\documentclass[12pt,a4paper]{article}

\usepackage[utf8]{inputenc}
\usepackage[T1]{fontenc}
\usepackage{graphicx,float}
\usepackage{amsmath,amssymb,amsfonts}
\usepackage{epsfig,color}
\usepackage{pdfpages,verbatim}
\usepackage{dsfont,float,booktabs}
\usepackage{accents}
\usepackage{amstext}
\usepackage{esint,mathtools}
\usepackage[margin=1.5cm]{geometry}
\usepackage[svgnames]{xcolor}
\usepackage{multirow}
\usepackage{tikz}
\usepackage{subfigure}
\usepackage{enumitem,colortbl}
\usepackage{adjustbox,pifont,amssymb}
\usepackage{appendix}
\newcommand{\cmark}{\ding{51}}%
\allowdisplaybreaks 

\usepackage[citecolor=blue,urlcolor=blue,unicode=true,pdfusetitle, bookmarks=true,bookmarksnumbered=true,bookmarksopen=true,bookmarksopenlevel=3, breaklinks=false,pdfborder={0 0 1},backref=false,colorlinks=true, linkcolor=blue]{hyperref}
\usepackage[acronym,toc,sort=use]{glossaries-extra}

\allowdisplaybreaks
\usetikzlibrary{arrows,arrows.meta,shapes,positioning,shadows,trees,calc,intersections,external}

\numberwithin{equation}{section}
\definecolor{gris1}{RGB}{246,246,246}
\definecolor{gris2}{RGB}{216,216,216}
\definecolor{gris3}{RGB}{224,224,224}
\definecolor{gris4}{RGB}{186,186,186}
\definecolor{gris5}{RGB}{135,135,135}
\makeglossaries

\renewcommand*{\glossaryentrynumbers}[1]{}

\newacronym{tg}{TG}{Teleparallel Gravity}
\newacronym{dof}{DoF}{Degrees of Freedom}
\newacronym{gr}{GR}{General Relativity}
\newacronym{gw}{GW}{Gravitational Waves}
\newacronym{tegr}{TEGR}{Teleparallel Equivalent of General Relativity}
\newacronym{flrw}{FLRW}{Friedmann–Lema{\^{i}}tre–Robertson–Walker}
\newacronym{tegb}{TEGB}{Teleparallel equivalent of the Gauss-Bonnet}
\newacronym{stegr}{STEGR}{Symmetric teleparallel equivalent of general relativity}
\newacronym{ngr}{NGR}{New General Relativity}
\newacronym{ppn}{PPN}{Parameterized post-Newtonian formalism}
\newacronym{lhs}{LHS}{Left hand side}
\newacronym{rhs}{RHS}{Right hand side}
\newacronym{wrt}{wrt}{with respect to}
\newacronym{cmb}{CMB}{Cosmic Microwave Background}
\newacronym{cdm}{CDM}{Cold Dark Matter}
\newacronym{eos}{EoS}{Equation of State}
\newacronym{de}{DE}{Dark energy}
\newacronym{bao}{BAO}{Baryonic Acoustic Oscillations}
\newacronym{wmap}{WMAP}{Wilkinson Microwave Anisotropy Probe}
\newacronym{bbn}{BBN}{Big Bang Nucleosynthesis}
\newacronym{qcd}{QCD}{Quantum Chromodynamics}
\newacronym{sn}{SN}{Supernova (we use abbreviations SNeIa and SNIa to refer to supernova type 1a)}
\newacronym{svt}{SVT}{Scalar-Vector-Tensor}
\newacronym{cc}{CC}{Cosmic Chronometers}
\newacronym{hde}{HDE}{Holographic Dark Energy}
\newacronym{plde}{PlDE}{Pilgrim Dark Energy}
\newacronym{shoes}{SH0ES}{Supernova H0 for the Equation of State}
\newacronym{des}{DES}{Dark Energy Survey}
\newacronym{tgrb}{TRGB}{Tip of the red-giant branch}
\newacronym{slacs}{SLACS}{Sloan Lens ACS}
\newacronym{tdcosmo}{TDCOSMO}{Time-delay Cosmography}
\newacronym{lisa}{LISA}{Laser Interferometer Space Antenna}
\newacronym{act}{ACT}{Atacama Cosmology Telescope}
\newacronym{spt}{SPT}{South Pole Telescope}
\newacronym{hst}{HST}{Hubble Space Telescope}
\newacronym{dr}{DR}{Data Release}
\newacronym{gaia3}{Gaia EDR3}{Gaia Early Data Release 3}
\newacronym{wl}{WL}{Weak Lensing}
\newacronym{sdss}{SDSS}{Sloan Digital Sky Survey}
\newacronym{df}{dF}{degree field}
\newacronym{boss}{BOSS}{Baryon Oscillation Spectroscopic Survey}
\newacronym{cmass}{CMASS}{Constant stellar MASS}
\newacronym{jla}{JLA}{Joint Light-curve Analysis}
\newacronym{kids}{KiDS}{Kilo-Degree Survey}
\newacronym{holicow}{H0LiCOW}{$H_0$ Lenses in COSMOGRAIL's Wellspring}
\newacronym{gp}{GP}{Gaussian processes}
\newacronym{mcmc}{MCMC}{Markov chain Monte Carlo}
\newacronym{lss}{LSS}{Large Scale Structure}
\newacronym{litebird}{LiteBIRD}{Lite (Light) satellite for the studies of B-mode polarization and Inflation from cosmic background Radiation Detection}
\newacronym{ccl}{CCL}{Core Cosmology Library}
\newacronym{core}{CORE}{Cosmic Origins Explorer}
\newacronym{prism}{PRISM}{Polarized Radiation Imaging and Spectroscopy Mission}
\newacronym{pixie}{PIXIE}{Primordial Inflation Explorer}
\newacronym{pico}{PICO}{Probe of Inflation and Cosmic Origins}
\newacronym{ska}{SKA}{Square Kilometre Array}
\newacronym{desi}{DESI}{Dark Energy Spectroscopic Instrument}
\newacronym{bingo}{BINGO}{Baryon acoustic oscillation In Neutral Gas Observations}
\newacronym{kvno}{KVNO}{Keck+VLT+UVES+Oklo measurements}
\newacronym{vlbi}{VLBI}{Very Long Baseline Interferometry}

\newglossarystyle{mylong}{%
  \setglossarystyle{long}%
     {\begin{longtable}[l]{@{}p{\dimexpr 2cm-\tabcolsep}p{0.8\hsize}}}
     {\end{longtable}}%
 }
\newcommand{\udt}[3]{#1^{#2}_{\phantom{#2}#3}}
\newcommand{\udut}[4]{#1^{#2\phantom{#3}#4}_{\phantom{#2}#3\phantom{#4}}}

\newcommand{\dut}[3]{#1_{#2}^{\phantom{#2}#3}}
\newcommand{\dudt}[4]{#1_{#2\phantom{#3}#4}^{\phantom{#2}#3}}
\newcommand{\st}[1]{\accentset{\Diamond}{#1}}
\newcommand{\lc}[1]{\accentset{\circ}{#1}}
\newcommand{\gc}[1]{\widehat{#1}}
\newcommand{\dd}{{\rm d}}
\newcommand{\DD}{{\rm D}}
\newcommand{\DDD}{{\mathcal{D}}}
\makeatletter
\long\def\dddddot#1{%
  {\mathop {#1}\limits ^{\vbox to-1.4\ex@ {\kern -\tw@ \ex@ \hbox {\normalfont .....}\vss }}}%
}
\long\def\multidots#1#2{%
  \count@=0
  {{\mathop {#2}\limits ^{\vbox to-1.4\ex@ {\kern -\tw@ \ex@ \hbox {\normalfont %
  \loop%
  \ifnum#1>\count@%
  .%
  \advance\count@ by1%
  \repeat%
  }\vss }}}}%
}
\makeatother
\usepackage{amsthm}
\theoremstyle{plain}

\theoremstyle{definition}

\theoremstyle{remark}

\newcommand{\xx}[0]{\mathrm{x}}

\newcommand{\HH}{\mathbf{H}}
\newcommand{\GGG}{\mathbf{G}}

\makeatletter
\renewcommand\paragraph{\@startsection{paragraph}{4}{\z@}%
	{-3.25ex\@plus -1ex \@minus -.2ex}%
	{1.5ex \@plus .2ex}%
	{\normalfont\normalsize\bfseries}}
\usepackage{tabularx}
\usepackage{booktabs}
\usepackage[raggedright]{titlesec}
\usepackage{siunitx}
\usepackage{mathtools}
\usepackage{footmisc}

\makeatother
\makeatletter
\renewcommand\subparagraph{\@startsection{subparagraph}{5}{\z@}%
	{-3.25ex\@plus -1ex \@minus -.2ex}%
	{1.5ex \@plus .2ex}%
	{\normalfont\normalsize\bfseries}}
\makeatother

\usepackage{pifont}
\newcommand{\xmark}{\ding{55}}

\newcommand{\order}[2]{\overset{\mathclap{\scriptscriptstyle #2}}{#1}\vphantom{#1}}
\DeclareSIUnit\parsec{pc}
\newcommand{\GG}{\mathbf{G}}

\newcommand{\sgn}{\text{sign}}

\renewcommand{\arraystretch}{1.15} 

\makeatletter
\renewcommand{\l@subsection}{\@dottedtocline{2}{1.5em}{3.4em}}
\makeatother

\usepackage{silence}
\newcommand{\midsepremove}{\aboverulesep = 0mm \belowrulesep = 0mm}
\midsepremove
\newcommand{\midsepdefault}{\aboverulesep = 0.605mm \belowrulesep = 0.984mm}
\midsepdefault

\begin{document}

\pagenumbering{roman}

\title{Teleparallel Gravity: From Theory to Cosmology}
\author{
  Sebastian Bahamonde\thanks{\href{mailto:sbahamonde@ut.ee}{sbahamonde@ut.ee}, \href{mailto:bahamonde.s.aa@m.titech.ac.jp}{bahamonde.s.aa@m.titech.ac.jp}}~\texorpdfstring{${}^{1,2}$}{},
  Konstantinos F. Dialektopoulos\thanks{\href{mailto:kdialekt@gmail.com}{kdialekt@gmail.com}}~\texorpdfstring{${}^{3,4,5}$}{},\\[2ex]
  Celia Escamilla-Rivera\thanks{\href{mailto:celia.escamilla@nucleares.unam.mx}{celia.escamilla@nucleares.unam.mx}}~\texorpdfstring{${}^{6}$}{},
  Gabriel Farrugia\thanks{\href{mailto:gfarr02@um.edu.mt}{gfarr02@um.edu.mt}}~\texorpdfstring{${}^{7,8}$}{},
  Viktor Gakis\thanks{\href{mailto:vgakis@central.ntua.gr}{vgakis@central.ntua.gr}}~\texorpdfstring{${}^{\phantom{i}7,8,9}$}{},\\[2ex]
  Martin Hendry\thanks{\href{mailto:martin.hendry@glasgow.ac.uk}{martin.hendry@glasgow.ac.uk}}~\texorpdfstring{${}^{10}$}{},
  Manuel Hohmann\thanks{\href{mailto:manuel.hohmann@ut.ee}{manuel.hohmann@ut.ee}}~\texorpdfstring{${}^{\phantom{ii}1}$}{},
  Jackson Levi Said\thanks{\href{mailto:jackson.said@um.edu.mt}{jackson.said@um.edu.mt}}~\texorpdfstring{${}^{\phantom{ii}7,8}$}{},\\[2ex]
  Jurgen Mifsud\thanks{\href{mailto:jurgen.mifsud@um.edu.mt}{jurgen.mifsud@um.edu.mt}}~\texorpdfstring{${}^{\phantom{ii}7,8}$}{},
  Eleonora Di Valentino\thanks{\href{mailto:eleonora.di-valentino@durham.ac.uk}{eleonora.di-valentino@durham.ac.uk}}~\texorpdfstring{${}^{\phantom{ii}{11}}$}{}\\[2ex]
  \texorpdfstring{${}^1$}{}Laboratory of Theoretical Physics, Institute of Physics,\\
  University of Tartu, W. Ostwaldi 1, 50411 Tartu, Estonia\\[1ex]
   \texorpdfstring{${}^2$}{}Department of Physics, Tokyo Institute of Technology,\\
  1-12-1 Ookayama, Meguro-ku, Tokyo 152-8551, Japan.\\[1ex]
  \texorpdfstring{${}^3$}{}Center for Gravitation and Cosmology, College of Physical Science and\\ Technology, Yangzhou University, Yangzhou 225009, China\\[1ex]
  \texorpdfstring{${}^4$}{}Laboratory of Physics, Faculty of Engineering, Aristotle University of\\Thessaloniki, 54124 Thessaloniki, Greece\\[1ex]
  \texorpdfstring{${}^5$}{}Department of Physics, Nazarbayev University,\\53 Kabanbay Batyr avenue, 010000 Astana, Kazakhstan\\[1ex]
  \texorpdfstring{${}^6$}{}Instituto de Ciencias Nucleares, Universidad Nacional Aut\'{o}noma de\\ M\'{e}xico, Circuito Exterior C.U., A.P. 70-543, M\'exico D.F. 04510, M\'{e}xico.\\[1ex]
  \texorpdfstring{${}^7$}{}Institute of Space Sciences and Astronomy, University of Malta,\\[1ex]
  \texorpdfstring{${}^8$}{}Department of Physics, University of Malta,\\[1ex]
  \texorpdfstring{${}^9$}{}Department of Physics, National Technical University of Athens,\\ Zografou Campus GR 157 73, Athens, Greece\\[1ex]
  \texorpdfstring{${}^{10}$}{}SUPA, School of Physics and Astronomy,\\ University of Glasgow, Glasgow G12 8QQ, UK\\[1ex]
  \texorpdfstring{${}^{11}$}{}Institute for Particle Physics Phenomenology, Department of Physics,\\ Durham University, Durham DH1 3LE, United Kingdom
}

\date{\today}
\thispagestyle{empty}
\maketitle

\begin{abstract}
Teleparallel gravity has significantly increased in popularity in recent decades, bringing attention to Einstein’s other theory of gravity. In this Review, we give a comprehensive introduction to how teleparallel geometry is developed as a gauge theory of translations together with all the other properties of gauge field theory. This relates the geometry to the broader metric-affine approach to forming gravitational theories where we describe a systematic way of constructing consistent teleparallel theories that respect certain physical conditions such as local Lorentz invariance. We first use teleparallel gravity to formulate a teleparallel equivalent of general relativity which is dynamically equivalent to general relativity but which may have different behaviors for other scenarios, such as quantum gravity. After setting this foundation, we describe the plethora of modified teleparallel theories of gravity that have been proposed in the literature. We attempt to connect them together into general classes of covariant gravitational theories. Of particular interest, we highlight the recent proposal of a teleparallel analogue of Horndeski gravity which offers the possibility of reviving all of the regular Horndeski contributions. In the second part of the Review, we first survey works in teleparallel astrophysics literature where we focus on the open questions in this regime of physics. We then discuss the cosmological consequences for the various formulations of teleparallel gravity. We do this at background level by exploring works using various approaches ranging from dynamical systems to Noether symmetries, and more. Naturally, we then discuss perturbation theory, firstly by giving a concise approach in which this can be applied in teleparallel gravity theories and then apply it to a number of important theories in the literature. Finally, we examine works in observational and precision cosmology across the plethora of proposal theories. This is done using some of the latest observations and is used to tackle cosmological tensions which may be alleviated in teleparallel cosmology. We also introduce a number of recent works in the application of machine learning to gravity, we do this through deep learning and Gaussian processes, together with discussions about other approaches in the literature.
\end{abstract}
\clearpage

\setcounter{tocdepth}{2}
\tableofcontents
\clearpage

\printglossaries

\clearpage
\setcounter{page}{1}
\pagenumbering{arabic}


\section{Introduction} \label{sec:introduction}

General relativity (\gls{gr}) is an astonishing theory both in its simplicity and its ability to retain a theoretically consistent framework while satisfying observational tests at almost all scales of physics. Together with quantum field theory, it forms one of the two pillars of modern physics. Its precise predictions and influence over physical processes in the Universe has set a standard for which all other physical theories aspire to.

After framing gravity through the perspective of his strong \textit{equivalence principle}, Einstein then formulated his field equations by adopting the Riemann tensor as the fundamental building block of geometric deformation, as suggested by Marcel Grossmann \cite{vanDongen:2018deh}. On the demand that the field equations be conserved, Einstein worked out their exact form through a process of reconstruction and elimination. Later on, it was noticed that they could have also been derived from the \emph{Bianchi identities}, from which one can derive an Einstein tensor $\lc{G}^{\mu\nu}$ that is covariantly conserved $\lc{\nabla}_{\mu}\lc{G}^{\mu\nu}=0$ (for a covariant derivative $\lc{\nabla}_{\mu}$ defined using the Levi-Civita connection $\udt{\lc{\Gamma}}{\rho}{\mu\nu}$ which is further explained in Sec.~\ref{ssec:Conventions}) \cite{misner1973gravitation}. However, Einstein did not have this information at the time. Using his approach, he was able to write down the field equation as \cite{Einstein:1917ce}
\begin{equation*}
    \lc{G}_{\mu\nu} = \frac{8\pi G}{c^4}\,\Theta_{\mu\nu}\,,
\end{equation*}
for an energy-momentum tensor $\Theta_{\mu\nu}$, and where $G$ is Newton's constant of gravity and $c$ the speed of light \cite{Sauer2007-SAUEUF}. These equations remain unchanged from their original form as presented by Einstein to the Prussian Academy of Sciences in November 1915 \cite{einstein1915allgemeinen}. It is these equations which have been used to explore the expansion and evolution of the Universe from its infinitesimal beginnings to its currently accelerating expansion phase, as well as the nature of black holes, the propagation of \gls{gw} \cite{Barack:2018yly} and the formation of galactic and compact structures in the Universe. There are even several attempts to extend the theory to meet the growing challenges from quantum theory and high energy particle physics.

The development of \gls{gr} has undergone astounding success over the decades. However, the establishment of quantum mechanics in the 1920s and then quantum field theory later on already started to expose some of the limitations of the theory early on. Observational constraints have also focused attention on some of the failings of \gls{gr} which not only occur on the very small scale of the physical spectrum but also on larger scales such as for galactic systems where modifications of the standard model of particles physics \cite{Glashow:1961tr,Weinberg:1967tq,Salam:1968rm} are necessary to preserve \gls{gr} as the underlying theory of gravity. This has led to the realization that dark matter plays an important role in galaxies and their clusters \cite{Rubin:1970zza,Zwicky:1937zza,Bertone:2010zza} in the astrophysical context but also has a crucial role in the cosmological context in the dynamics of the early Universe \cite{mukhanov_2005}. The best macroscopic version of this is \textit{cold dark matter} (\gls{cdm}). \gls{gr} also needs the addition of further corrections in the matter sector when considering the very early Universe where a rapid period of inflation \cite{Guth:1980zm,Linde:1981mu} is necessitated to explain the extremely flat and homogeneous Universe we observe at present. Saying that, there are other solutions to explain how inflation came about without any modification to the matter description of the standard model such as the Starobinsky's model \cite{Starobinsky:1980te} which modified the early Universe evolution in comparison to the standard model of cosmology. It may also be the case that modifications to the gravity sector may also explain this feature of galactic rotation curves such as in the case of Starobinsky gravity \cite{Sporea:2017eph} and conformal Weyl gravity \cite{Mannheim:2012qw}. In the same vein, the late-time Universe is undergoing a period of accelerated expansion \cite{Riess:1998cb,Perlmutter:1998np} which requires a cosmological constant $\Lambda$ \cite{RevModPhys.61.1} to adequately describe the observed expansion using \gls{gr} as the fundamental to describe gravitational phenomena.

Altogether, \gls{gr} can thus explain most phenomena provided adequate modifications are considered in the matter sector so that we consider the widely accepted concordance model in $\Lambda$\gls{cdm} plus inflation cosmology. However, the exotic nature of these particle species remains a total mystery in terms of observations despite significant theoretical advances in physics beyond the standard model of particle physics. In the same vein, it may also be the case that the standard model of particle physics does not require a significant a restructuring for meeting these observational challenges and it is that the gravitational section needs further revisiting. This may take the form of extensions from \gls{gr} or even modifications beyond \gls{gr} that may be an alternative to its original formulation. In the literature there have been numerous proposals for new theories of gravity \cite{Clifton:2011jh,Bull:2015stt,Capozziello:2011et,Saridakis:2021lqd,Nojiri:2017ncd} which are motivated either through phenomena or some theoretical approach, as well as for other reasons such as from quantum physics. One interesting possibility that has been gaining momentum in the literature in recent decades is that of \textit{teleparallel gravity} (\gls{tg}) where curvature is replaced by torsion as the mechanism by which geometric deformation produces a gravitational field. It does this by dislodging the uniquely curvature-based Levi-Civita connection with a torsion-based teleparallel connection. In fact, there are now thousands of publications related to the topic in the literature. One of the many theories that torsion-based approaches to gravity has produced is the \textit{teleparallel equivalent of general relativity} (\gls{tegr}) which is dynamically equivalent to \gls{gr}, meaning that they cannot be distinguished through classical experiments.

In fact, even Einstein himself, while not aware of the observational motivations for modifying \gls{gr}, was also one of the first proposers for modifications of \gls{gr} shortly after its inception. This took various forms with varying degrees of novelty. However, Einstein was captured by teleparallelism due to its reliance of the tetrad frames and its \textit{absolute teleparallelism} feature \cite{Goenner:2014mka}. Similar to the largely silent period between his papers on special and general relativity, Einstein devoted himself to teleparallelism and its potential for a unified field theory. He spent this time in correspondence with many of the main proponents for the new geometric framework at the time such as Roland Weitzenb\"{o}ck, \'{E}lie Cartan, Luther P. Eisenhart, Herman M\"{u}ntz, Jakob Grommer and Cornelius Lanczos to mention some. The main objective for Einstein was to use this new approach of absolute teleparallelism to unify gravity and electromagnetism since tetrad fields can support sixteen independent components representing ten degrees of freedom (\gls{dof}) that are related to the metric tensor while the remaining six would correspond to a separate connection that he wanted to show could be yielded to produce the electromagnetic sector. However, after three intense years of research between the 1928 summer and 1931 spring, together with eight publications on the topic, Einstein eventually abandoned the idea. He failed to establish a relationship between this new connection and electromagnetism which was his ultimate goal of having a unified field theory approach to gravity and electromagnetism. Instead he found that these seemingly extra \gls{dof} were related to the Lorentz group and simply a manifestation of the local Lorentz freedom of gravity.

The unified approach was largely abandoned after that time and only revived much later on by M\o{}ller in Ref.~\cite{Moller:1961} and then Pellegrini and Plebanski in Ref.~\cite{Pellegrini:1961} where they formulated a Lagrangian that satisfies the concepts behind absolute parallelism. This was also followed by Ref.~\cite{Moller:1968} where the first ideas of a well defined energy-momentum tensor for gravitation were conceptualized, and which led to significant advances in later decades as compared with the analogous case in \gls{gr}, which we now know to be quasi-local \cite{Brown:1992br}. Work on the topic remained in the foundations sector for the next couple of decades with compelling works such as the review by Hehl, Heyde and Kerlick in Ref.~\cite{RevModPhys.48.393}. Independently, Hayashi and Nakano in Ref.~\cite{Hayashi:1967se} also started to revive a form of Einstein's origin work. In their work, they formulated absolute teleparallelism through a gauge theory of translations on the Lorentz group. This was further extended to a fully fledged gravitational theory in Ref.~\cite{Hayashi:1979qx} where a \textit{new general relativity} (\gls{ngr}) formalism was presented together with its weak field limit. It was at this time that the motivations for looking for theories of gravity beyond \gls{gr} was starting to take off with many new toy model proposals tackling various theoretical and observational issues in the topic. In this background, absolute teleparallelism was not actively studied beyond very select groups. This was partly due to its very foundational nature as compared to other popular toy model theories at the time, which appealed to a much larger audience. This naturally leads us to the question of how would modern gravitational research look like if Einstein had continued to work on the concepts surrounding absolute teleparallelism, or if he had formulated \gls{gr} in terms of torsion from the beginning (namely, \gls{tegr})? This construction of gravity would have been written in terms of gauge theories thus eliminating the need for many future works on formulating \gls{gr} as a gauge theory within the curvature-based Levi-Civita connection. This is all the more prescient due to the fact that both \gls{gr} and \gls{tegr} both predict the same observational outcomes in their classical predictions (due to their dynamical equivalence).

Absolute teleparallelism was finally revived in its latest incarnation in the numerous works by Aldrovandi and Pereira (among others) which culminated in two pivotal books on the topic. The first is Ref.~\cite{aldrovandi1995introduction} where the underlying geometry of \gls{gr} is revisited and the foundational ideas behind various concepts in \gls{tg} are put on a concrete mathematical basis, while the second is Ref.~\cite{Aldrovandi:2013wha} where the physics on which \gls{tegr} is fully fleshed out together with potential future work on quantum gravity and quantum field theory. These works again resurrect Einstein's original goal of reconciling gravity with other branches of physics \cite{Maluf:2013gaa}. \gls{tegr} continues to be studied for potential benefits as compared with \gls{gr} such as in its quantum regime, and others. On the other hand, \gls{tg} has since undergone a renaissance with respect to (\gls{wrt}) potential modified theories of gravity. The first such modification was presented by Hayashi in Ref.~\cite{Hayashi:1979qx} where \gls{gr} was slightly modified to allow for small deviations from \gls{tegr} through a decomposition of its Lagrangian. However, the most impactful suggested modification came from Ferraro and Fiorini in Ref.~\cite{Ferraro:2006jd} where a form of $f(T)$ gravity first appeared in the literature, and later by Linder in Ref.~\cite{Linder:2010py}. The rationale here follows the same reasoning as in the very popular $f(\lc{R})$ theory of gravity \cite{Sotiriou:2008rp,Capozziello:2009nq,DeFelice:2010aj}. The spectrum of modified teleparallel theories has since drastically expanded to also include nonminimal couplings to matter, scalar fields as well as vector and tensor field additions, giving a plethora of teleparallel theories in which gravity can be formulated and studied.

The literature on \gls{tg} features a wide range of works with an increasing intensity of new additions to the growing body of publications on the topic. In this considerable body of works, we would like to point out two important reviews on the topic. Firstly, Ref.~\cite{Krssak:2018ywd} where the foundations of \gls{tg} and its particular application to \gls{tegr}, $f(T)$ and some other teleparallel theories was explored. This work built on foundational work represented in the book in Ref.~\cite{Aldrovandi:2013wha} and covered important advances on the issue of covariance in \gls{tg} \cite{Krssak:2015oua}, among other topics. Another hugely important work is Ref.~\cite{Cai:2015emx} where foundational aspects, such as the Poincar\'{e} gauge gravity origins of the theory, are explored but also many applications, both astrophysical and cosmology. These are important works in the community but the literature has since drastically evolved with even more clarity on the covariant formulation of \gls{tg} and its relationship to metric-affine theories being more laid out. \gls{tg} also now hosts a significantly expanded number of new theories of gravity, of particular interest we point out advances in the scalar-tensor section where an analogue of Horndeski gravity has also started gain popularity. In terms of observational astrophysics and cosmology, there has been serious new results such as in the regime of precision cosmology, observational constraints in new teleparallel models and novel approach to cosmology such as machine learning, as well as important progress in galactic and stellar physics. At the same time, observational physics, and particularly cosmology, is reaching the border of the region in which \gls{gr} can satisfy measurements such as with the late-time cosmological tensions as well as ambiguities in early Universe physics. Given these critical advances in the literature and the resolution of many more foundational issues in \gls{tg} as well as the growing pressure from observational cosmology for competitive new theories of gravity, we are motivated to reevaluate the topic and to bridge the gap between foundational theory and it observational consequences in astrophysical and cosmological systems.

\subsection{Original results and corrections to the literature} \label{sec:corrections}

This Review contains some original results that have not been reported in the literature yet. In addition, throughout the process of reviewing the most important literature of \gls{tg}, we have noticed some mistakes that we have corrected. We list them here for each section:
\begin{itemize}
    \item In Sec.~\ref{ssec:tggenprop} we provide a general description for any \gls{tg} theory in order to have a theory which certain physical properties such as having local Lorentz invariance of both the matter and gravitational action.
    \item In Sec.~\ref{sec5:extended} we re-derived the field equations for the most popular teleparallel theories and express them with the same notation.
    \item In Sec.~\ref{sec:spheri} we found the two most general good tetrad-spin connection pair in spherical symmetry for $f(T,B,\phi,X)$ gravity when one assumes that the teleparallel connection and the tetrad respect spherical symmetry.
	\item Eq.~\eqref{eq:torscalatv} corrects a sign mistake in Ref.~\cite{Bahamonde:2017wwk}.
    \item In Sec.~\ref{sec:cosmology_in_TG} we derived the flat Friedmann–Lema\^{i}tre–Robertson–Walker (\gls{flrw}) equations for each model in a consistent way. We also corrected the reported \gls{flrw} equations in higher order derivative teleparallel gravity (see Sec.~\ref{sec:HOT_back}).
    \item We verified the validity of the reconstructed solutions especially in cases where the modified gravity component behaves as a dark energy fluid. This appears in $f(T)$ in Sec.~\ref{sec:fT-reconstruction} and for $f(T,\Theta)$ in Sec.~\ref{sec:reconstruction-fTT-gravity}. Other reported solutions have been checked and corrected where applicable.
    \item In Sec.~\ref{sec:dyn_fTTG}, we have corrected the stability conditions for the Minkowski solution in the case of $f(T,T_G)$ gravity.
    \item We corrected the Noether symmetry conditions and the dynamical systems approach reported in the minimal $f(T)$-scalar coupling in Secs.~\ref{sec:noether_mincoupled} and \ref{sec:dynm_mincoupled}.
    \item We corrected the dynamical analysis for the non-local $f(T)$ gravity approach in Sec.~\ref{sec:dyn_nonlocal}.
    \item In the Supplementary annexes (Supplementary 1), the finding good tetrad-spin connection pairs approach for different anisotropic cosmological scenarios has been performed. Sets of good tetrad-spin connection pairs are also presented. We also corrected instances where bad tetrad-spin connection pairs have been used in literature.
    \item In anisotropic cosmologies, some reconstructed solutions such as in the Supplementary annexes (Supplementary 1) have been corrected. Moreover, in another part of the Supplementary annexes (Supplementary 1), the \gls{flrw} limiting behavior in the Noether symmetry approach has been corrected.
    \item A brief new discussion about the generalized Birkhoff's theorem in $f(T,B)$ gravity is provided in Sec.~\ref{sec:birkhoff}.
    \item In Sec.~\ref{sec:observation_status_precision} we provided a full compendium of \gls{tg} models and their analysis perspectives in precision and observational cosmology.
	\item In Sec.~\ref{subsection:ML_p} we analyzed \gls{tg} theories using machine learning techniques.
    \item In Sec.~\ref{subsec:cosmography_p} we corrected the kinetic equations for $f(T)$ cosmography. This is manifested in Eqs.~\eqref{f4T}--\eqref{f5T} where we corrected the snap parameters terms, which changes the current results already reported in the literature up to third order. For future analysis, this can change the kinetic evolution of the Universe at higher redshifts.
\end{itemize}
There are also some other minor new additions or corrected mistakes from the literature that we believe, are not so important to mention above.

\subsection{Review structure}

The Review is intended to cover the main developments in teleparallel gravity and its impact on astrophysics, gravitational waves (\gls{gw}) and cosmology. To facilitate this, we divide the Review into four main thematic areas covering the most important results coming from the last few decades while still comprehensively describing its origins. We show the Review breakdown graphically below and then go on to describe how each section is broken down.\\ \vspace{0.5cm}

\begin{tikzpicture}[scale=0.87,>=stealth,
topic0/.style={align=center,rectangle,minimum height=10mm,draw,rounded corners,fill=orange!30},
topic/.style={align=center,rectangle,minimum height=10mm,draw,rounded corners,fill=blue!20},
section/.style={align=center,rectangle,draw,outer sep=2mm,fill=gray!60,minimum size=35mm,minimum height=12mm}]
\node[topic0] (top) at (0,0) {Teleparallel Gravity\\ From Theory to Cosmology};
\node[topic] (found) at (-5,-2) {Foundations};
\node[topic] (app) at (5,-2) {Applications};
\node[topic] (math) at (-7.5,-4) {Geometry};
\node[topic] (phys) at (-2.5,-4) {Theories of\\ Gravity};
\node[topic] (astro) at (2.5,-4) {Astrophysics};
\node[topic] (cosmo) at (7.5,-4) {Cosmology};
\node[section,below of=math,yshift=-0.5cm] (metaff) {Metric-Affine Geometry\\ Sec.~\ref{sec2:affine}};
\node[section,below of=metaff,yshift=-0.5cm] (telegeo) {Teleparallel Geometry\\ Sec.~\ref{sec3:torsional}};
\node[section,below of=phys,yshift=-0.8cm] (ngr) {Teleparallel\\ Formulation and \\ TEGR\\ Sec.~\ref{sec4:TEGRNGR}};
\node[section,below of=ngr,yshift=-1.1cm] (ext) {Modified \\ Teleparallel Theories\\ Sec.~\ref{sec5:extended}};
\node[section,below of=astro,yshift=-0.5cm] (compact) {Gravitational Wave\\ Polarizations\\Sec.~\ref{sec9:GW}};
\node[section,below of=compact,yshift=-0.8cm] (gw) {Astrophysical Systems\\ Sec.~\ref{sec6:astrophysics}};
\node[section,below of=cosmo,yshift=-0.5cm] (flrw) {Background\\ Dynamics\\ Sec.~\ref{sec:cosmology_in_TG}};
\node[section,below of=flrw,yshift=-0.8cm] (pert) {Perturbation\\ Theory\\ Sec.~\ref{sec:cosmo-pert}};
\node[section,below of=pert,yshift=-0.8cm] (obs) {Precision\\ Observations\\ Sec.~\ref{sec10:Observation}};
\draw[->] (top) -- (found);
\draw[->] (top) -- (app);
\draw[->] (found) -- (math);
\draw[->] (found) -- (phys);
\draw[->] (app) -- (astro);
\draw[->] (app) -- (cosmo);
\end{tikzpicture}
\vspace{1cm}

We first open with an exploration of the geometric underpinnings of metric-affine geometry in Sec.~\ref{sec2:affine} where the so-called trinity of gravity is cemented as our starting off base. In Sec.~\ref{sec3:torsional}, we then focus on the specific foundations surrounding torsional teleparallel geometry. We here motivate teleparallel geometry as a gauge theory of translations and describe it as having an associated gauge field strength. We also discuss symmetries and transformations in teleparallel geometry and discuss how it couples to matter with the minimal coupling prescription.

\gls{gr} is built on the geometric foundations set by Riemann and others at the around the beginning of the twentieth century. In a similar vein, \gls{tg} is build on the work by Weitzenb\"{o}ck and many others who laid the geometric foundations for a torsional rather than curvature-based formulation of gravity. In Sec.~\ref{sec4:TEGRNGR}, we explore the foundational construction of teleparallel theories and its application to the formulation of the \gls{tegr}, which is dynamically equivalent to \gls{gr} while being born out of an entirely distinct geometric setting. Due to the tetrad-spin connection pair basis in which \gls{tg} is constructed, we revisit a number of tenets of this theory of gravity such as its ADM formalism and the possible effects it may have on quantum gravity (where it may no longer be equivalent to \gls{gr}). As in curvature-based theories of gravity, extensions to \gls{tegr} are interesting to study, as are modifications and alternative to \gls{tegr}. We explore the most promising of gravitational theories in Sec.~\ref{sec5:extended} where we introduce theories through their actions and determine many of their ensuing field equations.

After laying the foundations of \gls{tg} and exploring some of its most popular manifestations in theory, we move to the astrophysics sector in which we first describe the impact of \gls{tg} on gravitational radiation. We here focus on the polarization structure of \gls{gw} in Sec.~\ref{sec9:GW} where we reexamine some of the theories promoted in Sec.~\ref{sec5:extended}, in this context. In Sec.~\ref{sec6:astrophysics}, we then move on to other aspects of astrophysical systems in \gls{tg} such as the issue of spherically symmetric solutions and issues related to time-dependent solutions, as well as other symmetry settings. We use this information to discuss physical stars and to discuss open problems in this regime of the literature. Finally we close this area with a discussion of some work in the regime of galactic rotation curves in a particular setting of $f(T)$ gravity.

Finally, we tackle the cosmological consequences for some of the modified teleparallel theories of gravity explained in Sec.~\ref{sec5:extended}. We start with an analysis of the numerous background cosmology analyses in Sec.~\ref{sec:cosmology_in_TG} which forms the majority of works in the literature. This section delves into the many works on reconstruction methods, Noether symmetries, and dynamical systems. Following this lengthy exploration, we turn to the perturbative side of \gls{tg} in Sec.~\ref{sec:cosmo-pert}, where we explain how perturbations can be taken consistently in \gls{tg} and what form they should take. We also tackle the issue of perturbations in a cosmological background and some impacts on possible inflationary theory. Finally, Sec.~\ref{sec10:Observation} contains the core observational predictions from \gls{tg} in the cosmological regime. We here review works that constrain some of the most popular models within \gls{tg} through late-time, as well as give new perspectives on prospects of machine learning applications in teleparallel cosmology. We also discuss the state of the art in terms of the near-term prospects for observational cosmology and its possible impact in discriminating between models in gravity beyond $\Lambda$\gls{cdm}.

\subsection{Notation and conventions}\label{ssec:Conventions}

We lay out our conventions in this part of the Introduction. The literature on \gls{tg} contains numerous different combinations of conventions since there exist a number of non-competing choices one can take with respect to various definitions throughout the layers on which \gls{tg} is constructed. In the remainder of this work, we refer to the definitions declared here as the \textit{review convention}.

Firstly, we use a metric signature $(+1,-1,-1,-1)$ which is the dominant signature in the \gls{tg} literature. Also, we choose to write spacetime indices using lower case Greek alphabet letters ($\alpha,\,\beta,\,\dots$) and capital Latin letters ($A,\,B,\,\dots$) for the Minkowski spacetime (Lorentz indices). Capital Latin letters starting from ($I,\,J,\,\dots$) will denote purely spatial indices on the Minkowski spacetime (see Sec.~\ref{sec9:GW}), while lower case Latin letters from the middle of the alphabet ($i,\,j,\,\dots$) are reserved for purely spatial indices (see Secs.~\ref{sssec:metafftetspi},\,\ref{sssec:adm31split}). We also define the following commonly used terms related to the foundational formulation of \gls{tg}
\begin{itemize}
    \item Teleparallel connection -- The general torsional teleparallel connection that is composed of both the tetrad and spin connection defined in Eq.~\ref{eq:tg_connection};
    \item Weitzenb\"{o}ck connection -- The teleparallel connection in the Weitzenb\"{o}ck gauge (See Sec.~\ref{sec:LLTW}) where the spin connection vanishes (see Sec.~\ref{subsec:goodtetrad});
    \item Weitzenb\"{o}ck gauge -- The case where the spin connection vanishes (see Sec.~\ref{ssec:admformalism}).
\end{itemize}

The line element for flat spacetime in Cartesian coordinates is then described by
\begin{equation}
    \dd s^2=\eta_{\mu\nu}\dd x^\mu \dd x^\nu = \dd t^2-\dd x^2-\dd y^2-\dd z^2\,,
\end{equation}
where $\dd $ denotes the differential operator and $\eta_{\mu\nu}$ the Minkowski metric. All metric tensor line elements are produced by the tetrads $\udt{e}{A}{\mu}$ and inverse tetrads $\dut{E}{A}{\mu}$ (see Sec.~\ref{Grav_Coup_Prescrip})
\begin{equation}
    g_{\mu\nu}=\eta_{AB}\udt{e}{A}{\mu}\udt{e}{B}{\nu}\,, \quad \text{and} \quad g^{\mu\nu}=\eta^{AB}\dut{E}{A}{\mu}\dut{E}{B}{\nu}\,,\label{eq:def_metr_tetrad}
\end{equation}
where the tetrad is a fundamental dynamical variable in the theory. The tangent spacetime is described by trivial tetrads which are denoted by $\udt{h}{A}{\mu}$ with inverses $\dut{H}{A}{\mu}$. We will also define the one-forms $\mathbf{e}^A = \udt{e}{A}{\mu}\dd x^{\mu}$ and vector tetrads $\mathbf{E}_A = \dut{E}{A}{\mu}\partial_{\mu}$ (together with the trivial one-forms $\mathbf{h}^A = \udt{h}{A}{\mu}\dd x^{\mu}$ and trivial vector tetrads $\mathbf{H}_A = \dut{H}{A}{\mu}\partial_{\mu}$) (see Sec.~\ref{Sec:TEGR_gauge_theory}).

Tensor and operator quantities that have a dependence on the connection being employed are distinguished by the following definitions for the different possible connection options (see Sec.~\ref{sec2:affine})
\begin{itemize}
    \item The general connection $\udt{\gc{\Gamma}}{\alpha}{\mu\nu}$ e.g. $\udt{\gc{R}}{\alpha}{\beta{\mu\nu}}$;
    \item The Levi-Civita Connection $\udt{\lc{\Gamma}}{\alpha}{\mu\nu}$ e.g. $\udt{\lc{R}}{\alpha}{\beta{\mu\nu}}$;
    \item The symmetric teleparallel connection $\udt{\st{\Gamma}}{\alpha}{\mu\nu}$ e.g. $\udt{\st{R}}{\alpha}{\beta{\mu\nu}}$;
    \item The teleparallel connection $\udt{\Gamma}{\alpha}{\mu\nu}$ e.g. $\udt{R}{\alpha}{\beta{\mu\nu}}$,
\end{itemize}
which are used throughout the Review. We also define the following derivative operators
\begin{itemize}
    \item The covariant derivative $\nabla_{\mu}$ (with over-symbols for the various connection choices - as described above) which acts only on the general spacetime;
    \item The Fock-Ivanenko derivative $\DDD_{\mu}$ which acts on Lorentz indices only (see Sec.~\ref{sec:field_strength});
    \item The exterior covariant derivative $\DD$ and exterior derivative $\dd$ (see Sec.~\ref{sssec:diffform});
    \item The d'Alembertian operator $\Box = \nabla_\mu \nabla^ \mu = g^{\mu\nu} \nabla_\mu \nabla_\nu$.
\end{itemize}

For any connection, we define the Riemann curvature tensor \eqref{eq:spiconcurv} as \cite{misner1973gravitation}
\begin{equation}
    \gc{R}^\mu{}_{\nu\alpha\beta}:=\partial_\alpha \gc{\Gamma}^{\mu}{}_{\nu\beta}-\partial_\beta \gc{\Gamma}^{\mu}{}_{\nu\alpha}+\gc{\Gamma}^{\mu}{}_{\sigma\alpha} \gc{\Gamma}^{\sigma}{}_{\nu\beta} -\gc{\Gamma}^{\mu}{}_{\sigma\beta} \gc{\Gamma}^{\sigma}{}_{\nu\alpha}\,,\label{eq:riemann_ten_def}
\end{equation}
which defines the Levi-Civita Einstein tensor as
\begin{equation}
    \lc{G}_{\alpha\beta} := \lc{R}_{\alpha\beta}-\frac{1}{2}\lc{R}g_{\alpha\beta}\,,
\end{equation}
where $\lc{R} = g^{\mu\nu}\lc{R}_{\mu\nu}$. On the other hand, the teleparallel connection is defined through
\begin{equation}
    \Gamma^{\sigma}{}_{\nu\mu} := \dut{E}{A}{\sigma}\left(\partial_{\mu}\udt{e}{A}{\nu} + \udt{\omega}{A}{B\mu}\udt{e}{B}{\nu}\right)\,,
\end{equation}
where $\udt{\omega}{A}{B\mu}$ represents the components of the teleparallel spin connection (see Sec.~\ref{sec:tetradspin}). This can then be used to build the torsion \eqref{eq:torsion_tensor}, contortion \eqref{chp3_contortion_def} and superpotential \eqref{Eq:Superpotential_def} tensors as well as the torsion scalar \eqref{eq:torsion_def} defined through
\begin{subequations}
\begin{align}\label{torsion_tensor}
    T^{A}{}_{\mu\nu} & :=2\Gamma^{A}{}_{[\nu\mu]}=e^{A}{}_{\nu,\mu}-e^{A}{}_{\mu,\nu}+\omega^{A}{}_{B\mu}e^{B}{}_{\nu}-\omega^{A}{}_{B\nu}e^{B}{}_{\mu}=-2\Gamma^{A}{}_{[\mu\nu]}\,,\\[0.5ex] \label{contortion}
    K^{\rho}{}_{\mu\nu} & :=\Gamma^{\rho}{}_{\mu\nu}-\lc{\Gamma}^{\rho}{}_{\mu\nu}=\frac{1}{2}\left(T_{\mu}{}^{\rho}{}_{\nu}+T_{\nu}{}^{\rho}{}_{\mu}-T^{\rho}{}_{\mu\nu}\right)\,,\\[0.5ex] \label{superpotential}
    S_{\rho}{}^{\mu\nu} & :=K^{\mu\nu}{}_{\rho}-\delta_{\rho}^{\mu}T_{\sigma}{}^{\sigma\nu}+\delta_{\rho}^{\nu}T_{\sigma}{}^{\sigma\mu}=-S_{\rho}{}^{\nu\mu}\,,\\[0.5ex] \label{torsion_scalar}
    T & :=\frac{1}{2}S_{\rho}{}^{\mu\nu}T^{\rho}{}_{\mu\nu}\,.
\end{align}
\end{subequations}
where square brackets represent the anti-symmetric operator, i.e. $A_{[\mu\nu]}:=\frac{1}{2}\left(A_{\mu\nu} - A_{\nu\mu}\right)$. Also, all comma separated lower indices will refer to partial derivatives, such as $f_{,T}=\partial f/\partial T$. Unless otherwise stated, this Review observes the geometric units convention where
\begin{equation}\label{Eq:Con_units}
    c \equiv 1\,,\quad \kappa^2= 8\pi G\,,
\end{equation}
which will be mainly used for the foundations part of the Review, while the observational sections will mainly use SI units (as will be indicated).

As an example, consider the Einstein-Hilbert action \eqref{eq:einhilaction}
\begin{equation}
    \mathcal{S}_{\rm GR} := \frac{1}{2\kappa^2}\int \dd^4 x\, \sqrt{-g}\lc{R} + \mathcal{S}_{\rm m} = \frac{1}{2\kappa^2}\int \dd^4 x\, \sqrt{-g}\mathcal{L} + \mathcal{S}_{\rm m} = \frac{1}{2\kappa^2}\int \dd^4 x\, L + \mathcal{S}_{\rm m}\,,
\end{equation}
where the \gls{gr} subscript will be replaced for other theories, the Lagrangian $L$ is related to the Lagrangian density $\mathcal{L}$ through $L = \mathcal{L}\sqrt{-g}$, and where $\mathcal{S}_{\rm m}$ refers to the matter action. We take special note here for objects calculated using the Levi-Civita connection (such as $\udt{R}{\rho}{\sigma\mu\nu}$) in that we refer to these as \textit{standard gravity} objects since they are determined using the Levi-Civita connection. For minimally coupled matter system, the matter action will then be given by
\begin{equation}\label{eq:actionmatter}
    \mathcal{S}_{\rm m}:=\int \dd^4 x\, \sqrt{-g}\mathcal{L}_{\rm m}=\int \dd^4 x\, L_{\rm m}\,,
\end{equation}
which then defines the energy-momentum tensor through the variation
\begin{equation}\label{Eq:Con_EM_ten}
    \Theta_{\mu\nu} := \frac{-2}{\sqrt{-g}}\frac{\delta L_{\rm m}}{\delta g^{\mu\nu}}=e^{A}{}_{\mu}\left(\frac{1}{e}\frac{\delta L_{\rm m}}{\delta e^{A}{}_{\nu'}}\right)g_{\nu\nu'}:=e^{A}{}_{\mu}\Theta_{A}{}^{\nu'}g_{\nu\nu'}\,.
\end{equation}
The trace of the energy-momentum tensor will be denoted by $\Theta=\udt{\Theta}{\mu}{\mu}$.

When writing the \gls{flrw} line-element, we use $t$ to denote cosmic time (coordinate time as measured by a comoving observer in a homogeneous matter field) and $\tau$ (defined as $\dd \tau=N(t)\dd t/a(t)$) to denote conformal time (see Sec.~\ref{sec:power_spec_f_T}). Thus, the \gls{flrw} metric in spherical-like spatial coordinates is written as \eqref{eq:frwN}
\begin{equation} \label{eq:FLRW_metric}
    \dd s^2=N(t)^2\dd t^2-a(t)^2\Big[\frac{\dd r^2}{1-kr^2}+r^2(\dd \vartheta^2+\sin^2\vartheta \dd \varphi^2)\Big]\,,
\end{equation}
where $N(t)$ is the lapse function and $a(t)$ is the scale factor. We would like to point out that the lapse function cannot \textit{a priori} be absorbed into the definition of cosmic time since it may be dynamical in some branches of certain theories. We can then define the cosmic and conformal time derivatives
\begin{subequations}
\begin{align}
    \dot{a} (t) &:= \frac{\dd a}{\dd t}\,,\\[0.5ex]
    a^\prime (\tau) &:= \frac{\dd a}{\dd \tau}\,,
\end{align}
\end{subequations}
which then gives the Hubble parameter and the conformal Hubble parameter definitions respectively as
\begin{subequations}
\begin{align}
    H(t) &:= \frac{\dot{a}}{N\,a}\,,\\[0.5ex]
    \mathcal{H}(\tau) &:= \frac{a^\prime}{N\,a}\,,
\end{align}
\end{subequations}
where the lapse gauge (in which $N=1$) can be taken to recover the corresponding regular definitions, namely $H=\dot{a}/a$ and $\mathcal{H} = a'/a$. Throughout this Review we will refer to the energy density of a fluid as $\rho$, and its isotropic pressure as $p$. The equation of state is defined as
\begin{equation}
    p \equiv w \rho\,,
\end{equation}
where $w$ is the equation of state (\gls{eos}) parameter which often varies over the cosmic history of the Universe in modified theories of gravity.




\clearpage

\section{Metric-Affine Approaches}\label{sec2:affine}

We start our Review with a brief overview of metric-affine geometry and its fundamental properties, which lie at the foundation of \gls{tg} theories. In Sec.~\ref{ssec:metaffgeom}, we provide a summary of general metric-affine geometry in its most commonly used formulation in terms of a metric and an affine connection. In the restricted case of \gls{tg}, however, which is the main subject of this Review, a formulation in terms of a tetrad and spin connection is more common; this is Reviewed in Sec.~\ref{sec:tetradspin}. Finally, in Sec.~\ref{Sec:Geometric_Trinity}, we discuss how different classes of geometries can be used as alternative formulations of \gls{gr}, whose field equations agree with the usual, curvature-based formulation.

\subsection{Metric-affine geometry}\label{ssec:metaffgeom}

We now give a summary here of metric-affine geometry and its characteristic properties. In Sec.~\ref{sssec:metafffundvar} we summarize the fundamental fields that constitute the metric-affine geometry. Their characterizing, tensorial properties are displayed in Sec.~\ref{sssec:affcontens}. These tensorial quantities are related by the Bianchi identities, which we display in Sec.~\ref{sssec:affconbianchi}. Further, we discuss how they can be used to decompose the connection in Sec.~\ref{sssec:affcondec}. An overview of particular subclasses of metric-affine geometries is given in Sec.~\ref{sssec:metaffpartclass}.

\subsubsection{Fundamental variables}\label{sssec:metafffundvar}
In the most general metric-affine setting, which is more comprehensively reviewed in Ref.~\cite{Hehl:1994ue}, the fundamental variables are a metric (i.e., a symmetric rank-two tensor field) \(g_{\mu\nu}\) of Lorentzian signature, as defined in the conventions Sec.~\ref{ssec:Conventions}, as well as the coefficients \(\gc{\Gamma}^{\rho}{}_{\mu\nu}\) of an affine connection, with associated covariant derivative \(\gc{\nabla}\), both defined on a 4 dimensional spacetime manifold \(\mathcal{M}\). It thus follows that the metric has 10 independent components, while the connection has 64 independent components, and there is no \textit{a priori} relation between these two objects. Here and in the following we will use a hat $\gc{\,}$ whenever we refer to quantities which are related to this general affine connection. We have collected these notes on notation in Sec.~\ref{ssec:Conventions} for each of access.

\subsubsection{Characteristic tensors} \label{sssec:affcontens}

\begin{figure}[htbp]
\centering
\hspace{-1cm}\includegraphics[scale=1.2]{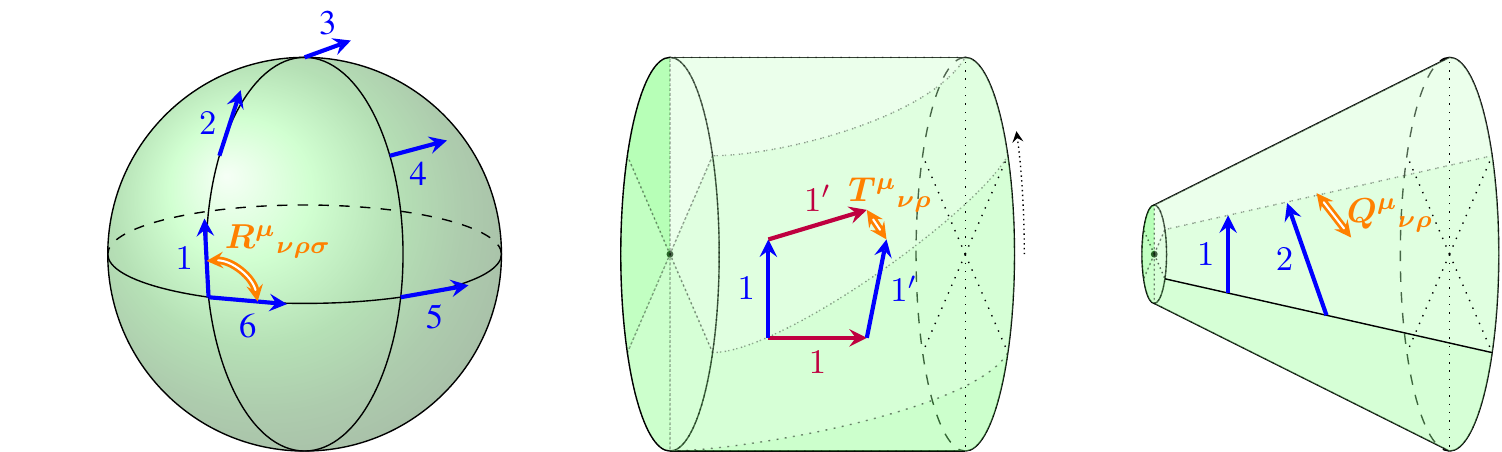}
\caption{Schematic geometrical representation of the curvature, torsion and non-metricity tensors by their effect on the parallel transport of vectors~\cite{Jarv:2020evp}. This figure was kindly provided by Laur J\"{a}rv.}
\label{fig:RTQ}
\end{figure}

The most general metric-affine geometry defined by a metric and affine connection is characterized by a number of tensorial quantities, whose definition and conventions we use in this Review are given below. From the connection alone one obtains the curvature tensor
\begin{equation}\label{eq:gennaffcurv}
\gc{R}^{\mu}{}_{\nu\rho\sigma} = \partial_{\rho}\gc{\Gamma}^{\mu}{}_{\nu\sigma} - \partial_{\sigma}\gc{\Gamma}^{\mu}{}_{\nu\rho} + \gc{\Gamma}^{\mu}{}_{\tau\rho}\gc{\Gamma}^{\tau}{}_{\nu\sigma} - \gc{\Gamma}^{\mu}{}_{\tau\sigma}\gc{\Gamma}^{\tau}{}_{\nu\rho}\,,
\end{equation}
as well as the torsion tensor
\begin{equation}\label{eq:genafftors}
    \gc{T}^{\mu}{}_{\nu\rho} = \gc{\Gamma}^{\mu}{}_{\rho\nu} - \gc{\Gamma}^{\mu}{}_{\nu\rho}\,.
\end{equation}
Using also the metric, one further defines the non-metricity tensor
\begin{equation}\label{eq:genaffnonmet}
\gc{Q}_{\mu\nu\rho} = \gc{\nabla}_{\mu}g_{\nu\rho} = \partial_{\mu}g_{\nu\rho} - \gc{\Gamma}^{\sigma}{}_{\nu\mu}g_{\sigma\rho} - \gc{\Gamma}^{\sigma}{}_{\rho\mu}g_{\nu\sigma}\,.
\end{equation}
These notions are illustrated in Fig.~\ref{fig:RTQ} by their effect on the parallel transport of a vector defined by the affine connection \(\gc{\Gamma}^{\mu}{}_{\nu\rho}\). Curvature causes the parallel transport along a closed curve to be non-trivial, i.e., to change the transported vector. With torsion, the parallel transport is not symmetric under exchanging the transported vector and the direction of transport. With non-metricity, the length of the vector, as measured by the metric, changes along the transport. Note that for the general affine connection \(\hat{\Gamma}^{\rho}{}_{\mu\nu}\), any of these tensorial quantities may be nonvanishing. More restricted classes of connections can be obtained by demanding that certain quantities vanish; we will discuss such cases in Sec.~\ref{sssec:metaffpartclass}.

\subsubsection{Bianchi identities} \label{sssec:affconbianchi}

The characterizing tensors of a general affine connection are not completely independent from each other. In particular, the curvature and torsion are related by the Bianchi identities~\cite{Kobayashi:1963fdg}
\begin{equation}\label{eq:bianchione}
\gc{R}^{\mu}{}_{[\nu\rho\sigma]} = \gc{\nabla}_{[\nu}\gc{T}^{\mu}{}_{\rho\sigma]} + \gc{T}^{\mu}{}_{\omega[\nu}\gc{T}^{\omega}{}_{\rho\sigma]}\,,
\end{equation}
and
\begin{equation}\label{eq:bianchitwo}
\gc{\nabla}_{[\omega}\gc{R}^{\mu}{}_{|\nu|\rho\sigma]} = -\gc{R}^{\mu}{}_{\nu\tau[\omega}\gc{T}^{\tau}{}_{\rho\sigma]}\,.
\end{equation}
This can easily be seen by direct calculation. For the first identity~\eqref{eq:bianchione}, one has
\begin{equation}
\gc{\nabla}_{[\nu}\gc{T}^{\mu}{}_{\rho\sigma]} = -2\partial_{[\nu}\gc{\Gamma}^{\mu}{}_{\rho\sigma]} - 2\gc{\Gamma}^{\mu}{}_{\omega[\nu}\gc{\Gamma}^{\omega}{}_{\rho\sigma]} + 2\gc{\Gamma}^{\omega}{}_{[\rho\nu}\gc{\Gamma}^{\mu}{}_{|\omega|\sigma]} + 2\gc{\Gamma}^{\omega}{}_{[\sigma\nu}\gc{\Gamma}^{\mu}{}_{\rho]\omega}\,,
\end{equation}
and
\begin{equation}
\gc{T}^{\mu}{}_{\omega[\nu}\gc{T}^{\omega}{}_{\rho\sigma]} = 2\gc{\Gamma}^{\mu}{}_{\omega[\nu}\gc{\Gamma}^{\omega}{}_{\rho\sigma]} - 2\gc{\Gamma}^{\mu}{}_{[\nu|\omega|}\gc{\Gamma}^{\omega}{}_{\rho\sigma]}\,.
\end{equation}
Combining these two terms, and sorting the indices, one obtains the expression
\begin{equation}
\gc{\nabla}_{[\omega}\gc{R}^{\mu}{}_{|\nu|\rho\sigma]} + \gc{T}^{\mu}{}_{\omega[\nu}\gc{T}^{\omega}{}_{\rho\sigma]} = 2\partial_{[\rho}\gc{\Gamma}^{\mu}{}_{\nu\sigma]} + 2\gc{\Gamma}^{\mu}{}_{\omega[\rho}\gc{\Gamma}^{\omega}{}_{\nu\sigma]} = \gc{R}^{\mu}{}_{[\nu\rho\sigma]}\,,
\end{equation}
which is the \gls{lhs} of the Bianchi identity~\eqref{eq:bianchione}. Similarly, the second identity~\eqref{eq:bianchitwo} can be proven. Expanding the covariant derivative on the \gls{lhs} yields
\begin{equation}\label{eq:curvderexpans}
\gc{\nabla}_{[\omega}\gc{R}^{\mu}{}_{|\nu|\rho\sigma]} = \partial_{[\omega}\gc{R}^{\mu}{}_{|\nu|\rho\sigma]} + \gc{\Gamma}^{\mu}{}_{\tau[\omega}\gc{R}^{\tau}{}_{|\nu|\rho\sigma]} - \gc{\Gamma}^{\tau}{}_{\nu[\omega}\gc{R}^{\mu}{}_{|\tau|\rho\sigma]} - \gc{\Gamma}^{\tau}{}_{[\rho\omega}\gc{R}^{\mu}{}_{|\nu\tau|\sigma]} - \gc{\Gamma}^{\tau}{}_{[\sigma\omega}\gc{R}^{\mu}{}_{|\nu|\rho]\tau}\,.
\end{equation}
Now it is easy to see that the last two terms are identical,
\begin{equation}
\gc{\Gamma}^{\tau}{}_{[\sigma\omega}\gc{R}^{\mu}{}_{|\nu|\rho]\tau} = -\gc{\Gamma}^{\tau}{}_{[\sigma\omega}\gc{R}^{\mu}{}_{|\nu\tau|\rho]} = \gc{\Gamma}^{\tau}{}_{[\rho\omega}\gc{R}^{\mu}{}_{|\nu\tau|\sigma]}\,.
\end{equation}
Further using the definition \eqref{eq:genafftors} of the torsion, as well as cyclically rearranging the indices within square brackets, the last two terms of the expansion~\eqref{eq:curvderexpans} can be combined to
\begin{equation}
-2\gc{\Gamma}^{\tau}{}_{[\rho\omega}\gc{R}^{\mu}{}_{|\nu\tau|\sigma]} = -\gc{R}^{\mu}{}_{\nu\tau[\omega}\gc{T}^{\mu}{}_{\rho\sigma]}\,,
\end{equation}
which is the \gls{rhs} of the Bianchi identity~\eqref{eq:bianchitwo}. It remains to show that the first three terms on the \gls{rhs} of the expansion~\eqref{eq:curvderexpans} vanish. Expanding the curvature in the derivative term using its definition~\eqref{eq:gennaffcurv} yields
\begin{equation}
\partial_{[\omega}\gc{R}^{\mu}{}_{|\nu|\rho\sigma]} = 2\partial_{[\omega}\partial_{\rho}\gc{\Gamma}^{\mu}{}_{|\nu|\sigma]} + 2\partial_{[\omega}\gc{\Gamma}^{\mu}{}_{|\tau|\rho}\gc{\Gamma}^{\tau}{}_{|\nu|\sigma]} + 2\gc{\Gamma}^{\mu}{}_{\tau[\rho}\partial_{\omega}\gc{\Gamma}^{\tau}{}_{|\nu|\sigma]}\,.
\end{equation}
The first term on the \gls{rhs} vanishes, since partial derivatives commute. In the remaining two terms one can substitute the derivatives on the connection coefficients by curvature tensors, which yields
\begin{subequations}
\begin{align}
2\partial_{[\omega}\gc{\Gamma}^{\mu}{}_{|\tau|\rho}\gc{\Gamma}^{\tau}{}_{|\nu|\sigma]} &= \gc{\Gamma}^{\tau}{}_{\nu[\sigma}\gc{R}^{\mu}{}_{|\tau|\omega\rho]} - 2\gc{\Gamma}^{\tau}{}_{\nu[\sigma}\gc{\Gamma}^{\mu}{}_{|\phi|\omega}\gc{\Gamma}^{\phi}{}_{|\tau|\rho]}\,,\\[0.5ex]
2\gc{\Gamma}^{\mu}{}_{\tau[\rho}\partial_{\omega}\gc{\Gamma}^{\tau}{}_{|\nu|\sigma]} &= \gc{\Gamma}^{\mu}{}_{\tau[\rho}\gc{R}^{\tau}{}_{|\nu|\omega\sigma]} - 2\gc{\Gamma}^{\mu}{}_{\tau[\rho}\gc{\Gamma}^{\tau}{}_{|\phi|\omega}\gc{\Gamma}^{\phi}{}_{|\nu|\sigma]}\,.
\end{align}
\end{subequations}
After permuting the indices in square brackets, one finds that the two curvature terms found here cancel the second and third term on the \gls{rhs} of the expansion~\eqref{eq:curvderexpans}. Finally, for the two terms cubic in the connection coefficients, one can exchange the dummy indices \(\tau \leftrightarrow \phi\) in the first term to find
\begin{equation}
\gc{\Gamma}^{\mu}{}_{\tau[\omega}\gc{\Gamma}^{\tau}{}_{|\phi|\rho}\gc{\Gamma}^{\phi}{}_{|\nu|\sigma]} + \gc{\Gamma}^{\mu}{}_{\tau[\rho}\gc{\Gamma}^{\tau}{}_{|\phi|\omega}\gc{\Gamma}^{\phi}{}_{|\nu|\sigma]} = 0\,,
\end{equation}
due to the antisymmetry in the indices \(\rho\) and \(\omega\). This completes the proof of the second Bianchi identity~\eqref{eq:bianchitwo}.

There are two notable special cases. For a symmetric connection, \(\gc{T}^{\mu}{}_{\nu\rho} = 0\), and so the Bianchi identities reduce to the well known relations
\begin{equation}\label{eq:bianchisymmetric}
\gc{R}^{\mu}{}_{[\nu\rho\sigma]} = 0\,, \quad
\gc{\nabla}_{[\omega}\gc{R}^{\mu}{}_{|\nu|\rho\sigma]} = 0\,.
\end{equation}
This is the case in particular for the Levi-Civita connection. For a flat connection one has \(\gc{R}^{\mu}{}_{\nu\rho\sigma} = 0\), and so the second Bianchi identity becomes trivial, while the first one reduces to
\begin{equation}\label{eq:bianchioneflat}
\gc{\nabla}_{[\nu}\gc{T}^{\mu}{}_{\rho\sigma]} + \gc{T}^{\mu}{}_{\omega[\nu}\gc{T}^{\omega}{}_{\rho\sigma]} = 0\,.
\end{equation}
If both torsion and curvature vanish, both Bianchi identities become trivial.

\subsubsection{Connection decomposition} \label{sssec:affcondec}

An important property of metric-affine geometry is the fact that in the presence of a metric, the coefficients \(\gc{\Gamma}^{\rho}{}_{\mu\nu}\) of the general affine connection can uniquely be decomposed into three parts in the form
\begin{equation}\label{eq:affcondec}
    \gc{\Gamma}^{\rho}{}_{\mu\nu} := \lc{\Gamma}^{\rho}{}_{\mu\nu} + \gc{K}^{\rho}{}_{\mu\nu} + \gc{L}^{\rho}{}_{\mu\nu}\,,
\end{equation}
where we have introduced the coefficients
\begin{equation}\label{eq:levicivita}
\lc{\Gamma}^{\mu}{}_{\nu\rho} := \frac{1}{2}g^{\mu\sigma}\left(\partial_{\nu}g_{\sigma\rho} + \partial_{\rho}g_{\nu\sigma} - \partial_{\sigma}g_{\nu\rho}\right)\,,
\end{equation}
of the Levi-Civita connection, the contortion tensor
\begin{equation}\label{eq:contor}
\gc{K}^{\mu}{}_{\nu\rho} := \frac{1}{2}\left(\gc{T}_{\nu}{}^{\mu}{}_{\rho} + \gc{T}_{\rho}{}^{\mu}{}_{\nu} - \gc{T}^{\mu}{}_{\nu\rho}\right)\,,
\end{equation}
as well as the disformation tensor
\begin{equation}\label{eq:disfor}
\gc{L}^{\mu}{}_{\nu\rho} := \frac{1}{2}\left(\gc{Q}^{\mu}{}_{\nu\rho} - \gc{Q}_{\nu}{}^{\mu}{}_{\rho} - \gc{Q}_{\rho}{}^{\mu}{}_{\nu}\right)\,.
\end{equation}
In this regime, the Christoffel symbols ($\lc{\Gamma}^{\mu}{}_{\nu\rho}$), the contortion tensor ($\gc{K}^{\mu}{}_{\nu\rho}$), and the disformation tensor ($\gc{L}^{\mu}{}_{\nu\rho}$) respectively quantify curvature, torsion, and non-metricity. Note that although the torsion~\eqref{eq:genafftors} is independent of the metric, this does not hold for the contortion~\eqref{eq:contor}, due to the fact that indices have been raised and lowered with the metric.

\subsubsection{Metric-affine subclasses} \label{sssec:metaffpartclass}

By demanding that certain of the tensorial quantities vanish, one may reduce the most general class of metric-affine geometries, which we introduced above and which is characterized by a general connection \(\hat{\Gamma}^{\mu}{}_{\nu\rho}\) next to the metric \(g_{\mu\nu}\), and thereby obtain particular subclasses. Various of these restricted geometries are relevant and used to model theories of gravity in the literature. By successive restriction, one finds the following special cases of metric-affine geometries:
\begin{enumerate}
\item $\gc{R}^{\mu}{}_{\nu\rho\sigma}\equiv0$: In the case of vanishing curvature, the connection is said to be \emph{flat}. This case is also known as (general) teleparallel geometry, if both torsion and non-metricity are present.
\item $\gc{T}^{\mu}{}_{\nu\rho}\equiv0$: If the torsion vanishes, the connection is called \emph{symmetric}, since the connection coefficients satisfy $\gc{\Gamma}^{\rho}{}_{[\mu\nu]}\equiv0$. This case is considered, e.g., in Weyl gravity.
\item $\gc{Q}_{\mu\nu\rho}\equiv0$: For vanishing non-metricity, the connection is called \emph{metric-compatible}, and the corresponding geometry is known as Riemann-Cartan geometry. This choice of the connection is used in Poincaré gauge theory.
\item $\gc{R}^{\mu}{}_{\nu\rho\sigma}\equiv0$, $\gc{Q}_{\mu\nu\rho}\equiv0$: If only torsion is nonvanishing, one obtains the case of \emph{metric} \emph{teleparallel geometry} (also called \emph{torsional geometry}), which is the main topic of this Review. For the metric teleparallel connection and its related quantities, we will omit the hat, and simply write $\Gamma^{\rho}{}_{\mu\nu}$. Moreover, from Sec.~\ref{sec3:torsional} we will just just the word teleparallel to refer to this geometry.
\item $\gc{R}^{\mu}{}_{\nu\rho\sigma}\equiv0$, $\gc{T}^{\mu}{}_{\nu\rho}\equiv0$: An alternative class of geometries is obtained by demanding that only non-metricity is nonvanishing. This class of geometries is known as \emph{symmetric teleparallelism}. We will denote the symmetric teleparallel connection and its associated quantities with a diamond on top, i.e., $\st{\Gamma}^{\rho}{}_{\mu\nu}$.
\item $\gc{T}^{\mu}{}_{\nu\rho}\equiv0$, $\gc{Q}_{\mu\nu\rho}\equiv0$: The most well known class of geometries, which is also most commonly used in the description of gravity theories, including \gls{gr} and its extensions, is the symmetric and metric-compatible connection. In this case the connection is uniquely determined as the \emph{Levi-Civita connection}, which we denote by a circle, i.e., $\lc{\Gamma}^{\rho}{}_{\mu\nu}$.
\item $\gc{R}^{\mu}{}_{\nu\rho\sigma}\equiv0$, $\gc{T}^{\mu}{}_{\nu\rho}\equiv0$, $\gc{Q}_{\mu\nu\rho}\equiv0$: Finally, if all tensorial quantities vanish, also the metric is constrained, and one finds the case of \emph{Minkowski space}. In this case the metric-affine geometry is fixed up to diffeomorphisms and does not carry any gravitational \gls{dof}.
\end{enumerate}

\begin{figure}[H]
\centering
\includegraphics[scale=0.8]{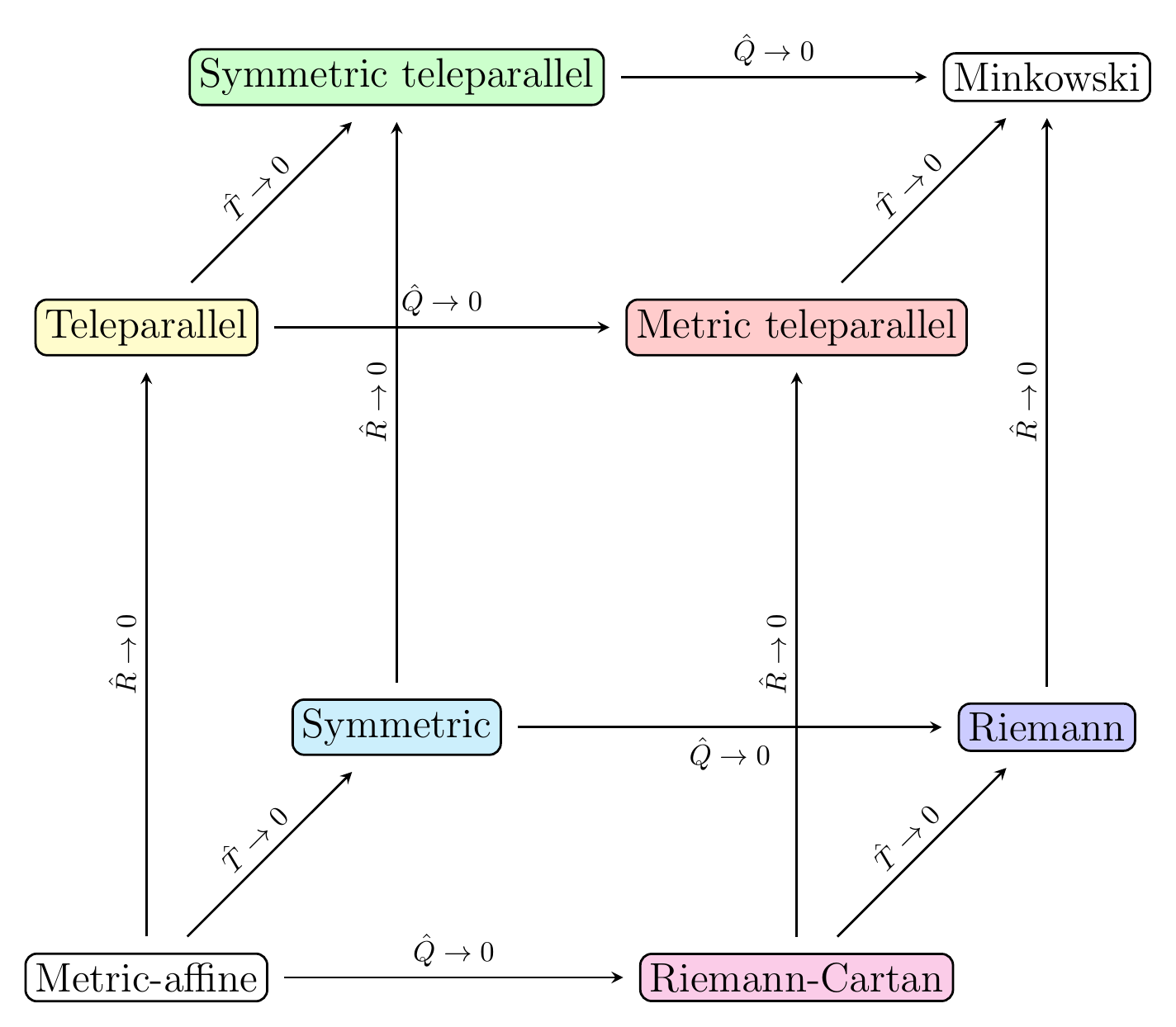}
\caption[Relation between different metric-affine geometries]{Relation between different metric-affine geometries. Starting from the most general class of metric-affine geometries in the lower, left, front corner, this diagram shows how more specific geometries are obtained by imposing that one or more of the tensorial quantities defined in Sec.~\ref{sssec:affcontens} vanish. In the most restricted case, which is Minkowski space in the upper, right, back corner, all three quantities vanish.}
\label{fig:metaffinecube}
\end{figure}

\begin{figure}[H]
\centering
\includegraphics[scale=0.8]{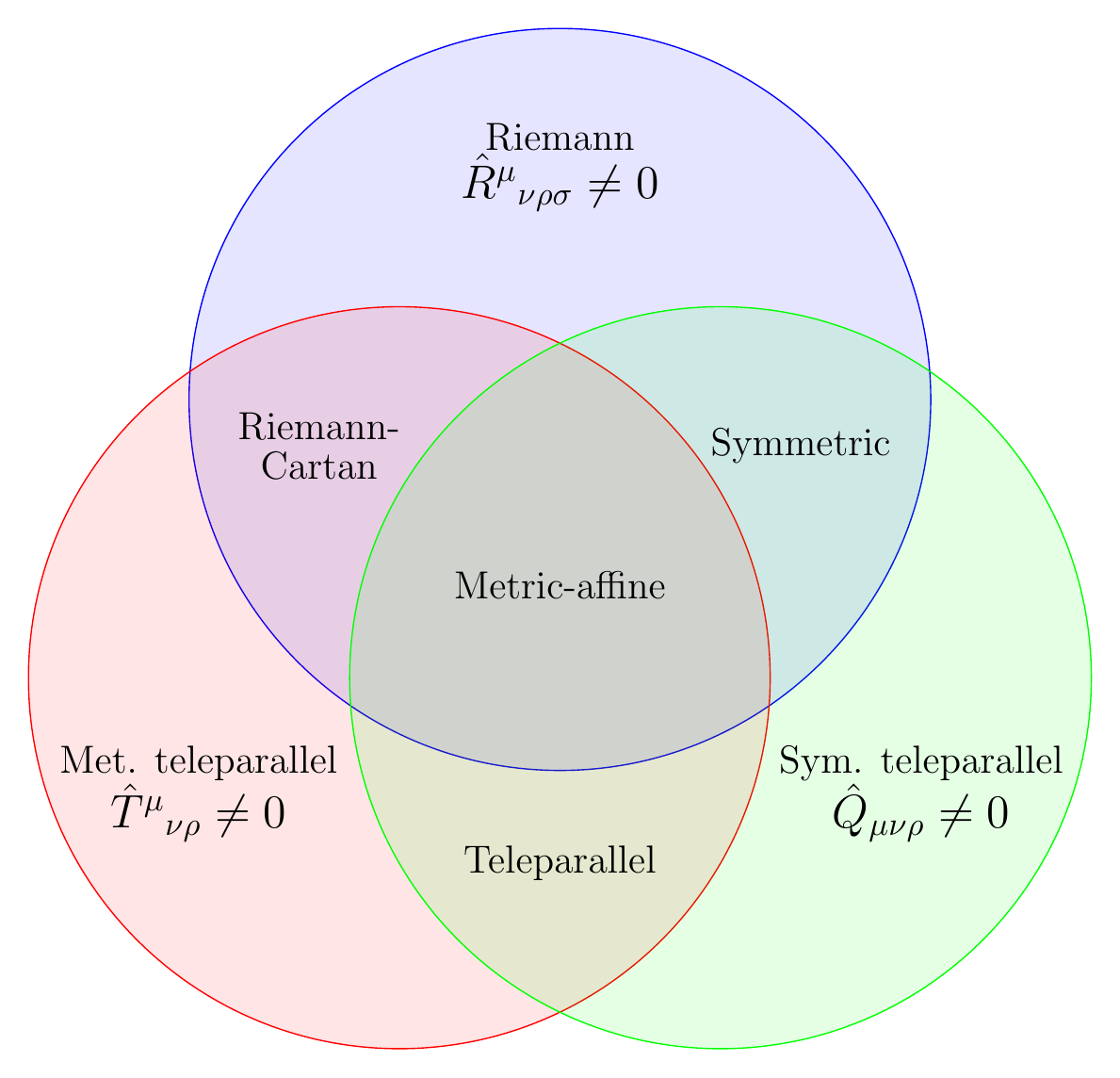}
\caption[Classification of metric-affine geometries]{Classification of metric-affine geometries. Every circle encloses those geometries in which one of the three tensorial quantities defined in Sec.~\ref{sssec:affcontens} is nonvanishing. On the intersections of the circles, more that one tensorial quantity is nonvanishing. Not shown is Minkowski space, where all three quantities vanish, and which is represented by the white area outside all three circles.}
\label{fig:metaffinevenn}
\end{figure}

The relations between the most general metric-affine geometry, in which none of the tensorial quantities vanish, and the special cases listed above, may be visualized in different ways. Fig.~\ref{fig:metaffinecube} shows how one may obtain the aforementioned special cases by imposing restrictions on the most general metric-affine geometry. Another visualization is given by the Venn diagram in Fig.~\ref{fig:metaffinevenn}. Each of the three colored circles represents one of the tensorial quantities, which is nonvanishing for all geometries inside the corresponding circle. The white outer region represents Minkowski space, where curvature, torsion and non-metricity vanish.

In the remainder of this Review we will mostly be working with a metric teleparallel connection \(\Gamma^{\rho}{}_{\mu\nu}\), and omit the accent on all quantities such as the torsion tensor \(T^{\mu}{}_{\nu\rho}\) which are derived from this connection. Any other connection, such as the general metric-affine connection \(\gc{\Gamma}^{\mu}{}_{\nu\rho}\), the Levi-Civita connection \(\lc{\Gamma}^{\mu}{}_{\nu\rho}\) and the symmetric teleparallel connection \(\st{\Gamma}^{\mu}{}_{\nu\rho}\), and the quantities derived from these connections, will be distinguished by denoting them with the corresponding accent.

\subsection{Tetrads and the spin connection} \label{sec:tetradspin}

In the field of \gls{tg}, which is the primary topic of this Review, the underlying geometry of spacetime is conventionally described in the tetrad-spin connection formulation, which we summarize here. In Sec.~\ref{sssec:metafftetspi}, we display how these quantities define the metric and the affine connection, which we had introduced as independent quantities in the previous section. The characteristic tensors describing the metric-affine geometry are then expressed in terms of the tetrad and spin connection in Sec.~\ref{sssec:spicontens}. An important property of this formulation is the local Lorentz invariance of the geometry, which we discuss in Sec.~\ref{sec:LLTW}, and which allows to choose a particular gauge for the spin connection in the teleparallel case. Finally, in Sec.~\ref{sssec:diffform}, we provide an equivalent mathematical description of the tetrad and spin connection in terms of differential forms.

\subsubsection{Relation to metric and connection}\label{sssec:metafftetspi}
\begin{figure}[htbp]
\centering
\includegraphics[scale=1]{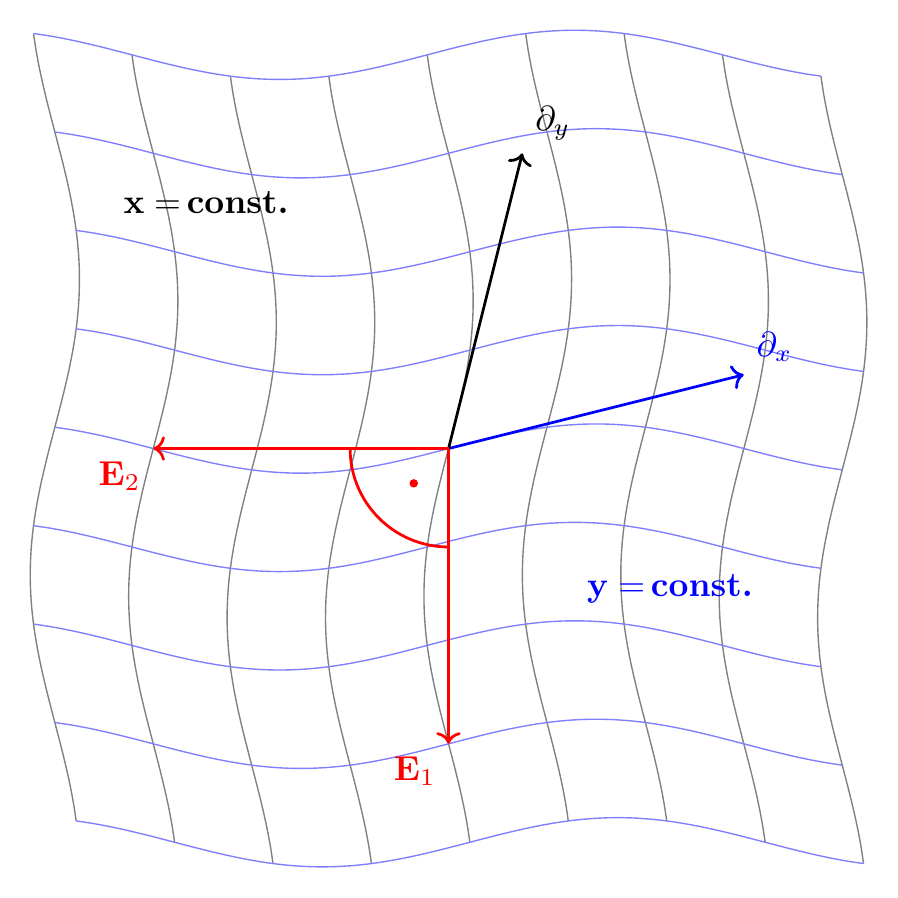}
\caption[Graphical representation of the tetrad]{Graphical representation of the tetrad, reduced to a 2 dimensional model manifold. A coordinate basis \((\partial_x, \partial_y)\) of the tangent space is, by definition, tangent to the coordinate lines at a given base point, but not necessary orthogonal or normalized with a given metric. In contrast, the tetrad corresponds to an orthonormal basis \((\mathbf{E}_1, \mathbf{E}_2) = (E_1{}^{\mu}\partial_{\mu}, E_2{}^{\mu}\partial_{\mu})\).}
\label{fig:tetrad}
\end{figure}

An alternative description of the metric-affine geometry outlined in the preceding section is to use a tetrad \(e^A{}_{\mu}\) and a spin connection \(\gc{\omega}^A{}_{B\mu}\), where capital Latin letters \(A, B = 0, \ldots, 3\) denote Lorentz indices, while small Greek letters $\mu, \nu = 0,...,3$ denote spacetime indices (where numbers refer to spacetime coordinates on the respective spaces). From the tetrad one constructs the Lorentzian metric via the relation
\begin{equation}\label{eq:metric}
g_{\mu\nu} = \eta_{AB}e^A{}_{\mu}e^B{}_{\nu}\,,
\end{equation}
where \(\eta_{AB} = \mathrm{diag}(1, -1, -1, -1)\) denotes the Minkowski metric. For the metric \(g_{\mu\nu}\) to be non-degenerate, one requires that the tetrad possesses an inverse \(E_A{}^{\mu}\), which satisfies \(E_A{}^{\mu}e^A{}_{\nu} = \delta^{\mu}_{\nu}\) and \(E_A{}^{\mu}e^B{}_{\mu} = \delta_A^B\), and defines the inverse metric as
\begin{equation}
g^{\mu\nu} = \eta^{AB}E_A{}^{\mu}E_B{}^{\nu}\,.
\end{equation}
Note that as a consequence, the vectors \((\mathbf{E}_A)\) form an orthonormal basis of the tangent space, i.e.,
\begin{equation}
g(\mathbf{E}_A, \mathbf{E}_B) = g_{\mu\nu}E_A{}^{\mu}E_B{}^{\nu} = \eta_{AB}\,.
\end{equation}
This is visualized in Fig.~\ref{fig:tetrad}. While the coordinate basis \((\partial_{\mu})\) is defined by the coordinate lines, and is in general neither orthogonal nor normalized, the opposite is true for the tetrad basis \((\mathbf{E}_A)\).

The frame coefficients \(E_A{}^{\mu}\) are also required in order to calculate the coefficients \(\gc{\Gamma}^{\mu}{}_{\nu\rho}\) of the affine connection from the spin connection \(\gc{\omega}^A{}_{B\mu}\) via
\begin{equation}\label{eq:genconnection}
\gc{\Gamma}^{\rho}{}_{\mu\nu} = E_A{}^{\rho}\left(\partial_{\nu}e^A{}_{\mu} + \gc{\omega}^A{}_{B\nu}e^B{}_{\mu}\right)\,,
\end{equation}
which is the unique affine connection satisfying the so-called ``tetrad postulate''
\begin{equation}\label{eq:tetradpost}
\partial_{\mu}e^A{}_{\nu} + \gc{\omega}^A{}_{B\mu}e^B{}_{\nu} - \gc{\Gamma}^{\rho}{}_{\nu\mu}e^A{}_{\rho} = 0\,.
\end{equation}
This condition necessitates that vectors remain parallel even at a distance, which is the source of the term \textit{teleparallel gravity}. It is this condition that forces the distinction between curvature and teleparallel based connections. Finally, note that for a given metric-affine geometry, a tetrad, in general, exists only locally, and that global tetrads exist only if the spacetime manifold \(\mathcal{M}\) is parallelizable; this fact can be understood by applying the theory of fiber bundles~\cite{Hohmann:2021dhr}.

\subsubsection{Characteristic tensors} \label{sssec:spicontens}

One advantage of the formulation in terms of a tetrad and spin connection, which makes it particularly useful for the description of teleparallel geometries, is the fact that the curvature and non-metricity defined in Sec.~\ref{sssec:affcontens}, and hence also the answer to the question whether they vanish or not, become properties of the spin connection only, and are independent of the choice of the tetrad. In this case, one defines the curvature tensor
\begin{equation}\label{eq:spiconcurv}
\gc{R}^A{}_{B\mu\nu} := \partial_{\mu}\gc{\omega}^A{}_{B\nu} - \partial_{\nu}\gc{\omega}^A{}_{B\mu} + \gc{\omega}^A{}_{C\mu}\gc{\omega}^C{}_{B\nu} - \gc{\omega}^A{}_{C\nu}\gc{\omega}^C{}_{B\mu}\,,
\end{equation}
the torsion tensor
\begin{equation}\label{eq:spicontors}
\gc{T}^A{}_{\mu\nu} := \partial_{\mu}e^A{}_{\nu} - \partial_{\nu}e^A{}_{\mu} + \gc{\omega}^A{}_{B\mu}e^B{}_{\nu} - \gc{\omega}^A{}_{B\nu}e^B{}_{\mu}\,,
\end{equation}
as well as the non-metricity tensor
\begin{equation}\label{eq:spiconnonmet}
\gc{Q}_{\mu AB} := -\eta_{AC}\gc{\omega}^C{}_{B\mu} - \eta_{CB}\gc{\omega}^C{}_{A\mu}\,.
\end{equation}
These are related to the previously defined quantities by
\begin{subequations}
\begin{align}
\gc{R}^A{}_{B\mu\nu} &= e^A{}_{\rho}E_B{}^{\sigma}\gc{R}^{\rho}{}_{\sigma\mu\nu}\,, &
\gc{R}^{\mu}{}_{\nu\rho\sigma} &= E_A{}^{\mu}e^B{}_{\nu}\gc{R}^A{}_{B\rho\sigma}\,,\\[0.5ex]
\gc{T}^A{}_{\mu\nu} &= e^A{}_{\rho}\gc{T}^{\rho}{}_{\mu\nu}\,, &
\gc{T}^{\mu}{}_{\nu\rho} &= E_A{}^{\mu}\gc{T}^A{}_{\nu\rho}\,,\\[0.5ex]
\gc{Q}_{\mu AB} &= E_A{}^{\nu}E_B{}^{\rho}\gc{Q}_{\mu\nu\rho}\,, &
\gc{Q}_{\mu\nu\rho} &= e^A{}_{\nu}e^B{}_{\rho}\gc{Q}_{\mu AB}\,,
\end{align}
\end{subequations}
so that spacetime and Lorentz indices are simply exchanged with the help of the tetrad and its inverse. The same convention may also be applied to any other tensor, by defining
\begin{equation}\label{eq:getrans}
Z^{A_1\cdots A_n}{}_{B_1\cdots B_m} := e^{A_1}{}_{\alpha_1} \cdots e^{A_n}{}_{\alpha_n}E_{B_1}{}^{\beta_1} \cdots E_{B_m}{}^{\beta_m}Z^{\alpha_1\cdots\alpha_n}{}_{\beta_1\cdots\beta_m}\,.
\end{equation}
This allows to decompose the spin connection in the form
\begin{equation}
\gc{\omega}^A{}_{B\mu} = \lc{\omega}^A{}_{B\mu} + \gc{K}^A{}_{B\mu} + \gc{L}^A{}_{B\mu}\,,
\end{equation}
in analogy to the decomposition~\eqref{eq:affcondec} of the affine connection displayed in Sec.~\ref{sssec:affcondec}. Here the Levi-Civita spin connection \(\lc{\omega}^A{}_{B\mu}\) is obtained from its affine connection counterpart \(\lc{\Gamma}^{\mu}{}_{\nu\rho}\) from definition~\eqref{eq:levicivita} by making use of the tetrad postulate~\eqref{eq:tetradpost}, with the general connection replaced by the Levi-Civita connection. It then follows that the contortion \(\gc{K}^A{}_{B\mu}\) and disformation \(\gc{L}^A{}_{B\mu}\) are again related to their counterparts with spacetime indices by the tetrad and its inverse using the transformation rule~\eqref{eq:getrans} for tensor indices.

\subsubsection{Local Lorentz transformations and Weitzenb{\"{o}}ck gauge} \label{sec:LLTW}

It is important to note that the tetrad \(e^A{}_{\mu}\) and spin connection \(\gc{\omega}^A{}_{B\mu}\) are not uniquely determined by the metric \(g_{\mu\nu}\) and affine connection \(\gc{\Gamma}^{\rho}{}_{\mu\nu}\). From the definition in Eq.~\eqref{eq:metric}, one can see that the metric is invariant, \(g_{\mu\nu} = g'_{\mu\nu}\), if one replaces the tetrad by
\begin{equation}\label{eq:tetlortrans}
    e^A{}_{\mu} \mapsto e'^A{}_{\mu} = \Lambda^A{}_Be^B{}_{\mu}\,,
\end{equation}
where \(\Lambda^A{}_B\) is a local (spacetime point dependent) Lorentz transformation, i.e., it must satisfy
\begin{equation}\label{eq:finlortranscond}
    \eta_{AB}\Lambda^A{}_C\Lambda^B{}_D = \eta_{CD}\,.
\end{equation}
Note, however, that the connection coefficients~\eqref{eq:genconnection} are not invariant under this transformation of the tetrad alone, unless the Lorentz transformation is global, \(\partial_{\mu}\Lambda^A{}_B \equiv 0\). To obtain invariance of the connection also under local Lorentz transformations, one must additionally transform the spin connection by the rule
\begin{equation}\label{eq:spclortrans}
    \gc{\omega}^A{}_{B\mu} \mapsto \gc{\omega}'^A{}_{B\mu} = \Lambda^A{}_C(\Lambda^{-1})^D{}_B\,\gc{\omega}^C{}_{D\mu} + \Lambda^A{}_C\,\partial_{\mu}(\Lambda^{-1})^C{}_B\,.
\end{equation}
It thus follows that the tetrad and spin connection which model a given metric-affine geometry are uniquely determined only up to a local Lorentz transformation, and that any tetrad and spin connection related by the transformations~\eqref{eq:tetlortrans} and~\eqref{eq:spclortrans} define the same metric-affine geometry. Hence, the tetrad and spin connection variables come together with a Lorentz gauge symmetry.

The Lorentz gauge freedom is of particular relevance for metric teleparallelism, with a flat, metric-compatible spin connection. In this case it is always possible to choose a gauge (at least locally, on a simply connected domain) such that the spin connection vanishes identically, \(\omega^A{}_{B\mu} \equiv 0\). This gauge is known as the Weitzenb\"{o}ck gauge. It further follows that in any other gauge the spin connection takes the simple form
\begin{equation}\label{eq:spclormatrep}
    \omega'^A{}_{B\mu} = \Lambda^A{}_C\partial_{\mu}(\Lambda^{-1})^C{}_B\,,
\end{equation}
and is thus a pure gauge \gls{dof}.

It is also helpful to note how the tetrad and spin connection transform under \emph{infinitesimal} Lorentz transformations, which are of the form
\begin{equation}
    \Lambda^A{}_B = \delta^A_B + \lambda^A{}_B\,,
\end{equation}
where the condition~\eqref{eq:finlortranscond} translates to
\begin{equation}
    \eta_{CD} = \eta_{AB}\left(\delta^A_C + \lambda^A{}_C\right)\left(\delta^B_D + \lambda^B{}_D\right) = \eta_{CD} + 2\lambda_{(CD)} + \mathcal{O}(\lambda^2)\,,
\end{equation}
and so \(\lambda_{(AB)} = 0\). In this case we find the transformation
\begin{equation}\label{eq:inftetlor}
\delta_{\lambda}e^A{}_{\mu} = e'^A{}_{\mu} - e^A{}_{\mu} = \lambda^A{}_Be^B{}_{\mu}\,,
\end{equation}
for the tetrad, and
\begin{equation}\label{eq:infspclor}
\delta_{\lambda}\gc{\omega}^A{}_{B\mu} = \gc{\omega}'^A{}_{B\mu} - \gc{\omega}^A{}_{B\mu} = -\partial_{\mu}\lambda^A{}_B - \omega^A{}_{C\mu}\lambda^C{}_B + \omega^C{}_{B\mu}\lambda^A{}_C = -\DDD_{\mu}\lambda^A{}_B\,,
\end{equation}
for the spin connection. We will make use of these relations later, when we consider the variation of the teleparallel spin connection and the notion of spacetime symmetries. $\DDD_{\mu}$ is called the Fock-Ivanenko derivative and is formally defined in Sec.~\ref{sec:field_strength}.

\subsubsection{Differential form formulation} \label{sssec:diffform}

A more compact description of the relevant objects in the tetrad and spin connection formulation, which can be found mostly in the more mathematically oriented literature on metric-affine and \gls{tg}, makes use of the language of differential forms~\cite{Kobayashi:1963fdg}. We give a brief guide to the reader on how to translate the most important quantities and relations from the differential form formulation to the component notation we use in this review. The starting point is the fact that the tetrad and the spin connection constitute the components of one-forms \(\mathbf{e}^A = e^A{}_{\mu}\,\dd x^{\mu}\) and \(\gc{\boldsymbol{\omega}}^A{}_B = \gc{\omega}^A{}_{B\mu}\,\dd x^{\mu}\), where the presence of Lorentz indices indicates that the former takes values in the Minkowski space, while the latter takes values in the Lie algebra \(\mathfrak{gl}(4)\) of the general linear group. In this language, also the tensorial quantities we introduced before are expressed as differential forms. In particular, we have the curvature
\begin{equation}
\gc{\mathbf{R}}^A{}_B := \frac{1}{2}\gc{R}^A{}_{B\mu\nu}\,\dd x^{\mu} \wedge \dd x^{\nu} = \dd\gc{\boldsymbol{\omega}}^A{}_B + \gc{\boldsymbol{\omega}}^A{}_C \wedge \gc{\boldsymbol{\omega}}^C{}_B\,,
\end{equation}
the torsion
\begin{equation}
\gc{\mathbf{T}}^A := \frac{1}{2}\gc{T}^A{}_{\mu\nu}\,\dd x^{\mu} \wedge \dd x^{\nu} = \DD\mathbf{e}^A = \dd\mathbf{e}^A + \gc{\boldsymbol{\omega}}^A{}_B \wedge \mathbf{e}^B\,,
\end{equation}
and finally the non-metricity
\begin{equation}
\gc{\mathbf{Q}}_{AB} := \gc{Q}_{\mu AB}\,\dd x^{\mu} = \DD\eta_{AB} = \dd\eta_{AB} - \gc{\boldsymbol{\omega}}^C{}_A \wedge \eta_{CB} - \gc{\boldsymbol{\omega}}^C{}_B \wedge \eta_{AC}\,,
\end{equation}
where we defined the covariant exterior derivative \(\DD\). Similarly, the spin connection decomposes in the form
\begin{equation}
\gc{\boldsymbol{\omega}}^A{}_B := \lc{\boldsymbol{\omega}}^A{}_B + \gc{\mathbf{K}}^A{}_B + \gc{\mathbf{L}}^A{}_B\,,
\end{equation}
using the contortion \(\gc{\mathbf{K}}^A{}_B\) and disformation \(\gc{\mathbf{L}}^A{}_B\). Finally, the Bianchi identities are more commonly encountered in the form
\begin{equation}
\DD\gc{\mathbf{T}}^A = \gc{\mathbf{R}}^A{}_B \wedge \mathbf{e}^B\,, \quad
\DD\gc{\mathbf{R}}^A{}_B = 0\,.
\end{equation}
However, in this Review we will mostly use the tensor component notation introduced earlier, which is more common in physical applications, and resort to differential forms only when referring to relevant literature.

\subsection{Geometric trinity of gravity} \label{Sec:Geometric_Trinity}

\begin{figure}[htbp]
\centering
\includegraphics[scale=1]{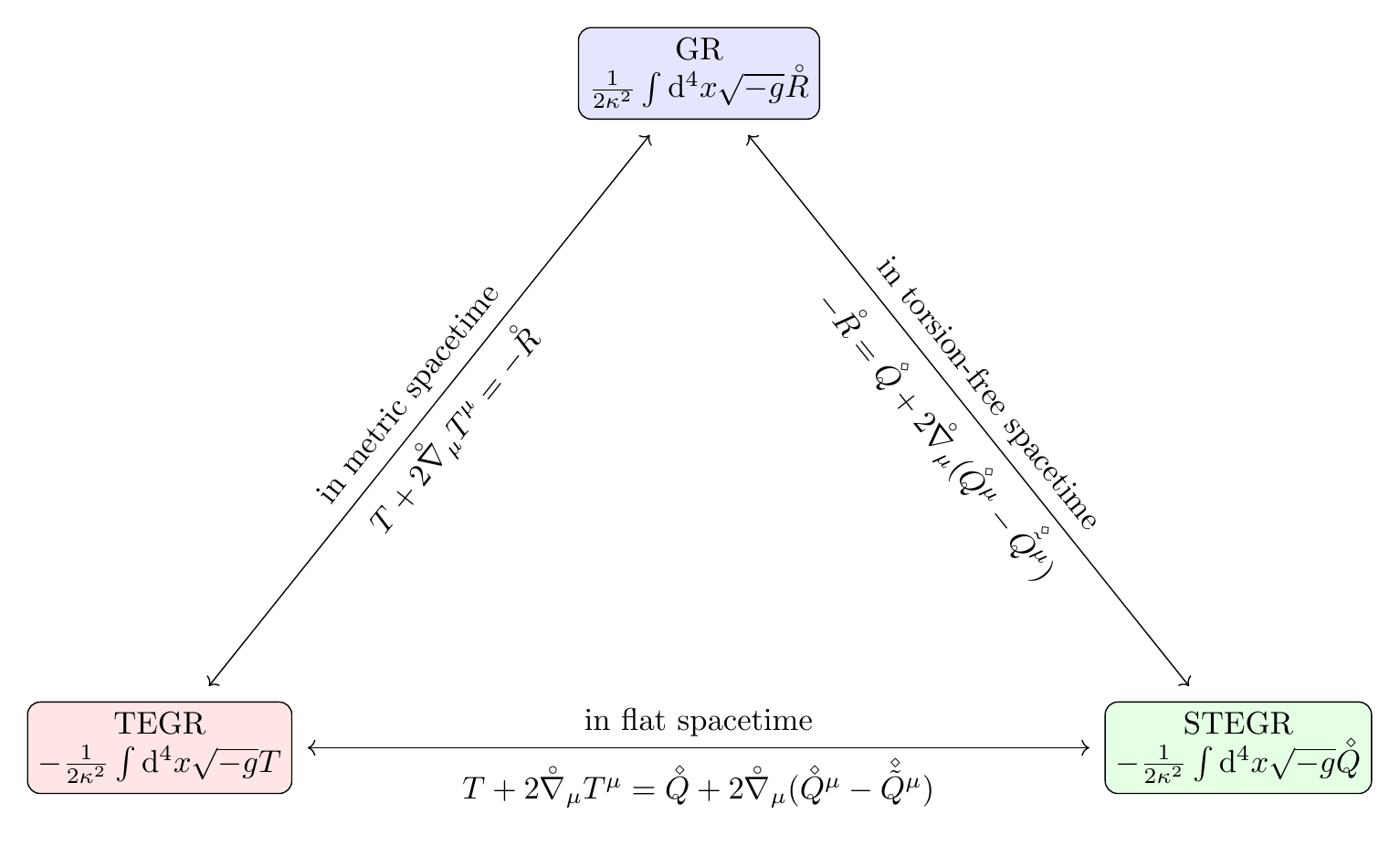}
\caption[Trinity of gravity]{Trinity of gravity. Arrows indicate the relations~\eqref{eq:ricsmtele} and~\eqref{eq:ricsstele} relating the Ricci scalar of the Levi-Civita connection to the torsion and non-metricity scalars.}
\label{fig:trinity}
\end{figure}

The existence of a unique decomposition~\eqref{eq:affcondec} of the connection coefficients \(\gc{\Gamma}^{\mu}{}_{\nu\rho}\) for every metric-affine geometry leads to an interesting consequence. It follows that the curvature tensor \(\gc{R}^{\mu}{}_{\nu\rho\sigma}\) allows for a similar decomposition
\begin{equation}\label{eq:gennaffcurvdec}
\gc{R}^{\mu}{}_{\nu\rho\sigma} = \lc{R}^{\mu}{}_{\nu\rho\sigma} + \lc{\nabla}_{\rho}\gc{D}^{\mu}{}_{\nu\sigma} - \lc{\nabla}_{\sigma}\gc{D}^{\mu}{}_{\nu\rho} + \gc{D}^{\mu}{}_{\tau\rho}\gc{D}^{\tau}{}_{\nu\sigma} - \gc{D}^{\mu}{}_{\tau\sigma}\gc{D}^{\tau}{}_{\nu\rho}\,,
\end{equation}
where we used the shorthand
\begin{equation}
\gc{D}^{\mu}{}_{\nu\rho} = \gc{\Gamma}^{\mu}{}_{\nu\rho} - \lc{\Gamma}^{\mu}{}_{\nu\rho} = \gc{K}^{\mu}{}_{\nu\rho} + \gc{L}^{\mu}{}_{\nu\rho}\,,
\end{equation}
called the distortion tensor. Two special cases of this relation deserve particular attention, which have in common that they lead to a vanishing curvature tensor \(\gc{R}^{\mu}{}_{\nu\rho\sigma} \equiv 0\). The first case is given by a metric teleparallel connection \(\gc{\Gamma}^{\mu}{}_{\nu\rho} \equiv \Gamma^{\mu}{}_{\nu\rho}\), which allows to express the Riemann tensor of the Levi-Civita connection as
\begin{equation}
\lc{R}^{\mu}{}_{\nu\rho\sigma} = K^{\mu}{}_{\tau\sigma}K^{\tau}{}_{\nu\rho} - K^{\mu}{}_{\tau\rho}K^{\tau}{}_{\nu\sigma} + \lc{\nabla}_{\sigma}K^{\mu}{}_{\nu\rho} - \lc{\nabla}_{\rho}K^{\mu}{}_{\nu\sigma}\,,
\end{equation}
in terms of the contortion tensor. The second special case is that of a symmetric teleparallel connection \(\gc{\Gamma}^{\mu}{}_{\nu\rho} \equiv \st{\Gamma}^{\mu}{}_{\nu\rho}\). In this case the Riemann tensor is expressed through the disformation tensor as
\begin{equation}
\lc{R}^{\mu}{}_{\nu\rho\sigma} = \st{L}^{\mu}{}_{\tau\sigma}\st{L}^{\tau}{}_{\nu\rho} - \st{L}^{\mu}{}_{\tau\rho}\st{L}^{\tau}{}_{\nu\sigma} + \lc{\nabla}_{\sigma}\st{L}^{\mu}{}_{\nu\rho} - \lc{\nabla}_{\rho}\st{L}^{\mu}{}_{\nu\sigma}\,.
\end{equation}
From these relations one further finds that the Ricci scalar can be expressed as
\begin{equation}\label{eq:ricsmtele}
\lc{R} = K^{\mu}{}_{\rho\nu}K^{\rho\nu}{}_{\mu} - K^{\mu}{}_{\rho\mu}K^{\rho\nu}{}_{\nu} - 2\lc{\nabla}_{\mu}K^{\mu\nu}{}_{\nu} = -T + 2\lc{\nabla}_{\mu}T_{\nu}{}^{\nu\mu}\,,
\end{equation}
in the case of metric teleparallelism, and as
\begin{equation}\label{eq:ricsstele}
\lc{R} = \st{L}^{\mu\nu\rho}\st{L}_{\rho\mu\nu} - \st{L}^{\mu}{}_{\mu\rho}\st{L}^{\rho\nu}{}_{\nu} + \lc{\nabla}_{\nu}\st{L}_{\mu}{}^{\mu\nu} - \lc{\nabla}_{\mu}\st{L}^{\mu\nu}{}_{\nu} = -\st{Q} + \lc{\nabla}_{\nu}\st{Q}_{\mu}{}^{\mu\nu} - \lc{\nabla}_{\mu}\st{Q}^{\mu\nu}{}_{\nu}\,,
\end{equation}
in the symmetric teleparallel case. Recall that the Ricci tensor is the central element of the gravitational part of the Einstein-Hilbert action
\begin{equation}\label{eq:einhilaction}
    \mathcal{S}_{\text{EH}} = \frac{1}{2\kappa^2}\int\dd^4x\sqrt{-g}\lc{R}\,,
\end{equation}
of \gls{gr}. Together with the previous findings, one thus sees that by assuming a metric or symmetric teleparallel geometry, one may transform the Einstein-Hilbert action into alternative formulations. For this purpose, note that the expressions~\eqref{eq:ricsmtele} and~\eqref{eq:ricsstele} for the Ricci scalar both involve a total divergence part, which turns into a boundary term when these relations are used in the Einstein-Hilbert action~\eqref{eq:einhilaction}. Omitting these boundary terms as well as the matter contribution, one arrives at the actions
\begin{equation}\label{eq:tegractiong}
    \mathcal{S}_{\text{TEGR}} = -\frac{1}{2\kappa^2}\int\dd^4x\sqrt{-g}T\,,
\end{equation}
of the \gls{tegr}, and
\begin{equation}
\mathcal{S}_{\text{STEGR}} = -\frac{1}{2\kappa^2}\int\dd^4x\sqrt{-g}\st{Q}\,,
\end{equation}
of the symmetric teleparallel equivalent of general relativity~\cite{Nester:1998mp} (\gls{stegr}), respectively. As the nomenclature suggests, these theories are equivalent to \gls{gr} in the sense that they lead to equivalent dynamics for the metric, which is the only fundamental field variable in the latter theory. This mutual equivalence, which is summarized in Fig.~\ref{fig:trinity}, has been coined the \emph{geometric trinity of gravity}~\cite{BeltranJimenez:2019tjy}.

A few comments regarding the mentioned notion of equivalence are nevertheless necessary. First, it must be emphasized that in the transition from the Einstein-Hilbert action to the \gls{tegr} or \gls{stegr} actions the flat, teleparallel connection has been employed only in the gravitational part of the action, where it is used to substitute the Levi-Civita curvature scalar by the torsion or non-metricity scalars. The matter action, however, remains unchanged, and no direct matter coupling to the teleparallel connections is introduced, in order to preserve the weak equivalence principle. This means that all matter fields retain their universal coupling to the metric and possibly its Levi-Civita connection (in the case of spinor fields), with the only possible substitution which arises from considering the metric as a derived quantity obtained from the tetrad instead of a fundamental field. It then follows that the dynamics of the matter fields in \gls{tegr} and \gls{stegr} is unaltered compared to GR, and also the energy-momentum tensor as the source of gravity is retained, which establishes the equivalence between these theories also at the matter side of the gravitational field equations. Other couplings between matter and the symmetric teleparallel connection may be conceived~\cite{BeltranJimenez:2020sih}; however, these would exceed the scope of this Review. The coupling between matter and the metric teleparallel framework will be discussed further in Sec.~\ref{Grav_Coup_Prescrip}.

Another remark must be made regarding the omission of the boundary terms in the \gls{tegr} and \gls{stegr} actions. While this does not influence the field equations, which are the Euler-Lagrange equations derived from the action, it does have an influence when spacetime boundaries are considered. This is the case, for example, for the Casimir effect, black hole entropy or Hamiltonian formulation. Hence, attention must be paid in these cases, and possible non-equivalence may arise. Here we will not further discuss the \gls{stegr} action; \gls{tegr}, however, will be discussed in detail in Sec.~\ref{ssec:tegr}.

Further, as discussed in Sec.~\ref{sec:tetradspin}, in the metric teleparallel case, in which only torsion is nonvanishing and which encompasses the \gls{tegr} action~\eqref{eq:tegractiong}, it is more common to formulate the geometry in terms of a tetrad and spin connection, so that the metric determinant is replaced by the tetrad determinant through the relation
\begin{equation}\label{eq:tetraddet}
e = \sqrt{-g}
\end{equation}
in the action functional of \gls{tg} theories.

Finally, we remark that also a more general teleparallel, i.e., flat connection may be employed in the relation in Eq.~\eqref{eq:gennaffcurvdec}, in order to express the Riemann tensor \(\lc{R}^{\mu}{}_{\nu\rho\sigma}\) in terms of the distortion tensor \(\gc{D}^{\mu}{}_{\nu\rho}\), leading to the notion of the \emph{general teleparallel equivalent of general relativity} (GTEGR) \cite{Jimenez:2019ghw}. In this case, imposing the flatness condition \(\gc{R}^{\mu}{}_{\nu\rho\sigma} \equiv 0\) implies that the Ricci scalar can be written as
\begin{equation}\label{eq:ricsgtele}
\lc{R} = \gc{D}^{\mu\rho\nu}\gc{D}_{\rho\nu\mu} - \gc{D}^{\mu}{}_{\rho\mu}\gc{D}^{\rho\nu}{}_{\nu} + \lc{\nabla}_{\nu}\gc{D}^{\mu\nu}{}_{\mu} - \lc{\nabla}_{\mu}\gc{D}^{\mu\nu}{}_{\nu} = -\gc{G} + \lc{\nabla}_{\mu}(2\gc{T}_{\nu}{}^{\nu\mu} + \gc{Q}_{\nu}{}^{\nu\mu} - \gc{Q}^{\mu\nu}{}_{\nu})\,,
\end{equation}
where
\begin{align}
    \gc{G} &= \frac{1}{4}\gc{T}_{\mu\nu\rho}\gc{T}^{\mu\nu\rho} + \frac{1}{2}\gc{T}_{\mu\nu\rho}\gc{T}^{\mu\rho\nu}-\udt{\gc{T}}{\alpha}{\mu\alpha}\udt{\gc{T}}{\beta\mu}{\beta}+\gc{Q}_{\mu\nu\rho}\gc{T}^{\rho\mu\nu} - \dut{\gc{Q}}{\mu\alpha}{\alpha}\udt{\gc{T}}{\beta}{\mu\beta} + \udt{\gc{Q}}{\alpha}{\alpha\mu}\udt{\gc{T}}{\beta\mu}{\beta}  \nonumber\\
    &+ \frac{1}{4}\gc{Q}_{\mu\nu\rho}\gc{Q}^{\mu\nu\rho} - \frac{1}{2}\gc{Q}_{\mu\nu\rho}\gc{Q}^{\nu\mu\rho} - \frac{1}{4} \dut{\gc{Q}}{\mu\alpha}{\alpha}\udut{\gc{Q}}{\mu}{\beta}{\beta} + \frac{1}{2}\dut{\gc{Q}}{\mu\alpha}{\alpha}\udt{\gc{Q}}{\beta}{\beta}{\mu}\,.
\end{align}
Omitting the boundary term, as for the \gls{tegr} and \gls{stegr} actions, as well as the matter contribution, one thus arrives at the GTEGR action
\begin{equation}
\mathcal{S}_{\text{GTEGR}} := -\frac{1}{2\kappa^2}\int\dd^4x\sqrt{-g}\gc{G}\,.
\end{equation}
Comparing with Fig.~\ref{fig:trinity}, this theory would be joining the lower two corners of the triangle. We end this section with a comment that also more general theories may be conceived, in which other tensorial quantities mediate the gravitational interaction, and which still reproduce the Einstein equations as the metric field equations obtained from the Einstein-Hilbert action in \gls{gr}, and that also for theories beyond \gls{gr} various equivalent and related teleparallel theories in their different flavors can be constructed~\cite{Bohmer:2021eoo}.

\clearpage

\section{Torsional Teleparallel Geometries}\label{sec3:torsional}

In this section, we discuss the foundations of \gls{tg} and how it connects with a fundamental theory of gauge translations. We also describe how \gls{tg} has been used in other branches of physics and its impact on the formation of gravitational scalar invariants.

\subsection{Teleparallel gravity as a gauge theory of translations} \label{Sec:TEGR_gauge_theory}

Gauge theory is a powerful framework in which to express physical theories as is well known. In \gls{gr}, the first attempt in Ref.~\cite{PhysRev.101.1597} which led to the forced introduction of tetrads and the Lorentz group being linked to an antisymmetric gauge current. Despite the coupling of gravity to the symmetric energy-momentum tensor in \gls{gr}, it is also possible to formulate \gls{gr} as a gauge theory of translations \cite{Koivisto:2019ejt,blagojevic2002gravitation}. However, \gls{tg} seems to find more benefits for this treatment as will be explored in this section. \gls{tg} introduces the tetrad fields in a much more natural way and can be interpreted as a native gauge theory of translations \cite{Aldrovandi:2013wha,Hayashi:1967se}. Through Noether's theorem \cite{ortin2004gravity} this results in a covariantly conserved energy-momentum tensor, which would not necessarily be the case for a gauge theory of rotation.

Through their existence, tetrads connect a general manifold to its local Minkowski space coordinates where a gauge translation transformation will appear as
\begin{equation}
	\tilde{x}^A \rightarrow x^A + \epsilon^A\,,
\end{equation}
in which $\epsilon^A=\epsilon^A (x^{\mu})$ characterizes the infinitesimal local transformation. The differential operators $P_A = \partial_{A}=\partial/\partial x^{A}$ will then be the generators of these infinitesimal transformations, which straightforwardly satisfy the commutation relations. In this way, the infinitesimal transformation can be written as
\begin{equation}
	\delta_{\epsilon} x^A=\epsilon^B \partial_B x^A\,.
\end{equation}
For a generic source field $\Psi(\tilde{x}^A(x^{\mu}))$, infinitesimal transformation on the local Minkowski space will be
\begin{equation}
	\delta_{\epsilon} \Psi(\tilde{x}^A(x^{\mu}))=\epsilon^{A}\partial_A \Psi(\tilde{x}^B(x^{\mu}))\,,
\end{equation}
where the Minkowski point $x^B$ will appear for every Minkowski space at point $x^{\mu}$. However, the derivative of this field will transform as
\begin{equation}\label{delta_eps_gauge}
	\delta_{\epsilon}\left( \partial_{\mu} \Psi\right) = \epsilon^A \partial_A \left(\partial_{\mu} \Psi \right) + \left( \partial_{\mu} \epsilon^A \right) \partial_A \Psi\,,
\end{equation}
which is only covariant when the parameter $\epsilon^A=\epsilon^A(x^{\mu})$ is a constant. The second term breaks the gauge invariance and is removed by the introduction of a gauge potential
\begin{equation}
B_{\mu}=\udt{B}{A}{\mu}\partial_{A}\,,
\end{equation}
which is a one-form taking values from the Lie algebra of the translation group. To remedy the gauge invariance violating term in Eq.~\eqref{delta_eps_gauge}, the gauge potential can be used to define the gauge derivative
\begin{equation}
	\mathbf{e}_{\mu} \Psi := \partial_{\mu} \Psi + \udt{B}{A}{\mu}\partial_{A} \Psi\,, \label{eq:inertial_cov_der}
\end{equation}
where inertial effects are initially ignored for clarity, they are incorporated into this description in the explanation that follows. In this description inertial is referring to Lorentz frames and particularly to the invariance of the theory under the action of boosts and rotations in a Lorentz framework setting which is a necessity for any eventual physical theory of gravity. This derivative directly leads to the correct gauge invariance relation under an infinitesimal gauge translation
\begin{equation}\label{inf_gaug_transl}
	\delta_{\epsilon} \left(\mathbf{e}_{\mu} \Psi\right) = \epsilon^A \partial_A \left(\partial_{\mu} \Psi \right)\,,
\end{equation}
where the transformation appears as $\delta_{\epsilon} \udt{B}{A}{\nu} = -\partial_{\nu} \epsilon^A(x^{\mu})$. Thus, the translational coupling prescription replaces the partial derivative with $\mathbf{e}_{\mu}$
through
\begin{equation}
	\partial_{\mu} \Psi \rightarrow \mathbf{e}_{\mu} \Psi\,,
\end{equation}
which observes the translational gauge invariance. For an arbitrary point $x^{\mu}$ on the manifold, the covariant derivative $\mathbf{e}_{\mu}$ can be written in terms of the local Minkowski space as
\begin{equation}\label{cov_der_for_zero_iner_system}
	\udt{e}{A}{\mu}\partial_A \Psi = \mathbf{e}_{\mu} \Psi = \left(\partial_A \Psi\right)\left(\partial_{\mu} x^A \right) + \udt{B}{A}{\mu} \partial_A \Psi\,,
\end{equation}
where $\udt{e}{A}{\mu} = \partial_{\mu}x^A + \udt{B}{A}{\mu}$ is a non-trivial tetrad field. The close relationship between the manifold and Minkowski space indices through the tetrad field $\udt{e}{A}{\mu}$ is called \textit{soldering}.

A trivial tetrad would mean $\mathbf{H}_{\mu}=\udt{h}{A}{\mu}\partial_A$ ($\udt{B}{A}{\mu}=\partial_{\mu} \epsilon^A$), and so the translational coupling prescription becomes
\begin{equation}\label{sec4:coupling}
	\udt{h}{A}{\mu} \partial_A \Psi \rightarrow \udt{e}{A}{\mu} \partial_A \Psi\,,
\end{equation}
where $\udt{B}{A}{\mu} \neq \partial_{\mu} \epsilon^A$ for the non-trivial tetrad field, and from which it follows that
\begin{equation}\label{tet_coupling_prescrip}
	\udt{h}{A}{\mu} \rightarrow \udt{e}{A}{\mu}\,,
\end{equation}
which is reminiscent of the Minkowski metric being raised to an arbitrary metric tensor for the general manifold
\begin{equation}
	\eta_{\mu\nu} \rightarrow g_{\mu\nu}\,.
\end{equation}
This is now a consequence of the prescription in Eq.~\eqref{tet_coupling_prescrip} due to the definitions $\eta_{\mu\nu} = \eta_{AB} \udt{h}{A}{\mu}\udt{h}{B}{\nu}$ and $g_{\mu\nu} = \eta_{AB}\udt{e}{A}{\mu}\udt{e}{B}{\nu}$.

Thus far, inertial effects have been ignored to make the argument clearer for the interpretation of \gls{tg} as a gauge theory of translations. We can now include inertial effects by considering arbitrary Lorentz transformations $\udt{\Lambda}{A}{B}$ from the Lorentz group $O(1,3)$. These transformations only apply to the local Minkowski space and their index representations, which means that local coordinates will transform as
\begin{equation}
	x^A \rightarrow \udt{\Lambda}{A}{B}(x) x^B = x'^A\,,
\end{equation}
so that scalar and tensor fields will similarly transform appropriately. Analogously, the gauge potential will also undergo local Lorentz transformations on its inertial index, meaning that
\begin{equation}
	\udt{B}{A}{\mu} \rightarrow \udt{\Lambda}{A}{B}(x) \udt{B}{B}{\mu}\,,
\end{equation}
will directly lead to the generalization of Eq.~\eqref{cov_der_for_zero_iner_system} to
\begin{equation}\label{trans_cov_der}
	\mathbf{e}_{\mu}\Psi = \udt{e}{A}{\mu} \partial_{A}\Psi = \partial_{\mu}\Psi + \udt{\omega}{A}{B\mu} x^B\partial_A \Psi + \udt{B}{A}{\mu}\partial_{A}\Psi\,,
\end{equation}
where $\udt{\omega}{A}{B\mu}$ is the teleparallel spin connection, which is a purely inertial Lorentz connection. By observing the local Lorentz invariance (such as through the transformation $\udt{e}{A}{\mu}=\udt{\Lambda}{A}{B}\udt{e'}{B}{\mu}$), it follows that the spin connection takes on components as we discussed in Sec.~\ref{sec:LLTW}
\begin{equation}\label{eq:TG_spin_connection}
	\udt{\omega}{A}{C\mu} = \udt{\Lambda}{A}{B}(x)\partial_{\mu} \dut{\Lambda}{C}{B}(x)\,,
\end{equation}
\gls{wrt} the Lorentz transformations, which indeed is totally inertial. The fuller tetrad can be reinterpreted as a local Lorentz invariant trivial and spin connection pair tetrad
\begin{equation}\label{chp3_iner_cov_der}
	\udt{h}{A}{\mu} = \partial_{\mu} x^A + \udt{\omega}{A}{B\mu} x^B = \mathcal{D}_{\mu} x^A\,,
\end{equation}
which raises the partial derivative to its local Lorentz invariant, while the non-trivial tetrad also includes the gauge potential
\begin{equation}
	\udt{e}{A}{\mu} = \udt{h}{A}{\mu} + \udt{B}{A}{\mu}\,,
\end{equation}
where $\mathcal{D}_{\mu}$ is the inertial covariant derivative, i.e. the covariant derivative \gls{wrt} the Lorentz group which is associated with the spin connection. Finally, to retain tetrad invariance under infinitesimal gauge translations ($\delta_{\epsilon} \left(\udt{e}{A}{\mu}\right) = 0$) implies that
\begin{equation}
\delta_{\epsilon} \udt{B}{A}{\mu} = -\mathcal{D}_{\mu} \epsilon^A\,,
\end{equation}
which can be viewed as a generalization of the partial derivative in Eq.~\eqref{inf_gaug_transl}.

\subsection{Gravitational coupling prescription} \label{Grav_Coup_Prescrip}

In \gls{gr}, through Einstein's choice of the Levi-Civita connection as the expression of geometric deformation (through Riemann geometry), the coupling prescription associated with gravity becomes
\begin{equation} \label{chp3_GR_coup_pres1}
    \partial_{\mu} \rightarrow \lc{\nabla}_{\mu}\,,
\end{equation}
by choice rather than a result of a gauge theory. This then results in the well known geodesic equation
\begin{equation}
    u^{\mu}\lc{\nabla}_{\mu}u^{\nu}=\frac{d^2x^\nu}{d\lambda^2}+\lc{\Gamma}^{\nu}{}_{\alpha\beta}\frac{d x^\alpha}{d\lambda}\frac{d x^\beta}{d\lambda}=0\,,
\end{equation}
from which it can be said that in \gls{gr} gravitation is not expressed in a force-like manner, i.e. its acceleration vanishes ($\lc{a}^{\nu}=0$). The coupling prescription described in Eq.~\eqref{chp3_GR_coup_pres1} is called the \textit{minimal coupling prescription} (namely, there the covariant derivative is coupled with the Levi-Civita connection) since it lays out a scenario where the spin connection of \gls{gr} is used, in all other cases such as that in teleparallel geometries, the coupling prescription is not necessarily minimal \cite{Aldrovandi:2013wha}. This point is important since the coupling prescription of teleparallel geometries can make use of the Levi-Civita connection components, while its spin connection will always be different to that of \gls{gr}.

The situation in \gls{tg} is entirely different in that gravitation emerges through a gauge theory of translations. In this setting, the inertial covariant derivative $\mathbf{h}_{\mu}$ is raised to $\mathbf{e}_{\mu}$ in the presence of gravity, leading directly to Eq.~\eqref{sec4:coupling} where the inertial covariant derivative is raised to its non-trivial tetrad field form in the presence of gravitation. The end result is the translational gauge coupling prescription
\begin{equation}\label{tetrad_coupling_prescrip}
	\udt{h}{A}{\mu} \rightarrow \udt{e}{A}{\mu}\,.
\end{equation}
The relationship with the inertial and general manifold metrics now appear as results of this prescription where \begin{equation}\label{tetrad_coup_prescript}
    \eta_{AB}\udt{h}{A}{\mu}\udt{h}{B}{\nu} \rightarrow \eta_{AB}\udt{e}{A}{\mu}\udt{e}{B}{\nu}\,,
\end{equation}
which is equivalent to saying that $\eta_{\mu\nu}$ is raised to the general manifold $g_{\mu\nu}$. In this way, we have a direct link between the metric tensor and the tetrad fields that can be used to relate the components of the two objects. Another important feature of these relations is that the mechanics of metric tensors (such as ansatz choices) can be used to solve these relations for the tetrad field components through Eq.~\eqref{tetrad_coup_prescript}.

\subsection{The field strength of gravity} \label{sec:field_strength}

In the context of the gauge structure of \gls{tg}, a gauge invariant field strength can be defined as a measure of the amount of geometric deformation which is then expressed as a gravitational or Maxwell-like force in the dynamical equations. Taking the example of classical electromagnetism, the vector potential $A^{\mu}$ induces a gauge field strength
\begin{equation}
    F_{\mu\nu} := \lc{\nabla}_{\mu} A_{\nu} - \lc{\nabla}_{\nu} A_{\mu}\,,
\end{equation}
on the general manifold, which also obeys field conservation laws \cite{jackson2007classical}. Now, taking the translational covariant derivative in Eq.~\eqref{trans_cov_der}, the gauge field strength can be similarly defined from the gauge derivatives $\mathbf{e}_{\mu}$ as
\begin{equation}
    \left[\mathbf{e}_{\mu},\mathbf{e}_{\nu}\right] = \udt{T}{A}{\mu\nu}\partial_{A}\,,
\end{equation}
where $\left[\mathbf{e}_{\mu},\mathbf{e}_{\nu}\right] := \frac{1}{2}\left(\mathbf{e}_{\mu}\mathbf{e}_{\nu} - \mathbf{e}_{\nu}\mathbf{e}_{\mu}\right)$, and the translational field strength is defined as
\begin{equation}\label{gauge_def_field_strength}
    \udt{T}{A}{\mu\nu} := \partial_{\mu}\udt{B}{A}{\nu} - \partial_{\nu}\udt{B}{A}{\mu} + \udt{\omega}{A}{B\mu} \udt{B}{B}{\nu} - \udt{\omega}{A}{B\nu} \udt{B}{B}{\mu}\,,
\end{equation}
and $\partial_{A}$ represents the generators of infinitesimal translations. While direct, Eq.~\eqref{gauge_def_field_strength} can also be recast in terms of the tetrad field $\udt{e}{A}{\mu}$ by identifying the vanishing antisymmetric identity
\begin{equation}
    \left[\mathcal{D}_{\mu},\mathcal{D}_{\nu}\right] x^{A} \equiv \mathcal{D}_{\mu}\left(\mathcal{D}_{\nu} x^{A}\right) - \mathcal{D}_{\nu} \left(\mathcal{D}_{\mu} x^{A}\right) = 0\,,
\end{equation}
so that the translational field strength turns out to be expressed as
\begin{equation}\label{eq:torsion_tensor}
    \udt{T}{A}{\mu\nu} = 2\left(\partial_{[\mu}\udt{e}{A}{\nu]} + \udt{\omega}{A}{B[\mu}\udt{e}{B}{\nu]}\right) = \partial_{\mu}\udt{e}{A}{\nu} - \partial_{\nu}\udt{e}{A}{\mu} + \udt{\omega}{A}{B\mu}\udt{e}{B}{\nu} - \udt{\omega}{A}{B\nu}\udt{e}{B}{\mu}\,,
\end{equation}
which will also inherit the property of invariance under gauge transformations $\udt{T'}{A}{\mu\nu} = \udt{T}{A}{\mu\nu}$, where square brackets are used to denote the antisymmetric operator. Thus, the teleparallel field strength turns out to be exhibited as an antisymmetric expression which can be written in terms of only general manifold indices as
\begin{equation}
    \udt{T}{\rho}{\mu\nu} = \dut{E}{A}{\rho}\udt{T}{A}{\mu\nu} = \udt{\Gamma}{\rho}{\nu\mu} - \udt{\Gamma}{\rho}{\mu\nu} = 2\udt{\Gamma}{\rho}{\left[\nu\mu\right]}\,,
\end{equation}
where $\udt{\Gamma}{\rho}{\mu\nu}$ is the teleparallel connection defined as
\begin{equation}\label{eq:tg_connection}
    \udt{\Gamma}{\rho}{\nu\mu} := \dut{E}{A}{\rho}\left( \partial_{\mu}\udt{e}{A}{\nu} + \udt{\omega}{A}{B\mu}\udt{e}{B}{\nu}\right) \equiv \dut{E}{A}{\rho}\mathcal{D}_{\mu}\udt{e}{A}{\nu}\,.
\end{equation}

In the class of frames in which the spin connection components vanish, the connection reduces to $\udt{\Gamma}{\rho}{\nu\mu} := \dut{E}{A}{\rho} \partial_{\mu}\udt{e}{A}{\nu}$ which will later be introduced as the Weitzenb\"{o}ck gauge connection, so that the torsion tensor becomes
\begin{equation}
    \udt{T}{A}{\mu\nu} = \partial_{\mu}\udt{e}{A}{\nu} - \partial_{\nu}\udt{e}{A}{\mu}\,,
\end{equation}
which acts as the field strength in \gls{tg}. Another important feature of this geometric formulation is that the Riemann measure of curvature vanishes for any frame in this regime, namely
\begin{equation}\label{chp3_rie_ten}
    \udt{R}{A}{B\nu\mu} \equiv \partial_{\nu} \udt{\omega}{A}{B\mu} - \partial_{\mu} \udt{\omega}{A}{B\nu} + \udt{\omega}{A}{C\nu} \udt{\omega}{C}{B\mu} - \udt{\omega}{A}{C\mu} \udt{\omega}{C}{B\nu} = 0\,,
\end{equation}
which can be seen as the substitution of the teleparallel spin connection \eqref{eq:TG_spin_connection} into the metric-affine Riemann tensor in Eq.~\eqref{eq:spiconcurv}.

\subsection{Inertial and gravitational effects} \label{sec:inertial_non_inertial_effs}

The concept of inertia played a crucial role in Einstein's formulation of \gls{gr} through the application of Mach's principle \cite{mach1960science,barbour1995mach}. To this end, Einstein's attempt was to remove the privileged role that inertial frames of reference play in classical mechanics by removing the dependence of measurements on the choice of coordinates. On the other hand, \gls{gr} does not fully resolve the foundational issues at the core of the definition of inertial such as the need, at times, to include absolute elements. For instance, consider rotating matter \gls{wrt} its local inertial system. At its core, this points to a spacetime structure that is not entirely governed by matter fields \cite{einstein1996meaning,iorio2007measurement}.

The dynamical variable in \gls{tg} for gravitation is the tetrad field which builds up to the metric tensor. Given that tetrad frames relate the general manifold with their Minkowski spaces, \gls{tg} offers a different perspective to this issue at the core of GR. Consider again the concept of trivial tetrad ($\udt{h}{A}{\mu}$) and spin connection pairs which relate inertial frames with their Minkowski spaces. Tetrads can also be represented by their tetrad fields which represent a basis of vectors for the tangent space of an arbitrary point $x^{\mu}$ on a manifold, i.e.
\begin{equation}
    \mathbf{H}_A = \dut{H}{A}{\mu} \partial_{\mu} \quad \text{and} \quad \mathbf{h}^A=\udt{h}{A}{\mu} \dd x^{\mu}\,,
\end{equation}
which conversely result in
\begin{equation}
    \partial_{\mu} = \udt{h}{A}{\mu}\mathbf{H}_{A} \quad \text{and} \quad \dd x^{\mu} = \dut{H}{A}{\mu}\mathbf{h}^{A}\,,
\end{equation}
when requiring that
\begin{equation}
    \textbf{h}^A\left(\textbf{h}_B\right) = \delta^A_B\,.
\end{equation}
This orthogonality condition directly leads to tetrad frame conditions
\begin{equation}
    \udt{h}{A}{\mu}\dut{H}{A}{\nu} = \delta_{\mu}^{\nu} \quad \text{and} \quad \udt{h}{A}{\mu}\dut{H}{B}{\mu} = \delta^{A}_{B}\,.\label{tetrad_ortho_cond}
\end{equation}
These frames and their associated conditions hold for any differentiable manifold. Another way of characterizing the trivial tetrad and spin connection pairs is through the so-called structure coefficients which are defined through the commutation relation \cite{sternberg1999lectures,aldrovandi1995introduction}
\begin{equation}
    \left[\mathbf{H}_A, \mathbf{H}_B\right] = \udt{f}{C}{AB}\textbf{h}_C\,,
\end{equation}
which can also be interpreted as the coefficients of the anholonomy of the tetrad fields. The dual expression, where the inverse tetrads are used, is found using the Cartan structure equation
\begin{equation}
    \dd \textbf{h}^C = -\frac{1}{2}\udt{f}{C}{AB} \mathbf{h}^{A} \wedge \mathbf{h}^{B} = \frac{1}{2} \left(\partial_{\mu} \udt{h}{C}{\nu} - \partial_{\nu} \udt{h}{C}{\mu}\right)\dd x^{\mu} \wedge \dd x^{\nu}\,,
\end{equation}
where the exterior (wedge) product is used. The structure coefficients can be related to the tetrad frames straightforwardly through
\begin{subequations}
\begin{align}
    \udt{f}{C}{AB} &\equiv \left[\mathbf{H}_A,\mathbf{H}_B\right]\mathbf{h}^C\\[0.5ex]
    &= \dut{H}{A}{\mu}\dut{H}{B}{\nu} \left(\partial_{\nu} \udt{h}{C}{\mu} - \partial_{\mu} \udt{h}{C}{\nu}\right)\,,\label{chp3_struc_coeff}
\end{align}
\end{subequations}
where the last equations is a result of the orthogonality conditions Eq.~\eqref{tetrad_ortho_cond}.

A special class of frames $\mathbf{h}'_A$ exist in which
\begin{equation}
    \udt{f'}{C}{AB} = 0\,, \label{eq:holonomic_condition}
\end{equation}
which directly sets $\dd \mathbf{h}'_A = 0$ meaning that $\mathbf{h}'_A$ can be expressed in closed differential form, and that for some $x'^A$, $\mathbf{h}'^A = \dd x'^A$ locally. In this case, the basis ${\mathbf{h}'^A}$ turns out to be integrable, or holonomic. However, $\mathbf{h}^A$ refers to holonomic frames meaning that all such frames observe this vanishing property. On the other hand, non-holonomic frames do not generally observe this vanishing condition, and indeed $\udt{f}{C}{AB} \neq 0$ in such cases.

In special relativity, the inertial effects within a frame can be represented by a so-called spin connection which is a Lorentz connection coming from the Lorentz group of transformations. As is well known, these effects do not occur in every inertial frame such as in Minkowski frames. Spin connection components can be generically induced by first considering a general coordinate system ${x^{\mu}}$ with the holonomic form
\begin{equation}
    \udt{h'}{A}{\mu} = \partial_{\mu} x'^A\,,
\end{equation}
where $x'^A = x'^A(x^{\mu})$ is position dependent. Taking an arbitrary Lorentz transformation $\udt{\Lambda}{A}{B}$ of vectors in this basis results in
\begin{equation}\label{chp3_iner_trans2}
    x^A=\udt{\Lambda}{A}{B}(x)x'^{B}\,,
\end{equation}
so that the frame gets transformed to
\begin{equation}\label{chp3_iner_trans}
    \udt{h}{A}{\mu} = \udt{\Lambda}{A}{B}(x) \udt{h'}{B}{\mu}\,.
\end{equation}
By writing this transformation in terms of the transformed $x^A$, the tetrad frame can be written in terms of the Lorentz matrices as
\begin{equation}
    \udt{h}{A}{\mu} = \udt{\Lambda}{A}{B}(x) \udt{h'}{B}{\mu} = \udt{\Lambda}{A}{B}(x) \partial_{\mu} \left( \dut{\Lambda}{C}{B}(x) x^C\right) = \partial_{\mu} x^A + \udt{\omega}{A}{C\mu} x^C\,,
\end{equation}
where the inverse transformation of Eq.~\eqref{chp3_iner_trans} and identity $\udt{\Lambda}{A}{B}\dut{\Lambda}{C}{B}=\delta^A_C$ was used, and where the spin connection is explicitly defined as
\begin{equation}\label{eq:TG_spin_conn}
    \udt{\omega}{A}{C\mu} := \udt{\Lambda}{A}{B}(x) \partial_{\mu} \dut{\Lambda}{C}{B}(x)\,,
\end{equation}
which has already been used in the general geometry framework around Eq.~\ref{eq:spclormatrep}. We can recognise this as the inertial covariant derivative from Eq.~\eqref{chp3_iner_cov_der} which lets us write $\udt{h}{A}{\mu} \equiv \mathcal{D} x^A$.

In this way, by initially considering an inertial frame, in which $\udt{\omega'}{A}{B\mu} = 0$, Lorentz transformations can be performed in a local (position-dependent) way through $\udt{\Lambda}{A}{B}$ which produces nonvanishing spin connection components. For each class of frames, a global (position-independent) Lorentz transformation ($\udt{\Lambda}{A}{B} = constant$) can relate the frames for every global point.

Noticing the property between Lorentz transformed trivial frames
\begin{equation}
    \udt{\Lambda}{A}{B} = \udt{h}{A}{\mu} \dut{H'{}}{B}{\mu}\,,
\end{equation}
which is derived through the transformation in Eq.~\eqref{chp3_iner_trans}. The coefficients of the anholonomy can then be written as Lorentz matrices by appropriately transforming some of the tetrad frames in Eq.~\eqref{chp3_struc_coeff} to give
\begin{equation}\label{chp3_anholo_express}
    \udt{f}{C}{AB} = \udt{\omega}{C}{BA} - \udt{\omega}{C}{AB}\,,
\end{equation}
where the general manifold index was raised with the inverse tetrad though $\udt{\omega}{C}{BA}=\dut{h}{A}{\mu}\udt{\omega}{C}{B\mu}$.

For inertial frames $\udt{h}{A}{\mu}$ which are related by the Lorentz connection, i.e.
\begin{equation}\label{chp3_tor_ten}
    \udt{T}{A}{\nu\mu} \equiv \partial_{\nu} \udt{h}{A}{\mu} - \partial_{\mu} \udt{h}{A}{\nu} + \udt{\omega}{A}{C\nu} \udt{h}{C}{\mu} - \udt{\omega}{A}{C\mu}\udt{h}{C}{\nu} = 0\,,
\end{equation}
which one would expect for such frames. This condition guarantees that gravity vanishes, however, frames do exist that exhibit nonvanishing torsion tensor components but which do not feature non-inertial frames.

Considering again the spin connection, one can show that it features an antisymmetry in its first two indices
\begin{equation}
    \omega_{AB\mu} = \eta_{AD}\udt{\omega}{D}{B\mu} = \eta_{AD} \udt{\Lambda}{D}{C}\partial_{\mu} \dut{\Lambda}{B}{C} = -\eta_{BE}\udt{\omega}{E}{A\mu} = -\omega_{BA\mu}\,,
\end{equation}
which leads to a number of properties. By taking three different combinations of the vanishing torsion tensor together with the antisymmetry property for the spin connection, it follows that
\begin{equation}
    0 = \dudt{f}{B}{C}{A} + \dudt{f}{A}{C}{B} - \udt{f}{C}{BA} = \dut{\omega}{BA}{C} - \dudt{\omega}{B}{C}{A} + \dut{\omega}{AB}{C} - \dudt{\omega}{A}{C}{B} - \left(\udt{\omega}{C}{AB} - \udt{\omega}{C}{BA}\right)\,,
\end{equation}
which directly gives the inverse relation to Eq.~\eqref{chp3_anholo_express} as
\begin{equation}\label{chp3_inver_rel_struc}
    \udt{\omega}{A}{BC} = \frac{1}{2}\left(\dudt{f}{B}{A}{C} + \dudt{f}{C}{A}{B} - \udt{f}{A}{BC}\right)\,.
\end{equation}
Through the spin connection, \gls{tg} observes local Lorentz invariance in all relations which is a fundamental symmetry of Nature. Local Lorentz invariance should not produce dynamics in a gravitational theory, but in \gls{tg} it does produce a second coupling prescription which will be related by the strong equivalence principle. This emerges as a correction term to the derivative terms for inertial frames which are raised to Lorentz covariant derivatives. To see this transition, consider a vector field $\phi'^C$. Using the tetrad field
\begin{equation}
    {\mathbf{h}'}_A = \delta^{\mu}_{A}\partial_{\mu}\,,
\end{equation}
which acts on the vector field as
\begin{equation}
    {\mathbf{h}'}_A {\phi'}^C = \delta^{\mu}_A \partial_{\mu} {\phi'}^C\,.
\end{equation}
Transforming to an unprimed frame through $\phi^C = \dut{\Lambda}{D}{C} (x) {\phi'}^D$ gives
\begin{subequations}
\begin{align}
    {\mathbf{h}'}_A {\phi'}^C &= \delta^{\mu}_{A} \partial_{\mu} \left(\dut{\Lambda}{D}{C} \phi^D\right)\\[0.5ex]
    &= \udt{\Lambda}{B}{A} \dut{\Lambda}{D}{C} \partial_B \phi^D + \left(\udt{\Lambda}{B}{A} \partial_{B} \dut{\Lambda}{D}{C}\right) \phi^D\\[0.5ex]
    &= \udt{\Lambda}{B}{A}\dut{\Lambda}{D}{C}\left(\partial_B \phi^D + \udt{\omega}{D}{EB}\phi^E\right)\\[0.5ex]
    &= \udt{\Lambda}{B}{A}\dut{\Lambda}{D}{C}\left(\partial_B \phi^D + \frac{1}{2}\left(\dudt{f}{E}{D}{B} + \dudt{f}{B}{D}{E} - \udt{f}{D}{EB}\right)\phi^E\right)\,.
\end{align}
\end{subequations}
In terms of the vector representation of the Lorentz generators
\begin{equation}\label{chp3_Lorentz_gen}
    \udt{(\udt{S}{B}{C})}{A}{D} = \delta^B_D \delta^A_C - \eta_{CD}\eta^{AB}\,,
\end{equation}
the coupling prescription takes the form
\begin{equation}
    \partial_B \phi^D + \frac{1}{4}\left(\dudt{f}{E}{D}{B} + \dudt{f}{B}{D}{E} - \udt{f}{D}{EB}\right)\udt{S}{E}{F}\phi^F\,,
\end{equation}
where an extra half appears due to the Lorentz generators. For a scalar field $\Psi$
\begin{subequations}
\begin{align}
    \partial_{\mu} \Psi &\rightarrow \mathcal{D}_{\mu} \Psi \\[0.5ex]
    &= \partial_{\mu} \Psi + \frac{1}{2}\udt{\omega}{C}{B\mu} \dut{S}{C}{B} \Psi \\[0.5ex]
    &= \partial_{\mu}\Psi + \frac{1}{2}\udt{h}{A}{\mu} \udt{\omega}{C}{BA} \dut{S}{C}{B} \Psi = \partial_{\mu} \Psi + \frac{1}{4}\udt{h}{A}{\mu} \left(\dudt{f}{B}{C}{A} + \dudt{f}{A}{C}{B} - \udt{f}{C}{BA}\right) \dut{S}{C}{B} \Psi\,,
\end{align}
\end{subequations}
where $\dut{S}{B}{C}$ now represents the Lorentz generators for the $\Psi$ field. The second term in the last relation is a compensating term that imposes the local Lorentz covariance of the derivative in the new inertial frame. This coupling prescription is hidden in the internal structure of \gls{gr} and does not play an active role in many of its considerations. Thus in \gls{tg}, the strong equivalence principle takes on a more active role in the construction of the theory. This concept is encapsulated in the idea behind the general covariance principle where Lorentz covariant formula in special relativity can be raised to its non-inertial form by the coupling prescription.

The Lorentz covariant derivative can then be raised to its non-inertial counterpart by the coupling principle in Eq.~\eqref{tetrad_coupling_prescrip} which gives
\begin{equation}\label{chp3_grav_cou_pres}
    \partial_{\mu} \Psi \rightarrow \mathcal{D}_{\mu}\Psi = \partial_{\mu} \Psi + \frac{1}{4}\udt{e}{A}{\mu} \left(\dudt{f}{B}{C}{A} + \dudt{f}{A}{C}{B} - \udt{f}{C}{BA}\right) \dut{S}{C}{B} \Psi\,,
\end{equation}
where the coefficients of the anholonomy are now raised to
\begin{equation}
    \udt{f}{C}{AB} = \dut{E}{A}{\mu}\dut{E}{B}{\nu} \left(\partial_{\nu} \udt{e}{C}{\mu} - \partial_{\mu} \udt{e}{C}{\nu}\right)\,,
\end{equation}
where we recall the definition in Eq.~\eqref{chp3_struc_coeff}.

A natural consequence of the coupling prescription is that the torsion no longer vanishes organically in Eq.~\eqref{chp3_tor_ten} meaning that the relation between the structure coefficients of the anholonomy and spin connection components in Eq.~\eqref{chp3_anholo_express} now takes the form
\begin{equation}
    \udt{f}{C}{AB} = \dut{E}{A}{\mu}\dut{E}{B}{\nu}\left(\udt{T}{C}{\nu\mu} - \udt{\omega}{C}{D\nu}\udt{e}{D}{\mu} + \udt{\omega}{C}{D\mu}\udt{e}{D}{\nu}\right)\,,
\end{equation}
or, in a more simplified form
\begin{equation}
    \udt{f}{C}{AB} + \udt{T}{C}{AB} = \udt{\omega}{C}{BA} - \udt{\omega}{C}{AB}\,,
\end{equation}
where $\udt{T}{C}{AB}$ is the torsion tensor in inertial indices. This gives a direct relationship between the structure coefficients, the torsion tensor and the spin connection. Similarly, the Riemann tensor in Eq.~\eqref{chp3_rie_ten} is found to be
\begin{equation}
    \udt{R}{A}{BCD} = \dut{h}{C}{\nu} \partial_{\nu} \udt{\omega}{A}{BD} - \dut{h}{D}{\mu} \partial_{\mu} \udt{\omega}{A}{BC} + \udt{\omega}{A}{EC} \udt{\omega}{E}{BD} - \udt{\omega}{A}{ED} \udt{\omega}{E}{BC} - \udt{\omega}{A}{BE} \udt{f}{E}{CD}\,.
\end{equation}
The same combination of indices as in inverse relation in Eq.~\eqref{chp3_inver_rel_struc} for the nonvanishing torsion tensor results in
\begin{equation}\label{chp3_Rel_struc_cont}
    \frac{1}{2}\left(\dudt{f}{B}{C}{A} + \dudt{f}{A}{C}{B} - \udt{f}{C}{AB}\right) = \udt{\omega}{C}{BA} + \udt{K}{C}{BA}\,,
\end{equation}
where the so-called contortion tensor is defined as
\begin{equation} \label{chp3_contortion_def}
    \udt{K}{C}{BA} := \frac{1}{2}\left(\dudt{T}{B}{C}{A} + \dudt{T}{A}{C}{B} - \udt{T}{C}{AB}\right)\,,
\end{equation}
which aligns with a \gls{tg} version of the general contortion tensor in Eq.~\eqref{eq:contor}. Thus, the coupling prescription in Eq.~\eqref{chp3_grav_cou_pres} can be written fully as
\begin{equation}\label{chp3_full_grav_coup_pres}
    \partial_{\mu} \Psi \rightarrow \partial_{\mu} \Psi + \frac{1}{2}\left(\udt{\omega}{C}{B\mu} + \udt{K}{C}{B\mu}\right) \dut{S}{C}{B} \Psi\,,
\end{equation}
where full gravitational coupling prescription is laid out and where the Lorentz part of this coupling prescription is separated from the gravitational sector.

In GR, the spin connection takes the form of the full expression of the coefficients of the anholonomy in Eq.~\eqref{chp3_Rel_struc_cont} as \cite{aldrovandi1995introduction}
\begin{equation}
    \udt{\lc{\omega}}{C}{AB} = \frac{1}{2} \left(\dudt{f}{B}{C}{A} + \dudt{f}{A}{C}{B} - \udt{f}{C}{AB}\right)\,,
\end{equation}
which is the Levi-Civita spin connection, and results in the fundamental identity
\begin{equation}\label{chp3_spin_iden}
    \udt{\lc{\omega}}{C}{AB} = \udt{\omega}{C}{AB} - \udt{K}{C}{AB}\,.
\end{equation}
The gravitational coupling prescription in Eq.~\eqref{chp3_full_grav_coup_pres} can equivalently be written as
\begin{equation}\label{chp3_GR_coup_pres2}
    \partial_{\mu} \Psi \rightarrow \partial_{\mu} \Psi + \frac{1}{2}\udt{\lc{\omega}}{C}{B \mu} \dut{S}{C}{B} \Psi\,,
\end{equation}
which is the coupling prescription for \gls{gr}. Thus, both curvature- and torsion-based approaches to gravity utilize the same coupling prescription, and are thus both consistent with the principles of covariance and strong equivalence.

The potential for separating inertial effects is revealed here as a fundamental difference between torsional- and curvature-based prescriptions for gravity. In the Levi-Civita spin connection $\udt{\lc{\omega}}{C}{B\mu}$ both gravitational and inertial are combined while in torsional gravity, the inertial effects of the spin connection $\udt{\omega}{C}{B\mu}$ is interpreted as being separated from the gravitational effects represented by the contortion tensor $\udt{K}{C}{B\mu}$. In fact, for a local frame in which the Levi-Civita spin connection vanishes, the identity implies that the inertial and gravitational effects exactly compensate for each other.

The separation of gravitational and inertial effects can immediately be seen by a free particle in Minkowski spacetime in which
\begin{equation}
    0 = \frac{\dd u^A}{\dd \tau} = \frac{\dd x^{\mu}}{\dd \tau} \frac{\partial u^A}{\partial x^{\mu}} = u^{\mu} \partial_{\mu} u^A\,,
\end{equation}
where the four-velocity $u^a$ is subject to the Lorentz coupling prescription, and $\tau$ is the proper time. In the \gls{gr} case, the coupling prescription in Eq.~\eqref{chp3_GR_coup_pres2} which together with the Lorentz generators in Eq.~\eqref{chp3_Lorentz_gen} tuns out to be
\begin{equation}
    u^{\mu} \left(\partial_{\mu} u^A + \frac{1}{2}\udt{\lc{\omega}}{C}{B\mu} \udt{(\dut{S}{C}{B})}{A}{D} u^D\right) = 0\,,
\end{equation}
giving
\begin{equation}\label{GR_geodesic_eq}
    u^{\mu} \left(\partial_{\mu} u^A + \udt{\lc{\omega}}{A}{B\mu} u^B\right) = 0\,,
\end{equation}
where the particle motion is dictated through the Levi-Civita spin connection rather than a separate force representing gravity. This again implies that the inertial and gravitational effects within \gls{gr} are compounded together. In \gls{tg}, the coupling prescription in Eq.~\eqref{chp3_full_grav_coup_pres} gives a different picture where
\begin{equation}
    u^{\mu}\left(\partial_{\mu} u^A + \frac{1}{2}\left(\udt{\omega}{C}{B\mu} - \udt{K}{C}{B\mu}\right) \udt{(\dut{S}{C}{B})}{A}{D} u^D \right)\,,
\end{equation}
simplifies to
\begin{equation}\label{Eq:Lorentz_force_like_eq}
    u^{\mu} \left(\partial_{\mu} u^A + \udt{\omega}{A}{B\mu} u^B\right) = \udt{K}{A}{D\mu} u^{\mu} u^{D}\,,
\end{equation}
where the effect of gravitation on the motion of a particle is now totally different in which the gravitational force is acting through the contortion tensor term. The geodesic equations in Eq.~\eqref{GR_geodesic_eq} and Eq.~\eqref{Eq:Lorentz_force_like_eq} are equivalent by the identity in Eq.~\eqref{chp3_spin_iden} but represent the fundamental difference between the effects of gravitation in \gls{gr} and \gls{tg}.

\subsection{Coupling to matter} \label{ssec:mattercoupling}

In \gls{tg}, the curvature associated with the Levi-Civita connection is replaced by torsion associated with the teleparallel connection. The gravitational sector can be readily reformulated in this framework, as is laid out in this work. However, the coupling to matter must remain consistent across these possible choices when generalizing the geometric underpinnings of gravitation. As pointed out in Ref.~\cite{BeltranJimenez:2019tjy} there are two important subtleties to bear in mind when considering the matter sector in a gravitational field, namely
\begin{itemize}
	\item Ambiguities in the coupling to matter may arise when changing the geometry;
	\item The treatment of bosonic and fermionic fields may lead to important differences that may lead to inconsistencies.
\end{itemize}

Considering a point particle in GR
\begin{equation}
	\mathcal{S} = mc^2 \int \dd \tau\,.
\end{equation}
For fixed parametrization \cite{Carroll:2004st}, this will follow the regular geodesic equation described in Eq.~\eqref{GR_geodesic_eq}, and will depend on the Levi-Civita connection. However, in other geometric theories of gravity this point needs to be revisited as discussed in Sec.~\ref{Grav_Coup_Prescrip}. This is important since observational constraints on the paths of test particles have become extremely precise, making this a crucial test for the formulation of any geometric theory of gravity.

The second difference to consider is the distinction between bosonic and fermionic fields since the former couples to the metric and is described by tensor representations while the latter also couples to the connection and is represented by spinors. Through the spinor representation, tetrads together with their respective gravitational spin connections are utilized to describe fermions.

In GR, the minimal coupling prescription is presumed where
\begin{equation}
    \eta_{\mu\nu} \rightarrow g_{\mu\nu}\,\quad {\rm and}\quad \partial_{\mu} \rightarrow \lc{\nabla}_{\mu}\,,
\end{equation}
where the Levi-Civita connection, $\udt{\lc{\Gamma}}{\sigma}{\mu\nu}$, is taken. Moreover, the condition of vanishing torsion and non-metricity forces the matter Lagrangian to take the regular form, $\Theta_{\mu\nu}=\frac{-2}{\sqrt{-g}}\frac{\delta L_{\rm m}}{\delta g^{\mu\nu}}$ (also given in Eq.~\eqref{Eq:Con_EM_ten}). If the Lagrangian had some dependence on the connection, $L_{\rm m}=L_{\rm m}(g_{\mu\nu}, \udt{\lc{\Gamma}}{\sigma}{\mu\nu}, \phi)$, would lead to a nonvanishing hypermomentum that would necessarily have to be equal to zero once the conditions of vanishing torsion and non-metricity are imposed \cite{BeltranJimenez:2019tjy}.

In \gls{tg}, the change in connection associated with the covariant derivative and thus mediation of the gravitational field imply a natural choice for the coupling to matter where the covariant derivative assumes the teleparallel connection which is torsionful and satisfies vanishing curvature and zero metricity. However, this is immediately problematic, for instance consider the case of bosons through the simple example of photons. In the absence of gravitation, the field strength is described by
\begin{equation}
    F_{\mu\nu} = 2\partial_{[\mu}A_{\nu]}\,,
\end{equation}
where $A_{\nu}$ is the vector potential associated with the electromagnetic field. If the teleparallel connection were taken for the coupling prescription then this would give
\begin{equation}
    F_{\mu\nu} \rightarrow 2\nabla_{[\mu}A_{\nu]} = 2\partial_{[\mu}A_{\nu]} + \udt{T}{\sigma}{\mu\nu}A_{\sigma}\,,
\end{equation}
in a gravitational field. This would mean that the photon is non-minimally coupled to torsion and does not observe the $U(1)$ gauge symmetry in its standard form which would be extremely problematic.

On the other hand, fermions would also pose a serious problem to a matter coupling prescription that takes on the teleparallel connection. Consider the Dirac equation~\cite{weinberg1995quantum}
\begin{equation}
    i\gamma^A \dut{E}{A}{\mu}{\rm D}_{\mu}\psi - mc \psi = 0\,,
\end{equation}
where $\gamma^A$ are the Dirac matrices, $m$ is the mass parameter, and $\psi$ is the fermionic field. The fermionic covariant derivative is given by
\begin{equation}\label{eq:fermionic_cov_der}
    {\rm D}_{\mu} = \partial_{\mu} - \udt{\omega}{AB}{\mu}\left[\gamma_A,\gamma_B\right]\,,
\end{equation}
which arises so that a well-defined Dirac equation that describes the dynamics of spinor fields on a general spacetime \cite{BeltranJimenez:2020sih}. It is this covariant derivative that is the source of the problem with regard to using the teleparallel connection for fermionic fields. In GR, the spin connection contains terms beyond the Lorentz connection and so provides the coupling required with the Levi-Civita connection. The spin connection of \gls{tg} is totally inertial and can be set to zero under an appropriate gauge condition which means that the connection will no longer be coupled to the fermionic fields. This last point will mean that the energy-momentum tensor will not always be conserved which will produce an inconsistent theory.

One option is to choose the coupling prescription $\partial_{\mu} \rightarrow \partial_{\mu}$ for bosons and $\partial_{\mu} \rightarrow {\rm D}_{\mu}$, where ${\rm D}_{\mu}$ is a fermionic covariant derivative that retains the connection coupling such that the energy-momentum tensor for matter is conserved. However, this would be quite arbitrary as a coupling prescription choice. It is for these reasons that the minimal coupling description for gravitation, as described in Sec.~\ref{Grav_Coup_Prescrip}, is identified for all matter coupling in \gls{tg} \cite{Huguet:2020ler,Fontanini:2018krt}. As examined below, it is this choice that most consistently provides a sound avenue for constructing teleparallel theories of gravity \cite{Mosna:2003rx}, i.e. we adopt the coupling prescription
\begin{subequations}\label{chp3_tele_coup_pres}
\begin{align}
    \udt{h}{A}{\mu} &\rightarrow \udt{e}{A}{\mu}\,,\\[0.5ex]
    \eta_{\mu\nu} &\rightarrow g_{\mu\nu}\,,
\end{align}
\end{subequations}
which is first introduced in Eq.~\eqref{sec4:coupling}. A consequence of this is that the contortion tensor identity presented in Eq.~\eqref{chp3_spin_iden} can now be written as
\begin{equation}\label{eq:cont_def}
    K^{\rho}{}_{\mu\nu} =\Gamma^{\rho}{}_{\mu\nu}-\lc{\Gamma}^{\rho}{}_{\mu\nu}=\frac{1}{2}\left(T_{\mu}{}^{\rho}{}_{\nu}+T_{\nu}{}^{\rho}{}_{\mu}-T^{\rho}{}_{\mu\nu}\right)\,,
\end{equation}
where the contortion tensor is recognised as the difference between the teleparallel and Levi-Civita connections respectively. Also, the definition in Eq.~\eqref{chp3_contortion_def} was used.

\subsubsection{Scalar fields} \label{chp3_scalar_field_coup}

Considering the simple but revealing case of a linear scalar field where the Lagrangian is given by
\begin{equation}
    \mathcal{L}_{\phi} := \frac{1}{2} \left(\eta^{\mu\nu} \partial_{\mu} \phi \partial_{\nu} \phi - \mu^2 \phi^2\right)\,,
\end{equation}
which is defined on the tangent space, $\mu = mc/\hbar$\footnote{We retain SI units for clarity here.} is the mass associated with the scalar field, and the $\phi=\phi(x)$ position dependence is suppressed for brevity's sake. The Lorentzian Klein-Gordon equation is produced by taking a variation \gls{wrt} $\phi$ giving \cite{weinberg1995quantum}
\begin{equation}
    \partial_{\mu}\partial^{\mu} \phi + \mu^2 \phi = 0\,.
\end{equation}

By adopting the minimal coupling presenting in Eq.~\eqref{chp3_tele_coup_pres}, the Lagrangian density is promoted to
\begin{equation}
    \mathcal{L}_{\phi} = \frac{1}{2}\left(g^{\mu\nu} \lc{\nabla}_{\mu} \phi \lc{\nabla}_{\nu} \phi - \mu^2 \phi^2\right)\,,
\end{equation}
where $\lc{\nabla}_{\mu} \phi=\partial_{\mu} \phi$ since $\phi$ is a scalar. Taking the variation of this Lagrangian \gls{wrt} the scalar field produces the gravitational Klein-Gordon equivalent
\begin{equation}
    \lc{\Box} \phi + \mu^2 \phi = 0\,,\label{eq:KGeqB}
\end{equation}
where $\lc{\Box} \phi = e^{-1} \partial_{\sigma} \left(e \partial^{\sigma} \phi \right)$ is the Laplace-Beltrami operator \cite{Aldrovandi:2013wha}, and $e = {\rm det} \left(\udt{e}{A}{\mu}\right) = \sqrt{-g}$. However, by noticing that
\begin{equation}
    \partial_{\mu} e = \partial_{\mu} \sqrt{-g} = \frac{1}{2\sqrt{-g}} \partial_{\mu} g = \frac{1}{2} \sqrt{-g} g^{\alpha\beta} \partial_{\mu} g_{\alpha\beta} = \sqrt{-g} \udt{\lc{\Gamma}}{\sigma}{\sigma\mu} = e \udt{\lc{\Gamma}}{\sigma}{\sigma\mu} = e\left(\udt{\Gamma}{\sigma}{\mu\sigma} - \udt{K}{\sigma}{\mu\sigma}\right)\,,
\end{equation}
implies that the Laplace-Beltrami operator on the scalar field can be written as
\begin{subequations}
\begin{align}
    \lc{\nabla}_{\mu}\lc{\nabla}^{\mu}\phi = \lc{\nabla}_{\mu}\partial^{\mu}\phi =\lc{\Box} \phi &= e^{-1} \left[e\partial_{\mu}\left(\partial^{\mu}\phi\right) + \left(\partial^{\mu}\phi\right) \partial_{\mu}e\right]\\[0.5ex]
    &= e^{-1}\left[e\partial_{\mu}\left(\partial^{\mu}\phi\right) + \left(\partial^{\mu}\phi\right) e \left(\udt{\Gamma}{\sigma}{\mu\sigma} - \udt{K}{\sigma}{\mu\sigma}\right)\right]\\[0.5ex]
    &= \left[\partial_{\mu} + \udt{\Gamma}{\sigma}{\mu\sigma} - \udt{K}{\sigma}{\mu\sigma}\right]\partial^{\mu}\phi \,.
\end{align}
\end{subequations}
Thus, the teleparallel Klein-Gordon Eq.~\eqref{eq:KGeqB} can be written as
\begin{equation}
    \left(\partial_{\mu} + \udt{\Gamma}{\sigma}{\mu\sigma} - \udt{K}{\sigma}{\mu\sigma} \right)\partial^{\mu}\phi + \mu^2 \phi = 0\,,
\end{equation}
from which it follows scalar fields couple to torsion through the contortion tensor.

\subsubsection{Fermion fields} \label{chp3_fermion_field_coup}

The half spin massive particles are described by the Dirac field which has an associated Lagrangian
\begin{equation} \label{chp_3_Dirac_Min_L}
    \mathcal{L}_{\psi} := \frac{ic\hbar}{2}\left(\bar{\psi} \gamma^A \dut{H}{A}{\mu} \partial_{\mu}\psi - \dut{H}{A}{\mu} \partial_{\mu}\bar{\psi} \gamma^A \psi\right) - mc^2 \bar{\psi}\psi \,,
\end{equation}
in the absence of gravity (where SI units are allowed so that the appearance of constants in the Dirac equation is clearer to see), and where $\bar{\psi}$ denotes he conjugate transpose of $\psi$. This produces the flat Dirac equation
\begin{equation}
    i\hbar \gamma^A \dut{H}{A}{\mu} \partial_{\mu} \psi - mc \psi = 0\,.
\end{equation}
Now using the spinor field covariant derivative defined in Eq.~\eqref{eq:fermionic_cov_der}, the Dirac field Lagrangian that is minimally coupled to torsion will be transformed to
\begin{subequations}
\begin{align}
    \mathcal{L}_{\psi} &= \frac{ic\hbar}{2}\left(\bar{\psi} \gamma^A \dut{H}{A}{\mu} \partial_{\mu}\psi - \dut{H}{A}{\mu} \partial_{\mu}\bar{\psi} \gamma^A \psi\right) - mc^2 \bar{\psi}\psi\\[0.5ex]
    &\rightarrow \frac{ic\hbar}{2}\left(\bar{\psi} \gamma^A \dut{E}{A}{\mu} {\rm D}_{\mu}\psi - \dut{E}{A}{\mu} {\rm D}_{\mu}\bar{\psi} \gamma^A \psi\right) - mc^2 \bar{\psi}\psi\,,
\end{align}
\end{subequations}
which conveniently separates the Levi-Civita and purely torsional terms, giving
\begin{equation}
    \mathcal{L}_{\psi} = \frac{ic\hbar}{2}\left(\bar{\psi} \gamma^A \dut{E}{A}{\mu} \lc{\nabla}_{\mu}\psi - \dut{E}{A}{\mu} \lc{\nabla}_{\mu}\bar{\psi} \gamma^A \psi - \udt{e}{a}{\bar{\mu}}g^{\bar{\mu}\mu}\udt{e}{b}{\lambda}\udt{e}{c}{\bar{\rho}}g^{\bar{\rho}\rho}\udt{K}{\lambda}{\rho\mu}\{\gamma_a,\left[\gamma_c,\gamma_c\right]\}\psi\right) - mc^2 \bar{\psi}\psi\,,
\end{equation}
where $\{\gamma_a,\gamma_b\} = 2\gamma_{(a}\gamma_{b)}$. Using several identities relating the Dirac matrices, it can be shown that \cite{BeltranJimenez:2020sih}
\begin{equation}
    \{\gamma_a,\left[\gamma_c,\gamma_c\right]\} = 4 i \dut{\epsilon}{abc}{d} \gamma_d \gamma^5\,,
\end{equation}
where $\epsilon_{ABCD}$ is the Levi-Civita symbol, and where the Lagrangian reduces to
\begin{equation}
    \mathcal{L}_{\psi} = \frac{ic\hbar}{2}\left(\bar{\psi} \gamma^A \dut{E}{A}{\mu} \lc{\nabla}_{\mu}\psi - \dut{E}{A}{\mu} \lc{\nabla}_{\mu}\bar{\psi} \gamma^A \psi - 2 i \epsilon^{\lambda\rho\mu\nu}T_{\lambda\rho\mu}\bar{\psi}\gamma^5\gamma_{\nu}\psi\right) - mc^2 \bar{\psi}\psi\,.
\end{equation}

The teleparallel Dirac Lagrangian can be recast in terms of the irreducible contributions to the torsion tensor. In Sec.~\ref{chp4_NGR_action}, the irreducible nature of the torsion tensor will be further probed. For the purpose of the present discussion, consider the decomposition of the torsion tensor
\begin{equation}
    \udt{T}{\lambda}{\mu\nu} = \frac{2}{3} \left(\udt{t}{\lambda}{\mu\nu} - \udt{t}{\lambda}{\nu\mu}\right) + \frac{1}{3} \left(\delta^{\lambda}_{\mu}v_{\nu} - \delta^{\lambda}_{\nu}v_{\mu}\right) + \udt{\epsilon}{\lambda}{\mu\nu\rho} a^{\rho}\,,
\end{equation}
where $a_{\mu} = \frac{1}{6}\epsilon_{\mu\nu\sigma\rho}T^{\nu\sigma\rho}$, $v_{\mu} = T^{\sigma}{}_{\sigma\mu}$ and $\udt{t}{\sigma}{\mu\nu} = \frac{1}{2}\left(T_{\sigma\mu\nu} + T_{\mu\sigma\nu}\right) + \frac{1}{6}\left(g_{\nu\sigma}v_{\mu} + g_{\nu\mu}v_{\sigma}\right) - \frac{1}{3}g_{\sigma\mu}v_{\nu}$ are the axial, vector and purely tensorial parts of the torsion tensor as discussed in Eq.~\eqref{eq:tortensatv123}. By considering the symmetries of the torsion tensor, the Dirac Lagrangian can be further reduced to \cite{BeltranJimenez:2020sih}
\begin{equation}
    \mathcal{L}_{\psi} = \frac{ic\hbar}{2}\left(\bar{\psi} \gamma^A \dut{E}{A}{\mu} \lc{\nabla}_{\mu}\psi - \dut{E}{A}{\mu} \lc{\nabla}_{\mu}\bar{\psi} \gamma^A \psi - 2 i \bar{\psi}\gamma^5\gamma^{\mu}a_{\mu}\psi\right) - mc^2 \bar{\psi}\psi\,,
\end{equation}
where it can be noted that the Dirac spinors couple to the axial part of the torsion tensor only, and so the other parts of the torsion tensor can vary freely without altering the fermionic sector. This is an important property to identify, it is also important to point out that ultimately, the Dirac equation will have the same solutions since \gls{tegr} is dynamically equivalent to GR, and so the differences we recognize will only pertain to the way that the TG contributions reproduce the standard equation.

\subsubsection{Boson fields} \label{chp3_bosonic_field_coup}

A large class of boson fields permeate through the standard model of particle physics with the most prominent being the photon under spin 1 representation of the Lorentz group. We consider this example to show the general matter coupling for bosonic fields in teleparallel geometries \cite{BeltranJimenez:2020sih,Weldon:2000fr}. Taking the four-potential $A_{\mu}$, the electromagnetic tensor can be written as \cite{jackson2007classical}
\begin{equation}
    F_{\mu\nu} = \partial_{\mu}A_{\nu} - \partial_{\nu} A_{\mu}\,,
\end{equation}
in the absence of gravity on the tangent space. By this construction, it immediately follows that the Bianchi equation below is satisfied by this field, i.e.
\begin{equation} \label{Maxwel_eq_1}
    \partial_{\mu} F_{\nu\sigma} + \partial_{\sigma} F_{\mu\nu} + \partial_{\nu} F_{\sigma\mu} = 0\,,
\end{equation}
while the electromagnetic Lagrangian
\begin{equation} \label{Maxwel_eq_2}
    \mathcal{L}_{\rm EM} := -\frac{1}{4} F_{\mu\nu}F^{\mu\nu}\,,
\end{equation}
produces the field equations
\begin{equation}
    \partial_{\mu} F^{\mu\nu} = 0\,,
\end{equation}
when varied \gls{wrt} the four-potential. Together Eq.~\eqref{Maxwel_eq_1} and Eq.~\eqref{Maxwel_eq_2} constitute the source-less Maxwell equations \cite{jackson2007classical}. Within the Lorentz gauge, where $\partial_{\mu}A^{\mu} = 0$, Eq.~\eqref{Maxwel_eq_2} directly leads to the wave equation
\begin{equation}
    \partial_{\sigma}\partial^{\sigma} A^{\mu} = 0\,.
\end{equation}
These Lorentz equations can be raised to their gravitational analogues by applying the coupling prescription in Eq.~\eqref{chp3_tele_coup_pres} so that
\begin{equation}
    \partial_{\sigma}A^{\mu} \rightarrow \lc{\nabla}_{\sigma}A^{\mu} = \partial_{\sigma}A^{\mu} + \left(\udt{\Gamma}{\mu}{\nu\sigma} - \udt{K}{\mu}{\nu\sigma}\right)A^{\nu}\,.
\end{equation}
Given the way that the electromagnetic tensor is defined, it can be written as
\begin{equation}
    F_{\mu\nu} = \lc{\nabla}_{\mu}A_{\nu} - \lc{\nabla}_{\nu} A_{\mu} = \partial_{\mu}A_{\nu} - \partial_{\nu} A_{\mu}\,,
\end{equation}
which retains the form of the electromagnetic tensor as in Minkowski space, and where the second equation is the result of the difference of the terms in the definition being symmetric. Naturally, this means that the same Bianchi identities will continue to be satisfied, namely $\partial_{\mu} F_{\nu\sigma} + \partial_{\sigma} F_{\mu\nu} + \partial_{\nu} F_{\sigma\mu} = 0$. However, the Lagrangian will be altered such that
\begin{equation}
    \mathcal{L}_{\rm EM} = -\frac{1}{4} F_{\mu\nu} F^{\mu\nu}\,,
\end{equation}
which produce
\begin{equation} \label{chp3_tele_maxwell_eqn_2}
    \lc{\nabla}_{\mu}F^{\mu\nu} = 0\,,
\end{equation}
when varied \gls{wrt} the four-potential. This constitutes the second part of Maxwell's equations in a teleparallel setting.

In this setting, the Lorentz gauge ($\lc{\nabla}_{\mu} A^{\mu} = 0$) can only be applied when the commutator on the four-potential is considered since the covariant derivatives are not commutative, i.e.
\begin{equation}
    \left[\lc{\nabla}_{\mu}, \lc{\nabla}_{\nu}\right] A^{\mu} = \lc{\nabla}_{\mu} \lc{\nabla}_{\nu} A^{\mu} - \lc{\nabla}_{\nu} \lc{\nabla}_{\mu} A^{\mu} = \lc{R}_{\mu\nu} A^{\mu} = - P_{\mu\nu}A^{\mu}\,,
\end{equation}
where $P_{\mu\nu} := \udt{P}{\sigma}{\mu\nu\sigma}$ and is defined by the Riemann tensor as calculated using both the teleparallel and Levi-Civita connections
\begin{equation} \label{chp3_Riemann_equiv}
    0 \equiv \udt{R}{\sigma}{\mu\nu\rho} = \udt{\lc{R}}{\sigma}{\mu\nu\rho} + \udt{P}{\sigma}{\mu\nu\rho}\,,
\end{equation}
where the contortion tensor in Eq.~\eqref{chp3_contortion_def} leads to a purely teleparallel definition of $\udt{P}{\sigma}{\mu\nu\rho}$ through
\begin{alignat}{2}
    \udt{P}{\sigma}{\mu\nu\rho} & :=\: & & \partial_{\nu} \udt{K}{\sigma}{\mu\rho} - \partial_{\rho} \udt{K}{\sigma}{\mu\nu} + \udt{\Gamma}{\sigma}{\theta\nu} \udt{K}{\theta}{\mu\rho} - \udt{\Gamma}{\sigma}{\theta\rho} \udt{K}{\theta}{\mu\nu} \nonumber\\[0.5ex]
   & \: & &- \udt{\Gamma}{\theta}{\mu\nu} \udt{K}{\sigma}{\theta\rho} + \udt{\Gamma}{\theta}{\mu\rho} \udt{K}{\sigma}{\theta\nu} + \udt{K}{\sigma}{\theta\rho} \udt{K}{\theta}{\mu\nu} - \udt{K}{\sigma}{\theta\nu} \udt{K}{\theta}{\mu\rho}\,.
\end{alignat}
In this context, the teleparallel Maxwell equation in Eq.~\eqref{chp3_tele_maxwell_eqn_2} reduces to \cite{Aldrovandi:2013wha}
\begin{equation}
    \lc{\nabla}_{\mu}\lc{\nabla}^{\mu} A^{\nu} + \udt{P}{\nu}{\mu}A^{\mu} = 0\,,
\end{equation}
where a coupling with the teleparallel connection appears while continuing to satisfy the gauge invariance of the Maxwell equations. Comparing to standard gravity through the equivalency in Eq.~\eqref{chp3_Riemann_equiv}, the same action is observed for the Levi-Civita connection in
\begin{equation}
    \lc{\nabla}_{\mu}\lc{\nabla}^{\mu} A^{\nu} - \lc{R}^{\mu\nu}A_{\mu} = 0\,,
\end{equation}
where the coupling also appears only in terms of the Levi-Civita connection.

More generally, the teleparallel coupling prescription in Eq.~\eqref{chp3_tele_coup_pres} has not been widely studied in the context of other bosonic fields. It would be interesting to further investigate how this coupling prescription may lead to further probes of \gls{tg}.

\subsubsection{Effects on the geodesic equation}\label{sec:geodesic_eqn}

The coupling prescription detailed in Eq.~\eqref{chp3_tele_coup_pres} describes how matter couples to the geometry of gravity. In Secs.~\ref{chp3_scalar_field_coup}--\ref{chp3_bosonic_field_coup}, the cases of scalar fields, fermions and bosonic fields are explored respectively through the \gls{tg} analogues of the Klein-Gordon, Dirac, and Maxwell equations. In these cases, the consistency of the theory is preserved by taking the Levi-Civita covariant derivative as the teleparallel coupling to matter.

An important nuance takes place when considering the paths of particles in gravitational theories based on non-Riemannian geometry. Given a Levi-Civita covariant derivative, $\lc{\nabla}_{\mu}$, a tangent vector field, $\xi^{\sigma} = \frac{\dd x^{\sigma}}{\dd \tau}$, is said to be an autoparallel of $\DD_{\mu}$ provided that it remains parallel to itself along the length of the path \cite{Vinckers:2020cmw,Wald:1984rg}
\begin{equation}
    \xi^{\sigma}\lc{\nabla}_{\sigma} \xi^{\mu} = 0\,.
\end{equation}
Autoparallels generalize the concept of a straight line to a general affine spacetime, while geodesics define the shortest path between two points. However, geodesics require a determination of length such as the metric tensor in Riemannian geometries. In \gls{gr}, the Levi-Civita covariant derivative
\begin{equation}
    \lc{\nabla}_{\mu}\xi_{\sigma} = \partial_{\mu} \xi_{\sigma} - \udt{\lc{\Gamma}}{\rho}{\mu\sigma} \xi_{\rho}\,,
\end{equation}
results in the same equation when considering either geodesic or autoparallel equations, given by
\begin{equation}\label{chp3_GR_geo_eq}
    \frac{\dd ^2 x^{\mu}}{\dd \tau^2} + \udt{\lc{\Gamma}}{\mu}{\alpha\beta} \frac{\dd x^{\alpha}}{\dd \tau} \frac{\dd x^{\beta}}{\dd \tau} = 0\,,
\end{equation}
which is the well known equation that describes the paths that test particles take.

The teleparallel coupling prescription means that the geodesic equation in Eq.~\eqref{chp3_GR_geo_eq} is transformed by Eq.~\eqref{eq:cont_def} to
\begin{equation}\label{chp3_test_part_eq}
    \frac{\dd ^2 x^{\mu}}{\dd \tau^2} + \udt{\Gamma}{\mu}{\alpha\beta} \frac{\dd x^{\alpha}}{\dd \tau} \frac{\dd x^{\beta}}{\dd \tau} = \udt{K}{\mu}{\alpha\beta} \frac{\dd x^{\alpha}}{\dd \tau} \frac{\dd x^{\beta}}{\dd \tau}\,,
\end{equation}
which is the result of the transformation from the curvature-based Levi-Civita connection to the torsion-based teleparallel connection. Thus, the geodesic equation in \gls{gr} is now transformed into a Lorentz-like equation where a force term appears on the \gls{rhs} featuring the contortion tensor which drives test particle paths on the \gls{lhs}. Importantly, both equations represent the same paths and so feature the same solutions albeit from different perspectives. This equation highlights an important property to the \gls{gr} geodesic equation in that it recovers a Lorentz force-like form in that instead of the geometry determining the path of test particles, these particles are acted upon by a force-like contribution, in the form of the contortion tension, which acts on the test particles as a force term would.

For generalized teleparallel theories, the free fall paths of particles continues to be determined by Eq.~\eqref{chp3_test_part_eq} where the theory determines the particular geometry of the spacetime.

\subsubsection{A note on the energy-momentum tensor}

Consider a matter action \eqref{eq:actionmatter}
\begin{equation}
    \mathcal{S}_{\rm m}:=\int \dd^4 x\, \sqrt{-g}\mathcal{L}_{\rm m}=\int \dd^4 x\, L_{\rm m}\,,
\end{equation}
which can be minimally coupled to any teleparallel theory of gravity, and where $\mathcal{L}_{\rm m}$ and $L_{\rm m}$ are the Lagrangian density and Lagrangian respectively. The matter section is treated using the minimal coupling description (see Sec.~\ref{Grav_Coup_Prescrip} \cite{BeltranJimenez:2020sih}). Thus, the general matter Lagrangian may depend on the matter contributions $\phi$ and both the tetrad $\udt{e}{A}{\mu}$ and connection $\udt{\lc{\Gamma}}{\rho}{\mu\nu}$. By taking variations \gls{wrt} the tetrad and connections leads to the definition of the regular energy-momentum tensor
\begin{equation}
    \Theta_{\mu\nu} :=\frac{-2}{\sqrt{-g}}\frac{\delta L_{\rm m}}{\delta g^{\mu\nu}}=e^{A}{}_{\mu}\left(\frac{1}{e}\frac{\delta L_{\rm m}}{\delta e^{A}{}_{\nu'}}\right)g_{\nu\nu'}:=e^{A}{}_{\mu}\Theta_{A}{}^{\nu'}g_{\nu\nu'}\,,
\end{equation}
from which we denote the energy-momentum trace as $\Theta=\udt{\Theta}{\mu}{\mu}$, and the hypermomentum as
\begin{equation}
    \dut{\Delta}{\rho}{\mu\nu} := \,\frac{\delta \mathcal{S}_{\rm m}}{\delta \udt{\Gamma}{\rho}{\mu\nu}}\,.
\end{equation}
In principle, the hypermomentum contributes to a connection field equation in the same way that the regular energy-momentum tensor contributes to the metrical (tetrad) variation field equation. As in the vast majority of the literature, we take a vanishing hypermomentum meaning that the matter fields are assumed to not couple to the connection. This will incur some constraints on the matter section (such as with fermions which do have nonzero hypermomentum due their coupling with the axial torsion). However, this effect is largely suppressed in real systems and is ignored in most works within and outside of \gls{tg}.

\subsection{Symmetries of teleparallel geometries} \label{sec:symmetries}

As introduced in Sec.~\ref{sec:tetradspin}, teleparallel theories have the tetrads and spin connection as their fundamental variables. Then, the notion of symmetries in \gls{tg} should be taken carefully since, differently to purely metric based theories (as \gls{gr}), now both quantities may or may not respect the symmetries of a certain spacetime. This study was fully described in Ref.~\cite{Hohmann:2019nat} where it was assumed that both the tetrads and the teleparallel connection satisfy these symmetries. A recent quick summary of these results as long as some applications were also presented in the recent paper Ref.~\cite{Pfeifer:2022txm}.

In the following, we explain how these symmetries can be connected to ansatz choices for the tetrad and spin connection. Let us start by defining what symmetries are in \gls{tg}. A geometric object (on the manifold $\mathcal{M}$) will be invariant under the group action $\varphi: G\times \mathcal{M}\rightarrow \mathcal{M}$, where $G$ being a Lie group on $\mathcal{M}$. Thus, the symmetry of \gls{tg} geometries is defined as a group action $\varphi$ such that both the metric and affine connection are invariant. For practical reasons, it is convenient to consider infinitesimal symmetries to find out the symmetries underlying a certain teleparallel theory. To do this, one can use the Lie derivatives \gls{wrt} a vector $X_{\xi}\in \textrm{Vect}\,\mathcal{M}$ acting on both quantities, which acting on the metric and the connection give us the following
\begin{subequations}
\begin{align}\label{eq:metsymcond}
(\mathcal{L}_{X_{\xi}}g)_{\mu\nu} &= X_{\xi}^{\rho}\partial_{\rho}g_{\mu\nu} + \partial_{\mu}X_{\xi}^{\rho}g_{\rho\nu} + \partial_{\nu}X_{\xi}^{\rho}g_{\mu\rho}\,,\\[0.5ex]
(\mathcal{L}_{X_{\xi}}\Gamma)^{\mu}{}_{\nu\rho} &= X_{\xi}^{\sigma}\partial_{\sigma}\Gamma^{\mu}{}_{\nu\rho} - \partial_{\sigma}X_{\xi}^{\mu}\Gamma^{\sigma}{}_{\nu\rho} + \partial_{\nu}X_{\xi}^{\sigma}\Gamma^{\mu}{}_{\sigma\rho} + \partial_{\rho}X_{\xi}^{\sigma}\Gamma^{\mu}{}_{\nu\sigma} + \partial_{\nu}\partial_{\rho}X_{\xi}^{\mu}\\[0.5ex]
&= \nabla_{\rho}\nabla_{\nu}X_{\xi}^{\mu}
- \nabla_{\rho}(X_{\xi}^{\sigma}T^{\mu}{}_{\nu\sigma})\,,
\end{align}
\end{subequations}
where we have used the flat condition $R^{\mu}{}_{\nu\rho\sigma}\equiv 0$ (see Eq.~\eqref{eq:spiconcurv}). It is worth mentioning that in the standard case where the connection is the Levi-Civita one, if one assumes that the metric is invariant under a group G, $(\mathcal{L}_{X_{\xi}}g)_{\mu\nu}=0$, it follows that the Levi-Civita connection is also invariant under the group G, meaning that $(\mathcal{L}_{X_{\xi}}\lc{\Gamma})^{\mu}{}_{\nu\rho}=0$. This property is true since the Levi-Civita connection depends directly on the metric. However, for the teleparallel connection, if we assume $(\mathcal{L}_{X_{\xi}}g)_{\mu\nu}=0$, it does not mean that $(\mathcal{L}_{X_{\xi}}\Gamma)^{\mu}{}_{\nu\rho}=0$. Thus, these two quantities are zero if the metric and spin connection are invariant under the action of the group $G$. Since the \gls{tg} fundamental variables are the tetrads and spin connection, one easily finds that the Lie derivatives acting on them are
\begin{equation}
(\mathcal{L}_{X_{\xi}}e)^A{}_{\mu} = X_{\xi}^{\nu}\partial_{\nu}e^A{}_{\mu} + \partial_{\mu}X_{\xi}^{\nu}e^A{}_{\nu}\,, \quad (\mathcal{L}_{X_{\xi}}\omega)^A{}_{B\mu} = X_{\xi}^{\nu}\partial_{\nu}\omega^A{}_{B\mu} + \partial_{\mu}X_{\xi}^{\nu}\omega^A{}_{B\nu}\,.
\end{equation}
To fulfill the conditions $(\mathcal{L}_{X_{\xi}}g)_{\mu\nu}=(\mathcal{L}_{X_{\xi}}\Gamma)^{\mu}{}_{\nu\rho}=0$ (invariance under $G$), one then requires
\begin{equation}\label{eq:infisymcondgen}
(\mathcal{L}_{X_{\xi}}e)^A{}_{\mu} = -\boldsymbol{\lambda}_{\xi}^A{}_B e^B{}_{\mu}\,, \quad
(\mathcal{L}_{X_{\xi}}\omega)^A{}_{B\mu} = \DD_{\mu}\boldsymbol{\lambda}_{\xi}^A{}_B\,,
\end{equation}
where $\DD_{\mu}\boldsymbol{\lambda}_{\xi}^A{}_B = \partial_{\mu}\boldsymbol{\lambda}_{\xi}^A{}_B + \omega^A{}_{C\mu}\boldsymbol{\lambda}_{\xi}^C{}_B - \omega^C{}_{B\mu}\boldsymbol{\lambda}_{\xi}^A{}_C$ was used and $\boldsymbol{\lambda}_{\xi}$ is the local Lie homomorphism (see Ref.~\cite{hamermesh2012group}) given by
\begin{equation}\label{eq:lochomo}
\boldsymbol{\lambda}_{\xi}(x) = \left.\frac{\dd}{\dd t}\boldsymbol{\Lambda}_{\hat{\xi}(t)}(x)\right|_{t = 0}\,.
\end{equation}
Now, one can use local Lorentz transformations (for both the tetrads and spin connection) and choose a specific gauge $\Lambda_A{}^{B}$ where the spin connection $\omega'^A{}_{B\mu}$ (Weitzenb\"{o}ck gauge) vanishes to simplify the above conditions. After doing this, one gets that the conditions~\eqref{eq:infisymcondgen} become
\begin{equation}\label{eq:infisymcondwb}
(\mathcal{L}_{X_{\xi}}e')^A{}_{\mu} = -{\boldsymbol{\lambda}'}_{\xi}^A{}_B e'^b{}_{\mu}\,, \quad
0 \equiv (\mathcal{L}_{X_{\xi}}\omega')^A{}_{B\mu} = \partial_{\mu}{\boldsymbol{\lambda}'}_{\xi}^A{}_B\,.
\end{equation}
One can notice that in the Weitzenb\"{o}ck gauge, it is much easier to solve the conditions in such a way that both the metric and the connection are invariant under the group action. We can define \textit{symmetric tetrads} as the tetrads satisfying the above conditions, leading to $(\mathcal{L}_{X_{\xi}}g)_{\mu\nu}=(\mathcal{L}_{X_{\xi}}\Gamma)^{\mu}{}_{\nu\rho}\equiv 0$, that are related by global Lorentz transformations
\begin{equation}
\boldsymbol{\bar{\Lambda}}'_{F\,B}{}^A = \Lambda_B{}^D\Lambda^A{}_C\,\boldsymbol{\Lambda}'_{F\,D}{}^C\,.
\end{equation}
This condition is very important since using it, one is able to find the symmetric tetrads that are compatible with the underlying symmetries. Since all computations are easier in the Weitzenb\"{o}ck gauge, we will omit the primes when we are computing the symmetries in this case and whenever we perform a Lorentz transformation, the transform quantities (tetrad and spin connection) will be denoted with the primes.

It is also important to mention that the condition $(\mathcal{L}_{X_{\xi}}\Gamma)^{\mu}{}_{\nu\rho}=0$ implies that
\begin{equation}\label{eq:torsym}
(\mathcal{L}_{X_\xi}\Gamma)^\sigma{}_{[\mu\nu]}=-\frac{1}{2}(\mathcal{L}_{X_\xi}T)^\sigma{}_{\mu\nu} = 0\,,
\end{equation}
which tells us that the torsion tensor also satisfies the symmetries. This means that symmetric tetrads will automatically impose that the torsion tensor also respect the symmetries assumed.

For any \gls{tg} theory of gravity constructed from a tetrad, the torsion tensor and their covariant derivatives (\gls{wrt} Levi-Civita or Weitzenb\"{o}ck) will also respect the symmetries, then, the field equations of this theory $W_{\mu\nu}$ will also satisfy the symmetry conditions. In principle, the conditions related to the connection can be relaxed and only impose the symmetries in the tetrads (or the metric).

We will now use the definitions above to derive the most general tetrads in the Weitzenb\"{o}ck gauge satisfying axial, spherical and cosmological symmetries.

\subsubsection{Axial symmetry - \texorpdfstring{$\mathrm{SO}(2)$}{}} \label{sec:axialsymmetry}

Let us first derive the symmetric tetrad for axial symmetry. We will work in spherical coordinates $(t,r,\vartheta,\varphi)$.
In this case, there is only one generator given by $X_z=-\partial_{\varphi}$ (only one of the generators of the case in Eq.~\eqref{eq:genvecz}) which is related to the $\mathrm{SO}(2)$ group. Then, for axial symmetry, we can choose a local homomorphism in the domain of the group $\mathrm{SO}(2)$. One canonical choice in this domain gives us that the differential \eqref{eq:lochomo} acts on the generator $X_z$, which in this case gives
\begin{equation}\label{eq:axalghomo2}
\quad 
{\boldsymbol{\lambda}}(X_z)= \begin{pmatrix}
0 & 0 & 0 & 0\\
0 & 0 & -1 & 0\\
0 & 1 & 0 & 0\\
0 & 0 & 0 & 0
\end{pmatrix}\,.
\end{equation}
We can solve the set of equations in Eq.~\eqref{eq:infisymcondwb} in the Weitzenb\"{o}ck gauge and find the most general symmetric tetrad satisfying axial symmetry. One notices that there are two possible branches which solve Eq,~\eqref{eq:infisymcondwb}.

The first branch that we label as \textit{regular branch}, reads as follows
\begin{equation}\label{eq:axtetradwb}
    e^0{}_{\mu} = C^0{}_{\mu}\,, \quad
    e^1{}_{\mu} = C^1{}_{\mu}\cos\varphi - C^2{}_{\mu}\sin\varphi\,, \quad
    e^2{}_{\mu} = C^1{}_{\mu}\sin\varphi + C^2{}_{\mu}\cos\varphi\,, \quad
    e^3{}_{\mu} = C^3{}_{\mu}\,,
\end{equation}
with $C^A{}_{\mu}$ being 16 functions of the coordinates $t, r, \vartheta$. It is possible to perform a Lorentz transformation for both the tetrad~\eqref{eq:tetlortrans} and the spin connection~\eqref{eq:spclortrans} in such a way that the tetrads do not depend on $\varphi$. This can be done by choosing the following Lorentz transformation
\begin{equation}\label{lorentz1}
    \Lambda^A{}_B = \begin{pmatrix}
    1 & 0 & 0 & 0\\
    0 & \cos\varphi & \sin\varphi & 0\\
    0 & -\sin\varphi & \cos\varphi & 0\\
    0 & 0 & 0 & 1
\end{pmatrix}\,,
\end{equation}
and obtain another tetrad-spin connection pair satisfying axial symmetry
\begin{subequations}\label{axial1a}
	\begin{align}
	e'^0{}_{\mu} &= C^0{}_{\mu}\,, \quad
	e'^1{}_{\mu} = C^1{}_{\mu}\,, \quad
	e'^2{}_{\mu} =C^2{}_{\mu}\,, \quad
	e'^3{}_{\mu} = C^3{}_{\mu}\,,\\[0.5ex]
	\omega'{}^1{}_{2\varphi} &= -\omega'{}^2{}_{1\varphi} = -1\,.
	\end{align}
\end{subequations}

The second branch that we label as \textit{solely axially symmetric branch} which also solve the symmetry conditions~\eqref{eq:infisymcondwb} and give us the following tetrad in the Weitzenb\"{o}ck gauge
\begin{equation}\label{eq:axtetradwC}
e^0{}_{\mu} = C^0{}_{\mu}\,, \quad
e^1{}_{\mu} = C^1{}_{\mu}\,, \quad
e^2{}_{\mu} =C^2{}_{\mu}\,, \quad
e^3{}_{\mu} = C^3{}_{\mu}\,,
\end{equation}
where again $C^A{}_{\mu}$ are 16 arbitrary functions of $t, r, \vartheta$, but now there is not a $\varphi$ dependant in this tetrad. This branch trivially solves Eq.~\eqref{eq:infisymcondwb}. We notice that this branch is related to the first branch via Eq.~\eqref{lorentz1}. However, the above tetrad has zero spin connection whereas in the first branch~\eqref{axial1a}, the spin connection is non-zero. Thus, these two branches are different solutions satisfying axial symmetry in \gls{tg}. One important remark here is that the second branch cannot recover spherical symmetry in \gls{tg} in any limit. Even though the metric can be spherically symmetric in a certain limit, the connection will always be axially symmetric. This is the reason why we label this branch as solely axially symmetric branch. On the other hand, the first branch (regular one), has a spherically symmetric limit which will be the same tetrad that we will find in the next section.

\subsubsection{Spherical symmetry - \texorpdfstring{$\mathrm{SO}(3)$}{}}

In this section, we derive the most general spherically symmetric tetrads in the Weitzenb\"ock gauge. It is convenient to start the computations using spherical coordinates as we did in the previous section. In these coordinates, one finds three generators related to the spherical symmetry
\begin{equation}\label{eq:genvecz}
    X_x = \sin\varphi\,\partial_{\vartheta} + \frac{\cos\varphi}{\tan\vartheta}\,\partial_{\varphi}\,, \quad
    X_y = -\cos\varphi\,\partial_{\vartheta} + \frac{\sin\varphi}{\tan\vartheta}\,\partial_{\varphi}\,, \quad X_z = -\partial_{\varphi}\,.
\end{equation}
These generators are associated to the rotational group $\mathrm{SO}(3)$. Now, we need to choose a local homomorphism $\boldsymbol{\Lambda}$ in the domain of the group $\mathrm{SO}(3)$. A canonical choice for $\boldsymbol{\Lambda}$ is given by
\begin{subequations}
\begin{align}\label{eq:axgrouphomo}
\boldsymbol{\Lambda} &= \begin{pmatrix}
1 & 0 \\
0 & \mathcal{R}(\alpha,\beta,\gamma)
\end{pmatrix}\\
&=\begin{pmatrix}
1&0&0&0 \\
0& \cos\alpha & -\sin\alpha & 0 \\
0&\sin\alpha & \cos\alpha & 0 \\
0&0 & 0 & 1
\end{pmatrix}
\begin{pmatrix}
1&0&0&0 \\
0&\cos\beta & 0 & -\sin\beta \\
0& 0 & 1 & 0 \\
0& \sin\beta & 0 & \cos\beta
\end{pmatrix}\begin{pmatrix}
1 & 0 & 0 & 0\\
0 & 1 & 0 & 0\\
0 & 0 & \cos\gamma & -\sin\gamma\\
0 & 0 & \sin\gamma & \cos\gamma
\end{pmatrix}\,,
\end{align}
\end{subequations}
where $\mathcal{R}(\alpha,\beta,\gamma) =R_z(\alpha)R_y(\beta)R_x(\gamma)$ are the rotation matrices about the Cartesian coordinates axis with angles $\alpha,\beta,\gamma$. Then, the differential \eqref{eq:lochomo} of $\boldsymbol{\Lambda}$ acts on the generators of rotations as
\begin{equation}\label{eq:axalghomo}
{\boldsymbol{\lambda}}(X_x)= \begin{pmatrix}
0 & 0 & 0 & 0\\
0 & 0 & 0 & 0\\
0 & 0 & 0 & -1\\
0 & 0 & 1 & 0
\end{pmatrix}\,, \quad
{\boldsymbol{\lambda}}(X_y) = \begin{pmatrix}
0 & 0 & 0 & 0\\
0 & 0 & 0 & 1\\
0 & 0 & 0 & 0\\
0 & -1 & 0 & 0
\end{pmatrix}\,, \quad {\boldsymbol{\lambda}}(X_z)= \begin{pmatrix}
0 & 0 & 0 & 0\\
0 & 0 & -1 & 0\\
0 & 1 & 0 & 0\\
0 & 0 & 0 & 0
\end{pmatrix}\,.
\end{equation}
In order to find the symmetric tetrad, we need to replace the quantities in Eq.~\eqref{eq:infisymcondwb} and then solve the resulting equations. It is sufficient to just take the first part of that equation, giving us the following form of the tetrad in the Weitzenb\"{o}ck gauge
\begin{subequations}\label{eq:sphertetradwb}
\begin{alignat}{2}
	\mathbf{e}^0 & =\: & & C_1\dd t + C_2\dd r\,,\\[0.5ex]
	\mathbf{e}^1 & =\: & & \sin\vartheta\cos\varphi(C_3\dd t + C_4\dd r) + (C_5\cos\vartheta\cos\varphi - C_6\sin\varphi)\dd\vartheta \nonumber\\[0.5ex]
	& \: & &- \sin\vartheta(C_5\sin\varphi + C_6\cos\vartheta\cos\varphi)\dd\varphi\,,\\[0.5ex]
	\mathbf{e}^2 & =\: & & \sin\vartheta\sin\varphi(C_3\dd t + C_4\dd r) + (C_5\cos\vartheta\sin\varphi + C_6\cos\varphi)\dd\vartheta\nonumber\\[0.5ex]
	& \: & &+ \sin\vartheta(C_5\cos\varphi - C_6\cos\vartheta\sin\varphi)\dd\varphi\,,\\[0.5ex]
	\mathbf{e}^3 & =\: & & \cos\vartheta(C_3\dd t + C_4\dd r) - C_5\sin\vartheta\dd\vartheta + C_6\sin^2\vartheta\dd\varphi\,,
\end{alignat}
\end{subequations}
where all the 6 functions $C_i$ depend only on $r$ and $t$, and the $\vartheta$ dependence is fully determined. We can here notice that this tetrad is a special case of the axially symmetric regular branch~\eqref{eq:axtetradwb} in the spherically symmetric limit. Since the symmetry condition~\eqref{eq:infisymcondwb} in spherical symmetry only has the above tetrad as a solution (in the Weitzenb\"{o}ck gauge), then, one can conclude that the second branch in axial symmetry (see Eq.~\eqref{eq:axtetradwC}) cannot become spherically symmetric in any limit.

The above symmetric tetrad is the most general tetrad satisfying spherical symmetry. The metric associated to this tetrad is
\begin{equation}
\dd s^2= (C_1^2-C_3^2)\dd t^2- (C_4^2-C_2^2)\dd r^2 - (C_5^2-C_6^2)(\dd \vartheta^2+\sin^2\varphi\, \dd \varphi^2)-2(C_3C_4-C_1C_2)\dd r \, \dd t\,,
\end{equation}
which is clearly non-diagonal.

We can simplify the form of the tetrad Eq.~\eqref{eq:sphertetradwb} by performing a Lorentz transformation in order to eliminate all the $\varphi$ dependence. To do this, let us take the following Lorentz transformation
\begin{equation}\label{eq:spherlt10}
\Lambda^A{}_B = \begin{pmatrix}
1 & 0 & 0 & 0\\
0 & \sin\vartheta\cos\varphi & \sin\vartheta\sin\varphi & \cos\vartheta\\
0 & \cos\vartheta\cos\varphi & \cos\vartheta\sin\varphi & -\sin\vartheta\\
0 & -\sin\varphi & \cos\varphi & 0
\end{pmatrix}\,,
\end{equation}
yielding the tetrad
\begin{equation}\label{eq:tetspher1}
\mathbf{e}'^0 = C_1\dd t + C_2\dd r\,, \quad
\mathbf{e}'^1 = C_3\dd t + C_4\dd r\,, \quad
\mathbf{e}'^2 = C_5\dd\vartheta - C_6\sin\vartheta\dd\varphi\,, \quad
\mathbf{e}'^3 = C_6\dd\vartheta + C_5\sin\vartheta\dd\varphi\,,
\end{equation}
and then also having the following non-zero components of the spin connection
\begin{equation}\label{eq:spiconspher11}
\omega'^1{}_{2\vartheta} = -\omega'^2{}_{1\vartheta} = -1\,, \quad
\omega'^1{}_{3\varphi} = -\omega'^3{}_{1\varphi} = -\sin\vartheta\,, \quad
\omega'^2{}_{3\varphi} = -\omega'^3{}_{2\varphi} = -\cos\vartheta\,.
\end{equation}
The pair of Eqs.~\eqref{eq:tetspher1}--\eqref{eq:spiconspher11} is equivalent as taking the tetrad in Eq.~\eqref{eq:sphertetradwb} with a zero spin connection (in the Weitzenb\"{o}ck gauge).

Let us now consider that the field equations of any arbitrary \gls{tg} theory is denoted by $W_{\mu\nu}$ and its antisymmetric part is $W_{[\mu\nu]}$. This raises the question of which components of $W_{[\mu\nu]}$ are nonvanishing if we consider that the field equations respect spherical symmetry?

To respond to this question, let us first take the vector $X_z$ from the rotations in Eq.~\eqref{eq:genvecz}. The invariance of $W_{[\mu\nu]}$ to this generator will be given by
\begin{equation}\label{eq:Eaxial}
0 = (\mathcal{L}_{X_z}W)_{[\mu\nu]} = \partial_{\varphi}W_{[\mu\nu]}\,,
\end{equation}
then, all the components of $W_{[\mu\nu]}$ do not depend on $\varphi$. Furthermore, by considering the other two rotational generators $X_x$ and $X_y$ (see~\eqref{eq:genvecz}), we can compute
\footnotesize{\begin{equation}
(\cos\varphi\mathcal{L}_{X_x}W + \sin\varphi\mathcal{L}_{X_y}W)_{[\mu\nu]} = \begin{pmatrix}
0 & 0 & -(\sin\vartheta)^{-2}W_{[t\varphi]} & W_{[t\vartheta]}\\
0 & 0 & -(\sin\vartheta)^{-2} W_{[r\varphi]} & W_{[r\vartheta]}\\
(\sin\vartheta)^{-2} W_{[t\varphi]} & (\sin\vartheta)^{-2} W_{[r\varphi]} & 0 & 0\\
-W_{[t\vartheta]} & -W_{[r\vartheta]} & 0 & 0
\end{pmatrix}\,,
\end{equation}}
\normalsize which gives us that
\begin{align}
W_{[t\vartheta]}=W_{[r\vartheta]}=W_{[t\varphi]}=W_{[r\varphi]}=0\,.\label{eq:antisym1}
\end{align}
Furthermore, from the generators $X_x$ and $X_y$ we also find that

\footnotesize{\begin{equation}
(\cos\varphi\mathcal{L}_{X_y}W- \sin\varphi\mathcal{L}_{X_x}W)_{[\mu\nu]} =
\begin{pmatrix}
	0 & -\partial_{\vartheta}W_{[tr]} & 0 & 0\\
	\partial_{\vartheta}W_{[tr]} & 0 & 0 & 0\\
	0 & 0 & 0 & \cot\vartheta\, W_{[\vartheta\varphi]} - \partial_{\vartheta}W_{[\vartheta\varphi]}\\
	0 & 0 & -\cot\vartheta\, W_{[\vartheta\varphi]} + \partial_{\vartheta}W_{[\vartheta\varphi]} & 0
	\end{pmatrix}\,,
\end{equation}}
\normalsize where we have also used Eq.~\eqref{eq:antisym1}. The above equation can be easily solved by setting
\begin{equation}\label{eq:Espherical}
W_{[tr]} = \tilde{W}_{[tr]}(t, r)\,, \quad
W_{[\vartheta\varphi]} = \tilde{W}_{[\vartheta\varphi]}(t, r)\sin\vartheta\,.
\end{equation}

From here, we can conclude that the field equations $W_{\mu\nu}$ of any \gls{tg} theory in spherical symmetry will have only two nonvanishing antisymmetric components which are $W_{[tr]}$ and $W_{[\vartheta\varphi]}$. This statement is valid for any \gls{tg}.

Let us finish this section by explicitly showing the form of the torsion tensor for the tetrad in Eq.~\eqref{eq:sphertetradwb} in the Weitzenb\"{o}ck gauge. As we pointed before, symmetric tetrads respecting spherical symmetry will impose that the torsion tensor $T^{\alpha}{}_{\mu\nu}$ will also respect this symmetries. The non-zero independent components for the torsion tensor $T^{\alpha}{}_{\mu\nu}$ for the symmetric tetrad~\eqref{eq:sphertetradwb} are
\begin{subequations}\label{chp3_torsion_comp}
	\begin{align}
	T^t{}_{tr} &= \frac{C_4 \left(C_{2,t}-C_{1,r}\right)+C_2 \left(C_{3,r}-C_{4,t}\right)}{C_1 C_4-C_2 C_3}\,, \\[0.5ex]
	T^r{}_{tr} &=\frac{C_3 \left(C_{1,r}-C_{2,t}\right)+C_1 \left(C_{4,t}-C_{3,r}\right)}{C_1 C_4-C_2 C_3}\,, \\[0.5ex]
	T^t{}_{\vartheta\varphi} &= \frac{2 C_2 C_6}{C_2 C_3-C_1 C_4}\sin\vartheta\,, \\[0.5ex]
	T^r{}_{\vartheta\varphi} &= \frac{2 C_1 C_6}{C_1 C_4-C_2 C_3}\sin\vartheta\,, \\[0.5ex]
	T^{\vartheta}{}_{t\vartheta} &= T^{\varphi}{}_{t\varphi} = \frac{-C_3 C_5+C_{5,t} C_5+C_6 C_{6,t}}{C_5^2+C_6^2}\,, \\[0.5ex]
	T^{\varphi}{}_{t\vartheta} &= -\frac{T^{\vartheta}{}_{t\varphi}}{\sin^2\vartheta} = \frac{1}{\sin\vartheta}\Big[\frac{C_3 C_6-C_{5,t} C_6+C_5 C_{6,t}}{C_5^2+C_6^2}\Big]\,,\\[0.5ex]
	T^{\vartheta}{}_{r\vartheta} &= T^{\varphi}{}_{r\varphi} = \frac{-C_4 C_5+C_{5,r} C_5+C_6 C_{6,r}}{C_5^2+C_6^2}\,, \\[0.5ex]
	T^{\varphi}{}_{r\vartheta} &= -\frac{T^{\vartheta}{}_{r\varphi}}{\sin^2\vartheta} = \frac{1}{\sin\vartheta}\Big[\frac{C_4 C_6-C_{5,r} C_6+C_5 C_{6,r}}{C_5^2+C_6^2}\Big]\,,\label{eq:torsionc}
	\end{align}
	where $C_{i,r}=\partial C_i/\partial r $ and $C_{i,t}=\partial C_i/\partial t $. Note that $C_1 C_4-C_2 C_3\neq0$ always, otherwise the inverse of the metric and the inverse of the tetrad is singular. We directly notice that the torsion tensor indeed also respects spherical symmetry.
\end{subequations}

\subsubsection{Friedmann–Lema\^{i}tre–Robertson–Walker cosmological symmetry} \label{subsymmFLRW}

Here, we derive the tetrads and spin connection satisfying cosmological symmetries in a \gls{flrw} spacetime. First, we will derive the tetrads in the Weitzenb\"{o}ck gauge and then as we did in previous sections, we will perform a Lorentz transformation to find an equivalent pair constructed from tetrads and a spin connection. The generators for cosmological symmetries are given by Eq.~\eqref{eq:genvecz} plus three additional vector fields related to the translator generators, namely
\begin{subequations}\label{eq:genvectrans}
\begingroup \addtolength{\jot}{3pt}
\begin{align}
X_{1} & =\chi\sin\vartheta\cos\varphi\,\partial_{r}+\frac{\chi}{r}\cos\vartheta\cos\varphi\,\partial_{\vartheta}-\frac{\chi\sin\varphi}{r\sin\vartheta}\,\partial_{\varphi}\,,\\[0.5ex]
X_{2} & =\chi\sin\vartheta\sin\varphi\,\partial_{r}+\frac{\chi}{r}\cos\vartheta\sin\varphi\,\partial_{\vartheta}+\frac{\chi\cos\varphi}{r\sin\vartheta}\,\partial_{\varphi}\,,\\[0.5ex]
X_{3} & =\chi\cos\vartheta\,\partial_{r}-\frac{\chi}{r}\sin\vartheta\,\partial_{\vartheta}\,,
\end{align}

\endgroup
\end{subequations}
where \(\chi = \sqrt{1 - kr^2}\) and $k$ is the spatial curvature describing flat \gls{flrw} ($k=0$), or open FLRW universe ($k=-1$) or closed \gls{flrw} universe ($k=1$).

Following the same question procedure asked before, which components of $W_{[\mu\nu]}$ are nonvanishing if we consider that the field equations respect cosmological symmetries? To respond to this question, we can first take the result found in the spherical symmetry case. We find that the only nonvanishing components are $W_{[tr]}$ and $W_{[\vartheta\varphi]}$. Besides the rotational generators in spherical symmetry, in cosmological symmetries we also have the extra symmetries given by the translational generators in Eq.~\eqref{eq:genvectrans}. If we apply $X_3$ to the remaining nonvanishing antisymmetric components of the field equations, we find
\begin{equation}
    (\mathcal{L}_{X_3}W)_{[t\vartheta]} = -\chi\sin\vartheta\tilde{W}_{[tr]}\,, \quad m(\mathcal{L}_{X_3}W)_{[r\varphi]} = \frac{\sin^2\vartheta\tilde{W}_{[\vartheta\varphi]}}{\chi r^2}\,.
\end{equation}
The above equations then gives us that $W_{[tr]}=W_{[\vartheta\varphi]}=0$. Thus, independent of the \gls{tg} considered, the antisymmetric part of its field equations is always zero. This a consequence of the strong condition imposed by the cosmological symmetry. This statement is also valid for any rank 2-tensor. Therefore, if one is able to find symmetric tetrads satisfying cosmological symmetries, they will automatically solve the antisymmetric field equations in any \gls{tg}. A recent paper~\cite{Hohmann:2020zre} performed a complete classification of cosmological teleparallel geometries.

In the following subsections, we will derive the tetrads and spin connection satisfying cosmological symmetry for the three possibles cases: $k=0$, $k>0$ and $k<0$.

\paragraph{Flat case ($k=0$) - ISO(3)}

We can take the result found in spherical symmetry as a first step~\eqref{eq:sphertetradwb} in the Weitzenb\"{o}ck gauge and now consider the generator $X_3$, giving us that the functions $C_i$ must have the following form
\begin{equation}
C_1 = N(t)\,, \quad
C_4 = a(t)\,, \quad
C_5 = a(t)r\,, \quad C_2 = C_3 = C_6 = 0\,.
\end{equation}
We can identify $N(t)$ as the lapse function and $a(t)$ as the scale factor of the Universe. Thus, by replacing the above form of the functions in the tetrad~\eqref{eq:sphertetradwb}, we find that the most general tetrad satisfying cosmological symmetries in the Weitzenb\"{o}ck gauge becomes
\begin{subequations}\label{eq:cosmoftetradwb}
	\begin{align}
	    \mathbf{e}^0 &= N(t)\dd t\,,\\[0.5ex]
	    \mathbf{e}^1 &= a(t)\left[\sin\vartheta\cos\varphi\dd r + r\cos\theta\cos\varphi\dd\vartheta - r\sin\vartheta\sin\varphi\dd\varphi\right]\,,\\[0.5ex]
	    \mathbf{e}^2 &= a(t)\left[\sin\vartheta\sin\varphi\dd r + r\cos\theta\sin\varphi\dd\vartheta + r\sin\vartheta\cos\varphi\dd\varphi\right]\,,\\[0.5ex]
	    \mathbf{e}^3 &= a(t)\left[\cos\vartheta\dd r - r\sin\vartheta\dd\vartheta\right]\,.
	\end{align}
\end{subequations}
As we have shown before, this tetrad will give $W_{[\mu\nu]}=0$ for any \gls{tg} theory. If we perform the Lorentz transformation~\eqref{eq:spherlt10} we find that the new tetrad can be written as follows
\begin{equation}\label{eq:diagtetcosmoflat}
\mathbf{e}'^0 = N(t)\dd t\,, \quad
\mathbf{e}'^1 = a(t)\dd r\,, \quad
\mathbf{e}'^2 = a(t)r\dd\vartheta\,, \quad
\mathbf{e}'^3 = a(t)r\sin\vartheta\dd\varphi\,,
\end{equation}
while having a non-zero spin connection with the same components as~\eqref{eq:spiconspher11}.

Another interesting result is that the tetrad~\eqref{eq:cosmoftetradwb} can be transformed to Cartesian coordinates $(t,x,y,z)$, and gives us that the tetrad in the Weitzenb\"{o}ck gauge in Cartesian coordinates reads
\begin{equation}\label{eq:carttetcosmoflat}
\mathbf{e}^0 = N(t)\dd t\,, \quad
\mathbf{e}^1 = a(t)\dd x\,, \quad
\mathbf{e}^2 = a(t)\dd y\,, \quad
\mathbf{e}^3 = a(t)\dd z\,,
\end{equation}
or equivalently
\begin{eqnarray}
e^A{}_\mu=\textrm{diag}(N(t),a(t),a(t),a(t))\,.
\end{eqnarray}
The above tetrad in Cartesian coordinates is diagonal and is valid in the Weitzenb\"{o}ck gauge. Since the form of this tetrad is very simple, we will largely use this tetrad in the cosmology part in Sec.~\ref{sec:cosmology_in_TG} in a more practical setting. Furthermore, since this tetrad respects cosmological symmetries, we will also have $W_{[\mu\nu]}=0$ for any \gls{tg} theory. It is also useful to note that the torsion tensor in the diagonal tetrad in Cartesian coordinates becomes very simple
\begin{equation}
T^{i}{}_{0j} = NH\delta^{i}{}_{j}\,,\label{Torsionflat}
\end{equation}
where $i,j=1,2,3$ are the spatial spacetime indices and $H=\dot{a}/(Na)$ is the Hubble parameter.

Let us finish this section by noting that we still have the freedom to choose a different time coordinate \(t \mapsto \tilde{t}\) due to the fact that the time diffeomorphisms commute with the cosmological symmetry. By doing this, one can always choose a constant lapse function, which usually is chosen to be $N(t)=1$.

\paragraph{Real tetrad for $k\geq0$ - SO(4) }\label{sec:realtetrad}
The first non-trivial case for a spatially curved FLRW cosmology has the following translation generators
\begin{equation}
\boldsymbol{\lambda}^{\pm}(X_1) = \pm\begin{pmatrix}
0 & 0 & 0 & 0\\
0 & 0 & 0 & 0\\
0 & 0 & 0 & 1\\
0 & 0 & -1 & 0
\end{pmatrix}\,, \quad
\boldsymbol{\lambda}^{\pm}(X_2) = \pm\begin{pmatrix}
0 & 0 & 0 & 0\\
0 & 0 & 0 & -1\\
0 & 0 & 0 & 0\\
0 & 1 & 0 & 0
\end{pmatrix}\,, \quad
\boldsymbol{\lambda}^{\pm}(X_3) = \pm\begin{pmatrix}
0 & 0 & 0 & 0\\
0 & 0 & 1 & 0\\
0 & -1 & 0 & 0\\
0 & 0 & 0 & 0
\end{pmatrix}\,.
\end{equation}
Using these translator generators and the rotational ones (see Eq.~\eqref{eq:genvecz}), we can solve the symmetric conditions~\eqref{eq:infisymcondwb}, which gives us the spherically symmetric tetrad~\eqref{eq:sphertetradwb} and some additional conditions on the functions, namely
\begin{equation}
C_1 = N(t)\,, \quad
C_4 = \frac{a(t)}{\chi}\,, \quad
C_5 = r\chi a(t)\,, \quad
C_6 = \mp \sqrt{k}\, r^2\,a(t)\,, \quad
C_2 = C_3 = 0\,,
\end{equation}
where $k\in \{0,\pm 1\}$, giving us that the final expression of the symmetric tetrad satisfying cosmological symmetries in the Wetizen\"ock gauge being equal to
\begin{subequations}\label{eq:cosmoptetradwb}
	\begin{align}
	\mathbf{e}_{\pm}^0 &= N(t)\dd t\,,\\[0.5ex]
	\mathbf{e}_{\pm}^1 &= a(t)\left[\frac{\sin\vartheta\cos\varphi}{\chi}\dd r + r(\chi\cos\theta\cos\varphi \pm \sqrt{k} r\sin\varphi)\dd\vartheta - r\sin\vartheta(\chi\sin\varphi \mp \sqrt{k} r\cos\vartheta\cos\varphi)\dd\varphi\right]\,,\nonumber\\[0.5ex]
	\mathbf{e}_{\pm}^2 &= a(t)\left[\frac{\sin\vartheta\sin\varphi}{\chi}\dd r + r(\chi\cos\theta\sin\varphi \mp \sqrt{k} r\cos\varphi)\dd\vartheta + r\sin\vartheta(\chi\cos\varphi \pm \sqrt{k} r\cos\vartheta\sin\varphi)\dd\varphi\right]\,,\nonumber\\[0.5ex]
	\mathbf{e}_{\pm}^3 &= a(t)\left[\frac{\cos\vartheta}{\chi}\dd r - r\chi\sin\vartheta\dd\vartheta \mp \sqrt{k} r^2\sin^2\vartheta\dd\varphi\right]\,.
	\end{align}
\end{subequations}
Here, \(\chi = \sqrt{1 - kr^2}\). It should be noted that $k=1$ ($k=-1$) represents a closed (open) universe whereas $k=0$, a flat universe. As expected, when $k=0$, we have $\chi=1$ and then we recover the flat \gls{flrw} symmetric tetrad~\eqref{eq:cosmoftetradwb}. Note that the above tetrad is real only if either $k=0$ or $k=1$, so that, the tetrad becomes complex for the negatively curved $k$ (see Sec.~\ref{sec:negativsym} for another tetrad which is real for $k\geq 0$).

In this case we have two different tetrads depending on the sign chosen. By taking the following Lorentz transformation
\begin{equation}\label{eq:diagltcosmopos2}
\Lambda_{\pm}^A{}_B = \begin{pmatrix}
1 & 0 & 0 & 0\\
0 & \sin\vartheta\cos\varphi & \sin\vartheta\sin\varphi & \cos\vartheta\\
0 & \chi\cos\vartheta\cos\varphi \pm \sqrt{k}\,r\sin\varphi & \chi\cos\vartheta\sin\varphi \mp \sqrt{k}\, r\cos\varphi & -\chi\sin\vartheta\\
0 & \pm \sqrt{k}\,r\cos\vartheta\cos\varphi - \chi\sin\varphi & \chi\cos\varphi \pm \sqrt{k}\, r\cos\vartheta\sin\varphi & \mp \sqrt{k}\,r\sin\vartheta
\end{pmatrix}\,,
\end{equation}
the tetrads~\eqref{eq:cosmoptetradwb} transform as follows
\begin{equation}\label{diagtetcosmopos}
\mathbf{e}_{\pm}'^0 = N(t)\dd t\,, \quad
\mathbf{e}_{\pm}'^1 = \frac{a(t)}{\chi}\dd r\,, \quad
\mathbf{e}_{\pm}'^2 = a(t)r\dd\vartheta\,, \quad
\mathbf{e}_{\pm}'^3 = a(t)r\sin\vartheta\dd\varphi\,,
\end{equation}
which then induces a non-zero spin connection with components,
\begin{subequations}\label{eq:spiconcosmopos}
	\begin{align}
	\omega_{\pm}'^1{}_{2\vartheta} 	&= -\omega_{\pm}'^2{}_{1\vartheta} = -\chi\,, &
	\omega_{\pm}'^1{}_{2\varphi} 	&= -\omega_{\pm}'^2{}_{1\varphi} = \pm \sqrt{k}\,r\sin\vartheta\,, &
	\omega_{\pm}'^1{}_{3\vartheta} 	&= -\omega_{\pm}'^3{}_{1\vartheta} = \mp \sqrt{k}\,r\,,\\[0.5ex]
	\omega_{\pm}'^1{}_{3\varphi} 	&= -\omega_{\pm}'^3{}_{1\varphi} = -\chi\sin\vartheta\,, &
	\omega_{\pm}'^2{}_{3r} 			&= -\omega_{\pm}'^3{}_{2r} = \pm\sqrt{k}\frac{1}{\chi}\,, &
	\omega_{\pm}'^2{}_{3\varphi} 	&= -\omega_{\pm}'^3{}_{2\varphi} = -\cos\vartheta\,.
	\end{align}
\end{subequations}
As expected, the Lorentz transformation~\eqref{eq:diagltcosmopos2} becomes the same one as the one performed in Eq.~\eqref{eq:spherlt10} for the flat case, and then the spin connection~\eqref{eq:spiconcosmopos} is also the same as the one found in Eq.~\eqref{eq:spiconspher11} when $k=0$.

For the symmetric tetrad~\eqref{eq:cosmoptetradwb} in the Weitzenb\"{o}ck gauge or equivalently, by considering the tetrad in Eq.~\eqref{diagtetcosmopos} with the spin connection in Eq.~\eqref{eq:spiconcosmopos}, we find that the non-zero components of the torsion tensor $T^{\alpha}{}_{\mu\nu}$ become
	\begin{equation}
	T^{i}{}_{tj}=NH \delta^i{}_j\,,\quad T^{r}{}_{\vartheta\varphi}=\mp 2 \sqrt{k} r^2 \sin \vartheta \chi\,,\quad T^{\varphi}{}_{r\vartheta}= -\frac{T^{\vartheta}{}_{r\varphi}}{\sin^2\vartheta}=\mp 2\frac{\sqrt{k}}{\chi}\csc\vartheta\,.
\end{equation}
As expected, the flat case $k=0$ coincides with the torsion found before in Eq.~\eqref{Torsionflat} and if $k=-1$, the torsion tensor becomes complex, meaning that one needs to be careful on interpreting the physical quantities for the negatively spatial curvature. Even though the torsion tensor is complex, it is important to mention that some important scalars as the torsion scalar $T$ and the boundary term $B$ (see Eqs.~\eqref{Torsion_scalar} and \eqref{Eq:boundary_term_def}, respectively) become
\begin{equation}\label{eq:torsionnonflat}
    T = \frac{6 k}{a^2}-6 H^2\,,\quad B=-\frac{6 \dot{H}}{N}-18 H^2
\end{equation}
are real for this tetrad for any $k$. Therefore, any action constructed from these scalars will be real for this tetrad.

\paragraph{Real tetrad for $k\leq0$ - SO(1,3)} \label{sec:negativsym}

Following the same algorithm mentioned in the previous sections, there is another non-trivial translator generators satisfying cosmological symmetries, namely
\begin{equation}
\boldsymbol{\lambda}^{\pm}(X_1) = \pm\begin{pmatrix}
0 & 1 & 0 & 0\\
1 & 0 & 0 & 0\\
0 & 0 & 0 & 0\\
0 & 0 & 0 & 0
\end{pmatrix}\,, \quad
\boldsymbol{\lambda}^{\pm}(X_2) = \pm\begin{pmatrix}
0 & 0 & 1 & 0\\
0 & 0 & 0 & 0\\
1 & 0 & 0 & 0\\
0 & 0 & 0 & 0
\end{pmatrix}\,, \quad
\boldsymbol{\lambda}^{\pm}(X_3) = \pm\begin{pmatrix}
0 & 0 & 0 & 1\\
0 & 0 & 0 & 0\\
0 & 0 & 0 & 0\\
1 & 0 & 0 & 0
\end{pmatrix}\,.
\end{equation}
Then, the spherically symmetric tetrad~\eqref{eq:sphertetradwb} for this case becomes
\begin{subequations}\label{eq:cosmontetradwb}
	\begin{align}
	\mathbf{e}_{\pm}^0 &= \pm N(t)\chi\dd t \mp a(t)\sqrt{-k}\,\frac{r}{\chi}\dd r\,,\\[0.5ex]
	\mathbf{e}^1 &= a(t)\left[\sin\vartheta\cos\varphi\left(\dd r - \frac{N(t)}{a(t)}\sqrt{-k} r\dd t\right) + r\cos\vartheta\cos\varphi\dd\vartheta - r\sin\vartheta\sin\varphi\dd\varphi\right]\,,\\[0.5ex]
	\mathbf{e}^2 &= a(t)\left[\sin\vartheta\sin\varphi\left(\dd r - \frac{N(t)}{a(t)}\sqrt{-k} r\dd t\right) + r\cos\vartheta\sin\varphi\dd\vartheta + r\sin\vartheta\cos\varphi\dd\varphi\right]\,,\\[0.5ex]
	\mathbf{e}^3 &= a(t)\left[\cos\vartheta\left(\dd r - \frac{N(t)}{a(t)}\sqrt{-k}r\dd t\right) - r\sin\vartheta\dd\vartheta\right]\,,
	\end{align}
\end{subequations}
where again \(\chi = \sqrt{1 - kr^2}\) and the functions were taken to be of the form of
\begin{equation}
C_1 = \pm\chi N(t)\,, \quad
C_2 = \mp \frac{r}{\chi}\sqrt{-k}\,a(t)\,, \quad
C_3 =-\sqrt{-k} r\,N(t)\,, \quad
C_4 = a(t)\,, \quad
C_5 = ra(t)\,, \quad
C_6 = 0\,,
\end{equation}
to fulfill the symmetry condition n Eq.~\eqref{eq:infisymcondwb}. As in the previous case, this symmetric tetrad can become the flat \gls{flrw} symmetric tetrad~\eqref{eq:cosmoftetradwb} but as it happened before, the above tetrad is complex for $k=1$. Similarly as we did in the previous section, we can perform the following Lorentz transformation
\begin{equation}\label{eq:diagltcosmoneg2}
\Lambda_{\pm}^A{}_B = \begin{pmatrix}
\pm\chi & \sqrt{-k}r\sin\vartheta\cos\varphi & \sqrt{-k}r\sin\vartheta\sin\varphi &\sqrt{-k} r\cos\vartheta\\
\pm \sqrt{-k}r & \chi\sin\vartheta\cos\varphi & \chi\sin\vartheta\sin\varphi & \chi\cos\vartheta\\
0 & \cos\vartheta\cos\varphi & \cos\vartheta\sin\varphi & -\sin\vartheta\\
0 & -\sin\varphi & \cos\varphi & 0
\end{pmatrix}\,,
\end{equation}
to rewrite the tetrad~\eqref{eq:cosmontetradwb} as
\begin{equation}\label{eq:diagtetcosmoneg}
\mathbf{e}_{\pm}'^0 = N(t)\dd t\,, \quad
\mathbf{e}_{\pm}'^1 = \frac{a(t)}{\chi}\dd r\,, \quad
\mathbf{e}_{\pm}'^2 = a(t)r\dd\vartheta\,, \quad
\mathbf{e}_{\pm}'^3 = a(t)r\sin\vartheta\dd\varphi\,,
\end{equation}
but now with a non-zero spin connection which has the following components,
\begin{subequations}\label{eq:spiconcosmoneg2}
	\begin{align}
	\omega_{\pm}'^0{}_{1r} 			&= \omega_{\pm}'^1{}_{0r} = -\sqrt{-k}\frac{1}{\chi}\,, &
	\omega_{\pm}'^0{}_{2\vartheta} 	&= \omega_{\pm}'^2{}_{0\vartheta} = -\sqrt{-k}r\,, &
	\omega_{\pm}'^0{}_{3\varphi} 	&= \omega_{\pm}'^3{}_{0\varphi} =-\sqrt{-k} r\sin\vartheta\,, \\[0.5ex]
	\omega_{\pm}'^1{}_{2\vartheta} 	&= -\omega_{\pm}'^2{}_{1\vartheta} = -\chi\,, &
	\omega_{\pm}'^1{}_{3\varphi} 	&= -\omega_{\pm}'^3{}_{1\varphi} = -\chi\sin\vartheta\,, &
	\omega_{\pm}'^2{}_{3\varphi} 	&= -\omega_{\pm}'^3{}_{2\varphi} = -\cos\vartheta\,.
	\end{align}
\end{subequations}
The nonvanishing components of the torsion tensor $T^{\lambda}{}_{\mu\nu}$ for the tetrad-spin connection pair~\eqref{eq:diagtetcosmoneg}--\eqref{eq:spiconcosmoneg2} (or equivalently with the Weitzenb\"{o}ck tetrad~\eqref{eq:cosmontetradwb}) can be expressed as
\begin{subequations}\label{eq:torsionFLRWb}
	\begin{equation}
T^i{}_{0j} =\Big(H+\frac{1}{a}\sqrt{-k}\Big)N\,\delta^i{}_j\,.
\end{equation}
\end{subequations}
Note that if $k=1$, the above torsion tensor is complex. Furthermore, the situation is slightly different than the previous section since the the torsion scalar and boundary term (again see Eqs.~\eqref{Torsion_scalar} and \eqref{Eq:boundary_term_def})
\begin{equation}\label{eq:torsion_nonflat_2ndbranch}
    T=-\frac{12 \sqrt{-k} H}{a }+\frac{6 k}{a^2}-6 H^2\,,\quad B=-\frac{12 \sqrt{-k} H}{a}-\frac{6 \dot{H}}{N}-18 H^2
\end{equation}
are complex for $k>0$, meaning that any action constructed from this tetrad will be complex for the positively curved case.

Let us finish the cosmological symmetry section by emphasising again that the tetrads with zero spin connection components, namely Eq.~\eqref{eq:cosmoptetradwb} and Eq.~\eqref{eq:cosmontetradwb}, which are valid for $k\in \{0,\pm 1\}$ will solve all the antisymmetric field equations for any modified teleparallel theory of gravity. Equivalently, the tetrad-spin connection pairs given in Eqs.~\eqref{diagtetcosmopos}--\eqref{eq:spiconcosmopos} and Eqs.~\eqref{eq:diagtetcosmoneg}--\eqref{eq:spiconcosmoneg2} also satisfy all the antisymmetric field equations for any teleparallel theory. Then, we will use these tetrads in forthcoming sections to study \gls{flrw} cosmology.

\subsubsection{Maximally symmetric spacetimes - ISO(1,3), SO(1,4), SO(2,3)} \label{sec:maxisym}

A spacetime with the maximum number of symmetries can be obtained by taking the following generators can be written as
\begin{subequations}\label{eq:genvecmax}
	\begin{align}
	X_0 &= \chi\partial_t - kr\chi T_k(t)\partial_r\,,\\[0.5ex]
	X_X &= r\sin\vartheta\cos\varphi\partial_t + \chi^2\sin\vartheta\cos\varphi T_k(t)\partial_r + \frac{\cos\vartheta\cos\varphi T_k(t)}{r}\partial_{\vartheta} - \frac{\sin\varphi T_k(t)}{r\sin\vartheta}\partial_{\varphi}\,,\\[0.5ex]
	X_Y &= r\sin\vartheta\sin\varphi\partial_t + \chi^2\sin\vartheta\sin\varphi T_k(t)\partial_r + \frac{\cos\vartheta\sin\varphi T_k(t)}{r}\partial_{\vartheta} + \frac{\sin\varphi T_k(t)}{r\cos\vartheta}\partial_{\varphi}\,,\\[0.5ex]
	X_Z &= r\cos\vartheta\partial_t + \chi^2\cos\vartheta T_k(t)\partial_r - \frac{\sin\vartheta T_k(t)}{r}\partial_{\vartheta}\,,
	\end{align}
\end{subequations}
where $\chi=\sqrt{1-kr^2}$ and we have defined
\begin{equation}
T_k(t) := \begin{cases}
\tanh t & \text{for } k = 1\,,\\
t & \text{for } k = 0\,,\\
\tan t & \text{for } k = -1
\end{cases}\,.
\end{equation}
 The metric reproduced by the above tetrad becomes
\begin{equation}\label{eq:maxsymmetric}
    \dd s^2= \dd t^2 - C_k^2(t)\left[\frac{\dd r^2}{1 - kr^2} + r^2\left(\dd\vartheta^2 + \sin^2\vartheta\dd\varphi^2\right)\right]\,,
\end{equation}
where
\begin{equation}
C_k(t) := \begin{cases}
\cosh t & \text{for } k = 1\,,\\
1 & \text{for } k = 0\,,\\
\cos t & \text{for } k = -1
\end{cases}\,.
\end{equation}
Thus, $k=0$ represents Minkowski and $k=1$ ($k=-1$) de-Sitter (anti de-Sitter) spacetimes. The Minkowski case is represented by the ISO$(1,3)$ group whereas for the de-Sitter (anti de-Sitter) the group is SO$(2,3)$ (SO$(1,4)$).

It is easy to notice that for Minkowski, the tetrad in the Weitzenb\"{o}ck gauge~\eqref{eq:cosmoftetradwb} with $N(t)=a(t)=\textrm{const.}=c$, is the solution of the symmetry condition under the generators~\eqref{eq:genvecmax}, which gives us the following tetrad
\begin{subequations}\label{eq:minkowski}
	\begin{align}
	\mathbf{e}^0 &=c\,\dd t\,,\\[0.5ex]
	\mathbf{e}^1 &= c\left[\sin\vartheta\cos\varphi\dd r + r\cos\theta\cos\varphi\dd\vartheta - r\sin\vartheta\sin\varphi\dd\varphi\right]\,,\\[0.5ex]
	\mathbf{e}^2 &= c\left[\sin\vartheta\sin\varphi\dd r + r\cos\theta\sin\varphi\dd\vartheta + r\sin\vartheta\cos\varphi\dd\varphi\right]\,,\\[0.5ex]
	\mathbf{e}^3 &= c\left[\cos\vartheta\dd r - r\sin\vartheta\dd\vartheta\right]\,.
	\end{align}
\end{subequations}
One can also perform a coordinate transformation and find that in Cartesian coordinates $(t,x,y,z)$, the tetrad satisfying Minkowski symmetries in the Weitzenb\"{o}ck gauge becomes a diagonal one, namely
\begin{equation}
e^A{}_\mu=c\,\textrm{diag}(1,1,1,1)\,.
\end{equation}

Finally, it is important to mention that for de-Sitter and anti de-Sitter, there are not symmetry solutions to the symmetry condition~\eqref{eq:infisymcondwb} under the generators~\eqref{eq:genvecmax}. This means that there are no de-Sitter or anti de-Sitter symmetries in \gls{tg}. Then, the metric can be de-Sitter or anti de-Sitter and then it can respect the symmetries but the teleparallel connection cannot respect this type of symmetry.

\subsection{Conformal and disformal transformations} \label{Sec:conformal_trans}

The concept of conformal and disformal transformations arises from the question which is the most general possibility to construct a metric \(\tilde{g}_{\mu\nu}\) from a given metric \(g_{\mu\nu}\) and a scalar field \(\phi\). If one demands that the new metric depends only on the values of the old metric and scalar field, but not on their derivatives, one finds that the most general form of the new metric is given by the conformal transformation~\cite{Weyl:1921rzm}
\begin{equation}\label{eq:conf_trans}
    \tilde{g}_{\mu\nu} = \mathfrak{A}(\phi)g_{\mu\nu}\,,
\end{equation}
where \(\mathfrak{A}\) is a free function of the scalar field. If one allows also for a dependence on first derivatives of the scalar field, the most general class of transformations is of the disformal type~\cite{Bekenstein:1992pj}
\begin{equation}\label{eq:conf_trans_metr}
    \tilde{g}_{\mu\nu} = \mathfrak{A}(\phi, X)g_{\mu\nu} + \mathfrak{B}(\phi, X)\partial_{\mu}\phi\partial_{\nu}\phi\,,
\end{equation}
with two free functions \(\mathfrak{A}, \mathfrak{B}\) of the scalar field and its kinetic term
\begin{equation}
    X = -\frac{1}{2}g^{\mu\nu}\partial_{\mu}\phi\partial_{\nu}\phi\,.
\end{equation}
If the metric is defined by a tetrad \(e^A{}_{\mu}\), it is more convenient to introduce a different parametrization of disformal transformations. In this case one may define~\cite{Ezquiaga:2017ner}
\begin{equation}\label{eq:tetdisfcomp}
    \tilde{e}^A{}_{\mu} = \mathfrak{C}(\phi, X)e^A{}_{\mu} + \mathfrak{D}(\phi, X)\partial_{\mu}\phi\partial_{\nu}\phi g^{\nu\rho}e^A{}_{\rho}\,,
\end{equation}
while the spin connection is left unchanged, \(\tilde{\omega}^A{}_{B\mu} = \omega^A{}_{B\mu}\), since any transformation preserving its flatness and antisymmetry could be absorbed into a local Lorentz transformation. The relation between the different parametrizations is given by
\begin{equation}
    \mathfrak{A} = \mathfrak{C}^2\,, \quad \mathfrak{B} = 2\mathfrak{D}(\mathfrak{C} - X\mathfrak{D})\,,
\end{equation}
where here, and in the remainder of this section, we will omit the arguments \(\phi, X\) for brevity's sake. Note that we recover the class of conformal transformations for \(\mathfrak{C}_{,X} = \mathfrak{D} = 0\), where the subscript denotes the derivative \gls{wrt} the argument \(X\).

In order to obtain an invertible transformation, one must demand \(\mathfrak{C} \neq 0\) and \(\mathfrak{C} - 2X\mathfrak{D} \neq 0\). This can also be seen from the transformation behavior of the inverse tetrads, which is given by
\begin{equation}\label{eq:invtetdisfcomp}
    \tilde{E}_A{}^{\mu} = \frac{1}{\mathfrak{C}}\left(E_A{}^{\mu} - \frac{\mathfrak{D}}{\mathfrak{C} - 2X\mathfrak{D}}g^{\mu\nu}E_A{}^{\rho}\partial_{\nu}\phi\partial_{\rho}\phi\right)\,,
\end{equation}
where is found using the inverse formula~\eqref{tetrad_ortho_cond} (which can also be used for the non-trivial tetrads), and likewise from the transformation of the determinant of the tetrad, given by
\begin{equation}
    \tilde{e} = \mathfrak{C}^3(\mathfrak{C} - 2X\mathfrak{D})e\,.
\end{equation}
It is a remarkable fact that the transformation of the metric determinant takes a similarly simple form
\begin{equation}
    \tilde{g} = \mathfrak{A}^3(\mathfrak{A} - 2X\mathfrak{B})g = \mathfrak{C}^6(\mathfrak{C} - 2X\mathfrak{D})^2g\,,
\end{equation}
in terms of the original parameter functions \(\mathfrak{A}, \mathfrak{B}\). It is straightforward to derive the corresponding transformation rules for other relevant quantities in the teleparallel geometry, such as the torsion,
\begin{equation}
    \tilde{T}^A{}_{\mu\nu} = \mathfrak{C}T^A{}_{\mu\nu} + 2\partial_{[\mu}\mathfrak{C}e^A{}_{\nu]} + 2\eta^{AB}E_B{}^{\rho}\left(\partial_{\rho}\phi\partial_{[\mu}\mathfrak{D}\partial_{\nu]}\phi + \mathfrak{D}\partial_{[\nu}\phi\nabla_{\mu]}\partial_{\rho}\phi\right)\,,
\end{equation}
and contortion
\begin{alignat}{2}
\tilde{K}_{AB\mu} & =\: & & K_{AB\mu} + 2E_{[A}{}^{\alpha}E_{B]}{}^{\beta}\Bigg(\frac{\mathfrak{D}\lc{\nabla}_{\mu}\lc{\nabla}_{\alpha}\phi\lc{\nabla}_{\beta}\phi}{\mathfrak{C} - 2X\mathfrak{D}} + \frac{\mathfrak{D}^2\lc{\nabla}_{\mu}\phi\lc{\nabla}_{\alpha}\phi\lc{\nabla}_{\beta}X}{\mathfrak{C}(\mathfrak{C} - 2X\mathfrak{D})}\nonumber\\[0.5ex]
& \: & &+ \frac{\mathfrak{C}_{,X}\mathfrak{D}g_{\alpha\mu}\lc{\nabla}_{\beta}\phi\lc{\nabla}_{\gamma}\phi\lc{\nabla}^{\gamma}X}{\mathfrak{C}(\mathfrak{C} - 2X\mathfrak{D})} + \frac{\mathfrak{D}_{,X}\lc{\nabla}_{\alpha}X\lc{\nabla}_{\beta}\phi\lc{\nabla}_{\mu}\phi}{\mathfrak{C}} + \frac{\mathfrak{C}_{,X}\lc{\nabla}_{\alpha}Xg_{\beta\mu}}{\mathfrak{C}} + \frac{\mathfrak{C}_{,\phi}\lc{\nabla}_{\alpha}\phi g_{\beta\mu}}{\mathfrak{C} - 2X\mathfrak{D}}\Bigg)\,,
 \end{alignat}
 which constitute a special case of a more general class of scale transformations of the teleparallel affine connection~\cite{Iosifidis:2018zwo}. Note that the latter is independent of the teleparallel spin connection \(\omega^A{}_{B\mu}\), and can thus be obtained directly from the transformation of the Levi-Civita spin connection \(\lc{\omega}^A{}_{B\mu}\) using the relation
\begin{equation}
    0 = \tilde{\omega}^A{}_{B\mu} - \omega^A{}_{B\mu} \equiv \lc{\tilde{\omega}}^A{}_{B\mu} - \lc{\omega}^A{}_{B\mu} + \tilde{K}^A{}_{B\mu} - K^A{}_{B\mu}\,.
\end{equation}
To obtain the corresponding forms given only with spacetime indices, one must take into account the transformation of the tetrad and its inverse, which relate these different representations. Explicit formulas can be found in Ref.~\cite{Golovnev:2019kcf}.

A more compact form of the disformal transformations can be obtained by using the language of differential forms. For this purpose one defines the one-forms~\cite{Hohmann:2019gmt}
\begin{equation}
\boldsymbol{\psi}_A = \phi_{,A}\dd\phi\,, \quad
\boldsymbol{\pi}_A = \DD\phi_{,A} = \dd\phi_{,A} - \boldsymbol{\omega}^B{}_A \wedge \phi_{,B}\,, \quad
\lc{\boldsymbol{\pi}}_A = \lc{\DD}\phi_{,A} = \dd\phi_{,A} - \lc{\boldsymbol{\omega}}^B{}_A \wedge \phi_{,B}\,,
\end{equation}
using the abbreviation
\begin{equation}
\phi_{,A} = E_A{}^{\mu}\partial_{\mu}\phi\,.
\end{equation}
In terms of these one-forms, the disformal transformation~\eqref{eq:tetdisfcomp} takes the simple form
\begin{equation}
    \tilde{\mathbf{e}}^A = \mathfrak{C}\mathbf{e}^A + \mathfrak{D}\boldsymbol{\psi}^A\,,
\end{equation}
while the corresponding dual vector fields obey the transformation
\begin{equation}
    \tilde{\mathbf{E}}_A = \frac{1}{\mathfrak{C}}\left(\delta_A^B - \frac{\mathfrak{D}}{\mathfrak{C} - 2X\mathfrak{D}}\phi_{,A}\phi_{,C}\eta^{BC}\right)\mathbf{E}_B\,,
\end{equation}
which is obtained from the corresponding formula in Eq.~\eqref{eq:invtetdisfcomp}. Finally, we note that the torsion two-form transforms as
\begin{equation}
    \tilde{\mathbf{T}}^A = \mathfrak{C}\mathbf{T}^A + \mathfrak{C}_{,\phi}\dd\phi \wedge \mathbf{e}^A + \mathfrak{C}_{,X}\dd X \wedge \mathbf{e}^A + \mathfrak{D}\boldsymbol{\pi}^A \wedge \dd\phi + \mathfrak{D}_{,X}\dd X \wedge \boldsymbol{\psi}^A\,,
\end{equation}
while for the contortion we have
\begin{alignat}{2}
\tilde{\mathbf{K}}_{AB} & =\: & & \mathbf{K}_{AB} + 2\Bigg(-\frac{\mathfrak{D}\phi_{,[A}\lc{\boldsymbol{\pi}}_{B]}}{\mathfrak{C} - 2X\mathfrak{D}} - \frac{\mathfrak{D}^2X_{,[A}\phi_{,B]}\,\dd\phi}{\mathfrak{C}(\mathfrak{C} - 2X\mathfrak{D})}\nonumber\\[0.5ex]
& \: & & - \frac{\mathfrak{C}_{,X}\mathfrak{D}g^{-1}(\dd\phi, \dd X)\phi_{,[A}\mathbf{e}_{B]}}{\mathfrak{C}(\mathfrak{C} - 2X\mathfrak{D})} + \frac{\mathfrak{D}_{,X}X_{,[A}\phi_{,B]}\,\dd\phi}{\mathfrak{C}} + \frac{\mathfrak{C}_{,X}X_{,[A}\mathbf{e}_{B]}}{\mathfrak{C}} + \frac{\mathfrak{C}_{,\phi}\phi_{,[A}\mathbf{e}_{B]}}{\mathfrak{C} - 2X\mathfrak{D}}\Bigg)\,.
\end{alignat}
Similar expressions may also be derived for the transformation behavior of the one-forms \(\boldsymbol{\psi}_A\) and \(\boldsymbol{\pi}_A\) introduced above~\cite{Hohmann:2019gmt}.

\subsection{The teleparallel Gauss-Bonnet invariant}\label{Sec3:TeleGB}

In $(3+1)-$dimensions, the Gauss-Bonnet term is a topological invariant quantity that does not contribute dynamically since it is a total divergence term in the gravitational action. For a standard gravity scenario, based on the Levi-Civita connection, this boundary term takes on the form \cite{nakahara2003geometry}
\begin{equation}
	\lc{\mathcal{G}} := \lc{R}^2 - 4\lc{R}_{\mu\nu}\lc{R}^{\mu\nu} + \lc{R}_{\mu\nu\kappa\lambda}\lc{R}^{\mu\nu\kappa\lambda}\,,
\end{equation}
where the variation $\delta\left[\int\,\dd ^4 x \sqrt{-g} \lc{\mathcal{G}}\right]/\delta g^{\mu\nu}$ naturally vanishes as a boundary term. In this setup, the Gauss-Bonnet invariant is derived from first principles through the Gauss-Bonnet theorem which connects the curvature of surfaces in differential geometry with their Euler characteristic in topology. The teleparallel equivalent of the Gauss-Bonnet (\gls{tegb}) invariant is found by transforming the Gauss-Bonnet term $\lc{\mathcal{G}}$ through the contortion tensor to its teleparallel equivalent $T_G$. This procedure will produce a dynamical scalar $T_G$ as well as a total divergence term $B_G$ which together will sum to the Gauss-Bonnet scalar $\lc{\mathcal{G}} = T_G + B_G$ \cite{Bahamonde:2016kba,delaCruz-Dombriz:2017lvj}. This is the fundamental difference between the scalar $\lc{\mathcal{G}}$ and $T_G$. A teleparallel analogue to the Gauss-Bonnet theorem may lead to a separate scalar invariant that does not depend on the results of standard gravity.

The \gls{tegb} term was first derived in Ref.~\cite{Kofinas:2014owa} using the language of forms, since the coordinate description leads to much more cumbersome equations. In standard gravity, the Gauss-Bonnet invariant is expressed in this context as
\begin{equation}\label{chp3_LC_GB_term}
    \lc{\mathcal{G}} = \epsilon_{abcd} \lc{R}^{ab} \wedge \lc{R}^{cd}\,,
\end{equation}
such that the variation of
\begin{equation}
    \lc{S}_{\lc{\mathcal{G}}} := \frac{1}{2\kappa^2} \int \lc{\mathcal{G}}+\mathcal{S}_{\rm m}\,,
\end{equation}
vanishes, where $\kappa^2 = 8\pi G$ (as described in Eq.~\ref{Eq:Con_units}) is the gravitational constant (whose value is not relevant here), and the exterior (wedge) product is again used. Using the relation between the Levi-Civita and teleparallel connections in Eq.~\eqref{chp3_spin_iden}, namely
\begin{equation}
    \udt{\lc{\Gamma}}{C}{AB} = \udt{\Gamma}{C}{BA} + \udt{K}{C}{BC}\,,
\end{equation}
together with the following properties
\begin{itemize}
    \item $\lc{\rm D} \lc{R}^{AB} = 0$ -- The covariant derivative of the Lorentz curvature, $\lc{\rm D} \lc{R}^{AB} = \dd R^{AB} + \udt{\omega}{A}{C}\wedge R^{CB} + \udt{\omega}{B}{C}\wedge R^{AC}$, vanishes which gives the second Bianchi identity in the language of differential forms \cite{Gasperini:1517921};
    \item $\lc{\rm D}\mathbf{e}^a = 0$ -- The Levi-Civita Lorentz covariant derivative is torsionless and satisfies metricity, and so the tetrad exterior derivative will identically vanish;
    \item $\lc{R}^{AB} + \lc{\rm D}K^{AB} = R^{AB} - \udt{K}{A}{C}\wedge K^{CB}$ -- This describes how the Ricci tensors as calculated with both connections are related together which can be observed from Eq.~\eqref{chp3_Riemann_equiv}. The Levi-Civita Lorentz derivative term can be expressed in terms of the teleparallel connection through $\lc{\rm D}\udt{K}{A}{B} = {\rm D}\udt{K}{A}{B} - 2\udt{K}{A}{C}\wedge\udt{K}{C}{B}$ which directly expresses the Levi-Civita Ricci tensor in terms of teleparallel quantities as
    \begin{equation}
        \udt{\lc{R}}{A}{B} = \udt{R}{A}{B} + \udt{K}{A}{C}\wedge\udt{K}{C}{B} - {\rm D}\udt{K}{A}{B}\,.
    \end{equation}
\end{itemize}
These properties combine to re-express the Levi-Civita based Gauss-Bonnet invariant in Eq.~\eqref{chp3_LC_GB_term} as \cite{Kofinas:2014owa,Kofinas:2014daa}
\begin{equation}
    S_{\rm TEGB} := \frac{1}{2\kappa^2}\int \left(\mathcal{T}_G - \dd \mathcal{B}_G \right)+\mathcal{S}_{\rm m}\,,
\end{equation}
where the integrands can be expressed as
\begin{subequations}
\begin{align}
    \mathcal{T}_G &:= T_G\,\mathbf{e}^1\wedge \mathbf{e}^2\wedge \mathbf{e}^3\wedge \mathbf{e}^4\,,\\[0.5ex]
    \mathcal{B}_G &:= \epsilon_{ABCD} \left(2K^{AB}\wedge \lc{R}^{CD} + K^{AB}\wedge \lc{\rm D} K^{CD}\right)\,,
\end{align}
\end{subequations}
in which the \gls{tegb} Lagrangian is given by
\begin{alignat}{2} \label{eq:gb_scalar}
    T_G & =\: & \delta^{\mu\nu\sigma\lambda}_{\alpha\beta\gamma\epsilon}K^\alpha{}_{\chi\mu}K^{\chi \beta}{}_\nu K^\gamma{}_{\xi \sigma}K^{\xi \epsilon}{}_\lambda+2\delta^{\mu\nu\sigma\lambda}_{\alpha\beta\gamma\epsilon}K^{\alpha\beta}{}_\mu K^\gamma{}_{\chi\nu}K^{\chi\epsilon}{}_{\xi}K^\xi{}_{\sigma\lambda}+2\delta^{\mu\nu\sigma\lambda}_{\alpha\beta\gamma\epsilon} K^{\alpha\beta}{}_\mu K^{\gamma}{}_{\chi\nu} D_\lambda K^{\chi \epsilon}{}_\sigma\,.
\end{alignat}
In terms of coordinates rather than differential forms, these scalar invariants can then be expressed as \cite{Bahamonde:2016kba}
\begin{equation}\label{chp3_GB_scalars}
    \mathcal{\lc{G}} = T_G + B_G\,,
\end{equation}
where the boundary term is given by~\cite{Bahamonde:2016kba}
\begin{equation}\label{eq:gb_boundary_term}
    B_G = \frac{1}{e}\partial_{\mu}
\Big[e\delta^{\mu\nu\sigma\lambda}_{\alpha\beta\gamma\epsilon}K^{\alpha\beta}{}_{\nu}\Big(  K^{\gamma}{}_{\xi \sigma}K^{\xi \epsilon}{}_\lambda-\frac{1}{2}\lc{R}^{\gamma\epsilon}{}_{\sigma\lambda}\Big)\Big]\,,
\end{equation}
with $D_\lambda K^{\chi \epsilon}{}_\sigma=\partial_\lambda K^{\chi \epsilon}{}_\sigma+(\mathring{\Gamma}+K)^\chi{}_{\beta\lambda}K^{\beta \epsilon}{}_\sigma+(\mathring{\Gamma}+K)^\epsilon{}_{\beta\lambda}K^{ \chi\beta}{}_\sigma-(\mathring{\Gamma}+K)^\beta{}_{\sigma\lambda}K^{\chi\epsilon}{}_\beta$. This term is a total divergence term. Thus, the total Levi-Civita Gauss-Bonnet invariant can be written in terms of the teleparallel connection through the scalars in Eq.~\eqref{chp3_GB_scalars} which contain a teleparallel scalar and a total divergence term. As in standard gravity, these scalars can be used to construct theories of gravity \cite{delaCruz-Dombriz:2017lvj,delaCruz-Dombriz:2018nvt,Zubair:2015yma}, and may also lead to a purely \gls{tegb} invariant constructed purely within \gls{tg}. Another important aspect of the Gauss-Bonnet term formulated in Eq.~\eqref{eq:gb_scalar} is that $T_G$ is a topological invariant in 4 dimensions (as $\lc{G}$) and $B_G$ is a boundary term in all dimensions. also would be a boundary term.

\subsection{Analogue with other branches of physics}\label{sec:crystals}

\gls{tg} offers a new perspective on interpreting gravitational interactions that revisits the foundations on which gravity is built. Similar to the introduction of curvature as the mediator of gravitation in \gls{gr}, attempts have been made in other branches of physics to utilize this description, and its geometric baggage, in advancing or uniting these field descriptions. One important direction in which the mechanics of geometric gravity has been used to describe other phenomena is that of lattice structures \cite{doi:10.1098/rspa.1955.0171,Jizba:2009qf,Klinkhamer:2018mmk} which can be applied within the trinity of geometric structures, namely Riemann curvature \cite{Kleinert:2003za}, torsion \cite{Nissinen:2019pbm,Nissinen:2018dnq,boehmer2012gauge} and non-metricity.

\begin{figure}[htp!]
    \centering
    \includegraphics[scale=0.8]{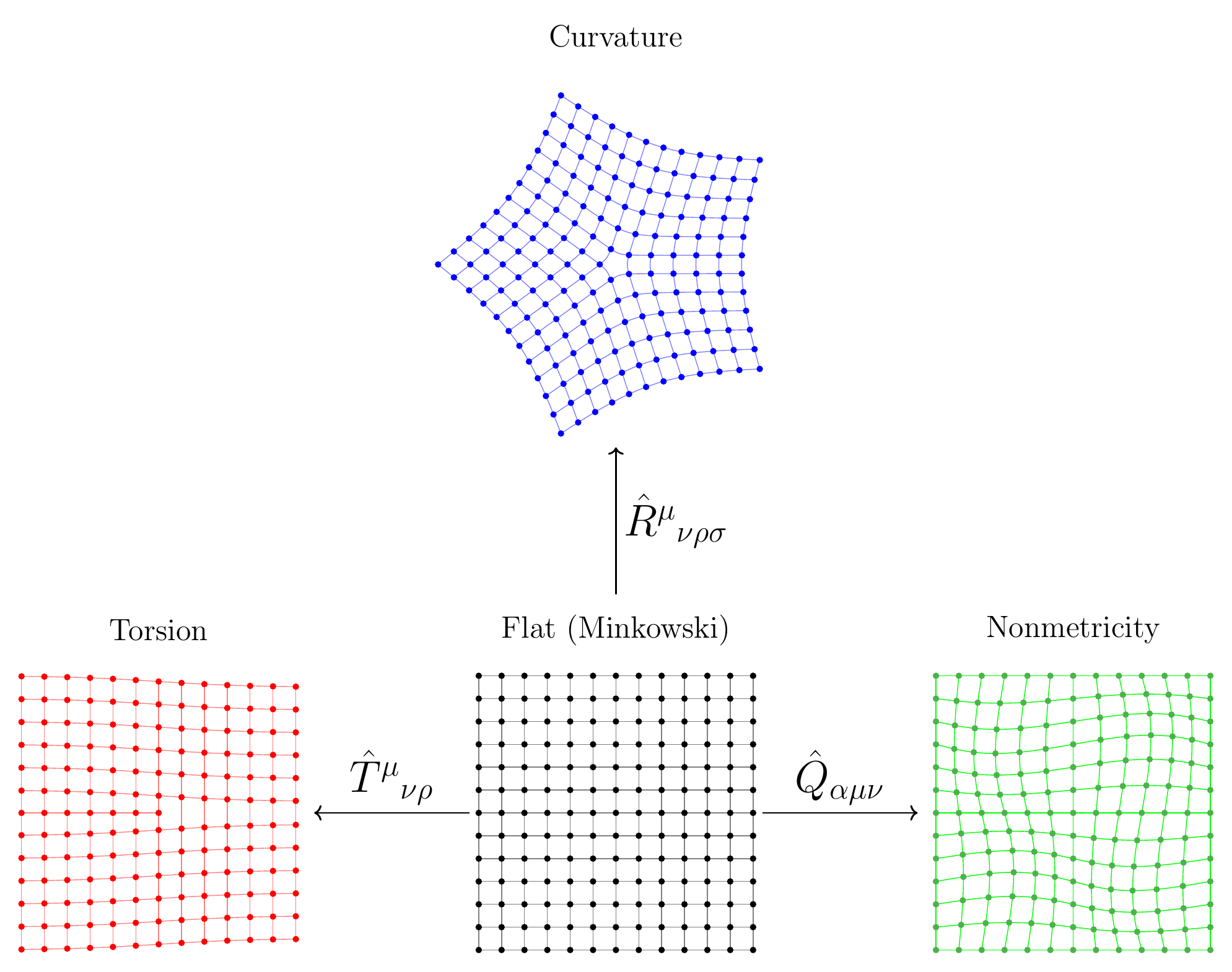}
    \caption{Crystalline structure and its analogy with curvature, torsion and non-metricity.}
    \label{fig:analogyCM}
\end{figure}

\begin{figure}[htp!]
    \centering
    \includegraphics[width=0.32\textwidth]{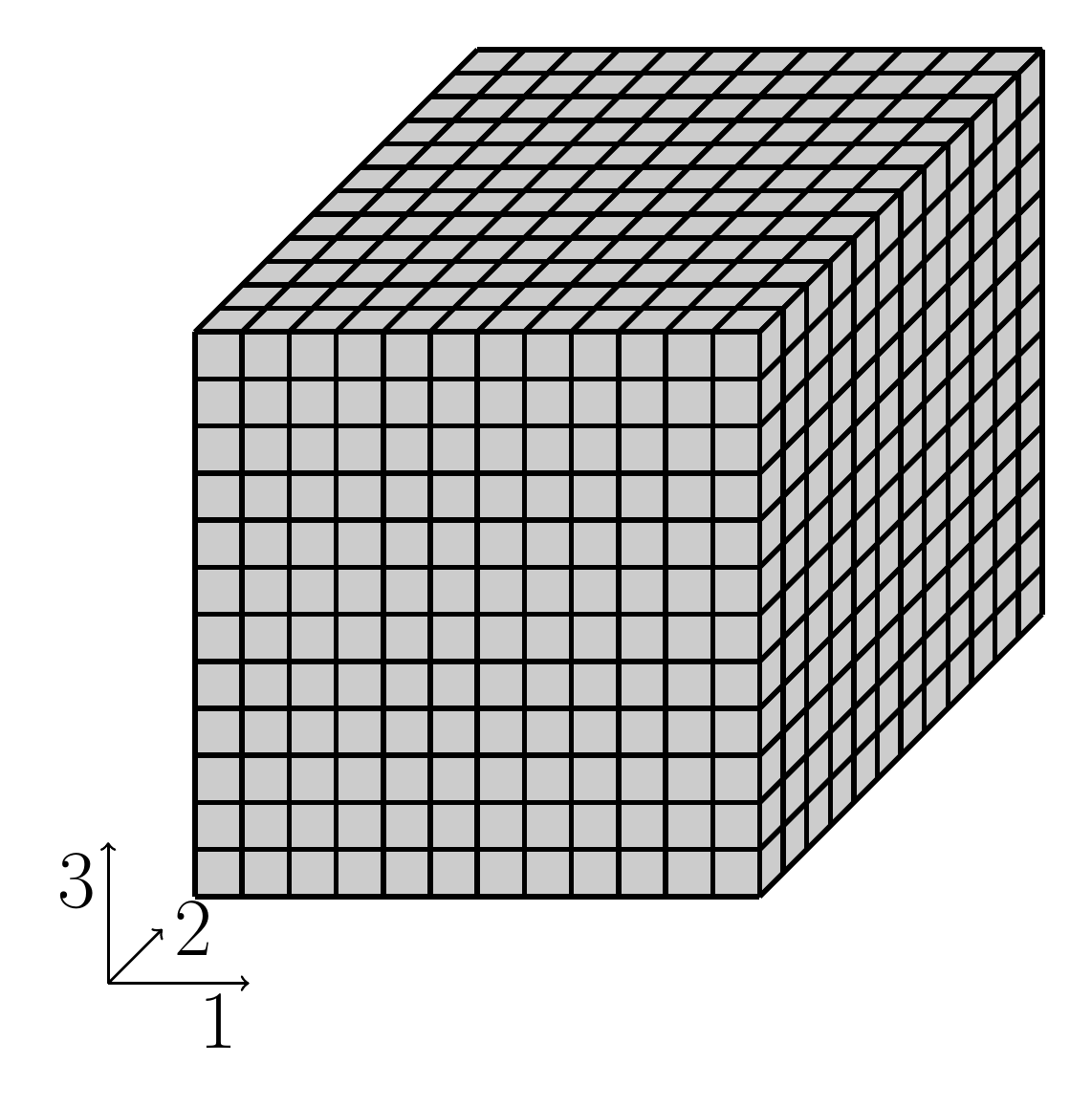}
    \includegraphics[width=0.32\textwidth]{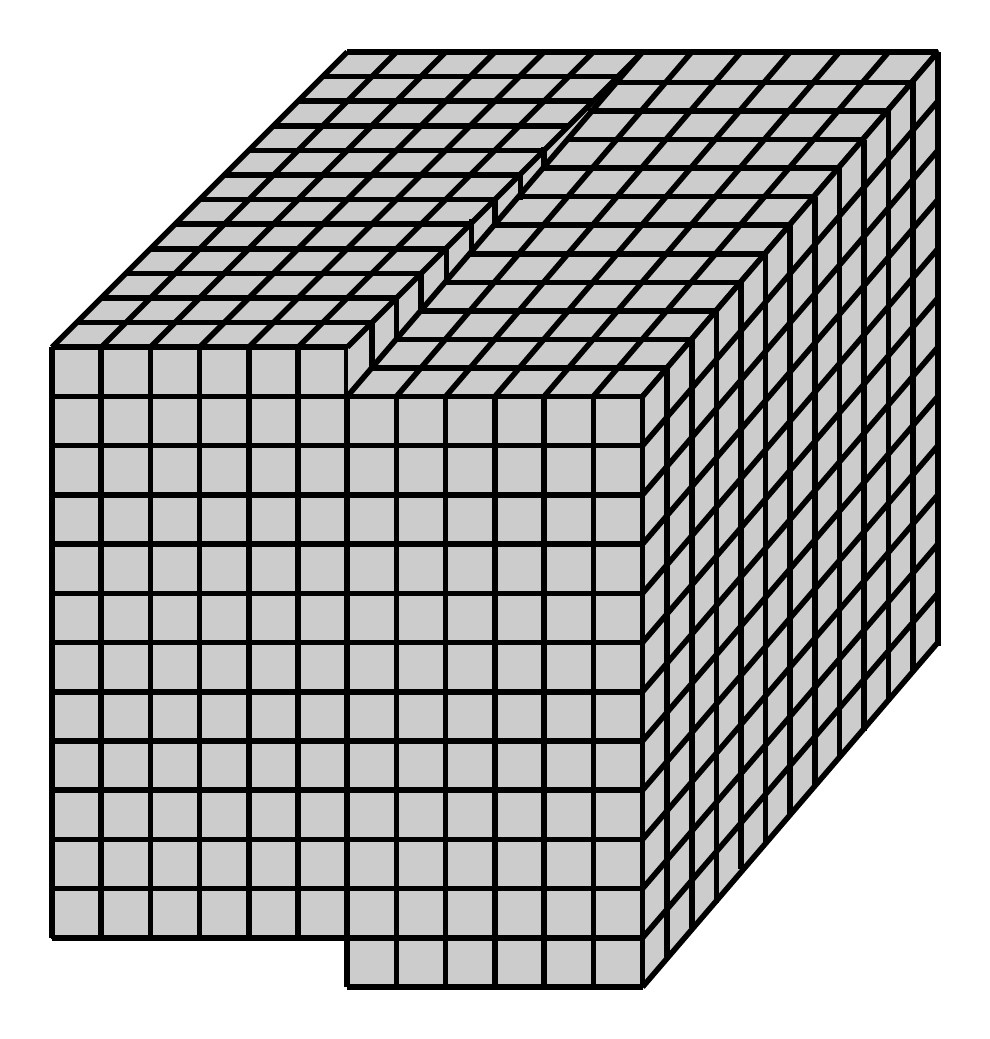}
    \includegraphics[width=0.32\textwidth]{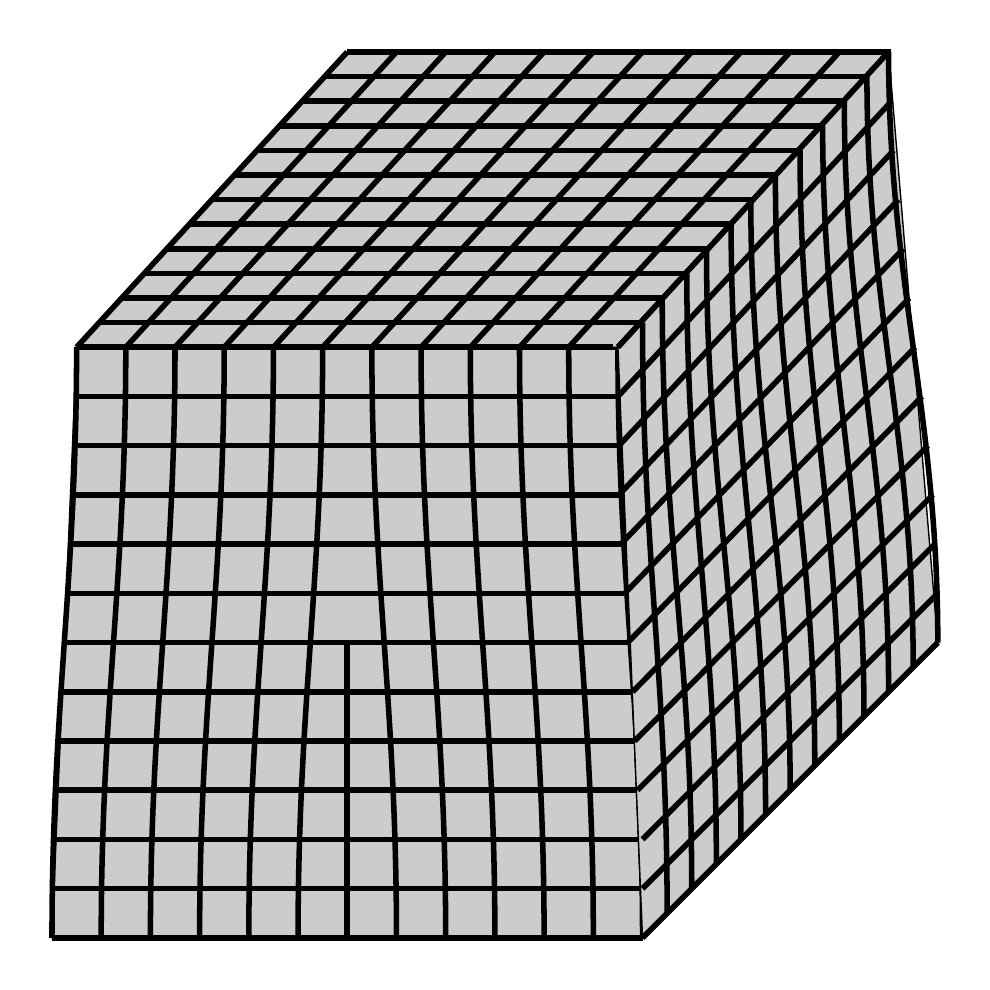}
    \caption{Crystal dislocations are shown against a regular crystal with no dislocation (left), and where \textit{screw} (middle) and \textit{edge} (right) dislocations are represented.}
    \label{fig:dislocation_cubes}
\end{figure}

In the context of continuum mechanics, the geometric tools of gravity have been used to describe various effects in continuum mechanics. The different geometric tools available within the trinity of gravity can describe different types of possible deformations within these structures as shown in Fig.~\ref{fig:analogyCM}, where deformations are made to an analogous Minkowski crystal. In Fig.~\ref{fig:analogyCM} curvature, torsion and metricity are shown to produce distinct and non-overlapping changes to crystal geometry. For continuum mechanics, deformations can be studied through this prism of a trinity of geometric tools.

For torsional geometry, the tools being developed can be used to describe dislocations which are crystallographic defects, or irregularities, in the crystal structure \cite{hull2011introduction}. Dislocation movements appear as microstructures glide over each other producing geometric information that can be interpreted through the tensorial formalism associated with gravitational theories. The creation and movement of a large number of dislocations can lead to plastic deformations, while more generally, the number and arrangement of dislocations can effect the properties associated with a material body. Thus, through the lens of geometry the impact of dislocations can be studied through the geometric tools being developed in gravitational physics.

The dislocations in these structures take on the form of particle-like bodies in that their behavior can be modeled in this way. As identified in Ref.~\cite{Frank1949}, this phenomenon can be described by a deformation field is represented
\begin{align}
	u_x = 0\,, \quad u_y = 0\,,\quad u_z = \frac{b}{2\pi}\arctan\frac{y}{x}\,,
\end{align}
which illustrates a screw dislocation in Cartesian coordinates, and where $b$ is the maximal displacement between the microstructures, and $u = (u_x, u_y, u_z)$ is the dislocation velocity. This is a solution to the Navier-Stokes equation and produces a longitudinal length contraction by a factor $\sqrt{1-\frac{v^2}{c_s^2}}$. The dislocation also has an associated energy of $E_0/\sqrt{1-\frac{v^2}{c_s^2}}$ ($E_0$ being the rest energy of the dislocation). The transverse speed of sound is represented by $c_s$, which when identified with the speed of light, renders the motion analogous to that of an electric field undergoing contractions or expansions. In this way, propagating dislocation solutions behave exactly as relativistic particles.

In Ref.~\cite{Kondo1952}, torsion is first related with dislocations in that their density and dynamics could be mapped by the torsion tensor. In Fig.~\ref{fig:dislocation_cubes}, the two main types of dislocations are represented, namely \textit{edge} and \textit{screw} type dislocations. As shown in the axis defined in the figure, we take $1$, $2$ and $3$ to point to the right, backwards and up respectively. Hence, an edge dislocation can be interpreted as the $1-3$ plane being shifted in the $1-$axis, and so giving a contribution to $\udt{T}{1}{13}$ component of the torsion tensor. Similarly, screw dislocations can be indicated as the $1-2$ plane being shifted in the $3-$axis direction, and thus giving a contribution to $\udt{T}{3}{12}$. In the context, torsion then becomes a measure of dislocation density for a vanishing lattice spacing scenario. There has been a number of important works that build on this approach to using torsion for dislocation theory \cite{10.2307/99752,Unzicker:2000gg,Zubkov:2019vhf,Lazar:2000gw,Nissinen:2018dnq}. On the other hand, it may also be the case that the torsion tensor can be used on a discrete space which may better model certain crystal interactions.

Lattice discretizations using the \gls{tg} framework has also been applied using Regge calculus as in Refs.~\cite{CASELLE1989457,Schmidt:2001vd} where dislocations are added with the curvature associated disclinations to form a fuller theory of discrete lattice structures using differential geometry. Another approach was attempted in Ref.~\cite{Pereira:2002ff} where the usual Regge calculus is used within a torsional framework in which the tetrads are used to replace the metric. This approach is then used to replace the Riemann scalar description with a \gls{tg} approach. More recently, a more direct quantum gravity approach was used in Ref.~\cite{Zubkov:2003jp} where the Abelian gauge group of translations is used immediately as the lattice variable without the use of additional structures.

In closing, \gls{tg} has reached into a number of seemingly unrelated topics but which share a geometric connection through which the mechanics developed in the theory can be more broadly applied. One other example of this is the body of work related to Cosserat mechanics where the geometric manifold is constrained to not express dislocations but only rotations which is a long studied topic in continuum mechanics \cite{cosserat1909}. This has been strongly connected to the mechanics of \gls{tg} by associating the tetrad again with the manifold deformations and using the torsion tensor to express actions on the manifold \cite{Burnett:2008bx,Burnett:2009pw,Chervova:2009kf,Chervova:2010ig,Burnett:2010fb,10.1093/qjmam/hbr011,doi:10.1098/rspa.2011.0718,10.1093/qjmam/hbr011,2020arXiv200606800B}. Another would be in condensed matter media where low-energy fermions act like an effective and non-relativistic torsional medium \cite{Nissinen:2019kld,Nissinen:2019mkw,Laurila:2020yll}. While the governing equations are different to those of gravity, it is another application of this geometric outlook in another setting. On the other hand, there has not been much progress of an analogous interpretation of non-metricity in terms of crystal structures, which may be an interesting area to focus future works on the topic.

\clearpage

\section{The Teleparallel Formulation of Gravity and its General Relativity Equivalent}\label{sec4:TEGRNGR}

In this section, we explore the theoretical foundations on which we can construct \gls{tg} theories. Ultimately, this is the goal of gravitational physics, namely, producing theories of gravity on which to probe Nature. Firstly, we describe some general properties of teleparallel theories of gravity in Sec.~\ref{ssec:tggenprop}, such as the local Lorentz structure of the gravitational and matter contributions together with other symmetry and conservation considerations. In Sec.~\ref{ssec:admformalism}, we expand on the $3+1$ decomposition of teleparallel theories and the procedure by which this can be developed. The \gls{tegr} is then extensively described in Sec.~\ref{ssec:tegr} where we derive this from first principles depending on only a few physical assumptions, after which we then analyze the properties of this theory and its dynamical equivalent to \gls{gr}. Finally, we discuss some quantum gravity attempts in Sec.~\ref{ssec:quantumhighdim} where we also highlight some future possibilities for the theory.

\subsection{General properties of teleparallel gravity}\label{ssec:tggenprop}
Before we come to a detailed study of specific \gls{tg} theories and their properties, we discuss a number of general properties which can be shown to hold for any \gls{tg} theory. The starting point for these considerations is a generic teleparallel action and its field equations, which we derive in Sec.~\ref{sssec:tggenactfield}. We then derive a number of properties from the local Lorentz invariance of the matter action in Sec.~\ref{sssec:genmatactlorinv}, as well as the gravitational action in Sec.~\ref{sssec:gengravactlorinv}. In Sec.~\ref{sssec:palatini}, we show how these properties appear within the Palatini approach. They are further used in Sec.~\ref{sssec:genenmomcons}, where we discuss the energy-momentum conservation. The latter is closely related to the Bianchi identities, which we discuss in Sec.~\ref{sssec:genbianchi}. Finally, we study the premetric approach in Sec.~\ref{sssec:genpremetric}.

\subsubsection{Action and field equations} \label{sssec:tggenactfield}

In \gls{tg} one most commonly assumes an action which of the form
\begin{equation}\label{eq:matgravaction}
    \mathcal{S}_{\text{TG}} := \mathcal{S}_{\text{g}}[e, \omega] + \mathcal{S}_{\text{m}}[e, \chi]\,,
\end{equation}
where the gravitational part \(\mathcal{S}_{\text{g}}\) of the action depends on the tetrad \(e^A{}_{\mu}\) and the spin connection~\(\omega^A{}_{B\mu}\), while the matter part depends on the tetrad \(e^A{}_{\mu}\) and arbitrary matter fields \(\chi^I\), but not on the spin connection; see Ref.~\cite{Koivisto:2018aip,BeltranJimenez:2020sih} for an extensive discussion. In brief the reason for this is that we assume that the hypermomentum vanishes which is one way of imposing that this coupling does not occur. If it were to depend on spin then we would be effective introducing a second matter tensor which would result from the variation of the matter Lagrangian \gls{wrt} the spin connection. The variation of the matter part of the action, after integration by parts to remove any derivatives acting on the field variations, can thus be written in the form
\begin{equation}\label{eq:matactvar}
    \delta\mathcal{S}_{\text{m}} = \int\dd^4x\,e\,(\Theta_A{}^{\mu}\delta e^A{}_{\mu} + \Omega_I\delta\chi^I)\,,
\end{equation}
where \(\Omega_I = 0\) are the matter field equations and \(\Theta_A{}^{\mu}\) is the energy-momentum tensor~\eqref{Eq:Con_EM_ten}. The corresponding variation of the gravitational action takes the form
\begin{equation}\label{eq:gravactvar}
    \delta\mathcal{S}_{\text{g}} = -\int\dd^4x\,e\,(W_A{}^{\mu}\delta e^A{}_{\mu} + Y_A{}^{B\mu}\delta\omega^A{}_{B\mu})\,,
\end{equation}
with tensors \(W_A{}^{\mu}\) and \(Y_A{}^{B\mu}\) arising from the variation and integration by parts, whose explicit form depends on the particular theory under consideration. The expression for \(W_A{}^{\mu}\) will be shown for a number of theories in sections~\ref{ssec:tegr} and~\ref{sec5:extended}; see~\eqref{eq:tegrfield} for a simple example. We do not show \(Y_A{}^{B\mu}\) here for brevity, since it turns out to be redundant in the derivation of the field equations, so that the latter are fully determined from $W_A{}^{\mu}$ alone. We will show this explicitly in the remainder of this section and in section~\ref{sssec:gengravactlorinv}.

From the variation \gls{wrt} the tetrad one obtains the field equations
\begin{equation}\label{eq:gentetradfieldlor}
W_A{}^{\mu} = \Theta_A{}^{\mu}\,.
\end{equation}
Another, more common form of the field equations is obtained by transforming the first index into a spacetime index with the tetrad, while lowering the second index, to obtain
\begin{equation}
W_{\mu\nu} = e^A{}_{\mu}g_{\rho\nu}W_A{}^{\rho}\,, \quad
\Theta_{\mu\nu} = e^A{}_{\mu}g_{\rho\nu}\Theta_A{}^{\rho}\,,
\end{equation}
so that the field equations take the form
\begin{equation}\label{eq:gentetradfield}
W_{\mu\nu} = \Theta_{\mu\nu}\,.
\end{equation}
The derivation of the field equations for the spin connection, however, is less trivial, since by definition it must be flat, $R^{\alpha}{}_{\beta\mu\nu} = 0$, and metric-compatible, $\nabla_{\alpha}g_{\mu\nu} = 0$. Different possibilities exist in order to maintain this property during the variation procedure \cite{Golovnev:2017dox,Hohmann:2021fpr}. One possibility is to explicitly restrict the variation of the spin connection such that its curvature and non-metricity are unaffected. This is the case if and only if the variation obeys the form
\begin{equation}
\delta\omega^A{}_{B\mu} = \DDD_{\mu}\xi^A{}_B = \partial_{\mu}\xi^A{}_B + \omega^A{}_{C\mu}\xi^C{}_B - \omega^C{}_{B\mu}\xi^A{}_C\,,
\end{equation}
where \(\xi^A{}_B\) are the components of an antisymmetric matrix, \(\xi^{(AB)} = 0\), where the index has been raised with the Minkowski metric. The variation \(\delta_{\omega}\mathcal{S}_{\text{g}}\) of the gravitational action \gls{wrt} the spin connection, while keeping the tetrad \(e^A{}_{\mu}\) fixed, then reads~\cite{Hohmann:2017duq}
\begin{equation}
\delta_{\omega}\mathcal{S}_{\text{g}} = -\int\dd^4x\,e\,Y_A{}^{B\mu}(\partial_{\mu}\xi^A{}_B + \omega^A{}_{C\mu}\xi^C{}_B - \omega^C{}_{B\mu}\xi^A{}_C)\,.
\end{equation}
After integration by parts, and applying the covariant divergence formula
\begin{equation}
\partial_{\mu}(eX^{\mu}) = e\lc{\nabla}_{\mu}X^{\mu} = e\left(\partial_{\mu}X^{\mu} + \lc{\Gamma}^{\mu}{}_{\nu\mu}X^{\nu}\right)\,,
\end{equation}
to the vector field \(X^{\mu} = Y_A{}^{B\mu}\xi^A{}_B\), this takes the form
\begin{equation}\label{eq:genactvarspicon}
\delta_{\omega}\mathcal{S}_{\text{g}} = \int\dd^4x\,e\left(\partial_{\mu}Y_A{}^{B\mu} + \lc{\Gamma}^{\mu}{}_{\nu\mu}Y_A{}^{B\nu} - \omega^C{}_{A\mu}Y_C{}^{B\mu} + \omega^B{}_{C\mu}Y_A{}^{C\mu}\right)\xi^A{}_B\,.
\end{equation}
Finally, recalling that the variation \(\xi^A{}_B\) must be antisymmetric, the resulting field equations therefore take the form
\begin{equation}\label{eq:genspiconfield2}
    0 = -\tilde{W}^{AB} = \partial_{\mu}Y^{[AB]\mu} + \lc{\Gamma}^{\mu}{}_{\nu\mu}Y^{[AB]\nu} - \omega^{C[A}{}_{\mu}Y_C{}^{B]\mu} + \omega^{[B}{}_{C\mu}Y^{A]C\mu}\,.
\end{equation}
This equation is more conveniently expressed in spacetime indices. For this purpose, note that the first term on the \gls{rhs} of the field equation~\eqref{eq:genspiconfield2} can be rewritten using
\begin{equation}
    \partial_{\mu}Y^{[AB]\mu} = \partial_{\mu}Y^{[\rho\sigma]\mu}e^A{}_{\rho}e^B{}_{\sigma} + Y^{\rho\sigma\mu}\partial_{\mu}e^{[A}{}_{\rho}e^{B]}{}_{\sigma} + Y^{\rho\sigma\mu}e^{[A}{}_{\rho}\partial_{\mu}e^{B]}{}_{\sigma}\,.
\end{equation}
After transforming the third term in~\eqref{eq:genspiconfield2} by using the antisymmetry of the spin connection, from which it follows
\begin{equation}
-\omega^{C[A}{}_{\mu}Y_C{}^{B]\mu} = \omega^{[A|C|}{}_{\mu}Y_C{}^{B]\mu} = \omega^{[A}{}_{C\mu}Y^{|C|B]\mu}\,,
\end{equation}
the derivatives acting on the tetrad can be combined with the spin connection to form the connection coefficients
\begin{subequations}
\begin{align}
Y^{\rho\sigma\mu}\partial_{\mu}e^{[A}{}_{\rho}e^{B]}{}_{\sigma} + \omega^{[A}{}_{C\mu}Y^{|C|B]\mu} &= \Gamma^{\nu}{}_{\rho\mu}Y^{\rho\sigma\mu}e^{[A}{}_{\nu}e^{B]}{}_{\sigma}\,,\\[0.5ex]
Y^{\rho\sigma\mu}e^{[A}{}_{\rho}\partial_{\mu}e^{B]}{}_{\sigma} + \omega^{[B}{}_{C\mu}Y^{A]C\mu} &= \Gamma^{\nu}{}_{\sigma\mu}Y^{\rho\sigma\mu}e^{[A}{}_{\rho}e^{B]}{}_{\nu}\,.
\end{align}
\end{subequations}
Finally, one is left with the task of transforming the second term in the field equation~\eqref{eq:genspiconfield2}. Here, we can use the relation
\begin{equation}
    \lc{\Gamma}^{\mu}{}_{\nu\mu} = \lc{\Gamma}^{\mu}{}_{\mu\nu} = \Gamma^{\mu}{}_{\mu\nu} - K^{\mu}{}_{\mu\nu} = \Gamma^{\mu}{}_{\mu\nu} = \Gamma^{\mu}{}_{\nu\mu} - T^{\mu}{}_{\mu\nu}\,,
\end{equation}
which consists of the following three steps. The first equality follows from the fact that the Christoffel symbol components \(\lc{\Gamma}^{\mu}{}_{\nu\rho}\) are symmetric in their lower two indices. In the second equality they are replaced by the teleparallel connection coefficients and the contortion using the definition~\eqref{eq:affcondec}. In the third equality the second term is omitted since the trace \(K^{\mu}{}_{\mu\nu}\) of the contortion over its first two indices vanishes due to antisymmetry in these indices. Finally, in the fourth and last equality the lower two indices of the teleparallel connection coefficients \(\Gamma^{\mu}{}_{\nu\rho}\) are switched by introducing a torsion tensor. This can also be obtained directly using
\begin{equation}
    \lc{\Gamma}^{\mu}{}_{\nu\mu} = \Gamma^{\mu}{}_{\nu\mu} - K^{\mu}{}_{\nu\mu} = \Gamma^{\mu}{}_{\nu\mu} - T^{\mu}{}_{\mu\nu}\,,
\end{equation}
by taking the trace over the outer indices of the contortion tensor~\eqref{eq:contor}. Combining the partial derivative and the connection coefficients into a teleparallel covariant derivative, the connection field equation finally reads
\begin{equation}\label{eq:genspiconfield}
0 = -\tilde{W}^{\mu\nu} = \nabla_{\rho}Y^{[\mu\nu]\rho} - T^{\sigma}{}_{\sigma\rho}Y^{[\mu\nu]\rho}\,.
\end{equation}
Note that by construction, it is antisymmetric in its two indices.

Another possibility to impose the properties of the spin connection is by introducing Lagrange multipliers~\cite{Golovnev:2017dox,Hohmann:2021fpr}. In this case one adds an additional term to the action, which is given by
\begin{equation}\label{eq:genactlagmul}
\mathcal{S}_{\text{LM}} := \int\dd^4x\,e\left(\tilde{q}^{(AB)\mu}\omega_{AB\mu} + \tilde{r}_A{}^{B\mu\nu}R^A{}_{B\mu\nu}\right)\,,
\end{equation}
where \(\tilde{q}^{AB\mu}\) and \(\tilde{r}_A{}^{B\mu\nu}\) are Lagrange multipliers. These equations are obtained by taking a variation \gls{wrt} these Lagrange multipliers, and then enforcing that the spin connection is flat, \(R^{\mu}{}_{\nu\rho\sigma} = 0\), and metric-compatible, \(Q_{\mu\nu\rho} = 0\). Now the spin connection field equation is obtained by the variation
\begin{equation}
    \delta_{\omega}\mathcal{S}_{\text{LM}} = \int\dd^4x\,e\left(-2\tilde{r}^{AB[\mu\nu]}\mathcal{D}_{\nu}\delta\omega_{AB\mu} + \tilde{q}^{(AB)\mu}\delta\omega_{AB\mu} - Y^{AB\mu}\delta\omega_{AB\mu}\right)\,,
\end{equation}
\gls{wrt} the full spin connection \(\delta\omega^A{}_{B\mu}\). Integration by parts yields
\begin{subequations}
\begin{alignat}{2}
\delta_{\omega}\mathcal{S}_{\text{LM}} & =\: & & \int\dd^4x\,e\left(2\lc{\nabla}_{\nu}\tilde{r}^{\rho\sigma[\mu\nu]} + 2K^{\rho}{}_{\omega\nu}\tilde{r}^{\omega\sigma[\mu\nu]} + 2K^{\sigma}{}_{\omega\nu}\tilde{r}^{\rho\omega[\mu\nu]} + \tilde{q}^{(\rho\sigma)\mu} - Y^{\rho\sigma\mu}\right)e^A{}_{\rho}e^B{}_{\sigma}\delta\omega_{AB\mu}\\[0.5ex]
& =\: & & \int\dd^4x\,e\left(2\nabla_{\nu}\tilde{r}^{\rho\sigma[\mu\nu]} - 2K^{\mu}{}_{\omega\nu}\tilde{r}^{\rho\sigma[\omega\nu]} - 2K^{\nu}{}_{\omega\nu}\tilde{r}^{\rho\sigma[\mu\omega]} + \tilde{q}^{(\rho\sigma)\mu} - Y^{\rho\sigma\mu}\right)e^A{}_{\rho}e^B{}_{\sigma}\delta\omega_{AB\mu}\,,
\end{alignat}
\end{subequations}
so that the field equations read
\begin{equation}
2\nabla_{\nu}\tilde{r}^{\rho\sigma[\mu\nu]} - 2K^{\mu}{}_{\omega\nu}\tilde{r}^{\rho\sigma[\omega\nu]} - 2K^{\nu}{}_{\omega\nu}\tilde{r}^{\rho\sigma[\mu\omega]} + \tilde{q}^{(\rho\sigma)\mu} - Y^{\rho\sigma\mu} = 0\,.
\end{equation}
Here the contribution which is symmetric in \(\rho\) and \(\sigma\) simply determines the value of the Lagrange multiplier \(\tilde{q}^{(\rho\sigma)\mu}\), and so it does not contribute to the field equations for the remaining fields. One is thus left with the antisymmetric part
\begin{equation}\label{eq:lagmulasymfield}
2\nabla_{\nu}\tilde{r}^{[\rho\sigma][\mu\nu]} - 2K^{\mu}{}_{\omega\nu}\tilde{r}^{[\rho\sigma][\omega\nu]} - 2K^{\nu}{}_{\omega\nu}\tilde{r}^{[\rho\sigma][\mu\omega]} - Y^{[\rho\sigma]\mu} = 0\,.
\end{equation}
In order to eliminate the remaining Lagrange multipliers from this equation, one may take its divergence with the teleparallel connection \(\nabla_{\mu}\), and use the relation
\begin{equation}
2\nabla_{\mu}\nabla_{\nu}\tilde{r}^{[\rho\sigma][\mu\nu]} = 2\nabla_{[\mu}\nabla_{\nu]}\tilde{r}^{[\rho\sigma]\mu\nu} = -T^{\omega}{}_{\mu\nu}\nabla_{\omega}\tilde{r}^{[\rho\sigma]\mu\nu}\,,
\end{equation}
which follows from the fact that the teleparallel connection has vanishing curvature, but nonvanishing torsion, to obtain the equation
\begin{equation}
\nabla_{\mu}\left(Y^{[\rho\sigma]\mu} + 2K^{\mu}{}_{\omega\nu}\tilde{r}^{[\rho\sigma][\omega\nu]} + 2K^{\nu}{}_{\omega\nu}\tilde{r}^{[\rho\sigma][\mu\omega]}\right) + T^{\omega}{}_{\mu\nu}\nabla_{\omega}\tilde{r}^{[\rho\sigma]\mu\nu} = 0\,.
\end{equation}
Expanding the contortion tensor in terms of its definition~\eqref{eq:contor}, and applying the product rule, this equation becomes
\begin{equation}
\nabla_{\mu}Y^{[\rho\sigma]\mu} - 2T^{\omega}{}_{\omega\mu}\nabla_{\nu}\tilde{r}^{[\rho\sigma][\mu\nu]} - 3\nabla_{[\omega}T^{\omega}{}_{\mu\nu]}\tilde{r}^{[\rho\sigma][\mu\nu]} = 0\,.
\end{equation}
On the last term, one can use the first Bianchi identity~\eqref{eq:bianchioneflat} for a flat connection, which yields
\begin{equation}
3\nabla_{[\omega}T^{\omega}{}_{\mu\nu]} = -3T^{\omega}{}_{\tau[\omega}T^{\tau}{}_{\mu\nu]} = T^{\omega}{}_{\omega\tau}T^{\tau}{}_{\mu\nu} + T^{\omega}{}_{\mu\tau}T^{\tau}{}_{\nu\omega} - T^{\omega}{}_{\nu\tau}T^{\tau}{}_{\mu\omega} = T^{\omega}{}_{\omega\tau}T^{\tau}{}_{\mu\nu}\,,
\end{equation}
where the last two terms cancel each other due to symmetry. This leads to the equation
\begin{equation}
    \nabla_{\mu}Y^{[\rho\sigma]\mu} - 2T^{\omega}{}_{\omega\mu}\nabla_{\nu}\tilde{r}^{[\rho\sigma][\mu\nu]} - T^{\omega}{}_{\omega\tau}T^{\tau}{}_{\mu\nu}\tilde{r}^{[\rho\sigma][\mu\nu]} = 0\,.
\end{equation}
Comparing this with the antisymmetric Eq.~\eqref{eq:lagmulasymfield}, we see that it matches the covariant derivative of the Lagrange multiplier \(\tilde{r}^{[\rho\sigma][\mu\nu]}\), contracted with the torsion vector \(T^{\omega}{}_{\omega\tau}\). Performing the same contraction on the antisymmetric Eq.~\eqref{eq:lagmulasymfield} yields
\begin{equation}
T^{\omega}{}_{\omega\tau}\left(2\nabla_{\nu}\tilde{r}^{[\rho\sigma][\tau\nu]} + T^{\tau}{}_{\mu\nu}\tilde{r}^{[\rho\sigma][\mu\nu]} - Y^{[\rho\sigma]\tau}\right) = 0\,,
\end{equation}
where we have once again used the definition~\eqref{eq:contor} of the contortion in terms of the torsion. Taking the sum of the last two equations, we find that the Lagrange multiplier cancels, and the remaining equation reproduces the spin connection Eq.~\eqref{eq:genspiconfield}. This equation has six independent components, and so it determines the six free components of the spin connection, which are left after fixing the metric compatibility and flatness constraints. The remaining components of Eq.~\eqref{eq:lagmulasymfield} determine the Lagrange multiplier. Hence, we have shown that the method of Lagrange multipliers yields the same field equations as the constrained variation which we used earlier in this section.

We finally remark that \textit{a priori}, there is no relation between the field equations obtained by variation \gls{wrt} the tetrad~\eqref{eq:gentetradfield} and the spin connection~\eqref{eq:genspiconfield}. However, this changes if one imposes a number of conditions, as will be shown below.

\subsubsection{Local Lorentz invariance of the matter action} \label{sssec:genmatactlorinv}

We first turn our attention to the matter action \(\mathcal{S}_{\text{m}}\), which does not depend on the spin connection, following the assumed decomposition~\eqref{eq:matgravaction}. We demand that the matter action is invariant under infinitesimal and local Lorentz transformations \(\lambda^A{}_B\) with \(\lambda^{(AB)} = 0\), which act on the tetrad following the transformation~\eqref{eq:inftetlor}, and which we assume to act trivially on matter fields, \(\delta_{\lambda}\chi^I = 0\). The resulting variation of the matter action is therefore given by
\begin{equation}\label{eq:matactlortrans}
    \delta_{\lambda}\mathcal{S}_{\text{m}} = \int\dd^4x\,e\,\Theta_A{}^{\mu}\lambda^A{}_Be^B{}_{\mu} = \int\dd^4x\,e\,\Theta_{\mu\nu}\lambda^{\mu\nu}\,,
\end{equation}
where
\begin{equation}
    \lambda^{\mu\nu} = g^{\nu\rho}E_A{}^{\mu}e^B{}_{\rho}\lambda^A{}_B\,.
\end{equation}
From the antisymmetry of \(\lambda^{\mu\nu}\) it thus follows that the antisymmetric part of the energy-momentum tensor vanishes, \(\Theta_{[\mu\nu]} = 0\). Note that this holds both on-shell and off-shell, i.e., independently of the matter field equations, since these have not entered the derivation. It does, however, depend on the assumption that there is no direct coupling between the matter fields and the spin connection. Given such a coupling, one would obtain a contribution from the spin current~\cite{Obukhov:2006ge}.

We also remark that the condition of local Lorentz invariance of the matter action is equivalent to demanding that the tetrad enters the matter action only through the metric and its derivatives (which appear through the Christoffel symbols in covariant derivatives of the matter tensor fields \(\chi^I\)). To see this, one uses the symmetry of the energy-momentum tensor to write the variation of the matter action \gls{wrt} the tetrad as
\begin{equation}\label{eq:matactmetvar}
    \delta_e\mathcal{S}_{\text{m}} = \int\dd^4x\,e\,\Theta_A{}^{\mu}\delta e^A{}_{\mu} = \int\dd^4x\,e\,\Theta^{\nu\mu}\eta_{AB}e^B{}_{(\nu}\delta e^A{}_{\mu)} = \frac{1}{2}\int\dd^4x\,e\,\Theta^{\mu\nu}\delta g_{\mu\nu}\,,
\end{equation}
which shows that it is fully described by the variation of the metric. Hence, the matter action does not depend on any other \gls{dof} arising from the tetrad besides the ones which determine the metric. The converse is obvious: if the tetrad enters into the matter action only through the metric, then local Lorentz invariance follows from the fact \(\delta_{\lambda}g_{\mu\nu} = 0\) that the metric is invariant under local Lorentz transformations.

\subsubsection{Local Lorentz invariance of the gravitational action} \label{sssec:gengravactlorinv}

Next, we turn our attention to the gravitational part \(\mathcal{S}_{\text{g}}\) of the action. Demanding invariance under local Lorentz transformations, as discussed for the matter action above, we must here also take into account the transformation~\eqref{eq:infspclor} of the spin connection. It then follows that the total induced variation of the action is given by
\begin{equation}\label{eq:gravactlortrans}
    \delta_{\lambda}\mathcal{S}_{\text{g}} = -\int\dd^4x\,e\,\left[W_A{}^{\mu}\lambda^A{}_Be^B{}_{\mu} - Y_A{}^{B\mu}(\partial_{\mu}\lambda^A{}_B + \omega^A{}_{C\mu}\lambda^C{}_B - \omega^C{}_{B\mu}\lambda^A{}_C)\right]\,.
\end{equation}
After integration by parts, we then find the expression
\begin{equation}
    \delta_{\lambda}\mathcal{S}_{\text{g}} = -\int\dd^4x\,e\,\left(W_A{}^{\mu}e^B{}_{\mu} + \partial_{\mu}Y_A{}^{B\mu} + \lc{\Gamma}^{\mu}{}_{\nu\mu}Y_A{}^{B\nu} - \omega^C{}_{A\mu}Y_C{}^{B\mu} + \omega^B{}_{C\mu}Y_A{}^{C\mu}\right)\lambda^A{}_B\,.
\end{equation}
Taking into account the antisymmetry of \(\lambda^A{}_B\), and comparing with the field Eqs.~\eqref{eq:gentetradfield} and~\eqref{eq:genspiconfield}, we see that local Lorentz invariance imposes that they are related by
\begin{equation}\label{eq:genfieldasym}
    W_{[\mu\nu]} = \tilde{W}_{\mu\nu}\,,
\end{equation}
and so the antisymmetric part of the tetrad field equation agrees with the spin connection equation~\cite{Hohmann:2017duq}. This reflects the fact that the spin connection is a pure gauge \gls{dof}, which is not restricted by an independent field equation. The linear dependence between the field equations exactly accounts for this gauge freedom. Hence, in the remainder of this Review we will make use of this fact, and omit deriving the spin connection field equations explicitly, as they are redundant.

As for the matter action, we can also obtain an equivalent description of local Lorentz invariance of the gravitational action by demanding that the gravitational action is constructed from the metric~\eqref{eq:metric}, the torsion~\eqref{eq:spicontors} and its covariant derivative. To see this, note first that the general variation of the metric and the torsion tensor can be obtained from those of the tetrad and the spin connection as
\begin{equation}\label{eq:mettorsvar}
    \delta g_{\mu\nu} = 2\eta_{AB}e^A{}_{(\mu}\delta e^B{}_{\nu)}\,, \quad \delta T^{\rho}{}_{\mu\nu} = E_A{}^{\rho}\left(2\delta\omega^A{}_{B[\mu}e^B{}_{\nu]} + 2\mathcal{D}_{[\mu}\delta e^A{}_{\nu]} - T^{\sigma}{}_{\mu\nu}\delta e^A{}_{\sigma}\right)\,.
\end{equation}
Two properties follow immediately from these variations. First, note that for a local Lorentz transformation given by the relations~\eqref{eq:inftetlor} and~\eqref{eq:infspclor}, both \(\delta g_{\mu\nu}\) and \(\delta T^{\rho}{}_{\mu\nu}\) vanish. This can easily be seen by direct calculation, which yields
\begin{equation}
    \delta_{\lambda}g_{\mu\nu} = 2\eta_{AC}e^A{}_{(\mu}e^B{}_{\nu)}\lambda^C{}_B = 2e^A{}_{\mu}e^B{}_{\nu}\lambda_{(AB)} = 0\,,
\end{equation}
as well as
\begin{equation}
    \delta_{\lambda}T^{\rho}{}_{\mu\nu} = E_A{}^{\rho}\left[2\mathcal{D}_{[\mu}(\lambda^A{}_Be^B{}_{\nu]}) - 2E_C{}^{\sigma}\mathcal{D}_{[\mu}e^C{}_{\nu]}\lambda^A{}_Be^B{}_{\sigma} - 2e^B{}_{[\nu}\mathcal{D}_{\mu]}\lambda^A{}_B\right] = 0\,.
\end{equation}
Secondly, one finds that the variation of the spin connection can be expressed in terms of the variations of the torsion and the tetrad. This relation can be derived in multiple steps. First, it follows from variation of the tetrad postulate~\eqref{eq:tetradpost} that the variation of the spin connection takes the form
\begin{equation}\label{eq:spicontorsvar}
    \delta\omega^A{}_{B\mu} = e^A{}_{\rho}E_B{}^{\nu}\left[\delta\Gamma^{\rho}{}_{\nu\mu} - \nabla_{\mu}(E_C{}^{\rho}\delta e^C{}_{\nu})\right]\,,
\end{equation}
in terms of the variations of the tetrad and the teleparallel affine connection. The coefficients of the latter can then be decomposed using the relation~\eqref{eq:affcondec}, where the disformation vanishes, to obtain
\begin{equation}
    \delta\Gamma^{\rho}{}_{\nu\mu} = \delta\lc{\Gamma}^{\rho}{}_{\nu\mu} + \delta K^{\rho}{}_{\nu\mu}\,,
\end{equation}
where the variation of the Levi-Civita connection coefficients is given by
\begin{equation}
    \delta\lc{\Gamma}^{\rho}{}_{\nu\mu} = \frac{1}{2}g^{\rho\sigma}\left(\lc{\nabla}_{\nu}\delta g_{\sigma\mu} + \lc{\nabla}_{\mu}\delta g_{\nu\sigma} - \lc{\nabla}_{\sigma}\delta g_{\nu\mu}\right).
\end{equation}
The variation of the contortion follows from its definition~\eqref{eq:contor} and reads
\begin{equation}
    \delta K^{\rho}{}_{\nu\mu} = \frac{1}{2}\delta T^{\rho}{}_{\mu\nu} - g^{\rho\sigma}g_{\tau(\mu}\delta T^{\tau}{}_{\nu)\sigma} + g^{\rho\sigma}T_{(\mu\nu)}{}^{\tau}\delta g_{\sigma\tau} - T^{\sigma}{}_{(\mu}{}^{\rho}\delta g_{\nu)\sigma}\,.
\end{equation}
Taking the sum of the latter two expressions, one finds that the variation of the teleparallel affine connection is given by
\begin{equation}
    \delta\Gamma^{\rho}{}_{\nu\mu} = \frac{1}{2}g^{\rho\sigma}\left(\nabla_{\nu}\delta g_{\sigma\mu} + \nabla_{\mu}\delta g_{\nu\sigma} - \nabla_{\sigma}\delta g_{\nu\mu}\right) + \frac{1}{2}\delta T^{\rho}{}_{\mu\nu} - g^{\rho\sigma}g_{\tau(\mu}\delta T^{\tau}{}_{\nu)\sigma}\,.
\end{equation}
Finally, substituting the metric variation \(\delta g_{\mu\nu}\) in terms of the tetrad variation given by the expression~\eqref{eq:mettorsvar}, we have expressed the variation of the teleparallel connection in terms of the variations of the tetrad and torsion.

With the help of the relations derived above, it is possible to rewrite the variation in Eq.~\eqref{eq:gravactvar} of the gravitational action in terms of the variations of the tetrad and the torsion. By successive substitution and integration by parts, one finds~\cite{Hohmann:2017duq,Hohmann:2018ijr}
\begin{subequations}
\begin{alignat}{2}
\delta\mathcal{S}_{\text{g}} & =\: & & -\int\dd^4x\,e\,\left\{W_A{}^{\mu}\delta e^A{}_{\mu} + Y_{\rho}{}^{\nu\mu}\left[\delta\Gamma^{\rho}{}_{\nu\mu} - \nabla_{\mu}(E_C{}^{\rho}\delta e^C{}_{\nu})\right]\right\}\\[0.5ex]
& =\: & & -\int\dd^4x\,e\,\bigg\{\left[W_{\nu}{}^{\mu} + T_{\sigma}{}^{\sigma\rho}(2Y^{\mu}{}_{(\nu\rho)} + Y_{\nu\rho}{}^{\mu}) + \nabla_{\rho}(2Y_{\nu}{}^{[\mu\rho]} + Y^{\rho\mu}{}_{\nu})\right]E_A{}^{\nu}\delta e^A{}_{\mu}\nonumber\\[0.5ex]
& \: & &+ \left[\frac{1}{2}Y^{\mu\nu}{}_{\rho} - Y_{\rho}{}^{\mu\nu}\right]\delta T^{\rho}{}_{\mu\nu}\bigg\}\\[0.5ex]
& =\: & & -\int\dd^4x\,e\,(U_A{}^{\mu}\delta e^A{}_{\mu} + Z_{\rho}{}^{\mu\nu}\delta T^{\rho}{}_{\mu\nu})\,,
 \end{alignat}
 \end{subequations}
where the variation terms are given by
\begin{equation}\label{eq:actvartorstoconn}
U_A{}^{\mu} := \left[W_{\nu}{}^{\mu} + T_{\sigma}{}^{\sigma\rho}(2Y^{\mu}{}_{(\nu\rho)} + Y_{\nu\rho}{}^{\mu}) + \nabla_{\rho}(2Y_{\nu}{}^{[\mu\rho]} + Y^{\rho\mu}{}_{\nu})\right]E_A{}^{\nu}\,, \quad
Z_{\rho}{}^{\mu\nu} := \frac{1}{2}Y^{\mu\nu}{}_{\rho} - Y_{\rho}{}^{[\mu\nu]}\,,
\end{equation}
where we have applied the antisymmetry of the torsion tensor in its last two indices.

To derive the converse relation, one starts from the variation of the action in the equivalent form
\begin{equation}\label{eq:gravactvar2}
\delta\mathcal{S}_{\text{g}} = -\int\dd^4x\,e\,(U_A{}^{\mu}\delta e^A{}_{\mu} + Z_{\rho}{}^{\mu\nu}\delta T^{\rho}{}_{\mu\nu})\,.
\end{equation}
Inserting the variation~\eqref{eq:mettorsvar} of the torsion into this variation, one obtains
\begin{subequations}
\begin{alignat}{2}
\delta\mathcal{S}_{\text{g}} & =\: & & -\int\dd^4x\,e\,\left[U_A{}^{\mu}\delta e^A{}_{\mu} + Z_{\rho}{}^{\mu\nu}E_A{}^{\rho}\left(2\delta\omega^A{}_{B[\mu}e^B{}_{\nu]} + 2\mathcal{D}_{[\mu}\delta e^A{}_{\nu]} - T^{\sigma}{}_{\mu\nu}\delta e^A{}_{\sigma}\right)\right]\\[0.5ex]
& =\: & & -\int\dd^4x\,e\,\Big[\left(U_A{}^{\mu} - E_A{}^{\nu}T^{\mu}{}_{\rho\sigma}Z_{\nu}{}^{\rho\sigma} - 2E_A{}^{\rho}K^{\sigma}{}_{\rho\nu}Z_{\sigma}{}^{\mu\nu} + 2E_A{}^{\rho}\lc{\nabla}_{\nu}Z_{\rho}{}^{\mu\nu}\right)\delta e^A{}_{\mu}\nonumber\\[0.5ex]
& \: & &+ 2E_A{}^{\rho}e^B{}_{\nu}Z_{\rho}{}^{\mu\nu}\delta\omega^A{}_{B\mu}\Big]\\[0.5ex]
& =\: & & -\int\dd^4x\,e\,\left[\left(U_A{}^{\mu} - 2E_A{}^{\rho}T^{\sigma}{}_{\sigma\nu}Z_{\rho}{}^{\mu\nu} + 2E_A{}^{\rho}\nabla_{\nu}Z_{\rho}{}^{\mu\nu}\right)\delta e^A{}_{\mu} + 2E_A{}^{\rho}e^B{}_{\nu}Z_{\rho}{}^{\mu\nu}\delta\omega^A{}_{B\mu}\right]\,,
\end{alignat}
\end{subequations}
where we have performed integration by parts, and omitted antisymmetrization in the last two indices of \(Z_{\rho}{}^{\mu\nu}\), since it is already antisymmetric due to the antisymmetry of the torsion in the variation~\eqref{eq:gravactvar2}. By comparison with the action variation~\eqref{eq:gravactvar}, one thus finds the relation
\begin{equation}\label{eq:actvarconntotors}
    W_A{}^{\mu} := U_A{}^{\mu} - 2E_A{}^{\rho}T^{\sigma}{}_{\sigma\nu}Z_{\rho}{}^{\mu\nu} + 2E_A{}^{\rho}\nabla_{\nu}Z_{\rho}{}^{\mu\nu}\,, \quad Y_A{}^{B\mu} := E_A{}^{\rho}e^B{}_{\nu}(Z_{\rho}{}^{\mu\nu} - Z^{\nu\mu}{}_{\rho})\,,
\end{equation}
between the occurring terms, where the antisymmetrization in the last term is due to the fact that the spin connection is, by definition, antisymmetric in its two Lorentz indices. One easily checks that the relations~\eqref{eq:actvartorstoconn} and~\eqref{eq:actvarconntotors} are indeed each other's inverses.

Writing the variation of the action in the form shown in Eq.~\eqref{eq:gravactvar2}, the local Lorentz transformation~\eqref{eq:gravactlortrans} of the gravitational part of the action can now simply be written as
\begin{equation}
\delta_{\lambda}\mathcal{S}_{\text{g}} = -\int\dd^4x\,e\,U_A{}^{\mu}\lambda^A{}_Be^B{}_{\mu} = \int\dd^4x\,e\,U_{\mu\nu}\lambda^{\mu\nu}\,,
\end{equation}
and thus it is of the same form as the corresponding local Lorentz transformation~\eqref{eq:matactlortrans} of the matter action. Hence, we see that the gravitational action is invariant under arbitrary local Lorentz transformations if and only if \(U_{[\mu\nu]} = 0\), since \(\lambda^{\mu\nu}\) is antisymmetric by definition of a local Lorentz transformation. With the relation~\eqref{eq:actvartorstoconn}, this translates to
\begin{equation}\label{eq:actvarlorinv}
0 = U_{[\mu\nu]} = W_{[\mu\nu]} - T^{\sigma}{}_{\sigma\rho}Y_{\mu\nu}{}^{\rho} + \nabla_{\rho}Y_{\mu\nu}{}^{\rho} = W_{[\mu\nu]} - \tilde{W}_{\mu\nu}\,,
\end{equation}
and thus reproduces the Lorentz invariance condition~\eqref{eq:genfieldasym} which we derived from the variation in Eq.~\eqref{eq:gravactlortrans}. Following the same routine as we did for the variation~\eqref{eq:matactmetvar} of the matter action, the gravitational action is locally Lorentz invariant if and only if its variation can be written as
\begin{equation}\label{eq:gravactvarmettor}
\delta\mathcal{S}_{\text{g}} = -\int\dd^4x\,\sqrt{-g}\,\left(\frac{1}{2}U^{\mu\nu}\delta g_{\mu\nu} + Z_{\rho}{}^{\mu\nu}\delta T^{\rho}{}_{\mu\nu}\right)\,,
\end{equation}
thus also expressing the determinant of the tetrad through that of the metric, which means that the action can be fully expressed through the metric and the torsion tensor (and their covariant derivatives). Note that this is equivalent to the statement that the gravitational action can be completely expressed in terms of variables which are invariant under local Lorentz transformations~\cite{Hohmann:2021fpr,Hohmann:2021dhr,Blixt:2022rpl}.

\subsubsection{Palatini approach} \label{sssec:palatini}

The aforementioned considerations regarding the local Lorentz invariance open the possibility for yet another formulation of \gls{tg} theories, which is known as the Palatini approach~\cite{BeltranJimenez:2018vdo}. This approach is similar to writing the gravitational action and its variation in the form on Eq.~\eqref{eq:gravactvarmettor}, using the metric \(g_{\mu\nu}\) instead of the tetrad as one of the fundamental field variables. For the second field variable, the teleparallel affine connection with coefficients \(\Gamma^{\mu}{}_{\nu\rho}\) is chosen. This allows one to write the variation of the gravitational part of the action in the form
\begin{equation}\label{eq:gravactvarpalat}
    \delta\mathcal{S}_{\text{g}} = -\int\dd^4x\,\sqrt{-g}\,\left(\frac{1}{2}U^{\mu\nu}\delta g_{\mu\nu} + Y_{\mu}{}^{\nu\rho}\delta\Gamma^{\mu}{}_{\nu\rho}\right)\,,
\end{equation}
while the variation of the matter action is expressed in the form~\eqref{eq:matactmetvar} through the symmetric energy-momentum tensor \(\Theta^{\mu\nu}\) and the metric variation \(\delta g_{\mu\nu}\). Instead of \textit{a priori} assuming that the connection is flat and metric-compatible, one imposes these conditions by introducing an additional term
\begin{equation}
    \mathcal{S}_{\text{LM}} = \int(\tilde{r}_{\mu}{}^{\nu\rho\sigma}R^{\mu}{}_{\nu\rho\sigma} + \tilde{q}^{\mu\nu\rho}Q_{\mu\nu\rho})\sqrt{-g}\dd^4x\,,
\end{equation}
into the action, where \(\tilde{r}_{\mu}{}^{\nu\rho\sigma}\) and \(\tilde{q}^{\mu\nu\rho}\) are Lagrange multipliers, which is similar to the corresponding term~\eqref{eq:genactlagmul} in the tetrad and spin connection formulation. The equations obtained by taking a variation \gls{wrt} these Lagrange multipliers, which then yields the conditions that the connection represented by \(\Gamma^{\mu}{}_{\nu\rho}\) is flat, \(R^{\mu}{}_{\nu\rho\sigma} = 0\), and metric-compatible, \(Q_{\mu\nu\rho} = 0\). The remaining field equations are obtained by variation \gls{wrt} the metric and to the affine connection, essentially following the same procedure as shown in Sec.~\ref{sssec:tggenactfield} using the tetrad and spin connection as fundamental variables. Variation \gls{wrt} the affine connection, and raising one index for convenience, yields the equation
\begin{equation}\label{eq:genactvaraffcon}
    2\nabla_{\sigma}\tilde{r}^{\mu\nu[\rho\sigma]} - 2\tilde{r}^{\mu\nu[\rho\sigma]}K^{\tau}{}_{\sigma\tau} - 2\tilde{r}^{\mu\nu[\sigma\tau]}K^{\rho}{}_{\sigma\tau} - 2\tilde{q}^{\rho(\mu\nu)} - Y^{\mu\nu\rho} = 0\,.
\end{equation}
As in the tetrad and spin connection formulation, one finds that the symmetric part of this equation only determines the value of the Lagrange multiplier \(\tilde{q}^{\rho(\mu\nu)}\), so that one only retains the antisymmetric part of the equation in order to derive a field equation for the physical fields. Eliminating also the Lagrange multiplier \(\tilde{r}_{\mu}{}^{\nu\rho\sigma}\) by taking the covariant divergence of the antisymmetric equation, one derives the connection field equation
\begin{equation}
\nabla_{\rho}Y^{[\mu\nu]\rho} - Y^{[\mu\nu]\rho}T^{\tau}{}_{\tau\rho} = 0\,,
\end{equation}
thereby reproducing the field Eq.~\eqref{eq:genspiconfield} for the spin connection (see Ref.~\cite{Hohmann:2021fpr} for a detailed derivation). By taking a variation \gls{wrt} the metric one obtains the equation
\begin{equation}
U^{\mu\nu} + 2\nabla_{\rho}\tilde{q}^{\rho(\mu\nu)} + 2\tilde{q}^{\sigma(\mu\nu)}K^{\rho}{}_{\sigma\rho} = \Theta^{\mu\nu}\,,
\end{equation}
which now contains the Lagrange multiplier \(\tilde{q}^{\rho(\mu\nu)}\). This must be eliminated by substituting its value from the previously found in Eq.~\eqref{eq:genactvaraffcon}, which contains the Lagrange multiplier \(\tilde{r}_{\mu}{}^{\nu\rho\sigma}\); the latter is eliminated using the same procedure as above, finally leading to the field equation
\begin{equation}
U^{\mu\nu} - \nabla_{\rho}Y^{(\mu\nu)\rho} + Y^{(\mu\nu)\rho}T^{\tau}{}_{\tau\rho} = \Theta^{\mu\nu}\,,
\end{equation}
which now contains only the physical fields. In order to relate this equation to the field equations in the tetrad and spin connection formulation, one substitutes \(U^{\mu\nu}\) from the relation~\eqref{eq:actvartorstoconn}, keeping only its symmetric part, as mandated by the Lorentz invariance condition~\eqref{eq:actvarlorinv}, and finally also uses the connection field Eq.~\eqref{eq:genspiconfield}, to reproduce the tetrad field Eq.~\eqref{eq:gentetradfield}. Hence, the Palatini approach yields the same field equations as the more common approach we discussed before, albeit using different fundamental fields.

\subsubsection{Energy-momentum conservation} \label{sssec:genenmomcons}

Another important relation can be derived from the demand that the action is invariant under diffeomorphisms. Infinitesimally, these are generated by vector fields \(X = X^{\mu}\partial_{\mu}\), and the induced change on any tensor field is given by its Lie derivative. For the tetrad and spin connection these Lie derivatives are given by the relations~\eqref{eq:infisymcondgen}. For the matter action this yields the variation
\begin{equation}
\delta_X\mathcal{S}_{\text{m}} = \int\dd^4x\,e\,\left[\Theta_A{}^{\mu}(X^{\nu}\partial_{\nu}e^A{}_{\mu} + \partial_{\mu}X^{\nu}e^A{}_{\nu}) + \Omega_I\mathcal{L}_X\chi^I\right]\,.
\end{equation}
One then imposes that the matter field equations \(\Omega_I = 0\) are satisfied. Hence, it must be emphasized that everything which follows holds \emph{only on-shell}. Integrating by parts, as well as a few transformations of the resulting terms, lead to
\begin{subequations}
\begin{alignat}{2}
\delta_X\mathcal{S}_{\text{m}} & =\: & & \int\dd^4x\,e\,\Theta_A{}^{\mu}(X^{\nu}\partial_{\nu}e^A{}_{\mu} + \partial_{\mu}X^{\nu}e^A{}_{\nu})\\[0.5ex]
& =\: & & \int\dd^4x\,e\,\left[\Theta_A{}^{\mu}\partial_{\nu}e^A{}_{\mu} - \lc{\Gamma}^{\rho}{}_{\mu\rho}\Theta_A{}^{\mu}e^A{}_{\nu} - \partial_{\mu}(\Theta_A{}^{\mu}e^A{}_{\nu})\right]X^{\nu}\\[0.5ex]
& =\: & & \int\dd^4x\,e\,\left(\lc{\Gamma}^{\rho}{}_{\nu\mu}\Theta_{\rho}{}^{\mu} - \lc{\Gamma}^{\rho}{}_{\mu\rho}\Theta_{\nu}{}^{\mu} - \partial_{\mu}\Theta_{\nu}{}^{\mu}\right)X^{\nu}\\[0.5ex]
& =\: & & -\int\dd^4x\,e\,X^{\nu}\lc{\nabla}_{\mu}\Theta_{\nu}{}^{\mu}\,,
\end{alignat}
\end{subequations}
from which follows the energy-momentum conservation
\begin{equation}\label{eq:enmomconserv}
\lc{\nabla}_{\nu}\Theta^{\mu\nu} = 0\,,
\end{equation}
since the vector field is arbitrary. A crucial step in this derivation is the substitution
\begin{subequations}
\begin{alignat}{2}
\lc{\Gamma}^{\rho}{}_{\nu\mu}\Theta_{\rho}{}^{\mu} & =\: & & \frac{1}{2}g^{\rho\sigma}(\partial_{\nu}g_{\sigma\mu} + \partial_{\mu}g_{\nu\sigma} - \partial_{\sigma}g_{\nu\mu})\Theta_{\rho}{}^{\mu}\\[0.5ex]
& =\: & & \frac{1}{2}\Theta^{\mu\sigma}(\partial_{\nu}g_{\sigma\mu} + \partial_{\mu}g_{\nu\sigma} - \partial_{\sigma}g_{\nu\mu})\\[0.5ex]
& =\: & & \frac{1}{2}\Theta^{\mu\sigma}\partial_{\nu}g_{\sigma\mu}\\[0.5ex]
& =\: & & \Theta^{\mu\sigma}\eta_{AB}e^B{}_{\mu}\partial_{\nu}e^A{}_{\sigma}\\[0.5ex]
& =\: & & \Theta_A{}^{\sigma}\partial_{\nu}e^A{}_{\sigma}\,,
 \end{alignat}
 \end{subequations}
which makes explicit use of the symmetry \(\Theta^{[\mu\nu]} = 0\) of the energy-momentum tensor. This shows that the Lorentz invariance of the matter action is a necessary ingredient in deriving the conservation Eq.~\eqref{eq:enmomconserv}. Note that energy-momentum conservation is a consequence of the invariance of the matter action under diffeomorphisms, and that its common form~\eqref{eq:enmomconserv} is a consequence of coupling to the metric only, which is independent of the choice of the gravitational action.

\subsubsection{Bianchi identities} \label{sssec:genbianchi}

Also, for the gravitational action \(\mathcal{S}_{\text{g}}\) the invariance under diffeomorphisms is imposed. In this case the induced variation is given by
\begin{equation}\label{eq:gengravdiffvar}
\delta_X\mathcal{S}_{\text{g}} = -\int\dd^4x\,e\,\left[W_A{}^{\mu}(X^{\nu}\partial_{\nu}e^A{}_{\mu} + \partial_{\mu}X^{\nu}e^A{}_{\nu}) + Y_A{}^{B\mu}(X^{\nu}\partial_{\nu}\omega^A{}_{B\mu} + \partial_{\mu}X^{\nu}\omega^A{}_{B\nu})\right]\,.
\end{equation}
This variation can be decomposed in a number of steps. We start with the second part of this term, which contains the contribution from the spin connection variation \(Y_A{}^{B\mu}\). After integration by parts this contribution reads
\begin{equation}
-\int\dd^4x\,e\,\left(Y_A{}^{B\mu}\partial_{\nu}\omega^A{}_{B\mu} - Y_A{}^{B\mu}\partial_{\mu}\omega^A{}_{B\nu} - \partial_{\mu}Y_A{}^{B\mu}\omega^A{}_{B\nu} - \lc{\Gamma}^{\rho}{}_{\mu\rho}Y_A{}^{B\mu}\omega^A{}_{B\nu}\right)X^{\nu}\,,
\end{equation}
where the last term originates from the derivative of the determinant \(e\) of the tetrad. Note that the first two terms constitute an antisymmetric derivative of the spin connection. Using the fact that its curvature~\eqref{eq:spiconcurv} vanishes, one thus finds the expression
\begin{equation}
-\int\dd^4x\,e\,\left(Y_A{}^{B\mu}\omega^A{}_{C\mu}\omega^C{}_{B\nu} - Y_A{}^{B\mu}\omega^A{}_{C\nu}\omega^C{}_{B\mu} - \partial_{\mu}Y_A{}^{B\mu}\omega^A{}_{B\nu} - \lc{\Gamma}^{\rho}{}_{\mu\rho}Y_A{}^{B\mu}\omega^A{}_{B\nu}\right)X^{\nu}\,.
\end{equation}
After renaming indices, one can extract a common factor to obtain
\begin{equation}
-\int\dd^4x\,e\,\left(Y_C{}^{B\mu}\omega^C{}_{A\mu} - Y_A{}^{C\mu}\omega^B{}_{C\mu} - \partial_{\mu}Y_A{}^{B\mu} - \lc{\Gamma}^{\rho}{}_{\mu\rho}Y_A{}^{B\mu}\right)\omega^A{}_{B\nu}X^{\nu}\,.
\end{equation}
Comparing with the expression~\eqref{eq:genactvarspicon}, and using the fact that the spin connection is antisymmetric, we see that this is simply
\begin{equation}
-\int\dd^4x\,e\,\tilde{W}^{AB}\omega_{AB\tau}X^{\tau} = -\int\dd^4x\,e\,W^{[AB]}\omega_{AB\tau}X^{\tau}\,,
\end{equation}
where we used the local Lorentz invariance, which implies the relation~\eqref{eq:genfieldasym} between the tetrad and spin connection field equations. Inserting this in the variation~\eqref{eq:gengravdiffvar}, we thus find the expression
\begin{equation}
    \delta_X\mathcal{S}_{\text{g}} = -\int\dd^4x\,e\,\left[W_A{}^{\mu}(X^{\nu}\partial_{\nu}e^A{}_{\mu} + \partial_{\mu}X^{\nu}e^A{}_{\nu}) + W^{AB}\omega_{AB\tau}X^{\tau}\right]\,,
\end{equation}
where we omitted the antisymmetrization brackets on the last term, since it is contracted with the antisymmetric spin connection. Once again, we perform integration by parts to obtain
\begin{equation}
\delta_X\mathcal{S}_{\text{g}} = -\int\dd^4x\,e\,\left(W_A{}^{\mu}\partial_{\nu}e^A{}_{\mu} - W_A{}^{\mu}\partial_{\mu}e^A{}_{\nu} - \partial_{\mu}W_A{}^{\mu}e^A{}_{\nu} - \lc{\Gamma}^{\rho}{}_{\mu\rho}W_A{}^{\mu}e^A{}_{\nu} + W^{AB}\omega_{AB\nu}\right)X^{\nu}\,.
\end{equation}
Note that the second and third term can be combined by transforming a Lorentz index into a spacetime index, which yields
\begin{equation}
\delta_X\mathcal{S}_{\text{g}} = -\int\dd^4x\,e\,\left(W_A{}^{\mu}\partial_{\nu}e^A{}_{\mu} - \partial_{\mu}W_{\nu}{}^{\mu} - \lc{\Gamma}^{\rho}{}_{\mu\rho}W_A{}^{\mu}e^A{}_{\nu} + W^{AB}\omega_{AB\nu}\right)X^{\nu}\,.
\end{equation}
This can also be done with the remaining Lorentz indices, so that one obtains
\begin{equation}
\delta_X\mathcal{S}_{\text{g}} = -\int\dd^4x\,e\,\left(W_{\rho}{}^{\mu}E_A{}^{\rho}\partial_{\nu}e^A{}_{\mu} - \partial_{\mu}W_{\nu}{}^{\mu} - \lc{\Gamma}^{\rho}{}_{\mu\rho}W_{\nu}{}^{\mu} + W^{\rho\mu}e^A{}_{\rho}e^B{}_{\mu}\omega_{AB\nu}\right)X^{\nu}\,.
\end{equation}
Now the first and the last term can be combined to yield the teleparallel affine connection
\begin{equation}
\delta_X\mathcal{S}_{\text{g}} = -\int\dd^4x\,e\,\left(W_{\rho}{}^{\mu}\Gamma^{\rho}{}_{\mu\nu} - \partial_{\mu}W_{\nu}{}^{\mu} - \lc{\Gamma}^{\rho}{}_{\mu\rho}W_{\nu}{}^{\mu}\right)X^{\nu}\,.
\end{equation}
After splitting the teleparallel connection in the first term into a contortion and Levi-Civita connection
\begin{equation}
\delta_X\mathcal{S}_{\text{g}} = -\int\dd^4x\,e\,\left(W_{\rho}{}^{\mu}K^{\rho}{}_{\mu\nu} + W_{\rho}{}^{\mu}\lc{\Gamma}^{\rho}{}_{\mu\nu} - \partial_{\mu}W_{\nu}{}^{\mu} - \lc{\Gamma}^{\rho}{}_{\mu\rho}W_{\nu}{}^{\mu}\right)X^{\nu}\,,
\end{equation}
the remaining terms combine into a covariant derivative
\begin{equation}
\delta_X\mathcal{S}_{\text{g}} = -\int\dd^4x\,e\,\left(W_{\rho}{}^{\mu}K^{\rho}{}_{\mu\nu} - \lc{\nabla}_{\mu}W_{\nu}{}^{\mu}\right)X^{\nu}\,,
\end{equation}
using the fact that the Levi-Civita connection is symmetric. The resulting equations
\begin{equation}\label{eq:genbianchi}
W^{\rho\nu}K_{\rho\nu}{}^{\mu} - \lc{\nabla}_{\nu}W^{\mu\nu} = 0\,,
\end{equation}
are the Bianchi identities for a general \gls{tg} theory. Note that although we have used the relation~\eqref{eq:genfieldasym} between the tetrad and spin connection field equations, which follows from the Lorentz invariance of the gravitational action, we have not imposed that either of these field equations hold. Hence, the Bianchi identities~\eqref{eq:genbianchi} hold both on-shell and off-shell, i.e., they are geometric identities. Using the antisymmetry \(K_{(\mu\nu)\rho} = 0\) of the contortion, one can decompose them into their symmetric and antisymmetric parts
\begin{equation}\label{eq:decbianchi}
\lc{\nabla}_{\nu}W^{(\mu\nu)} = 0\,, \quad
W^{[\rho\nu]}K_{\rho\nu}{}^{\mu} - \lc{\nabla}_{\nu}W^{[\mu\nu]} = 0\,.
\end{equation}
Together with a corresponding decomposition
\begin{equation}\label{eq:Wsym_Theta_Wantisym}
W_{(\mu\nu)} = \Theta_{\mu\nu}\,, \quad
W_{[\mu\nu]} = 0\,,
\end{equation}
of the field Eqs.~\eqref{eq:gentetradfield}, which follows from the symmetry \(\Theta_{[\mu\nu]} = 0\) of the energy-momentum tensor. We see that inserting the field equations into the Bianchi identity~\eqref{eq:decbianchi} yields the energy-momentum conservation~\eqref{eq:enmomconserv}. Hence, we find that the latter is indeed imposed by the field equations.

Thus, for any \gls{tg} theory we can always write its field equations as the symmetric and antisymmetric $W^{\mu\nu}$ (which is the variation of the action \gls{wrt} the tetrad). This then accounts for both the tetrad and spin connection field equations.

\subsubsection{Premetric approach} \label{sssec:genpremetric}

To conclude the preceding discussion of the general properties of \gls{tg} theories, we finally show how these can be found from the premetric point of view~\cite{Itin:2016nxk,Hohmann:2017duq,Itin:2018dru}. For this purpose, we write the gravitational part \(\mathcal{S}_{\text{g}}\) of the general action in yet another form
\begin{equation}
\mathcal{S}_{\text{g}} = \frac{1}{2}\int\mathbf{T}^A \wedge \mathbf{H}_A = -\frac{1}{8}\int\dd^4x\,e\,T^A{}_{\mu\nu}H_{A\rho\sigma}\epsilon^{\mu\nu\rho\sigma}\,,
\end{equation}
where we introduced the gravitational \emph{excitation two-form}
\begin{equation}
\mathbf{H}_A = \frac{1}{2}H_{A\mu\nu}\dd x^{\mu} \wedge \dd x^{\nu}\,.
\end{equation}
Any \gls{tg} theory can be defined by specifying a \emph{constitutive relation}, i.e. a functional dependence of \(H_{A\mu\nu}\) on the tetrad \(e^A{}_{\mu}\) and the torsion tensor \(T^A{}_{\mu\nu}\). It follows that its variation can be expressed in the form
\begin{equation}
\delta H_{A\mu\nu} = \mathcal{E}_{AB\mu\nu}{}^{\rho}\delta e^B{}_{\rho} + \frac{1}{2}\mathcal{T}_{AB\mu\nu}{}^{\rho\sigma}\delta T^B{}_{\rho\sigma}\,,
\end{equation}
where the quantities \(\mathcal{E}_{AB\mu\nu}{}^{\rho}\) and \(\mathcal{T}_{AB\mu\nu}{}^{\rho\sigma}\) depend on the choice of the constitutive relation. The full variation of the gravitational part of the action is then expressed as
\begin{equation}\label{eq:gravactvarpre}
\delta\mathcal{S}_{\text{g}} = \int(\boldsymbol{\Upsilon}_A \wedge \delta\mathbf{e}^A + \boldsymbol{\Pi}_A \wedge \delta\mathbf{T}^A)\,,
\end{equation}
in terms of a three-form and a two-form
\begin{equation}
\boldsymbol{\Upsilon}_A := \frac{1}{6}\Upsilon_{A\mu\nu\rho}\dd x^{\mu} \wedge \dd x^{\nu} \wedge \dd x^{\rho}\,, \quad
\boldsymbol{\Pi}_A := \frac{1}{2}\Pi_{A\mu\nu}\dd x^{\mu} \wedge \dd x^{\nu}\,, \quad
\end{equation}
whose components are given by
\begin{subequations}
\begin{align}
\Upsilon_{A\mu\nu\rho} &:= \frac{1}{8}T^B{}_{\alpha\beta}\mathcal{E}_{BA\gamma\delta}{}^{\tau}\epsilon^{\alpha\beta\gamma\delta}\epsilon_{\tau\mu\nu\rho} = 3T^B{}_{[\mu\nu}\mathcal{E}_{BA\rho\tau]}{}^{\tau}\,,\\[0.5ex]
\Pi_{A\mu\nu} &:= \frac{1}{2}H_{A\mu\nu} - \frac{1}{16}T^B{}_{\alpha\beta}\mathcal{T}_{BA\gamma\delta}{}^{\omega\tau}\epsilon^{\alpha\beta\gamma\delta}\epsilon_{\omega\tau\mu\nu} = \frac{1}{2}H_{A\mu\nu} + \frac{3}{2}T^B{}_{[\mu\nu}\mathcal{T}_{BA\omega\tau]}{}^{\omega\tau}\,.
\end{align}
\end{subequations}
This variation is complemented with the variation of the matter action, written as
\begin{equation}\label{eq:matactvarpre}
\delta\mathcal{S}_{\text{m}} = \int\boldsymbol{\Sigma}_A \wedge \delta\mathbf{e}^A\,,
\end{equation}
in terms of the energy-momentum three-form
\begin{equation}
\boldsymbol{\Sigma}_A := \frac{1}{6}\Sigma_{A\mu\nu\rho}\dd x^{\mu} \wedge \dd x^{\nu} \wedge \dd x^{\rho}\,.
\end{equation}
This approach is related to the preceding discussion through the torsion variation
\begin{equation}
\delta\mathbf{T}^A = \DD\delta\mathbf{e}^A + \delta\boldsymbol{\omega}^A{}_B \wedge \mathbf{e}^b\,,
\end{equation}
from which follows that one can equivalently rewrite the variation~\eqref{eq:gravactvarpre} of the gravitational action as~\cite{Hohmann:2017duq,Hohmann:2018vle}
\begin{equation}
\delta\mathcal{S}_{\text{g}} = \int\left[(\boldsymbol{\Upsilon}_A - \DD\boldsymbol{\Pi}_A) \wedge \delta\mathbf{e}^A - \boldsymbol{\Pi}_A \wedge \mathbf{e}^B \wedge \delta\boldsymbol{\omega}^A{}_B\right]\,.
\end{equation}
By comparing with the variation in Eq.~\eqref{eq:gravactvar}, one therefore obtains the identifications
\begin{equation}
    W_A{}^{\mu} = \epsilon^{\mu\nu\rho\sigma}\left(\frac{1}{2}\mathcal{D}_{\nu}\Pi_{A\rho\sigma} - \frac{1}{6}\Upsilon_{A\nu\rho\sigma}\right)\,, \quad
Y_A{}^{B\mu} = \frac{1}{2}\epsilon^{\mu\nu\rho\sigma}\Pi_{A\nu\rho}e^B{}_{\sigma}\,.
\end{equation}
Similarly, comparing the variations in Eqs.~\eqref{eq:matactvarpre} and~\eqref{eq:matactvar}, one obtains the relation
\begin{equation}
\Theta_A{}^{\mu} = \frac{1}{6}\Sigma_{A\nu\rho\sigma}\epsilon^{\mu\nu\rho\sigma}\,.
\end{equation}
With these identifications in place, we see that the Lorentz invariance of the matter action, which is manifest in the symmetry of the energy-momentum tensor, takes the form
\begin{equation}
\boldsymbol{\Sigma}^{[A} \wedge \mathbf{e}^{B]} = 0\,,
\end{equation}
while the Lorentz invariance of the gravitational action implies
\begin{equation}
\boldsymbol{\Upsilon}^{[A} \wedge \mathbf{e}^{B]} + \boldsymbol{\Pi}^{[A} \wedge \mathbf{T}^{B]} = 0\,.
\end{equation}
Further, these two relations can be used to derive the field equation
\begin{equation}\label{eq:prespiconfield}
\DD\left(\boldsymbol{\Pi}^{[A} \wedge \mathbf{e}^{B]}\right) = \DD\boldsymbol{\Pi}^{[A} \wedge \mathbf{e}^{B]} + \boldsymbol{\Pi}^{[A} \wedge \mathbf{T}^{B]} = 0\,,
\end{equation}
which is obtained from variation \gls{wrt} the spin connection, from the antisymmetric part of the tetrad field equation
\begin{equation}\label{eq:pretetradfield}
\DD\boldsymbol{\Pi}_A - \boldsymbol{\Upsilon}_A = \boldsymbol{\Sigma}_A\,.
\end{equation}
Finally, the covariant energy-momentum conservation takes the form
\begin{equation}
\lc{\DD}\boldsymbol{\Sigma}_A = 0\,,
\end{equation}
while the Bianchi identities for the gravitational sector read
\begin{equation}
\lc{\DD}\left(\DD\boldsymbol{\Pi}_A - \boldsymbol{\Upsilon}_A\right) = 0\,.
\end{equation}
The main advantage of this premetric approach lies in its formal analogy to electrodynamics and the possibility to write the gravitational field equations in a form which is reminiscent of gauge theories. This becomes apparent by writing the tetrad field Eq.~\eqref{eq:pretetradfield} in the form
\begin{equation}
\DD\boldsymbol{\Pi}_A = \boldsymbol{\Upsilon}_A + \boldsymbol{\Sigma}_A\,.
\end{equation}
The \gls{rhs} of this equation can be interpreted as the total energy-momentum current, which contains both a gravitational contribution \(\boldsymbol{\Upsilon}_A\) and a matter contribution \(\boldsymbol{\Sigma}_A\). This total current is covariantly conserved \gls{wrt} the teleparallel connection,
\begin{equation}
0 = \DD^2\boldsymbol{\Pi}_A = \DD(\boldsymbol{\Upsilon}_A + \boldsymbol{\Sigma}_A)\,,
\end{equation}
since \(\DD^2 = 0\) due to the flatness of this connection. Together with the torsion relations
\begin{equation}
    \DD\mathbf{e}^A = \mathbf{T}^A\,, \quad
    \DD\mathbf{T}^A = 0\,,
\end{equation}
one easily recognizes the similarity to the equations
\begin{equation}
    \dd\mathbf{A} = \mathbf{F}\,, \quad
    \dd\mathbf{F} = 0\,, \quad
    \dd\star\mathbf{F} = \mathbf{J}\,, \quad
    \dd\mathbf{J} = 0\,,
\end{equation}
in Lorentz-Maxwell electrodynamics, where \(\mathbf{A}\) is the electromagnetic vector potential, \(\mathbf{F}\) is its field strength and \(\mathbf{J}\) is the conserved electromagnetic current density. Also here more general theories of electrodynamics, such as effects of media, can be formulated by replacing the excitation \(\star\mathbf{F}\) with a more general constitutive relation~\cite{Hehl:2001ced}. This analogy is particularly visible in the case that the gravitational excitation \(\mathbf{H}_A\) depends linearly on the torsion \(\mathbf{T}^A\), in which case \(\boldsymbol{\Pi}_A = \mathbf{H}_A\)~\cite{Itin:2016nxk}. In the non-linear case, this relation becomes more involved; a number of constitutive relations for various modified \gls{tg} theories are given in Ref.~\cite{Hohmann:2017duq}.

We finally remark that in other, curvature-based gravity theories, similar formal analogies with electrodynamics or Yang-Mills theories can be constructed. This is the case, e.g., for MacDowell-Mansouri gravity~\cite{Wise:2006sm} or Chern-Simons gravity~\cite{Wise:2009fu}.

\subsection{The ADM formalism}\label{ssec:admformalism}
A very useful tool to study gravity theories is the ADM formalism, named after Arnowitt, Deser and Misner. Before discussing its application to particular \gls{tg} theories, we now discuss a few general aspects of its use in this context. In Sec.~\ref{sssec:adm31split}, we discuss the $3+1$ decomposition of the teleparallel geometry into space and time components. This leads to the introduction of lapse and shift variables in Sec.~\ref{sssec:lapseshift}. The question of Lorentz gauge fixing is discussed in Sec.~\ref{sssec:admlorgauge}. In Sec.~\ref{sssec:genadmcanvar}, we introduce canonical variables, and discuss their irreducible decomposition in Sec.~\ref{sssec:velmomirrdec}.

\subsubsection{\texorpdfstring{$3+1$}{3 + 1} decomposition of teleparallel geometry}\label{sssec:adm31split}

In order to perform a Hamiltonian analysis, one usually starts with a split of all dynamical variables in their space and time components. Such a decomposition is done relative to a foliation of spacetime with timelike hypersurfaces, i.e., one identifies the spacetime manifold \(\mathcal{M}\), which is assumed to be globally hyperbolic, with a direct product of the form \(\mathbb{R} \times \Sigma\). This identification allows essentially two different possibilities to split the dynamical variables into space and time components, which we will write out explicitly for the tetrads and their duals:

\begin{enumerate}
\item
In the first approach, one uses coordinates \((x^{\mu}) = (x^0 = t, x^i)\), where lowercase Latin letters denote spatial indices following the convention~\ref{ssec:Conventions}, which are adapted to the foliation, such that the time coordinate \(t\) parametrizes the different spatial slices, while \((x^i)\) are spatial coordinates on each spatial slice \(\Sigma_t \subset \mathcal{M}\). In these coordinates the tetrad naturally decomposes as
\begin{equation}
\mathbf{e}^A = e^A{}_{\mu}\dd x^{\mu} = e^A{}_0\dd t + e^A{}_i\dd x^i\,,
\end{equation}
while its inverse decomposes as
\begin{equation}
\mathbf{E}_A = E_A{}^{\mu}\partial_{\mu} = E_A{}^0\partial_0 + E_A{}^i\partial_i\,.
\end{equation}
Notable works using this decomposition are Refs.~\cite{Maluf:1994ji,Maluf:1998ae,Maluf:2000ah,Maluf:2001rg,daRochaNeto:2011ir,Okolow:2011np,Okolow:2011nq,Okolow:2013lwa,Ferraro:2016wht,Ferraro:2018tpu,Guzman:2020kgh,Blixt:2018znp,Blixt:2019mkt,Hohmann:2019sys}.

\item
Another approach does not make use of adapted coordinates, but employs the spacetime metric \(g_{\mu\nu}\). In this approach, one starts from the unique normalized, future pointing, hypersurface-orthogonal vector field \(n^{\mu}\partial_{\mu}\), which satisfies
\begin{equation}
    n^{\mu}n_{\mu} = g_{\mu\nu}n^{\mu}n^{\nu} = 1\,, \quad X^{\mu}n_{\mu} = g_{\mu\nu}X^{\mu}n^{\nu} = 0\,,
\end{equation}
for any vector field \(X^{\mu}\partial_{\mu}\) which is tangent to the spatial hypersurfaces. The metric then decomposes into the form
\begin{equation}
    g_{\mu\nu} = n_{\mu}n_{\nu} - h_{\mu\nu}\,,
\end{equation}
where \(h_{\mu\nu}\) restricts to a Riemannian metric on every spatial hypersurface and \(n^{\mu}h_{\mu\nu} = 0\). Raising one index, one obtains two projectors
\begin{equation}\label{eq:orparprojectors}
\delta^{\mu}_{\nu} = (P_{\perp})^{\mu}_{\nu} + (P_{\parallel})^{\mu}_{\nu}\,, \quad
(P_{\perp})^{\mu}_{\nu} = n^{\mu}n_{\nu}\,, \quad
(P_{\parallel})^{\mu}_{\nu} = -h^{\mu}_{\nu}\,,
\end{equation}
where it is convenient to write the indices directly on top of each other, as for the Kronecker symbol \(\delta^{\mu}_{\nu}\). These projectors can be applied to any covariant or contravariant index
\begin{subequations}
\begin{align}
V^{\mu} &= (P_{\perp})^{\mu}_{\nu}V^{\nu} + (P_{\parallel})^{\mu}_{\nu}V^{\nu} = n^{\mu}n_{\nu}V^{\nu} - h^{\mu}_{\nu}V^{\nu}\,,\\[0.5ex]
V_{\mu} &= (P_{\perp})_{\mu}^{\nu}V_{\nu} + (P_{\parallel})_{\mu}^{\nu}V_{\nu} = n_{\mu}n^{\nu}V_{\nu} - h_{\mu}^{\nu}V_{\nu}\,.
\end{align}
\end{subequations}
A common notation for the obtained projected components is given by
\begin{equation}
n_{\mu}V^{\mu} = V^{\perp}\,, \quad
n^{\mu}V_{\mu} = V_{\perp}\,, \quad
-h^{\mu}_{\nu}V^{\nu} = V^{\bar{\mu}}\,, \quad
-h_{\mu}^{\nu}V_{\nu} = V_{\bar{\mu}}\,,
\end{equation}
where a bar on top of an index is a shorthand notation to indicate that the corresponding index has been contracted with a spatial projector. Hence, for the tetrad one has the decomposition
\begin{equation}
e^A{}_{\mu} = n_{\mu}e^A{}_{\perp} + e^A{}_{\bar{\mu}}\,,
\end{equation}
while its dual decomposes as
\begin{equation}
E_A{}^{\mu} = n^{\mu}E_A{}^{\perp} + E_A{}^{\bar{\mu}}\,.
\end{equation}
Notable works using this decomposition are Refs.~\cite{Cheng:1988zg,Blagojevic:2000qs,Blagojevic:2000xd,Mitric:2019rop,Blagojevic:2020dyq}.
\end{enumerate}

Finally, one can analogously decompose also the teleparallel spin connection \(\omega^A{}_{B\mu}\). However, this step can be avoided by working in the Weitzenb\"{o}ck gauge, which is always possible, as we shall see below.

\subsubsection{Lapse and shift variables}\label{sssec:lapseshift}

Before performing the Hamiltonian analysis, it turns out to be more convenient to introduce a different set of variables in order to replace the temporal tetrad components \(e^A{}_0\) or \(e^A{}_{\perp}\) introduced in the $3+1$ decomposition. For this purpose one writes the components of the unit normal vector field in the form
\begin{equation}\label{eq:normveccoordls}
n^0 = \frac{1}{N}\,, \quad
n^i = -\frac{N^i}{N}\,,
\end{equation}
introducing the lapse \(N\) and the components \(N^i\) of the shift vector field. They satisfy the relation
\begin{equation}
\partial_0 = Nn^{\mu}\partial_{\mu} + N^i\partial_i\,,
\end{equation}
and therefore relate the two different possibilities to split tensor components in the previous section. Using these variables, it is possible to express the temporal tetrad components in the form
\begin{equation}\label{eq:tetradlapseshift}
e^A{}_0 = Nn^A + N^ie^A{}_i\,,
\end{equation}
where \(n^A = e^A{}_{\mu}n^{\mu} = e^A{}_{\perp}\) is the expression of the unit normal vector field in the tetrad basis. A crucial insight in the transition to the new variables is the fact that the latter can be fully expressed in terms of the spatial tetrad components \(e^A{}_i\), without using the time components \(e^A{}_0\). To see this, note first that the condition that \(n^{\mu}\) is normal to the spatial hypersurfaces can be expressed as
\begin{equation}\label{eq:normveclorortho}
0 = g(n, \partial_i) = n^{\mu}g_{\mu i} = n^A\eta_{AB}e^B{}_i = n_Ae^A{}_i = n_i\,,
\end{equation}
from which likewise follows
\begin{equation}
n_0 = n_Ae^A{}_0 = n_A(Nn^A + N^ie^A{}_i) = N\,.
\end{equation}
The second condition is the normalization, which in the tetrad basis simply reads
\begin{equation}
1 = n^An_A = \eta_{AB}n^An^B\,,
\end{equation}
and so involves only the Minkowski metric. These two conditions uniquely determine \(n^A\) up to a sign difference, and leads to the expression~\cite{Okolow:2011np}
\begin{equation}
n^A{} = \pm\frac{1}{6}\epsilon^A{}_{BCD}\varepsilon^{ijk}e^B{}_ie^C{}_je^D{}_k\,,
\end{equation}
where \(\varepsilon_{ijk}\) denotes the totally antisymmetric tensor field defined by the induced metric \(h_{ij} = \eta_{AB}e^A{}_ie^B{}_j\) on the spatial hypersurfaces. The sign is finally fixed by the demand that the normal vector field is future pointing, which means positive lapse, \(N > 0\), following the decomposition~\eqref{eq:normveccoordls}. This can be conducted by realizing that the four Lorentz vectors \((n^A, e^A{}_i)\) with \(i = 1,2,3\) are linearly independent; this follows from the fact that only if the \(e^A{}_i\) are linearly independent, the metric \(h_{ij}\) is non-degenerate, while the orthogonality condition~\eqref{eq:normveclorortho} implies that also \(n^A\) is linearly independent. Hence, these four Lorentz vectors constitute a basis of Minkowski space. Assuming that the three spatial tetrad components \(e^A{}_i\) are related to an oriented coordinate basis \(\partial_i\) of the spatial hypersurfaces, then translates the condition that \(n^A\) is future pointing into a condition on the orientation of the basis \((n^A, e^A{}_i)\). This basis is oriented if the determinant
\begin{subequations}
\begin{alignat}{2}
e & =\: & & \epsilon_{ABCD}e^A{}_0e^B{}_1e^C{}_2e^D{}_3\\[0.5ex]
& =\: & & N\epsilon_{ABCD}n^Ae^B{}_1e^C{}_2e^D{}_3\\[0.5ex]
& =\: & & \pm\frac{1}{6}N\epsilon_{ABCD}\epsilon^A{}_{EFG}\varepsilon^{ijk}e^B{}_1e^C{}_2e^D{}_3e^E{}_ie^F{}_je^G{}_k\\[0.5ex]
& =\: & & \mp N\varepsilon^{ijk}h_{i1}h_{j2}h_{k3}\\[0.5ex]
& =\: & & \mp N\varepsilon_{123}\\[0.5ex]
& =\: & & \mp N\sqrt{h}\,,
 \end{alignat}
 \end{subequations}
is positive, and so the lower sign must be chosen. This determines the normal vector to be given by
\begin{equation}
n^A{} = -\frac{1}{6}\epsilon^A{}_{BCD}\varepsilon^{ijk}e^B{}_ie^C{}_je^D{}_k\,.
\end{equation}
Together with the relation~\eqref{eq:tetradlapseshift}, it is thus possible to express the time component \(e^A{}_0\) of the tetrad through its spatial components, the lapse and the shift. In summary, this means that the 16 components \(e^A{}_{\mu}\) of the tetrad are equivalently expressed by the 16 variables
\begin{equation}\label{eq:admvariables}
N\,, \quad N^i\,, \quad e^A{}_i\,.
\end{equation}
This is the ADM decomposition of the tetrad in \gls{tg}. With its help, the inverse tetrad is written as
\begin{equation}
E_A{}^0 = \frac{1}{N}n_A\,, \quad
E_A{}^i = -\eta_{AB}h^{ij}e^B{}_j - \frac{N^i}{N}n_A\,,
\end{equation}
where \(h^{ij}\) denotes the inverse of the spatial metric \(h_{ij} = -\eta_{AB}e^A{}_ie^B{}_j\). Further, for the metric this yields the usual ADM decomposition
\begin{equation}
g_{00} = N^2 - N^iN^jh_{ij}\,, \quad
g_{0i} = -N^jh_{ij}\,, \quad
g_{ij} = -h_{ij}\,,
\end{equation}
using the fact that \(n_i = 0\), together with the inverse metric
\begin{equation}
g^{00} = \frac{1}{N^2}\,, \quad
g^{0i} = -\frac{N^i}{N^2}\,, \quad
g^{ij} = \frac{N^iN^j}{N^2} - h^{ij}\,.
\end{equation}
Transforming the indices of the metric to Lorentz indices with the help of the tetrad, this yields the helpful relations
\begin{equation}
\eta_{AB} = n_An_B - \eta_{AC}\eta_{BD}h^{ij}e^C{}_ie^D{}_j\,,
\end{equation}
and
\begin{equation}
\eta^{AB} = n^An^B - h^{ij}e^A{}_ie^B{}_j\,.
\end{equation}
These allow us to express the orthogonal and parallel projectors~\eqref{eq:orparprojectors} as
\begin{equation}\label{eq:orparprojlor}
    (P_{\perp})^A_B = n^An_B\,, \quad (P_{\parallel})^A_B = -\eta_{BC}h^{ij}e^A{}_ie^C{}_j\,,
\end{equation}
in the tetrad basis, which we will use later for the decomposition of further quantities into the spatial and normal components.

\subsubsection{Lorentz gauge fixing} \label{sssec:admlorgauge}

As discussed in Sec.~\ref{ssec:tggenprop}, the covariant formulation of \gls{tg} generically involves a Lorentz gauge freedom, which is implemented via the flat, metric-compatible spin connection \(\omega^A{}_{B\mu}\). Since the latter represents a pure gauge \gls{dof}, the question arises how to treat this variable in the Hamiltonian formalism. The treatment of the spin connection can essentially be classified into three approaches:
\begin{enumerate}
\item
Using Lagrange multipliers: As discussed in Sec.~\ref{sssec:tggenactfield}, one possibility to include the spin connection into the teleparallel action and enforce its flatness and compatibility with the metric is by introducing Lagrange multipliers, which impose the conditions of vanishing curvature and non-metricity. In the Hamiltonian formalism, these lead to constraints on the spin connection, thus implementing its flatness and metric compatibility. Further primary constraints arise from the Lorentz gauge freedom, linking the spin connection and the tetrad through their common change under Lorentz transformations, which leave the action invariant. Finally, also for the Lagrange multipliers, constraints arise which implement the condition that their canonical momenta vanish identically. This large number of variables and associated constraints therefore makes this approach rather cumbersome. It is used in Refs.~\cite{Maluf:1994ji,Blagojevic:2000qs,Blagojevic:2000xd}.

\item
Integrating the spin connection: Another approach to treat the spin connection, which does not necessitate the use of Lagrange multipliers, arises from the fact that locally it may be integrated in the form~\eqref{eq:spclormatrep} and thus expressed by a local Lorentz transformation \(\Lambda^A{}_B\). Instead of \(\omega^A{}_{B\mu}\), one thus considers \(\Lambda^A{}_B\) as the fundamental field variable which implements the Lorentz gauge freedom. As a consequence, the derived connection is flat and metric-compatible by construction, and therefore does not have to be enforced by constraints. Together with the absence of Lagrange multipliers, the number of variables and associated constraints is therefore significantly reduced compared to the aforementioned approach. Nevertheless, constraints still appear from the gauge invariance of the teleparallel action under local Lorentz transformations, and these constraints intertwine the conjugate momenta of the tetrad and spin connection \gls{dof}. However, this approach is not common for the Hamiltonian analysis.

\item
Fixing the gauge: Finally, it is also possible to eliminate the spin connection completely prior to performing the Hamiltonian analysis. This can be achieved starting from the preceding approach which integrates the spin connection locally to a variable \(\Lambda^A{}_B\), which constitutes a local Lorentz transformation, by making a change of variables according to
\begin{equation}\label{eq:wbgfixfieldvar}
\tilde{e}^A{}_{\mu} = (\Lambda^{-1})^A{}_Be^B{}_{\mu}\,, \quad
\tilde{\Lambda}^A{}_B = \Lambda^A{}_B\,,
\end{equation}
which is equivalent to the change of ADM variables given by
\begin{equation}
\tilde{N} = N\,, \quad
\tilde{N}^i = N^i\,, \quad
\tilde{e}^A{}_i = (\Lambda^{-1})^A{}_Be^B{}_i\,, \quad
\tilde{\Lambda}^A{}_B = \Lambda^A{}_B\,.
\end{equation}
One finds that the new variable \(\tilde{e}^A{}_{\mu}\) simply represents the Weitzenb\"{o}ck tetrad which is associated to the original tetrad \(e^A{}_{\mu}\). Since the Weitzenb\"{o}ck tetrad is uniquely determined from the original tetrad (up to a global Lorentz transformation) by the condition that \(\Lambda^A{}_B\) relates the original spin connection \(\omega^A{}_{B\mu}\) to the vanishing Weitzenb\"{o}ck spin connection, it follows that the new variable \(\tilde{e}^A{}_B\) is invariant under any further local Lorentz transformation \(\hat{\Lambda}^A{}_B\), which acts on the original variables as
\begin{equation}
e'^A{}_{\mu} = \hat{\Lambda}^A{}_Be^B{}_{\mu}\,, \quad
\Lambda'^A{}_B = \hat{\Lambda}^A{}_C\Lambda^C{}_B\,.
\end{equation}
Note that here the Lorentz transformation \(\hat{\Lambda}^A{}_B\) is applied only to the first index of the variable \(\Lambda^A{}_B\), but not to the second. This is due to the fact that the first index of \(\Lambda^A{}_B\) represents the variable, non-Weitzenb\"{o}ck gauge, in which the tetrad is given by \(e^A{}_{\mu}\), while the second index refers to the fixed Weitzenb\"{o}ck gauge, from which the spin connection \(\omega^A{}_{B\mu}\) is obtained through the transformation \(\Lambda^A{}_B\). This can also be seen by explicitly calculating the transformation
\begin{subequations}
\begin{alignat}{2}
\omega'^A{}_{B\mu} & =\: & & \Lambda'^A{}_C\partial_{\mu}(\Lambda'^{-1})^C{}_B\\[0.5ex]
& =\: & & \hat{\Lambda}^A{}_D\Lambda^D{}_C\partial_{\mu}\left[(\Lambda^{-1})^C{}_E(\hat{\Lambda}^{-1})^E{}_B\right]\\[0.5ex]
& =\: & & \hat{\Lambda}^A{}_D\left[(\hat{\Lambda}^{-1})^E{}_B\Lambda^D{}_C\partial_{\mu}(\Lambda^{-1})^C{}_E + \partial_{\mu}(\hat{\Lambda}^{-1})^D{}_B\right]\\[0.5ex]
& =\: & & \hat{\Lambda}^A{}_D(\hat{\Lambda}^{-1})^E{}_B\omega^D{}_{E\mu} + \hat{\Lambda}^A{}_D\partial_{\mu}(\hat{\Lambda}^{-1})^D{}_B\,,
 \end{alignat}
 \end{subequations}
which is the behavior~\eqref{eq:spclortrans} of a spin connection under local Lorentz transformations. From the change of variables~\eqref{eq:wbgfixfieldvar} one can see that these transformations cancel each other for \(\tilde{e}^A{}_{\mu}\), as can be seen by explicitly calculating
\begin{equation}
\tilde{e}'^A{}_{\mu}
= (\Lambda'^{-1})^A{}_Be'^B{}_{\mu}g
= (\Lambda^{-1})^A{}_B(\hat{\Lambda}^{-1})^B{}_C\hat{\Lambda}^C{}_De^D{}_{\mu}g
= (\Lambda^{-1})^A{}_Be^B{}_{\mu}g
= \tilde{e}^A{}_{\mu}\,.
\end{equation}
Hence, Lorentz transformations act exclusively on \(\tilde{\Lambda}^A{}_B\), so that the new variables \((\tilde{e}^A{}_{\mu}, \tilde{\Lambda}^A{}_B)\) provide a split into a gauge-invariant and a pure gauge variable. Further, one finds that the metric and teleparallel affine connection, which enter \gls{tg} actions, are expressed in the new variables as
\begin{equation}
g_{\mu\nu} = \eta_{AB}e^A{}_{\mu}e^B{}_{\nu} = \eta_{AB}\Lambda^A{}_C\Lambda^B{}_D\tilde{e}^C{}_{\mu}\tilde{e}^D{}_{\nu} = \eta_{AB}\tilde{e}^A{}_{\mu}\tilde{e}^B{}_{\nu}\,,
\end{equation}
as well as
\begin{subequations}
\begin{alignat}{2}
\Gamma^{\mu}{}_{\nu\rho} & =\: & & E_A{}^{\mu}\left[\partial_{\rho}e^A{}_{\nu} + \Lambda^A{}_B\partial_{\rho}(\Lambda^{-1})^B{}_Ce^C{}_{\nu}\right]\\[0.5ex]
& =\: & &(\Lambda^{-1})^D{}_A\tilde{E}_D{}^{\mu}\left[\partial_{\rho}(\Lambda^A{}_B\tilde{e}^B{}_{\nu}) + \Lambda^A{}_B\partial_{\rho}(\Lambda^{-1})^B{}_C\Lambda^C{}_E\tilde{e}^E{}_{\nu}\right]\\[0.5ex]
& =\: & &(\Lambda^{-1})^D{}_A\tilde{E}_D{}^{\mu}\left[\Lambda^A{}_B\partial_{\rho}\tilde{e}^B{}_{\nu} + \tilde{e}^B{}_{\nu}\partial_{\rho}\Lambda^A{}_B - \partial_{\rho}\Lambda^A{}_E\tilde{e}^E{}_{\nu}\right]\\[0.5ex]
& =\: & & \tilde{E}_A{}^{\mu}\partial_{\rho}\tilde{e}^A{}_{\nu}\,,
\end{alignat}
\end{subequations}
where the latter confirms the previous statement that \(\tilde{e}^A{}_{\mu}\) represents the tetrad in the Weitzenb\"{o}ck gauge. Since any Lorentz invariant \gls{tg} action can fully be expressed in terms of the metric and teleparallel affine connection through its torsion, as we have shown in Sec.~\ref{sssec:gengravactlorinv}, it follows that the action depends only on the gauge-invariant variable \(\tilde{e}^A{}_{\mu}\), while the pure Lorentz gauge variable \(\tilde{\Lambda}^A{}_B\) does not contribute to the action. Its canonical momenta therefore vanish identically, and it does not need to be considered further in the Hamiltonian approach, so that one is left with working with \(\tilde{e}^A{}_{\mu}\) as the only teleparallel field variable. Note that this approach is equivalent to imposing the Weitzenb\"{o}ck gauge \(\omega^A{}_{B\mu} \equiv 0\) from the beginning~\cite{Blixt:2019mkt,Golovnev:2021omn}, and thus to working in the non-covariant, pure tetrad formulation of \gls{tg}. This is employed in Refs.~\cite{Cheng:1988zg,Maluf:1998ae,Maluf:2000ah,Maluf:2001rg,daRochaNeto:2011ir,Okolow:2011np,Okolow:2011nq,Okolow:2013lwa,Ferraro:2016wht,Ferraro:2018tpu,Guzman:2020kgh,Blixt:2018znp,Blixt:2019mkt,Hohmann:2019sys,Mitric:2019rop,Blagojevic:2020dyq}.
\end{enumerate}
In the following, we will follow the latter, gauge-fixed approach, and use the Weitzenb\"{o}ck tetrad \(\tilde{e}^A{}_{\mu}\) and its ADM decomposition as the only teleparallel field variable. For convenience, we will drop the tilde on all variables, hence assuming the Weitzenb\"{o}ck gauge from the beginning.

\subsubsection{Canonical variables} \label{sssec:genadmcanvar}

In the following we will assume that the gravitational dynamics is obtained from a teleparallel action of the form~\eqref{eq:matgravaction}, where we further assume that the gravitational action \(\mathcal{S}_{\text{g}}\) is of first derivative order in the gravitational field variables. From the Lorentz invariance of the gravitational action it then follows that it is constructed from the metric \(g_{\mu\nu}\) and the torsion tensor \(T^{\mu}{}_{\nu\rho}\) only, while derivatives thereof are excluded, as they would introduce at least second order derivatives of the tetrad. Hence, the action depends only on the tetrads \(e^A{}_{\mu}\) and their first order derivatives \(\partial_{\mu}e^A{}_{\nu}\), where the latter enter only through the torsion. Splitting these torsion into space and time components,
\begin{equation}\label{eq:tor31split}
T^A{}_{00} = 0\,, \quad
T^A{}_{0i} = -T^A{}_{i0} = \partial_0e^A{}_i - \partial_ie^A{}_0\,, \quad
T^A{}_{ij} = \partial_ie^A{}_j - \partial_je^A{}_i\,,
\end{equation}
it follows from the fact that the torsion is antisymmetric in its last two indices, that the only terms which enter the action are
\begin{equation}
e^A{}_0\,, \quad e^A{}_i\,, \quad \partial_0e^A{}_i\,, \quad \partial_ie^A{}_0\,, \quad \partial_ie^A{}_j\,,
\end{equation}
while the time derivative \(\partial_0e^A{}_0\) does not contribute. Equivalently, one may use the relation~\eqref{eq:tetradlapseshift} to replace \(e^A{}_i\) by the ADM variables~\eqref{eq:admvariables}, so that the action is expressed by the fields
\begin{equation}
N\,, \quad N^i\,, \quad e^A{}_i\,, \quad \partial_0e^A{}_i\,, \quad \partial_iN\,, \quad \partial_iN^j\,, \quad \partial_ie^A{}_j\,.
\end{equation}
The time derivatives (or \emph{velocities}) of the spatial tetrad components are conventionally denoted
\begin{equation}
v^A{}_i = \dot{e}^A{}_i = \partial_0e^A{}_i\,.
\end{equation}
Variation of the gravitational action \(\mathcal{S}_{\text{g}}\) \gls{wrt} the velocities defines their conjugate momenta \(\pi_A{}^i\) as
\begin{equation}\label{eq:canmomdef}
\delta_{\dot{e}}\mathcal{S}_{\text{g}} = \int\dd^4x\,\pi_A{}^i\,\delta\dot{e}^A{}_i\,.
\end{equation}
Note that by definition, they are tensor densities, i.e., they include the density factor \(e\) given by the determinant of the tetrad. It follows that also the Hamiltonian
\begin{equation}\label{eq:hamiltondef}
H = \pi_A{}^iv^A{}_i - L\,,
\end{equation}
like the Lagrangian \(L\), is a scalar density.

\subsubsection{Irreducible decomposition} \label{sssec:velmomirrdec}

As argued in Sec.~\ref{sssec:genadmcanvar}, varying the action of a generic \gls{tg} theory \gls{wrt} the velocities \(\dot{e}^A{}_i\) leads to the canonical momenta \(\pi_A{}^i\), which are conjugate to the spatial tetrad components \(e^A{}_i\). These momenta are obtained as functions which depend on the teleparallel field variables and their time derivatives. In order to proceed with the Hamiltonian analysis, this relation needs to be inverted and the velocities determined as functions of the momenta. This task can become rather involved, since the relation between momenta is, in general, non-trivial. However, it can be simplified by choosing a decomposition of the momenta and velocities into irreducible components under the rotation group acting on the spatial hypersurfaces. For the momenta, one defines the objects
\begin{subequations}\label{eq:canmomirrdec}
\begin{align}
\accentset{\mathcal{V}}{\pi}^i &= n^A\pi_A{}^i\,,\\[0.5ex]
\accentset{\mathcal{A}}{\pi}^{ij} &= e^A{}_kh^{k[i}\pi_A{}^{j]}\,,\\[0.5ex]
\accentset{\mathcal{S}}{\pi}^{ij} &= e^A{}_kh^{k(i}\pi_A{}^{j)} - \frac{1}{3}h^{ij}e^A{}_k\pi_A{}^k\,,\\[0.5ex]
\accentset{\mathcal{T}}{\pi} &= \frac{1}{3}e^A{}_i\pi_A{}^i\,,
\end{align}
\end{subequations}
where the letters \(\mathcal{V}, \mathcal{A}, \mathcal{S}, \mathcal{T}\) are chosen to indicate the vector, antisymmetric, symmetric trace-free and trace components. From these irreducible components, the original components of the momenta are recovered as
\begin{equation}\label{eq:canmomirrcomp}
\pi_A{}^i = \eta_{AB}\left[n^B\accentset{\mathcal{V}}{\pi}^i + e^B{}_j\left(\accentset{\mathcal{A}}{\pi}^{ij} + \accentset{\mathcal{S}}{\pi}^{ij}\right) + e^B{}_jh^{ij}\accentset{\mathcal{T}}{\pi}\right]\,.
\end{equation}
A similar decomposition is introduced for the velocities \(v^A{}_i = \dot{e}^A{}_i\), and which can be defined as
\begin{subequations}\label{eq:canvelirrdec}
\begin{align}
\accentset{\mathcal{V}}{v}_i &= n_Av^A{}_i\,,\\[0.5ex]
\accentset{\mathcal{A}}{v}_{ij} &= \eta_{AB}e^B{}_{[i}v^A{}_{j]}\,,\\[0.5ex]
\accentset{\mathcal{S}}{v}_{ij} &= \eta_{AB}e^B{}_{(i}v^A{}_{j)} - \frac{1}{3}\eta_{AB}h_{ij}h^{kl}e^B{}_lv^A{}_k\,,\\[0.5ex]
\accentset{\mathcal{T}}{v} &= \frac{1}{3}\eta_{AB}h_{ij}h^{kl}e^B{}_lv^A{}_k\,,
\end{align}
\end{subequations}
and from which the original velocity components are recovered as
\begin{equation}\label{eq:canvelirrcomp}
v^A{}_i = n^A\accentset{\mathcal{V}}{v}_i + e^A{}_kh^{kj}\left(\accentset{\mathcal{A}}{v}_{ij} + \accentset{\mathcal{S}}{v}_{ij}\right) + e^A{}_i\accentset{\mathcal{T}}{v}\,.
\end{equation}
The virtue of this decomposition becomes apparent, for example, in calculating the product \(\pi_A{}^iv^A{}_i\), which enters the Hamiltonian. It follows from the irreducibility of the decomposition above that this term reduces to
\begin{equation}\label{eq:velmomirrdec}
\pi_A{}^iv^A{}_i = \accentset{\mathcal{V}}{\pi}^i\accentset{\mathcal{V}}{v}_i + \accentset{\mathcal{A}}{\pi}^{ij}\accentset{\mathcal{A}}{v}_{ij} + \accentset{\mathcal{S}}{\pi}^{ij}\accentset{\mathcal{S}}{v}_{ij} + 3\accentset{\mathcal{T}}{\pi}\accentset{\mathcal{T}}{v}\,,
\end{equation}
which can be seen explicitly as follows. First, note that the last two terms in the decomposition~\eqref{eq:velmomirrdec} combine as
\begin{subequations}
\begin{alignat}{2}
\accentset{\mathcal{S}}{\pi}^{ij}\accentset{\mathcal{S}}{v}_{ij} + 3\accentset{\mathcal{T}}{\pi}\accentset{\mathcal{T}}{v} & =\: & & \left(e^A{}_kh^{k(i}\pi_A{}^{j)} - \frac{1}{3}h^{ij}e^A{}_k\pi_A{}^k\right)\left(\eta_{BC}e^C{}_{(i}v^B{}_{j)} - \frac{1}{3}\eta_{BC}h_{ij}h^{lm}e^C{}_mv^B{}_l\right)\nonumber\\[0.5ex]
 & \: & & + \frac{1}{3}\eta_{BC}h^{jk}e^A{}_ie^C{}_k\pi_A{}^iv^B{}_j\\[0.5ex]
 & =\: & &\eta_{BC}e^A{}_kh^{k(i}\pi_A{}^{j)}e^C{}_{(i}v^B{}_{j)} + \frac{2}{3}\eta_{BC}h^{jk}e^A{}_ie^C{}_k\pi_A{}^iv^B{}_j\nonumber\\[0.5ex]
 & \: & & - \frac{1}{3}\eta_{BC}h^{ij}e^A{}_k\pi_A{}^ke^C{}_{(i}v^B{}_{j)} - \frac{1}{3}\eta_{BC}h_{ij}h^{lm}e^C{}_mv^B{}_le^A{}_kh^{k(i}\pi_A{}^{j)}\\[0.5ex]
 & =\: & & \eta_{BC}e^A{}_kh^{k(i}\pi_A{}^{j)}e^C{}_{(i}v^B{}_{j)}\,.
 \end{alignat}
\end{subequations}
Together with the second term in the decomposition~\eqref{eq:velmomirrdec} this yields
\begin{equation}
\accentset{\mathcal{A}}{\pi}^{ij}\accentset{\mathcal{A}}{v}_{ij} + \accentset{\mathcal{S}}{\pi}^{ij}\accentset{\mathcal{S}}{v}_{ij} + 3\accentset{\mathcal{T}}{\pi}\accentset{\mathcal{T}}{v} = \eta_{BC}h^{ki}e^A{}_ke^C{}_i\pi_A{}^jv^B{}_j = (P_{\parallel})^A_B\pi_A{}^iv^B{}_i\,,
\end{equation}
since cross-terms arising from the contraction of symmetric and antisymmetric tensors cancel, and we obtain the parallel projector~\eqref{eq:orparprojlor}. Finally, the remaining term is now simply the orthogonal part
\begin{equation}
\accentset{\mathcal{V}}{\pi}^i\accentset{\mathcal{V}}{v}_i = n^An_B\pi_A{}^iv^B{}_i = (P_{\perp})^A_B\pi_A{}^iv^B{}_i\,,
\end{equation}
so that their sum indeed yields \(\pi_A{}^iv^B{}_i\). An alternative approach to this result is by introducing the notation
\begin{equation}
\accentset{\mathcal{V}}{\pi}_A{}^i = n_A\accentset{\mathcal{V}}{\pi}^i\,, \quad
\accentset{\mathcal{A}}{\pi}_A{}^i = \eta_{AB}e^B{}_j\accentset{\mathcal{A}}{\pi}^{ij}\,, \quad
\accentset{\mathcal{S}}{\pi}_A{}^i = \eta_{AB}e^B{}_j\accentset{\mathcal{S}}{\pi}^{ij}\,, \quad
\accentset{\mathcal{T}}{\pi}_A{}^i = \eta_{AB}e^B{}_jh^{ij}\accentset{\mathcal{T}}{\pi}
\end{equation}
and
\begin{equation}
    \accentset{\mathcal{V}}{v}^A{}_i = n^A\accentset{\mathcal{V}}{v}_i\,, \quad
    \accentset{\mathcal{A}}{v}^A{}_i = e^A{}_kh^{kj}\accentset{\mathcal{A}}{v}_{ij}\,, \quad
    \accentset{\mathcal{S}}{v}^A{}_i = e^A{}_kh^{kj}\accentset{\mathcal{S}}{v}_{ij}\,, \quad
    \accentset{\mathcal{T}}{v}^A{}_i = e^A{}_i\accentset{\mathcal{T}}{v}\,,
\end{equation}
for the terms in the decompositions \eqref{eq:canmomirrcomp} and~\eqref{eq:canvelirrcomp}. Using this notation, the decomposition~\eqref{eq:velmomirrdec} equivalently reads
\begin{equation}
\pi_A{}^iv^A{}_i = \accentset{\mathcal{V}}{\pi}_A{}^i\accentset{\mathcal{V}}{v}^A{}_i + \accentset{\mathcal{A}}{\pi}_A{}^i\accentset{\mathcal{A}}{v}^A{}_i + \accentset{\mathcal{S}}{\pi}_A{}^i\accentset{\mathcal{S}}{v}^A{}_i + \accentset{\mathcal{T}}{\pi}_A{}^i\accentset{\mathcal{T}}{v}^A{}_i\,,
\end{equation}
which is proven in a similar way. Another advantage of the irreducible decomposition is that the different components in the relations of the momenta and velocities generically decouple, which simplifies inverting these relations. This will be shown explicitly when we discuss the application of the ADM formalism to specific \gls{tg} theories.

\subsection{The teleparallel equivalent of general relativity} \label{ssec:tegr}

In the context and circumstances of the points raised in this section thus far, a gravitational analogue to \gls{gr} can be written down that is dynamically equivalent. It was Einstein who first proposed the \gls{tegr} \cite{2005physics...3046U}, which was initially part of his attempt to unify gravitational and electromagnetic interactions. An attempt that ultimately failed but which produced an independent theoretical formalism of gravitation that was dynamically equivalent to GR. In practical terms, this means that the field equations and so the classical predictions of the theory, would be equivalent. However, given the differences in their action formulations, these theories are not bound to the same fate when quantum considerations are taken into account, among other areas where it is the action rather than the dynamical equations that play the more important role.

\subsubsection{The TEGR action and field equations} \label{sec:TEGR_acion}

In \gls{gr}, the Einstein-Hilbert action is ultimately dependent on a single dynamical variable, namely the metric tensor \cite{misner1973gravitation}. This occurs through the metric tensor itself and the Riemann tensor expressed through
\begin{equation}
    \udt{\lc{R}}{\rho}{\lambda\nu\mu} = \udt{\lc{\Gamma}}{\rho}{\lambda\mu,\nu} - \udt{\lc{\Gamma}}{\rho}{\lambda\nu,\mu} + \udt{\lc{\Gamma}}{\rho}{\sigma\nu} \udt{\lc{\Gamma}}{\sigma}{\lambda\mu} - \udt{\lc{\Gamma}}{\rho}{\sigma\mu} \udt{\lc{\Gamma}}{\sigma}{\lambda\nu}\,,
\end{equation}
which gives a measure of the curvature associated with the Levi-Civita connection. If this is exchanged with the teleparallel connection, the resulting Riemann tensor
\begin{equation}
    \udt{R}{\rho}{\lambda\nu\mu} = \udt{\Gamma}{\rho}{\lambda\mu,\nu} - \udt{\Gamma}{\rho}{\lambda\nu,\mu} + \udt{\Gamma}{\rho}{\sigma\nu} \udt{\Gamma}{\sigma}{\lambda\mu} - \udt{\Gamma}{\rho}{\sigma\mu} \udt{\Gamma}{\sigma}{\lambda\nu} \equiv 0\,,
\end{equation}
will identically vanish due to the connection being curvature-less property. Using Eq.~\eqref{eq:affcondec} the two connections can be related together through the contortion tensor $\udt{K}{\rho}{\mu\nu}$ with the relation
\begin{equation}
    \udt{\Gamma}{\rho}{\mu\nu} = \udt{\lc{\Gamma}}{\rho}{\mu\nu} + \udt{K}{\rho}{\mu\nu}\,,\label{Eq:Contortion_def}
\end{equation}
which means that the two forms of the Riemann tensor can be related through \cite{Aldrovandi:2013wha}
\begin{equation}
    0 \equiv \udt{R}{\rho}{\lambda\nu\mu} = \udt{\lc{R}}{\rho}{\lambda\nu\mu} + \udt{P}{\rho}{\lambda\nu\mu}\,,
\end{equation}
where
\begin{alignat}{2}
    \udt{\lc{R}}{\rho}{\lambda\nu\mu} & =\: & &-\Big( \udt{K}{\rho}{\lambda\mu,\nu} - \udt{K}{\rho}{\lambda\nu,\mu} + \udt{\lc{\Gamma}}{\rho}{\sigma\nu} \udt{K}{\sigma}{\lambda\mu} - \udt{\lc{\Gamma}}{\rho}{\sigma\mu} \udt{K}{\sigma}{\lambda\nu} + \udt{\lc{\Gamma}}{\sigma}{\lambda\mu} \udt{K}{\rho}{\sigma\nu} - \udt{\lc{\Gamma}}{\sigma}{\lambda\nu} \udt{K}{\rho}{\sigma\mu} \nonumber\\[0.5ex]
     & \: & &+ \udt{K}{\rho}{\sigma\nu} \udt{K}{\sigma}{\lambda\mu} - \udt{K}{\rho}{\sigma\mu} \udt{K}{\sigma}{\lambda\nu}\Big) =: -\udt{P}{\rho}{\lambda\nu\mu}\,.
\end{alignat}
It is important to note that the $\udt{P}{\rho}{\lambda\nu\mu}$ tensor is expressed only in terms of the teleparallel connection. Naturally, this will lead to a relation for the Ricci tensor
\begin{equation}
    0 \equiv R_{\lambda\mu} = \udt{R}{\rho}{\lambda\rho\mu} = \udt{\lc{R}}{\rho}{\lambda\rho\mu} + \udt{P}{\rho}{\lambda\rho\mu}\,,
\end{equation}
that ultimately produces the Ricci scalar associated with standard gravity resulting in
\begin{equation}
    0 = R = \lc{R} + P\,,
\end{equation}
where $R = g^{\lambda\mu}R_{\lambda\mu}$, and
\begin{equation}
    P = g^{\lambda\mu} \udt{P}{\rho}{\lambda\rho\mu} = \frac{2}{e} \partial_\rho\left(e \udt{T}{\mu\rho}{\mu}\right) + K^{\rho\sigma\mu} K_{\mu\sigma\rho} - \udt{K}{\rho}{\sigma\rho} \udt{K}{\mu\sigma}{\mu}\,,
\end{equation}
where the following identities were used
\begin{equation}
    \udt{K}{\mu}{\rho\mu} = \udt{T}{\mu}{\mu\rho}\,,\quad \udt{T}{\sigma}{\mu\nu} = - \udt{T}{\sigma}{\nu\mu}\,.
\end{equation}
The first term in $P$ is a total divergence term written as
\begin{equation}\label{Eq:boundary_term_def}
    B := \frac{2}{e}\partial_\rho \left(e T^{\mu}{}_{\mu}{}^\rho\right) \equiv -\frac{2}{e}\partial_\rho \left(e \udt{T}{\mu\rho}{\mu}\right)\,,
\end{equation}
while the remainder can be simplified to
\begin{equation}
    K^{\rho\sigma\mu} K_{\mu\sigma\rho} - \udt{K}{\rho}{\sigma\rho} \udt{K}{\mu\sigma}{\mu} = \frac{1}{4} T^{\rho\sigma\mu} T_{\rho\sigma\mu} + \frac{1}{2} T^{\mu\sigma\rho} T_{\rho\sigma\mu} - \udt{T}{\rho}{\rho\sigma} \udut{T}{\mu}{\mu}{\sigma}\,,
\end{equation}
which is the original definition of the torsion scalar, namely
\begin{equation}\label{Torsion_scalar}
    T = \frac{1}{4} T^{\rho\sigma\mu} T_{\rho\sigma\mu} + \frac{1}{2}T^{\mu\sigma\rho} T_{\rho\sigma\mu} - \udt{T}{\rho}{\rho\sigma} \udut{T}{\mu}{\mu}{\sigma}\,.
\end{equation}
Hence, the Ricci scalar differs by a boundary term \gls{wrt} the torsion scalar and then, it can be written as
\begin{equation} \label{Eq:Ricci_broken_tor_bound}
    \lc{R} = - P = - T + B\,,
\end{equation}
which guarantees the dynamical equivalence between the Einstein-Hilbert and \gls{tegr} action \eqref{eq:tegractiong} constructed from the torsion scalar. In this way, \gls{gr} and \gls{tegr} must produce identical field equations, which may only be distinguishable in appearance rather than in any dynamical features. The reason a divergence term arises is because both the \gls{gr} and \gls{tegr} Lagrangians depend on the first and second derivatives of the metric and tetrad respectively, however, the second derivative elements in the tetrad instance simplify to a total divergence term. In fact, the boundary term $B$ is related to the origin of the Lovelock theorem in that it forces even minor modifications of the Einstein-Hilbert action to produce high order field equations. Thus, using the tetrad as the fundamental variable of the theory, instead of the metric, gives the possibility of obtaining second order field equations using more general actions from a local Lagrangian, in 4 dimensions, which is consistent with the Lovelock theorem.

On the boundary term, its definition has a noteworthy feature which is based on a well known theorem that related partial and covariant derivatives using the metric tensor determinant, namely
\begin{equation}
    \lc{\nabla}_{\mu} A^{\mu} = \frac{1}{\sqrt{-g}} \partial_{\mu}\left(\sqrt{-g} A^{\mu}\right)\,,
\end{equation}
where $A^{\mu}$ is an arbitrary tensor. Using the tetrad determinant relation in Eq.~\eqref{eq:tetraddet}, this can be directly used to rewrite the boundary term definition \cite{Bahamonde:2015zma,Bahamonde:2020lsm} in the following way
\begin{equation}
    B = 2\lc{\nabla}_{\mu}\left(\udut{T}{\rho}{\rho}{\mu}\right)\,,
\end{equation}
from which the total divergence nature of the term becomes much more evident. In terms of making calculations, this expression of the boundary term can be useful in determining the value of the quantity. However, we highlight that despite the calculation relying on the Levi-Civita connection, it does not mean that this is not a genuine torsion term, its just an algebraic way to re-express the term in this way.

On the other hand, an interesting property of the torsion scalar is that it can be written in a so-called superpotential format in which \cite{deAndrade:1997gka}
\begin{equation}\label{eq:torsion_def}
    T = \frac{1}{2} \dut{S}{\rho}{\sigma\mu}\udt{T}{\rho}{\sigma\mu}\,,
\end{equation}
where the superpotential takes on the form
\begin{equation}
    \dut{S}{\rho}{\sigma\mu} = \udt{K}{\sigma\mu}{\rho} - \delta^{\sigma}_{\rho} \udut{T}{\nu}{\nu}{\mu} + \delta^{\mu}_{\rho} \udut{T}{\nu}{\nu}{\sigma}\,, \label{Eq:Superpotential_def}
\end{equation}
which observes the anti-symmetry $\dut{S}{\rho}{\sigma\mu} = -\dut{S}{\rho}{\mu\sigma}$.

Consider the following action for \gls{tegr} in which the matter contribution is also included (to the gravitational component in Eq.~\eqref{eq:tegractiong}) and the boundary term is already removed since it does not contribute dynamically
\begin{equation}\label{eq:tegraction}
    \mathcal{S}_{\rm TEGR} := -\frac{1}{2\kappa^2} \int \dd^4x e \, T + \int \dd^4 x e \mathcal{L}_{\rm m}\,,
\end{equation}
where $\kappa^2 = 8\pi G$ as defined in Eq.~\eqref{Eq:Con_units}, and the negative sign for the gravitational sector appears so that the exact same dynamics are obtained when compared with the Einstein-Hilbert action. Using the variation identities in Appendix~\ref{App:variations}, and the energy-momentum definition in Eq.~\eqref{Eq:Con_EM_ten}, we can write the \gls{tegr} field equations as
\begin{equation}\label{eq:tegrfield}
    \dut{W}{A}{\mu} = e^{-1}\partial_{\sigma}(e \dut{S}{A}{\mu\sigma}) - \udt{T}{\sigma}{\nu A} \dut{S}{\sigma}{\nu\mu} + \frac{1}{2} \dut{E}{A}{\mu} T + \,\udt{\omega}{B}{A\nu} \dut{S}{B}{\nu\mu} = \kappa^2 \dut{\Theta}{A}{\mu}\,,
\end{equation}
after some lengthy simplification. By contracting with $\udt{e}{A}{\beta}g_{\mu\alpha}$, we can rewrite the field equations as
\begin{equation}
    \udt{e}{A}{\beta}\,g_{\mu\alpha}\, e^{-1}\partial_{\sigma} (e\dut{S}{A}{\mu\sigma}) - \udt{T}{\sigma}{\nu\beta}\dudt{S}{\sigma}{\nu}{\alpha} + \frac{1}{2} g_{\alpha\beta} T + \udt{\omega}{B}{\beta\nu} \dudt{S}{\beta}{\nu}{\alpha} = \kappa^2 \Theta_{\alpha\beta}\,.
\end{equation}
The field equation components satisfy both symmetric and antisymmetric
equations
\begin{equation}
W_{(\mu\nu)}=\Theta_{\mu\nu},\quad\text{and}\quad W_{[\mu\nu]}\equiv0\,,\label{eq:TEGR_antisym_eq}
\end{equation}
where we stress that the antisymmetric equations $W_{[\mu\nu]}$ for \gls{tegr} are actually identically satisfied $W_{[\mu\nu]}\equiv0$ due to the Bianchi identities (see Sec.~\ref{sssec:genbianchi}), and where the energy-momentum tensor is expressed as $\Theta_{\alpha\beta}=\Theta_{\beta\alpha}=\udt{e}{A}{\beta}\Theta_{A\alpha}=\udt{e}{A}{\beta}g_{\mu\alpha}\dut{\Theta}{A}{\mu}$. These field equations organically admit all solutions that appear in \gls{gr} since they are dynamically equivalent to each other. This can be seen after noticing that the Einstein tensor can be written in term of teleparallel quantities as follows
\begin{equation}\label{eq:einsteintensor}
    \lc{ G}_{\alpha\beta} = -\Big( T^{B}{}_{\nu\beta}S_{B}{}^{\nu}{}_{\alpha} -  \omega^B{}_{\beta\nu} S_B{}^{\nu}{}_{\alpha} -\frac{1}{e} g_{\mu\alpha}e^{A}{}_{\beta}\partial_{\nu}(e S_{A}{}^{\mu\nu})-\frac{T}{2} g_{\alpha\beta}\Big)\,.
\end{equation}
Also, these equations are invariant under local Lorentz transformations and diffeomorphisms since each of the terms in the torsion scalar in Eq.~\eqref{Torsion_scalar} are invariant under these symmetries. The tetrad and spin connection represent independent \gls{dof} and thus are determined through extra equations. In the case of \gls{tegr} the spin connection field equations are found to be identically satisfied \cite{Krssak:2018ywd} which is not the case for more general theories as will be explored in Sec.~\ref{sec5:extended}. This is consistent with the spin connection implementation in \gls{tegr} where it represents \gls{dof} associated with the Lorentz group and so cannot contribute to the number of dynamical equations for an arbitrary gravitational system \cite{Golovnev:2017dox,Blagojevic:2000qs,Krssak:2015lba}.

It is interesting to note that these field equations can be reformulated as a current equation using the gauge current $\dut{J}{A}{\mu}$ (see Sec.~\ref{ssec:gravenergy} for more details), which is derived in Appendix~\ref{App:Gauge_Curr_def}, namely \cite{deAndrade:2000kr,deAndrade:1997cj}
\begin{equation}
    e^{-1}\partial_{\sigma} (e \dut{S}{A}{\mu\sigma}) - \frac{1}{2}\dut{J}{A}{\mu} = \kappa^2 \dut{\Theta}{A}{\mu}\,,
\end{equation}
where in this case $\dut{J}{A}{\mu}$ becomes the Noether energy-momentum density of gravitation \cite{Moller:1961,deAndrade:2000kr,Krssak:2015rqa}.

Using this formulation of the field equations \cite{Shirafuji:1996im}, the vacuum field equations can be written with the gauge current as a source term
\begin{equation}
    \partial_{\sigma}(e \dut{S}{A}{\mu\sigma}) = \frac{e}{2} \dut{J}{A}{\mu}\,,
\end{equation}
so that the gravitational field can take on a Noether gauge current structure. Also, the anti-symmetry of the superpotential gives rise directly to the conservation equation
\begin{equation}
    \partial_{\mu} \left(\frac{e}{2}\dut{J}{A}{\mu} + e\dut{\Theta}{A}{\mu}\right) = 0\,,
\end{equation}
which can be shown to produce the \gls{tegr} field equations.

The motivation for the particular choice of \gls{tegr} action~\eqref{eq:tegraction} was to produce a \gls{gr} equivalent theory using torsional terms. However, we can also arrive at this point using a perspectives from gauge theory as our motivation. As discussed in Sec.~\ref{Sec:TEGR_gauge_theory}, TG can be written as the gauge theory of translations and the field strength is the torsion tensor. As with any gauge theory, the action is constructed from the trace of the torsion tensor squared, and since there are three possible contractions of the torsion tensor, it is natural to consider the action~\cite{Aldrovandi:2013wha}
\begin{equation}
    \mathcal{S}_{\rm TEGR}=\frac{1}{2\kappa^2}\int \textrm{tr}\Big(\boldsymbol{T}\wedge\star \boldsymbol{T} \Big)\,,
\end{equation}
where $\boldsymbol{T}=(1/2)T^{A}{}_{\mu\nu}P_A \dd x^\mu\wedge \dd x^\nu$ is the torsion 2-field and $P_A = \partial_A$ is the translation generators (see Sec.~\ref{Sec:TEGR_gauge_theory}). The last term appearing in the integral is the dual 2-form of the torsion tensor, i.e., $\star \boldsymbol{T}=(1/2)\star T^{A}{}_{\mu\nu}P_A \dd x^\mu\wedge \dd x^\nu$ with $\star T^{A}{}_{\mu\nu}\equiv (e/2)\epsilon_{\mu\nu\alpha\beta}S^{A\alpha\beta}$. This is in analogy to Yang-Mills theory where a gauge theory of the non-abelian symmetry group is considered and its action is constructed from the trace of the field strength. It turns out that after some identities, one can show that the above action is identical to the \gls{tegr} action~\eqref{eq:tegraction}. This means that even without the \textit{a priori} knowledge of \gls{gr}, one can still formulate \gls{tegr} using information coming from gauge theory and produce the dynamics of \gls{gr} using the standard formulation of gauge theory.

\subsubsection{The gravitational energy problem} \label{ssec:gravenergy}

A direct consequence of the fact that the field Eqs.~\eqref{eq:tegrfield}, being a special case of the general field Eqs.~\eqref{eq:gentetradfieldlor} arise as the Euler-Lagrange equations arising from the action~\eqref{eq:tegraction}
\begin{equation}
    eW_A{}^{\mu} = \partial_{\nu}\frac{\partial L}{\partial(\partial_{\nu}e^A{}_{\mu})} - \frac{\partial L}{\partial e^A{}_{\mu}} = \frac{\delta L_{\text{m}}}{\delta e^A{}_{\mu}} = e\Theta_A{}^{\mu}\,,
\end{equation}
of a first order Lagrangian, is the fact that they can be written in the so-called potential form
\begin{equation}\label{eq:tegrfieldpot}
\partial_{\nu}(eS_A{}^{\mu\nu}) - \kappa^2eJ_A{}^{\mu} = \kappa^2e\Theta_A{}^{\mu}\,.
\end{equation}
Here the two terms on the \gls{lhs} are given by
\begin{equation}
    S_A{}^{\mu\nu} = \frac{1}{e}\frac{\partial L}{\partial(\partial_{\nu}e^A{}_{\mu})} = K^{\mu\nu}{}_A - E_A{}^{\mu}T_{\rho}{}^{\rho\nu} + E_A{}^{\nu}T_{\rho}{}^{\rho\mu}\,,
\end{equation}
and
\begin{equation}\label{eq:gravenmompseudo}
    J_A{}^{\mu} := \frac{1}{\kappa^2e}\frac{\partial L}{\partial e^A{}_{\mu}} = \frac{1}{\kappa^2}\left(E_A{}^{\lambda}S_B{}^{\nu\mu}T^B{}_{\nu\lambda} - \frac{1}{2}E_A{}^{\mu}T + \omega^B{}_{A\nu}S_B{}^{\mu\nu}\right)\,.
\end{equation}
One finds that the expression \(J_A{}^{\mu}\) formally appears in analogy to the matter energy-momentum tensor \(\Theta_A{}^{\mu}\). However, there is a fundamental difference between these two objects. It is well known that \(\Theta_A{}^{\mu}\) is a tensor, and as such can only satisfy a covariant conservation equation
\begin{equation}
\lc{\DDD}_{\mu}(e\Theta_A{}^{\mu}) = e(\partial_{\mu}\Theta_A{}^{\mu} - \lc{\omega}^B{}_{A\mu}\Theta_B{}^{\mu} + \lc{\Gamma}^{\nu}{}_{\nu\mu}\Theta_A{}^{\mu}) = 0\,.
\end{equation}
The quantity \(J_A{}^{\mu}\), in contrast, is not a tensor, which can already be seen from the explicit appearance of the spin connection \(\omega^A{}_{B\mu}\) in its definition~\eqref{eq:gravenmompseudo}. It follows from the antisymmetry \(S_A{}^{(\mu\nu)} = 0\) of the superpotential that in the absence of matter it satisfies the conservation equation \cite{Combi:2017crv}
\begin{equation}
    \partial_{\mu}\partial_{\nu}(eS_A{}^{\mu\nu}) - \kappa^2\partial_{\mu}(eJ_A{}^{\mu}) = -\kappa^2\partial_{\mu}(eJ_A{}^{\mu}) = 0\,.
\end{equation}
In order to compensate for this difference, note that the last term in the expression~\eqref{eq:gravenmompseudo} can be combined with the first term in the \gls{tegr} field Eq.~\eqref{eq:tegrfieldpot} to form a Fock-Ivanenko derivative,
\begin{equation}
    \partial_{\nu}(eS_A{}^{\mu\nu}) - \omega^B{}_{A\nu}eS_A{}^{\mu\nu} = \DDD_{\nu}(eS_A{}^{\mu\nu})\,.
\end{equation}
This allows for us to write the \gls{tegr} field equations in the form
\begin{equation}\label{eq:tegrfieldpot2}
    \DDD_{\nu}(eS_A{}^{\mu\nu}) - \kappa^2e\, t_A{}^{\mu} = \kappa^2e\Theta_A{}^{\mu}\,.
\end{equation}
In this equation,
\begin{equation}
t_A{}^{\mu} = \frac{1}{\kappa^2}\left(E_A{}^{\lambda}S_B{}^{\nu\mu}T^B{}_{\nu\lambda} - \frac{1}{2}E_A{}^{\mu}T\right)\,,
\end{equation}
is a tensor. In the absence of matter, it satisfies the covariant conservation equation
\begin{equation}
\DDD_{\mu}\DDD_{\nu}(eS_A{}^{\mu\nu}) - \kappa^2\DDD_{\mu}(et_A{}^{\mu}) = -\kappa^2\DDD_{\mu}(et_A{}^{\mu}) = 0\,,
\end{equation}
where also in this case the first term vanishes since \(S_A{}^{\mu\nu}\) is antisymmetric it its last two indices, while the Fock-Ivanenko derivatives commute due to the flatness of the teleparallel connection. It is common to interpret \(t_A{}^{\mu}\) as the energy-momentum tensor of gravity, while the difference
\begin{equation}
    i_A{}^{\mu} = J_A{}^{\mu} - t_A{}^{\mu} = \frac{1}{\kappa^2}\omega^B{}_{A\nu}S_B{}^{\mu\nu}\,,
\end{equation}
is attributed to inertial effects. However, as can be seen from the expression on the \gls{rhs}, this term is not invariant under local Lorentz transformations, and even vanishes in the Weitzenb\"{o}ck gauge. It follows from this lack of Lorentz invariance that it cannot be an observable quantity in a theory in which Lorentz invariance is preserved, so that all physically observable quantities must also be invariant under local Lorentz transformations; hence, and the physical relevance of the split into \(i_A{}^{\mu}\) and \(J_A{}^{\mu}\) is questionable \cite{Aldrovandi:2013wha,BeltranJimenez:2021kpj,Golovnev:2017dox}.

\subsubsection{The degrees of freedom of TEGR} \label{sssec:tegrhamilton}

The number and nature of the \gls{dof} of any field theory is a core question by which to assess its consistency as well as the wellposedness of the Cauchy problem\footnote{The Cauchy problem is related to the existence and uniqueness of solutions to the field equations given boundary conditions (for more information see Ref.~\cite{ringstrom2009cauchy})}. These are central to our understanding of any formulation of gravity since hidden \gls{dof} can infiltrate every numerical problem by introducing some uncertainty in the necessary number of initial conditions to determine a system. In field theories in which higher-order field equations appear, Ostrogradsky instability is well known to lead to ghost-like \gls{dof} \cite{Motohashi:2014opa} which has created a favourable environment for second-order formulations of gravity. However, even in second-order theories of gravity, its important to identify whether \gls{dof} are propagating or non-propagating since the latter would not have an effect on observations. This is not a trivial task and remains an open question in many formulations of gravity.

The most popular way to count the number of \gls{dof} in \gls{tegr} is to use a Hamiltonian analysis approach (see Sec.~\ref{ssec:admformalism}) which gives the total number of localized \gls{dof} associated with a field theory \cite{ortin2004gravity}. In this setup, the \gls{dof} are associated with the pairs of conjugate variables, namely the dynamical system generalized coordinates and their conjugate momenta. The total number of \gls{dof} can then be calculated by finding the number of irreducible first-class constraints $a$ and second-class constraints $b$ and using the formula
\begin{equation}
    \text{Number of \gls{dof}s} = \frac{2N-2a-b}{2}\,,
\end{equation}
where $2N$ is the number of canonical coordinates \cite{10.1007/3-540-58339-414}. In this terminology, each first-class constraint eliminates a \gls{dof} while it takes two second-class constraints to achieve the same result. This is consistent with classical mechanics where it is known that first-class constraints and their associated gauge fixing conditions form a system of second-class constraints.

\gls{tegr} is dynamically equivalent to \gls{gr} and so produces the same classical predictions in their field equations such as metric solutions and energy conditions. However, these theories stem from starkly different Lagrangians as evidenced in Sec.~\ref{sec:TEGR_acion} where the Ricci scalar in Eq.~\eqref{Eq:Ricci_broken_tor_bound} is separating into a contributing second-order torsion scalar and a total divergence term. The result is a drastically different Einstein-Hilbert and \gls{tegr} Lagrangian which naturally leads to the question of how many \gls{dof} appear. Another important property of \gls{tegr} is its dependence on the tetrad rather than the metric tensor which has a significant impact on the Hamiltonian analysis implementation.

The formulation of the Hamiltonian structure of \gls{tegr} has had a varied history in the literature. In Ref.~\cite{Maluf:2000ag} the \gls{tegr} Hamiltonian was formulated using a strategy in which an auxiliary field representing a $3+1$ decomposition of the torsion tensor was used to reduce the over-all order the Euler-Lagrange equations \cite{Maluf:1994ji,Maluf:1998ae,daRochaNeto:2011ir,Maluf:2013gaa}. The approach is altered in Ref.~\cite{Blagojevic:2000qs} where curvature is eliminated through the use of Lagrange multipliers and the Weitzenb\"{o}ck gauge is not immediately assumed. This larger set of dynamical variables leads to a more intricate analysis that eventually produces 2 physical \gls{dof} corresponding to a massless graviton. The canonical formulation that makes up the Hamiltonian analysis has also been put in the language of one-forms in Refs.~\cite{Okolow:2011nq,Okolow:2013lwa} where it is shown that the constraint algebra is indeed closed.

The previous works on the topic of the number of \gls{dof} of \gls{tegr} left some questions open which were then tackled in Ref.~\cite{Ferraro:2016wht} where an exhaustive study of the first- and second-class constraints is conducted resulting in \gls{tegr} having the same number of \gls{dof} as \gls{gr} in $n-$dimensions, namely $n(n-3)/2$. In the context of \gls{tegr}, this point is again independently verified in Ref.~\cite{Blagojevic:2020dyq} where the Hamiltonian analysis is revisited. It is worth noting that both analyses were done in the Weitzenb\"{o}ck gauge.

\subsection{Outlook toward quantum teleparallel gravity and other dimensions} \label{ssec:quantumhighdim}

The prospect of a viable quantum theory of gravity has come into sharp focus in the context of community efforts in recent years. Given the expanded framework of alternative formulations of \gls{gr} and modified theories of gravity, the amount of work in the literature has drastically increased with several promising avenues of research. The added contribution of possible detections through experimental and observational advances in recent years has also highlighted the vital nature of this subject and brought to the fore questions of quantum gravity properties in many reformulations of gravitation.

Ultimately quantum gravity phenomena may be expressed in a variety of possibles ways. The recent detection of \gls{gw} makes the likelihood of astrophysical or cosmological signatures of quantum gravity a real possibility with the next generation of \gls{gw} observatories \cite{Calcagni:2019ngc}. Another crucial outlet for quantum phenomena in a gravitational context is through gamma rays which may reveal some of the effects of quantum gravity corrections within classical theories of gravity such as energy-dependent dispersion relations \cite{AmelinoCamelia:1997gz}. Quantum gravity could also have a strong impact on our measurements of beyond the standard model of particle physics which has mostly been directed at neutrino physics in the realm of gravitation \cite{Formaggio:2016cuh}.

\gls{tegr} has a number of attractive properties which make it might be more amenable to a quantization procedure. For instance, the strong equivalence principle has been shown not to be a fundamental requirement of the theory (as explained further in Sec.~\ref{sec3:torsional}) unlike in \gls{gr} where the theory necessitates the principle \textit{a priori}. The strong equivalence principle continues to satisfy all observational constraints \cite{Will:2001mx}, while on the other hand, the fundamental nonlocality of quantum theory may eventually require the violation of this principle where in \gls{tg} may offer an alternative avenue that largely preserve the dynamics of GR. Another crucial aspect to quantizing gravity is the fact that the spin connection is solely related to the local Lorentz transformation while the teleparallel connection includes no inertial effects and so represents a genuine gravitational connection \cite{Aldrovandi:2005mz,Aldrovandi:2006cy}. In \gls{gr} the connections share these properties and so pose a serious issue for such procedures. \gls{tg} offers a scenario that is more akin to the regular field theory approach to quantum gravity. While not enough in and of itself, these properties may help in the formation of a quantum theory of gravity in \gls{tegr} where some of the core issues of \gls{gr} are eliminated \textit{ab initio}. \gls{tegr} also has a number of other properties which makes it more malleable for a quantum setting such as its likeness to Yang-Mills theories \cite{Ho:2015bia} giving \gls{tg} a strong similarity to a particle physics theory. Another notable property is that \gls{tg} naturally has a Gibbons-Hawking-York boundary term embedded in its action giving the \gls{tegr} Hamiltonian a more well-defined expression \cite{Oshita:2017nhn}. Despite all this further work is needed on both the theoretical development in terms of quantum predictions from \gls{tegr} \cite{Aldrovandi:2003pd} as well as more information on the what observational signatures to expect from a quantum theory of gravity.

Another interesting proposed route for a quantum \gls{tegr} approach comes from Ref.~\cite{Pereira:2020yzi}, where the Poincar\'{e} group is replaced with the de Sitter group which both retain the Lorentz group of symmetries as a subgroup. This approach has some attractive features which naturally define a particular length scale $l$ which may be related to the cosmological constant ($\Lambda = 3/l_{\rm pl}^2$)~\cite{Jennen:2015bxa,Koivisto:2021ofz}. The length scale in this scenario may coincide with the Planck length scale $l_{\rm pl}$. This is different to the scenario in the Poincar\'{e} group where no invariant length scale is possible at all. Thus, by coupling the field theoretical approach of \gls{tg} with a special relativity that replaces the Poincar\'{e} with the de Sitter group provides an interesting regime to produce a field theoretic theory of gravity that has a naturally associated length scale (possibly relating the cosmological constant with the Planck scale).

The literature already contains a number of intriguing works on quantum aspects of \gls{tg}. For instance, in Ref.~\cite{Dupuis:2019unm} the first quantization suggests a loop quantum gravity nature to the discretization of \gls{tg} similar to the development of regular loop quantum gravity from the Einstein-Hilbert metric description. Renormalization is tackled in Ref.~\cite{Krssak:2015rqa} where the resemblance of renormalization to the process of finding an appropriate spin connection conjugate to a tetrad choice is pointed out and used to highlight the possibility of defining local energy and momentum densities in \gls{tg}. Another important contribution to the literature on the quantization of \gls{tg} is the series of works in Refs.~\cite{Okolow:2013cua,Okolow:2013dua,Okolow:2013lba,Okolow:2013ifa} where projective techniques are used to construct kinematic quantum states of \gls{tegr}. However, the work needs more development to fully quantize the theory. More generally, the quantum regime has also been studied in the context of black hole solutions of \gls{tegr} such as Refs.~\cite{Ulhoa:2014eka,Ulhoa:2019ibd} where the well known Weyl quantization procedure is applied to the Schwarzschild and Kerr solutions, as well as Ref.~\cite{Ulhoa:2011py} in which an initial second quantization procedure is taken on the Schwarzschild solution. Of note as well is some initial work that has started in handling brane cosmology, namely Refs.~\cite{Nozari:2012qi,Behboodi:2014tda}, but it is still in very early stages when compared with standard gravity.

\clearpage

\section{Modified Teleparallel Gravity Theories}\label{sec5:extended}
This section is devoted in the so-called extensions or modifications of \gls{tg}. The motivation is the same as in \gls{gr}: the difficulty of finding a consistent quantum theory of gravity, the nature of singularities, but mostly the unknown, \textit{dark} Universe led cosmology to the pursuit of a better description for the gravitational interactions. There have been a great many proposals for modifying the \gls{tegr} and in this section we will try to present the features of most of them. In Fig.~\ref{fig:modifiedgravity} one can see a bird's eye view of the spectrum of theories, deviating from \gls{gr}. Together with the different ways of modifying \gls{gr}, some of the most important example theories are depicted as well. The figure is drawn by thinking on breaking some of the conditions in the Lovelock's theorem. It turns out that some parts of the figure are connected. For example, some theories which add invariants can be rewritten as scalar-tensor theories. Furthermore, theories can be part of multiple branches of the figure. An example of this could be teleparallel theories, which is a framework beyond \gls{gr} which modifies the geometry. As we will see in this section, modified teleparallel theories can also break other conditions in the Lovelock's theorem, as it happens in teleparallel scalar-tensor theories (see Sec.~\ref{sec:scalartensor}) where besides modifying the geometry, one adds a scalar field in the Lagrangian.
In what follows, we will summarize different proposed teleparallel theories that have been studied in the literature.

\tikzset{
  font={\fontsize{10pt}{12}\selectfont}}

\begin{figure}[H]
        \centering
        \resizebox{\columnwidth}{!}{
\begin{tikzpicture}[thick,rounded corners,block/.style={align=center,rectangle,fill=white},lv/.style={draw=green!50!black,fill=green!20!white},lv2/.style={draw=black!50!black,fill=black!20!white},hs/.style={draw=yellow!50!black,fill=yellow!20!white},ms/.style={draw=blue!50!black,fill=blue!20!white},mf/.style={draw=red!50!black,fill=red!20!white},mx/.style={draw=purple!50!black,fill=purple!20!white},mN/.style={draw=cyan!50!black,fill=cyan!20!white}]
\draw (-7.5,3.5) rectangle (0.75+0.1,0.5);
\draw (-7.5+12,3.5) rectangle (0.75+12-0.1,0.5+0.1);
\draw (-7.5,3.5-8) rectangle (0.75,0.5-8+0.2-0.1-0.1);
\draw (-7.5+12,3.5-8) rectangle (0.75+12+0.1,0.5-8-0.1);
\draw (-7.5+12+3.5-1,3.5-8+4+0.5) rectangle (0.75+12-4+4+2+1.5-1-2,0.5-8+4-0.6+0.3-0.3-0.1);
\draw (-7.5+12+3.5+0.4-16-2+1.2-0.2,3.5-8+4) rectangle (0.75+12-4+4-16+0.5,0.5-8+4);
\node at (-3,3.25) {\hspace{-1cm}\textbf{Non-Riemannian geometry}};
\node[block,lv] at (-5.5,2.5-0.2) {Metric-affine\\ gravity~\cite{Hehl:1994ue}};
\node[block,lv] at (-5.5-0.1,1.25-0.1) {Non-commutative\\ geometry~\cite{Aschieri:2005yw}};
\node[block,lv] at (-1-0.3,2.5) {Einstein-Cartan~\cite{Hehl:1994ue}};
\node[block,lv] at (-1.2-0.2-0.2,1.75) {Poincaré gauge gravity~\cite{Hehl:1994ue}};
\node[block,lv] at (-1-0.2-0.1,1) {Teleparallel theories~\cite{Aldrovandi:2013wha}};
\node at (3,3.25) {\hspace{11cm}\textbf{Higher-order theories}};
\node[block,hs] at (5.7+2-1-1+5.2+0.1,1.5) {Lovelock \\ theories~\cite{Deruelle:1989fj}};
\node[block,hs] at (5.7+2-2+2+0.2+0.2+0.1,2.5) {$f(\lc{R})$~\cite{DeFelice:2010aj}};
\node[block,hs] at (5.7+2-2+4+0.1+0.4+0.2+0.2,2.5) {$f(\lc{R},\Theta)$~\cite{Harko:2011kv}};
\node[block,hs] at (5.7+2-2+0.1+0.1,2.5) {$f(\lc{R},\lc{\mathcal{G}})$~\cite{Nojiri:2005jg}};
\node[block,hs] at (5.7+2-1-1+0.1,1.5) {Non-local\\ theories~\cite{Deser:2007jk}};
\node[block,hs] at (5.7+2-1+2-0.5-0.1+0.1+0.1,1.5+0.1) {Conformal\\ Weyl~\cite{Mannheim:1988dj}};
\node at (3,3.25-8.1) {\hspace{11cm} \textbf{Tensor-vector-scalar theories}};
\node[block,ms] at (4.25+2-0.2,-1.75-4) {Einstein-{\AE}ther~\\ \cite{Jacobson:2008aj}};
\node[block,ms] at (4.25+2+2+1-0.2-0.1,-1.75-4) {Proca theories\\ \cite{DeFelice:2016yws}};
\node[block,ms] at (4.25+2+2+1+2+0.5-0.1-0.1,-1.75-4+0.1) {Bimetric \\ gravity~\cite{Hassan:2011zd}};
\node[block,ms] at (4.25+2-0.5,-1.75-4-1-0.1) {Horndeski~\\ \cite{Kobayashi:2019hrl}};
\node[block,ms] at (4.25+2+2+1-0.5-0.1-0.1,-1.75-4-1-0.1-0.1) {Beyond Horndeski~\\ \cite{Gleyzes:2014dya}};
\node[block,ms] at (4.25+2+2+1+2+1-0.5-0.1,-1.75-4-1-0.2) {Massive\\ gravity~\cite{deRham:2010kj}};
\node at (3-6,3.25-8.1) { \textbf{$D$-dimensional theories}};
\node[block,mf] at (4.25+2-12.5+0.2,-1.75-4+0.1) {Braneworld~\cite{Sahni:2002dx}};
\node[block,mf] at (4.25+2+2+1-12.8+0.4,-1.75-4-0.1+0.2) {Randall\\
Sundrum~\cite{Randall:1999ee}};
\node[block,mf] at (4.25+2+2+1+2+0.5-12.3+0.1,-1.75-4) {DGP~\cite{Dvali:2000xg}};
\node[block,mf] at (4.25+2-0.5-12+0.08+0.1+0.2,-1.75-4-1-0.1) {Kaluza-Klein~\cite{Klein:1926tv}};
\node[block,mf] at (4.25+2+2+1-0.5-12+0.4-0.15,-1.75-4-1+0.1-0.2) {EdGB\\ gravity~\cite{Kanti:1995vq}};
\node[block,mf] at (4.25+2+2+1+2+1-0.5-12+0.1-0.05+0.08-0.3-0.1,-1.75-4-1) {M-theory~\cite{Maldacena:1997re}};
\node at (3+1+3+2+2-1+0.4+1-1-0.3,3.25-3-1-0.3+0.2+0.5) { \textbf{Quantum gravity theories}};
\node[block,mx] at (4.25+2-12.5+15.5+0.2-0.9-0.1+0.2-0.1,-1.75-4+4.2+0.5-0.1) {Ho\v{r}ava-Lifschitz~\\ \cite{Mukohyama:2010xz}};
\node[block,mx] at (4.25+2-12+2.5+15.5+0.1-0.9+0.2,-1.75-4+4.2+0.5) {String theory~\cite{polchinski_1998}};
\node[block,mx] at (4.25+2-12+2.5+15.5+0.2+2.6-1-2.7+0.1,-1.75-4+4.2-1.8-0.1) {Rainbow gravity~\\ \cite{Amelino-Camelia:2013wha}};
\node[block,mx] at (4.25+2-12+2.5+15.5+0.2-3-0.8,-1.75-4+4.2-1+0.5-0.2-0.1) {Loop quantum\\ gravity~\cite{Sahlmann:2010zf}};
\node[block,mx] at (4.25+2-12+2.5+15.5-0.8,-1.75-4+4.2-1+0.5-0.2) {Asymptotic \\ safety~\cite{Niedermaier:2006wt}};
\node[block,mx] at (4.25+2-12+2.5+15.5+0.5+2-1-4-1-0.2-0.1-0.1,-1.75-4+4.2-1-0.8-0.1) {Supergravity~\\ \cite{Yamaguchi:2011kg}};
\node at (3+1+3+2+2-1+0.4+1-17.6+0.5,3.25-3-1-0.3+0.2) {\textbf{Other approaches}};
\node[block,mN] at (4.25+2-12.5+15.5+0.2-17.6+0.2+1.2-0.1-0.18-0.1,-1.75-4+4.2-0.1) {Padmanabhan\\ thermod.~\cite{Padmanabhan:2009vy}};
\node[block,mN] at (4.25+2-12+2.5+15.5-17.6+0.2+1-0.08-0.1,-1.75-4+4.2) {Holography~\cite{Witten:1998qj}};
\node[block,mN] at (4.25+2-12.5+15.5+0.2-17.6+0.2+1.1-0.1+0.3,-1.75-4+4.2-1-0.3) {Entropic \\gravity~\cite{Verlinde:2010hp}};
\node[block,mN] at (4.25+2-12+2.5+15.5-17.6+0.2+0.3+0.8,-1.75-4+4.2-1) {Analogue \\ gravity~\cite{Barcelo:2005fc}};
\node[block,lv2] (GR) at (2.5,-2) {\large \textbf{General} \\ \large \textbf{Relativity}};
\draw[->] (GR) to node [sloped,above] {Changing geometry}    node [above] {} (-3,0.3);
\draw[->] (GR) to node [sloped,above] {Adding invariants}    node [above] {} (7.5-1,0.3);
\draw[->] (GR) to node [sloped,below] {Adding new fields}    node [above] {} (8.5-1.7,-4.3);
\draw[->] (GR) to node [sloped,below] {Changing dimension}    node [below] {} (8.5-12+1.7,-4.3);
\draw[->] (GR) to node [sloped,above] {Other approaches}    node [below] {} (2.5-5-0.19,-2);
\draw[->] (GR) to node [sloped,above] {Quantization}    node [below] {} (2.5+5.3-0.8-0.1,-2);
\end{tikzpicture}
	}\caption{Representation of some possible ways of modifying \gls{gr} through breaking the Lovelock's theorem along with some examples.}\label{fig:modifiedgravity}
\end{figure}
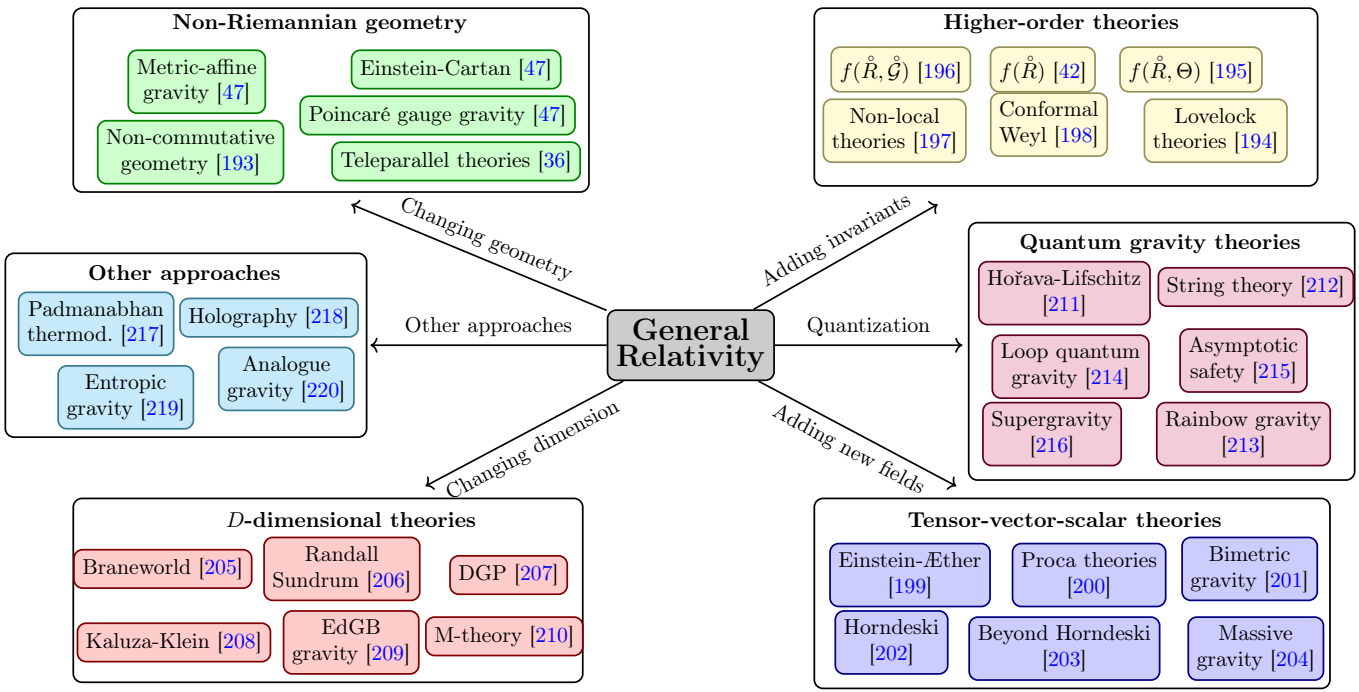

\subsection{Foundations of modified teleparallel theories and how to make them fully invariant}\label{subsec:goodtetrad}

One of the main motivations of this Review, has been the fact that \gls{tg} gained a lot of interest in the astrophysics and cosmology community in the last few years. However, since it is a not so well studied framework (yet), a lot of ambiguities and misunderstandings have arisen through the years; especially in the modifications sector of the theory.

\gls{tg}, being a theory built on the tangent space, must be invariant under general coordinate transformations, as well as under local Lorentz transformations, because special relativity should be recovered in locally inertial frames \cite{1972gcpa.book.....W}. As already known from special relativity, local Lorentz transformations define two classes of frames: the inertial frames, being the ones in which inertial effects vanish, and the non-inertial ones, where inertial effects are present. Quantitatively, these frames can be distinguished by a purely inertial connection, also known as spin connection, which at inertial frames vanishes, while at non-inertial frames is different from zero.

Before going in more depth, let us briefly explain the issue of local Lorentz invariance. The first formulation of teleparallel theories assumed that the spin connection was always zero, and then the torsion tensor would only depend on the tetrads, namely
\begin{equation}\label{eq:torsion111}
    T^{\lambda}{}_{\mu\nu} =E_A{}^\lambda(e^{A}{}_{\nu,\mu}-e^{A}{}_{\mu,\nu})\,.
\end{equation}
As we have explained in Sec.~\ref{sec:LLTW}, the above torsion tensor is a particular one which is computed in the so-called Weitzenb\"{o}ck gauge (where the spin connection vanishes). Now, let us take a local Lorentz transformation only in the tetrads $e'{}^A{}_{\mu}=\Lambda^A{}_B e^B{}_\mu $. By doing this transformation, the above torsion tensor transforms as
\begin{equation}
    T'{}^{\lambda}{}_{\mu\nu}=
    T^\lambda{}_{\mu\nu}+\Lambda_A{}^B E_B{}^\lambda(e^C{}_\nu \partial_\mu \Lambda^A{}_C-e^C{}_\mu \partial_\nu \Lambda^A{}_C)\,,
\end{equation}
where one immediately notices that this quantity is non-covariant under local Lorentz transformations. Further, the torsion scalar in this particular gauge will transform as~\cite{Cho:1975dh}
\begin{equation}
    T'=T+\frac{8}{e}\partial_\mu(e \,\eta^{CB} \, E_B{}^\mu \partial_A \Lambda^A{}_C )\,.
\end{equation}
From the above equation, we get that in the old version of \gls{tg} where one only has the tetrads as fundamental variables and the spin connection is always zero, \gls{tegr} is a pseudo-local Lorentz invariant theory. Since this extra term only appears as a boundary term in the above equation, independently of the choice of the spin connection (or local Lorentz transformation), the manifestation of breaking the local Lorentz invariance does not appear in \gls{tegr} (at least at the level of the field equations - since it is just \gls{gr}). For more information, please check Sec.~\ref{sec:LLTW} and Sec.~\ref{sec:field_strength}. However, the situation is very different when one modifies the action and studies modified teleparallel theories of gravity. In these theories, for example such as in $f(T)$ gravity, the boundary term appearing in the above equation is not a boundary term anymore, meaning that theory will depend on $\Lambda^A{}_B$. This means that when one is modifying \gls{tegr}, the theories would break local Lorentz invariance if one assumes that the spin connection vanishes and construct theories with the torsion tensor~\eqref{eq:torsion111} only depending on the tetrads. This problem was not well understood until the authors in Ref.~\cite{Krssak:2015oua} noticed that the problem of breaking the local Lorentz invariance was only related to choosing the particular Weitzenb\"{o}ck gauge. This can be easily seen by noticing that the torsion tensor~\eqref{torsion_tensor}
transforms covariantly under local Lorentz transformations if we take the simultaneous transformations in the tetrads and the spin connection (see Eqs.~\eqref{eq:tetlortrans} and \eqref{eq:spclortrans}), which indeed gives us that
$T'{}^{\lambda}{}_{\mu\nu} =T^{\lambda}{}_{\mu\nu}\,.$
In conclusion, are teleparallel theories fully invariant (diffeomorphisms and local Lorentz)? The answer is yes, they are if we perform the simultaneous local Lorentz transformations for the tetrad and the flat spin connection. In addition, this quantity is a tensor and thus transforms invariantly under diffeomorphisms. This means that, any action (and consequently field equations) constructed from the torsion tensor will be fully invariant.

The spin connection in \gls{gr} describes both gravitational and inertial effects, and because of the equivalence principle, there exists a local frame, where this connection vanishes. However, the picture in \gls{tg} becomes different: gravitational effects are described by a translational gauge potential, since \gls{tg} is a gauge theory of the translation group (see discussion in Sec.~\ref{Sec:TEGR_gauge_theory}), while the spin connection represents purely inertial effects. That is why, when one sets the spin connection to zero local Lorentz invariance is broken.

Much of the work on \gls{tg} has been performed in the Weitzenb\"{o}ck gauge. When the community realized that modified theories (and specifically $f(T)$ gravity, as we will see below) were not local Lorentz invariant, and since they were (still) ignoring the fact that a nonvanishing spin connection could solve the problem, they introduced the notion of ``good'' and ``bad'' tetrads \cite{Tamanini:2012hg}. \textit{Good} were those tetrads that led to non-trivial solutions of the field equations (of $f(T)$ gravity), with a vanishing spin connection, and \textit{bad} were those tetrads were the theory trivially reduced to \gls{tegr}.
As expected, these notions tend to become obsolete since the realization that teleparallel theories are local Lorentz invariant, but for old times' sake, it is nice to keep a part of them and slightly modify it. For this reason, and since it concerns all theories built on teleparallel geometry, we thought it would be better to discuss what a \textit{good tetrad-spin connection pair} would be, before we dive into the realm of modified teleparallel theories.

In Sec.~\ref{sec:symmetries} we defined the concept of symmetric tetrads, that are specific tetrads satisfying certain symmetries related to teleparallel geometries. These tetrads are important to study astrophysical systems and cosmology since in them, one always has certain symmetries associated to the physical situation, such as spherical or axial symmetry. In \gls{tg}, one has the additional complication that these type of tetrads do not necessarily solve the antisymmetric part of the field equations of a certain theory. Moreover, one also needs to be careful that the choice of the symmetric tetrad is compatible with the choice of the spin connection. Due to the Lorentz transformations, one can always choose a specific gauge where the spin connection vanishes (Weitzenb\"{o}ck gauge), but one can also find a pair of tetrad with a non-zero spin connection which respects the symmetries and also solves the antisymmetric field equations of a theory as we will see later in this section.
Then, it is convenient to define a subclass of symmetric tetrads being compatible with the choice of the spin connection satisfying the symmetries. We will label the ``good tetrad-spin connection pair'' as the pair satisfying the following three conditions:
\begin{enumerate}
	\item The tetrad must be a symmetric tetrad compatible with the choice of the spin connection. This means that it must fulfill Eq.~\eqref{eq:infisymcondwb} (Weitzenb\"{o}ck gauge) or Eq.~\eqref{eq:infisymcondgen} (non-zero spin connection).

	\item It must solve the antisymmetric field equations without constraining the underlying theory.
\end{enumerate}
This definition was put together in order to match and extend the first ideas related to \gls{tg} in different symmetries~\cite{Tamanini:2012hg}, where the authors defined the good tetrads. These first ideas were introduced in the non-invariant theory, so it was not really clear that the spin connection plays an important role in \gls{tg}. Let us here explain all the conditions in more detail. The first condition ensures that the good tetrad-spin connection pair satisfy the underlying symmetries of the problem. \textit{A priori}, there is no fundamental reason to assume that the connection must also respect the symmetries of the physical situation, but, it is a sensible choice to impose this condition to then have the torsion tensor (and field equations) satisfying the same symmetries as the tetrad (or the metric). The second condition is the most important characteristic of the good pair, meaning that they are some specific pair which solve the antisymmetric field equations of a certain set of equations of a \gls{tg}. For practical reasons, these are the pairs that one needs to use to compute the field equations for different physical situations with a certain type of symmetry. It can also be inferred that the pair depends on the theory studied and it is possible to find that a good tetrad-spin connection pair in one theory is not a good tetrad-spin connection pair in another theory. It is of great importance though, that a good pair does not constrain the theory itself. If for example, the antisymmetric field equations, say in $f(T)$ gravity, yield $d^2f/dT^2 = 0,$ the associated tetrad-spin connection pair would not be considered as a good one. This condition is ignored in many studies concerning \gls{tg}.

Then, a good tetrad-spin connection pair satisfies the symmetry conditions that are compatible with either a zero spin connection or a non-zero spin connection and also solve the antisymmetric part of the field equations of the theory. Thus, in any modified \gls{tg}, one would first need to find a symmetric tetrad (with or without zero spin connection) satisfying a certain the symmetry of the physical situation and then find the good tetrad-spin connection pair which solve the antisymmetric field equations of this theory, with the important condition that the symmetric field equations cannot be the trivial ones (GR$+\Lambda$). To match the existing literature, apart from the good tetrad-spin connection pair, we will refer to \textit{good tetrads} as the ones that satisfy the conditions above and are computed in the Weitzenb\"{o}ck gauge.

In the following, we will give a comprehensive review of the most important modified teleparallel theories explored in the literature, along with their most important theoretical descriptions.

\subsection{New general relativity}
\label{sec:NGR}

We start the discussion on modifications of \gls{tegr} with a class of theories, dubbed \emph{New General Relativity} (\gls{ngr})~\cite{Hayashi:1979qx}. This was the first modification of \gls{tegr} considered in the literature and it was formulated in 1979 by Hayashi and Shirafuji. In their original use of the name, \gls{ngr} refers to a one-parameter subclass of a larger, three-parameter class of models, while nowadays the term \gls{ngr} is also used to refer to the whole three-parameter class. In this section we will discuss the structure, action and field equations of \gls{ngr} in section~\ref{chp4_NGR_action}, as well as its Hamiltonian analysis and the number of \gls{dof} in section~\ref{sssec:ngrhamilton}.

\subsubsection{Action and field equations} \label{chp4_NGR_action}
A straightforward generalization of the \gls{tegr} action~\eqref{eq:tegraction} can be obtained by realizing that the torsion scalar can be decomposed into a sum of three terms. Two different decompositions of this type are commonly used, for which we introduce the notation
\begin{equation}\label{eq:torscalar_decomposition}
T = \frac{1}{4}T_1 + \frac{1}{2}T_2 - T_3 = \frac{3}{2}T_{\text{axi}} + \frac{2}{3}T_{\text{ten}} - \frac{2}{3}T_{\text{vec}}\,.
\end{equation}
Here the terms in the first parametrization read
\begin{equation}
T_1 := T^{\mu\nu\rho}T_{\mu\nu\rho}\,, \quad
T_2 := T^{\mu\nu\rho}T_{\rho\nu\mu}\,, \quad
T_3 := T^\mu{}_{\mu\rho}T_\nu{}^{\nu\rho}\,.
\end{equation}
Alternatively, one may use the decomposition of the torsion tensor into parts which transform under irreducible representations of the Lorentz group, and which can be written in two equivalent ways. First, one may define
\begin{subequations}\label{eq:tortensatv123}
\begin{align}
a_{\mu} &:= \frac{1}{6}\epsilon_{\mu\nu\sigma\rho}T^{\nu\sigma\rho}\,,\label{eq:toraxidef}\\[0.5ex]
v_{\mu} &:= T^{\sigma}{}_{\sigma\mu}\,,\label{eq:torvecdef}\\[0.5ex]
t_{\sigma\mu\nu} &:= \frac{1}{2}\left(T_{\sigma\mu\nu} + T_{\mu\sigma\nu}\right) + \frac{1}{6}\left(g_{\nu\sigma}v_{\mu} + g_{\nu\mu}v_{\sigma}\right) - \frac{1}{3}g_{\sigma\mu}v_{\nu}\,,\label{eq:tortendef}
\end{align}
\end{subequations}
where \(\epsilon_{\mu\nu\sigma\rho}\) represents the totally antisymmetric Levi-Civita tensor associated to the metric \(g_{\mu\nu}\). Note that the tensor part \(t_{\sigma\mu\nu}\) is symmetric in its first two indices, while its totally symmetric part and any trace vanish
\begin{equation}
t_{\alpha\mu\nu} = t_{\mu\alpha\nu}, \quad
t_{\alpha\mu\nu} + t_{\nu\alpha\mu} + t_{\mu\nu\alpha} = 0, \quad
t^{\alpha\mu}{}_{\alpha} = t_{\alpha}{}^{\alpha\mu} = t^{\mu\alpha}{}_{\alpha} = 0\,.
\end{equation}
In terms of these one finds the scalar invariants\footnote{Note a sign mistake in the definition of \(T_{\text{axi}}\) in Ref.~\cite[Eq. (12)]{Bahamonde:2017wwk}.}
\begin{subequations}\label{eq:torscalatv}
\begin{align}
T_{\text{axi}} &:= a_{\mu}a^{\mu} = \frac{1}{18}\left(2T_{\sigma\mu\nu}T^{\mu\sigma\nu} - T_{\sigma\mu\nu}T^{\sigma\mu\nu}\right)\,,\\[0.5ex]
T_{\text{vec}} &:= v_{\mu}v^{\mu} = T^{\sigma}{}_{\sigma\mu}T_{\rho}{}^{\rho\mu}\,,\\[0.5ex]
T_{\text{ten}} &:= t_{\sigma\mu\nu}t^{\sigma\mu\nu} = \frac{1}{2}\left(T_{\sigma\mu\nu}T^{\sigma\mu\nu} + T_{\sigma\mu\nu}T^{\mu\sigma\nu}\right) - \frac{1}{2}T^{\sigma}{}_{\sigma\mu}T_{\rho}{}^{\rho\mu}\,.
\end{align}
\end{subequations}
Another possibility is to start from a decomposition of the torsion tensor in the form
\begin{equation} \label{chp4_tor_ten_decom}
T^{\mu}{}_{\nu\rho} = \accentset{\text{axi}}{T}^{\mu}{}_{\nu\rho} + \accentset{\text{vec}}{T}^{\mu}{}_{\nu\rho} + \accentset{\text{vec}}{T}^{\mu}{}_{\nu\rho}\,,
\end{equation}
where the three components are given by
\begin{equation}
\accentset{\text{axi}}{T}_{\mu\nu\rho} := \epsilon_{\mu\nu\rho\sigma}a^{\sigma}\,, \quad
\accentset{\text{vec}}{T}_{\mu\nu\rho} := \frac{1}{3}(g_{\mu\nu}v_{\rho} - g_{\mu\rho}v_{\nu})\,, \quad
\accentset{\text{ten}}{T}_{\mu\nu\rho} := \frac{2}{3}(t_{\mu\nu\rho} - t_{\mu\rho\nu})\,.
\end{equation}
These tensors vanish when mutually contracted with each other due to the symmetries of the torsion tensor, and yield the scalar invariants in the form
\begin{equation}
T_{\text{axi}} = -\frac{1}{6}\accentset{\text{axi}}{T}^{\mu}{}_{\nu\rho}\accentset{\text{axi}}{T}_{\mu}{}^{\nu\rho}\,, \quad
T_{\text{ten}} = \frac{3}{4}\accentset{\text{ten}}{T}^{\mu}{}_{\nu\rho}\accentset{\text{ten}}{T}_{\mu}{}^{\nu\rho}\,, \quad
T_{\text{vec}} = \frac{3}{2}\accentset{\text{vec}}{T}^{\mu}{}_{\nu\rho}\accentset{\text{vec}}{T}_{\mu}{}^{\nu\rho}\,.
\end{equation}
Hence, the different parametrizations are related by
\begin{equation}\label{eq:torscalatv123}
T_{\text{axi}} = \frac{1}{18}(2T_2 - T_1)\,, \quad
T_{\text{ten}} = \frac{1}{2}(T_1 + T_2 - T_3)\,, \quad
T_{\text{vec}} = T_3\,,
\end{equation}
or conversely by
\begin{equation}\label{eq:torscal123atv}
T_1 = \frac{2}{3}T_{\text{vec}} + \frac{4}{3}T_{\text{ten}} - 6T_{\text{axi}}\,, \quad
T_2 = \frac{1}{3}T_{\text{vec}} + \frac{2}{3}T_{\text{ten}} + 6T_{\text{axi}}\,, \quad
T_3 = T_{\text{vec}}\,.
\end{equation}
With these definitions in place, one generalizes the \gls{tegr} action~\eqref{eq:tegraction} to an arbitrary linear combination of the three scalar invariants, hence
\begin{equation}\label{eq:ngrlagrangian}
    \mathcal{L}_{\rm NGR} := c_1T_1 + c_2T_2 + c_3T_3 + \mathcal{L}_{\text{m}} = c_{\text{axi}}T_{\text{axi}} + c_{\text{ten}}T_{\text{ten}} + c_{\text{vec}}T_{\text{vec}} + \mathcal{L}_{\text{m}}\,.
\end{equation}
The relation between the different constants follows from the relations~\eqref{eq:torscalatv123} and~\eqref{eq:torscal123atv} to be
\begin{equation}
c_{\text{axi}} := 6(c_2 - c_1)\,, \quad
c_{\text{ten}} := \frac{4}{3}c_1 + \frac{2}{3}c_2\,, \quad
c_{\text{vec}} := \frac{2}{3}c_1 + \frac{1}{3}c_2 + c_3\,,
\end{equation}
or equivalently
\begin{equation}
c_1 := \frac{1}{2}c_{\text{ten}} - \frac{1}{18}c_{\text{axi}}\,, \quad
c_2 := \frac{1}{2}c_{\text{ten}} + \frac{1}{9}c_{\text{axi}}\,, \quad
c_3 := c_{\text{vec}} - \frac{1}{2}c_{\text{ten}}\,.
\end{equation}
Obviously, for the particular values
\begin{equation}\label{eq:ngrtegrvalues}
c_1 = \frac{1}{4}\,, \quad
c_2 = \frac{1}{2}\,, \quad
c_3 = -1 \quad \Leftrightarrow \quad
c_{\text{axi}} = \frac{3}{2}\,, \quad
c_{\text{ten}} = \frac{2}{3}\,, \quad
c_{\text{vec}} = -\frac{2}{3}\,,
\end{equation}
the Lagrangian~\eqref{eq:ngrlagrangian} reduces to that of \gls{tegr}. It was recently shown that the Lagrangian~\eqref{eq:ngrlagrangian} can be renormalized at one-loop order without introducing higher-order terms~\cite{Casadio:2021zai}. However, the theory with the three parameters contain ghosts.

The field equations are easily derived by variation of the action~\eqref{eq:ngrlagrangian} \gls{wrt} the tetrad and read
\begin{alignat}{2}\label{eq:ngrfieldeq}
\kappa^2\Theta_{\mu\nu} & =\: & & \frac{1}{2}\left(c_1T_1 + c_2T_2 + c_3T_3\right)g_{\mu\nu} + 2\lc{\nabla}^{\rho}\left(c_1T_{\nu\mu\rho} + c_2T_{[\rho\mu]\nu} + c_3T^{\sigma}{}_{\sigma[\rho}g_{\mu]\nu}\right)\nonumber\\[0.5ex]
& \: & &+ c_1T^{\rho\sigma}{}_{\mu}\left(T_{\nu\rho\sigma} - 2T_{[\rho\sigma]\nu}\right) - \frac{1}{2}c_3T^{\sigma}{}_{\sigma\rho}\left(T^{\rho}{}_{\mu\nu} + 2T_{(\mu\nu)}{}^{\rho}\right)\nonumber\\[0.5ex]
& \: & &+ \frac{1}{2}c_2\left[T_{\mu}{}^{\rho\sigma}\left(2T_{\rho\sigma\nu} - T_{\nu\rho\sigma}\right) + T^{\rho\sigma}{}_{\mu}\left(2T_{[\rho\sigma]\nu} - T_{\nu\rho\sigma}\right)\right]\,,
\end{alignat}
in terms of the torsion tensor and the energy-momentum tensor~\eqref{Eq:Con_EM_ten}, and can equivalently be expressed through the decomposition~\eqref{eq:tortensatv123} as
\begin{alignat}{2}
\kappa^2\Theta_{\mu\nu} & =\: & & c_{\text{axi}}\left(\frac{1}{2}a^{\rho}a_{(\rho}g_{\mu\nu)} - \frac{4}{9}\epsilon_{\nu\alpha\beta\gamma}a^{\alpha}t_{\mu}{}^{\beta\gamma} - \frac{2}{9}\epsilon_{\mu\nu\rho\sigma}a^{\rho}v^{\sigma} - \frac{1}{3}\epsilon_{\mu\nu\rho\sigma}\lc{\nabla}^{\rho}a^{\sigma}\right)\nonumber\\[0.5ex]
& \: & &+ c_{\text{vec}}\left(\frac{1}{2}v^{\rho}v_{(\rho}g_{\mu\nu)} + \frac{4}{3}t_{\mu[\rho\nu]}v^{\rho} + 2g_{\mu[\nu}\lc{\nabla}^{\rho}v_{\rho]} - \frac{1}{2}\epsilon_{\mu\nu\rho\sigma}a^{\rho}v^{\sigma}\right)\nonumber\\[0.5ex]
& \: & &+ c_{\text{ten}}\left(\frac{2}{3}t_{\alpha[\beta\gamma]}t^{\alpha\beta\gamma}g_{\mu\nu} - \frac{4}{3}t_{\mu[\rho\sigma]}t_{\nu}{}^{\rho\sigma} + 2\lc{\nabla}^{\rho}t_{\mu[\nu\rho]} - \frac{2}{3}t_{\nu[\mu\rho]}v^{\rho} + \frac{1}{2}\epsilon_{\mu\alpha\beta\gamma}a^{\alpha}t_{\nu}{}^{\beta\gamma}\right)\,.
\end{alignat}
The energy-momentum tensor \(\Theta_{\mu\nu}\) is symmetric by assuming a minimal, Lorentz invariant matter coupling as discussed in Sec.~\ref{ssec:mattercoupling}. The field equations are then decomposed into their symmetric and antisymmetric parts, which are given by
\begin{alignat}{2}\label{eq:ngrfieldeqsym}
\kappa^2\Theta_{\mu\nu} & =\: & & \lc{\nabla}^{\rho}\left[(2c_1 + c_2)T_{(\mu\nu)\rho} + c_3(T^{\sigma}{}_{\sigma\rho}g_{\mu\nu} - T^{\sigma}{}_{\sigma(\mu}g_{\nu)\rho})\right]\nonumber\\[0.5ex]
& \: & & + \frac{1}{2}\left(c_1T_1 + c_2T_2 + c_3T_3\right)g_{\mu\nu} + c_1\left(T^{\rho\sigma}{}_{(\mu}T_{\nu)\rho\sigma} - 2T_{[\rho\sigma]\mu}T^{\rho\sigma}{}_{\nu}\right)\nonumber\\[0.5ex]
&\: & & + \frac{1}{2}c_2\left(T^{\rho\sigma}{}_{(\mu}T_{\nu)\rho\sigma} + 2T_{[\rho\sigma]\mu}T^{\rho\sigma}{}_{\nu} - T_{\mu}{}^{\rho\sigma}T_{\nu\rho\sigma}\right) - c_3T^{\sigma}{}_{\sigma\rho}T_{(\mu\nu)}{}^{\rho}\,,
 \end{alignat}
and
\begin{equation}\label{eq:ngrfieldeqasym}
0 = \frac{1}{2}(2c_2 + c_3)\lc{\nabla}^{\rho}T_{\rho\mu\nu} - (2c_1 - c_2)\lc{\nabla}^{\rho}T_{[\mu\nu]\rho} + \frac{1}{2}(2c_1 - 3c_2 - c_3)T^{\rho\sigma}{}_{[\mu}T_{\nu]\rho\sigma}\,,
\end{equation}
using the geometric identity
\begin{equation}
    3\lc{\nabla}_{[\rho}T^{\rho}{}_{\mu\nu]} + T^{\rho}{}_{\mu\nu}T^{\sigma}{}_{\sigma\rho} - T^{\rho\sigma}{}_{[\mu}T_{\nu]\rho\sigma} = 0\,,
\end{equation}
in the second equation. Recall that in the covariant formulation of \gls{tg}~\cite{Krssak:2018ywd} we use here, in which an arbitrary, flat spin connection is assumed, the antisymmetric part~\eqref{eq:ngrfieldeqasym} is equivalently obtained by variation \gls{wrt} the spin connection, and vanishes identically if the parameter values satisfy
\begin{equation}\label{eq:ngrasymvancond}
2c_1 - c_2 = 2c_2 + c_3 = 0 \quad \Leftrightarrow \quad
c_{\text{ten}} + c_{\text{vec}} = 9c_{\text{ten}} - 4c_{\text{axi}} = 0\,,
\end{equation}
and are hence a multiple of the \gls{tegr} values~\eqref{eq:ngrtegrvalues}.

The theory based on general form of the Lagrangian~\eqref{eq:ngrlagrangian} discussed above is known as (the three-parameter class of) \gls{ngr}~\cite{Hayashi:1979qx}. It can also be obtained from the premetric approach~\cite{Itin:2016nxk}. However, as mentioned in the introduction of this section, in the original work the name \gls{ngr} is used to refer to the more restricted one-parameter class. This smaller class is obtained by fixing \(c_{\text{vec}}\) and \(c_{\text{ten}}\) to their \gls{tegr} values \eqref{eq:ngrtegrvalues}, and allowing only \(c_{\text{axi}}\) to deviate, so that the parameters take the form
\begin{equation}\label{eq:ngroneparvalues}
c_1 = \frac{1}{4} - \frac{\epsilon}{18}\,, \quad
c_2 = \frac{1}{2} + \frac{\epsilon}{9}\,, \quad
c_3 = -1 \quad \Leftrightarrow \quad
c_{\text{axi}} = \epsilon + \frac{3}{2}\,, \quad
c_{\text{ten}} = \frac{2}{3}\,, \quad
c_{\text{vec}} = -\frac{2}{3}\,,
\end{equation}
with a free parameter \(\epsilon\). This restricted class has the interesting and distinguishing property that observational discriminators, such as the Parameterized post-Newtonian (\gls{ppn}) limit, agree with \gls{tegr}, and hence with \gls{gr}~\cite{Hayashi:1979qx,Ualikhanova:2019ygl}. Its field equations take the form
\begin{equation}\label{eq:ngroneparfieldeq}
\kappa^2\Theta_{\mu\nu} = \lc{G}_{\mu\nu} + \epsilon\left(\frac{1}{2}a^{\rho}a_{(\rho}g_{\mu\nu)} - \frac{4}{9}\epsilon_{\nu\alpha\beta\gamma}a^{\alpha}t_{\mu}{}^{\beta\gamma} - \frac{2}{9}\epsilon_{\mu\nu\rho\sigma}a^{\rho}v^{\sigma} - \frac{1}{3}\epsilon_{\mu\nu\rho\sigma}\lc{\nabla}^{\rho}a^{\sigma}\right)\,,
\end{equation}
showing that they reduce to \gls{gr} in the absence of axial torsion \(a_{\mu}\).

Note that the covariant, three-parameter approach to \gls{ngr} discussed above differs in two further aspects from the originally proposed theory~\cite{Hayashi:1979qx}. In the latter, a non-covariant formulation is adopted, in which the spin connection is assumed to vanish identically, and the tetrad is the only dynamical field. Further, a different matter coupling is assumed, in which Dirac fields couple to the (vanishing) teleparallel spin connection instead of the Levi-Civita connection. As a consequence, their energy-momentum tensor receives an antisymmetric contribution, which acts as a nonvanishing source for the antisymmetric Eq.~\eqref{eq:ngrfieldeqasym}. Hence, in this original approach is was assumed that the parameter values strictly differ from the condition~\eqref{eq:ngrasymvancond}, since otherwise the theory would not have allowed for fermionic fields.

\subsubsection{Hamiltonian analysis and degrees of freedom} \label{sssec:ngrhamilton}

An important question which arises in the context of modifications of \gls{gr} is the appearance and nature of additional \gls{dof}. Of particular importance is the question whether such new \gls{dof} are pathological, such as to behave as ghosts or tachyons \cite{VanNieuwenhuizen:1973fi,Kuhfuss:1986rb}, whether their time evolution is uniquely defined, such that the theory has a well-defined Cauchy problem, and whether there is a mismatch between the linearized and the full, non-linear theory~\cite{Jimenez:2019tkx}. Specifically, in Ref.~\cite{Jimenez:2019tkx} the authors found that the gauge symmetry that appears in the quadratic Lagrangian, realized as a massless 2-from field, i.e. Kalb-Ramond field, is necessarily broken at cubic order, signaling a strong coupling problem around Minkowski spacetime. This means that the Minkowski solution is very unstable against perturbations and thus it will be impossible to be reached from an aritrary point in phase space. For \gls{ngr} in its most general form this is still an open question. The most comprehensive approach to answering this question is the Hamiltonian analysis (see Ref.~\cite{Blixt:2020ekl} for a recent review).

Already, for the original one-parameter \gls{ngr}~\cite{Hayashi:1979qx}, possible issues concerning a non-unique time evolution, and hence an ill-defined Cauchy problem, have been noted, with the help of a simple example~\cite{Kopczynski:1982}. Consider the diagonal tetrad \(e^A{}_{\mu} = \mathrm{diag}(1,1,1,1)\) and nonvanishing spin connection \(\omega^A{}_{B\mu}\dd x^{\mu} = \Omega^A{}_B(t)\dd t\), where the spatial components \(\Omega^I{}_J(t) = 0\) vanish, with \(I, J = 1, \ldots, 3\). One finds that this spin connection is indeed flat, \(R^A{}_{B\mu\nu} = 0\), while the only nonvanishing torsion components are given by
\begin{equation}
T^0{}_{0I} = -T^0{}_{I0} = \Omega^0{}_I\,.
\end{equation}
Note that, in particular, the axial torsion~\eqref{eq:toraxidef} vanishes, due to the contraction with the totally antisymmetric tensor. Further, the metric \(g_{\mu\nu}\) obtained from the diagonal tetrad \(e^A{}_{\mu}\) is simply the Minkowski metric, and so the curvature of the Levi-Civita connection vanishes. Hence, this tetrad and spin connection constitute a solution to the field Eqs.~\eqref{eq:ngroneparfieldeq}. Since the functions \(\Omega^0{}_I(t)\) can be chosen arbitrarily, it follows that their time evolution, and hence also the time evolution of the nonvanishing torsion components, is not determined by the field equations and initial conditions. While this has no observable consequences if the matter couples only to the metric, it does lead to an undetermined influence on matter if one allows for a direct coupling to the spin connection. Note that similar statements can be made also about other \gls{tg} theories, such as \(f(T)\) gravity~\cite{Jimenez:2020ofm,Golovnev:2020nln}.

In order to obtain a well-defined Cauchy problem also for the torsion, different modifications of \gls{ngr} have been suggested. The most straightforward extension is to consider the full action~\eqref{eq:ngrlagrangian}, with three free parameters, where the deviation from their \gls{tegr} values must be chosen sufficiently small so their \gls{ppn} limit is compatible with solar system observations~\cite{Kopczynski:1982}. Other possibilities are given by adding higher order terms~\cite{Moller:1978} or parity-violating terms~\cite{MuellerHoissen:1983vc} to the action. However, it has been pointed out that the non-unique time evolution occurs only for particular solutions and is not generic~\cite{nester1988there}, and that tetrads which are related by the remaining partial Lorentz invariance which does not alter the axial torsion may be regarded as physical equivalent~\cite{PhysRevD.24.3312}.

\begin{figure}[htbp]
\centering
\includegraphics[width=\textwidth]{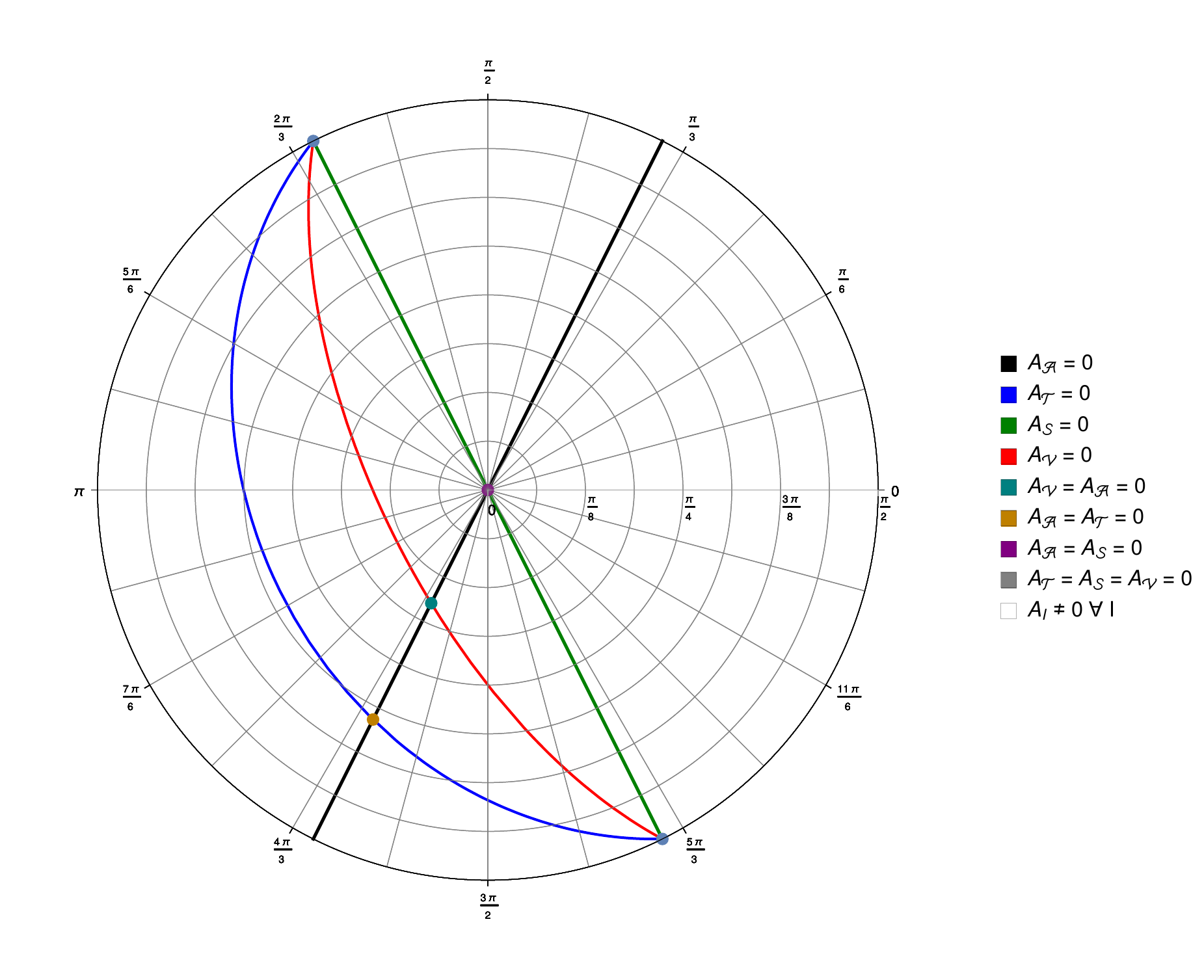}
\caption[Primary constraints of \gls{ngr}]{Primary constraints of \gls{ngr}, depending on the values of the parameters in the action. The radial axis shows \(0 \leq \alpha \leq \frac{\pi}{2}\), while the polar axis shows \(0 \leq \beta \leq 2\pi\), using the definition~\eqref{eq:ngrpolarcoord}. The red line \(A_{\mathcal{V}} = 0\), except for the two points where it meets the plot boundary \(\alpha = \frac{\pi}{2}\), constitutes the one-parameter class of \gls{ngr}, with \gls{tegr} being marked by the dot at the intersection of the red and black lines, \(A_{\mathcal{V}} = A_{\mathcal{A}} = 0\). Permission for use of this figure was kindly provided by the authors of Ref.~\cite{Blixt:2018znp}.}
\label{fig:ngrprimcons}
\end{figure}

In order to address the aforementioned issues, and assess the consistency of \gls{ngr}, various approaches towards a full Hamiltonian description have been carried out. For the most general, three-parameter class of \gls{ngr} governed by the Lagrangian~\eqref{eq:ngrlagrangian}, the canonical momenta defined by the variation~\eqref{eq:canmomdef} are given by Ref.~\cite{Blixt:2018znp}
\begin{equation}\label{eq:ngrcanmom}
\pi_A{}^i = \frac{\sqrt{h}}{N}\left\{T^B{}_{0j}M^i{}_A{}^j{}_B + T^B{}_{kl}\left[M^i{}_A{}^k{}_BN^l + 2Nh^{ik}(c_2n_Be_A{}^l + c_3n_Ae_B{}^l)\right]\right\}\,,
\end{equation}
using the abbreviation
\begin{equation}\label{eq:ngrvelmommat}
M^i{}_A{}^j{}_B = -2(2c_1h^{ij}\eta_{AB} - (c_2 + c_3)n_An_Bh^{ij} + c_2e_A{}^je_B{}^i + c_3e_A{}^ie_B{}^j)\,.
\end{equation}
Note that \(e_A{}^i\) is \emph{not} the inverse tetrad, but defined from the tetrad by raising and lowering indices with the Minkowski and induced metrics, \(\eta_{AB}\) and \(h^{ij}\), so that
\begin{equation}\label{eq:notinvtetrad}
e_A{}^i = \eta_{AB}h^{ij}e^B{}_j \neq E_A{}^i = e_A{}^i + n_A\frac{N^i}{N}\,.
\end{equation}
Further, from the decomposition~\eqref{eq:tor31split} of the torsion follows that the velocities enter the momenta~\eqref{eq:ngrcanmom} only via the component \(T^A{}_{0i}\) in the first term, while the remaining terms depend only on the lapse \(N\), shift \(N^i\), spatial tetrad components \(e^A{}_i\) and their spatial derivatives. Hence, in order to solve for the velocities \(v^A{}_i = \partial_0e^A{}_i\) in terms of the momenta, it is convenient to isolate the term which is linear in the velocities on one side of the equation, which then takes the form
\begin{equation}\label{eq:ngrvellinrel}
M^i{}_A{}^j{}_Bv^B{}_j = S_A{}^i\,,
\end{equation}
where the term \(S_A{}^i\) on the \gls{rhs} is defined as
\begin{equation}
S_A{}^i = \frac{N}{\sqrt{h}}\pi_A{}^i + \left[\partial_k(Nn^B + N^me^B{}_m) - T^B{}_{kl}N^l\right]M^i{}_A{}^k{}_B - 2NT^B{}_{kl}h^{ik}(c_2n_Be_A{}^l + c_3n_Ae_B{}^l)\,,
\end{equation}
and does not contain any velocities. In order to solve the Eq.~\eqref{eq:ngrvellinrel} for the velocities \(v^A{}_i\), it is necessary to invert the object in Eq.~\eqref{eq:ngrvelmommat}, viewed as a matrix. This can be achieved by applying the irreducible decomposition shown in Sec.~\ref{sssec:velmomirrdec} to both sides of the Eq.~\eqref{eq:ngrvellinrel}. On the \gls{rhs}, proceeding in analogy to the momentum decomposition~\eqref{eq:canmomirrdec} yields the decomposition
\begin{subequations}
\begin{align}
\accentset{\mathcal{V}}{S}^i &= \frac{N}{\sqrt{h}}\accentset{\mathcal{V}}{\pi}^i - 2Nc_3T^B{}_{kl}h^{ik}e_B{}^l + 2(2c_1 + c_2 + c_3)\left[\partial_k(Nn^B + N^me^B{}_m) - T^B{}_{kl}N^l\right]n_Bh^{ik}\,,\\[0.5ex]
\accentset{\mathcal{A}}{S}^{mp} &= \frac{N}{\sqrt{h}}\accentset{\mathcal{A}}{\pi}^{mp} - 2Nc_2h^{lm}h^{pk}T^B{}_{kl}n_B - 2(2c_1 - c_2)\left[\partial_k(Nn^B + N^se^B{}_s) - T^B{}_{kl}N^l\right]e_B{}^{[m}h^{p]k}\,,\\[0.5ex]
\accentset{\mathcal{S}}{S}^{mp} &= \frac{N}{\sqrt{h}}\accentset{\mathcal{S}}{\pi}^{mp} - 2(2c_1 + c_2)\left[\partial_k(Nn^B + N^se^B{}_s) - T^B{}_{kl}N^l\right]\left(e_B{}^{(m}h^{p)k} - \frac{1}{3}h^{pm}e_B{}^k\right)\,,\\[0.5ex]
\accentset{\mathcal{T}}{S} &= \frac{N}{\sqrt{h}}\accentset{\mathcal{T}}{\pi} - \frac{2}{3}(2c_1 + c_2 + 3c_3)\left[\partial_k(Nn^BN^me^B{}_m) - T^B{}_{kl}N^l\right]e_B{}^k\,.
\end{align}
\end{subequations}
Applying the same decomposition to the \gls{lhs} of the velocity Eq.~\eqref{eq:ngrvellinrel}, and substituting the velocity components~\eqref{eq:canvelirrdec}, one finds that the matrix~\eqref{eq:ngrvelmommat} greatly simplifies, and the Eq.~\eqref{eq:ngrvellinrel} decomposes into the components
\begin{equation}
\accentset{\mathcal{V}}{S}^i = -2A_{\mathcal{V}}\accentset{\mathcal{V}}{v}^i\,, \quad
\accentset{\mathcal{A}}{S}^{mp} = -2A_{\mathcal{A}}\accentset{\mathcal{A}}{v}^{mp}\,, \quad
\accentset{\mathcal{S}}{S}^{mp} = -2A_{\mathcal{S}}\accentset{\mathcal{S}}{v}^{mp}\,, \quad
\accentset{\mathcal{T}}{S} = -2A_{\mathcal{T}}\accentset{\mathcal{T}}{v}\,,
\end{equation}
where the constants \(A_{\mathcal{V}, \mathcal{A}, \mathcal{S}, \mathcal{T}}\) are given by
\begin{equation}
A_{\mathcal{V}} = 2c_1 + c_2 + c_3\,, \quad
A_{\mathcal{A}} = 2c_1 - c_2\,, \quad
A_{\mathcal{S}} = 2c_1 + c_2\,, \quad
A_{\mathcal{T}} = 2c1 + c_2 + 3c_3\,,
\end{equation}
or equivalently,
\begin{equation}
A_{\mathcal{V}} = c_{\text{ten}} + c_{\text{vec}}\,, \quad
A_{\mathcal{A}} = \frac{1}{2}c_{\text{ten}} - \frac{2}{9}c_{\text{axi}}\,, \quad
A_{\mathcal{S}} = \frac{3}{2}c_{\text{ten}}\,, \quad
A_{\mathcal{T}} = 3c_{\text{vec}}\,.
\end{equation}
If all of these constants are nonvanishing, one can solve all velocity components for the corresponding momentum components. In general, however, this is not the case, as certain constants may vanish, depending on the values of the parameters \(c_{1,2,3}\), thereby leading to the appearance of primary constraints, which are given by
\begin{subequations}\label{eq:ngrprimcons}
\begin{align}
A_{\mathcal{V}} = 0 \quad &\Rightarrow \quad 0 \approx \accentset{\mathcal{V}}{C}^i = \frac{\accentset{\mathcal{V}}{\pi}^i}{\sqrt{h}} - 2c_3T^B{}_{kl}h^{ik}e_B{}^l\,,\\[0.5ex]
A_{\mathcal{A}} = 0 \quad &\Rightarrow \quad 0 \approx \accentset{\mathcal{A}}{C}^{mp} = \frac{\accentset{\mathcal{A}}{\pi}^{mp}}{\sqrt{h}} - 2c_2h^{lm}h^{pk}T^B{}_{kl}n_B\,,\\[0.5ex]
A_{\mathcal{S}} = 0 \quad &\Rightarrow \quad 0 \approx \accentset{\mathcal{S}}{C}^{mp} = \frac{\accentset{\mathcal{S}}{\pi}^{mp}}{\sqrt{h}}\,,\\[0.5ex]
A_{\mathcal{T}} = 0 \quad &\Rightarrow \quad 0 \approx \accentset{\mathcal{T}}{C} = \frac{\accentset{\mathcal{T}}{\pi}}{\sqrt{h}}\,,
\end{align}
\end{subequations}
which defines the constraint expressions \(C_A{}^i\) in terms of their irreducible decomposition. The different possible cases can most easily be visualized by introducing normalized parameters \(\tilde{c}_i\) and polar coordinates \((\alpha, \beta)\) using the definition
\begin{equation}\label{eq:ngrpolarcoord}
\tilde{c}_i = \frac{c_i}{\sqrt{c_1^2 + c_2^2 + c_3^2}}\,, \quad
\tilde{c}_1 = \sin\alpha\cos\beta\,, \quad
\tilde{c}_2 = \sin\alpha\sin\beta\,, \quad
\tilde{c}_3 = \cos\alpha\,.
\end{equation}
These coordinates parametrize a sphere, one hemisphere of which is shown in Fig.~\ref{fig:ngrprimcons}; the other hemisphere is obtained by a point reflection at the origin. A different parametrization, based on the axial-vector-tensor decomposition of the torsion, is given in Ref.~\cite{Blixt:2019ene}. The same result is presented in the language of differential forms in Ref.~\cite{Hohmann:2019sys} and in the premetric approach in Ref.~\cite{Guzman:2020kgh}.

Several particular subclasses of \gls{ngr} have been studied, which can be characterized by their primary constraints. The most prominent example is \gls{tegr} given by the parameter values~\eqref{eq:ngrtegrvalues}, which imply \(A_{\mathcal{V}} = A_{\mathcal{A}} = 0\); this is discussed in Sec.~\ref{sssec:tegrhamilton}. A toy model with \(c_1 = 1/2, c_2 = c_3 = 0\) with no primary constraints has been studied in Ref.~\cite{Okolow:2011np} in the language of differential forms, in order to present the general formalism of Hamiltonian analysis in \gls{tg}. The one-parameter class of \gls{ngr}, with parameters~\eqref{eq:ngroneparvalues}, has been studied for \(\epsilon \neq 0\), so that the only primary constraint \(\accentset{\mathcal{V}}{C}^i \approx 0\) arises from \(A_{\mathcal{V}} = 0\), in Ref.~\cite{Cheng:1988zg}. Finally, also an extension of \gls{ngr} by a parity-violating term has been considered, whose Lagrangian is given by \cite{Chen:1987}
\begin{equation}
\mathcal{L} = \frac{1}{2}\left(1 - \frac{\lambda}{2} - \nu\right)T_1 + \frac{\lambda}{2}T_2 - T_3 + 6\sigma v_{\mu}a^{\mu}\,,
\end{equation}
which depends on the constants \(\lambda, \nu, \sigma\). Note the appearance of a parity-violating term \(v_{\mu}a^{\mu}\), governed by the parameter \(\sigma\). In the absence of this term, choosing \(\sigma = 0\), the theory reduces to \gls{ngr} with the parameters given by
\begin{equation}
c_1 = \frac{1}{2}\left(1 - \frac{\lambda}{2} - \nu\right)\,, \quad
c_2 = \frac{\lambda}{2}\,, \quad
c_3 = -1\,,
\end{equation}
or equivalently,
\begin{equation}
c_{\text{axi}} = 3\left(\frac{3}{2}\lambda + \nu - 1\right)\,, \quad
c_{\text{ten}} = \frac{2}{3}(1 - \nu)\,, \quad
c_{\text{vec}} = \frac{1}{3}(\nu - 2)\,.
\end{equation}
Further restricting to \(\nu = 0\), the theory reduces to the one-parameter \gls{ngr} values~\eqref{eq:ngroneparvalues} with \(\epsilon = 2(\lambda - 1)/9\), and giving \gls{tegr} for \(\lambda = 1\). This class of theories avoids the undetermined evolution of the torsion mentioned above if the parameters satisfy the inequality
\begin{equation}
\nu(1 - \nu - \lambda) - 2\sigma^2 \neq 0\,.
\end{equation}
For pure \gls{ngr}, where \(\sigma = 0\), this condition reduces to \(\nu \neq 0\) and \(1 - \nu - \lambda \neq 0\). By comparison with the conditions for the appearance of primary constraints, one finds that this corresponds to \(A_{\mathcal{V}} \neq 0\) and \(A_{\mathcal{A}} \neq 0\), so that no vector and antisymmetric constraints appear. Further, consistency of the \gls{ppn} limit with solar system observations, \(|\nu| \leq 1\) is required, so that also \(A_{\mathcal{S}} \neq 0\) and \(A_{\mathcal{T}} \neq 0\). Hence, this corresponds to the generic class of \gls{ngr} theories, in which there are no primary constraints.

Further studies in the Hamiltonian approach to \gls{ngr} have so far been restricted to such special cases. In the case that none of the constants \(A_{\mathcal{V}, \mathcal{A}, \mathcal{S}, \mathcal{T}}\) vanish, so that none of the primary constraints~\eqref{eq:ngrprimcons} appear. In this case all velocities can be expressed in terms of canonical momenta, and the full Hamiltonian is identical to the kinematic Hamiltonian~\eqref{eq:hamiltondef} and reads~\cite{Blixt:2018znp}
\begin{alignat}{2}\label{eq:ngrhamiltongeneric}
H & =\: & & N\sqrt{h}\left(\frac{\accentset{\mathcal{V}}{C}_i\accentset{\mathcal{V}}{C}^i}{4A_{\mathcal{V}}} - \frac{\accentset{\mathcal{A}}{C}_{ij}\accentset{\mathcal{A}}{C}^{ij}}{4A_{\mathcal{A}}} - \frac{\accentset{\mathcal{S}}{C}_{ij}\accentset{\mathcal{S}}{C}^{ij}}{4A_{\mathcal{S}}} - \frac{3\accentset{\mathcal{T}}{C}^2}{4A_{\mathcal{T}}} - \tilde{T} - \frac{n^A\tilde{\nabla}_i\pi_A{}^i}{\sqrt{h}}\right)\nonumber\\[0.5ex]
& \: & & - N^k(T^A{}_{jk}\pi_A{}^j + e^A{}_k\tilde{\nabla}_i\pi_A{}^i) + \tilde{\nabla}_i[\pi_A{}^i(Nn^A + N^je^A{}_j)]\,,
\end{alignat}
where we used the abbreviation
\begin{equation}
\tilde{T} = c_1\eta_{AB}T^A{}_{ij}T^B{}_{kl}h^{ik}h^{jl} + c_2e_A{}^ie_B{}^jT^A{}_{kj}T^B{}_{li}h^{kl} + c_3e_A{}^ie_B{}^jh^{kl}T^A{}_{ki}T^B{}_{lj}\,,
\end{equation}
while \(\tilde{\nabla}_i\) denotes the purely spatial Levi-Civita covariant derivative of the induced metric \(h_{ij}\), and \(e_A{}^i\) are the spatial tetrad components~\eqref{eq:notinvtetrad}. Note that in this case none of the expressions \(C_A{}^i\) given by the \gls{rhs} of the expressions~\eqref{eq:ngrprimcons} are constraints, i.e., they are not imposed to vanish; they simply turn out to be convenient for expressing the Hamiltonian. Furthermore, note that all derivatives of the lapse \(N\) and shift \(N^i\) are gathered into a total divergence term, while they appear as Lagrange multipliers in the remaining terms. Hence, one finds that they lead to the further constraints
\begin{equation}
\frac{\accentset{\mathcal{V}}{C}_i\accentset{\mathcal{V}}{C}^i}{4A_{\mathcal{V}}} - \frac{\accentset{\mathcal{A}}{C}_{ij}\accentset{\mathcal{A}}{C}^{ij}}{4A_{\mathcal{A}}} - \frac{\accentset{\mathcal{S}}{C}_{ij}\accentset{\mathcal{S}}{C}^{ij}}{4A_{\mathcal{S}}} - \frac{3\accentset{\mathcal{T}}{C}^2}{4A_{\mathcal{T}}} - \tilde{T} - \frac{n^A\tilde{\nabla}_i\pi_A{}^i}{\sqrt{h}} \approx 0\,,
\end{equation}
and
\begin{equation}
T^A{}_{jk}\pi_A{}^j + e^A{}_k\tilde{\nabla}_i\pi_A{}^i \approx 0\,,
\end{equation}
which correspond to the Hamiltonian and momentum constraints in \gls{gr}. Further, it follows that imposing these constraints leads to a vanishing Hamiltonian on the constraint surface, up to a total divergence, as it is also the case in \gls{gr}.

In the case that any of the constants \(A_{\mathcal{V}, \mathcal{A}, \mathcal{S}, \mathcal{T}}\) vanishes, the corresponding term in the Hamiltonian~\eqref{eq:ngrhamiltongeneric} would diverge, and so it must be replaced, taking into account the arising primary constraints. The most straightforward case in which this can be seen is the non-trivial (\(\epsilon \neq 0\)) one-parameter \gls{ngr}~\eqref{eq:ngroneparvalues}. In this case one has \(A_{\mathcal{V}} = 0\), leading to the constraint \(\accentset{\mathcal{V}}{C}^i \approx 0\). It then follows that the kinematic Hamiltonian is given by~\cite{Cheng:1988zg}
\begin{alignat}{2}
H_0 & =\: & & N\sqrt{h}\Bigg\{\frac{9}{8\epsilon}\left[\frac{\pi_{[ij]}}{\sqrt{h}} - \left(1 - \frac{2}{9}\epsilon\right)n_AT^A{}_{ij}\right]\left[\frac{\pi^{[ij]}}{\sqrt{h}} - \left(1 - \frac{2}{9}\epsilon\right)n_AT^{Aij}\right] - \tilde{T} - \frac{n^A\tilde{\nabla}_i\pi_A{}^i}{\sqrt{h}}\nonumber\\[0.5ex]
& \: & & - \frac{\pi_{(ij)}\pi^{(ij)}}{4h} + \frac{(\pi_A{}^ie^A{}_i)^2}{8h}\Bigg\} - N^k(T^A{}_{jk}\pi_A{}^j + e^A{}_k\tilde{\nabla}_i\pi_A{}^i) + \tilde{\nabla}_i[\pi_A{}^i(Nn^A + N^je^A{}_j)]\,,
\end{alignat}
to which a Lagrange multiplier term
\begin{equation}
    H_1 = V_i\left(\accentset{\mathcal{V}}{\pi}^i + 2\sqrt{h}T^B{}_{kl}h^{ik}e_B{}^l\right)\,,
\end{equation}
must be added, in order to implement the primary constraint \(\accentset{\mathcal{V}}{C}_i \approx 0\). In order to obtain the full Hamiltonian, further terms must be added, which arise from the time evolution of the constraint \(\accentset{\mathcal{V}}{C}_i\), given by its Poisson bracket with the kinematic Hamiltonian~\cite{Cheng:1988zg}.

\subsection{\texorpdfstring{$f(T)$}{fT} gravity}
\label{sec:f(T)gravity}

A relatively recent modification of \gls{tegr} is the so-called $f(T)$ theory of gravity. In the same spirit as $f(\lc{R})$ gravity, that is a straightforward generalization of the Einstein-Hilbert action, $f(T)$ theory was proposed almost more than a decade ago as a generalization to the \gls{tegr} action, by R.~Ferraro and F.~Fiorini \cite{Ferraro:2006jd}. In this section, we will study its action and field equations together with its behavior under conformal and disformal transformations.

\subsubsection{Action and field equations}\label{ssec:Action_and_field_equations}
\begin{figure}[H]
    \centering
    \includegraphics[scale=1.15]{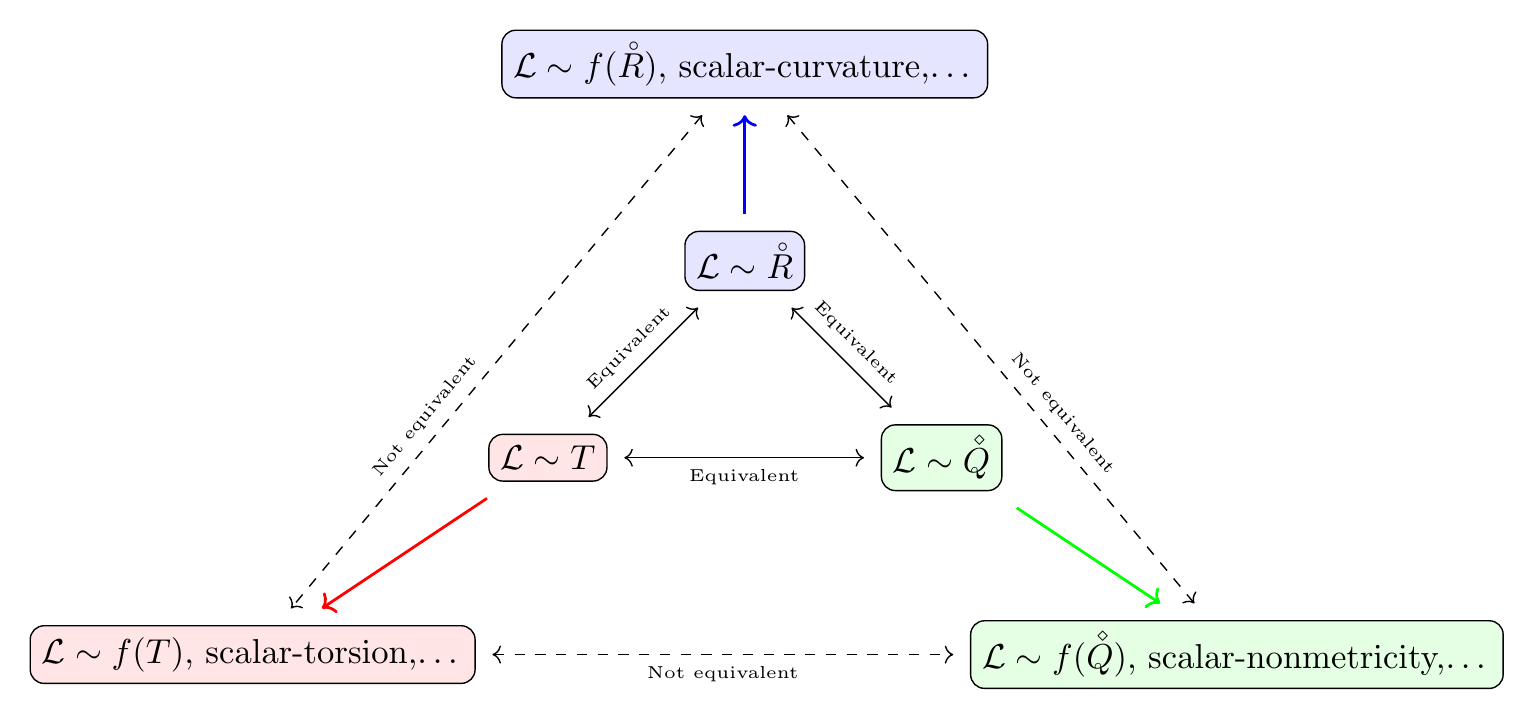}
    \caption{Equivalency between GR, \gls{tegr} and \gls{stegr}, but not equivalency between their modifications.}
    \label{fig:Chapter5-1}
\end{figure}

The action of $f(T)$ theory reads
\begin{equation}\label{f(T)action}
    \mathcal{S}_{f(T)} := \frac{1}{2 \kappa ^2} \int \dd ^4 x \, e \, f(T) + \mathcal{S}_{\rm m}\,,
\end{equation}
where $\mathcal{S}_{\rm m}$ is the action related to the matter sector defined as in Eq.~\eqref{eq:actionmatter}. Let us take a step back and consider the geometric trinity of gravity discussed in Sec.~\ref{Sec:Geometric_Trinity}. There we saw from Eq.~\eqref{eq:ricsmtele} and Eq.~\eqref{eq:ricsstele}, that the relation between the Ricci, the torsion and the non-metricity scalars is
\begin{equation}\label{eq:RTQ}
\lc{R} = -T+B = -\st{Q}-B_{\st{Q}}\,,
\end{equation}
where we defined the two boundary terms, $B$ and $B_{\st{Q}}$ as
\begin{subequations}
\begin{align}\label{eq:boundaryB}
    B &= 2 \lc{\nabla}_\mu T^\mu = \frac{2}{e}\partial _{\mu}(eT^\mu)\,, \\[0.5ex]
    B_{\st{Q}} &= \lc{\nabla}_{\nu}\st{Q}_{\mu}{}^{\mu\nu} - \lc{\nabla}_{\mu}\st{Q}^{\mu\nu}{}_{\nu}\,.\label{eq:boundaryB_b}
\end{align}
\end{subequations}
It is easy to conclude from Eq.~\eqref{eq:RTQ} that since the total derivatives, $B$ and $B_{\st{Q}}$ will not contribute in the action integral, the field equations, resulting from their variations, will be equivalent in all three theories, GR, \gls{tegr} and \gls{stegr}, even though the geometric background is different in each one of them. The same does not happen though when one generalizes to arbitrary functions of these scalars \cite{Jarv:2018bgs} and this is described in the Fig.~\ref{fig:Chapter5-1}. Recently, a different decomposition of the curvature scalar has been utilized,
\begin{equation}
    \lc{R} = \lc{G} + \lc{\mathcal{B}}\,,
\end{equation}
where $\lc{G} = g^{\mu\nu}\left(\lc{\Gamma}^{\lambda}{}_{\nu\sigma}\lc{\Gamma}^{\sigma}{}_{\lambda \mu} + \lc{\Gamma}^{\sigma}{}_{\mu\nu}\lc{\Gamma}^{\lambda}{}_{\lambda\sigma}\right)$ and $\lc{\mathcal{B}} = \partial_{\nu}\left(\partial_{\mu}(gg^{\mu\nu})/ \sqrt{-g} \right)/\sqrt{-g}$, that are pseudo-scalars (in contrast with $T$ and $B$) and a new generalized theory was proposed \cite{Bohmer:2021eoo} that has the form
\begin{equation}
    \mathcal{S}_{f(\lc{G},\lc{\mathcal{B}})} := \frac{1}{2\kappa ^2}\int \dd^4 x \sqrt{-g}f(\lc{G},\lc{\mathcal{B}}) + \mathcal{S}_{\rm m}\,,
\end{equation}
with $f$ being an arbitrary function of $\lc{G}$ and $\lc{\mathcal{B}}$.

In greater detail, $f(\lc{R})$ theories result in fourth order field equations and that is because, the Ricci scalar consists of second order derivatives of the metric. On the other hand, both the torsion and the non-metricity scalar are built of only first derivatives of the tetrad and the metric respectively, yielding second order field equations. Actually, because of the similarity of these two scalars, $f(\st{Q})$ and $f(T)$ theories are indistinguishable at cosmological background level \cite{Jarv:2018bgs}; that is why if one wants to look for differences between the two theories, they have to study perturbations \cite{Jimenez:2019ovq}. If one wants to imitate $f(\lc{R})$ theory in the teleparallel geometries, they would have to consider the respective boundary term to be a variable of the arbitrary function, meaning $f(T,B)$ and $f(\st{Q},B_{\st{Q}})$ theories.

As already discussed in a previous section, modified teleparallel theories employ as their field variables both the tetrad and the spin connection, meaning that both of them should be described by respective field equations. These two variations are closely connected, since the field equations of the spin connection are the antisymmetric part of the field equations of the tetrad.

Specifically, varying the action \eqref{f(T)action} \gls{wrt} the tetrad $e^A{}_{\mu}$ we get its field equations that read
\begin{equation}\label{f(T)field_equations_tetrad}
-\frac{1}{e}f_T \partial _{\nu}\left(e S_A{}^{\mu\nu}\right) - S_A{}^{\mu\nu}\partial _{\nu} f_T +
f_T T^B{}_{\nu A}S_B{}^{\nu\mu} - f_T \omega ^B{}_{A\nu}
S_B{}^{\nu\mu} - \frac{1}{2}f E_A {}^{\mu} = \kappa ^2 \Theta _A
{}^{\mu}\,,
\end{equation}
where $f_T$ is the first derivative of $f$ \gls{wrt} $T$, $S_A{}^{\mu\nu}$ is the superpotential, $\omega ^B{}_{A\nu}$ the spin connection and $\Theta _A{}^{\mu}$ is the energy-momentum tensor of the matter fields; details regarding the variations can be found in the Appendix~\ref{App:variations}. Moreover, varying the action \eqref{f(T)action} \gls{wrt} the spin connection \cite{Golovnev:2017dox,Krssak:2017nlv,Hohmann:2017duq} we get
\begin{equation}\label{f(T)field_equations_spin connection}
    \delta _\omega \mathcal{S}_{f(T)} =S_{[AB]}{}^{\nu}\partial_{\nu} f_T = 0\,.
\end{equation}
It is worth noticing that the \gls{rhs} vanishes, because the energy-momentum tensor, $\Theta_A{}^{\mu}$, has no antisymmetric part. This occurs since we have assumed that matter is not coupled to the spin connection. As has already been discussed in a general framework in Sec.~\ref{sssec:gengravactlorinv}, Eq.~\eqref{f(T)field_equations_tetrad} and \eqref{f(T)field_equations_spin connection} are related, in the sense that the second is the antisymmetric part of the first. Let us see why: it is helpful to write down the \gls{lhs} of Eq.~\eqref{f(T)field_equations_tetrad} as
\begin{subequations}
\begin{align}
    W_{AB} &:= -\eta_{BC}e^C{}_{\mu} \left(\frac{1}{e}f_T \partial _{\nu}\left(e S_A{}^{\mu\nu}\right) + S_A{}^{\mu\nu}\partial _{\nu} f_T -
f_T T^D{}_{\nu A}S_D{}^{\nu\mu} + f_T \omega ^D{}_{A\nu}
S_D{}^{\nu\mu} + \frac{1}{2}f E_A {}^{\mu}\right) \\[0.5ex] \label{eq:f(T)contracted}
&=-S_{AB}{}^{\nu}\partial_\nu f_T - f_T\lc{G}_{AB}-\frac{1}{2}\eta _{AB}(f-Tf_T)\,,
\end{align}
\end{subequations}
where $\lc{G}_{AB}$ is the symmetric Einstein tensor of the Levi-Civita connection expressed with tetrads. Note that we used Eq.~\eqref{eq:einsteintensor} to obtain an Einstein term $\lc{G}_{AB}$. Since the last term is symmetric as well, the only antisymmetric part of the tetrad field equations is the first term,
\begin{equation}
    W_{[AB]} =- S_{[AB]}{}^{\nu}\partial _{\nu}f_T\,,
\end{equation}
which coincides with Eq.~\eqref{f(T)field_equations_spin connection}.
This is imposed by the local Lorentz invariance and indicates the fact that the spin connection is nothing more but a pure gauge \gls{dof}, and not a St\"{u}ckelberg field.

Apart from the local Lorentz invariance of $f(T)$ theories, there have been other controversial issues as well. These theories apparently suffer from strong coupling problems, since in some backgrounds, the kinetic term of some physical \gls{dof} disappears at quadratic order, and thus perturbation theory, at linear order, breaks down at these backgrounds.

Specifically, in Ref.~\cite{Jimenez:2020ofm} the authors perturb a trivial tetrad and they find explicitly a new mode at fourth order in perturbations, that goes beyond the Minkowski background. Apart from that, cosmological perturbations up to second order \cite{Izumi:2012qj,Golovnev:2018wbh} show that no extra modes appear around a flat Friedmann background, signalling again a strong coupling problem. In Ref.~\cite{Bahamonde:2022ohm} though, the authors performed perturbations in non-flat cosmological backgrounds; it turns out that the tensor and vector modes are well determined just like in the flat case. Unlike the flat case though, the scalar perturbations are fully determined by the perturbed equations at linear order. Unfortunately, this does not mean that the strong coupling is not present in non-flat backgrounds, since some of these scalar modes are still non-propagating. It could be seen as progress though, towards limiting the strongly coupled behaviour in such backgrounds. What's more, it could mean that the theory is not ill-behaved, but the way we approach the perturbation analysis is not the ideal one, thus leading to strong coupling problems.

Up to now, there have been three different Hamiltonian analyses of $f(T)$ gravity in the literature. There is indeed a consensus that there are extra \gls{dof}, however, the number of them is still under investigation. A recent review on these analyses is Ref.~\cite{Blixt:2020ekl}, while in Ref.~\cite{Golovnev:2020zpv} the authors review the open problems in $f(T)$ gravity.

In the first paper that appeared almost immediately after the introduction of $f(T)$ theory, \cite{Li:2011rn}, the authors claim to find three new dynamical modes, meaning that in total there are five \gls{dof} in $3+1$ dimensions. Nevertheless, it is claimed \cite{Golovnev:2020zpv, Blagojevic:2020dyq} that, not only their arguments lack justification, but also part of the results are incorrect. In particular, in many cases it seems that all Lorentz constraints become second class, which would mean breaking of the Lorentz symmetry. However, there are known backgrounds with no new \gls{dof} in linear perturbations, thus rendering their results controversial.

Some years later, a new perspective was given to the field \cite{Ferraro:2018tpu,Ferraro:2018axk,Guzman:2019ozl,Ferraro:2020tqk}, by claiming that there exist only three \gls{dof} in a 4 dimensional spacetime, and thus making $f(T)$ theories comparable to $f(\lc{R})$ in this regard. However, it was shown in Ref.~\cite{Blagojevic:2020dyq}, that the Poisson brackets of the Lorentz constraints are incomplete, and as a consequence their main conclusion is wrong.

Even though, the last study that came our recently, \cite{Blagojevic:2020dyq}, points out several ``mistakes'' of the previous approaches, it does not seem to lack problems itself. The authors claim, among other things, that the total number of \gls{dof} in 4 dimensions is five. Yet, the matrix of Poisson brackets of Lorentz constraints does not seem to be non-degenerate, since it contains a non-trivial term \cite{Golovnev:2020zpv}. They claim that this non-trivial term proves that the constraints are second class, however, since the matrix could be degenerate, they could be a combination of first and second class.

Last but not least, apart from pure $f(T)$ theories, there have been similar attempts to generalize \gls{tegr} as well, that could be considered as special cases. One of them is the so-called unimodular gravity, where one considers the determinant of the tetrad to be constant. The Hamiltonian formulation of unimodular \gls{tegr} was studied in Ref.~\cite{daRochaNeto:2011ir}, reconstruction of unimodular $f(T)$ models was studied in Ref.~\cite{Nassur:2016yhc}, inflationary cosmology was studied in Ref.~\cite{Bamba:2016wjm} and holographic dark energy models of $f(T)$ were studied in Ref.~\cite{Godonou:2017ugt}.

\subsubsection{Conformal and disformal transformations in \texorpdfstring{$f(T)$}{fT} gravity} \label{sec:con_dis_trans}
As discussed in Sec.~\ref{Sec:conformal_trans}, if one wants to construct a quantity, say a metric $\tilde{g}_{\mu\nu}$ from a given metric $g_{\mu\nu}$ and a scalar field $\phi$, they come across with the notion of conformal and disformal transformations. The conformal transformation of the metric is given by
\begin{equation}\label{eq:conformal_metric}
    \tilde{g}_{\mu\nu} = \Omega^2(\phi)g_{\mu\nu}\,,\quad \tilde{g}^{\mu\nu} = \Omega ^{-2}(\phi)g^{\mu\nu}\,,
\end{equation}
where $\Omega^2 = \mathfrak{A}$ from Sec.~\ref{Sec:conformal_trans}. Under the same conformal transformation, the tetrad and its inverse must transform as \cite{Yang:2010ji,Bamba:2013jqa,Obukhov:1982zn}
\begin{equation}\label{eq:conformal_tetrad}
    \tilde{e}^A{}_{\mu} = \Omega(\phi)e^A{}_{\mu}\,,\quad \tilde{E}_A{}^{\mu} = \Omega ^{-1}(\phi)E_A{}^{\mu}\,.
\end{equation}
Respectively, the volume element $e$ transforms as
\begin{equation}
    \tilde{e} = \Omega ^4 (\phi)e\,.
\end{equation}

Having the above relations in mind, we can easily calculate how the rest of the teleparallel quantities transform under \eqref{eq:conformal_tetrad}. Hence, the torsion tensor transforms as
\begin{equation}
    \tilde{T}^{A}{}_{\mu\nu} = \Omega T^{A}{}_{\mu\nu} + 2\partial _{[\mu}\Omega \, e^A{}_{\nu]}\,,
\end{equation}
and the contortion tensor as
\begin{equation}
    \tilde{K}_{AB\mu} = K_{AB\mu} + 2E_{[A}{}^{\alpha}E_{B]}{}^{\beta}g_{\beta\mu}\frac{\Omega_{,\phi}}{\Omega}\lc{\nabla} _{\alpha}\phi \,.
\end{equation}
The torsion scalar becomes\begin{equation}
    \tilde{T} = \Omega^{-2}T + 4 \Omega^{-3}g^{\mu\nu}T_{\mu}(\partial_{\nu}\Omega) - 6 \Omega ^{-4}g^{\mu\nu} (\partial_{\mu}\Omega)( \partial _{\nu}\Omega) \,,
\end{equation}
from which we can derive the inverse transformation that reads
\begin{equation}
    T = \Omega ^2 \tilde{T} - 4\Omega \tilde{g}^{\mu\nu}\tilde{T}_\nu(\partial _{\mu}\Omega) - 6 \tilde{g}^{\mu\nu}(\partial _{\mu} \Omega)( \partial_{\nu} \Omega)\,.\label{eq:tor_scal_conf_trans}
\end{equation}
Furthermore, the boundary term transforms as
\begin{equation}
    B = \Omega ^2 \tilde{B} - 4 \Omega \tilde{T}^{\mu} \partial_{\mu}\Omega - 18 \tilde{\partial}_{\mu}\Omega + 6 \Omega \tilde{\lc{\Box}} \Omega\,.
\end{equation}

After introducing how the teleparallel quantities behave under conformal transformations, let us proceed by studying what happens with $f(T)$ gravity. If one introduces two auxiliary fields $\chi$ and $\phi$ we can rewrite the gravitational part of the action \eqref{f(T)action} as
\begin{equation}\label{f(T)_action_scalar-tensor_1}
    \mathcal{S}_{f(T)} =\frac{1}{2\kappa ^2}\int \dd ^4x \,e \,f(T) = -\frac{1}{2\kappa ^2}\int \dd ^4x \,e \left[ \chi (T-\phi)- f(\phi)\right]\,.
\end{equation}
Varying this action \gls{wrt} $\chi$ yields $\phi = T$ which shows that the above action is equivalent to \eqref{f(T)action}, unless $f''(T)\equiv f_{TT} =0$; in this case, the theory reduces to \gls{tegr}. Varying the action \gls{wrt} the other scalar $\phi$ it yields $\chi = -f'(\phi)$. Setting $F(\phi) = - f'(\phi)$, we can rewrite the above action as
\begin{equation}\label{f(T)_action_scalar-tensor}
    \mathcal{S}_{f(T)} = \frac{1}{2\kappa ^2} \int \dd ^4x\, e \left[ -F(\phi)T - V(\phi)\right]\,,
\end{equation}
with $V(\phi) = \phi f'(\phi)- f(\phi)$. This formulation produces a scalar-tensor type of theory without a kinetic term for the scalar field, as is the case for $f(\lc{R})$ gravity.

Applying the conformal transformation \eqref{eq:conformal_tetrad} in Eq.~\eqref{f(T)_action_scalar-tensor} the action transforms into
\begin{equation}
    \mathcal{S}_{f(T)} = \frac{1}{2\kappa^2} \int\, \dd^4 x \tilde{e} \left[- F(\phi)\left(\Omega^{-2}\tilde{T} - 4 \Omega ^{-3}\tilde{g}^{\mu\nu}\tilde{T}^{\nu}(\partial_{\mu}\Omega) - 6 \Omega^{-4}\tilde{g}^{\mu\nu}(\partial_{\mu}\Omega)( \partial_{\nu}\Omega) \right) - \Omega^{-4}V(\phi)\right]\,.
\end{equation}
Setting $\Omega^2 = F(\phi)$ the above action becomes
\begin{equation}\label{eq: conformal_f(T)}
    \mathcal{S}_{f(T)} = \frac{1}{2\kappa ^2}\int \dd ^4x\,\tilde{e}\left[-\tilde{T} +\frac{2}{F(\phi)}\tilde{g}^{\mu\nu}\tilde{T}_{\nu} \partial _{\mu}F(\phi)+ \frac{3F'(\phi)^2}{2F(\phi)^2}\tilde{g}^{\mu\nu}(\partial _{\mu}\phi)( \partial_{\nu}\phi) - \frac{V(\phi)}{F(\phi)^2} \right]\,.
\end{equation}
If we introduce a new scalar field $\psi$ so that
\begin{equation}\label{eq: scalar_redefintion}
    \frac{d\psi}{d\phi} = \sqrt{3}\frac{F'(\phi)}{F(\phi)} \Rightarrow \psi = \sqrt{3} \ln F(\phi)\,,
\end{equation}
the action \eqref{eq: conformal_f(T)} becomes
\begin{equation}\label{eq:conforal1}
    \mathcal{S}_{f(T)} = \frac{1}{2\kappa ^2} \int \dd ^4x\, \tilde{e} \left[-\tilde{T} + \frac{2}{F}\tilde{T}^{\mu}\partial_{\mu}F + \frac{1}{2}\tilde{g}^{\mu\nu}\tilde{\lc{\nabla}}_{\mu}\psi \tilde{\lc{\nabla}}_{\nu}\psi -U(\psi)\right]\,,
\end{equation}
where $U(\psi) = V(\phi)/F(\phi)^2$. As can be seen from Eq.~\eqref{eq:conforal1}, conformal transformations cannot lead to the Einstein frame in $f(T)$ gravity because of the fact that the torsion tensor is not conformally invariant. Further manipulation of the second term in Eq.~\eqref{eq: conformal_f(T)} yields
\begin{equation}
    \frac{1}{F}\partial_{\mu}F = \partial _{\mu}(\ln F)\,,
\end{equation}
and by substituting it back and integrating by parts we get
\begin{equation}
    \mathcal{S}_{f(T)} = \frac{1}{2\kappa ^2} \int \dd ^4x\, \tilde{e} \left[-\tilde{T} -\frac{\psi}{\sqrt{3}}\tilde{B} + \frac{1}{2}\tilde{g}^{\mu\nu}\tilde{\lc{\nabla}}_{\mu}\psi \tilde{\lc{\nabla}}_{\nu}\psi -U(\psi)\right]\,,
\end{equation}
because of Eq.~\eqref{eq: scalar_redefintion}. We thus see that a conformal transformation of $f(T)$ gravity leads to \gls{tegr} with a non-minimal coupling of a phantom scalar field, meaning that its kinetic term has the opposite sign, to the boundary term $\tilde{B}$. More details on conformal transformations in $f(T)$ gravity can be found in Refs.~\cite{Wright:2016ayu, Yang:2010ji} and references therein.

Before we move to other theories, it is interesting to see what happens if one allows the new tetrad in Eq.~\eqref{eq:conformal_tetrad} to depend on first derivatives of the scalar field as well. Can we get rid of the above symmetry-breaking term? In this case, the transformations are called disformal and have the form
\begin{equation}\label{eq:conf_trans_tetrad}
    \tilde{e}^A{}_{\mu} = \mathfrak{C}(\phi,X)e^A{}_{\mu} + \mathfrak{D}(\phi,X)g^{\nu\rho}e^A{}_{\rho}(\partial_{\mu}\phi)( \partial_{\nu}\phi) \,.
\end{equation}
The inverse tetrad transforms as
\begin{equation}
\tilde{E}_A{}^{\mu} = \frac{1}{\mathfrak{C}}\left( E_A{}^{\mu} - \frac{\mathfrak{D}}{\mathfrak{C}-2X\mathfrak{D}}g^{\mu\nu}E_A{}^{\rho}(\partial_{\nu}\phi)( \partial_{\rho}\phi)\right)\,,
\end{equation}
and the volume element as
\begin{equation}\label{eq:disformal_determinant}
    \tilde{e} = \mathfrak{C}^3\left(\mathfrak{C}-2X\mathfrak{D}\right)e\,.
\end{equation}
One can thus immediately obtain the transformation of the metric, that reads
\begin{equation}
    \tilde{g}_{\mu\nu} = \mathfrak{C}^2g_{\mu\nu} + 2\mathfrak{D}(\mathfrak{C}-X\mathfrak{D})(\partial_{\mu}\phi)(\partial_{\nu}\phi)\,.
\end{equation}
The disformal transformations of the rest of the teleparallel quantities are given in Sec.~\ref{Sec:conformal_trans}.

Considering now the scalar-torsion representation of $f(T)$ gravity, given in Eq.~\eqref{f(T)_action_scalar-tensor_1} and setting (for simplicity) $\phi = f'(\phi)$ the action is rewritten as
\begin{equation}
    \mathcal{S}_{f(T)} = -\frac{1}{2\kappa^2}\int \dd ^4 x e \left(\phi T - V(\phi)\right)\,,
\end{equation}
where $V(\phi) = f(T)-\phi T$. Using the first of Eq.~\eqref{eq:RTQ} to replace $T$ in the above action we can see, as previously, that the term $\phi B$, integrated by parts, will become $-2\int \dd ^4 x e T_{\mu}\partial ^{\mu}\phi$. This term takes the following form under disformal transformations
\begin{equation}\label{Tmu_conformal}
    \tilde{T}_{\mu}\tilde{\partial} ^{\mu}\phi = \frac{1}{\left(\mathfrak{C}-2\mathfrak{D}\right)^2}\left[\left(1-2X\frac{\mathfrak{D}}{\mathfrak{C}}\right)T_{\mu}\partial^{\mu}\phi + 3 \frac{\partial _{\mu} \mathfrak{C}}{\mathfrak{C}}\partial^{\mu}\phi + \frac{\mathfrak{D}}{\mathfrak{C}}\left(\partial^{\alpha}\phi\partial^{\beta}\phi\lc{\nabla} _{\alpha} \partial _{\beta}\phi +2 X \lc{\Box}\phi\right)\right]\,.
\end{equation}
Replacing this together with Eq.~\eqref{eq:disformal_determinant} into the action we readily see that it is not possible to remove the symmetry-breaking terms with disformal transformations. More details on disformal transformations in $f(T)$ gravity can be found in Refs.~\cite{Golovnev:2019kcf, Hohmann:2019gmt} and references therein.

\subsection{\texorpdfstring{$f(T,B)$}{fTB} gravity}\label{sec:f_T_B_gravity}

Another very interesting and well-studied generalization of \gls{tegr} is the so-called $f(T,B)$ gravity, where apart from the torsion scalar one considers the boundary term $B$, Eq.~\eqref{eq:boundaryB}, that is the difference between the Ricci scalar, computed from the Levi-Civita connection and the torsion scalar, Eq.~\eqref{eq:RTQ}. The action of this theory reads
\begin{equation}\label{eq:f(T,B)action}
\mathcal{S}_{f(T,B)}:= \frac{1}{2\kappa ^2}\int \dd^4x \, e \,f(T,B) +
 \mathcal{S}_{\text{m}}\,.
\end{equation}

This theory was firstly proposed in Ref.~\cite{Bahamonde:2015zma} by S.~Bahamonde, C.~B\"{o}hmer and M.~Wright, and the motivation was, mostly, to fully incorporate the dynamics of $f(\lc{R})$ theory. However, as one could immediately realize, $f(\lc{R})$ is just a limiting case of Eq.~\eqref{eq:f(T,B)action} (only when the specific combination $-T+B$ appears as an argument to the function), and thus $f(T,B)$ has offers much richer phenomenology.

As in any teleparallel theory, since the torsion scalar depends both on the frame field and the spin connection, $f(T,B)$ theory will result non-trivial equations for both of them. Thus, the antisymmetric part of the tetrad field equations, will coincide with the field equations for the inertial connection. This means that, if one chooses to work on the Weitzenb\"{o}ck gauge, where the spin connection vanishes, the respective field equations will be solely solved by the tetrad.

Let us go a step further, in order to examine quantitatively the above picture. Varying the action \eqref{eq:f(T,B)action} \gls{wrt} the tetrad we obtain its field equations that read
\begin{alignat}{2}\label{fieldequationsfTB}
& \: & & 	E_A{}^{\mu} \lc{\square} f_B - E_A {}^{\nu} \lc{\nabla} ^{\mu} \lc{\nabla}_{\nu} f_B +
	\frac{1}{2} B f_B E_A{}^{\mu} - \left(\partial _{\nu}f_B + \partial
	_{\nu}f_{T} \right)S_A{}^{\mu\nu} \nonumber \\[0.5ex]
	& \: & & -\frac{1}{e} f_{T}\partial _{\nu} (eS_A{}^{\mu\nu})
	+ f_{T}
	T^{B}{}_{\nu A}S_{B}{}^{\nu\mu}- f_T \omega ^B{}_{A\nu}
	S_B{}^{\nu\mu} -\frac{1}{2} f E_{A}{}^{\mu} = \kappa ^2 \Theta _A{}^{\mu} \,,
\end{alignat}
where $f_B$ and $f_T$ are partial derivatives of $f$ \gls{wrt} the boundary term and the torsion scalar. Details on the variations can be found in the Appendix~\ref{App:variations}.
As one would expect, by choosing the form of the theory to be $f(T,B) = f(-T+B)$, the above equations coincide with the equations of motion in $f(\lc{R})$ gravity, expressed in the tetrad formulation. As already shown in full generality in Sec.~\ref{sssec:gengravactlorinv} and also discussed in the previous section of $f(T)$ gravity, the antisymmetric part of the tetrad's field equations will coincide with the equations of motion of the spin connection.

We will focus on specific applications of this theory, as well as on constraints of its functional form in the following sections. In particular, in Sec.~\ref{Sec:fTB cosmology} we will study its cosmological solutions, in Sec.~\ref{f_TB_matter_dens} we will see the matter density perturbation equation, its polarization modes will be presented in Sec.~\ref{Sec:fTB_pol-modes} and precision cosmology will be discussed in Sec.~\ref{sec:f(TB)_test}. Conformal transformations of $f(T,B)$ gravity are studied in Ref.~\cite{Wright:2016ayu}; the authors show that generalized non-minimally coupled ``teleparallel dark energy models'' (see Sec.~\ref{Sec5:scalar-torsion_models}) are conformally equivalent to $f(T, B)$ gravity.

One particular interesting case is when $f(T,B)=-T+F(B)$ since this theory is conformally equivalent to~\eqref{eq:TDEconformal0} with the coupling function $A(\phi)=\beta^2(1+\frac{\phi}{2\beta\sqrt{3}})^2$. This theory only has a non-minimally coupling between the torsion scalar and the scalar field and the kinetic term has the correct sign (canonical); this will be studied in detail in Sec.~\ref{Sec:conf-trans-fTB}. The existence and the stability of specific relativistic solutions, such as de Sitter and scaling solutions, are studied in Ref.~\cite{Paliathanasis:2017flf}; the authors claim that if one uses a Lagrange multiplier, then $f(T,B)$ gravity becomes equivalent to \gls{gr} plus a minimally coupled non-canonical scalar field. In Ref.~\cite{Sahlu:2020mij} the authors consider two specific models of the theory, one exponential and one power-law, and they find solving the field equations numerically, that both of them are capable of reproducing the late-time acceleration of the Universe. In Ref.~\cite{Bhattacharjee:2020chn,Zubair:2018wyy} the authors constrain the function form of $f$ using energy conditions. Thermodynamics is used in Refs.~\cite{Bahamonde:2016cul,Pourbagher:2019zhq}, in order to constrain and cosmologically reconstruct $f(T,B)$ models. The $H_0$ tension was studied in Ref.~\cite{Escamilla-Rivera:2019ulu}, together with cosmologically viable models; weak field limit tests were studied in Refs.~\cite{Farrugia:2020fcu,Capozziello:2020vil} and stability analysis in Ref.~\cite{Franco:2020lxx}. Further, bouncing cosmological solutions are considered in Ref.~\cite{Caruana:2020szx}, Noether symmetries in \cite{Bahamonde:2016grb}. In \cite{Paliathanasis:2021kuh} the minisuperspace quantization is considered in cosmology. Last but not least, some other approaches of the form $f(T,\lc{R})$ have been considered in \cite{Myrzakulov:2012qp,SALTI:2018vqq}, however, they are considered controversial because of the mixed framework they use, i.e. both curvature and torsion are non-zero. A complete analysis about $f(T,B)$ cosmology will be provided in Secs.~\ref{sec:fTBbackgroundcosmo} and \ref{sec:f(TB)_test}.

\subsection{Extensions of new general relativity} \label{Sec:Ext_NGR}

Following Sec.~\ref{sec:NGR}, \gls{ngr} is described by the Lagrangian density \eqref{eq:ngrlagrangian}. There are actually two more quadratic scalars that one can construct from the torsion tensor~\cite{Bahamonde:2017wwk}
\begin{equation}\label{eq:quadratic_torsion_odd-parity}
    P_1 := u_{\mu}a^{\mu}\,,\quad P_2 := \epsilon _{\mu\nu\rho\sigma}t^{\lambda\mu\nu}t_{\lambda}{}^{\rho\sigma}\,,
\end{equation}
however, they are both parity violating. The interested reader should read Ref.~\cite{MuellerHoissen:1983vc} where the authors argue that parity violating terms play a significant role in the wellposedness of the Cauchy problem. It is not so clear thought that such terms can play a fully consistent role in gravitational theories.

Theories beyond \gls{ngr} have been proposed as well. In Ref.~\cite{Bahamonde:2017wwk} S.~Bahamonde, C.~B\"{o}hmer and M.~Kr\v{s}\v{s}\'{a}k consider the theory which assumes the Lagrangian density
\begin{equation}\label{lagr_f(Tax,Tvec,Tten)}
    \mathcal{L} = f(T_{\rm axi},T_{\rm ten},T_{\rm vec})\,,
\end{equation}
i.e. an arbitrary function of the irreducible parts of the torsion tensor. One can also use the alternative decomposition of the torsion scalar, described by the equation \eqref{eq:torscalar_decomposition}, and thus consider theories of the form $f(T_1,T_2,T_3)$. The equations of motion for the above theory \eqref{lagr_f(Tax,Tvec,Tten)} in vacuum are given by
\begin{equation}
    E_A{}^{\mu} f+ \frac{\delta f}{\delta e^A{}_{\mu}} = 0\,,
\end{equation}
where\begin{equation} \frac{\delta f}{\delta e^A{}_{\mu}} = \frac{\delta f}{\delta e^A{}_{\mu}}|_{\rm axi} + \frac{\delta f}{\delta e^A{}_{\mu}}|_{\rm ten} + \frac{\delta f}{\delta e^A{}_{\mu}}|_{\rm vec}\,,\end{equation} and each of the terms on the \gls{rhs} are
\begin{subequations}
\begin{alignat}{2}
    \frac{\delta f}{\delta e^A{}_{\mu}}|_{\rm axi} & =\: & & -\frac{2}{3}\left[ \epsilon _{BC}{}^{DJ}f_{T_{\rm axi}}a^B\left(E_D{}^{\mu}T^C{}_{AJ}-E_J{}^{\mu}\omega^C{}_{AD}\right)+e^{-1}\partial _{\nu}\left(e\,\epsilon _{BA}{}^{CD}f_{T_{\rm axi}}a^B E_C{}^{\nu}E_D{}^{\mu}\right)\right]\,, \\[0.5ex]
    \frac{\delta f}{\delta e^A{}_{\mu}}|_{\rm vec} & =\: & & 2f_{T_{\rm vec}} \left(v^{\mu}\omega ^{\rho}{}_{A\rho} - v^B \omega ^{\mu}{}_{AB} - T^{\mu}{}_{AB}v^{B} - v^{\mu}v_A\right) - 2 e^{-1}\partial _{\nu}\left[e\,f_{T_{\rm vec}} \left(v^{\mu}E_A{}^{\nu} - v^{\nu}E_A{}^{\mu}\right)\right]\,, \\[0.5ex]
    \frac{\delta f}{\delta e^A{}_{\mu}}|_{\rm ten} & =\: & & f_{T_{\rm ten}}\Big[-2 T^B{}_{A\sigma}T_B{}^{\mu\sigma} - T^{\mu}{}_{\rho\sigma}T^{\rho}{}_A{}^{\sigma} - T^B{}_{\rho A}T^{\rho}{}_{B}{}^{\mu} +T^{\mu}{}_{AB}v^B +v^{\mu}v_A -v^{\mu}\omega^\rho{}_{A\rho} + \nonumber\\[0.5ex]
    & \: & &+v^B\omega^{\mu}{}_{AB}+2T_B{}^{\rho\mu} + T^{\rho}{}_{B}{}^{\mu}- T^{\mu}{}_B{}^{\rho}\Big] -\nonumber \\[0.5ex]
    & \: & &-e^{-1}\partial _{\nu}\left[e\,f_{T_{\rm ten}} \left(-2T_A{}^{\mu\nu}+T^{\mu\nu}{}_A -T^{\nu\mu}{}_A - v^{\mu}E_A{}^{\nu} + v^{\nu}E_A{}^{\mu}\right) \right]\,.
\end{alignat}
\end{subequations}
The theory expressed in the Lagrangian in Eq.~\eqref{lagr_f(Tax,Tvec,Tten)} is manifestly covariant and invariant under local Lorentz transformations, since the spin connection is nonvanishing. Alternatively, one could work on those frames where the spin connection vanishes (Weitzenb\"{o}ck gauge) but in that case they should employ only with good tetrads \cite{Tamanini:2012hg,Ferraro:2011us,Bengochea:2008gz,Ferraro:2008ey}; more details can be found in Sec.~\ref{sec:f(T)gravity}. In the same spirit, inclusion of parity violating terms and the boundary term was proposed \cite{Bahamonde:2017wwk}, in theories of the form
\begin{equation}
    \mathcal{L} = f(T_{\rm axi},T_{\rm ten},T_{\rm vec},B,P_1,P_2)\,,\label{eq:LagrangianGen}
\end{equation}
as well as higher order contractions of the torsion tensor such as
\begin{equation}
    S_1 := t^{\lambda\mu\nu}v_{\lambda}a_{\mu}v_{\nu}\,,\quad S_2 := t^{\lambda\mu\nu}a_{\lambda}v_{\mu}a_{\nu}\,.
\end{equation}
In principle, because of the nature of the torsion tensor and the fact that it depends only on first derivatives of the tetrad (in contrast with the Ricci scalar), one can construct infinite-order contractions of the torsion tensor and still obtain second order field equations. However, as we will discuss in Sec.~\ref{sec:BDLS} we do not know how physical such theories would be.

In Ref.~\cite{Koivisto:2018loq} the authors considered a generalization of \gls{ngr} by adding nine functions of the d'Alembertian operator, showing that it can accommodate the ghost- and singularity-free structure that was realized in the metric theories \cite{Conroy:2017yln,Heisenberg:2018vsk,Biswas:2011ar}. Furthermore, in Ref.~\cite{Geng:2016yke} the authors study the extensions of \gls{ngr} as a reduction of the 5 dimensional Kaluza-Klein theory and in Ref.~\cite{Formiga:2014rsa} the equivalence between \gls{ngr} and a Weyl geometry is presented.

\subsection{Gauss-Bonnet theories}
\label{Sec:GB_theories}

The Gauss-Bonnet invariant, depending only in higher order curvature terms
\begin{equation}
    \lc{\mathcal{G}} = \lc{R}^2 - 4 \lc{R}_{\mu\nu}\lc{R}^{\mu\nu} +
\lc{R}^{\alpha\beta\mu\nu}\lc{R}_{\alpha\beta\mu\nu}\,,
\end{equation}
could be considered as a subclass of higher order theories. In the teleparallel framework its form becomes
\begin{equation}
    \lc{\mathcal{G}} = T_{G} + B_{G}\,,
\end{equation}
where $T_{G}$ is the \gls{tegb} term and
$B_{\mathcal{G}}$ its boundary term, as shown in Sec.~\ref{Sec3:TeleGB}.

Theories that have been studied in the literature have the form
\begin{equation}\label{fTGaction}
    \mathcal{S}_{f(T,T_{G})} := \frac{1}{2 \kappa ^2} \int
\dd^4x \,e \,f(T,T_G)+ \mathcal{S}_{\rm m}\,.
\end{equation}
The equations of motion in the Weitzenb\"{o}ck gauge are presented in Ref.~\cite{Kofinas:2014owa} by G.~Kofinas and E.~N.~Saridakis. In Ref.~\cite{Capozziello:2016eaz} cosmological solutions using Noether symmetries were found. More general theories such as $f(T,B,T_{G},B_G)$ were studied in Ref.~\cite{Bahamonde:2016kba}. In curvature-based gravity, the Gauss-Bonnet term is intrinsically related to the Lovelock theorem which defines the conditions by which an action for gravity can be written such that it has field equation derivatives that are at most second order in the metric. Naturally, the scalar is also related to the teleparallel analogue of this theorem in \gls{tg}. The Lovelock analogue in \gls{tg} as well as cosmological solutions are presented and studied in Refs.~\cite{Gonzalez:2015sha,Gonzalez:2019tky}; while couplings with a scalar field and its kinetic term are considered in Ref.~\cite{Bahamonde:2020vfj}. Varying the action \eqref{fTGaction} we get the following field equations
\begin{alignat}{2}
& \: & &-\frac{1}{e}f_{T}\partial_{\nu}(e S_{A}\,^{\mu\nu})- S_{A}\,^{\mu\nu} (\partial_{\nu}f_{T})+f_{T}T^{B}\,_{\nu A}S_{B}\,^{\nu\mu} -\frac{1}{2}fE_A{}^\mu-f_{T_G}\delta^{MBCD}_{IJKL}E_{D}{}^{\mu}K^{IJ} {} _M K^{K} {} _ {EB}\partial_A K^{EL} {} _C  \nonumber\\[0.5ex]
& \: & &+\frac{1}{2e}\partial_{\beta}\Big(\eta_{AL}(Y^{B[LH]}-Y^{H[LB]}+Y^{L[BH]})E_{H}{}^{\beta}E_{B}{}^{\mu}\Big)+\frac{1}{2e}T_{IAB}E_{H}{}^{\mu}(Y^{B[IH]}-Y^{H[IB]}+Y^{I[BH]})\nonumber\\[0.5ex]
& \: & & \,\,\,\,\,\,\,\,\,\,\,\,\,\,\,\,\,\,\,\,\,\,\,\,\,\,\,\,\,\,\,\,\,\,\,\,\,\,\,\,\,\,\,\,\,\,\,\,\,\,\,\,\,\,\,\,\,\,\,\,\,\,\,\,\,\,\,\,\,\,\,\,\,\,\,\,\,\,\,\,\,\,\,\,\,\,\,\,\,\,\,\,\,\,\,\,\,\,\,\,\,\,\,\,\,\,\,\,\,\,\,\,\,\,\,\,\,\,\,\,\,\,\,\,\,\,\,\,\,\,\,\,\,\,\,\,\,\,\,\,\,\,\,\,\,\,\,\,\,\,\,\,\,\,\,\,\,\,\,\,\,\,\,\,\,\,\,\,\,\,\,\,\,\,\,\,\,\,\,\,\,\,\,\,\,\,\,\,\,\,\,\,\,\,\,\,\,\,\,\,\,\,\,\,\,\,\,\,\,\,\,\,\,\,\,\,\,\,\,\,\,\,\,\,=\kappa^2\Theta_{A}{}^\mu\,,\nonumber\\
\label{eq:fTTG}
\end{alignat}
where we introduced the following tensors
\begin{align}
    Y^{B}{}_{IJ}& := ef_{T_{G}}X^{B}{}_{IJ}-2\delta^{CABD}_{ELKJ}\partial_{\mu}\big(ef_{TG}E_{D}{}^{\mu}K^L{}_{C}{}^{E}K_{IA}{}^{K}\big)\,, \label{eq:defXY}
\end{align}
and
\begin{alignat}{2}
X^{A}{}_{IJ}& :=\: & &K_J {}^E {} _B K^K {} _ {FC} K^{FL} {} _D\delta^{ABCD} _ {IEKL} +  
 K^E {} _ {IB} K^K {} _ {FC} K^{FL} {} _D \delta^{BACD} _ {EJKL} + 
 K^K {} _ {EC} K^{EF} {} _B K_J {}^L {} _D \delta^{CBAD} _ {KFIL} \nonumber\\[0.5ex]
& \: & &+ 
 K^F {} _ {ED} K^{EL} {} _B K^K {} _ {IC}\delta^{DBCA} _ {FLKJ}+2 K^K {} _ {EB} K^{EL} {} _F K^F {} _ {CD}\delta^{ABCD} _ {IJKL} + 
 2 K^{KE} {} _B K_J {}^L {} _F K^F {} _ {CD}\delta^{BACD} _ {KEIL} \nonumber\\[0.5ex]
& \: & &+ 
 2 K^{EL} {} _F K^K {} _ {IB} K^A {} _ {CD}\delta^{FBCD} _ {ELKJ} + 
 2 K^{FC} {} _D K^
   D {} _ {EB} K^{EL} {} _I \eta_ {MJ} \delta^{DBMA} _ {FCKL} +2 K^K {} _ {EB}\delta^{ABCD} _ {IJKL}\partial_D K^{EL} {} _C \nonumber\\[0.5ex]
& \: & &
+ 
 2 K^{KE} {} _B \delta^{BACD} _ {KEIL}\partial_D K_J{}^L {} _C\,.\label{X}
\end{alignat}
Details about the variations can be found in the Appendix~\ref{App:variations}. Applications of this theory will be discussed in the following sections.

\subsection{Higher-order derivatives}\label{sec:HOT}

Inspired mostly by the curvature case where theories with higher order derivative terms were studied, such as $f(\lc{R},\lc{R}_{\mu\nu}\lc{R}^{\mu\nu},\lc{R}^{\alpha\beta\mu\nu}\lc{R}_{\alpha\beta\mu\nu},\lc{\Box} \lc{R}, (\lc{\nabla} \lc{R})^2,...)$ and more, together with the fact that such terms arise naturally in dimensional reductions of higher-dimensional theories, such as Kaluza-Klein, and in general in quantum corrections \cite{Addazi:2021xuf}, the higher-order teleparallel (HOT) theory described by the action
\begin{equation}
    \mathcal{S}_{\rm HOT} := \frac{1}{2\kappa ^2} \int \dd^4x \, e\, f(T,(\lc{\nabla} T)^2, \lc{\Box} T)+\mathcal{S}_{\rm m}\,,
\end{equation}
where
\begin{equation}
    (\lc{\nabla} T)^2 \equiv g^{\mu\nu}(\lc{\nabla} _{\mu}T)(\lc{\nabla}_{\nu}T) = \eta ^{AB}E_A{}^{\mu}E_B{}^{\nu}(\lc{\nabla}_{\mu}T)(\lc{\nabla}_{\nu}T)\,,
\end{equation}
and
\begin{equation}
    \lc{\Box} T \equiv g^{\mu\nu}\lc{\nabla}_{\mu}\lc{\nabla}_{\nu}T= \eta ^{AB}E_A{}^{\mu}E_B{}^{\nu} \lc{\nabla}_{\mu}\lc{\nabla}_{\nu}T\,,\label{eq:higher_order_terms}
\end{equation}
was proposed by G.~Otalora and E.~N.~Saridakis in Ref.~\cite{Otalora:2016dxe}. The field equations in the Weitzenb\"ock gauge for this theory become
\begin{alignat}{2}
   & \: & & -\frac{1}{e}f_T\partial_{\nu}\left( e S_{A}{}^{\mu\nu}\right)-S_{A}{}^{\mu\nu}(\partial_{\nu}f_T)+f_T T^{B}{}_{\nu A}S_{B}{}^{\nu\mu}-\frac{1}{2}E_A{}^{\mu}f\nonumber \\[0.5ex]
	& \: & & -\frac{1}{2}\sum _{i=1}^2 \left\{f_{X_i} \frac{\partial X_i}{\partial e^A{}_{\mu}}- \frac{1}{e}\left[ \partial _{\alpha}\left(ef_{X_i}\frac{\partial X_i}{\partial (\partial _{\alpha}e^A{}_{\mu})}\right) - \partial _{\alpha}\partial _{\nu} \left( ef_{X_i}\frac{\partial X_i}{\partial (\partial _{\alpha}\partial _{\nu}e^A{}_{\mu})}\right)\right]\right\} \nonumber\\[0.5ex]
	& \: & & + \frac{1}{2e}\partial _{\sigma}\partial_{\alpha}\partial _{\nu}\left(ef_{X_2}\frac{\partial X_2}{\partial(\partial_{\sigma}\partial_{\alpha}\partial_{\nu}e^A{}_{\mu})}\right)- \frac{1}{2e}\partial_{\beta}\partial _{\sigma}\partial_{\alpha}\partial _{\nu}\left(ef_{X_2}\frac{\partial X_2}{\partial(\partial_{\beta}\partial_{\sigma}\partial_{\alpha}\partial_{\nu}e^A{}_{\mu})}\right)=\kappa^2 \Theta _A{}^{\mu}\,,
\end{alignat}
where $X_i=((\lc{\nabla} T)^2, \lc{\Box} T)$ and then $f_{X_i}=\partial f/\partial X_i$. In principle, such theories that contain higher order derivatives in the field equations are vulnerable to Ostrogradsky ghosts. However, as stated in Ref.~\cite{Otalora:2016dxe} this is not necessarily the case here since the theory is formulated on different foundations it may be an indication of extra \gls{dof} rather than ghosts, as in $f(\lc{R})$ gravity.

\subsection{Scalar-tensor theories}\label{sec:scalartensor}

Another very well motivated class of theories, capable of describing the accelerating expansion of the Universe at present and early times has been the scalar-tensor theories. The first and probably the simplest scalar-tensor theory introduced, was the Brans-Dicke theory \cite{Brans:1961sx} which describes a linear non-minimal coupling of a scalar field to gravity that introduces a new, scalar \gls{dof} to the theory and the gravitational constant becomes dynamical. In that spirit, gravity is described both by the dynamics of the metric and the scalar field. In principle there are no restrictions in the number of scalar fields introduced.

\subsubsection{Scalar-tensor theories with couplings with $T$ and $B$}
\label{Sec5:scalar-torsion_models}
In the same spirit, instead of coupling the scalar field to the metric and invariants constructed from the Levi-Civita connection, it is interesting to see what happens in the teleparallel framework, meaning if we couple it to the tetrad and invariants constructed from the teleparallel connection. These theories have been known as scalar-torsion theories and in this section we will present interesting and well-studied models, their field equations as well as how they transform under conformal transformations.

\paragraph{Action and field equations}
Before we study specific scalar-torsion models, we will start from a more general action that includes an arbitrary function of the torsion scalar $T$, the boundary term $B$, a scalar field $\phi$ and its kinetic term $X\equiv -(1/2) \, g^{\mu\nu}\partial_{\mu}\phi\partial_{\nu}\phi=-(1/2)(\lc{\nabla}\phi)^2$. Its action reads
\begin{align}
\mathcal{S}_{ f(T,B,\phi,X)} := \int
\dd^4x\, e\, \left[
\frac{1}{2\kappa^2}f(T,B,\phi,X) + \mathcal{L}_{\rm m}
\right] \,.\label{action}
\end{align}
Variations of Eq.~\eqref{action} \gls{wrt} the tetrad yield the field equations
\begin{alignat}{2}
 	& \: & & \delta_{\beta}^{\mu}\lc{\Box} f_{B}-\lc{\nabla}^{\mu}\lc{\nabla}_{\beta}f_{B}+
\frac{1}{2}	B f_{B}\delta_{\beta}^{\mu} -
	\Big[(\partial_{\nu}f_{B})+(\partial_{\nu}f_{T})\Big]S_{\beta}{}^{\mu\nu}
	\nonumber\\[0.5ex]
& \: & & 	-\frac{1}{e}f_{T}e^{A}{}_{\beta}\partial_{\nu}(e S_{A}{}^{\mu\nu}) +
	f_{T}T^{B}{}_{\nu\beta}S_{B}{}^{\nu\mu} - f_T \omega^B{}_{\beta\nu}
	S_B{}^{\nu\mu}-
	\frac{1}{2}f \delta_{\beta}^{\mu}-\frac{1}{2}\, f_{X}\partial^{\mu}\phi \partial_{\beta}\phi =\kappa^2 \Theta_{\beta}^{\mu}\,, \label{fieldeqgeneral2}
\end{alignat}
while the field equations of the spin connection will be the antisymmetric part of Eq.~\eqref{fieldeqgeneral2},
\begin{equation}
    W_{[\mu\nu]}= 2\Big[(\partial_{\rho}f_{B})+(\partial_{\rho}f_{T})\Big]S_{[\mu}{}^{\rho}{}_{\nu]}=3T^{\rho}{}_{[\mu\nu}\partial_{\rho]}(f_T+f_B)\,.\label{antifTB}
\end{equation}
Additionally, the field equations for the scalar field read
\begin{eqnarray}
\partial_{\mu}\Big(ef_{X}g^{\mu\nu}\partial_{\nu}\phi\Big)+e f_{\phi}=0\,.\label{fieldeq2}
\end{eqnarray}
Details about the variations can be found in the Appendix~\ref{App:variations}. An important theory that can be constructed from the above action is the following
\begin{equation}\label{action:scalar-torsion-coupling-T-and-B}
    \mathcal{S}=\int \dd^4x \, e\, \Big[\frac{1}{2\kappa^2}\Big(f(T)+F_1(\phi)T
+F_2(\phi)B\Big)-\frac{1}{2}\omega(\phi)\lc{\nabla}_\mu \phi \lc{\nabla}^\mu \phi+V(\phi)\Big] + \mathcal{S}_{\rm m}\,,
\end{equation}
where there are non-minimal couplings between the torsion scalar $T$ and the boundary term $B$. For this specific theory, the field equation~\eqref{fieldeqgeneral2} can be rewritten as
\begin{alignat}{2}
 & \: & &-(F_1(\phi)+f_T)\lc{G}_\beta^\mu+\delta_{\beta}^{\mu}\lc{\Box} F_{2}(\phi)-\lc{\nabla}^{\mu}\lc{\nabla}_{\beta}F_{2}(\phi) -
		\partial_\nu\Big(f_T+F_1(\phi)+F_2(\phi)\Big)S_{\beta}{}^{\mu\nu}	-
	\frac{1}{2} \delta_{\beta}^{\mu}(f-Tf_T)
\nonumber	\\[0.5ex]
 & \: & &- f_T e^A{}_\beta\omega^B{}_{A\nu}
	S_B{}^{\nu\mu}-\kappa^2\omega(\phi)(\partial^\mu\phi)(\partial_\beta\phi)+\frac{1}{2}\kappa^2\omega(\phi)(\partial_\nu \phi) (\partial^\nu \phi)\delta^\mu_\beta-\kappa^2V(\phi) =\kappa^2 \Theta_{\beta}^{\mu}\,,
	\label{fieldeqgeneralscalar}
	\end{alignat}
whereas the modified Klein-Gordon equation~\eqref{fieldeq2} is reduced to
\begin{eqnarray}
\omega(\phi)\lc{\Box} \phi+\omega'(\phi)(\partial_\alpha \phi)(\partial^\alpha \phi) +\frac{dV}{d\phi}+\frac{1}{2\kappa^2}(F_1'(\phi)T+F_2'(\phi)B)=0\,.\label{fieldeq2KG}
\end{eqnarray}

As already mentioned, the action~\eqref{action} is a very general one and incorporates a lot of interesting scalar-torsion models.
One of them is the ``Teleparallel dark energy'' model \cite{Geng:2011aj}. Motivated mainly by the quintessence model in \gls{gr} \cite{Capozziello:2003gx,Copeland:2003cv,Capozziello:2002rd}, its action reads
\begin{equation}\label{action:tele_dark_energy}
    \mathcal{S}_{\rm Tel. DE} := \int \dd^4x e\left[-\left(\frac{1}{\kappa^2} + \xi\phi^2 \right)\frac{T}{2} - \frac{1}{2}\partial_{\mu}\phi \partial^{\mu}\phi + V(\phi) \right] + \mathcal{S}_{\rm m}\,,
\end{equation}
where we have set $\omega(\phi) = 1, f(T) = -T$ and $F_1 = \kappa^2 \xi \phi,\,F_2 = 0$ in Eq.~\eqref{action:scalar-torsion-coupling-T-and-B}. Some Teleparallel dark energy models were found to be invariant under the Gasperini-Veneziano duality transformation \cite{Paliathanasis:2021gfq,Paliathanasis:2021nqa}.

Obviously, when the coupling of the scalar field is minimal, the theory coincides with the known quintessence models; but once the non-minimal coupling is switched on, the theory presents a much richer phenomenology. In particular, one can obtain a dark energy sector that behaves not only as quintessence, but also as phantom or phantom-divide crossing during evolution. This means that phantom behavior appears naturally without the need of ghost fields which come with ambiguous quantum behavior. The theory \eqref{action:tele_dark_energy} was further extended to include other irreducible parts of the torsion scalar as well \cite{Otalora:2014aoa}. A specific class of theories of Eq.~\eqref{action:scalar-torsion-coupling-T-and-B} with $f(T) = -T$ was considered in Ref.~\cite{Bahamonde:2015hza,Zubair:2016uhx}
\begin{equation}\label{scalar-torsion_action}
    \mathcal{S} = \int \dd^4x e \left[-\left(\frac{1}{\kappa^2} + F(\phi)\right)\frac{T}{2} -\frac{1}{2}G(\phi)B -\frac{\omega(\phi)}{2} \partial_{\mu}\phi \partial^{\mu}\phi + V(\phi)\right] + \mathcal{S}_{\rm m}\,.
\end{equation}
Another interesting class of theories are the so-called tachyonic teleparallel scalar-torsion theories (TTST), where a generalized non-minimal coupling of the scalar field with the torsion scalar and the boundary term is considered \cite{Bahamonde:2019gjk}. Their action reads
\begin{equation}\label{eq:action_TTST}
    \mathcal{S}_{\rm TTST} := \int \dd ^4 x e \left[-\frac{T}{2\kappa ^2} +\frac{1}{2}f(\phi)T+\frac{1}{2}g(\phi)B + V(\phi)\sqrt{1-\frac{2X}{V(\phi)}} + \mathcal{L}_{\rm m}\right] \,,
\end{equation}
where $V(\phi)$ is the potential. As one can immediately see, considering $f(\phi) = - g(\phi)$, a non-minimal coupling between the scalar field and the Ricci scalar is recovered because of the first relation in Eq.~\eqref{eq:RTQ}; further, the boundary term does not play any role when $g(\phi) = \textrm{constant}$ and it can be shown that such theories allow the crossing of the phantom divide line \cite{Banijamali:2012nx,Fazlpour:2014qla}. The coupling of the scalar field with the boundary term emanates from the fact that without fine-tuning, the system will evolve to a late-time accelerating attractor solution \cite{Bahamonde:2015hza,Zubair:2016uhx}. This theory is a sub-class of Eq.~\eqref{action}, and thus the equations of motion will be exactly the same with \eqref{fieldeqgeneral2} and \eqref{fieldeq2}, if we set
\begin{equation}
    f(T,B,\phi,X) = -T +2\kappa^2 \left[\frac{1}{2}f(\phi)T+\frac{1}{2}g(\phi)B + V(\phi)\sqrt{1-\frac{2X}{V(\phi)}}\right]\,.
\end{equation}
A dynamical systems analysis of this theory is presented in Ref.~\cite{Bahamonde:2019gjk} and it shown that the tachyonic field might be a good candidate for the solution of the cosmic coincidence and the dark energy problems.

In a series of papers \cite{Hohmann:2018rwf,Hohmann:2018vle,Hohmann:2018dqh,Hohmann:2018ijr} the most general teleparallel scalar-torsion theories of gravity in their covariant formulation are discussed. The only two necessary restrictions are invariance of the theory under diffeomorphisms and local Lorentz invariance as well as no coupling between the matter fields and the spin connection. The general formalism and the relation between different classes of theories with conformal transformations are discussed in Ref.~\cite{Hohmann:2018rwf}; a discussion on multiple scalar fields is also presented, as is also done in Refs.~\cite{Bahamonde:2018miw,Hohmann:2018vle}. In the second paper of the series \cite{Hohmann:2018dqh} a Lorentz invariant theory is presented that consists of the torsion scalar, a scalar field, its kinetic term and a derivative coupling between the torsion
and the scalar field and its Lagrangian reads
\begin{equation}\label{action:L(T,X,Y,phi)}
    \mathcal{S}_{f(T,Y,\phi,X)} := \frac{1}{2\kappa^2} \int \dd ^4x\, e \, f(T,Y,\phi,X) + S_{\rm m}\,,
\end{equation}
where $X$ is the kinetic term of the scalar field $\phi$ and $Y$ is the coupling between the torsion and the derivative of the scalar field
\begin{equation}
    Y = g^{\mu\nu} T^{\rho}{}_{\rho\mu}\partial_{\nu}\phi\,.
\end{equation}
Part of this action, without the derivative coupling has been studied in Ref.~\cite{Abedi:2018lkr}. It is interesting to note that linear terms of $Y$ in the action, either minimally or non-minimally coupled, would lead to a $B-$like contribution after an integration by parts; however, non-linear terms that could be possibly be part of the theory lead to more interesting phenomenology. Variations of the action \eqref{action:L(T,X,Y,phi)} \gls{wrt} to the tetrad yield the field equations
\begin{alignat}{2}
    & \: & &\frac{1}{2}\lc{\nabla}^{\mu}\left(f_Y\partial_\beta\phi\right) - \frac{1}{2}\delta_{\beta}^{\mu}\lc{\nabla}^{\rho}\left(f_Y\partial_\rho \phi\right) + \frac{1}{2}f_Y\left(g^{\nu\mu}T_{(\beta\nu)}{}^{\rho}\partial_\rho\phi + \frac{1}{2}T^{\rho}{}_{\beta}{}^{\mu}\partial_\rho\phi + T^{\rho}{}_{\rho\beta}\partial^\mu \phi\right) -(\partial_{\nu}f_{T})S_{\beta}{}^{\mu\nu}-	\nonumber\\[0.5ex]
    & \: & &	-\frac{1}{e}f_{T}e^{A}{}_{\beta}\partial_{\nu}(e S_{A}{}^{\mu\nu}) + f_{T}T^{B}{}_{\nu\beta}S_{B}{}^{\nu\mu} - f_T \omega^B{}_{\beta\nu} S_B{}^{\nu\mu} - \frac{1}{2}f \delta_{\beta}^{\mu}-\frac{1}{2}  f_{X}\partial^{\mu}\phi \partial_{\beta}\phi =\kappa^2 \Theta_{\beta}^{\mu}\,. \label{tetrad_equat:f(T,X,Y,phi}
\end{alignat}
Further, variations of the action \gls{wrt} to the spin connection coincide with the antisymmetric part of the tetrad equations \eqref{tetrad_equat:f(T,X,Y,phi}, meaning that only those pairs of tetrads and spin connection that solve both field equations are considered solutions of the theory. The spin connection equations read
\begin{align}
    3 \partial _{[ \rho} f_T T^{\rho}{}_{\mu\nu]} + \partial _{[\mu} f_Y \partial _{\nu]}\phi - \frac{3}{2}f_Y T^{\rho}{}_{[\mu\nu} \partial _{\rho]}\phi = 0\,.
\end{align}
Finally, variations \gls{wrt} the scalar field yields the Klein-Gordon equation
\begin{equation}
    g^{\mu\nu}\lc{\nabla}_{\mu}\left( f_YT^{\rho}{}_{\rho\nu} - f_X \partial _{\nu}\phi\right) - f_{\phi} = \kappa^2 \Theta\,,
\end{equation}
where $\Theta$ is the trace of the energy-momentum tensor. Conformal transformations as well as relation of scalar-torsion to scalar-tensor theories are discussed in Refs.~\cite{Hohmann:2018dqh,Hohmann:2018ijr}. In addition, disformal transformations in scalar-torsion gravity are discussed in Ref.~\cite{Hohmann:2019gmt} as well as in Sec.~\ref{Sec:conformal_trans}.

\paragraph{Conformal transformations teleparallel dark energy}\label{Sec:conf-trans-fTB}
Considering a teleparallel dark energy model of the form
\begin{equation}\label{eq:TDEconformal0}
    \mathcal{S}_{\rm TDE} := \frac{1}{2\kappa ^2} \int \dd ^4 x e \left[ -A(\phi) T - \frac{1}{2}\partial_{\mu}\phi\partial^{\mu}\phi + V(\phi) \right]+ \mathcal{S}_{\rm m}\,,
\end{equation}
and applying a conformal transformation of the form \eqref{eq:conformal_tetrad} in an attempt to remove the kinetic term of the scalar field, we end up with
\begin{equation}\label{eq:TDEconformal}
    \mathcal{S}_{\rm TDE} = \frac{1}{2\kappa ^2} \int \dd ^4 x \tilde{e} \left[ - A(\phi)\left(\Omega^2\tilde{T} - 4 \Omega \tilde{T}^{\mu}\partial_{\mu}\Omega - 6 \tilde{g}^{\mu\nu}(\partial_{\mu}\Omega)(\partial_{\nu}\Omega) \right) - \frac{1}{2}\Omega ^2 \tilde{g}^{\mu\nu}\partial_{\mu}\phi\partial_{\nu}\phi +V(\phi) \right]\,,
\end{equation}
where we considered only the gravitational part of the action \eqref{eq:TDEconformal0}. In order for the kinetic term of the scalar field to vanish, we have to set
\begin{equation}
    \Omega = \exp\left(\int \frac{1}{2\sqrt{3A(\phi)}}\dd\phi\right)\,,
\end{equation}
and expressing it in its inverse form, meaning $\phi = \phi(\Omega)$ we can substitute it in Eq.~\eqref{eq:TDEconformal} and get
\begin{equation}\label{eq:TDEconformal2}
    \mathcal{S}_{\rm TDE} = \frac{1}{2\kappa^2} \int \dd^4 x \tilde{e}\left[ -A(\Omega)\Omega^{-2}\tilde{T} + 4\Omega^{-3}A(\Omega)\tilde{T}^{\mu}\partial_{\mu}\Omega + U(\Omega)\right]\,,
\end{equation}
with $U(\Omega) = \Omega^{-4}V(\phi)$. We can rewrite the second term in the above action \eqref{eq:TDEconformal2} introducing a function $G(\Omega)$ such that
\begin{equation}
    \Omega^{-3}A(\Omega)\partial_{\mu}\Omega = \partial _{\mu}G(\Omega)\,.
\end{equation}
Integrating Eq.~\eqref{eq:TDEconformal2} by parts we get
\begin{equation}\label{eq:TDEconformal3}
    \mathcal{S}_{\rm TDE} = \frac{1}{2\kappa^2}\int \dd^4 x \tilde{e} \left[ -A(\Omega)\Omega^{-2}\tilde{T}-2G(\Omega)\tilde{B}+U(\Omega)\right]\,.
\end{equation}
We can see now that there is no kinetic term for the scalar field $\Omega$, meaning that it is just an auxiliary field. Varying the action \eqref{eq:TDEconformal3} and solving the respective equations for $\Omega$, we obtain a function of the form $\Omega = \Omega (\tilde{T},\tilde{B})$. This means that the integrand in Eq.~\eqref{eq:TDEconformal3} is just a function $f(\tilde{T},\tilde{B})$, thus making the teleparallel dark energy model \eqref{eq:TDEconformal0} conformally equivalent to $f(T,B)$ gravity. A more general class of disformally invariant theories can be found in Ref.~\cite{Hohmann:2019gmt}.

\subsubsection{Teleparallel Horndeski gravity - BDLS theory}
\label{sec:BDLS}

The motivation for Horndeski's theory was to write down the most general Lagrangian with a single scalar field, that leads to second order field equations. Almost thirty years after its publication, it was realized that this theory is completely equivalent to generalized covariant galileons \cite{Deffayet:2009wt,Capozziello:2018gms,Bhattacharya:2015chc}. However, part of it was severely constrained after the GW 170817 and GRB 170817A events, which was the motivation to study if the constraints persisted in the teleparallel framework as well.

Bahamonde-Dialektopoulos-Levi Said (BDLS) theory of gravity is nothing but the teleparallel analogue of the known Horndeski gravity, i.e. the most general scalar-tensor theory with a single scalar field in 4 dimensions leading to second order field equations, satisfying some necessary conditions to retain a finite class of theories. The theory itself was formulated in Ref.~\cite{Bahamonde:2019shr}.

The necessary criteria for the theory to be formulated are the following:
\begin{itemize}
    \item
    The field equations both for the tetrad and for the scalar field must be of second order,
    \item
    the scalar invariants should not be parity-violating,
    \item
    and contractions of the torsion tensor must be at most quadratic.
\end{itemize}
The first condition appears in order to avoid ghost instabilities while the second one refers to contractions of the scalar field (through its derivatives) with the irreducible parts of the torsion tensor. Those scalars should be invariant under parity transformations. The last one is a particular feature of \gls{tg}. In \gls{gr} and its extensions, thus in Horndeski gravity as well, Lovelock's theorem prohibits other terms to be part of these Lagrangians. In \gls{tg} though, this is not the case; the torsion tensor consists of first derivatives in the tetrad, in contrast with the Riemann tensor that consists of second derivatives in the metric. This means that, in principle, infinite contractions of the torsion tensor could be considered and still the field equations would be second order. However, it is unclear how physical such higher order contributions will be. For this reason, in Ref.~\cite{Bahamonde:2019shr}, it was demanded that, the contributing scalar invariants of the theory be at most quadratic contractions of the torsion tensor.

In order to form the most general Lagrangian that satisfies the above criteria, we should formulate first the one in the tangent space and then raise it to the general manifold through the coupling prescription. The theory considers also the following conditions for the scalar field \cite{Deffayet:2011gz}:
\begin{enumerate}
    \item
    The Lagrangian contains up to second order derivatives of the scalar field,
    \item
    the Lagrangian is a polynomial in second order derivatives of the scalar field,
    \item \label{sec5:BDLS-3rd crit}
    the corresponding field equations are at most second order in derivatives of the scalar field (which coincides with the first criterion above).
\end{enumerate}

The Lagrangian for a scalar field $\phi$ with a kinetic term $X = -\frac{1}{2}\partial_{\mu}\phi \partial ^{\mu}\phi$, that is invariant under the Galilean transformation $\phi \rightarrow \phi + b_{\mu}x^{\mu} + c$, in the tangent space has the form
\begin{equation}\label{flat_galileon_action}
    \mathcal{L} = \sum _{i=1} ^5 c_i \mathcal{L}_i\,,
\end{equation}
where $c_i$ are arbitrary constants and the subscript refers to the number of appearances of the scalar field in each component of the Lagrangians which are
\begin{subequations}
\begin{alignat}{2} \label{flat_galileon_action_L1}
    \mathcal{L}_1 & :=\: & & \phi \,, \\[0.5ex] \label{flat_galileon_action_L2}
    \mathcal{L}_2 & :=\: & & X \,,\\[0.5ex] \label{flat_galileon_action_L3}
    \mathcal{L}_3 & :=\: & & X \bar{\Box} \phi \,,\\[0.5ex] \label{flat_galileon_action_L4}
    \mathcal{L}_4 & :=\: & & -X (\bar{\Box}\phi)^2 + (\bar{\Box}\phi) \phi,_{\mu}\phi ,_{\nu} \phi ^{,\mu\nu} + X  \phi^{,\mu\nu} \phi,_{\mu\nu} - \phi,_{\mu} \phi^{,\mu\nu} \phi,_{\nu\rho}\phi^{,\rho}\,,\\[0.5ex] \label{flat_galileon_action_L5}
    \mathcal{L}_5 & :=\: & & -2X (\bar{\Box}\phi)^3 - 3(\bar{\Box}\phi)^2 \phi,_{\mu}\phi,_{\nu}\phi^{,\mu\nu} + 6 X \bar{\Box}\phi \phi,_{\mu\nu}\phi^{,\mu\nu} \nonumber \\[0.5ex]
    & \: & &+6\bar{\Box}\phi \phi,_{\mu}\phi^{,\rho} \phi^{,\mu\nu} \phi,_{\nu\rho} -4X\phi,_{\mu}{}^{\nu}\phi,_{\nu}{}^{\rho} \phi,_{\rho}{}^{\mu} \nonumber \\[0.5ex]
   & \: & &+3 \phi,_{\mu\nu}\phi^{,\mu\nu}\phi,_{\rho}\phi,_{\lambda}\phi^{,\lambda\rho} - 6 \phi,_{\mu}\phi^{,\mu\nu}\phi,_{\nu\rho}\phi^{,\lambda\rho}\phi,_{\lambda}\,.
\end{alignat}
\end{subequations}
The d'Alembertian is given by $\bar{\Box} = \partial _{\mu}\partial^{\mu}$ which changes in any gravitational theory since it is a derivative operator, comma ``,'' denotes partial derivative and barred quantities refer to Minkowski spacetime.

Covariantizing the action \eqref{flat_galileon_action}, i.e considering it in a general non-flat manifold, we first need to replace the arbitrary constants $c_i$ to general functions of the form $G_i(\phi,X)$.
In this way, the Lagrangians \eqref{flat_galileon_action_L1}--\eqref{flat_galileon_action_L5} take the form
\begin{subequations}
\begin{align}
\mathcal{L}_2 &:= G_2(\phi,X)\,,\label{HG_2}\\[0.5ex]
\mathcal{L}_3 &:= G_3(\phi,X)\lc{\Box}\phi\,,\label{HG_3}\\[0.5ex]
\mathcal{L}_4 &:= G_4(\phi,X)\left(-T+B\right) + G_{4,X}(\phi,X)\left[\left(\lc{\Box}\phi\right)^2 - \phi_{;\mu\nu}\phi^{;\mu\nu}\right]\,,\label{HG_4}\\[0.5ex]
\mathcal{L}_5 &:= G_5(\phi,X)\lc{G}_{\mu\nu}\phi^{;\mu\nu} - \frac{1}{6}G_{5,X}(\phi,X)\Big[\left(\lc{\Box}\phi\right)^3 + 2\dut{\phi}{;\mu}{\nu}\dut{\phi}{;\nu}{\alpha}\dut{\phi}{;\alpha}{\mu} - 3\phi_{;\mu\nu}\phi^{;\mu\nu}\left(\lc{\Box}\phi\right)\Big]\,,\label{HG_5}
\end{align}
\end{subequations}
where $G_{i,X}=\partial G_i/\partial X$ and $\phi_{;\mu\nu}=\lc{\nabla}_\mu\lc{\nabla}_\nu\phi\,.$
In the $\mathcal{L}_4$ and $\mathcal{L}_5$ Lagrangians there are two new terms, $G_4(\phi,X) (-T+B)$ and $G_5(\phi,X)\lc{G}_{\mu\nu}\phi^{;\mu\nu}$, where $\lc{G}_{\mu\nu}$ is the Einstein tensor formulated in teleparallel geometry, i.e., Eq.~\eqref{eq:einsteintensor}. That is because the covariant derivatives do not commute, in order to compensate for the higher order terms that would appear in the field equations, we have to add these extra terms.

However, the theory is not complete yet. In the teleparallel geometry there exist more terms that satisfy the aforementioned criteria. Specifically, if we consider up to quadratic contractions of the torsion tensor, criterion~\ref{sec5:BDLS-3rd crit}, the most general Lagrangian satisfying the conditions explained above of \gls{tg} (without a scalar field) is $f(T_{\rm axi}, T_{\rm vec},T_{\rm ten})$ that was discussed in Sec.~\ref{Sec:Ext_NGR}, which means that (at least) these invariants, i.e. the irreducible decomposition of the torsion tensor should be part of our theory.

Indeed, in \gls{tg} there exists a set of scalars that should be included in a new component of the Lagrangian, $\mathcal{L}_{\rm Tele}$, which consist of the irreducible parts of the torsion tensor, contracted with derivatives of the scalar field. Linear contractions of the torsion tensor are
\begin{equation}
    I_1 := t^{\mu\nu\sigma} \phi_{;\mu}\phi_{;\nu}\phi_{;\sigma}\,,\quad I_2 := v^{\mu} \phi_{;\mu}\,,\quad I_3 := a^{\mu} \phi_{;\mu}\,.
\end{equation}
This is the full set of scalar that can be constructed considering linear contractions of the torsion tensor because of the symmetry of $t_{\mu\nu\sigma}$ in its first two indices and the fact that $t^{\sigma\mu}{}_{\sigma} = t^{\sigma}{}_{\sigma}{}^{\mu} = t^{\mu\sigma}{}_{\sigma} = 0$. However, due to the the fact that $t_{(\mu\nu\rho)} = 0$, it can be easily shown that $I_1$ vanishes. Moreover, conforming with the second criterion, $I_3$ cannot be considered, together with any scalar terms that contains an odd number of the axial part of the torsion tensor, because it is not parity invariant. Contracting the tensorial part of the torsion tensor with second order derivatives of the scalar field would result higher order derivatives in the field equations.

Following the third criterion mentioned above, the theory includes also quadratic contractions of the torsion tensor with derivatives of the scalar field. Those scalars that are not parity-violating are
\begin{subequations}
\begin{align}
    J_1 &:= a^{\mu}a^{\nu} \phi_{;\mu}\phi_{;\nu}\,,\quad J_2 := v^{\mu}v^{\nu} \phi_{;\mu}\phi_{;\nu}\,, \quad J_3 := v_{\sigma}t^{\sigma\mu\nu} \phi_{;\mu}\phi_{;\nu}\,,\quad J_4 := v_{\mu}t^{\sigma\mu\nu} \phi_{;\sigma}\phi_{;\nu}\,,\\[0.5ex]
    J_5 &:= t^{\sigma\mu\nu}t_{\sigma}{}^{\bar{\mu}}{}_{\nu} \phi_{;\mu}\phi_{;\bar{\mu}}\,, \quad J_6 := t^{\sigma\mu\nu}t_{\sigma}{}^{\bar{\mu}\bar{\nu}} \phi_{;\mu}\phi_{;\nu}\phi_{;\bar{\mu}}\phi_{;\bar{\nu}}\,,\quad J_7 := t^{\sigma\mu\nu} t^{\bar{\sigma}\bar{\mu}}{}_{\sigma} \phi_{;\mu}\phi_{;\nu}\phi_{;\bar{\sigma}}\phi_{;\bar{\mu}}\,, \\[0.5ex]
    J_8 &:= t^{\sigma\mu\nu}t_{\sigma\mu}{}^{\bar{\nu}} \phi_{;\nu}\phi_{;\bar{\nu}}\,,\quad J_9 := t^{\sigma\mu\nu}t^{\bar{\sigma}\bar{\mu}\bar{\nu}} \phi_{;\sigma}\phi_{;\mu}\phi_{;\nu}\phi_{;\bar{\sigma}}\phi_{;\bar{\mu}}\phi_{;\bar{\nu}}\,,\quad J_{10} := \epsilon^{\mu}{}_{\nu\rho\sigma}a^{\nu}t^{\alpha\rho\sigma} \phi_{;\mu}\phi_{;\alpha}\,.
\end{align}
\end{subequations}
Nevertheless, these scalars are not all independent with each other. In particular, one can notice that $J_2 = I_2^2,\,J_3=J_4$ and $J_7 = -2 J_6$. In addition, as with $I_1$, the vanishing of the totally symmetric part of the tensor torsion makes $J_9$ vanish. In addition, contractions of quadratic order torsion tensor with second order derivatives of the scalar field, lead to higher order field equations. This means that the most general theory that satisfies the above criteria ends up having seven independent scalars in total and apart from the sum of \eqref{HG_2}-\eqref{HG_5} it includes also an extra Lagrangian that has the form
\begin{align}\label{Ltele}
    \mathcal{L}&_{\text{Tele}} :=G_{\text{Tele}}(\phi,X,T,T_{\text{axi}},T_{\text{vec}},I_2,J_1,J_3,J_5,J_6,J_8,J_{10})\,,
\end{align}
where, instead of choosing its tensorial decomposition, we used the torsion scalar itself, which is possible due to the relationship in Eq.~\eqref{eq:torscalar_decomposition}. Notice that instead of the three irreducible parts of the torsion tensor, we considered the torsion scalar with its axial and its vectorial parts. What is also interesting to notice is that the boundary term $B$ does not contribute in Eq.~\eqref{Ltele}. Even though $B$ itself is second order in the Lagrangian, it produces the fourth order elements of the theory, thus not satisfying the first criterion\footnote{These contributions make $f(\lc{R})$ theory have fourth order field equations}. This is similar to the standard Horndeski theory case where $f(\lc{R})$ does not appear in the action since that term contains higher order derivatives. Thus, the Lagrangian of BDLS theory, i.e. the teleparallel analogue of Horndeski gravity is given by
\begin{equation}\label{BDLS_Action}
    \mathcal{S}_{\text{BDLS}} := \frac{1}{2\kappa^2}\int \dd^4x\, e \mathcal{L}_{\rm Tele} + \frac{1}{2\kappa^2} \sum_{i=2}^{5}\int \dd^4x\, e\mathcal{L}_{i} + \mathcal{S}_{\rm m}\,.
\end{equation}
Once one sets $\mathcal{L}_{\rm Tele} = 0$, the action takes the standard Horndeski form, formulated in the teleparallel geometry. Variations of the scalars and the arbitrary functions, together with the equations of motion of the tetrad and the scalar field can be found in the Appendix~\ref{App:variations}.

Before we move on, we would like to remark the motivation for BDLS theory. Apart from a competitive exercise and an analytically interesting theory, that is why Horndeski formulated his theory in the first place, it is an interesting terrain to test the possible supremacy of teleparallel model over their curvature-based counterparts. BDLS not only contains the standard Horndeski theory, as well as most of the extended teleparallel theories of gravity (e.g. $f(T)$, $f(T_{\rm axi},T_{\rm vec},T_{\rm ten})$, teleparallel dark-energy models, Gauss-Bonnet theory in the conformal frame and more) but it extends it just by changing the background geometry. The Fig.~\ref{fig:BDLS} shows some important theories that are part of the BDLS theory along with some important references. This results in richer phenomenology because it contains new terms through the extra Lagrangian $\mathcal{L}_{\rm Tele}$ and because of this extra terms it revives several Horndeski models that were severely constrained in the curvature formulation after the GW 170817 and GRB 170817A events. More details can be found in Refs.~\cite{Bahamonde:2019shr,Bahamonde:2019ipm} and also in Sec.~\ref{sec:cosmo-pert} were cosmological perturbations are discussed. Recently, it was found that the most general BDLS theory propagates up to 7 \gls{dof}. In cosmology, it was also found that this theory can generate a late-time, low energy de-Sitter vacuum state by using the well-tempered method~\cite{Bernardo:2021izq,Bernardo:2021bsg}. Keeping with recent developments in cosmology, in Ref.~\cite{Dialektopoulos:2021ryi} Noether's theorem was used to classify the many models that can be produced in this new class of theories. We conclude this section by mentioning a similar theory introduced in Ref.~\cite{Nicosia:2020egv} which considers a teleparallel vector-tensor theory constructed as an analogue of the so-called generalized Proca theory. This theory assumes a vector $A_\mu$ field coupled to gravity which only propagates up to 3 \gls{dof} in order to avoid Ostrogradsky ghosts and comply also with the representation of the massive spin 1 Lorentz group. This theory contains up to second order derivatives in the field equations (for the tetrads and the vector) and only parity preserving scalars were considered. The field equations of the theory are of second order for both the tetrad and the vector field and only parity preserving scalars were considered. One important difference \gls{wrt} teleparallel Horndeski is the fact that the scalars constructed from the torsion tensor were considered to be only linear.

\begin{figure}[htp!]\label{Tele_Map}
	\tikzset{
  font={\fontsize{9pt}{11}\selectfont}}
\footnotesize{\begin{tikzpicture}
[auto,
decision/.style={diamond, draw=blue, thick, fill=blue!20,
	text width=8em,align=flush center,
	inner sep=1pt},
block/.style ={rectangle, draw=blue, thick, fill=blue!20,
	text width=8em,align=center, rounded corners,
	minimum height=3em},
line/.style ={draw, thick, -latex',shorten >=2pt},
cloud/.style ={draw=red, thick, ellipse,text width=8em,fill=red!20,	text width=6em,align=center,	minimum height=3em}]
\matrix [column sep=12mm,row sep=30mm]
{&	&\node [cloud] (GR) {GR};\\[-11ex]
&
	\node [cloud] (expert4) {Kinetic braiding~\cite{Deffayet:2010qz}}; &
	\node [cloud] (expert) {Generalized Brans-Dicke\\ \cite{Brans:1961sx,Perrotta:1999am}}; &
	\node [cloud] (init) {k-essence, quintessence\\ \cite{Copeland:2006wr}}; \\[-8ex]
	\node [cloud] (expert2) {Quartic couplings};&	& \node [cloud] (identify) {\ \ \ Horndeski\newline~\cite{Horndeski:1974wa,Kobayashi:2011nu}}; & \\[-15ex]
		\node [cloud] (GR2) {GR};
		&\node [block] (TEGR2) {TEGR};
		& \node [decision] (evaluate) {Teleparallel Horndeski};\\[-16ex]
	\node [block] (decide8) {Cubic Teleparallel \newline\cite{Gonzalez-Espinoza:2019ajd}};	&	& \node (evaluate2) {}; & \node [block] (decide5) {Conformal Teleparallel~\cite{Maluf:2011kf}};\\[6ex]
	\node [block] (decide3) {Kinetic Teleparallel~\cite{Hohmann:2018dqh,Abedi:2015cya}};	& \node [block] (decide) {Non-minimally couplings between $\phi$ and $T$ and $B$~\cite{Zubair:2016uhx}}; &\node [block] (decide4) {$f(T,T_{\rm axi},T_{\rm vec})$~\cite{Bahamonde:2017wwk}}; & \\[-4ex]
	\node [block] (stop2) {Couplings with $T$~\cite{Geng:2011aj}};	& \node [block] (stop) {Couplings with $B$~\cite{Bahamonde:2015hza}}; & \node [block] (decide6) {$f(T)$~\cite{Ferraro:2006jd,Cai:2015emx}}; & \node [block] (decide7) {New General Relativity~\cite{Hayashi:1979qx}};\\[-11ex]
&	&\node [block] (TEGR) {TEGR};\\
};
\begin{scope}[-stealth,every path/.style]
\path (identify) -- (init);
\path (GR2) edge node [right] {} (TEGR2);
\path (TEGR2) edge node [right] {} (GR2);
\draw [rounded corners=0.8cm] (-8.5-0.2-0.2-0.2-0.2-0.2+0.6,0.5+0.4-0.2) rectangle +(8+0.2+0.4-0.1+0.2+0.2+0.2-0.2,2.5+0.4-0.2) node [midway] {};
\node[anchor=west,text width=6cm] (note1) at (-7.5,1.1+0.2) { \gls{gr} and \gls{tegr} are equivalent at the level of their field equations};
\path (evaluate) edge node [right] {$\mathcal{L}_{\rm Tele}=0$} (identify);
\path (evaluate) edge node [right] {$\mathcal{L}_{\rm Tele}\neq0$} (evaluate2);
\path (evaluate2) edge node [above,sloped] {\scriptsize{$\hspace{-1cm}\tilde{G}_{\rm Tele}=(\tilde{G}_4(\phi)+F(\phi))T$}} (decide);
\path (evaluate2) edge node [below,sloped] {\scriptsize{$G_5=G_3=0,G_4=\tilde{G}_4(\phi)$}} (decide);
\path (evaluate2) edge node [above] {$G_4=G_5=0, G_{\rm Tele}=F(\phi)T$ } (decide8);
	\path (evaluate2) edge node [sloped,above] {$G_3=G_4=G_5=0$ and $G_{\rm Tele}=F(\phi,X)T$} (decide3);
	\path (evaluate2) edge node [above,sloped] {$G_2=G_3=G_4=G_5=0$} (decide4);

	\path (decide8) edge[bend right=40] node [below,sloped] {$G_3=0$} (stop2);

		\path (evaluate2) edge node [below,sloped] { $G_{\rm Tele}=f(T,T_{\rm axi},T_{\rm vec})$} (decide4);
	\path (decide4) edge node [below,sloped] {$ f=(\frac{3}{2}T_{\rm axi}+\frac{2}{3}T_{\rm ten})^2$} (decide5);
	\path (decide4) edge node [sloped,below] {\hspace{-0.8cm}$f=a_1 T_{\rm axi}+a_2 T_{\rm ten}+a_3 T_{\rm vec}$} (decide7);
	\path (decide4) edge node [below,sloped] {$f=f(T)$} (decide6);

		\path (decide7) edge node [sloped,above] {\scriptsize{\hspace{-0.5cm}$a_1=3/2, a_2=2/3$}} (TEGR);
			\path (decide7) edge node [sloped,below] {\scriptsize{$a_3=-2/3$}} (TEGR);
				\path (decide6) edge node [right] {$f=T$} (TEGR);
					\path (stop) edge node [sloped,above] {\hspace{0.5cm}$G_2=0,\tilde{G}_4=1$} (TEGR);
					\path (stop2) edge node [sloped,below] {$G_2=0,F=1$} (TEGR);
	\path (decide) edge node [right] {$F(\phi)=1$} (stop);
		\path (decide3) edge node [below,sloped] {$F=F(\phi)$} (stop2);
	\path (decide) edge node [above,sloped] {$\tilde{G}_4(\phi)=0$} (stop2);
	\path (identify) edge node [right] {$G_3=G_5=0,G_4=\tilde{G}_4(\phi)$} (expert);
	\path (identify) edge node [below,sloped] {$G_5=0$} (expert2);
	\path (identify) edge node [below,sloped] {$G_4=1,G_5=0$} (expert4);
	\path (expert) edge node [below] {\scriptsize{$\tilde{G}_4(\phi)=1$}} (init);
	\path (expert4) edge node [above,sloped] {$G_2=G_3=0$} (GR);
	\path (expert) edge node [above,sloped] {$G_2=0,$} (GR);
		\path (expert) edge node [below,sloped] {$\tilde{G}_4=1$} (GR);
		\path (init) edge node [above,sloped] {$G_2=0$} (GR);
			\path (expert2) edge[bend left=40] node [above,sloped] {$G_2=G_3=0,\,G_4=1$} (GR);
	\end{scope}
	\end{tikzpicture}}
	\caption{Relationship between the Teleparallel Horndeski analogue and various theories.}
	\label{fig:BDLS}
\end{figure}

\subsection{Theories non-minimally coupled with matter} \label{sec:modifiedmatter}

Another route that became popular in modified gravity is to construct theories with non-minimally couplings between the matter sector and gravity. The most popular of them is the so-called $f(\lc{R},\Theta)$ where $\Theta:=\Theta^\mu{}_\mu$ is the trace of the energy-momentum tensor and $\lc{R}$ the Ricci scalar~\cite{Harko:2011kv}. Due to the nature of these theories, the continuity equation may also be modified leading to further generalizations and richer phenomenology. In \gls{tg}, a similar theory was introduced in Ref.~\cite{Kiani:2013pba,Harko:2014aja} by considering the following action
\begin{equation}
\mathcal{S}_{f(T,\Theta)} := \frac{1}{2\kappa^2}\int \dd ^4 x\, e\, f(T,\Theta)+\mathcal{S}_{\rm m}\,,
\end{equation}
where $T$ is the torsion scalar instead of the Ricci scalar. By taking variations \gls{wrt} the tetrads, we find the following field equations
\begin{alignat}{2}
	-
	(\partial_{\nu}f_{T})S_{\beta}{}^{\mu\nu}
	-\frac{1}{e}f_{T}e^{A}{}_{\beta}\partial_{\nu}(e S_{A}{}^{\mu\nu}) +
	f_{T}T^{B}{}_{\nu\beta}S_{B}{}^{\nu\mu}& \: & & - f_T \omega^B{}_{\beta\nu}
	S_B{}^{\nu\mu}-
	\frac{1}{2}f \delta_{\beta}^{\mu}\nonumber\\
& \: & &	=\kappa^2 \Theta_{\beta}^{\mu}-f_{\Theta}\Big(\Theta_{\beta}^\mu+e^A{}_\nu g^{\alpha\beta}\frac{\delta \Theta_{\alpha\beta}}{\delta e^A{}_\mu}\Big)\,,\label{eq:fTtheta}
 \end{alignat}
where $f_{\Theta}=\partial{f}/\partial{\Theta}$ and the variation appearing on the \gls{rhs} of the equation can be expanded as
\begin{equation}
e^A{}_\lambda g^{\mu\nu}\frac{\delta \Theta_{\mu\nu}}{\delta e^A{}_\rho}=-2\Theta_{\lambda}{}^\rho+\delta_{\lambda}{}^\rho \mathcal{L}_{\rm m}-2g^{\nu\rho }g^{\alpha\beta}\frac{\partial^2 \mathcal{L}_{\rm m}} {\partial g^{\lambda \nu}\partial g^{\alpha\beta}}\,.
\end{equation}
Here $\mathcal{L}_{\rm m}$ is the density matter Lagrangian. It is common to consider that the matter Lagrangian does not depend on the derivatives of the metric (or equivalently derivatives of the tetrads), therefore, the last term in the above equation is usually neglected. If one assumes that the matter is described by a perfect fluid with energy density $\rho$ and pressure $p$, then $\Theta_{\mu\nu}=(\rho+p)u_\mu u_\nu-p g_{\mu\nu}$ with $u_\mu$ being the 4-velocity of the fluid which is normalized as $u_\mu u^\mu=+1$. For this particular case, we can take $\mathcal{L}_{\rm m}=-p$ and the above term becomes
\begin{equation}
e^A{}_\lambda g^{\mu\nu}\frac{\delta \Theta_{\mu\nu}}{\delta e^A{}_\rho}=-2\Theta_{\lambda}{}^\rho-p \delta_\lambda{}^\rho\,.
\end{equation}
This particular choice was assumed in Ref.~\cite{Harko:2014aja} too in order to analyze the cosmology of this theory filled with a standard perfect fluid. It can be shown than in this theory, the standard energy-momentum tensor is non conserved~\cite{Saez-Gomez:2016wxb}
\begin{eqnarray}\label{eq:continuity_fTT}
\lc{\nabla}_{\mu}\Theta^{\mu\nu}=\frac{f_{\Theta}}{\kappa^2-f_{\Theta}}\left[\left(\Theta^{\mu\nu}+g^{\mu\nu}p\right)\lc{\nabla}_{\mu}\log f_{\Theta}+g^{\mu\nu}\lc{\nabla}_{\mu}\left(p+\frac{1}{2}\Theta\right)\right]\,,
\end{eqnarray}
which can be interpreted as an additional force which is usually labelled as the fifth force~\cite{Harko:2011kv}. One can further include the boundary term $B$ in the action. Further, for that extended theory, the process of quantum fluctuations induced by matter-gravity couplings were studied~\cite{Chen:2021oal}. Further, it was found that these fluctuations are compatible with cosmological observations~\cite{Bernardo:2021ynf}.

Another route for constructing theories with couplings between matter and the gravitational sector is by considering actions with couplings between the Lagrangian density and the teleparallel scalars. The first theory considering this was formulated in Ref.~\cite{Harko:2014sja} where the authors considered an action based on $f_1(T)+(1+\lambda f_2(T))\mathcal{L}_{\rm m}$ (without $\mathcal{S}_{\rm m}$) where $\lambda$ is a constant. This theory was further extended including the boundary term in Ref.~\cite{Bahamonde:2017ifa} with the action
\begin{equation}
\mathcal{S}_{f(T,B,\mathcal{L}_{\rm m})} := \frac{1}{2\kappa^2}\int \dd ^4 x \,e\, f(T,B,\mathcal{L}_m)\,.\label{actionfTBL}
\end{equation}
The above theory also contains the theory $f(T,\mathcal{L}_{\rm m})$ which was firstly presented in Ref.~\cite{Feng:2015awr}. These theories are the teleparallel versions of other non-minimally matter-gravity coupled theories constructed from a Riemannian case with the Ricci scalar $\lc{R}$ being the quantity which is coupled with $\mathcal{L}_{\rm m}$~\cite{Bertolami:2007gv,Harko:2010mv}. Furthermore, the $f(\lc{R},\mathcal{L}_{\rm m})$ case can be derived from Eq.~\eqref{actionfTBL} by considering the case $f(-T+B,\mathcal{L}_{\rm m})$. The variations of the above action \gls{wrt} the tetrads yields
	\begin{alignat}{2}
& \: & &	\delta_{\beta}^{\mu}\lc{\Box} f_{B}-\lc{\nabla}^{\mu}\lc{\nabla}_{\beta}f_{B}+
\frac{1}{2}	B f_{B}\delta_{\beta}^{\mu} -
	\Big[(\partial_{\nu}f_{B})+(\partial_{\nu}f_{T})\Big]S_{\beta}{}^{\mu\nu}
\nonumber	\\
& \: & &	-\frac{1}{e}f_{T}e^{A}{}_{\beta}\partial_{\nu}(e S_{A}{}^{\mu\nu}) +
	f_{T}T^{B}{}_{\nu\beta}S_{B}{}^{\nu\mu} - f_T \omega^B{}_{\beta\nu}
	S_B{}^{\nu\mu}-
	\frac{1}{2}f \delta_{\beta}^{\mu}-\frac{1}{2}f_{L}\mathcal{L}_{\rm m}\delta^{\mu}_{\beta} =\frac{1}{2}f_{L}\Theta_{\beta}^{\mu}\,,
	\label{fieldeqgenerafTLm}
	\end{alignat}
where $f_L=\partial f/\partial \mathcal{L}_{\rm m}$. The \gls{tegr} case is then recovered when $f=-T+\mathcal{L}_{\rm m}$ and further modifications can be made by other forms of $f$. For this theory, one also finds that the energy-momentum tensor is non-conserved. The process of induced particle production was recently studied in Ref.~\cite{Harko:2021bdi} without including the boundary term.

\subsection{Non-local theories} \label{sec:nonlocal}

One of the most well known proposals for a non-local modification in gravity, and specifically of the Einstein-Hilbert action, was suggested in \cite{Deser:2007jk}, where their action reads
\begin{equation}\label{Deser_Non-local}
    \mathcal{S}_{\rm NL} := \frac{1}{2\kappa^2}\int \dd^4x\sqrt{-g}\lc{R}\left(1+f(\lc{\Box} ^{-1}\lc{R})\right) + \mathcal{S}_{\rm m}\,,
\end{equation}
where $f$ is an arbitrary function that depends on the retarded Green function evaluated at the Ricci scalar, i.e.
\begin{equation}\label{Green_function}
    \lc{ \Box} ^{-1}F (x)\equiv \int \dd^4x' e(x')F(x')G(x,x')\,.
\end{equation}
Such corrections arise naturally when one considers quantum loop effects, while they are considered as a possible solution to the black hole information paradox \cite{Amadei:2019ssp,Stoica:2018uli}. Since then, non-locality theories have been studied in different contexts and have given some interesting results in open problems of cosmology, see more details in Ref.~\cite{Bahamonde:2017sdo} and references therein. Some interesting non-singular and ghost-free higher order theories that include torsion are studied in Refs.~\cite{delaCruz-Dombriz:2018aal,delaCruz-Dombriz:2019tge}.

For the same reasons, the non-local teleparallel gravity (TNL) was proposed in Ref.~\cite{Bahamonde:2017bps}. Its action has the form
\begin{align}
    \mathcal{S}_{\rm TNL} &:= - \frac{1}{2\kappa ^2}\int \dd^4x \,e \, T + \frac{1}{2\kappa ^2} \int \dd^4x\, e\, Tf\left(\lc{\square}^{-1}T\right)+\mathcal{S}_{\rm m} \,, \label{nonlocaltele}
\end{align}
where the non-local operator, i.e. the retarded Green function is evaluated at the torsion scalar $T$. The authors show that the theory is consistent with cosmological data from \gls{sn}eIa+\gls{bao}+\gls{cc}+$H_0$ observations. Varying this action \gls{wrt} the tetrad we get the field equations of the theory that read
\begin{alignat}{2}
	& \: & & -\left[S_A{}^{\beta\mu}\partial _{\mu} +\frac{1}{e}\partial _{\mu}(eS_A{}^{\beta\mu}) - T^{\sigma}{}_{\mu A} S_{\sigma}{}^{\mu\beta} - \frac{1}{2}TE_A{}^{\beta}\right]\left[f(\lc{\Box}^{-1}T)+\lc{\Box} ^{-1}(Tf'(\lc{\Box}^{-1}T)) \right] \nonumber \\
	& \: & & + \lc{G}_A{}^\beta -\frac{1}{2} \partial _{\rho}\left[\lc{\Box}^{-1} (Tf'(\lc{\Box}^{-1}T))\right] \partial_{\sigma}T \left( g^{\sigma\rho}E_A{}^{\beta}-2g^{\beta(\rho} E_A{}^{\sigma)}\right) = \kappa^2\Theta_A{}^\beta\,.
\end{alignat}

A generalization of Eq.~\eqref{nonlocaltele} was proposed in Ref.~\cite{Bahamonde:2017sdo} where the authors introduced also the effect of the d'Alembertian operator on the boundary term $B$. Specifically, the action of the generalized teleparallel non-local theory (GTNL) has the form
\begin{align}
    \mathcal{S}_{\rm GTNL} &:= - \frac{1}{2\kappa ^2}\int \dd^4x\, e\, T + \frac{1}{2\kappa ^2} \int \dd^4x \, e \left(\xi T + \chi B \right) f\left(\lc{\square} ^{-1}T,\lc{\square} ^{-1}B\right)+\mathcal{S}_{\rm m} \,. \label{generalizednonlocaltele}
\end{align}
In the second term $\xi$ and $\chi$ are two constant coupling terms. It can be easily seen that by choosing $\xi = - \chi = -1$ one can obtain the Ricci scalar, since $\lc{R} = - T + B$. From the definition of the inverse d'Alembertian operator \eqref{Green_function} we directly see that
\begin{equation}
    \lc{\Box} ^{-1}\lc{R} = - \lc{\Box} ^{-1}T+ \lc{\Box} ^{-1}B\,.
\end{equation}
Effectively, this means that in the case when $f(\lc{\Box}^{-1}T,\lc{\Box}^{-1}B) = f(-\lc{\Box}^{-1}T + \lc{\Box}^{-1}B)$ and simultaneously $\xi = - \chi = -1$ we recover the action \eqref{Deser_Non-local}. With the equations of motion being too complicated to study the theory, it is less cumbersome to study a localized version of Eq.~\eqref{generalizednonlocaltele}, introducing scalar fields in the form of Lagrange multipliers. This method was also used for the theory \eqref{Deser_Non-local} in Ref.~\cite{Nojiri:2007uq}.

Rewriting the action \eqref{generalizednonlocaltele} in a more suitable way using four scalar fields $\phi,\psi,\theta,\zeta$, it takes the following form
\begin{subequations}
\begin{align}
	\mathcal{S}_{\rm GTNL} &= \frac{1}{2\kappa^2}\int \dd^{4}x\, e\left[-T+ \, (\xi T+\chi B)f(\phi,\varphi) +\theta (\lc{\square} \phi - T) + \zeta (\lc{\square} \varphi -B)\right] +\mathcal{S}_{\rm m} \\
	&=\frac{1}{2\kappa^2}\int \dd^{4}x\, e\Big[-T+(\xi T+\chi B)f(\phi,\varphi) - (\partial_{\mu} \theta)( \partial^{\mu} \phi) - \theta T - (\partial _{\mu} \zeta)( \partial ^{\mu} \varphi) -\zeta B\Big]+\mathcal{S}_{\rm m} \,, \label{action2}
\end{align}
\end{subequations}
where the last two terms in the first line act as Lagrange multipliers. The mixed kinetic terms of the auxiliary fields introduce ghost-like terms that cannot be gauge transformed away \cite{DeFelice:2014kma}. This would in principle lead to non-viability of the theory, however, one can choose the mass of these modes to be larger than the cut-off of the theory. A ghost-free non-local theory has also been presented in Ref.~\cite{Nojiri:2019dio}. Varying this action \gls{wrt} $\theta$ and $\zeta$ we get $\phi= \lc{\square}^{-1}T$ and $\varphi= \lc{\square}^{-1}B$ respectively, while varying the action \gls{wrt} $\phi$ and $\varphi$ we get
\begin{align}
\lc{\square} \theta &= -(\xi T + \chi B) \frac{\partial f(\phi,\varphi)}{\partial\phi}\,,\quad
\lc{\square} \zeta = -(\xi T + \chi B) \frac{\partial f(\phi,\varphi)}{\partial\varphi}\,,
\end{align}
which are the field equations for the dynamical scalar fields. The corresponding field equations for the tetrad in the Weitzenb\"{o}ck gauge read
\begin{alignat}{2}
    & \: & &(1-\xi f(\phi,\varphi)+\theta)E_A{}^\beta\lc{G}_{\beta}{}^\mu+\frac{1}{2}(\partial^\lambda\theta)(\partial_\lambda\phi)E_A{}^\beta-\frac{1}{2}(\partial^\beta\theta)(\partial_\alpha \phi)E_A{}^\alpha-\frac{1}{2}(\partial^\beta\phi)(\partial_\alpha \theta)E_A{}^\alpha\nonumber\\
    & \: & &+\frac{1}{2}(\partial^\lambda\zeta)(\partial_\lambda\psi)E_A{}^\beta-\frac{1}{2}(\partial^\beta\zeta)(\partial_\alpha \psi)E_A{}^\alpha-\frac{1}{2}(\partial^\beta\psi)(\partial_\alpha \zeta)E_A{}^\alpha+E_A{}^\rho S_{\rho}{}^{\mu\beta}\partial_\mu\Big[(\xi+\chi)f(\phi,\psi)-\theta-\zeta\Big]\nonumber\\
    & \: & &+(E_A{}^\mu \lc{\nabla}_{\mu}\lc{\nabla}^\beta-E_A{}^\beta \lc{\Box})(\zeta-\chi f(\phi,\varphi))=\kappa^2 \Theta_A{}^\beta\,.
\end{alignat}
It would be interesting to study the behavior of the theory at lower scales, e.g. in astrophysical systems, since cosmologically it seems very promising.

\subsection{Other approaches}

There are other teleparallel theories that are less popular in the literature but it is worthwhile to mention them. For example, there are some approaches to the so-called Kaluza-Klein theory in the context of \gls{tg}. The standard version of Kaluza-Klein theory formulates a 5 dimensional theory which unifies electromagnetism with gravity with the Lagrangrian being composed by a 5 dimensional Einstein-Hilbert action. In the teleparallel equivalent of Kaluza-Klein theory~\cite{deAndrade:1999vq,Barbosa:2002mg}, only the internal (or tangent) space is 5 dimensional but the spacetime remains 4 dimensional. This formulation is achieved by considering a 5 dimensional translational gauge theory which turns out to unify gravity with electromagnetism. Later, it was found that the Kaluza-Klein theory reduction of \gls{tegr} at low energies in the absence of electromagnetism introduces a non-minimally coupling between a scalar field and the torsion tensor~\cite{Geng:2014nfa}. The Lagrangian of this reduced theory becomes $\mathcal{L}=\phi T+2T^\mu\partial_\mu \phi$, which can be integrated out to find that the scalar field is non-minimally coupled to both the torsion scalar and the boundary term $B$. This reduced theory was then generalized to the case of $f(T)$ gravity in the Kaluza-Klein reduced formulation.

Another approach that has been explored in \gls{tg} is the so-called mimetic gravity. The standard version of mimetic gravity is a modified version of \gls{gr} which respects conformal symmetries as an internal \gls{dof}. This is achieved by rewriting the metric as $g_{\mu\nu}=-\tilde{g}_{\mu\nu}\tilde{g}^{\alpha\beta}\partial_\alpha\partial_\beta \phi$ where $\tilde{g}_{\mu\nu}$ is an auxiliary metric and $\phi$ is a scalar field~\cite{Sebastiani:2016ras,Chamseddine:2013kea}. This parametrization is invariant under Weyl transformations. To be consistent, one must impose that $g^{\mu\nu}\partial_\mu \partial_\nu \phi=-1$. These types of theories are usually formulated using Lagrange multipliers. For example, the simplest modification is constructed with the Ricci scalar as $\mathcal{L}=\lc{R}+\lambda(g^{\mu\nu}\partial_\mu \partial_\nu \phi+1)$. These theories have the interesting feature that they can unify dark matter with dark energy. In Ref.~\cite{Mirza:2017afs}, the authors formulated a teleparallel version of mimetic gravity by considering the Lagrangian $\mathcal{L}=f(T)+\lambda(g^{\mu\nu}\partial_\mu \partial_\nu \phi+1)$. This is achieved by considering that the tetrads are unchanged and the conformal transformations are performed in the Minkowski metric.

Last but not least, an interesting theory has been proposed in Ref.~\cite{Hohmann:2020dgy}, which is a generalization of \cite{Li:2020xjt}, where the authors study a general teleparallel theory that is a linear combination of the five scalar invariants, quadratic in the torsion tensor. Two of these scalars have odd parity and thus they allow coupling to pseudo-scalar fields, in order for the Lagrangian to be even under parity. Then, one can formulate a \gls{tg} theory which is similar to the well-known Nieh-Yan parity-violating theory~\cite{Nieh:1981ww}. In the same spirit as in gauge theories, these fields are called \textit{teleparallel axions} and they present some interesting phenomenology at cosmological scales, especially in the early Universe. Finally, a local Lorentz covariant version of \cite{Li:2020xjt} was considered in \cite{Li:2021wij} and non-flat universes were also taken into account. It was recently reported that some \gls{tg} parity-violating models suffer from ghost instabilities but the so-called modified \gls{tg} Nieh-Yan is free of them~\cite{Li:2022mti}.

\subsection{Good tetrad spin connection examples} \label{subsec:goodtetradsph}

In general, finding a tetrad spin connection pair which satisfy the antisymmetric field equations is a difficult task. In this section, we will give an overview of different good tetrad-spin connection pairs found in the literature for different spacetimes satisfying different symmetries. As discussed above, a good tetrad-spin connection pair depends on the theory, therefore, some results provided below will be also theory dependent.

\subsubsection{Static spherically symmetric case } \label{sec:spheri}

As a first glimpse to understanding how important is to choose the correct tetrad-spin connection pair, we will perform a very simple calculation to show that one needs to be very careful when one is considering teleparallel theories of gravity. Let us start by considering a static spherically symmetric metric
\begin{equation}\label{metric222}
    \dd s^2=\mathcal{A}(r)\dd t^2-\mathcal{B}(r)\dd r^2- \mathcal{M}(r)^2(\dd \theta^2+\sin\theta^2\, \dd\varphi^2)\,,
\end{equation}
and choose a tetrad which is diagonal in the Weitzenb\"ock gauge:
\begin{equation}
    e^A{}_\mu=\textrm{diag}(\sqrt{\mathcal{A}(r)},\sqrt{\mathcal{B}(r)},\mathcal{M}(r),\mathcal{M}(r)\sin\vartheta)\,.\label{tetrad00}
\end{equation}
Clearly, the above tetrad reproduces the metric~\eqref{metric222} but the teleparallel connection does not respect spherical symmetry (if the spin connection vanishes) since $(\mathcal{L}_{X_{\xi}}\Gamma)^{\mu}{}_{\nu\rho}\neq0$ (see Sec.~\ref{sec:symmetries} for more details). In principle, there is no a problem if the connection does not have the same symmetries as the tetrad. However, in this case, is it correct to choose this tetrad-spin connection pair? The answer is simple: the choice of Eq.~\eqref{tetrad00} is incompatible with a vanishing spin connection. This can be easily seen in an example. If we use this tetrad in $f(T)$ gravity~\eqref{f(T)field_equations_tetrad}, we end up with a nonvanishing antisymmetric field equation being equal to
\begin{equation}
    f_{TT}\,\frac{d}{dr}T=0\,,
\end{equation}
which means that either $f_{TT}=0$ (which is $f(T)=-T+\Lambda$) or $T=\textrm{const.}$, which again has the same dynamics as \gls{tegr} plus a cosmological constant. This small example shows that one needs to be careful on how to choose the correct tetrad-spin connection pair, otherwise the theory will be trivially \gls{gr}. This important result was totally omitted in many papers and due to this, many incorrect results can be found in the literature.

The way to circumvent this issue is to always consider a good tetrad-spin connection, which is related to the symmetries of the problem and also to solve the antisymmetric field equations of the studied theory. We will now derive the correct pair for spherical symmetry.
Let us consider the most general tetrad satisfying spherical symmetry in the Weitzenb\"{o}ck gauge where the spin connection is zero, which is given by Eq.~\eqref{eq:sphertetradwb}. In this section, we will further assume stationarity. As a first step, we will just consider the case in $f(T,B,\phi,X)$ gravity but then, one can conclude that the result can be easily extended this to other teleparallel theories. For this theory, the antisymmetric field equation are given by Eq.~\eqref{antifTB}. For the static spherically symmetric tetrad, there are only two nonvanishing equations
\begin{equation}
    W_{[tr]}\propto C_3 C_5 (f'_T+f'_B)=0\,,\quad W_{[\vartheta\varphi]}\propto C_1 C_6(f_T'+f_B')=0\,,\label{anti2}
\end{equation}
where primes are derivatives \gls{wrt} the radial coordinate, and the $C_i$ functionals are first shown in Eq.~\eqref{eq:sphertetradwb}. The easiest way to solve the system is by taking $(f_T+f_B)'=0$, but this case is only true for either $f(\lc{R})$ gravity (which trivially satisfies the antisymmetric field equations for any spacetime) or $T,B$ constants (which is equivalent to \gls{tegr} plus a cosmological constant). Since we are interested in the non-trivial cases, we can assume $(f_T+f_B)'\neq 0$. Then, to solve the system we notice that there are two possibles branches, i) $C_3=0$; ii) $C_3\neq0$. For the first branch, one uniquely solves the system for $C_3=C_6=0$, and then, due to the remaining metric \gls{dof}, without loosing generality, we can eliminate the cross terms by further setting $C_2=0$ and also setting $\mathcal{A}(r)=\sqrt{C_1(r)}, \ \mathcal{B}(r)=\sqrt{C_4(r)}$ and $C_5(r)=\xi \mathcal{M}(r)$, where $\xi=\pm 1$. Note that the sign of $C_5$ does not affect the metric but it affects the form of the tetrad. By taking all of these forms of $C_i=C_i(r)$ in Eq.~\eqref{eq:sphertetradwb}, we find that the first branch that solves the antisymmetric field equations in $f(T,B,\phi,X)$ gravity in the Weitzenb\"ock gauge reads
\begin{equation}\label{goodspher}
e^A{}_\mu=\left(
\begin{array}{cccc}
    \sqrt{\mathcal{A}(r)} & 0 & 0 & 0 \\
    0 & \sqrt{\mathcal{B}(r)} \sin \vartheta \cos \varphi &\xi\mathcal{M}(r) \cos \vartheta \cos \varphi & -\xi\mathcal{M}(r)\sin \vartheta \sin \varphi \\
    0 & \sqrt{\mathcal{B}(r)} \sin \vartheta \sin \varphi & \xi \mathcal{M}(r)\cos \vartheta \sin \varphi &\xi \mathcal{M}(r)\sin \vartheta \cos \varphi \\
    0 & \sqrt{\mathcal{B}(r)} \cos \vartheta & - \xi\mathcal{M}(r)\sin \vartheta & 0 \\
\end{array}
\right)\,,\quad \xi=\pm 1\,,
\end{equation}
which reproduces the metric in the standard form
\begin{equation}
\dd s^2= \mathcal{A}(r)\dd t^2- \mathcal{B}(r)\dd r^2 - \mathcal{M}(r)^2(\dd \vartheta^2+\sin^2\varphi\, \dd \varphi^2)\,.
\end{equation}
We can switch on the spin connection by performing the following Lorentz transformation $e'^{A}{}_\mu=\Lambda^A{}_B e^B{}_\mu$, with $\Lambda^A{}_B $ being,
\begin{equation}\label{eq:spherlt1}
\Lambda^A{}_B = \begin{pmatrix}
1 & 0 & 0 & 0\\
0 & \sin\vartheta\cos\varphi & \sin\vartheta\sin\varphi & \cos\vartheta\\
0 & \cos\vartheta\cos\varphi & \cos\vartheta\sin\varphi & -\sin\vartheta\\
0 & -\sin\varphi & \cos\varphi & 0
\end{pmatrix}\,,
\end{equation}
which gives us another good tetrad-spin connection pair which is equivalent to Eq.~\eqref{goodspher} with $\omega^A{}_{B\mu}=0$ having a diagonal tetrad
\begin{equation}\label{sphericdiag}
e'^A{}_\mu=\textrm{diag}(\sqrt{\mathcal{A}(r)},\sqrt{\mathcal{B}(r)},\xi \mathcal{M}(r),\xi\mathcal{M}(r)\sin\vartheta)\,,\quad \xi=\pm 1\,,
\end{equation}
but now having a non-zero spin connection $\omega^{A}{}_{B\mu}$ with the following non-zero components
\begin{equation}\label{eq:spiconspher1}
\omega'^{1}{}_{2\vartheta}=-\omega'^{2}{}_{1\vartheta}=-1\,,\quad \omega'^{1}{}_{3\varphi}=-\omega'^{3}{}_{1\varphi}=-\sin\vartheta\,,\quad \omega'^{2}{}_{3\varphi}=-\omega'^{3}{}_{2\varphi}=-\cos\vartheta\,.
\end{equation}
One important remark is that the Lorentz transformation must be carried out for both the tetrads and the spin connection. It is interesting to note that the scalar-torsion $T$ and the boundary term $B$ become
\begin{subequations}
\begin{alignat}{2}
T& =\: & &-\frac{2 \left(\xi \sqrt{\mathcal{B}}-\mathcal{M}'\right) \left(\mathcal{M} \mathcal{A}'+\mathcal{A} (\mathcal{M}'-\xi \sqrt{\mathcal{B}})\right)}{\mathcal{A} \mathcal{B} \mathcal{M}^2}\,,\\
B& =\: & &\frac{\mathcal{M}' \left(4 \mathcal{B} \mathcal{M} \mathcal{A}'-2 \mathcal{A} \left(\mathcal{M} \mathcal{B}'+2 \xi \mathcal{B}^{3/2}\right)\right)}{\mathcal{A} \mathcal{B}^2 \mathcal{M}^2}-\frac{\mathcal{B} \mathcal{M} \mathcal{A}'^2+\mathcal{A} \left(-2 \mathcal{B} \mathcal{M} \mathcal{A}''+\mathcal{M} \mathcal{A}' \mathcal{B}'+4 \xi \mathcal{B}^{3/2} \mathcal{A}'\right)}{2 \mathcal{A}^2 \mathcal{B}^2 \mathcal{M}}\nonumber\\
& \: & &+\frac{4 \mathcal{M}''}{\mathcal{B} \mathcal{M}}+\frac{4 \mathcal{M}'^2}{\mathcal{B} \mathcal{M}^2}\,.
\end{alignat}
\end{subequations}
 Since these quantities are related to gravitational effects, one should expect that they should vanish in the Minkowski limit. One notices that in the limit $\mathcal{A},\mathcal{B}\rightarrow 1,\,\mathcal{M}\rightarrow r$, $T=B=0$ are zero for $\xi=1$ but non-zero for $\xi=-1$. However, after redefining $\sqrt{\mathcal{A}}=\mathcal{\tilde{A}},\, \sqrt{\mathcal{B}}=\mathcal{\tilde{B}}$, one can have the possibility of obtaining the Minkowski metric for $\mathcal{\tilde{A}},\mathcal{\tilde{B}}\rightarrow -1,\,\mathcal{M}\rightarrow r$ and giving us the opposite result, which is that $T=B=0$ for $\xi=-1$ and they are non-zero for $\xi=1$. It is still not clear in the literature what is the physical interpretation of having these scalars different to zero in Minkowski. Some authors argue that the teleparallel connection is not satisfying the Minkowski symmetries and therefore, the torsion scalar is non-zero even though one assumes the metric being the Minkowski one. To avoid this issue, several studies~\cite{Bahamonde:2020vpb,Bahamonde:2020bbc,Bahamonde:2019jkf,Bahamonde:2019zea} assumed $\xi=1$. However, there are other papers ignoring this issue~\cite{Ruggiero:2015oka,Finch:2018gkh,Farrugia:2016xcw,Iorio:2015rla} and studied the $\xi=-1$ where both $T,B$ do not vanish in the Minkowski limit unless we take the limit with the negative functions as we just discussed. Interestingly, the sign of $\xi $ drastically changes the form of the field equations of $f(T,B,\phi,X)$ gravity in spherical symmetry. This will be explored further in Sec.~\ref{sec6:astrophysics}.

Let us now explore the second non-trivial branch appearing in Eq.~\eqref{anti2} which is the one when $C_3\neq0 $. For this case, the system~\eqref{anti2} is solved uniquely if $C_1(r)=C_5(r)=0$. For this branch, one needs to be careful since the metric change its signature unless we assume that the functions $C_2(r)=i \,\mathcal{B}(r)\,,\quad C_3(r)=i \,\mathcal{A}(r)$ are imaginary. Furthermore, one still has some freedom to choose $C_4=0$ to eliminate the cross term in the metric. After doing this
\begin{equation}\label{tetrad2new}
e^{A}{}_\mu=\left(
\begin{array}{cccc}
0 & i \mathcal{B}(r) & 0 & 0 \\
i \mathcal{A}(r) \sin\vartheta \cos\varphi & 0 &- \mathcal{M}(r) \sin\varphi & - \mathcal{M}(r) \sin\vartheta \cos\vartheta \cos\varphi \\
i \mathcal{A}(r) \sin\vartheta \sin\varphi & 0 & \mathcal{M}(r) \cos\varphi & - \mathcal{M}(r) \sin\vartheta \cos\vartheta \sin\varphi \\
i \mathcal{A}(r) \cos\vartheta & 0 & 0 & \mathcal{M}(r) \sin^2\vartheta \\
\end{array}
\right)\,,
\end{equation}
where we have set $C_6=\mathcal{M}(r)$. It should be noted that this tetrad is complex but the metric is real and its signature remains unchanged, i.e.,
\begin{equation}\label{metric22}
    \dd s^2=\mathcal{A}(r)^2\dd t^2-\mathcal{B}(r)^2\dd r^2- \mathcal{M}(r)^2\dd \Omega^2\,.
\end{equation}
For this tetrad the sign of $\mathcal{M}(r)$ does not play any role. Thus, the complex tetrad in~\eqref{tetrad2new} also satisfy the antisymmetric field equations. It is worth noticing that even though this quantity is complex, both the torsion scalar and the boundary term
\begin{subequations}
\begin{align}
    T&=\frac{4 \mathcal{A}' \mathcal{M}'}{\mathcal{A} \mathcal{B}^2 \mathcal{M}}+\frac{2 \mathcal{M}'{}^2}{\mathcal{B}^2 \mathcal{M}^2}+\frac{2}{\mathcal{M}^2}\,,\\
    B&=\frac{\mathcal{M}' \left(8 \mathcal{B} \mathcal{A}'-4 \mathcal{A} \mathcal{B}'\right)}{\mathcal{A} \mathcal{B}^3 \mathcal{M}}+\frac{2 \left(\mathcal{B} \mathcal{A}''-\mathcal{A}' \mathcal{B}'\right)}{\mathcal{A} \mathcal{B}^3}+\frac{4 \mathcal{M}''}{\mathcal{B}^2 \mathcal{M}}+\frac{4 \mathcal{M}'{}^2}{\mathcal{B}^2 \mathcal{M}^2}\,,
\end{align}
\end{subequations}
are real, so that, the action will be well-defined under this tetrad. It is still unclear if other important physical quantities can be complex for this tetrad. In the Minkowski limit, $T,B\neq0$ unless we take the Minkowski limit for $\mathcal{A}\,,\mathcal{B}=-1\,,\mathcal{M}=r$.

Similarly as we did above, one can perform the following local Lorentz transformation
\begin{equation}\label{eq:spherlt1B}
\Lambda^A{}_B =\left(
\begin{array}{cccc}
0 & -i \sin\vartheta\cos\varphi& -i \sin\vartheta\sin\varphi& -i \cos\vartheta \\
-i & 0 & 0 & 0 \\
0 & -\sin\varphi& \cos\varphi& 0 \\
0 & -\cos\vartheta \cos\varphi& -\cos\vartheta \sin\varphi& \sin\vartheta\\
\end{array}
\right)\,,
\end{equation}
simultaneously for the tetrad and the spin connection, to gives us a spin connection with the following non-zero components
\begin{equation}\label{spinc1}
\omega'{}^{0}{}_{2\varphi}=\omega'{}^{2}{}_{0\varphi}=i \sin\vartheta\,,\quad \omega'{}^{0}{}_{3\vartheta}=\omega'{}^{3}{}_{0\vartheta}=-i\,,\quad \omega'{}^{2}{}_{3\varphi}=-\omega'{}^{3}{}_{2\varphi}=-\cos\vartheta\,,
\end{equation}
and a diagonal tetrad in the form
\begin{equation}\label{tetraddiag1}
e^A{}_\mu=\textrm{diag}(\mathcal{A}(r),\mathcal{B}(r),\mathcal{M}(r),\mathcal{M}(r)\sin\vartheta)\,.
\end{equation}
The tetrad in~\eqref{tetrad2new} in the Weitzenb\"{o}ck gauge is equivalent as the pair~\eqref{spinc1}--\eqref{tetraddiag1}. It is straightforward to show that all the good tetrad-spin connection presented in this section for $f(T,B,\phi,X)$ gravity, also satisfy the antisymmetric field equations of more general teleparallel theories such as the generalized \gls{ngr} theory (see Sec.~\ref{sec:NGR}), and also for \gls{tegb} extensions such as the ones discussed in Sec.~\ref{Sec:GB_theories}.

\subsubsection{Time-dependent spherically symmetric case} \label{sec:time-dependent_tetrad_nonflat}

In this section, we will perform a similar analysis as in the previous section for $f(T,B,\phi,X)$ gravity but we will assume that $\phi=\phi(t),\, T=T(t),\, B=B(t)$. This means that we have $\frac{\dd}{\dd r}\,(f_T+f_B)=0$ but $\frac{\dd}{\dd t}\, (f_T+f_B)\neq 0$. This case would be important for cosmological scenarios, for example. Consider the \gls{flrw} cosmological scenario, where the cosmological principle will impose that these quantities must be time-dependent only. Using the most general time-dependent spherically symmetric tetrad~\eqref{eq:sphertetradwb}, the antisymmetric field equation~\eqref{antifTB} for this case is reduced to the following system
\begin{subequations}
\begin{align}
W_{[tr]}&\propto (\dot{f}_T+\dot{f}_B)(C_4 C_5-C_5'C_5-C_6 C'_{6})=0\,,\label{time1}\\
W_{[\vartheta\varphi]}&\propto (\dot{f}_T+\dot{f}_B)C_2 \, C_6=0 \,,\label{time2}
\end{align}
\end{subequations}
where dots and primes denote derivatives \gls{wrt} time and the radial coordinate, respectively. We have two possible branches: i) $C_6\neq 0$; ii) $C_6=0$. Let us first explore the first branch. For this case we need to impose ($f_B\neq-f_T$)
\begin{equation}
    C_4(t,r)=\frac{1}{C_5}\Big(C_5C_5'+C_6 C_6'\Big)\,,
\end{equation}
to solve Eq.~\eqref{time1}. We have further assumed $C_5\neq0$ since that branch is very restrictive and for \gls{flrw}, this case only becomes the flat \gls{flrw} case. Moreover, due to the remaining gauge freedom in the metric we can further set $C_3=0$ without loosing generality. Let us emphasise here that $T$ and $B$ were assumed to depend only on $t$ but after assuming the above assumptions, this condition is not true yet. To continue, we will focus on the non-flat \gls{flrw} case, where the functions are
\begin{subequations}
\begin{align}
C_1(t,r)&=\xi_1\, N(t)\,,\quad C_5(t,r)=\xi_2\sqrt{r^2a(t)^2-C_6(t,r)^2}\,,\\
    C_6(t,r)&=\xi_3 \sqrt{k}r^2a(t)\,,\quad \xi_1=\pm 1\,, \xi_2=\pm 1\,, \xi_3=\pm 1\,.
\end{align}
\end{subequations}
Then, for \gls{flrw} cosmology one can write down the following good tetrad in the Weitzenb\"ock gauge in $f(T,B,\phi,X)$ gravity
\begin{subequations}\label{eq:FRW1}
	\begin{alignat}{2}
	\mathbf{e}^0 & =\: & & \xi_1 N(t)\,\dd t \,,\\
\mathbf{e}^1 & =\: & & \frac{a(t)\cos\varphi\sin\vartheta}{\sqrt{1-kr^2}\xi_2} \,\dd r+r a(t) \left(- \xi_3\sqrt{k} r \sin \varphi + \xi_2\sqrt{1-k r^2}\cos \vartheta \cos \varphi \right)\dd \vartheta\nonumber\\
	& \: & & - r a(t) \sin \vartheta \left(\xi_3\sqrt{k} r \cos \vartheta \cos \varphi + \xi_2 \sqrt{1-k r^2} \sin \varphi \right)\dd\varphi\,,\\
	\mathbf{e}^2 & =\: & & \frac{a(t)\sin\varphi\sin\vartheta}{\sqrt{1-kr^2}} \,\dd r+r a(t) \left(\xi_3\sqrt{k} r \cos \varphi + \xi_2 \sqrt{1-k r^2} \cos \vartheta\sin \varphi \right)\dd \vartheta\nonumber\\
	& \: & & +r a(t) \sin \vartheta \left(-\xi_3 \sqrt{k} r \cos \vartheta \sin \varphi + \xi_2\sqrt{1-k r^2} \cos \varphi \right)\dd\varphi\,,\\
	\mathbf{e}^3 & =\: & & \frac{a(t)\cos\vartheta}{\sqrt{1-kr^2}}\, \dd r-\xi_2 a(t)r\sqrt{1-kr^2}\sin\vartheta\,\dd \vartheta +\xi_3 \sqrt{k}r^2a(t)\sin^2\vartheta\,\dd \varphi\,,
\end{alignat}
\end{subequations}
where $\xi_1=\pm 1$, $\xi_2=\pm 1$ and $\xi_3=\pm 1$. It should be noted that this result is compatible with the one found in Sec.~\ref{subsymmFLRW} where it was shown that the same above good tetrad for non-flat \gls{flrw} applies for any generic modified teleparallel theory. Independently of the sign chosen for the tetrad, the scalars $T$ and $B$ become
\begin{equation}\label{eq:Sec6_scalars}
    T=-6H^2 +\frac{6 k}{a^2}\,,\quad
    B=-18H^2-6\frac{\dot{H}}{N}\,,
\end{equation}
where we have introduced the Hubble parameter $H(t)=\dot{a}/(Ha)$. We can again perform a Lorentz transformation for the tetrad-spin connection pair \eqref{eq:FRW1} and find that under the Lorentz transformation,
\footnotesize{\begin{equation}\label{eq:diagltcosmopos}
	\Lambda^A{}_B = \begin{pmatrix}
	\xi_1^{-1} & 0 & 0 & 0\\
	0& \xi_2^{-1}\cos\varphi\sin\vartheta & \xi_2^{-1}\sin\varphi\sin\vartheta & \xi_2^{-1}\cos\vartheta\\
	0 & \xi_2\cos\vartheta \cos \varphi \sqrt{1-k r^2} -\xi_3 \sqrt{k} r \sin \varphi & \xi_2\cos\vartheta \sin \varphi \sqrt{1-k r^2} +\xi_3 \sqrt{k} r \cos \varphi & -\xi_2\sqrt{1-kr^2}\sin\vartheta \\
	0 & -\xi_2\sin \varphi \sqrt{1-k r^2} - \xi_3\sqrt{k} r \cos\vartheta\cos \varphi& \xi_2\cos \varphi \sqrt{1-k r^2} -\xi_3 \sqrt{k} r \cos \vartheta \sin\varphi & \xi_3\sqrt{k}r\sin\vartheta
	\end{pmatrix}\,.
	\end{equation}}\normalsize
the tetrad~\eqref{eq:FRW1} (independently of the sign) becomes diagonal
\begin{equation}
e'^A{}_\mu=\textrm{diag}\Big(N(t),\frac{a(t)}{\sqrt{1-kr^2}},a(t)r,a(t)r\sin\vartheta\Big)\,,\label{FRWdiag}
\end{equation}
and the spin connection $\omega'{}^{A}{}_{B\mu}$ have the following non-zero components,
\begin{subequations}
\begin{align}\label{FRWdiagw}
\omega'{}^1{}_{2\vartheta} 	&=-\omega'{}^2{}_{1 \vartheta} =- \sqrt{1-kr^2}\,,\quad \omega'{}^1{}_{3 \varphi} =-\omega'{}^3{}_{1 \varphi} = \sqrt{1-kr^2}\sin\vartheta\,,\\
 \omega'^1{}_{3 \vartheta} &=-\frac{}{}\omega'^3{}_{1 \vartheta} = \frac{\xi_3}{\xi_2} \sqrt{k}r\,, \quad \omega_{\pm}'^2{}_{1 \varphi} 	=-\omega'^1{}_{2 \varphi} =\xi_2 \xi_3 \sqrt{k}r\sin\vartheta\,,\\
\omega'^2{}_{3 r} 	&=-\omega'^3{}_{2 r} =\xi_2\xi_3\frac{\sqrt{k}}{\sqrt{1-k r^2}}\,, \quad \omega'^2{}_{3 \varphi} 	=-\omega'^3{}_{2 \varphi} =-\cos\vartheta\,.
\end{align}
\end{subequations}
Note that we have used $\xi_i=1/\xi_i$ since $\xi_i=\pm 1$. As we have remarked before, the pair \eqref{eq:FRW1} with zero spin connection is identical to the pair \eqref{FRWdiag} with non-zero spin connection~\eqref{FRWdiagw}, so both pairs are good tetrad-spin connection pair which solve the antisymmetric field equations for $f(T,B,\phi,X)$ gravity while respecting the symmetries.

Let us now explore the second branch (ii) where $C_6=0$. For this case, the first antisymmetric equation~\eqref{time1} will be zero if $C_4=C_5'$. To further express the metric in its diagonal form, we must take $C_1=C_3 C_5'/C_2$ with $C_2\neq0$ (the case $C_2=0$ is also more restrictive).
Furthermore, by taking the \gls{flrw} example, we need to set
\begin{equation}
C_2(t,r)=\frac{\xi_3\sqrt{k}r\,a(t)}{\sqrt{kr^2-1}}\,,\quad C_3(t,r)= \xi_1\sqrt{-k}\, r N(t)\,,\quad C_5=\xi_2 r\,a(t)\,,\quad \xi_1=\pm\,, \xi_2=\pm 1 \,, \xi_3=\pm 1\,.
\end{equation}
Thus, another good tetrad-spin connection for \gls{flrw} cosmology is given by
\begin{subequations}\label{eq:FRW2}
\begin{alignat}{2}
	\mathbf{e}^0 & =\: & & \frac{\xi_1\xi_2}{\xi_3}N(t)\sqrt{1-k r^2}\,\dd t+\xi_3\frac{\sqrt{-k}ra(t)}{\sqrt{1-kr^2}} \, \dd r \,,\\
	\mathbf{e}^1 & =\: & & \xi_1 N(t)\sqrt{-k} r \sin \vartheta \cos \varphi \, \dd t+\xi_2 a(t) \sin \vartheta \cos \varphi \,\dd r+\xi_2 ra(t) \cos \vartheta \cos \varphi \,\dd \vartheta\nonumber\\
	& \: & & -\xi_2 r a(t) \sin \vartheta \sin \varphi \, \dd\varphi\,,\\
	\mathbf{e}^2 & =\: & & \xi_1 N(t)\sqrt{-k} r \sin\vartheta \sin \varphi \,\dd t+\xi_2 a(t) \sin \vartheta \sin \varphi \,\dd r+\xi_2 r a(t) \cos \vartheta \sin \varphi \, \dd \vartheta\nonumber\\
	& \: & & +\xi_2 r a(t) \sin \vartheta \cos \varphi \, \dd\varphi\,,\\
\mathbf{	e}^3 & =\: & & \xi_1 N(t)\sqrt{-k} r \cos \vartheta \, \dd t+\xi_2 a(t) \cos \vartheta \, \dd r-\xi_2 r a(t) \sin \vartheta \,\dd \vartheta \,,
\end{alignat}
\end{subequations}
with a zero spin connection, which has the scalars given by
\begin{equation}
T=-6H^2+ 12\xi_1\xi_2\frac{ \sqrt{-k} H}{a }+\frac{6 k}{a^2}\,,\quad B=-18H^2-\frac{6\dot{H}}{N}+\frac{12\xi_1}{\xi_2}\frac{\sqrt{-k} H}{a}\,.
\end{equation}
This tetrad reproduces the standard non-flat \gls{flrw} metric and is real only for $k=-1$ or $k=0$. For the $k=1$ case, the scalars become complex (as explained in~\ref{sec:negativsym}). It is easy to see that both tetrads~\eqref{eq:FRW1} and \eqref{eq:FRW2} are equivalent for the flat case $k=0$. There is one important difference in this tetrad which is that depending on the sign chosen of the tetrad (where $\xi_1=\pm 1$ and $\xi_2= \pm 1$), one could have different $T$ and $B$. Finally, for the above tetrad in \gls{flrw} cosmology~\eqref{eq:FRW2}, one can perform the following Lorentz transformation
\begin{equation}\label{eq:diagltcosmoneg}
\Lambda^A{}_B = \begin{pmatrix}
\frac{\xi_1 \xi_2}{\xi_3}\sqrt{1-k r^2} & -\xi_1 \sqrt{-k}\,r\sin\vartheta\cos\varphi & -\xi_1 \sqrt{-k}\,r\sin\vartheta\sin\varphi & -\xi_1 \sqrt{-k}\,r\cos\vartheta\\
-\xi_3 \sqrt{-k}\, r & \xi_2\sqrt{1-k r^2}\sin\vartheta\cos\varphi & \xi_2\sqrt{1-k r^2}\sin\vartheta\sin\varphi & \xi_2\sqrt{1-k r^2}\cos\vartheta\\
0 & \xi_2\cos\vartheta\cos\varphi & \xi_2\cos\vartheta\sin\varphi & -\xi_2\sin\vartheta\\
0 & -\xi_2\sin\varphi & \xi_2\cos\varphi & 0
\end{pmatrix}\,,
\end{equation}
to get the same diagonal \gls{flrw} tetrad~\eqref{FRWdiag} with the following non-zero components for the spin connection
\begin{subequations}
\begin{align}
\omega_{\pm}'^0{}_{1r} 	&=\omega_{\pm}'^1{}_{0r} = \xi_1 \xi_2 \frac{\sqrt{-k}}{\sqrt{1-kr^2}}\,,\quad \omega_{\pm}'^0{}_{2\vartheta} =\omega_{\pm}'^2{}_{0\vartheta} = \frac{\xi_1}{\xi_2}\sqrt{-k}r\,,\quad \omega_{\pm}'^0{}_{3\varphi} =\omega_{\pm}'^3{}_{0\varphi} = \frac{\xi_1}{\xi_2} \sqrt{-k}r\sin\vartheta\,,\\
\omega_{\pm}'^1{}_{2\vartheta} 	&=-\omega_{\pm}'^2{}_{1\vartheta} = -\sqrt{1-kr^2}\,,\quad \omega_{\pm}'^1{}_{3\varphi} 	=-\omega_{\pm}'^3{}_{1\varphi} = -\sqrt{1-kr^2}\sin\vartheta\,,\quad \omega_{\pm}'^2{}_{3\varphi} 	=-\omega_{\pm}'^3{}_{2\varphi} =-\cos\vartheta\,.\nonumber\\ \label{eq:spiconcosmoneg}
\end{align}
\end{subequations}
Therefore, the above spin connection with the diagonal tetrad components~\eqref{FRWdiag} are also a good tetrad-spin connection pair. Let us remark here that the scalars $T$ and $B$ (or any other constructed from the torsion tensor) will have the same value after making the Lorentz transformation since both the spin connection and the tetrad compensates the transformation.

It is also important to mention that as we have pointed out before, the good tetrad-spin connection pair is field equation dependent. However, for the \gls{flrw} cosmological pair found in this section, it has been proved in Ref.~\cite{Hohmann:2019nat} that both \eqref{eq:FRW1} and \eqref{eq:FRW2} with zero spin connection are good tetrad-spin connection pairs valid for any modified teleparallel theory of gravity. This is not the same for the static spherically symmetric good tetrad-spin connection pair~\eqref{goodspher} which may not be a good pair for some specific theories. Depending on the teleparallel theory, one must check if this pair is good or not.

\subsubsection{Good tetrad-spin connection pair: axial symmetry} \label{sec:axials}

The situation is very complicated in axial symmetry since it is hard to find a good tetrad-spin connection pair in modified \gls{tg} due to the impact of the time dependence. This pair is really essential in performing calculations in axial symmetry in modified \gls{tg}. There are some works in \gls{tegr} such as in Ref.~\cite{Gonzalez:2011dr}, but they are trivial since in this case any choice would be a good tetrad-spin connection pair (all would be identical to \gls{gr} in the end). As was discussed in Sec.~\ref{sec:axialsymmetry}, there are two different branches solving the symmetry condition which respects axial symmetry, which we labelled as the regular branch and the solely axially symmetric branch. In Ref.~\cite{Bejarano:2014bca}, the authors studied the Kerr case in $f(T)$ gravity and they found a good tetrad-spin connection pair such that for this spacetime $T=0$. This tetrad choice is also consistent with the fact that the Ricci scalar is $\lc{R}=0$ for the Kerr geometry, and then solves the antisymmetric and the symmetric field equations in $f(T)$ gravity. This tetrad is part of the solely axially symmetric branch, so that, it cannot respect spherical symmetry in the teleparallel point of view in any limit. This means that the metric can become spherically symmetric (for the Schwarzschild limit) but the teleparallel connection cannot become spherically symmetric, meaning that $\mathcal{L}_{X_\zeta}\Gamma^\sigma{}_{\mu\nu}\neq 0$. The failure of this condition for this tetrad is manifest in the nonvanishing component of the torsion tensor $T^{\varphi}{}_{\vartheta \varphi}=\cot\vartheta$. Moreover, the tetrad found in Ref.~\cite{Bejarano:2014bca} is time-dependent whereas the Kerr geometry is known to be static. Another point to remark about this work is that $T=0$ but the three scalars $T_{\rm vec}, T_{\rm axi}$ or $T_{\rm ten}$ are not vanishing. Then, only in $f(T,B)$ gravity, this tetrad will solve the antisymmetric field equations and this occurs in a trivial way since $T=B=0$ generates a theory which is equivalent to \gls{gr} plus a cosmological constant.

The first non-trivial known good tetrad-spin connection pair for a more general axial symmetric case was derived in Ref.~\cite{Jarv:2019ctf} for $f(T,\phi)$ gravity, but it can be proved that this pair is also a good pair for many \gls{tg} theories (similarly as the pair found before in spherical symmetry). This solution does not contain any Pleba\'{n}ski–Demia\'{n}ski spacetime as special cases (except from Schwarzschild). Later, in Ref.~\cite{Bahamonde:2020snl} the authors analyzed in detail the possibility of finding good tetrad-spin connection pairs in axial symmetry for a very general theory known as $f(T,B,\phi,X)$ gravity which includes the boundary term and a scalar field with its kinetic term. They focused mainly in the regular branch in order to obtain a smooth limit from axial symmetry to spherical symmetry. First, they started with the tetrad~\eqref{eq:axtetradwb} in the Weitzenb\"{o}ck gauge satisfying axial symmetry that can be also written in the following form
\small{\begin{equation}
e^A{}_\mu = \left(
\begin{array}{cccc}
H_{00} & H_{01} & -H_{02} & H_{03} \\
H_{10}\cos\varphi-H_{20}\sin\varphi &H_{11}\cos\varphi - H_{21}\sin\varphi & H_{22}\sin\varphi +H_{12}\cos\varphi &H_{13}\cos\varphi -H_{23}\sin\varphi \\
H_{10}\sin\varphi +H_{20}\cos\varphi & H_{11}\sin\varphi +H_{21} \cos\varphi & H_{12}\sin\varphi -H_{22}\cos\varphi & H_{13}\sin\varphi +H_{23}\cos\varphi \\
H_{30} & H_{31} & -H_{32} & H_{33} \\
\end{array}
\right)\,,
\label{eq: general axial tetrad}
\end{equation}}\normalsize
where $C^A{}_{\mu}$ where replaced by $H_{ij}$ and they denote 16 different functions which depend only on $t,r,\vartheta$. The metric reproduced by the above tetrad contains all the possible cross terms $(\dd t,$ $\dd r,$ $\dd t,$ $\dd \varphi,$ $\dd t,$ $\dd \vartheta,$ $\dd r,$ $\dd \varphi,$ $\dd r,$ $\dd \vartheta,$ $\dd \vartheta,$ $\dd \varphi)$. Then, there is some gauge freedom related to the metric. For example, one can choose a gauge such that only the $\dd t\, \dd \varphi $ cross term differs from zero. This choice is the standard one chosen for the Kerr metric in Boyer-Lindquist coordinates. The authors in Ref.~\cite{Bahamonde:2020snl} studied the stationary case and set
\begin{equation}\label{eq:H1choice}
    H_{01}=H_{02}=H_{20}=H_{33}=H_{30}=H_{10}=H_{21}=H_{22}=H_{13}=0\,,\quad H_{31}=\frac{H_{11}H_{12}}{H_{32}}\,,
\end{equation}
giving the following metric
\begin{equation}\label{metric0}
    \dd s^2=H_{00}^2 \dd t^2-H_{11}^2 \left(\frac{H_{12}^2}{H_{32}^2}+1\right)\dd r^2-(H_{12}^2+H_{32}^2)\dd \vartheta^2-(H_{23}^2-H_{03}^2)\dd \varphi^2+2 H_{00} H_{03}\dd t\, d\varphi\,.
\end{equation}
This choice was inspired by the spherically symmetric good tetrad in the Weitzenb\"{o}ck gauge~\eqref{goodspher} since the the tetrad~\eqref{eq: general axial tetrad} with \eqref{eq:H1choice} gives us
\begin{equation}
    e^A{}_\mu = \left(
    \begin{array}{cccc}
    H_{00} & 0 & 0 &H_{03} \\
    0 & H_{11}\cos \varphi & H_{12}\cos \varphi & -H_{23}\sin \varphi \\
    0 & H_{11}\sin \varphi & H_{12}\sin \varphi & H_{23}\cos \varphi \\
    0 & H_{11}H_{12}/H_{32} & -H_{32} & 0 \\
    \end{array}
    \right)\,,\label{tetradaxial}
\end{equation}
which reproduces~\eqref{goodspher} in the limit
\begin{subequations}
\begin{align}\label{spherical}
H_{00} &= \sqrt{\mathcal{A}(r)}\,, \quad H_{11}=\sqrt{\mathcal{B}(r)}\sin\vartheta\,,\quad H_{12}=\sqrt{\mathcal{C}(r)}\cos\vartheta\,, \quad \\
H_{23} & = H_{32}=\sqrt{\mathcal{C}(r)}\sin\vartheta\,,\quad H_{03}=0\,.
\end{align}
\end{subequations}
If we assume stationarity and replace~\eqref{tetradaxial} in the $f(T,B,\phi,X)$ field equations, one notices that there is only one nonvanishing antisymmetric equation, that reads
\begin{equation}
    W_{[r\vartheta]}=\frac{1}{2}\left[ (f_{T,\vartheta}+f_{B,\vartheta}) Q_\vartheta + (f_{T,r}+f_{B,r}) Q_r\right]=0\,,\label{antiGG}
\end{equation}
where commas denote differentiation and
\begin{eqnarray}
Q_{\vartheta}=\frac{H_{00,r}}{H_{00}}-\frac{H_{11}}{H_{23}}+\frac{H_{23,r}}{H_{23}}\,,\quad Q_{r}=\frac{H_{12}-H_{23,\vartheta}}{H_{23}}-\frac{H_{00,\vartheta}}{H_{00}}\,.
\end{eqnarray}
The tetrad~\eqref{tetradaxial} contains one additional free function compared to the metric~\eqref{metric0}. This extra free function can be found by solving the remaining antisymmetric field equation~\eqref{antiGG}. Thus, in order to find a good tetrad in the Weitzenb\"{o}ck gauge, one would need to solve Eq.~\eqref{antiGG} for one of the six functions $H_{ij}$. However, doing this in general is a very complicated task since $T$ and $B$ become very cumbersome in axial symmetry. In order to tackle this problem, the authors in Ref.~\cite{Bahamonde:2020snl} studied four different particular cases where they were able to find good tetrads. The most important good tetrads in the Weitzenb\"{o}ck gauge found are:
\begin{enumerate}
	\item \textbf{Case ($Q_r=Q_\vartheta=0$):} A good tetrad that does not contain any Pleba\'{n}ski–Demia\'{n}ski spacetime which is valid for any form of $f(T,B,\phi,X)$, and also generalizes the tetrad found in Ref.~\cite{Jarv:2019ctf}, which reads
\footnotesize{\begin{align}\label{goodtetradaxial1}
e^A{}_\mu =
 \left(
\begin{array}{cccc}
H_{00} & 0 & 0 &H_{03} \\
0 & \left(\frac{H_{00,r}H_{23}}{H_{00}}+H_{23,r}\right)\cos \varphi & \left(\frac{H_{00,\vartheta}H_{23}}{H_{00}}+H_{23,\vartheta}\right)\cos \varphi & -H_{23}\sin \varphi \\
0 & \left(\frac{H_{00,r}H_{23}}{H_{00}}+H_{23,r}\right)\sin \varphi & \left(\frac{H_{00,\vartheta}H_{23}}{H_{00}}+H_{23,\vartheta}\right)\sin \varphi &H_{23} \cos \varphi \\
0 &H_{32}^{-1}\left(\frac{H_{00,\vartheta}H_{23}}{H_{00}}+H_{23,\vartheta}\right) \left(\frac{H_{00,r}H_{23}}{H_{00}}+H_{23,r}\right) & -H_{32} & 0 \\
\end{array}
\right) \,.\nonumber \\
\end{align}}\normalsize
\item \textbf{Case ($Q_r=0$ and $f_{T,\vartheta}+f_{B,\vartheta} = 0$):} A good tetrad behaving as a family of Taub-NUT-like spacetimes which needs $\phi=\phi(r)$
\begin{equation}
e^A{}_\mu=\left(
\begin{array}{cccc}
    \sqrt{\mathcal{A}(r)} & 0 & 0 & \sqrt{\mathcal{A}(r)} \left(C_2+C_1\cos \vartheta\right) \\
    0 & \sqrt{\mathcal{B}(r)} \sin \vartheta \cos \varphi & \sqrt{\mathcal{C}(r)}\cos \vartheta \cos \varphi & - \sqrt{\mathcal{C}(r)}\sin \vartheta\sin \varphi \\
    0 & \sqrt{\mathcal{B}(r)} \sin \vartheta \sin \varphi & \sqrt{\mathcal{C}(r)}\cos \vartheta \sin\varphi & \sqrt{\mathcal{C}(r)}\sin \vartheta \cos \varphi \\
    0 & \sqrt{\mathcal{B}(r)} \cos \vartheta & -\sqrt{\mathcal{C}(r)}\sin \vartheta & 0 \\
\end{array}
\right)\,,\label{tetradgoodTAUB}
\end{equation}
where $C_1$ and $C_2$ are constants. The corresponding metric is then
\begin{alignat}{2}\label{metrictaublike}
    \dd s^2 & =\: & & \mathcal{A}(r)\dd t^2 -\mathcal{B}(r)\dd r^2- \mathcal{C}(r)\dd\vartheta^2- \left[\mathcal{C}(r)\sin ^2\vartheta -\mathcal{A}(r) \left(C_1 \cos \vartheta +C_2\right)^2\right]\dd\varphi^2\nonumber\\
    & \: & &+\,2 \mathcal{A}(r) \left(C_1 \cos \vartheta +C_2\right)\dd t\,,
\end{alignat}
so that, the Taub-NUT spacetime is recovered by setting
\begin{equation}
    \mathcal{A}(r)=1/\mathcal{B}(r)=\frac{(r-r_{+})(r-r_{-})}{r^2+b^2}\,, \quad \mathcal{C}(r)=r^2+b^2\,,\quad \mathcal{D}(r,\vartheta)=2b \cos\vartheta\,,
\end{equation}
where $r_{\pm}=M\pm\sqrt{M^2+b^2}$ and $b$ is the Taub-NUT parameter and $M$ the mass. It should be noted that when $C_2=C_1=0$, this tetrad exactly matches the spherically symmetric good tetrad~\eqref{goodspher}.
\item \textbf{Case ($Q_\vartheta=0$ and $f_{T,r}+f_{B,r} = 0$):} A good tetrad having a metric which is always axially symmetric (unless a trivial case is considered) needing that $\phi=\phi(\vartheta)$
\begin{equation}\label{goodtetradaxial3}
e^A{}_\mu=\left(
\begin{array}{cccc}
\mathcal{B}(\vartheta) & 0 & 0 &\mathcal{B}(\vartheta) \\
0 &\mathcal{D}_1(\vartheta)\mathcal{D}_{2}'(r) \cos \varphi & 0 & -\mathcal{D}_1(\vartheta) \mathcal{D}_2(r) \sin \varphi \\
0 &\mathcal{D}_1(\vartheta)\mathcal{D}_{2}'(r) \sin \varphi & 0 &\mathcal{D}_1(\vartheta) \mathcal{D}_2(r) \cos \varphi \\
0 & 0 & -\mathcal{C}(\vartheta) & 0 \\
\end{array}
\right)\,.
\end{equation}
\item \textbf{Case ($(f_{T,r}+f_{B,r}) Q_r \neq0$ and $(f_{T,\vartheta}+f_{B,\vartheta})Q_{\vartheta}\neq0$):} A good tetrad in $f(T,B)$ gravity for a slowly rotating Kerr spacetime which is given by the tetrad~\eqref{tetradaxial} with
\begin{subequations}
\begin{align}
    H_{00}&=\sqrt{1-\frac{2Mr}{\Sigma}}\,,\ \ H_{11}=\frac{H_{32}}{\sqrt{\Delta}}\,, \ \ H_{12}=\sqrt{\Sigma-H_{32}^2}\,,\label{kerrA}\\
    H_{23}&=\sqrt{\sin^2\vartheta\left(\frac{2a^{2}Mr\sin^{2}\vartheta}{\Sigma}+a^2+r^2\right)+H_{03}^2}\,,\\
    H_{03}&=-\frac{2aMr\sin ^2\vartheta}{\sqrt{\Sigma(\Sigma-2Mr)}} \,,\label{kerrB}\\
    H_{32}(r,\vartheta) &= r \sin\vartheta+a^2\, \mathcal{A}(r,\vartheta)+\mathcal{O}(a^4)\,,\\
    \mathcal{A}(r,\vartheta)&=\frac{\sin \vartheta \cos ^2\vartheta \left(4 \mu ^5+6 \mu ^2 r^{3/2}+r^{5/2}+4 \mu r^2+\mu ^4 \sqrt{r}-16 \mu ^3 r\right)}{2 \mu ^2 r^{3/2} \left(-\mu ^2-4 \mu \sqrt{r}+r\right)}\,,
\end{align}
\end{subequations}
where $\Sigma=r^2+a^2\cos^2\vartheta,\, \Delta=r^2-2M r+a^2,\ \mu=\sqrt{r-2M}$ and $a$ is the angular momentum (per unit mass) which is assumed to be small ($a\ll 1$)\,. Note that $H_{00}, H_{11},H_{23}$ and $H_{03}$ need to be expanded up to second order in $a$.
\end{enumerate}
All the above tetrads were computed in the Weitzenb\"{o}ck gauge but one can also perform a local Lorentz transformation and then one could have a non-zero spin connection. The three first cases solve all the antisymmetric field equations for $f(T,B,\phi,X)$ gravity whereas the last case is only valid for $f(T,B)$ gravity. These results can be used to further analyze the symmetric field equations and explore axial symmetry in modified \gls{tg}. For example, as was done in Ref.~\cite{Bahamonde:2020bbc} for Schwarzschild, the slowly rotating Kerr good tetrad can be used to find perturbative teleparallel modifications of slowly rotating axially symmetric solutions around Kerr.

\subsubsection{Cylindrical and other symmetries}

There are few works in \gls{tg} dealing with other symmetries such as cylindrical or planar ones. There are no works dealing with the most general form of the tetrad respecting these symmetries. In Ref.~\cite{Houndjo:2012sz}, the authors found that for cylindrical symmetries (stationary case), it is convenient to work with Weyl-type coordinates $x^\mu=(t,r,\varphi,z)$ such that the metric can be written as
\begin{equation}
    \dd s^2=e^{2u(r)}\dd t^2-e^{2(k(r)-u(r))}\dd r^2-w(r)^2e^{-2u(r)}\dd \varphi^2-e^{2(k(r)-u(r))}\dd z^2\,,
\end{equation}
where $u(r), k(r)$ and $w(r)$ are arbitrary functions depending only on the radial coordinate. It can be shown that a good tetrad-spin connection pair for $f(T,B)$ gravity is the one in which the spin connection vanishes and has the following diagonal tetrad
\begin{equation}
    e^A{}_\mu=\textrm{diag}\Big(e^{u(r)},e^{k(r)-u(r)},w(r) e^{-u(r)},e^{k(r)-u(r)}\Big)\,,\label{cylindrical}
\end{equation}
which gives the following scalar-torsion and boundary term
\begin{subequations}
\begin{align}
    T&=\frac{e^{2 u-2 k}}{w} \left(2 k' w'-2 w u'^2\right)\,,\\
    B&=\frac{2 e^{2 u-2 k} }{w}\left(w \left(k''-u''\right)+w' \left(k'-u'\right)+w''\right)\,.
\end{align}
\end{subequations}
Here primes denotes derivatives \gls{wrt} $r$. It can be shown that the tetrad~\eqref{cylindrical} satisfied the antisymmetric equations for $f(T,B)$ gravity and also for other more general theories. For the \gls{gr} case in vacuum ($f=-T$), one finds the following solutions
\begin{equation}
    w(r)=c_0+c_1r\,,\quad k(r)=\frac{c_2}{c_1}\log(c_0+c_1 r)\,,\quad u(r)=\pm \sqrt{\frac{c_2}{c_1}}\log(c_0+c_1r)\,,
\end{equation}
which after making some coordinate transformations, it can be recast as the well known Levi-Civita cylindrical solution~\cite{Stephani:2003tm}. In Ref.~\cite{Houndjo:2012sz}, the authors found perturbed solutions for torsion squared gravity behaving similarly as the above metric. In Ref.~\cite{Nurbaki:2020dgw}, the authors used the Noether's symmetry approach to find solutions for $f(T)$ gravity but they are just written implicitly and not in an exact form for the metric coefficients. In Refs.~\cite{Jawad:2015eey,Jawad:2018oze}, the authors studied the collapse for cylindrical stars but they unfortunately, used a dynamical cylindrical tetrad that is not compatible with the choice of setting the spin connection to zero, and then, the antisymmetric equations presented there are not satisfied.

Other more exotic symmetries such as planar ones have also been explored in the literature. In Ref.~\cite{Rodrigues:2013uua}, these geometries was studied in $f(T)$ gravity. It can be shown that for the coordinates $x^\mu=(t,r,x,y)$, a diagonal tetrad of the form $e^A{}_\mu=\textrm{diag}(\mathcal{A}(r),\mathcal{B}(r),\mathcal{C}(r),\mathcal{C}(r))$ which reproduces a planar metric $\dd s^2=\mathcal{A}(r)^2\dd t^2-\mathcal{B}(r)^2\dd r^2-\mathcal{C}(r)^2(\dd x^2+\dd y^2)$, is compatible with the choice of a zero spin connection, leading to the conclusion that~\cite{Rodrigues:2013uua} correctly chose the tetrad ensuring that the antisymmetric field equations in $f(T)$ are satisfied. In this paper, in the end, they only studied the case of \gls{gr} since they always chose either $T=0$ or $T=\textrm{const}$. This is the reason why they found the same exact solutions known in \gls{gr} for these geometries which are the M\o{}ller, Kattler-Wittaker and planar de-Sitter
solutions.

\clearpage

\section{Cosmology in Teleparallel Gravity}\label{sec:cosmology_in_TG}

The relatively recent observational discovery of the accelerating expansion of the Universe~\cite{Riess:1998cb,Perlmutter:1998np} has led to an intense few decades of theoretical and observational work to determine the origin and properties of the so-called \textit{dark energy} (\gls{de}). In terms of numerical analysis, the most consistent model that fulfil these requirements is $\Lambda$\gls{cdm} where \textit{cold dark matter} (\gls{cdm}) acts as a gravitational well in galaxies while a cosmological constant $\Lambda$ dominates on cosmological scales, and \gls{gr} acts at all scales as the fundamental theory of gravitation. However, theoretically this has led to numerous problems \cite{Bull:2015stt} which may ultimately be addressed by revisiting this approach to gravity. By investigating possible cosmology beyond the concordance model, we may design a better formulation of gravitation that requires less physics beyond the standard model of particle physics in the dark matter sector, as well as having a more nature emergence of late-time acceleration in the Universe, and possibly resolving some issues of the inflationary epoch and its origins.

Cosmology is dominated by the two assumptions of homogeneity and isotropy which are supported to a high degree by unprecedented observational evidence \cite{dodelson2003modern,Aghanim:2018eyx} resulting in the widely accepted cosmological principle \cite{peacock1999cosmological}. This has led to the well known \gls{flrw} metric which is a universal solution in cosmology. Considering its form in cosmic time $t$ gives it the 4 dimensional representation
\begin{equation} \label{eq:frwN}
    \dd s^2 = N(t)^2\dd t^2-a(t)^2\Big[\frac{\dd r^2}{1-kr^2}+r^2(\dd \vartheta^2+\sin^2\vartheta \dd \varphi^2)\Big]\,,
\end{equation}
where $N(t)$ and $a(t)$ represent the lapse function and scale factor respectively. In many cases, the lapse function is found to not be dynamical, to determine this feature the metric \ref{eq:frwN} has to be substituted into the dynamical equations to determine is the lapse function plays a role in the equations of motion of the theory. In those cases where the lapse function is found to be nondynamical, it can be absorbed into the definition of cosmic time. Another important property to point out is that this is being defined in spherical polar coordinates in which the Lorentz frame is not represented by the Minkowski metric which directly leads to a nonvanishing appearance of the spin connection components, as explained in Sec.~\ref{subsec:goodtetradsph}. Finally, the curvature parameter $k$ corresponds to open ($k=-1$), flat ($k=0$), or closed ($k=+1$) cosmologies.

As we can notice from Eq.~\eqref{eq:frwN}, the scale factor $a(t)$ is a function of time $t$, where $a(t_0) = a_0$ at the present time $t_0$. It is standard to use the redshift $z \equiv (a_0/a) - 1$ as a proxy for the age or scale factor. The redshift can be measured for distant sources; it is the fractional amount by which the wavelength of a photon has been stretched by the expansion between the time the photon is emitted and the time it is received. The expansion rate $H \equiv \dot{a}/(Na)$ is a function of time, with the value $H_0$ as the Hubble-Lema\^{i}tre constant and where the dot denotes a derivative \gls{wrt} $t$. The cosmological deceleration parameter is then $q \equiv - (\ddot{a}/a)/H^2 = (1 + z)\dot{H}/H - 1$, where dots refer to derivatives with cosmic time.

Once this is considered, a \gls{gr} equation of motion (Friedmann and acceleration equations) for $a(t)$ for a flat universe filled with fluids \textit{i} (e.g., nonrelativistic matter, radiation, and dark energy) of energy densities $\rho_i$ can be written as
\begin{subequations}
\begin{align}
    H^2 &=\left(\frac{\dot{a}}{a}\right)^2 = \frac{8\pi G}{3} \rho\,, \\[0.5ex]
    \frac{\ddot{a}}{a} &= -4\pi G (\rho + 3 p)\,.
\end{align}
\end{subequations}
In these equations $\rho$ and $p$ are the total density and pressure and dots represent differentiation \gls{wrt} cosmic time. Hence, they can be written as sums of the contributions of the individual components as $\rho\equiv \sum_{i} \rho_i$ and $p\equiv \sum_{i} p_i$. If the fluids have pressures $p_i$, then the change $\dd (\rho a^3)$ in the total energy $(\rho = \sum_i \rho_i)$ per comoving volume is equal to the work $-p \dd (a^3)$, where $p = \sum_i p_i$, done by the fluid. If we define equation-of-state parameters $w_i \equiv p_i/\rho_i$ (e.g., $w_{\rm m} = 0$ for matter and $w_{\rm r} = 1/3$ for radiation), then the second form of the Friedmann equation can be written as $q_0 = (1+3w_{\rm Tot})/2$, where $w_{\rm Tot} \equiv \sum_i p_i/\rho_i$ is the total equation-of-state parameter. Thus, if \gls{gr} is correct, the observations require that the Universe has $w_{\rm Tot} < - 1/3$. Thus, some \gls{de}, a negative-pressure fluid, is postulated to account for cosmic acceleration. Another way to represent this, and other, contributions is through the density parameter which we define in terms of the critical density within the $\Lambda$\gls{cdm} Friedmann equations, namely $\rho_c = 3H^2/8\pi G$ (where we retain SI units for convenience). We can then define density parameters for each contribution $\rho_i$ as $\Omega_i \coloneqq \rho_i/\rho_c$. 

An analogous description can be prescribed for certain \gls{tg} theories of gravity, those which \gls{flrw} equations can be recast in a \gls{gr}-like format, such as in $f(T)$ and $f(T,B)$ gravity. When this formulation is permitted, the gravitational source acts as an effective fluid (with energy density $\rho_{\text{eff}}$ and pressure $p_{\text{eff}}$) with \gls{eos} $w_{\text{eff}}$. Consequently, the Universe's acceleration is achieved when the total \gls{eos} $w_{\rm Tot} = \frac{p + p_{\text{eff}}}{\rho + \rho_{\text{eff}}} < -1/3$.

Moreover, the overwhelming historical evidence has pointed to a flat universe \cite{Melchiorri:1999br,Hinshaw:2012aka,Aiola:2020azj,Balkenhol:2021eds} with further support from inflationary theories \cite{Linde:1981mu,Guth:1980zm}. However, some recent observations have prompted a reexamination of the closed universe model \cite{Aghanim:2018eyx,DiValentino:2019qzk,Handley:2019tkm,DiValentino:2020hov}, which provides a substantial improvement of the fit of early-time data, but is in disagreement with external additional data sets as acoustic oscillations seen in cosmic microwave background anisotropy data.

Saying that, a non-flat cosmology would require a radical overhaul of our standard approaches to cosmology and the interpretation of data and so should be taken with caution. In the context of \gls{tg}, the vast majority of works have been based on a flat cosmological background while some works have investigated the possibility of a non-flat cosmology, as will be explored in this section.

Considering again the general metric in Eq.~\eqref{eq:frwN} which describes a universe that observes an isotropic and homogeneous cosmology. The tetrads feature a branching effect in which the positive curvature parameter ($k=+1$) can be described by the good tetrad in Eq.~\eqref{eq:FRW1} while the negative curvature parameter ($k=-1$) branch are described by Eq.~\eqref{eq:FRW2} (otherwise one would have a complex tetrad). For the case of a flat \gls{flrw} cosmology, both good tetrads given in Eqs.~\eqref{eq:FRW1} and \eqref{eq:FRW2} coincide with each other. It is important to say again that this is not the only choice that describes the \gls{flrw} background but it is the only once in which the spin connection is allowed to vanish while the teleparallel connection respects the cosmological symmetries.

Another important tetrad to highlight is used particularly for flat cosmologies ($k=0$), where the metric can be written as
\begin{equation}\label{FLRW_metric}
    \mathrm{d}s^2 = N(t)^2\mathrm{d}t^2 - a^2(t) \left(\mathrm{d}x^2+\mathrm{d}y^2+\mathrm{d}z^2\right)\,,
\end{equation}
and the diagonal tetrad results in the tetrad
\begin{equation}\label{FLRW_tetrad}
    \udt{e}{A}{\mu}=\text{diag}\left(N(t),\,a(t),\,a(t),\,a(t)\right)\,,
\end{equation}
which turns out to be in the Weitzenb\"{o}ck gauge for the extensions to \gls{tegr} explored here. An important remark here is that the above tetrad (with vanishing spin connection) is the only one that has the property that both the tetrad and the teleparallel connection obey cosmological symmetries for flat \gls{flrw}. One can also relax the condition that the teleparallel connection enjoys the symmetries of cosmology, but then, the corresponding cosmological equations would not respect the symmetries of cosmology.

For the diagonal tetrad~\eqref{FLRW_tetrad} with a vanishing spin connection, one finds that by using the definitions of the torsion scalar \eqref{Torsion_scalar} and the boundary term \eqref{Eq:boundary_term_def}, they become
\begin{equation}\label{eq:Tor_sca_flrw}
    T = -6H^2\,, \quad B = -18H^2-\frac{6\dot{H}}{N}\,.
\end{equation}
Here, we again used that $H=\dot{a}/(aN)$. One naturally reproduces the standard Ricci scalar in this scenario
\begin{equation}
    \lc{R} = -T + B = -12H^2-6\frac{\dot{H}}{N}\,.
\end{equation}

In the following sections, the different manifestations of this cosmological setup are explored in the dynamics of the various extensions beyond $\Lambda$\gls{cdm}, which are investigated together with their cosmological evolution, and properties.

In this section, we explore the predominant cosmological constructions in \gls{tg}. Given the enormity of the task, we first describe the background on the different tools that have been employed to explore the background cosmology of the different \gls{tg} theories, and then delve into the specific theories themselves. In the Supplementary annexes (Supplementary 1) we provide some basic cosmological methods that will be used in the following sections such as the reconstruction method, the Noether symmetry approach and dynamical systems in cosmology (see Supplementary annexes (Supplementary 1)). Additionally, some definitions regarding bouncing solutions are also explained in the Supplementary annexes (Supplementary 1). Those definitions and methods will be applied for several \gls{tg} theories in the next sections for the case of \gls{flrw} cosmology. A more advanced topic regarding anisotropic cosmology is also presented in the Supplementary annexes (Supplementary 1) as a complementary study for the interested reader.

\subsection{\texorpdfstring{$f(T)$}{f(T)} cosmology}\label{sec:f_T_cosmo_back}

In this section we will analyze $f(T)$ cosmology in flat \gls{flrw} using the techniques/methods explained in the previous sections. If we use the diagonal tetrad~\eqref{FLRW_tetrad} in Cartesian coordinates and replace it in the $f(T)$ field equations~\eqref{f(T)field_equations_tetrad}, we arrive at the modified \gls{flrw} equations described by
\begin{subequations}
	\begin{align}
	-6H^2f_T- \frac{1}{2}f &= \kappa^2\rho\label{Friedmann_1B}\,, \\[0.5ex]
	-2f_T(3H^2+\dot{H}) - 2H\dot{f}_T - \frac{1}{2}f &= -\kappa^2 p\label{Friedmann_2B}\,,
	\end{align}
\end{subequations}
where dots are derivatives \gls{wrt} the time, so that $\dot{f}_T=f_{TT}\dot{T}$. It is important to mention that the most general teleparallel theory constructed from of up to quadratic contractions of torsion,\newline $f(T_{\rm axi},T_{\rm ten},T_{\rm vec},P_1,P_2)$ gravity (see Sec.~\ref{Sec:Ext_NGR}), is dynamically equivalent to $f(T)$ gravity in flat \gls{flrw} since $T_{\rm ten}=T_{\rm axi}=P_1=P_2=0$ and $T_{\rm vec}=9H^2$, meaning that only one scalar is nonvanishing and due to~\eqref{eq:torscalar_decomposition}, the theory can be always represented as a $f(T)$ gravity theory. This conclusion is only valid for flat \gls{flrw} at the background level. These equations will be analyzed in detail in the following sections.

\subsubsection{Reconstruction method}\label{sec:fT-reconstruction}

In the trivial extension of $f(T)$ gravity, the reconstruction procedure has been explored extensively in the literature. One common approach involves describing the matter sector as a manifestation of torsion usually by taking the ansatz $f(T) = -T + F(T)$. The latter's $F(T)$ role acts as the source for the matter sector as can be seen from Eq.~\eqref{Friedmann_1B}
\begin{equation}
    3H^2 = \kappa^2 \rho + \frac{F(T)}{2} -T F_T\,,
\end{equation}
where we recall that $T=-6H^2$. A torsional fluid $\rho_{\text{eff}}$ can therefore be defined which can be used to reconstruct the $F(T)$ function \cite{Dent:2011zz}
\begin{equation}\label{eq:fT-reconstruction-density}
    \kappa^2 \rho_{\text{eff}} = \frac{F(T)}{2}-T F_T \, \implies F(T) = \kappa^2 \sqrt{-T} \int\frac{\rho_{\text{eff}}}{(-T)^{\frac{3}{2}}} \, \textrm{d}T.
\end{equation}
This approach has been applied for various fluids including Ricci \gls{de}, \gls{hde} and $(m,n)$ \gls{hde}, Tsallis \gls{hde} including its power-law and logarithmic corrected variants, \gls{plde} and $(m,n)$ \gls{plde} \cite{Daouda:2011yf,Farooq:2013ava,Debnath:2014wga,Huang:2013xca,Chattopadhyay:2019bgm,Jawad:2017hvv,Waheed:2020cxw}. However, in the aforementioned works, a further assumption on the cosmological expansion behavior leads to a disagreement with the Friedmann equations. Meanwhile, in \cite{Said:2017nti}, the reconstruction procedure can be applied in the absence of fluids $\rho=p=0$ through an effective \gls{eos}
\begin{equation}
    w_{\text{eff}} = -1-\frac{16T\dot{H}f_{TT}}{f}\,,
\end{equation}
provided the evolution of $H(t)$ is set. Meanwhile, in \cite{Nojiri:2020wmh}, a correspondence between \gls{hde} and $f(T)$ gravity is produced via a suitable identification of the IR cut-off length.

An alternative proposal is to make use of the matter component in order to reconstruct the Lagrangian as opposed to an \textit{a priori} desired form for $\rho_{\text{eff}}$ \cite{Myrzakulov:2013wza}. This leads to an alternative relation
\begin{equation}\label{eq:fT-reconstruction-matterfluids}
    f(T) = -\kappa^2 \sqrt{-T}\int \frac{\rho_{\text{m}}}{(-T)^\frac{3}{2}} \, \dd T\,.
\end{equation}
For the $\Lambda$\gls{cdm} cosmology in the presence of dust matter, the Lagrangian reduces to the standard $\Lambda$\gls{cdm} Lagrangian \cite{Salako:2013gka,ElHanafy:2019zhr}. However, it can be generalized for arbitrary perfect fluids with \gls{eos} $w$ to give~\cite{Myrzakulov2011}
\begin{equation}\label{eq:arbPerFluidsF(T)}
    f(T) = T_0 \Omega_{w0} \left(-\frac{\Omega_\Lambda}{\Omega_{\rm m0}}\right)^{1+w}\left[\, _2F_1\left(-\frac{1}{2},-w;\frac{1}{2};\frac{T}{T_0 \Omega_\Lambda}\right)+ \frac{T}{T_0 \Omega_\Lambda} \, _2F_1\left(\frac{1}{2},-w;\frac{3}{2};\frac{T}{T_0 \Omega_\Lambda}\right)\right]\,,
\end{equation}
where $\Omega_\Lambda = 1-\Omega_{\rm m0}$, $T_0$ represents the torsion scalar evaluated at current times, and
\begin{equation}
    _{2}F_{1}(a,b;c;z) \equiv \frac{\Gamma(c)}{\Gamma(a) \Gamma(b)} \sum\limits_{k=0}^{\infty} \frac{\Gamma(a+k)\Gamma(b+k)}{\Gamma(c+k) \, k!} z^k
\end{equation}
represents Gauss's hypergeometric function which is convergent for $|z| < 1$ \cite{buchholz2013confluent}. Naturally, the standard result is recovered for $w = 0$. If the fluid is instead assumed to be composed of a spinor field \cite{Myrzakulov:2013wza}, the Lagrangian becomes a rescaled version $\Lambda$\gls{cdm}, i.e. \gls{tegr} with the addition of a constant. In the case of \gls{hde} and \gls{qcd} \gls{de} \cite{Chattopadhyay2013,Chattopadhyay:2014jfa}, the reconstructed solution does not satisfy the Friedmann equations and hence are not reported. 

An alternative proposal is to reconstruct through a specification of the deceleration parameter $q(z)$. During periods where the matter content is mostly composed of baryonic matter, the reconstructed function is obtained from the relation \cite{ElHanafy:2019zhr}
\begin{equation}\label{eq:fT-reconstruction-Hanafy}
    f(z) = -6\Omega_{\rm m0} {H_0}^2 e^{\int_0^z \frac{1+q(\bar{z})}{1+\bar{z}} \, \dd\bar{z}} \int_0^z \frac{(1+\bar{z})^2(1+q(\bar{z}))}{e^{\int_0^z \frac{1+q(\bar{z})}{1+\bar{z}} \, \dd\bar{z}}} \, \dd\bar{z}\,.
\end{equation}
This is derived from Eq.~\eqref{eq:fT-reconstruction-matterfluids} by expressing it in terms of redshift $z$ and the deceleration parameter $q(z)$ via the relations $H(z) = H_0 e^{\int_0^z \frac{1+q(\bar{z})}{1+\bar{z}} \, \dd\bar{z}}$ and $T = -6H^2$. In particular, the parametrization $q(z) = q_0 + q_1 X(z)$ \cite{ElHanafy:2019zhr} for constants $q_0, q_1$ and for two functional choices of $X(z)$ which determines the cosmological evolution was investigated.

\begin{figure}[!ht]
	\centering
	\includegraphics[scale=0.5]{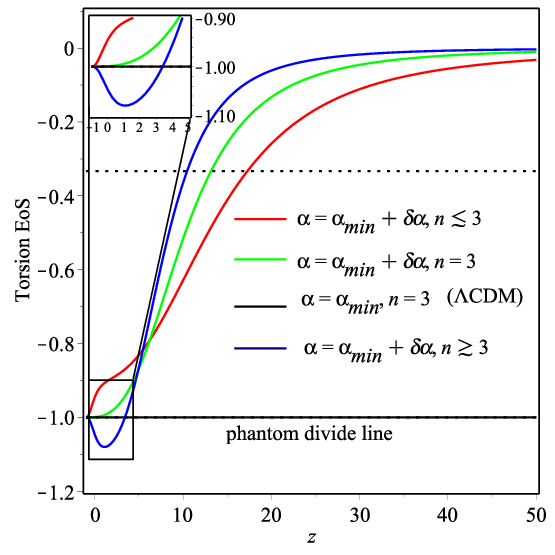}
	\caption{The torsional fluid \gls{eos} behavior for the model $w_{\rm eff} = -\frac{1}{1 + \alpha(1+z)^n}$ for different choices of $n$ as presented in Ref.~\cite{ElHanafy:2019zhr}. Here, $\Omega_{\rm m0} = 0.297$ and $\delta\alpha = 10^{-3}$. Permission for use of this figure was kindly provided by the authors of Ref.~\cite{ElHanafy:2019zhr}.}
	\label{fig:fT_EoS_recon}
\end{figure}

The first two models $X(z) = \frac{z(1+z)}{1+z^2}$ \cite{Mamon:2015osa} and $X(z) = \frac{\ln(\widetilde{N}+z)}{1+z} - \ln \widetilde{N}$, $\widetilde{N} > 1$ \cite{Mamon:2016dlv} do not yield a viable cosmology. On the other hand, the final model explores the effective fluid \gls{eos} $w_\text{eff} = -\frac{1}{1 + \alpha(1+z)^n}$ for some constants $\alpha, n$ \cite{Mukherjee:2016eqj}, or equivalently in terms of $q(z)$, 
\begin{equation}
    q(z) = -1 + \frac{3\alpha(1+z)^n}{2\left[1+\alpha(1+z)^n\right]}.
\end{equation}
The model describes an early matter domination and late de Sitter phases. In particular, $n = 3$ yields the standard $\Lambda$\gls{cdm} cosmology, hence $n$ is expected to deviate slightly from this value. In this case, the reconstructed function is
\begin{equation}
    f(z) = -9\Omega_{\rm m0} {H_0}^2 \left[1+\alpha(1+z)^n\right]^{\frac{3}{2n}} \int_0^z \frac{(1+\bar{z})^{n+2}}{\left[1+\alpha(1+\bar{z})^n\right]^{1+\frac{3}{2n}}} \, \dd\bar{z}\,.
\end{equation}
Here, $\Omega_{\rm m} \leq 1$ provided $\alpha > \alpha_\text{min} = \frac{{\Omega_{\rm m0}}^{\frac{n}{3}}}{1-{\Omega_{\rm m0}}^{\frac{n}{3}}}$. For $\alpha$ slightly larger than this minimum value leads to the following torsional fluid behaviors:
\begin{enumerate}[label = (\alph*{})]
    \item $n \lesssim 3$: Quintessence with $w_{\text{eff}} \gtrsim -1$ at present;
    \item $n \gtrsim 3$: Quintom (quintessence $\to$ phantom transition at $z \sim 4$) with $w_{\text{eff}} \lesssim -1$ at present;
    \item $n = 3$: Quintessence with $w_{\text{eff}} \sim -1$ at present.
\end{enumerate}
For all cases, $w_{\text{eff}} \to 0$ at early-times and asymptotically approaches \gls{de} in the future (see Fig.~\ref{fig:fT_EoS_recon}). Overall, this produces a viable cosmological model. 

A similar reconstruction procedure uses the combination of the Eqs.~\eqref{Friedmann_1B} and \eqref{Friedmann_2B} \cite{Myrzakulov2011,Chakrabarti:2019bed},
\begin{equation}
    f - 2Tf_T + \frac{4\dot{H}}{1+w} \left[f_T+2Tf_{TT}\right] = 0\,,
\end{equation}
and the specification of the perfect fluid's \gls{eos}. Due to the presence of the $\dot{H}$ terms, solutions are obtained only in special limited scenarios such as specification of the evolutionary behavior and setting $w$ to take constant values. For instance, for a power law cosmology $a(t) \propto t^n$ and $w = 0$, $f(T) \propto (-T)^{\frac{3n}{2}}$ which, however, does not host \gls{gr} as a limit unless $n = \frac{2}{3}$. In Ref.~\cite{Chakrabarti:2019bed}, the evolutionary behavior is specified through the jerk parameter $j(t) \equiv \dddot{a}/(aH^3)$ where three models were considered (see the Supplementary annexes (Supplementary 1)):
\begin{enumerate}[label = (\alph*{})]
    \item $j = 1$: A de Sitter accelerating behaviour can be produced. At early-times, the torsional fluid varies in nature $(-1 < w_{\text{eff}} < 1)$ whereas it always approaches phantom behavior at late-times. 
    \item $j = \frac{s^2}{H^2}$: late-time acceleration when $m \gg n$ provided $s > 0$. A best-fit value $\mu \sim 0.39$ where $\mu \coloneqq \frac{s(1-p)^2}{s(1-p)^2 + 4mn}$ was obtained using \gls{sn}Ia Union 2.1 luminosity distance data \cite{Suzuki2012}. Here, the nature of the matter fluid influences the behavior of $w_{\rm eff}$:
\begin{enumerate}[label = (\roman*{})]
    \item $w = 0$: $w_{\text{eff}} >0$ at all times with $w_{\text{eff}} \to 0$ at late-times;
    \item $w = \pm \frac{1}{2}$: $w_{\text{eff}} \to \pm \frac{1}{2}$ at late-times;
    \item $w = 1$: quintom behavior; $w_{\text{eff}} \sim 1$ at early-times, crosses the phantom-divide line and approaches $w_{\text{eff}} \to -1$ at late-times.
\end{enumerate}
    \item $j = 1 - \epsilon f(a)$ with slowly varying function $f(a) = \frac{9}{8\epsilon} + \frac{\epsilon_0}{a}$, $\epsilon, \epsilon_{0} \ll 1$: A decelerating universe $a(t) \sim t^{\frac{4}{5}}$ results and thus not explored in further detail. 
\end{enumerate} 
As a final remark, for all the aforementioned jerk parameter considerations, the models are asymptotically stable against homogeneous perturbations. 

The reconstruction procedure can also be applied to generate the spatially flat \gls{flrw} cosmology through non-flat \gls{flrw} spacetimes \cite{Paliathanasis:2021uvd}. Accounting for the spatial curvature $k$, the Friedmann equations are obtained following the first branch solution described in Sec.~\ref{sec:realtetrad} with tetrad Eq.~\eqref{eq:cosmoptetradwb}, leading to
\begin{subequations}
\begin{align}
\kappa^2\rho &=-\frac{f}{2} - 6H^2 f_T\,, \label{eq:fT-00Friedmann-non-flat}\\[0.5ex]
-\kappa^2 p &= -2f_T(3H^2+\dot{H}) + \frac{2k}{a^2}f_T - 2H\dot{f}_T -\frac{f}{2}\,, \label{eq:fT-iiFriedmann-non-flat}
\end{align}
\end{subequations}
with $T = -6H^2 + \frac{6k}{a^2}$ as given in Eq.~\eqref{eq:torsionnonflat}. In the avenue that the second branch solution (Sec.~\ref{sec:negativsym}) is used, the form of $T$ changes to Eq.~\eqref{eq:torsion_nonflat_2ndbranch} which is complex for $k > 0$. In the following works, the former description has been employed. Following Refs.~\cite{Hanafy:2014ica,Hanafy:2015lda}, an almost flat universe is constructed by eliminating the explicit spatial curvature dependence in Eq.~\eqref{eq:fT-iiFriedmann-non-flat} which leads to the reconstructed $f(T)$ function to be
\begin{equation}\label{eq:reconstruction-fT-flat-like-reconstructedB}
    f(t)\sim\int^{t}\exp\left[\int^{\tilde{t}}\frac{k^{2}+(3\ddot{a}a-5\dot{a}^{2})k+2\ddot{a}^{2}a^{2}+4\dot{a}^{4}-7\dot{a}^{2}\ddot{a}a+\dot{a}\dddot{a}a^{2}}{\dot{a}a\left(\ddot{a}a-\dot{a}^{2}+k\right)}\,\dd t'\right]\dd\tilde{t}\,.
\end{equation}
Here, $k$ reappears by virtue of the torsion scalar. The dependence of $k$ can be further suppressed by setting $3\ddot{a}a -5\dot{a}^2 = 0$, i.e. $a(t) \propto \left[1-\frac{2H_0}{3}(t-t_0)\right]^{-\frac{3}{2}}$ with $t_0$ representing the cut-off time and $H_0 \coloneqq H(t_0)$. Thus, the evolutionary behavior is independent of $k$ and therefore flat-like. In addition, it has been shown to be consistent with early inflation. The associated reconstructed function for this flat-like scenario takes on the series expansion $f(T) = \sum\limits_{n=0}^\infty \alpha_n (-T)^{-\frac{n}{2}}$ for some coefficients $\alpha_n$.

Use of spatial curvature also appears in Refs.~\cite{Nashed:2014lva,Hanafy:2014bsa} to generate an accelerating de Sitter cosmology. In the limit when the matter content behavior approaches that of \gls{de} yields\footnote{The reconstructed function obtained in \cite{Nashed:2014lva} is not a solution. Substituting Eqs.~(3.5) and (3.6) in (3.21) does not lead to Eq.~(3.22); instead, the equation identically reduces to zero.}~\cite{Nashed:2014lva}
\begin{equation}
    f(T) \sim \exp \left[\frac{T}{12}\frac{a^2(\ddot{a}a - \dot{a}^2 - k)}{\dot{a}^2(\ddot{a}a - \dot{a}^2 + k)}\right]\,,
\end{equation}
provided that $\frac{a^2(\ddot{a}a - \dot{a}^2 - k)}{\dot{a}^2(\ddot{a}a - \dot{a}^2 + k)}$ is constant. If a vacuum is assumed, $f(T)$ then takes on the form $f(T) \propto \exp \left(-\frac{T+6{H_0}^2}{12{H_0}^2}\right)$ \cite{Hanafy:2014bsa}. Contrary to the standard spatially flat cases, $f(T)$ no longer reduces to a cosmological constant, instead it becomes dynamical which asymptotically approaches the latter behavior.

\subsubsection{Noether's symmetry approach} \label{sec:cosmo-Noether-fT}

In the first trivial extension of teleparallelism, $f(T)$ gravity, Noether symmetry has been studied in numerous works. Generally, the configuration space is taken to be $\mathcal{Q} = \lbrace t, a, T\rbrace$ from which the Noether condition is then solved (see the Supplementary annexes (Supplementary 1)). Due to the simplicity of the Lagrangian, irrespective of whether the time symmetry generator is included in the analysis, the Noether condition imposes the Lagrangian constraint $f(T) \propto (-T)^n$ for some constant $n$ \cite{Wei:2011aa,Myrzakulov:2012sp,Sk:2017ucb,Basilakos:2013rua,Atazadeh:2011aa,Sadjadi:2012xa,Dong:2013rea,Fazlollahi:2020doj,Bajardi:2021tul}. This generates a power-law expansion behavior, $a(t) \sim t^{\frac{3}{2n}}$, and henceforth includes a number of cosmological solutions. However, according to Ref.~\cite{Sk:2017ucb}, $n$ can only take a singular value of $n = \frac{3}{2}$.

This restriction arises once the Euler-Lagrange field equations are solved in conjuction with the resulting conserved charge arising from the Noether symmetry. Together, they restrict $n = \frac{3}{2}$ leading to a coasting cosmology. An analogous restriction appears in $f(\lc{R})$ theories where the model reduces to $f(\lc{R}) \propto \lc{R}^{\frac{3}{2}}$ \cite{Sarkar:2012ug,Capozziello:2008ima}, a result which also holds true in the Palatini formulation \cite{Sk:2016dkj}. Despite this similarity, the resulting cosmological behaviors are distinctively different from one another due to the fundamental differences between the two theories.

In Ref.~\cite{Sadjadi:2012xa}, the Noether symmetry approach was further extended to include velocities in the Noether symmetry vector. In the absence of these velocities, the model reduces to the $f(T) \propto (-T)^n$ result. Otherwise, further solutions are obtained, including $f(T) = \pm \sqrt{2c_1 T + 2c_2}$ for integration constants $c_{1,2}$. This leads to an entirely different cosmological behavior where it now describes a cosmological turnaround and acceleration/deceleration phases during \gls{cdm} domination. Furthermore, the total \gls{eos} parameter does not cross the phantom-divide line and is always non-phantom. Nevertheless, these models do not have a \gls{tegr} background, making them less popular.

Meanwhile, in Ref.~\cite{Dong:2013rea}, the Noether condition is not generally solved but only after specific $f(T)$ ansatz are considered. Provided a symmetry exists for the considered ansatz, the model is then constrained against \gls{sn}+$H(z)$+\gls{bao} data. In summary, only the $f(T) = \alpha T + \beta T^n$ ansatz leads to meaningful results as it can match with observations with $n = -1$ being the best-fitting parameter. The $f(T) \propto (-T)^n, e^{nT}$ or $f(T) \sim \sum_n (-T)^n$ ansatz are insufficient to match with observations.

Finally, Ref.~\cite{Fazlollahi:2020doj,Bajardi:2021tul} investigate the correspondence with the Wheeler-de Witt equation where the resulting classical universe trajectories lead back to the power-law scale factor solution $a \sim t^{\frac{3}{2n}}$.

\subsubsection{Dynamical system approach}\label{sec:dynamics-fT}

The dynamical systems approach has been extensively investigated in $f(T)$ gravity \cite{Jamil:2012yz,Hohmann:2017jao,ElHanafy:2017sih,Awad:2017ign,Awad:2017yod,Awad:2017sau,Jamil:2012nma}. Given the simple nature of the field equations \gls{flrw} equations~\eqref{Friedmann_1B}-\eqref{Friedmann_2B} as they can be solely expressed in terms of the Hubble parameter and its time derivative, allows for the straightforward construction of the one dimensional autonomous system
\begin{equation}\label{eq:dynamic-fT-equation}
\dot{H} = -\dfrac{1+w}{4}\dfrac{f - 2Tf_T}{f_T + 2T f_{TT}} = 3(1+w) \dfrac{f(H)-Hf_H}{f_{HH}} = -\dfrac{3}{2}\left(1+w_{\rm Tot}\right)H^2 \equiv \mathcal{F}(H)\,,
\end{equation}
where $w$ represents the perfect fluid \gls{eos} and $w_{\rm Tot}$ is the total \gls{eos}, and where $f_{H}$ and $f_{HH}$ represent first and second derivatives of $f(H)$ \gls{wrt} $H$. Clearly, the critical points always correspond to Minkowski or de Sitter points which stability is identified by $\mathcal{F}_H$. This approach has been considered in Refs.~\cite{Jamil:2012yz,Awad:2017ign,Awad:2017yod,Awad:2017sau,Jamil:2012nma} for various choices of the $f(T)$ Lagrangian.

In Refs.~\cite{Jamil:2012yz,Jamil:2012nma}, the model $f(T) = \alpha T + \beta \sqrt{-T} + \gamma$, where $\alpha,$ $\beta$ and $\gamma$ are constants, was considered. However, this is simply rescaled \gls{tegr} with a cosmological constant hence not leading to other new dynamics beyond the $\Lambda$\gls{cdm} framework.

Meanwhile, Ref.~\cite{Awad:2017yod} considers a dust matter fluid for different $f(T)$ model ansatz:
\begin{enumerate}[label={(\roman*)}]
	\item $-T + \alpha (-T)^b$: Semi-stable Minkowski and accelerating (decelerating) stable (unstable) de Sitter critical points. For $b \ll 1$, the sequence Big Bang singularity $\to$ matter domination $\to$ de Sitter acceleration is recovered. Similar conclusions are drawn in Ref.~\cite{Awad:2017sau};
	\item $-T + \alpha T_0 (1-e^{-p \sqrt{T/T_0}})$: Semi-stable Minkowski and stable or unstable de Sitter critical points. A cosmological turnaround can be realized while a correct matter era is recovered for $p > 0$;
	\item $-T + \alpha T_0 \left(1- e^{-\beta T/T_0}\right)$: Similar critical points to model (ii). Cosmological evolution changes depending on $p$, but a deceleration $\to$ acceleration sequence can be realized for $p = 1/2$. Larger values of $p$ yield better agreements with observations;
	\item $T e^{\beta T_0/T}$: The number of critical points depend on the parameter $\beta$.
	\begin{enumerate}
		\item $\beta < 0$: Only a semi-stable Minkowski solution arises. A future Big Crunch singularity $(H < 0)$ or deceleration $\to$ acceleration which ends at the Minkowski point $(H > 0)$ can be realized;
		\item $\beta > 0$: Semi-stable Minkowski and accelerating (decelerating) stable (unstable) de Sitter point. The cosmological evolutionary behavior depends on the magnitude of $H$. In particular, $H > \sqrt{2\beta}|H_0|$ appears to realize a viable cosmological scenario.
	\end{enumerate}
\end{enumerate}

For arbitrary $f(T)$ functions, the complete dynamical nature of the theory was obtained in Ref.~\cite{Hohmann:2017jao} in the presence of dust and radiation fluids. To account for the two-fluid system, a further phase-space variable $X = \frac{\rho_\text{r}}{\rho_\text{m}+\rho_\text{r}}$ was defined, leading to a 2 dimensional system. This also modifies Eq.~\eqref{eq:dynamic-fT-equation} to
\begin{equation}
\dot{H} = -(X+3)H\frac{W}{W_H}\,,
\end{equation}
where $W(H) = f(H) - H f_H$. The critical points are listed in Table~\ref{table:dynamics-critpts-fT}.

\begin{figure}[!ht]
	\centering
	\includegraphics[width=\textwidth]{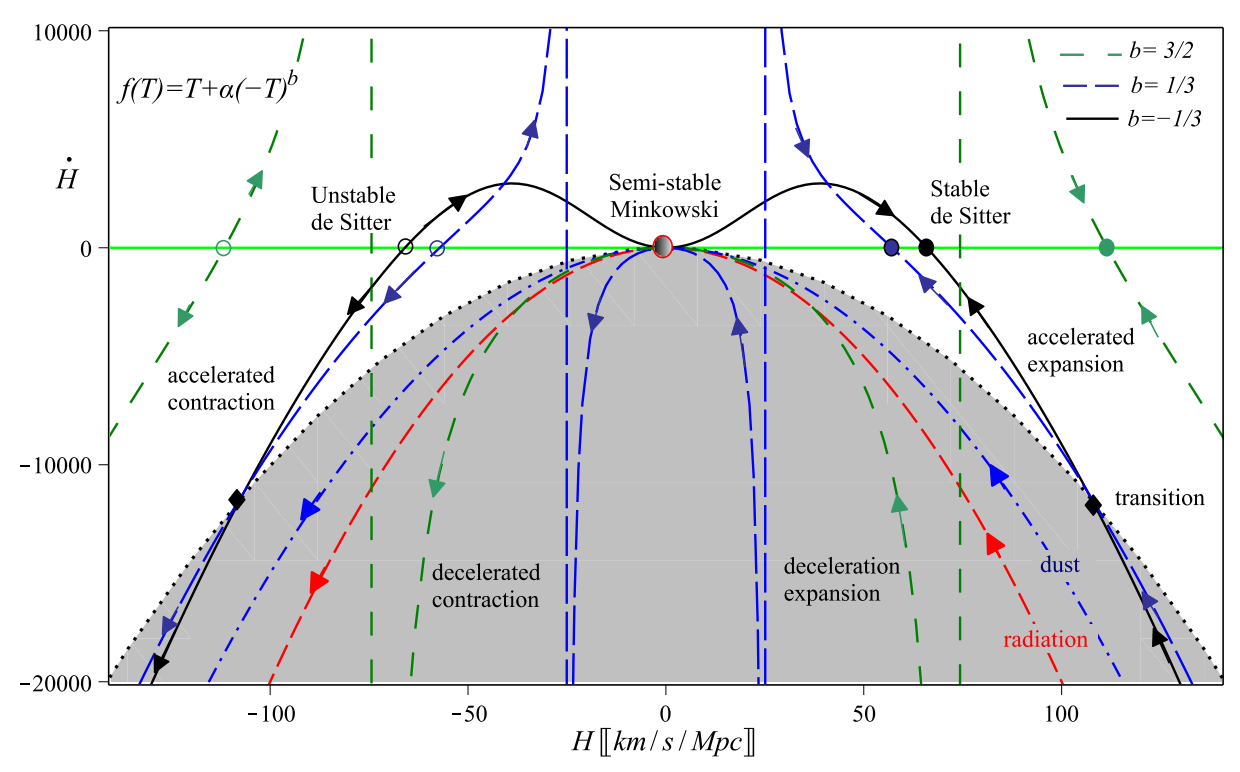}
	\caption{The dynamical behavior for the $f(T) = -T + \alpha (-T)^b$ power-law model\protect\footnotemark{} for different choices of $b$. Here, the gray (white) region represents decelerated (accelerated) expansion with the dotted line representing the transition. Circles represent the critical points with empty (filled) being repeller (attractor) and grey being semi-stable. Focusing on the $H > 0$ plane, a matter domination $\to$ late-time de Sitter is achieved for $b = -\frac{1}{3}$. Permission for use of this figure was kindly provided by the authors of Ref.~\cite{Awad:2017yod}.}
	\label{fig:fT_dynamics}
\end{figure}

Evidently, the existence conditions depend on the behavior of $W$ and $W_H$. Meanwhile, attractor solutions are $P_1$ (vacuum de Sitter) and $P_3$ (non-empty static Minkowski geometry). Therefore, a late-time acceleration can be recovered through $P_1$. Furthermore, for any $f(T)$ model, no cyclic and oscillating solutions are possible. Moreover, the sequence where the Universe starts with an initial accelerating phase (inflationary), become decelerating and finally evolves to an accelerating de Sitter phase cannot be realized. In the particular case when $f(T) = -T + \alpha(-T)^b$, the critical points as found in Refs.~\cite{Awad:2017yod,Awad:2017sau} are recovered with the extra critical point $P_2$ which cannot appear in the aforementioned works due to the different dynamical phase-space analysis. A simple dynamical representation for different choices of $b$ are highlighted in Fig.~\ref{fig:fT_dynamics}.

\begin{table}[!ht]
	\centering
	\midsepremove
	\begin{tabularx}{\textwidth}{lcp{3cm}XX}
		\toprule
		\cellcolor{gris3}& \cellcolor{gris3}\boldmath{$(H^\star, X^\star)$} &\cellcolor{gris3} \textbf{Existence} & \cellcolor{gris3}\textbf{Stability} & \cellcolor{gris3}\textbf{Properties} \\ \midrule
		\cellcolor{gris1}& \cellcolor{gris1} & \cellcolor{gris1} & \cellcolor{gris1}Attractor $H^\star > 0$& \cellcolor{gris1} \\
		\multirow{-2}{*}{\cellcolor{gris1}$P_1$} & \multirow{-2}{*}{\cellcolor{gris1}$(H^\star \neq 0,0)$} & \multirow{-2}{*}{\cellcolor{gris1}$W^\star = 0$} &\cellcolor{gris1}Repeller for $H^\star < 0$ & \multirow{-2}{*}{\cellcolor{gris1}de Sitter vacuum} \\
		\cellcolor{gris3}$P_2$ & \cellcolor{gris3}$(H^\star \neq 0,1)$ & \cellcolor{gris3}$W^\star = 0$ & \cellcolor{gris3}Saddle & \cellcolor{gris3}Same behavior as $P_1$ \\
		\cellcolor{gris1} & \cellcolor{gris1} & \cellcolor{gris1} & \cellcolor{gris1}Attractor $W_H^\star > 0$ & \cellcolor{gris1}Non-empty \\
		\multirow{-2}{*}{\cellcolor{gris1}$P_3$} & \multirow{-2}{*}{\cellcolor{gris1}$(0,X^\star)$} & \multirow{-2}{*}{\cellcolor{gris1}$W^\star > 0, W_H^\star \neq 0$} & \cellcolor{gris1}Repeller $W_H^\star < 0$ &\cellcolor{gris1} Static Minkowski geometry \\
		\cellcolor{gris3}$P_4$ & \cellcolor{gris3}$(0,X^\star)$ & \cellcolor{gris3}$W^\star = 0$ & \cellcolor{gris3}Not determined & \cellcolor{gris3}Minkowski vacuum solution \\
		\cellcolor{gris1} & \cellcolor{gris1} & \cellcolor{gris1}$W^\star > 0\,,$ & \cellcolor{gris1} & \cellcolor{gris1} \\
		\multirow{-2}{*}{\cellcolor{gris1}$P_5$} & \multirow{-2}{*}{\cellcolor{gris1}$(0,X^\star)$} & \cellcolor{gris1}$W_H^\star \to \pm \infty$ & \multirow{-2}{*}{\cellcolor{gris1}Not determined} & \multirow{-2}{*}{\cellcolor{gris1}Non-vacuum static universe} \\
		\cellcolor{gris3} & \cellcolor{gris3} & \cellcolor{gris3}$W^\star > 0\,,$& \cellcolor{gris3} &\cellcolor{gris3} \\
		\multirow{-1}{*}{\cellcolor{gris3}$P_6$} & \multirow{-1}{*}{\cellcolor{gris3}$(0,X^\star)$} & \cellcolor{gris3}$W_H^\star=0\,,$ & \multirow{-1}{*}{\cellcolor{gris3}Not determined} &\multirow{-1}{*}{\cellcolor{gris3}Type IV finite singularity} \\
		\cellcolor{gris3} & \cellcolor{gris3} & \cellcolor{gris3}$H/W_H \to 0$ & \cellcolor{gris3} &\cellcolor{gris3} \\
		\bottomrule
	\end{tabularx}
	\midsepdefault
	\caption{Summary of the critical points $(H=H^\star,X=X^\star)$ for an arbitrary $f(T)$ model in the presence of dust and radiation fluids as found in Ref.~\cite{Hohmann:2017jao}. Here, $W^\star$ and $W_H^\star$ represent the function $W(H)$ and its derivative evaluated at $H = H^\star$.}
	\label{table:dynamics-critpts-fT}
\end{table}

\footnotetext{To match with the review convention, the \gls{tegr} term's signature is reversed compared to \cite{Awad:2017yod}.}

For the remaining works~\cite{Wu:2010xk,Setare:2012ry,Setare:2013xh,Feng:2014fsa,Mirza:2017vrk,Ganiou:2018dta}, instead of a 1 dimensional system, a larger autonomous system is constructed which allows for a more general investigation of non-de Sitter critical points. Of note, two notable approaches were employed. The first follows Refs.~\cite{Wu:2010xk,Setare:2012ry,Setare:2013xh,Feng:2014fsa,Mirza:2017vrk} which consider dust and radiation fluids with density $\Omega_{\rm m}$ and $\Omega_{\rm r}$, respectively, and the phase-space variables\footnote{With the exception of Ref.~\cite{Feng:2014fsa} where the combination of phase-space variables $\tilde{X} = x + y$ is considered.}
\begin{equation}
\tilde{x} := \frac{f(T) + T}{T}\,, \quad\tilde{y} := -2(f_T +1)\,, \quad\tilde{z} := 2T f_{TT}\,, \quad\Omega_{\rm m} := \frac{\kappa^2 \rho_{\rm m}}{3H^2}\,, \quad\Omega_\text{r} := \frac{\kappa^2 \rho_\text{r}}{3H^2}\,.
\end{equation}

Following Refs.~\cite{Setare:2012ry,Setare:2013xh}, for arbitrary choices of the gravitational Lagrangian, a stable de Sitter attractor, a stable or saddle scaling solution with matter and a possibly unstable scaling radiation solution were found whereas an inflationary solution was not observed. For particular $f(T)$ models, the cosmological sequence starting with radiation domination $\to$ matter domination $\to$ de Sitter acceleration was investigated.

As summarized in Table~\ref{table:fT-dynamics-Sertare}, a viable model which realizes the sequence is obtained either in the power-law model $T^a$ or in the logarithmic model $T^a \left[\ln (\gamma T)\right]^b$ while the exponential models are able to realize the matter to \gls{de} domination sequence leading to a stable de Sitter phase but do not host dust and radiation domination epochs. Of further note, in the power-law model, a stable de Sitter solution is realized for $a = \frac{1}{2}$ which corresponds to the case where a $\sqrt{-T}$ term appears in the Lagrangian. As this term does not contribute to the field equations, the stability of the solution is to be treated lightly.

\begin{table}[!ht]
	\centering
	\midsepremove
	\begin{tabular}{cccc}
		\toprule
		\cellcolor{gris3}\boldmath{$F(T)$} \cellcolor{gris3}\textbf{Model} &\cellcolor{gris3} \boldmath{$m = 0,\,\, r = -1$} & \cellcolor{gris3}\boldmath{$m_{r} > -1,\,\, r = -1$} & \cellcolor{gris3}\boldmath{$m = -\frac{1}{2},\,\, r = -\frac{1}{2}$} \\
		\cellcolor{gris3}\boldmath{$F(T)$} & \cellcolor{gris3}\boldmath{$\Omega_\text{m}$/$\Omega_\text{r}$} \textbf{Domination} &\cellcolor{gris3}\boldmath{$\Omega_\text{m} \to$} \textbf{DE Domination} & \cellcolor{gris3}\textbf{Stable de Sitter} \\ \midrule
		\cellcolor{gris1}$(-T)^a$ & \cellcolor{gris1}Always & \cellcolor{gris1}$a < 1$ & \cellcolor{gris1}$a = \frac{1}{2}$ \\
		\cellcolor{gris3}$(-T)^a e^{bT}$ & \cellcolor{gris3}$a = 1$ & \cellcolor{gris3}$a > -2$ & \cellcolor{gris3}$a = \frac{1}{8}$ \\
		\cellcolor{gris1}$(-T)^a e^{\frac{b}{T}}$ & \cellcolor{gris1}$a = 1$ & \cellcolor{gris1}$a > 0$ & \cellcolor{gris1}$a = \frac{1}{8}$ \\
		\cellcolor{gris3}$(-T)^a \left[\ln (\gamma T)\right]^b$ & \cellcolor{gris3}$a = \frac{11 \pm 6\sqrt{3}}{13}$ & \cellcolor{gris3}$a > -2$ & \cellcolor{gris3}$a = \frac{11 \pm 6\sqrt{3}}{13}$ \\
		\bottomrule
	\end{tabular}
	\midsepdefault
	\caption{Summary of the existence of matter and radiation domination epochs, the matter to \gls{de} domination and finally the existence of a stable de Sitter cosmology, for various $f(T)$ models as carried out in Ref.~\cite{Setare:2012ry}. Here, $m \coloneqq \frac{Tf_{TT}}{f_T}$ and $r \coloneqq -\frac{Tf_T}{f}$.}
	\label{table:fT-dynamics-Sertare}
\end{table}

\begin{figure}[!ht]
	\centering
	\includegraphics[width=0.5\textwidth]{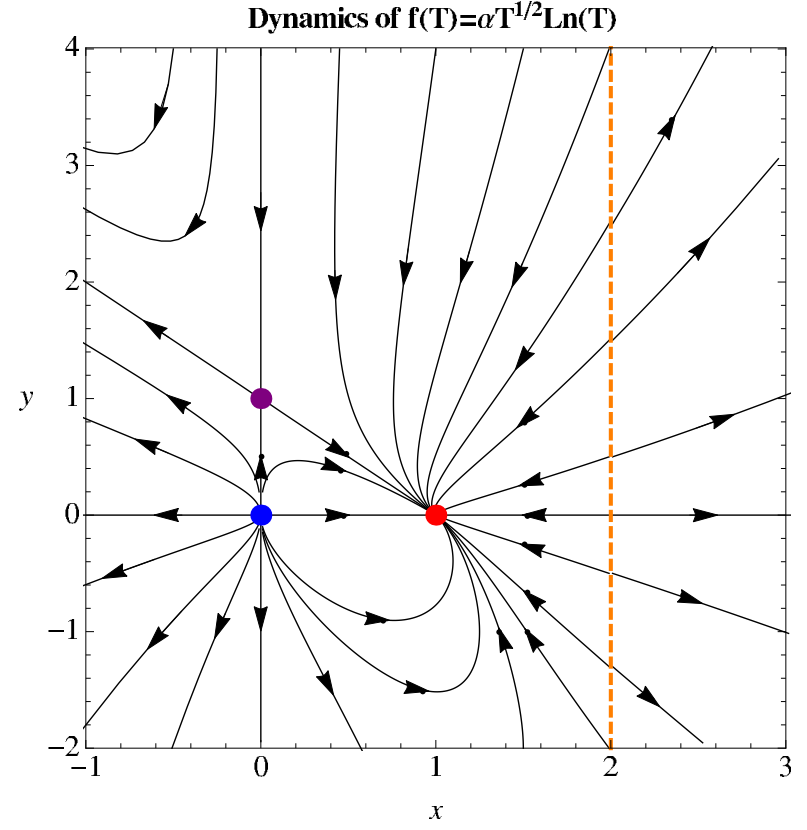}
	\caption{The dynamical behavior for the logarithmic $f(T) = -T + \alpha \sqrt{-T} \ln T$ model with the critical points marked with circles (blue - radiation, violet - matter, red - de Sitter). The orange line $(x = 2)$ represents a singularity present in the dynamical system. In fact, the evolutionary behavior is distinct between $x < 2$ and $x > 2$. To match our conventions, $x \to \tilde{x}$, $y \to \Omega_{\rm m}$ and $f(T) \to -T + f(T)$. This figure first appeared in Ref.~\cite{Feng:2014fsa}.} \label{fig:fT_dynamics_log}
\end{figure}

In Ref.~\cite{Feng:2014fsa}, a \gls{de} dominated de Sitter sink is found which suggests the resulting late-time behavior of the theory. In addition, the critical points $(\tilde{x}^\star,{\Omega_\text{m}}^\star) = (0,0)$ and $(0,1)$ are found which correspond to radiation and matter dominated epochs respectively, are also obtained for the models $f(T) = -T + \alpha (-T)^n$ and $f(T) = -T + \alpha (-T)^\beta \ln T$, $\beta = \lbrace \frac{1}{2}, 1\rbrace$. In the power-law and the $\beta = \frac{1}{2}$ cases, the domination phases correspond to their expected decelerating expansion rate. However, for the logarithmic $\beta = 1$ case, the rate of deceleration (or possibly acceleration) depends on the magnitude of $\alpha$, hereby in agreement to the bifurcation observed in the dynamical analysis. Overall, a suitable radiation $\to$ matter $\to$ de Sitter evolutionary process can be realized for suitable parameter choices (see Fig.~\ref{fig:fT_dynamics_log} for the $\beta = \frac{1}{2}$ logarithmic model phase-space representation). 

\begin{figure}[!t]
	\centering
	\includegraphics[width=\textwidth]{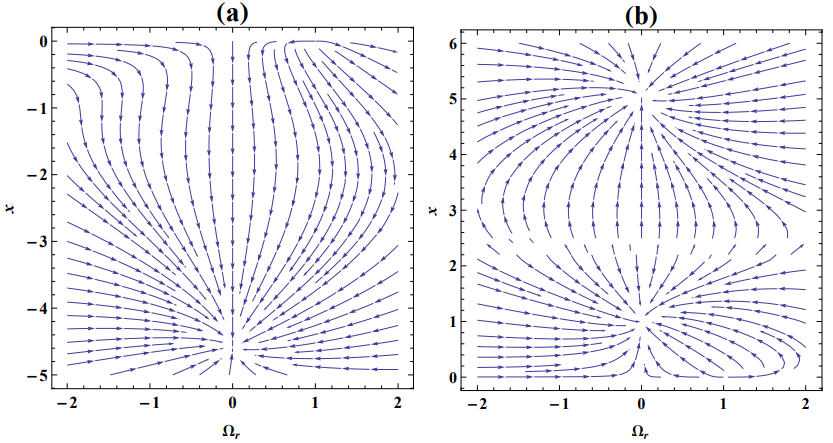}
	\caption{The dynamical behavior for the exponential $f(T) = -T + T_0\big(1 - e^{-p\frac{T}{T_0}}\big)$ model for (a)~$p = 1$ and (b)~$p = 1.5$. For this model, $x_0 = 0$ meaning the critical points $P_1 \, (0,1)$ (radiation) and $P_2 \, (0,0)$ (matter) appear in both plots. The critical point $P_3$ (de Sitter) also appears with (a)~$x_1 = -4.59$ while in (b)~$x_1 = 1$ and 5.04. Thus, for both $p$ choices, the sequence radiation $\to$ matter $\to$ de Sitter can be achieved. Note the identification $x \to \tilde{x}$ for this figure. Permission for use of this figure was kindly provided by the authors of Ref.~\cite{Mirza:2017vrk}.}
	\label{fig:fT_dynamics_exp}
\end{figure}

Following Ref.~\cite{Wu:2010xk} and the improved results in Ref.~\cite{Mirza:2017vrk}, critical points of the form $(\tilde{x}^\star, {\Omega_\text{r}}^\star)$ were obtained for arbitrary choices of $f(T)$, as summarized in Table~\ref{table:fT-dynamics-Mirza}. Overall, the sequence starting from an inflationary period all the way to a late-time acceleration can be realized provided the following conditions are met: (i) $\tilde{x}_0$ and $\tilde{x}_1$ exist, (ii) $-1 < \tilde{x}_0 \leq 0$ and (iii) $\tilde{y}_{\tilde{x}}(\tilde{x}_0) > -2$. These conditions were examined for a selected number of $f(T) = -T + F(T)$ models with $F(T)$ being equal to: 

{\setlength{\jot}{-8pt}
\begin{align*}
& \text{(a)} \quad T^n & & n < 1\,, & & \text{(d)} \quad T_0\left(1 - e^{-p\sqrt{\frac{T}{T_0}}}\right) & & p \geq 0.5\,, \\
& \text{(b)} \quad \sqrt{\frac{T}{q T_0}} \ln \left(\frac{qT_0}{T}\right) & & q > 0\,, & & \text{(e)} \quad T_0\left(1 - e^{-p\frac{T}{T_0}}\right) & & 0.203 \leq p \leq 1.977, \\
& & & & & & & p \geq 6.110, \, p \neq 1.256, \\
& \text{(c)} \quad T^n \tanh \left(\frac{T_0}{T}\right) & & n < 1\,,
\end{align*}}
\hspace*{-0.275cm} which highlight the possibility of realizing the desired sequence of cosmological behaviors (see, for instance, the exponential model dynamical behavior in Fig.~\ref{fig:fT_dynamics_exp}). However, this appears to be in disagreement with the results obtained in Ref.~\cite{Hohmann:2017jao} where it is claimed that inflation $\to$ deceleration $\to$ accelerating de Sitter phase cannot be realized for arbitrary $f(T)$ models. 

\begin{table}[!t]
	\centering
	\begin{tabularx}{\textwidth}{lclX}
		\toprule
		\cellcolor{gris3}& \cellcolor{gris3}\boldmath{$(\tilde{x}^\star, {\Omega_\text{r}}^\star)$} & \cellcolor{gris3}\textbf{Stability} & \cellcolor{gris3}\textbf{Cosmological Description} \\ \midrule
		\cellcolor{gris1}$P_1$ & \cellcolor{gris1}$\left(\tilde{x}_0, 1+\tilde{x}_0\right)$ & \cellcolor{gris1}Saddle or unstable & \cellcolor{gris1}Decelerating radiation or \gls{de} dominated solution \\
		\cellcolor{gris3}$P_2$ & \cellcolor{gris3}$\left(\tilde{x}_0, 0\right)$ & \cellcolor{gris3}Stable or saddle & \cellcolor{gris3}Decelerating matter or dark energy dominated solution \\
		\cellcolor{gris1}$P_3$ & \cellcolor{gris1}$\left(\tilde{x}_1, 0\right)$ & \cellcolor{gris1}Stable &\cellcolor{gris1} Dark energy dominated de Sitter accelerating phase \\
		\cellcolor{gris3}$P_4$ & \cellcolor{gris3}$\left(\tilde{x}_2, 0\right)$ & \cellcolor{gris3}Stable or saddle & \cellcolor{gris3}Inflationary phase (without graceful exit) \\
		\bottomrule
	\end{tabularx}
	\caption{The critical points along with their stability and cosmological description for an arbitrary $f(T)$ model based on the analysis carried out in Ref.~\cite{Mirza:2017vrk}. Here, $\tilde{x}_0$ is a solution to the equation $\tilde{y}(\tilde{x}) + 2\tilde{x} = 0$, $\tilde{x}_1$ for $\tilde{y}(\tilde{x}) + \tilde{x} = 1$ and $x_2$ when $1-\frac{1}{2}\tilde{y} + \tilde{y}_{\tilde{x}} \left(\tilde{x}+\frac{1}{2}\tilde{y}\right)$ diverges.}
	\label{table:fT-dynamics-Mirza}
\end{table}

Finally, in the second consideration examined in Ref.~\cite{Ganiou:2018dta}, the phase-space variables
\begin{equation}
\tilde{x} = -\frac{\ddot{f}}{H\dot{f}}\,, \quad\tilde{y} = \frac{f}{4H^2\dot{f}}\,, \quad\tilde{z} = \frac{3H^2+\dot{H}}{H^2}\,,
\end{equation}
were chosen and two scenarios were considered: (i) absence of matter fluids and (ii) presence of dust and radiation components. Furthermore, the case when the parameter $m = -\frac{\ddot{H}}{H^3}$ takes on constant values $m = 0$ (quasi-de Sitter evolution) and $m = -\frac{9}{2}$ (matter dominated evolution) was explored.

Starting with the vacuum case, an asymptotic stable de Sitter late-time cosmology can be reached while the matter domination is necessarily an unstable point. On the other hand, in the presence of matter, dust dominated unstable critical points and an asymptotic \gls{de} dominated de~Sitter critical point are obtained. However, it is unclear whether a radiation domination epoch can be generated.

Another interesting work which uses dynamical system is~\cite{Bohmer:2019qfi} where the authors studied the cosmology of $D$-dimensional $f(T)$ gravity. They focused on eleven dimensions (with seven dimensions being compactified) and found that in this case it is possible to find an early inflationary epoch without introducing any extra matter source. In another work~\cite{Fiorini:2013hva}, the compactification of extra dimensions in flat \gls{flrw} cosmology for $f(T)$ was further studied obtaining a correct tetrad in six and seven dimensions which can be used as the starting point to compactify the extra dimensions.

\subsection{\texorpdfstring{$f(T,B)$}{f(T,B)} cosmology} \label{sec:fTBbackgroundcosmo}

If we use the \gls{flrw} diagonal tetrad in Eq.~\eqref{FLRW_tetrad} in the $f(T,B)$ gravity equations~\eqref{fieldequationsfTB}, we arrive at the following modified Friedmann equations~\cite{Bahamonde:2016cul,Paliathanasis:2017flf,Zubair:2018wyy}
\begin{subequations}
\begin{align}
	3 H\dot{f}_B -3H^2 ( 3f_B+2f_T)-3 f_B\dot{H}-\frac{1}{2} f(T,B)&=\kappa ^2\rho\,,\label{eq:fTBFRW1}\\[0.5ex]
	-(3H^2+\dot{H})(2f_T+3f_B)-2 H\dot{f}_T+\ddot{f}_B-\frac{1}{2} f(T,B)&=-\kappa ^2 p\,.\label{eq:fTBFRW2}
\end{align}
\end{subequations}
The equations can be recast into a \gls{tegr}-like form when the \gls{tegr} contribution is maintained, i.e. $f(T,B) = -T + F(T,B)$, where the gravitational component $F(T,B)$ acts as a gravitational fluid with associated energy density $\rho_{\text{eff}}$ and pressure $p_{\text{eff}}$. In other words, Eqs.~\eqref{eq:fTBFRW1} and \eqref{eq:fTBFRW2} are rewritten as
\begin{subequations}
\begin{align}
    3H^2 &= \kappa^2 \left(\rho +\rho_{\text{eff}}\right)\,,\\[0.5ex]
    3H^2+2 \dot{H} &= -\kappa^2\left(p+p_{\text{eff}}\right)\,,
\end{align}
\end{subequations}
with the gravitational fluid defined as
\begin{subequations}
\begin{align}
\kappa^2 \rho_{\text{eff}} &= 3H^2\left(3F_B + 2F_T\right) - 3H\dot{F}_B + 3\dot{H}F_B + \frac{F(T,B)}{2}\,,\label{eq:fluid1}\\[0.5ex]
\kappa^2 p_{\text{eff}} &= -\frac{F(T,B)}{2}-(3H^2+\dot{H})\left(3F_B + 2F_T\right) -2H\dot{F}_T+\ddot{F}_B\,.\label{eq:fluid2}
\end{align}
\end{subequations}
Consequently, the effective \gls{eos} is \cite{Escamilla-Rivera:2019ulu,EscamillaRivera:2010py,Escamilla-Rivera:2021xql}
\begin{equation}\label{eq:eos_fTB}
    w_{\text{eff}} := \frac{p_{\text{eff}}}{\rho_{\text{eff}}} = -1+\frac{\ddot{F}_B-3H\dot{F}_B-2\dot{H}F_T-2H\dot{F}_T}{3H^2\left(3F_B+2F_T\right)-3H\dot{F}_B+3\dot{H}F_B+\frac{1}{2}F}\,.
\end{equation}
Under the assumption that the matter fluids do not interact with the gravitational fluid, the latter satisfies the standard fluid conservation equation
\begin{equation} \label{eq:conservations}
\dot{\rho}_{\text{eff}} + 3H(\rho_{\text{eff}} + p_{\text{eff}}) = 0\,.
\end{equation}
As discussed in Sec.~\ref{subsymmFLRW}, for the non-flat \gls{flrw} case there are two branches. Interestingly, in the first branch~\eqref{eq:cosmoptetradwb}, the boundary term does not depend on $k$ (see Eq.~\eqref{eq:torsionnonflat}). This means that for this branch the Friedman equations become
\begin{subequations}
\begin{align}
	3 H\dot{f}_B -3H^2 ( 3f_B+2f_T)-3 f_B\dot{H}-\frac{1}{2} f(T,B)&=\kappa ^2\rho\,,\label{eq:fTBFRW1non}\\[0.5ex]
	-(3H^2+\dot{H})(2f_T+3f_B)-2 H\dot{f}_T+\ddot{f}_B+\frac{2 k f_T}{a^2}-\frac{1}{2} f(T,B)&=-\kappa ^2 p\,.\label{eq:fTBFRW2non}
\end{align}
\end{subequations}
Recently, this branch was studied using dynamical systems finding that Milne, Milne-like and de-Sitter solutions exist~\cite{Paliathanasis:2022pgu}.
It is worth mentioning that the second branch~\eqref{eq:cosmontetradwb} can only be used for $k\leq0$ (since $T,B$ become complex) has not been used in the literature yet.

Following the same idea as we did in the previous section, we will analyze these equations further in the next sections.

\subsubsection{Reconstruction method}\label{sec:fTB-reconstruction-background}

One notices from the $f(T,B)$ cosmological equations~\eqref{eq:fTBFRW1}-\eqref{eq:fTBFRW2} that reconstructing for a general $f(T,B)$ function is not generally possible; thus, the ansatz choices $f(T,B) = f_1(T) + f_2(B)$ and its sub-case with $f_1(T) = -T$ were considered, for the (i) power-law $a(t) \propto t^h, h > 0$, (ii) de Sitter, (iii) $\Lambda$\gls{cdm}, and (iv) phantom dominated, cosmological behaviors \cite{Bahamonde:2016cul,Paliathanasis:2017flf,Zubair:2018wyy}. The respective solutions are summarized in the Supplementary annexes (Supplementary 1). A noticeable feature appears for $\Lambda$\gls{cdm} where the sub-case scenario offers a much simpler Lagrangian compared to the separable ansatz as it negates the existence of Gauss's hypergeometric function.

Beyond the reconstruction procedure, the solutions were then tested against the validity of the generalized second law of thermodynamics as well as their stability against homogeneous perturbations. In the former, each model is able to satisfy the constraint. In the latter, only the power-law and de Sitter behaviors were explored where both are able to exhibit stability. More importantly, however, is the fact that the power-law cosmology is only stable for $h > 1$, i.e., an expanding behavior.

\subsubsection{Noether's symmetry approach} \label{sec:cosmo-Noether-fTB}

Following Refs.~\cite{Bahamonde:2016grb,Bahamonde:2018ibz}, various $f(T,B)$ ansatz have been considered to examine the Noether symmetries, namely
\begin{align*}
\text{(i)} \quad b_0 B^k + t_0 (-T)^m\,, & & \text{(ii)} \quad B^l (-T)^n\,, & & \text{(iii)} \quad -T + F(B)\,, & &\text{(iv)} \quad B^n\,,
\end{align*}
where $b_0, t_0, k, m, l$ and $n$ are constants and $F(B)$ is some arbitrary function not linear in $B$. Of note, the first model ansatz only satisfies the Noether condition provided $k = 1$ reducing the model to the power-law $f(T)$ model investigated in Sec.~\ref{sec:cosmo-Noether-fT}. For the remaining models, further restrictions and cosmological implications are summarized in Table~\ref{table:cosmo-noethersym-fTB}. It is noted that for the $-T + F(B)$ ansatz, a similar $F(B)$ functional form appears in Refs.~\cite{Bahamonde:2016cul,Paliathanasis:2017flf}.

For the $f(\lc{R},T)$ gravity sub-case, the Noether symmetry has been explored in Ref.~\cite{Capozziello:2014bna}. As the Noether condition yields an intractable system of equations to solve generally, two $f(\lc{R},T)$ ansatz were considered which reduce to: (i)~$f(\lc{R},T) = g_0 \lc{R} + h_0 T$ and (ii)~$f(\lc{R},T) = f_0\lc{R}^n (-T)^{1-n}$ where $f_0, g_0, h_0$ and $n$ are arbitrary integration constants. As the first model is a \gls{tegr} rescaling, the second model offers a richer cosmology provided that $n \neq 1$ and is absent in both $f(\lc{R})$ and $f(T)$ formulations.

\begin{table}[!ht]
\centering
\begin{tabularx}{\textwidth}{lX}
\toprule
\cellcolor{gris3}\boldmath{$f(T,T_B)$} & \cellcolor{gris3}\textbf{Symmetry and Cosmological Implications} \\ \midrule
\cellcolor{gris1} & \cellcolor{gris1}Condition enforces $m = \frac{1-l}{2}$. Depending on the symmetry, power-law $a(t) \propto t^{\frac{1+l}{3}}$ or\\
\multirow{-2}{*}{\cellcolor{gris1}$B^l (-T)^n$}&\cellcolor{gris1}exponential $a(t) = \left[\alpha_0 e^{\frac{\beta t}{\sqrt{6}}} + \alpha_1\right]^{\frac{1}{3}}$ behaviors can be recovered. \\
\cellcolor{gris3} & \cellcolor{gris3}$F(B)$ constrained to be $F(B) = -\frac{B}{3} \ln B$. \\
\multirow{-1}{*}{\cellcolor{gris3}$-T + F(B)$}&\cellcolor{gris3}When dust is considered, $a(t) = \left[\alpha_2 e^{\alpha_3 t} +\alpha_4 t+\alpha_5 \right]^{1/3}$.\\
\cellcolor{gris3}&\cellcolor{gris3}In the case of dark energy, the solution is identical except $\alpha_4 = 0$. \\
\cellcolor{gris1}$B^n$ & \cellcolor{gris1}Symmetries exist for dust and $w = \frac{1}{2n-1}$. However, no cosmology has been reported. \\
\bottomrule
\end{tabularx}
\caption{A summary of the existence of symmetries and the cosmological implications of the $f(T,T_B)$ ansatz considered in Refs.~\cite{Bahamonde:2016grb,Bahamonde:2018ibz}. Here, $\alpha_0,\dots{},\alpha_5$ and $\beta$ are integration constants.}
\label{table:cosmo-noethersym-fTB}
\end{table}

\subsubsection{Dynamical system approach} \label{Sec:fTB cosmology}

The dynamics of the specific class $f(T,B) = -T + F(B)$, for some unknown function $F(B)$, has been investigated via its scalar-tensor equivalent form by introducing an auxiliary scalar field $\phi = F_B$ with an associated potential $V(\phi) = B F_B - F(B)$ \cite{Paliathanasis:2017flf}. This scalar field (or equivalently the $F(B)$ contribution) is assumed to act as the source for \gls{de}. In particular, the case when \gls{de} is absent during certain phases of evolution was studied, which leads to matter dominated phases. Here, the matter fluids are assumed to have an \gls{eos} $w_\text{m} \in [0,1)$.

One avenue is to set the \gls{de} energy density and pressure to be vanishing which leads to $V(\phi) \propto \phi^{\frac{2-\gamma}{2+\gamma}}$ (or equivalently, $F(B) \propto B^{-\frac{2-\gamma}{2\gamma}}$). However, according to dynamical analysis, the cosmology is unable to evolve past the matter domination phase. Instead, \gls{de} was then considered with non-vanishing energy density and pressure while still being magnitudes smaller than that of matter. In this case, the scalar field evolution equation becomes\footnote{The matter contribution Eq.~(74) in Ref.~\cite{Paliathanasis:2017flf} was not expressed correctly.}
\begin{equation}
    \ddot{\phi} + \dfrac{V}{2} + \dfrac{(1-\gamma)}{3\gamma^2 t^2} \left(4-3\gamma^2 \kappa^2 \rho_{\rm m0}\right) = 0\,,
\end{equation}
where $\gamma = 1 + \omega_\text{m}$ is the barotropic index. Naturally, during matter domination, the bracketed term is vanishing reducing the system to the previous scenario. In this case, a solution exists for the power-law potential $V(\phi) = V_0 \phi^\mu$ with $\mu \neq 1$ without the necessary constraint that $\mu = \frac{2-\gamma}{2+\gamma}$. Additionally, $p_\phi \simeq 0$ for this potential meaning that the scalar fluid behaves as dust at leading order. The dynamical systems analysis reveals the existence of three critical points for $\mu \neq -1$ with deviation from matter domination occurring only for $\mu > 1$.

An extension of the above approach is presented in Ref.~\cite{Paliathanasis:2017efk} where the dynamical system procedure was applied with a different set of phase-space variables which permit a more general description of the resulting cosmology. In order to conform with the signature notation used in the Review, the auxiliary scalar field $\phi$ relations are modified to $\phi = -F_B$ and $V(\phi) = F(B) - BF_B$. Thus, the chosen phase-space variables are
\begin{equation}
    \tilde{x} := \frac{\dot{\phi}}{H} \,, \quad \tilde{y} := \frac{V(\phi)}{6H^2} \,, \quad \lambda := -\frac{V_\phi}{V(\phi)} \,.
\end{equation}
Here, the choice of $\lambda$ determines the nature of the potential (and hence, the choice of $F(B)$). In particular, two scenarios were explored: (a) $\lambda$ = constant leading to $V(\phi) \propto e^{-\lambda \phi}$, and (b) $\lambda \neq 0$, i.e. a general potential. As the critical points of (a) are contained in the general case (b), we highlight the latter results. For the above phase-space variables, five critical points are obtained which nature and existence are summarized in Table~\ref{table:dynamics-critpts-fTB-2}.

\begin{table}[!ht]
	\centering
	\midsepremove
	\begin{tabularx}{\textwidth}{lcp{2cm}Xp{7cm}}
		\toprule
		\cellcolor{gris3}& \cellcolor{gris3}\boldmath{$(\tilde{x}^\star, \tilde{y}^\star,\lambda^\star)$}& \cellcolor{gris3}\textbf{Existence} & \cellcolor{gris3}\textbf{Stability} & \cellcolor{gris3}\textbf{Properties} \\ \midrule
		\cellcolor{gris1}$P_1$ & \cellcolor{gris1}$(1,0,\lambda_0)$& \cellcolor{gris1} $\bar{\Gamma}(\lambda_0) = 0$ & \cellcolor{gris1}Unstable & \cellcolor{gris1}Matter fluid absent; effective fluid \newline behaves as a stiff fluid $(w_{\text{eff}} = 1)$ \\
		\cellcolor{gris3}$P_2$ & \cellcolor{gris3}$(2\lambda_0^+,\lambda_0^-,\lambda_0)$& \cellcolor{gris3}$\lambda_0 \geq \tilde{\lambda},$ \newline $\bar{\Gamma}(\lambda_0) = 0$ & \cellcolor{gris3}$\lambda_0 > \tilde{\lambda}$ & \cellcolor{gris3}Effective fluid behaves as dust \newline $(w_{\text{eff}} = 0)$ \\
		\cellcolor{gris1}$P_3$ & \cellcolor{gris1}$(-\frac{6}{\lambda_0}+2,\frac{6}{\lambda_0}-1,\lambda_0)$& \cellcolor{gris1}$\bar{\Gamma}(\lambda_0) = 0$ & \cellcolor{gris1}$\lambda_0 < \tilde{\lambda}, (-3+\lambda_0)\bar{\Gamma}_\lambda(\lambda_0)>0$ & \cellcolor{gris1}Matter fluid absent; effective fluid \gls{eos} $w_{\text{eff}} = \frac{2\lambda_0-9}{3}$ (accelerating for $\lambda_0 < \frac{9}{2}$ and de~Sitter for $\lambda = 3$) \\
		\cellcolor{gris3}$P_4$ & \cellcolor{gris3}$(0,1,3)$& \cellcolor{gris3}Always & \cellcolor{gris3}$\Re(\bar{\Gamma}(3)) > 0$ & \cellcolor{gris3}Matter fluid absent; de Sitter behavior (effective fluid behaves as \gls{de}) \\
		\cellcolor{gris1}$P_5$ & \cellcolor{gris1}$(1,0,0)$ & \cellcolor{gris1}Always & \cellcolor{gris1}Unstable & \cellcolor{gris1}Same as $P_1$\\
		\bottomrule
	\end{tabularx}
	\midsepdefault
	\caption{The critical points for $f(T,B) = -T + F(B)$ together with their existence conditions viewed from the introduction of an auxiliary scalar field $\phi = -F_B$ \cite{Paliathanasis:2017efk}. Here, $\lambda_0$ is an arbitrary value, $\bar{\Gamma}(\lambda) \equiv \frac{V_{\phi\phi}}{{V_\phi}^2}-1$, $\lambda^{\pm}_0 \equiv \frac{3}{2\lambda_0}(1\pm w)$ and $\tilde{\lambda} \equiv \frac{3}{2}(3+w)$.}
	\label{table:dynamics-critpts-fTB-2}
\end{table}
	
Evidently, the model hosts stiff fluid behaviors $(P_1,P_5)$ and two possible stable de~Sitter behaviors $(P_2, P_4)$. Meanwhile, the matter domination epoch is achieved via $P_2$. In the instance when $\lambda =$ constant, only the first critical points $P_1$--$P_3$ remain. For constant $\lambda$, matter domination $\to$ late-time acceleration can be achieved through $P_2 \to P_3$ whereas for the general model, a similar behavior is achieved through $P_2 \to P_3/P_4$.

Recently, the scalar-tensor approach has also been explored for the $f(T,B) = F_1(T) + F_2(B)$ model ansatz where $F_1(T) = -T + f_0 (-T)^n$ and $F_2(B) = \frac{B \ln B}{\lambda}$ where $f_0, \lambda$ and $n$ are constants \cite{Paliathanasis:2021ysb}. In particular, the $F_2(B)$ function leads to a scalar potential $V(\phi) \propto e^{-\lambda \phi}$. Thus, this model serves as an extension of the previous $\lambda$ = constant scenario where a torsion power-law contribution has now been included. For this choice, in the absence of matter fluids, the chosen phase-space variables are
\begin{equation}
    \tilde{x} := \frac{\dot{\phi}}{\sqrt{1+H^2}} \,, \quad \tilde{y} := \frac{V(\phi)}{6H^2} \,, \quad \tilde{z} := \frac{(-T)^n}{1+H^2}\,, \quad \eta := \frac{H}{\sqrt{1+H^2}} \,.
\end{equation}
Since $\tilde{y}$ is constrained by the Friedmann constraint \eqref{eq:fTBFRW1} and $\tilde{z} = \tilde{z}(\eta)$ by virtue of $T = -6H^2$, the system's dimensionality reduces to a 2 dimensional one. This leads to six finite critical points $(\tilde{x},\eta)$ being:
\begin{enumerate}[label = (\alph*{})]
    \item two unstable critical points where the effective fluid behaves as stiff matter. Only the kinetic term of the auxiliary field contributes to the fluid's behavior;
    \item two critical points where the effective fluid behaves either as a perfect fluid with $w_{\text{eff}} = \frac{2\lambda}{3}-3$ for $\lambda \neq 3$ or as a cosmological constant for $\lambda = 3$. These points are attractors for $\lambda < 3$ and $n < 1$, otherwise, they act as sources/saddle points;
    \item a de Sitter critical point which can serve as a late-time attractor;
    \item a Minkowskian spacetime in which stability can be determined through centre manifold theory. Nonetheless, for the considered numerical analysis, it is observed to be unstable.
\end{enumerate}
Meanwhile, the critical points lying at the infinite boundary cannot be attractors. Thus, a viable cosmology can be realized for suitable parameter choices. However, to confirm the existence of matter and radiation dominated phases, the contribution from matter fluids still needs to be incorporated in the analysis.

Scenarios where the contributions of \gls{de} without a scalar field are assumed to be the source has not been studied widely yet. In Ref.~\cite{Franco:2020lxx} a dynamical analysis was presented where, contrary to the previous approach, the cosmic acceleration effect is analyzed without the introduction of an auxiliary scalar field. In the following, an $f(T,B)$ \gls{de} which is fluid-like is used to obtain a richer population of stability points that can be constrained by observational surveys \cite{Karpathopoulos:2017arc}. The models can also be tested against thermodynamics and cosmological stability \cite{Bahamonde:2016cul,Pourbagher:2019zhq} as well as energy conditions \cite{Zubair:2018wyy}. 

Taking the ansatz $f(T,B) = -T + F(T,B)$, in order to construct the dynamical system, the \gls{flrw} equations \eqref{eq:fTBFRW1}--\eqref{eq:fTBFRW2} are first rewritten as 
\begin{subequations}
\begin{align}
\Omega + \Omega_{\text{eff}} &= 1\,, \label{eq:friedmann_f} \\[0.5ex]
3 + 2\left(\frac{H'}{H}\right) &= \frac{F}{2H^2} + 9F_B +6F_T + \left(\frac{H'}{H}\right)(3F_B + 2F_T - F'_B) + 2F'_T - F''_B - \frac{\kappa^2 p}{H^2}\,, \label{eq:friedmann_f2} \\[0.5ex]
\Omega_{\text{eff}} &\equiv 3F_B + 2F_T - F'_B + \frac{F}{6H^2} + \left(\frac{H'}{H}\right)F_B\,,
\end{align}
\end{subequations}
which each $i$ denotes the component density parameter $\Omega_i = \kappa ^2 \rho_i/(3H^2)$. The prime $(\prime)$ denotes derivatives \gls{wrt} $N_{\rm f} =\ln{a}$, with a chain rule given by $\dd/\dd t = H (\dd/\dd N_{\rm f})$. These set of equations impose a condition over the form of the derivative $f'(T,B)$. Taking the derivative \gls{wrt} $N_{\rm f}$ in Eq.~\eqref{eq:friedmann_f} and making use of the \gls{flrw} equations, the relation
\begin{equation}
6\left(\frac{H'}{H}\right)F_B + 2\left(\frac{H'}{H}\right)F_T + \left(\frac{H'^2}{H^2} + \frac{H''}{H}\right)F_B + \frac{F'}{6H^2} = 0\,,
\end{equation}
is obtained which can be used to impose a condition, at least at a background level, over the form of the derivative $F^\prime(T,B)$. Naturally, further constraints can be obtained once cosmological perturbations are considered (see Sec.~\ref{sec:cosmo-pert}). 

Following Eqs.~\eqref{eq:friedmann_f}--\eqref{eq:friedmann_f2}, a set of conveniently specified dimensionless variables are chosen to construct the autonomous system \cite{Shah:2019mxn,Mirza:2017vrk}. First, the parameter \cite{Odintsov:2018uaw}
\begin{equation}\label{eq:ansatz_lambda}
\lambda = \frac{\ddot{H}}{H^3} = \frac{H'^2}{H^2} + \frac{H''}{H}\,,
\end{equation}
is introduced. For cases where $\lambda$ = constant, some cosmological solutions can be recovered including de~Sitter/quasi-de~Sitter $(\lambda=0)$ and matter domination $(\lambda = 9/2)$. This ansatz, which shall be assumed in the following, shows cosmological viable scenarios as analogous to models with barotropic fluids with \gls{eos} $w$. However, this choice restricts the number of critical points which in turn leaves an incomplete picture of the whole cosmological evolution for the chosen $F(T,B)$ model. To remove this imposition, either a different dynamical system approach or suitable choice of the $F(T,B)$ function must be adopted. In fact, the latter case is discussed in \cite{Rave-Franco:2021yvu} for the mixed power-law model $F(T,B) \propto B^k T^m$ for constants $k, m$. Following this, the dynamical phase-space variables 
\begin{equation}\label{eq:aut-system}
\tilde{x} := F_B\,, \quad
\tilde{y} := F'_B\,, \quad
\tilde{z} := \frac{H'}{H} = \frac{\dot{H}}{H^2}\,, \quad
g := \frac{F}{6H^2}
\end{equation}
are introduced. The energy constriction, Eq.~\eqref{eq:friedmann_f}, in terms of the above variables becomes
\begin{equation}
\Omega + 3\tilde{x} +2F_T - \tilde{y} +g +\tilde{z} \tilde{x}=1\,, \label{eq:constrain}
\end{equation}
where $\Omega = \kappa^2\rho/(3H^2)$ depends on the other dynamical variables. We can write the autonomous system for this theory using \eqref{eq:friedmann_f} as \cite{Franco:2020lxx,Rave-Franco:2021yvu} 
\begin{subequations}
\begin{align} 
\tilde{x}' &= \tilde{y}\,, \label{eq:variables_system} \\[0.5ex]
\tilde{y}' &= 3g + (9 + 3\tilde{z})\tilde{x} + F_T(6 + 2\tilde{z}) + 2 F'_T - \tilde{z}\tilde{y} -3 - 2\tilde{z} -\frac{\kappa^2 p}{H^2}\,, \label{eq:variables_system2} \\[0.5ex]
\tilde{z}' &= \lambda - 2\tilde{z}^2\,, \label{eq:variables_system4}\\[0.5ex]
g' &= -6\tilde{z}\tilde{x} - 2\tilde{z} F_T - \lambda \tilde{x} - 2\tilde{z}\tilde{g}\,. \label{eq:variables_system3}
\end{align}
\end{subequations}
To follow the constraint of the system in Eqs.~\eqref{eq:constrain}--\eqref{eq:variables_system3}, we need to write $F_T$ either as a dynamical variable or expressed in terms of the described variables. The latter can be done by considering a specific form of $F(T,B)$. Finally, we can rewrite the \gls{eos} \eqref{eq:eos_fTB} as 
\begin{equation}\label{eq:eos_dynamicalfTB}
    w_{\text{eff}} = \frac{H^2 \tilde{z} (2\tilde{z}+3)-\kappa ^2 p \tilde{z}}{3 H^2 \left[g'+\tilde{z} (g-\tilde{x}\tilde{z}+3\tilde{x}+\tilde{y})+\lambda x\right]}\,.
\end{equation}
From Eqs.~\eqref{eq:variables_system} and \eqref{eq:variables_system4}, we notice that there is not an explicit dependency of $F(T,B)$; therefore, for the critical points, we require that
\begin{equation}\label{eq:fTB-y-and-z-critpts_constraint}
\tilde{z} = \pm \sqrt{\frac{\lambda}{2}}\,, \quad \tilde{y}=0\,.
\end{equation}
As $\tilde{z} = \dot{H}/H^2$, this variable becomes an important determining factor for the stability and cosmological nature (power-law $\tilde{z} \neq 0$ or de Sitter $\tilde{z} = 0$) of the critical points. To achieve this, we are going to consider three cosmologically viable models, determine the critical points (by determining the remaining critical values $\tilde{x}$ and $g$), then consider the eigenvalues of the stability matrix and the constriction equation Eq.~\eqref{eq:constrain}. In the following, the results reported from \cite{Franco:2020lxx} are for the case when the perfect fluid is sourced by dust $w = 0$.

\begin{enumerate}
	\item \textbf{General Taylor Expansion model} -- The form for this model was presented in Ref.~\cite{Farrugia:2018gyz}, given by
	{\small
		\begin{alignat}{2}
		F(T,B) & =\: & & F(T_0, B_0) + F_T(T_0,B_0) (T-T_0) + F_B(T_0,B_0) (B-B_0) + \frac{1}{2!}F_{TT}(T_0,B_0) (T-T_0)^2 \nonumber\\[0.5ex]
		& \: & &+ \frac{1}{2!}F_{BB}(T_0,B_0) (B-B_0)^2 + F_{TB}(T_0,B_0) (T-T_0)(B-B_0) + \mathcal{O}(T^3,B^3)\,, \label{taylor_model}
		\end{alignat}
	}
	\hspace*{-3mm} which gives the general Taylor expansion of the $f(T,B)$ Lagrangian about its Minkowski values for the torsion scalar $T$ and boundary term $B$. The Minkowski space is chosen as the background for this model since all models should contain this spacetime in some limit. We notice from here that we need to take into account beyond linear approximations since $B$ is a boundary term at linear order. Where locally spacetime appears to be Minkowski, we can consider $T_0 = B_0 = 0$. Labelling the constant coefficients as $A_i$, the Lagrangian can be rewritten as
	\begin{equation}\label{taloy}
	F(T,B)\simeq A_0+A_1 T + A_2 T^2 + A_3 B^2 + A_4 TB\,,
	\end{equation}
	where the linear boundary term does not contribute so we have omitted it. We notice from this specific form that the first term can be seen as $A_0 \approx \Lambda$, therefore we are dealing with a cosmic acceleration as a consequence of the $F(T,B)$. Thus, the form of this model can be written in terms of the dynamical variables as
	\begin{equation} \label{eq:taylor_system}
	    F_T = -(3+\tilde{z})\tilde{x} - 2g - A_1\,,
	\end{equation}
	at linear order in torsion and with $A_0=0$, i.e we are switching-off the cosmological constant. This can be done since an explicitly time-dependent factor appears and then a different approach has to be taken. The critical points for this model are
	\begin{equation}\label{taylor_crit}
	g= -A_1\,, \quad 
	\tilde{x} = \frac{A_1 -1}{3 \pm \sqrt{\frac{\lambda}{2}}}\,,
	\end{equation}
	where the case $3 \pm \sqrt{\frac{\lambda}{2}} \neq 0$ was not explored. The constriction evolution equation \eqref{eq:constrain} sets $\Omega=0$ hereby denoting contributions only related to $F(T,B)$ gravity which imply that the constriction evolution equation in Eq.~\eqref{eq:constrain} is now $\Omega=0$ which denotes only contributions related to $F(T,B)$ gravity. According to these points, the following eigenvalues result 
	\begin{equation}\label{taylor_eigenvalues}
	\omega_{1} =-3 \mp \sqrt{\frac{\lambda}{2}}\,, \quad \omega_{2} =-3 \mp 2\sqrt{\frac{\lambda}{2}}\,,
	\quad \omega_{3} =\mp 4\sqrt{\frac{\lambda}{2}}\,, \quad \omega_{4} =\pm 2\sqrt{\frac{\lambda}{2}}\,.
	\end{equation}
	Considering values $\lambda \neq 0$, we get that $\text{Re}(\omega_3) = -\text{Re}(\omega_4) \neq 0$ implying that, for this system, all the critical points are saddle-like. The solutions for this case are in agreement with the cosmological constraints found in Ref.~\cite{Escamilla-Rivera:2019ulu}. According to these results, the critical points behave as $A_{i}< A_{i+1}$ (with $A_0=0, \, A_1=1$), show a decelerating behavior with $0 \leq w_\text{eff} < 1$ when $B$ dominates at redshift $z\approx 1$ followed by a $\Lambda$\gls{cdm} behavior.
	\item \textbf{Power Law model} -- Considering a Lagrangian of separated power law style models for the torsion and boundary scalars, we can write a model like \cite{Bahamonde:2016grb}
	\begin{equation}
	F(T,B) = b_0 B^k + t_0 (-T)^m\,, \label{eq:powerlaw}
	\end{equation}
	for constants $b_0$ and $t_0$. This is an interesting model since it was already been shown in Ref.~\cite{Said:2017nti} that for $m<0$, the Friedmann equations will be effected mostly in the accelerating late-time Universe while for $m>0$, this impact will take place for the early Universe, assuming no input from the boundary contribution. By incorporating the boundary term, this analysis will reveal an effect of $B$ on the combined evolution within $f(T,B)$ cosmology. The form for this model can be written in terms of the dynamical variables as
	\begin{equation}\label{eq:powerlaw_dynamical}
	F_T = -mg - \frac{m}{k}(3+\tilde{z})\tilde{x}\,.
	\end{equation}
	The critical points for this scenario are
	\begin{equation}
	g = \frac{k-m}{m(1-k)}\,, \quad 
	\tilde{x} = -\frac{k}{m} \Big(\frac{m-1}{k-1}\Big) \frac{1}{3 \pm \sqrt{\frac{\lambda}{2}}}\,,
	\end{equation}
where, once more, the case $3 \pm \sqrt{\frac{\lambda}{2}} \neq 0$ was not explored. In this scenario, we find attractors or saddle points for the positive branch whereas repellent or saddle points for the negative branch. Additionally, for the former case, keeping $b_0, t_0 > 0$, $\Lambda$\gls{cdm} and late cosmic acceleration can be recovered.

	\item \textbf{Mixed Power Law Model} -- In order to reproduce several important power law scale factors relevant for several cosmological epochs, in Ref.~\cite{Bahamonde:2016grb} a form of $F(T,B)$ given by
	\begin{equation} \label{eq:mixed_power}
	F(T,B) = f_0 B^k (-T)^m\,,
	\end{equation}
	was presented, where the second and fourth order contributions will now be mixed, and $f_0,k,m$ are arbitrary constants. We can recover $\Lambda$\gls{cdm} when the index powers vanish, i.e. when $k=m=0$. For this case, the model can be written in terms of the dynamical variables through
	\begin{equation}\label{eq:mixed_case}
	F_T = -mg\,.
	\end{equation}
	In comparison to the latter $F(T,B)$ scenarios, this case implies
	\begin{equation}
	\tilde{x} = F_B = k f_0 B^{k-1} (-T)^m = \frac{k}{B}F = -\frac{k}{6(3H^2 + \dot{H})}F = -\frac{F}{6H^2}\frac{k}{3 +\frac{\dot{H}}{H^2}}=-\frac{g k}{3+\tilde{z}}\,,
	\end{equation}
	from which we can notice that $\tilde{x}$ is not an independent variable of the dynamical system. In the same way, as $\tilde{y} = \tilde{x}'$, we directly obtain that $\tilde{y} = \tilde{y}(g,\tilde{z})$. With these conditions, the autonomous system can be reduced to a 2 dimensional dynamical phase-space
	\begin{equation}
	\tilde{z}' = \lambda - 2 \tilde{z}^2\,, \quad
	g' = g\left[\frac{6\tilde{z}(k+m-1) + \lambda k + 2\tilde{z}^2(m-1)}{3+\tilde{z}}\right]\,,
	\end{equation}
	with critical points
	\begin{equation}
	\tilde{z} = \pm \sqrt{\frac{\lambda}{2}}\,, \quad \text{and} \quad g=0\,.
	\end{equation}
	Under these values, the constriction of the system is given by $\Omega = 1$. Once more, $\tilde{z}$ turns out to be the determining factor in the behavior of the dynamics of the system. On the other hand, the system turns out to be relatively straightforward to analyze with clear cut results which tally with the general results of the power law model. 
	
	However, the more general approach in Ref.~\cite{Rave-Franco:2021yvu}, which includes both radiation $\rho_{\rm r}$ and a perfect fluid with \gls{eos} $w$, $\rho_w$, leads to a different conclusion. Most notably, $\lambda$ is no longer constrained to be constant as it is expressed in terms of the dynamical variables. This alters the former matter dominated critical point to a late-time de Sitter one with $\Omega_{\rm r} = \Omega_w = 0, \Omega_{\text{eff}} = 1$ which can be stable under suitable parameter conditions. Investigation of the cosmological evolution close to the critical point was then used to best-fit $H_0$ using $H(z)$ observations. The obtained values can match with both early $(\Omega_{\rm r} \neq 0)$ and late-time $(\Omega_{\rm r} = 0)$ observations. Such $H_0$ values can be derived for three cosmological scenarios which links their behavior together with the error propagation on the free variables of the system. A full description of this matter can be seen in Fig.~3 in \cite{Rave-Franco:2021yvu}.
\end{enumerate}

The above results can be linked in a more straightforward manner if we consider directly the form for the \gls{eos} Eq.~\eqref{eq:eos_fTB}. At the critical point, for the first two cases (Taylor and Power law model),
\begin{equation}
w_{\text{eff}} = -1 \mp \frac{2}{3} \sqrt{\frac{\lambda}{2}}\,.
\end{equation}
Meanwhile, for the Mixed Power Law model
\begin{equation}
\text{\cite{Franco:2020lxx} $\lambda =$ constant:} \quad w_{\text{eff}} = -1 \mp \frac{2}{3}(k + m)\sqrt{\frac{ \lambda}{2}}\,, \quad \text{\cite{Rave-Franco:2021yvu} non-constant $\lambda$:} \quad w_{\text{eff}} = -1\,.
\end{equation}
Notice that we recover $\Lambda$\gls{cdm} in the \gls{gr} limit. In both \gls{eos} scenarios, we recover a $\Lambda$\gls{cdm} model when $\lambda$ vanishes as $\tilde{z} = 0$. Notice that we can rewrite the $\tilde{z}$-variable from Eq.~\eqref{eq:aut-system} using the definition of the second cosmographic parameter, the deceleration parameter $q$, as $\tilde{z}=-(q+1)$. Thus, $\lambda = 0$ leads to $q=-1$. We can also rewrite the ansatz for $\lambda$ given in Eq.~\eqref{eq:ansatz_lambda} in terms of the third cosmographic parameter, the jerk, as $j = \lambda \mp 3\sqrt{\frac{\lambda}{2}} +1$. Notice that when $\lambda=0$, we recover the standard value $j=1$.

\begin{figure}[!ht]
	\centering
	\includegraphics[width=0.8\textwidth]{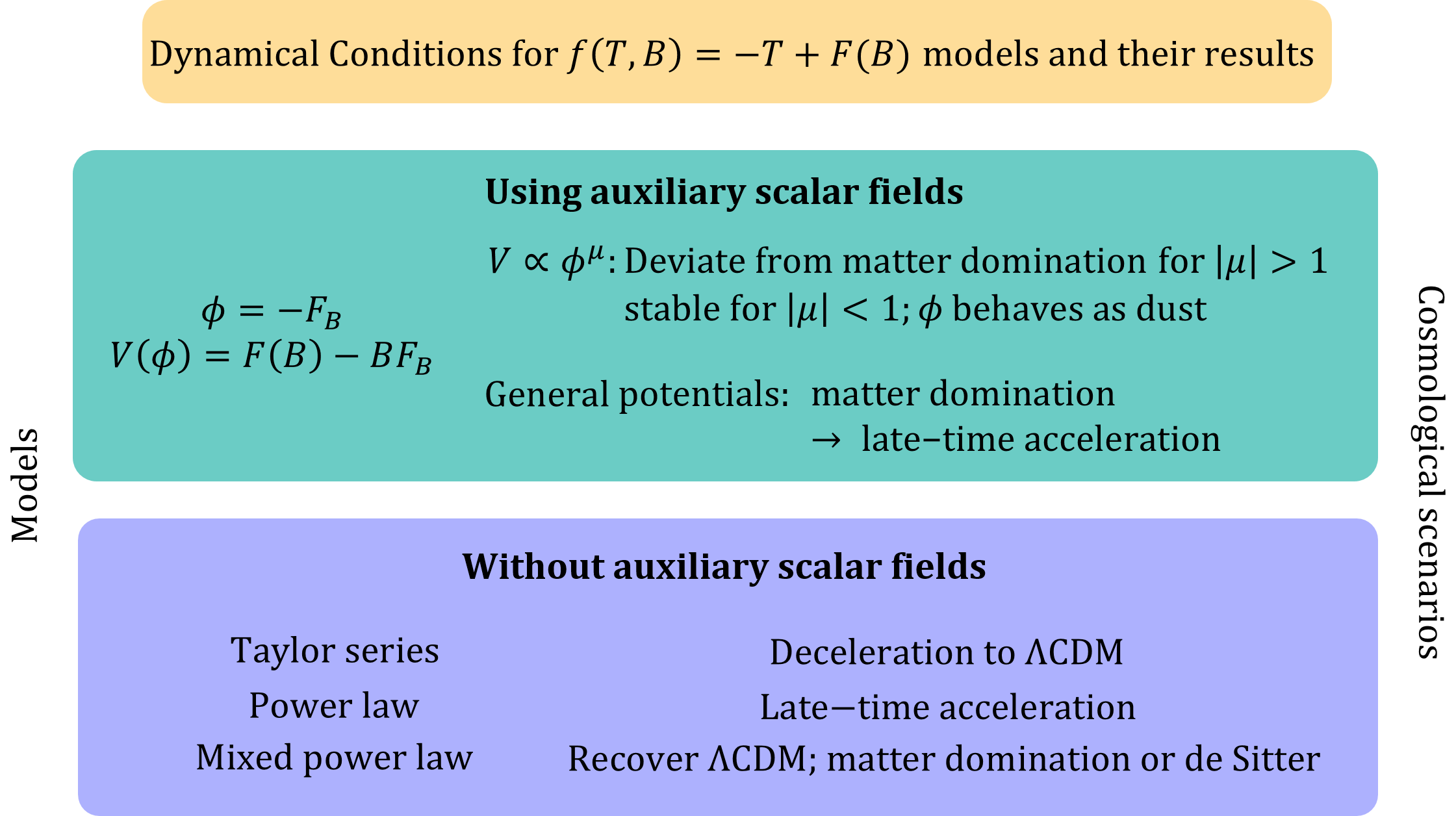}
	\caption{Dynamical behavior for $f(T,B) = -T + F(B)$ models and their results.}
	\label{fig:dynamical_fTB}
\end{figure}

Notice that an important feature of this analysis (as summarized in Fig.~\ref{fig:dynamical_fTB}) is the role of couplings between the torsion tensor $T$ and the boundary term $B$. These represent the second order and fourth order contributions to the field equations respectively, and in a particular combination, forming $f(\lc{R})=f(-T+B)$ gravity. In $f(\lc{R})$ gravity, these couplings do appear but in a very prescribed format.

\subsection{Teleparallel Gauss-Bonnet cosmology}

In the case of $f(T,T_G)$ gravity, since the tensorial field equations are very involved, it is easier to use the minisuperspace for flat \gls{flrw} cosmology. To do this, we require the torsion scalar and the teleparallel Gauss-Bonnet scalar in the minisuperspace, which are given by
\begin{align}
    T_G = \frac{24H^2}{N}\dot{H}+24 H^4\,,\quad T=-6H^2\,,
\end{align}
and then the Friedmann equations for $f(T,T_G)$ gravity can be directly found by using the minisuperspace analysis and then fixing the gauge such that $N=1$, giving us
\begin{subequations}
\begin{align}
-6 f_T H^2-12 H^3 \dot{f}_{T_G}+12f_{T_G} H^2 \left(\dot{H}+H^2\right)-\frac{1}{2} f(T,T_G)&=\kappa^2\rho\,,\label{eq:FRWfTG1}\\[0.5ex]
-2 H\dot{f}_T-2 f_T \left(\dot{H}+3 H^2\right)+12f_{T_G} H^2 \left(\dot{H}+H^2\right)-8 H \dot{f}_{T_G} \left(\dot{H}+H^2\right)&\nonumber\\
-4 H^2 \ddot{f}_{T_G}-\frac{1}{2} f(T,T_G)&=-\kappa^2p\,,\label{eq:FRWfTG2}
\end{align}
\end{subequations}
where $f_{T_G}=\partial f/\partial T_G$. It is worth mentioning that $B_G=0$ in flat \gls{flrw} and then the standard Gauss-Bonnet term is just $\lc{\mathcal{G}}=T_G$. 

\subsubsection{Reconstruction method}\label{sec:reconstruction-fTTG}

The \gls{tegb} extension has been investigated in Refs.~\cite{delaCruz-Dombriz:2017lvj,delaCruz-Dombriz:2018nvt} for various late-time behaviors. Due to the introduction of the \gls{tegb} scalar, a number of model ansatz were considered
\begin{align*}
    \text{(i)} &\quad f(T,T_G) = g(T) + h(T_G)\,, & \text{(iv)} &\quad f(T,T_G) = -T + T_G g(T)\,, \\[0.5ex]
    \text{(ii)} &\quad f(T,T_G) = T g(T_G)\,, & \text{(v)} &\quad f(T,T_G) = -T + \mu (-T)^\beta {T_G}^\gamma\,. \\[0.5ex]
    \text{(iii)} &\quad f(T,T_G) = T_G g(T)\,, & &
\end{align*}
Of note, models (iii) and (iv) are distinct despite their apparent similarity. Nonetheless, the resulting reconstructed solutions are related via:
\begin{equation}
    \text{Model (iv)} \qquad f(T,T_G) = -T + \frac{3T_G}{4T} + \text{Solution of model (iii)}\,.
\end{equation}
As such, only the results of model (iii) are summarized. Although a majority of the ansatz and cosmological behaviors considered do yield an analytical solution (see Supplementary annexes (Supplementary 1)), some scenarios do not host a solution. Note that for the power-law cosmology $a(t) \sim t^\alpha$ reconstructed solutions, the coasting cosmology $(\alpha = 1)$ has been investigated separately as $T_G = 0$ in this case.

Certain reconstructed solutions constitute relatively more complicated forms including a lack of a closed form or are expressed in terms of Gauss's hypergeometric function. Despite this apparent complexity, the nature of the perfect fluid may yield simpler results. For instance, the $\Lambda$\gls{cdm} cosmology with $w = 0$ reduces to the standard $\Lambda$\gls{cdm} Lagrangian for model (i). However, non-trivial models are also obtained, such as
\begin{equation}\label{eq:fTTG-reconstruction-LCDM-nontrivial-I}
f(T,T_G) = \frac{3 T_G (\Omega_\Lambda T_0 -2 T)}{8 T^2}\,,
\end{equation}
by virtue of model (iii). This could be resourceful in other sectors such as cosmological perturbations as a different choice of Lagrangian may lead to different conclusions even though the background evolution is still that of $\Lambda$\gls{cdm}.

The reconstructed models were then examined and checked if they are able to host vacuum solutions i.e. whether $f(0,0) = 0$. The reconstructed $\Lambda$\gls{cdm} model behavior cannot host vacuum solutions (see Supplementary annexes (Supplementary 1)). As shown in Table~\ref{table:reconstruct_sols-vacuum}, the reconstructed $\Lambda$\gls{cdm} model behavior cannot host vacuum solutions. On the contrary, the power-law can satisfy the vacuum constraint subject to conditions. While the de Sitter scenario is not listed, the vacuum condition can be satisfied by suitable Lagrangian choices.

\begin{table}[!htb]
\centering
\midsepremove
\begin{tabularx}{0.8\textwidth}{rlX}
\toprule
\cellcolor{gris3}\textbf{Ansatz} & \cellcolor{gris3}\textbf{\boldmath{$\Lambda$}CDM} & \cellcolor{gris3}\textbf{Power-law} \\ \midrule
 \cellcolor{gris1}& \cellcolor{gris1} & \cellcolor{gris3}$\alpha \neq 1$: $w <-1, \, \alpha < 0$ or $w >-1, \, 0<\alpha <1$ \\
\multirow{-2}{*}{\cellcolor{gris1}(i)}&\multirow{-2}{*}{\cellcolor{gris1}\xmark}&$\alpha = 1$: $w > -1$ or vacuum, and depending on $h(T_G)$ \\
\cellcolor{gris2} & \cellcolor{gris2} & \cellcolor{gris3}$\alpha \neq 1$: $3\alpha(1+w) > 1, \, -3 < \alpha < 7$ \\
 \multirow{-2}{*}{\cellcolor{gris2}(ii)}& \multirow{-2}{*}{\cellcolor{gris2}Unknown}& \cellcolor{gris1}$\alpha = 1$: $g(0)$ is finite \\
\cellcolor{gris3}(iii) & \cellcolor{gris3}\xmark & \cellcolor{gris3}$4 + 3\alpha(1+w) > 0$ \\
\cellcolor{gris1}(iv) & \cellcolor{gris1}\xmark & \cellcolor{gris1}Vacuum or $4 + 3\alpha(1+w) > 0$ \\
\cellcolor{gris3}(v) & \cellcolor{gris3}\xmark & \cellcolor{gris3}Always \\
\bottomrule
\end{tabularx}
\midsepdefault
\caption{The conditions for the reconstructed models found in Refs.~\cite{delaCruz-Dombriz:2017lvj,delaCruz-Dombriz:2018nvt} to obey the vacuum condition $f(0,0) = 0$ for the $\Lambda$\gls{cdm} and power-law cosmology subject that all individual contributions are maintained. In the latter cosmology, the distinction between $\alpha = 1$ and $\alpha \neq 1$ is necessary due to branching solutions.}
\label{table:reconstruct_sols-vacuum}
\end{table}

On a different framework, the reconstruction procedure has been applied to generate models yielding \gls{plde} with the IR cutoff length taken as the future event horizon in Ref.~\cite{Jawad2015b} and Hubble horizon in Ref.~\cite{Chattopadhyay:2014xaa}. While analytical solutions for the power-law and de Sitter cosmologies were obtained along with numerical solutions for bouncing cosmologies and the unification of matter-accelerated cosmological phases, no solution behaves as \gls{plde} throughout the cosmic phase. Primarily, only the power-law cosmology leads to the desired phantom behaviour at late times.

\subsubsection{Noether's symmetry approach}\label{sec:cosmo-Noether-fTTG}

Following Refs.~\cite{Capozziello:2016eaz,Bahamonde:2018ibz,Bajardi:2021tul}, the following $f(T,T_G)$ model ansatz were chosen to simplify the Noether condition system of equations, being
\begin{align*}
    \text{(i)} \quad g_0 {T_G}^k + t_0 (-T)^m\,, & & \text{(ii)} \quad {T_G}^l (-T)^m\,, & & \text{(iii)} \quad {T_G}^n\,,
\end{align*}
where $g_{0}, t_0, k, m, l$ and $n$ are constants. Model~(i) imposes the constraint $k = 1$ reducing the Lagrangian to the $f(T)$ power-law case. For the remaining cases, symmetries can be found along with their associated cosmological behaviors, a summary of which is provided in Table~\ref{table:cosmo-noethersym-fTTG}. The ${T_G}^l (-T)^m$ ansatz leads to interesting phenomenology as besides the standard power-law cosmology, trigonometric and hypergeometric behaviors can also be recovered. Additionally, the power-law behavior is recovered as a classical trajectory following the Wheeler-de Witt formulation \cite{Bajardi:2021tul}. Meanwhile, there is no clear evolutionary behavior for model~(iii) although a power-law behavior is observed if $a_{1,2} = 0$.

\begin{table}[!ht]
\centering
\midsepremove
\begin{tabularx}{\textwidth}{p{2cm}X}
\toprule
\cellcolor{gris3}\boldmath{$f(T,T_G)$} & \cellcolor{gris3}\textbf{Symmetry and Cosmological Implications} \\ \midrule
\cellcolor{gris1} & \cellcolor{gris1}Noether condition constrains $m = \frac{1-l}{2}$. Depending on the symmetry considered,\\
\multirow{-1}{*}{\cellcolor{gris1}${T_G}^l (-T)^m$}&\cellcolor{gris1}power-law, trigonometric $a(t) \sim \tan t$ or hypergeometric $a(t) \sim \tanh t$ behaviors \\
\cellcolor{gris1}&\cellcolor{gris1}are achieved. \\
\cellcolor{gris3}& \cellcolor{gris3}Symmetries exist for dust and $w = \frac{1}{4n-1}$. A partial analytical solution is obtained \\
\multirow{-2}{*}{\cellcolor{gris3}${T_G}^n$} &\cellcolor{gris3}for $n \neq 1, \frac{1}{4}$, being $\left[4(n-1)\right]^{\frac{n-1}{4n-5}} \int \left[(4n-5)a_0 a^4-a_1\right]^{\frac{n-1}{4n-5}} \, da = t + a_2$. \\
\bottomrule
\end{tabularx}
\midsepdefault
\caption{Summary of conditions and cosmological behaviors obtained for the model $f(T,T_G)$ ansatz considered and analyzed in Refs.~\cite{Capozziello:2016eaz,Bahamonde:2018ibz,Bajardi:2021tul}. Here, $a_{0,1,2}$ are integration constants.}
\label{table:cosmo-noethersym-fTTG}
\end{table}

For the extended generalized teleparallel class $f(T,B,T_G,B_G)$ gravity \cite{Bahamonde:2018ibz}, since for the \gls{flrw} cosmology $B_G = 0$, the theory reduces to that of $f(T,B,T_G)$ gravity. Despite this simplification, given its general complexity, specific ansatz choices were considered for the Noether symmetry approach
\begin{align*}
\text{(i)} \quad f_0 (-T)^m + f_1 B^n + f_2 {T_G}^q\,, & & \text{(ii)} \quad f_0 (-T)^m B^n {T_G}^q\,,
\end{align*}
where $f_{0,1,2}, m, n$ and $q$ are constants. The Noether condition reveals that the first model imposes the constraint $n = q = 1$ reducing it to the $f(T)$ power-law model (Sec.~\ref{sec:cosmo-Noether-fT}). Meanwhile, the second model contains two Noether symmetries but its general cosmological behavior was not explored.

\subsubsection{Dynamical system approach}\label{sec:dyn_fTTG}

The cosmological dynamics of $f(T,T_G)$ gravity was investigated in Ref.~\cite{Kofinas:2014aka}. In particular, the model $f(T,T_G) = -T + \alpha_1 \sqrt{T^2 + \alpha_2 T_G}$, where $\alpha_{1,2} \neq 0$ are constants, was studied. In this case, the presence of a perfect dust fluid was assumed and the following dimensionless phase-space parameters were defined
\begin{equation}
\tilde{x} = \sqrt{1+\frac{2\alpha_2}{3}\left(1+\frac{\dot{H}}{H^2}\right)}\,, \quad \Omega_{\rm m} = \frac{\kappa^2 \rho_{\rm m}}{3H^2}\,.
\end{equation}
As the phase-space $(\tilde{x}, \Omega_{\rm m})$ is noncompact, the boundary can be studied with Poincar\'{e}'s projection method. Two new phase-space variables, $(r, \theta)$, having domain $r \in [0,1)$ and $\theta \in [0,\frac{\pi}{2}]$, are introduced to determine any critical points lying on the boundary. These are related to the original parameters via
\begin{equation}
\tilde{x} := \frac{r}{1-r}\cos \theta\,, \quad \Omega_{\rm m} := \frac{r}{1-r}\sin \theta\,.
\end{equation}
In this way, the boundary is reached by taking $r \to 1^-$. The critical points can therefore be determined, and are summarized in Table~\ref{table:dynamics-critpts1-fTTG}.

	\begin{table}[!ht]
		\centering
		\midsepremove
		\begin{tabularx}{\textwidth}{lcp{2.6cm}X}
			\toprule
		\cellcolor{gris3}	& \cellcolor{gris3}\boldmath{$(\tilde{x}^\star, \Omega_{\rm m}^\star)$} & \cellcolor{gris3}\textbf{Stability} & \cellcolor{gris3}\textbf{Cosmological Description} \\ \midrule
		\cellcolor{gris1}$P_1$ & \cellcolor{gris1}$\big(\sqrt{1-\frac{\alpha_2}{3}},$ $\Omega_{\rm{m}1}\big)$ & \cellcolor{gris1}Stable/saddle & \cellcolor{gris1}Decelerating scaling solution $a(t) \propto t^{2/3}$ \\
			%
				\cellcolor{gris3}& 	\cellcolor{gris3} & 	\cellcolor{gris3}Stable node, & 	\cellcolor{gris3}Dark energy dominated universe, fluid can behave as \\
					\cellcolor{gris3} & 	\cellcolor{gris3}& 	\cellcolor{gris3}unstable node & 	\cellcolor{gris3}phantom, cosmological constant, quintessence or\\
		 \multirow{-3}{*}{\cellcolor{gris3}$P_2$}& 	 \multirow{-3}{*}{\cellcolor{gris3}$\left(x_-,0\right)$} & 	\cellcolor{gris3}or saddle & 	\cellcolor{gris3}non-phantom depending on parameter constraints \\
				\cellcolor{gris1} & 	\cellcolor{gris1} & 	\cellcolor{gris1}Stable node, & 	\cellcolor{gris1} \\
					\cellcolor{gris1} & 	\cellcolor{gris1} & 	\cellcolor{gris1}unstable node& 	\cellcolor{gris1} \\
				 \multirow{-3}{*}{\cellcolor{gris1}$P_3$} & 	 \multirow{-3}{*}{\cellcolor{gris1}$\left(x_+,0\right)$} & 	\cellcolor{gris1}or saddle & 	 \multirow{-3}{*}{\cellcolor{gris1}Same behavior as the critical point $P_2$} \\
				\cellcolor{gris3} & 	\cellcolor{gris3} & 	\cellcolor{gris3} & 	\cellcolor{gris3}Dark energy dominated universe, decelerating for $\alpha_2 > 0$ \\
				\cellcolor{gris3}& 	\cellcolor{gris3} & 	 \multirow{-2}{*}{\cellcolor{gris3}Unstable node} & 	\cellcolor{gris3}and accelerating for $\alpha_2 < 0$. Fluid can be phantom, \\
		 \multirow{-3}{*}{\cellcolor{gris3}$P_4$} & \multirow{-3}{*}{\cellcolor{gris3}$\left(0,0\right)$} & 	 \multirow{-2}{*}{\cellcolor{gris3}\cellcolor{gris3}or saddle} & 	\cellcolor{gris3}a cosmological constant or quintessence \\
			\cellcolor{gris1} & \cellcolor{gris1} & \cellcolor{gris1} & 	\cellcolor{gris1}Super-accelerated phantom solution ($\alpha_2 > 0$) or an \\
				\multirow{-2}{*}{\cellcolor{gris1}$Q_1$} & \multirow{-2}{*}{\cellcolor{gris1}$\left(\infty,0\right)$} & \multirow{-2}{*}{\cellcolor{gris1}Saddle} & 	\cellcolor{gris1}an asymptotically decelerating universe ($\alpha_2 < 0$) \\
			\cellcolor{gris3} & \cellcolor{gris3} & \cellcolor{gris3}Unstable & \cellcolor{gris3}\\
				\multirow{-2}{*}{\cellcolor{gris3}$Q_2$} & \multirow{-2}{*}{\cellcolor{gris3}$\left(\infty,\infty\right)$} & \cellcolor{gris3}or stable & \multirow{-2}{*}{\cellcolor{gris3}Future singularity for \makebox{$\alpha_2 < 0$}, past singularity for $\alpha_2 > 0$}\\
			\cellcolor{gris1}$Q_3$ & \cellcolor{gris1}$\left(0,\infty\right)$ & \cellcolor{gris1}Undetermined & \cellcolor{gris1}Future, past or intermediate singularity\\
			\bottomrule
		\end{tabularx}
		\midsepdefault
		\caption{The critical points for the $f(T,T_G) = -T + \alpha_1 \sqrt{T^2 + \alpha_2 T_G}$, $\alpha_1, \alpha_2 \neq 0$, model as obtained in Ref.~\cite{Kofinas:2014aka}, together with their stability nature and cosmological description. The variables $\Omega_{\rm{m}1} \coloneqq 1-\frac{\alpha_1 (6-5\alpha_2)\sqrt{9-3\alpha_2}}{6(\alpha_2-3)}$, $x_{\pm} \coloneqq \frac{3 \pm \sqrt{3{\alpha_1}^2(4\alpha_2-3)+9}}{3\alpha_1}$ have been defined for simplicity.}
		\label{table:dynamics-critpts1-fTTG}
	\end{table}

\begin{figure}[!ht]
	\centering
	\subfigure[$\alpha_1 = -\sqrt{33}$ and $\alpha_2 = 4$]{\includegraphics[width=0.325\textwidth]{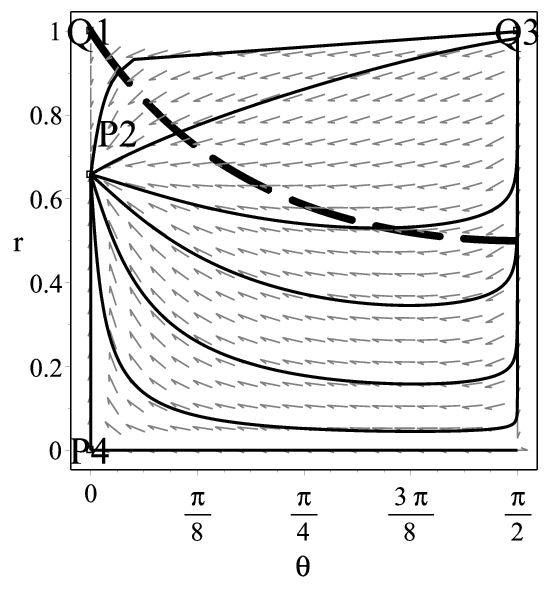}}
    \subfigure[$\alpha_1 = -\alpha_2 = \frac{1}{2}$]{\includegraphics[width=0.325\textwidth]{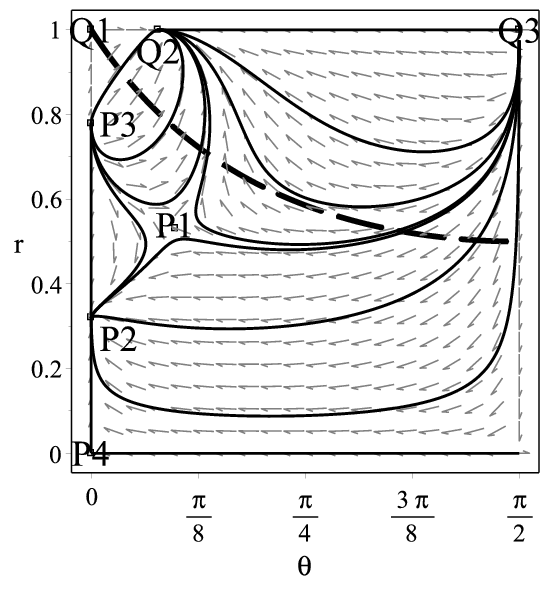}}
    \subfigure[$\alpha_1 = 2\alpha_2 = 3$]{\includegraphics[width=0.325\textwidth]{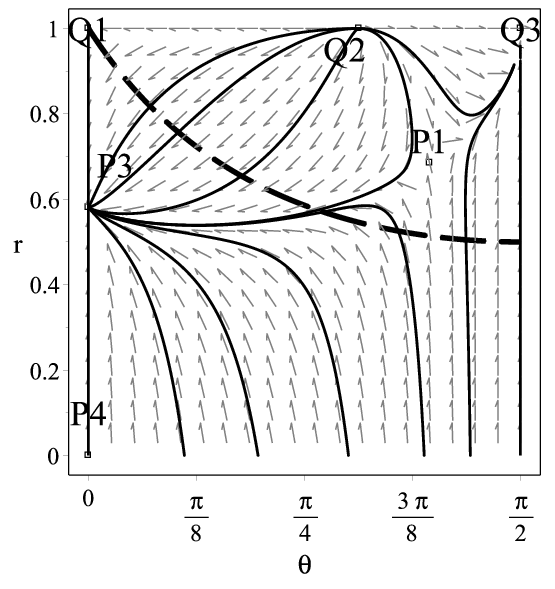}}
	\caption{Pioncar\`{e} phase-space representations of the $f(T,T_G) = -T + \alpha_1 \sqrt{T^2 + \alpha_2 T_G}$ model for different choices of $\alpha_1$ and $\alpha_2$. Depending on the latter's magnitudes, certain critical points appear. Observe that in (b) (and (c)), the universe can evolve towards a future singularity $Q_2$ ($Q_3$) as opposed to the phantom $P_2$ (quintessence $P_3$) behavior. Permission for use of this figure was kindly provided by the authors of Ref.~\cite{Kofinas:2014aka}.}
	\label{fig:fTTG_dynamics}
\end{figure}

Immediately, one notes that the critical points $Q_2, Q_3$ are unlikely to be physically viable as $\Omega_\text{m} > 1$. Focusing on the remaining points, a realistic cosmology can only be obtained if a transition from $P_1$ (akin to a matter dominated solution) to $P_2/ P_3$ is realized provided that the latter points describe an accelerating cosmology. Meanwhile, the critical points $P_4$ and $Q_1$ are saddle or unstable points which could represent intermediary phases. Additionally, Ref.~\cite{Kofinas:2014aka} also investigate the behavior for certain parameter choices which are illustrated in Fig.~\ref{fig:fTTG_dynamics}. For $\alpha_1 = -\sqrt{33}$ and $\alpha_2 = 4$, a late-time de Sitter attractor $(P_2)$ without formation of any singularities results. On the other hand, late-time phantom $(P_2)$ for $\alpha_1 = -\alpha_2 = \frac{1}{2}$ and quintessence $(P_3)$ for $\alpha_1 = 2\alpha_2 = 3$ behaviors can be recovered. 

On the other hand, Ref.~\cite{Tretyakov:2016uvv} focuses attention on the stability of the Minkowski solution by examining the de Sitter space behavior (similar approaches appear in \cite{Tretyakov:2014kra,Faraoni:2004dn}). Here, the phase-space variables are taken to be $(H, \dot{H})$ which results in an autonomous system yielding critical points of the form $(H^\star,0)$, namely de Sitter behavior. Thus, the matter density component approaches a constant value, $\Lambda$. In this way, a Minkowski solution can be investigated in the limit when $H^\star, \Lambda \to 0$. Additionally, however, a true Minkowski vacuum exists provided the conditions $f(0,0) = 0, \, f_T(0,0) < 0$ are also satisfied. For the previous model $f(T,T_G) = -T + \alpha_1 \sqrt{T^2 + \alpha_2 T_G}$, a stable Minkowski solution is obtained when\footnote{The stability conditions are significantly different from Ref.~\cite{Tretyakov:2016uvv}. This is due to evaluation of the derivatives $F_H, F_D$ of the function $F(H,D)$ (Eq.~(12)) at the critical point, where the expressions Eqs.~(17) and (18) have not been recovered.}
\begin{equation}
\alpha_1 < 0\,, \quad 0 < \alpha_2 < \frac{3}{{\alpha_1}^2} \left[1+{\alpha_1}^2 + 3\sqrt{1+3{\alpha_1}^2}\right]\,.
\end{equation}
Combined with the results in \cite{Kofinas:2014aka}, the existence of a Minkowski solution together with a viable cosmological sequence can be achieved. For instance, the choice $\alpha_1 = -\sqrt{33}$ and $\alpha_2 = 4$ hosts a stable Minkowski solution and a saddle matter domination behavior which later evolves towards a stable de Sitter behavior.

\subsection{Teleparallel cosmology with non-minimal couplings to matter} \label{sec:fTT_cosmo_back}

Different theories with non-minimally couplings between the matter sector and gravity have been considered (see Sec.~\ref{sec:modifiedmatter}). One popular one is described by the trace of the energy-momentum tensor $\Theta:=\Theta^\mu{}_\mu$. For this theory, considering a perfect fluid, and that the density Lagrangian does not depend on derivatives of the tetrads, the field equations~\eqref{eq:fTtheta} in flat \gls{flrw} cosmology are reduced to
\begin{subequations}
\begin{align}
    -6 f_T H^2-(\rho+p)f_{\Theta}-\frac{1}{2} f(T,\Theta)&=\kappa^2\rho\,,\label{eq:frwtrace1}\\[0.5ex]
	-2H\dot{f}_T - 2f_T (\dot{H}+3H^2) - \frac{1}{2}f(T,\Theta) &= -\kappa^2 p\,,\label{eq:frwtrace2}
\end{align}
\end{subequations}
where $f_{\Theta}=\partial f/ \partial \Theta$. Along with this theory, $f(T,B,\mathcal{L}_{\rm m})$ gravity has been studied in cosmology. Using the field equation~\eqref{fieldeqgenerafTLm} for flat \gls{flrw} and a perfect fluid, we arrive at
	\begin{subequations}
	\begin{align}
	3 H\dot{f}_B -3H^2 ( 3f_B+2f_T)-3 f_B\dot{H}-\frac{1}{2} f(T,B,\mathcal{L}_{\rm m})&=0\,,\label{eq:fTBLmFRW1}\\[0.5ex]
	-(3H^2+\dot{H})(2f_T+3f_B)-2 H\dot{f}_T+\ddot{f}_B-\frac{1}{2} f(T,B,\mathcal{L}_{\rm m})&=-f_L(\rho+p)\,.\label{eq:fTBLmFRW2}
	\end{align}
	\end{subequations}
	where $f_L=\partial f/\partial \mathcal{L}_{\rm m}$. In the following, we will mainly concentrate on analyzing these two set of \gls{flrw} equations.

\subsubsection{Reconstruction method}\label{sec:reconstruction-fTT-gravity}

In Ref.~\cite{Momeni:2014jja}, the authors reconstruct the $f(T,\Theta)$ Lagrangian for a diverse number of scenarios (see the Supplementary annexes (Supplementary 1)). Overall, a reconstructed solution is obtained with some cases containing exceptions and further extensions. In the case of \gls{hde} and \gls{plde} \cite{Shahzad:2019jdc,Salako:2015xra}, the reconstructed $f(T,\Theta)$ solutions do not satisfy the Friedmann equations and thus are not reported. It is remarked that a $\Lambda$\gls{cdm} reconstructed solution appears in Ref.~\cite{Junior:2015bva} although the result is incorrect as the spatial components of $\nabla_\mu \Theta^{\mu\nu}$ are identically zero as opposed to constituting a condition for $f(T,\Theta)$.

In Ref.~\cite{Nassur:2015zba}, the reconstruction procedure in the presence of a perfect fluid and the ansatz choice $f(T,\Theta) = f_1(T) + f_2(\Theta)$ was explored via the equivalent scalar-tensor formulation by introducing an auxiliary scalar field $\Phi$ and proper functions $P_{1,2}, Q$ satisfying
\begin{subequations}
\begin{align}
f(T,\Theta) &= P_1(\Phi) T + P_2(\Phi) \Theta + Q(\Phi)\,, \label{eq:fTT-recon-scalar-relation} \\[0.5ex]
f_1(T) &= P_1 T + Q_1(\Phi)\,, \quad f_2(\Theta) = P_2 \Theta + Q_2(\Phi)\,, \quad Q = Q_1 + Q_2\,.
\end{align}
\end{subequations}
The crucial relation for this approach is
\begin{equation}
8H\dot{P}_1 + 4\dot{H}P_1 = -\rho(1+w)(1+2P_2)\,.
\end{equation}
For simplicity, \cite{Nassur:2015zba} apply the method of separation of variables by setting the above equal to some constant $\lambda$ whose magnitude alters the form of the solutions:
\begin{enumerate}
    \item $\lambda = 0$: This leads to $f(T,\Theta) = 2\widetilde{k}(-T)^{\frac{3}{4}} - \frac{w \Theta}{1-3w}$ for some integration constant $\widetilde{k}$, a solution which contrasts with that reported in Ref.~\cite{Nassur:2015zba}. By virtue of the Friedmann equations, the power-law cosmology $a(t) \sim t^{\frac{1}{2-2w}}$ results. 
    \item $\lambda \neq 0$: Here, a reconstructed solution is obtained provided the cosmological behaviour is specified. The authors explore the unification of matter and \gls{de} domination eras and matter-\gls{de} transition phase which reconstructed solutions are listed in  the Supplementary annexes (Supplementary 1).
\end{enumerate}

\subsubsection{Dynamical system approach}

In Refs.~\cite{Carloni:2015lsa,Bahamonde:2017ifa}, the dynamics of the general torsion-matter coupling model $f(T,B,\mathcal{L}_{\rm m})$ (see Sec.~\ref{sec:modifiedmatter})) was investigated. Given the complexity of the field equations, particular functional ansatz were chosen for the dynamical analysis, namely:
\begin{enumerate}
	\item $f(T,B,\mathcal{L}_{\rm m}) = f(\lc{R},\mathcal{L}_{\rm m})$ which is the curvature-matter coupling scenario previously explored in various works \cite{Ribeiro:2014sla,Azizi:2014qsa,An:2015mvw,Azevedo:2016ehy};
	\item $f(T,B,\mathcal{L}_{\rm m}) = f(T,\mathcal{L}_{\rm m})$;
	\begin{enumerate}[label = ({\alph*})]
		\item $f(T,\mathcal{L}_{\rm m}) = f_1(T) + 2\kappa^2 f_2(T) \mathcal{L}_{\rm m}$: \cite{Carloni:2015lsa} investigates six model ansatz of which three (i) $f_1 = -T + \alpha_1 T^2 + 2\Lambda, \, f_2 = -1 + \beta_1 T\,,$ (ii) $f_1 = -T + 2\Lambda, \, f_2 = -1 + \beta_2 T + \beta_3 T^2\,,$ (iii) $f_1 \propto e^{\alpha_2 T^2}, \, f_2 \propto e^{-T}$ host a repeller or saddle power-law behavior (such as matter domination) critical points which evolves towards a \gls{de} dominated de Sitter attractor. Thus, an appropriate late-time cosmology can be realized;
		\item $f(T,\mathcal{L}_{\rm m}) = -\Lambda \exp\left[-\dfrac{1}{\Lambda} \left(T + \mathcal{L}_{\rm m}\right)\right]$, $\Lambda > 0$ and $f(T,\mathcal{L}_{\rm m})= M^{-\epsilon}\left(T + \mathcal{L}_{\rm m}\right)^{1+\epsilon}$, $\epsilon, M$ are constants. Here, contrary to their curvature analogues investigated in Refs.~\cite{Harko:2010mv,Azevedo:2016ehy}, require no dynamical system analysis since the scale factor can be directly solved;
	\end{enumerate}
	\item $f(T,B,\mathcal{L}_{\rm m}) = c_1 T + c_2 B^s + \left(c_3 + c_4 B^r\right)\mathcal{L}_{\rm m}$ with $c_4 \neq 0$, $r$ and $s$ are constants. Here, the critical points cannot be determined for arbitrary choices of the parameters. To this end, the following three scenarios were considered \cite{Bahamonde:2017ifa}: (a) $r = 1$, (b) $c_2 = 0$ and (c) $r = 1-s$, $c_2 \neq 0$, $s = \{ -2,-1,2,3,4,5\}$. While all sub-cases yield stable critical points (for sub-case (c), it is only for $s = -2$) and is able to describe an accelerating universe (with sub-case (a) describing a de Sitter cosmology), it is unclear whether they are able to host the desired sequence of cosmological epochs.
\end{enumerate}

\subsection{Teleparallel scalar-tensor cosmology}

This section will be devoted to studying cosmology for different scalar-tensor teleparallel theories (see Sec.~\ref{sec:scalartensor} for more details). One particular class of theory that has been studied in cosmology is the one described by the action~\eqref{action} which is constructed from the torsion scalar and the boundary term. For this theory, the field equation~\eqref{fieldeqgeneral2} in flat \gls{flrw} becomes
	\begin{subequations}
	\begin{align}
	-\frac{1}{2} f(T,B)+3 H\dot{f}_B -3H^2 ( 3f_B+2f_T)-3 f_B\dot{H}-\frac{1}{2}f_X\dot{\phi}^2&=\kappa ^2\rho\,,\label{eq:fTBphiXFRW1}\\[0.5ex]
	-(3H^2+\dot{H})(2f_T+3f_B)-2 H\dot{f}_T+\ddot{f}_B-\frac{1}{2} f(T,B)&=-\kappa ^2 p\,,\label{eq:fTBphiXFRW2}
	\end{align}
	\end{subequations}
whereas the modified Klein-Gordon equation~\eqref{fieldeq2} yields
\begin{equation}
\dot{f}_X\dot{\phi} +3 f_X H\dot{\phi}+f_X \ddot{\phi}+f_{\phi}=0\,,
\end{equation}
where $f_X=\partial f/\partial X$ and $f_\phi=\partial f/\partial \phi$.
A particular case which has been extensively used in the literature for cosmology is the one described by~\eqref{action:scalar-torsion-coupling-T-and-B} where the kinetic term appears linearly and the scalar field is non-minimally coupled to both $T$ and $B$. For this theory, the flat \gls{flrw} are
\begin{subequations}
\begin{alignat}{2}\label{eq:nonmin_coupling_fTB}
    -6 f_T H^2-\frac{1}{2} f(T)-3H^2F_1(\phi)+3 H\dot{F}_2(\phi)& =\: & &\kappa^2\Big(\rho+\frac{1}{2}\omega(\phi)\dot{\phi}^2+V(\phi)\Big)\,,\\[0.5ex]
    -2 H\dot{f}_T-2 f_T (\dot{H}+3 H^2)-\frac{1}{2} f(T)-F_1(\phi)(3H^2+2\dot{H})& \: & &\nonumber\\[0.5ex]
    -2H\dot{F}_1(\phi)+\ddot{F}_2(\phi)& =\: & &-\kappa^2\Big(p+\frac{1}{2}\omega(\phi)\dot{\phi}^2-V(\phi)\Big)\,, \label{eq:nonmin_coupling_fTB_2}
\end{alignat}
\end{subequations}
whereas the modified Klein-Gordon equation becomes
\begin{equation}
    - \frac{3}{\kappa ^2}\Big(H^2 (F_1'(\phi)+3F_2'(\phi))+ \dot{H}F_2'(\phi)\Big)+3 H \omega(\phi)\dot{\phi}+\omega(\phi)\ddot{\phi}+\frac{1}{2} \dot{\phi}^2 \omega'(\phi)+V'(\phi)=0\,, \label{eq:nonmin_coupling_fTB_KG}
\end{equation}
where primes denote differentiation \gls{wrt} the scalar field. On the other hand, a further extension of the $f(T,B,X,\phi)$ action is the quintom model investigated in Ref.~\cite{Bahamonde:2018miw} where two canonical scalar fields $\phi$ and $\sigma$ are introduced, both being non-minimally coupled with the torsion $T$ and boundary $B$ scalars. The gravitational Lagrangian is thus generalized to
\begin{equation}\label{eq:non-min-Lagrangian-dynamics}
\mathcal{L} = -\dfrac{1}{2\kappa^2}\left[T + \left(f_1(\phi) + f_2(\sigma)\right)T + \left(g_1(\phi) + g_2(\sigma)\right) B\right] - \dfrac{\xi}{2} \partial_\mu \phi \partial^\mu \phi - \dfrac{\chi}{2} \partial_\mu \sigma \partial^\mu \sigma + V(\phi,\sigma)\,,
\end{equation}
where $\xi$ and $\chi$ are constants and $V(\phi,\sigma)$ represents the combined potential of both scalar fields. For this model, the flat \gls{flrw} equations are
\begin{subequations}
\begin{alignat}{2}
    3H^2(1+f_1(\phi) + f_2(\sigma)) & =\: & & \kappa^2 \left[\rho + V(\phi,\sigma) + \frac{\xi}{2}\dot{\phi}^2 + \frac{\chi}{2}\dot{\sigma}^2 + 3H(\dot{g}_1(\phi) + \dot{g}_2(\sigma))\right] \,, \label{eq:FLRW_quintom_1}\\[0.5ex]
    (3H^2+2\dot{H})(1+f_1(\phi) + f_2(\sigma))& =\: & & -\kappa^2\Big[p - V(\phi,\sigma) + \frac{\xi}{2}\dot{\phi}^2 + \frac{\chi}{2}\dot{\sigma}^2 \nonumber\\[0.5ex]
    &\:& & +2H(\dot{f}_1(\phi) + \dot{f}_2(\sigma)) - \ddot{g}_1(\phi) - \ddot{g}_2(\sigma)\Big] \,, \label{eq:FLRW_quintom_2}
\end{alignat}
\end{subequations}
whereas the modified Klein-Gordon equations for both scalar fields are
\begin{subequations}
\begin{align}
    \xi(\ddot{\phi} + 3H\dot{\phi}) + \frac{3H^2}{\kappa^2} f^\prime_1(\phi) + \frac{3}{\kappa^2}g^\prime_1(\phi)(3H^2 + \dot{H}) + V_\phi &= 0 \,, \\[0.5ex]
    \chi(\ddot{\sigma} + 3H\dot{\sigma}) + \frac{3H^2}{\kappa^2} f^\prime_2(\phi) + \frac{3}{\kappa^2}g^\prime_2(\phi)(3H^2 + \dot{H}) + V_\sigma &= 0 \,,
\end{align}
\end{subequations}
where $V_\phi=\partial V/\partial \phi$ and $V_\sigma=\partial V/\partial \sigma$. In the following, we will mainly concentrate on these kind of theories.

\subsubsection{Reconstruction method}

In Ref.~\cite{Chakrabarti:2017moe}, the reconstruction procedure was applied for the minimal coupling scenario $F_1(\phi) = F_2(\phi) = 0$ where the matter component is solely described by the scalar field with kinetic coupling $\omega(\phi) = 1$ and potential $V(\phi) = \frac{\phi^{n+1}}{n+1}$, where $n \neq -3, \pm 1, 0$ is a constant. In particular, the authors explore the case when the Klein-Gordon equation, generally expressed in Euler's form \cite{Euler1997}
\begin{equation}
    \ddot{\phi} + f_1(t) \dot{\phi} + f_2(t)\phi + f_3(t) \phi^n = 0\,,
\end{equation}
for time-dependent functions $f_{1,2,3}(t)$ satisfy the relation \cite{Harko2013ic}
\begin{equation}
    \frac{1}{n+3}\frac{\ddot{f}_3}{f_3} - \frac{n+4}{(n+3)^2}\left(\frac{\dot{f}_3}{f_3}\right)^2 + \frac{n-1}{(n+3)^2} \frac{f_1 \dot{f}_3}{f_3} + \frac{2\dot{f}_1}{n+3} + \frac{2(n+1){f_1}^2}{(n+3)^2} = f_2\,.
\end{equation}
This condition guarantees an analytical solution for $\phi(t)$. As $f_1 = 3H, f_2 = 0$ and $f_3 = 1$ for this particular case, the above relation leads to $a(t) \propto t^{\frac{n+3}{3(n+1)}}$ leading to $\phi(t) \propto t^{\frac{2(1+3n)}{1-n^2}}$. Thus, the reconstructed form of $f(T)$ is obtained via the Friedmann Eq.~\eqref{eq:nonmin_coupling_fTB} to be
\begin{equation}
    f(T) \sim (-T)^{n^2-1} + (-T)^{\frac{2(1+3n)}{n-1}}\,.
\end{equation}
Other scalar potentials were also investigated, namely power law (mixed and Higgs), exponential and logarithmic. Except for the exponential case, which leads to a rescaled \gls{tegr} solution, no solutions were obtained for the remaining cases.

In Ref.~\cite{Bamba:2013jqa}, the conformal $f(T)$ theory was studied having non-minimal couplings $F_1(\phi) = \frac{C\phi^2}{2}$, $F_2(\phi) = \frac{D\phi^2}{2}$, kinetic coupling $\omega(\phi) = 2\kappa^2$ and potential $V(\phi) = \frac{2\kappa^2 V_0 \phi^{m+1}}{m+1}$, where $C$, $D$, $V_0$ and $m$ are constants. To ensure conformal invariance, $C = \frac{1}{6}, D = \frac{1}{3}$ and $m = 3$ which is assumed in the reported results. For simplicity, only simple cosmological scenarios were considered for reconstruction. 

Starting with a power-law cosmology $a(t) \propto t^\alpha$ for some constant $\alpha$, in the absence of the potential $V(\phi)$, $f(T)$ takes the form
\begin{alignat}{2}
f(T) & =\: & & - \frac{y^{-2 \alpha } \left({c_1}^4 + 2 (\alpha -1)^2 {c_2}^2\right)}{4 (4 \alpha -1)}-\frac{{c_1}^3 {c_2} y^{\frac{1-5 \alpha }{2}}}{2-5 \alpha }+\frac{{c_1}^2 {c_2}^2 y^{1-3 \alpha}}{2 \alpha -1}+\frac{{c_1} {c_2}^3 y^{\frac{3-7 \alpha }{2}}}{7 \alpha -4} \nonumber \\[0.5ex]
& \: & & -\frac{{c_2}^4 y^{2-4 \alpha}}{4(5-8 \alpha )}+\frac{2 \kappa^2 \rho_0 y^{-\frac{3}{2}\alpha (1+w)}}{3 \alpha (1+w) - 1}\,, \quad y \equiv -\frac{6\alpha^2}{T}\,,
\end{alignat}
where $c_{1,2}$ are integration constants originating from the modified Klein-Gordon equation. It is remarked that the reconstructed solution is invalid for certain choices of $\alpha$ and $w$, such as radiation domination ($\alpha = \frac{1}{2}$). 

For a de Sitter cosmology, $f-2T f_T$ becomes constant in time (see Sec.~\ref{sec:fT-reconstruction}) making the scalar field act as the source for the accelerated behavior. The latter's evolution, obtained from the Klein-Gordon equation \eqref{eq:nonmin_coupling_fTB_KG}, is dependent on the choice of potential. For the power-law potential, $\phi$ approaches a constant value\footnote{An extra factor of 4 appears in Ref.~\cite{Bamba:2013jqa}.} $\phi = \pm \sqrt{\frac{-2H^2}{V_0}}$ whereas in its absence $(V_0 = 0)$, the scalar field decays with time as $\phi(t) \propto e^{-H t}$. 

Finally, for a $\Lambda$\gls{cdm} cosmology in the absence of the scalar potential, $f(T)$ reduces to the standard $\Lambda$\gls{cdm} Lagrangian with $\phi = e^{-\int H \, dt}$ making the evolution independent of the scalar field's behavior. 

In view of the above observations, the scalar field can be made to act only as a source for the late-time acceleration by constructing a potential which is predominantly absent during earlier phases (such as matter domination) but becomes dominant as the Universe transitions to accelerated phases (such as de Sitter).

\subsubsection{Noether's symmetry approach}

In the following, the Noether symmetry approach has also been explored within teleparallel scalar-tensor theories. Generally, the symmetries also constrain the form of the scalar-tensor coupling as well as the scalar potentials. For simplicity, the minimally and non-minimally coupled cases shall be discussed separately. 

\paragraph{Minimally coupled scalar fields}\label{sec:noether_mincoupled}

In Ref.~\cite{Aslam:2012tj}, the authors investigate the teleparallel analogue of Saez-Ballester theory \cite{Saez:1986dil}, which has been extensively explored in curvature-based models \cite{Jamil:2012zm,Rao:2008zza,Adhav:2007zzb,Ram:2009zzb,Rao:2007zz}. Following the action given in Eq.~\eqref{action:scalar-torsion-coupling-T-and-B}, the theory is recovered when $F_1(\phi) = F_2(\phi) = 0$. For this case, the theory exhibits one Noether symmetry together with the functional constraints
\begin{equation}
f(T) = t_0 T^2 + C\,, \quad \omega(\phi) \propto \phi^{-4}\,, \quad V(\phi) = V_0 \phi^{-4} + C\,,
\end{equation}
where $C, t_0$ and $V_0$ are integration constants. Ref.~\cite{Aslam:2012tj} then fixes the latter parameters to investigate the resulting cosmology. In particular, a $\Lambda$\gls{cdm}-like behavior is observed with a scalar field which is phantom in nature and approaches \gls{de} at late-times. Despite this, it has been noted that the $C$ term disappears from the Lagrangian and hence has no physical consequences. However, its contribution still appeared in the cosmological equations which may lead to a different cosmological behavior.

A further extension is explored in Ref.~\cite{Jamil:2012fs} which includes the contribution of dust matter for the particular choice $\omega(\phi) = \epsilon$, where $\epsilon = +1, -1$ correspond to a quintessence or phantom field respectively. In this case, the Noether condition imposes the following functional constraints
\begin{equation}
    f(T) = \dfrac{4}{3}c_1 (-T)^{3/4} + c_3\,, \quad V(\phi) = c_4 + c_5\left(c_1 \phi + c_2\right)^2\,,
\end{equation}
where $c_1,\dots{},c_5$ are integration constants. For $c_1 = \frac{3}{4}$, $c_2 = c_4 = 0$ and $c_5 = 1$, the following distinct evolutionary behaviors were obtained:
\begin{enumerate}[label = (\alph*{})]
\item \textbf{Quintessence} -- The scalar field grows with time while its \gls{eos} becomes close to the phantom-divide line. The number of \textit{e}-foldings increases exponentially with time;
\item \textbf{Phantom} -- The scalar field decays in an oscillatory fashion with its \gls{eos} rapidly decreasing into the phantom regime. The number of \textit{e}-foldings becomes asymptotically flat.
\end{enumerate}

Finally, the minimal coupling case consisting of \gls{tegr}, a scalar field with $\omega(\phi) = 1$ coupled with an electromagnetic tensor scalar field coupling was explored in Ref.~\cite{Tajahmad:2016bjs}. Namely, this introduces the contribution $\frac{1}{4} f(\phi)^2 F_{\mu\nu} F^{\mu\nu}$ to the gravitational action, i.e.
\begin{equation}
    \mathcal{L} = -\frac{T}{2\kappa^2} - \frac{1}{2}\phi_{;\mu}\phi^{;\mu} + V(\phi) + \frac{1}{4} f(\phi)^2 F_{\mu\nu} F^{\mu\nu}\,,
\end{equation}
where $f(\phi)$ represents the coupling strength, to the gravitational action which role is to explain inflationary phases \cite{Maleknejad:2012fw} as well as to generate late-time acceleration \cite{Vakili:2014sda}. Since \gls{tegr} is considered, the field equations are identical to its curvature analogue meaning the Noether symmetries match with those obtained, for instance, in Ref.~\cite{Vakili:2014sda}. However, there appear to be differences in the Noether equations. For instance, neglecting time symmetries for simplicity, the condition $2\alpha_a = 0$ appearing in Ref.~\cite{Tajahmad:2016bjs} should be $\alpha + 2a \alpha_a = 0$ as correctly reported in Ref.~\cite{Vakili:2014sda}\footnote{The typo in Ref.~\cite{Tajahmad:2016bjs} did not effect the main results which are in a different convention to other works.}.

\paragraph{Non-Minimally coupled scalar fields}

Non-minimal scalar field couplings with the torsion and boundary term scalars have been explored in Refs.~\cite{Kucukakca:2013mya,Gecim:2017hmn,Bajardi:2021tul} with $f(T) = 0$, $\omega(\phi) = 1$\footnote{The Lagrangians between Refs.~\cite{Kucukakca:2013mya,Gecim:2017hmn} differ by a factor of 2. The necessary factors have been included for sake of consistency.}. Starting with Refs.~\cite{Kucukakca:2013mya,Gecim:2017hmn}, in order to simplify the Noether symmetry analysis, the following models were considered:
\begin{align*}
\textbf{Model I } \text{\cite{Kucukakca:2013mya}} & \quad F_2(\phi) = 0, \, V(\phi) = \lambda \phi^{\frac{6}{2n+3}} \text{ where } n \neq -3/2\,, \\[0.5ex]
\textbf{Model II } \text{\cite{Gecim:2017hmn}} & \quad F_2(\phi) \neq 0, \, V(\phi) = \lambda \phi^2\,,
\end{align*}
with $\lambda \geq 0$. The first model describes the absence of the boundary term coupling while a general power-law potential is considered whereas the second model maintains both couplings while the potential is taken to be quadratic.

Through Noether symmetry, both $F_1(\phi)$ and $F_2(\phi)$ coupling functions turn out to be constrained, as summarized in Table~\ref{table:noethersym-nonmin-couplingI}. Evidently, both coupling terms generally take on a power-law form and each consist of a quadratic coupling which is reminiscent of teleparallel \gls{de} \cite{Geng:2011aj}.

\begin{table}[!ht]
\centering
\midsepremove
\begin{tabularx}{\textwidth}{p{2.5cm}X}
\toprule
\cellcolor{gris3}\textbf{Model} & \cellcolor{gris3}\boldmath{$F_1(\phi)$} \textbf{and} \boldmath{$F_2(\phi)$} \\ \midrule
\cellcolor{gris1}\textbf{I} & \cellcolor{gris1} $F_1(\phi) = -\frac{(2n+3)^2}{24}\phi^2$, $F_2(\phi) = 0$ \\
\multicolumn{2}{l}{\cellcolor{gris3}\textbf{II}} \\
\cellcolor{gris1}$(a) \; \tilde{k} \neq -3$ & \cellcolor{gris1}$F_1(\phi) = \frac{3}{8}\phi^2 + \frac{3c_1}{2} \phi^{\frac{2(\tilde{k}+3)}{3}}\,,$ \ \ \ $F_2(\phi) = -\frac{\phi^2}{4} - \frac{3c_1}{2(\tilde{k}+3)}\phi^{\frac{2(\tilde{k}+3)}{3}}$ \\
\cellcolor{gris2}$(b) \; \tilde{k} = -3$ & \cellcolor{gris2}$F_1(\phi) = \frac{3}{8}\phi^2 - \frac{3c_2}{\phi^2} + \frac{3c_1}{2}\,,$ \ \ \ \ $F_2(\phi) = -\frac{\phi^2}{4} - c_1 \ln \phi$ \\
\bottomrule
\end{tabularx}
\midsepdefault
\caption{The resulting functional constraints imposed by Noether symmetry for the two models considered in Refs.~\cite{Kucukakca:2013mya,Gecim:2017hmn}. The integration constants $c_{1,2}$ and $\tilde{k}$ arise from the Noether constraint. Observe that in Model~II, when $\tilde{k} = 0$ and $c_1 = -\frac{1}{2}$, $F_2(\phi) = 0$, which matches the Model~I result when $n = 0$ as expected.}
\label{table:noethersym-nonmin-couplingI}
\end{table}

Prior to examining the resulting cosmological behaviors, the coordinates were first transformed to a new set of coordinates $(u,z)$ with $z$ being a cyclic coordinate\footnote{The Model~II point-like Lagrangian in terms of these new coordinates is $\mathcal{L} = -\frac{\dot{u}^2}{2} + 2c_1 \alpha_1 u^{\frac{2\tilde{k}+3}{3}} \dot{u}\dot{z} + \lambda u^2 + \frac{4c_2 {\alpha_1}^2 \dot{z}^2}{u^2}$ with the $c_2$ contribution being absent for $\tilde{k} \neq -3$, and $\alpha_1$ is an integration constant. This is different from that reported in Ref.~\cite{Gecim:2017hmn}, hereby affecting the nature of the cosmological solutions. As $c_2$ only appears for the case when $\tilde{k} = -3$, the presented solutions shall hold true only for $c_2 = 0$.}. Thus, this yields a constant of motion $I_0$ which influences the nature of the cosmology as shown in Table~\ref{table:noethersym-nonmin-couplingII}.

\begin{table}[!ht]
\centering
\midsepremove
\begin{tabular}{ll}
\toprule
\cellcolor{gris3}\textbf{Model} &\cellcolor{gris3} \textbf{Scale Factor} \\ \midrule
\multicolumn{2}{l}{\textbf{Case I:} \cellcolor{gris3}$I_0 = 0$} \\ \midrule
\cellcolor{gris1}Model I & \cellcolor{gris1}$a \propto \left\lbrace \begin{array}{ll} \left(z_1 + z_2 t\right)^{-\frac{1}{2n}}, & \cellcolor{gris1}\text{for } n \neq 0 \\ e^{-\frac{2\alpha_0}{3} \left(z_1 + z_2 t\right)}, & \text{for } n = 0 \end{array}\right.$ \\
Model II & $a \propto \left(z_1 + z_2 t\right)^{-\frac{1}{\tilde{k}}}$ \\ \midrule
\multicolumn{2}{l}{\cellcolor{gris3}\textbf{Case II:} $I_0 \neq 0$} \\ \midrule
\cellcolor{gris1}Model I & \cellcolor{gris1}\\
\cellcolor{gris1}(a) $n \neq 0, \, -3, \, -6$ & \cellcolor{gris1}$a \propto \left(b t + c_1\right)^{-\frac{3}{n(n+3)}}$ \\
\cellcolor{gris1}(b) $n = 0$ &\cellcolor{gris1}$a = \left(\alpha_0 u_1+I_0 t\right)^{\frac{1}{6}} e^{\frac{2}{3} \left(-\alpha_0 z_3+\frac{2 \alpha_0 u_1 \lambda t}{I_0}+\lambda t^2\right)}$ \\
\cellcolor{gris1}(c) $n = -3$ & \cellcolor{gris1}$a \propto e^{-\frac{I_0 t}{6 \alpha_0}} \left(c_2+\frac{2 \alpha_0 \lambda {u_2}^3}{{I_0}^2} e^{-\frac{I_0 t}{\alpha_0}}\right)^{\frac{1}{6}}$ \\
Model II & \\
(a) $\tilde{k} \neq -3, \, -6$ & $a \propto \left[\left(\frac{3\lambda \alpha_1 c_1 \left(u_3 + \ell t\right)^2}{{I_0}^2(\tilde{k}+6)} -\frac{3}{8\alpha_1 c_1 \tilde{k}}\right)\left(u_3 + \ell t\right)^{-\frac{\tilde{k}}{\tilde{k}+3}} + z_4\right]^{-\frac{1}{\tilde{k}}}$ \\
(b) $\tilde{k} = -3$ & $a \propto \left[z_5 {u_4}^2 e^{-\frac{I_0 t}{c_1 \alpha_1}} + z_6\right]^{\frac{1}{3}}$ \\
\bottomrule
\end{tabular}
\midsepdefault
\caption{The corresponding cosmological behaviors for each Noether symmetry depending on the constant of motion $I_0$ as obtained in Refs.~\cite{Kucukakca:2013mya,Gecim:2017hmn}. Here, $b \coloneqq \frac{I_0(n+3)}{\alpha_0(2n+3)}$, $\ell \coloneqq -\frac{I_0(n+3)}{3\alpha_0 c_1}$ and $u_1,\dots{}, u_4$, $z_1,\dots{}, z_6$ are integration constants. It is remarked that the cases $n = -6$ and $\tilde{k} = -6$ were not studied.}
\label{table:noethersym-nonmin-couplingII}
\end{table}

Starting with $I_0 = 0$, the potential becomes zero in both models. Depending on the choice of integration constants, both models can contain static $(z_1 \neq 0, z_2 = 0)$ and power-law $(z_1 = 0, z_2, n \neq 0)$ behaviors while Model~I can also host (anti)-de Sitter $(z_2 \neq 0, n = 0)$ behavior. Transition periods may be realized for more general parameter considerations.

Moving on to the $I_0 \neq 0$ case, each model exhibits distinct behaviors. In Model~I, $n \neq 0, \, -3, \, -6$ leads to a power-law behavior which can result into an ever-accelerating phantom or quintessence cosmology. For $n = 0$, the scale factor yields an evolution which can be associated with a multi-fluid scenario. Furthermore, a late-time de Sitter phase is achieved since the scalar field component approaches \gls{de} behavior. Lastly, for $n = -3$, the scalar field component can behave as a non-phantom fluid which approaches a \gls{de} state at late-times. Similar to the $n = 0$ case, this universe becomes accelerating and approaches a de Sitter-like state.

For Model II, for $\tilde{k} \neq -3, \, -6$, various cosmological behaviors can be obtained. For instance, this universe may experience accelerating and decelerating phases whereas the scalar field fluid can behave as either a phantom or a non-phantom fluid. Therefore, this solution can be used to explain transitioning periods. In particular, a late-time de Sitter-like behavior can be achieved for small $\tilde{k}$ values. Finally, $\tilde{k} = -3$ can host a de Sitter solution which is expanding provided $\frac{I_0}{c_1 \alpha_1} < 0$ and leads to a growing potential which serves as the main driving force for the expansion.

In the absence of the boundary term coupling $(F_2(\phi) = 0)$, four Noether symmetries are obtained which constrain $F_1(\phi)$ and $V(\phi)$ to be \cite{Bajardi:2021tul}:
\begin{align*}
    \text{(I)} &\quad F_1(\phi) = -\frac{(2n+3)^2}{24} \phi^2, \, V(\phi) \propto \phi^{\frac{6}{2n+3}}, & \text{(II)} & \quad F_1(\phi) \propto \phi^2, \, V(\phi) \propto \phi^2, \\[0.5ex]
    \text{(III)} &\quad F_1(\phi) = \text{constant}, \, V(\phi) \propto e^{\alpha\phi}, & \text{(IV)} &\quad F_1(\phi) \propto c_1 + c_2 \phi + c_3 \phi^2, \\[0.5ex] 
    & & &\quad V(\phi) \propto (c_1 + c_2 \phi + c_3 \phi^2)^2 \,,
\end{align*}
for constants $n$ and $\alpha$. Observe that the first two models are precisely the cases described previously in \cite{Kucukakca:2013mya,Gecim:2017hmn} whereas the third and fourth model are new symmetries which could not be derived in the previous analysis. While these scenarios were not explored in further detail, the quantum cosmological description using the Wheeler-de Witt equation and Hartle's criterion was applied for the second model. In this case, the Universe's wave function satisfies Hartle's criterion and thus the classical trajectories reduce to the cosmological behaviors as obtained in \cite{Kucukakca:2013mya,Gecim:2017hmn}.

In the case of generalized couplings, the results obtained in Refs.~\cite{Tajahmad:2017ywa,Myrzakulov:2019yqt} are reviewed. Ref.~\cite{Tajahmad:2017ywa} considers the $f(T,\phi,X)$ sub-class 
\begin{equation}\label{lagra:behzad}
    \mathcal{L} = f(\phi)T - U(\phi, X)g(T) - \frac{\omega(\phi)}{2}\phi_{,\mu} \phi^{,\mu} + V(\phi)\,,
\end{equation}
where $U(\phi, X)$ is a coupling function chosen be $U = h(\phi) \dot{\phi}$. Applying the Noether condition leads to the following functional constraints
\begin{equation}
    f(\phi) \propto \phi^2\,, \quad g(T) \propto \sqrt{-T}\,, \quad h(\phi) \propto \phi\,, \quad V(\phi) = V_0 \phi^2\,, \quad \omega(\phi) \text{ is constant}\,,
\end{equation}
where $V_0$ is an integration constant. The corresponding cosmological behavior was investigated for the particular choices $f(\phi) = \frac{3}{32}\phi^2$ and $\omega(\phi) = 1$, leading to the scale factor
\begin{equation}
    a(t) = (c_1 t + c_2)^{2/3} \exp\left(\frac{8}{3} V_0 t^2 + \left[c_3 - \frac{16}{3}\frac{V_0 c_2}{c_1}\right]t+c_4\right)\,,
\end{equation}
with $c_1,\dots{}, c_4$ representing integration constants. However, as both $f(\phi), \omega(\phi) > 0$, this may lead to a non-canonical scalar field, depending on the form of the second term in the Lagrangian \eqref{lagra:behzad}.

On the other hand, Refs.~\cite{Myrzakulov:2019yqt,Kucukakca:2014vja,Gecim:2016gqx} consider a fermionic coupling in the form
\begin{equation}
    \mathcal{L} = h(u)f(T) - \frac{i}{2}\left[\bar{\psi}\Gamma^\mu D_\mu \psi - \left(\bar{D}_\mu \bar{\psi}\right)\Gamma^\mu \psi\right] + V(u)\,,
\end{equation}
where $\psi$ is the fermion function, $\Gamma^\mu = \dut{E}{A}{\mu} \gamma^A$ with $\gamma^a$ representing the Dirac matrices, and $u = \psi\bar{\psi}$. Overbars represent the respective adjoint operators. The TEGR coupling case $f(T) = T$ has been explored in 3+1 \cite{Kucukakca:2014vja} and 2+1 \cite{Gecim:2016gqx} spacetimes. Starting with the former, the Noether condition imposes the constraint on the coupling functions $h(u) \propto u^{\frac{2n+1}{3}}$ and $V(u) \propto u$, where $n$ is some constant which greatly influences the nature of the cosmology. This can be summarised as follows:
\begin{enumerate}[label = (\alph*{})]
    \item $n = -\frac{1}{2}$ yields $a(t) \sim t^{\frac{2}{3}}$ with the fermionic field behaving as dust;
    \item $n = 1$ gives de Sitter and hence can describe inflationary epochs. Here, the fermionic field behaves as a cosmological constant;
    \item $n \neq 1, - \frac{1}{2}$ leads to $a(t) \sim t^{\frac{1}{1-n}}$. For $0 < n < 1$, the fermionic field has phantom behaviour while $n > 1$ leads to quintessence behaviour.
\end{enumerate}
In 2+1 spacetime, the functional constraints take on various forms depending on whether the gauge term $g(t,q^i)$ in the Noether symmetry condition is maintained (see the Supplementary annexes (Supplementary 1)). When neglected $(g = 0)$, the resulting cosmology becomes identical to the 3+1 scenario, else we find $h(u) \propto u^{\frac{2k}{k+1}}$ and $V(u) \propto u^{\frac{2}{k+1}}$ for some constant $k$, which magnitude alters the cosmological behaviour, namely:
\begin{enumerate}[label = (\alph*{})]
    \item $k \neq 1$: $a(t) \sim t^{\frac{k+1}{2(1-k)}}$ which can be decelerating $(-1 < k < \frac{1}{3})$, coasting $(k = \frac{1}{3})$ or otherwise accelerating. Meanwhile, the fermionic field can behave as a quintessence field $(\frac{1}{2} < k < 1)$ or a phantom field $(k < -1 \text{ or } k > 1)$;
    \item $k = 1$: de Sitter with fermionic field behaving as a cosmological constant.
\end{enumerate}

Finally, for the general case \cite{Myrzakulov:2019yqt}, the Noether condition imposes the constraints $f(T) \propto T^{\frac{m-1}{n}}$, $V(u) \propto u$ and $h(u) \propto u^m$, leading to a power-law cosmology $a \sim t^{\frac{3n}{2}}$. Interestingly, the parameter $m$ does not affect the evolution. As only a power-law behavior is obtained, the model appears to be insufficient to describe the whole Universe's history but it may be viable for specific epochs.

\subsubsection{Dynamical system approach}

Similar to the Noether symmetry approach, the dynamical system approach is explored for a number of teleparallel scalar-tensor theories, categorised under minimally and non-minimal couplings. These shall be discussed separately with a comparison made in cases where the non-minimal dynamics reduce to the minimal ones.

\paragraph{Minimally coupled scalar fields} \label{sec:dynm_mincoupled}

In Refs.~\cite{Biswas:2015cva,Awad:2017ign}, the $f(T)$ gravity minimal coupling model having $F_1(\phi) = F_2(\phi) = 0$ and $\omega(\phi) = 1$ was studied. In Ref.~\cite{Biswas:2015cva}, an interaction between a dust fluid and the \gls{de} scalar fluid was considered for the model $f(T) = -T + \alpha \sqrt{-T}$. As this reduces to standard minimally coupled \gls{tegr} with a scalar field, the results are not explored further.

Instead, the deviation from \gls{tegr} is explored in Ref.~\cite{Awad:2017ign} in the absence of matter fluids with the scalar (inflaton) field obeying the constant-roll approximation $\ddot{\phi} = \beta H \dot{\phi}$, where $\beta$ represents the constant-roll parameter. This, in turn, leads to the assumption that the Hubble parameter takes on the form
\begin{equation}
H(\phi) = M \cos\left(\kappa \sqrt{\frac{\beta}{2}}\phi\right)\,,
\end{equation}
with $M \leq \kappa^{-1}$. For this model, the $f(T)$ dynamical system approach discussed in Sec.~\ref{sec:dynamics-fT} can be directly applied by using Eq.~\eqref{eq:dynamic-fT-equation} with the inflaton's \gls{eos}
\begin{equation}
    w_\phi = \frac{\frac{1}{2}\dot{\phi}^2 - V(\phi)}{\frac{1}{2}\dot{\phi}^2 + V(\phi)}\,,
\end{equation}
in conjunction with the \gls{flrw} equations~\eqref{eq:nonmin_coupling_fTB}--\eqref{eq:nonmin_coupling_fTB_2}. This leads to the 1-dimensional autonomous relation
\begin{equation}\label{eq:dynamics-fT-scalarmatterfield}
\dot{H} = \dfrac{6}{f_{HH}}\left[f- Hf_H + 2\kappa^2 V(\phi) \right]\,.
\end{equation}
The $f(T)$ function can be reconstructed by combining the \gls{flrw} equations~\eqref{eq:nonmin_coupling_fTB}-\eqref{eq:nonmin_coupling_fTB_2} whereas the potential is obtained from the modified Klein-Gordon equation~\eqref{eq:nonmin_coupling_fTB_KG}. The resulting functions take on the forms
\begin{subequations}
\begin{align}
f(\Phi) &= 3 M^2 \cos (2 \Phi ) + c_1 \left[\Phi \cos (\Phi )- \sin (\Phi )\right]+c_2\,, \\[0.5ex]
V(\Phi) &= \frac{M^2 [(\beta +3) \cos (2 \Phi )-\beta +6]}{2 \kappa ^2} + \frac{(\beta +3) c_1 \sin (\Phi )}{6 \kappa ^2}-\frac{\beta {c_1}^2}{144 \kappa ^2 M^2} -\frac{c_2}{2 \kappa ^2} \,,
\end{align}
\end{subequations}
where $\Phi \coloneqq \kappa \sqrt{\frac{\beta}{2}}\phi$, $c_1$ and $c_2$ being integration constants. It is remarked that the $f(T)$ solution can be expressed in terms of the torsion scalar by inverting the relation $T = -6M^2 \cos^2(\Phi)$. Substituting back into the autonomous relation Eq.~\eqref{eq:dynamics-fT-scalarmatterfield} leads to
\begin{equation}
    \dot{H} = \beta\left(H^2-M^2\right) + \frac{\beta c_1}{12M} \sqrt{M^2-H^2}\,,
\end{equation}
which is different than the one given in \cite{Awad:2017ign}. This leads to a different set of critical points and hence, a possible different dynamical behavior. Nonetheless, we highlight the reported results. 

For $c_1 < 0$, the universe interpolates between a Type~IV initial singularity (which replaces the initial Big Bang singularity) and a late-time de Sitter cosmology. Throughout the cosmic evolution, the universe experiences a decelerated expansion and a late-time behaviour compatible with $\Lambda$CDM especially for small $c_1$ values. Meanwhile, for $c_1 > 0$, a cyclic universe is obtained, one which interpolates between two Type~II singularities.

\paragraph{Non-minimally coupled scalar fields}

The dynamical system analysis for the non-minimally coupled scalar field case has been extensively studied in numerous works. Here, we review Refs.~\cite{Jarv:2015odu,Wei:2011yr,Skugoreva:2014ena,Bahamonde:2018miw,Xu:2012jf,Jamil:2012vb,DAgostino:2018ngy,Bahamonde:2015hza}.

Starting with the $T$-coupling cases ($F_2(\phi) = 0$) with $f(T) = -T$, in Ref.~\cite{Xu:2012jf}, the coupling $F_1(\phi) = \xi \phi^2$ with kinetic coupling $\omega(\phi) = 1$ and potential $V(\phi) \propto e^{-\lambda \phi}$ was explored using the dimensionless phase-space variables
\begin{equation}
\tilde{x}^2 := \dfrac{\kappa^2 \dot{\phi}^2}{6H^2}\,, \quad \tilde{y}^2 := \dfrac{\kappa^2 V(\phi)}{3H^2}\,, \quad \tilde{z} := \sqrt{|\xi|}\phi\,, \quad \Omega_{\rm{m}} := \dfrac{\kappa^2\rho_\text{m}}{3H^2} \label{eq:dynamics_nonmincoupling_variables}
\end{equation}
in the presence of dust. Solving the autonomous system and use of Poincar\'{e}'s projection method yields a number of critical points as summarized in Table~\ref{table:dynamics-critpts-nonmincoupling-Xu}.
\begin{table}[!b]
	\centering
	\midsepremove
	\begin{tabularx}{\textwidth}{p{1cm} >{\centering\arraybackslash}p{3.45cm} p{2.55cm} p{4.4cm} X}
		\toprule
		\cellcolor{gris3}& 	\cellcolor{gris3}$(\tilde{x}^\star, \tilde{y}^\star, \tilde{z}^\star)$ & 	\cellcolor{gris3}\textbf{Existence} & 	\cellcolor{gris3}\textbf{Stability} & 	\cellcolor{gris3}\textbf{Properties} \\ \midrule
		\cellcolor{gris1}A & 	\cellcolor{gris1}$\left(0,0,0\right)$ & 	\cellcolor{gris1}Always & 	\cellcolor{gris1}Saddle &	\cellcolor{gris1} Matter dominated \newline decelerating universe \\
		\cellcolor{gris3}B$^{(+)}$/ \newline C$^{(-)}$ & 	\cellcolor{gris3}$(\pm 1, 0, 0)$ & \cellcolor{gris3}$\xi = 0$ & 	\cellcolor{gris3}Unstable $(\lambda^{(+)} < \sqrt{6}$, \newline $\lambda^{(-)} > 6)$, \newline saddle otherwise & 	\cellcolor{gris3}\gls{de} dominated dec. \newline universe behaving as a stiff fluid \\
		\cellcolor{gris1}D & 	\cellcolor{gris1}$\Big(\frac{\lambda}{\sqrt{6}}, \sqrt{1-\frac{\lambda^2}{6}}, 0\Big)$ & 	\cellcolor{gris1}$\xi = 0,$ $\lambda^2 \leq 6$ & 	\cellcolor{gris1}Stable node $(\lambda^2 < 3)$, \newline saddle otherwise & 	\cellcolor{gris1}\gls{de} (quintessence) \newline dominated universe with \newline $a(t) \propto t^{\frac{2}{\lambda^2}}$ \\
		\cellcolor{gris3}E & 	\cellcolor{gris3}$\left(\sqrt{\frac{3}{2}}\frac{1}{\lambda}, \sqrt{\frac{3}{2}}\frac{1}{\lambda}, 0\right)$ & 	\cellcolor{gris3}$\xi = 0,$ $\lambda^2 \geq 3$ & 	\cellcolor{gris3}Stable node \newline $(3 < \lambda^2 < \frac{24}{7})$, \newline stable spiral $(\lambda^2 > \frac{24}{7})$ &	\cellcolor{gris3} Decelerating matter \newline dominated phase with \newline \gls{de} behaving as dust \\
		\cellcolor{gris1}F$^{(-)}$/ \newline G$^{(+)}$ & 	\cellcolor{gris1}$\Big(0, \sqrt{\frac{2(\xi \pm \sqrt{\xi(\xi-\lambda^2)})}{\lambda^2}},$ \newline $\frac{(\xi - \sqrt{\xi(\xi-\lambda^2)})\sqrt{|\xi|}}{\lambda \xi}\Big)$ & 	\cellcolor{gris1}$0 < \lambda^2 \leq \xi$, \newline or $\xi < 0$ (G) & 	\cellcolor{gris1}F: Stable node $(\lambda^2 < \xi)$; \newline G: Saddle & 	\cellcolor{gris1}Accelerating de Sitter \newline \gls{de} domination \\
		\cellcolor{gris3}J &	\cellcolor{gris3} $\left(0, 1, 0\right)$ & 	\cellcolor{gris3}$\xi \neq 0,$ $\lambda = 0$ & 	\cellcolor{gris3}Stable spiral $(\frac{3}{8} < \xi)$, \newline stable node $(0 < \xi < \frac{3}{8})$, \newline saddle otherwise & \cellcolor{gris3}Accelerating de Sitter \newline \gls{de} domination \\ \midrule
		\cellcolor{gris1}& 	\cellcolor{gris1}$({\tilde{x}_r}^\star, {\tilde{y}_r}^\star, {\tilde{z}_r}^\star)$ & 	\cellcolor{gris1}\textbf{Existence} & 	\cellcolor{gris1}\textbf{Stability} & 	\cellcolor{gris1}\textbf{Properties} \\ \midrule
		\cellcolor{gris3}$Q_{1,\pm}$ & \cellcolor{gris3}$\left(\mp \frac{1}{\sqrt{2}}, 0, \pm \frac{1}{\sqrt{2}}\right)$ &	\cellcolor{gris3} $\xi > 0$ & 	\cellcolor{gris3}Unstable $(0 < \xi < \frac{3}{8})$, \newline saddle otherwise & 	\cellcolor{gris3}Accelerating $(\xi > \frac{1}{6})$ \newline with phantom behavior $(\xi > \frac{3}{8})$. Arbitrary \gls{de} \gls{eos} \\
		\cellcolor{gris1}$Q_{2,\pm}$ & 	\cellcolor{gris1}$\left(\pm \frac{1}{\sqrt{2}}, 0, \pm \frac{1}{\sqrt{2}}\right)$ & 	\cellcolor{gris1}$\xi > 0$ & 	\cellcolor{gris1}Saddle & 	\cellcolor{gris1}Non-accelerating \newline universe. Arbitrary \gls{de} \gls{eos} \\
		\bottomrule
	\end{tabularx}
	\midsepdefault
	\caption{Summary of the critical points investigated in Ref.~\cite{Xu:2012jf} for the non-minimal torsion scalar coupling model. A--J represent the finite boundary critical points whereas $Q_i$ are the critical points found at the boundary derived using Poincar\'{e}'s projection method. Here, the Poincar\`{e} variables are $\tilde{x}_r = \rho \cos \theta \sin \psi$, $\tilde{y}_r = \rho \cos \psi$, $\tilde{z}_r = \rho \sin \theta \sin \psi$ where $\rho = \frac{r}{\sqrt{1-r^2}}$ with $r = \sqrt{\tilde{x}^2+\tilde{y}^2+\tilde{z}^2}$, $\theta \in [0, 2\pi]$ and $\psi \in [-\frac{\pi}{2}, \frac{\pi}{2}]$.}
	\label{table:dynamics-critpts-nonmincoupling-Xu}
\end{table}

\begin{figure}[!ht]
	\centering
	\subfigure[Minimal coupling $\lambda = 1$, $\xi = 0$]{\includegraphics[width=0.49\textwidth]{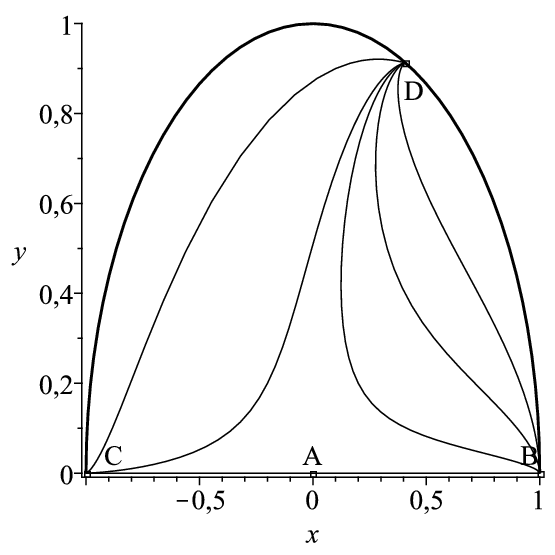}}
    \subfigure[Non-minimal coupling $\lambda = 2$, $\xi = -10^{-3}$]{\includegraphics[width=0.49\textwidth]{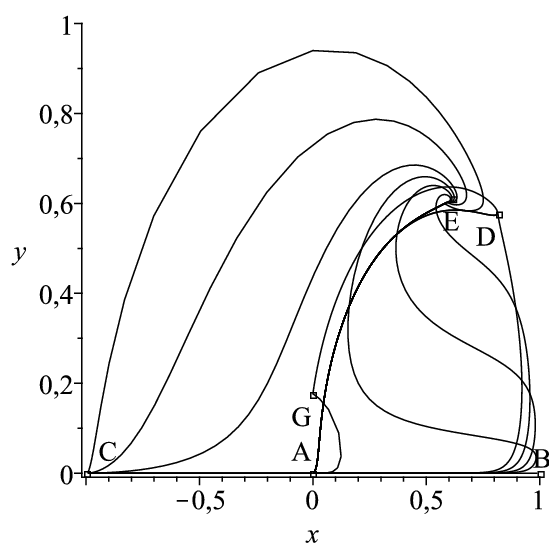}}
    \caption{Projection of the finite phase-space on the $x$-$y$ plane for the $T$-coupling (a) minimal and (b) non-minimal scalar field cases. In the former, the evolution approaches the late-time quintessence dominated behavior (D). Once the non-minimal coupling is introduced, the behavior changes considerably. While the critical points B--E no longer exist, the cosmology can remain for a long finite time near E. Then, the system evolves to the matter dominated state A followed by the de Sitter critical point G. However, as this point is saddle, it does not represent the final evolutionary state. Here, $x \to \tilde{x}$ and $y \to \tilde{y}$ to match our conventions. This figure first appeared in Ref.~\cite{Xu:2012jf}.}
	\label{fig:nonmin_dynamics_Xu}
\end{figure}

The critical points D, E, F, J can be dynamically stable, corresponding to a quintessence universe (D), decelerating matter dominated universe with \gls{de} behaving as matter (E), and \gls{de} dominated de Sitter accelerating universes (F, J). Furthermore, the critical points (B--E) only appear if the scalar field is minimally coupled $(\xi = 0)$. Meanwhile, the boundary critical points $Q_{1,2}$ are unstable and leave an arbitrary \gls{de} fluid behavior meaning that these points could describe transition, inflationary or de Sitter phases. For minimally coupled scenarios, a viable cosmology can be realized following the trajectory A $\to$ D while for non-minimally coupled cases A $\to$ F/G/J provided that the relevant existence and stability criteria are satisfied. A finite phase-space representation of the latter for both minimally and non-minimally coupled cases is illustrated in Fig.~\ref{fig:nonmin_dynamics_Xu}.

An extension of the previous model, explored in Ref.~\cite{Wei:2011yr}, takes into account the interaction between the scalar field and the matter content via an interaction term $Q$ for a number of interactions and potentials, namely (i)~$Q = 0$, (ii) $Q \propto \rho_\text{m} \dot{\phi}$, (iii) $Q \propto H \rho$, (iv) $Q \propto H \rho_\text{m}$, and (\textbf{A}) $V(\phi) \propto e^{-\lambda \phi}$, (\textbf{B}) $V(\phi) \propto \phi^{n}$. Here, $\lambda$ and $n$ are constants. Interestingly, the interaction models (iii) and (iv) yield no critical points. Meanwhile, only a few selected models can realize a viable cosmology. \textbf{A}(i) can realize a matter $\to$ de Sitter sequence whereas only the late-time de Sitter phase is apparent in \textbf{B}(i). On the other hand, \textbf{A}(ii) and \textbf{B}(ii) yield a scaling solution to de Sitter.

Meanwhile, the special non-interacting case $Q = 0$ in the presence of a dust fluid was investigated in Ref.~\cite{DAgostino:2018ngy} but for a linear $V(\phi) \propto \phi$ and exponential $V(\phi) \propto e^{-\phi}$ potentials. The critical points therefore reduce to the ones given in \cite{Wei:2011yr} by setting $n = \lambda = 1$. In this case, the exponential model is still able to generate a viable cosmology from matter domination $\to$ de Sitter.

In Ref.~\cite{Skugoreva:2014ena}, the $T$-coupling is investigated for the generalized power-law form $F_1(\phi) = \xi \phi^{\tilde{N}}$ for some $\tilde{N} > 0$ and power-law potential $V(\phi) = V_0 \phi^n$ with $V_0 > 0$ and $n$ is some constant. Here, the dynamics have been explored in vacuum using a new set of dimensionless phase-space variables
\begin{equation}
m := \dfrac{\dot{\phi}}{H \phi}\,, \quad A := \dfrac{\tilde{N} \phi^{\tilde{N}}}{1-\xi \phi^{\tilde{N}}}\,.
\end{equation}
Depending on whether $\tilde{N} = 2$ or $\tilde{N} \neq 2$, different critical points are obtained. In the latter, only a de Sitter behavior appears. On the other hand, for $\tilde{N} = 2$, the choice of $\xi$ determines the cosmological behavior:
\begin{enumerate}
    \item $\xi > 0$: no stable points and therefore unable to describe any late-time behavior;
    \item $\xi < 0$: this universe can evolve from de Sitter $\to$ power-law provided $n \neq 2$ and $n$ is odd. To maintain a late-time accelerating behavior, $(n^2-4)\xi < 2$.
\end{enumerate}

Finally, Ref.~\cite{Jamil:2012vb} considers $f(T) = \alpha T + c_1 \sqrt{-T} + c_2$, $F_1(\phi) \propto \phi^2$, $\omega(\phi) = \pm 1$ and potential $V(\phi) \propto e^{\beta \phi}$ where $c_{1,2}, \alpha$ and $\beta > 0$ are constants. For the phase-space variables Eq.~\eqref{eq:dynamics_nonmincoupling_variables} leads to two unstable critical points which their physical description was not explored. Nonetheless, through numerical analysis, a late-time phantom behavior was found.

\begin{figure}[!ht]
    \centering
    \includegraphics[width = 0.6\textwidth]{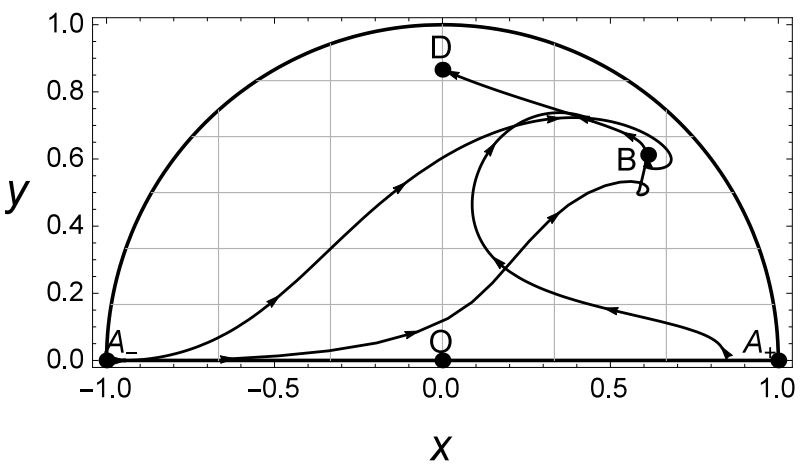}
    \caption{Dynamical behavior of the $B$-non-minimal coupling scenario described via the projection onto the $x$-$y$ phase-space plane. Clearly, the evolution evolves towards a late-time de Sitter attractor (D). Note the identification $x \to \tilde{x}$ and $y \to \tilde{y}$ to match the review convention. Permission for use of this figure was kindly provided by the authors of Ref.~\cite{Bahamonde:2015hza}.}
    \label{fig:nonmin_dynamics_Bahamonde}
\end{figure}

Moving on to the case which explores the $B$-coupling, Ref.~\cite{Bahamonde:2015hza} is reviewed which investigates the model $F_1(\phi) = 0$, $F_2(\phi) = \chi \phi^2$, $\omega(\phi) = 1$ and $V(\phi) \propto e^{-\lambda \phi}$. In this case, the phase-space variables are
\begin{equation}
\tilde{x}^2 := \dfrac{\kappa^2\dot{\phi}^2}{6H^2}\,, \quad \tilde{y}^2 := \dfrac{\kappa^2 V(\phi)}{3H^2}\,, \quad \tilde{z} := 2\sqrt{6}\chi\phi\,, \quad \Omega_{\rm{m}} := \dfrac{\kappa^2\rho_\text{m}}{3H^2}\,.
\end{equation}
To maintain physical solutions, the following additional conditions were imposed: $\tilde{x}^2 + \tilde{y}^2 + \tilde{x}\tilde{z} \leq 1$ (ensuring that the matter density is positive) and $y > 0$ (the potential is assumed to be positive). Similar to Ref.~\cite{Xu:2012jf}, Poincar\'{e}'s projection method is necessary to complete the analysis. However, as the boundary critical points turn out to be non-physical, only the finite critical points shall be discussed which are summarized in Table~\ref{table:dynamics-critpts-nonmincoupling-Bahamonde1}. It is remarked that the stability and physical behavior has been carried out only for the dust matter case $(w = 0)$ but can be easily extended for more general fluids. In addition, the special case $\chi = 0$ leads to the identical critical points obtained in Ref.~\cite{Xu:2012jf} for the $\xi = 0$ model as they lead to to the same gravitational Lagrangian, the only difference being the generalization to an arbitrary fluid constant \gls{eos}. When $\chi \neq 0$, only the critical points O (decelerating matter domination) and D, E (accelerating de Sitter) remain which can constitute a viable cosmological evolution (see Fig.~\ref{fig:nonmin_dynamics_Bahamonde}). 

\begin{table}[!ht]
	\centering
	\midsepremove
	\begin{tabularx}{\textwidth}{lc p{2.25cm}p{4.95cm}X}
		\toprule
		\cellcolor{gris3}	&	\cellcolor{gris3} \boldmath{$(\tilde{x}^\star, \tilde{y}^\star, \tilde{z}^\star)$} & 	\cellcolor{gris3}\textbf{Existence} & 	\cellcolor{gris3}\textbf{Stability} & 	\cellcolor{gris3}\textbf{Properties} \\ \midrule
		\cellcolor{gris1}O & 	\cellcolor{gris1}$\left(0,0,0\right)$ & 	\cellcolor{gris1}Always & 	\cellcolor{gris1}Saddle & 	\cellcolor{gris1}Matter dominated \newline decelerating universe \\
		\cellcolor{gris3}A$_{\pm}$ & 	\cellcolor{gris3}$(\pm 1, 0, 0)$ & 	\cellcolor{gris3}$\chi = 0$ & 	\cellcolor{gris3}$P_{2,+}:$ Unstable $(\lambda < \sqrt{6})$, \newline saddle otherwise; \newline $P_{2,-}:$ Unstable $(\lambda > \sqrt{6})$, \newline saddle otherwise &	\cellcolor{gris3} \gls{de} dominated dec. \newline universe behaving as a stiff fluid \\
		\cellcolor{gris1}B &	\cellcolor{gris1} $\Big(\sqrt{\frac{3}{2}}\frac{1+w}{\lambda}, \sqrt{\frac{3}{2}}\frac{\sqrt{1-w^2}}{\lambda}, 0\Big)$ & 	\cellcolor{gris1}$\chi = 0,$ $\lambda^2 \geq$ \newline $3(1+w)$ & 	\cellcolor{gris1}Stable node $(3 < \lambda^2 < \frac{24}{7})$, \newline stable spiral $(\lambda^2 > \frac{24}{7})$ & 	\cellcolor{gris1}Scaling solution \\
		\cellcolor{gris3}C &	\cellcolor{gris3} $\left(\frac{\lambda}{\sqrt{6}}, \sqrt{1-\frac{\lambda^2}{6}}, 0\right)$ & 	\cellcolor{gris3}$\chi = 0,$ \newline $\lambda^2 < 6$ &	\cellcolor{gris3} Stable node $(\lambda^2 < 3)$, \newline saddle otherwise$^a$ &	\cellcolor{gris3} Quintessence \gls{de} \newline domination with \newline $a(t) \propto t^{\frac{2}{\lambda^2}}$ \\
		\cellcolor{gris1}D & 	\cellcolor{gris1}$\left(0, 1, \sqrt{\frac{2}{3}}\lambda \right)$ & 	\cellcolor{gris1}Always & 	\cellcolor{gris1}Stable spiral $(48\chi > \lambda^2+6)$, \newline stable node \newline $(0 < 48\chi < \lambda^2+6)$, \newline saddle otherwise & 	\cellcolor{gris1}Accelerating de Sitter \gls{de} domination \\
		\cellcolor{gris3}E &	\cellcolor{gris3} $\left(0, 1, 0\right)$ & 	\cellcolor{gris3}$\chi \neq 0,$ \newline $\lambda = 0$ & 	\cellcolor{gris3}Stable spiral $(\frac{1}{8} < \chi)$, \newline stable node $(0 < \chi < \frac{1}{8})$, \newline saddle otherwise & 	\cellcolor{gris3}Accelerating de Sitter \gls{de} domination \\
		\bottomrule
		\multicolumn{5}{@{}p{0.95\textwidth}}{\footnotesize $^a$A parameter $\beta$ appears in Ref.~\cite{Bahamonde:2015hza}, however no source for its significance was found. Nonetheless, the point corresponds to the same behavior found in Ref.~\cite{Xu:2012jf}.}
	\end{tabularx}
	\midsepdefault
	\caption{Summary of the critical points for the boundary-scalar coupling model considered in Ref.~\cite{Bahamonde:2015hza}. The stability and cosmological implications only reflect the dust matter case.}
	\label{table:dynamics-critpts-nonmincoupling-Bahamonde1}
\end{table}

Lastly, the generalized quintom case model which considers both $T$ and $B$-couplings is discussed in Ref.~\cite{Bahamonde:2018miw}. In particular, the following non-minimal couplings and potentials have been considered
\begin{subequations}
\begin{align}
f_1(\phi) &= c_1 \phi^2\,, \quad f_2(\sigma) = c_2 \sigma^2\,, \quad g_1(\phi) = c_3 \phi^2\,, \quad g_2(\sigma) = c_4 \sigma^2\,,\\[0.5ex]
V(\phi,\sigma) &= \tilde{V}(\phi) + \bar{V}(\sigma) = V_1 e^{-\lambda_1 \phi} + V_2 e^{-\lambda_2 \sigma}\,,
\end{align}
\end{subequations}
\noindent where $c_1,\dots{}, c_4$, $V_1, V_2$ and $\lambda_1, \lambda_2$ are constants. To build the dynamical autonomous system, the following dimensionless variables were defined
\begin{subequations}
\begin{align}
\Omega_{\rm m} &= \dfrac{\kappa^2\rho_\text{m}}{3H^2}\,, \quad \tilde{x}^2 = \dfrac{\kappa^2\dot{\phi}^2}{6H^2}\,, \quad \tilde{y}^2 = \dfrac{\kappa^2 V_1(\phi)}{3H^2}\,, \quad \tilde{z} = 2\sqrt{6}\xi \phi\,, \\[0.5ex]
\tilde{u}^2 &= \dfrac{\kappa^2 \dot{\sigma}^2}{6H^2}\,, \quad \tilde{v}^2 = \dfrac{\kappa^2 V_2(\sigma)}{3H^2}\,, \quad \tilde{w} = 2\sqrt{6}\chi \sigma\,,
\end{align}
\end{subequations}
\noindent where the dynamical variable $s$ can be determined via the Friedmann equation \eqref{eq:FLRW_quintom_1} hereby reducing the dimensionality of the autonomous system. Here, the matter content is assumed to behave as a perfect fluid with \gls{eos} $w \in \left[-1, 1\right]$ whereas the constraints $\tilde{y}, \tilde{v} \geq 0$ (to ensure non-negative potentials as in standard quintessence theories) and $\Omega_{\rm m} > 0$ (to ensure a positive matter energy density) were considered. Solving the autonomous systems yields 13 critical points characterizing a matter dominated phase and \gls{de} dominated phases, the latter of which consists of:
\begin{enumerate}[label = (\alph*{})]
    \item one of the scalar fields being absent with the other frozen in time with either zero or non-zero potential energy;
    \item both scalar fields frozen in time while the Universe accelerates in a de Sitter phase.
\end{enumerate}
Based on their dynamical stability, a matter $\to$ \gls{de} evolutionary sequence can be achieved under suitable parameter choices.

\subsection{Other theories} \label{sec:gen_cosmo_back}

In the following, we explore other theories investigated in the context of cosmology. These approaches have not yet been extensively explored as the previous \gls{tg} models especially under the approaches considered in this section. For this reason, a brief summary of the results is provided together with associated \gls{flrw} equations where applicable.

\subsubsection{Non-local theories}

The \gls{flrw} for non-local given by~\eqref{generalizednonlocaltele} are
\begin{subequations}
\begin{alignat}{2}
    \kappa^2\rho & =\: & & 3 H^2 (\theta+1-\xi f(\phi,\varphi))+3 \chi H \dot{f}-\dot{\zeta} (3 H+\frac{1}{2}\dot{\varphi})-\frac{1}{2} \dot{\theta} \dot{\phi} \,,\\[0.5ex]
    -\kappa^2 p & =\: & & (2 \dot{H}+3 H^2) (1+\theta-\xi f(\phi,\varphi))+\frac{1}{2} \dot{\phi} (\dot{\theta}-4 \xi H f_{\phi})-2 \xi H \dot{\varphi} f_{\varphi}+2 H \dot{\theta}\nonumber\\[0.5ex]
    & \: & &-\ddot{\zeta}+\frac{1}{2} \dot{\zeta} \dot{\varphi} + \chi\ddot{f} \,,
\end{alignat}
\end{subequations}
while the scalar field equations are
\begin{subequations}
\begin{align}
0&=-6 \chi \dot{H}f_{\phi}-6 H^2 (\xi +3 \chi ) f_{\phi}+3 H\dot{\theta}+\ddot{\theta}\,,\\[0.5ex]
0&=-6 \chi \dot{H}f_{\varphi}-6 H^2 (\xi +3 \chi ) f_{\varphi}+3 H\dot{\zeta}+\ddot{\zeta}\,,\\[0.5ex]
0&=3 H\dot{\phi}+6 H^2+\ddot{\phi}\,,\quad 0=\dot{H}+3 H\dot{\varphi}+18 H^2+\ddot{\varphi}\,.
\end{align}
\end{subequations}
In the above, $\phi= \lc{\square}^{-1}T$, $\varphi= \lc{\square}^{-1}B$, $\lc{\square} \theta = -(\xi T + \chi B) \frac{\partial f(\phi,\varphi)}{\partial\phi}$ and
$\lc{\square} \zeta = -(\xi T + \chi B) \frac{\partial f(\phi,\varphi)}{\partial\varphi}$.

\paragraph{Noether's symmetry approach}

The study of Noether symmetries in GTNL gravity has been investigated in Refs.~\cite{Bahamonde:2017sdo,Channuie:2017txg} using its scalar-tensor equivalent form Eq.~\eqref{action2}. Taking the configuration space $\mathcal{Q} = (t, a, \phi, \varphi, \theta, \zeta)$ yields 7 distinct symmetries which stem from the generalized Noether vector
\begin{alignat}{3} 
\textbf{X} & =\: & & \left(c_1 t + c_2, \dfrac{1}{3}(c_1-c_3)a, c_4 + c_5 \left(6 \ln a + \psi\right), c_6 + c_7\left(6 \ln a + \varphi\right) + c_9, \right. \nonumber \\[0.5ex]
& \: & & c_{10} + c_3 \theta, (c_3-c_7) \zeta - c_5 \theta + c_8\bigg)\,,
\end{alignat}
with $c_1, \dots, c_{10}$ being integration constants and $\psi \equiv \varphi-\phi$\footnote{Following Ref.~\cite{Bahamonde:2017sdo}, some integration constants do not appear for particular symmetries (for example, $c_{10}$ in \textbf{S1}). However, as these constants do not appear in both the Noether vector and the function $f(\phi,\psi)$, they are taken to be zero.}. Each symmetry constraints the form of $f(\phi, \psi)$ (and ultimately the non-local theory) as summarized in Table~\ref{table:noethersym-nonlocal} and, correspondingly, the cosmological behavior. The obtained results match with the non-local $f(T)$ ($\chi = \varphi = \zeta = 0$) and $f(\lc{R})$ ($\xi = -\chi = -1$, $\theta = -\zeta$ and $f(\phi,\varphi) = f(\psi)$) gravity limiting cases \cite{Channuie:2017txg,Bahamonde:2017sdo}.

 Interestingly, some forms of $f(\phi,\psi)$ host exponential contributions which are known to make the theory renormalizable (at least in the non-local $f(\lc{R})$ cases) \cite{Tomboulis:2015esa,Modesto:2017sdr,Modesto:2017hzl}. In fact, a similar observation is found in the study of Noether symmetries in non-local $f(\lc{R})$ gravity \cite{Bajardi:2020mdp}.

\begin{table}[!ht]
	\centering
	\midsepremove
	\begin{tabularx}{\textwidth}{lXX}
		\toprule
		\multicolumn{2}{l}{\cellcolor{gris3}\textbf{Symmetry}} & \cellcolor{gris3}\boldmath{$f(\phi,\psi)$} \\ \midrule
		\cellcolor{gris1}\textbf{S1} & \cellcolor{gris1}$c_{10} = 0$, $c_{3,7} \neq 0$, $c_4 c_7 \neq c_5(c_6 + c_9)$ & \cellcolor{gris1}$\frac{1}{\xi} + \frac{c_{11} c_{12}}{c_3} \exp\left[\frac{c_3}{c_{12}}(c_5 \varphi - c_7 \phi)\right]$ \\
		\cellcolor{gris3}\textbf{S2} & \cellcolor{gris3}$c_{3,10} = 0$, $c_7 \neq 0$, $c_4 c_7 = c_5(c_6 + c_9)$ & \cellcolor{gris3}$c_{11} + F(-c_7 \phi + c_5 \varphi)$ \\
		\cellcolor{gris1}\textbf{S3} & \cellcolor{gris1}$c_{7,10} = 0$, $c_{3,5} \neq 0$, $c_5 \neq - c_6$ & \cellcolor{gris1}$\frac{c_3 - c_{10}}{\xi c_3} + c_{11} \exp\left(\frac{c_3}{c_6 c_9}\varphi\right)$ \\
		\cellcolor{gris3}\textbf{S4} & \cellcolor{gris3}$c_{3,7,10} = 0$, $c_5 \neq 0$, $c_5 = - c_6 = c_9$ & \cellcolor{gris3}$c_{11} + G(\varphi)$ \\
		\cellcolor{gris1}\textbf{S5} & \cellcolor{gris1}$c_{5,7,10} = 0$, $c_{3,4} \neq 0$ & \cellcolor{gris1}$\frac{c_3 - c_{10}}{\xi c_3} + H\left(-\frac{c_6 + c_9}{c_4} \phi + \varphi\right) e^{c_3 \phi/c_4}$ \\
		\cellcolor{gris3}\textbf{S6} & \cellcolor{gris3}$c_{3,4,5,7} = 0$, $c_6 \neq - c_7$ & \cellcolor{gris3}$\frac{c_{10}\psi}{(c_6+c_9)\xi} + I(\phi)$ \\
		\cellcolor{gris1}\textbf{S7} & \cellcolor{gris1}$c_{3,4,5,7,10} = 0$, $c_6 = - c_7 = -c_9$ & \cellcolor{gris1}Arbitrary \\
		\bottomrule
	\end{tabularx}
	\midsepdefault
	\caption{A summary of the distinct Noether symmetries \textbf{S}$i$ together with the associated constrained $f(\phi,\psi)$ function derived in Refs.~\cite{Bahamonde:2017sdo,Channuie:2017txg}. Here, $c_1,\dots{}, c_{11}$ are integration constants, $c_{12} \equiv c_5 c_6 - c_4 c_7 + c_5 c_9$, while $F(x), G(x), H(x)$ and $I(x)$ are arbitrary functions.}
	\label{table:noethersym-nonlocal}
\end{table}

Only the cosmological expansion behavior of \textbf{S1} has been investigated in detail where it hosts both de Sitter and power-law behaviors \cite{Bahamonde:2017sdo}. Nonetheless, it is claimed that these behaviors also appear in the remaining symmetries. For the non-local $f(T)$ gravity sub-case, the de Sitter behavior appears as a solution to the field equations \cite{Channuie:2017txg}. As the general analytic scale factor behavior is not obtained, use of statefinder parameters showed the cosmological sequence consisting of deceleration $\to$ acceleration $\to$ asymptotically de Sitter with periods where this universe lies in a phantom domination epoch which may suggest quintom behavior. However, the deceleration phase may persist for a sufficiently long time leading to a fine tuning problem of the initial conditions \cite{Channuie:2017txg}.

\paragraph{Dynamical system approach} \label{sec:dyn_nonlocal}

The cosmological dynamics of the non-local $f(T) = T(\mathcal{A} \lc{\Box}^{-1} T + \mathcal{B}- 1)$ gravity model, where $\mathcal{A}, \mathcal{B}$ are constants, has been investigated in Ref.~\cite{Bamba:2017ufh}. Here, the presence of interacting dust (matter), radiation, \gls{de} and scalar field fluids are considered. This interaction is described through the interaction coefficients $\Gamma_a$
\begin{equation}
\sum\limits_a \Gamma_a = 0\,, \quad \Gamma_a = 3H^3 \sum\limits_{b} \alpha_{ab}\Omega_b\,,
\end{equation}
where $-\frac{1}{3}\alpha_{\rm{m}\Lambda} = \alpha_{\Lambda\Lambda} = \alpha_{\text{r}\Lambda} = \alpha_{\phi\Lambda} = 2\beta$ for some constant $\beta$. The phase-space variables are taken to be the density parameters $\Omega_a$, leading to a 4 dimensional autonomous system. However, careful re-examination illustrates the following key observation.

The Friedmann equations yield the constraint $\sum_a \Omega_a = 1-2\mathcal{B}$ which, in principle, allows for an elimination of one of the dynamical phase-space variables reducing the dimensionality of the system. Solving the 4 dimensional system instead fixes the value of $\mathcal{B}$. More importantly, a realistic critical point is obtained provided all the density parameters are non-negative. This ultimately leaves the critical point $(\Omega_{\rm{m}}^\star, \Omega_\Lambda^\star, \Omega_\text{r}^\star, \Omega_\phi^\star) = (0,0,0,0)$ (Point~C) as the only viable point as the remaining critical points do not appear for the following reasons:
\begin{enumerate}
	\item Point A becomes non-physical as the matter density becomes negative;
	\item Points B, D and E reduce to the critical point C.
\end{enumerate}
This critical point represents an empty static universe which is either stable or saddle. Thus, it is not suited to describe the late-time behavior.

Finally, the stability of a de Sitter solution for the special case when $\mathcal{A} = -1$ and $\mathcal{B} = 0$ was investigated. In the absence of matter fluids, the de Sitter cosmology turns out to be unstable, while inclusion of matter fluids yields an asymptotically stable static universe critical point. Thus, the model does not appear to be a viable alternative in describing a late-time, de Sitter behavior.

\subsubsection{Higher-order derivative teleparallel gravity} \label{sec:HOT_back}

The higher-order derivative theory described in Sec.~\ref{sec:HOT} adds two scalars that contain depend on the torsion scalar. These scalars for flat \gls{flrw} become
\begin{subequations}
\begin{align}
    (\lc{\nabla} T)^2 &=\frac{144H^2}{N^2}\dot{H}^2\,,\\[0.5ex]
    \lc{\Box} T &=-\frac{12 \dot{H}^2}{N^2}-\frac{12 H \ddot{H}}{N^2}-\frac{12 H\dot{ H} }{N^3}\left(3 HN^2-\dot{N}\right)\,,
\end{align}
\end{subequations}
and then, by performing variations \gls{wrt} the lapse and the scale factor (and then setting $N=1$), we arrive at the modified \gls{flrw} given by
\begin{subequations}
\begin{alignat}{2} 
    \kappa^2\rho& =\: & &-\frac{1}{2} f(T,X_1, X_2)-6 H^2 \ddot{f}_{X_2}+6 H \dot{f}_{X_2} (\dot{H}-3 H^2)-6 f_{X_2} (H \ddot{H}+3 H^2 \dot{H}+\dot{H}^2)\nonumber\\[0.5ex]
    & \: & &-144 H^3 \dot{f}_{X_1} \dot{H}-144 f_{X_1} H^3 (\ddot{H}+3 H \dot{H})-6 f_T H^2\,, \label{eq:HOD-friedmann-1} \\[0.5ex]
    -\kappa^2 p& =\: & &-\frac{1}{2} f(T,X_1, X_2)-2 \dddot{f}_{X_2} H-2 \ddot{f}_{X_2} (\dot{H}+6 H^2)-18 H \dot{f}_{X_2} (\dot{H}+H^2)\nonumber\\[0.5ex]
    &\: & &-6 f_{X_2} (H \ddot{H}+3 H^2 \dot{H}+\dot{H}^2)-48 H^2 \ddot{f}_{X_1} \dot{H}-48 H \dot{f}_{X_1} (2 H \ddot{H}+6 H^2 \dot{H}+3 \dot{H}^2)\nonumber\\[0.5ex]
    & \: & &-48 f_{X_1} \left(6 H^3 \ddot{H}+9 H^4 \dot{H}+\dot{H}^3+H^2 (\dddot{H}+12 \dot{H}^2)+4 H \dot{H} \ddot{H}\right)-2 H \dot{f}_T-2 f_T (\dot{H}+3 H^2)\,, \label{eq:HOD-friedmann-2}
\end{alignat}
\end{subequations}
where $f_{X_1}=\partial f/\partial (\lc{\nabla} T)^2$ and $f_{X_2}=\partial f/\partial \lc{\Box} T$. It should be noted that the second \gls{flrw}~\eqref{eq:HOD-friedmann-2} is not the same as the one reported in~\cite{Otalora:2016dxe}. This means that the study performed in that reference could contain some mistakes.

\paragraph{Noether's symmetry approach}

In Ref.~\cite{Bajardi:2021tul}, the Noether symmetry approach for the $f(T, \lc{\Box}T)$ sub-class has been investigated for the configuration space $\mathcal{Q} = (a, T, \lc{\Box}T)$. However, the derived \gls{flrw} equations do not match with the ones given previously. It appears that the issue stems from the determination of the Lagrange multipliers $\lambda_1, \lambda_2$. Considering the minisuperspace higher-order theory Lagrangian 
\begin{equation}
    \mathcal{L} = a^3 f(T,\lc{\Box}T) -\lambda_1 (T + 6H^2) - \lambda_2 \left(\lc{\Box}T - \ddot{T} - 3H \dot{T}\right)
\end{equation}
and taking variations \gls{wrt} $T$ yields
\begin{equation}
    \frac{\partial \mathcal{L}}{\partial T} - \frac{\dd }{\dd t}\left(\frac{\partial \mathcal{L}}{\partial \dot{T}}\right) + \frac{\dd^2}{\dd t^2}\left(\frac{\partial \mathcal{L}}{\partial \ddot{T}}\right) = 0 \implies a^3 f_T - \lambda_1 + 3\dot{H} \lambda_2 + 3H\dot{\lambda}_2 + \ddot{\lambda}_2 = 0.
\end{equation}
Contrary to that reported, it does not constrain $\lambda_1$ directly due to the terms emerging from the $\dot{T}$ and $\ddot{T}$ contributions. Re-expressing the minisuperspace Lagrangian in terms of the scale factor $a(t)$ instead would resolve this issue leading to the correct set of field equations. Thus, the Noether symmetry results may no longer hold and thus are not reported.

\paragraph{Dynamical system approach}
For the general higher-order theory theory, the dynamical system approach has been explored in Ref.~\cite{Otalora:2016dxe}. Generally, an autonomous system can be easily constructed following the definition of the dynamical variables
\begin{equation}
Z_1 := H, \; Z_2 := \dfrac{\dot{H}}{H^2}, \; \dots, \; Z_{l+1} := \dfrac{\stackrel{l}{H}}{H^{l+1}}\,,
\end{equation}
where overhead indices represent the number of time derivatives. Any critical point of the system would then be classified under two distinct cases:
\begin{enumerate}[label = (\alph*{})]
	\item ${Z_1}^\star > 0, {Z_{2, \, \dots{}, \, l+2}}^\star = 0$: de Sitter cosmology;
	\item ${Z_1}^\star = , {Z_{l+2}}^\star = l!(l+2){Z_2}^{\star(l+1)}$: power law evolution $a(t) \propto t^{-\frac{1}{{Z_2}^\star}}$.
\end{enumerate}
Although the general dynamical behavior of the theory was not explored, two ansatz forms were considered:
\begin{align}
\textrm{(i)} -T - \frac{\alpha_1 (\lc{\nabla} T)^2}{T^2} -\alpha_2 e^{\frac{\delta (\lc{\nabla} T)^2}{T^4}}\,, \quad \textrm{(ii)} -T - \frac{\beta_1 \lc{\Box} T}{T} - \frac{\beta_2 (\lc{\Box} T)^2}{T^3} - \beta_3 e^{\frac{\sigma \lc{\nabla} T}{T^3}}\,,
\end{align}
where $\alpha_{1,2}$, $\beta_{1,2,3}$ with $\beta_2 = \frac{7\beta_1}{34}$, $\delta$ and $\sigma$ are constants. For simplicity, the dynamics have been investigated in the presence of a perfect matter fluid having constant \gls{eos}. However, as Eq.~\eqref{eq:HOD-friedmann-2} does not match with the one reported in \cite{Otalora:2016dxe}, the resulting dynamical behaviors may be different from those reported. Here, we summarise the results for completeness.

The first model contains quintessence dominated and de Sitter phases as late-time attractors. Furthermore, the cosmological sequence radiation $\to$ matter $\to$ \gls{de} can be realized for suitable choices of $\alpha_1, \alpha_2$ and $\delta$. In the second model, only a stable de Sitter critical point appears while intermediary phases are not well realized making the model not viable.

\subsubsection{Mimetic and unimodular teleparallel cosmology}

In the unimodular formulation \cite{Bamba:2016wjm,Nassur:2016yhc}, the $f(T)$ Lagrangian has been reconstructed for the power law cosmology $a(t) \propto t^q$ in the presence of a perfect fluid with constant \gls{eos}, leading to\footnote{The torsion scalar Eq.~(40) in Ref.~\cite{Nassur:2016yhc} is incorrect. Correcting yields the solution found in Ref.~\cite{Bamba:2016wjm}.}
\begin{equation}
    f(T) = \Lambda + \frac{(3q-1) \kappa^2 \rho_{\rm m0}}{1-3q(2+w)}\left(\dfrac{T}{T_0}\right)^{-\frac{3q(1+w)}{2(3q-1)}}\,, \quad \lambda = \Lambda - 2\kappa^2\rho_{\rm m0} \left(\dfrac{T}{T_0}\right)^{-\frac{3q(1+w)}{2(3q-1)}}\,,
\end{equation}
where $\lambda$ represents the Lagrange multiplier which imposes the unimodularity constraint and $\Lambda$ is an integration constant. For the case of \gls{hde}, the Lagrangian reduces to rescaled \gls{tegr} \cite{Godonou:2017ugt}.

Meanwhile, the cosmological dynamical behavior of mimetic $f(T)$ gravity has been investigated in Ref.~\cite{Mirza:2017afs} in the presence of dust, radiation and dark matter fluids. Generally, the model hosts both decelerating and accelerating \gls{de} dominated phases, radiation and matter dominated periods, and an early inflationary epoch. In particular, for the power-law model $f(T) = -T + \alpha (-T)^b$ with $b < 1$, the cosmological sequence
\begin{center}
	radiation $\to$ matter $\to$ (decelerating) dark matter $\to$ (accelerating) dark energy
\end{center}
is achieved. However, the inflationary phase cannot be realized.

\subsubsection{Teleparallel Axions}\label{sec:cosmo_axions}

The background cosmological dynamics of teleparallel axion theory has been investigated in Refs.~\cite{Li:2020xjt,Li:2021wij,Hohmann:2020dgy}. In particular, the particular sub-case where the vector, axial and tensor quadratic contributions give rise to the \gls{tegr} contribution is considered, namely the gravitational action
\begin{equation}
    \mathcal{L} = -\frac{1}{2\kappa^2}\left(T + b \phi P_1 + \tilde{b} \phi P_2 \right) - \frac{\omega(\phi)}{2\kappa^2} g^{\mu\nu}\partial_\mu \phi \partial_\nu \phi + V(\phi)\,,
\end{equation}
where $P_1 = u_{\mu}a^{\mu}$ and $P_2 = \epsilon_{\mu\nu\rho\sigma}t^{\lambda\mu\nu}t_{\lambda}{}^{\rho\sigma}$ are the odd-parity quadratic torsion contributions (see Sec.~\ref{Sec:Ext_NGR}), $\phi$ is a pseudo-scalar field, $b$ and $\tilde{b}$ are constants. For the \gls{flrw} geometry in the presence of spatial curvature $k$, there are two distinct tetrad branches which lead to the same metric as discussed in Sec.~\ref{sec:time-dependent_tetrad_nonflat}. Due to the presence of the odd-parity terms, the chosen branch affects the forms of the \gls{flrw} equations:
\begin{subequations}
\begin{align}
    3H^2 + \frac{3k}{a^2} &= \kappa^2\left(\rho + \frac{\omega(\phi)}{2\kappa^2} \dot{\phi}^2 + V(\phi)\right) \,, \\[0.5ex]
    2\dot{H} + 3H^2 + \frac{k}{a^2} &= -\kappa^2 p - \frac{\omega(\phi)}{2} \dot{\phi}^2 + \kappa^2 V(\phi) + \begin{cases} 
    \frac{bu\dot{\phi}}{a}, & \text{for the first (axial) branch}\,, \\
    0, & \text{for the second (vector) branch}\,, 
    \end{cases}
\end{align}
\end{subequations}
where $u = \pm \sqrt{k}$. Observe that in both cases, the contribution from $P_2$ is absent since $t_{\lambda\mu\nu} = 0$ for both branches. Meanwhile, the $P_1$ contribution is absent in the vector branch as $a^\mu = 0$ reducing the model to a minimally coupled scenario whereas in the axial branch, the contribution is retained leading to a non-minimal coupling. Furthermore, the axion coupling does not influence the late-time dynamics due to the $u/a$ dependence but it will affect the pre-inflation cosmology. It is also convenient to define the effective axion field energy density $\rho_\phi$ and pressure $p_\phi$:
\begin{equation}
    \rho_\phi = \frac{\omega(\phi)}{2\kappa^2} \dot{\phi}^2 + V(\phi)\,, \quad p_\phi = \frac{\omega(\phi)}{2\kappa^2} \dot{\phi}^2 - V(\phi) - \begin{cases} 
    \frac{bu\dot{\phi}}{\kappa^2 a}, & \text{for the axial branch}\,, \\
    0, & \text{for the vector branch}\,.
    \end{cases}
\end{equation}

\begin{figure}[!ht]
	\centering
	\subfigure[$b = 0$]{\includegraphics[width=0.49\textwidth]{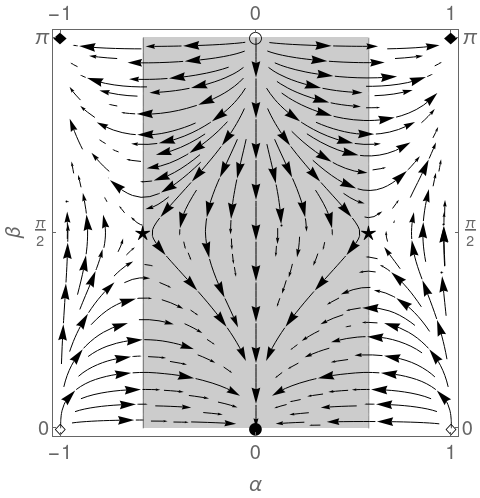}}
    \subfigure[$b = \sqrt{8/3}$]{\includegraphics[width=0.49\textwidth]{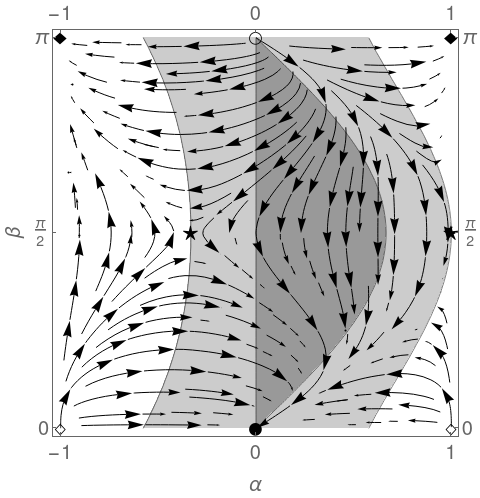}}
    \caption{Dynamical behavior for the teleparallel axion model in vacuum for two choices of $b$ illustrated in the $\alpha$-$\beta$ phase-space. Here, filled (open) circles denote stable, expanding (unstable, contracting) de Sitter critical points whereas filled (empty) diamonds denote Big Crunch (Big Bang) critical points. Stars denote the saddle transition points. Shades from dark to white indicate phantom behavior $w_{\rm eff} < -1$, acceleration $-1 < w_{\rm eff} < -1/3$ and deceleration $w_{\rm eff} > -1/3$, where $w_{\rm eff} = p_\phi/\rho_\phi$, respectively. It can be noted that no phantom behavior arises for $b = 0$ but appears for $b = \sqrt{8/3}$. In both cases, the late-time attractor is either de Sitter or Big Crunch. Permission for use of this figure was kindly provided by the authors of Ref.~\cite{Hohmann:2020dgy}.}
	\label{fig:dynamics_axion}
\end{figure}

In view of the above, \cite{Hohmann:2020dgy} focus their attention to explore the cosmological dynamics of the axial scenario in vacuum $(p = \rho = 0)$ by introducing the phase-space variables $(\alpha, \beta)$ satisfying the relations
\begin{equation}
    \dot{\phi} = \sqrt{\frac{2\kappa^2 V(\phi)}{\omega(\phi)}} \frac{\alpha}{\sqrt{1-\alpha^2}}, \quad H = \sqrt{\frac{\kappa^2 V(\phi)}{3(1-\alpha^2)}} \cos \beta, \quad a = u \left(\sqrt{\frac{\kappa^2 V(\phi)}{3(1-\alpha^2)}} \sin \beta\right)^{-1}\,,
\end{equation}
which are constrained by $-1 < \alpha < 1$ and $\sgn \, u = \sgn \sin \beta$. As pre-inflationary epochs were considered, the constant roll approximation was assumed with $\omega(\phi) = 1$ and constant potential $V(\phi)$. In turn, critical points describing a Big Bang/Big Crunch singularity ($\alpha = \pm 1$), an infinitely expanding or contracting de~Sitter universe $(\alpha = 0, \beta/\pi \in \mathbb{Z})$ and a saddle transition phase between acceleration and deceleration $(\alpha_\pm = \frac{b \pm \sqrt{b^2+8}}{2\sqrt{6}}, \frac{\beta}{\pi} - \frac{1}{2} \in \mathbb{Z})$ were obtained. A representation of the dynamical behavior is given in Fig.~\ref{fig:dynamics_axion}. 

\subsection{Inflation}

Cosmic inflation was proposed to solve the horizon, flatness and cosmic magnetic-monopole problems \cite{Guth:1980zm,Linde:1981mu} which had plagued theories of early Universe cosmology for a long time. The number of inflationary scenario proposals \cite{Baumann:2014nda,Romania:2012av,Nojiri:2017ncd} has since drastically increased where various slow- and fast-roll possible decay mechanisms for the inflationary scalar field(s) have been put forward. Despite all this, direct observational evidence of an inflationary era in the cosmic history of the Universe continues to evade measurement keeping the possibility of other exotic solutions to the early Universe puzzles as a possibility.

$\Lambda$\gls{cdm} contains no natural mechanism to produce an inflationary scenario on its own. This may of course be resolved by some exotic particle physics beyond the standard model in the matter sector. One of the interesting properties of gravity beyond \gls{gr} is that it allows for designer models to naturally induce an inflationary mechanism in the very early Universe which may give a more natural source to such a scenario.

In this section, we explore some of the possible teleparallel inflationary scenarios proposed in the literature. While interesting, there are still many possible realizations of cosmic inflation that are yet to be explored, as well as more precise predictions that these models may produce. On the other hand, there are already interesting possible scenarios in which inflation is either induced naturally or in the same way of as in $\Lambda$\gls{cdm}.

\subsubsection{Born-Infeld inspired inflation}

The pioneering work by Born and Infeld \cite{doi:10.1098/rspa.1934.0010,1933Natur.132.1004B} assumes a finiteness for physical quantities at all times in order to avoid the divergences that were emerging in physics at the time. Their initial work concerned the self-energy of point-like particles \cite{Born:1933qff}, but has since been expanded as the root inspiration into several different branches. In gravity theory, there have been numerous manifestations of Born-Infeld inspire gravity theories starting with Ref.~\cite{Deser:1998rj} where the first proposal turned out to be plagued by ghosts \cite{BeltranJimenez:2017doy}. Since this time, Born-Infeld theories have been incorporated into metric-affine formalisms \cite{Vollick:2003qp}, Eddington-inspired Born-Infeld gravity \cite{Banados:2008fj}, and many others. These have been used to eliminate a range of singularities from black holes to the big bang singularity.

Given the status of black hole physics in \gls{tg} (as explained further in Sec.~\ref{sec6:astrophysics}), the core work in the \gls{tg} literature has been in the realm of early Universe cosmology. In this background, Refs.~\cite{Ferraro:2006jd,Ferraro:2008ey,Fiorini:2009ux} explore a \gls{tg} model \textit{\`{a} la} Born-Infeld within the context of an $f(T)$ gravity Lagrangian in the spirit of the action in Eq.~\eqref{f(T)action}\footnote{Note that the convention of the torsion scalar in this Review is twice the torsion scalar in the above references} in which
\begin{equation}
    \mathcal{L}_{\rm BI-1} = \lambda_{\rm BI} \left[\sqrt{1+\frac{2\left(T + 2 \Lambda\right)}{\lambda_{\rm BI}}}-1\right]\,,
\end{equation}
where $\lambda_{\rm BI}$ is a scaling parameter which recovered $\Lambda$\gls{cdm} in the limit where $\lambda_{\rm BI} \rightarrow \infty$. For a flat \gls{flrw} Universe, for very early-times ($a \rightarrow 0$), the scale factor and Hubble parameter are given, respectively, as
\begin{equation}
    a(t) \sim \exp\left[\rule{0cm}{0.9cm} \right.\sqrt{\frac{\lambda_{\rm BI}}{12}\left(1+\frac{4\Lambda}{\lambda_{\rm BI}}\right)}\,\,\,t \left.\rule{0cm}{0.9cm}\right]\,, \quad H(t) \sim \sqrt{\frac{\lambda_{\rm BI}}{12}\left(1 + \frac{4\Lambda}{\lambda_{\rm BI}}\right)}\,,
\end{equation}
which regularizes the divergence of \gls{tegr} at very early-times. This naturally induces an inflationary era from geometry without the need of an inflaton field. The redshift relation $z=1/a(t) - 1$ implies that $z$ tends to a constant for the early Universe which produces a divergence in the particle horizon
\begin{equation}
    \eta = \int_0^1 \frac{\dd a}{a\dot{a}} \rightarrow \infty\,,
\end{equation}
giving a universe that is entirely causally connected, in agreement with the isotropic part of the cosmological principle and the \gls{cmb}. It is important to point out that despite these properties at early-times in the Universe, this model tends to regular $\Lambda$\gls{cdm} for late-times (provided that $\lambda_{\rm BI} \gg \Lambda$) which leads to a cosmic history that is extremely similar to that of $\Lambda$\gls{cdm} up to some very fine differences. In this way, the scaling parameter $\lambda_{\rm BI}$ acts takes on the role of an effective initial vacuum energy driver for the inflationary stage. In Ref.~\cite{Jana:2014aca}, very preliminary results are shown where the $\lambda_{\rm BI}$ is constrained using \gls{sn} data from Ref.~\cite{Nesseris:2004wj}. Here, it is found that $\lambda_{\rm BI} \lesssim 10^{-11}\,{\rm Mpc}^{-2}$.

In Refs.~\cite{Fiorini:2013kba,Bouhmadi-Lopez:2014tna}, a slightly more generalized scenario is considered in which
\begin{equation}\label{eq:BI_mod_fiorini}
    \mathcal{L}_{\rm BI-2} = \lambda_{\rm BI} \left[\sqrt{\det\left(g_{\mu\nu}+\frac{2\mathcal{F}_{\mu\nu}}{\lambda_{\rm BI}}\right)} - \sqrt{\det g_{\mu\nu}}\right]\,,
\end{equation}
where $\mathcal{F}_{\mu\nu}$ contains only first derivatives of the tetrad and is chosen to be
\begin{equation}
    \mathcal{F}_{\mu\nu} := \alpha_1 \dut{S}{\mu}{\rho\sigma}T_{\nu\rho\sigma} + \alpha_2 \dut{S}{\rho\mu}{\sigma}\udt{T}{\rho}{\nu\sigma} + \alpha_3 g_{\mu\nu} T\,,
\end{equation}
where $\alpha_i$ are dimensionless constants. The only meaningful condition on $\mathcal{F}_{\mu\nu}$ (and thus its constants) is that it must limit to \gls{gr} in the low energy limit ($\mathcal{F} \ll \lambda_{\rm BI}$), which in this setting means $\textrm{Tr}(\dut{\mathcal{F}}{\mu}{\nu})=T$. For the case where $\alpha_1 = 0 = \alpha_2$, the Lagrangian in Eq.~\eqref{eq:BI_mod_fiorini} becomes again an $f(T)$ model, but is distinct to $f(T)$ in all other cases.

In a flat \gls{flrw} setting and a perfect fluid source term $\rho$, Ref.~\cite{Fiorini:2013kba} finds the Friedmann equation
\begin{equation}
    \frac{\sqrt{1 - A_2H^2}}{\sqrt{1 - A_1 H2}} \left(1+ 2 A_2 H^2 - 3 A_1 A_2 H^4\right) - 1 = \frac{2 \kappa^2 \rho}{\lambda_{\rm BI}}\,,
\end{equation}
where $A_1 = 6\left(\alpha_2+2\alpha_3\right)/\lambda_{\rm BI}$ and $A_2 = 2\left(2\alpha_1 + \alpha_2 + 6\alpha_3\right)/\lambda_{\rm BI}$. The study also finds that the fluid conservation equation is observed at all scales. The authors study an intriguing form of the $A_i$ parameter, namely the case where $A_2 = 0$ which implies that $A_1 = 12/\lambda_{\rm BI}$. For the flat \gls{flrw} case they find again that in this limit
\begin{equation}
    a(t) \sim \exp\left[\sqrt{\frac{\lambda_{\rm BI}}{12}}\,\,\,t\right]\,,
\end{equation}
for the very early Universe, giving $H \sim \sqrt{\lambda_{\rm BI}/12}$ which is a maximum in this regime and which drives the early Universe to a de Sitter phase. The work also goes on to the study the evolution in nonflat scenarios but the tetrad that is adopted is not compatible with the Weitzenb\"{o}ck gauge tetrad we derive in Eqs.~\eqref{eq:FRW1} and \eqref{eq:FRW2} (for the $k=+1$ and $k=-1$ choices, respectively). While the results in this case are interesting, the antisymmetric field equations (see Sec.~\ref{subsec:goodtetrad}) will render an incompatibility with this gauge choice.

Finally, in Ref.~\cite{Fiorini:2015hob} another more exotic branch of these parameter settings is explored, namely where $A_1 = A_2$ which reduces the Friedmann equation to
\begin{equation}
    6H^2 \left(1-\frac{9H^2}{2\lambda_{\rm BI}}\right) = 2\kappa^2 \rho\,,
\end{equation}
which produces two disconnected solutions for $H^2$ with the positive sign being ignored since it is detached from \gls{gr}. To retain strong enough resemblance to \gls{gr}, the $\lambda_{\rm BI}$ must be positive, in which case, the scale factor near the very early Universe gives
\begin{equation}
    a^4(t) = \frac{\rho_0}{6\kappa^2\lambda_{\rm BI}\left(1 \pm 4\lambda_{\rm BI} t\right)} + \mathcal{O}(\lambda_{\rm BI} t^2)\,,
\end{equation}
and a Hubble parameter $H(t) = \mp \lambda_{\rm BI}/\left(1 \pm 4\lambda_{\rm BI} t\right) + \mathcal{O}(\lambda_{\rm BI} t^2)$, where $\rho_0$ is the energy density at present time for the perfect fluid. This produces a minimum for the scale factor at $t=0$ where
\begin{equation}
    a_{\rm Min}^4 = \frac{\rho_0}{6\kappa^2 \lambda_{\rm BI}}\,,
\end{equation}
and where $H_{\rm Min} = \lambda_{\rm BI}$. This is called a \textit{brusque bounce}, and represents a spacetime in which finite objects do not crash into zero volume.

\subsubsection{Power spectra in LQC \texorpdfstring{$f(T)$}{f(T)} Gravity}\label{sec:power_spec_f_T}

Loop quantum cosmology within the \gls{flrw} framework hosts the matter bounce scenario \cite{WilsonEwing:2012pu}, described by the Friedmann equation
\begin{equation}
    H^2 = \frac{\rho}{3}\left(1-\frac{\rho}{\rho_\text{cr}}\right)\,,
\end{equation}
where $\rho$ is the matter fluid energy density and $\rho_\text{cr} = \frac{3}{\lambda^2 \gamma^2}$ is a critical density originating from quantum considerations with $\lambda = \sqrt{\frac{\sqrt{3}}{2}\gamma}$ and $\gamma \approx 0.2375$ the Barbero-Immirzi parameter. Assuming a dust perfect isotropic fluid, the scale factor and Hubble parameter take on the forms
\begin{equation}
    a(t) = \left(\frac{3}{4}\rho_\text{cr} t^2 + 1\right)^{\frac{1}{3}}\,, \qquad H(t) = \frac{2 \rho_\text{cr} t}{4+3 \rho_\text{cr} t^2}\,.
\end{equation}
Clearly, the bounce occurs at $t = 0$.

In contrast to curvature-based theories, the teleparallel formulation allows for a direct reformulation of the LQC Lagrangian purely in terms of the torsion scalar by virtue of the direct correspondence between $T$ and $H$, namely $T = -6H^2$. In this way, the LQC Lagrangian can be recast in terms of $f(T)$ gravity \cite{Cai:2011tc,Bamba:2012ka,Amoros:2013nxa}
\begin{equation}
    f(T) = -\frac{3}{\lambda \gamma}\sqrt{-\frac{T}{6}} \arcsin\left(2\lambda \gamma \sqrt{-\frac{T}{6}}\right) + \frac{3}{\gamma^2 \lambda^2}\sin^2\left[\frac{1}{2}\arcsin\left(2\lambda \gamma \sqrt{-\frac{T}{6}}\right)\right]\,.
\end{equation}
Alternatively, the Lagrangian can be obtained via reconstruction of the Friedmann equations \cite{Amoros:2013nxa,Caruana:2020szx}. An extension to the standard LQC formulation is considered in Ref.~\cite{Casalino:2020kdr} where the critical density is instead parametrised as $\rho_{\text{cr}} = \frac{12}{\alpha^2}$ where $\alpha$ is a constant which measures the deviation from \gls{tegr}. This modifies the teleparallel Lagrangian to be recast in the following form
\begin{equation}
    f(T) = -\frac{12}{\alpha^2}\left[1-\sqrt{1+\frac{\alpha^2 T}{6}} - \alpha \sqrt{-\frac{T}{6}}\arcsin\left(\alpha \sqrt{-\frac{T}{6}}\right)\right]\,,
\end{equation}
which recovers \gls{tegr} $f(T) = -T$ in the limit $|\alpha| \to 0$.

Despite this apparent equivalence, there are crucial differences between the LQC and teleparallel formulations. Most notably, the critical density is quantum mechanically derived in LQC but becomes a free parameter in the teleparallel formulation. This has important consequences when matching with spectra observations as the freedom in $\rho_\text{cr}$ may be sufficient to reconcile different data sets. Meanwhile, equivalence only holds at a background level with deviations appearing at perturbative regimes.

In this setting, the matter sector is sourced by some scalar field $\phi$ having an associated potential $V(\phi)$. The latter's nature is crucial for the matter bounce cosmology as it affects both spectral indices and power spectra. A simple choice for such scalar field is one which effective fluid acts as dust with $\rho = \frac{\dot{\phi}^2}{2} + V(\phi)$ and $p = \frac{\dot{\phi}^2}{2} - V(\phi) = 0$. However, this leads to a perfectly scalar-invariant spectrum $(n_s = 1)$, contradicting \gls{cmb} observations $n_s = 0.9649 \pm 0.0042$ \cite{Akrami:2018odb}. Additionally, this does not lead to a reheating phase. As such, the scalar potential must provide \cite{Haro:2014wha,deHaro:2014tla,Haro:2015zta}:
\begin{enumerate}[label = (\alph*{})]
    \item Matching with spectral indices and power spectra observations;
    \item Account for reheating processes;
    \item Stability against small perturbations.
\end{enumerate}
Some works also include a late-time accelerating expansion as a criterion but this is not necessarily the case. Many such scalar fields cannot achieve both goals of an early-time inflationary epoch coupled with late-time accelerating expansion. For instance, a quadratic potential does not meet both this extra criterion. However, it may be modified to include a small cosmological constant provided instant preheating and matter dominated fields are obtained at low and large $\phi$ values respectively. Similar considerations can be achieved via quintessence type potentials.

Another possibility focuses on the quasi-matter contraction scenario which focuses on the matter fluid's \gls{eos} to be $\omega \ll 1$ as opposed to being identically zero. In this case, the potential $V(\phi) \propto e^{-\sqrt{3(1+\omega)}\phi}$ and the scalar spectral index changes to $n_s = 1 + 12\omega$ allowing for a non-invariant spectrum. A further generalization considers $V(\phi) \propto e^{\sqrt{3}\phi (1+f(\phi))}$ with $|f(\phi)| \ll 1$ and $|\phi f_\phi| \ll 1$. Despite such models do not allow for a transition between the contraction and expansion phases in \gls{gr}, this does not appear in the $f(T)$ formulation \cite{deHaro:2015wda}.

We now turn our attention to the scalar and tensor power spectra. To investigate the scalar power spectrum, the longitudinal gauge is considered with Bardeen potentials $\Phi$ and $\Psi$. Perturbing the scalar field about a background value $\order{\phi}{0}$ as $\phi=\order{\phi}{0}+\delta\phi$, for the potentials considered, no anisotropic contributions arise. From the perturbed $f(T)$ field equations, the $\Phi$ and $\Psi$ modes are equal according to Ref.~\cite{Haro:2013bea} and in follow up works \cite{deHaro:2014kxa,deHaro:2014tla,Haro:2014wha,Haro:2015zta,deHaro:2015wda,Haro:2017mir}. However, this is not the case as a proper account of the perturbed tetrad was not considered.

The perturbed tetrad introduces a further \gls{dof} which has not been accounted for in the aforementioned works. As this missing term appears in the anisotropic parts of the field equations, the $\Phi$ and $\Psi$ potentials are no longer equal despite being of the same order at least within sub-horizon length scales. Even if the latter scales are considered, only one of the antisymmetric $f(T)$ field equations has been accounted for despite the existence of a second antisymmetric field equation. Neglect of this secondary equation in conjunction with having equal potentials leads to having a greater number of field equations than \gls{dof} which is only reconciled once the extra scalar \gls{dof} is taken into account.

On the other hand, the tensor power spectrum is obtained by exploring the tensor perturbations which yields the relation (expressed in Fourier space, details provided in Sec.~\ref{sec:ten_pert})
\begin{equation}
    \ddot{h}_{\lambda} + \left(3H+\frac{\dot{f}_T}{f_T}\right) \dot{h}_{\lambda} + \frac{k^2}{a^2} h_{\lambda} = 0\,,
\end{equation}
which is a subset of the general Eq.~\eqref{eq:GWPE_general}, and where $\lambda$ denotes the two polarization states of the tensor perturbations. Introducing the Mukhanov-Sasaki variables $z_t = a\sqrt{f_T}$ and $v = hz_t$ recasts the equation in the form
\begin{equation}
    v^{\prime\prime} - k^2 v - \frac{z^{\prime \prime}}{z} v = 0\,,
\end{equation}
where primes denote derivatives \gls{wrt} conformal time (defined through $d\eta = dt/a$ \cite{Peebles2020}). This leads to the computation of the tensor power spectrum
\begin{equation}
    P_T := \frac{k^3}{2\pi^2} \left| \frac{v(\eta)}{z(\eta)} \right|^2 = \frac{{\rho_\text{cr}}^2}{\rho} {R_T}^2 = \frac{16\rho_\text{cr}}{3\rho}\text{Si}^2\left(\frac{\pi}{2}\right)\,, \quad R_T \coloneqq \int_{-\infty}^{\infty} \frac{\dd\eta}{{z_t}^2}\,,
\end{equation}
where $\text{Si}(x)$ is the sine integral function.

Under the assumption that the scalar power spectrum result is correct, the tensor-scalar ratio $r \approx 6.7187$, considerably larger than currently observed bounds $r < 0.056$ \cite{Akrami:2018odb}. However, the magnitude of this ratio together with the magnitudes of the power spectra changes depending on the choice of potential \cite{Haro:2014wha} as well as the magnitude of the background scalar field $\bar{\phi}$ at the bounce \cite{deHaro:2014kxa,Haro:2014wha}. Another important point to highlight is that this results depends on a particular solution of the scalar field equation of motion where a vanishing pressure for matter is assumed, which may not be the case. Ref.~\cite{Haro:2014wha} goes on to find numerical solutions which do satisfy current bounds on the tensor-scalar ratio.

\subsubsection{Higgs inflation}

In the standard model of particle physics all the matter fields couple minimally with the gravitational section except for the case of a Higgs field where no suppression needs to take place \cite{Rubio:2018ogq}. Thus, the Higgs field is an interesting avenue in which to explore differences between \gls{gr}, \gls{tegr}, and \gls{stegr}. In \gls{gr}, Higgs inflation then emerges as quantum corrections between the Einstein-Hilbert action and the Higgs coupling field \cite{Bezrukov:2013fka}. By considering direct couplings with \gls{tg} quantities like the torsion scalar instead of the Ricci scalar, we can probe possible differences in these different inflationary scenarios.

Taking a generic scalar that can then be non-minimally coupled to the scalar torsion through \cite{Raatikainen:2019qey,MohseniSadjadi:2016ukp,MohseniSadjadi:2015yex}
\begin{equation}
	\mathcal{S}_{\rm Higgs} = -\frac{1}{2\kappa^2}\int d^4 e\left[ F(\phi) T + 2G(\phi)\lc{\nabla}_{\alpha}T^{\alpha} + K(\phi) g^{\alpha\beta} \partial_{\alpha} \phi \partial_{\beta} \phi + V(\phi)\right] + \mathcal{S}_{\rm m} \left(\phi,\Psi,\udt{e}{A}{\mu},\udt{\lc{\omega}}{A}{B\mu}\right)\,,
\end{equation}
where all other matter fields $\Psi$ are minimally coupled to the torsion scalar, and $K$ and $V$ are kinetic and potential functions, and which is a particular form of the action introduced in Eq.~\eqref{action:L(T,X,Y,phi)}. In Ref.~\cite{Raatikainen:2019qey}, this model is investigated for Higgs inflation in which the potential is given by
\begin{equation}
    V = \frac{\lambda}{4}\left(\phi^2 - v^2\right)^2\,,
\end{equation}
where $\lambda$ and $v$ are constants, and where
\begin{equation}
    K = K_0\,,\quad F = F_0 \left(1+\zeta\phi^2\right)\,,\quad G = G_1 \phi^2\,,
\end{equation}
where $K_0,\,F_0$ and $G_1$ are constants. In this work, the tetrad postulate of \gls{tg} is used to expand the particular terms in the action. This is done so that the problematic term in Sec.~\ref{sec:con_dis_trans} (namely, the vector torsion coupling term) can be isolated and set to zero for one branch of possible function settings. This study is, however, not purely contained to \gls{tg}, and also includes a broader contribution from non-metricity. Despite this, for the branch where an Einstein frame exists, regular Higgs inflation ($\zeta \phi \gg 1$) produces a very high tensor-to-scalar ratio which is well constrained by observations \cite{Aghanim:2018eyx}. In the other branch, in which an Einstein frame is not recovered by a conformal (or disformal) transformation, a working inflationary model is not recovered due to linear perturbations not being dominant. This is conducted using an \gls{flrw} metric in conformal time.

Thus, in Ref.~\cite{Raatikainen:2019qey}, Higgs inflation appears problematic. However, the particular prescription being explored is very specific and may yield more realistic results. Moreover, other scalar field coupling terms may also produce more practical inflationary scenarios such as those explored in Refs.~\cite{Benisty:2020vvm,Hamada:2013mya,Gundhi:2018wyz,Schlogel:2015ota}.

The question of the naturalness of inflation in \gls{tg} remains an open question and poorly tackled in the literature. An interesting recent work on the topic is Ref.~\cite{Jarv:2021ehj} where inflation is explored through a nonminimally coupled scalar field in $f(T)$ gravity \cite{Farrugia:2016qqe} (see Sec.~\ref{sec:scalartensor}) in the same way that previous works have approach the same setting in $f(\lc{R})$ gravity \cite{Capozziello:2011et,Clifton:2011jh}. They find the slow-roll conditions for a near scale invariant power spectrum in this general setup and explore the dynamical systems for quadratic and quartic potentials. In both cases, inflation is very difficult or impossible to produce in the scalar-torsion approach. However, there are many other possible options for the scalar field potential which may produce an inflationary epoch that is more consistent with cosmic evolution in the early Universe. Other works also consider the impact of a coupled scalar field in generalized $f(T,\phi)$ gravity such as Refs.~\cite{Gonzalez-Espinoza:2020azh,Gonzalez-Espinoza:2021qnv} where power spectra for scalar and tensor modes are calculated (which have led to determinations of the spectral index and scalar-to-tensor ratio). In Ref.~\cite{Goodarzi:2018feh}, a power-law model of $f(T)$ gravity was probed for its possible inflationary scenarios where the spectral index and power spectrum is again calculated but where the model was not totally related to possible late-time accelerated expansion, for consistency. Finally, Ref.~\cite{RezaeiAkbarieh:2018ijw} explores possible tachyonic inflationary scenarios which has some interesting explanations for recent observational measurements.

\subsection{Essential conditions for viability in cosmological models} \label{sec:pros_ghosts_instab}

In this section a number of prominent modified theories within \gls{tg} have been explored through the works that appear in the literature in the cosmological setting at background level. However, cosmology can also yield theoretical limits on models in the context of their behavior against various kinds of instabilities such as ghosts and other forms of instability. Ghost instabilities represent unstable propagating \gls{dof} where a wrong sign in the propagator gives rise to unbounded states in the phase space of the theory \cite{Woodard:2015zca,Motohashi:2014opa,DeFelice:2006pg}. This became a notable feature due to the popularity in fourth-order gravitational theories in recent decades but is also a property that can arise in second-order gravitational theories. The issue can also be expressed through a model exhibiting the correct sign in its kinetic term which can be used to put limits on potential models that can be realized within such classes of theories \cite{Barth:1983hb,1978GReGr...9..353S}. There has been recent work on producing stable models for theories that feature a ghost \cite{Deffayet:2021nnt} but this has not yet been realized in a \gls{tg} setting.

Another possible problematic property in modified gravitational theories is gradient or Laplacian instabilities \cite{DeFelice:2016ucp,Rubakov:2014jja} where even small wavelength perturbations grow are potentially exponential rates unless otherwise tamed. The instability arises out of an imaginary propagation velocity in one of the perturbation sectors, and is only avoided by demanding positive propagation speeds at all scales. Similarly, if a massive field is admitted, tachyonic instabilities can appear for regions of negative effective mass \cite{Perivolaropoulos:2020uqy}. These instabilities can cause problems at background level which are related to instabilities in the vacuum. A special case of this instability is the scenario where a negative effective mass appears but which grows slower than the Hubble rate. This issue appears in $\Lambda$CDM \cite{Gsponer:2021obj} and may not gives grounds to discard a particular theory. However, it is something that must be accounted for and studied for potential problematic features.

A healthy theoretical model or class of models must be exorcised of any such instabilities which may produce unnatural expressions of a theory that are incompatible with the real Universe. Another crucial condition on which to assess whether a theoretical framework is healthy or not is the possibility of strongly coupled solutions. This was introduced in Secs.~\ref{subsymmFLRW}, \ref{sssec:ngrhamilton} and will be discussed further in the remainder of the Review. In essence, strong coupling occurs for specific solutions where the \gls{dof} do not all appear in the linearization of the theory. This means that, for those solutions, that possibly unbounded modes may coupling different order perturbations in an irregular way. Moreover, given that solutions are approximations to real systems such as in cosmology but also for astrophysical solutions, it means that such spacetimes may not approximate well these setups since even minor perturbations away from background assumptions may have a large impact at background level. This will pollute the background solution since perturbations will no longer be contained to perturbative order level. Hence, it is crucial that any theoretical setup not exhibit strong coupling for the most physical solutions such as those that express the cosmological principle, i.e. the \gls{flrw} spacetime. A viable theory should first be based on a healthy theoretical framework in which instabilities are adequately addressed and that physically realistic spacetimes are not strongly coupled.

\subsection{Perspectives of teleparallel cosmology} \label{sec:cosmo_back_conclu}

The reconstruction procedure is a viable alternative approach as a means to explore the gravitational teleparallel action at least from a background cosmological viewpoint. Briefly, $f(T)$ gravity is the simplest class of models to reconstruct and can provide solutions which match with observations as well as reproduce non-flat spacetime cosmologies purely from a gravitational standpoint. However, no $f(T)$ Lagrangian can perfectly reproduce the $\Lambda$\gls{cdm} cosmology unless it is $\Lambda$\gls{cdm} Lagrangian. This is not the case for other teleparallel extensions such as $f(T,B)$ and $f(T,T_G)$ gravity. For these theories, the Lagrangian reproduces the background $\Lambda$\gls{cdm} cosmology with the possibility of generating differing behaviors at perturbative regimes. The $\Lambda$\gls{cdm} cosmology can also be produced via a conformal scalar theory by requiring the scalar field to be the driving field to cause late-time acceleration. Introducing the trace of the stress-energy tensor $\Theta$ increases the complexity of the solutions which may make them less appealing for broader use. Nonetheless, this formulation provides an interesting alternative to the Einstein static universe as the closed spacetime geometry requirement is lifted. Additionally, unification and transitional eras can be purely described from this torsional-matter coupling viewpoint.

The dynamical system approach serves as an interesting alternative to explore the Universe's cosmological dynamics prescribed by a gravitational model. Through critical points and their stability, the corresponding cosmological sequence can be determined, although, differing phase-space variables choices influence the ensuing analysis especially in $f(T)$ gravity. Overall, the models under review here together with associated tables and figures of the relevant dynamical analysis are summarized in Figs.~\ref{fig:dynamic-overview-noscalar} and \ref{fig:dynamic-overview-scalar}. 

\begin{figure}[!ht]
	\centering
	\begin{tikzpicture}[node distance=1.75cm, >=stealth,
	topic0/.style={align=center,rectangle,minimum height=10mm,draw,rounded corners,fill=orange!30},
	topic/.style={align=center,rectangle,minimum height=10mm,draw,rounded corners,fill=blue!20},
	subtopic/.style={align=center,rectangle,draw,minimum size=35mm,minimum height=12mm}]
	\node[topic0] (top) at (-2.65,0) {Dynamical Systems Without Scalar Fields};
	\node[topic, below of = highder] (tormatt) at (-5, -0.5) {Torsion-Matter Coupling \\ \cite{Carloni:2015lsa,Bahamonde:2017ifa}};
	\node[topic, below of = tormatt] (fTB) {$f(T,B)$ \\ \cite{Paliathanasis:2017flf,Franco:2020lxx,Rave-Franco:2021yvu} \\ 
	Table~ \ref{table:dynamics-critpts-fTB-2}, Fig.~\ref{fig:dynamical_fTB}};
	\node[topic, right of = fTB, xshift = 4 cm] (fTTG) {$f(T,T_G)$ \\ \cite{Kofinas:2014aka,Tretyakov:2016uvv} \\ Table~\ref{table:dynamics-critpts1-fTTG}, Fig.~\ref{fig:fTTG_dynamics}};
	\node[topic, below of = fTB] (fT) {$f(T)$};
	\node[subtopic, right of = fT, xshift = 6 cm] (fT-m1) {$\dot{H} = \mathcal{F}(H)$ Method \\ \cite{Jamil:2012yz,Hohmann:2017jao,ElHanafy:2017xsm,ElHanafy:2017sih,Awad:2017ign,Awad:2017yod,Bamba:2016gbu,Awad:2017sau,Jamil:2012nma} \\ Table~\ref{table:dynamics-critpts-fT}, Fig.~\ref{fig:fT_dynamics}};
	\node[subtopic, below of = fT-m1] (fT-m2) {Different Phase-Space Variables \\ \cite{Wu:2010xk,Setare:2012ry,Setare:2013xh,Feng:2014fsa,Mirza:2017vrk,Ganiou:2018dta} \\ Tables~\ref{table:fT-dynamics-Sertare}, \ref{table:fT-dynamics-Mirza}, Figs.~\ref{fig:fT_dynamics_log}, \ref{fig:fT_dynamics_exp}};
	\node[subtopic, below of = fT-m2] (mim-fT) {Mimetic Approach \cite{Mirza:2017afs}};
	\node[topic, right of = tormatt, xshift = 10.5cm] (nonloc) {Non-Local \cite{Bamba:2017ufh}};
	\node[topic, below of = nonloc] (highder) {Higher-Order Derivative \\ \cite{Otalora:2016dxe}};
	\draw[->] (top) -- (tormatt);
	\draw[->] (tormatt) -- (fTB);
	\draw[->] (fTB) -- (fT);
	\draw[->] (fTTG) -- (fT);
	\draw[->] (top) -- (fTTG);
	\draw[->] (fT) -- (fT-m1);
	\draw[->] (-2.5,-5.75) -- (-2.5,-7.5) -- (fT-m2);
	\draw[->] (-2.5,-7.5) -- (-2.5,-9.25) -- (mim-fT);
	\draw[->] (top) -- (4.25,0) -- (4.25,-2.25) -- (nonloc);
	\draw[->] (4.25,-2.25) -- (4.25,-4) -- (highder);
	\end{tikzpicture}
	\caption{Breakdown summary of the reviewed works involving the use of the dynamical system approach in the absence of scalar fields.}
	\label{fig:dynamic-overview-noscalar}
\end{figure}
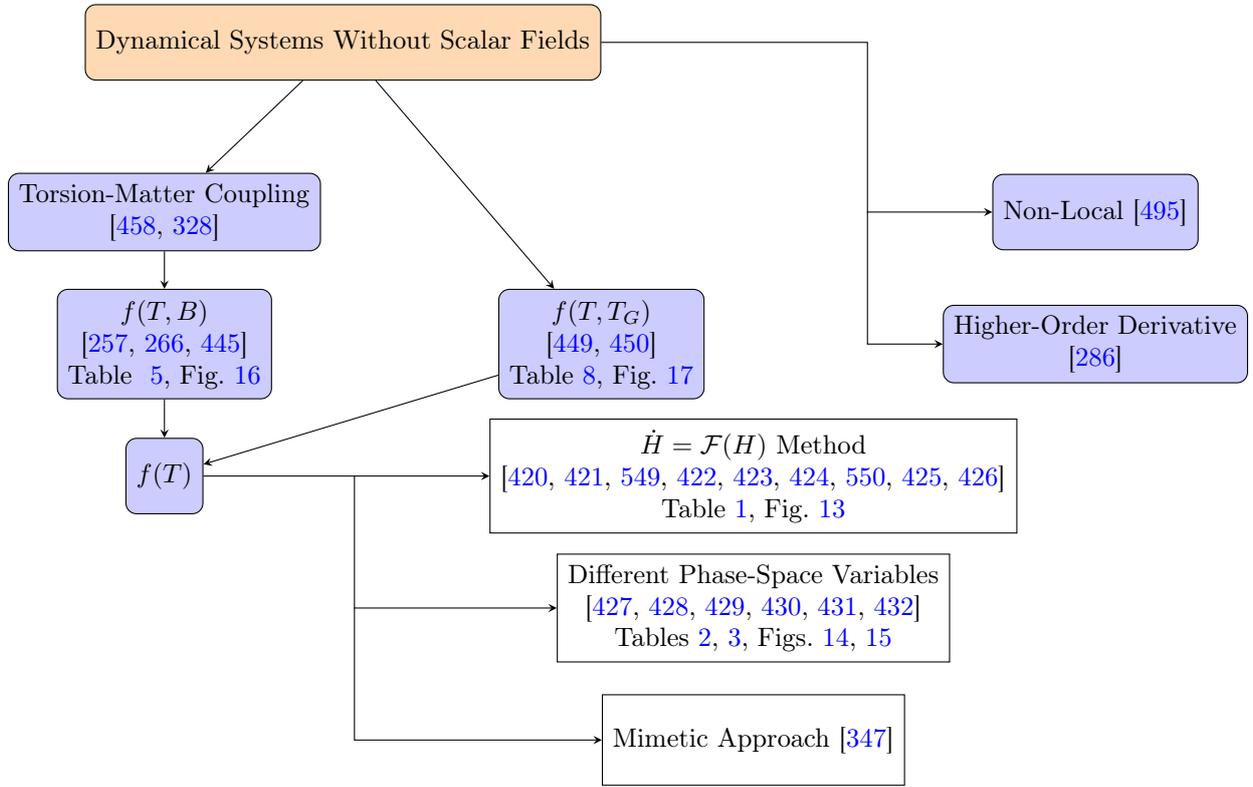

\begin{figure}[!ht]
	\centering
	\begin{tikzpicture}[node distance=1.75cm, >=stealth,
	topic0/.style={align=center,rectangle,minimum height=10mm,draw,rounded corners,fill=orange!30},
	topic/.style={align=center,rectangle,minimum height=10mm,draw,rounded corners,fill=blue!20},
	subtopic/.style={align=center,rectangle,draw,minimum size=35mm,minimum height=12mm}]
	\node[topic0] (top) {Dynamical Systems With Scalar Fields};
	\node[topic] (mincoup) at (-3.5, -1.5) {Minimally Coupled \cite{Biswas:2015cva,Awad:2017ign}};
	\node[topic, right of = mincoup, xshift = 5 cm] (nonmincoup) {Non-Minimally Coupled};
	\node[topic, below of = mincoup] (gencoup) {Generalized $T$ and $B$ Couplings \cite{Bahamonde:2018miw}};
	\node[topic, below of = nonmincoup] (fT-coup) {$f(T)$ Coupling \cite{Jamil:2012vb}};
	\node[topic, right of = fT-coup, xshift = 3.5 cm] (telax) {Teleparallel Axions \\ \cite{Hohmann:2020dgy} Fig.~\ref{fig:dynamics_axion}};
	\node[subtopic, below of = gencoup] (T-coup) {$T$ coupling \cite{Xu:2012jf,DAgostino:2018ngy,Skugoreva:2014ena} \\ (incl. matter interaction \cite{Wei:2011yr}) \\ Table~\ref{table:dynamics-critpts-nonmincoupling-Xu}, Fig.~\ref{fig:nonmin_dynamics_Xu} };
	\node[subtopic, below of = fT-coup] (B-coup) {$B$ coupling \cite{Bahamonde:2015hza} \\ Table~\ref{table:dynamics-critpts-nonmincoupling-Bahamonde1}, Fig.~\ref{fig:nonmin_dynamics_Bahamonde}};
	\draw[->] (top) -- (mincoup);
	\draw[->] (top) -- (nonmincoup);
	\draw[->] (nonmincoup) -- (gencoup);
	\draw[->] (nonmincoup) -- (fT-coup);
	\draw[->] (nonmincoup) -- (telax);
	\draw[->] (gencoup) -- (T-coup);
	\draw[->] (gencoup) -- (B-coup);
	\end{tikzpicture}
	\caption{Breakdown summary where the dynamical system approach has been applied in the presence of scalar fields.}
	\label{fig:dynamic-overview-scalar}
\end{figure}
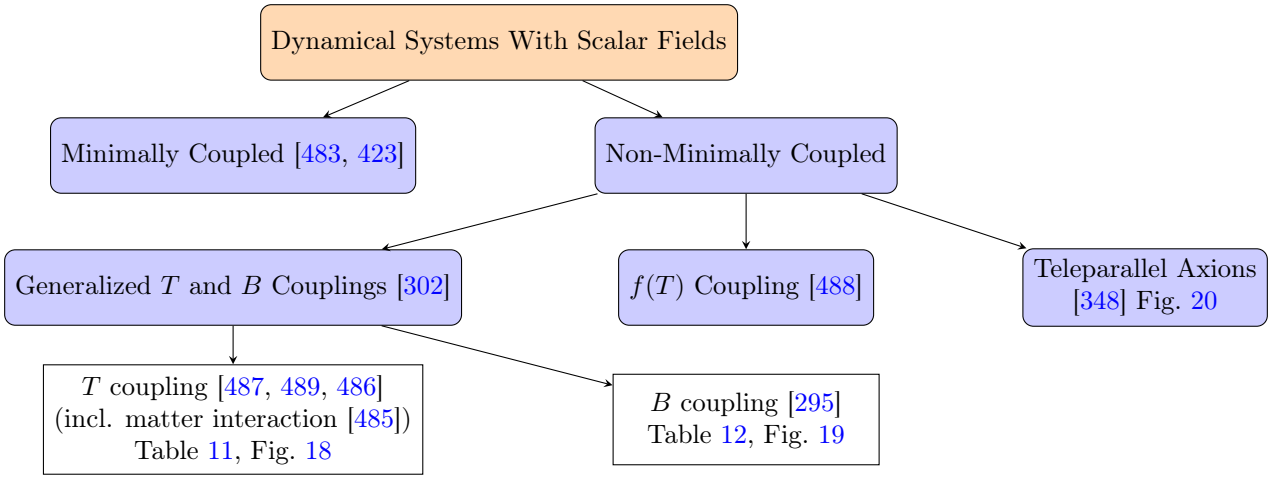

As the dynamical analysis can provide a general overview of the cosmology, a selection criterion to distinguish between suitable models has been adopted, one which hosts the cosmological sequence described by an unstable inflationary period $\to$ saddle radiation/matter domination epochs $\to$ (stable) late-time accelerated expansion. In hindsight, the teleparallel description offers a vast selection of models whose evolutionary sequence are summarized in Figs.~\ref{fig:fT-dynamic-summary} and \ref{fig:dynamic-summary}. However, only a few choices of \gls{tg} models appear to host the complete cosmological sequence while the majority of models start from matter domination. Even so, such \gls{tg} models may indeed host the full cosmological sequence needed to describe the Universe. The absence of such behavior is also imparted by the choice of the dynamical phase-space variables which greatly influence the number of critical points, and hence the evolutionary behavior, of the theory. Of note, a late-time accelerating attractor is present in practically all \gls{tg} theories. Additionally, other behaviors including bouncing scenarios, phantom and quintessence regimes can be realized. In some cases, such as in $f(T)$ gravity, absence of cosmological sequences or epochs including cyclic or oscillatory solutions can also be determined through this approach. Moreover, only the dynamical behavior of $f(T)$ gravity has been explored for arbitrary functional choices, making the study on other \gls{tg} extensions to be of further interest.

\begin{figure}[!ht]
	\centering
	\begin{tikzpicture}[node distance=1.5cm, >=stealth,
	topic/.style={align=center,rectangle,minimum height=10mm,draw,rounded corners,fill=blue!20},
	subtopic/.style={rectangle,draw,minimum size=35mm,minimum height=12mm}]
	\node[topic] (infl) {Inflation};
	\node[topic, right of = infl, xshift = 3cm] (rad) {Radiation \\ Domination};
	\node[topic, right of = rad, xshift = 3cm] (matt) {Matter \\ Domination};
	\node[topic, right of = matt, xshift = 3cm] (acc) {Late-time \\ Acceleration};
	\node[subtopic, align = justify, above of = infl, xshift = 2.4cm, yshift = 0.5cm] (infl-models) {Uncertain \cite{Wu:2010xk,Mirza:2017vrk,Hohmann:2017jao}; can produce \\ Big Bang $\to$ matter $\to$ de Sitter};
	\node[align = center, subtopic, xshift = 0.5 cm, yshift = -0.5cm, below of = rad] (rad-models) {e.g. $-T + \alpha (-T)^n$ and $-T + \alpha \sqrt{-T} \ln (-T)$};
	\node[subtopic, align = justify, above of = matt, xshift = 1.7cm, yshift = 0.5cm] (matt-models) {e.g. $-T + (-T)^a e^{bT}$ and $-T + (-T)^a e^{b/T}$ $(a > 0)$};
	\draw[->] (infl) -- (rad);
	\draw[->] (rad) -- (matt);
	\draw[->] (matt) -- (acc);
	\draw[->] ([xshift=-2.4cm]infl-models.south) -- (infl);
	\draw[->] ([xshift=-0.5cm]rad-models.north) -- (rad);
	\draw[->] ([xshift=-1.7cm]matt-models.south) -- (matt);
	\end{tikzpicture}
	\caption{Representation of how $f(T)$ gravity realizes different cosmological sequences based on the investigated model. The choice of Lagrangian also invokes the existence of certain phases and development towards future phases. Some model examples have been included.}
	\label{fig:fT-dynamic-summary}
\end{figure}
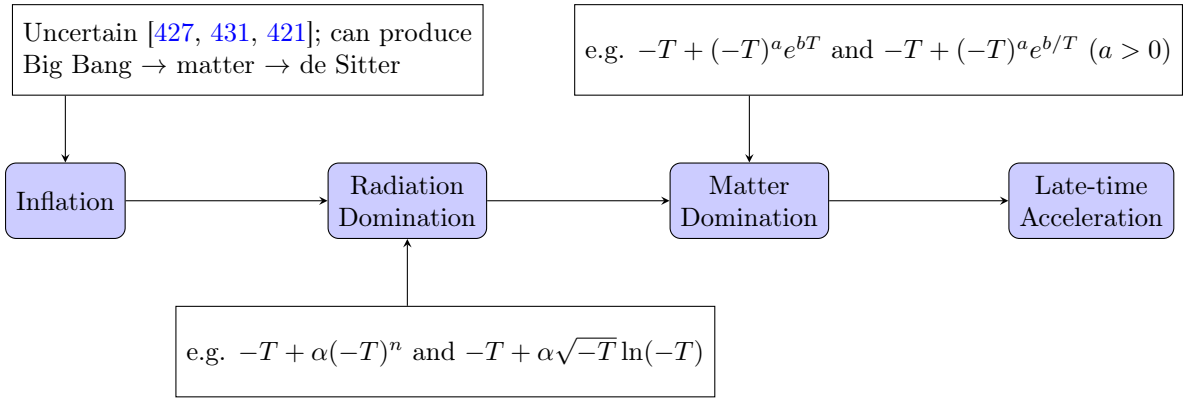

\begin{figure}[!ht]
	\centering
	\begin{tikzpicture}[node distance=1.5cm, >=stealth,
	topic/.style={align=center,rectangle,minimum height=10mm,draw,rounded corners,fill=blue!20},
	subtopic/.style={rectangle,draw,minimum size=35mm,minimum height=12mm}]
	\node[topic] (infl) {Inflation};
	\node[topic, right of = infl, xshift = 3cm] (rad) {Radiation \\ Domination};
	\node[topic, right of = rad, xshift = 3cm] (matt) {Matter \\ Domination};
	\node[topic, right of = matt, xshift = 3cm] (acc) {Late-time \\ Acceleration};
	\node[subtopic, align = justify, below of = infl, xshift = 4.2cm, yshift = -0.5cm] (infl-models) {Mimetic $f(T)$ \\ Teleparallel Axions: Big Bang $\to$ transition phases $\to$ de Sitter};
	\node[subtopic, align = justify, above of = matt, xshift = -2cm, yshift = 1cm, text width = 150mm] (matt-models) {\vspace*{-\baselineskip}\begin{align*}
	    f(T,T_G): & \text{ e.g. } -T + \alpha_1 \sqrt{T^2 + \alpha_2 T_G} \text{ for } \alpha_1 = -\sqrt{33}, \, \alpha_2 = 4 \\ 
	    f(T,B): & \text{ e.g. } -T + \alpha B^{\frac{\mu}{\mu-1}} \, (\mu > 1), \text{ Taylor expanded model} \\ 
	    \text{Torsion-Matter:} & \text{ e.g. } f_1(T) + 2\kappa^2 f_2(T)\mathcal{L}_\text{m} \\ 
	    \text{Scalar-Tensor:} & \text{ e.g. quintom model } (c_4 > 0, \, c_2 < -2c_4); \\ 
	    & \quad \quad \text{ minimal } (\xi = 0, \lambda^2 \leq 6) \text{ and non-minimal quadratic $T$~couplings } (\lambda^2 \leq \xi)
	    \end{align*}};
	\draw[->] (infl) -- (rad);
	\draw[->] (rad) -- (matt);
	\draw[->] (matt) -- (acc);
	\draw[->] ([xshift=-4.2cm]infl-models.north) -- (infl);
	\draw[->] ([xshift=2cm]matt-models.south) -- (matt);
	\end{tikzpicture}
	\caption{Representation of how other teleparallel gravitational models (including possible Lagrangian~choices) realize different cosmological sequences.}
	\label{fig:dynamic-summary}
\end{figure}
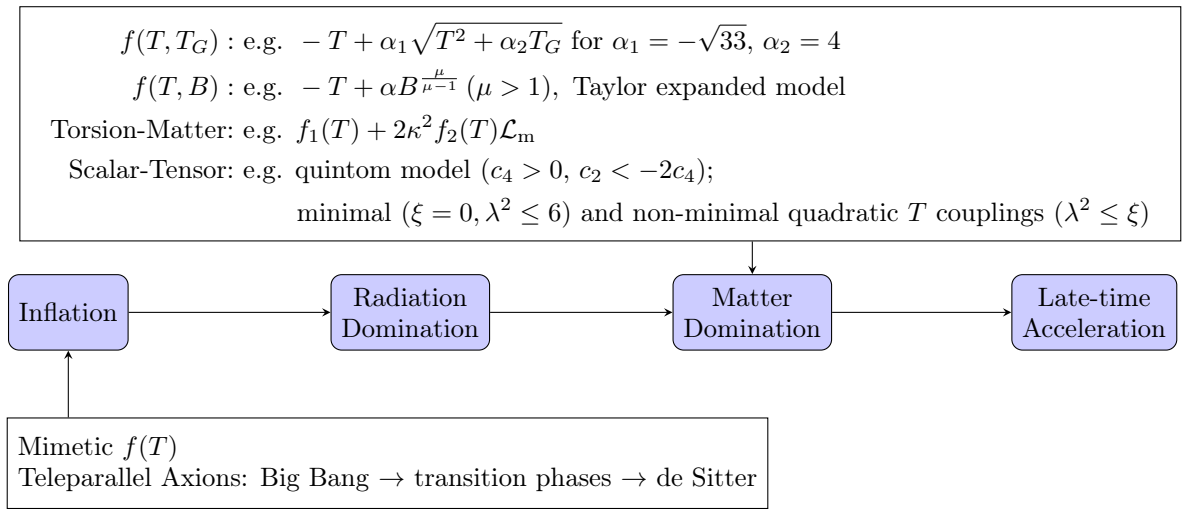

Focusing our attention to the topic of inflation, the different approaches revealed the potential existence of an inflationary period in \gls{tg}, namely in scalar-tensor formulations and from cosmological dynamics. In the latter, a viable cosmological sequence can be realised for $f(T)$ gravity and its mimetic equivalent. Other models mainly the \gls{tg} equivalents of Born-Infeld, LQC and Higgs inflation have focussed their attention in the realisation of this inflationary epoch. For instance, in the Born-Infeld formulation, having $\lambda_{\text{BI}} \lesssim \SI{e-11}{\mega\parsec^{-2}}$ ensures casual connectedness, and realisation of an inflationary epoch with $\Lambda$CDM-like behaviour at late times. However, the main issue outlined in the reviewed works lies in the cosmological perturbative regime. In the \gls{tg} LQC and Higgs formulations, they both suffer from large tensor-scalar ratios, marking it an important viability requirement for \gls{tg} models.

In view of the above, it must still be emphasized that generating the background cosmology is insufficient to confirm the viability of the model. For instance, the model could also ensure the generation of the desired cosmological sequence through dynamical systems but unable to conform with local or cosmological observations. Ensuring existence of vacuum solutions is another important criterion, a property which has in fact helped to greatly reduce the number of models in the reconstructed solutions obtained in $f(T,B)$ and $f(T,T_G)$ gravity. Ultimately, all these different approaches help in providing a set of necessary criteria to build a viable model. The next step is the exploration of the perturbative regime which holds another realm of important observations and conditions crucial to ensure the validity of the model.
\clearpage

\section{Cosmological Perturbations in Teleparallel Gravity}\label{sec:cosmo-pert}

The background geometry of the Universe, as also discussed in depth in Sec.~\ref{sec:cosmology_in_TG}, has been considered as being spatially flat \gls{flrw} geometry. The general class of \gls{flrw} geometries is generated by assuming that the Universe is homogeneous and isotropic (cosmic principle \cite{peacock1999cosmological}). Observations related to the distribution of structure formation at large scales, due to the isotropic nature of \gls{cmb} render such spatially flat cosmological geometries well motivated and founded \cite{Melchiorri:1999br,Hinshaw:2012aka,Aiola:2020azj,Balkenhol:2021eds,dodelson2003modern}. This is the case even if a few percent non-zero spatial curvature cannot be ruled out at high precision~\cite{Aghanim:2018eyx,DiValentino:2019qzk,Handley:2019tkm,DiValentino:2020hov}.

Dropping the weak field approximation of Minkowski spacetime, and considering full perturbation theory in more general backgrounds (such as the \gls{flrw} background solution), these deviations from the cosmic principle can be tackled in a more direct way. This is the origin of cosmological perturbation theory and was first introduced by Lifshitz in 1946 \cite{Lifshitz:1945du}. See~\cite{Malik:2008im} for a modern exposition of the theory of cosmological perturbations. In this framework, we attribute the inhomogeneous and anisotropic structure to the perturbations of the flat \gls{flrw} metric. These perturbations turn out to describe the observed \gls{cmb} anisotropy and large scale structure very well. Although, one needs to be very careful with the perturbative framework since it is only meaningful in very specific limits \cite{Kodama:1985bj}. For example, at late-times the matter overdensities become significantly more dominant to higher perturbative levels than the linear one.

In the perturbative framework, up to linear order, the metric assumes the expansion \cite{Malik:2008im}
\begin{align}
    g_{\mu\nu}=\order{g}{0}{}_{\mu\nu}+\delta g_{\mu\nu}\,,\label{eq:perturbed_metric_abstract}
\end{align}
where $\order{g}{0}{}_{\mu\nu}$ is the background value of the metric and $\delta g_{\mu\nu}$ are the small inhomogeneous and anisotropic correction to the background spacetime, which satisfies $|\delta g_{\mu\nu}|\ll1$. As the background evolution is exhaustively reviewed in Sec.~\ref{sec:cosmology_in_TG}, our main focus in this section will be to study the perturbation of the metric. In a very similar manner one can formulate the perturbation of the tetrad as
\begin{align}
    e^{A}{}_{\mu}=\order{e}{0}^{A}{}_{\mu}+\delta e^{A}{}_{\mu},\label{eq:perturbed_tetrad_abstract}
\end{align}
where $\order{e}{0}^{A}{}_{\mu}$ denotes the background value of the tetrad while $\delta e^{A}{}_{\mu}$ is the first order perturbation of the tetrad obeying $|\delta e^{A}{}_{\mu}|\ll1$.

In general, in order to have proper perturbative analysis the background solution of any variable under consideration ($\order{e}{0}^{A}{}_{\mu}$ or $\order{g}{0}{}_{\mu\nu}$) must not lead to trivialization of the Hamiltonian constraints. This would mean that some of the fields may assume anomalous propagation or completely lose their kinetic term thus signaling potentially fatal problems in the analysis. The most common issue of this sort is the so called strong coupling issue as discussed also in Sec.~\ref{sssec:ngrhamilton}, where some constraint in the Hamiltonian analysis is trivialized due to specific (bad) background solution thus stripping away constraints some \gls{dof} loses its dynamics. Studies in the direction for the Minkowski background in $f(T)$ theory has been carried out in \cite{Jimenez:2020ofm,BeltranJimenez:2021dce}. Furthermore, in cosmological backgrounds where a background tetrad solution $\order{e}{0}^{A}{}_{\mu}$ was such that $f_{T}\equiv f_{T}(t)$, it was shown in \cite{Blagojevic:2020dyq} to be linked with a strong coupling issue. Whether or not this is the case, the overall nature of the strong coupling problem is traced back to the non-linearity of the constraints in the velocities which is introduced by modifying the theory $T\rightarrow f(T)$ \cite{Chen:1998ad,Ong:2013qja,Chen:2014qtl}. Still we need more studies in other \gls{tg} theories such as $-T+F(B)$ gravity (or other extensions) to understand if this situation also appears in other theories.

We also mention, that there is a class of perturbations which are called homogeneous \cite{Boehmer:2007tr,Boehmer:2009yz,Bohmer:2009fc}. In this class of perturbations, one directly perturbs the background quantities and field equations. For instance one perturbs the scale factor $a(t)$ as $a(t)=\order{a}{0}(t)+\delta a(t)$ and then one can find the field equations of $\delta a(t)$ by perturbing the background Friedman equations of the theory. In this approach one could perturb the matter density and pressure to determine the behavior of the effective \gls{eos} at perturbative level. We can also use this approach to find the deceleration parameter and how it behaves under specific choices of the perturbation variables along with particular parameters arising from choosing specific models. Considering the effective \gls{eos} and the deceleration parameter, one is able to obtain stability conditions, as it was done for $f(T)$ in Refs.~\cite{Wu:2011xa,Salako:2013gka} and for $f(T,B)$ in Ref.~\cite{Bahamonde:2016cul}. This method of perturbations in completely different than the standard one, for example outlined in \cite{Malik:2008im}, since we perturb already background evaluated quantities as indicated with the scale factor. This only affects the background \gls{dof} which for a metric in flat \gls{flrw} background are just 2 whereas in the general perturbative scheme they are 10. One of the main uses of the homogeneous perturbations is to probe the background stability of a theory in contrast to the general perturbations which probe the stability of the perturbed background.

\subsection{\texorpdfstring{$3+1$}{3+1} split in cosmology}\label{sec:cos_3_plus_1}

We start by reviewing the geometry of the \gls{flrw} spacetime, as presented in Ref.~\cite{Hohmann:2020vcv} and in Sec.~\ref{sssec:adm31split}, in order to facilitate the introduction of perturbations, and to lay out the notation to be used. Assuming that the spacetime is modeled after a globally hyperbolic manifold $\mathcal{M}$, which is diffeomorphic to the product manifold $\mathbb{R}\times\Sigma$, where $\Sigma$ is a maximally symmetric, 3 dimensional manifold \cite{Gourgoulhon:2007ue,bojowald_2010}. We will denote by $i:\mathbb{R}\times\Sigma\to\mathcal{M}$ the diffeomorphism relating these manifolds and we define the time coordinate $t$ on $\mathcal{M}$ as the projection $t=\mathrm{pr}_{1}\circ i^{-1}:\mathcal{M}\to\mathbb{R}$. In addition, we equip $\Sigma$ with local coordinates $(x^{i})$, thus inducing the coordinate system $(x^{\mu})=(t,x^{i})$ for $\mathcal{M}$. For the details regarding the conventions, see Sec.~\ref{ssec:Conventions}.

The space manifold $\Sigma$ is taken to be maximally symmetric, which means that it is equipped with a metric
\begin{equation}
    \gamma=\gamma_{ij}\mathrm{d}x^{i}\otimes\mathrm{d}x^{j}\,,
\end{equation}
where $\otimes$ denotes the usual tensor product, admitting the maximal number of Killing vector fields. Hence the metric can be split as
\begin{equation}
    g_{\mu\nu}=n_{\mu}n_{\nu}-\bar{g}_{\mu\nu}\,,\label{eq:metricsplit}
\end{equation}
into the hypersurface (co-)normal
\begin{equation}
    n^{\mu}\partial_{\mu}=N^{-1}\partial_{t}\,,\quad n_{\mu}\mathrm{d}x^{\mu}=N\,\mathrm{d}t\,,
\end{equation}
and the spatial metric
\begin{equation}
    \bar{g}_{\mu\nu}\mathrm{d}x^{\mu}\otimes\mathrm{d}x^{\nu}=a^{2}\gamma_{ij}\mathrm{d}x^{i}\otimes\mathrm{d}x^{j}\,,
\end{equation}
where $N=N(t)$ is the lapse and $a=a(t)$ the scale factor. They satisfy the normalization $n_{\mu}n^{\mu}=1$ and orthogonality $n^{\mu}\bar{g}_{\mu\nu}=0$. The metric is thus given by
\begin{equation}
    g=g_{\mu\nu}\mathrm{d}x^{\mu}\otimes\mathrm{d}x^{\nu}=N^{2}\mathrm{d}t\otimes\mathrm{d}t-a^{2}\gamma_{ij}\mathrm{d}x^{i}\otimes\mathrm{d}x^{j}\,.\label{eq:metric_3_1_FLRW}
\end{equation}
This metric expressed in spherical coordinates was previously introduced in Eq.~\eqref{eq:frwN}. In case one chooses Cartesian coordinates, and a flat \gls{flrw} geometry for the metric \eqref{eq:metric_3_1_FLRW}, we obtain the form of Eq.~\eqref{FLRW_metric}. Both of these cases were discussed in detail in Sec.~\ref{sec:cosmology_in_TG}. Further, we denote by $\epsilon_{\mu\nu\rho\sigma}$ the totally antisymmetric tensor defined by the spacetime metric $g_{\mu\nu}$. Its spatial part is denoted as $\varepsilon_{\mu\nu\rho}$ and defined by
\begin{equation}
    \varepsilon_{\mu\nu\rho}=n^{\sigma}\epsilon_{\sigma\mu\nu\rho}\,,\quad\epsilon_{\mu\nu\rho\sigma}=4\varepsilon_{[\mu\nu\rho}n_{\sigma]}\,.
\end{equation}
We note that by construction, $n^{\mu}\varepsilon_{\mu\nu\rho}=0$ and we remark that $\varepsilon_{\mu\nu\rho}$ is related to the totally antisymmetric tensor $\upsilon_{ijk}$ of the metric $\gamma_{ij}$ via
\begin{equation}
    \varepsilon_{\mu\nu\rho}\mathrm{d}x^{\mu}\otimes\mathrm{d}x^{\nu}\otimes\mathrm{d}x^{\rho}=a^{3}\upsilon_{ijk}\mathrm{d}x^{i}\otimes\mathrm{d}x^{j}\otimes\mathrm{d}x^{k}\,.
\end{equation}
We will make use of these relations in the following sections.

We split the arbitrary tensors into spatial and temporal components by first introducing the spatial tensor fields
\begin{equation}
    \Pi_{\mu}^{c}\partial_{c}\otimes\mathrm{d}x^{\mu}=a\delta_{b}^{c}\,\partial_{c}\otimes\mathrm{d}x^{b}\,,\quad\Pi_{c}^{\mu}\partial_{\mu}\otimes\mathrm{d}x^{c}=a^{-1}\delta_{c}^{b}\,\partial_{b}\otimes\mathrm{d}x^{c}\,.
\end{equation}
It follows that they are related to the unit (co-)normal $n_{\mu}$ and induced spatial metric $\bar{g}_{\mu\nu}$ by
\begin{equation}
    n_{\mu}\Pi_{i}^{\mu}=0\,,\quad n^{\mu}\Pi_{\mu}^{i}=0\,,\quad\bar{g}_{\mu\nu}\Pi_{i}^{\mu}\Pi_{j}^{\nu}=\gamma_{ij}\,,\quad\gamma_{ij}\Pi_{\mu}^{i}\Pi_{\nu}^{j}=\bar{g}_{\mu\nu}\,.
\end{equation}
Consequently, we can write the temporal and spatial projectors in the form
\begin{equation}
    \delta_{\nu}^{\mu}=n^{\mu}n_{\nu}-\bar{g}_{\nu}^{\mu}=n^{\mu}n_{\nu}-\Pi_{j}^{\mu}\Pi_{\nu}^{j}\,,\quad\Pi_{\mu}^{i}\Pi_{j}^{\mu}=\delta_{j}^{i}\,.
\end{equation}
For a vector field $X=X^{\mu}\partial_{\mu}$, we then introduce the notation for the temporal and spatial components:
\begin{equation}
    X=N^{-1}\hat{X}^{0}\partial_{t}+a^{-1}\hat{X}^{a}\partial_{a}\,,\quad\hat{X}^{0}=n_{\mu}X^{\mu}=NX^{0}\,,\quad\hat{X}^{b}=-\Pi_{\mu}^{b}X^{\mu}=-aX^{b}\,.
\end{equation}
 Conversely, for a covector field $\Xi=\Xi_{\mu}\mathrm{d}x^{\mu}$ we write
\begin{equation}
    \Xi=N\hat{\Xi}_{0}\,\mathrm{d}t+a\hat{\Xi}_{b}\,\mathrm{d}x^{b}\,,\quad\hat{\Xi}_{0}=n^{\mu}\Xi_{\mu}=N^{-1}\Xi_{0}\,,\quad\hat{\Xi}_{b}=\Pi_{b}^{\mu}\Xi_{\mu}=a^{-1}\Xi_{b}\,.
\end{equation}
Hence we see that the temporal and spatial components are related by
\begin{equation}
    \hat{X}^{0}=\hat{X}_{0}\,,\quad\hat{X}^{a}=-\gamma^{ab}\hat{X}_{b}\,,
\end{equation}
and analogously for $\Xi$. Hence, the decomposed tensor fields indices are raised and lowered with the background metric $\dd s^{2}=g=\mathrm{d}t\otimes\mathrm{d}t-\gamma_{ij}\mathrm{d}x^{i}\otimes\mathrm{d}x^{j}$. Note that the components of the latter are independent of the time coordinate $t$, in contrast to the components of the spacetime metric $g_{\mu\nu}$.

We now make use of the tensor decomposition introduced above to decompose the covariant derivative of the Levi-Civita connection $\lc\nabla$ (as defined in Secs.~\ref{ssec:Conventions}, \ref{sec2:affine} and \ref{ssec:admformalism}). By considering $X^{\mu}$ and $Y^{\mu}$ as spatial, we have that $n_{\mu}X^{\mu}=n_{\mu}Y^{\mu}=0$, and then, the covariant derivative
\begin{equation}
    \lc{\nabla}_{X}Y=(X^{\mu}\lc{\nabla}_{\mu}Y^{\nu})\partial_{\nu}\,,
\end{equation}
is not necessarily spatial. Its decomposition
\begin{equation}
    \lc{\nabla}_{X}Y=\lc{\mathrm{D}}_{X}Y+nK(X,Y)\,,
\end{equation}
where $n := n^{\alpha} \partial_{\alpha}$ is the unit normal vector, defines the spatial covariant derivative
\begin{equation}
    \lc{\mathrm{D}}_{X}Y=(X^{\mu}\lc{\mathrm{D}}_{\mu}Y^{\nu})\partial_{\nu}=(X^{\mu}\lc{\nabla}_{\mu}Y^{\nu})\bar{g}_{\nu}^{\rho}\partial_{\rho}\,,
\end{equation}
as well as the extrinsic curvature
\begin{equation}\label{eq:extrinsic_curvature}
    K(X,Y)=K_{\mu\nu}X^{\mu}Y^{\nu}=-n_{\nu}X^{\mu}\lc{\nabla}_{\mu}Y^{\nu}\,,
\end{equation}
which acts as a measure geometric deformation \gls{wrt} a hypersurface \cite{Wald:1984rg}, and where Eq.~\eqref{eq:metricsplit} was used. The latter is related to the acceleration \cite{Giulini:2015qha} vector field $A_{\mu}$ by
\begin{equation}
    K_{\mu\nu}=\lc{\nabla}_{\mu}n_{\nu}+n_{\mu}A_{\nu}\,,\quad A_{\mu}=n^{\nu}\lc{\nabla}_{\nu}n_{\mu}\,.
\end{equation}

In order to relate the introduced decomposition of covariant derivatives on model spacetimes $\mathcal{M}$ to those projected on the spatial manifold $\Sigma$, we denote by $\mathrm{d}_{i}$ the Levi-Civita derivative of the metric $\gamma_{ij}$, which acts on vector fields $\xi^{i}$ on $\Sigma$ as
\begin{equation}
    \mathrm{d}_{i}\xi^{j}=\partial_{i}\xi^{j} + \bar{\Gamma}^{j}{}_{ki}\xi^{k}\,,
\end{equation}
with the spatial Christoffel symbols of the spatial metric $\gamma_{ij}$
\begin{equation}
    \bar{\Gamma}^{i}{}_{jk}=\frac{1}{2}\gamma^{il}(\partial_{j}\gamma_{lk}+\partial_{k}\gamma_{jl}-\partial_{l}\gamma_{jk})\,.
\end{equation}
In the case of the \gls{flrw} metric, the acceleration and extrinsic curvature are given by
\begin{equation}
    A_{\mu}=0\,,\quad K_{\mu\nu}=H\bar{g}_{\mu\nu}\,,\quad H=\frac{1}{N}\frac{\partial a}{a}\,,
\end{equation}
where the dot denotes differentiation \gls{wrt} time and $H$ is usual Hubble function already introduced in various places as Sec.~\ref{ssec:Conventions} and Sec.~\ref{sec:cosmology_in_TG}. This just signifies the geometrical origin of the Hubble function as derived from a fully geometrical point of view. Having laid out explicitly the underlying geometrical framework, we will introduce the cosmological perturbation scheme in detail in the next sections.

\subsection{Irreducible scalar-vector-tensor decomposition} \label{sssec:Irreducible_decomposition}

Expanding on the form of the perturbations, the general first order expansion of the perturbed metric of Eq.~\eqref{eq:perturbed_metric_abstract} can be explicitly determined in specific backgrounds. Fixing the background to be spatially flat \gls{flrw}, the perturbation of the metric $\delta g_{\mu\nu}$ can be parametrized in a irreducible form as
\begin{equation}
    \delta g_{\mu\nu}:=\left[\begin{array}{cc} - 2\varphi & a\left(\partial_{i}\mathcal{B}+\mathcal{B}_{i}\right)\\
    a\left(\partial_{i}\mathcal{B}+\mathcal{B}_{i}\right) & 2a^{2}\left(-\psi\delta_{ij}+\partial_{i}\partial_{j}h+2\partial_{(i}h_{j)}+\frac{1}{2}h_{ij}\right)
\end{array}\right]\,.\label{eq:hS-1}
\end{equation}
The constituent fields that have an index are all divergenceless, i.e, $\partial^{i}h_{ij}=0$ and also for any of the vectors $\partial_{i}X^{i}=0$. In addition $h_{ij}$ is symmetric and traceless $h_{ij}\delta^{ij}=0$. Since the metric is a symmetric (0,2) tensor in general, it assumes 10 \gls{dof} which are distributed as 4 scalars $\left\{ \varphi,\mathcal{B},\psi,h\right\} $ (account for 1 \gls{dof} each), 2 vectors $\left\{ \mathcal{B}_{i},h_{j}\right\} $ (account for 2 \gls{dof} each) and 1 tensor $h_{ij}$ (accounts for 2 \gls{dof}). This is the most general form of the perturbation of the metric in a flat \gls{flrw} background and it was first introduced by Lifshitz \cite{Carroll:2004st,Malik:2008im,Mukhanov:1990me,Dodelson:2003ft,Kodama:1985bj}. On the other hand, the tetrad can be represented as a generic 4 by 4 matrix with no symmetries which means that it enjoys the full set of 16 \gls{dof} contrary to the metric which can be also represented as symmetric matrix assuming only 10 \gls{dof}. Hence we distribute these 16 \gls{dof} of the perturbation of the tetrad $\delta e^{I}{}_{\mu}$ as follows
\begin{equation}
    \delta e^{I}{}_{\mu}:=\left[\begin{array}{cc} - \varphi & -a\left(\partial_{i}\beta+\beta_{i}\right)\\
    \delta^{I}{}_{i}\left(\partial^{i}b+b^{i}\right) & a\delta^{Ii}\left(-\psi\delta_{ij}+\partial_{i}\partial_{j}h+2\partial_{(i}h_{j)}+\frac{1}{2}h_{ij}+\epsilon_{ijk}\left(\partial^{k}\sigma+\sigma^{k}\right)\right)
\end{array}\right]\,,\label{eq:tetrad perturbation-1}
\end{equation}
where the capital Latin letters indicate 3 dimensional spatial Minkowski indices (see Sec.~\ref{ssec:Conventions}). Here, again all the vectors are solinoidal $\partial_{i}X^{i}=0$ and to be compatible with the metric $\mathcal{B}\equiv b-\beta$, $\mathcal{B}_{i}\equiv b_{i}-\beta_{i}$. In this case, the 16 \gls{dof} are split as follows: 6 scalars $\left\{ \varphi,b,\beta,\psi,h,\sigma\right\} $, 4 vectors $\left\{ b_{i},\beta_{i},h_{i},\sigma_{i}\right\} $ and 1 tensor $h_{ij}$. The counting is again the same as the in case of the metric. The main differences schematically are the introduction of a pseudo-scalar $\sigma$ and a pseudo-vector $\sigma_{i}$ which gives us $\mathcal{B}\rightarrow\{b,\beta\}$, $\mathcal{B}_{i}\rightarrow\{b_{i},\beta_{i}\}$. The extra 6 \gls{dof} are due to the Lorentz group which the metric is already invariant under. The antisymmetric part, that introduces the terms $\partial^{k}\sigma$ and $\sigma^{k}$, was elusive at first by some early works like Ref.~\cite{Chen:2010va} and were later correctly included in Ref.~\cite{Zheng:2010am,Wu:2012hs,Izumi:2012qj}.

We note also that one can introduce the spacetime indexed tetrad
\begin{equation}
    \tau_{\mu\nu}:=\eta_{AB}\order{e}{0}^{A}{}_{\mu}\delta e^{B}{}_{\nu}\,,\label{eq:perttens}
\end{equation}
which calculated explicitly for Eq.~\eqref{eq:tetrad perturbation-1} reads as
\begin{equation}
    \tau_{\mu\nu}=\left[\begin{array}{cc} - \varphi & -a\left(\partial_{i}\beta+\beta_{i}\right)\\
    a(\partial_{i}b+b_{i}) & a^{2}\left(-\psi\delta_{ij}+\partial_{i}\partial_{j}h+2\partial_{(i}h_{j)}+\frac{1}{2}h_{ij}+\epsilon_{ijk}\left(\partial^{k}\sigma+\sigma^{k}\right)\right)
\end{array}\right]\,,\label{eq:deltatetradgreek}
\end{equation}
and turns out to be more suitable for calculations since the Lorentz indices can be dropped in the process. Just as the inverse metric perturbation, the inverse tetrad perturbation can be calculated using the relations $e^{A}{}_{\mu}E_{A}{}^{\nu}=\delta_{\mu}^{\nu}$ and $g_{\mu\alpha}g^{\alpha\nu}=\delta_{\mu}^{\nu}$ which were introduced in Secs.~\ref{ssec:Conventions} and \ref{sssec:metafftetspi}.

The following convention for the perturbation of the energy-momentum tensor of a perfect fluid will be employed
\begin{equation}
    \delta\Theta_{\mu}{}^{\nu}:=\left[\begin{array}{cc} \delta\rho & (\rho+p)(v^{i}+\partial^{i}v)\\
    -a^{2}(\rho+p)(v_{i}+\partial_{i}v) & -\delta p\delta_{j}^{i}
\end{array}\right]\,,\label{eq:deltaenergy}
\end{equation}
where $\rho$ is the matter density, $p$ denotes the pressure, and $v,v^{i}$ denote the scalar and vector parts of the perturbation of the velocity field. In general, the spatial part $\delta\Theta_{i}{}^{j}$ can include an anisotropic stress piece $\Pi_{i}{}^{j}$ which can be further split into a scalar-vector-tensor (\gls{svt}) decomposition as\begin{subequations}
\begin{align}
\Pi_{ij} & =\Pi_{ij}^{S}+\Pi_{ij}^{V}+\Pi_{ij}^{T}\,,\label{eq:PiAS}\\[0.5ex]
\Pi_{ij}^{S} & :=\partial_{i}\partial_{j}\Pi^{S}\,,\label{eq:PiASS}\\[0.5ex]
\Pi_{ij}^{V} & :=-\frac{1}{2}(\partial_{j}\Pi_{i}^{V}+\partial_{i}\Pi_{j}^{V})\,,\label{eq:PiASV}\\[0.5ex]
\Pi_{ij}^{T} & \phantom{:=}\textrm{\textrm{is the tensor part\,,}}\label{eq:PiAST}
\end{align}
\end{subequations} where $\partial^{i}\Pi_{i}^{V}\equiv0\equiv\partial^{i}\Pi_{ij}^{T}$,
$\Pi_{ij}^{T}\equiv\Pi_{(ij)}^{T}$ and $\delta^{ij}\Pi_{ij}^{T}\equiv0$.

We close off by introducing our convention for the Fourier transform of a perturbation $X$ as
\begin{align}
X\left(t,x\right) & =\int\frac{\dd \omega}{\left(2\pi\right)^{1/2}}\int\frac{d^{3}k}{\left(2\pi\right)^{3/2}}\left[X(\omega,k_{j})e^{-i\omega t+ik_{j}x^{j}}+X^{\dagger}(\omega,k_{j})e^{i\omega t-ik_{j}x^{j}}\right]\,,\label{eq:fourier_trans_con}
\end{align}
where $\omega$ is the time component and $k_{j}$ is the spatial wave covector of the Fourier transform, and $X^{\dagger}$ is the conjugate of $X$. We also raise and lower the indices as $X_{0}=X^{0}$, $X_{j}=-X^{j}$ and use the conventions $\Box:=\partial_{\mu}\partial^{\mu}=\partial_{0}^{2}-\partial^{2}$, where the spatial Laplacian is defined as $\partial^{2}:=-\eta^{ij}\partial_{j}\partial_{i}=\delta^{ij}\partial_{i}\partial_{j}$. Along these lines, the norm of the wave covector $k_{i}$ is defined as $k^{2}:=-\eta^{ij}k_{i}k_{j}=\delta^{ij}k_{i}k_{j}$. In the Supplementary annexes (Supplementary 2), we provide some useful expressions for the \gls{svt} decomposition for different important tensors in \gls{tg}.

\subsection{Gauge transformations} \label{sec:gauge_trans}

We will utilize the active approach of Refs.~\cite{Bruni:1996im,Malik:2008yp} in setting up the gauge transformations of tensor fields. In this approach the perturbation of a tensor field $Z$ along the direction of a vector field $Y$ at any order is given by
\begin{equation}
    \widetilde{Z}=e^{\mathcal{L}_{Y}}Z\,,
\end{equation}
where $\widetilde{Z}$ is the transformed tensor field and $\mathcal{L}_{Y}$ is the Lie derivative along $Y$. The vector field $Y$ that generates the transformation and the tensor field $Z$ can be further expanded up to second order as
\begin{subequations}
\begin{align}
    Y & =Y_{1}+\frac{1}{2}Y_{2}\,,\\[0.5ex]
    Z & =Z_{1}+\frac{1}{2}Z_{2}+\mathcal{O}(\left(\mathcal{L}_{Y}\right)^{3})\,,
\end{align}
\end{subequations}
and the exponential map can be expanded up to second order as
\begin{equation}
    e^{\mathcal{L}_{Y}}=1+\mathcal{L}_{Y_{1}}+\frac{1}{2}\left(\mathcal{L}_{Y_{1}}\right)^{2}+\frac{1}{2}\left(\mathcal{L}_{Y_{2}}\right)^{2}+\mathcal{O}(\left(\mathcal{L}_{Y}\right)^{3})\,.
\end{equation}
Combining the equations above, we finally arrive at
\begin{subequations}
\begin{align}
\widetilde{Z}_{0} & =Z_{0}\,,\\[0.5ex]
\widetilde{Z}_{1} & =Z_{1}+\mathcal{L}_{Y_{1}}Z_{0}\,,\\[0.5ex]
\widetilde{Z}_{2} & =Z_{2}+\mathcal{L}_{Y_{2}}Z_{0}+\left(\mathcal{L}_{Y_{1}}\right)^{2}Z_{0}+2\mathcal{L}_{Y_{1}}Z_{1}\,,
\end{align}
\end{subequations}
where in general, for a scalar field $\phi$, a vector field $X_{\mu}$ and a (0,2) tensor $t_{\mu\nu}$ we have
\begin{subequations}
\begin{align}
\mathcal{L}_{Y}\phi & =Y^{\lambda}\phi_{,\lambda}\,,\\[0.5ex]
\mathcal{L}_{Y}X_{\mu} & =X_{\mu,\alpha}Y^{\alpha}+X_{\lambda}Y^{\lambda}{}_{,\mu}\,,\\[0.5ex]
\mathcal{L}_{Y}t_{\mu\nu} & =t_{\mu\nu,\lambda}Y^{\lambda}+t_{\mu\lambda}Y^{\lambda}{}_{,\nu}+t_{\lambda\nu}Y^{\lambda}{}_{,\mu}\,.
\end{align}
\end{subequations}
Without digressing, let us state that the passive approach, counterpart to the active we just presented is just stating that an infinitesimal change of coordinates along the vector field $Y^{\mu}$ is described by $\widetilde{x}^{\mu}=x^{\mu}+Y_{1}^{\mu}$ which is usually further split as $\widetilde{t}=t+Y_{1}^{0}$ and $\widetilde{x}^{i}=x^{i}+Y_{1}^{i}$. This is the usual first order approach followed for example in Refs.~\cite{bardeen1980gauge,Kodama:1985bj,Mukhanov:1990me}, but that method is rather cumbersome if one needs to consider higher orders.

Next, we will calculate the gauge transformation of the components of the perturbed tetrad. In what follows we will only determine the first order perturbations so that, we drop the subscripts $Y_{1}^{\mu}\rightarrow Y^{\mu}$. The transformed perturbed tetrad according to the active approach is
\begin{equation}
    \widetilde{\delta e}^{A}{}_{\mu}=\delta e^{A}{}_{\mu}+\mathcal{L}_{Y}e^{A}{}_{\mu}\,.
\end{equation}
We further split $Y^{\mu}$ as $Y^{\mu}=\left\{ Y^{0},\zeta(Y^{i}+\delta{}^{ij}\partial{}_{j}Y)\right\} $ where $\zeta\in\{1,\frac{1}{a}\}$, which incorporates both conventions (in terms of how $Y^{\mu}$ can be split in the $3+1$ decomposition), $\partial{}_{i}Y^{i}=0$ and $Y$ the scalar part coming from splitting $Y^{i}$ \begin{subequations}\label{eq:gauge_transformations}
\begin{eqnarray}
\widetilde{\varphi} & = & \varphi-\dot{\xi}{}^{0}\,,\quad\widetilde{\psi}=\psi+H\xi^{0}\,,\quad\widetilde{\beta}=\beta-\frac{1}{a}\xi{}^{0}-\xi{}^{0}\,,\quad\widetilde{\beta_{i}}=\beta_{i}\,,\\[0.5ex]
\widetilde{b} & = & b-a\left(\dot{\zeta}\xi+\zeta\dot{\xi}\right)\,,\quad\widetilde{b}_{i}=b_{i}+a\left(\dot{\zeta}\xi_{i}+\zeta\dot{\xi_{i}}\right)\,,\quad\widetilde{\sigma}=\sigma\,,\quad\widetilde{\sigma}_{i}=\sigma^{i}-\frac{1}{2}\epsilon^{i}{}_{jk}\partial^{j}\xi^{k}\,,\\[0.5ex]
\widetilde{h} & = & h-\zeta\xi\,,\quad\widetilde{h}_{i}=h_{i}+\frac{1}{2}\zeta\xi_{i}\,,\quad\widetilde{h}_{ij}=h_{ij}\,.
\end{eqnarray}
\end{subequations}
These transformations suggest the grouping $\left\{ \varphi,\psi,\beta\right\} $, $\left\{ b,h\right\} $ and $\left\{ b_{i},h_{i},\sigma_{i}\right\} $ for the fields. This grouping is generated by gathering all fields that transform similarly. In this way, we can set to zero only one element from each group due to the similarity in their transformation. For example we can always set the pseudo-vector $\sigma_{i}$ to zero but not the pseudo-scalar $\sigma$ since it is gauge invariant. Using the transformations from Eq.~\eqref{eq:gauge_transformations} and the grouping of the fields in order to gather a few popular gauge fixing choices in Table~\ref{table:gauge_choices_tetrad_metric}.

\begin{table}[H]
\centering
\midsepremove
\begin{tabular}{lll}
\toprule
\cellcolor{gris3}\textbf{Gauge} & \cellcolor{gris3}\textbf{Metric} & \cellcolor{gris3}\textbf{Tetrad}\tabularnewline
\midrule
\cellcolor{gris1}Flat & \cellcolor{gris1}$\psi=0,h=0$ & \cellcolor{gris1}$\psi=0,h=0$\tabularnewline
\cellcolor{gris3}Unitary & \cellcolor{gris3}$\delta\Phi=0,h=0$ & \cellcolor{gris3}$\delta\Phi=0,h=0$\tabularnewline
\cellcolor{gris1}Newtonian & \cellcolor{gris1}$\mathcal{B}=0,h=0$ & \cellcolor{gris1}$b=\beta,h=0$\tabularnewline
\cellcolor{gris3}Synchronous & \cellcolor{gris3}$\varphi=0,\mathcal{B}=0$ & \cellcolor{gris3}$\varphi=0,b=\beta$\tabularnewline
\bottomrule
\end{tabular}
\midsepdefault
\caption{Common gauge choices for the metric and corresponding tetrad}
\label{table:gauge_choices_tetrad_metric}
\end{table}

For completeness we introduce a generic scalar field $\Phi$ and a commonly used scalar called the shear potential $\chi:=a(\beta-b+a\dot{E})$ which transform as \begin{subequations}
\begin{align}
\widetilde{\chi} & =\chi+Y^{0}\,,\\[0.5ex]
\widetilde{\delta\Phi} & =\delta\Phi+\dot{\Phi}Y^{0}\,.
\end{align}
\end{subequations}
From all the scalar \gls{dof} we can construct
the following gauge invariant variables found in the literature but
adapted to the perturbed tetrad, \begin{subequations}
\begin{align}
\Phi_{B} & =\frac{d}{dt}\left(a(b-\beta)\right)-\frac{d}{dt}(a^{2}\dot{h})\,,\\[0.5ex]
\Psi_{B} & =-\psi+a(-\beta+b)H-a{}^{2}H\dot{h}\,,\\[0.5ex]
\delta\Phi_{f} & =\delta\Phi+\frac{\dot{\Phi}}{H}\psi\,,\\[0.5ex]
\delta\Phi_{N} & =\delta\Phi+a(-\beta+b)\dot{\Phi}-a{}^{2}\dot{\Phi}\dot{h}\,,
\end{align}
\end{subequations}
where $\Phi_{B},\Psi_{B}$ are the usual Bardeen potentials \cite{bardeen1980gauge}, $\delta\Phi_{f}$ is called the Mukhanov-Sasaki variable and $\delta\Phi_{N}$ is a Bardeen type variable built from a scalar field and the metric/tetrad \gls{dof}. Finally, it should be noted that in Ref.~\cite{Hohmann:2020vcv} there is an extensive and quite general perturbative framework for \gls{tg} that one could follow. This framework was also applied to \gls{tegr} using gauge invariant variables.

\subsection{Tensor perturbations} \label{sec:ten_pert}

Starting with the most important and simple perturbations (the tensor ones), one finds that the most general case of \gls{gw} propagation equation for this sector yields~\cite{Saltas:2014dha}
\begin{equation}
    \ddot{h}_{ij}+(3+\alpha_{M})H\dot{h}_{ij}+\biggl((1+\alpha_{T})\frac{k^{2}}{a^{2}}+m^{2}\biggr)h_{ij}=\Pi_{ij}^{T}\,,\label{eq:GWPE_general}
\end{equation}
where $\alpha_{T}:=c_{T}^{2}-1$ is the \textit{tensor speed excess} with $c_{T}:=c_{\rm g}/c$ (here revert to SI units for clarity's sake). This quantity describes the modification in the \gls{gw} propagation speed. In addition, the \textit{Planck-mass run rate} $\alpha_{M}$ enters as a friction term and it is related to the \textit{cosmological strength of gravity} $M_{*}^{2}$ (the kinetic term of tensor perturbations) by $\alpha_{M}:=\frac{\dd\log(M_{*}^{2})}{\dd\log a}$, $m$ denotes the effective graviton mass and $\Pi_{ij}^{T}$ is the tensor part of the anisotropic stress \eqref{eq:PiAST}. For the case of \gls{gr}, Eq.~\eqref{eq:GWPE_general} is reproduced by setting $\alpha_{M}=\alpha_{T}=m=0$. For the rest of the analysis, only cases where $\Pi_{ij}^{T}=0$ are considered. The detection of the \gls{gw} event GW170817 and the $\gamma$-ray burst GRB 170817A place a strong constraint on the speed at which \gls{gw} propagate, leading to the following constrain \cite{Monitor:2017mdv}
\begin{equation}
    -3\times10^{-15}<\Big|\frac{c_{\rm g}}{c}-1\Big|<7\times10^{-16}\,.\label{GW_constraint}
\end{equation}

In addition, observations for the mass of the graviton set an upper bound of $m_{\rm g}<1.2\times10^{-22}$ eV/$c^{2}$ from Ref.~\cite{Abbott:2016blz}. There is also a stronger constraint coming directly from Solar System tests from which we have the stronger upper bound $m_{\rm g}<10^{-23}$ eV/$c^{2}$ \cite{Will:2018gku,Will:2018mcj}. Hence suggesting that our current understanding of the graviton regarding its mass was not really changed from the GW170817 event \cite{deRham:2016nuf}.

In what follows we will present the results of two major classes of \gls{tg} theories, namely $f(T,B)$ (see Sec.~\ref{sec:f_T_B_gravity}) and BDLS (Teleparallel analogue of Horndeski) gravity (see Sec.~\ref{sec:BDLS}) since these are the only results currently reported in the literature. In order to consider the tensor perturbations we only need the tensor part of the perturbed tetrad in Eq.~\eqref{eq:tetrad perturbation-1} which is
\begin{equation}
    \delta e^{I}{}_{j}=\frac{a}{2}\delta^{Ii}h_{ij}\,.\label{ten_pert}
\end{equation}
Starting with the $f(T,B)$ case, by inserting the above perturbation into the second order action in Fourier space \eqref{eq:fourier_trans_con}, one can derive directly the \gls{gw} propagation equation which is just Eq.~\eqref{eq:GWPE_general} by setting $\alpha_{T}=0$ and $m^{2}=0$. From this equation, the stability condition $f_{T}<0$ is implied\footnote{The stability condition $f_{T}<0$ comes from the fact that the Eq.~\eqref{eq:GWPE_general} with $\alpha_{T}=0$ and $m^{2}=0$ can also be generated from an action analogous to the form $\int\frac{\dd^{3}k}{\left(2\pi\right)^{3/2}}\dd t(-a^{3}f_{T})\left[\dot{h}_{ij}\dot{h}^{ij}+\frac{k^{2}}{a^{2}}h_{ij}h^{ij}\right]$. This type of action in order to be healthy $(-a^{3}f_{T})>0\Rightarrow f_{T}<0$ needs to be imposed.}. The Planck mass run rate, representing a frictional term \cite{Ezquiaga:2017ekz,Saltas:2014dha,Riazuelo:2000fc} turns out to be in this case
\begin{equation}
    \alpha_{M}=\frac{1}{H}\frac{\dot{f}_{T}}{f_{T}}\,.\label{ten_pert_stab}
\end{equation}
It should be noted here that $\dot{f}_{T}=f_{TT}\dot{T}+f_{TB}\dot{B}$. AS in \gls{gr}, the propagation of the tensor waves in $f(T,B)$ gravity is also the speed of light \cite{Copeland:2018yuh}
\begin{equation}
    c_{T}^{2}=1\,,
\end{equation}
and it is in agreement with \cite{Monitor:2017mdv}.

In this context, $f(T,B)$ gravity is not that strongly constrained by present observations related to Eq.~\eqref{GW_constraint}. However a full stability analysis of the perturbations needs to be performed, i.e., also including the scalar sector.

Let us point out that exactly the same results hold true for the class of $f(T)$ theories and hence one can deduce that the boundary term only influences the functional form of the results only through the time derivative $\dot{f}_{T}={f}_{TT}\dot{T}+{f}_{TB}\dot{B}$. Another very famous theory that enjoys the same functional picture in the sense of $f(T)$, is $f(\lc{R})$ gravity \cite{DeFelice:2010aj} since the \gls{gw}s propagation at the speed of light and the the Planck mass run rate is given as $\alpha_{M}=\dot{f}_{R}/(Hf_{R})$. These similarities were to be expected to some degree since both $f(T)$ and $f(\lc{R})$ are subclasses of $f(T,B)$.

The next major class of theories we consider is the BDLS theory, which encapsulates all possible theories constructed with up to second order derivatives of the tetrad and a scalar field (see Sec.~\ref{sec:BDLS}). Following the same procedure, one can find the gravitational propagation equation for the tensor perturbations as Eq.~\eqref{eq:GWPE_general} by setting $m^{2}=0$, then
\begin{equation}
    \alpha_{T}=\frac{2X}{M_{\ast}^{2}}\left(2G_{4,X}-2G_{5,\phi}-G_{5,X}(\ddot{\phi}-\dot{\phi}H)-2G_{{\rm Tele,J_{8}}}-\frac{1}{2}G_{{\rm Tele,J_{5}}}\right)\,,\label{alpha_T}
\end{equation}
and the effective Planck mass is given by
\begin{align}
M_{\ast}^{2} & =2\Big(G_{4}-2XG_{4,X}+XG_{5,\varphi}-\dot{\phi}XHG_{5,X}+2XG_{{\rm Tele,J_{8}}}+\frac{1}{2}XG_{{\rm Tele,J_{5}}}-G_{{\rm Tele,T}}\Big)\,.\label{eff_Planck_mass}
\end{align}
Let us stress that at this point the only nonvanishing contributing background scalars to the $G_{\text{Tele}}$ term are $T=-6H^{2}$, $T_{\text{vec}}=-9H^{2}$, and $I_{2}=3H\dot{\phi}$, while all the other scalars vanish up to first order.

The fact that the $G_{\text{Tele}}$ terms enter Eqs.~\eqref{alpha_T} and \eqref{eff_Planck_mass} means that in contrast to the standard Horndeski theory there is the potential of revised speed of waves instead of trivialization of most interesting classes of models. Hence one can still find a wide variety of modes that respect the light of speed propagation constraint by solving $\alpha_{T}=0$. For the case of $G_{\text{Tele}}=0$ we directly recover the standard results for the \gls{gw} propagation equation \cite{Kobayashi:2011nu}.

Both of the above major classes correctly reproduce the most trivial result of \gls{tegr} and $f(T)$ theories which corresponds to $\alpha_{M}=\alpha_{T}=0$, i.e,
\begin{equation}
    \ddot{h}_{ij}+3H\dot{h}_{ij}+\frac{k^{2}}{a^{2}}h_{ij}=0\,,\label{TEGR_GWPE}
\end{equation}
which is also identical to the result of \gls{gr}. For other theories like the ones containing the teleparallel Gauss-Bonnet terms (see Sec.~\ref{Sec:GB_theories}), it is expected that $c_{T}\neq1$ as it was found for similar theories as in the case of the modified Gauss-Bonnet one~\cite{DeFelice:2010sh}. This study has not been analyzed yet. Recently, two \gls{tg} parity-violating models have found that some of those theories can produce a gravitational wave birefringence and dispersion meaning that the dispersion relation could depend on the wave number and their polarization~\cite{Hohmann:2022wrk,Wu:2021ndf}.

\subsection{Vector perturbations} \label{sec:vector_pert}

In order to study the vector perturbations at linear order, the tetrad perturbation of Eq.~\eqref{eq:tetrad perturbation-1} reduces to
\begin{align}
    \left[\delta e^{A}{}_{\mu}\right] & =\left[\begin{array}{cc}
    0 & a\beta_{i}\\
    \delta^{I}{}_{i}b^{i} & a\delta^{Ii}\epsilon_{ijk}\sigma^{k}
\end{array}\right]\,.\label{=00003D0003B4evector}
\end{align}
In the literature, so far, there have not been reported explicitly any theories with dynamical vector perturbations in flat \gls{flrw}. It is instructive to go through the result \cite{Bahamonde:2020lsm} of $f(T,B)$ gravity which also includes \gls{gr}, $f(T)$ and also $f(\lc R)$ as sub-cases. For that calculation the gauge can been fixed by setting $h_{i}\equiv0$ (see Sec.~\ref{sec:gauge_trans} for details). Perturbing the field equations which we now denote as $W_{\mu\nu}$ and for $\dot{f}_{B}+\dot{f}_{T}\neq0$ (which is a theory different to $f(\lc{R})$) we get that $\beta_{i}=0$ and $\sigma_{i}=0$. Hence one eventually arrives at
\begin{alignat}{2}
    W_{ij}(i\neq j):\quad & 0=\: & \dot{b}_{j}+b_{j}\left(2H+\frac{\dot{f}_{T}}{f_{T}}\right)\,.\label{Wijvector}
\end{alignat}
The above equation is just a constraint equation since only first order derivatives on time appear and hence the vector perturbations are not propagating. One can also read off the same stability condition $f_{T}<0$ as in the tensor case. It so happens that $f(T)$ gravity~\cite{Golovnev:2018wbh} is again portrayed by the same functional picture as $f(T,B)$.

In the case of $\dot{f}_{B}+\dot{f}_{T}\equiv0$ which means $f_{T}=-f_{B}=-f_{R}$ or in other words, $f(T,B)\rightarrow f(\lc{R})$, all antisymmetric field equations are trivialized ($W_{[\mu\nu]}\equiv0$). By further introducing $Y_{i}:=b_{i}-\beta_{i}$, we end up with
\begin{alignat}{2}
    W_{ij}(i\neq j):\quad & 0=\: & \dot{Y}_{j}+Y_{j}\left(2H+\frac{\dot{f}_{R}}{f_{R}}\right)\,,\label{WijvectorfR}
\end{alignat}
which is a well known result for $f(\lc{R})$ theories \cite{DeFelice:2010aj}. Notice again that this result has the exactly the same functional form as the one for $f(T,B)$ above.

Lastly, for the simplest case of \gls{tegr} one recovers the constraint equation
\begin{alignat}{2}
    W_{ij}(i\neq j):\quad & 0=\: & \dot{b}_{j}+2Hb_{j}\,,\label{WijvectorTEGR}
\end{alignat}
and thus there are no propagating vectorial \gls{dof} which agrees with the literature.

\subsection{Scalar perturbations}

\label{sec:perturbationsfTB}

Focusing on the scalar sector which is the most involved, the analysis
will be confined to the scalar part of Eq.~\eqref{eq:tetrad perturbation-1}
that reads
\begin{align}
\left[\delta e^{A}{}_{\mu}\right] & =\left[\begin{array}{cc}
\varphi & a\partial_{i}\beta\\
\delta^{I}{}_{i}\partial^{i}b & a\delta^{Ii}\left(-\psi\delta_{ij}+\partial_{i}\partial_{j}h+\epsilon_{ijk}\partial^{k}\sigma\right)
\end{array}\right]\,,\label{eq:scalar_pert_tetrad}
\end{align}
in which the Newtonian gauge ($b=\beta$ and $h=0$) will be adopted. In the following, the field equations for $f(T,B)$ and $f(T,\Theta)$ gravity theories are reported since they are quite general including a wide range of theories. One can find more details about the derivations in the Supplementary annexes (Supplementary 2).

\subsubsection{\texorpdfstring{$f(T,B)$}{B} gravity\label{ssec:f_T_B_scalar_pert}}

The linearized field equations of the scalar perturbations for $f(T,B)$
are given by
\begin{subequations}
\begin{alignat}{2}
W_{00}:\; \kappa^{2}\delta\rho & =\: & & 3H\delta\dot{f}_{B}+\Big(\frac{k^{2}}{a^{2}}+\frac{B}{2}\Big)\delta f_{B}-6H^{2}\delta f_{T}-\frac{1}{2}f_{T}\delta T-\frac{2Hk^{2}f_{T}}{a}b+\dot{\psi}(12Hf_{T}-3\dot{f}_{B})\nonumber \\[0.5ex]
& \: & & + \frac{2k^{2}f_{T}}{a^{2}}\psi+6H\varphi(2Hf_{T}-\dot{f}_{B})\,,\label{eq:f_T_B_scalar_pert00}\\[0.5ex]
W_{ij}(i\neq j):\; \psi-\varphi & =\: & &  \frac{1}{f_{T}}(a(\dot{f_{T}}+\dot{f}_{B})b-\delta f_{B})\,,\label{Wij}\\[0.5ex]
W_{i}^{i}:\; -\kappa^{2}\delta p & =\: & & \delta\ddot{f}_{B}+\delta f_{B}\left(\frac{2k^{2}}{3a^{2}}+\frac{B}{2}\right)-2(3H^{2}+\dot{H})\delta f_{T} -\frac{2k^{2}}{3a}(\dot{f}_{B}+3Hf_{T}+\dot{f}_{T})b\nonumber \\[0.5ex]
 & \: & & -2H\delta\dot{f}_{T}-\frac{1}{2}f_{T}\delta T+\frac{2k^{2}f_{T}}{3a^{2}}\psi+2f_{T}\ddot{\psi}+2\dot{\psi}(6Hf_{T}+\dot{f}_{T})+\dot{\varphi}(2Hf_{T}-\dot{f}_{B})\nonumber \\[0.5ex]
& \: & & +\varphi\left(4f_{T}\left(-\frac{2k^{2}f_{T}}{3a^{2}}+3H^{2}+\dot{H}-2\ddot{f}_{B}\right)+4H\dot{f}_{T}\right)\,,\label{Wii}\\[0.5ex]
W_{0i}:\; \kappa^{2}av(p+\rho) & =\: & & \delta\dot{f}_{B}-3H\delta f_{B}+2f_{T}\dot{\psi}-2H\delta f_{T}+(2f_{T}H-\dot{f}_{B})\varphi\,,\\[0.5ex]
W_{i0}:\; \kappa^{2}av(p+\rho) & =\: & & \delta\dot{f}_{B}-H\delta f_{B}+2f_{T}\dot{\psi}+2(\dot{f}_{T}+\dot{f}_{B})\psi+(2f_{T}H-\dot{f}_{B})\varphi\,,
\end{alignat}
\end{subequations}
where $\delta f_{T}=f_{TT}\delta T+f_{TB}\delta B$ and $\delta f_{B}=f_{BT}\delta T+f_{BB}\delta B$,
while the antisymmetric part of the field equations reads as
\begin{equation}
    W_{i0}-W_{0i}:\; 0 = H(\delta f_{T}+\delta f_{B})+\psi(\dot{f_{T}}+\dot{f}_{B})\,,\label{anti}
\end{equation}
and the energy-momentum conservation in the case of dust (for the
general case, see the Supplementary annexes (Supplementary 2) is given by
\begin{subequations}
\begin{alignat}{2}
\lc{\nabla}_{\mu}\Theta_{0}{}^{\mu}:\; \delta\dot{\rho}+3H\delta\rho & =\: & & \frac{\rho}{a}k^{2}v+3\dot{\psi}\rho\,,\\[0.5ex]
\lc{\nabla}_{\mu}\Theta_{i}{}^{\mu}:\; a\dot{v}+aHv & =\: & & -\varphi\,.\label{W=00005B0i=00005D}
\end{alignat}
\end{subequations}
Notice that $\sigma$ completely drops out from the field equations, just like in $f(T)$ gravity. On top of this, also the antisymetric part of the field equations is not trivialized \eqref{anti}, just like in $f(T)$ \cite{Zheng:2010am,Izumi:2012qj,Golovnev:2018wbh}. In the limit of $f(T,B)\rightarrow f(-T+B)=f(\lc{R})$ where ($f_{T}\rightarrow-f_{R},\, f_{B}\rightarrow f_{R}$), one can recover after a few trivial substitutions the usual equations reported in Ref.~\cite{DeFelice:2010aj}. In addition, in this limit the antisymmetric part of the field equations in Eq.~\eqref{anti} is trivialized and the scalar field $b$ completely drops off from the field equations\footnote{After expanding $\delta T,\,\delta B\,,\delta f_{T}$ and $\delta f_{B}$ as prescribed in Supplementary annexes (Supplementary 2).}. This is to be expected since for $f(\lc R)$ there are no antisymmetric field equations and there should be only 4 scalar fields in the equations \cite{DeFelice:2010aj}.

\subsubsection{\texorpdfstring{$f(T,\Theta)$}{G} gravity \label{ssec:f_T_Theta}}

For the case of $f(T,\Theta)$ gravity, as also introduced in Sec.~\ref{sec:modifiedmatter}, we use the model ansatz $f(T,\Theta)=-T-F(T,\Theta)$ for some arbitrary function $F(T,\Theta)$ which is formulated as \gls{tegr} plus a modification. The analysis follows from Ref.~\cite{Farrugia:2016pjh} where the field equations are first presented in the flat \gls{flrw} background (see Eqs.~\eqref{eq:frwtrace1}-\eqref{eq:frwtrace2})
\begin{subequations}
\begin{alignat}{2}
\frac{\kappa^{2}}{2}\rho & =\: & & \left(1+F_{T}\right)3H^{2}+\dfrac{T+F}{4}+\dfrac{F_{\Theta}}{2}\left(\rho+p\right)\,,\label{eq:fTT-Friedmann-tt}\\[0.5ex]
-\frac{\kappa^{2}-F_{\Theta}}{2}(p+\rho) & =\: & & \left(1+F_{T}\right)\dot{H}-12H^{2}\dot{H}F_{TT}+H\left(\dot{\rho}-3\dot{p}\right)F_{T\Theta}\,,\label{eq:fTT-Friedmann-trace}
\end{alignat}
\end{subequations}
 where we have used $\Theta=\rho-3p$ and $\dot{f}_{T}=\dot{F}_{T}=F_{TT}\dot{T}+F_{T\Theta}\dot{\Theta}=-12H\dot{H}F_{TT}+(\dot{\rho}-3\dot{p})F_{T\Theta}$.
Combining these equations (or using Eq.~\eqref{eq:continuity_fTT}) yields the modified continuity equation
\begin{align}
\left(\kappa^{2}-F_{\Theta}\right)\left[\dot{\rho}+3H\left(\rho+p\right)\right] & =\dfrac{F_{\Theta}}{2}\big(\dot{\rho}-\dot{p}\big)+(\rho+p)\left[-12H\dot{H}F_{T\Theta}+\left(\dot{\rho}-3\dot{p}\right)F_{\Theta\Theta}\right]\,.\label{eq:fTT-continuity}
\end{align}
Special care is needed in the presence of the trace of the energy momentum-tensor since it modifies the continuity equation, which leads to the non-conservation of the stress-energy tensor. Nevertheless we impose the conservation as a constraint in order to limit the possible functional form. This constraints reads as
\begin{align}
    0 & =F_{\Theta}\left(\dot{p}-\dot{\rho}\right)+6(\rho+p)\left[4H\dot{H}F_{T\Theta}+\left(\dot{p}-\frac{\dot{\rho}}{3}\right)F_{\Theta\Theta}\right]\,.
\end{align}
On the other hand, the perturbed field equations of Eq.~\eqref{f(T)field_equations_tetrad}, in the Newtonian gauge, read as.
\begin{subequations}
\begin{alignat}{2}
W_{00}:\; \frac{\kappa^2}{2}\delta\rho & =\: & & \left(1+F_{T}\right)\left[a^{-2}\partial^{2}\psi-3H\dot{\psi}-3H^{2}\varphi\right]+3H^{2}\bigg[F_{TT}\delta T+F_{T\Theta}\delta\Theta\bigg]\nonumber \\[0.5ex]
 & \: & & +\frac{F_{\Theta}}{4}\left(3\delta\rho-\delta p-\partial^{2}\Pi^{\text{S}}\right)+\frac{\rho+p}{2}\bigg[F_{T\Theta}\delta T+F_{\Theta\Theta}\delta\Theta\bigg]\,,\label{eq:00-first}\\[0.5ex]
W_{ij}(i\neq j):\; \left(\kappa^2-F_{\Theta}\right)\partial_{j}\partial^{i}\Pi^{\text{S}}& =\: & & a^{-1}\partial_{j}\partial^{i}\beta\left[-12H\dot{H}F_{TT}+\left(\dot{\rho}-3\dot{p}\right)F_{T\Theta}\right]\nonumber\\[0.5ex]
& \: & &+a^{-2}\left(1+f_{T}\right)\partial_{j}\partial^{i}\left(\psi-\varphi\right)\,,\label{eq:ij-first}\\[0.5ex]
W_{i}^{i}:\; \frac{\kappa^2}{2}\left(\delta p+\dfrac{\partial^{2}\Pi^{\text{S}}}{3}\right)& =\: & & \left(1+F_{T}\right)\left[H\dot{\phi}+3H^{2}\phi+3H\dot{\psi}+2\dot{H}\varphi+\ddot{\psi}-\frac{1}{3}a^{-2}\partial^{2}\left(\psi-\varphi\right)\right]\nonumber \\[0.5ex]
 & \: & & -F_{TT}\Big(3H^{2}\delta T+2\dot{H}\delta T+H\delta\dot{T}+12H^{2}\dot{H}\varphi\Big)\nonumber \\[0.5ex]
 & \: & & +F_{T\Theta}\bigg[-\left(3H^{2}+\dot{H}\right)\delta\Theta-H\delta\dot{\Theta}+\left(\dot{\rho}-3\dot{p}\right)\left(\frac{\delta T}{12H}+H\phi\right)\bigg]\nonumber \\[0.5ex]
& \: & & +12H^{2}\dot{H}\left(F_{TTT}\delta T+F_{TT\Theta}\delta\Theta\right)\nonumber\\[0.5ex]
& \: & &-H\left(\dot{\rho}-3\dot{p}\right)\left[F_{TT\Theta}\delta T+F_{T\Theta\Theta}\delta\Theta\right]-\dfrac{F_{\Theta}}{4}\bigg(\delta\Theta-\frac{2}{3}\partial^{2}\pi^{\text{S}}\bigg)\,,\label{eq:trace-first}\\[0.5ex]
W_{0i} :\; \frac{a^{2}}{2}\left(\rho+p\right)\left(\kappa^2-F_{\Theta}\right)\partial_{i}v& =\: & & \left(1+F_{T}\right)\partial_{i}\left(\dot{\psi}+H\varphi\right)-H\left(F_{TT}\partial_{i}\delta T+F_{T\Theta}\partial_{i}\delta\Theta\right)\,,\label{eq:0i-first}\\[0.5ex]
W_{i0}:\; -\frac{a^{2}}{2}\left(\rho+p\right)\left(\kappa^2-F_{\Theta}\right)\partial_{i}v& =\: & & -\left(1+F_{T}\right)\partial_{i}\left(\dot{\psi}+H\varphi\right)-\partial_{i}\psi\left[-12H\dot{H}F_{TT}+\left(\dot{\rho}-3\dot{p}\right)F_{T\Theta}\right]\,,\label{eq:i0-first}
\end{alignat}
\end{subequations}
and the antisymmetric part
\begin{equation}
    W_{i0}+W_{0i}:\; H\left(F_{TT}\partial_{i}\delta T+F_{T\Theta}\partial_{i}\delta\Theta\right) = -\partial_{i}\psi\left[-12H\dot{H}F_{TT}+\left(\dot{\rho}-3\dot{p}\right)F_{T\Theta}\right]\,,\label{eq:anti_f_T_theta}
\end{equation}
where $\udt{W}{A}{\rho}$ denotes the linearized field equations, $\delta T$ can be found in the Supplementary annexes (Supplementary 2) and $\delta\Theta=\delta\rho-3\left(p+\delta p\right)-\partial^{2}\Pi^{\text{S}}$. These equations properly reproduce the results reported in Refs.~\cite{Zheng:2010am,Harko:2014aja}. Furthermore, the scalar perturbation $\sigma$ does not appear in the field equations, leaving its effect to be negligible \cite{Zheng:2010am,Golovnev:2018wbh}.

In addition to the field equations, we also need to perturb the conservation laws for the stress-energy tensor and the velocity perturbation. Thus, the perturbed continuity equation is
\begin{alignat}{2}
    \lc{\nabla}_{\mu}\Theta_{0}{}^{\mu}: & \: & & \left(\kappa^2-F_{\Theta}\right)\left[\delta\dot{\rho}+3H\left(\delta\rho+\delta p+\frac{\partial^{2}\pi^{\text{S}}}{3}\right)-3\left(\rho+p\right)\dot{\psi}+\left(\rho+p\right)\partial^{2}v\right]=\nonumber \\[0.5ex]
    & \: & & \frac{F_{\Theta}}{2}\left(\delta\dot{\rho}-\delta\dot{p}-\partial^{2}\dot{\pi^{\text{S}}}\right)+F_{T\Theta}\left[-2a^{-2}\partial^{2}\psi\left(\dot{\rho}-3\dot{p}\right)+\left(\rho+p\right)\left(3H\delta T+\delta\dot{T}\right)\right.\nonumber \\[0.5ex]
    & \: & & \left.-12H\dot{H}\left(\delta\rho+\delta p\right)+6H\left(H\varphi+\dot{\psi}\right)\left(\dot{\rho}-3\dot{p}\right)+\left(\dot{\rho}+\dot{p}\right)\delta T\right]\nonumber \\[0.5ex]
    & \: & & +F_{\Theta\Theta}\bigg[\left(\rho+p\right)\left(3H\delta\Theta+\dot{\Theta}\right)+\dfrac{1}{2}\big(3\delta\rho-\delta p-\partial^{2}\pi^{\text{S}}\big)\left(\dot{\rho}-3\dot{p}\right)+\left(\dot{\rho}+\dot{p}\right)\delta\Theta\bigg]\nonumber \\[0.5ex]
    & \: & & -12H\dot{H}\left(\rho+p\right)\left(F_{TT\Theta}\delta T+F_{T\Theta\Theta}\delta\Theta\right)+\left(\rho+p\right)\left(\dot{\rho}-3\dot{p}\right)\left(F_{T\Theta\Theta}\delta T+F_{\Theta\Theta\Theta}\delta\Theta\right)\,.\label{eq:continuity-first}
\end{alignat}
and the linearized conservation of momentum equation equation
\begin{alignat}{2}
\lc{\nabla}_{\mu}\Theta_{i}{}^{\mu}:-\dfrac{1}{2}F_{\Theta}\left[a^{2}\left(\dot{\rho}-\dot{p}\right)\partial_{i}v+\partial_{i}\left(\delta\rho-\delta p-\partial^{2}\pi^{\text{S}}\right)\right] & =\: & & \left(\kappa^2-F_{\Theta}\right)\Big\lbrace\left(\rho+p\right)\left[a^{2}\partial_{i}\dot{v}+2a^{2}H\partial_{i}v+\partial_{i}\varphi\right]\nonumber \\[0.5ex]
 & \: & & +a^{2}\dot{p}\partial_{i}v+\partial_{i}\delta p+\partial_{i}\partial^{2}\pi^{\text{S}}\Big\rbrace\,.\label{eq:velocity}
\end{alignat}

The scalar field $\sigma$ completely drops out from the field equations and in addition a non-trivial antisymmetric part of the field equations is obtained \eqref{eq:anti_f_T_theta}. This is also case in $f(T,B)$ gravity \eqref{ssec:f_T_B_scalar_pert}. In this instance the scalar perturbations are coupled with the perturbations of the energy-momentum components and so this is not enough information to determine the impact of these cosmological perturbations on observational parameters.

\subsection{Growth index and the \texorpdfstring{$f\sigma_{8}$}{} parameter} \label{sssec:Growth-index}

A viable way to carefully explore the phenomenology of proposed models is to compare with observations. To achieve this goal, we need to find tools that can help efficiently in discriminating between different classes of models. The growth index $\gamma$ (see Eq.~\ref{eq:growth-gamma}), which describes a way of parameterizing the growth of density perturbations in a nonrelativistic matter component is a good example of such a tool. In recent years, the parameter $\gamma$ has been used to discriminate spatially open from spatially flat universes. As an extension, $\gamma$ has been used in the context of \gls{de} in order to study deviations from standard \gls{gr} models. The reason why this parameter is important is due the fact that it has a clear quasi-constant signature at very low redshifts. At this point, it is important to study (theoretically) in which models the growth index $\gamma$ can be \textit{exactly} constant, $\gamma\approx3(w-1)/(6w-5)$ which is \gls{gr} plus \gls{de} fluid or $\gamma\approx6/11$ in the case of $\Lambda$\gls{cdm}. Also some indirect values for modified gravity are reported in Ref.~\cite{Basilakos:2012uu}. Hence, the value of $\gamma$ offers a way to distinguish between modified gravity models.

Let us start by considering the evolution of linear scalar perturbations $\delta_{\text{m}}$ as
\begin{equation}
    \delta_{\text{m}}:=\frac{\delta\rho}{\rho}+3HV\,,\label{delta_m_def}
\end{equation}
in the matter component in the Universe with \gls{de} components (we neglect radiation at the matter and \gls{de} dominated stages). Inside the Hubble radius dynamics of $\delta_{\text{m}}$ is given by
\begin{equation}
    \ddot{\delta}_{\text{m}}+2H\dot{\delta}_{\text{m}}-4\pi G\rho\delta_{\text{m}}=0\,.\label{del}
\end{equation}
In the absence of spatial curvature, the evolution of the $H$ as a function of the redshift $z=\frac{a_{0}}{a}-1$ at $z\ll z_{{\rm eq}}$ reads
\begin{equation}
    h^{2}(z)=\Omega_{\text{m0}}~(1+z)^{3}+(1-\Omega_{\text{m0}})~\exp\left[3\int_{0}^{z}dz'~\frac{1+w_{{\rm DE}}(z')}{1+z'}\right]\,,
\end{equation}
with $h(z):=\frac{H}{H_{0}}$ and $w_{{\rm DE}}(z):=p_{{\rm DE}}(z)/\rho_{{\rm DE}}(z)$. This equation will hold for all \gls{flrw} models inside \gls{gr}. Now, if we consider that the relative density of matter component in terms of the critical one $\Omega_{\text{m}}=\Omega_{\text{m0}}\left(\frac{a^{3}}{a_{0}^{3}}h^{2}\right)^{-1}$, we can write the effective \gls{de} \gls{eos} as
\begin{equation}
    w_{{\rm DE}}=-\frac{1}{3(1-\Omega_{\text{m}})}\,\frac{\dd\ln\Omega_{\text{m}}}{\dd\ln(1+z)}\,.\label{wDE_general}
\end{equation}
In this representation, the effective Newton's parameter $G_{\text{eff}}$ can be parametrized as
\begin{equation}
    G_{\text{eff}}=GQ\,,\quad\text{where}\quad Q=\frac{2+4\Omega_{\text{m}}^{2}}{3+3\Omega_{\text{m}}^{2}}\,,
\end{equation}
which allows the matter perturbation equation~\eqref{del} to retain it form while absorbing modified terms into this effective parameter. While we can work with $\delta_{m}$, it is convenient to use the growth function $f:=\frac{\dd\ln\delta_{\text{m}}}{\dd\ln a}$ and the \gls{de} \gls{eos} \eqref{wDE_general} in order to arrive at the non-linear equation
\begin{equation}
    \frac{\dd f}{\dd\ln a}+f^{2}+\frac{1}{2}\left(1-\frac{\dd\ln\Omega_{\text{m}}}{\dd\ln a}\right)f=\frac{3}{2}~\Omega_{\text{m}}\,,
\end{equation}
where we can recover $\delta_{\text{m}}$ from $f$. Notice that $f=f_{0}$ (with $f_{0}$ constant) if $\delta_{\text{m}}\propto a^{f_{0}}$. In particular, $f\to1$ for the $\Lambda$\gls{cdm} case and for large $z$, while for $f=1$ we obtain the Einstein-de Sitter universe. As is standard, we can derive the growth of perturbations by parameterizing $f$ in terms of the \gls{de} related quantities as
\begin{equation}
    f=\Omega_{\text{m}}(z)^{\gamma(z)}\,.\label{eq:growth-gamma}
\end{equation}
We note that in this case it is important to study whether $\gamma$ can be exactly constant or not. In such cases, the the modified evolution of matter perturbations assumes the form of Eq.~\eqref{meszaroseq} which is
\begin{equation}
    \ddot{\delta}_{\text{m}}+2H\dot{\delta}_{\text{m}}-4\pi\rho G_{{\rm eff}}\delta_{\text{m}}=0\,,\label{eq:matterdens}
\end{equation}
where $G_{{\rm eff}}$, as the effective gravitational constant which depends on the choice of the theory. As an example, for effectively massless scalar-tensor models $G_{{\rm eff}}$ varies \gls{wrt} $t$.

The galaxies and quasars observed in spectroscopic observations are biased tracers of such structure formation, leading to degeneracy between the amplitude of matter fluctuations and biasing parameters. This issue could be a consequence of the complicated calculation of the growth function from the isotropic power spectrum derived from clustering of cosmic sources. However, spectroscopic observations contain extra information about the velocity field arising from gravitational collapse by separately measuring the power spectrum along and perpendicular to the line-of-sight. Measurements of the velocity field can help to differentiate between the effect of \gls{de} and modified gravity as the source of the accelerating universe through measurements of Redshift-Space Distortions (RSD) \cite{Colless:2003wz}. These distortions emerge due to the gravitational pull of matter over-densities which causes velocity deviations from the smooth Hubble flow. These peculiar velocities are imprinted in galaxy redshift surveys in which recessional velocity is used as the line-of-sight coordinate for galaxy positions. This in turn leads us to an apparent compression of radial clustering.

The resulting anisotropy in the clustering of galaxies is correlated with the speed at which structure grows; deviations from \gls{gr} which cause slower or faster growth result in smaller or larger anisotropic distortions in the observed RS clustering. The rate of change of the amplitude of the clustering is given by $f\sigma_{8}(a)=\dd\sigma_{8}(a)/\dd\ln a$, where $a=1/(1+z)$ and $f\equiv f(a)$ has been defined in Eq.~\eqref{eq:growth-gamma}. Because RSD measurements are sensitive to the product of the growth rate and the amplitude of matter fluctuations, a wide range in redshift coverage is important to constrain the evolution in clustering amplitude and directly probe \gls{gr}. On the other hand, if one were to assume a $\Lambda$\gls{cdm} model where \gls{gr} properly explains gravitational collapse, the growth rate can be predicted to high precision and RSD results can be used to constrain $\sigma_{8}(a)$, thereby provide insight into other fundamental physics such as neutrino masses.

\subsubsection{\texorpdfstring{$f(T)$ gravity}{G}}

In Ref.~\cite{Chen:2010va}, an analysis of the scalar cosmological perturbations in $f(T)$ gravity was presented. In particular, they chose to investigate a power-law model scenario where it is possible to find constraints on the growth matter factor in sub-horizon scales around $\mathcal{O}$(100 Mpc). This is illustrated by the field equation for the perturbation of the additional scalar field of the theory, $\delta\phi$ read as at first order perturbations when $f(T)$ is fixed to a constant. In Ref.~\cite{Dent:2011zz}, in comparison to Ref.~\cite{Chen:2010va}, a reconstruction of $f(T)$ was considered as a quintessence model, where it was noticed that strong deviation of the matter over density evolution at small scales and weak otherwise. In this direction, it was shown that at linear perturbative order it is possible to break the degeneracy between a family of $f(T)$ models. As an extension of the aforementioned work on $f(T)$, this quintessence analysis showed some deviations for large scales. According to Eq.~\eqref{eq:growth-gamma}, it is possible to analyze cases where the growth index $\gamma$ can be parameterized. In such cases, as in Ref.~\cite{Fu:2011zze}, a parameterization was given in the form $\gamma(z)=\gamma_{0}+\gamma_{1}\frac{z}{1+z}$ in order to study if a power law model can be constrained using observations.

Including a new scalar field, $\alpha_{{\rm m}}$ in the linear equations, as in Ref.~\cite{Geng:2012vn}, could contribute further to the anisotropic effects between the gravitational potentials $\psi$ and $\varphi$. This new scalar degree of freedom vanishes in the sub-horizon scales of $f(T)$ gravity and it is constrained with observations related to the flux power spectrum of the Lyman-alpha forest in quasar absorption spectra. According to this analysis it was found that there is a faster growth rate if the tracker solution is followed by the usual quintessence-like evolution. While proposing $f(T)$ power law models can help avoiding degeneracies at linear order, in Ref.~\cite{Basilakos:2016xob}, it was found that an asymptotic form of this particular $f(T)$ model can be given by $\gamma\approx\frac{6}{11-6b}$. This served a generalization to the described analyses in the regime where $\gamma$ can vary with redshift. This allows for an accurate determination of $b$ and also the ability of testing it in the range of validity of the observations. Furthermore, in Ref.~\cite{Gonzalez-Espinoza:2018gyl}, the evolution of the weighted growth rate, denoted by $f\sigma_{8}(a)$, was investigated for several different values of the parameter $n$ in the power law model. The results were based in observations derived from redshifts distortions reported by Planck 2015 Collaboration in a sample detailed in \cite{Kazantzidis:2018rnb}.

Using machine learning techniques, as will be further introduced in Sec.~\ref{subsection:ML_p} and detailed in Ref.~\cite{LeviSaid:2021yat} the role of the growth data was studied by reconstructing observations from RSD $f\sigma_{8}$ data together with the Hubble data that comes from cosmic chronometer and \gls{sn} type Ia data. The advantage of this reconstruction, is that it acquires more simulated data points in order to constrain modified gravity models with better precision. In addition, in Ref.~\cite{Nunes:2018xbm,Nunes:2018evm} a $f(T)$ power law model was constrained using the \gls{cmb} temperature power spectrum from early-time estimations that includes baryonic acoustic oscillations and also, local Hubble constant measurements. By including this analysis, it was possible to set a playground where the evolution of tensor modes in $f(T)$ gravity could influence the power spectrum and the B-modes polarization.

In Ref.~\cite{Nunes:2018xbm,Nunes:2018evm}, the power model was further constrained by considering \gls{gw} multi-messenger astronomy, especially on the effects of primordial \gls{gw} on the \gls{cmb} anisotropies and the BB spectrum. It was shown that only an amplitude modification could be compared with the standard $\Lambda$\gls{cdm} cosmology, i.e the more the parameter $n$ in the model deviates from \gls{gr}, the larger the \gls{gw} amplitude decay is compared to the canonical $\Lambda$\gls{cdm}. Additionally, one can constrain the growth factor by including detections from neutron stars mergers that include an EM counterpart.

\subsubsection{\texorpdfstring{$f(T,B)$ gravity}{G}\label{f_TB_matter_dens}}

Considering dust for the perfect fluid, and following Refs.~\cite{DeFelice:2010aj,Tsujikawa:2007gd}, the variable $V:=av$ is introduced and along with the gauge invariant variable dubbed density contrast $\delta_{\text{m}}$ as in Eq.~\eqref{delta_m_def}. In order to determine the time derivative of this parameter, we need to utilize the continuity equation to obtain the density parameter time derivative, which is
\begin{equation}
    \delta\dot{\rho}+3H\delta\rho=\frac{k^{2}\rho V}{a^{2}}+3\rho\dot{\psi}\,.\label{eq:Continuity EQ}
\end{equation}
The time derivative of the density contrast parameter can then be written as
\begin{subequations}
\begin{alignat}{2}
\dot{\delta}_{\text{m}} & =\: & & -\frac{\lc\nabla^{2}V}{a^{2}}+3\dot{\psi}+3\frac{d}{dt}(HV)\,,\label{eq:=00003D00003D0003B4mdot-1}\\[0.5ex]
\dot{V} & =\: & & -\varphi\,,\label{eq:Vdot-1}
\end{alignat}
\end{subequations}
where the time derivative of $V$ is also presented. By combining both derivatives, we obtain
\begin{equation}
    \ddot{\delta}_{\text{m}}+2H\dot{\delta}_{\text{m}}=\frac{\lc{\nabla}^{2}\varphi}{a^{2}}+3\ddot{\psi}+3\frac{d^{2}}{dt^{2}}(HV)+6H\dot{\psi}+6H\frac{d}{dt}(HV)\,.\label{matter_pert_equation}
\end{equation}

In the sub-horizon approximation $k\gg aH$, $k$ being well inside the Hubble radius, the dominant terms are $k$ and $\delta\rho$. Now that we have all the prerequisites we need to proceed, let us first summarize the dominant terms in this limit
\begin{equation}
    \left\{ \frac{k^{2}}{a^{2}}|\varphi|,\frac{k^{2}}{a^{2}}|\psi|,\frac{k^{2}}{a^{2}}|\beta|,\frac{k^{2}}{a^{2}}|\delta f_{T}|,\frac{k^{2}}{a^{2}}|\delta f_{B}|\right\} \gg\left\{ H^{2}|\varphi|,H^{2}|\psi|,H^{2}|\beta|,H^{2}|\delta f_{T}|,H^{2}|\delta f_{B}|\right\} \,,\label{spacederaprox}
\end{equation}
and
\begin{equation}
    \dot{|X}|\lesssim|HX| \quad {\textstyle {\rm where}}\quad X\in\left\{ \varphi,\psi,\beta,\delta f_{T},\delta f_{B},\dot{\varphi},\dot{\psi},\dot{\beta},\delta\dot{f_{T}},\delta\dot{f_{B}}\right\} \,.\label{timederaprox}
\end{equation}
Thus, it follows directly that in Fourier space \eqref{eq:fourier_trans_con} of the sub-horizon limit of Eq.~\eqref{matter_pert_equation} that it reduces to \eqref{eq:matterdens} as
\begin{equation}
    \ddot{\delta}_{\text{m}}+2H\dot{\delta}_{\text{m}}\simeq-\frac{k^{2}\varphi}{a^{2}}=4\pi\rho G_{{\rm eff}}\delta_{\text{m}}=\frac{\kappa^{2}}{2}\rho G_{{\rm eff}}\delta_{{\rm m}}\,,\label{meszaroseq}
\end{equation}
from which it follows that the only contributing scalar is $\varphi$. Along a similar vein, $\Sigma_{{\rm def}}$ is a parameter sensitive to weak lensing which appears when we write the lensing potential $-\left(\varphi+\psi\right)$ in terms of the matter density contrast $\delta_{\text{m}}$, so $\Sigma_{{\rm def}}$ plays a similar role to $G_{{\rm eff}}$ but between the lensing potential and $\delta_{\text{m}}$ specifically. This parameter is defined as
\begin{equation}
    \Sigma:=\frac{1}{2}\frac{G_{{\rm eff}}}{G}\left(1+\frac{\psi}{\varphi}\right)\,,\label{Sigmadef}
\end{equation}
which is calculated in along with $G_{{\rm eff}}$ in the sub-horizon approximation. Due to the nature of the $f(T,B)$ there is branching in various sectors by solving the field equations \eqref{eq:f_T_B_scalar_pert00}-\eqref{anti}. By introducing the useful notation
\begin{subequations}
\begin{align}
\Pi & :=f_{B}+f_{T}\,,\label{PI}\\[0.5ex]
\Upsilon & :=f_{BB}+2f_{TB}+f_{TT}=\Pi_{T}+\Pi_{B}\,,\label{YPSILON}\\[0.5ex]
\Xi & :=f_{TB}^{2}-f_{TT}f_{BB}=-\Pi_{T}\Pi_{B}+f_{TB}\Upsilon\,,\label{KSI}
\end{align}
\end{subequations}
induced branches are classified as:
\begin{description}
\item [{1.}] $\left\{ \Pi\neq\text{const},\Upsilon\neq0\right\} $ \\
 Which can further be classified using $\Xi=-\Pi_{T}\Pi_{B}+f_{TB}\Upsilon$
\begin{description}
\item [{1.a}] $\left\{ \Pi\neq\text{const},\Upsilon\neq0,\Xi\neq0\right\} $
most general case of $f(T,B)$
\item [{1.b}] $\left\{ \Pi\neq\text{const},\Upsilon\neq0,\Xi=0\right\} $
includes $f(T)$
\end{description}
\item [{2.}] $\left\{ \Pi\neq\text{const},\Upsilon\equiv0\right\} $ \\
 Which can further be classified using $\Upsilon\equiv0\Rightarrow\Pi_{B}\equiv-\Pi_{T}$
into Eq.~\eqref{KSI} as $\Xi=\Pi_{T}^{2}=\Pi_{B}^{2}$
\begin{description}
\item [{2.a}] $\left\{ \Pi\neq\text{const},\Upsilon=0,\Xi\neq0\right\} $
\item [{2.b}] $\left\{ \Pi={\rm const},\Upsilon=0,\Xi=0\right\} $ the
unique $f(\lc{R})$ case
\end{description}
\end{description}

\begin{table}[H]
\begin{centering}
\midsepremove
\resizebox{18cm}{!}{
\global\long\def\arraystretch{1.5}
\begin{tabular}{ccccc}
\toprule
\cellcolor{gris3}\textbf{Class} & \cellcolor{gris3}\textbf{Conditions} & \cellcolor{gris3} \textbf{Models} & \cellcolor{gris3}\boldmath{$G_{{\rm eff}}$} & \cellcolor{gris3}\boldmath{$\Sigma$}\tabularnewline
\midrule
\cellcolor{gris1}1 & \cellcolor{gris1}\boldmath{$\left\{ \Pi\neq0,\Upsilon\neq0\right\} $} & \cellcolor{gris1} & \cellcolor{gris1} & \cellcolor{gris1}\tabularnewline
\cellcolor{gris3}1a & \cellcolor{gris3}$\Xi\neq0$ & \cellcolor{gris3}General $f(T,B)$ & \cellcolor{gris3}$-G\frac{4\Upsilon}{36H^{2}(f_{BB}f_{TT}+2\Xi)+3\Upsilon f_{T}}$ & \cellcolor{gris3}$-\frac{\Upsilon}{\Upsilon f_{T}+12H^{2}\left(f_{BB}f_{TT}+2\Xi\right)}$\tabularnewline
\cellcolor{gris1}1b & \cellcolor{gris1}$\Xi=0$ & \cellcolor{gris1}Includes $f(T)$ & \cellcolor{gris1}$G\frac{A_{2}}{A_{6}}$ & \cellcolor{gris1}$\frac{\Delta_{4}}{\Delta_{10}}=-\frac{A_{2}}{A_{6}}$\tabularnewline
\midrule
\cellcolor{gris3}2 & \cellcolor{gris3}\boldmath{$\left\{ \Pi\neq0,\Upsilon\equiv0\right\} $} & \cellcolor{gris3} & \cellcolor{gris3} & \cellcolor{gris3}\tabularnewline
\cellcolor{gris1}2a & \cellcolor{gris1}$\Xi\neq0$ & \cellcolor{gris1}Less general $f(T,B)$ & \cellcolor{gris1}$-G\frac{4}{3(f_{T}+12H^{2}f_{TB})}$ & \cellcolor{gris1}$-\frac{1}{f_{T}+12H^{2}f_{TB}}$\tabularnewline
\cellcolor{gris3}2b & \cellcolor{gris3}$\Xi=0$ & \cellcolor{gris3}Only $f(\lc{R})$ & \cellcolor{gris3}$G\left(\frac{4}{3f_{R}}+\frac{1}{3(-f_{R}+3\frac{k^{2}}{a^{2}}f_{RR})}\right)$ & \cellcolor{gris3}$\Sigma\sim\frac{1}{f_{R}}$\tabularnewline
\bottomrule
\end{tabular}}
\par\end{centering}
\midsepdefault
\caption{Summarizing the cases of all the subclasses of $f(T,B)$ in the sub-horizon
approximation (recalling that definitions for $A_2,\,A_6\,,\Delta_4\,,\Delta_{10}$ are in Appendix~\ref{sec:Geff-appendix}). For more details see \cite{Bahamonde:2020lsm}}
\label{tab:f_T_B_branches}
\end{table}

These branches may also be indicators of variable \gls{dof}, since we know for sure that $f(\lc{R})$ has 3 \gls{dof} and $f(T)$ has either 3 or 5 \gls{dof} \cite{Blagojevic:2020dyq,Li:2011rn,Ong:2013qja,Ferraro:2018tpu} as also discussed in Sec.~\ref{sec:f(T)gravity}. The classification induced by these branches is illustrated in Table~\ref{tab:f_T_B_branches} and the relevant coefficients are included in the Appendix~\ref{sec:Geff-appendix}.

\subsubsection{\texorpdfstring{$f(T,\Theta)$ gravity}{G}}

Investigating the growth formation for epochs well within the matter domination periods, we set the pressure components $p=\delta p=0$ along with the absence of anisotropic stress. In addition, models that obey the conservation of the stress-energy tensor will be assumed. Even when the models do not obey the conservation law, the result reported in Ref.~\cite{Farrugia:2016pjh} is still obtained. Also, the conclusions regarding the resulting growth evolution remain the same.

Following the procedure of Sec.~\ref{f_TB_matter_dens}, and combining Eqs.~\eqref{eq:i0-first} and \eqref{eq:0i-first}, the relation
\begin{equation}
    F_{TT}\left(12H\dot{\psi}+12H^{2}\phi-12\dot{H}\psi-4a^{-2}k^{2}\xi\right)=-F_{T\Theta}\left(\delta\rho-3\rho\psi\right)\,,\label{eq:fTT-0i-and-i0-combined}
\end{equation}
is obtained where we define $\xi\coloneqq aH\beta$. Here, we have eliminated $\varphi$ using Eq.~\eqref{eq:ij-first}, where we have obtained
\begin{equation}
    \varphi=\psi-\frac{3\xi}{1+F_{T}}\left(4\dot{H}F_{TT}+\rho F_{T\Theta}\right)\,,\label{eq:fTT-phi-psi-relation}
\end{equation}
which, then combined with Eq.~\eqref{eq:fTT-0i-and-i0-combined} results in
\begin{equation}
    4\xi F_{TT}=\frac{\rho F_{T\Theta}\left(\delta_{\text{m}}-3Ha^{2}v-3\psi\right)+F_{TT}\left[12H\dot{\psi}+12\psi\left(H^{2}-\dot{H}\right)\right]}{\frac{k^{2}}{a^{2}}+\frac{9H^{2}}{1+F_{T}}\left(4\dot{H}F_{TT}+\rho F_{T\Theta}\right)}\,,
\end{equation}
which in the in the sub-horizon approximation assumes the form
\begin{equation}
    4\xi F_{TT}\approx\frac{a^{2}}{k^{2}}\left\lbrace \rho F_{T\Theta}\left(\delta_{\text{m}}-3Ha^{2}v-3\psi\right)+F_{TT}\left[12H\dot{\psi}+12\psi\left(H^{2}-\dot{H}\right)\right]\right\rbrace \,.
\end{equation}
Approximating by leading order, we obtain
\begin{equation}
    \xi\sim\frac{a^{2}H^{2}}{k^{2}}\left(\delta_{\text{m}}-a^{2}Hv+\psi\right)\,,\label{eq:fTT-xi-subhorizon}
\end{equation}
which can be further reduced using equation Eq.~\eqref{eq:velocity} as
\begin{equation}
    \left(\dot{v}+2Hv+\dfrac{\varphi}{a^{2}}\right)(\kappa^2-F_{\Theta})=-\frac{F_{\Theta}}{2a^{2}}\delta_{\text{m}}\,.\label{eq:fTT-euler-dustcase}
\end{equation}
and from Eq.~\eqref{eq:fTT-phi-psi-relation}
\begin{equation}
    \xi\sim\frac{a^{2}H^{2}}{k^{2}}\left(\varphi + \psi\right) \ll \psi\,.
\end{equation}
Hence, we find that $\varphi\simeq\psi$. We note that in this analysis we implicitly imposed $F_{\Theta},\,F_{T\Theta},\,F_{TT}\neq0$ but nevertheless the conclusion $\varphi\simeq\psi$ remains the same for the two gravitational potentials. This is important since these potentials play a dominant role in the calculation of the deflection parameter \eqref{Sigmadef}.

Using the sub-horizon limit approximation, one can then proceed to derive the evolution equation for the gauge-invariant fractional overdensity $\delta_{\text{m}}$. Using Eqs.~\eqref{eq:00-first}, \eqref{eq:0i-first} and the definition of $\delta_{\text{m}}$ in Eq.~\eqref{delta_m_def}
\begin{alignat}{2}
    \left(\frac{\kappa^2}{2}-\frac{3F_{\Theta}}{4}-\frac{\rho F_{\Theta\Theta}}{2}\right)\delta_{\text{m}} & =\: & &\left(1+F_{T}\right)\dfrac{k^{2}\psi}{a^{2}\rho}+\dfrac{1}{2}\bigg\lbrace F_{T\Theta}\left[12H\left(\dot{\psi}+H\varphi\right)-4a^{-1}Hk^{2}w\right]\nonumber \\[0.5ex]
    &\: & & +3Ha^{2}v\rho F_{\Theta\Theta}\bigg\rbrace+\frac{3F_{\Theta}}{4}a^{2}Hv\,,
\end{alignat}
which in conjunction with $\psi\simeq\varphi$ and $\frac{k^{2}\psi}{a^{2}H^{2}}\gg\psi$ reduces to
\begin{equation}
    A\delta_{\text{m}}=\left(1+F_{T}\right)\dfrac{k^{2}\psi}{a^{2}\rho}+\frac{3Ha^{2}v}{2}\rho F_{\Theta\Theta}+\frac{3F_{\Theta}}{4}a^{2}Hv\,,\label{test1}
\end{equation}
where $A\coloneqq\frac{\kappa^2}{2}-\frac{3}{4}F_{\Theta}-\frac{1}{2}F_{\Theta\Theta}\rho$. In addition, From Eq.~\eqref{eq:i0-first}, in the sub-horizon limit, we obtain the relation $H\varphi\sim\rho a^{2}v$ which can be further used to simplify \eqref{test1} as
\begin{equation}
    A\delta_{\text{m}}=\left(1+F_{T}\right)\dfrac{k^{2}\psi}{a^{2}\rho}\,,\label{eq:delta-relation-1}
\end{equation}
from which one generates the generalized matter density equation as
\begin{alignat}{2}
 & \: & & \delta''_{m}(a)+\frac{1}{aAH}\left[aAH'+2A'aH+3AH-\frac{3AH}{\kappa^2-F_{\Theta}}\left(4H'aHF_{T\Theta}+\rho F_{\Theta\Theta}\right)\right]\delta'_{\text{m}}(a)\nonumber \\[0.5ex]
 & \: & & +\frac{1}{Aa^{2}H^{2}}\left[a^{2}H^{2}A''+3A'aH^{2}+A'a^{2}HH'-\frac{3A'aH^{2}}{\kappa^2-F_{\Theta}}\left(4H'aHF_{T\Theta}+\rho F_{\Theta\Theta}\right)\right.\nonumber \\[0.5ex]
 & \: & & \left.-\frac{\kappa^2-F_{\Theta}}{2(1+f_{T})}A\rho-\frac{k^{2}F_{\Theta}}{4a^{2}}\right]\delta_{\text{m}}(a)=0\,.\label{eq:fTT-subhorizon-growth}
\end{alignat}
It is straightforward to check that the limits $A=\frac{\kappa^2}{2}$ and $F(T,\Theta)=0$ are recovered as expected. Usually, at this stage, one can read off some $G_{{\rm eff}}$ but this is rather not the case since we have a modified continuity equation as can be seen from Eqs.~\eqref{eq:continuity-first}-\eqref{eq:velocity}. Regarding its solubility, this system of equations would yield two different solutions for $\delta_{\text{m}}$ dubbed the growing and decaying modes. However, in most cases the interest lies in the formation of structure and thus only the growing solution is probed. A very similar model from the curvature-based gravity framework with very similar behavior is $f(\lc{R},\Theta)$ \cite{Alvarenga:2013syu}.

Henceforth, for the above reasons, extra caution while using first order perturbations is needed. The use of the correct perturbed tetrad is quite straightforward at this stage. However, in general one should carry out a full Hamiltonian analysis to fully understand if strong coupling appears in a theory. Then try to probe a specific background in order to extract information about linear perturbations and their dynamics.

\subsubsection{Closing remarks}

As closing remarks we would like to point out that in most of these works, and in general, one needs to be very careful considering (the scalar) perturbations. There are actually two three major reasons for that, namely:
\begin{enumerate}
    \item \textbf{A correct choice of perturbed tetrad} -- \label{enu:A-correct-choice} In most of the references discussed in Sec.~\ref{sssec:Growth-index} a wrong perturbed tetrad was used that only assumed 13 \gls{dof}. The missing \gls{dof} is traced back to the antisymmetric part of the perturbation of the tetrad which is described by the scalar field field $\sigma$ and the vector field $\sigma_{i}$. The tetrad is neither symmetric, like the metric tensor, nor antisymmetric and thus it must assume 16 \gls{dof} in general, thus not including the antisymmetric contributions one is left with 3 \gls{dof} less. This is illustrated in a more clear way via the full spacetime indexed perturbation of the tetrad dubbed as $\tau_{\mu\nu}$ defined in Eq.~\eqref{eq:deltatetradgreek}. Having said that, it was proven in Refs.~\cite{Izumi:2012qj,Bahamonde:2020lsm}, that the pseudo-scalar field $\sigma$ is not dynamical for neither $f(T)$ nor $f(T,B)$ in a flat \gls{flrw} background. Once one extends these theories, by adding more scalar fields or more contributions to the actions, then a full analysis using the correct perturbed tetrad Eq.~\eqref{eq:tetrad perturbation-1} must be performed.
    \item \textbf{Overfixing the gauge} -- One needs to be very careful first to use the correct perturbed tetrad assuming the correct number of \gls{dof} as discussed in Sec.~\ref{enu:A-correct-choice} and then fix the gauge properly as discussed in Sec.~\ref{sec:gauge_trans}. In some instances but not properly understanding the gauge transformation of the fields one can be mislead in overfixing the gauge thus artificially removing \gls{dof} which lead to a false result in principle.
    \item \textbf{The strong coupling issue} -- The next and maybe the most important reason is related to the strong coupling issue in modified theories like $f(T)$ seem to be haunted from, as also discussed in Sec.~\ref{ssec:Action_and_field_equations}. This phenomenon is directly linked with the existence of some (scalar) \gls{dof} which in general is propagating but in specific (highly symmetric) backgrounds becomes non-dynamical, at least it seems that way in first order perturbations. If one goes beyond first order perturbations then eventually one will find these missing \gls{dof} in higher perturbative orders. Hence if there are hidden \gls{dof} in higher perturbative orders, the linear order perturbations might not be dominant anymore and a full analysis of all orders until the missing \gls{dof} is located will be needed. This is directly linked to the fact that linear perturbations are not sensitive to non-linear velocity constraints in general. Finally, we stress that, specifically for $f(T)$ this is a problem of choosing a proper background solution, not a problem of the actual theory \cite{Blagojevic:2020dyq}. Further, we do not know if this problem will also appear in other modified \gls{tg} different to $f(T)$ gravity.
\end{enumerate}

Henceforth, for the above reasons, extra caution while using first order perturbations is needed. The use of the correct perturbed tetrad and avoid over-fixing the gauge is quite straightforward at this stage. Regarding the strong coupling issue, it would be wise to first make sure that a proper Hamiltonian analysis of the action has been carried out and the number of \gls{dof} is known. Then try to probe a specific background solution and its perturbations. Although it is expected that $f(T)$ is strongly coupled for the available background solutions used so far, it should be stressed that this is purely due to the non-linearity $f(T)$ introduces. In the case of $-T+\Lambda$ for example there is no such issue since the dependence on $T$ is linear.

\clearpage

\section{Polarization of Gravitational Waves in Teleparallel Gravity}\label{sec9:GW}

In this section, the polarization status of some major classes of \gls{tg} theories will be reviewed. The \gls{ngr} will be the starting point for calculating the polarizations by utilizing the Newman-Penrose formalism. In this case it will be evident that the polarizations depend heavily on the parameters of \gls{ngr}. Subsequently, the case of $f(T,B)$ gravity will be illustrated where the polarizations are calculated with a more traditional method using the geodesic deviation equation. This case differs significantly from \gls{ngr} since there is a massive scalar on top of the tensor modes. Finally, the polarization modes of the BDLS theory, the most general scalar-tensor theory built on \gls{tg}, will be exhibited. For this case there is yet another distinct method, of calculating the polarization modes, used by employing the \gls{svt} decomposition. First the number of \gls{dof} are calculated and then the various results are applied for the calculation of polarizations.

\subsection{Polarization modes in metric theories of gravity}\label{subsec:Polarization_modes_in_general}

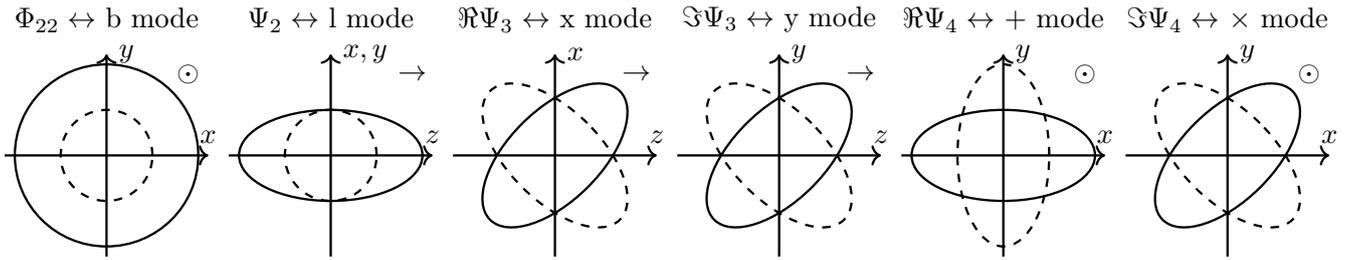
\begin{figure}[H]
\resizebox{\textwidth}{!}{
\begin{tikzpicture}[thick,scale=0.25]

\foreach \x in {0,11,...,55} \draw[->] (\x,5) to +(10,0);
\foreach \x in {5,16,...,60} \draw[->] (\x,0) to +(0,10);

\node[above] at (5,10.5) {$\Phi_{22}\leftrightarrow$ b mode};
\node[above] at (10,5) {$x$};
\node[right] at (5,10) {$y$};
\node at (9,9) {$\odot$};
\draw[dashed] (5,5) circle [radius=2.25];
\draw (5,5) circle [radius=4.5];

\node[above] at (16,10.5) {$\Psi_{2}\leftrightarrow$ l mode};
\node[above] at (21,5) {$z$};
\node[right] at (16,10) {$x,y$};
\node at (20,9) {$\rightarrow$};
\draw (16,5) ellipse [x radius=4.5,y radius=2.25];
\draw[dashed] (16,5) circle [radius=2.25];

\node[above] at (27,10.5) {$\Re\Psi_{3}\leftrightarrow$ x mode};
\node[above] at (32,5) {$z$};
\node[right] at (27,10) {$x$};
\node at (31,9) {$\rightarrow$};
\draw[rotate around={45:(27,5)}] (27,5) ellipse [x radius=4.5,y radius=2.25];
\draw[dashed,rotate around={45:(27,5)}] (27,5) ellipse [x radius=2.25,y radius=4.5];

\node[above] at (38,10.5) {$\Im\Psi_{3}\leftrightarrow$ y mode};
\node[above] at (43,5) {$z$};
\node[right] at (38,10) {$y$};
\node at (42,9) {$\rightarrow$};
\draw[rotate around={45:(38,5)}] (38,5) ellipse [x radius=4.5,y radius=2.25];
\draw[dashed,rotate around={45:(38,5)}] (38,5) ellipse [x radius=2.25,y radius=4.5];

\node[above] at (49,10.5) {$\Re\Psi_{4}\leftrightarrow$ $+$ mode};
\node[above] at (54,5) {$x$};
\node[right] at (49,10) {$y$};
\node at (53,9) {$\odot$};
\draw (49,5) ellipse [x radius=4.5,y radius=2.25];
\draw[dashed] (49,5) ellipse [x radius=2.25,y radius=4.5];

\node[above] at (60,10.5) {$\Im\Psi_{4}\leftrightarrow$ $\times$ mode};
\node[above] at (65,5) {$x$};
\node[right] at (60,10) {$y$};
\node at (64,9) {$\odot$};
\draw[rotate around={45:(60,5)}] (60,5) ellipse [x radius=4.5,y radius=2.25];
\draw[dashed,rotate around={45:(60,5)}] (60,5) ellipse [x radius=2.25,y radius=4.5];

\end{tikzpicture}
}%
\caption{All possible polarizations of \gls{gw} travelling in z-direction, starting with the scalar modes breathing, longitudinal, the vector modes x,y, and the tensor modes $+,\times$. The \gls{gw} deforms a sphere of freely falling test particles. The symbols $\rightarrow$ and $\odot$ denote propagation parallel and perpendicular to the paper plane accordingly.}
\label{fig:polarizations}
\end{figure}

In metric theories of gravity, due to the nature of the Riemann tensor, only six \gls{gw} polarizations are allowed \cite{Eardley:1973br,Eardley:1974nw,Barack:2018yly}. These six polarizations can be also classified in terms of their helicity content as two tensor (helicity $\pm 2$) modes plus (+) and cross ($\times$), two vector (helicity $\pm 1$) modes called x and y and two scalar (helicity 0) modes named breathing and longitudinal modes. This classification is illustrated in Fig.~\ref{fig:polarizations}. In general, we can explore the polarizations of the \gls{gw}s by measuring their amplitudes with \gls{gw} detectors \cite{Nakao:2000ug,Takeda:2020tjj,Will:2001mx,Tobar:1999tf}.

For a \gls{gw} incident toward the Earth, the polarization content of the \gls{gw} determines the relative amplitudes that are measured at different detector locations on the Earth's surface, modulated also by each detectors' antenna pattern~\cite{Will:2001mx}. The current ground-based advanced interferometer network - comprising LIGO Hanford, LIGO Livingston and Virgo - is therefore not yet sufficient to fully reconstruct all the \gls{gw} polarization of detected signals \cite{Takeda:2020tjj}. However, the existing three-detector network can already be used to compare the evidence for some specific subsets of all the possible polarization combinations \cite{LIGOScientific:2021sio}.

In Ref.~\cite{Abbott:2017oio,PhysRevLett.123.011102,LIGOScientific:2020tif}, the LIGO and Virgo collaborations carried out a Bayesian model comparison between some extreme alternative hypotheses -- specifically, comparing full tensor with full vector or full scalar -- for the polarization content of the \gls{gw} signals detected from three events in the second LIGO-Virgo Observing Run: GW170814, GW170817 and GW170818. The first and third events were binary black hole mergers while GW170817 was a binary neutron star merger with an explicitly identified electromagnetic counterpart in its host galaxy NGC4993. Consequently the constraints obtained from these analyses were strongest for GW170817, which favoured the full tensor hypothesis over both full vector and full scalar -- in each case with a log Bayes factor greater than 20. For the two binary black hole mergers also, however, the full tensor hypothesis was also modestly favoured. To further clarify, therefore, this means \gls{gw}s are highly more like to have only tensor polarization modes compared to only vector or only scalar ones. However, we do not yet have any information about a mixture of polarizations such as tensor and scalar modes or vector and scalar modes etc. Hence we cannot yet exclude theories that predict such mixtures of polarization pairs.

More recently, the LIGO and Virgo collaborations have carried out a further Bayesian analysis that again compares full tensor with full vector and full scalar hypotheses, but without assuming a specific waveform model. (The analyses carried out in Ref.~\cite{Abbott:2017oio,PhysRevLett.123.011102,LIGOScientific:2020tif} had adopted \gls{gr} template waveforms). This analysis was applied in Ref.~\cite{LIGOScientific:2020tif} to a subset of the \gls{gw} detections reported in the GWTC-2 catalog, which included events observed during the first part of the 3rd LIGO-Virgo Observing Run, denoted O3a~\cite{abbott2021gwtc2}. Again, none of the mergers studied showed significant evidence favoring a non-\gls{gr} hypothesis.

Notwithstanding the limitations of the current three detector global network, the near-future prospects for conducting high-precision polarization tests of general metric theories of gravity are very good. By the late 2020s, the Advanced LIGO detectors will have undergone significant further improvements, enhancing their sensitivity to the so-called ``A+" configuration: a mid-scale, incremental upgrade to Advanced LIGO that is scheduled to happen between 2026-2028 \cite{Miller:2014kma}. This should increase their reach by a factor of several. Similar enhancements to the Advanced Virgo detector \cite{2021arXiv210509247N} are anticipated on the same timescale, and more importantly by that time the KAGRA \cite{Michimura:2020xnj} and LIGO India \cite{Unnikrishnan:2013qwa} detectors will also have joined the global network, with LIGO India planned to begin operations at the ``A+" configuration. Hence we can expect that by the end of this decade an extended ``Big Five" global network of advanced interferometers will be fully operational, with significantly higher event rates and improved sky localization capability~\cite{2018LRR....21....3A}. Data from these five detectors will be able to fully characterize the polarization content of \gls{gw} sources, significantly reducing or eliminating degeneracies, and hence will provide an excellent means for testing the predictions of teleparallel theories of gravity.

The six aforementioned polarizations are directly linked with the symmetries of the electric components of the Riemann tensor denoted as $\lc{R}_{i0j0}$, which assumes strictly only 6 independent components due to its symmetry. These components are responsible for the evolution of the geodesic deviation equation \cite{Carroll:2004st}
\begin{equation}
    \ddot{x}_i = -\lc{R}_{i0j0} x^j\,,\label{eq:geodesic_dev}
\end{equation}
where dots represent coordinate time derivatives, $(t,x,y,z) = (0,1,2,3)$, $i = \lbrace 1,2,3 \rbrace$ and $x^j = (x, y, z)$. Since in \gls{tg} we mostly assume that matter is coupled to the metric (and the Levi-Civita connection), the force-like equation which describes the motion of particles in \gls{tg} coincides with the geodesic equation (see Sec.~\ref{Grav_Coup_Prescrip}). This means that the information related to the polarizations would be only encoded in the Riemann tensor (as in \gls{gr}). The tetrad contains more \gls{dof} than the metric, but the extra ones cannot be directly measured due to the matter coupling prescription adopted. On the other hand, the extra information that could carry the tetrad could be found by indirect \gls{gw} observations. Restricting to massless \gls{gw}s there is the $\mathrm{E}(2)$ classification \cite{Eardley:1973br,Eardley:1974nw,Will:2001mx}, that utilizes the Newman-Penrose variables in order to facilitate the calculation of the polarizations in a given gravitational theory. Within this framework the six components of the electric Riemann tensor can be represented as
\begin{align}
    \lc{R}{}_{0i0j} & =\left(\arraycolsep=2pt\def\arraystretch{1.5}\begin{array}{ccc}
    \frac{1}{2}(\Re\Psi_{4}+\Phi_{22}) & \frac{1}{2}\Im\Psi_{4} & -2\Re\Psi_{3}\\
    \frac{1}{2}\Im\Psi_{4} & -\frac{1}{2}(\Re\Psi_{4}-\Phi_{22}) & 2\Im\Psi_{3}\\
    -2\Re\Psi_{3} & 2\Im\Psi_{3} & -6\Psi_{2}
    \end{array}\right)\,.\label{eq:RiemannELECTRIC_E2}
\end{align}
where $\Phi_{22},\Psi_{2},\Psi_{3},\Psi_{4}$ are Newman-Penrose variables, $\Re$ represents the real part and $\Im$ the imaginary one. These Newman-Penrose variables can also be classified \gls{wrt} their helicity states through
\begin{equation}
\arraycolsep=2pt\def\arraystretch{1.5}\begin{array}{lllllc}
\Psi_{2} & : & s=0\,, & \Phi_{22} & : & s=0\,,\\
\Psi_{3} & : & s=-1\,, & \overline{\Psi}_{3} & : & s=1\,,\\
\Psi_{4} & : & s=-2\,, & \overline{\Psi}_{4} & : & s=2\,,
\end{array}\label{eq:POL_AMPL_helicities}
\end{equation}
where the overbar denotes complex conjugation. From
Eq.~\eqref{eq:POL_AMPL_helicities} we can deduce that $\Phi_{22},\Psi_{2}$
are related to scalar \gls{dof}, $\Psi_{3}$ is related to vectorial \gls{dof}
and $\Psi_{4}$ is related to tensorial \gls{dof}. Finally, we can visualize this parametrization of Eq.~\eqref{eq:RiemannELECTRIC_E2} in Fig.~\ref{fig:polarizations}.

We will use the notation of Ref.~\cite{will2018theory} and denote the basis vectors by \(l^{\mu}, n^{\mu}, m^{\mu}, \bar{m}^{\mu}\) as
\begin{equation}\label{NP_basis}
l = \partial_0 + \partial_3\,, \quad
n = \frac{1}{2}(\partial_0 - \partial_3)\,, \quad
m = \frac{1}{\sqrt{2}}(\partial_1 + i\partial_2)\,, \quad
\bar{m} = \frac{1}{\sqrt{2}}(\partial_1 - i\partial_2)\,,
\end{equation}
and we consider a plane wave propagating in the positive \(x^3\) direction, which corresponds to a single Fourier mode \eqref{eq:fourier_trans_con}. The wave covector then takes the form \(k_{\mu} = -\omega l_{\mu}\), where $\omega$ characterizes the norm of the wave covector, and any generic perturbation $p_{\mu\nu}$ will be expanded as
\begin{equation}
p_{\mu\nu}=P_{\mu\nu}e^{i\omega u}\,,\label{eqn:zwave}
\end{equation}
where we introduced the retarded time \(u = x^0 - x^3\) and the wave amplitude is denoted \(P_{\mu\nu}\). As shown in Ref.~\cite{Eardley:1974nw}, the Riemann tensor of a null plane wave is determined completely by the six so-called electric components. For the wave in Eq.~\eqref{eqn:zwave}, these can be written in the Newman-Penrose basis as
\begin{subequations}
\begin{align}
\Psi_2 &= -\frac{1}{6}\lc{R}_{nlnl} = \frac{1}{12}\ddot{h}_{ll}\,, \quad
\Psi_3 = -\frac{1}{2}\lc{R}_{nln\bar{m}} = -\frac{1}{2}\overline{\lc{R}_{nlnm}} = \frac{1}{4}\ddot{h}_{l\bar{m}} = \frac{1}{4}\overline{\ddot{h}_{lm}}\,,\\[0.5ex]
\Psi_4 &= -\lc{R}_{n\bar{m}n\bar{m}} = -\overline{\lc{R}_{nmnm}} = \frac{1}{2}\ddot{h}_{\bar{m}\bar{m}} = \frac{1}{2}\overline{\ddot{h}_{mm}}\,,
\quad \Phi_{22} = -\lc{R}_{nmn\bar{m}} = \frac{1}{2}\ddot{h}_{m\bar{m}}\,,\label{eqn:riemcomp}
\end{align}
\end{subequations}
where dots denote derivatives \gls{wrt} the retarded time \(u\). This representation of Newman-Penrose variables serves as the standard way of calculating the polarization content of massless \gls{gw}.

However, note the amplitudes in Eq.~\eqref{eq:POL_AMPL_helicities} are not observer-independent quantities. For example, if in one frame $\Psi_{4}\neq0$ but $\Psi_{2}\neq0$, then there exists a frame in which $\Psi_{4}=0$. Hence, the absence or presence of the components of various helicities depends upon the frame chosen. Nonetheless, there are still some frame invariant statements for the amplitudes
which include a set of quasi-Lorentz invariant classes of \gls{gw}. Each class is labeled by the Petrov type of its nonvanishing Weyl tensor and the maximum number of nonvanishing amplitudes as seen by any observer. These labels observer independent quantities. These classes are illustrated in Table~\ref{table:E2classes}.

\begin{table}[H]
\midsepremove
\centering \resizebox{\textwidth}{!}{
\begin{tabular}{lll}
\toprule 
\cellcolor{gris3}\textbf{Class} & \cellcolor{gris3}\textbf{Conditions} & \cellcolor{gris3}\textbf{Description}\tabularnewline
\midrule 
\multicolumn{1}{l}{\cellcolor{gris1}} & \multicolumn{1}{l}{\cellcolor{gris1}} & \cellcolor{gris1}All standard observers measure the same value for
$\Psi_{2}$,\tabularnewline
\multirow{-2}{*}{\cellcolor{gris1}II$_6$} & \multirow{-2}{*}{\cellcolor{gris1}$\Psi_{2}\neq0$} & \cellcolor{gris1}but disagree on the presence or absence of all other
modes.\tabularnewline
\multicolumn{1}{l}{\cellcolor{gris3}} & \multicolumn{1}{l}{\cellcolor{gris3}} & \cellcolor{gris3}All standard observers agree on the absence of $\Psi_{2}$
and on the presence \tabularnewline
\multirow{-2}{*}{\cellcolor{gris3}III$_{5}$} & \multirow{-2}{*}{\cellcolor{gris3}$\Psi_{2}=0$, $\Psi_{3}\neq0$} & \cellcolor{gris3}of $\Psi_{3}$, but disagree on the presence or
absence of $\Psi_{4}$ and $\Phi_{22}$.\tabularnewline
\cellcolor{gris1}N$_{3}$ & \cellcolor{gris1}$\Psi_{2}=0$, $\Psi_{3}=0$, $\Phi_{22}\neq0$,
$\Psi_{4}\neq0$ & \cellcolor{gris1}Presence or absence of all modes is observer-independent.\tabularnewline
\cellcolor{gris3}N$_{2}$ & \cellcolor{gris3}$\Psi_{2}=0$, $\Psi_{3}=0$, $\Phi_{22}=0$, $\Psi_{4}\neq0$ & \cellcolor{gris3}Independent of observer.\tabularnewline
\cellcolor{gris1}O$_{1}$ & \cellcolor{gris1}$\Psi_{2}=0$, $\Psi_{3}=0$, $\Phi_{22}\neq0$,
$\Psi_{4}=0$ & \cellcolor{gris1}Independent of observer.\tabularnewline
\cellcolor{gris3}O$_{0}$ & \cellcolor{gris3}$\Psi_{2}=\Psi_{3}=\Phi_{22}=\Psi_{4}=0$ & \cellcolor{gris3}Independent of observer: No wave.\tabularnewline
\bottomrule
\end{tabular}}
\midsepdefault
\caption{E(2) classes as labeled by the Petrov type of its nonvanishing Weyl
tensor and the maximum number of nonvanishing amplitudes as seen by
any observer. These classes are independent of observer \cite{will2018theory}.}
\label{table:E2classes}
\end{table}

This is exactly the so called $E(2)$ classification for \gls{gw}s and each of these classes of a particular metric theory is defined to be the class of its most general wave. More details also including almost null waves can be found in Ref.~\cite{will2018theory}.

We present yet another way to probe the polarizations via the use of the \gls{svt} decomposition as introduced in Sec.~\ref{sssec:Irreducible_decomposition}. The idea is to use the \gls{svt} decomposition of the tetrad (and consequently the metric) on Minkowski spacetime and then calculate the electric components of the Riemann tensor \cite{Bahamonde:2021dqn}, $\lc{R}{}_{0i0j}$ which reads as
\begingroup
\begin{align}
\lc{R}{}_{0i0j} & =\left(
\arraycolsep=2pt\def\arraystretch{1.5}\begin{array}{ccc}
\ddot{\psi}-\tfrac{1}{2}\ddot{h}_{+} & -\tfrac{1}{2}\ddot{h}_{\times} & -\tfrac{1}{2}ik(\dot{\beta}_{1}+\dot{\Lambda}_{1})\\
-\tfrac{1}{2}\ddot{h}_{\times} & \ddot{\psi}+\tfrac{1}{2}\ddot{h}_{+} & -\tfrac{1}{2}ik(\dot{\beta}_{2}+\dot{\Lambda}_{2})\\
-\tfrac{1}{2}ik(\dot{\beta}_{1}+\dot{\Lambda}_{1}) & -\tfrac{1}{2}ik(\dot{\beta}_{2}+\dot{\Lambda}_{2}) & \ddot{\psi}-k^{2}(\dot{\chi}+\Phi)
\end{array}\right)\,,\label{eq:ERiemSVT}
\end{align}
\endgroup
where we introduced the gauge invariant variables
\begin{subequations}
\begin{align}
\chi & :=b-\dot{h}\,,\label{eq:gauge_inv1}\\[0.5ex]
\Phi & :=\varphi-\dot{\beta}\,,\\[0.5ex]
\Sigma_{j} & :=h_{j}k{}^{2}+i\varepsilon{}_{jlp}k{}^{l}\sigma{}^{p}\,,\\[0.5ex]
\Xi_{j} & :=ik^{2}b{}_{j}-2\varepsilon{}_{jlp}k{}^{l}\dot{\sigma}{}^{p}\,,\\[0.5ex]
\Lambda_{j} & :=-b_{j}+2\dot{h}{}_{j}\,.\label{eq:gauge_inv2}
\end{align}
\end{subequations}
The representation of the electric components of the Riemann tensor in Eq.~\eqref{eq:ERiemSVT} actually holds for any case, irrespectively if the \gls{gw} is massless or massive and hence is more flexible.

\subsection{Polarization modes in new general relativity} \label{subsec:Polarization modes in NGR}

We review the analysis of polarizations of \gls{gw} in \gls{ngr}, introduced in Sec.~\ref{sec:NGR}, as developed in Ref.~\cite{Hohmann:2018xnb,Hohmann:2018jso}.
The Lagrangian of \gls{ngr} \cite{Hayashi:1979qx} reads as
\begin{align}\label{eq:LNGR}
	L
	&= e \big(c_1 T^\rho{}_{\mu\nu}T_{\rho}{}^{\mu\nu} + c_2T^\rho{}_{\mu\nu} T^{\nu\mu}{}_\rho + c_3 T^\rho{}_{\mu\rho}T^{\sigma\mu}{}_{\sigma}\big)
	= e G_{\alpha\beta}{}^{\mu\nu\rho\sigma}T^{\alpha}{}_{\mu\nu}T^\beta{}_{\rho\sigma}\,,
\end{align}
where the three real parameters $c_1, c_2$ and $c_3$ define different \gls{ngr} theories and the supermetric \cite{Ferraro:2016wht} or constitutive tensor \cite{Itin:2016nxk,Hohmann:2017duq}
\begin{align}\label{eq:G}
	G_{\alpha\beta}{}^{\mu\nu\rho\sigma} =c_1 g_{\alpha\beta}g^{\rho[\mu}g^{\nu]\sigma} - c_2 \delta_\beta^{[\mu}g^{\nu][\rho}\delta^{\sigma]}_\alpha - c_3\delta_\alpha^{[\mu}g^{\nu][\rho}\delta^{\sigma]}_\beta\,,
\end{align}
which serves in compactifying the calculations later on.

\subsubsection{Linearized regime and field equations}

We start by linearizing the field equations by introducing the perturbations of the tetrad, inverse tetrad and the Lorentz transformation defining the spin connection
\begin{subequations}\label{eqn:pertfields}
\begin{align}
    e^A{}_\mu &= \delta^A_\mu + \epsilon\ u^A{}_\mu\,,\\[0.5ex]
    E_A{}^\mu &= \delta_A^\mu + \epsilon\ v_A{}^\mu\,,\\[0.5ex]
    \Lambda^A{}_B &= \delta^A_B + \epsilon\ w^A{}_B\,,
\end{align}
\end{subequations}
where $\epsilon \ll 1$ acts as the perturbation parameter. One can easily show, at first order in $\epsilon$, from $e^{A}{}_{\mu}E_{A}{}^{\nu}=\delta^\nu_\mu$ that $v_A{}^\mu \delta_\mu^B = - u^B{}_\nu \delta^\nu_A$ and from $\eta_{AB}\Lambda^{A}{}_{C}\Lambda^{B}{}_{D}=\eta_{CD}$ implies that $w_{AB} = -w_{BA}$. We consider a tetrad $\hat{e}^A{}_{\mu}$ that is related to the original tetrad $e^A{}_\mu$, by a local Lorentz transformation $\tilde{\Lambda}^A{}_B$, i.e. $\hat{e}^A{}_\mu = \tilde{\Lambda}^A{}_B e^B{}_\mu$. Consequently, the torsion tensors of the respective tetrads will be $\hat{T}^A = \tilde{\Lambda}^A{}_B T^B$, and the connections are given in terms of two further Lorentz transformations $\hat{\Lambda}$ and $\Lambda$ by
\begin{equation}
    \hat{\omega}^A{}_B = \hat{\Lambda} ^A{}_C \dd (\hat{\Lambda}^{-1})^C{}_B\,,\,\,\, \omega ^A{}_B = (\tilde{\Lambda}^{-1})^A{}_C \hat{\Lambda}^C{}_D \dd (\tilde{\Lambda}^E{}_B(\hat{\Lambda}^{-1})^D{}_E) = \Lambda ^A{}_D \dd (\Lambda^{-1})^D{}_B\,.  \end{equation}
In the case when $\tilde{\Lambda} = \hat{\Lambda}$ the spin connection of the tetrad vanishes, $\omega ^A{}_B = 0$, meaning we choose the Weitzenb\"ock gauge. Do not confuse the $\,\hat{}\,$ symbol used here to denote the transformed tetrad, with the $\,\widehat{}\,$ one used in Sec.~\ref{sec2:affine}, \ref{sec3:torsional}, \ref{sec4:TEGRNGR} and \ref{sec5:extended} and denotes the quantities calculated with the symmetric teleparallel connection.

The gauge transformations regarding the perturbations in this framework are
\begin{subequations}
\begin{alignat}{5}
    \hat{\mathbf{e}}^{A} & =\: & & \tilde{\Lambda}^{A}{}_{B}\mathbf{e}^{B} & \:\Rightarrow\: & \hat{u}^{A}{}_{\mu} & & =\: & \tilde{w}^{A}{}_{\mu}+u^{A}{}_{\mu}\,,\\[0.5ex]
    \Lambda^{A}{}_{D} & =\: & & (\tilde{\Lambda}^{-1})^{A}{}_{C}\hat{\Lambda}^{C}{}_{D} & \:\Rightarrow\: & \hat{w}^{A}{}_{B} & & =\: & \tilde{w}^{A}{}_{B}+w^{A}{}_{B}\,.\label{eq:pertgt}
\end{alignat}
\end{subequations}
We change indices using $\delta^A_\mu$ and $\delta^\mu_A$ while raising/lowering any kind of index is done with the Minkowski metrics $\eta_{AB}$ and $\eta_{\mu\nu}$ or its inverse.

The torsion tensor up to first order reads
\begin{align}
	T^A{}_{\mu\nu} = 2 \partial_{[\mu} e^A{}_{\nu]} + 2\omega^A{}_{B[\mu}e^B{}_{\nu]} =2 \epsilon \big( \partial_{[\mu} u^A{}_{\nu]} - \partial_{[\mu} w^A{}_{\nu]} \big) + \mathcal{O}(\epsilon^2)\,.
\end{align}
This perturbative expansion of the torsion tensor does not have a zeroth-order contribution since we assumed that the background is described by the Minkowski metric with the diagonal background tetrad $\delta^A_\mu$ which indeed ensures that the teleparallel connection and the metric respect maximally symmetric spacetimes (see Sec.~\ref{sec:maxisym}). For further convenience we will transform all the indices of the torsion tensor into Greek indices, up to first order, as $T^{\lambda}{}_{\mu\nu}=E_{A}{}^{\lambda}T^{A}{}_{\mu\nu}$. We expand the \gls{ngr} Lagrangian~\eqref{eq:LNGR} up to the first non-trivial order
\begin{align}\label{eq:pertL}
	L^{(2)}=\epsilon^2 G_{\alpha\beta}{}^{\mu\nu\rho\sigma}\big( \partial_\mu u^\alpha{}_\nu - \partial_\mu w^\alpha{}_\nu \big)\big( \partial_\rho u^\beta{}_\sigma - \partial_\rho w^\beta{}_\sigma \big)
	+ \mathcal{O}(\epsilon^3)\,,
\end{align}
and note that the determinant $e$ and $G_{\alpha\beta}{}^{\mu\nu\rho\sigma}$ contribute only their zeroth order since the quadratic torsion terms are already second order. From this Lagrangian we can derive the field equations \gls{wrt} the variables $u$ and $w$ as
\begin{subequations}
\begin{alignat}{4}
0 & =\: & & \partial_{\lambda}\frac{\partial L^{(2)}}{\partial\partial_{\lambda}u^{\tau}{}_{\kappa}} & \Leftrightarrow\: & 0\: & =\: & G_{\tau\beta}{}^{\lambda\kappa\rho\sigma}\partial_{\lambda}\big(\partial_{\rho}u^{\beta}{}_{\sigma}-\partial_{\rho}w^{\beta}{}_{\sigma}\big)\,,\\[0.5ex]
0 & =\: & & \partial_{\lambda}\frac{\partial L^{(2)}}{\partial\partial_{\lambda}w^{\tau}{}_{\kappa}} & \Leftrightarrow\: & 0\: & =\: & (G_{\tau\beta}{}^{\lambda\kappa\rho\sigma}-\eta_{\gamma\tau}\eta^{\xi\kappa}G_{\xi\beta}{}^{\lambda\gamma\rho\sigma})\partial_{\lambda}\big(\partial_{\rho}u^{\beta}{}_{\sigma}-\partial_{\rho}w^{\beta}{}_{\sigma}\big)\,.
\end{alignat}
\end{subequations}
It is clear that $u$ and $w$ are not independent variables of the theory thus we introduce the new gauge invariant with Eq.~\eqref{eq:pertgt}) variable $x_{\beta\sigma} = u_{\beta\sigma} - w_{\beta\sigma}$ gives us
\begin{align}\label{eq:fieldeq}
	W^{\tau\kappa} &:= G^{\tau\beta\lambda\kappa\rho\sigma}\partial_\lambda\partial_\rho \xx_{\beta\sigma}=0\,.
\end{align}
We can further split $\xx_{\beta\sigma}$ as $\xx_{\beta\sigma} = s_{\beta\sigma} + a_{\beta\sigma}$ ($s_{\beta\sigma}$ being the symmetric part and $a_{\beta\sigma}$ being the anti-symmetric part) and thus replacing it back into Eq.~\eqref{eq:fieldeq} we get
\begin{alignat}{2}
W^{\tau\kappa} & =\: & & \partial_{\rho}\big[(2c_{1}-c_{2})\partial^{\rho}a^{\tau\kappa}-(2c_{1}-c_{2})\partial^{\kappa}a^{\tau\rho}+(2c_{2}+c_{3})\partial^{\tau}a^{\rho\kappa}\big]\nonumber \\[0.5ex]
& \: & & +\partial_{\rho}\big[(2c_{1}+c_{2})\partial^{\rho}s^{\tau\kappa}-(2c_{1}+c_{2})\partial^{\kappa}s^{\tau\rho}+c_{3}\big(\eta^{\tau\kappa}(\partial^{\rho}s^{\beta}{}_{\beta}-\partial_{\beta}s^{\rho\beta})-\eta^{\tau\rho}(\partial^{\kappa}s^{\beta}{}_{\beta}-\partial_{\beta}s^{\kappa\beta})\big)\big]\,.\label{eqn:linvaceom}
\end{alignat}
In the same manner we split also the field equations $W^{\tau\kappa}$ into symmetric and antisymmetric parts thus arriving at $ W^{(\tau\kappa)}$ and $W^{[\tau\kappa]}$ as
\begin{subequations}
\begin{alignat}{2}
W^{(\tau\kappa)} & =\: & & \partial_{\rho}\big[-(2c_{1}+c_{2}+c_{3})\partial^{(\tau}a^{\kappa)\rho}\big]\nonumber \\[0.5ex]
 & \: & & +\partial_{\rho}\big[(2c_{1}+c_{2})\partial^{\rho}s^{\tau\kappa}-(2c_{1}+c_{2}+c_{3})\partial^{(\tau}s^{\kappa)\rho}+c_{3}\big(\eta^{\tau\kappa}(\partial^{\rho}s^{\beta}{}_{\beta}-\partial_{\lambda}s^{\rho\lambda})-\eta^{\rho(\tau}\partial^{\kappa)}s^{\beta}{}_{\beta}\big)\big]\,,\label{eq:sym}\\[0.5ex]
W^{[\tau\kappa]} & =\: & & \partial_{\rho}\big[(2c_{1}-c_{2})\partial^{\rho}a^{\tau\kappa}+(2c_{1}-3c_{2}-c_{3})\partial^{[\tau}a^{\kappa]\rho}\big]+\partial_{\rho}\big[(2c_{1}+c_{2}+c_{3})\partial^{[\tau}s^{\kappa]\rho})\big)\big]\,.\label{eq:anti}
\end{alignat}
\end{subequations}
A few remarks on the structure of these fields equations:
\begin{enumerate}
\item It is possible to fully decouple $W^{(\tau\kappa)}$ and $W^{[\tau\kappa]}$
by demanding $(2c_{1}+c_{2}+c_{3})=0$.
\item If one demands that Eq.~\eqref{eq:anti} vanishes identically, in addition to the decoupling conditions $(2c_{1}+c_{2}+c_{3})=0$, $(2c_{1}-c_{2})=0$ also $(2c_{1}-3c_{2}-c_{3})=0$ has to be satisfied, which implies $c_{1}=-\frac{1}{4}c_{3}$ and $c_{2}=-\frac{1}{2}c_{3}$. It turns out that the family of Lagrangians satisfying these conditions are multiples of \gls{tegr}. In \gls{tegr} the antisymmetric part of the field equations is satisfied trivially because the spin connection enters only as a boundary term in the action (see Sec.~\ref{sec5:extended} for an extended discussion)
up to a boundary term~\cite{Cho:1975dh}.
\item The weak field limit of \gls{tegr} has already been studied in Ref.~\cite{Obukhov:2002hy} and the fully general case, albeit in a different representation, in Ref.~\cite{Hohmann:2018jso,Kuhfuss:1986rb}.
\end{enumerate}

In the next step, the principal polynomial will be used in order to investigate the propagation properties of the \gls{gw}. After that the polarizations of the \gls{gw} will be obtained.

\subsubsection{Principal symbol approach}\label{sec:principle_sym_app}

The principal polynomial is a tool that helps us understand the propagation properties of a field from its equation of motion. More specifically the principal polynomial is defined as the determinant of the principal symbol \cite{hormander2015analysis,hormander2007analysis}. The vanishing of the principal polynomial defines the wave covectors $k_{\mu}$ of the propagating \gls{dof} of the theory. If these
covectors are known then one can calculate the speed of propagation and the mass (if not massless) of the fields under examination.

The principal symbol is defined as the highest derivative operator of the partial differential equation in Fourier space Eq.~\eqref{eq:fourier_trans_con}. In order to derive the principal polynomial we need to rewrite the field Eq.~\eqref{eq:fieldeq} as
\begin{align}
	\hat{W}^{\tau\kappa} = G^{\tau\beta\lambda\kappa\rho\sigma}k_\lambda k_\rho \hat x_{\beta\sigma} = P^{\tau\beta\kappa\sigma}(k) \hat x_{\beta\sigma}\,,
\end{align}
where $\hat x_{\beta\sigma}$ is the Fourier transformation of our original field variable $x_{\beta\sigma}$ (and not the hat accent that was used in the previous section) and
\begin{alignat}{2}
    P^{\tau\beta\kappa\sigma}(k) & =\: & & \frac{c_{1}}{2}\eta^{\tau\beta}(K\eta^{\kappa\sigma}-k^{\kappa}k^{\sigma})-\frac{c_{2}}{4}(k^{\beta}k^{\kappa}\eta^{\sigma\tau}-k^{\beta}k^{\tau}\eta^{\kappa\sigma}+k^{\sigma}k^{\tau}\eta^{\beta\kappa}-K\eta^{\beta\kappa}\eta^{\sigma\tau})\nonumber \\[0.5ex]
    & \: & & -\frac{c_{3}}{4}(k^{\tau}k^{\kappa}\eta^{\sigma\beta}-k^{\beta}k^{\tau}\eta^{\kappa\sigma}+k^{\sigma}k^{\beta}\eta^{\tau\kappa}-K\eta^{\tau\kappa}\eta^{\sigma\beta})\,,
\end{alignat}
where $K=\eta(k,k)=\eta^{\mu\nu}k_{\mu}k_{\nu}$ is the norm of the wave covector. The principal polynomial $P(k)$ is given by the determinant of the principal symbol, which is interpreted as a metric on the space of fields $y^{\tau\kappa} = P^{\tau\beta\kappa\sigma}(k) \hat x_{\beta\sigma}$.

From the anti-symmetry of the field equations in the indices $\lambda\kappa$ and $\rho\sigma$, it is immediately clear that the principal symbol is degenerate, since fields of the form $\hat x_{\beta\sigma} = k_{\sigma}V_{\beta}(k)$ solve the field equations trivially. The principal symbol being degenerate indicates that there are gauge \gls{dof} in the theory. Thus we need to derive a non-degenerate version of the principal symbol by restricting the field equations to the subspace of fields, on which they are non-degenerate\footnote{This feature is common in field theories with gauge \gls{dof} and appears also in general premetric theories of electrodynamics~\cite{Pfeifer:2016har} for example.}. In order to do that we just perform the following split
\begin{align}\label{eq:xdecomp}
	\hat x_{\beta\sigma} = k_\beta k_\sigma U + V_\beta k_\sigma + k_\beta X_\sigma + Q_{\beta\sigma}\,,
\end{align}
where the scalar $U$, the $1$-form components $V_\alpha$ and $X_\alpha$ and the $(0,2)$-tensor $Q_{\beta\sigma}$ satisfy the constraints
\begin{align}
    k_\alpha V^\alpha = 0,\quad k_\alpha X^\alpha = 0,\quad k_\alpha Q^\alpha{}_\beta = 0\,, \quad k_\alpha Q_\beta{}^\alpha = 0\,.
\end{align}
The $4$ \gls{dof} from $U$ and $V^\alpha$ cannot be dynamical since they identically solve the field equations and so do not propagate. The remaining $12$ \gls{dof} are divided with $4-1=3$ in $X_\alpha$ and $16-7=9$ in $Q_{\alpha\beta}$, which span the subspace $\mathcal{V}$. Expanding~$Q^{\tau\kappa}$ further into its symmetric traceless and antisymmetric part as well as its trace by writing $Q^{\tau\kappa} = S^{\tau\kappa} + A^{\tau\kappa} + \frac{1}{3} (\eta^{\tau\kappa} - \frac{k^\tau k^\kappa}{K}) Q^\sigma{}_\sigma$, and using Eq.~\eqref{eq:G}, the Fourier space \eqref{eq:fourier_trans_con} field equations become
\begin{alignat}{2}
\hat{W}^{\tau\kappa} & =\: & & (2c_{1}+c_{2}+c_{3})Kk^{\tau}X^{\kappa}+(2c_{1}+c_{2})KS^{\tau\kappa}+(2c_{1}-c_{2})KA^{\tau\kappa}\nonumber \\[0.5ex]
& \: & & +\frac{1}{3}Q^{\sigma}{}_{\sigma}K\big(2c_{1}+c_{2}+3c_{3}\big)\big(\eta^{\kappa\tau}-\frac{1}{K}k^{\tau}k^{\kappa}\big)\,.
\end{alignat}
To analyze them further we observe that they decompose into their contractions with $k$, their trace, their symmetric traceless and antisymmetric part as follows
\begin{subequations}
\begin{alignat}{2}
    \hat{W}^{\tau\kappa}k_{\tau}k_{\kappa} & =\: & & 0\,,\label{eq:FE_ngr_eq1}\\[0.5ex]
    \hat{W}^{\tau\kappa}k_{\kappa} & =\: & & 0\,,\\[0.5ex]
    \hat{W}^{\tau\kappa}k_{\tau} & =\: & & (2c_{1}+c_{2}+c_{3})K^{2}W^{\kappa}=0\,,\label{eq:vec}\\[0.5ex]
    \hat{W}^{\tau}{}_{\tau} & =\: & & (2c_{1}+c_{2}+3c_{3})KQ^{\tau}{}_{\tau}=0\,,\label{eq:tr}\\[0.5ex]
    \hat{W}^{[\tau\kappa]}-\frac{1}{K}k^{[\tau}\hat{W}^{|\sigma|\kappa]}k_{\sigma} & =\: & & (2c_{1}-c_{2})KA^{\tau\kappa}=0\,,\label{eq:asym}\\[0.5ex]
    \hat{W}^{(\tau\kappa)}-k^{(\tau}\hat{W}^{|\sigma|\kappa)}k_{\sigma}-\tfrac{1}{3}\big(\eta^{\tau\kappa}-\frac{1}{K}k^{\tau}k^{\kappa}\big)\hat{W}^{\sigma}{}_{\sigma} & =\: & & (2c_{1}+c_{2})KS^{\tau\kappa}=0\,,\label{eq:symf}
\end{alignat}
\end{subequations}
The first two equations are satisfied trivially for any choice of parameters $c_1, c_2$ and $c_3$. The remaining four non-trivial field equations can be represented by a block diagonal matrix acting on a field space vector which is an element of $\mathcal{V}$
\begin{align}
    &\left(
    \begin{array}{cccc}
        K(2c_{1}+c_{2}+3c_{3}) & 0 & 0 & 0\\
        0 & K^{2}(2c_{1}+c_{2}+c_{3})I_{3} & 0 & 0\\
        0 & 0 & K(2c_{1}-c_{2})I_{2} & 0\\
        0 & 0 & 0 & K(2c_{1}+c_{2})I_{2}
    \end{array}\right)\left(
    \begin{array}{c}
        Q^{\tau}{}_{\tau}\\
        X^{\kappa}\\
        \hat{A}^{\tau\kappa}\\
        \hat{S}^{\tau\kappa}
    \end{array}\right)=\left(
    \begin{array}{c}
        0\\
        0\\
        0\\
        0
    \end{array}\right)\,,\nonumber\\
    &\label{eq:matrixf}
\end{align}
where we condensed the matrices corresponding to $X^{\kappa}$(3 \gls{dof}) $,\hat{A}^{\tau\kappa}$(2 \gls{dof}) and $\hat{S}^{\tau\kappa}$(2 \gls{dof}) by writing them in terms of the identity matrix $I_{n}$ where $n$ stands for the dimension. For example
\begin{align}
    K(2c_{1}-c_{2})I_{2} & =K(2c_{1}-c_{2})
    \left(\begin{array}{cc}
        1 & 0\\
        0 & 1
    \end{array}\right)=\left(
    \begin{array}{cc}
        K(2c_{1}-c_{2}) & 0\\
        0 & K(2c_{1}-c_{2})
\end{array}\right)\,.
\end{align}
Due to their simple nature the principal polynomial is now easily obtained as determinant of the above matrix
\begin{align}\label{eq:pp}
	P(k) = (2 c_1 + c_2 + c_3)^3(2 c_1 + c_2 + 3 c_3) (2c_1 - c_2)^3(2 c_1 + c_2)^5 K^{15}\,.
\end{align}
In order to solve the field Eq.~\eqref{eq:pp} in a non-trivial way, a dispersion relation needs to be found from $P(k)=0$ such that $P(k)$ is non-degenerate. This solution is $K=0$ which means that the fields $X^{\kappa},Q^{\tau}{}_{\tau},S^{\tau\kappa}$ and $A^{\tau\kappa}$ travel at the speed of light and they are massless. Hence we find that for \gls{ngr} theories of gravity, perturbations propagate with the speed of light. On the other hand, for $K\neq0$ the only solution of the field equations is that the fields themselves vanish identically, that is, no combination of coefficients can consistently solve each field equation expressed in Eqs.~\eqref{eq:FE_ngr_eq1}--\eqref{eq:symf} simultaneously. Note that For the $W^{\kappa}$ mode a double pole in its propagator was found, which is consistent with Refs.~\cite{Kuhfuss:1986rb,VanNieuwenhuizen:1973fi}.

\subsubsection{Newman-Penrose formalism and polarizations}\label{sec:polar}

We can now study the polarizations of \gls{ngr}. The whole preparation we need in the previous sections was performed to show whether the norm of the wave covector $K$ is zero or not. As a matter of fact, $K=0$ for our system and thus \gls{gw}s are massless and travel at the speed of light independently of the parameters of the theory \(c_1, c_2, c_3\). This directly allows us to make use of the well known Newman-Penrose formalism which was introduced in Sec.~\ref{subsec:Polarization_modes_in_general} in order to link directly the linearized field equations to particular polarizations.
Afterwards we will see which polarizations are acceptable by using the classification scheme detailed in Ref.~\cite{Eardley:1973br,Eardley:1974nw}.

Recall that we consider minimal coupling between gravity and matter, i.e., coupling only through the metric seen as function of the tetrad, but not through the flat spin connection. As also discussed in Sec.~\ref{sssec:genmatactlorinv}, this is the most straightforward way of making the energy-momentum tensor local Lorentz covariant. This is the usual coupling prescription for non-spinning matter, which we will henceforth assume. It follows from this choice of the matter coupling that test particles follow the geodesics of the metric, and hence the autoparallel curves of its Levi-Civita connection. The effect of a \gls{gw} on an ensemble of test particles, or any other type of \gls{gw} detector, such as the mirrors of an interferometer, is therefore described by the corresponding geodesic deviation equation. The observed \gls{gw} signal hence depends only on the Riemann tensor derived from the Levi-Civita connection.

The metric perturbation components \(\delta g_{\mu\nu}\) are derived from the perturbation ansatz in Eq.~\eqref{eqn:pertfields} which takes the form
\begin{equation}
    g_{\mu\nu} =\order{g}{0}{}_{\mu\nu}+\delta g_{\mu\nu} = \eta_{\mu\nu} + \epsilon(\eta_{\mu\rho}u^{\rho}{}_{\nu} + \eta_{\nu\rho}u^{\rho}{}_{\mu}) = \eta_{\mu\nu} + 2\epsilon s_{\mu\nu}\,.
\end{equation}
Note that they depend only on the symmetric perturbation of the tetrad, so that these are the only components whose presence or absence we must determine. We now examine which of the components~\eqref{eqn:riemcomp} may occur for \gls{gw} satisfying the linearized field Eqs.~\eqref{eqn:linvaceom}.

\begin{figure}[ht]
\centerline{\includegraphics[width=0.8\textwidth]{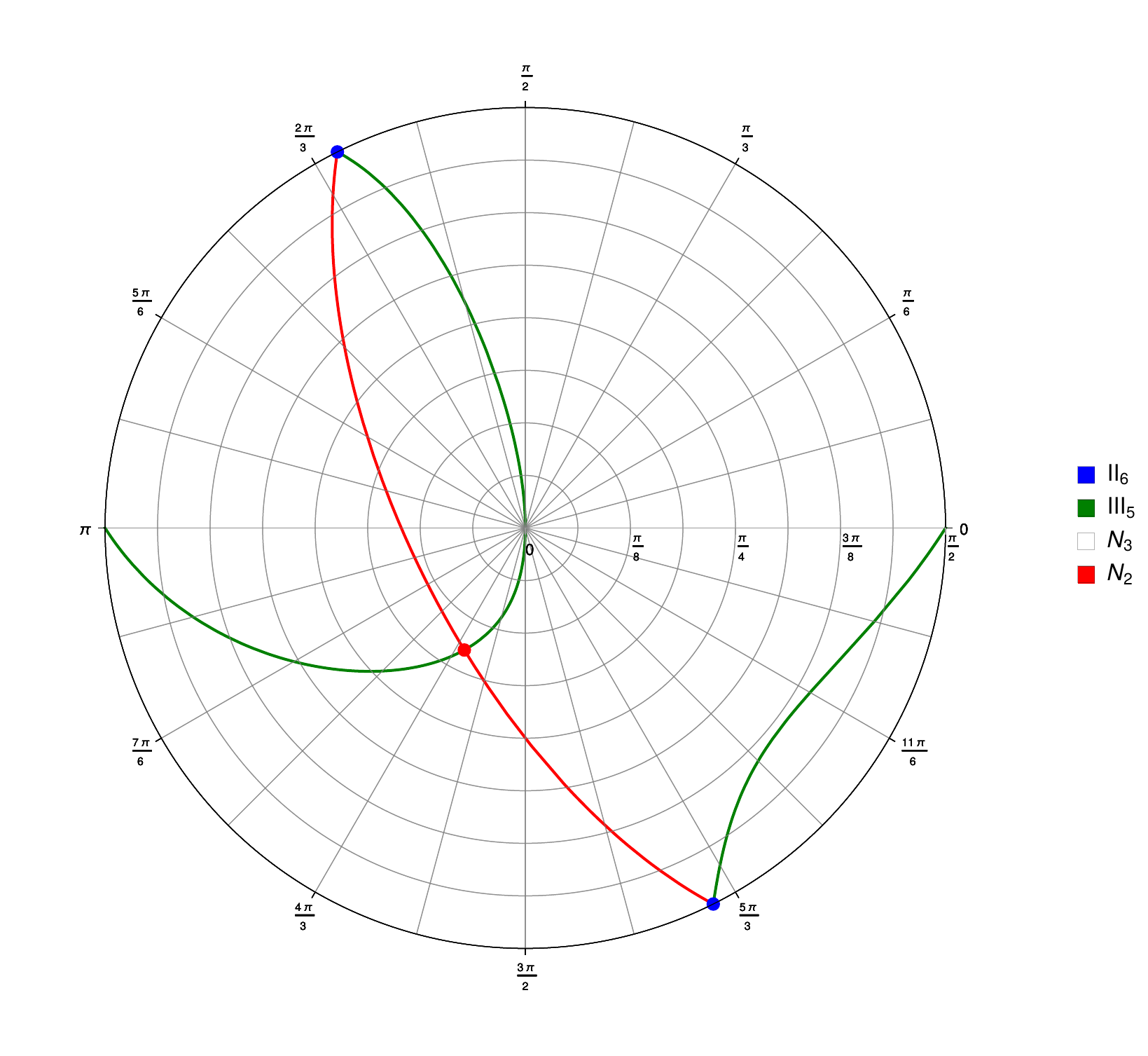}}
\caption{Visualization of the parameter space. Permission for use of this figure was kindly provided by the authors of Refs.~\cite{Hohmann:2018xnb,Hohmann:2018jso}.}
\label{fig:ngrpol}
\end{figure}

Inserting the wave ansatz Eq.~\eqref{eqn:zwave} and projecting the field equation tensor \(W_{\mu\nu}\) in the Newman-Penrose basis Eq.~\eqref{NP_basis}, we find the only non trivially satisfied equations are
\begin{subequations}\label{eqn:tnp_wave} 
\begin{alignat}{5}
0 & =\: & & W_{\mu\nu}n^{\mu}n^{\nu} & =\: & W_{nn} & & & \:=\: & (2c_{1}+c_{2}+c_{3})\ddot{s}_{nl}+2c_{3}\ddot{s}_{m\bar{m}}+(2c_{1}+c_{2}+c_{3})\ddot{a}_{nl}\,,\\[0.5ex]
0 & =\: & & W_{\mu\nu}m^{\mu}n^{\nu} & =\: & W_{mn} & & & \:=\: & \overline{W_{\bar{m}n}}=(2c_{1}+c_{2})\ddot{s}_{ml}+(2c_{1}-c_{2})\ddot{a}_{ml}\,,\\[0.5ex]
0 & =\: & & W_{\mu\nu}n^{\mu}m^{\nu} & =\: & W_{nm} & & & \:=\: & \overline{W_{n\bar{m}}}=-c_{3}\ddot{s}_{ml}+(2c_{2}+c_{3})\ddot{a}_{ml}\,,\\[0.5ex]
0 & =\: & & W_{\mu\nu}m^{\mu}\bar{m}^{\nu} & =\: & W_{m\bar{m}} & & & \:=\: & W_{\bar{m}m}=-c_{3}\ddot{s}_{ll}\,,\\[0.5ex]
0 & =\: & & W_{\mu\nu}l^{\mu}n^{\nu} & =\: & W_{ln} & & & \:=\: & (2c_{1}+c_{2})\ddot{s}_{ll}\,,
\end{alignat}
\end{subequations}
where the dots denote derivatives \gls{wrt} retarded time \eqref{eqn:zwave}. In order for the system of Eqs.~\eqref{eqn:tnp_wave} to be satisfied the conditions in Table \ref{table:summarizingcases} must be met. Specifically, by setting
\begin{equation}
    c_1 = C \sin \theta \cos \phi\,\quad c_2 = C \sin \theta \sin \phi\,\quad c_3 = C \cos \theta\,,
\end{equation}
with $C = (c_1^2 + c_2^2 + c_3 ^2)^{1/2}$ being non-zero, we can plot the above classes in polar coordinates as in Fig.~\ref{fig:ngrpol}. The zenith angle $\theta$ (in polar coordinates) is shown by the radial axis, while the azimuth angle $\phi$ is shown by the polar axis.
\begin{table}[htp!]
\centering
\midsepremove
\begin{tabular}{lll}
\toprule
\cellcolor{gris3}\textbf{Conditions} & \cellcolor{gris3}\textbf{Class} & \cellcolor{gris3}\textbf{Fig.~\ref{fig:ngrpol}} \\ \midrule
\cellcolor{gris1}\(2c_1 + c_2 = c_3 = 0\) & \cellcolor{gris1}$\mathrm{II}_{6}$ (6 polarizations) & \cellcolor{gris1}\tikz{\filldraw[fill=blue,draw=black](-0.15,-0.15) rectangle (0.15,0.15);}\\
\cellcolor{gris3}$2c_{1}(c_{2}+c_{3})+c_{2}^{2}=0$ and $2c_{1}+c_{2}+c_{3}\neq0$ & \cellcolor{gris3}$\mathrm{III}_{5}$ (5 polarizations) & \cellcolor{gris3}\tikz{\filldraw[fill=green!50!black,draw=black](-0.15,-0.15) rectangle (0.15,0.15);}\\
\cellcolor{gris1}$2c_{1}(c_{2}+c_{3})+c_{2}^{2}\neq0$ and $2c_{1}+c_{2}+c_{3}\neq0$ & \cellcolor{gris1}$\mathrm{N}_{3}$ (3 polarizations) & \cellcolor{gris1}\tikz{\filldraw[fill=white,draw=black](-0.15,-0.15) rectangle (0.15,0.15);}\\
\cellcolor{gris3}$2c_{1}+c_{2}+c_{3}=0$ and $c_{3}\neq0$ & \cellcolor{gris3}$\mathrm{N}_{2}$ (2 polarizations) & \cellcolor{gris3}\tikz{\filldraw[fill=red,draw=black](-0.15,-0.15) rectangle (0.15,0.15);}\\
 \bottomrule
\end{tabular}
\midsepdefault
\caption{Summarizing the cases appearing in Fig.~\ref{fig:ngrpol}.}
\label{table:summarizingcases}
\end{table}

This concludes our discussion of \gls{gw} polarizations in \gls{ngr}. More details regarding the construction details of the Fig.~\ref{fig:ngrpol} can be found in Refs.~\cite{Hohmann:2018xnb,Hohmann:2018jso}. We have seen that depending on the parameters \(c_1, c_2, c_3\) we obtain the \(\mathrm{E}_2\) class \(\mathrm{II}_6\), \(\mathrm{III}_5\), \(\mathrm{N}_3\) or \(\mathrm{N}_2\), with \(\mathrm{N}_3\) filling most of the parameter space. We have also seen that there exists a family of theories besides \gls{tegr} which is of class \(\mathrm{N}_2\) and thus exhibits the same two tensor modes as in \gls{gr}. Theories in this class therefore cannot be distinguished from \gls{gr} by observing the polarizations of \gls{gw} alone.

Finally let us state that, the exact same results hold true for the case of generalized \gls{ngr} described by the Lagrangian $\mathcal{L} = f(T_{1},T_{2},T_{3})$ as introduced in Eq.~\eqref{lagr_f(Tax,Tvec,Tten)}. The only difference is that now the constants $c_{i}$ are generated from the function $f$ as $c_{1}\rightarrow f_{1}:=\left.(\partial f/\partial T_{1})\right|_{T_{1}=0}$,
$c_{2}\rightarrow f_{2}:=\left.(\partial f/\partial T_{2})\right|_{T_{2}=0}$
and $c_{3}\rightarrow f_{3}:=\left.(\partial f/\partial T_{3})\right|_{T_{3}=0}$. Thus at the level of polarizations in a Minkowski background \gls{ngr} and generalized \gls{ngr} cannot be distinguished.

\subsection{Polarization modes in \texorpdfstring{$f(T,B)$}{} gravity} \label{Sec:fTB_pol-modes}

The polarization content of $f(T,B)$ gravity closely resembles that of $f(\lc{R})$ gravity \cite{Liang:2017ahj,Yang:2011cp,Capozziello:2008rq,Capozziello:2009nq,Berry:2011pb}. This is due to the fact that both theories in first order perturbations in Minkowski, predict one massive dynamical scalar field on top of the usual tensor perturbations. The exposition closely follows Ref.~\cite{Farrugia:2018gyz} in what follows. Since the theory does predict a massive \gls{gw} the E(2) classification cannot really be used as it is only valid for small masses. On the other hand, some promising recent progress has been made in using the Newman-Penrose approach for massive particles \cite{Hyun:2018pgn}. Thus, this can be properly studied by using the traditional approach where the geodesic deviation equation \eqref{eq:geodesic_dev} is directly used, which holds for any case. A similar calculation was carried out in Ref.~\cite{Capozziello:2019msc}.
The Weitzenb\"{o}ck gauge is employed both in the background and all perturbative
orders. The first order perturbation of the torsion scalar is found to be
\begin{equation}
T^{(1)}=0\,.
\end{equation}
The torsion scalar assumes no first order perturbation which as a
consequence has 
\begin{equation}
\lc{R}^{(1)}=B{}^{(1)}=\eta^{\mu\nu}\partial_{\rho}\partial_{\nu}\udt{h}{\rho}{\mu}-\lc\Box h\,,
\end{equation}
where $h:=\udt{h}{\mu}{\mu}$ and $\lc\Box:=\partial_{\mu}\partial^{\mu}$.
We expand the $f(T,B)$ function in Taylor series assuming it is expandable
around the background values values $T_{0}$ and $B_{0}$, namely
\begin{alignat}{2}
f(T,B) & =\: & & f(T_{0},B_{0})+f_{T}(T_{0},B_{0})(T-T_{0})+f_{B}(T_{0},B_{0})(B-B_{0})\nonumber\\[0.5ex]
 & \: & & +\frac{1}{2!}f_{TT}(T_{0},B_{0})(T-T_{0})^{2}+\frac{1}{2!}f_{BB}(T_{0},B_{0})(B-B_{0})^{2}\nonumber\\[0.5ex]
 & \: & & +f_{TB}(T_{0},B_{0})(T-T_{0})(B-B_{0})+\mathcal{O}(T^{3},B^{3})\,.
\end{alignat}
Inserting all the above into the field equations of $f(T,B)$ gravity,
namely Eq.~\eqref{fieldequationsfTB}, order by order we obtain 
\begin{subequations}
\begin{alignat}{2}
0 & =\: & & \eta_{\mu\nu}f(0,0)\,,\\[0.5ex]
0 & =\: & & -f_{T}(0,0)\lc{G}_{\mu\nu}^{(1)}+f_{BB}(0,0)\left(\eta_{\mu\nu}\lc\Box-\partial_{\mu}\partial_{\nu}\right)\lc{R}^{(1)}\,,\label{eq:first-order-perturbation-TB}
\end{alignat}
\end{subequations}
where we have used the fact that $\lc{R}^{(1)}=B^{(1)}$, and that $f(0,0)=0$ from the zeroth order condition. The latter condition is another statement for the fact that the arbitrary Lagrangian function does not include a cosmological constant.

We proceed following Ref.~\cite{Abedi:2017jqx} and we define an effective mass by considering the trace of the first-order equation. This is also similar to the $f(\lc{R})$ gravity case. However, our effective mass is different to that of Ref.~\cite{Abedi:2017jqx}. Indeed, by taking the trace 
\begin{equation}
    f_{T}(0,0)\lc{R}^{(1)}+3f_{BB}(0,0)\lc\Box\lc{R}^{(1)}=0\,,\label{eq:trace-TB}
\end{equation}
we identify the effective mass $m$ by bringing the equation in the form $(\lc\Box-m^{2})\delta\lc{R}^{(1)}=0$, which turns out to be \cite{Farrugia:2018gyz}\footnote{This is slightly different to the effective mass that appears in Ref.~\cite{Abedi:2017jqx} where a typo appears which was corrected in Ref.~\cite{Farrugia:2018gyz}.}
\begin{equation}
    m^{2}:=-\dfrac{f_{T}(0,0)}{3f_{BB}(0,0)}\,.
\end{equation}
We remark that in the $|m^{2}|\rightarrow\infty$ limit (for instance when $f_{BB}(0,0)=0$ and $f_{T}(0,0)\neq0$), the equation reduces to that of \gls{gr}. Since it is known that $f(T)$ gravity yields no further \gls{gw} modes~\cite{Bamba:2013ooa}, this special condition leads to a broader class of theories in which at first order yield the \gls{gw} solutions. In the case where $f_{BB}(0,0)\neq0$ we can follow the procedure of $f(\lc{R})$ gravity \cite{Liang:2017ahj,Yang:2011cp,Capozziello:2008rq,Capozziello:2009nq,Berry:2011pb} since $f(\lc{R})$ acts as a particular subclass of $f(T,B)$ gravity, namely $f(-T+B)$ gravity. Firstly, we introduce the tensor $\bar{h}_{\mu\nu}$ to be 
\begin{equation}
    h_{\mu\nu}=\bar{h}_{\mu\nu}-\dfrac{1}{2}\bar{h}\eta_{\mu\nu}+\dfrac{f_{BB}(0,0)}{f_{T}(0,0)}\eta_{\mu\nu}\lc{R}^{(1)}\,,\label{eq:trace-reversed}
\end{equation}
where $\bar{h}$ represents the trace of $\bar{h}_{\mu\nu}$. Similarly to the previous section we consider the non-trivial case of $f_{T}(0,0)\neq0$ (otherwise \gls{gr} cannot be obtained at any limit). This simplifies Eq.~\eqref{eq:first-order-perturbation-TB} to 
\begin{equation}
    \partial^{\rho}\partial_{\nu}\bar{h}_{\rho\mu}+\partial^{\rho}\partial_{\mu}\bar{h}_{\nu\rho}-\eta_{\mu\nu}\partial^{\rho}\partial^{\alpha}\bar{h}_{\rho\alpha}-\lc\Box\bar{h}_{\mu\nu}=0\,.
\end{equation}
As shown in Ref.~\cite{Liang:2017ahj}, it is possible to consider the Lorenz gauge condition $\partial^{\mu}\bar{h}_{\mu\nu}=0$, which simplifies the wave equation to 
\begin{equation}
    \lc\Box\bar{h}_{\mu\nu}=0\,,
\end{equation}
as well as the traceless condition $\bar{h}=0$. This allows for the solution 
\begin{equation}
    \bar{h}_{\mu\nu}=A_{\mu\nu}\exp\left(ik_{\rho}x^{\rho}\right)\,,
\end{equation}
where $k_{\rho}$ is the four-wavevector, $A_{\mu\nu}$ are constant coefficients, $k_{\rho}k^{\rho}=0$, $k^{\mu}A_{\mu\nu}=0$ and $\udt{A}{\mu}{\mu}=0$. The last conditions are the Lorenz gauge and traceless conditions respectively. On the other hand, the solution for Eq.~\eqref{eq:trace-TB} is 
\begin{equation}
    \lc{R}^{(1)}=F\exp\left(ip_{\mu}x^{\mu}\right)\,,\label{eq:Ricci-sol}
\end{equation}
where $F$ is a constant and $p_{\mu}$ is another four-wavevector such that $p_{\mu}p^{\mu}=-m^{2}$. Hence, the full solution for $h_{\mu\nu}$ is constructed as 
\begin{align}
    h_{\mu\nu} & =A_{\mu\nu}\exp\left(ik_{\rho}x^{\rho}\right)+\dfrac{f_{BB}(0,0)}{f_{T}(0,0)}\eta_{\mu\nu}F\exp\left(ip_{\mu}x^{\mu}\right)\,.
\end{align}
Note that from Eq.~\eqref{eq:first-order-perturbation-TB}- \eqref{eq:trace-TB},
the Ricci tensor is found to be 
\begin{equation}
\lc{R}^{(1)}{}_{\mu\nu}=\dfrac{1}{6}\eta_{\mu\nu}\lc{R}^{(1)}-\dfrac{f_{BB}(0,0)}{f_{T}(0,0)}\partial_{\mu}\partial_{\nu}\lc{R}^{(1)}\,,
\end{equation}
from which the solution of the Ricci scalar \eqref{eq:Ricci-sol}
simplifies to 
\begin{equation}
\lc{R}^{(1)}{}_{\mu\nu}=\left(\dfrac{1}{6}\eta_{\mu\nu}-\dfrac{1}{3m^{2}}p_{\mu}p_{\nu}\right)\lc{R}^{(1)}\,.
\end{equation}
Hence, it is trivial to verify that taking the trace part of the above equation, it yields a consistent
relation for the Ricci scalar, as expected.

We proceed by analyzing the polarization states of the \gls{gw} by using the geodesic deviation Eq.~\eqref{eq:geodesic_dev}. Assuming propagation in the $z$ direction we find the linearized components of the Riemann tensor as
\begin{equation}
    \lc{R}_{i0j0}=\dfrac{1}{2}k_{0}^{2}\bar{h}_{ij}-\dfrac{1}{6m^{2}}\left[\eta_{ij}p_{0}^{2}\lc{R}^{(1)}+p_{i}p_{j}\lc{R}^{(1)}\right]\,.
\end{equation}
Therefore, the geodesic deviation becomes 
\begin{subequations}
\begin{alignat}{2}
\ddot{x} & =\: & & \left[\dfrac{1}{2}k_{0}^{2}\bar{h}_{+}+\dfrac{1}{6m^{2}}p_{0}^{2}\lc{R}^{(1)}\right]x+\dfrac{1}{2}k_{0}^{2}\bar{h}_{\times}y\,,\\[0.5ex]
\ddot{y} & =\: & & \left[-\dfrac{1}{2}k_{0}^{2}\bar{h}_{+}+\dfrac{1}{6m^{2}}p_{0}^{2}\lc{R}^{(1)}\right]y+\dfrac{1}{2}k_{0}^{2}\bar{h}_{\times}x\,,\\[0.5ex]
\ddot{z} & =\: & & \dfrac{1}{6m^{2}}\left(p_{0}^{2}-p_{3}^{2}\right)\lc{R}^{(1)}z=-\dfrac{1}{6}\lc{R}^{(1)}z\,,
\end{alignat}
\end{subequations}
where in the last equation we have used that $p_{\mu}p^{\mu}=-m^{2}$. Additionally, since the wave propagates in the $z$-direction, we have used and defined $\bar{h}_{11}=-\bar{h}_{22}\equiv\bar{h}_{+}$ and $\bar{h}_{12}=\bar{h}_{21}\equiv\bar{h}_{\times}$, which represent the massless $+$ and $\times$ polarizations.

As we observe, in the \gls{tegr} and $f(T)$ limit, namely at $|m^{2}|\rightarrow\infty$, the remaining modes are the $+$ and $\times$ polarizations as expected. However, in the case $|m^{2}|<\infty$ we find the presence of the longitudinal and breathing modes in the geodesic deviation equations. The fact that we find a combination of scalar polarizations is due to the fact the scalar component is massive field. Thus $f(T,B)$ gravity includes more polarization modes than $f(T)$ gravity. The reason for this behavior is directly attributed to the boundary term that is not trivialised at first order $B^{(1)}\neq0$.

\subsection{Polarization modes in the teleparallel analogue of Horndeski gravity}

We thoroughly review the calculation of the polarization states in BDLS theory as presented in Ref.~\cite{Bahamonde:2021dqn}. This is the most complex and inclusive case since it contains the most general scalar-tensor form on top of any teleparallel theory that can be build from scalars quadratic to torsion coupled to a scalar field that leads to second order field equations (see Sec.~\ref{sec:BDLS}). The layout will contain two parts, in the first one we will find the \gls{dof} and how the overall theory branches. In the second part we will adapt the polarization analysis using the first as a guide.

\normalsize The starting point is considering the linearized field equations for the BDLS theory, which were calculated by expanding the action in Eqs.~\eqref{HG_2}-\eqref{HG_5} and \eqref{Ltele} up to second order in the perturbation of the tetrad $\delta e^{A}{}_{\mu}$ and the perturbation of scalar field $\delta \phi$. We then perform a change of variables for the tetrad as indicated in Eq.~\eqref{eq:perttens} in order to simplify the notation and facilitate the calculation process. Thus our new perturbative variables are $\tau_{\mu\nu}$, $\delta \phi$ for which we vary the second order action to obtain their field equations as
\begin{subequations}
\begin{alignat}{2}
    W_{\mu\nu} & =\: & & -2G_{\text{Tele},T_{\text{vec}}}\left(\partial^{\lambda}\partial_{\mu}\tau_{\lambda\nu}-\partial_{\mu}\partial_{\nu}\tau-\partial_{\lambda}\partial_{\sigma}\tau^{\lambda\sigma}\eta_{\mu\nu}+\Box\tau\,\eta_{\mu\nu}\right)\nonumber \\[0.5ex]
    & & & +\left(-2G_{\text{Tele},T}+2G_{4}\right)\left(\Box\tau_{(\mu\nu)}-\partial^{\lambda}\partial_{\mu}\tau_{(\nu\lambda)}-\partial^{\lambda}\partial_{\nu}\tau_{(\mu\lambda)}+\partial_{\mu}\partial_{\nu}\tau+\partial_{\lambda}\partial_{\sigma}\tau^{\lambda\sigma}\eta_{\mu\nu}-\Box\tau\,\eta_{\mu\nu}\right)\nonumber \\[0.5ex]
    & & & +\frac{4}{9}G_{\text{Tele},T_{\text{ax}}}\left(\Box\tau_{[\mu\nu]}-\partial^{\lambda}\partial_{\nu}\tau_{[\mu\lambda]}+\partial^{\lambda}\partial_{\mu}\tau_{[\nu\lambda]}\right)+\left(-G_{\text{Tele},I_{2}}+2G_{4,\phi}\right)\left(\partial_{\mu}\partial_{\nu}\delta\phi-\eta_{\mu\nu}\Box\delta\phi\right)\,,\label{Wmunutotal}\\[0.5ex]
    \widehat{W} & =\: & & \left(G_{\text{Tele},I_{2}}-2G_{4,\phi}\right)\left(\Box\tau-\partial_{\lambda}\partial_{\sigma}\tau^{\lambda\sigma}\right)+\left(G_{\text{Tele},\phi\phi}+G_{2,\phi\phi}\right)\,\delta\phi+\left(G_{\text{Tele},X}+G_{2,X}-2G_{3,\phi}\right)\Box\delta\phi\,,\label{Wscalartotal}
\end{alignat}
\end{subequations}
where we have already applied the background field equations
\begin{equation}
    0 = G_{\text{Tele}} + G_{2}\,,\quad 0 = G_{\text{Tele},\phi} + G_{2,\phi}\,.
\end{equation}
In the next step we perform a \gls{svt} decomposition of the field equations \eqref{Wmunutotal} and \eqref{Wscalartotal} by inserting Eq.~\eqref{eq:deltatetradgreek} and splitting them into $3+1$, transforming into Fourier space Eq.~\eqref{eq:fourier_trans_con} and using the gauge invariants variables Eqs.~\eqref{eq:gauge_inv1}-\eqref{eq:gauge_inv2}
we arrive at
\begin{subequations}
\begin{alignat}{2}
    W_{00} & =\: & & k^{2}\Big((G_{\text{Tele},I_{2}}-2G_{4,\phi})\delta\phi+2G_{\text{Tele},T_{\text{vec}}}\Phi-4(G_{4}-G_{\text{Tele},T}+G_{\text{Tele},T_{\text{vec}}})\psi\Big)\,,\label{eq:Scalareq1}\\[0.5ex]
    k^{j}W{}_{j0} & =\: & & ik^{2}\bigl((G_{\text{Tele},I_{2}}-2G_{4,\phi})\delta\dot{\phi}+2G_{\text{Tele},T_{\text{vec}}}\chi k{}^{2}-2(2(G_{4}-G_{\text{Tele},T})+3G_{\text{Tele},T_{\text{vec}}})\dot{\psi}\bigr)\,,\label{eq:Scalareq2}\\[0.5ex]
    \eta_{jl}W{}^{jl} & =\: & & 3(G_{\text{Tele},I_{2}}-2G_{4,\phi})\delta\ddot{\phi}+2k{}^{2}\Big((2(G_{4}-G_{\text{Tele},T})+3G_{\text{Tele},T_{\text{vec}}})\dot{\chi}+(G_{\text{Tele},I_{2}}-2G_{4,\phi})\delta\phi\nonumber \\[0.5ex]
    & & & +2(G_{4}-G_{\text{Tele},T}+G_{\text{Tele},T_{\text{vec}}})\Phi-2(G_{4}-G_{\text{Tele},T}+2G_{\text{Tele},T_{\text{vec}}})\psi\Big)\nonumber \\[0.5ex]
    & & & -6\Big(2(G_{4}-G_{\text{Tele},T})+3G_{\text{Tele},T_{\text{vec}}}\Big)\ddot{\psi}\,,\label{eq:Scalareq3}\\[0.5ex]
    k_{l}\epsilon^{ljk}W_{jk} & =\: & & -\frac{8}{3}iG_{\text{Tele},T_{\text{ax}}}k{}^{2}\bigl(\ddot{\sigma}+k{}^{2}\sigma\bigr)\,,\label{eq:Scalareq4}\\[0.5ex]
    \widehat{W} & =\: & & (G_{\text{Tele},X}+G_{2,X}-2G_{3,\phi})\delta\ddot{\phi}+\delta\phi\Big(G_{\text{Tele},\phi\phi}+G_{2,\phi\phi}+(G_{\text{Tele},X}+G_{2,X}-2G_{3,\phi})k{}^{2}\Big)\nonumber \\[0.5ex]
    & & & +(G_{\text{Tele},I_{2}}-2G_{4,\phi})\bigl(-k{}^{2}(\dot{\chi}+\Phi-2\psi)+3\ddot{\psi}\bigr)\,.\label{eq:Scalareq5}
\end{alignat}
\end{subequations}
On the other hand, the vector part consists of 3 linearly independent equations for the gauge invariant variables $\left(\beta{}_{i},\Sigma{}_{i},\Lambda{}_{i}\right)$
\begin{subequations}
\begin{alignat}{2}
    W_{0j} & =\: & & \frac{1}{9}\Big\{18G_{\text{Tele},T_{\text{vec}}}\ddot{\beta}{}_{j}+k{}^{2}\Big[\Big(2G_{\text{Tele},T_{\text{ax}}}-9(G_{4}-G_{\text{Tele},T})\Big)\Lambda{}_{j}\nonumber \\[0.5ex]
    & & & -\Big(9(G_{4}-G_{\text{Tele},T})+2G_{\text{Tele},T_{\text{ax}}}\Big)\beta{}_{j}\Big]-2\Big(9G_{\text{Tele},T_{\text{vec}}}+2G_{\text{Tele},T_{\text{ax}}}\Big)\dot{\Sigma}{}_{j}\Big\}\,,\label{eq:Vectoreq1}\\[0.5ex]
    W_{j0} & =\: & & \frac{1}{9}\Big\{ k{}^{2}\Big[\Big(2G_{\text{Tele},T_{\text{ax}}}-9(G_{4}-G_{\text{Tele},T})\Big)\beta{}_{j}-\Big(9(G_{4}-G_{\text{Tele},T})+2G_{\text{Tele},T_{\text{ax}}}\Big)\Lambda{}_{j}\Big]\nonumber\\[0.5ex]
    & \: & &+4G_{\text{Tele},T_{\text{ax}}}\dot{\Sigma}{}_{j}\Big\}\,,\label{eq:Vectoreq2}\\[0.5ex]
    k^{l}W{}_{lj} & =\: & & -\frac{1}{9}i\Big\{ k{}^{2}\Big[\Big(9(G_{4}-G_{\text{Tele},T})+2(9G_{\text{Tele},T_{\text{vec}}}+G_{\text{Tele},T_{\text{ax}}})\Big)\dot{\beta}{}_{j}\nonumber \\[0.5ex]
    & & &+\Big(9(G_{4}-G_{\text{Tele},T})-2G_{\text{Tele},T_{\text{ax}}}\Big)\dot{\Lambda}{}_{j} -18G_{\text{Tele},T_{\text{vec}}}\Sigma{}_{j}\Big]+4G_{\text{Tele},T_{\text{ax}}}\ddot{\Sigma}{}_{j}\Big\}\,,\label{eq:Vectoreq3}
\end{alignat}
\end{subequations}
while the tensor field is described by one equation
\begin{align}
    W_{ij} & =(G_{4}-G_{\text{Tele},T})(\ddot{h}_{ij}+k{}^{2}h_{ij})\,.\label{eq:Tensoreq}
\end{align}
In order to study the system as a whole, we can combine all the \gls{svt} sectors into one master matrix as done in \cite{Bahamonde:2021dqn}. The determinant of this master matrix, which is also the principal polynomial $P(k)$ is the most important quantity and it reads as
\begin{align}
P(k) & =-2^{14}3^{-5}(G_{4}-G_{\text{Tele},T}){}^{5}G_{\text{Tele},T_{\text{vec}}}{}^{3}G_{\text{Tele},T_{\text{ax}}}{}^{3}k{}^{12}\bigl(\omega^{2}-k{}^{2}\bigr)^{8}\Bigl(\tilde{c}_{1}+\tilde{c}_{2}\bigl(\omega^{2}-k{}^{2}\bigr)\Bigr)\,,\label{eq:Pk}
\end{align}
where
\begin{subequations}
\begingroup \addtolength{\jot}{2pt}
\begin{alignat}{2}
    \tilde{c}_{1} & :=\: & & 2(G_{\text{Tele},\phi\phi}+G_{2,\phi\phi})(2(G_{4}-G_{\text{Tele},T})+3G_{\text{Tele},T_{\text{vec}}})\,,\label{eq:c1tdef}\\[0.5ex]
    \tilde{c}_{2} & :=\: & & -3(G_{\text{Tele},I_{2}}-2G_{4,\phi})^{2}-2(G_{\text{Tele},X}+G_{2,X}-2G_{3,\phi})(2(G_{4}-G_{\text{Tele},T})+3G_{\text{Tele},T_{\text{vec}}})\,.\label{eq:c2tdef}\\[0.5ex]
    \tilde{c}_{3} & :=\: & & -G_{\text{Tele},\phi\phi}-G_{2,\phi\phi}\,,\label{eq:c3tdef}\\[0.5ex]
    \tilde{c}_{4} & :=\: & & G_{\text{Tele},X}+G_{2,X}-2G_{3,\phi}\,,\label{eq:c4tdef}\\[0.5ex]
    Z_{1} & :=\: & & -\frac{(G_{4}-G_{\text{Tele},T})G_{2,\phi\phi}}{-3G_{4,\phi}{}^{2}+(G_{4}-G_{\text{Tele},T})(2G_{3,\phi}-G_{2,X})}\,,\label{Z1}\\[0.5ex]
    Z_{2} & :=\: & & \frac{\bigl(3(G_{\text{Tele},I_{2}}-2G_{4,\phi})^{2}+2(2(G_{4}-G_{\text{Tele},T})+3(G_{\text{Tele},T_{\text{vec}}}))(G_{\text{Tele},X}+G_{2,X})\bigr)}{4(2(G_{4}-G_{\text{Tele},T})+3(G_{\text{Tele},T_{\text{vec}}}))}\,.\label{Z2}
\end{alignat}
\endgroup
\end{subequations}

In order to properly solve the system we need to find all the cases where $P(k)$ in Eq.~\eqref{eq:Pk} is non-degenerate. This was performed in a exhaustive manner in Ref.~\cite{Bahamonde:2021dqn} and all the results are gathered in the Table~\ref{tab:GW_DoF}.

In the Case 0, the standard Horndeski gravity was explicitly studied which also matches the results reported in Ref.~\cite{Hou:2017bqj,Gong:2018ybk} but a new branch that only entails a massless sector was further discovered. In this sector the only propagating \gls{dof} are the tensorial ones.
For the Case 1, which is the most general version of BDLS, 7 propagating \gls{dof} where found which can be though of as just two extra massless scalars and one massless vector on top of the standard Horndeski \gls{dof}.

The rest of the cases are simpler versions of Case 1 and include various combinations from just tensor modes to combinations of scalar,vector and tensor modes. In Table \ref{tab:dof_classes} there is a link between well known theories of the scalar-tensor form and the full analysis included in Table \ref{tab:GW_DoF}. Hence, in general there is a plethora of sub branches including most combinations of \gls{svt} propagating \gls{dof}.

Using the results regarding the \gls{dof} detailed in Table \ref{tab:GW_DoF} and using the representation of the electric components of the Riemann tensor \eqref{eq:ERiemSVT}, the polarizations of the respective cases were exhaustively studied. The resulting cases~\cite{Bahamonde:2021dqn} of \gls{gr} and $f(T)$ gravity are first shown as the straightforward instances of the analysis. Next, there is the standard Horndeski Case 0.I which assumes tensor polarizations for the massless sector and and both the scalar polarizations for the massive sector. For Case 0.II, the new branch of Horndeski gravity, only tensor polarizations where found as expected, since only tensorial \gls{dof} were found as shown in Table~\ref{tab:GW_DoF}.

In Case 1 (the full BDLS theory), there were found the breathing mode along tensor modes for the massless sector and both scalar modes for the massive sector. One could compare this to standard Horndeski (Case 0.I) as having an extra breathing mode in the massless sector. This exact polarization imprint is, also, only shared with Case 3 despite the fact that the \gls{dof} in these two cases are different. Regarding Cases 2.I.a, 2.I.b, 2.II.b, 5 and 6 in general, although there is a massless scalar \gls{dof}, it does not leave a polarization imprint. This also happens for Cases 2.I.a and 4.I.a where instead, there is a massive scalar. There is also a similar behavior vectorial \gls{dof} for the Cases 1, 2.I.a, 2.I.b, 2.II.a and 2.II.b where although there is a vectorial \gls{dof} it is invisible to the detectors. This phenomenon is directly linked to the fact that these \gls{dof} are not purely metrical or that they do not couple strong enough with the metric.

Hence we see that even in the full BDLS theory Case 1, there are all sorts of scalar-vectorial-tensorial \gls{dof} found, only the scalar and tensorial leave a polarization imprint in end. Obviously there is a plethora of possible combinations described by the various subclasses of how this happens but nevertheless this is a very interesting illustration of how metrical and non-metrical \gls{dof} behave within this system.


\begin{center}
\begin{table}[H]
\begin{centering}
\setlength{\tabcolsep}{7pt}{ 
\global\long\def\arraystretch{1.8}
\begin{adjustbox}{width=\textwidth}
\begin{tabular}{c|c|ccc|cc|c}
\hline 
\textbf{\cellcolor{gris3}Cases} & \textbf{\cellcolor{gris3}Conditions} & \multicolumn{3}{c}{\textbf{\cellcolor{gris3}Sectors}} & \textbf{\cellcolor{gris3}} & \textbf{\cellcolor{gris3}} & \textbf{\cellcolor{gris3}PDoF}\tabularnewline
\hline 
\textbf{\cellcolor{gris1}} & \textbf{\cellcolor{gris1}} & \multicolumn{3}{c|}{\textbf{\cellcolor{gris1}\cellcolor{gris1}}Massless $\left(\omega^{2}=k{}^{2}\right)$\textbf{\cellcolor{gris1}}} & \multicolumn{2}{c|}{\textbf{\cellcolor{gris1}}Massive $\left(\omega^{2}-k{}^{2}=m^{2}\right)$\textbf{\cellcolor{gris1}}} & \textbf{\cellcolor{gris1}}\tabularnewline
\hline 
\textbf{\cellcolor{gris3}} & \textbf{\cellcolor{gris3}} & \textbf{\cellcolor{gris3}}Scalar & \textbf{\cellcolor{gris3}}Vector & \textbf{\cellcolor{gris3}}Tensor & \textbf{\cellcolor{gris3}}Scalar & \textbf{\cellcolor{gris3}}$m^{2}$ & \textbf{\cellcolor{gris3}}\tabularnewline
\textbf{\cellcolor{gris1}}- & \textbf{\cellcolor{gris1}}$G_{4}-G_{\text{Tele},T}\neq0$ & \textbf{\cellcolor{gris1}}- & \textbf{\cellcolor{gris1}}- & \textbf{\cellcolor{gris1}}1 & \textbf{\cellcolor{gris1}}- & \textbf{\cellcolor{gris1}}- & \textbf{\cellcolor{gris1}}2\tabularnewline
\textbf{\cellcolor{gris3}} & \textbf{\cellcolor{gris3}}$G_{\text{Tele},T_{\text{vec}}}=0,\,G_{\text{Tele},T_{\text{ax}}}=0,$ & \textbf{\cellcolor{gris3}} & \textbf{\cellcolor{gris3}} & \textbf{\cellcolor{gris3}} & \textbf{\cellcolor{gris3}} & \textbf{\cellcolor{gris3}} & \textbf{\cellcolor{gris3}}\tabularnewline
\multirow{-2}{*}{\cellcolor{gris3}0} & \textbf{\cellcolor{gris3}}$G_{\text{Tele},I_{2}}=0,G_{\text{Tele},X}=0,G_{\text{Tele},\phi\phi}=0$ & \multirow{-2}{*}{\cellcolor{gris3}} & \multirow{-2}{*}{\cellcolor{gris3}} & \multirow{-2}{*}{\cellcolor{gris3}} & \multirow{-2}{*}{\cellcolor{gris3}} & \multirow{-2}{*}{\cellcolor{gris3}} & \multirow{-2}{*}{\cellcolor{gris3}}\tabularnewline
\textbf{\cellcolor{gris1}} & \textbf{\cellcolor{gris1}}$G_{2,\phi\phi}\neq0$ and & \textbf{\cellcolor{gris1}} & \textbf{\cellcolor{gris1}} & \textbf{\cellcolor{gris1}} & \textbf{\cellcolor{gris1}} & \textbf{\cellcolor{gris1}} & \textbf{\cellcolor{gris1}}\tabularnewline
\multirow{-2}{*}{\cellcolor{gris1}0.I} & \textbf{\cellcolor{gris1}}$-3G_{4,\phi}{}^{2}+(G_{4}-G_{\text{Tele},T})\left(2G_{3,\phi}-G_{2,X}\right)\neq0$ & \multirow{-2}{*}{\cellcolor{gris1}-} & \multirow{-2}{*}{\cellcolor{gris1}-} & \multirow{-2}{*}{\cellcolor{gris1}1} & \multirow{-2}{*}{\cellcolor{gris1}1} & \multirow{-2}{*}{\cellcolor{gris1}$Z_1$} & \multirow{-2}{*}{\cellcolor{gris1}3}\tabularnewline
\textbf{\cellcolor{gris3}} & \textbf{\cellcolor{gris3}}$G_{2,\phi\phi}=0$ and & \textbf{\cellcolor{gris3}} & \textbf{\cellcolor{gris3}} & \textbf{\cellcolor{gris3}} & \textbf{\cellcolor{gris3}} & \textbf{\cellcolor{gris3}} & \textbf{\cellcolor{gris3}}\tabularnewline
\multirow{-2}{*}{\cellcolor{gris3}0.II} & \textbf{\cellcolor{gris3}}$-3G_{4,\phi}{}^{2}+(G_{4}-G_{\text{Tele},T})\left(2G_{3,\phi}-G_{2,X}\right)=0$ & \multirow{-2}{*}{\cellcolor{gris3}-} & \multirow{-2}{*}{\cellcolor{gris3}-} & \multirow{-2}{*}{\cellcolor{gris3}1} & \multirow{-2}{*}{\cellcolor{gris3}-} & \multirow{-2}{*}{\cellcolor{gris3}-} & \multirow{-2}{*}{\cellcolor{gris3}2}\tabularnewline
\textbf{\cellcolor{gris1}} & \textbf{\cellcolor{gris1}}$G_{\text{Tele},T_{\text{vec}}}\neq0,\,G_{\text{Tele},T_{\text{ax}}}\neq0$ & \textbf{\cellcolor{gris1}} & \textbf{\cellcolor{gris1}} & \textbf{\cellcolor{gris1}} & \textbf{\cellcolor{gris1}} & \textbf{\cellcolor{gris1}} & \textbf{\cellcolor{gris1}}\tabularnewline
\multirow{-2}{*}{\cellcolor{gris1}1} & \textbf{\cellcolor{gris1}}$\tilde{c}_{1}\neq0,\tilde{c}_{2}\neq0$ & \multirow{-2}{*}{\cellcolor{gris1}2} & \multirow{-2}{*}{\cellcolor{gris1}1} & \multirow{-2}{*}{\cellcolor{gris1}1} & \multirow{-2}{*}{\cellcolor{gris1}1} & \multirow{-2}{*}{\cellcolor{gris1}$-\tilde{c}_{1}/\tilde{c}_{2}$} & \multirow{-2}{*}{\cellcolor{gris1}7}\tabularnewline
\textbf{\cellcolor{gris3}}2 & \textbf{\cellcolor{gris3}}$G_{\text{Tele},T_{\text{vec}}}\neq0,\,G_{\text{Tele},T_{\text{ax}}}\neq0,\tilde{c}_{1}=0,\tilde{c}_{2}=0$ & \textbf{\cellcolor{gris3}} & \textbf{\cellcolor{gris3}} & \textbf{\cellcolor{gris3}} & \textbf{\cellcolor{gris3}} & \textbf{\cellcolor{gris3}} & \textbf{\cellcolor{gris3}}\tabularnewline
\textbf{\cellcolor{gris1}}2.I & \textbf{\cellcolor{gris1}}$G_{\text{Tele},T_{\text{ax}}}\neq0
$& \textbf{\cellcolor{gris1}} & \textbf{\cellcolor{gris1}} & \textbf{\cellcolor{gris1}} & \textbf{\cellcolor{gris1}} & \textbf{\cellcolor{gris1}} & \textbf{\cellcolor{gris1}}\tabularnewline
\textbf{\cellcolor{gris3}}2.I.a & \textbf{\cellcolor{gris3}}$
\tilde{c}_{3}\neq0,\tilde{c}_{4}\neq0$ & \textbf{\cellcolor{gris3}}1 & \textbf{\cellcolor{gris3}}1 & \textbf{\cellcolor{gris3}}1 & \textbf{\cellcolor{gris3}}1 & \textbf{\cellcolor{gris3}}$-\tilde{c}_{3}/\tilde{c}_{4}$ & \textbf{\cellcolor{gris3}}6\tabularnewline
\textbf{\cellcolor{gris1}}2.I.b & \textbf{\cellcolor{gris1}}$
\tilde{c}_{3}=0,\tilde{c}_{4}=0$ & \textbf{\cellcolor{gris1}}1 & \textbf{\cellcolor{gris1}}1 & \textbf{\cellcolor{gris1}}1 & \textbf{\cellcolor{gris1}}- & \textbf{\cellcolor{gris1}}- & \textbf{\cellcolor{gris1}}5\tabularnewline
\textbf{\cellcolor{gris3}}2.II & \textbf{\cellcolor{gris3}}$G_{3,\phi}=Z_{2},\,G_{2,\phi\phi}=-G_{\text{Tele},\phi\phi}$ & \textbf{\cellcolor{gris3}} & \textbf{\cellcolor{gris3}} & \textbf{\cellcolor{gris3}} & \textbf{\cellcolor{gris3}} & \textbf{\cellcolor{gris3}} & \textbf{\cellcolor{gris3}}\tabularnewline
\textbf{\cellcolor{gris1}}2.II.a & \textbf{\cellcolor{gris1}}$G_{\text{Tele},I_{2}}-2G_{4,\phi}\neq0$ & \textbf{\cellcolor{gris1}}2 & \textbf{\cellcolor{gris1}}1 & \textbf{\cellcolor{gris1}}1 & \textbf{\cellcolor{gris1}}- & \textbf{\cellcolor{gris1}}- & \textbf{\cellcolor{gris1}}6\tabularnewline
\textbf{\cellcolor{gris3}}2.II.b & \textbf{\cellcolor{gris3}}$G_{\text{Tele},I_{2}}-2G_{4,\phi}=0$ & \textbf{\cellcolor{gris3}}1 & \textbf{\cellcolor{gris3}}1 & \textbf{\cellcolor{gris3}}1 & \textbf{\cellcolor{gris3}}- & \textbf{\cellcolor{gris3}}- & \textbf{\cellcolor{gris3}}5\tabularnewline
\textbf{\cellcolor{gris1}}3 & \textbf{\cellcolor{gris1}}$G_{\text{Tele},T_{\text{vec}}}\neq0,\,G_{\text{Tele},T_{\text{ax}}}=0,\tilde{c}_{1}\neq0,\tilde{c}_{2}\neq0$ & \textbf{\cellcolor{gris1}}1 & \textbf{\cellcolor{gris1}}- & \textbf{\cellcolor{gris1}}1 & \textbf{\cellcolor{gris1}}1 & \textbf{\cellcolor{gris1}}$-\tilde{c}_{1}/\tilde{c}_{2}$ & \textbf{\cellcolor{gris1}}4\tabularnewline
\textbf{\cellcolor{gris3}}4 & \textbf{\cellcolor{gris3}}$G_{\text{Tele},T_{\text{vec}}}\neq0,\,G_{\text{Tele},T_{\text{ax}}}=0,\tilde{c}_{1}=0,\tilde{c}_{2}=0$ & \textbf{\cellcolor{gris3}} & \textbf{\cellcolor{gris3}} & \textbf{\cellcolor{gris3}} & \textbf{\cellcolor{gris3}} & \textbf{\cellcolor{gris3}} & \textbf{\cellcolor{gris3}}\tabularnewline
\textbf{\cellcolor{gris1}}4.I & \textbf{\cellcolor{gris1}}$G_{\text{Tele},T_{\text{vec}}}=-\tfrac{2}{3}(G_{4}-G_{\text{Tele},T}),\,G_{4,\phi}=\tfrac{1}{2}G_{\text{Tele},I_{2}}$ & \textbf{\cellcolor{gris1}} & \textbf{\cellcolor{gris1}} & \textbf{\cellcolor{gris1}} & \textbf{\cellcolor{gris1}} & \textbf{\cellcolor{gris1}} & \textbf{\cellcolor{gris1}}\tabularnewline
\textbf{\cellcolor{gris3}}4.I.a & \textbf{\cellcolor{gris3}}$\tilde{c}_{3}\neq0,\tilde{c}_{4}\neq0$ & \textbf{\cellcolor{gris3}}- & \textbf{\cellcolor{gris3}}- & \textbf{\cellcolor{gris3}}1 & \textbf{\cellcolor{gris3}}1 & \textbf{\cellcolor{gris3}}$-\tilde{c}_{3}/\tilde{c}_{4}$ & \textbf{\cellcolor{gris3}}3\tabularnewline
\textbf{\cellcolor{gris1}}4.I.b & \textbf{\cellcolor{gris1}}$\tilde{c}_{3}=0,\tilde{c}_{4}=0$ & \textbf{\cellcolor{gris1}}- & \textbf{\cellcolor{gris1}}- & \textbf{\cellcolor{gris1}}1 & \textbf{\cellcolor{gris1}}- & \textbf{\cellcolor{gris1}}- & \textbf{\cellcolor{gris1}}2\tabularnewline
\textbf{\cellcolor{gris3}}4.II & \textbf{\cellcolor{gris3}}$G_{3,\phi}=Z_{2},\,G_{2,\phi\phi}=-G_{\text{Tele},\phi\phi}$ & \textbf{\cellcolor{gris3}} & \textbf{\cellcolor{gris3}} & \textbf{\cellcolor{gris3}} & \textbf{\cellcolor{gris3}} & \textbf{\cellcolor{gris3}} & \textbf{\cellcolor{gris3}}\tabularnewline
\textbf{\cellcolor{gris1}}4.II.a & \textbf{\cellcolor{gris1}}$G_{\text{Tele},I_{2}}-2G_{4,\phi}\neq0$ & \textbf{\cellcolor{gris1}}1 & \textbf{\cellcolor{gris1}}- & \textbf{\cellcolor{gris1}}1 & \textbf{\cellcolor{gris1}}- & \textbf{\cellcolor{gris1}}- & \textbf{\cellcolor{gris1}}3\tabularnewline
\textbf{\cellcolor{gris3}}4.II.b & \textbf{\cellcolor{gris3}}$G_{\text{Tele},I_{2}}-2G_{4,\phi}=0$ & \textbf{\cellcolor{gris3}}- & \textbf{\cellcolor{gris3}}- & \textbf{\cellcolor{gris3}}1 & \textbf{\cellcolor{gris3}}- & \textbf{\cellcolor{gris3}}- & \textbf{\cellcolor{gris3}}2\tabularnewline
\textbf{\cellcolor{gris1}}5 & \textbf{\cellcolor{gris1}}$G_{\text{Tele},T_{\text{vec}}}=0,\,G_{\text{Tele},T_{\text{ax}}}\neq0,\tilde{c}_{1}\neq0,\tilde{c}_{2}\neq0$ & \textbf{\cellcolor{gris1}}1 & \textbf{\cellcolor{gris1}}- & \textbf{\cellcolor{gris1}}1 & \textbf{\cellcolor{gris1}}1 & \textbf{\cellcolor{gris1}}$-\tilde{c}_{1}/\tilde{c}_{2}$ & \textbf{\cellcolor{gris1}}4\tabularnewline
\textbf{\cellcolor{gris3}}6 & \textbf{\cellcolor{gris3}}$G_{\text{Tele},T_{\text{vec}}}=0,G_{\text{Tele},T_{\text{ax}}}\neq0,\tilde{c}_{1}=0,\tilde{c}_{2}=0$ & \textbf{\cellcolor{gris3}}1 & \textbf{\cellcolor{gris3}}- & \textbf{\cellcolor{gris3}}1 & \textbf{\cellcolor{gris3}}- & \textbf{\cellcolor{gris3}}- & \textbf{\cellcolor{gris3}}3\tabularnewline
\textbf{\cellcolor{gris1}}7 & \textbf{\cellcolor{gris1}}$G_{\text{Tele},T_{\text{vec}}}=0,\,G_{\text{Tele},T_{\text{ax}}}=0,\,\tilde{c}_{1}\neq0,\tilde{c}_{2}\neq0$ & \textbf{\cellcolor{gris1}}- & \textbf{\cellcolor{gris1}}- & \textbf{\cellcolor{gris1}}1 & \textbf{\cellcolor{gris1}}1 & \textbf{\cellcolor{gris1}}$-\tilde{c}_{1}/\tilde{c}_{2}$ & \textbf{\cellcolor{gris1}}3\tabularnewline
\textbf{\cellcolor{gris3}}8 & \textbf{\cellcolor{gris3}}$G_{\text{Tele},T_{\text{vec}}}=0,\,G_{\text{Tele},T_{\text{ax}}}=0,\tilde{c}_{1}=0,\tilde{c}_{2}=0$ & \textbf{\cellcolor{gris3}}- & \textbf{\cellcolor{gris3}}- & \textbf{\cellcolor{gris3}}1 & \textbf{\cellcolor{gris3}}- & \textbf{\cellcolor{gris3}}- & \textbf{\cellcolor{gris3}}2\tabularnewline
\hline 
\end{tabular}\end{adjustbox}}
\par\end{centering}
\caption{All branches of the BDLS theory represented according to their PDoF (propagating degree of freedom). To each of the scalar, vector and tensor components correspond 1,2 and 2 \gls{dof} respectively. The quantities $\tilde{c}_{i}$ are defined in~\eqref{eq:c1tdef},~\eqref{eq:c2tdef},~\eqref{eq:c3tdef} and \eqref{eq:c4tdef} while $Z_{1}$ and $Z_{2}$ are defined in Eqs.~\eqref{Z1}-\eqref{Z2}) respectively.}
\label{tab:GW_DoF}
\end{table}
\par\end{center}

\begin{center}
\begin{table}[H]
\begin{centering}
\midsepremove
\setlength{\tabcolsep}{7pt}
\global\long\def\arraystretch{1.7}
{\resizebox{\columnwidth}{!}{ 
\begin{tabular}{lccc}
\toprule
\multicolumn{1}{c}{\textbf{Theory}\cellcolor{gris3}} & \multicolumn{1}{c}{\cellcolor{gris3}\textbf{Case}} & \multicolumn{1}{c}{\cellcolor{gris3}\textbf{PDoF}} & \multicolumn{1}{c}{\cellcolor{gris3}\textbf{Lagrangian Density $\mathcal{L}_{i}$ $\left(\mathcal{S}_{i}=\frac{1}{2\kappa^{2}}\int d^{4}x\,e\,\mathcal{L}_{i}\right)$}}\tabularnewline
\midrule
\gls{gr} (or \gls{tegr}) or $f(T)$ (Sec.~\ref{sec:f(T)gravity}) \cellcolor{gris1} & \cellcolor{gris1}- & \cellcolor{gris1}2 & \cellcolor{gris1}$\lc R$ or $f(T)$\tabularnewline
Horndeski (Sec.~\ref{sec:BDLS}) \cellcolor{gris3} & \cellcolor{gris3}0.I & \cellcolor{gris3}3 & \cellcolor{gris3}Eqs.~\eqref{HG_2}-\eqref{HG_5}\tabularnewline
$G_{{\rm Tele}}$ (Sec.~\ref{sec:BDLS}) \cellcolor{gris1} & \cellcolor{gris1}1 & \cellcolor{gris1}7 & \cellcolor{gris1}$G_{\text{Tele}}\left(\phi,X,T,T_{\text{ax}},T_{\text{vec}},I_{2},J_{1},J_{3},J_{5},J_{6},J_{8},J_{10}\right)$\tabularnewline
Extented \gls{ngr} (Sec.~\ref{Sec:Ext_NGR}) \cellcolor{gris3} & \cellcolor{gris3}2.I.b & \cellcolor{gris3}5 & \cellcolor{gris3}$f(T,T_{{\rm ax}},T_{{\rm vec}})$\tabularnewline
Generalized teleparallel dark energy (Sec.~\ref{Sec:conf-trans-fTB}) \cellcolor{gris1} & \cellcolor{gris1}7 & \cellcolor{gris1}3 & \cellcolor{gris1}$-A(\phi)T-\frac{1}{2}\partial_{\mu}\phi\partial^{\mu}\phi-V(\phi)$\tabularnewline
Generalized Teleparallel Scalar-Tensor (Sec.~\ref{Sec5:scalar-torsion_models}) \cellcolor{gris3} & \cellcolor{gris3}7 & \cellcolor{gris3}3 & \cellcolor{gris3}$F(\phi)T+P(\phi,X)-G_{3}(\phi,X)\Box\phi$\tabularnewline
Tachyonic teleparallel gravity (Sec.~\ref{Sec5:scalar-torsion_models}) \cellcolor{gris1} & \cellcolor{gris1}7 & \cellcolor{gris1}3 & \cellcolor{gris1}$f(T,X,\phi)$\tabularnewline
\bottomrule
\end{tabular}}}
\par\end{centering}
\midsepdefault
\caption{Common models in the literature (see Sec.~\ref{sec5:extended}) shown against the analysis presented in Table~\ref{tab:GW_DoF}.}
\label{tab:dof_classes}
\end{table}
\par\end{center}

\clearpage

\section{Astrophysical Systems}\label{sec6:astrophysics}
This section will be devoted to studying the most important results and problems in astrophysics in \gls{tg}. We will focus on Solar System tests and review the most important results regarding compact objects such as stars, black holes and wormholes. Since this is an active field of research, there are many important issues that have not been addressed yet. Some of them will be listed at the end of this section.

\subsection{Spherical symmetry and solutions}\label{sub:sphsolu}
There are several studies aiming to find spherically symmetric exact or perturbed solutions in different \gls{tg}. There are no good tetrad-spin connection pairs in \gls{tg} for the most general time-dependent spherically symmetric metric $\dd s^2=\mathcal{A}(r,t)\dd t^2-\mathcal{B}(r,t)\dd r^2-\mathcal{M}(r,t)\dd \Omega^2$ reported in the literature. At a first ingredient at this, one needs to consider a good tetrad-spin connection pair in spherical symmetry (see Sec.~\ref{subsec:goodtetradsph}) and since this pair depends on the theory, one should be careful in checking if Eq.~\eqref{goodspher} is still a good tetrad-spin connection pair in the studied theory at hand.

The first paper studying spherical symmetry in modified \gls{tg} was Ref.~\cite{Ferraro:2011ks} where the authors found an exact solution in $f(T)$ gravity which behaves as the standard Schwarzschild solution. The problem of this solution is that the torsion scalar becomes zero ($T=0$), therefore, it is trivial since $f(T)$ becomes \gls{tegr}. There are other vacuum exact solutions in $f(T)$ in the literature having a similar property with $T=\textrm{const}.$~\cite{Paliathanasis:2014iva} or also with $T=0$~\cite{Bejarano:2017akj}, which again are trivial since those solutions become \gls{tegr} plus a cosmological constant (or just \gls{tegr}). In Ref.~\cite{Daouda:2012nj}, the authors found different solutions with $T\neq \textrm{const.}$, but the analysis was not complete since they did not consider one of the field equations. It is also important to mention that there are other several incorrect works in the literature that also did not consider a good tetrad-spin connection pair, and then, they ignored the appearance of a nonvanishing antisymmetric field equation~\cite{HamaniDaouda:2011iy,Meng:2011ne,Capozziello:2012zj,Houndjo:2013qna,Kofinas:2015hla}. As discussed in Sec.~\ref{subsec:goodtetradsph}, one cannot use any arbitrary spin connection-tetrad pair since they must solve the antisymmetric field equations of the theory. In several earlier works, different authors found exact solutions in $f(T)$ gravity but they used the incorrect tetrad in the Weitzenb\"{o}ck gauge (not a good tetrad) which in the end is manifested in a non-zero antisymmetric field equation of the form of $f_{TT} T'=0$ which is only satisfied for the \gls{gr} case.

The first non-trivial vacuum exact spherically symmetric solutions in \gls{tg} (different to just \gls{gr}+$\Lambda$) was derived in Ref.~\cite{Bahamonde:2019jkf}. In this paper, several solutions were found in $f(T,B)$ gravity using the Noether's symmetry approach. The problem of these solutions is that all of them have a metric with a power-like form $\mathcal{A}(r)\propto r^m$ with $\mathcal{B}(r)=\textrm{const}$. Hence, all of these solutions might not be so physically interesting since they cannot describe black holes nor standard astrophysical systems. Two recent papers found the first non-trivial exact black hole solutions in $f(T,B)$ gravity and also in teleparallel scalar-tensor theories~\cite{Bahamonde:2021srr,Bahamonde:2022lvh}. These solutions are based on the complex tetrad in Eq.~\eqref{tetrad2new}. Still, these solutions have not been analysed in detail to know if they might provide realistic black hole solutions. From these works, one can then state that for a generic $f(T,B)$ theory, the Birkhoff's theorem is not satisfied.

Since it is hard to find exact solutions, it is also helpful to use perturbation techniques to find perturbed solutions in \gls{tg}. The main idea behind this is to use \gls{gr} in the background and add small perturbations related to the modification of \gls{tegr}. Regarding perturbed spherically symmetric solutions, the first study in \gls{tg} was done in Ref.~\cite{Ruggiero:2015oka}. The authors used the tetrad in Eq.~\eqref{goodspher} with $\xi=-1$ and found perturbed solutions around Minkowski. It is important to recall here that for this tetrad, $T$ does not become zero for Minkowski unless one assumes a non-trivial limit in the metric functions. This means that even though the metric is Minkowski, the connection (teleparallel one) is non-trivial. This idea was then further explored for $\xi=1$ in Ref.~\cite{DeBenedictis:2018wkp}, where the authors found perturbed solutions around Schwarzschild in $-T-\epsilon (\alpha/2) T^p$ with $p=2$. This study was then generalized in Ref.~\cite{Bahamonde:2019jkf} for different $p>1$ with $p\in \mathbb{Z} $. Later, in Ref.~\cite{Bahamonde:2020bbc,Bahamonde:2021srr} the authors found more perturbed solutions in $f(T,B)$ gravity for the two cases and also for the complex tetrad. We will now provide some of the most important results for the case $\xi=1$ since this study does not have the issue of $T$ and $B$ being nonvanishing in the Minkowski limit.

The $f(T,B)$ gravitational field equations~\eqref{fieldequationsfTB} in spherical symmetry with the good tetrad-spin connection pair given by the tetrad~\eqref{goodspher} with $\xi=1$ and $\mathcal{M}(r)=r$, which gives the metric
\begin{equation}
    \dd s^2= \mathcal{A}(r)\dd t^2- \mathcal{B}(r)\dd r^2 - r^2(\dd \vartheta^2+\sin^2\varphi\, \dd \varphi^2)\,,\label{metricspher}
\end{equation}
and with a zero spin connection become
\begin{subequations}
\begin{align}
     \kappa ^2 \rho & =-\frac{1}{2} f-\frac{ r \mathcal{B} (\sqrt{\mathcal{B}}-1) \mathcal{A}'+\mathcal{A} (r \mathcal{B}'+2 \mathcal{B}^{3/2}-2 \mathcal{B})}{ r^2 \mathcal{A} \mathcal{B}^2}f_T+\frac{r \mathcal{B}' f'_B-4 \mathcal{B}^{3/2} \left(f'_B+f'_T\right)+4 \mathcal{B} f'_T}{2 r \mathcal{B}^2}\nonumber\\[0.5ex]
    & -\frac{r^2 \mathcal{B} \mathcal{A}'^2+r \mathcal{A} \Big[r \mathcal{A}' \mathcal{B}'+4 \mathcal{B}^{3/2} \mathcal{A}'-2 \mathcal{B} (r \mathcal{A}''+4 \mathcal{A}')\Big]+4 \mathcal{A}^2 (r \mathcal{B}'+2 \mathcal{B}^{3/2}-2 \mathcal{B})}{4 r^2 \mathcal{A}^2 \mathcal{B}^2}f_B-\frac{f''_B}{ \mathcal{B}}\,,\label{Eq1}\\[0.5ex]
    - \kappa ^2 p_r &= -\frac{1}{2} f-\frac{r \mathcal{A}'+4 \mathcal{A} }{2 r \mathcal{A} \mathcal{B}}f'_B-\frac{ r (\sqrt{\mathcal{B}}-2) \mathcal{A}'+2 \mathcal{A} (\sqrt{\mathcal{B}}-1)}{ r^2 \mathcal{A} \mathcal{B}}f_T\nonumber\\[0.5ex]
    &+\frac{ -r^2 \mathcal{B} \mathcal{A}'^2+r \mathcal{A} \left[-r \mathcal{A}' \mathcal{B}'-4 \mathcal{B}^{3/2} \mathcal{A}'+2 \mathcal{B} (r \mathcal{A}''+4 \mathcal{A}')\right]-4 \mathcal{A}^2 \left(r \mathcal{B}'+2 \mathcal{B}^{3/2}-2 \mathcal{B}\right)}{4r^2 \mathcal{A}^2 \mathcal{B}^2}f_B\,,\label{Eq2}\\[0.5ex]
    - \kappa ^2 p_l &= -\frac{1}{2} f+\frac{r \mathcal{A}'-2 \mathcal{A} (\sqrt{\mathcal{B}}-1)}{2 r \mathcal{A} \mathcal{B}}f'_T+\frac{r \mathcal{B}'-2 \mathcal{B}^{3/2} }{2 r \mathcal{B}^2}f'_B-\frac{f''_B}{ \mathcal{B}}\nonumber\\[0.5ex]
    & +\frac{-r^2 \mathcal{B} \mathcal{A}'^2+r \mathcal{A} \left[-r \mathcal{A}' \mathcal{B}'-4 \mathcal{B}^{3/2} \mathcal{A}'+2 \mathcal{B} (r \mathcal{A}''+3 \mathcal{A}')\right]+\mathcal{A}^2 (-2 r \mathcal{B}'-8 \mathcal{B}^{3/2}+4 \mathcal{B}^2+4 \mathcal{B})}{4 r^2 \mathcal{A}^2 \mathcal{B}^2}f_T\nonumber\\[0.5ex]
    & +\frac{-r^2 \mathcal{B} \mathcal{A}'^2+r \mathcal{A} \left[-r \mathcal{A}' \mathcal{B}'-4 \mathcal{B}^{3/2} \mathcal{A}'+2 \mathcal{B} (r \mathcal{A}''+4 \mathcal{A}')\right]-4 \mathcal{A}^2 (r \mathcal{B}'+2 \mathcal{B}^{3/2}-2 \mathcal{B})}{4 r^2 \mathcal{A}^2 \mathcal{B}^2}f_B \,,\label{Eq3}
\end{align}
\end{subequations}
where we have assumed an anisotropic fluid with energy density $\rho$ and lateral and radial pressures given by $p_{l}$ and $p_r$ respectively. Here, primes denote differentiation \gls{wrt} the radial coordinate, i.e., $f_T'=f_{TT}T'+f_{TB}B'$ and the scalar-torsion and boundary term yield
\begin{subequations}
\begin{align}
T&=-\frac{2 \left(\sqrt{\mathcal{B}}-1\right) \left(r \mathcal{A}'-\mathcal{A} \sqrt{\mathcal{B}}+\mathcal{A}\right)}{r^2 \mathcal{A} \mathcal{B}}\,,\\[0.5ex]
B&=\frac{-r^2 \mathcal{B} \mathcal{A}'^2+r \mathcal{A} \left(-r \mathcal{A}' \mathcal{B}'-4 \mathcal{B}^{3/2} \mathcal{A}'+2 \mathcal{B} \left(r \mathcal{A}''+4 \mathcal{A}'\right)\right)-4 \mathcal{A}^2 \left(r \mathcal{B}'+2 \mathcal{B}^{3/2}-2 \mathcal{B}\right)}{2 r^2 \mathcal{A}^2 \mathcal{B}^2}\,.
\end{align}
\end{subequations}
These two quantities are zero for the Minkowski case $\mathcal{A}=\mathcal{B}=1$. As discussed above, there are non-trivial vacuum exact solutions but all of them have $\mathcal{B}=\textrm{const}$. One example is the following vacuum exact solution~\cite{Bahamonde:2019jkf}
\begin{equation}
    \mathcal{A}(r)^2=\left( \displaystyle\frac{r}{r_0} \right)^{\frac{4 n (n-1)(2n -3)}{4 n^2 -8 n +5}}\,,\quad \mathcal{B}(r)^2=\textrm{const}=\frac{ (2n-1)(4n -5)}{4 n^2 - 8n +5}\,,\quad f(T,B)=f_0 T^{n}\,.\label{solsph}
\end{equation}
It is important to mention that the standard conservation equation for the fluid holds. This can be seen by differentiating Eq.~\eqref{Eq2} \gls{wrt} $r$ and then using Eqs.~\eqref{Eq1}--\eqref{Eq3} accordingly, finding the standard conservation equation
\begin{equation}
    p'_r=-(\rho+p_r)\frac{\mathcal{A}'}{2\mathcal{A}}+\frac{2}{r}(p_l-p_r)\,,\label{TOV}
\end{equation}
which is the standard conservation equation obtained in \gls{gr}. This equation follows from the the fact that the theory is formulated within the standard teleparallel description of gravity which is diffeomorphism invariant (see Sec.~\ref{sssec:genenmomcons}). Thus, one finds that the energy-momentum tensor satisfies the standard conservation equation $\lc{\nabla}^{\mu}\Theta_{\mu\nu} = 0$.

Let us emphasise here that the system~\eqref{Eq1}--\eqref{Eq3} only has two independent equations for the vacuum case. In Ref.~\cite{Golovnev:2021htv,Golovnev:2020las}, the authors noticed that for $f_T\neq 0$ and $r (\mathcal{B}-2) \mathcal{A}'+\mathcal{A} (\mathcal{B}-1)\neq0$, these equations can be rewritten in the case of vacuum $f(T)$ gravity as
\begin{subequations}
\begin{align}
    0&=-r^2 \mathcal{A}'^2-r^2 \mathcal{A} \left(\mathcal{A}' \mathcal{B}'-(\mathcal{B}-1) \mathcal{A}''\right)+\mathcal{A}^2 (\mathcal{B}-1)^2 (\mathcal{B}+1)\label{eq:eqsymm}\,,\\[0.5ex]
    0&=-\frac{1}{2} f-\frac{ r (\sqrt{\mathcal{B}}-2) \mathcal{A}'+2 \mathcal{A} (\sqrt{\mathcal{B}}-1)}{r^2 \mathcal{A} \mathcal{B}}f_T\,,\label{eq:eqsymm2}
\end{align}
\end{subequations}
where the first equation does not depend on the form of $f$. This can be achieved by solving Eqs.~\eqref{Eq1}--\eqref{Eq2} for $f_T'$ and $f_T$ and then by replacing them in Eq.~\eqref{Eq3}. By doing this, one can eliminate all the $f$ dependence and then one finds that for any generic form of $f$, the metric functions must satisfy~\eqref{eq:eqsymm}. This equation is not easily solvable for $\mathcal{A}$ or $\mathcal{B}$ unless one assumes something for those functions. One interesting result that one finds is that when we assume the ansatz $\mathcal{B}=1/\mathcal{A}$, which is the form of many well known black hole solutions (like Schwarzschild), one gets that the unique solution of Eq.~\eqref{eq:eqsymm} is a Schwarzschild de-Sitter metric $\mathcal{A}=1-2M/r+\Lambda r^2$. Plugging this result into the second equation~\eqref{eq:eqsymm2}, we find that $f(T)=c_1 T-6\Lambda$. This means that the unique solution in $f(T)$ gravity which behaves as $\mathcal{B}=1/\mathcal{A}$ is a trivial one when the theory becomes \gls{tegr} plus a cosmological constant. This statement was shown for the two possible tetrads (the real and complex ones) in Ref.~\cite{Bahamonde:2021srr}. In other words, this means that any solution in $f(T)$ gravity beyond \gls{gr} must have $\mathcal{B}\neq 1/\mathcal{A}$.

Let us now find perturbed solutions by assuming a Schwarzschild background
\begin{subequations}
\begin{align}
    \mathcal{A}(r)&=1-\frac{2M}{r}+\epsilon\, a(r)\,,\label{AA}\\[0.5ex]
    \mathcal{B}(r)&=\Big(1-\frac{2M}{r}\Big)^{-1}+\epsilon \,b(r)\,,\label{BB}
\end{align}
\end{subequations}
where $M$ is the Schwarzschild mass that can be arbitrary large and $\epsilon\ll 1$ is a small tracking parameter. Up to first order in $\epsilon$, there are not solutions around Minkowski ($M=0$). Moreover, to get solutions around Minkowski, one would need to go beyond first order in $\epsilon$. It can be proved that for power-law $f(T)$ gravity, there is only contributions in $a(r)$ and $b(r)$ if one makes perturbations up to fourth order in $\epsilon$. On the contrary, when one uses the other choice of the tetrad~\eqref{goodspher} (with $\xi=-1$) and also the complex tetrad~\eqref{tetrad2new} which have nonvanishing torsion scalars at Minkowski, one finds non-trivial perturbed solutions around Minkowski (see Sec.~\ref{sec:galaxy} to see the real case for $\xi=-1$).

For the $M\neq0$ one can get perturbed solutions around Schwarzschild by only considering perturbations up to first order in $\epsilon$ for the following form of $f(T,B)$ as
\begin{equation}\label{model_choice}
    f(T,B)= -T-\frac{1}{2} \epsilon \left(\alpha T^q+\beta B^m+\gamma B^s T^w+\zeta (\xi T+\chi B )^u\right)\,.
\end{equation}
Different solutions were found in Ref.~\cite{Bahamonde:2020bbc} depending on the parameters chosen. As an example, for the case $q=m=2$, $\zeta=0$ and $w=-s+1$, one gets the following perturbed solution
\begin{subequations}
\begin{alignat}{2}
    \mathcal{A}(r)& =\: & &\mu^2+\epsilon\Big[C_2-\frac{1}{r^2(\mu^2-1)^2}\Big(3 \beta  \mu^7-\frac{1}{2} (\alpha +13 \beta ) \mu^6-4 \beta \mu^5+\frac{1}{2} \mu^4 (15 \alpha +43 \beta +2 C_1r) \nonumber\\[0.5ex]
    & \: & &-\frac{2}{3}\mu^3 (32 \alpha +35 \beta )-\frac{1}{2} \mu^2 (31 \alpha +51 \beta +4 C_1r)+4 \beta \mu+\frac{1}{2} (17 \alpha +21 \beta +2 C_1r)\nonumber\\[0.5ex]
    & \: & &-\frac{\beta }{\mu}+2 (\alpha +\beta ) \left(3 \mu ^2-1\right) \log\mu \Big)\Big]\,,\label{sec:solper1}\\[0.5ex]
    \mathcal{B}(r)& =\: & &\mu^{-2}+\epsilon\Big[\frac{C_1}{r \mu^4}+\frac{C_2 \left(\mu^2-1\right)}{\mu^4}+\frac{1}{r^2(\mu^2-1)}\Big(\frac{1}{2} (25 \alpha +37 \beta )-4 (\alpha +2 \beta ) \mu\nonumber\\[0.5ex]
    &\: & &-\frac{2 (16 \alpha +13 \beta )}{3 \mu}-\frac{2 (\alpha +3 \beta )}{\mu^2}+\frac{4 (\alpha +\beta )}{\mu^3}+\frac{-21 \alpha -25 \beta }{2 \mu^4}+\frac{2 \beta }{\mu^5}+\frac{2 (\alpha +\beta ) }{\mu^4}\log\mu\Big)\Big]\,,\nonumber\\[0.5ex]\label{sec:solper2}
\end{alignat}\label{eq:solper_com}
\end{subequations}
where $\mu^2=1-2M/r$ and $C_i$ are integration constants. One important remark is that $C_i$ can be set by imposing different physical situations. One can see this by expanding the above solutions up to $1/r$, yielding
\begin{subequations}
\begin{align}
    a(r)&\sim \Big(C_2+\frac{16 (\alpha +\beta )}{3 M^2}\Big)-\Big(C_1+\frac{16 (\alpha +\beta )}{M}\Big)\frac{1}{r}+\mathcal{O}\Big(\frac{1}{r^2}\Big) \,,\\[0.5ex]
    b(r)&\sim \Big(C_1-2 C_2 M+\frac{16 (\alpha +\beta )}{3 M} M\Big)\frac{1}{r}+\mathcal{O}\Big(\frac{1}{r^2}\Big)\,.
\end{align}
\end{subequations}
In Ref.~\cite{Bahamonde:2020bbc}, they chose the constants as
\begin{equation}
    C_1=-\frac{16\alpha +\beta}{M}\,,\quad C_2=-\frac{16 \alpha +\beta}{3 M^2}\label{CCcase1}
\end{equation}
to recover the Schwarzschild metric at infinity. Since these constants are related to the boundary conditions of the system, one can also choose them in other ways. The above solution is also a generalization of the solution found in Ref.~\cite{DeBenedictis:2018wkp} where a power-law $f(T)$ gravity was considered with $\beta=0$. For more details about the specific form of $a(r)$ and $b(r)$ for other solutions, see Ref.~\cite{Bahamonde:2020bbc}. Taking a similar approach, another spherically symmetric solution around Schwarzschild was found~\cite{Bahamonde:2020vpb} in a theory concerning the five scalars that one can construct from contractions of torsion, which is an extension of new \gls{gr} (see Sec.~\ref{Sec:Ext_NGR}). 

Let us here finish this section by mentioning two recent papers where exact solutions were found for the complex tetrad~\eqref{tetrad2new}. The first interesting solution presented in Ref.~\cite{Bahamonde:2021srr} reads
\begin{eqnarray}
    \dd s^2=\Big(1-\frac{2M}{r}+\frac{Q}{r^2}\Big)\dd t^2-\Big(\frac{2 Mr-Q-r^2}{2 Q-r^2}\Big)^{-1}\dd r^2-r^2d\Omega^2\,,
\end{eqnarray}with 
\begin{eqnarray}
f(T)=4 f_0 \, \frac{\left(2\pm\sqrt{Q^2 T^2-2 Q T+4}\right)}{\left(Q T+2\pm\sqrt{Q^2 T^2-2 Q T+4}\right) \sqrt{8-2 Q T\pm 4 \sqrt{Q^2 T^2-2Q T+4}}}\,.
\end{eqnarray} This exact solution is similar to Reissner–Nordstr\"om one but with $g_{rr}\neq -1/g_{tt}$ and hence, it describes a black hole with two event horizons. A second interesting black hole solution was found for the Born-Infeld $f(T)$ gravity $f(T)=\lambda\Big(\sqrt{1+\frac{2T}{\lambda}}-1\Big)$ and is given by
\begin{equation}
    \dd s^2=\frac{a_1^2 }{r}\Big[\sqrt{\lambda } (a_0 \lambda +r)-2 \tan ^{-1}\left(\frac{\sqrt{\lambda } r}{2}\right)\Big]\dd t^2-\frac{\lambda ^{5/2} r^5}{(4 + r^2 \lambda)^2}\Big[\sqrt{\lambda } (a_0 \lambda +r)-2 \tan ^{-1}\left(\frac{\sqrt{\lambda } r}{2}\right)\Big]^{-1}\dd r^2-r^2\dd \Omega^2\,,\label{eq:com_f_T_metric}
\end{equation}
where $a_0,a_1$ are integration constants. By fixing the constants accordingly, this spacetime is asymptotically flat and it describes a black hole with one horizon. Another recent paper~\cite{Bahamonde:2022lvh} found that for the teleparallel scalar-tensor theory presented in Sec.~\ref{action:scalar-torsion-coupling-T-and-B}, it is possible to find exact scalarized black hole solutions. One example of them are the so-called Bocharova–Bronnikov–Melnikov–Bekenstein (BBMB) provided by a theory with a coupling between the scalar field and the torsion scalar. For more details about other solutions, we refer the reader to see Ref.~\cite{Bahamonde:2022lvh}. Further new analysis are needed to understand the nature of these solutions as long as if they are healthy or not since they were constructed from the complex tetrad.

\subsection{Generalized Birkhoff's theorem} \label{sec:birkhoff}

The Birkhoff's theorem states that in \gls{gr} (in vacuum), the spherically symmetric spacetime must be static and its solution is described by the Schwarzschild metric. Since we know that $f(T,B)$ gravity has exact spherically symmetric solutions that differs from Schwarzschild (see for example Eq.~\eqref{solsph}), then we know that the generalized Birkhoff's theorem will not hold for all forms of $f$. It is obvious to mention that if $T,B$ are constants, then the dynamics of $f(T,B)$ will be just \gls{tegr} plus a cosmological constant. Then, for any $f$, the unique vacuum spherically symmetric solution is Schwarzschild de-Sitter with $\Lambda$ being related to the constants $T_0$ and $B_0$. Then, when $T_0=B_0=0$, we will recover the standard Schwarzschild solution and the generalized Birkhoff's theorem will hold. This statement will also be valid for more general teleparallel theories having other scalars. However, this case is not so interesting since it is just \gls{gr} plus $\Lambda$. For example, in $f(T)$ gravity, if $T=T_0$ (constant), the field equation~\eqref{eq:f(T)contracted} contracted with two tetrads becomes
\begin{equation}
    \lc{G}_{\alpha\beta}-\frac{1}{2}\frac{1}{f_T}(f-T_0f_T)g _{\alpha\beta}=\kappa^2 \frac{1}{f_T}\Theta_{\alpha\beta}\,,
\end{equation}
where we have used $\partial_\nu f_T=0$ and we have divided by $f_T$. Now, since $T=T_0$, all the functions evaluated at $T_0$ will be just constants, meaning that one define $\frac{1}{f_T}(f-T_0f_T)=\textrm{const.}:=\Lambda_{\rm eff}$. Thus, the $f(T)$ field equations with $T$ constant yields
\begin{equation}
    \lc{G}_{\alpha\beta}-\frac{1}{2}\Lambda_{\rm eff} g _{\alpha\beta}=\kappa^2 \frac{1}{f_T}\Theta_{\alpha\beta}=:\tilde{\kappa}^2\Theta_{\alpha\beta}\,,
\end{equation}
which clearly is the Einstein field equations with an extra effective cosmological constant.

In the following, we will ask the following question: can one have time-dependent vacuum spherical symmetry in a general modified teleparallel theory? To explore this, let us first take the most general time-dependent spherically symmetric tetrad given by Eq.~\eqref{eq:sphertetradwb} and consider $f(T,B,\phi,X)$ gravity. The antisymmetric field equation of this theory reads
\begin{equation}
    0=\partial_{[\mu}( f_T+f_B)T^{\mu}{}_{\lambda\nu]}=\frac{1}{3}\Big(T_{\lambda} \partial_\nu (f_T+ f_B)-T_{\nu} \partial_\lambda (f_T+f_B)+T^{\mu}{}_{\lambda\nu}\partial_\mu (f_T+f_B)\Big)\,.\label{antifTBphiX}
\end{equation}
In general, this equation depends on the form of $f$ and also on the form of the tetrad that can give a dependency on $r$ and $t$ in $T,B$ and $\phi$. Let us study this equation without assuming an explicit form of the function $f$. This strategy of course does not consider all the possible models constructed from $f$ since it could be that for some models, the equation is solved by just replacing the form of $f$ and also $T$ and $B$ in the above antisymmetric field equation.

Further, let us assume that $f_T+f_B$ depends on both $t$ and $r$, meaning that the torsion scalar, the boundary term and the scalar field $T=T(r,\vartheta)$ $B=B(r,\vartheta)$, $\phi=\phi(r,\vartheta)$ (or at least one of them depend on the two variables), then the only way to solve the antisymmetric field equation without choosing $f$ would be to constrain some components of the torsion tensor and also the torsion vector. For the most general spherically symmetric tetrad satisfying spherical symmetry in teleparallel geometries in the Weitzenb\"{o}ck gauge~\eqref{eq:sphertetradwb}, one finds out that for this case, we must have
\begin{equation}
    T_t=-T^r{}_{tr}\,,\quad T_r=-T^{t}{}_{rt}\,,\quad T^{t}{}_{\phi\theta}=T^{r}{}_{\phi\theta} = 0 \label{antieqcond}
\end{equation}
to solve the antisymmetric equation \eqref{antifTBphiX}. These equations
\begin{subequations}
\begin{align}
    0&=2 C_3 C_5-2 \dot{C}_{5}C_5-2 C_6 \dot{C}_{6}\,,\\[0.5ex]
    0&=2 C_4 C_5-2 C'_{5}C_5-2 C_6 C'_{6}\,,\\[0.5ex]
    0&=C_2 C_6\,,\quad 0=C_1 C_6\,,\label{confanti}
\end{align}
\end{subequations}
where primes and dots are derivatives \gls{wrt} $r$ and $t$, respectively. One notices that the only way to solve these equations is by imposing
\begin{equation}
    C_6=0\,,\quad C_3=\dot{C}_{5}\,,\quad C_4=C_{5}'\,.
\end{equation}
The first condition is always needed to solve Eq.~\eqref{confanti} since the inverse of the tetrad diverges when $C_2=C_1=0$. This solution observes that $T=0$ (which can be seen by calculating the torsion scalar using the tetrad in~\eqref{eq:sphertetradwb} and the torsion components in~\eqref{chp3_torsion_comp}) but the boundary term $B=B(r,\vartheta)$, meaning that the Ricci scalar $\lc{R}=B$ and then $f(T,B,\phi,X)$ gravity becomes $f(\lc{R},\phi,X)$ gravity. One can then conclude that if $f_T+f_B$ depends on both $t$ and $r$ (and we do not choose or replace $f$), the only solution of the antisymmetric field equations is the one which matches with the curvature-based theory $f(\lc{R},\phi,X)$ gravity by having the metric
\begin{equation}
    \dd s^2=(C_1^2-\dot{C}_{5}^2)\dd t^2+2 (C_1 C_{2}-C'_{5} \dot{C}_{5})\dd t\, \dd r-(C_{5}'{}^2-C_{2}^2)\dd r^2-C_5d\Omega^2\,.
\end{equation}
Without loss of generality, due to the remaining gauge freedom in the metric, we can choose $C_5=C_5(r)=r^2$ and also set $C_2=0$ to convert our metric into its diagonal form. After doing all of this, one finds that the above metric is highly constrained since $g_{rr}=-1$. Then, for this particular case, the Birkhoff theorem does not hold but the spacetime is static.

Let us finish this discussion by emphasizing that the above computation is only valid in the theory $f(T,B,\phi,X)$ and in the specific case when one does not know the specific form of $f$ and assumes that all the scalars depend on the two coordinates $t,r$, which means that the only way to solve the antisymmetric equation~\eqref{antifTBphiX} is by imposing that the terms multiplying $\dot{f}_T+\dot{f}_B$ and $f'_T+f_B'$ vanish which is achieved if Eq.~\eqref{antieqcond} is true. Still the question about the generalized Birkhoff's theorem can be thought off as being open the context of modified \gls{tg}.

\subsection{Solar System tests}

Many Solar System observations have been performed, and they have confirmed that \gls{gr} can explain these effects in a good agreement with observations~\cite{Will:2001mx}. Thus, any meaningful modified gravity must pass the so-called Solar System tests.
In this section we will explore two main routes to constrain theories of gravity: either using solutions of the studied theory and then analyze them by performing the standard particle phenomenology of geodesics using observables (i.e. the classical tests); or to use the so-called parametrized post-Newtonian formalism, that basically uses post-Newtonian expansions and express the possible deviations from \gls{gr} in different parameters.

\subsubsection{Solar System tests using particle motion observables}

One interesting route for testing teleparallel theories in astrophysical systems is to find exact or perturbed solutions and then constrain them using particle phenomenology, by which we mean both classical astrophysical tests of gravity as well as more recent ones such as gravitomagnetic effects. The first paper doing this for Solar System in $f(T)$ gravity was Ref.~\cite{Farrugia:2016xcw,Iorio:2015rla}, which was performed using a perturbed solution around Minkowski found in Ref.~\cite{Ruggiero:2015oka}. These results are correct but they assumed a tetrad having a nonvanishing torsion scalar in the absence of gravity (see Eq.~\eqref{goodspher} with $\xi=-1$). Later in Ref.~\cite{DeBenedictis:2016aze}, the authors analyzed the tetrad \eqref{goodspher} with $\xi=1$ (which has the correct limit in the scalars) and they found the corrections to the perihelion shift predicted by a power-law $f(T)=-T-(\alpha/2)T^n$ gravity with $n=2$. This work was extended for $n={2,3,...,10}$ in Ref.~\cite{Bahamonde:2019zea} and also finding the photon sphere and perihelion shift of the perturbed solutions around Schwarzschild to then constrain the models accordingly with the perihelion of Mercury. Then, in Ref.~\cite{Bahamonde:2020bbc} studied $f(T,B)$ gravity, finding new perturbed solutions around Schwarzschild (Eqs.~\eqref{sec:solper1}--\eqref{sec:solper2} is one solution) and then computing different observables such as the perihelion shift, deflection of light and the Shapiro delay. After finding them, they also constrained the parameters of the models using Solar System observations. Using a similar approach, in Ref.~\cite{Farrugia:2020fcu} the authors also studied $f(T,B)$ gravity in a weak field limit and found the geodetic and Lense-Thirring effects predicted by this theory, noticing that the modifications play an important role and then, these effects may be also use to constrain models.

\begin{figure}[H]
	\begin{center}
		\includegraphics[scale=0.7]{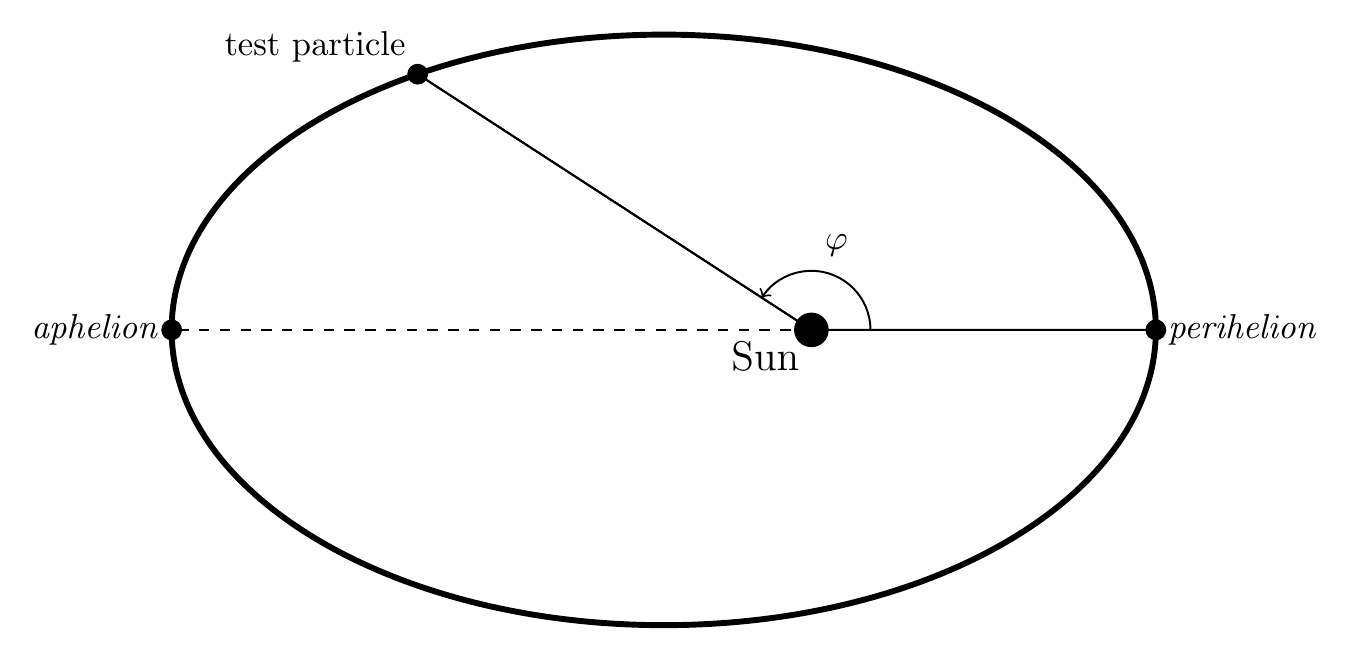}
	\end{center}
	\caption{\label{fig:class_solar_fig}
		The Sun causes gravitational effects to be exhibited by a test mass in orbit. The solar orbit is characterized by a closest point \textit{perihelion} and a furthest point \textit{aphelion}}
\end{figure}

Let us here compute the most important observables that are usually used to constrain models for the real tetrad and $\xi=+1$. In general, the observables that we will derive are also valid for any physical situation when we have a test body orbiting a massive compact object. This happens in the Solar System (where the Sun is the massive compact object) as depicted in Fig.~\ref{fig:class_solar_fig}, but also can be used in other astrophysical scenarios such as modelling the movements of the stars around the supermassive black hole Sgr A* in our galaxy. When the masses of the system have the same order of magnitude, then, the geodesic equation cannot be used and one needs to use the machinery of the two-body problem. It is interesting to mention that it turns out that for some observables, the final result coincides with the one that it will be presented here for the geodesic motion by just replacing the mass of the central massive compact to the sum of the two bodies ($M\rightarrow M_1+M_2$).

Let us here consider the case of geodesics of test particles orbiting a certain compact object. The worldline $q(\tau)=(t(\tau), r(\tau), \theta(\tau), \phi(\tau))$ of a test particle, with $\tau$ being an affine parameter, can be expressed as
\begin{equation}
2\mathcal{L}=g_{\mu\nu}\dot{q}^{\mu}\dot{q}^{\nu}=\mathcal{A}\,\dot{t}^2-\mathcal{B}\,\dot{r}^2-r^2\dot{\vartheta}^2-r^2\sin^2{\vartheta}\,\dot{\varphi}^2\,,\label{ELL}
\end{equation}
where dots are derivatives \gls{wrt} $\tau$. Due to spherical symmetry, we can consider the motion of the test particle at the equatorial plane $\vartheta=\pi/2$, without loss of generality. The energy $k$ and angular momentum $h$ are the conserved quantities that are given by
\begin{equation}\label{kh}
    k = \frac{\partial \mathcal{L}}{\partial \dot t} =\mathcal{A}\dot{t}\,,\quad h = -\frac{\partial \mathcal{L}}{\partial \dot \varphi} = r^2 \dot \varphi\,.
\end{equation}
Particles must obey the normalization condition so that by setting if $2\mathcal{L}=\sigma$, then massive (massless) particles would be described by $\sigma=1$ ($\sigma=0$). $\sigma$ is a constant that characterizes the massive or massless behavior of the type of particle being investigated. By using this normalization in Eq.~\eqref{ELL}, we arrive at the following expression
\begin{equation}\label{17}
    \dot{r}^2=\mathcal{B}^{-1}\Big(\frac{k^2}{\mathcal{A}}-\frac{ h^2}{r^2}-\sigma\Big)\,,
\end{equation}
that can be further rewritten in term of an effective potential giving us
\begin{equation}
\dot r^2 =-2V(r)\,,\label{potentialEq}
\end{equation}
where we have defined the potential as
\begin{equation}\label{eq:pot1}
    V(r) =-\frac{1}{2}\mathcal{B}^{-1}\Big(\frac{k^2}{\mathcal{A}}-\frac{ h^2}{r^2}-\sigma\Big)\,.
\end{equation}
Up to this point, we have not assumed any form of the metric. Then, Eq.~\eqref{potentialEq} is valid for any spacetime and any test particle. As discussed in Sec.~\ref{sub:sphsolu}, there are not many exact spherically symmetric solutions in modified teleparallel theories of gravity. However, there are some perturbed solutions around Schwarzschild, for example, the ones expressed in Eq.~\eqref{eq:solper_com} for $f(T,B)$ gravity. If we then assume that our metric functions can be expressed as in Eqs.~\eqref{AA}--\eqref{BB}, the potential can be expanded up to first order of $\epsilon$, yielding
\begin{alignat}{2}\label{eq:pot}
    V(r) & =\: & & - \frac{1}{2} k^2 + \frac{1}{2} \left(1-\frac{2M}{r}\right) \left( \frac{h^2}{r^2} + \sigma \right) \nonumber\\[0.5ex]
    & \: & & + \frac{\epsilon}{2} \left[ k^2 \left( \frac{a(r)}{1-\frac{2M}{r}} + b(r)\left(1-\frac{2M}{r}\right)\right) - b(r)\left( \sigma + \frac{h^2}{r^2} \right)\left(1-\frac{2M}{r}\right)^2\right]\,.
\end{alignat}
Now, we have all the ingredients to compute some interesting observables. For their more detailed derivation and schematic representations, see Ref.~\cite{Wald:1984rg}.
For the perihelion shift, which measures how much the perihelion is shifted after an orbit, we need to assume massive objects and a nearly circular orbit such that
\begin{equation}
    \textrm{\textbf{perihelion shift:}}\quad \sigma=1\,,\quad \textrm{and}\quad V(r)=V'(r)=0\,,
\end{equation}
where now $r=r(\varphi)=r_c+r_\varphi(\varphi)$, giving us that the potential equation~\eqref{potentialEq} can be rewritten in the following form
\begin{equation}
    \left(\frac{\dd r_\varphi}{\dd\varphi}\right)^2 = - 2 \frac{(r_c + r_\varphi)^4}{h^2} V(r_c + r_\varphi)\,.
\end{equation}
Now, if we assume a nearly circular orbit, we can expand $r_\varphi/r_c$ up to second order to approximate the above equation as
\begin{equation}
    \left(\frac{\dd r_\varphi}{\dd\varphi}\right)^2 = - \frac{r_c^4}{h^2} V''(r_c)r_\varphi^2 + \mathcal{O}\left(\tfrac{r_\varphi^3}{r_0^3}\right) \,,
\end{equation}
where $V(r_c) = 0$ and $V'(r_c)=0$ were used. The above equation for $r_\phi$ is in the form of an oscillator with a wave number $K=\sqrt{\frac{r_c^4}{h^2} V''(r_c)}$, giving us that the final expression for the perihelion shift becomes
\begin{equation}
    \Delta \varphi =2\pi\Big(\frac{1}{K}-1\Big) =2\pi \left(\frac{h}{r_c^2\sqrt{V''(r_c)}}-1\right)\,.
\end{equation}
For any perturbed solution around Schwarzschild, we find that the perihelion shift is
\begin{subequations}
\begin{align}\label{eqn:perihelionshift}
    \Delta \varphi &= \Delta\varphi_{\rm GR}+ \epsilon\, \Delta\varphi_{\epsilon} \\[0.5ex]
    &=6 \pi q + 27 \pi q^2 +135 \pi q^3+ \mathcal{O}(q^4) + \epsilon \, \Delta\varphi_{\epsilon}\,,
\end{align}
\end{subequations}
where $q=M/r_c$ and $\Delta\varphi_{\epsilon}$ denotes the first order correction from the perturbed solution ($a(r)$ and $b(r)$ in Eq.~\eqref{AA} and \eqref{BB}). In Ref.~\cite{DeBenedictis:2016aze,Bahamonde:2019zea,Bahamonde:2020bbc,Bahamonde:2020vpb}, the authors derived the first order correction of different observables such as the photon sphere and the perihelion shift for different perturbed solutions found in $f(T)$, $f(T,B)$ and $f(T_{\rm ax},T_{\rm vec},T_{\rm ten},P_1,P_2)$ gravity, respectively. Then, they used Solar System observations to constrain the corresponding free parameters of their model. For example, for a squared-power law $f(T)$ model
\begin{equation}
f(T)=-T-\frac{1}{2}\epsilon \, \alpha T^2\,,\label{fTpowerlaw}
\end{equation}
the first order correction to the perihelion shift is $\Delta\varphi_{\epsilon}=8 \pi \alpha q^2/r_c^2$, respectively~\cite{DeBenedictis:2016aze,Bahamonde:2019zea}. It is important to emphasize again here that these corrections are different depending on how one sets the integration constants $C_i$ in~\eqref{sec:solper1}-\eqref{sec:solper2} (with $\beta=0$).

In the Solar System, one can use the observations for the perihelion shift of Mercury which is $42,98\pm 0.040\, \textrm{''/cen} $~\cite{Will:2001mx}. This means that any deviations in the perihelion shift predicted by the modifications of \gls{tegr} should have the following maximum value for Mercury
\begin{equation}
    \Delta\varphi_{\epsilon,\rm max}\Big|_{\rm Mercury}\approx 0.18 \, \textrm{''/cen}\,.
\end{equation}
For the model~\eqref{fTpowerlaw}, the maximum value that the constant should take is $\alpha_{\rm max}\sim 10^{20}\, \textrm{km}^2$ to match the Mercury observations. Furthermore, there are other astrophysical systems where perihelion shift has been measured. For instance, the perihelion shift produced in the system composed by the supermassive black hole in the center of our galaxy, namely Sagittarius A* (Sgr A*), and the S stars which are orbiting it~\cite{Abuter:2020dou}. One can also use these observations to constrain teleparallel theories, but this has not been reported in the literature yet.

Another interesting observable is the deflection of light $ \Delta\varphi$ which is the angle characterizing the difference between the light trajectory at infinity (with or without the presence of a gravitating central mass). This quantity can be easily found by taking Eq.~\eqref{potentialEq} and rewriting it as
\begin{equation}
    \frac{\dd\varphi}{\dd r} = \frac{\dot \varphi}{\dot r} = \frac{h}{\sqrt{-2 V(r)} r^2}\,,
\end{equation}
then integrating this equation from $r_0$ (minimal distance) to infinity. It should be noted that the deflection of light for non-asymptotically flat spacetimes (such as in Schwarzschild de-Sitter) cannot be obtained by using the above formula since it goes beyond the event horizon. Instead, one needs to use the method used in Refs.~\cite{Ishak:2010zh,Azreg-Ainou:2017obt}. Besides these cases, by integrating the above equation and using $\dot r(r_0)=0$ ($V(r_0)=0$), we get
\begin{equation}
    \Delta\varphi = \pm 2\int_{r_0}^\infty \dd\bar r \frac{\mathcal{B}(\bar r)^{1/2}}{\bar r^2}\Big(\frac{\mathcal{A}(r_0)}{r_0^2 \mathcal{A}(\bar r)}-\frac{1}{\bar r^2}\Big)^{-1/2}-\pi\,,\label{phiis}
\end{equation}
where $r_0^2=\Big(\frac{h}{k}\Big)^2\mathcal{A}(r_0)$ was used and one can choose the + sign when further evaluating. Similarly as we did before, one can find the deflection of light predicted by \gls{gr} (the Schwarzschild contribution) plus a new contribution coming from the perturbed solution. After considering the approximation $r\gg M$, one finds
\begin{equation}
    \Delta \varphi\approx \varphi_{\rm GR}+\epsilon \, \varphi_{\epsilon} = \frac{4 M}{r_0}+\frac{ M^2}{r_0^2}\left(\frac{15 \pi }{4}-4\right)+\frac{ M^3}{r_0^3}\Big(\frac{244-45 \pi}{6}\Big)+\epsilon \, \varphi_{\epsilon}\,.\label{deflectionlight}
\end{equation}
This observable was explicitly found for different theories in Refs.~\cite{Bahamonde:2019zea,Bahamonde:2020bbc,Bahamonde:2020vpb}. As an example, for the theory~\eqref{fTpowerlaw}, the first order correction of the deflection of light is $\varphi_{\epsilon}=64 \alpha M^3 /(45 r_0^5)$. One can use observations from the Very Long Baseline Interferometry (\gls{vlbi}), which uses radio-telescopes on Earth~\cite{robertson1991new}, to constrain the teleparallel corrections. In this case, \gls{gr} predicts that the deflection of light (see the above equation) produced by the Sun is $\vartheta_{\rm GR}\approx 1.756''$. Then, the observed value over the \gls{gr} predicted value is~\cite{Shapiro:2004zz,Will:2001mx}
\begin{equation}
    \frac{\vartheta_{\rm obs}}{ \vartheta_{\rm GR}}\approx 1.0001\pm 0.0001\,,
\end{equation}
which tells us that for the model \eqref{fTpowerlaw}, the maximum value that $\alpha$ can take is $\alpha_{\rm max}\sim 10^{19}\, \textrm{km}^2$.

Another interesting experiment that can be used to put bounds in modified theories is the Cassini-Huygens experiment which is an experiment composed by a system of Earth-spacecraft-Earth. In this experiment, the fractional frequency shift $y$ of a ray of light after passing this system is measured~\cite{Bertotti:2003rm}, and its definition is
\begin{equation}
    y=2\frac{v_{\rm Cassini-Huygens} l_{\rm Earth}+v_{\rm Earth}l_{\rm Cassini-Huygens}}{l_{\rm Earth}+l_{\rm Cassini-Huygens}}\Delta \varphi\,,
\end{equation}
where $\Delta \varphi$ is the deflection of light (Eq.~\eqref{deflectionlight}), $v_{\rm Earth}$ and $v_{\rm Cassini-Huygens}$ are the transverse velocities of the Earth and the Cassini-Huygens spacecraft, respectively, and $l_{\rm Earth}$ and $l_{\rm Cassini-Huygens}$ are the distances from the Earth to the Sun and from the Cassini-Huygens spacecraft to the Sun, respectively. After assuming that the distance Cassini-Huygens-Sun is much larger than the distance Earth-Sun ($l_{\rm Cassini-Huygens}\gg l_{\rm Earth}$), we get
\begin{equation}
    y=y_{\rm GR}+\epsilon \,y_{\epsilon}\approx \frac{8M}{r_0}v_{\rm Earth}+\epsilon y_{\epsilon}\,.
\end{equation}
For the power-law model~\eqref{fTpowerlaw}, the first order fractional frequency shift was found to be equal to $y_\epsilon=128 M^3\alpha/(15 r_{0}^5)$~\cite{Bahamonde:2020bbc}. In this experiment, it has been measured that $y_{\rm obs}\sim 10^{-10}\pm 10^{-14}$~\cite{Bertotti:2003rm} and since the \gls{gr} predicted value (see above equation) is around $y_{\rm GR}\approx 1.690\times 10^{-9}$, the parameter $\alpha$ in \eqref{fTpowerlaw} cannot exceed $\alpha_{\rm max}\sim 10^{23}\, \textrm{km}^2$.

\begin{figure}[h]
	\centering
	\includegraphics[width=0.5\textwidth]{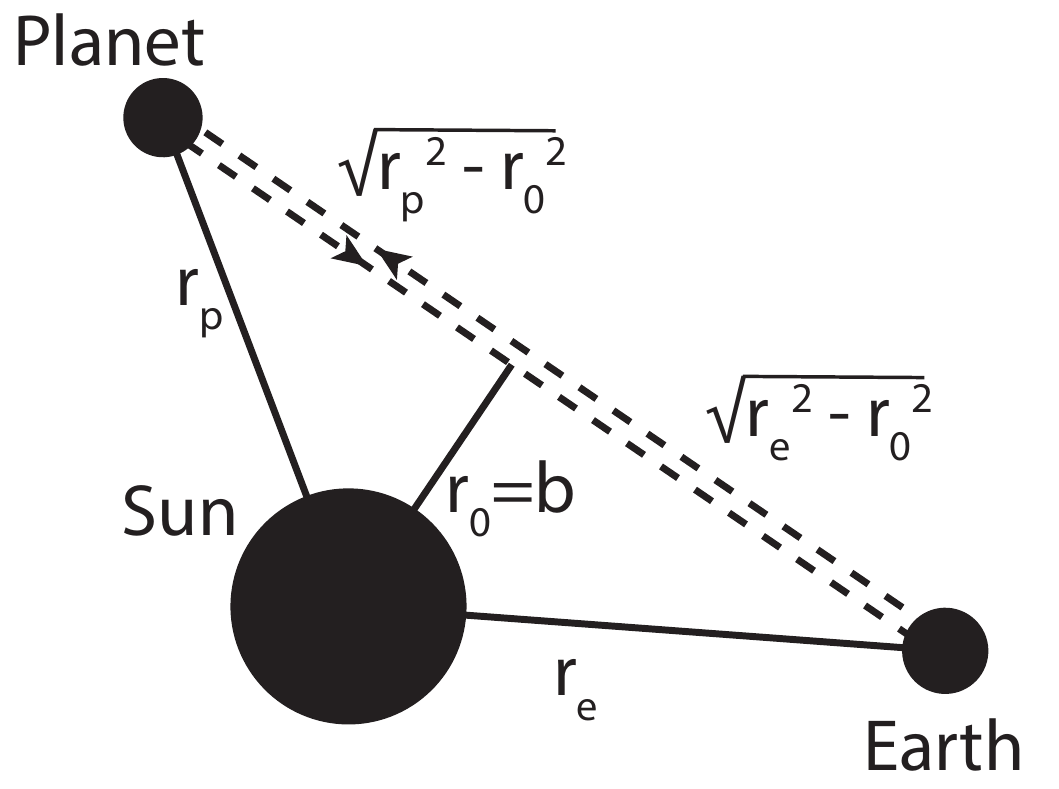}
	\caption{The Shapiro delay setup is shown in the context of the calculation that is performed. Here, the central mass is shown as the Sun, the mirror is shown as the planet and the emission point is taken to be from Earth.}
	\label{fig:shapiro}
\end{figure}

Another quantity that can be used to constrain models is the Shapiro delay or the retardation of light. By integrating Eq.~\eqref{potentialEq}, we find that the time required for a signal to be emitted from $r_0$ to $r$ is ($\sigma=0$)
\begin{equation}\label{timeshapiro}
    t(r,r_0) = \int\limits_{r_0}^r \dd\bar{r}\Big[\left(1-\frac{r_0^2 \mathcal{A}(\bar{r})}{\bar{r}^2 \mathcal{A}(r_0)}\right)\frac{\mathcal{A}(\bar{r})}{\mathcal{B}(\bar{r})}\Big]^{-1/2}\,,
\end{equation}
where $r_0^2=\Big(\frac{h}{k}\Big)^2\mathcal{A}(r_0)$ was used. The Shapiro delay is defined as the time
required for a signal to propagate from a radius $r_e$ to a point of closest encounter to the central mass $r_0$, and then from there to a mirror at radius $r_m$ where it gets reflected and returns on the same path to the emitter, as shown in Fig.~\ref{fig:shapiro} for a specific example case. This quantity is then defined as
\begin{equation}
    \Delta t_{\rm Shapiro}(r_{\rm e},r_{\rm m},r_0)=\frac{1}{2}\left(t(r_{\rm e},r_0)+t(r_{\rm m},r_0)-\sqrt{r_{\rm e}^2-r_0^2}-\sqrt{r_{\rm m}^2-r_0^2} \right)\,.
\end{equation}
By assuming a small central object $r \gg M$, we arrive at the following expression for the Shapiro delay
\begin{subequations}
\begin{align}
    \Delta t_{\rm Shapiro}(r,r,r_0)&=\Delta t_{\rm Shapiro,GR}(r,r,r_0)+\epsilon\, \Delta t_{\rm Shapiro,\epsilon}(r,r,r_0)\\[0.5ex]
    &\approx \frac{M \sqrt{r^2-r_0^2}}{r+r_0}+2 M \log \left(\frac{\sqrt{r^2-r_0^2}+r}{r_0}\right)+\epsilon\,\, \Delta t_{\rm Shapiro,\epsilon}(r,r,r_0)\,.\label{eq:shap}
\end{align}
\end{subequations}
For the theory~\eqref{fTpowerlaw}, the first order correction becomes
\begin{equation}
    \Delta t_{\rm Shapiro,\epsilon}(r,r,r_0)=\frac{4\alpha M^3 \sqrt{r^2-r_0^2}}{30(r+r_0)} \Big[\frac{7}{ r^3 r_0}+\frac{7}{ r^2 r_0^2}+\frac{17}{ r r_0^3}+\frac{20}{ r_0^4}\Big]\,.
\end{equation}
One example of an experiment that has measured the Shapiro delay is the Viking mission on Mars~\cite{Reasenberg:1979ey}, which essentially observed the time travel of a signal measured on the Earth after performing the trajectory Earth-Mars-Sun-Earth. It has been found that $\frac{\Delta t_{\rm obs}}{\Delta t_{\rm GR}}=1.000\pm 0.001$ where $\Delta t_{\rm GR} \approx 2.664\times 10^{-4}\, \textrm{s}$. Thus, for the power-law form in Eq.~\eqref{fTpowerlaw}, one finds that the maximum constant $\alpha_{\rm max}\sim 10^{20}\, \textrm{km}^2$. There are other measurements regarding the Shapiro delay outside the Solar System that can be used to constrain the corrections of \gls{gr}. For instance, there is an important effort in the field of pulsars to measure these effects (see for example~\cite{Demorest:2010bx,Cromartie:2019kug}). These constrains have not been used in the context of \gls{tg} yet.

The gravitational redshift can be also used to constrain models, for example by using experiments with a hydrogen-maser clock on a rocket. For a light ray propagating from altitudes $r_1$ and $r_2$ (with $r_1<r_2$), the gravitational redshift can be expressed as
\begin{eqnarray}
    z\equiv\frac{\nu_2}{\nu_1}-1=\sqrt{\frac{\mathcal{A}(r_2)}{\mathcal{A}(r_1)}}-1\,,
\end{eqnarray}
where $\nu_i$ are the frequencies measured at $r_i$. Then, the fractional frequency between these two points becomes
\begin{equation}
    \Big(\frac{\nu_2}{\nu_1}\Big)\approx \Big(\frac{\nu_2}{\nu_1}\Big)_{\rm GR}+\epsilon\, \Big(\frac{\nu_2}{\nu_1}\Big)_{\epsilon} = 1+M(r_1^{-1}-r_2^{-1})+\epsilon\, \Big(\frac{\nu_2}{\nu_1}\Big)_{\epsilon}\,,
\end{equation}
where we have assumed that $r_i\gg M$. The leading order contribution for the power-law case~\eqref{fTpowerlaw} is found to be equal to $\Big(\frac{\nu_2}{\nu_1}\Big)_{\epsilon}=\frac{2}{5}M^3\alpha \left(r_1^{-5}-r_2^{-5}\right)$. There are several experiments performed on the Earth which measures this effect. The most common one is the one which uses hydrogen-maser clock on a rocket launched to an altitude of about $10^{7}$ m~\cite{Vessot:1980zz}, that found that $\frac{\Delta \nu_{\rm obs}}{\Delta \nu_{\rm GR}}=1.000\pm0.0002$, which effectively constrain the parameter $\alpha$ in~\eqref{fTpowerlaw} to have the maximum value of the order of $\alpha_{\rm max}\sim 10^{22}\, \textrm{km}^2$.

Another important effect that has been measured is the so-called geodetic effect, which appears when a central body (a source) is rotating and one has a gyroscope which is orbiting it. In this situation, the gyroscope starts to precess. To obtain this effect, one can follow the procedure as it is explained in~\cite{rindler2003relativity}. By starting from a spherically symmetric metric, we can consider that the system is rotating with an angular frequency $\omega$, such that the angular coordinate $\varphi\rightarrow \tilde{\varphi}-\omega \, t$. By sitting the gyroscope in the circular polar point $\vartheta=\pi/2$, one finds that the metric becomes
\begin{equation}
    \dd s^2= e^{2\Psi(r)}(\dd t-e^{-2\Psi(r)}\omega r^2\dd \tilde{\varphi})^2-\mathcal{A}(r)r^2e^{-2\Psi(r)}\dd \tilde{\varphi}^2\,.
\end{equation}
Here, $e^{2\Psi(r)}=\mathcal{A}(r)-r^2\omega^2$. One can further find that the angular frequency of the gyroscope would be given by~\cite{rindler2003relativity}
\begin{equation}
    \Omega = \frac{e^\Psi}{2\sqrt{2}}\left[k^{ik} k^{jl} (\omega_{i,j} - \omega_{j,i})(\omega_{k,l} - \omega_{l,k})\right]^{1/2}\,,
\end{equation}
where $\omega_i = e^{-2\Psi} \omega r^2 \delta^3_i$ and $k^{ij}$ is the spatial 3-metric. By using several identities, one finds that the precession over one orbit (per year) becomes
\begin{equation}\label{geodetic_precess}
\Omega_{\text{dS}} = \sqrt{\frac{A'}{2r}}\left[1-\sqrt{\frac{1}{\mathcal{B}}\left(1-\frac{r\mathcal{A}'}{2\mathcal{A}}\right)}\right]\,,
\end{equation}
where $\mathcal{B}$ is the $g_{rr}$ component of the metric. The geodetic effect has been measured by the Gravity Probe B experiment which was a satellite that orbited the Earth with a gyroscope precessing due to the rotation of the Earth. It was found a geodetic precession of $-6601.8\pm 18.3$ mas$/$yr at around $642$ km from the Earth. In~\cite{Farrugia:2020fcu}, the authors calculated this effect for the power-law~\eqref{fTpowerlaw} finding that $\alpha_{\rm max}\sim 10^{32}\, \textrm{km}^2$.

The last observable that we will show here is the Lense-Thirring effect which measures a relativistic precession of a rotating body (e.g. a gyroscope) near a massive compact body. This effect can be found by analyzing the geodesic equation of a freely falling gyroscope which is initially at rest and has a spin characterized by the spin vector $S^\mu$. By using the force-like equation~\eqref{Eq:Lorentz_force_like_eq} which is equivalent as the geodesic equation~\eqref{GR_geodesic_eq}, we arrive at
\begin{equation}
    \frac{\dd S^\mu}{\dd\tau} + \udt{\Gamma}{\mu}{\sigma\rho} S^\sigma u^\rho = \udt{K}{\mu}{\sigma\rho} S^\sigma u^\rho\,.
\end{equation}
Here we have assumed that $u^\mu$ is the gyroscope's velocity measured when it is at rest, i.e., $u^\mu S_\mu=0$. If we again assume that the metric can be expressed as $g_{\mu\nu}=\eta_{\mu\nu}+\delta g_{\mu\nu}$ and $|h_{\mu\nu}|\ll 1$ (as in Eq.~\eqref{eq:perturbed_metric_abstract}), which is a weak field approximation, the above equation can be written as
\begin{equation}\label{lense_thirring_GPB}
    \frac{\dd S_i}{\dd\tau} = \epsilon_{ikl} \Omega^k S^l\,,
\end{equation}
where $\Omega^k \equiv -\frac{1}{2}\epsilon^{kmn}\partial_m h_{0n}$. Physically speaking, the quantity $\Omega^k$ represents the angular velocity precession vector of the gyroscope. In Ref.~\cite{Farrugia:2020fcu}, it was found that the Lense-Thirring effect for $f(T,B)$ gravity in the weak field regime is the same as the one predicted by \gls{gr}. There is only a Newtonian rescaling for this theory, therefore, one cannot constrain this theory by using weak field approximations. It would be interesting to extend this study by not assuming the weak field limit and study some specific models of $f(T,B)$ with its respective perturbed solutions.

An important comment that should be taken into account is that the value of all of these observables depend on the choice of the integration constants in the perturbed solutions. In~\cite{Pfeifer:2021njm}, the authors used other boundary conditions (to obtain a black hole) as the ones considered in the results provided above, and then they found that the maximum value that the parameters of the theory can reach, have a different order of magnitude. Furthermore, the expressions presented above assumed the real tetrad with $\xi=+1$. When one chooses $\xi=-1$ or the complex tetrad, these observables are different. It was found that they lead to corrections to
general relativity at lower order than the the real tetrad with $\xi=+1$~\cite{Bahamonde:2021srr}.

Let us stress again that all of these observables can be used in a direct way with any solution (exact or perturbed) to constrain gravitational theories by using, for example, Solar System observations or any other astrophysical system where these effects occur. There are also other ways to contrast Solar System observations with modified gravitational models. One very important method is the so-called \gls{ppn} analysis, that we will explain in the next section.

\subsubsection{Parametrized post-Newtoninan formalism}

One of the most useful analytical tools to test modified theories of gravity on smaller scales is by the so-called ``Parametrized Post-Newtoninan'' (\gls{ppn}) formalism~\cite{Will:2001mx} which essentially is encoded in ten parameters. These parameters have been tested with high precision (for example in the Solar System) and essentially they are obtained by considering the lowest orders deviations from the \gls{ppn} expansion of a given gravitational theory (weak field limit). The standard formalism has been developed in curvature-based theories containing a metric as a fundamental field. Since the fundamental variable in \gls{tg} is the tetrads, the formalism needs to be adapted by assuming perturbations in the tetrad. This study was first performed in Ref.~\cite{Ualikhanova:2019ygl} and we will review its most important results here.

One key ingredients to the \gls{ppn} formalism is assuming that the matter acting as a source of a gravitational field is given by a perfect fluid whose velocity is small compared with the velocity of light in a specific frame. Then, one can expand in orders of velocity. The energy-momentum tensor is then
\begin{equation}\label{eqn:tmunu}
    \Theta^{\mu\nu} = (\rho + \rho\Pi + p)u^{\mu}u^{\nu} - pg^{\mu\nu}\,,
\end{equation}
where $u_{\nu}$ is the 4-velocity and is normalized by $u^{\mu}u_{\mu} = +1$ and $\rho,p$ and $\Pi$ are the energy density, pressure and specific internal energy, respectively. To proceed, we need to expand all the important fields in orders of velocity $\mathcal{O}(n) \propto |\vec{v}|^n$. For the tetrads, the expansion can be written as
\begin{equation}\label{eqn:tetradexp}
    e^A{}_{\mu} = \order{e}{0}^A{}_{\mu} + \order{e}{1}^A{}_{\mu} + \order{e}{2}^A{}_{\mu} + \order{e}{3}^A{}_{\mu} + \order{e}{4}^A{}_{\mu} + \mathcal{O}(5)\,,
\end{equation}
 where we have assumed that the expansion is taken around a flat diagonal background tetrad $\order{e}{0}^A{}_{\mu} = \mathrm{diag}(1, 1, 1, 1)$. Here, overscripts denote the velocity order considered, i.e., $\order{e}{n}^A{}_{\mu}\sim \mathcal{O}(n)$. We will only consider expansions up to the fourth order in the velocity. It is useful to lower the Lorentz index, and this can be done by contracting the above perturbation with the Minkowski metric, yielding
 \begin{equation}
e_{\mu\nu} = \order{e}{0}^A{}_{\mu}\eta_{AB}e^B{}_{\nu}\,, \quad
\order{e}{n}_{\mu\nu} = \order{e}{0}^A{}_{\mu}\eta_{AB}\order{e}{n}^B{}_{\nu}\,.
\end{equation}
One important remark is that after doing the expansions one finds that not all the tetrad perturbation components are relevant. If we assume the Newtonian energy conservation and a time reversal symmetry, one finds that the only non-trivial components are
\begin{equation}\label{eqn:ppnfields}
\order{e}{2}_{00}\,, \quad
\order{e}{2}_{ij}\,, \quad
\order{e}{3}_{0i}\,, \quad
\order{e}{3}_{i0}\,, \quad
\order{e}{4}_{00}\,.
\end{equation}
Since the tetrads are the fundamental variables of all teleparallel theories (we assume the Weitzenböck gauge), we can then use the above perturbation in a specific theory and expand all the geometrical objects in the underlying theory in the velocity orders. If in our studied theory we have a scalar field $\phi$ (for example in teleparallel Horndeski, see Sec.~\ref{sec:BDLS}), we also need to expand it in velocity orders, namely,
\begin{equation}\label{eqn:scalarexp}
\phi = \Phi + \psi = \Phi + \order{\psi}{1} + \order{\psi}{2} + \order{\psi}{3} + \order{\psi}{4} + \mathcal{O}(5)\,.
\end{equation}
Moreover, if one has arbitrary functions of a geometrical variable, as in the teleparallel analogue of Horndeski gravity, one would need to expand the corresponding Lagrangian function. For example for this theory, this can be done by taking the following Taylor series expansion
\begin{subequations}
\begin{alignat}{2}
    G_i(\phi, X) & =\: & & G_i(\Phi, 0) + G_{i,\phi}(\Phi, 0)\psi + G_{i,X}(\Phi, 0)X + \frac{1}{2}G_{i,\phi\phi}(\Phi, 0)\psi^2 + G_{i,\phi X}(\Phi, 0)\psi X \nonumber\\[0.5ex]
    & \: & & + \frac{1}{2}G_{i,XX}(\Phi, 0)X^2 + \ldots\\[0.5ex]
    & =\: & & \GG_i + \GG_{i,\phi}\psi + \GG_{i,X}X + \frac{1}{2}\GG_{i,\phi\phi}\psi^2 + \GG_{i,\phi X}\psi X + \frac{1}{2}\GG_{i,XX}X^2 + \ldots\,,
\end{alignat}
\end{subequations}
where $\Phi=\order{\phi}{0}$, bold letters denote the constant Taylor coefficients at the background level and $G_i$ are the functions related to teleparallel Horndeski (see Eq.~\eqref{HG_2}--\eqref{HG_5}). This is just an example where the Lagrangian involves a function of an arbitrary scalar, but the expansion should be in the same way for other theories different to the one written above.

These are all the ingredients needed to start the computations. Now, one needs to insert all of these expansions in the field equations, expand each quantity accordingly in the \gls{ppn} approximation and then solve the equations order by order. To do this easily, one introduces some suitable potential ansatz consisting of \gls{ppn} potentials and some constant coefficients. One can take the following ansatz for the tetrad at each velocity order:
\begin{subequations}
\begin{align}
\order{e}{2}_{00} &= a_1U\,, \qquad
\order{e}{2}_{ij} = a_2U\delta_{ij} + a_3U_{ij}\,,\\[0.5ex]
\order{e}{3}_{i0} &= b_1V_i+b_2W_i\,, \qquad \order{e}{3}_{0i} = b_3V_i+b_4W_i\,,\\[0.5ex]
\order{e}{4}_{00} &= c_1\Phi_1 + c_2\Phi_2 + c_3\Phi_3 + c_4\Phi_4 + c_5U^2+c_6\Phi_W+c_7\mathfrak{A}\,,
\end{align}
\end{subequations}
with the potentials satisfying~\cite{will2018theory,Will:2001mx}
\begin{subequations}
\begin{align}
\nabla^2 U&=-4\pi\rho\,,\qquad \nabla^2 \chi = -2U\,, \qquad
U_{ij} = \chi_{,ij} + U\delta_{ij}\,, \\[0.5ex]
\nabla^2 V_{i} &= -4\pi\rho v_i\,, \qquad
\nabla^2 W_{i} = -4\pi\rho v_i + 2 U_{,0i}\,,\\[0.5ex]
\nabla^2\Phi_1&= -4\pi\rho v^2\,, \qquad
\nabla^2\Phi_2= -4\pi\rho U\,, \qquad
\nabla^2\Phi_3= -4\pi\rho \Pi\,, \qquad
\nabla^2\Phi_4= -4\pi p\,,
\end{align}
\end{subequations}
where $\nabla^2 = \delta^{ij}\partial_i\partial_j$. Using these ansatz, the tetrad can be identified with the standard \gls{ppn} parameters as follows~\cite{Hohmann:2019qgo}
\begin{subequations}\label{eq:standardppnt}
\begin{align}
\order{e}{2}_{00} &= U\,,\\[0.5ex]
\order{e}{2}_{(ij)} &= \gamma U\delta_{ij}\,,\\[0.5ex]
\order{e}{3}_{(0i)} &= -\frac{1}{4}(3 + 4\gamma + \alpha_1 - \alpha_2 + \zeta_1 - 2\xi)V_i - \frac{1}{4}(1 + \alpha_2 - \zeta_1 + 2\xi)W_i\,,\\[0.5ex]
\order{e}{4}_{00} &= \frac{1}{2}(1 - 2\beta)U^2 + \frac{1}{2}(2 + 2\gamma + \alpha_3 + \zeta_1 - 2\xi)\Phi_1 + (1 + 3\gamma - 2\beta + \zeta_2 + \xi)\Phi_2\nonumber\\[0.5ex]
&\phantom{=}+ (1 + \zeta_3)\Phi_3 + (3\gamma + 3\zeta_4 - 2\xi)\Phi_4 - \xi\Phi_W - \frac{1}{2}(\zeta_1 - 2\xi)\mathfrak{A}\,.
\end{align}
\end{subequations}
where
\begin{subequations}
\begin{align}
\Phi_W&=\int \int \rho' \rho'' \frac{\vec{x}-\vec{x}'}{|\vec{x}-\vec{x}'|^3}\cdot \Big[\frac{\vec{x}'-\vec{x}''}{|\vec{x}-\vec{x}''|}- \frac{\vec{x}-\vec{x}''}{|\vec{x}'-\vec{x}''|}\Big]\dd^3x' \dd^3x'' \,,\\[0.5ex]
\mathfrak{A}&=\int \dd^3x'\frac{\rho(t,\vec{x}')\left[v_i(t,\vec{x}')(x_i - x_i')\right]^2}{|\vec{x} - \vec{x}'|^3}\,.
\end{align}
\end{subequations}
For more details regarding these potentials, see Sec.~4.2.1 in Ref.~\cite{will2018theory}.
The so-called \gls{ppn} parameters are the ten parameters appearing in the above equation
\begin{equation}
    \gamma,\beta,\alpha_1,\alpha_2,\alpha_3,\zeta_1,\zeta_2,\zeta_3,\zeta_4,\xi\,.
\end{equation}
The parameter $\xi$ measures preferred-location effects, $\alpha_i$ is related to preferred-frame effects while $\zeta_i$ measure possible violations of the conservation of the total momentum. Finally, the parameters $\gamma$ and $\beta$ are related to how much spacetime curvature is produced (by unit rest mass) and the amount of non-linearity in the gravitational model, respectively. For \gls{gr} (or equivalently \gls{tegr}), we have
\begin{equation}
    \alpha_i=\zeta_i=\xi=0\,, \qquad \gamma=\beta=1\,.
\end{equation}
The strongest observations bounds for $\gamma$ and $\beta$ are set by the Cassini-Huygens tracking experiment~\cite{Bertotti:2003rm} and measurements provided by the perihelion precession of Mercury~\cite{Will:2001mx}, which are
\begin{equation}
\gamma - 1 = 4(\beta-1) \leq (2.1 \pm 2.3) \cdot 10^{-5}\,.
\end{equation}
\gls{gr} works very well at Solar System scales, so using these bounds one can set constrains on modified theories of gravity on smaller scales. Some scalar-tensor theories can have screening mechanisms which allow us to have a non-trivial density
profile for the scalar field and then, modifications of gravity become negligible at smaller scales and only become important on other scales such as at cosmological scales~\cite{Burrage:2017qrf}. Still, the screening mechanisms for teleparallel theories have not been studied in the literature.

There are few studies in teleparallel concerning the \gls{ppn} analysis explained above. The first study doing preliminary analysis under these directions was done in Ref.~\cite{Li:2013oef} for teleparallel \gls{de} (see Sec.~\ref{sec:scalartensor}). The authors concluded that all the \gls{ppn} parameters of this theory are the same as \gls{gr}. The same result in a slightly more general teleparallel scalar-tensor theory was obtained in Ref.~\cite{Chen:2014qsa}. Later, the case of a teleparallel scalar-tensor theory with a scalar field non-minimally coupled to both the boundary term $B$ and the torsion scalar $T$ was studied, finding some deviations for $\gamma$ and $\beta$~\cite{Sadjadi:2016kwj}. Some years later U. Ualikhanova and M.~Hohmann presented a formal and general way to construct the \gls{ppn} formalism in \gls{tg} in a more rigorous way  \cite{Ualikhanova:2019ygl}, which is the one presented above. They specifically studied a teleparallel theory constructed from an arbitrary function $f(T_1,T_2,T_3)$, where $T_i$ are the 3 non-parity violating scalars that can be constructed from the torsion tensor (see Sec.~\ref{Sec:Ext_NGR}). They found that only $\gamma$ and $\beta$ are different to the \gls{gr} value and for the specific case of $f(T)$ gravity, the \gls{ppn} parameters are identical to the \gls{gr} ones. Therefore, $f(T)$ gravity has the same predictions as \gls{gr} at smaller scales such as at Solar Systems. Later in Ref.~\cite{Flathmann:2019khc,Emtsova:2019qsl}, the authors studied a more general teleparallel scalar-tensor theory composed of $f(T,X,Y,\phi)$ where $Y=g^{\mu\nu}T^{\rho}{}_{\rho\mu}\phi_{,\nu}$ (see Sec.~\ref{sec:scalartensor}). This theory can be understood as a generalization of the model studied in Ref.~\cite{Sadjadi:2016kwj}. They also found that only $\gamma$ and $\beta$ differ from their \gls{gr} value. Finally, recently this study was further generalized to the teleparallel Horndeski case, again finding that only $\gamma$ and $\beta$ differ from \gls{gr}~\cite{Bahamonde:2020cfv}. One interesting remark is the fact that there are some classes of theories in teleparallel Horndeski having $\gamma=\beta=1$, and then, all of those model will have the same predictions as \gls{gr} at smaller scales. Two recent papers Refs.~\cite{Rao:2021azn,Gonzalez-Espinoza:2021nqd} found that for a \gls{tg} parity violating model which is similar to a Nieh-Yan model and also for a higher-order derivative teleparallel gravity (see Sec.~\ref{sec:HOT}), all the \gls{ppn} parameters are the same as \gls{gr}. All of the most important results are summarized in Table~\ref{table:PPN}.

\begin{table}[H]
\centering
\resizebox{18cm}{!}{\begin{tabular}{lcc}
\toprule
    \cellcolor{gris3}\textbf{Theory}& \cellcolor{gris3}\boldmath{$\beta-1$} &\cellcolor{gris3} \boldmath{$\gamma-1$ } \\ \midrule
    \cellcolor{gris1}$f(T)$/GR~\cite{Ualikhanova:2019ygl} &\cellcolor{gris1}0 & \cellcolor{gris1}0\\
    \cellcolor{gris3}$f(T_1,T_2,T_3)$~\cite{Ualikhanova:2019ygl} &\cellcolor{gris3}$-\frac{2f_{,1} + f_{,2} + f_{,3}}{4(2f_{,1} + f_{,2} + 2f_{,3})}$ & \cellcolor{gris3}$-2\frac{2f_{,1} + f_{,2} + f_{,3}}{2f_{,1} + f_{,2} + 2f_{,3}}$\\
    \cellcolor{gris1}&\cellcolor{gris1}$\frac{f_Y}{8\left(4f_Tf_X-3f_Y^2\right)\left(f_Y^2-f_Tf_X\right)^2}\Big[f_Tf_Xf_Y^2\left(16f_{T\phi}-7f_Y\right)+3f_Y^4\left(f_Y-2f_{T\phi}\right)$ & \cellcolor{gris1}\\
    \multirow{-2}{*}{$f(T,\phi,X,Y)$~\cite{Flathmann:2019khc} \cellcolor{gris1}}& \cellcolor{gris1}$-8f_T^2f_X^2f_{T\phi}+2f_T^2f_Y\left(2f_X^2+f_Yf_{X\phi}-2f_Xf_{Y\phi}\right)\Big]$& \multirow{-2}{*}{ \cellcolor{gris1}$\frac{f_Y^2}{2f_Tf_X-2f_Y^2}$}\\
    \cellcolor{gris3}Teleparallel Horndeski~\cite{Bahamonde:2020cfv}&\cellcolor{gris3}$-\frac{\tilde{\beta}}{8 \left(\HH_{,1} \HH_{,5}+\HH_{,3}^2\right){}^2 \left(3 \HH_{,3}^2-2 \HH_{,1} \left(\HH_{,4}-2 \HH_{,5}\right)\right)}$& \cellcolor{gris3}$-\frac{2 \HH_{,1} \HH_{,4}+\HH_{,3}^2}{2 \left(\HH_{,1} \HH_{,5}+\HH_{,3}^2\right)}$\\
 \bottomrule
\end{tabular}}
\caption{Summary of \gls{ppn} parameters for different teleparallel theories reported in the literature. In all of these theories, all the other \gls{ppn} parameters are zero (the same as \gls{gr}). The terms $\HH_i$ and $\tilde{\beta}$ are written in the appendix~\ref{app:PPN}.}
\label{table:PPN}
\end{table}

Let us finish here by mentioning that a viable modified theory of gravity must pass all the Solar system tests, and one possible route that has not been explored in teleparallel theories are by considering some screening mechanisms~\cite{Burrage:2017qrf}.

\subsection{Astrophysical compact objects}

Soon after the development of $f(T)$ gravity, different papers studied different compact objects in \gls{tg}. The main conclusion of the first works is that the majority of them suffered from the lack of understanding the covariant formulation of \gls{tg}, leading some authors to totally ignore the fact that they were choosing the incorrect good-tetrad spin connection pair. This created an enormous number of papers with mistakes since many authors ignored the fact that their choice contained some non-trivial antisymmetric field equation. The standard mistake was to consider a diagonal tetrad $e^A{}_\mu=(\mathcal{A},\mathcal{B},r,r\sin\vartheta)$ with a vanishing spin-connection. As was explained in Sec.~\ref{subsec:goodtetradsph}, this choice is not correct since it constrain theories to be just \gls{gr}. Thus, any paper using that diagonal tetrad in spherical coordinates with a vanishing spin connection would be incorrect. Taking this into account, in the following we will review some references which applied correctly the covariant formulation, without forgetting the role of the spin-connection (and the antisymmetric equations).

\subsubsection{Stars}\label{sec:stars}

\begin{figure}[H]
	\begin{center}
		\includegraphics{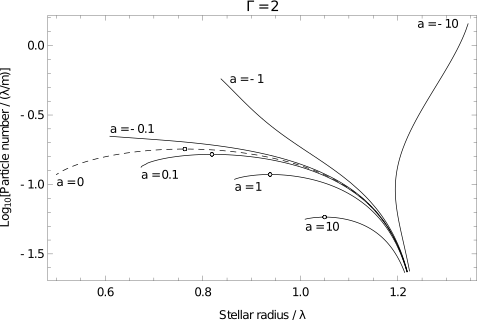}
	\end{center}
	\caption{\label{fig:g2_NR}
		Particle number--to--stellar radius curves for $f(T)=-T-(a/2)T^2$ gravity with a polytropic index equal to $\Gamma=2$. The dashed line $a=0$ corresponds to \gls{gr}. Permission for use of this figure was kindly provided by the authors of Ref.~\cite{Ilijic:2018ulf}.}
\end{figure}

The majority of the works related to stars in modified \gls{tg} have been carried out in $f(T)$ gravity. The first paper where these astrophysical systems were studied properly was done in Ref.~\cite{Boehmer:2011gw} for relativistic spherically symmetric stars in $f(T)$ gravity. Essentially, they used the tetrad-spin connection pair given by Eq.~\eqref{goodspher} for $f(T)$ gravity. The equations for this case are written in Eqs.~\eqref{Eq1}--\eqref{Eq3} if one sets $f(T,B)=f(T)$. Then, the set of equations~\eqref{Eq1},~\eqref{Eq2},~\eqref{Eq3} and the conservation equation~\eqref{TOV} describe the internal structure of self-gravitating fluids in $f(T,B)$ gravity. As happens in \gls{gr}, we need to choose an equation of state to close the system. Later in Ref.~\cite{Ilijic:2018ulf}, the authors studied compact stars in $f(T)=-T-(a/2)T^2$ gravity for an isotropic fluid ($p_r=p_l=p$) with a polytropic equation of state given by
\begin{equation}
    \rho = K p^{1/\Gamma}+\frac{p}{\Gamma-1}=K_2 n+\frac{K_1 n^{\Gamma}}{\Gamma-1}\,,\label{rhon}
\end{equation}
with the adiabatic exponent being $\Gamma=2$, $K_1$ and $K_2$ are constants and we have introduced the particle number density $n$. Then, the star would be sourced similarly as it is usually described by \gls{gr} since its source will be a fluid with a polytropic equation of state. However, since the field equations of the $f(T)$ (or other modifications) are different to \gls{gr}, the structure of the star could be different (for example its mass and radius). As has been discussed in the review in Ref.~\cite{Olmo:2019flu}, the definition of the mass of the star in modified gravity has been vastly debated. In \gls{gr}, this mass coincides with the ADM mass of the spacetime defined as
\begin{equation}
    M := 4\pi\int \rho r^2\dd r\,,
\end{equation}
but in a more general gravitational theory, this equation is not known to be true. Therefore, in Ref.~\cite{Ilijic:2018ulf}, the authors used a more general way to understand the matter contained in a stellar system which is related with particle number density defined as
\begin{equation}
    n:=\frac{dN}{dV}\,,
\end{equation}
where $N$ is the total particle number enclosed in the differential volume $dV=4\pi r^2\sqrt{\mathcal{B}}\dd r$. Then, we can integrate this equation from its centre $r=0$ to its radius $R$, yielding that the particle number is
\begin{equation}
    N=4\pi\int n(r)\sqrt{\mathcal{B}}r^2 \dd r\,.
\end{equation}
This equation must be used with $p=kn^\Gamma$, \eqref{rhon} and the field equations~\eqref{Eq1}--\eqref{Eq3} to describe the internal structure of the star. From these equations, one can also derive a modified version of the standard Tolman-Oppenheimer-Volkoff (TOV) equation that helps us describing compact stars in an easier way. The numerical procedure used in Ref.~\cite{Ilijic:2018ulf} works as follows: there are three field equations in spherical symmetry for $f(T)$ gravity and four variables (two metric functions $\mathcal{A},\mathcal{B}$ and the energy density $\rho$ and pressure $p$ for the fluid). To close the system one must assume an equation of state for the fluid, that as we pointed out above, one can assume to be a polytropic one. Then, some normalized variables are introduced in order to have all functions and constants dimensionless. To have a realistic star, one must impose the correct boundary conditions. At the center of the star, one can impose that the pressure satisfies $p(r=0)=p_0$ where $p_0$ is a constant and the metric functions $\mathcal{B}(r=0)=1$ and $(\mathcal{A}'/(2\mathcal{A}))\|_{r=0}=0$. At the radius of the star, one imposes that $p(r=R)=0$. Having all of that, one can numerically solve the equations. Fig.~\ref{fig:g2_NR} shows numerical solutions for the system which are the particle-to-stellar radius curves for different values of the modifications of \gls{gr} coming from $-(a/2) T^2$. This graph would be the analogous of the well known mass--radius plot usually displayed in \gls{gr} and its extensions. Then, it can be seen from this figure that it is possible to construct compact stars in squared-$f(T)$ gravity having different properties as \gls{gr}. One can notice that if $a>0$($a<0$), then the amount of matter supported against gravity is less(more) than in \gls{gr}.

Moreover, in Ref.~\cite{Ilijic:2018ulf} they also claimed that the energy density of the star over the central region is roughly constant ($n$ incompressible core) for positive values of $a$. It can be also conclude that for large positive values of $a$, there is a phase transition where the energy density and pressure change its sign within the interior of the star. Recently, this study was further generalized for boson stars~\cite{Ilijic:2020vzu} by adding a non-interacting complex scalar field. They found that the mass of the boson stars increases when $a$ is negative and large enough in absolute value. This means that in $f(T)$ gravity, the mass is no longer bounded as in \gls{gr} and then, the situation is quite different to \gls{gr} since the boson stars will not be necessarily a mini boson star. A similar study related to boson stars was done in Ref.~\cite{Horvat:2014xwa} in a theory where the torsion scalar is non-minimally coupled to a complex scalar field. In this case, the authors found that for a sufficiently large coupling constant, the energy density increases outwardly, surrounded by a thick shell, following an abrupt decrease in its profile ending in the so-called asymptotic tail. It is interesting to mention that this feature does not occur in a theory where one has the Ricci scalar non-minimally (or minimally) coupled to a complex scalar field. Later, the same authors~\cite{Horvat:2015qva} generalized this study by adding a $U(1)$ gauge field, leading to an increase in the maximal mass and the particle number. Recently, in a series of papers~\cite{deAraujo:2021pni,Fortes:2021ibz}, a similar result for $f(T)$-power law gravity was found.

In Ref.~\cite{Chakrabarti:2020ngy}, with the Raychaudhuri equation, collapsing stellar distributions were analyzed in $f(T)$ gravity. The authors considered a star with an interior behaving as a \gls{flrw} metric. They found a critical condition for $f(T)$, which tells us that depending on its form, one could model collapsing stars or not.

Regarding other modified teleparallel theories, some studies have been performed with stars in $f(T,\Theta)$ where $\Theta$ is the trace of the energy-momentum tensor. In Ref.~\cite{Pace:2017dpu}, the authors used a perturbative approach and derived the TOV equation for this model and later, in Ref.~\cite{Pace:2017aon}, the same authors analyzed the case of quark stars finding the corresponding mass-radius and mass-central density relations. More studies regarding neutron stars are needed. Using pulsars observations, it was recently reported in Ref.~\cite{Lin:2021ijx} that in power-law $f(T)$ gravity one can construct realistic neutron stars by considering certain bounds for the parameter of the theory.

Other works that are important to mention are related to junction conditions which are needed to have a correct matching of different regions of spacetime. These conditions are important to construct a viable collapsing star model. In Ref.~\cite{delaCruz-Dombriz:2014zaa}, they analyzed the junction conditions for $f(T)$ gravity by using the ADM decomposition and the thin shell formulation finding that one requires more junction conditions than in \gls{gr}. In this formulation, one imposes conditions in such a way to remove delta-like distributions in the field equations. The extra conditions in $f(T)$ gravity are related to the tetrads and also with the continuity in the torsion scalar. This study was then generalized in Ref.~\cite{Velay-Vitow:2017odc} in a more general approach by considering the variational principle for a covariant version of $f(T)$ gravity. The conclusion of this work states that for highly symmetric spacetimes such as in spherical symmetry, one has the same standard junction conditions as in \gls{gr} but for more general situations, one obtains different junction conditions that are the same as the ones obtained in Ref.~\cite{delaCruz-Dombriz:2014zaa} in a specific limit. Recently, in~\cite{Fiorini:2021mps} it was shown that the remnant group of local Lorentz transformations for teleparallel theories play a crucial role to match tetrads in order to ensure that scalars constructed from teleparallel quantities are continuous.

As a final remark, it is interesting to mention that there are no works related to the thermodynamics or Hayashi tracks of stars~\cite{Wojnar:2020txr} in modified teleparallel theories or in any other theory different to $f(T)$. Further, the complex tetrad~\eqref{tetrad2new} has not been studied to construct stars. This could lead to some new phenomenology that could be important to study in the future.

\subsubsection{Black holes}\label{sec:blackholes}

The study of black holes in teleparallel theories is an on-going research stream in the literature. Several studies like Refs.~\cite{Sousa:2003sx,CastelloBranco:2012fy,Castello-Branco:2013iza} have analyzed different types of black holes but in \gls{tegr} which in the end, is equivalent to \gls{gr}. Different attempts have been made in order to find exact black hole solutions in modifications of \gls{tegr}, like Noether's symmetry approach~\cite{Bahamonde:2019jkf,Paliathanasis:2014iva}, or directly by assuming different forms of the function for $f(T)$~\cite{Junior:2015dga}, but none of them are neither correct nor were able to find exact black hole solutions for the real tetrad. Recall that this tetrad provides a smooth limit to Minkowski for the scalars when we switch off gravity. Unfortunately, only perturbed methods (as the one mentioned in Sec.~\ref{sub:sphsolu}) or numerical approaches have been used to study the real tetrad. For example, in Ref.~\cite{Debnath:2014yya}, the authors studied the evolution of primordial black holes in $f(T)$ gravity by assuming a Chaplygin gas. They studied different properties of such black holes like its accretion mass by using a numerical approach. In Ref.~\cite{Aftergood:2014wla}, the authors studied the matter conditions for regular black holes in $f(T)$ gravity. To do this, they used a power-law form of $f$ and study the equations of motion of the interior of the black hole by determining the structure of matter in the vicinity of a possible singular point. Basically, the authors used perturbed approaches to determine the conditions for which it is possible to find non-singular black holes. Later, in Ref.~\cite{Boehmer:2019uxv}, the authors studied a Born-infield like form of $f(T)= \lambda(\sqrt{1+2T/\lambda}-1)$ and find the existence of a regular black hole interior which has a new length scale related to the Born-Infeld parameter $\lambda$. Due to this new length scale, they showed that the central curvature singularity is replaced by an infinitely long cosmic string (with constant curvature invariants). This means that the interior of the black hole is free from singularities. They used numerical and dynamical system techniques to show this, and then they used perturbation techniques to obtain an analytical perturbed solution for the interior of the black hole. Later on, the same authors noticed that this perturbed solution is of the form of a non-rotating BTZ black hole~\cite{Boehmer:2020hkn}. Using numerical techniques it was found that a squared power law $f(T)$ gravity case does for the real tetrad with $\xi=+1$ does not contain asymptotically flat black hole solutions beyond \gls{tegr}~\cite{DeBenedictis:2022sja}. Also in Ref.~\cite{Emtsova:2021ehh} the study for Schwarzschild in \gls{tegr} was reanalyzed to understand the role of the spin connection.

As we discussed in Sec.~\ref{sub:sphsolu}, perturbation theory allows us to obtain spherically symmetric solutions such as for example the ones expressed in Eqs.~\eqref{sec:solper1}--\eqref{sec:solper2}. Since this solution contains a logarithmic term, it is not so easy to check if this solution can represent a black hole solution since we need $\mathcal{A}(r_h)=\mathcal{B}^{-1}(r_h)=0$ and $\textrm{det}g_{\mu\nu}|_{r=r_h}\neq 0$, with $r_h$ being the event horizon, to describe a black hole perturbed solution. A recent paper~\cite{Pfeifer:2021njm} showed that when the $C_i$ are given by Eq.~\eqref{CCcase1} and $\beta=0$, the solution is not a black hole. However, one can also set $C_i$ in a different way, which is
\begin{subequations}
\begin{alignat}{2}\label{eq:C2}
 C_2 & =\: & &\frac{\alpha}{3 \mu_{\text{h}}^2}\Big(-6+12 \mu_{\text{h}} - 21 \mu_{\text{h}}^2 - 108 \mu_{\text{h}}^3 + 66 \mu_{\text{h}}^4 + 20 \mu_{\text{h}}^5 - 39 \mu_{\text{h}}^6 + 12 \mu_{\text{h}}^7 \nonumber\\[0.5ex]
 & \: & &+ 12 \mu_{\text{h}}^2 \ln(\mu_{\text{h}})\Big) + \zeta_1\,,\\[0.5ex]
C_1& =\: & &2MC_2-\frac{32}{3}\alpha\,,\label{eq:C1} \\[0.5ex]
 \zeta_1 & =\: & & -\frac{\mu_{\text{h}}-1}{3 \mu_{\text{h}}^2}(6 - 6 \mu_{\text{h}} - 49 \mu_{\text{h}}^2 + 59 \mu_{\text{h}}^3 - 7 \mu_{\text{h}}^4 - 27 \mu_{\text{h}}^5 + 12 \mu_{\text{h}}^6) - 4 \ln(\mu_{\text{h}})\,,
\end{alignat}
\end{subequations}
where $\mu_h=(1-2M/r_{\text{h}})^{1/2}$, to have a black hole solution. After showing that by choosing the above constants one has a perturbed black hole solution, the authors in Ref.~\cite{Pfeifer:2021njm} analyzed several observables such as perihelion shift, Shapiro delay or light deflection finding that the influence of the perturbed solution is much smaller than the one derived from the solution with the constants~\eqref{CCcase1}. Then, they analyzed several thermodynamics quantities such as the surface gravity and the sparsity finding corrections of the Hawking effect.

An important remark about black holes in $f(T)$ gravity is that it is not possible to have any solution beyond \gls{tegr} plus a cosmological constant with the metric functions satisfying $\mathcal{A}(r)=\frac{1}{\mathcal{B}(r)}$. In other words, the only solution with $\mathcal{A}(r)=\frac{1}{\mathcal{B}(r)}$ in $f(T)$ gravity is the Reissner–Nordstr\"om solution and the form of $f$ is always constrain to be $-T+\Lambda$. This conclusion is valid for any of the two tetrads.

One interesting feature that black holes produce is the so-called photon sphere, which is the radius that defines a region where photons are forced to travel in orbits, that can be derived by taking circular photon orbits, which is obtained by setting
\begin{equation}
    \textrm{\textbf{photon sphere:}}\quad \sigma=0\,,\quad \textrm{and}\quad V(r=r_c)=V'(r=r_c)=0\,,
\end{equation}
where $r_c$ is the radial circular orbit. For the solution Eqs.~\eqref{AA}--\eqref{BB}, we can find the photon sphere predicted by $f(T,B)$ gravity. To do this, we need to expand $r_c=r_0+\epsilon r_1$ and solve $V(r=r_c)=V'(r=r_c)=0$ order by order. By doing this, we get that the zeroth-order which is related to the Schwarzschild contribution, gives the standard photon sphere result $r_0=3M$ and the momentum $h_{0\pm} = \pm 3 \sqrt{3} k_0 M$. This observable is related to the shadow of the black hole. Then, the first order correction of the photon sphere will depend on the perturbed solution found for the specific teleparallel model. It is interesting to mention that the prediction of the photon sphere is equivalent for the integration constants \eqref{eq:C2}-\eqref{eq:C1} than for \eqref{CCcase1} for $f(T)$ gravity, and it becomes $r_{\rm ph}\approx (1.5+0.282675\, \beta)\frac{2M}{r}$. 

Another point that is worth mentioning is that black hole singularities in \gls{tg} might have a different description as it is usually described in theories like \gls{gr}. To analyze this, one usually uses the Hawking-Penrose singularities theorems which essentially defines singularities as being the points/paths where the geodesic is incomplete for moving particles. In \gls{tg} the so-called Raychaudhuri's equation might be different, so that, it is still unclear if the evolution equations for the congruence of curves would be the same. Moreover, a useful way to realize if one has a coordinate singularity is by constructing several scalars such as the Kretschmann scalar. It is also unclear which teleparallel scalars would give a hint whether one has a coordinate singularity. 

As pointed out in Sec.~\ref{sub:sphsolu}, if one allows the torsion scalar to be nonzero for the Minkowski limit, one can take the results of the complex tetrad~\eqref{tetrad2new} as being the first attempts to construct black hole configurations in modified teleparallel gravity. The recent solutions found in Ref.~\cite{Bahamonde:2021srr,Bahamonde:2022lvh} might help us to understand more about black holes in these theories as well as the role of the torsion scalar being nonzero for the Minkowski limit. One remarkable aspect of the complex tetrad is that it is that it is much simpler to obtain exact spherically symmetric black hole solutions. Further, there is a much richer family of exact scalarized asymptotically flat black hole solutions in \gls{tg} than in the standard Riemannian case. Let us finish this section by emphasizing that we are still very far away of understanding how black holes will be in modified teleparallel theories.

\subsubsection{Wormholes}\label{sec:wormholes}

A wormhole is usually described by the well known Morris-Thorne metric which is given by~\cite{Morris:1988cz}
\begin{equation}
    \dd s^{2}=e^{2\Phi(r)} \dd t^{2}-\frac{1}{1-\frac{b(r)}{r}}\dd r^{2}-r^{2}(\dd \vartheta^{2}+\sin^{2}\vartheta \dd \varphi^{2})\,,
\label{Morristhorne}
\end{equation}
where $\Phi(r)$ and $b(r)$ are the redshift and shape function, respectively. One then needs to have different properties for these functions to be able to describe a wormhole, like the non-monotonically decrease of $b(r)$ from infinity to a minimal value $r_{0}$ at the throat ($b(r_{0})=r_{0}$), and the so-called flaring-out condition stating that $(b-b'r)/b^{2}>0$ at the throat. For a more detailed description about wormholes, see Refs.~\cite{Visser:1989kh,Morris:1988tu,Visser:1995cc}.

In order to study wormholes in \gls{tg}, one needs to use a good tetrad-spin connection pair. Since they are usually described as spherically symmetric (as the metric above), one can then choose a zero spin connection but a tetrad being of the form
\begin{equation}\label{goodspher3}
e^A{}_\mu=\left(
\begin{array}{cccc}
    e^{\Phi(r)} & 0 & 0 & 0 \\
    0 & \Big(1-\frac{b(r)}{r}\Big)^{-1/2} \sin \vartheta \cos \varphi &r \cos \vartheta \cos \varphi & -r\sin \vartheta \sin \varphi \\
    0 & \Big(1-\frac{b(r)}{r}\Big)^{-1/2} \sin \vartheta \sin \varphi & r\cos \vartheta \sin \varphi & r\sin \vartheta \cos \varphi \\
    0 & \Big(1-\frac{b(r)}{r}\Big)^{-1/2} \cos \vartheta & - r\sin \vartheta & 0 \\
\end{array}
\right)\,,
\end{equation}
which is the same tetrad~\eqref{goodspher} with the specific metric coefficients chosen accordingly to reproduce the Morris-Thorne metric.

The great majority of studies related to wormholes in teleparallel theories have been done in $f(T)$ gravity. The first paper studying this using a good tetrad-spin connection pair found that spherically symmetric wormholes can be supported in $f(T)$ gravity by matter satisfying the standard energy conditions~\cite{Bohmer:2011si}. This conclusion cannot be made in \gls{gr} since in this case, one requires exotic matter to maintain the form of the throat of the wormhole. The main point here is that the modification of gravity, which in this case is related to the form of $f(T)$, can create a repulsive gravitational effect to then leads to the possibility of having wormholes supported by standard matter without violating the energy conditions. This result is not unique in modifications of \gls{tg}, since in many modified theories of gravity one can achieve this situation~\cite{Lobo:2009ip,Harko:2013yb}. In Ref.~\cite{Bohmer:2011si}, they assumed a squared power-law form of $f(T)$, and later, other authors assumed other models to studying wormholes. For example in Ref.~\cite{Rani:2016zbd} they used exponential and logarithmic forms of $f$ finding that for the first case, the null energy condition must be violated to construct wormholes and for the later, one can get wormholes satisfying all the energy conditions. Similar studies like in Refs.~\cite{Sharif:2014rda,Jawad:2015uea,Rani:2016gnl,Mustafa:2019ops} also found similar results but in a non-commutative background by setting a specific form of the energy density. In these different papers, different authors found that it is possible to construct stable wormholes exhibiting a conformal motion.

Concerning other teleparallel theories different to $f(T)$, there is only one paper where the authors studied a scalar field model non-minimally coupled with the scalar-torsion and the boundary term~\cite{Bahamonde:2016jqq}. In this paper, they found exact wormhole solutions using the Noether's symmetry approach.

The main conclusion that one could point out then is that in modified \gls{tg}, it is possible to construct wormholes satisfying the standard matter energy conditions.

\subsection{Galaxy rotation curves}\label{sec:galaxy}

Besides the problem of explaining the late-time acceleration of the Universe in a theoretically satisfactory way, \gls{gr} also suffers from the problem in explaining galaxy rotation curves. The only way to cure this in \gls{gr} is to introduce dark matter, which is a type of matter that has not been detected yet. Modified gravity also has tried to tackle this issue with the aim of not introducing dark matter but to fully describe the motion of galaxies using the new corrections appearing in the underlying theory. We here distinguish between the astrophysical dark matter responsible for galactic dynamics and the cosmological dark matter which is related to the stability of structure formation in the early Universe and the impact that has on the energy budget of the Universe in the \gls{cmb}. In this context, we can attempt to explain astrophysical dark matter in galactic systems but leave open the problem of cosmological dark matter \cite{Peebles2020} which we tackle in Sec.~\ref{sec10:Observation}. 

In $f(T)$ gravity, this can be achieved in a satisfactory way as it was explored in Ref.~\cite{Finch:2018gkh}. In this paper, the authors used a spherically symmetric perturbed solution around Minkowski for power-law $f(T)=-T-\alpha T^n$ gravity by assuming the tetrad~\eqref{goodspher} with $\xi=-1$, which is the branch where $T$ does not vanish in the Minkowski limit, which may be theoretically problematic. This does not happen for the $\xi=1$ branch which is defined in Eq.~\eqref{goodspher}, and may be an interesting analysis to perform and compare. In this case, the solution is explicitly given by the metric~\eqref{metricspher} with
\begin{subequations}\label{eqffffff}
\begin{align}
    \mathcal{A}(r) &= 1-\epsilon\, \Big(\frac{2M}{r}+\alpha \, \frac{r^{2-2n}}{2n-3}2^{3n-1} \Big)\,,\\[0.5ex]
    \mathcal{B}(r) &= 1+\epsilon\,\Big(\frac{2M}{r}+\alpha(1-3n+2n^2)\, \frac{r^{2-2n}}{2n-3}2^{3n-1}\Big)\,,
\end{align}
\end{subequations}
with $n\neq 3/2$ and $\epsilon\ll 1$ being the tracking parameter. In essence, this solution behaves as a weakly Schwarzschild plus an additional small correction coming from $\alpha$. To determine the rotational curve profile, one can consider a test particle following orbits around the galactic core, which is described by the effective potential~\eqref{eq:pot}. In this work, the authors used the perturbed solution around a Minkowski background in Eq.~\eqref{eq:pot} which would mean setting $M$ to being a small parameter, and choosing $a(r)$ and $b(r)$ described by the perturbations around this background, and with the corresponding integration constant related to a small mass. Furthermore, they assumed $h=0$, which effectively means that there are no meaningful angular momentum contributions from the galactic system model. Note that the definition used in Ref.~\cite{Finch:2018gkh} differs from ours. In our definitions, we absorbed the energy contribution in the potential whereas in this work the potential does not have the energy. Let us consider their form of the potential (Eq.~(23) in Ref.~\cite{Finch:2018gkh})
\begin{equation}
    V_{\rm eff}=\frac{1}{2}-\epsilon\, \Big(\frac{M}{r}+\alpha \frac{r^{2-2n}}{2n-3}2^{3n-2}\Big)\,.
\end{equation}
Next, we need to find the velocity profile that can be obtained by considering the following centripetal and gravitational acceleration
\begin{equation}
    a_c=\frac{v^2}{r}\,,\quad a_g=-\frac{dV_{\rm eff}}{dr}\,,
\end{equation}
which gives us
\begin{equation}
    v_{\rm eff}^2=-r \frac{dV_{\rm eff}}{dr}\,.
\end{equation}

\begin{figure}[H]
    \centering
    \includegraphics[scale=0.3]{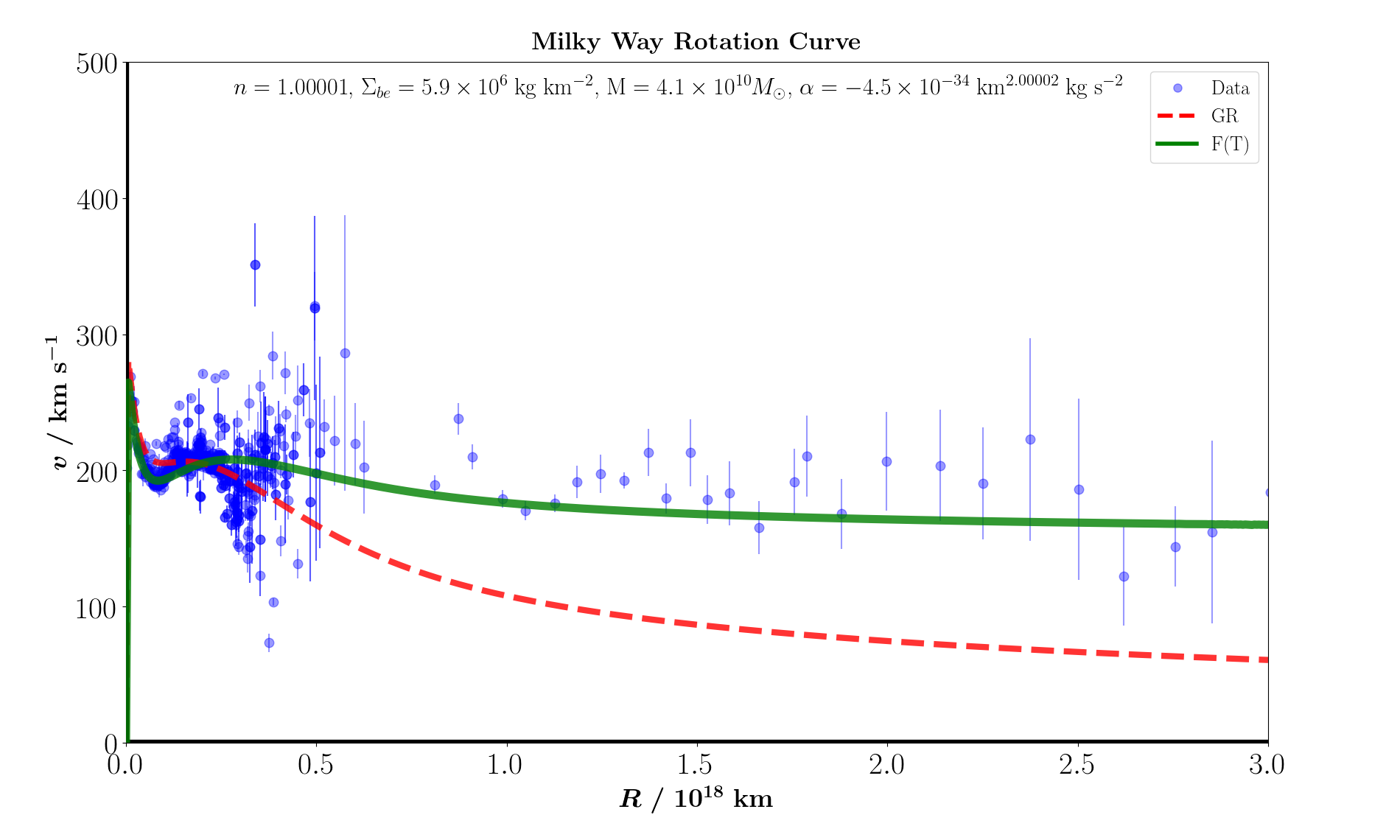}
    \caption{Best fit for a power-law $f(T)=-T-\alpha T^n$ with the constants $n=1.00001$ and $\alpha=$\num{-4.50573152405E-34}$\;\text{km}^{2n}\;\text{kg}^{-1}\;\text{s}^{-2}$, bulge surface mass density of $5.94\times10^6\;\text{kg}\;\text{km}^{-2}$ and disk mass $4.08\times10^{10}M_\odot$ (these are the best fit parameter presented in Ref.~\cite{Finch:2018gkh}). Permission for use of this figure was kindly provided by the authors of Ref.~\cite{Finch:2018gkh}.}
    \label{fig:galaxy}
\end{figure}

The simplest model of a galaxy is composed by two regions, namely, the disk and the bulge and each of them must be taken in a separated way. The standard approach for the disk is to use a velocity profile composed by the sum of the combined potential of the individual sources within the galaxy as described in Ref.~\cite{casertano1983rotation}. For the bulk, one assumes that this region is described as a spherical mass with a Vaucouleurs profile~\cite{Sofue:2008wt}. Following this analysis, the authors in Ref.~\cite{Finch:2018gkh} used data from the Milky way and three other galaxies to fit the parameters $\alpha$ and $n$ finding different sets of constants in such a way to describe the motion of the galaxy rotation curves in a satisfactory way without introducing any dark matter component. The main constraint that they found is that one requires $n$ to be small to fit the observations. Fig.~\ref{fig:galaxy} shows the best-fit model for the $\alpha$ parameter for the rotation profile for the Milky way for a power-law $f(T)=-T-\alpha T^n$ where one can notice that the profile fits the data with a good accuracy.

As a final remark, let us again mention that the authors used a tetrad with $T\neq0$ for Minkowski, which may be problematic foundationally. If one uses the same tetrad in Eq.~\eqref{goodspher} with $\xi=1$, one avoids this issue but there are no perturbations around Minkowski for $f(T)$ for this tetrad. Thus, to analyze the behavior of galaxy rotation curves in the case $\xi=1$, one would need to use the perturbed solutions around Schwarzschild (see Eqs.~\eqref{sec:solper1}--\eqref{sec:solper2}) and assume $M\ll1$ which is a valid approximation in this astrophysical scenario.

\subsection{Overview and future work}

As was stressed through this entire section, there are many unsolved and unexplored problems in modified teleparallel theories of gravity in the context of astrophysics. Some important open problems are listed below:
\begin{enumerate}
    \item \textbf{Exact or numerical black hole solutions} -- As was pointed out in Sec.~\ref{sec:blackholes}, for the real tetrad, there are only perturbed solutions or numerical solutions, but there are no analytical exact solutions describing black holes. Furthermore, numerical black holes solutions have not been analyzed in detail yet for theories that differ from $f(T)$ gravity. It would be interesting to analyze the solutions for the complex tetrad in more depth.
    \item \textbf{Quasi-normal modes} -- One interesting characteristic of the waveform of a merger of black holes is its final stage which is known as its ringdown phase. In this phase, the system is radiating \gls{gw} in the form of quasi-normal radiation. A recent paper~\cite{Maluf:2021ccx} studied the case of \gls{tegr} for Schwarzschild which can help as a tool to extend it for modified teleparallel gravity.
    \item \textbf{Time-dependent spherically symmetric good tetrad-spin connection pair} -- Thus far, it has been quite hard to perform a similar approach as in Sec.~\ref{subsec:goodtetradsph} for a generic time-dependent spherically symmetric spacetime beyond \gls{tegr}. This result could be very important for describing collapse of stars or if there is a way of finding theories beyond \gls{gr} satisfying the Birkhoff's theorem. Furthermore, one could also study time-dependent spacetimes such as cosmological wormholes.
    \item \textbf{Most general axially symmetric good tetrad-spin connection pair} -- As discussed in Sec.~\ref{sec:axials}, only some specific cases were derived. For example, a Taub-NUT-like form or a slowly rotating Kerr case has been found but, what is missing is a general axially symmetric good tetrad spin-connection pair which can include non-slowly rotating Kerr as a special case. If one has this, one could study axial symmetry in a generic way for any astrophysical system.
    \item \textbf{Galaxy rotation curves} -- As discussed in the previous section, in Ref.~\cite{Finch:2018gkh}, the authors investigated how galactic rotation curves are described in $f(T)$ gravity. They considered two different characteristics of galaxies: disk and the bulge. It would be interesting to re-analyze this problem by using other perturbed solutions (see Eqs.~\eqref{sec:solper1}--\eqref{sec:solper2}) for $f(T,B)$ gravity (and other theories or models) and then use Milky way data to fit the model with the observations which will add to the consistency of theory. More ambitiously, it would be ideal to constrain these models again a number of galactic rotation curve data sets.
    \item \textbf{White dwarfs and neutron stars} -- As discussed in Sec.~\ref{sec:stars}, there are some studies regarding compact stars in \gls{tg} but the majority of them are only focused on solving the TOV equations numerically to then find the mass-radius plot. There are no studies regarding other aspects of compact stars such as their early evolution that can gives us information about other characteristics of stars as 
    their effective temperature, luminosity, radiative core or how stars appear in the Hayashi tracks~\cite{Wojnar:2020txr}). In addition, all the works in \gls{tg} have been only focused on regular stars and then, more studies regarding neutron stars or white dwarfs are needed. Further, the complex tetrad have not been used to study such configurations.
    \item \textbf{Exotic objects and \gls{gw} echos} -- As we discussed in Sec.~\ref{sec:wormholes}, there are some studies about wormholes in \gls{tg}. Nevertheless, there are no studies about other exotic objects such as quark stars, gravastars or boson stars. Further, the \gls{gw} echos of exotic objects are important observational features (see~\cite{Cardoso:2019rvt}) that have not been computed in \gls{tg}.
    \item \textbf{Black hole singularities} -- As discussed in Sec.~\ref{sec:blackholes}, the Raychaudhuri's equation in \gls{tg} might have a different behaviour as the standard one. This might lead to a different interpretation of singularities in \gls{tg}. Further, coordinate-like singularities are usually understood by constructing several scalars such as the Kretschmann scalars. In \gls{tg}, one would need to construct other scalars to understand these types of singularities.
    \item \textbf{$3+1$ numerical simulations for \gls{gw} observations} -- The $3+1$ split formalism helps us to write down the Einstein's field equations in a particular form (Gauss-like equations) which is useful for numerical studies~\cite{Gourgoulhon:2005ng}. We know that \gls{tegr} has the same field equations as \gls{gr}. However, as we have seen in Sec.~\ref{sssec:tggenactfield}, they are expressed in terms of teleparallel quantities. What can happen if one uses the $3+1$ formalism for the \gls{tegr} field equations? could it be that this formalism leads to an easier numerical description of the Einstein's field equations? A first attempt towards understanding these questions were addressed in Ref.~\cite{Capozziello:2021pcg} and further followups studied are needed.
    \item \textbf{Weak lensing for galaxy distribution} -- There are no studies performing analyzes of weak lensing for galaxy distributions. This might gives us some hints towards understanding dark matter in a more satisfactory way and to understand better the physics of galaxies. 
\end{enumerate}
The above open problems are just some possible examples that have not been addressed in the literature. To fully describe astrophysical objects in \gls{tg} beyond~\gls{tegr}, one would need to start solving some of them.

\clearpage

\section{Observational and Precision Cosmology in Teleparallel Gravity} \label{sec10:Observation}

This section explored the pivotal regime of observational and precision cosmology of \gls{tg} and the various models that emerge from the earlier analyses. Here, we review the plethora of works in the literature on the topic, and explain how future works may open new possibilities for the most promising teleparallel models being explored in this Review.

More than using data to constrain the cosmological parameters inherent in these theories, when we discuss about precision cosmology, more than a qualitative (and statistics) endeavour, this is about a deeper understanding of the physics behind these theories of gravity (i.e. we want to go towards the inverse cosmology problem, that is to say, we want to know if the observations at hand can give us information about the nature of gravity). The $\Lambda$\gls{cdm} model has been extremely successful in meeting most of the observational data up to now. Assuming only \gls{gr} and the well-understood linear perturbations about a homogeneous and isotropic background cosmological model, with just six parameters, this model accounts for all cosmological observations on a wide range of scales. To achieve this we require the cosmological constant $\Lambda$, a constant usually associated with the vacuum energy, and dark matter, yet-undetected particle (or particles) which are predicted \cite{AlvesBatista:2021gzc}, for example, by supersymmetric extensions of the standard model of particle physics~\cite{Kazakov:2000ra}. While the theoretical backbone for both the cosmological constant and dark matter may be justly questioned, the $\Lambda$\gls{cdm} model has, hard-headed, maintained its hegemony; this is due to its performance in reproducing the observational data along with its numerical simplicity.

However, since the first release of the \gls{cmb} observations by the \texttt{Planck} Collaboration in 2013 in Ref.~\cite{Ade:2013zuv}, the determination of the Hubble constant $H_0$ based on the concordance model of cosmology started to be in tension with the cosmology-independent determination of this constant, e.g. via local Supernovae Ia (\gls{sn}) calibrated by the Hubble Space telescope in 2011 in Ref.~\cite{Riess2011}. The initial tension of 2.4$\sigma$ has been exacerbated over the years. On the one hand, systematic uncertainties have become better understood and \gls{cmb} data has accumulated. On the other hand, the survey of local \gls{sn} has increased and the anchors used to calibrate them significantly improved. The current status is that the two inferences of the Hubble-Lema{\^{i}}tre constant, one by the \texttt{Planck} Collaboration in 2018~\cite{Aghanim:2018eyx} assuming the $\Lambda$\gls{cdm} model, and the other by the \gls{shoes} Collaboration in 2019~\cite{Riess:2019cxk}, are in tension at the 4.4$\sigma$ level. Furthermore, using complementary sources of Cepheid calibration~\cite{Riess:2020fzl} we now reach a $4.2\sigma$ difference with the prediction from \texttt{Planck} and recently, a 5$\sigma$ using observations from \gls{hst} of Cepheids \cite{Riess:2021jrx}.

Many articles have been written trying to understand the implications of this tension: so far, it is the most severe problem confronting the standard model. The effect of local structure -- the so-called cosmic variance on $H_0$ -- has been thoroughly analyzed, along with thorough inspections of the error budget. Furthermore, physics beyond the concordance model has been investigated, with the aim that this tension could reveal alternatives to the highly tuned cosmological constant and the yet-undetected dark matter. So far, there is no well-posed panorama beyond the standard model that could solve this issue~\cite{DiValentino:2021izs}. Moreover, the $\Lambda$\gls{cdm} model successfully describes a range of observations from several epochs of the Universe: from primordial nucleosynthesis to the accelerating late-time cosmic expansion. As the standard cosmological parameters of this model are being inferred with increasing precision, there is no guarantee that the same model will fit more precise surveys from different cosmic epochs.

The current cosmological field set on the $\Lambda$\gls{cdm} model is supported by observational evidence as a description of the cosmic evolution on all cosmological scales~\cite{misner1973gravitation,Clifton:2011jh}, which is achieved by considering matter, the so-called dark matter, beyond the standard model of particle physics. Modelling itself this way, dark matter plays an important role in galactic structures~\cite{Baudis:2016qwx,Bertone:2004pz} and is manifest as cold dark matter particle(s), while \gls{de} in its simplest description is denoted by the cosmological constant~\cite{Peebles:2002gy,Copeland:2006wr} capable of producing the late-time cosmic accelerated expansion~\cite{Riess:1998cb,Perlmutter:1998np}. However, despite great efforts, some consistency problems persist with this constant~\cite{RevModPhys.61.1}, as well as a strong lack of direct observations of dark matter particle(s)~\cite{Schumann:2019eaa,Billard:2021uyg}.

Furthermore, this standard cosmological model was designed to
describe the Hubble flow, but the so-called $H_0$ tension problem now casts a shadow over this -- given that the observational discrepancy between cosmology-independent measurements~\cite{Riess:2019cxk,Wong:2019kwg,Breuval:2020trd,Reid:2019tiq,Kourkchi:2020iyz,Schombert:2020pxm,deJaeger:2020zpb,Shajib:2019toy,Blakeslee:2021rqi} and predicted ones~\cite{Aghanim:2018eyx,Ade:2015xua,Aiola:2020azj,Dutcher:2021vtw} from the early-time Universe (assuming the $\Lambda$\gls{cdm} model) seems to be growing. While measurements based on, for example, the tip of the red giant branch~\cite{Freedman:2020dne} (\gls{tgrb}, Carnegie-Chicago Hubble Program) or strong lensing measurements \gls{tdcosmo}+\gls{slacs}~\cite{Birrer:2020tax} have a relatively small mean value of $H_0$, their much larger error bars mean that they are currently unable to distinguish between the two faces of the tension. This issue could be resolved by future observations which involve new measuring techniques such as the use of \gls{gw} sources~\cite{Graef:2018fzu,Abbott:2017xzu} observed with ground-based detectors (such as the Einstein Telescope~\cite{Maggiore:2019uih}) or by the \gls{lisa} mission~\cite{Baker:2019nia,Audley:2017drz}.

\begin{figure}
\centering
    \includegraphics[width= 0.75\textwidth]{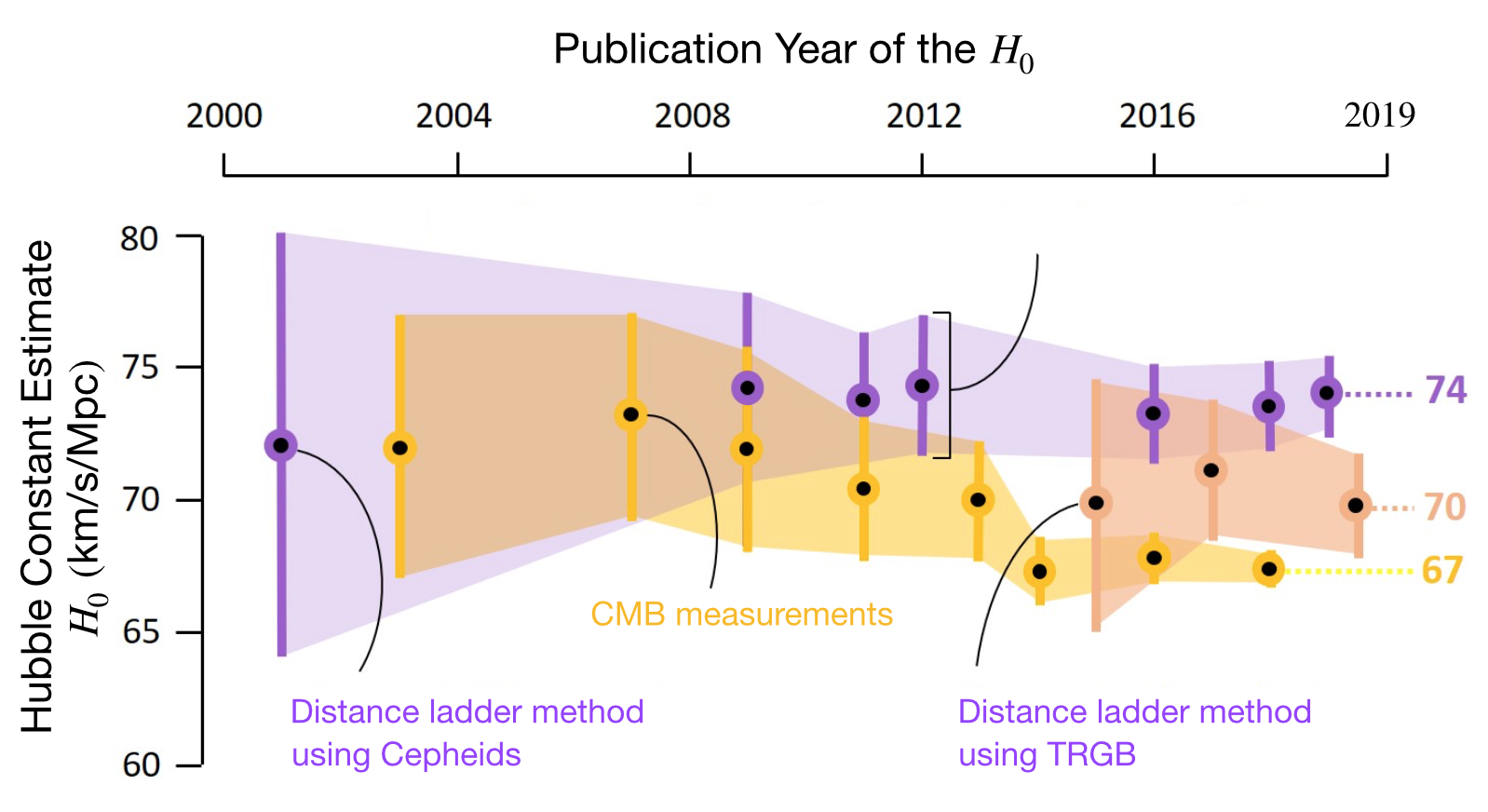}
    \caption[Hubble Constant estimates according to the publication year]{Hubble Constant estimates according to the publication year. The \gls{cmb}-based, early Universe mean value for $H_0$ is 67 km/s/Mpc. The Cepheid-based, late Universe mean value is 74 km/s/Mpc. A new alternative to Cepheids, i.e red giant stars that flare with a known intrinsic brightness, has only complicated the tension. They indicate a $H_0$ mean value of about 70 km/s/Mpc, a value that is midway between the other two, but in agreement with both the measurements within $2\sigma$. Other measurements, e.g. \gls{act}Pol 2020 (with $H_0 = 67.9 \pm 1.5\, \text{km/s/Mpc}$)~\cite{Aiola:2020azj}, \gls{act} \gls{dr}4 + \texttt{Planck} 2018 (with $H_0 = 67.53 \pm 0.56 \,\text{km/s/Mpc}$)~\cite{Naess:2020wgi}, \gls{hst} photometry + \gls{gaia3} (with $H_0 = 73.2 \pm 1.3 \,\text{km/s/Mpc}$)~\cite{Riess:2020fzl} have also been considered. Permission for use of this figure was kindly provided by the author of Ref.~\cite{Escamilla-Rivera:2021jkj}.
}
    \label{fig:tension_H0}
\end{figure}

Let us consider first two questions for the early-time Universe probes: {\it i)} what kind of smoking-gun signature can be used to identify unknown systematic errors that could affect the predictions for $H_0$ for the late Universe that is independent of the model of cosmology? {\it ii)} does there exist any indication of tension in the early-time data that could reveal systematic errors of the standard six-parameter $\Lambda$\gls{cdm} model that is exhibited purely in the early Universe? Aside from the well known small difference between the inference of $H_0$ from low and high angular resolution \texttt{Planck} data, all of the early-time data seem to be consistently predicting a low value of $H_0$. Experiments like \gls{act} and \gls{spt}, which use \gls{cmb} data to calibrate the sound horizon and baryonic acoustic oscillations (\gls{bao}), are in agreement with \texttt{Planck}. These methods lead to low $H_0$ values of $\sim$ 67-68.5 km s$^{-1}$ Mpc $^{-1}$. Another independent, consistent value of $H_0$ can also be obtained by using light element abundances to calibrate the sound horizon, \gls{bao}s and lower redshift probes. A compilation of the $H_0$ estimates up to 2019 is detailed in Fig.~\ref{fig:tension_H0}.

Our second question concerns anomalies with significant tension among high-redshift probes \cite{Abdalla:2022yfr}. Some of them appear as a departure from unity, at $\sim2.8\sigma$, of the inferred parameter A$_{\rm lens}$, a phenomenological parameter firstly introduced in~\cite{Calabrese:2008rt}, which is the amplitude of the gravitational lensing in the angular power spectra and can be used to rescale the lensing effect in the \gls{cmb} 2-point correlation function. If this deviation is confirmed, it will be clear evidence that something is not well understood in the relation between \gls{cmb} anisotropies and the growth of structure.

\begin{figure}
\centering
    \includegraphics[width= 0.85\textwidth]{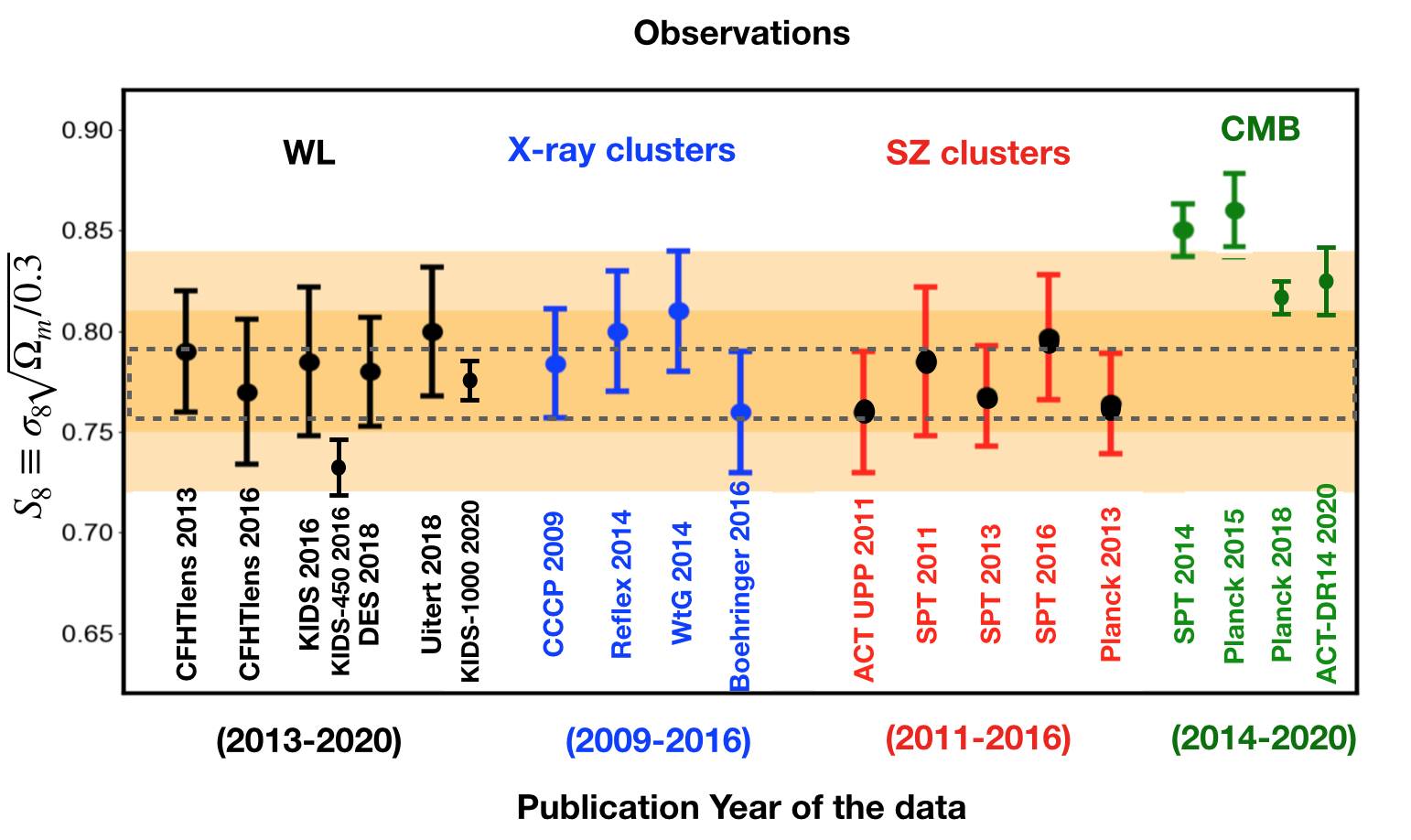}
    \caption{$S_8$ estimates according to the publication year from four different sources: \gls{wl}, X-ray clusters, Sunyaev-Zel'dovich (SZ) clusters and \gls{cmb} surveys. The error bars denote 1$\sigma$ CL and the year of publication for each analysis is denoted below the estimate, respectively: \cite{Heymans:2013fya,Choi:2015mnp,Hildebrandt:2016iqg,Heymans:2020gsg,vanUitert:2017ieu, Feigelson:2011ug,Bohringer:2014ooa,vonderLinden:2014haa,Boehringer:2016kqe, Hilton:2017gal,Bleem:2014iim,Aiola:2020azj,Aghanim:2018eyx}. The orange bars denote constraints from \gls{des} Y1 2018 with \texttt{Planck}, \gls{bao} from \gls{sdss}, 6\gls{df}, and \gls{boss} and \gls{jla} \gls{sn}Ia up to 2$\sigma$~\cite{Abbott:2017wau}. The dashed lines denote the 2$\sigma$ limit of \gls{kids}-1000~\cite{Heymans:2020gsg}. Notice that for the case $S_8=0.759^{+0.024}_{-0.021}$ from \gls{kids}-1000 according to Ref.~\cite{Asgari:2020wuj}, while the quoted number in Ref.~~\cite{Heymans:2020gsg} is a combination with \gls{boss}.
    }
    \label{fig:sigma_8}
\end{figure}

Up to now, we also have other $3\sigma$ tensions: {\it i)} the tension between the $\sigma_8$-$\Omega_{\rm m0}$ contours inferred from the \gls{cmb} in a $\Lambda$\gls{cdm} scenario and those inferred by cosmic shear data~\cite{Joudaki:2016kym,Hildebrandt:2018yau,Asgari:2020wuj,Abbott:2017wau,Abbott:2021bzy,Nunes:2021ipq}; {\it ii)} the tension between the cosmological parameters from the \gls{bao} signal in galaxies at $z<1$ and those of Ly$\alpha$ at higher redshift \cite{Cuceu:2019for} and {\it iii)} time derivative of the cosmological parameters with the \gls{cmb} fluctuation scale used to determine it \cite{Lombriser:2019ahl}. A compilation of the $S_8 := \sigma_8 \sqrt{\Omega_{\rm m0} /0.3}$, discussed in Sec.~\ref{sssec:Growth-index} and estimates up to 2020 is detailed in Fig.~\ref{fig:sigma_8}. The statistical errors of the above methods are expected to decrease in the next few years, and will reveal whether the tension is a statistical accident that we can expect when considering of order dozens of nuisance parameters, or whether it is hinting at yet to be discovered new physics.
For example, the latest DES-y3 release~\cite{Abbott:2021bzy,Amon:2021kas,Pandey:2021eex} is in agreement with the \texttt{Planck} $\Lambda$\gls{cdm} predictions, but it is showing an inconsistency of the amplitudes of the galaxy clustering with the galaxy–galaxy lensing. Even so, many have questioned if a solution to the late versus early-times discrepancy could be more credible if it also relaxed one or more of these additional tensions~\cite{DiValentino:2020vhf,DiValentino:2020zio,DiValentino:2020vvd,DiValentino:2020srs,Abdalla:2022yfr}.

\subsection{State-of-the-art}
\label{sec:observation_status_precision}

Over the last two years, \gls{tg} has started to take advantage of its analytical and physical description to gravity in order to provide a solution to the $H_0$ and $S_8$ tension problems (see Ref.~\cite{Verde:2019ivm,DiValentino:2020zio,DiValentino:2020vvd,DiValentino:2021izs,Perivolaropoulos:2021jda} for an overview of the tensions) and describe the observed and well-tested late-time dynamics with current cosmological data~\cite{Escamilla-Rivera:2020ges,Briffa:2020qli}. As we have discussed, in these kind of theories the curvature description is torsion-based one by the replacement of the Levi-Civita connection with a curvatureless connection labeled as the teleparallel connection~\cite{Weitzenbock1923}. It turns out, as discussed in detail in Sec.~\ref{Sec:Geometric_Trinity}, at the level of the action \gls{gr} and \gls{tegr} differ just by a boundary term which renders their field equations identical.
There are several ways to modify the \gls{tegr} proposal (see Sec.~\ref{sec5:extended} for a review of them).
With this methodology, we can generalize the Lagrangian to arbitrary functions of the torsion scalar in order to produce $f(T)$ gravity~\cite{Ferraro:2006jd,Linder:2010py,Chen:2010va}, which follows the same idea as $f(\lc{R})$ gravity. However, unlike $f(\lc{R})$ theory, $f(T)$ gravity produces generally second-order field equations, which makes them more easily open to test with observations.

Furthermore, given the importance of the boundary term in relating \gls{gr} and \gls{tegr}, $f(T,B)$ gravity has also been well-studied in Ref.~\cite{Escamilla-Rivera:2019ulu,Bahamonde:2015zma,Farrugia:2018gyz} as a possible extension to \gls{tegr}. This theory has been studied throughout this Review to describe \gls{flrw} cosmology at the background level and at the perturbation level (see Secs.~\ref{sec:fTBbackgroundcosmo} and~\ref{sec:cosmo-pert}).

Moreover, applying consistency tests with current cosmological data allows one to identify a viable gravity theory and deal at the same time with systematic effects in the data or any problems with the underlying cosmological model. Some data samples are sensitive to the theory and the dynamics of the Universe and some other samples are sensitive to the growth of large-scale structure~\cite{Cruz:2020cje}, e.g. in the early-time route of the distance ladder, \gls{bbn} allows a technique to measure precise abundances of hydrogen, helium, lithium and deuterium, and hence to test the cosmological model through being extremely sensitive at early-times~\cite{Aver:2015iza}. In these cases, these two sets of observations must be consistent with one another in order to solve the cosmological tension inside a viable theory of gravity. At late-times, any deviations between different theories can be measured through an effective \gls{eos} (mimicking a \gls{de} component) close to the standard $\Lambda$\gls{cdm} \gls{eos} value ($w=-1$).
Although the $\Lambda$\gls{cdm} model is very successful in explaining \textit{almost} all observations~\cite{Heymans:2013fya,Choi:2015mnp,Hildebrandt:2016iqg,Heymans:2020gsg,vanUitert:2017ieu, Bohringer:2014ooa,vonderLinden:2014haa,Boehringer:2016kqe, Bleem:2014iim,Aiola:2020azj,Aghanim:2018eyx},
it has some theoretical issues within its backbone structure. These include the mysterious physical origin of the two largest contributions to the energy content of the late-time Universe: cold dark matter (\gls{cdm}) and the cosmological constant, together with the too varied forms of inflation in the literature and their as yet unobserved nature.

Furthermore, through their mapping of the distance-redshift relation up to a redshift of $z \sim 1$, \gls{sn}Ia measurements have been a sensitive probe of the late-time cosmic acceleration. Another very powerful probe are \gls{gw}, e.g. those emitted from merging binary black holes and neutron stars, since the propagation of gravitons and photons can change at a fundamental level~\cite{LIGOScientific:2020tif}. To achieve this, \gls{gw} observations are sometimes coupled with a multimessenger signal through an electromagentic (EM) counterpart. These kind of sources are treated as \textit{standard sirens} (gravitational analogues of standard candles such as \gls{sn}) in the sense that the \gls{gw}s emitted by a compact binary contains information about the \gls{gw} luminosity of the binary system, and thus its luminosity distance~\cite{Schutz:1986gp}. On the other hand, massive binary black holes as standard sirens
have the disadvantage that the redshift is difficult to measure without an independently EM signal of the event. However, recently some progress has been made in determining the redshift from \gls{gw} observations alone \cite{LIGOScientific:2021aug}. If we identify the host galaxy then it should be possible to determine with joint events whether long-wavelength \gls{gw} and short-wavelength EM radiation undergo the same number of spacetime dimensions~\cite{2018JCAP...07..048P}. In Sec.~\ref{sec9:GW} we gave an overview of \gls{gw} in teleparallel theories. In this setting, another interesting feature of \gls{gw} astronomy is its potential in probing higher dimensional theories where spacetime \textit{leaks} into the extra dimensions, leading to the effect that cosmologically distant sources appear dimmer than they truly are, resulting in a systematic error in the inferred distance to the \gls{gw} source~\cite{Corman:2020pyr}. Also, there have been studies considering the propagation from the inspiraling of compact binary systems within the context of \gls{tg} theories in Ref.~\cite{Nunes:2018evm,Nunes:2019bjq}, meanwhile it is important to mention that these studies are in the context of a scenario where the \gls{gr} cosmological model is considered as an effective fluid in the standard perturbations equations. These studies lead open the idea of explaining these effects using non-standard extensions of \gls{gr}, as it is in \gls{tg} theories.

Moreover, unless higher precision probes of the Hubble flow $H_0$ and the growth of structure $S_8$ will be available without tension issues, these theories cannot be discriminated. The issue about the theoretical playground suggests that an approach to the problem of cosmic acceleration should rely on as few model-dependent quantities as possible. Since the number of viable gravitational theories at hand is large, on both astrophysical and cosmological scales, as well as in strong-gravity regimes, some of them could be good tools to compare theoretical predictions with survey observations, extracted from the \gls{cmb}, cosmological expansion, large-scale structures, astrophysical configurations and \gls{gw} signals. Moreover, these comparisons suffer from the so-called \textit{degeneracy problem}, which means that several gravitational theories are capable of describing certain observations with the same statistical confidence. A standard way of relaxing this issue consists of combining several observational measurements to reduce the phase space of free parameters. The search for \textit{viable} priors may result in numerical issues, where a possible solution could be to consider a cosmography approach rather than determining model-specific solutions from Friedmann equations and confronting them with observations. In this approach, cosmographic solutions are expressed, in the standard way, as derivatives of the scale factor, introducing the so-called cosmographic parameters which makes possible to fit the observations to the distance-redshift relation without prior theoretical assumptions.

In the standard approach to cosmography, we start with the \gls{flrw} metric and extend its parameters beyond the value of $H_0$, which evokes, at least, a fifth order Taylor expansion of the $a(t)$ in order to obtain a reliable approximation of the distance-redshift relation~\cite{Demianski:2016dsa,delaCruz-Dombriz:2016rxm,Luongo:2015zgq,Dunsby:2015ers}. This is needed when one is dealing with certain theories containing higher-order derivatives in the Hubble parameter.

A similar issue was tackled in Ref.~\cite{Escamilla-Rivera:2019aol}, where the cosmographic equations can be written down in an inverted version to study the dynamics through kinematics. In this approach, a homogeneous \gls{sn} observational sample was used to perform the statistical analysis of specific \gls{de} parameterizations, to directly obtain the kinematics of the cosmographic parameters. This approach relies on the fact that it is possible to set constraints over the cosmographic parameter values using statistical analyzes of the dynamics imposed in \gls{de}. In the literature, there are many studies that attempt to use these \gls{sn} observational samples to constrain the cosmography of models beyond $\Lambda$\gls{cdm}. However, due to the low redshift nature of the data sets in question, it may be misleading to consider such cases since it is unlikely to reveal any new dynamics without higher corrective terms.

A study in that direction was developed in Ref.~\cite{Capozziello:2019cav}, where an improved version of the standard cosmography approach was shown to be a viable solution to these issues. However, the approach is plagued with several shortcomings which limit its applicability in certain circumstances, and even more in extended theories of gravity. The main problems concern the inability of the proposed polynomials to be constrained with the currently available cosmological data so as to fix the kinematic expansion of the Universe, especially at early-times. Also, the arbitrary order of truncation of polynomials might compromise the predictive power of cosmography. The impossibility to measure them separately but only their sum, leads to different results depending on the probability distribution associated with each coefficient. To relax these issues, in Ref.~\cite{Escamilla-Rivera:2019aol,Munoz:2020gok} a mathematical expression to derive the \gls{eos} for a specific dynamical model was proposed. This expression allows us to obtain the cosmographic parameters without \textit{directly} assuming a cosmography-dependent polynomial series over them and avoid dealing with the aforementioned problems.

To apply these ideas to extended theories of gravity, a first step was performed in Ref.~\cite{Escamilla-Rivera:2019hqt}, where a novel computational tool based on machine learning for \gls{sn}, called Recurrent-Bayesian Neural Networks, was proposed to deal with different kinds of modified gravity models. A deep learning architecture can develop a trained homogeneous \gls{sn} sample, where the resulting dynamics of \gls{de} could lead to the necessity of another cosmological model different from $\Lambda$\gls{cdm}. In this scenario, the \textit{inverse cosmography approach} can fit statistically very well since it is not necessary to consider higher-order corrections of the cosmography series to obtain a convergent best-fit in comparison to the standard \gls{de} \gls{eos}.

Over the last few years an avalanche of observations and numerical techniques have been bringing new ways to test extended theories of gravity \cite{Bernardo:2021cxi,Dialektopoulos:2021wde,Bernardo:2021mfs}. Our purpose in this Section is to set the optimal conditions and recipes to treat them at the same level as the cosmological viable models derived from \gls{gr}.


\subsection{Viability of teleparallel cosmology} \label{sec:test an optimal TG}

In order to handle the analyses related to specific \gls{tg} models we require a strict recipe of the tools, theoretical and numerical, to study the cosmology derived from these models. To achieve this goal in the following sections (including information on standard statistical cosmology in the Supplementary annexes (Supplementary 3-4), we present several discussions on how to adapt \gls{tg} models at numerical level and study their cosmological constraints.
We can enlist these tools as follows:

\begin{itemize}
\item \textbf{Numerical codes and gravity model algorithms.} -- In the Supplementary annexes (Supplementary 3) we present three general algorithms to explain in detail how to implement some important \gls{tg} models using the standard numerical codes rewritten to analyze background and perturbative cosmology. Furthermore, we describe how to proceed with the Boltzmann code in order to adapt \gls{tg} models.

\item \textbf{Surveys and cosmological data} -- In the Supplementary annexes (Supplementary 4) we describe the current observational compilations available in the literature and how to handle the statistical analyses using $\chi^2$ methods and Bayesian inference. We will use these compilations in the description and computation of the best-fit cosmological parameters for several \gls{tg} scenarios.

\item \textbf{Cosmological compendium of Teleparallel scenarios} -- Using the information described in the Supplementary annexes (Supplementary 3-4)
we are ready to present in Sec.~\ref{subsec:compendium} the cosmologically viable \gls{tg} scenarios in order to study the current cosmic issues as described in the previous section.
In Table~\ref{tab:compendium} we present the most important extended theories where observational cosmology was studied, along with their references. As an extension, in Tables~\ref{Table-TG-physics}-\ref{Table-TG-physics2} we describe in detail the motivation, viability, maturity and late/early tests related to each theory under consideration.

\item \textbf{Precision cosmology for $f(T)$ and $f(T,B)$ gravity theories} -- Using the viable models described in the compendium, we present complete analyses for both theories using the respective modified version of Boltzmann codes and late-time data in Sec.~\ref{sec:f(T)_f(TB)_tests}. We further analyze the possibility of \textit{relaxing} the $H_0$ tension at the background level.

\item \textbf{Cosmography for \gls{tg}} -- This approach can be considered as a possible model-independent approach to tackle extensions and modifications of \gls{gr} in view of the current observational constraints, while breaking the degeneracy among models. In Sec.~\ref{subsec:cosmography_p} we will revise the proposals already available in the literature and confront them with precision cosmology tools, to present an update of the results.

\item \textbf{Beyond data-driven analysis for \gls{tg}} -- Reconstruction techniques that are model independent have since become more robust and better understood in terms of cosmological data sets. These nonparametric approaches, as they are commonly called in the literature, allow us to extract from the data the best cosmological model.
Using this nonparametric view, and as a perspective for future \gls{tg} scenarios analysis, in Sec.~\ref{subsec:gaussian_ML} we describe two current approaches on that path: \textit{(i)}
Reconstruction of data using Gaussian processes (\gls{gp}) that can offer an avenue by which gravitational theories can be reconstructed, guided by observational data without imposing stringent physical model assumptions. In modified \gls{tg}, this method has been applied to $f(T)$ models using compilations of galaxy ages and observations of \gls{bao} calibrated with Hubble data. \textit{(ii)}
Machine learning tools for \gls{tg} through deep learning techniques, which is a field of machine learning that uses several layers of non-linear processing neurons to obtain and transform at each successive layer an output from the previous layer.
A new deep learning method to classify $f(T)$ models is presented in this section.
\end{itemize}


\subsection{Compendium of teleparallel cosmologies} \label{subsec:compendium}

In order to confront different cosmological models with observations, and in view of the diversity of cosmological models to be confronted with observations, in this section we present a theoretical and observational state-of-art of the current \gls{tg} scenarios.

An overview of different cosmological models considered in the literature for different \gls{tg} theories is presented in Table~\ref{Table-TG-criteria}. This includes a compilation of constrained cosmological models discussed in the references indicated at the first column. The second column denotes the cosmological models already constrained by observations in the references mentioned. All these studies involve statistical tests on the family of models labeled and analytically detailed there. Notice that these equations were rewritten in comparison to the expressions in the reconstruction tables in the Supplementary annexes (Supplementary 1), in order to reduce the free parameter phase-space to be constrained. This is to ensure the convergence of the statistical test over the cosmological parameters of relevance and related directly with observable, e.g. the modulus distance luminosity for supernovae.

Together with the analysis of the cosmological models described, Table~\ref{Table-TG-physics} is detailed the following 2 different characteristics for each theory:
\begin{itemize}
    \item \textbf{Cosmological behaviours} -- These involve theories that can reproduce scenarios as early/late accelerated expansion, the dark sector and phantom/quintessence.
    \item \textbf{Maturity} -- Models that are already constrained by observations. It is indicated also which of them are not analyzed with observations yet.
\end{itemize}

All the models described in each theory are cosmological viable since they can reproduce several epoch of the Universe: from non-accelerating, matter dominated and late-time expansion era.

To complete the description of both Table~\ref{Table-TG-criteria} and Table~\ref{Table-TG-physics}, in Table~\ref{Table-TG-physics2} we included the observations used to constraint the models described. Notice that the analyses were made with late-time samples and early-time samples. For this we include the following catalogues described in detail in the Supplementary annexes (Supplementary 4):
\begin{itemize}
    \item \textbf{SNeIa} -- Pantheon compilation. Also, \gls{jla} 2015 from Ref.~\cite{Betoule_2014} and Union 2 (2010) compilation from Ref.~\cite{Suzuki_2012}.
    \item \textbf{BAO} -- Sample calibrated as inverse distance ladder. Also, it includes: \gls{sdss} from Ref.~\cite{Betoule:2014frx}, \gls{boss} from Ref.~\cite{Zhao:2016das}, the 2\gls{df} Galaxy Redshift Survey from Ref.~\cite{Colless:2003wz} and \gls{boss}, \gls{cmass}, Lyman-$\alpha$ \cite{Archidiacono:2019wdp}.
    \item \textbf{$H(z)$-CC} -- This includes the samples described in the Supplementary annexes (Supplementary 4) and also other one from Ref.~\cite{Moresco:2016mzx}.
    \item \textbf{CMB} -- From \texttt{Planck} legacy 2018.
    \item \textbf{WL} -- Optical and infrarred samples from \gls{kids} 450 + Viking (see Fig.~\ref{fig:sigma_8}).
\end{itemize}
Also we noted which of the theories are not tested with these observations yet. While there are reported constraints of these theories using early-time measurements, it is relevant to mention that all of them are performed by assuming the cosmological model derived from \gls{tg} as an effective dark energy fluid. Therefore, for future precision cosmology analyses of them it should be consider also the modified perturbed version of Boltzmann in the Supplementary annexes (Supplementary 3).

One important remark to note here is that the results reviewed in Table~\ref{Table-TG-physics2} depend on the cosmological perturbations of the \gls{tg} theory. As it was pointed out in Sec.~\ref{sec:cosmo-pert}, there is a strong coupling problem in $f(T)$ gravity that makes the conclusions regarding perturbations unclear or incomplete. Therefore, one needs to take these results with caution. However, for more general \gls{tg} theories, this issue has not been reported or studied in the literature since the Hamiltonian analysis has not been performed for them.

{
	\renewcommand{\arraystretch}{1.0}\begin{table}[htbp!]
	\centering
	\midsepremove
	\begin{tabularx}{\textwidth}{lX}
			\toprule
		\cellcolor{gris3}\textbf{Theory}	& 	\cellcolor{gris3}\textbf{Cosmological models constrained by observations}
	 \\
		\cellcolor{gris1} &	\cellcolor{gris1}
			1) Exponential Law: $ -T+1- e^{-T}$ \\
		\cellcolor{gris1} &	\cellcolor{gris1}	2) Power Law $T^{\alpha}$
	\\
		\cellcolor{gris1} &	\cellcolor{gris1}	3) Linear-Power: $-T+\beta (-T)^{n}$
		\\
		\cellcolor{gris1} &	\cellcolor{gris1}4) Logarithmic: $-T+\gamma \log{(\delta T)}$
	\\
		\multirow{-6}{*}{\cellcolor{gris1}$\ \ \ \ \ \ \ f(T)$} &	\cellcolor{gris1}	5) Hybrid: $-T+e^{\epsilon T}(-T)^{\alpha}$
	\\
		\multirow{-6}{*}{\cellcolor{gris1}
	\cite{Shaikh:2020fne,Sahlu:2019jmy,2020arXiv200507043M,Li:2018ixg,DAgostino:2020dhv}	} &	\cellcolor{gris1}	6) Linder: $1- e^{\sqrt{T/\zeta}}$
	\\
		\multirow{-6}{*}{	\cellcolor{gris1}\cite{DAgostino:2018ngy,2019arXiv190204406K, ElHanafy:2017sih,Capozziello:2017bxm}} &	\cellcolor{gris1}	7) Trigonometric: $-T+(-T)^n \tanh{\Big(\frac{T_0}{T}\Big)}$
	\\
			\multirow{-6}{*}{\cellcolor{gris1} \cite{Capozziello:2017uam,Farrugia:2016xcw,Chakrabarti:2019bed,Benetti:2020hxp}} &	\cellcolor{gris1}	8) Model independent: $\eta T + (T-T_0)\cosh{(T-T_0)} +\sinh{(T-T_0)}$
	\\
		\cellcolor{gris1}	\multirow{-6}{*}{\cellcolor{gris1}} &	\cellcolor{gris1}	9) Specific $w$-EoS (Hypergeometric):$\alpha\times \Big[\beta~ {}_2 F_{1} \Big(\frac{1}{2},-\frac{-T}{6}\Big) +\gamma~ {}_2 F_{1} \Big(\frac{1}{2},-\frac{-T}{6}\Big)\Big] $ \\
	\cellcolor{gris3}& \cellcolor{gris3}	1) Power Law: $-T+\alpha B^n +\beta (-T)^m$\\
			\multirow{-2}{*}{\cellcolor{gris3}$\ \ \ \ \,\,\, f(T,B)$} & \cellcolor{gris3}2) Mixed Power Law: $-T+\alpha B^n (-T)^m$\\
			\cellcolor{gris3}		\multirow{-2}{*}{\ \cite{Bhattacharjee:2020jfk,Capozziello:2019msc,Escamilla-Rivera:2019ulu}} & \cellcolor{gris3} 	3) Logarithmic: $-T+\alpha \log{B} $\\
				\cellcolor{gris3}\multirow{-2}{*}{} & \cellcolor{gris3}	4) Taylor polynomial: $f(0) +f_T (0) T+f_B(0) B+ f_{TB}(0)TB + f_{TT}(0)T^2 + f_{BB}(0) B^2 +\mathcal{O}(TB,T^2,B^2)$\\
				\cellcolor{gris1}$\ \ \ \ \ f(T,\phi)$ & \cellcolor{gris1} 	\\
				\cellcolor{gris1}\ \ \ \cite{Gonzalez-Espinoza:2020azh,Geng:2013uga,Gu:2012ww} & 	\cellcolor{gris1}\\
	\cellcolor{gris1}\ \ \ \cite{Ferraro:2006jd,Geng:2011ka} & \multirow{-3}{*}{\cellcolor{gris1}Mixed Power Law: $-T - G(T)G(\phi) -V(\phi)$, with $G(T)=T^s$,}	 	\\
				\cellcolor{gris1}	 & \multirow{-3}{*}{\cellcolor{gris1}$F(\phi)=\xi \phi^c$ and $V(\phi)=\lambda \phi^d$}\\
				\cellcolor{gris3}	$\, \, \, \,\,\,f(T,\Theta)$ &
					\cellcolor{gris3}1) Power Law: $-T+\alpha (-T)^n \Theta + \Lambda$
					\\
				\cellcolor{gris3}$\,\,\,\,\,\,$\cite{Harko:2014aja,Cai:2015emx}	 & \cellcolor{gris3}2) Power Law (second order): $-T+\alpha \Theta + \gamma T^2$\\
					\cellcolor{gris3}	$f(T,(\lc{\nabla} T)^2, \lc{\Box} T)$ &	\cellcolor{gris3} Model I: $T+\frac{\alpha_1 (\lc{\nabla} T)^2}{T^2} + \alpha_2 e^{\frac{\delta (\lc{\nabla} T)^2}{T^4}}$ \\
					\cellcolor{gris3}\ \ \ \ \ \ $\,\,$ \cite{Otalora:2016dxe}	&
						\cellcolor{gris3}Model 2: $T + \frac{\beta_1 \lc{\Box} T}{T} +\frac{\beta_2 (\lc{\Box} T)^2}{T^3} +\beta_3 e^{\frac{\sigma \lc{\Box} T}{T^3}}$	\\
		\cellcolor{gris1}$\,\, Tf(\lc{\Box}^{-1} T,\lc{\Box}^{-1}B)$ & 	\cellcolor{gris1}Model I: $f=-1-\lc{\Box}^{-1} T $\\
	\cellcolor{gris1}$\,\,\,\,\,\,\,\,\,\,\,$\cite{Bahamonde:2017sdo,Bamba:2017ufh} &	\cellcolor{gris1}Model II: $f=-1-\lc{\Box}^{-1} T + \lc{\Box}^{-1} B$\\
	\bottomrule
		\end{tabularx}
		\midsepdefault
		\caption{\label{Table-TG-criteria}
		Compilation of cosmological models from \gls{tg} theories. The cosmological models labeled at the second column correspond to the ones discussed in the references indicated at the first column. All these models involve statistical tests and constraining cosmological parameters analyses with observations. The type of observables and cosmological behaviours are detailed specifically in Tables~\ref{Table-TG-physics} and \ref{Table-TG-physics2}.	}
	\label{tab:compendium}
	\end{table}
}

\begin{table}[ht!]
	\centering
	\midsepremove
\resizebox{18cm}{!}{	\begin{tabular}{lcccccc}
		\toprule
	\multicolumn{1}{c}{\cellcolor{gris3}\textbf{Theories}}	&	\multicolumn{4}{c}{\cellcolor{gris3}\textbf{Cosmological behaviours}} &\multicolumn{2}{c}{\cellcolor{gris3}\textbf{Maturity}} 	\\ \midrule
		\cellcolor{gris1} & \cellcolor{gris1}\textbf{Early/Late} & \cellcolor{gris1}\textbf{Dark} & \cellcolor{gris1}\textbf{Phantom/} & \cellcolor{gris1}\textbf{Perturbations} & \cellcolor{gris1}\textbf{Early} & \cellcolor{gris1}\textbf{Late}
	\\
		\cellcolor{gris1} & \cellcolor{gris1}\textbf{accel.} & \cellcolor{gris1}\textbf{sector} & \cellcolor{gris1}\textbf{Quintessence} & \cellcolor{gris1}\textbf{(scalar/tensor)} & \cellcolor{gris1}\textbf{constraints} & \cellcolor{gris1}\textbf{constraints}
	\\\midrule
\cellcolor{gris3}$f(T)$	& \cellcolor{gris3}\cmark &\cellcolor{gris3}\cmark & \cellcolor{gris3}\cmark& \cellcolor{gris3}\cmark & \cellcolor{gris3}\cmark & \cellcolor{gris3}\cmark\\
\cellcolor{gris1}$f(T,B)$	& \cellcolor{gris1}\cmark &\cellcolor{gris1}\cmark & \cellcolor{gris1}\cmark& \cellcolor{gris1}\cmark & \cellcolor{gris1}\cmark & \cellcolor{gris1}\cmark\\
\cellcolor{gris3}$f(T,\phi)$	& \cellcolor{gris3}\cmark &\cellcolor{gris3}\cmark & \cellcolor{gris3}\cmark& \cellcolor{gris3}\cmark & \cellcolor{gris3}\cmark & \cellcolor{gris3}\cmark\\
\cellcolor{gris3}$f(T,\Theta)$	& \cellcolor{gris3}\cmark &\cellcolor{gris3}\cmark & \cellcolor{gris3}\cmark& \cellcolor{gris3}\cmark & \cellcolor{gris3}\cmark & \cellcolor{gris3}\cmark\\
\cellcolor{gris1}$f(T,\lc{\nabla}T,\lc{\Box}T)$	& \cellcolor{gris1}\cmark &\cellcolor{gris1}\xmark & \cellcolor{gris1}\cmark& \cellcolor{gris1}\xmark & \cellcolor{gris1}\xmark & \cellcolor{gris1}\xmark\\
\cellcolor{gris3}$Tf(\lc{\square}^{-1}T)$	& \cellcolor{gris3}\cmark &\cellcolor{gris3}\xmark & \cellcolor{gris3}\cmark& \cellcolor{gris3}\xmark & \cellcolor{gris3}\xmark & \cellcolor{gris3}\xmark\\
				\bottomrule
	\end{tabular}}
	\midsepdefault
\caption{A table of cosmological behaviour, which is a base for further observational constraints, and maturity, that indicates whether the particular theory can describe both early- and late-time behaviors in principle for the models reported in Table~\ref{Table-TG-criteria}. To that end, we list: (i) \textbf{Early/Late accel.} means that the theory models can explain early inflation and late-time accelerated expansion; (ii) \textbf{Dark sector} refers to the ability of this theory to explain both the cosmological dark matter and dark energy sector in terms of an effective dark energy; (iii) \textbf{Phantom/Quintessence} indicates whether the models can produce \gls{eos} values above and below $-1$; (iv) \textbf{Perturbations (scalar/tensor)} means that it produces reasonable scalar perturbations and tensor perturbations that predict gravitational wave propagation speeds equal or very close to that of light; (v/vi) \textbf{Early/Late constraints} indicates whether these models have been constrained against late-time data sets (e.g. SN, BAO or $H(z)$) and early-time data sets (e.g. CMB, BBN, BAO or WL) (See Table \ref{Table-TG-physics2} for further details).}

\label{Table-TG-physics}
\end{table}
\newpage

\begin{table}[ht!]
	\centering
	\midsepremove
	\begin{tabular}{lccccccc}
		\toprule
	\cellcolor{gris3}\textbf{Theories}	&	\multicolumn{7}{c}{\cellcolor{gris3}\textbf{Observables}}	\\ \midrule
		\cellcolor{gris1} & \multicolumn{3}{c}{\cellcolor{gris1}\textbf{Late-time}} & \multicolumn{4}{c}{\cellcolor{gris1}\textbf{Early-time}}
	\\ \midrule
\cellcolor{gris3}	& \cellcolor{gris3}SN &\cellcolor{gris3}BAO & \cellcolor{gris3}$H(z)$ & \cellcolor{gris3}CMB & \cellcolor{gris3}BBN & \cellcolor{gris3}BAO & \cellcolor{gris3}WL \\ \midrule
\cellcolor{gris1}$f(T)$	& \cellcolor{gris1}\cmark\,[Pantheon] &\cellcolor{gris1}\cmark & \cellcolor{gris1}\cmark\,[CC-2018]& \cellcolor{gris1}\cmark$^{*}$ & \cellcolor{gris1}\cmark & \cellcolor{gris1}\cmark & \cellcolor{gris1}\cmark\\
\cellcolor{gris3}$f(T,B)$	& \cellcolor{gris3}\cmark\,[Pantheon] &\cellcolor{gris3}\cmark & \cellcolor{gris3}\cmark\,[CC-2018]& \cellcolor{gris3}\cmark$^{*}$ & \cellcolor{gris3}NAY & \cellcolor{gris3}NAY & \cellcolor{gris3}NAY\\
\cellcolor{gris1}$f(T,\phi)$	& \cellcolor{gris1}\cmark\,[Union 2] &\cellcolor{gris1}\cmark & \cellcolor{gris1}NAY& \cellcolor{gris1}\cmark$^{*}$ & \cellcolor{gris1}NAY & \cellcolor{gris1}NAY & \cellcolor{gris1}NAY\\
\cellcolor{gris1}$f(T,\Theta)$	& \cellcolor{gris1}NAY &\cellcolor{gris1}NAY & \cellcolor{gris1}NAY& \cellcolor{gris1}NAY & \cellcolor{gris1}NAY & \cellcolor{gris1}NAY & \cellcolor{gris1}NAY\\
\cellcolor{gris3}$f(T,\lc{\nabla}T,\lc{\Box}T)$	& \cellcolor{gris3}NAY &\cellcolor{gris3}NAY & \cellcolor{gris3}NAY& \cellcolor{gris3}NAY & \cellcolor{gris3}NAY & \cellcolor{gris3}NAY & \cellcolor{gris3}NAY\\
\cellcolor{gris1}$Tf(\lc{\square}^{-1}T)$	& \cellcolor{gris1}\cmark\,[JLA] &\cellcolor{gris1}\cmark & \cellcolor{gris1}\cmark\,[CC-2016]& \cellcolor{gris1}NAY & \cellcolor{gris1}NAY & \cellcolor{gris1}NAY & \cellcolor{gris1}NAY\\
				\bottomrule
	\end{tabular}
	\midsepdefault
\caption{A table of early- and late-time constraints in the literature for the models reported in Table~\ref{Table-TG-criteria}. Here, NAY means \textit{not analyzed yet} in the literature which is different to the indication provided by the theoretical work highlighted in Table \ref{Table-TG-physics}. The asterisk indicates that the theories are constrained using CMB at background level which refers to whether the corrected Boltzmann equations have been used to analyze observational data in the literature.}\label{Table-TG-physics2}
\end{table}


\subsubsection{Precision cosmology for teleparallel scenarios}
\label{sec:f(T)_f(TB)_tests}

There are numerous alternative models to $\Lambda$\gls{cdm}, as there are numerous problems with which they tend to struggle with. Some models are plagued by theoretical instabilities and others require at least some degree of fine-tuning of the model parameters in order to meet observational constraints. However, numerical analyses have their own limitations in practice. A major drawback is their lack of efficiency, as is evident in the effort poured into making existing numerical codes faster to cope with the ever increasing demand for numbers of independent realizations, volume and resolution. This is performed in order to match the characteristics of current and future observational surveys. Unlike the standard cosmological model, which is unique and is widely accepted as a concordance model, there are many modified theories, which makes it more difficult to allocate much effort to forecasting for individual models or performing a continuous parameter search. Furthermore, simulations are often used as a black box, making the underlying physics untreatable, which does not help in developing reliable theoretical templates used in model constraints. Although a simulation can predict structure formation at very small scales, this information is not entirely trustworthy due to the uncertainties in modelling baryonic physics. In such cases we need accurate model predictions to further analyze the non-linear regime and when the physics governing the structure formation needs to considered. However, if approximation methods can be developed for the gravity models, and demonstrated to be valid in certain regimes, then it would be realistic to use observations in those regimes to determine model constraints much more efficiently. On this path, in this Review we will devoted to possible \gls{tg} models that can offers a possibility to solve (or alleviate) the cosmological problems described.

To proceed with the implementation of a specific \gls{tg} model in the numerical cosmological codes, we start by describing the required cosmological evolution equations. In $f(T)$ and $f(T,B)$ theories it was found in Sec.~\ref{sec:cosmology_in_TG} that the \gls{eos} can be our starting equation to employ in constriction analyses. These \gls{eos} equations for each theory are linked to a specific form of $f(T)$ and $f(T,B)$, whose can be related to specific cosmological models in order to investigate the effects of a late-time cosmic accelerated expansion without the influence of an exotic \gls{de} or extra fields.

With the background evolution and equation of state equations, we can proceed with their implementation in the numerical code for each specific theory. The architecture structures to achieve this task are detailed in the Supplementary annexes (Supplementary 3), where for the background analysis we need to modify a Markov Chain Monte Carlo (\gls{mcmc}) code and the information detailed in the Supplementary annexes (Supplementary 3-5).

\begin{figure}
\centering
    \includegraphics[width= 0.45\textwidth]{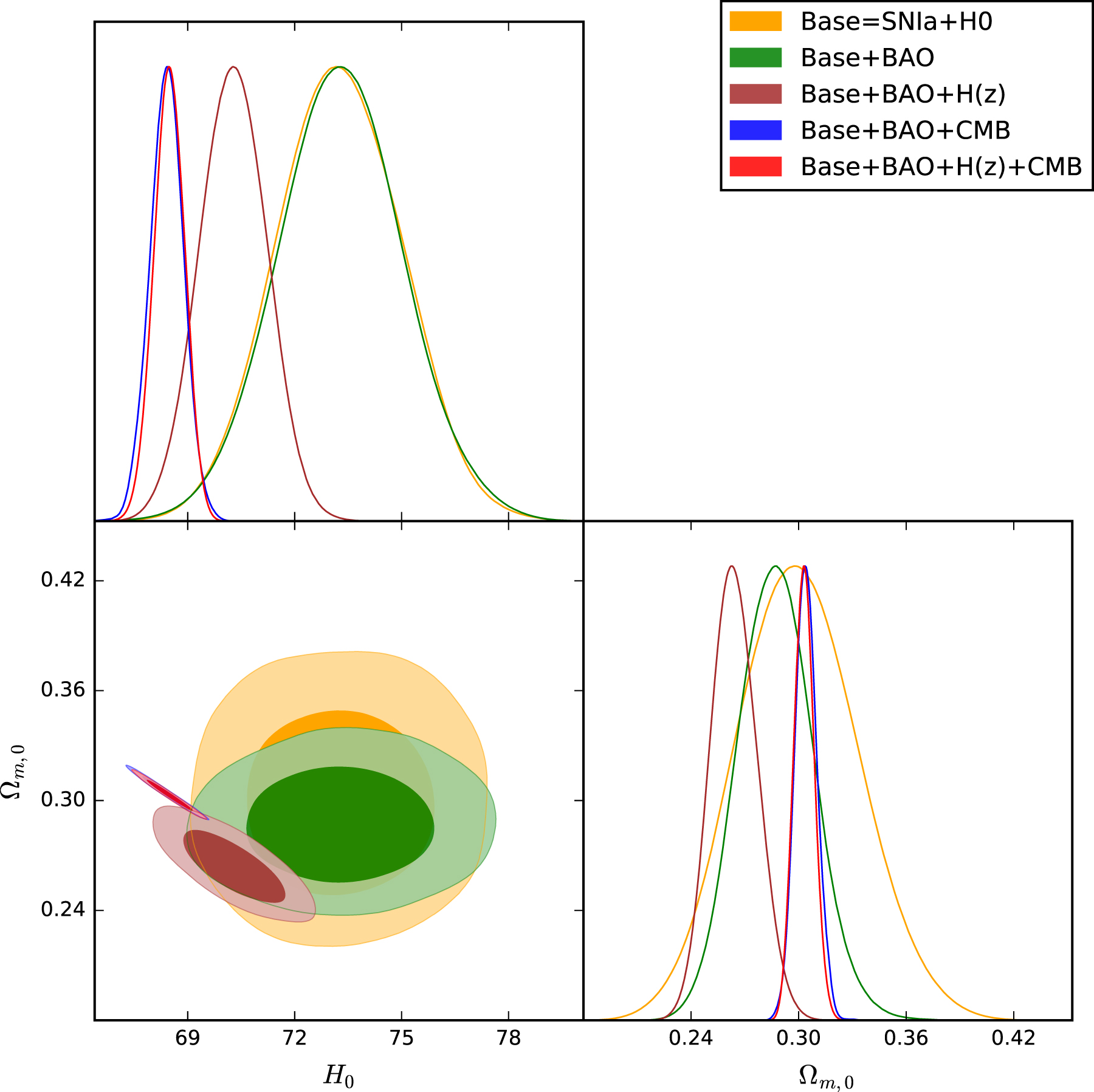}
    \includegraphics[width= 0.45\textwidth]{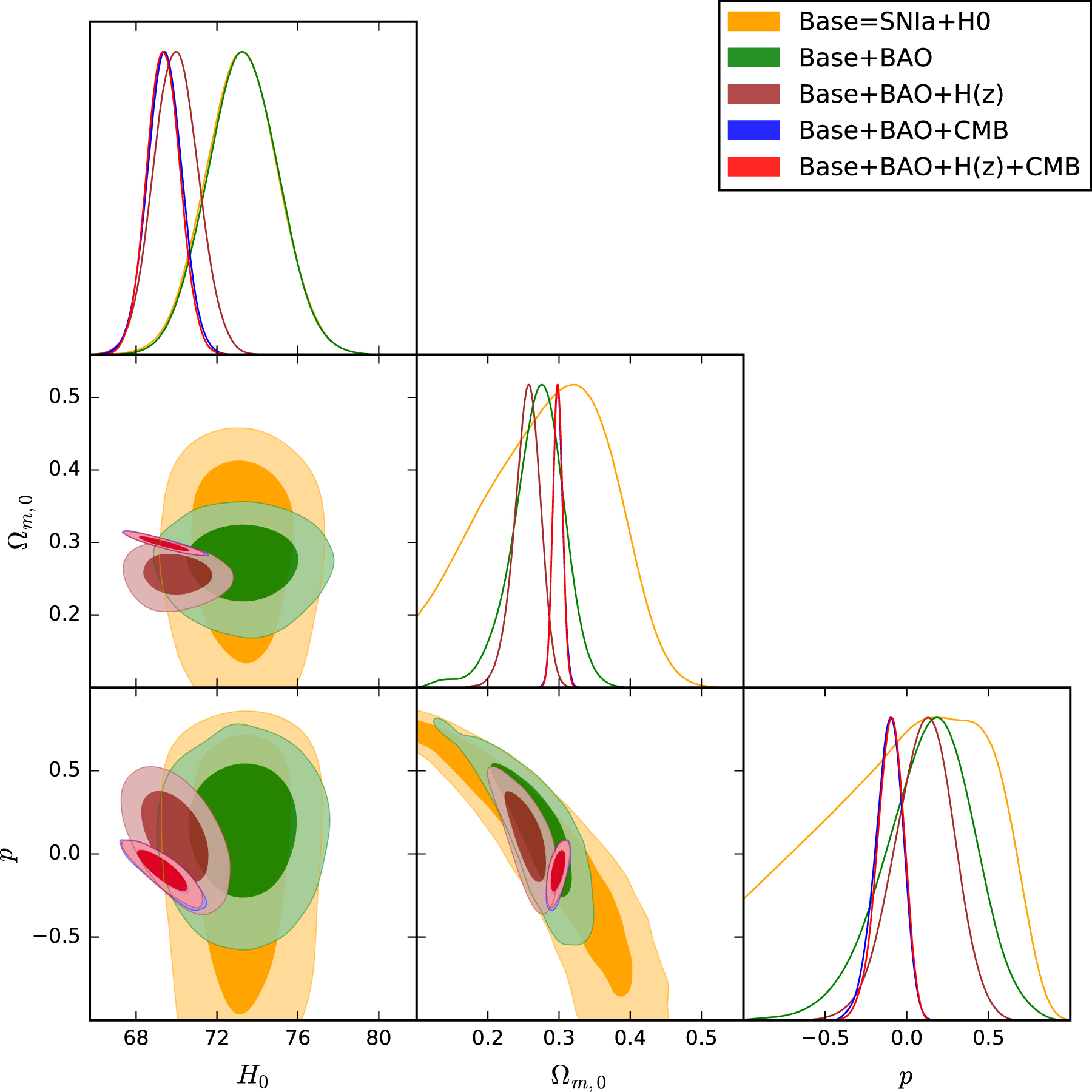}
    \caption{\textit{Left:} CL plots of $H_0$ and $\Omega_{\rm m0}$ in the $\Lambda$\gls{cdm}. Base denotes the combination of the \gls{jla} \gls{sn}eIa and the local value of the Hubble constant. \textit{Right:} Confidence regions of $H_0$, $\Omega_{\rm m0}$ and $p\equiv b$ in the $f_1(T)$ model. Base denotes the combination of the \gls{jla} \gls{sn}eIa and the local value of the Hubble constant. Permission for use of this figure was kindly provided by the authors of Ref.~\cite{Xu:2018npu}.}
    \label{fig:test_fT22}
\end{figure}


\subsubsection{Precision Cosmology for $f(T)$}\label{sec:f_T_models}

Observing the form of the first Friedmann Eqs.~\eqref{Friedmann_1B}, we deduce that in $f(T)$ cosmology we acquire an effective \gls{de} sector of gravitational origin. In particular, we can define the effective \gls{de} density as~\cite{Cai:2015emx,Santos:2021ypw}
\begin{equation}
    \rho_{\rm DE}\equiv\frac{3}{\kappa^2} \left[-\frac{f}{6}+\frac{TF_T}{3} \right]\,.
\end{equation}
Now, we are ready to analyze a set of viable $f(T)$ cosmological models. According to Tables~\ref{Table-TG-physics}-\ref{Table-TG-physics2}, there are several $f(T)$ models proposed to explain the present accelerating cosmic expansion. In this Review, we consider specific forms of $f(T)$ and confront them with \gls{bao}, \gls{cmb}, \gls{sn}eIa, and Hubble expansion data by using the publicly available \gls{mcmc} code \texttt{CosmoMC} \cite{cosmomcodes,Lewis:2002ah,Lewis:2013hha} which is interfaced with the Boltzmann integrator \texttt{CAMB} \cite{cambcodes,Lewis:1999bs}. For the analyses considered in this section, the \texttt{Planck} 2015 \gls{cmb} distance priors \cite{Huang:2014gta} are adopted, while the $H(z)$ data consists of 38 data points from Ref.~\cite{Farooq:2016zwm} and the \gls{sn}eIa data corresponds to the SNLS-\gls{sdss} \gls{jla} \cite{Betoule:2014frx}. This information is managed using the concepts described in the Supplementary annexes (Supplementary 4-5). The results described in the following sections are based on Ref.~\cite{Xu:2018npu}, were an adopted $H_0$ prior corresponds to $H_0^{R16}=73.24\pm1.74\,\mathrm{km}\,\mathrm{s}^{-1}\mathrm{Mpc}^{-1}$ \cite{Riess:2016jrr}. It is important to mention that such results, tighter to a prior, could have a larger significance which can derive in a worse constrain on the parameters in the $f(T)$ models. However, we should also point out that the methodology described here will allow a better control on the flat prior in future analysis on this matter. On the issue of priors in the MCMC analysis, there are strong concerns about the statistical coherence of such analyses since they may put undue bias toward such values. This is especially poignant when putting priors on the Hubble constant. Putting prior values on in these analysis may spoil the posterior outputs and thus the estimates on the model parameters \cite{Efstathiou:2021ocp}. On the other hand, there have been numerous studies in the literature that assume a prior on the Hubble constant. These analyses may prove useful in understanding viable parameter values. However, caution should be exercised when interpreting these studies.

In the standard $f(T)$ literature we deal with the following models:

\begin{table}[t]
    \centering
    \midsepremove
    \begin{tabular}{ l c c c }
       \toprule
       \cellcolor{gris3}\textbf{Data set} & \cellcolor{gris3}\boldmath{$H_0\,[\text{km/s/Mpc}]$} & \cellcolor{gris3}\boldmath{$\Omega_{\rm m0}$} & \cellcolor{gris3}\boldmath{$b$} \\
	 \midrule
		 \cellcolor{gris1}Base & \cellcolor{gris1}$73.2\pm1.8$ & \cellcolor{gris1}$0.284^{+0.100}_{-0.077}$ & \cellcolor{gris1}$-0.02^{+0.64}_{-0.37}$ \\
	 \cellcolor{gris3}Base+\gls{bao} & \cellcolor{gris3}$73.3\pm1.7$ & \cellcolor{gris3}$0.269^{+0.040}_{-0.030}$ & \cellcolor{gris3}~~$0.13^{+0.30}_{-0.24}$ \\
    	 \cellcolor{gris1}Base+\gls{bao}+$H(z)$ & \cellcolor{gris1}$70.0\pm1.1$ & \cellcolor{gris1}$0.255^{+0.020}_{-0.016}$ & \cellcolor{gris1}~~$0.10^{+0.19}_{-0.17}$ \\
		 \cellcolor{gris3}Base+\gls{bao}+\gls{cmb} & \cellcolor{gris3}$69.4\pm0.9$ & \cellcolor{gris3}$0.298\pm0.007$ & \cellcolor{gris3}$-0.11^{+0.10}_{-0.08}$~\\
		 \cellcolor{gris1}Base+\gls{bao}+$H(z)$+\gls{cmb} & \cellcolor{gris1}$69.4\pm0.8$ & \cellcolor{gris1}$0.298\pm0.007$ & \cellcolor{gris1}$-0.10^{+0.09}_{-0.07}$ \\[.3em]
	\bottomrule
    \end{tabular}
    \midsepdefault
    \caption{The mean value and the corresponding 68\% limits~\cite{Xu:2018npu} of the model parameters of the $f_1(T)$ model. Base refers to the JLA \gls{sn}eIa data and $H_0$ prior.}
    \label{tab:f_1_T}
\end{table}

\begin{figure}
\centering
    \includegraphics[width= 0.45\textwidth]{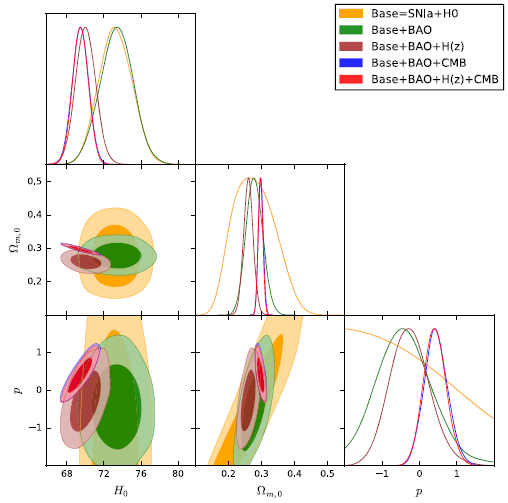}
    \includegraphics[width= 0.45\textwidth]{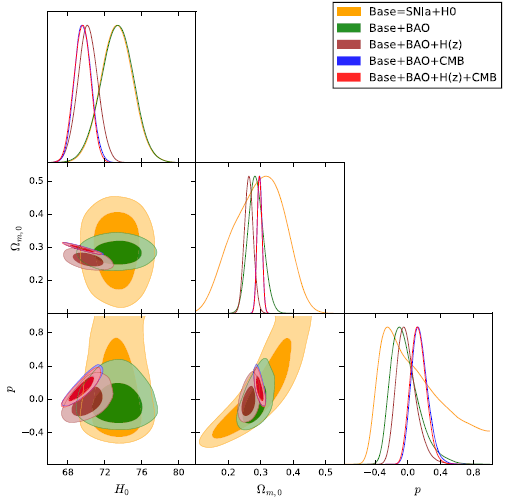}
    \caption{\textit{Left:} CL plots of $H_0$, $\Omega_{\rm m0}$ and $p$ in the square--root--exponential $f_2(T)$ model. \textit{Right:} Confidence regions of $H_0$, $\Omega_{\rm m0}$ and $p\equiv1/q$ in the $f_3(T)$ model. Base denotes the combination of the \gls{jla} \gls{sn}eIa and the local value of the Hubble constant. Permission for use of this figure was kindly provided by the authors of Ref.~\cite{Xu:2018npu}.}
    \label{fig:test_fT3}
\end{figure}

\begin{enumerate}
\item Power law model.
\label{sec:f_T_power_law}
We consider the power law model~\cite{Bengochea:2008gz} $(f_1(T))$ as the first $f(T)$ model, specified by
\begin{equation}
\label{fTmodel1}
    f_1(T)=-T+\alpha_1\,(-T)^b\,,
\end{equation}
where $\alpha_1$ and $b$ are constant model parameters, such that
    $ \alpha_1=\left(6H_0^2\right)^{1-b}\frac{\Omega_{\rm m0}-1}{2b-1}\,$ (this is found by evaluating the Friedmann equation at current times).
We depict the inferred confidence regions for the power law model in Fig.~\ref{fig:test_fT22} and the corresponding model parameter constraints in Table~\ref{tab:f_1_T}. It is clear that the $\Lambda$\gls{cdm} model is within the 1$\sigma$ confidence level (CL) when the \gls{cmb} data sets are not considered, while the $f_1(T)$ model becomes consistent with the concordance model of cosmology at 2$\sigma$ when including the \gls{cmb} data.


\item Square--root--exponential model.
\label{sec:f_T_square_root_exp}
The second $f(T)$ model (labelled as $(f_2(T))$) is the square--root--exponential model given by~\cite{Linder:2010py}
\begin{equation} \label{fTmodel2}
    f_2(T)=-T+\alpha_2\, T_0\left(\,1-e^{-p\sqrt{T/T_0}}\,\right)\,,
\end{equation}
with model parameters $\alpha_2$ and $p$, while
    $ \alpha_2=-\frac{1-\Omega_{\rm m0}}{1-(1+p)e^{-p}}\,.$
From the left panel of Fig. \ref{fig:test_fT3} as well ad from the inferred model parameter constraints of Table \ref{tab:f_2_T}, the Base data set was not able to constrain the model parameter $p$. However, the inclusion of the other data sets led to a constraint on $p$, with the tightest constraint being derived when considering the \gls{cmb} data set.
\begin{table}[t]
    \centering
    \midsepremove
    \begin{tabular}{ l c c c }
       \toprule
       \cellcolor{gris3}\textbf{Data set} & \cellcolor{gris3}\boldmath{$H_0\,[\text{km/s/Mpc}]$} & \cellcolor{gris3}\boldmath{$\Omega_{\rm m0}$} & \cellcolor{gris3}\boldmath{$p$} \\
	 \midrule
		 \cellcolor{gris1}Base & \cellcolor{gris1}$73.2\pm1.7$ & \cellcolor{gris1}$0.278^{+0.056}_{-0.070}$ & \cellcolor{gris1}$<0.23$ \\
	 \cellcolor{gris3}Base+\gls{bao} & \cellcolor{gris3}$73.4\pm1.7$ & \cellcolor{gris3}$0.279\pm0.025$ & \cellcolor{gris3}$-0.39^{+0.67}_{-0.80}$ \\
    	 \cellcolor{gris1}Base+\gls{bao}+$H(z)$ & \cellcolor{gris1}$70.0\pm1.1$ & \cellcolor{gris1}$0.260\pm0.015$ & \cellcolor{gris1}$-0.26^{+0.49}_{-0.57}$ \\
		 \cellcolor{gris3}Base+\gls{bao}+\gls{cmb} & \cellcolor{gris3}$69.5\pm0.9$ & \cellcolor{gris3}$0.298\pm0.007$ & \cellcolor{gris3}$0.44\pm0.32$~\\
		 \cellcolor{gris1}Base+\gls{bao}+$H(z)$+\gls{cmb} & \cellcolor{gris1}$69.5\pm0.8$ & \cellcolor{gris1}$0.297\pm0.007$ & \cellcolor{gris1}$0.41\pm0.31$ \\[.3em]
	\bottomrule
    \end{tabular}
    \midsepdefault
    \caption{The mean value and the corresponding 68\% limits~\cite{Xu:2018npu} of the model parameters of the $f_2(T)$ model. Base refers to the JLA \gls{sn}eIa data and $H_0$ prior.}
    \label{tab:f_2_T}
\end{table}

\item Exponential model.
\label{sec:f_T_exponential}
We now consider the exponential model~\cite{Linder:2010py} $(f_3(T))$, which is also motivated by $f(\lc{R})$ gravity~\cite{Linder:2009jz}, and is specified by
\begin{equation} \label{fTmodel3}
    f_3(T)=-T+\alpha_3\, T_0\left(1-e^{-qT/T_0}\right)\,,
\end{equation}
with
    $ \alpha_3=\frac{1-\Omega_{\rm m0}}{-1+(1+2q)e^{-q}}\,,$
and $q$ is the remaining model parameter. Similar to the previously considered model. The derived model parameter constraints are reported in Table~\ref{tab:f_3_T}, where the model parameter $p\equiv1/q$ has always been constrained by the data sets being considered. The corresponding confidence regions are illustrated in the right panel of Fig.~\ref{fig:test_fT3}.
\begin{table}[!ht]
    \centering
    \midsepremove
    \begin{tabular}{ l c c c }
       \toprule
       \cellcolor{gris3}\textbf{Data set} & \cellcolor{gris3}\boldmath{$H_0\,[\text{km/s/Mpc}]$} & \cellcolor{gris3}\boldmath{$\Omega_{\rm m0}$} & \cellcolor{gris3}\boldmath{$p$} \\
	 \midrule
		 \cellcolor{gris1}Base & \cellcolor{gris1}$73.3\pm1.7$ & \cellcolor{gris1}$0.295^{+0.084}_{-0.070}$ & \cellcolor{gris1}~~$0.10^{+0.22}_{-0.51}$ \\
	 \cellcolor{gris3}Base+\gls{bao} & \cellcolor{gris3}$73.3\pm1.8$ & \cellcolor{gris3}$0.285^{+0.022}_{-0.025}$ & \cellcolor{gris3}$-0.03^{+0.11}_{-0.21}$ \\
    	 \cellcolor{gris1}Base+\gls{bao}+$H(z)$ & \cellcolor{gris1}$70.2\pm1.1$ & \cellcolor{gris1}$0.263\pm0.014$ & \cellcolor{gris1}$-0.01^{+0.09}_{-0.14}$ \\
		 \cellcolor{gris3}Base+\gls{bao}+\gls{cmb} & \cellcolor{gris3}$69.6\pm0.9$ & \cellcolor{gris3}$0.296\pm0.008$ & \cellcolor{gris3}~~$0.15^{+0.08}_{-0.11}$\\
		 \cellcolor{gris1}Base+\gls{bao}+$H(z)$+\gls{cmb} & \cellcolor{gris1}$69.6\pm0.9$ & \cellcolor{gris1}$0.296\pm0.007$ & \cellcolor{gris1}~~$0.13^{+0.09}_{-0.11}$ \\[.3em]
	\bottomrule
    \end{tabular}
    \midsepdefault
    \caption{The mean value and the corresponding 68\% limits~\cite{Xu:2018npu} of the model parameters of the $f_3(T)$ model. Base refers to the \gls{jla} \gls{sn}eIa data and $H_0$ prior.}
    \label{tab:f_3_T}
\end{table}

\end{enumerate}

\begin{figure}[t!]
\centering
\includegraphics[width=0.48\textwidth,origin=c,angle=0]{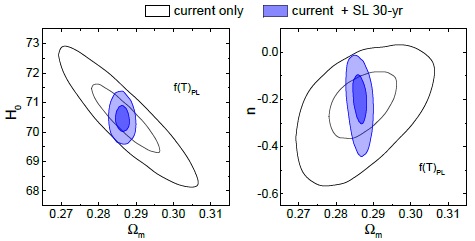}
\includegraphics[width=0.48\textwidth,origin=c,angle=0]{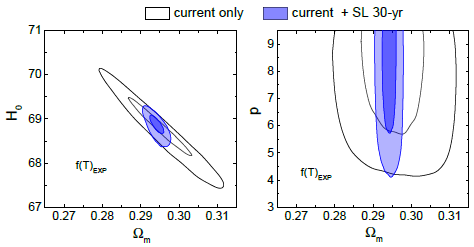}
\caption{Constraints in the $\Omega_{\rm m0}-H_0$ plane and in the $\Omega_{\rm m0}-n/p$ plane for power law $f(T)$ (left) and exponential $f(T)$ (right) models with current only and current + Sandage--Loeb 30--year data. This figure first appeared in Ref.~\cite{Geng:2015hen}.Permission for use of this figure was kindly provided by the authors of Ref.~\cite{Geng:2015hen}.}
\label{fig:SL_f(T)}
\end{figure}

There have been other studies of a similar nature in the literature such as Ref.~\cite{Briffa:2021nxg}. Upcoming constraints on the model parameters of $f(T)$ gravity are expected to be improved, for instance, Ref.~\cite{Geng:2015hen} explores the confrontation of $f(T)$ with forthcoming redshift drift data via the Sandage--Loeb test. It is expected that the thirty--nine metre European Extremely Large Telescope (E-ELT) will be equipped with a high--resolution spectrograph called CODEX (COsmic Dynamics EXperiment), which is designed to collect such Sandage--Loeb test signals. It was found that the redshift drift data alone cannot tightly constrain \gls{de} models because of the lack of low redshift data. However, it has been repeatedly shown that when the forthcoming redshift drift data is jointly combined with any other existing cosmological data set, this enhances the current model parameter constraints and leads to fewer parameter degeneracies.

In Ref.~\cite{Geng:2015hen}, the authors first constrain the power law and exponential $f(T)$ models using a combination of current data sets from \gls{sn}eIa, \gls{cmb}, \gls{bao}, and a direct measurement of the Hubble constant $H_0$, and then choose the best-fit models as fiducial models by producing 30 mock redshift drift data points, which are then used to jointly constrain the $f(T)$ models with this simulated data. The inferred constraints are illustrated in the panels of Fig.~\ref{fig:SL_f(T)} for the power law $f(T)$ and exponential $f(T)$ models.

Furthermore, advances related to the current scenario have been done using early-time data. In Refs.~\cite{Hashim:2020sez,Hashim:2021pkq} it has been reported that perturbed exponential infrared $f(T)$ produces a \gls{cmb} spectrum identical to $\Lambda$\gls{cdm}.


\subsubsection{Precision cosmology for \texorpdfstring{$f(T,B)$}{fTB} gravity}
\label{sec:f(TB)_test}

The generic \gls{eos} provided in Eq.~\eqref{eq:eos_fTB} which we now consider for four possible models in an $f(T,B)$ gravity setting. Our goal is to study the effects of dynamical \gls{de} for each models.

\begin{enumerate}
\item General Taylor expansion model.
Taking the general Taylor model that is introduced in Eq.~\eqref{taloy}, the \gls{eos} for the (background) effective fluid component can be determined through Eq.~\eqref{eq:eos_fTB} giving\footnote{We represented here $a^{(4)} = d^4 t/dt^4$.}:
\begin{subequations}
\begin{align}
    w(a) &= \frac{-24 A_3 + 6(H'+H^2) + 24H^3w_{x_1} -6w_{x_2} -24H^5 w_{x_7} +w_{x_9}}{-6(H'+H^2) + 72H^3 w_{x_4} +w_{x_5} +w_{x_6}}\\[0.5ex]
    &= \frac{-24w_{x_7} \dot{a}(t)^5+6 a(t)^4 \left[\ddot{a}(t)-4 A_3 a^{(4)}(t)\right]+24 a(t) \dot{a}(t)^3 w_{x_1} -6 a(t)^3 w_{x_2} + w_{x_9}} {-6 a(t)^4 \ddot{a}(t) + w_{x_5} + 72 a(t) \dot{a}(t)^3 w_{x_4} + w_{x_6}}\,,
\end{align}
\end{subequations}
where the $w_{x_i}$ definitions can be found in Appendix~\ref{sec:f_T_B_quantities}.
To perform the numerical analysis, we rewrite the above expression in terms of redshift $z=a_0 /a -1$, where $z=0$ corresponds to the present time. The \gls{eos} in this setting can be expressed
\begin{align} \label{eq:eosz_taylor}
    w(z)&= \frac{w_1(z) + 6\left[w_2(z) -w_6(z)\right] + 24\left[w_3(z)-w_4(z) -w_5(z)\right]} {-w_1(z) - 6 w_7(z) + 72 \left[w_8(z)- w_9(z) -w_{10}(z)\right] - \frac{12}{(z+1)^7}}\,,
\end{align}
where each $w_i(z)$ function is also defined in Appendix~\ref{sec:f_T_B_quantities}.
To solve this equation we analyze the behavior of the \eqref{eq:eosz_taylor} \gls{eos} associated with the fluid representation of $f(T,B)$ denoted by $w$ considering four kind of cases (c.f. Fig.~\ref{evolution_powerlaw}).

The specific cases that we can consider in order to develop a constraint analysis in the redshift range of observations include aspects when the contribution from $T$ is larger that $B$ ($T\gg B$), and vice versa ($T\ll B$):
\begin{itemize}
\item Case 1.1: Domination of the boundary term over the torsion scalar.
\item Case 1.2: Domination of the torsion scalar over the boundary term.
\item Case 2.1: Domination of the boundary term over the torsion scalar.
\item Case 2.2: Domination of the torsion scalar over the boundary term).
\end{itemize}
For Eq.~\eqref{eq:eosz_taylor}, we perform the fitting using the completed compilation of the observations described. In Fig.~\ref{contour_taylor_case1}, we see that the model has a preference for a phantom solution in agreement with \texttt{Planck} data for the density of matter.
\begin{figure}
\centering
\includegraphics[width=0.3\textwidth,origin=c,angle=0]{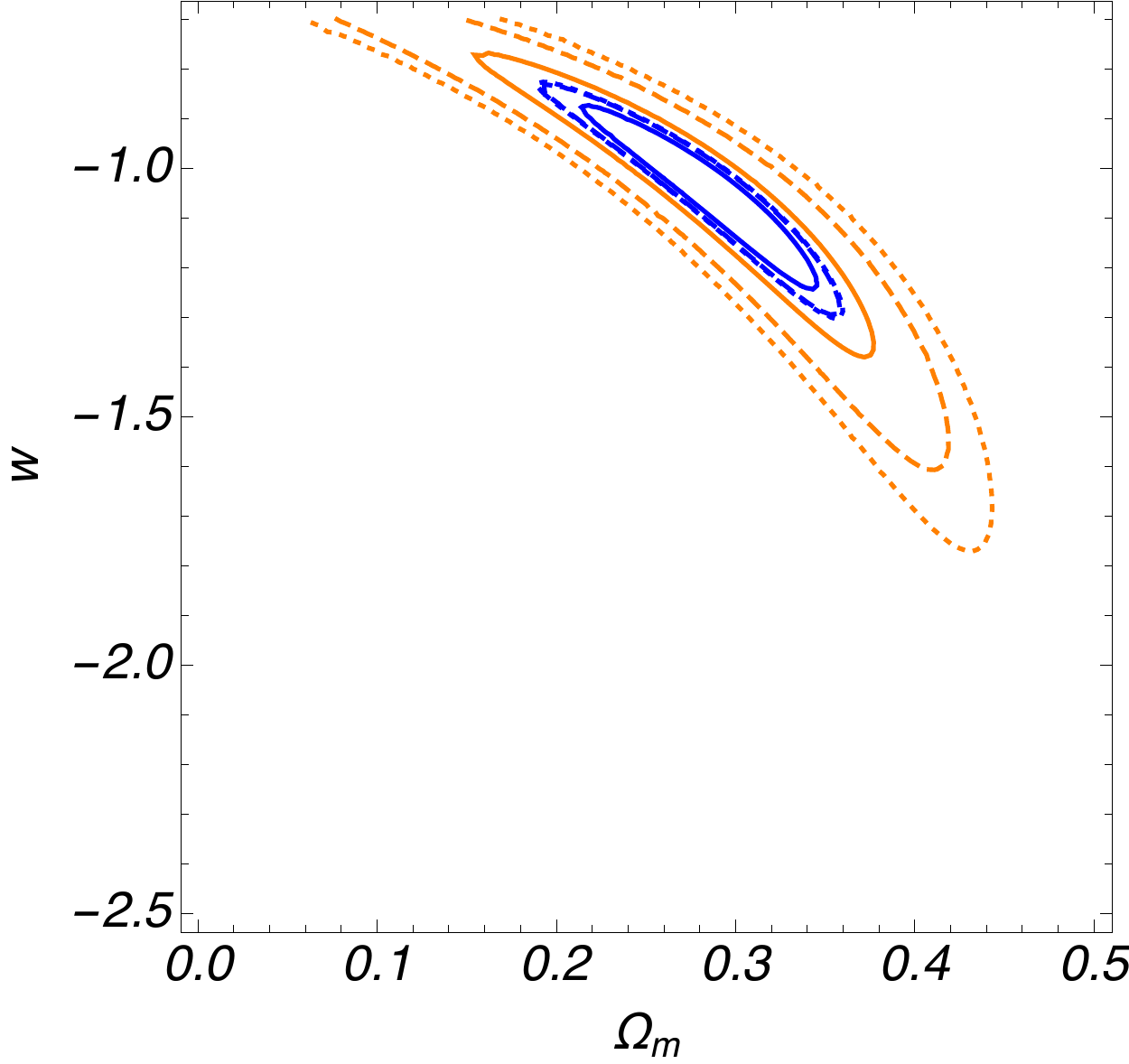}
\includegraphics[width=0.288\textwidth,origin=c,angle=0]{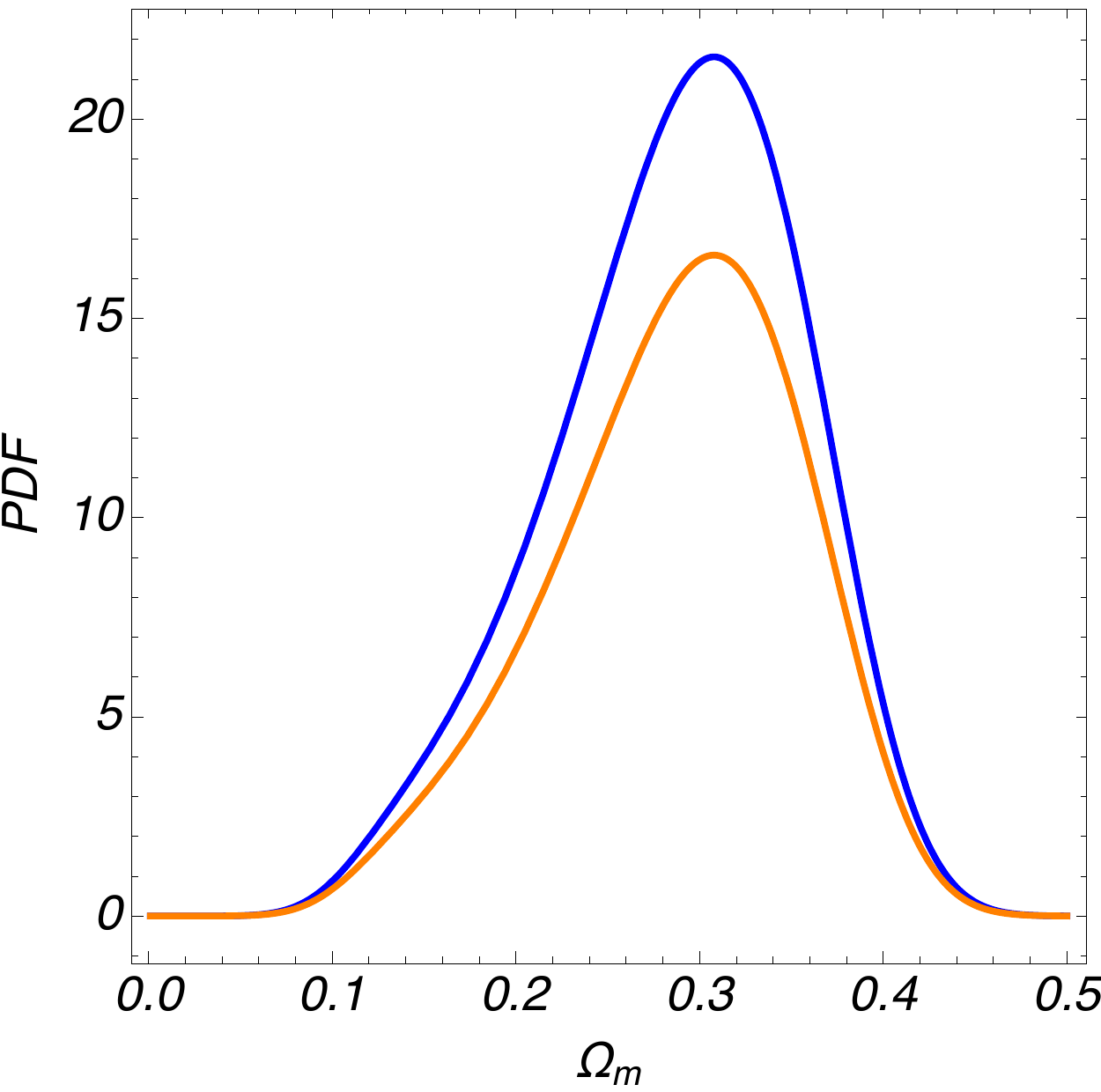}
\caption{General Taylor Expansion model contours plots for Cases 1.1 (orange color) and 1.2 (blue color) using \gls{cc}+Pantheon+\gls{bao} samplers. Permission for use of this figure was kindly provided by the authors of Ref.~\cite{Escamilla-Rivera:2019ulu}.}
\label{contour_taylor_case1}
\end{figure}

\item Power law model.
Now we consider a Lagrangian of separated power law style models for the torsion and boundary scalars such that
\begin{equation}
    f(T,B) = b_0 B^k + t_0 (-T)^m\,. \label{eq:powerlaw_model}
\end{equation}
Since we are interested in understanding whether this power law model can reproduce a \gls{de}-like behavior, we compute the \gls{eos} for the model in Eq.~\eqref{eq:powerlaw_model} and obtain
\begin{subequations}
\begin{align}
    w(a)&=-1+\frac{\frac{6^k b_0(2H^2 +H'+H^2)^k (k-1)k w_{x_1}}{(2H^2 +H'+H^2)^3} -2^{2+m}3^{m} (H^2)^{m+1} (m-1)m t_0 w_{x_2}} {3(-B^k b_0 +6H' -T^m t_0 +w_{x_4}+ w_{x_5})}, \\[0.5ex]
    &= -1+\frac{\frac{b_0 6^k (k-1) k \left[\frac{a(t) \ddot{a}(t)+2 \dot{a}(t)^2}{a(t)^2}\right]^k w_{x_1}} {\left[a(t) \ddot{a}(t)+2 \dot{a}(t)^2\right]^3}-\frac{t_0 2^{m+2} 3^m (m-1) m a(t) \left[\frac{\dot{a}(t)^2}{a(t)^2}\right]^{m+1} w_{x_2}}{\dot{a}(t)^5}} {3 \left\{\frac{6 \left[a(t) \ddot{a}(t)-\dot{a}(t)^2\right]}{a(t)^2}+ w_{x_4} + w_{x_5}-b_0 B^k-t_0 T^m\right\}}\,,
\end{align}
\end{subequations}
where the functions $w_{x_i}$ are given in Appendix~\ref{sec:f_T_B_quantities}.
To perform the numerical analysis, we rewrite the above expression in terms of the redshift $z=a_0 /a -1$, where $z=0$ corresponds to the present time. The power law model \gls{eos} can be expressed as
\begin{equation}\label{eq:eosz_powerlaw}
    w(z)=-1 + \frac{b_0 3^k 8^{k-2} (k-1) k (z+1)^{12} \left[\frac{1}{(z+1)^2}\right]^k w(z)_1 w(z)_4} {3 \left\{-b_0 B^k-(z+1)^3 w(z)_3 - b_0 2^{3 k-1} 3^k (k-1) k \left[\frac{1}{(z+1)^2}\right]^k-t_0 T^m+\frac{6}{(z+1)^2}\right\}}\,,
\end{equation}
where the functions $w(z)_{i}$ are given in Appendix~\ref{sec:f_T_B_quantities}.

\begin{figure}
\centering
\includegraphics[width=0.32\textwidth,origin=c,angle=0]{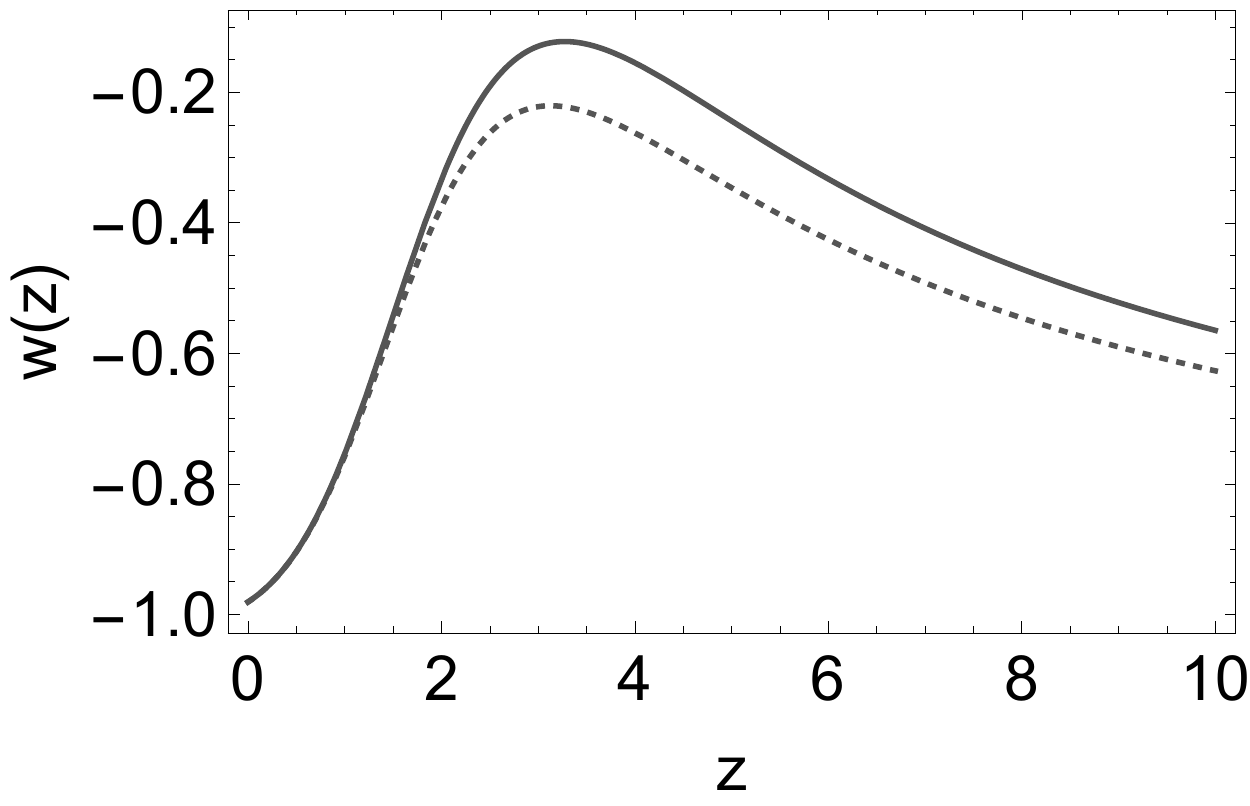}
\includegraphics[width=0.32\textwidth,origin=c,angle=0]{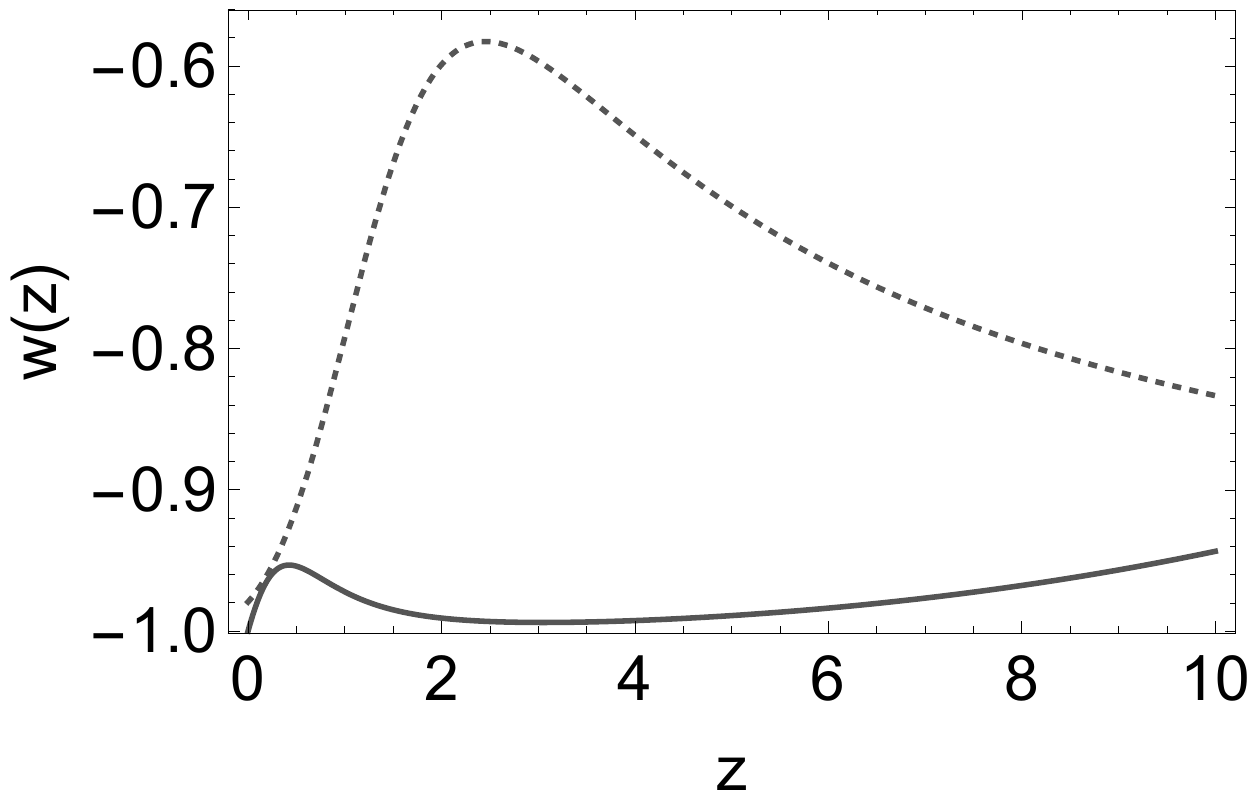}\\
\includegraphics[width=0.32\textwidth,origin=c,angle=0]{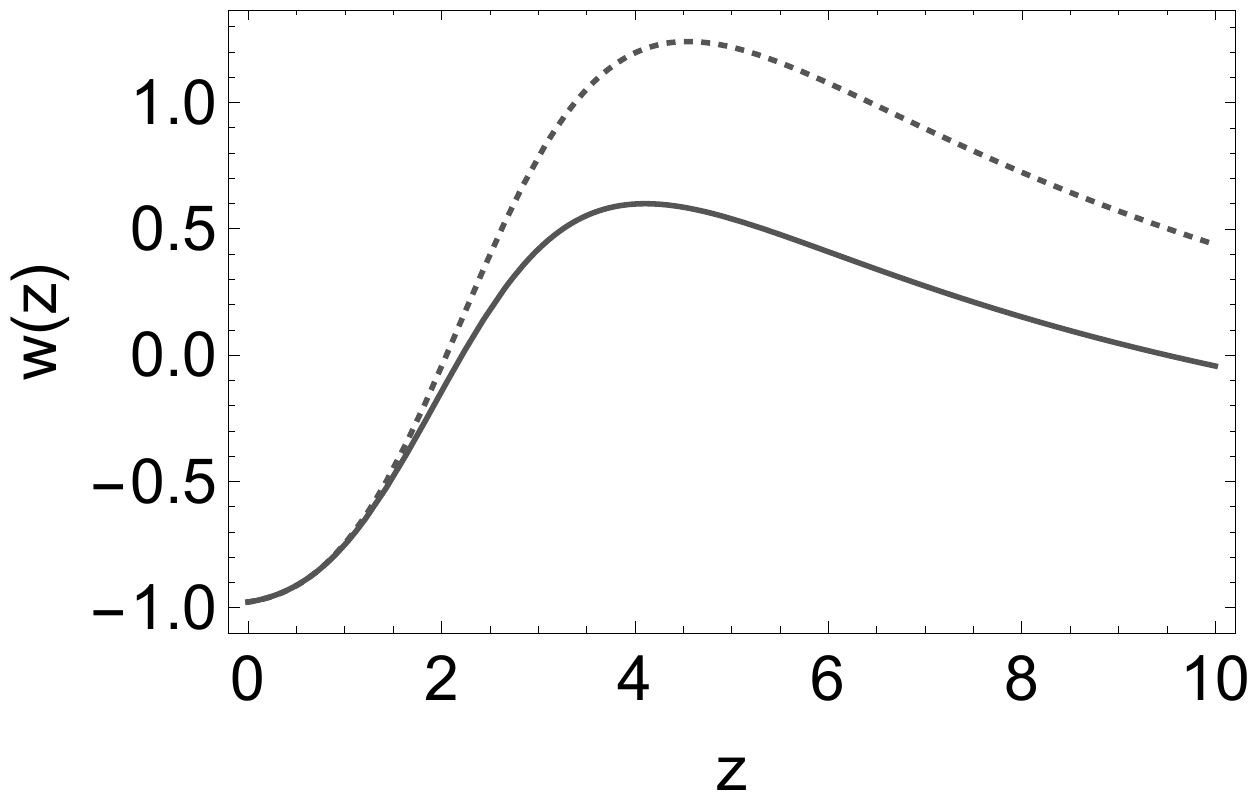}
\includegraphics[width=0.32\textwidth,origin=c,angle=0]{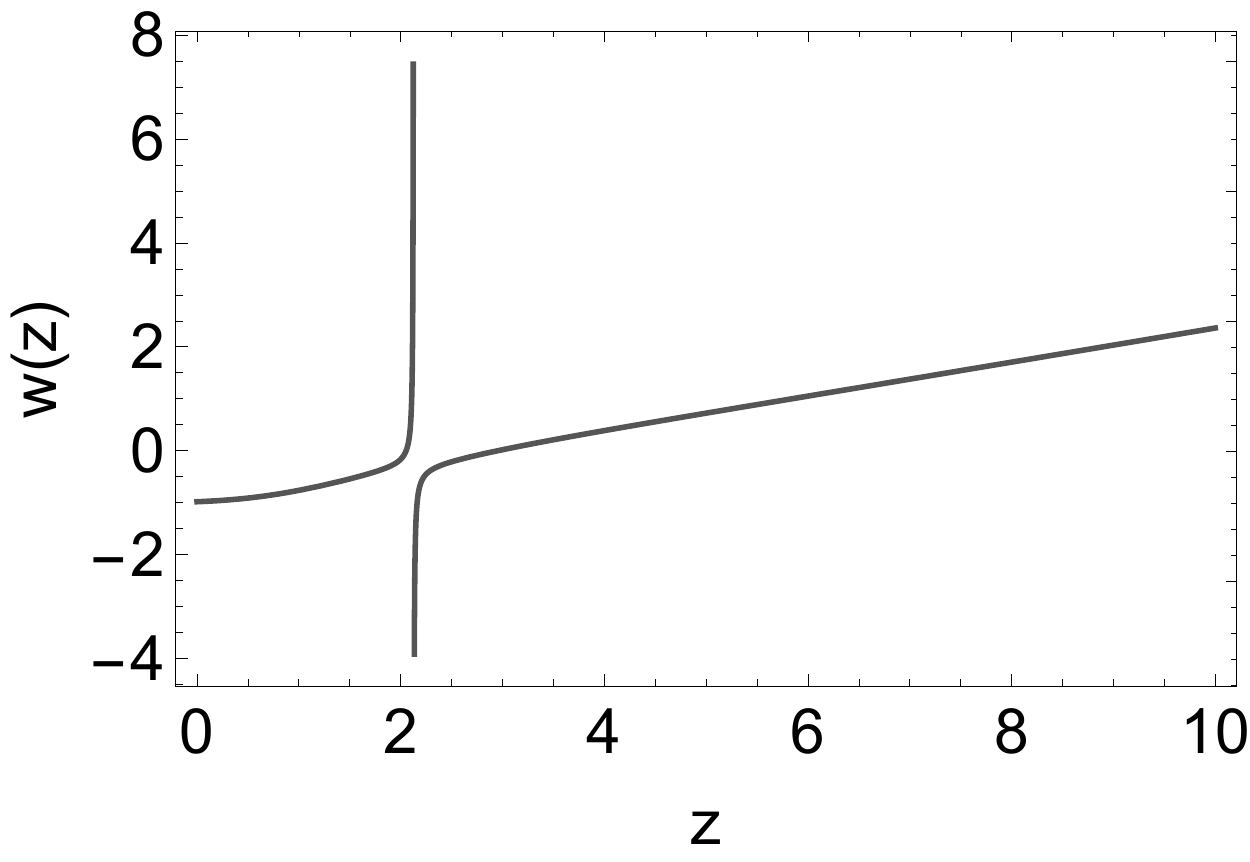}
\caption{Evolution of Power Law \gls{eos} \eqref{eq:eosz_powerlaw}. \textit{Top Left:} Case 1: solving $T$ and $B$, with $T\ll B$ (solid line) and $T\gg B$ (dashed line). \textit{Top Right:} Case 2: $m$ and $k$ as free parameters, with $m<k$ (solid line) and $m>k$ (dashed line). \textit{Bottom Left:} Case 3: $b_0$ and $t_0$ as free parameters, with $b_0$ negative and $f_0$ positive (solid line) and vice versa (dashed line). \textit{Bottom Right:} Case 4: $t_0$ and $m$ as free parameters and with negative values. The values considered are: $m=78.93$, $k=49.62$, $b_0, t_0 \approx 10^{15}$.
Permission for use of this figure was kindly provided by the authors of Ref.~\cite{Escamilla-Rivera:2019ulu}.}
\label{evolution_powerlaw}
\end{figure}

Notice that Eq.~\eqref{eq:eosz_powerlaw} reduces to the standard $\Lambda$\gls{cdm} model $w=-1$ when $k,\,m$ are zero, as expected. As a first strategy, we are going to analyze the behavior of the \gls{eos} in Eq.~\eqref{eq:eosz_powerlaw} which is associated with the fluid representation of $f(T,B)$ considering seven cases (c.f. Fig.~\ref{evolution_powerlaw}) as:
\begin{itemize}
    \item Case 1.1: Domination of the boundary term over the torsion scalar.
    \item Case 1.2: Domination of the torsion scalar over the boundary term.
    \item Case 2.1: $m$ and $k$ as free parameters with the condition that $m>k$.
    \item Case 2.2: $m$ and $k$ as free parameters with the condition that $k>m$.
    \item Case 3.1: $b_0$ and $t_0$ as free parameters with the condition that $b_0 < t_0$.
    \item Case 3.2: $b_0$ and $t_0$ as free parameters with the condition that $b_0 > t_0$.
    \item Case 4: $t_0$ and $m$ as free parameters and negative values.
\end{itemize}
Cases 1.1 and 1.2 (cf. with Fig.~\ref{evolution_powerlaw} - \textit{top left}) at $z<2$ show an accelerating cosmic solution. Case 1.1 decelerate at $z=3$, while Case 1.2 preserves this acceleration with \gls{eos} $w<-1/3$. It is important to point that that since the two scalars, namely the torsion scalar and boundary term, can interchange dominance over the evolution of the Universe, these cases will turn out to be mutually exclusive in most models. Cases 2.1 and 2.2 (cf. with Fig.~\ref{evolution_powerlaw} - \textit{top right}) have an \gls{eos} with $w<-1/3$, but shows an asymptotic behavior to $\Lambda$\gls{cdm} between $z=2$ and $z=4$. Afterwards starts to grow asymptotically to the first model at large z. Cases 3.1 and 3.2 cross the phantom divided-line, and below $z=2$ they are similar. Below $z=2.5$ both models can start with an \gls{eos} with $w<-1/3$, where both have an asymptotic behavior at large $z$, which can mimic a matter phase with $w=0$ (cf. with Fig.~\ref{evolution_powerlaw} - \textit{bottom left}). This is a case of an oscillating fluid representation of $f(T,B)$ \gls{eos} below $z=6$. Case 4 (cf. with Fig.~\ref{evolution_powerlaw} - \textit{bottom right}) has an oscillating particularity, but it experiences a divergence point due to the corresponding energy-density becoming zero. Thus, the possible cases of the different phases of the cosmic evolution are shown individually with all possible dominance terms being shown.

The confidence regions of this model are shown in Fig.~\ref{contour_powerlaw}, where we can infer the constraints on the theory against (\gls{cc}+Pantheon+\gls{bao}) observational data for free model parameters. The precise values are shown in Table~\ref{tab:f_T_B_power_law}. the very large value of these parameters may be problematic for other sectors of phenomenology.

\begin{table}[ht!]
\midsepremove
\begin{center}
\begin{tabular}{cllll}
\toprule
\cellcolor{gris3}\textbf{Parameter} & \cellcolor{gris3}\textbf{Best-fit} & \cellcolor{gris3}\textbf{Mean} \boldmath{$\pm\sigma$} & \cellcolor{gris3}\textbf{95\% lower} & \cellcolor{gris3}\textbf{95\% upper} \\ \midrule
\cellcolor{gris1}$H_{0 }\,[\text{km/s/Mpc}]$ & \cellcolor{gris1}$67.74$ & \cellcolor{gris1}$67.74_{-1.1}^{+1.1}$ & \cellcolor{gris1}$65.54$ & \cellcolor{gris1}$69.89$ \\
\cellcolor{gris3}$m$ & \cellcolor{gris3}$78.93$ & \cellcolor{gris3}$79.19_{-6.1}^{+4}$ & \cellcolor{gris3}$70.17$ & \cellcolor{gris3}$88.64$ \\
\cellcolor{gris1}$k$ & \cellcolor{gris1}$49.62$ & \cellcolor{gris1}$49.81_{-1}^{+0.73}$ & \cellcolor{gris1}$47.82$ & \cellcolor{gris1}$51.73$ \\
\cellcolor{gris3}$b_{0 }$ & \cellcolor{gris3}$8.16e+15$ & \cellcolor{gris3}$1.099e+16_{-1.1e+16}^{+7e+14}$ & \cellcolor{gris3}$9.981e+11$ & \cellcolor{gris3}$1.640e+15$ \\
\cellcolor{gris1}$c_{0 }$ &\cellcolor{gris1}$8.949e+15$ & \cellcolor{gris1}$7.974e+15_{-6.3e+15}^{+2.8e+15}$ & \cellcolor{gris1}$1.169e+16$ & \cellcolor{gris1}$1.074e+16$ \\
\bottomrule
 \end{tabular}
 \midsepdefault
 \caption{Parameters and mean values for the Power law model. \label{tab:f_T_B_power_law}}
 \end{center}
\end{table}
\begin{figure}
\centering
\includegraphics[width=0.55\textwidth,origin=c,angle=0]{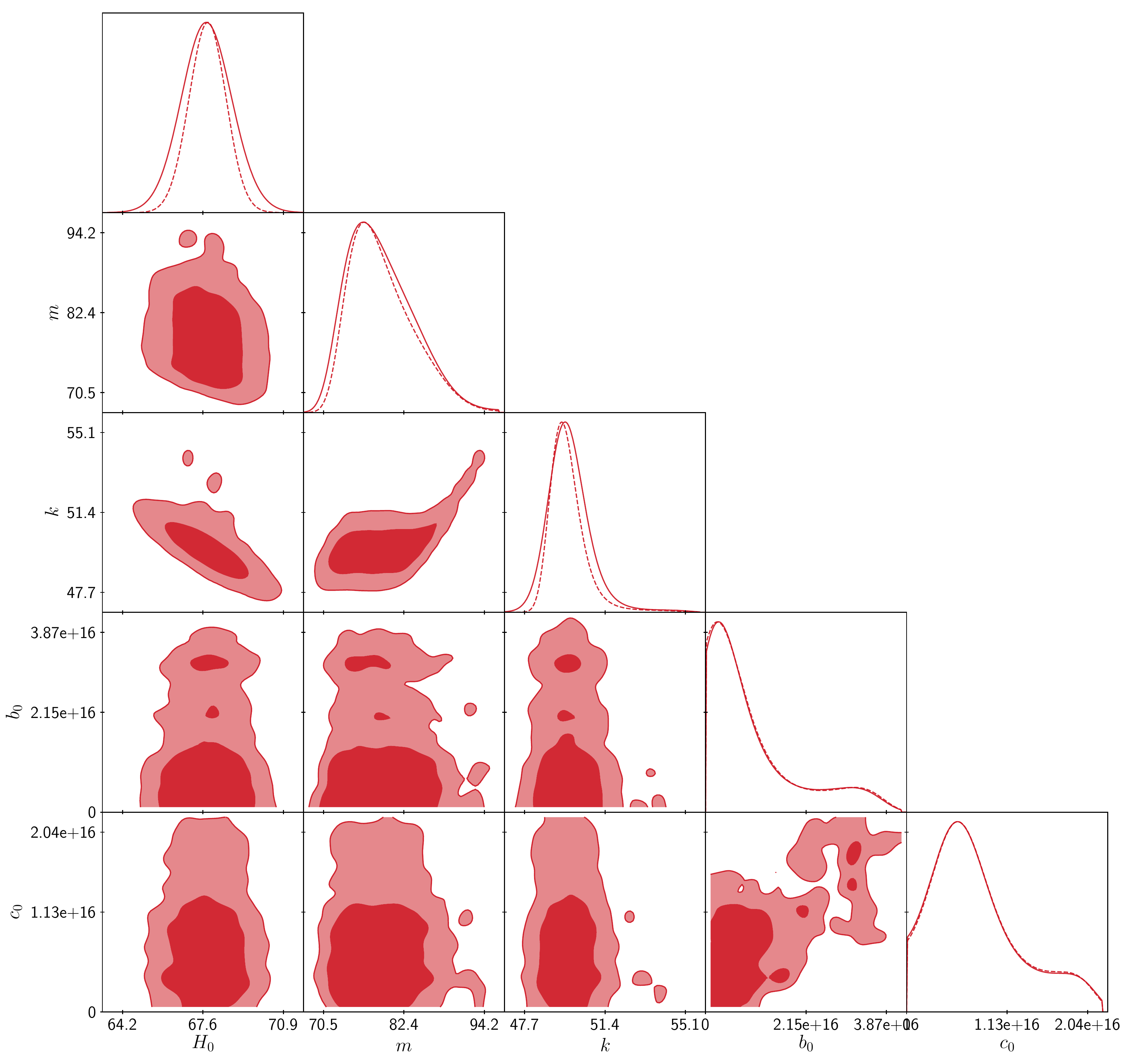}
\caption{1 dimensional marginalized distribution, and 2 dimensional contours with $68\%$ and $95\%$ confidence level for the free parameters of the Power Law model using the constrained solutions for $T$ and $B$ scalars and \gls{cc}+Pantheon+\gls{bao} total sampler. Permission for use of this figure was kindly provided by the authors of Ref.~\cite{Escamilla-Rivera:2019ulu}.}
\label{contour_powerlaw}
\end{figure}
\end{enumerate}

\subsubsection{Cosmography for TG gravity} \label{subsec:cosmography_p}

Observations from \gls{sn}eIa and gamma ray bursts are optimal tools in precision cosmology since with them it is possible to reconstruct the Hubble flow. Moreover, due to the degeneracy of model parameterizations, a viable path to overcome these issues is by using cosmography. To achieve this goal we expand the scale factor in Taylor series \gls{wrt} the cosmic time. Such an expansion leads to a distance-redshift relation which only relies on the assumption of the \gls{flrw} metric being model-independent since it does not depend on the specific form of the solution of the cosmic equations. It is convenient to introduce the following definitions
\begin{equation}\label{parB}
    H =\frac{1}{a}\frac{\dd a}{\dd t}\,,\quad   q = -\frac{1}{a}\frac{\dd^2a}{\dd t^2}H^{-2}\,,\quad j =\frac{1}{a}\frac{\dd^3a}{\dd t^3}H^{-3}\,, \quad
    s =\frac{1}{a}\frac{\dd^4a}{\dd t^4}H^{-4}\,,\quad l = \frac{1}{a}\frac{\dd^5a}{\dd t^5}H^{-5}\,,
\end{equation}
which denotes the Hubble, deceleration, jerk, snap and lerk parameters, respectively. Their present day values (denoted with a subscript 0) can be used to characterize the cosmic evolution, e.g. if $q_0 < 0$, this denotes an accelerated expansion, while $j_0$ allows to discriminate among different accelerating models.

Using these definitions straightforward we can obtain the following equations \cite{Capozziello:2010zz,Capozziello:2019cav}
\begin{subequations}
\begin{align}\label{dotH}
    \dot H&=-H^2(1+q)\,,\\[0.5ex]
    \label{ddotH}
    \ddot H&=H^3(j+3q+2)\,,\\[0.5ex]
\label{dddotH}
    \dddot H&=H^4[s-4j-3q(q+4)-6]\,,\\[0.5ex]
\label{ddddotH}
    H^{(4)}&=H^5[l-5s+10(q+2)j+30(q+2)q+24]\,,
\end{align}
\end{subequations}
where a dot denotes derivative \gls{wrt} the cosmic time $t$ and $H^{(4)}=\dd^4H/\dd t^4$. Eqs.~\eqref{dotH} -- \eqref{ddddotH} make it possible to relate the derivative of the Hubble parameter to the other cosmographic parameters.

Instead of choosing a parameterized form for a $f(T)$ model and then numerically solving the modified version of Friedmann equations for given values of the boundary conditions, it is possible to relate the present day values ($z=0$) of its derivatives to the cosmographic parameters $(q_0, j_0, s_0, l_0)$ in order to constraint them in a model-independent way and obtain a form of $f(T)$ that can be able to fit with the observational data at hand.

As a preliminary step, it is worth considering again that $T=-6H^2$. Differentiating \gls{wrt} $t$, we get the following expressions
\begin{subequations}
\begin{align}
    \dot T &= -12H\dot H\,, \label{dif1T} \\[0.5ex]
    \ddot T &= -12(\dot H^2+H\ddot H)\,, \label{dif2T}\\[0.5ex]
    \dddot T &= -12(3\dot H\ddot H+H\dddot H)\,,\label{dif3T} \\[0.5ex]
    T^{(4)} &= -12(3\ddot H^2+4\dot H\dddot H+HH^{(4)})\,. \label{dif4T}
\end{align}
\end{subequations}
The modified Friedmann given by Eqs.~\eqref{Friedmann_1B} and \eqref{Friedmann_2B} can be rewritten for the pressureless case $p=0$ as
\begin{subequations}
\begin{align}\label{friedfried}
    H^2&=-\frac{1}{12f_T}[-T\Omega_{m}+f(T)]\,,\\[0.5ex]
\label{acceacce}
    \dot H&=\frac{1}{4f_T}[-T\Omega_{m}-4H\dot{T}f_{TT}]\,.
\end{align}
\end{subequations}
Once with this quantities, we can derive up to fourth order in $H$ to derive our cosmographic parameters as
\begin{subequations}
\begin{alignat}{2} \label{expddH}
    \ddot H& =\: & &\frac{\Omega_{\rm m}}{4Hf_T}[H\dot T+T(3H^2+2\dot H)]-\frac{1}{f_T}[(2\dot H\dot T+H\ddot T)f_{TT}+H\dot T^2f_{TTT}]\,,\\[0.5ex]
    \dddot H& =\: & &\frac{\Omega_{\rm m}}{4H^2f_{T}}[-T(9H^4+6H^2\dot H+4\dot H^2)-H\dot T(3\dot H+6H^2)+H(H\ddot T-2\ddot H T)] \nonumber \\[0.5ex] & \: & &-\frac{1}{Hf_{T}}[\dot H\ddot Hf_{T}+(2\dot H^2\dot T+3H\ddot H\dot T+4H\dot H\ddot T+H^2\dddot T)f_{TT}+H^2\dot T^3f_{TTTT} \nonumber \\[0.5ex]& \: & &+H\dot T(4\dot H\dot T+3H\ddot T)f_{TTT}]\,, \label{expdddH}\\[0.5ex]
    H^{(4)}& =\: & &\frac{\Omega_{\rm m}}{4H^3f_T}[-T(10H\dot H\ddot H+12H^3\ddot H-27H^6-12H^2\dot H^2-8\dot H^3-2H^2\dddot H)+H^3\dddot T \nonumber \\[0.5ex] & \: & &+H^2\dot T(9H\dot H+27H^3-5\ddot H)-3H^2\ddot T(3H^2+\dot H)+7H\dot H^2\dot T]\nonumber \\[0.5ex] & \: & &-\frac{1}{H^2f_T}[(3H\dot H\dddot H+\dot H^2\ddot H+H\ddot H^2)f_T+H^2\dot T^2(7\dot H\dot T+6H\ddot T)f_{TTTT} \nonumber \\[0.5ex] & \: & &+(4H^2\dddot H\dot T+2\dot H^3\dot T+7H^2\dot H\dddot T+10H\dot H^2\ddot T+7H^2\ddot H\ddot T+11H\dot H\ddot H\dot T+H^3T^{(4)})f_{TT} \nonumber \\[0.5ex] & \: & &+H(10\dot H^2\dot T^2+7H\ddot H\dot T^2+21H\dot H\dot T\ddot T+3H^2\ddot T^2+4H^2\dot T\dddot T)f_{TTT} \nonumber \\[0.5ex] & \: & &+H^3\dot T^4f_{TTTT}]\,, \label{expddddH}
\end{alignat}
\end{subequations}
where the fourth and fifth derivative are denoted by the quantities $f^{(4)}(T)=\dd^4f(T)/\dd T^4$ and $f^{(v)}(T)=\dd^5f(T) / \dd T^5$, respectively. At this level, it is possible to derive a generic cosmography, moreover, the restriction of a specific $f(T)$ is still present. In Ref.~\cite{Capozziello:2011hj}, an $f(T)$ Taylor series model was studied up to fifth order to investigate \textit{a priori} a given $f(T)$ theory by simply comparing the theoretically predicted $\left(f_2,f_3\right)$ models with the observed ones. Unfortunately, this analysis was unable to predict the observationally values of $\left(f_2,f_3\right)$ models due the degeneracy over the cosmographic parameters and the difficulties concurrently constraining with current surveys. Furthermore, Ref.~\cite{Capozziello:2019cav} explored cosmographic constraints following the same methodology of exotic polynomials, e.g. Pad\'e polynomial. Moreover, the current state-of-the-art suggests that it is not yet possible to falsify the $\Lambda$\gls{cdm} model by robust statements adopting only this kind of cosmography.

At present day ($z=0$) values of $f(T)$ and its derivatives up to the fifth order. After some algebra, we have
\begin{subequations}
\begin{alignat}{2} \label{fT}
    \frac{f(T_0)}{6H_0^2}& =\: & &\Omega_{\rm m0}-2\,,\quad f_{T}(T_0)=1\,,\quad \frac{f_{TT}(T_0)}{(6H_0^2)^{-1}}=-\frac{3\Omega_{\rm m0}}{4(1+q_0)}+\frac{1}{2}\,,\\[0.5ex]
    \frac{f_{TTT}(T_0)}{(6H_0^2)^{-2}}& =\: & &\frac{-3\Omega_{\rm m0}(3q_0^2+6q_0+j_0+2)}{8(1+q_0)^3}+\frac{3}{4}\,,\\[0.5ex]
\label{f4T}
    \frac{f_{TTTT}(T_0)}{(6H_0^2)^{-3}}& =\: & &-\frac{3\Omega_{\rm m0}}{16(1+q_0)^5}[-s_0(1+q_0)+j_0(6q_0^2+17q_0+3j_0+5) \nonumber \\[0.5ex]& \: & &+3q_0(5q_0^3+20q_0^2+29q_0+16)+9]+\frac{15}{8}\,,\\[0.5ex]
\label{f5T}
    \frac{f_{TTTT}(T_0)}{(6H_0^2)^{-4}}& =\: & &-\frac{3\Omega_{\rm m0}}{32(1+q_0)^7}[l_0(1+q_0)^2+s_0(10q_0^3+43q_0^2+46q_0+13)-10j_0s_0(1+q_0)\nonumber \\[0.5ex]& \: & &+5j_0^2(6q_0^2+22q_0+3j_0+7)+j_0(45q_0^4+225q_0^3+412q_0^2+219q_0+32)\nonumber\\[0.5ex]& \: & &+3q_0(35q_0^5+210q_0^4+518q_0^3+666q_0^2
    +448q_0+150)+60]+\frac{105}{16}\,.
\end{alignat}
\end{subequations}
As we can notice from the latter equations, there is still a dependency of the model directly of the $f(T)$ form, which lead us to a model-dependent cosmography related to the Taylor $f(T)$ series form. As a perspective in this direction, the consideration of $f(T)$ models can help to understand the convergence of the series imposed by cosmography, even thought some of these models should be aligned to a particular form of Taylor series or extension of them. Due to the excessive number of free parameters in the equations of motion, \cite{Capozziello:2019cav} a degeneracy creeks into the system and so it becomes difficult to find a unique best-fit curve for these parameters. As a result, the order of the expansion does not have statistical relevance, which is a consequence of the low density of data points used to get this conclusion. Moreover, increasing the order of the expansion shows a deviation from the standard cosmological values.

\subsubsection{Probing \texorpdfstring{$f(T)$}{fT} gravity with the induced variation of fundamental constants} \label{sec:fT_varying_alpha}

Another cosmological probe of \gls{tg} involves measurements on the variation of the fine--structure constant $\alpha$, denoted by $\Delta\alpha/\alpha$. It predominantly consists of an archival astrophysical data set of quasar absorption lines observed at the Keck observatory \cite{Murphy:2003mi} and with the VLT (V) \cite{King:2012id}, along with a set of 21 dedicated new measurements (N) \cite{Agafonova:2011sp,Molaro:2013saa,Songaila:2014fza,Evans:2014yva,Kotus:2016xxb,Murphy:2016yqp,Martins:2017yxk,Bainbridge2017,Alves:2018mef}, and the constraint from the Oklo (O) natural nuclear reactor at an effective redshift of $z=0.14$ \cite{Petrov:2005pu}. The measurements contained in the N data set were reported from the ESO Ultraviolet and Visual Echelle Spectrograph (UVES) Large Program which was specifically developed for such measurements. For simplicity, we will henceforth be denoting the joint data set of $\mathrm{K}+\mathrm{V}+\mathrm{N}+\mathrm{O}$ by \gls{kvno}.

It is well--known that a non-minimal coupling between a scalar field and matter fields would break the Einstein equivalence principle, and would further lead to the variation of fundamental constants of Nature \cite{Uzan:2010pm,Uzan:2002vq,Hees:2014lfa}. For instance, a scalar field coupling with the electromagnetic Lagrangian would lead to a variation of the fine--structure constant, or Sommerfeld's constant, which characterizes the strength of the electromagnetic field and appears as a coupling constant in the electromagnetic action. A variation in the fundamental constants of Nature \cite{2001RPPh...64.1191F}, which could be conservatively defined as those theoretical free parameters that could not be calculated with our present knowledge of physics, has been a long--established intriguing question \cite{Dirac:1937ti,Dirac:1938mt} with pertinent consequences for fundamental physics and cosmology (see, for instance, Refs.~\cite{Uzan:2002vq,Uzan:2010pm,Martins:2017yxk}). Interestingly, when Dirac's numerological principle~\cite{Dirac:1937ti,Dirac:1938mt} was encapsulated in a field--theoretical framework, this led to the birth of the Jordan--Fierz--Brans--Dicke scalar--tensor theory of gravitation~\cite{jordan1937physikalischen,Fierz:1956zz,Brans:1961sx}.

A number of theoretical models have been proposed in order to explore the possibility of a dynamical fine--structure constant $\alpha\equiv e^2/\hbar c$. These models have been primarily formulated as Lagrangian theories with explicit
variation of the velocity of light $c$ (varying speed of light theories) \cite{Barrow:1999is}, or of the charge on the electron $e$ (varying electric charge theories) \cite{Bekenstein:1982eu,Livio:1998pp}. These models have been first formulated by Bekenstein~\cite{Bekenstein:1982eu} from a generalization of Maxwell's equations, that led to the construction of the cosmological varying--$e$ Bekenstein--Sandvik--Barrow--Magueijo theory of varying $\alpha$~\cite{Sandvik:2001rv,Barrow:2001iw,Barrow:2002db,Barrow:2002ed,Barrow:2011kr,Barrow:2013uza}. Other frameworks include, for instance, a runaway dilaton~\cite{Damour:2002mi,Damour:2002nv}, supersymmetric generalization of Bekenstein's model~\cite{Olive:2001vz} and a disformally coupled electromagnetic sector~\cite{vandeBruck:2015rma}. Each theoretical framework has its own cosmological signatures, and we should remark that the spacetime dependence of fundamental constants has also been linked~\cite{Hart:2019dxi} with the currently reported Hubble tension via the inferred effects in the ionisation history and profile of \gls{cmb} anisotropies. For example, in $f(T)$ gravity, the form of this fine--structure constant dependence can be obtained by a conformal transformation (refer to Sec.~\ref{sec:con_dis_trans} for further details) of the tetrad where
    $ \udt{\tilde{e}}{A}{\mu} = \Omega\,\udt{e}{A}{\mu}\,,$
    and $\dut{\tilde{E}}{A}{\mu} = \Omega^{-1}\,\dut{E}{A}{\mu}\,,$
which results in the regular conformal transformation $\tilde{g}_{\mu\nu}=\Omega^2 g_{\mu\nu}$, as expanded upon in Ref.~\cite{Wright:2016ayu}, where $\Omega^2 = -f_T=|f_T|$ (note that since in our conventions $T>0$ and $f_T<0$,
we have replaced $-f_T$ by $|f_T|$), and that tilde denotes conformally transformed quantities. It is well known that $f(T)$ gravity cannot be written in the Einstein frame through conformal transformations, which implies that it will induce a dependence in its associated fine--structure constant \cite{Brax:2012gr} (as will be shown later in Eq.~\eqref{B_F_Var_order}) . In fact, this produces an extra $2\Omega^{-2}\tilde{\partial}^{\mu}\left(\Omega^2\right)\udt{\tilde{T}}{\nu}{\nu\mu}$ term which cannot be removed (see Eq.~\eqref{eq:conforal1}). The remainder of the scalar field becomes a phantom field with the choice of $\phi=\sqrt{3}\ln f_T$~\cite{Wright:2016ayu}, which is partially favored by the 2018 \texttt{Planck} data~\cite{Aghanim:2018eyx}.

The result of a conformal transformation is the introduction of a new \gls{dof}, $\phi$, that arises from the transformation $\udt{A}{a}{\mu} \rightarrow \udt{\tilde{e}}{A}{\mu}$, where $\udt{\tilde{e}}{A}{\mu}$ represents the Einstein frame tetrad. This then induces an electromagnetic coupling which takes on the form
\begin{equation}\label{emag_action}
    \mathcal{S}_{\text{EM}} = \frac{1}{4} \int \mathrm{d}^4 x\, e B_F\left(\phi\right) F_{\mu\nu}F^{\mu\nu}\,,
\end{equation}
where $F_{\mu\nu}=A_{\nu,\mu} - A_{\mu,\nu}$ is the standard Faraday tensor, and $B_F\left(\phi\right)$ represents the nonvanishing $\phi-$coupling. The consequence of this induced coupling is that the fine--structure constant and the luminosity distance will be altered comparing to \gls{gr}~\cite{misner1973gravitation,PhysRevD.90.023017}. As in Refs.~\cite{Brax:2012gr,Olive:2001vz,Copeland:2003cv,Nunes:2016plz}, this can be expanded about $\phi(t=t_0)$, which is suitably small, to give
\begin{equation}\label{B_F_Var_order}
    B_F\left(\phi\right) \simeq 1+\beta_F\kappa^2\phi\,,
\end{equation}
where $\beta_F=\mathcal{O}(1)$ is a constant ($\beta_F\phi\ll \kappa^{-2}$).
Given an initially uncoupled Jordan--frame electromagnetic action, the fine--structure constant turns out to be given by~\cite{Olive:2001vz}
\begin{equation}
    \alpha_E (\phi) = \frac{\alpha_J(\phi)}{B_F(\phi)}\,,
\end{equation}
where $\alpha_E$ and $\alpha_J$ are the fine--structure constants in the Einstein and Jordan frames respectively.

\begin{table}[!ht]
    \centering
    \midsepremove
    \setlength\extrarowheight{0.8pt}
    \resizebox{\columnwidth}{!}{\begin{tabular}{ c c c c }
       \toprule
        \multicolumn{4}{c}{\cellcolor{gris3}\boldmath{$f_1^{}(T)$} \textbf{Model}}\\
       \cellcolor{gris3} \textbf{Parameter} & \cellcolor{gris3}\textbf{\gls{sn} + \gls{cc}+} \boldmath{$H_0^{\rm R}$} & \cellcolor{gris3}\textbf{\gls{sn} + \gls{cc} + \gls{kvno}} & \cellcolor{gris3}\textbf{\gls{sn} + \gls{cc} + \gls{kvno} +} \boldmath{$H_0^{\rm R}$} \\
	 \midrule
		 \cellcolor{gris1}$b$ & \cellcolor{gris1}$-0.16^{+0.24}_{-0.49}$ & \cellcolor{gris1} ~~$0.003^{+0.053}_{-0.059}$ & \cellcolor{gris1}$-0.001^{+0.050}_{-0.048}$ \\
	 \cellcolor{gris3}	$\Omega_{\rm m0}$ & \cellcolor{gris3}~~$0.281^{+0.036}_{-0.035}$ & \cellcolor{gris3}~~$0.300^{+0.026}_{-0.024}$ & \cellcolor{gris3}~~$0.283^{+0.023}_{-0.021}$ \\
		 \cellcolor{gris1}$H_0^{}\,[\text{km/s/Mpc}]$ & \cellcolor{gris1}~$72.8^{+1.4}_{-1.3}$ & \cellcolor{gris1}~$68.9^{+2.0}_{-1.9}$ & \cellcolor{gris1}~$72.2^{+1.2}_{-1.2}$ \\
		 \cellcolor{gris3}$\beta_F$ & \cellcolor{gris3}~~$0.28^{+0.32}_{-0.32}$ & \cellcolor{gris3}$-0.003^{+0.063}_{-0.056}$ & \cellcolor{gris3}$-0.003^{+0.074}_{-0.067}$~\\
		 \cellcolor{gris1}$\chi^2_\mathrm{min}$ & \cellcolor{gris1}$1042.578$ & \cellcolor{gris1}$1365.427$ & \cellcolor{gris1}$1370.014$ \\[.3em]
        \toprule
        \multicolumn{4}{c}{\cellcolor{gris3}\boldmath{$f_2^{}(T)$} \textbf{Model}}\\
        \cellcolor{gris3} \textbf{Parameter} & \cellcolor{gris3}\textbf{\gls{sn} + \gls{cc}+} \boldmath{$H_0^{\rm R}$} & \cellcolor{gris3}\textbf{\gls{sn} + \gls{cc} + \gls{kvno}} & \cellcolor{gris3}\textbf{\gls{sn} + \gls{cc} + \gls{kvno} +} \boldmath{$H_0^{\rm R}$} \\
	 \midrule
	\cellcolor{gris1}	$1/p$ & \cellcolor{gris1}~~$0.093^{+0.171}_{-0.079}$ & \cellcolor{gris1}$\left( 10.8^{+35.9}_{-4.9} \right) \times 10^{-3}$ & \cellcolor{gris1}$\left( 41.7^{+9.3}_{-30.8} \right) \times 10^{-3}$ \\
		\cellcolor{gris3}$\Omega_{\rm m0}$ & \cellcolor{gris3}~~$0.279^{+0.025}_{-0.031}$ & \cellcolor{gris3}~~$0.300^{+0.021}_{-0.020}$ & \cellcolor{gris3}~~$0.283^{+0.020}_{-0.019}$ \\
		\cellcolor{gris1}$H_0^{}\,[\text{km/s/Mpc}]$ & \cellcolor{gris1}~$72.2^{+1.3}_{-1.2}$ & \cellcolor{gris1}~$69.0^{+1.8}_{-1.9}$ & \cellcolor{gris1}~$72.2^{+1.3}_{-1.2}$ \\
	\cellcolor{gris3}	$\beta_F$ & \cellcolor{gris3}$-0.10^{+0.49}_{-0.56}$ & \cellcolor{gris3}$-0.01^{+0.45}_{-0.75}$ & \cellcolor{gris3}$-0.07^{+0.58}_{-0.55}$ \\
		\cellcolor{gris1}$\chi^2_\mathrm{min}$ & \cellcolor{gris1}$1045.741$ & \cellcolor{gris1}$1366.626$ & \cellcolor{gris1}$1371.221$\\[.3em]
	\toprule
        \multicolumn{4}{c}{\cellcolor{gris3}\boldmath{$f_3^{}(T)$} \textbf{Model}}\\
       \cellcolor{gris3} \textbf{Parameter} & \cellcolor{gris3}\textbf{\gls{sn} + \gls{cc}+} \boldmath{$H_0^{\rm R}$} & \cellcolor{gris3}\textbf{\gls{sn} + \gls{cc} + \gls{kvno}} & \cellcolor{gris3}\textbf{\gls{sn} + \gls{cc} + \gls{kvno} +} \boldmath{$H_0^{\rm R}$} \\
	 \midrule
	\cellcolor{gris1} $1/q$ & \cellcolor{gris1}~~$0.065^{+0.088}_{-0.045}$ & \cellcolor{gris1}$\left( 15.7^{+28.7}_{-9.5} \right) \times 10^{-3}$ & \cellcolor{gris1}~$0.029^{+0.018}_{-0.020}$ \\
	\cellcolor{gris3}	$\Omega_{\rm m0}$ & \cellcolor{gris3}~~$0.279^{+0.021}_{-0.020}$ & \cellcolor{gris3}~~$0.302^{+0.022}_{-0.023}$ & \cellcolor{gris3}~~$0.283^{+0.019}_{-0.019}$~ \\
		\cellcolor{gris1}$H_0^{}\,[\text{km/s/Mpc}]$ & \cellcolor{gris1}~$72.2^{+1.3}_{-1.2}$ & \cellcolor{gris1}~~$69.0^{+2.0}_{-1.9}$~ & \cellcolor{gris1}~$72.2^{+1.2}_{-1.3}$~ \\
	\cellcolor{gris3}	$\beta_F$ & \cellcolor{gris3}$-0.22^{+0.76}_{-0.40}$ & \cellcolor{gris3}$-0.05^{+0.61}_{-0.53}$ & \cellcolor{gris3}~~$0.00^{+0.61}_{-0.52}$~\\
	\cellcolor{gris1}	$\chi^2_\mathrm{min}$ & \cellcolor{gris1}$1046.074$ & \cellcolor{gris1}$1366.661$ & \cellcolor{gris1}$1371.241$ \\[.3em]
	\bottomrule
    \end{tabular}}
    \midsepdefault
    \caption{The mean value and the corresponding 68\% limits~\cite{LeviSaid:2020mbb} of the model parameters of three $f_{i}(T)$ models ($i\in\{1,2,3\}$), as described in Sec.~\ref{sec:fT_varying_alpha}. We also report the minimum $\chi^2$ value.}
    \label{tab:f_modelsA}
\end{table}

Given that $\Omega^2 = |f_T|$, this would imply that the $\tilde{\partial}^{\mu}\left(\Omega^2\right)$ would be very small rendering the additional term negligible. We will revisit this reasoning against the results of the analysis. With this approximation to the Einstein frame, the variation of the fine--structure constant takes the form
\begin{equation}\label{f_T_fine_struc_const}
    \frac{\Delta \alpha}{\alpha} = \frac{\kappa^{-2} + \sqrt{3}\beta_F [\ln |f_T(T_0)|]}{\kappa^{-2} + \sqrt{3}\beta_F [\ln |f_T|]} - 1\,,
\end{equation}
which vanishes for the $\Lambda$\gls{cdm} case of $f(T)=-T+\Lambda$, as expected. Eq.~\eqref{f_T_fine_struc_const} embodies the redshift dependence of the fine--structure constant in \gls{tg}, since the torsion scalar depends on redshift. Another consequence of a nonvanishing scalar field coupling to the electromagnetic action is that the luminosity distance will be altered~\cite{Hees:2014lfa}. By considering the standard derivation of luminosity distance~\cite{misner1973gravitation} with this new action, Ref.~\cite{PhysRevD.90.023017} shows that this leads to
\begin{subequations}
\begin{align}\label{eq:dL_f_T}
    d_L &= c\left(1+z\right)\sqrt{\frac{B_{F_0}}{B_F}} \int_0^z
\frac{\mathrm{d}z'}{H(z')}\\
    &=c\left(1+z\right) \sqrt{\frac{\kappa^{-2} + \sqrt{3}\beta_F [\ln
|f_T(T_0)|]}{\kappa^{-2} + \sqrt{3}\beta_F [\ln
|f_T|]}} \int_0^z
\frac{\mathrm{d}z'}{H(z')}\,,
\end{align}
\end{subequations}
as the luminosity distance for $f(T)$ gravity, which limits to the \gls{gr} formula for $B_F=1$.

We now consider the following $f(T)$ models, in which we primarily focus on the respective induced variation in the electromagnetic coupling and the inferred cosmological parameter constraints. These models were confronted with the \gls{kvno} varying fine--structure constant data set, along with the \gls{sn}eIa Pantheon compilation \cite{Scolnic:2017caz} and a \gls{cc} data set \cite{Moresco:2016mzx,Moresco:2012jh,Simon:2004tf,Stern:2009ep,Zhang:2012mp,Moresco:2015cya}. Occasionally, a prior on the Hubble constant is assumed in the following analyses, which is specified by $H_0^{\rm R}=74.03\pm1.42\,\mathrm{km}\,\mathrm{s}^{-1}\mathrm{Mpc}^{-1}$ \cite{Riess:2019cxk}.

\begin{figure}[t]
\begin{center}
    \includegraphics[width=0.48\columnwidth]{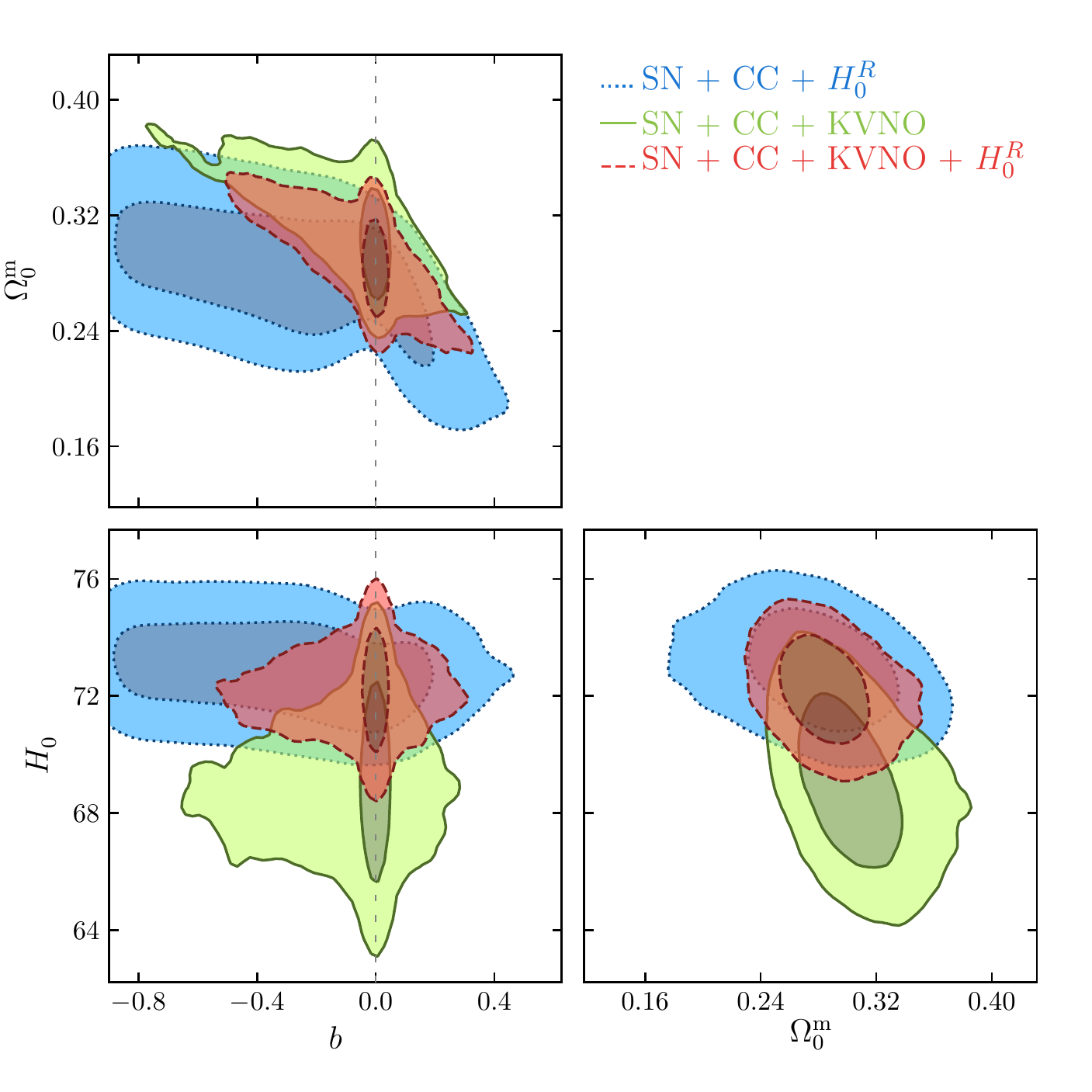}
    \includegraphics[width=0.48\columnwidth]{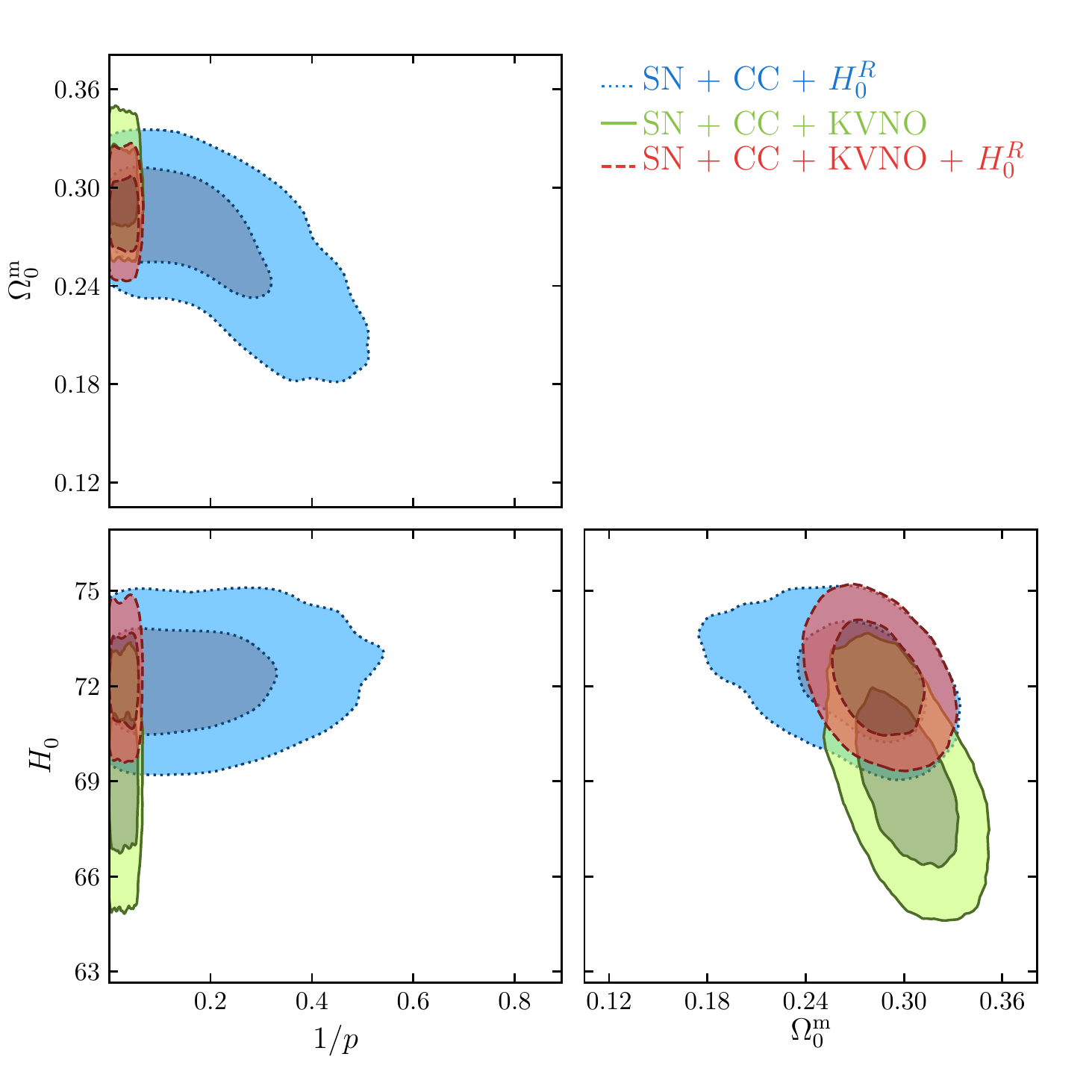}
    \caption{\label{fig:f1CDM_posteriors}{Marginalized 2 dimensional likelihood constraints on the parameters of the $f_1^{}(T)$ (\textit{left}) and $f_2^{}(T)$ (\textit{right}) models of Eq.~\eqref{fTmodel1} and Eq.~\eqref{fTmodel2}, respectively. Permission for use of these figures was kindly provided by the authors of Ref.~\cite{LeviSaid:2020mbb}.}}
\end{center}
\end{figure}

\begin{enumerate}

\item Power law model.
The derived constraint on the power law model~\cite{Bengochea:2008gz} $(f_1(T))$ parameter $b$ of Sec.~\ref{sec:f_T_power_law} was found to be $b=-0.16^{+0.24}_{-0.49}$ \cite{LeviSaid:2020mbb} when adopting the \gls{sn} + \gls{cc} + $H_0^{\rm R}$ joint data set, which is consistent with $\Lambda$\gls{cdm} and in line with the findings in previous studies~\cite{Nesseris:2013jea,Qi:2017xzl,Xu:2018npu,Anagnostopoulos:2019miu}. As illustrated in the left panel of Fig.~\ref{fig:f1CDM_posteriors}, the \gls{sn} + \gls{cc} + \gls{kvno} and \gls{sn} + \gls{cc} + \gls{kvno} + $H_0^{\rm R}$ joint data sets also improve the constraints on the model parameter $b$, which was found to be consistent with zero. Similar constraints have been reported in Ref.~\cite{Nunes:2016plz}, although the presented constraints are tighter. The reported constraints of Table~\ref{tab:f_modelsA}, also show that $\beta_F^{}$ is compatible with zero. Consequently, there is a negligible deviation from the $f(T)$ distance--duality relation in this model.


\item Square--root--exponential model.
In the square--root--exponential model~\cite{Linder:2010py} $(f_2(T))$ of Sec.~\ref{sec:f_T_square_root_exp}, the derived constraints on $1/p$ from the \gls{sn} + \gls{cc} + $H_0^{\rm R}$ and \gls{sn} + \gls{cc} + \gls{kvno} + $H_0^{\rm R}$ data sets were found to be consistent with zero at around 1$\sigma$ \cite{LeviSaid:2020mbb}. This model was also found to be in agreement with the $\Lambda$\gls{cdm}
model at around 2$\sigma$ when the \gls{sn} + \gls{cc} + \gls{kvno} data set was adopted (see, for instance, Refs.~\cite{Nesseris:2013jea,Qi:2017xzl,Nunes:2016plz,Nunes:2016qyp,Xu:2018npu,Anagnostopoulos:2019miu} for similar conclusions). The marginalized confidence contours are depicted in the right panel of Fig.~\ref{fig:f1CDM_posteriors}, while the list of all the derived parameter constraints is reported in the second panel of Table~\ref{tab:f_modelsA}.

\begin{figure}[t!]
\begin{center}
    \includegraphics[width=0.48\columnwidth]{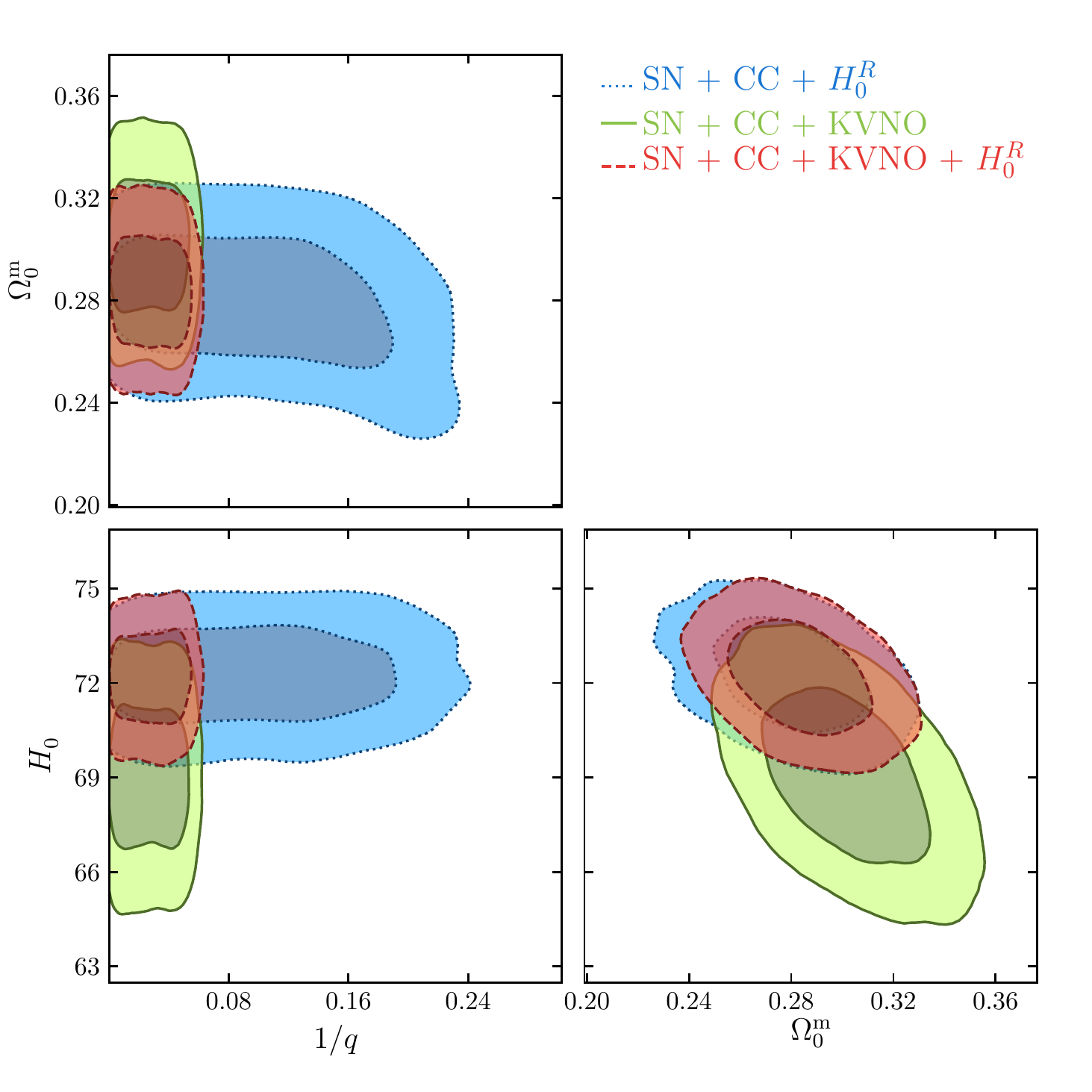}
    \caption{\label{fig:f3CDM_posteriors}{Marginalized 2 dimensional likelihood constraints on the parameters of the $f_3^{}(T)$ model of Eq.~\eqref{fTmodel3}.
    Permission for use of these figures was kindly provided by the authors of Ref.~\cite{LeviSaid:2020mbb}.}}
\end{center}
\end{figure}


\item Exponential model.
Similar to the previously considered model, in the $f_3(T)$~\cite{Linder:2010py} model we observe that the $\Lambda$\gls{cdm} model is recovered when $q\rightarrow+\infty$, or equivalently $1/q\rightarrow0^+$. The derived parameter constraints~\cite{LeviSaid:2020mbb} are reported in the third panel of Table~\ref{tab:f_modelsA}, whereas the marginalized confidence contours are depicted in top-left panel of Fig.~\ref{fig:f3CDM_posteriors}. Similar to the previous exponential $f(T)$ model, the model parameter $1/q$ is found to be consistent with the $\Lambda$\gls{cdm} limit at around 1$\sigma$, where such a result is compatible with the results of Refs.~\cite{Nesseris:2013jea,Nunes:2016qyp,Xu:2018npu,Anagnostopoulos:2019miu}. Furthermore, this exponential $f(T)$ model was found to be characterized by a null variation in the fine--structure constant, since $\beta_F^{}$ was always found to be consistent with zero.

\end{enumerate}

\subsection{Beyond data-driven analysis for TG cosmologies}
\label{subsec:gaussian_ML}

We have been discussing several proposals that can be cosmologically viable and well constrained by the current observational data. However, due the avalanche of these observational catalogues from different species, we require the implementation of new numerical technique to perform reconstruction that can be model independent in order to solve the inverse cosmology problem, i.e. going from the data to the best (or optimal) theory/model. These nonparametric approaches, as they are commonly called in the literature can be seen as a perspective for future \gls{tg} scenarios analysis, one of the can lies in the classification of reconstruction of data using Gaussian processes (\gls{gp}), and the second one as machine learning tools for \gls{tg} through deep learning techniques \cite{2020arXiv200301926S}. This section is devoted to explain how we can implement \gls{tg} scenarios in these ambiance. While the methodology behind each approach is wide, we present in the Supplementary annexes (Supplementary 6 and 7) the main ideas behind the numerical architecture.


\subsubsection{Gaussian processes and the reconstruction of f(T) gravity }

The discussed \gls{gp} approach in the Supplementary annexes (Supplementary 6) can be applied to a number of different $H(z)$ data sources, from which it is possible to reconstruct $H_0$. In the considered analyses, three sources of $H(z)$ data have been considered, namely \gls{cc}, \gls{sn}eIA and \gls{bao}. From the presented $H(z)$ data in the Supplementary annexes (Supplementary 4), only the \gls{cc} data as reported in Ref.~\cite{Moresco:2016mzx} are adopted in the following analyses \cite{Briffa:2020qli}.
Regarding the \gls{sn}eIa data which was introduced in the Supplementary annexes (Supplementary 4), the combination of the compressed Pantheon compilation is used \cite{Scolnic:2017caz} together with the CANDELS and CLASH Multi-cycle Treasury data \cite{Riess:2017lxs}. For this data set, the Hubble rate parameter measurements of $E(z)=H(z)/H_0$ were added along with the corresponding correlation matrix, such that only five of the reported six data points could be taken into consideration, since the $z=1.5$ data point is not Gaussian--distributed.
Moreover, a set of eight \gls{bao} data points from the \gls{sdss} (Data Release 12 and 14) \cite{Alam:2016hwk,Bourboux:2017cbm,Zhao:2018gvb} along with the corresponding correlation matrices were adopted in the following GP analyses. We refer the reader to Ref.~\cite{Briffa:2020qli} for further details on the inclusion of the \gls{bao} data set and \gls{sn}eIa measurements via an iterative technique.

\begin{figure}[t]
\begin{center}
    \includegraphics[width=0.48\columnwidth]{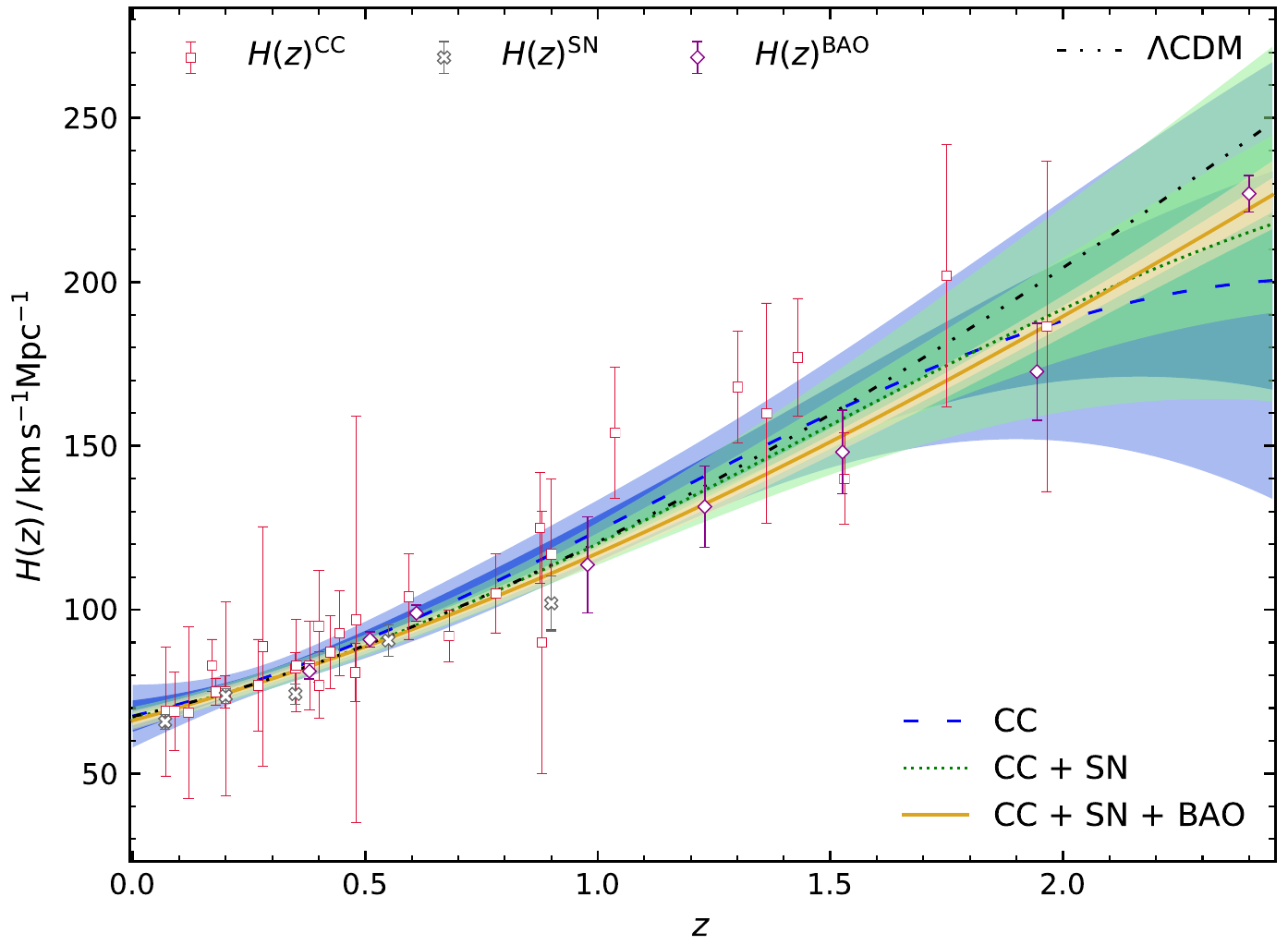}
    \includegraphics[width=0.48\columnwidth]{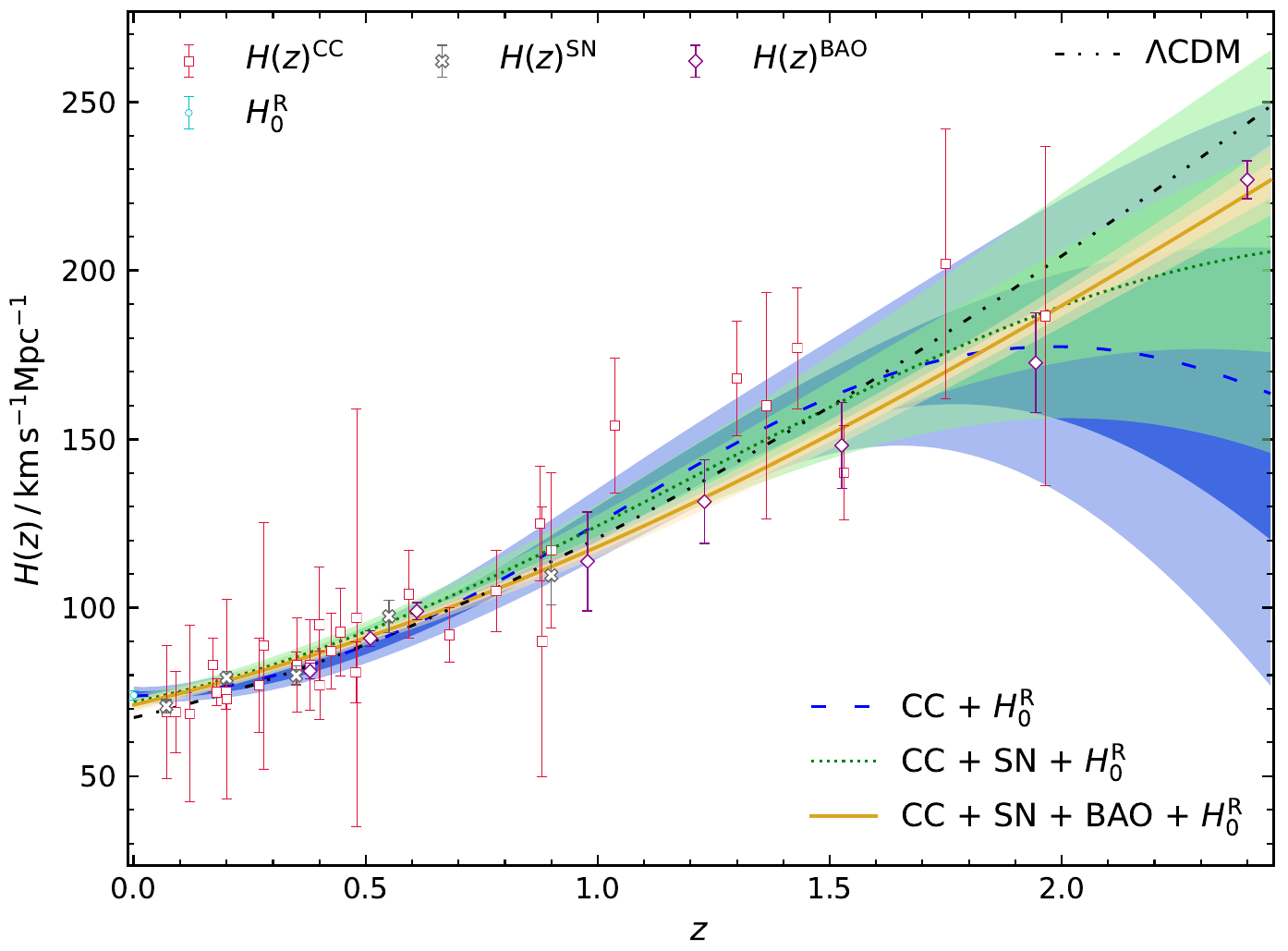}
    \includegraphics[width=0.48\columnwidth]{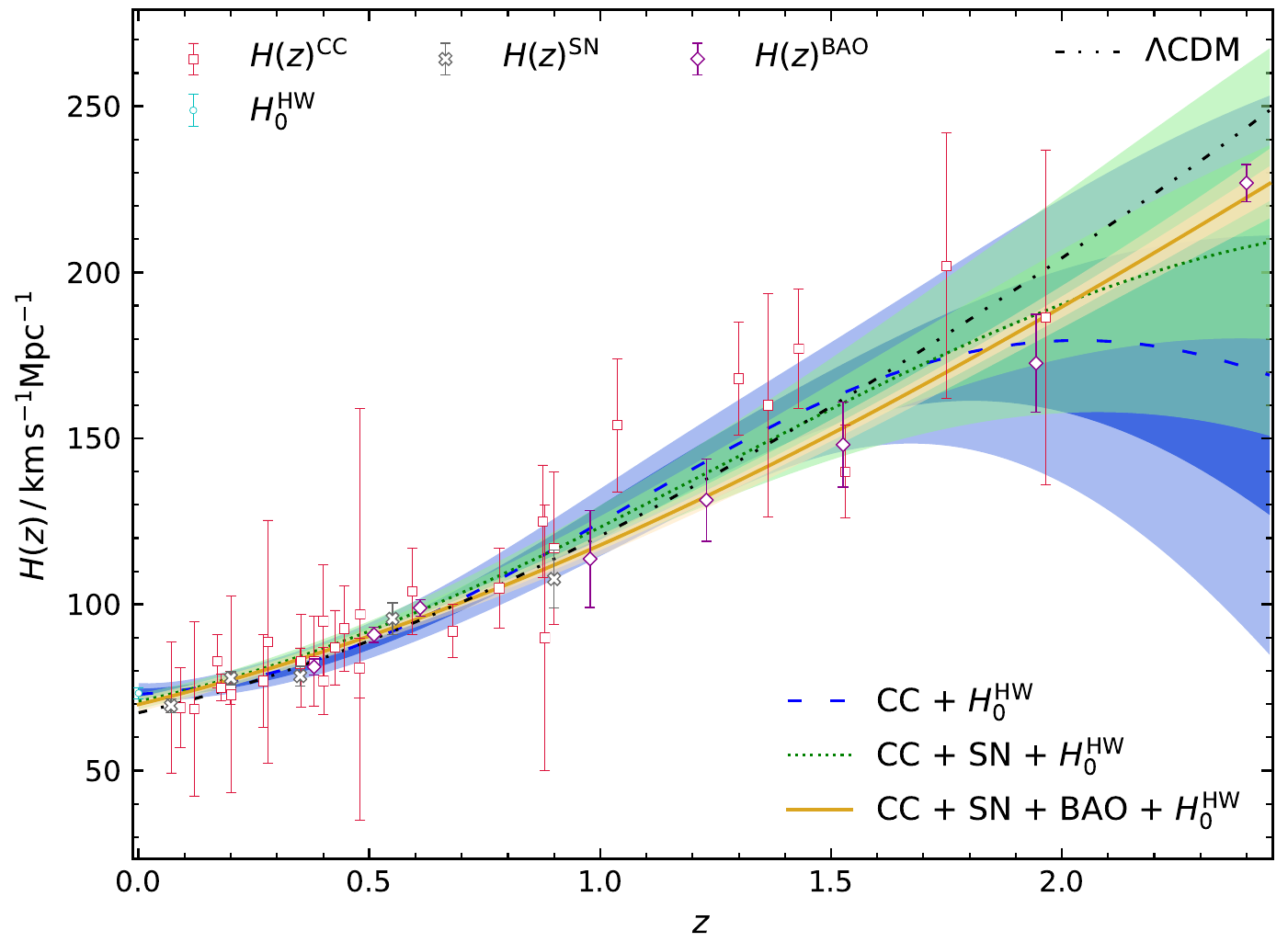}
    \includegraphics[width=0.475\columnwidth]{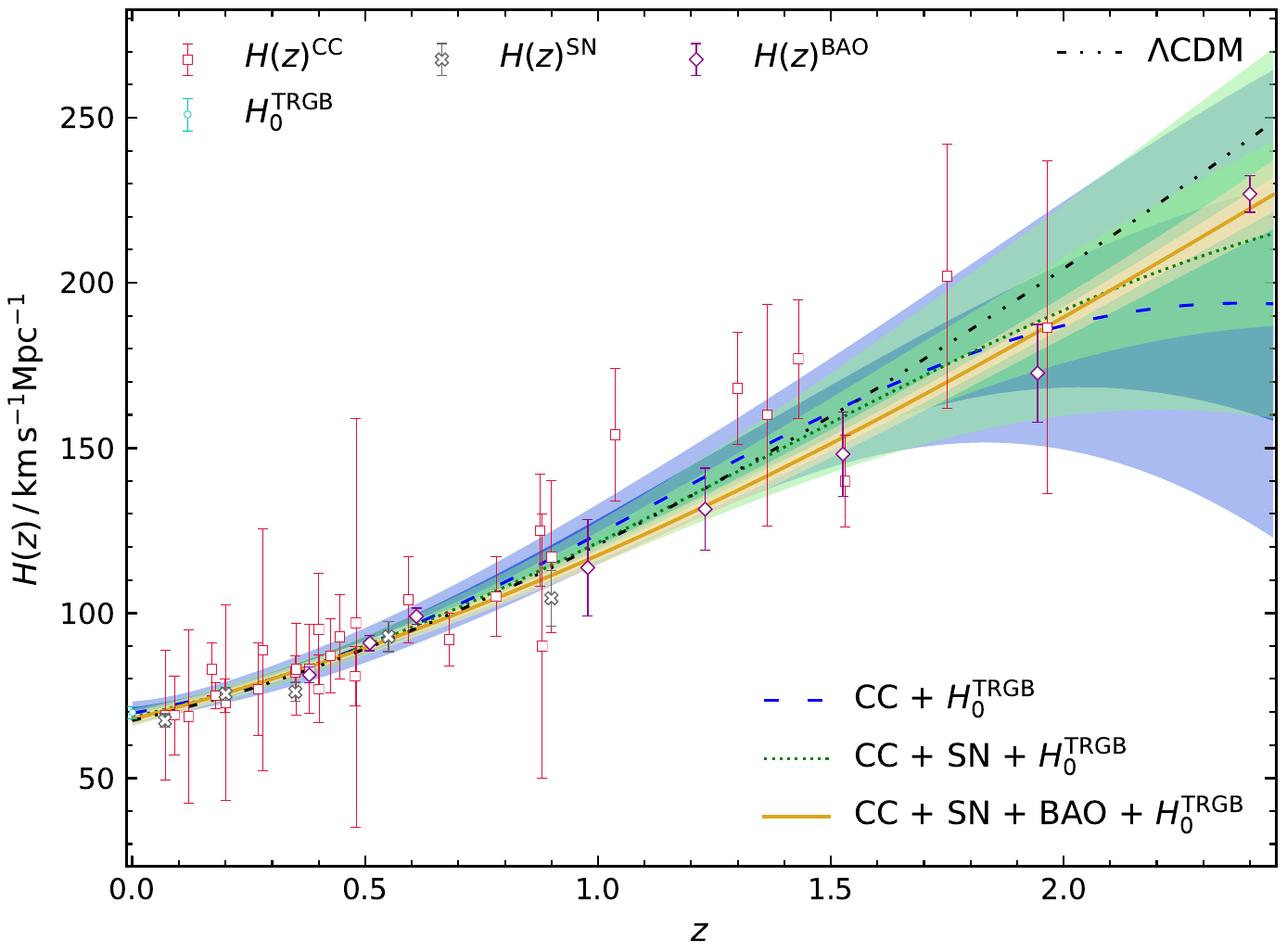}
    \caption{\label{fig:H_squaredexp}
    GP reconstructions of $H(z)$ with the squared exponential kernel function expressed in the Supplementary annexes (Supplementary 6). The data sets along with the different $H_0^{}$ priors are indicated in each respective panel. Permission for use of this figure was kindly provided by the authors of Ref.~\cite{Briffa:2020qli}.
    }
\end{center}
\end{figure}

\begin{table}[t]
\midsepremove
\setlength{\tabcolsep}{2.5pt}
\centering
\begin{tabular}{c c c c c c}
    \toprule
    \cellcolor{gris3} {\small\textbf{ Data set(s)}} & \cellcolor{gris3}{\small \boldmath{ $H_0\,[\text{km/s/Mpc}]$} }& \cellcolor{gris3}{\small \boldmath{ $d(H_0,H_0^{\rm R})$}} & \cellcolor{gris3}{\small \boldmath{ $d(H_0,H_0^{\rm TRGB})$}} & \cellcolor{gris3}{\small \boldmath{$d(H_0,H_0^{\rm HW})$}} & \cellcolor{gris3}{\small \boldmath{ $d(H_0,H_0^{\rm P18})$}} \\ \midrule
    \cellcolor{gris1} {\footnotesize \gls{cc}} & \cellcolor{gris1}$67.539 \pm 4.7720$ & \cellcolor{gris1}-1.3037 & \cellcolor{gris1}-0.4408 & \cellcolor{gris1}-1.1334 & \cellcolor{gris1}0.0290 \\
    \cellcolor{gris3} {\footnotesize \gls{cc}+\gls{sn}} & \cellcolor{gris3}$67.001 \pm 1.6531$ & \cellcolor{gris3}-3.2253 & \cellcolor{gris3}-1.1183 & \cellcolor{gris3}-2.6165 & \cellcolor{gris3}-0.2309 \\
    \cellcolor{gris1}{\footnotesize \gls{cc}+\gls{sn}+\gls{bao}} & \cellcolor{gris1}$66.197 \pm 1.4639$ & \cellcolor{gris1}-3.8407 & \cellcolor{gris1}-1.5127 & \cellcolor{gris1}-3.1132 & \cellcolor{gris1}-0.7776 \\
    \cellcolor{gris3} {\footnotesize \gls{cc}+$H_0^{\rm R}$} & \cellcolor{gris3}$73.782 \pm 1.3743$ & \cellcolor{gris3}-0.12556 & \cellcolor{gris3}1.7106 & \cellcolor{gris3}0.2166 & \cellcolor{gris3}4.3640 \\
    \cellcolor{gris3} {\footnotesize \gls{cc}+\gls{sn}+$H_0^{\rm R}$} & \cellcolor{gris3}$72.022 \pm 1.0756$ & \cellcolor{gris3}-1.1271 & \cellcolor{gris3}1.0265 & \cellcolor{gris3}-0.6220 & \cellcolor{gris3}3.8969 \\
    \cellcolor{gris1} {\footnotesize \gls{cc}+\gls{sn}+\gls{bao}+$H_0^{\rm R}$} & \cellcolor{gris1}$71.180 \pm 1.0245$ & \cellcolor{gris1}-1.6279 & \cellcolor{gris1}0.6447 & \cellcolor{gris1}-1.0457 & \cellcolor{gris1}3.3155 \\
    \cellcolor{gris3}{\footnotesize \gls{cc}+$H_0^{\rm TRGB}$} & \cellcolor{gris3}$69.604 \pm 1.7557$ & \cellcolor{gris3}-1.9599 & \cellcolor{gris3}-0.0760 & \cellcolor{gris3}-1.4908 & \cellcolor{gris3}1.2076 \\
    \cellcolor{gris1} {\footnotesize \gls{cc}+\gls{sn}+$H_0^{\rm TRGB}$} & \cellcolor{gris1}$68.468 \pm 1.2212$ & \cellcolor{gris1}-2.9695 & \cellcolor{gris1}-0.5942 & \cellcolor{gris1}-2.2641 & \cellcolor{gris1}0.8096 \\
    \cellcolor{gris3} {\footnotesize \gls{cc}+\gls{sn}+\gls{bao}+$H_0^{\rm TRGB}$} & \cellcolor{gris3}$67.811 \pm 1.1470$ & \cellcolor{gris3}-3.4070 & \cellcolor{gris3}-0.9036 & \cellcolor{gris3}-2.6233 & \cellcolor{gris3}0.3284 \\
    \cellcolor{gris1} {\footnotesize \gls{cc}+$H_0^{\rm HW}$} & \cellcolor{gris1}$72.966 \pm 1.6636$ & \cellcolor{gris1}-0.4863 & \cellcolor{gris1}1.2617 & \cellcolor{gris1}-0.1382 & \cellcolor{gris1}3.2043 \\
    \cellcolor{gris3}{\footnotesize \gls{cc}+\gls{sn}+$H_0^{\rm HW}$} & \cellcolor{gris3}$70.850 \pm 1.1991$ & \cellcolor{gris3}-1.7111 & \cellcolor{gris3}0.4710 & \cellcolor{gris3}-1.1550 & \cellcolor{gris3}2.6555 \\
    \cellcolor{gris1} {\footnotesize \gls{cc}+\gls{sn}+\gls{bao}+$H_0^{\rm HW}$} & \cellcolor{gris1}$69.911 \pm 1.1276$ & \cellcolor{gris1}-2.2717 & \cellcolor{gris1}0.0506 & \cellcolor{gris1}-1.6280 & \cellcolor{gris1}2.0355 \\
\bottomrule
\end{tabular}
\midsepdefault
\caption{\label{tab:se_kernel} Different GP reconstructions of $H_0$ \cite{LeviSaid:2020mbb} with the square exponential kernel function expressed in the Supplementary annexes (Supplementary 6). The reconstructed values of $H_0$ are complemented by their distance (in units of $\sigma$) from literature priors.}
\end{table}

The \gls{gp} approach is now applied to the various sources of Hubble data together with the priors described. The results pertaining to the value of $H_0$ are presented in Table~\ref{tab:se_kernel} which contains the principal results for the square exponential kernel. We here present the \gls{gp} inferred value of $H_0$ for the case of taking no prior, and the $H_0^{\rm R}$, $H_0^{\rm TRGB}$ and $H_0^{\rm HW}$ priors. In every case, it can be determined the distance (in units of $\sigma$) between the \gls{gp} determined value against the literature priors discussed above, so that this distance is defined as follows
\begin{equation}
    d\left(H_{0,i},H_{0,j}\right) = \frac{H_{0,i} - H_{0,j}}{\sqrt{\sigma_i^2 + \sigma_j^2}}\,,
\end{equation}
where $H_{0,i}$ and $H_{0,j}$ are two respective values of the present value of the Hubble parameter together with their respective $1\sigma$ uncertainties $\sigma_i$ and $\sigma_j$.

The derived results are presented in Fig.~\ref{fig:H_squaredexp}. The \gls{gp} approach is used for each set of priors for $H_0$ as shown in the sub-figures. Moreover, for every \gls{gp} reconstruction, the $1\sigma$ and $2\sigma$ regions are illustrated. As a reference point, it is present the $\Lambda$\gls{cdm} behavior in all instances.

\begin{figure}[t!]
\begin{center}
    \includegraphics[width=0.495\columnwidth]{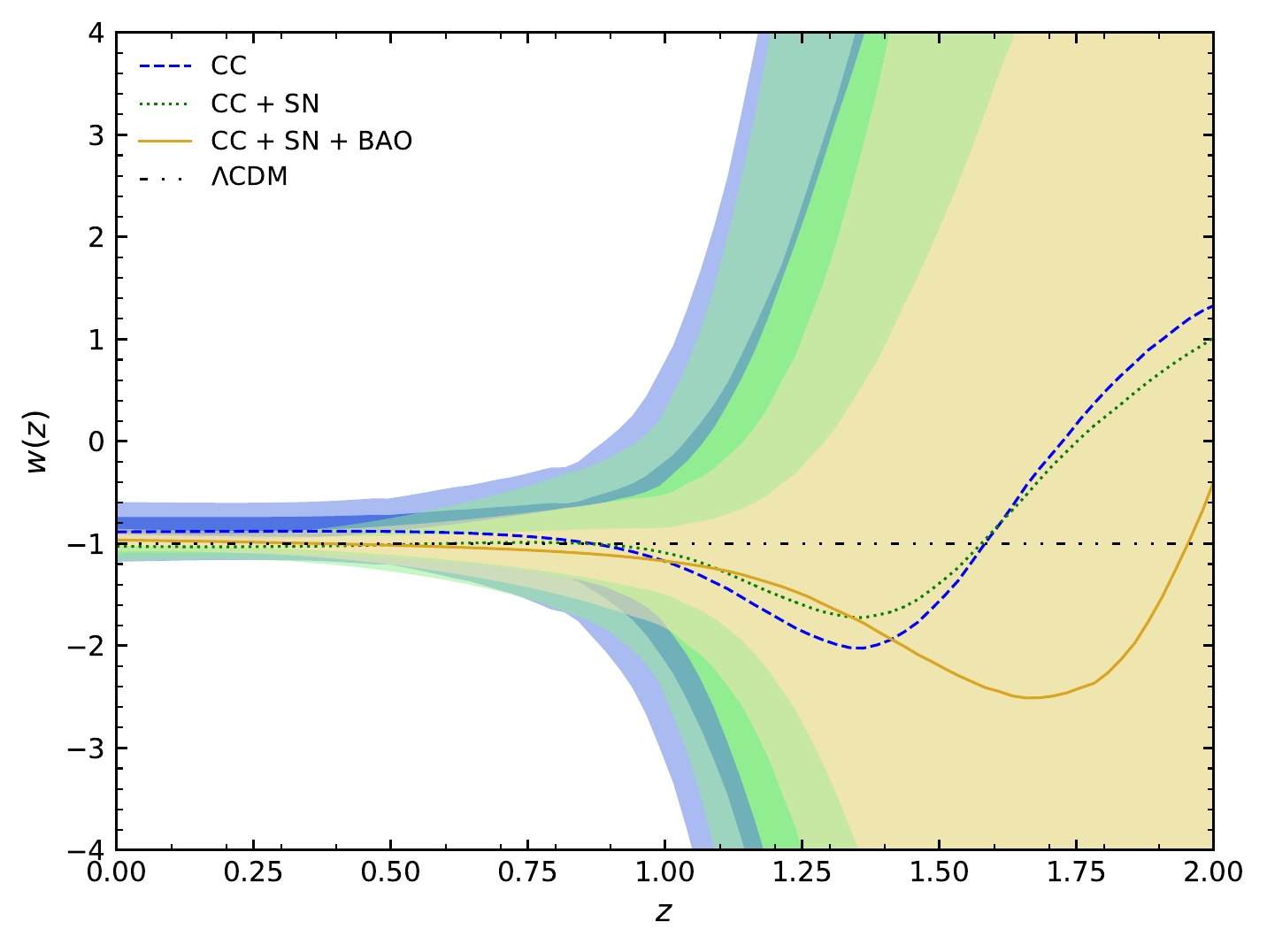}
    \includegraphics[width=0.495\columnwidth]{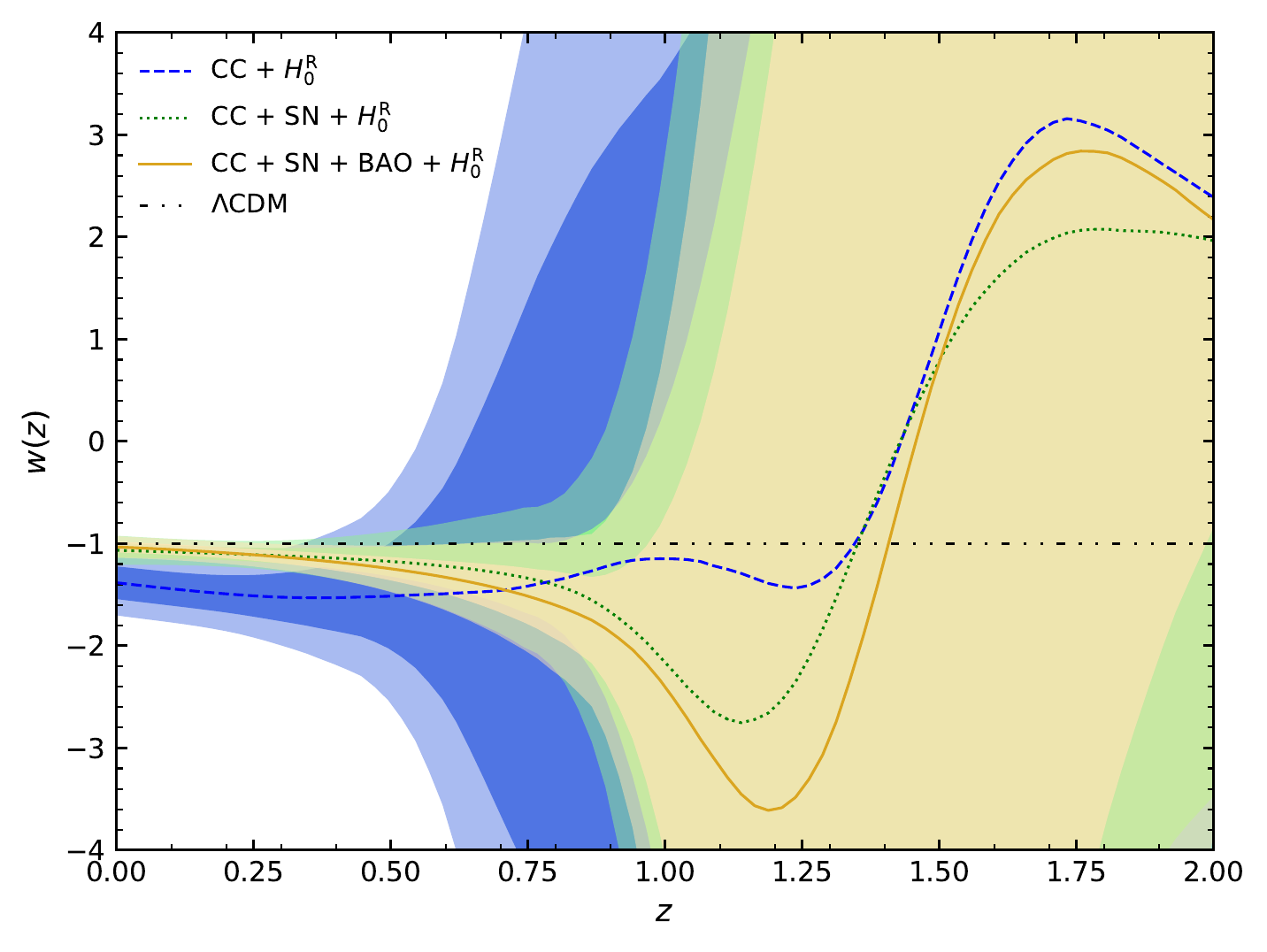}
    \includegraphics[width=0.495\columnwidth]{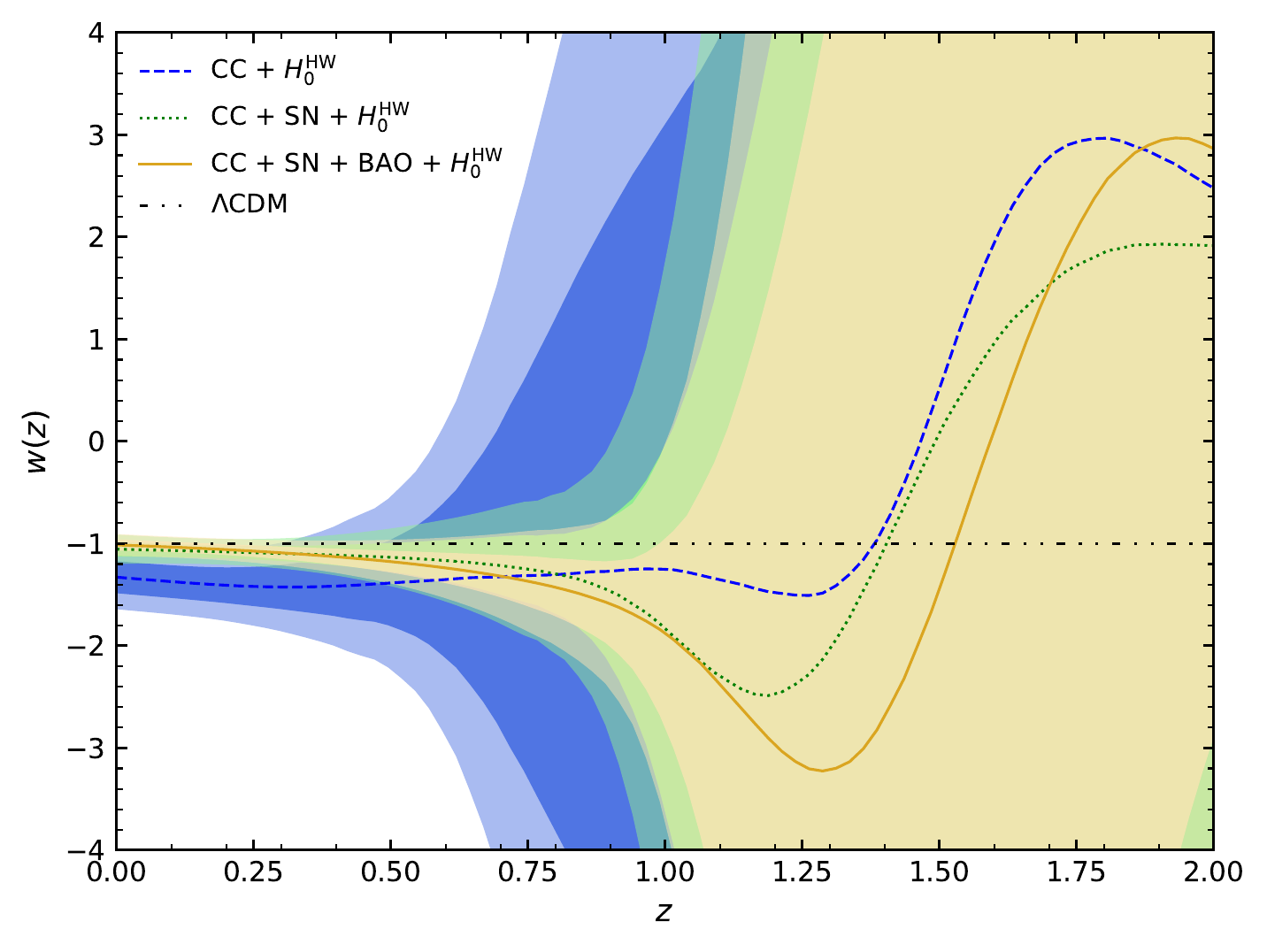}
    \includegraphics[width=0.495\columnwidth]{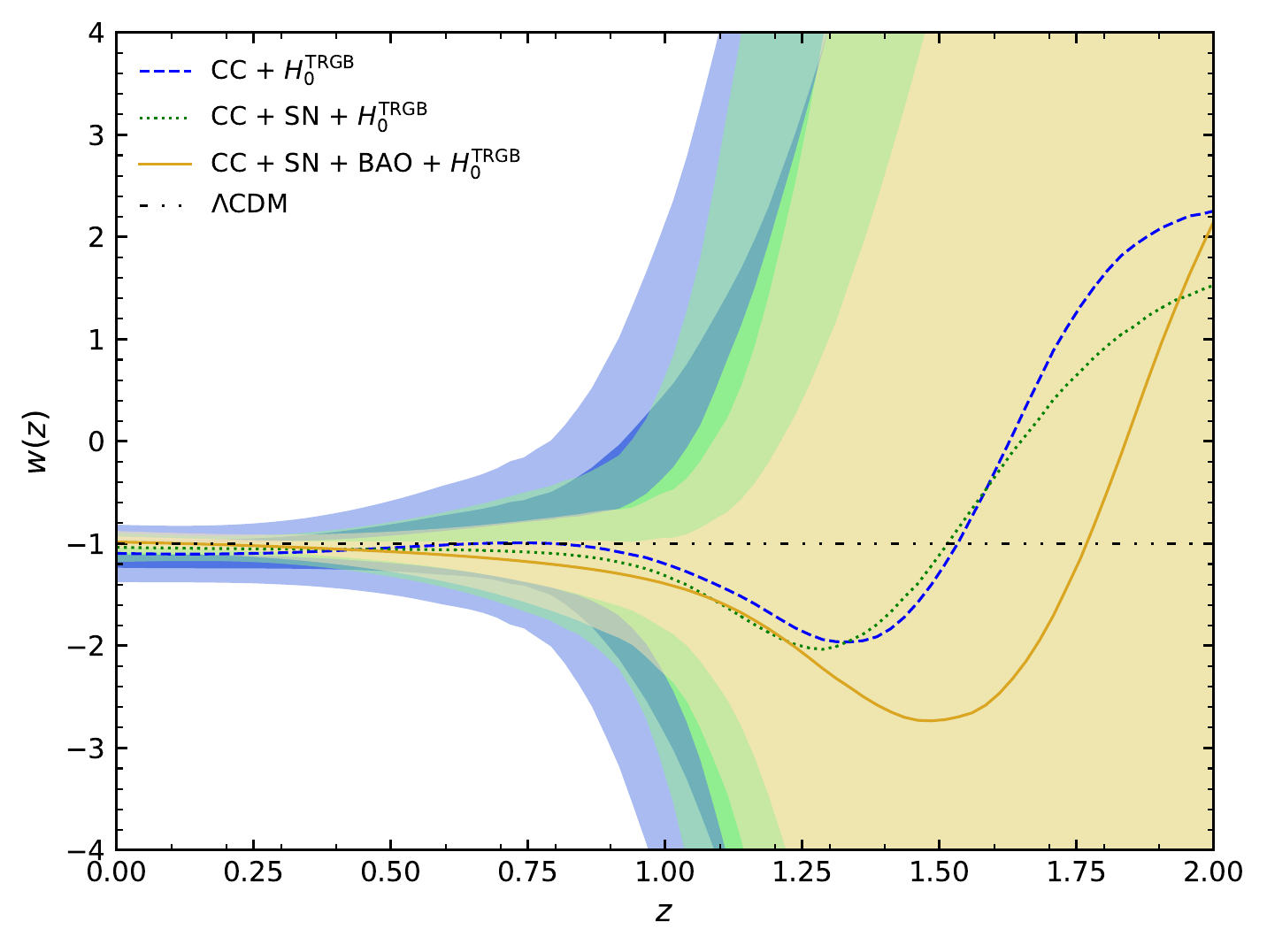}
    \caption{\label{fig:wz_squaredexp_cauchy}
    GP reconstructions of $w(z)$ with the squared exponential kernel functions, along with the $\Lambda$\gls{cdm} prediction. Permission for use of this figure was kindly provided by the authors of Ref.~\cite{Briffa:2020qli}.}
\end{center}
\end{figure}

There also exist diagnostic tools to assess preferences in the reconstructions toward deviations from $\Lambda$\gls{cdm} \cite{Yahya:2013xma,Zunckel:2008ti}. Considering the \gls{gr} first Friedmann equation for a cosmos filled with a dark fluid \gls{eos} $w(z)$
\begin{equation}
    E^2(z) := \frac{H^2(z)}{H_0^2} = \Omega_{\rm m0} \left(1+z\right)^3 + \Omega_{\rm k0} \left(1+z\right)^2 + \Omega_{\Lambda0}\exp\left[3\int_0^z \frac{1+w(z')}{1+z'} \mathrm{d}z'\right]\,,
\end{equation}
where $E(z)$ is the reduced Hubble parameter, and which can be rearranged to reconstruct the \gls{eos} of the dark fluid via
\begin{equation}
    w(z)=\frac{2(1+z)E(z)E'(z)-3E^2(z)}{3\left[E^2(z)-\Omega_{\rm m0}(1+z)^3\right]}\,,
\end{equation}
where it has been assumed that the Universe observes spatial flatness. The \gls{gp} reconstructions of $w(z)$ are reported in Fig.~\ref{fig:wz_squaredexp_cauchy}, where it is clear that $w=-1$ is not excluded by the currently available data. However, notice that this reconstruction is dependent on the matter density parameter, which restricts us from constructing physical models.

\begin{figure}[t!]
\begin{center}
    \includegraphics[width=0.495\columnwidth]{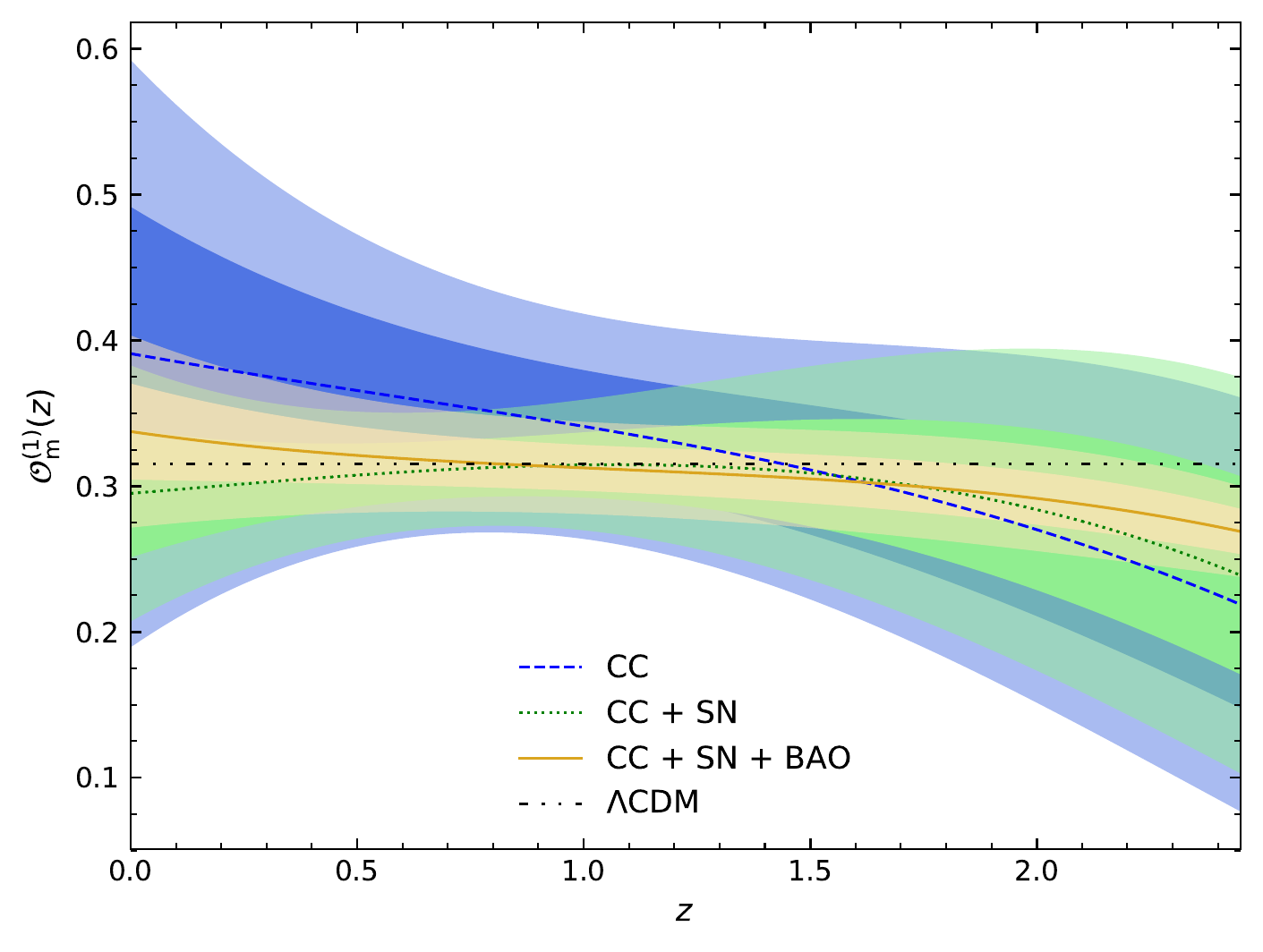}
    \includegraphics[width=0.495\columnwidth]{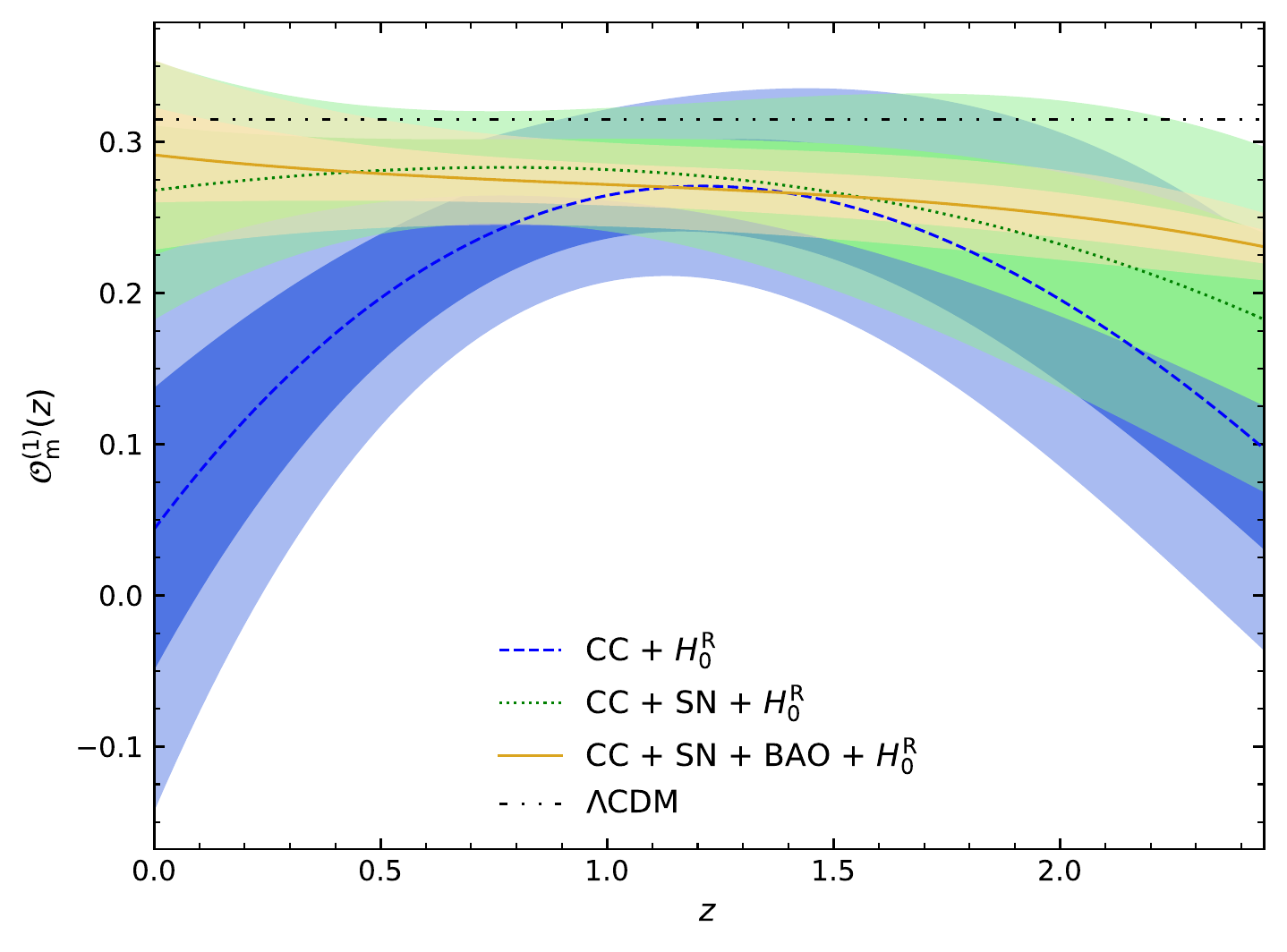}
    \includegraphics[width=0.495\columnwidth]{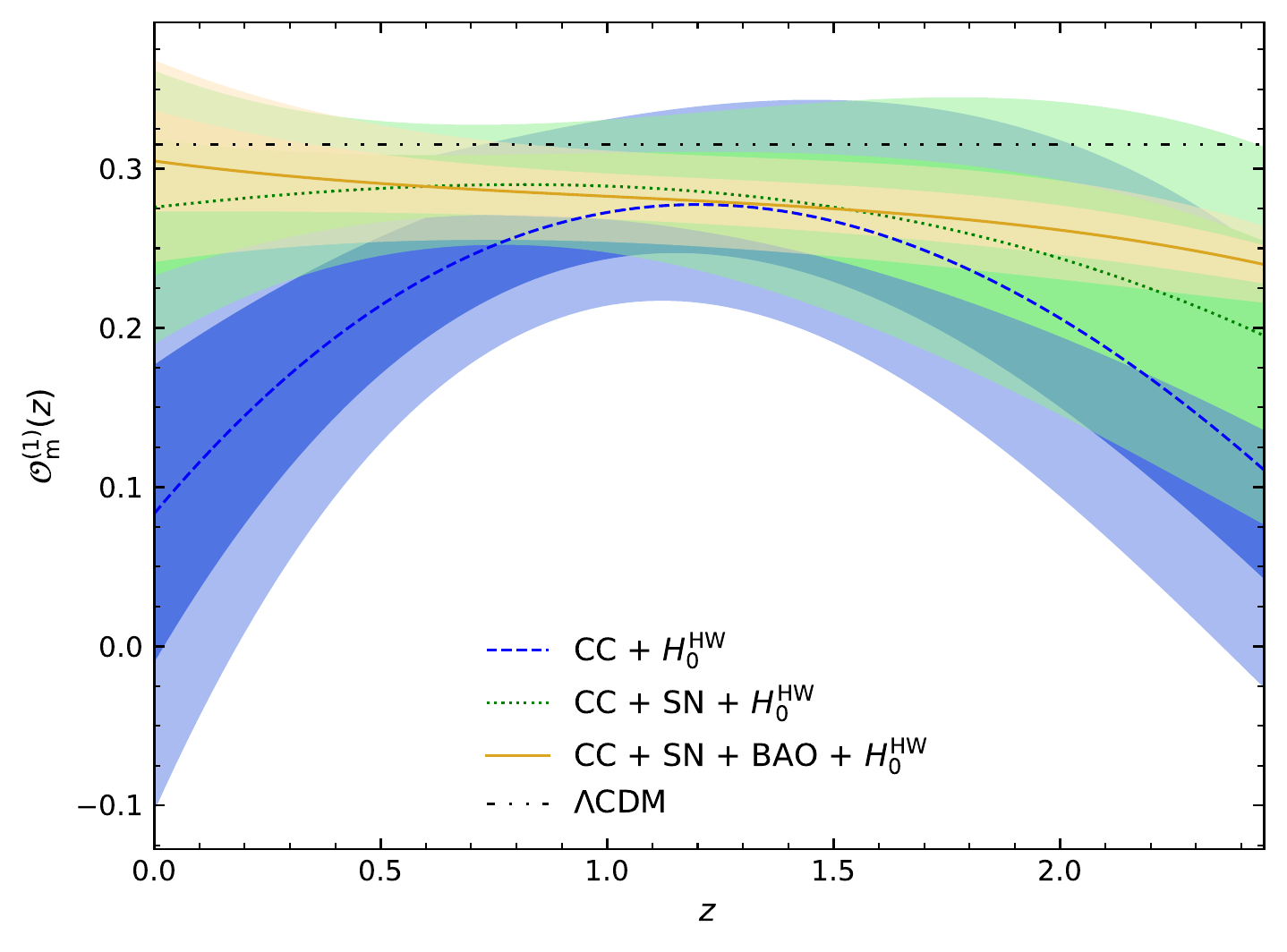}
    \includegraphics[width=0.495\columnwidth]{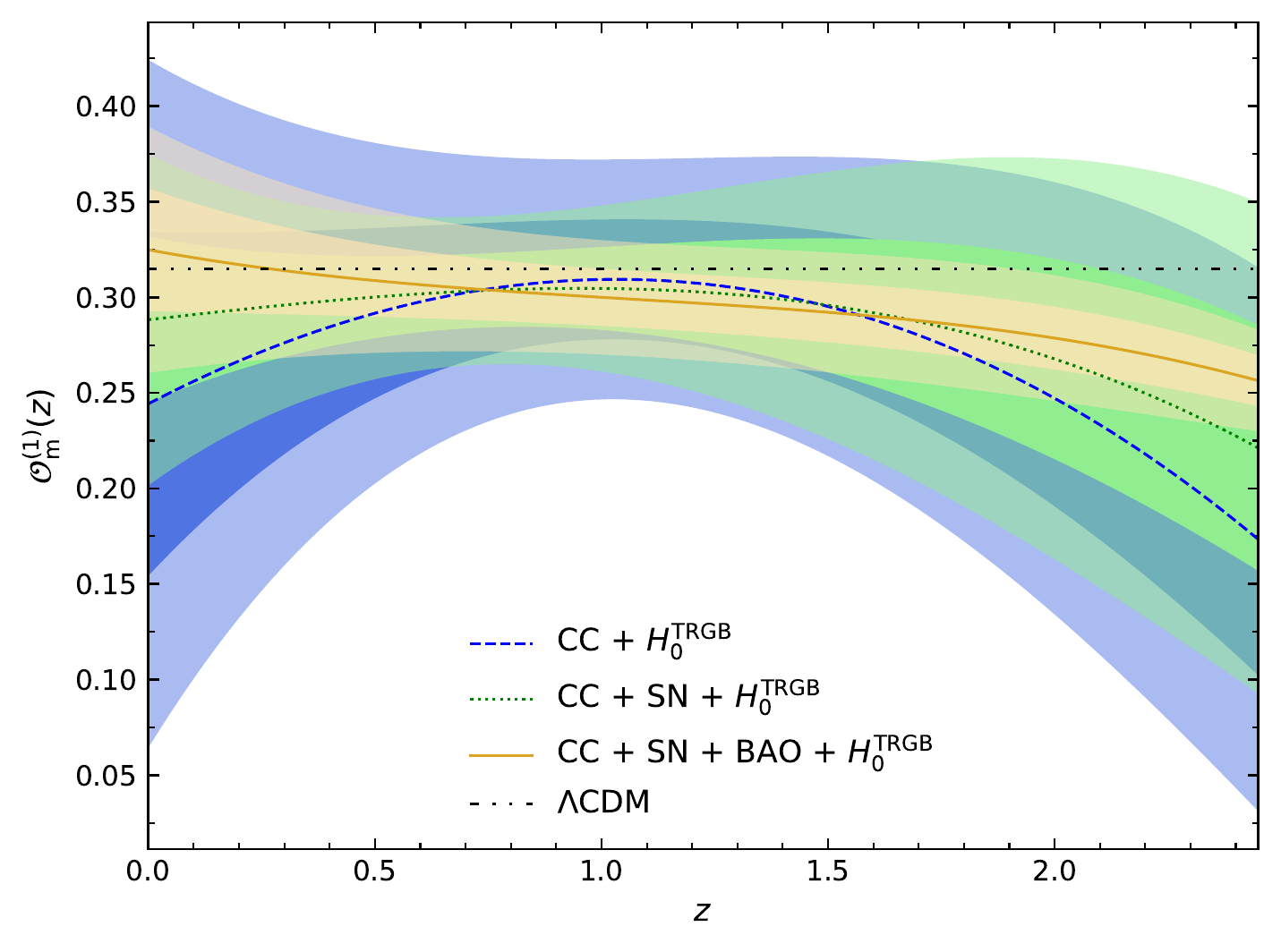}
    \caption{\label{fig:Om1_squaredexp_cauchy}
    GP reconstructions of $\mathcal{O}_m^{(1)}(z)$ with the squared exponential kernel function, along with the $\Lambda$\gls{cdm} prediction. Permission for use of this figure was kindly provided by the authors of Ref.~\cite{Briffa:2020qli}.
    }
\end{center}
\end{figure}

On the other hand, one could test the flat $\Lambda$\gls{cdm} model by considering the following diagnostic redshift function~\cite{LeviSaid:2020mbb}
\begin{equation}\label{consis_test}
    \mathcal{O}_m^{(1)} (z) :=\frac{E^2(z)-1}{z(3+3z+z^2)}\,,
\end{equation}
which reduces to $\mathcal{O}_m^{(1)} (z)=\Omega_{\rm m0}$ in the $\Lambda$\gls{cdm} scenario. Thus, this diagnostic can measure how close a data set is to \gls{gr}. We should remark that $\mathcal{O}_m^{(1)} (z)$ is not so much dependent on $\Lambda$\gls{cdm} but a characteristic parameter that can test how consistent a data set is with this model. As reported in Fig.~\ref{fig:Om1_squaredexp_cauchy}, one could clearly notice that the considered data sets are in a very good agreement with the $\Lambda$\gls{cdm} predictions, although the $H_0$ prior has a significant effect on the reconstruction and hence the viability of the concordance model of cosmology.

Complementary to the $\mathcal{O}_m^{(1)} (z)$, its derivative can also offer a useful way to quantify deviations from $\Lambda$CDM, defined through
\begin{equation}\label{diagnostic_for_H}
    \mathcal{L}^{(1)}(z)=3(1-E^2(z))(1+z)^2+2z(3+3z+z^2)E(z)E'(z)\,.
\end{equation}

\begin{figure}[t]
\begin{center}
    \includegraphics[width=0.495\columnwidth]{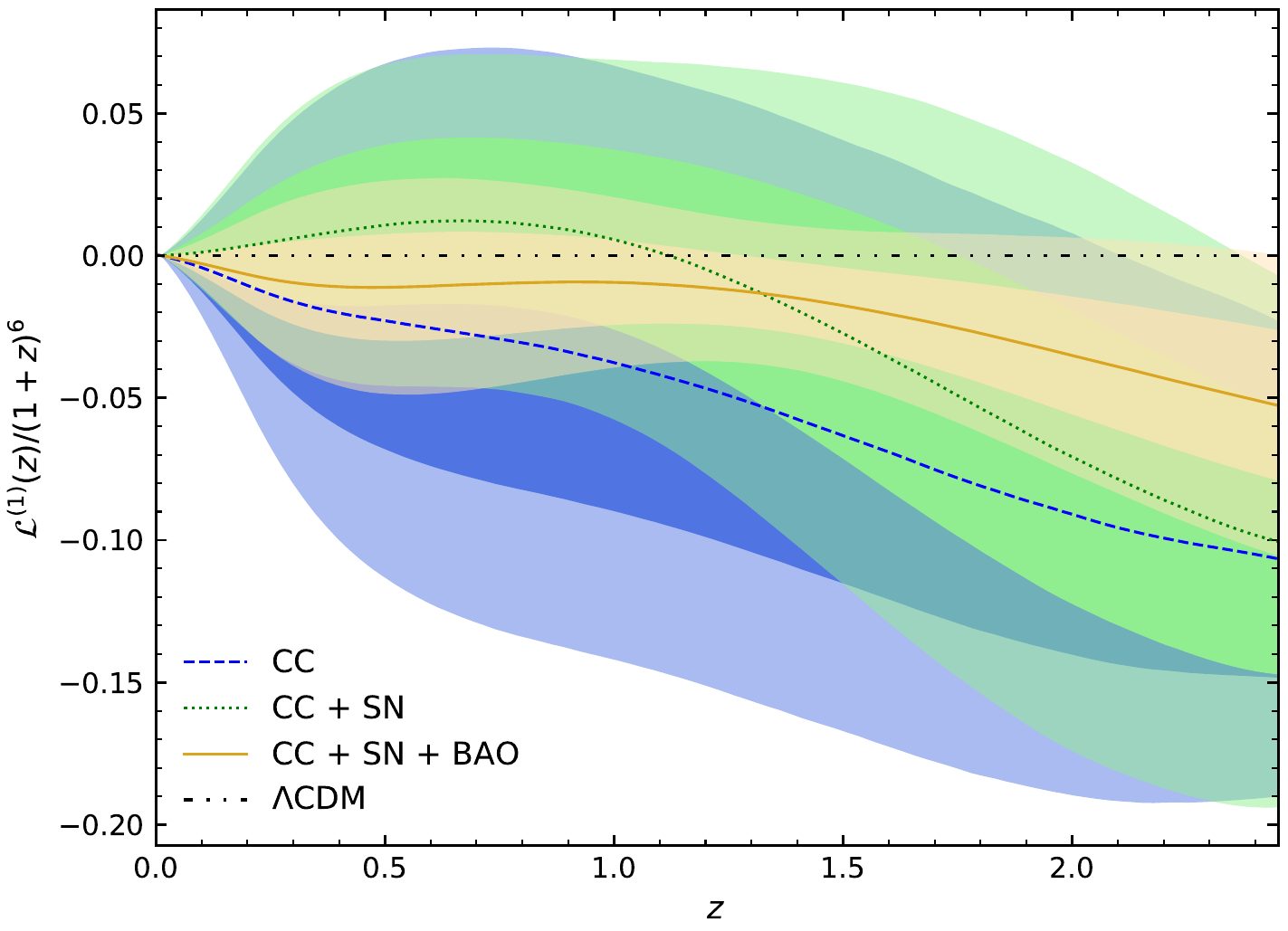}
    \includegraphics[width=0.495\columnwidth]{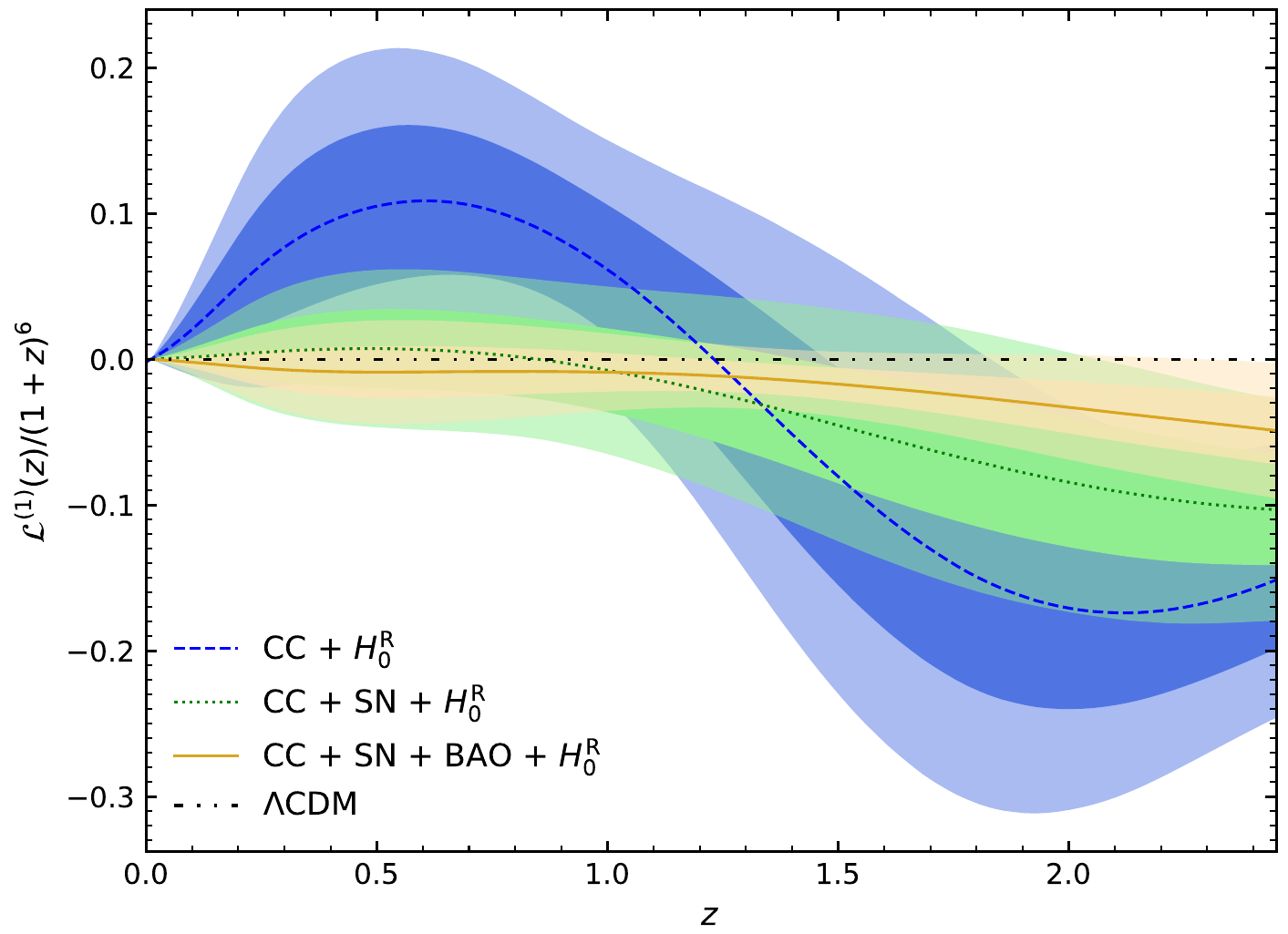}
    \includegraphics[width=0.495\columnwidth]{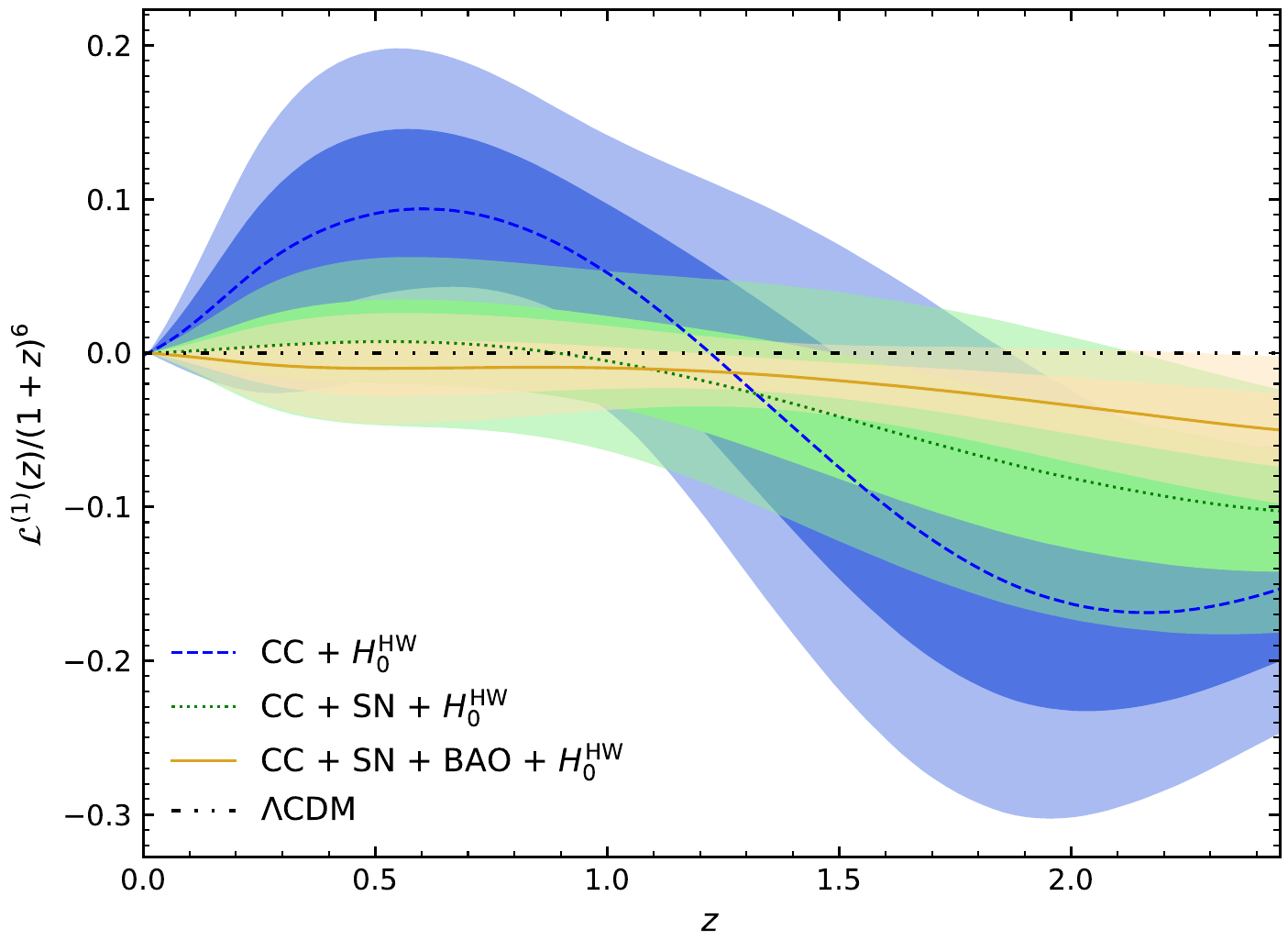}
    \includegraphics[width=0.495\columnwidth]{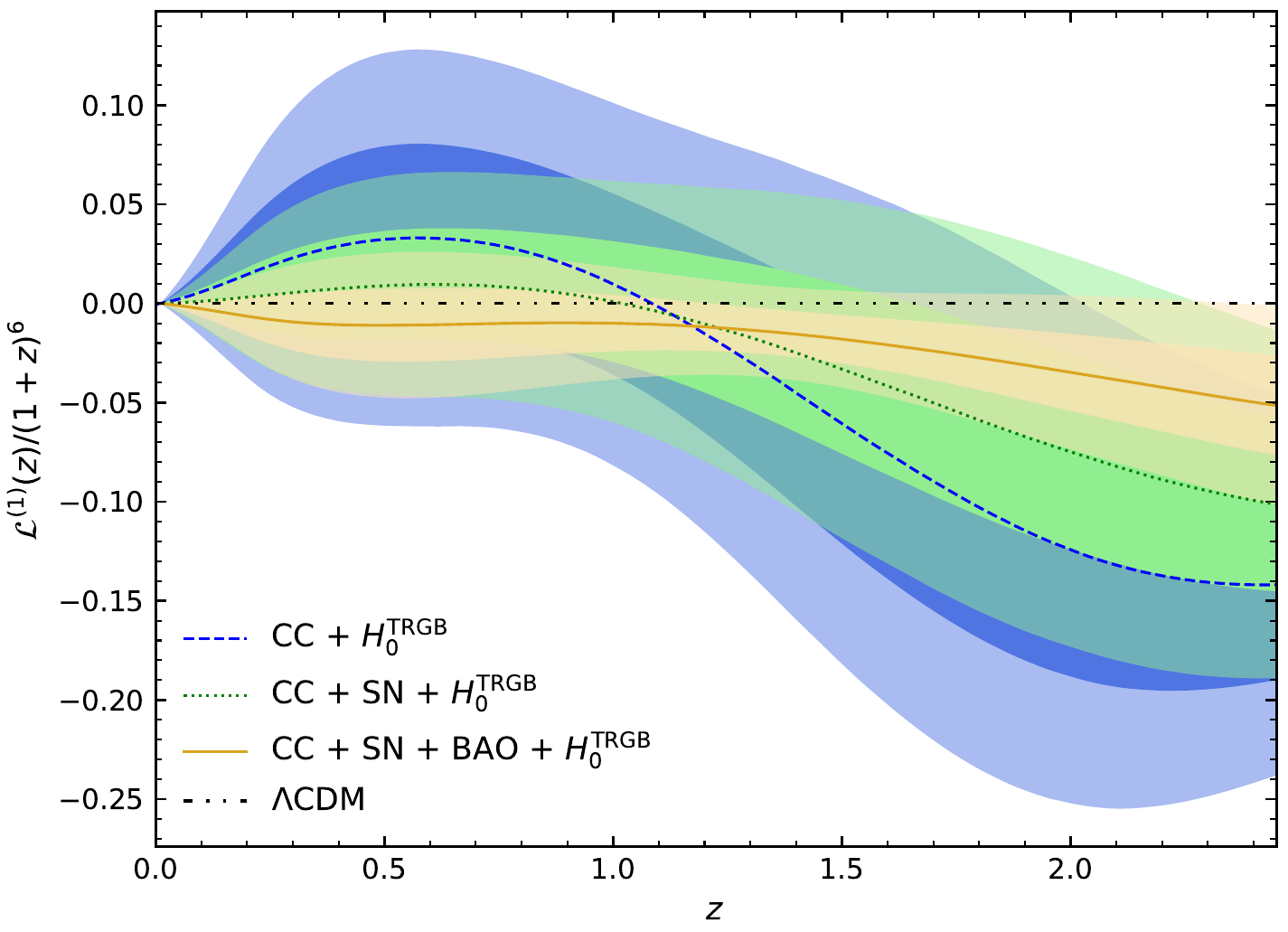}
    \caption{\label{fig:L1_squaredexp}
    GP reconstructions of $\mathcal{L}^{(1)}(z)/(1+z)^6$ with the squared exponential kernel function expressed in the Supplementary annexes (Supplementary 6). The data sets along with the different $H_0^{}$ priors are indicated in each respective panel. Permission for use of this figure was kindly provided by the authors of Ref.~\cite{Briffa:2020qli}.
    }
\end{center}
\end{figure}

\begin{figure}[t!]
\begin{center}
    \includegraphics[width=0.495\columnwidth]{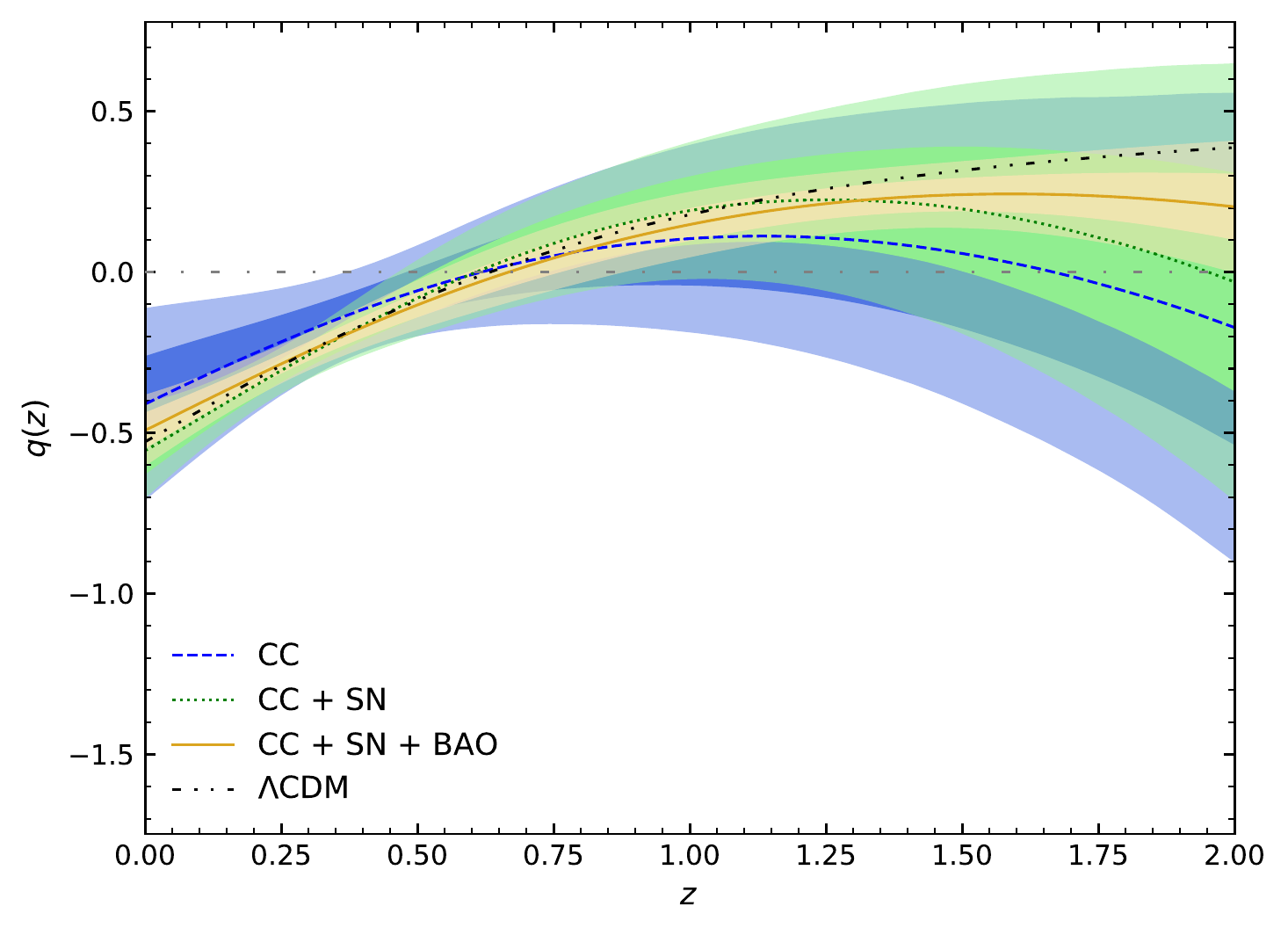}
    \includegraphics[width=0.495\columnwidth]{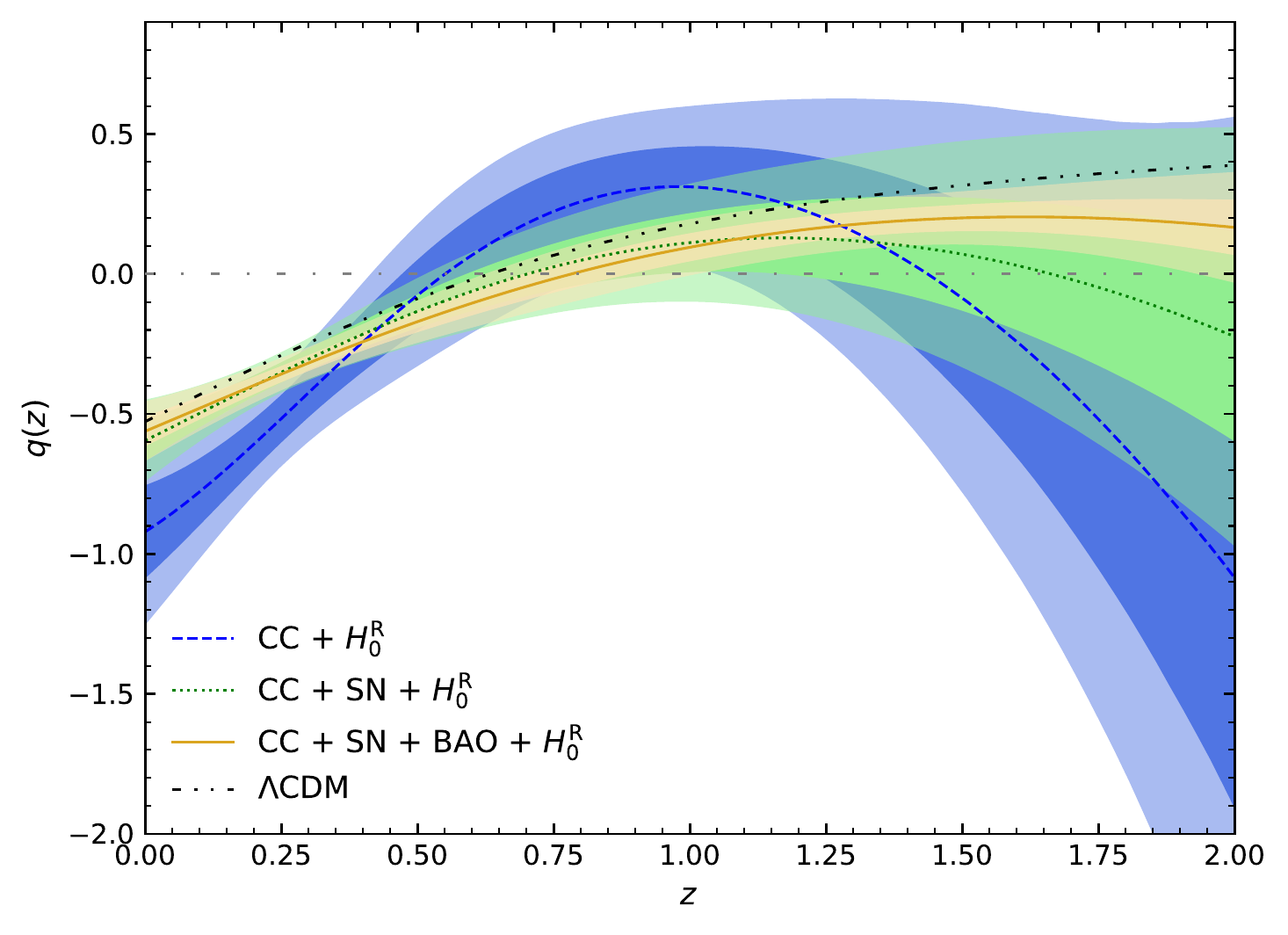}
    \includegraphics[width=0.495\columnwidth]{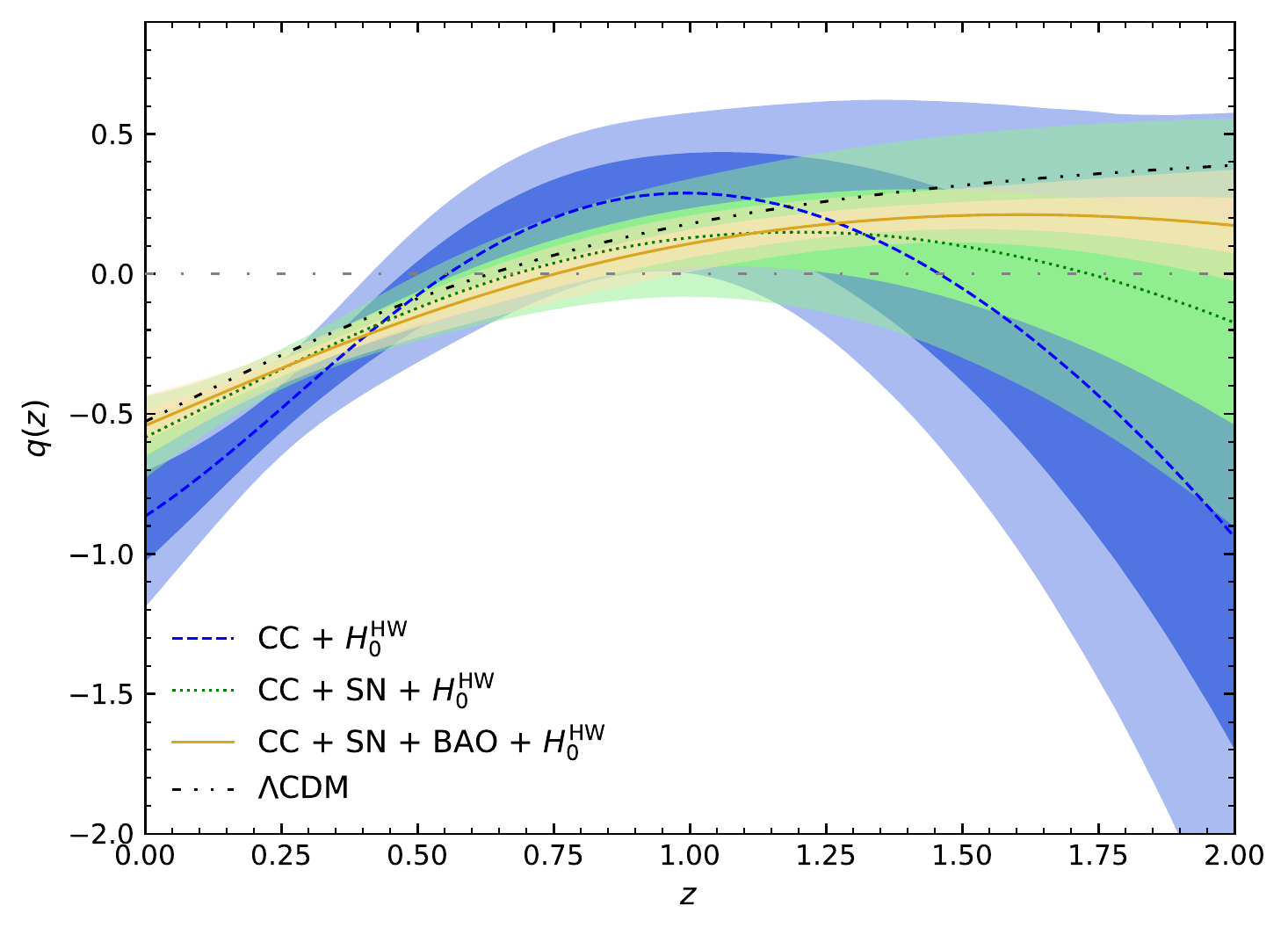}
    \includegraphics[width=0.495\columnwidth]{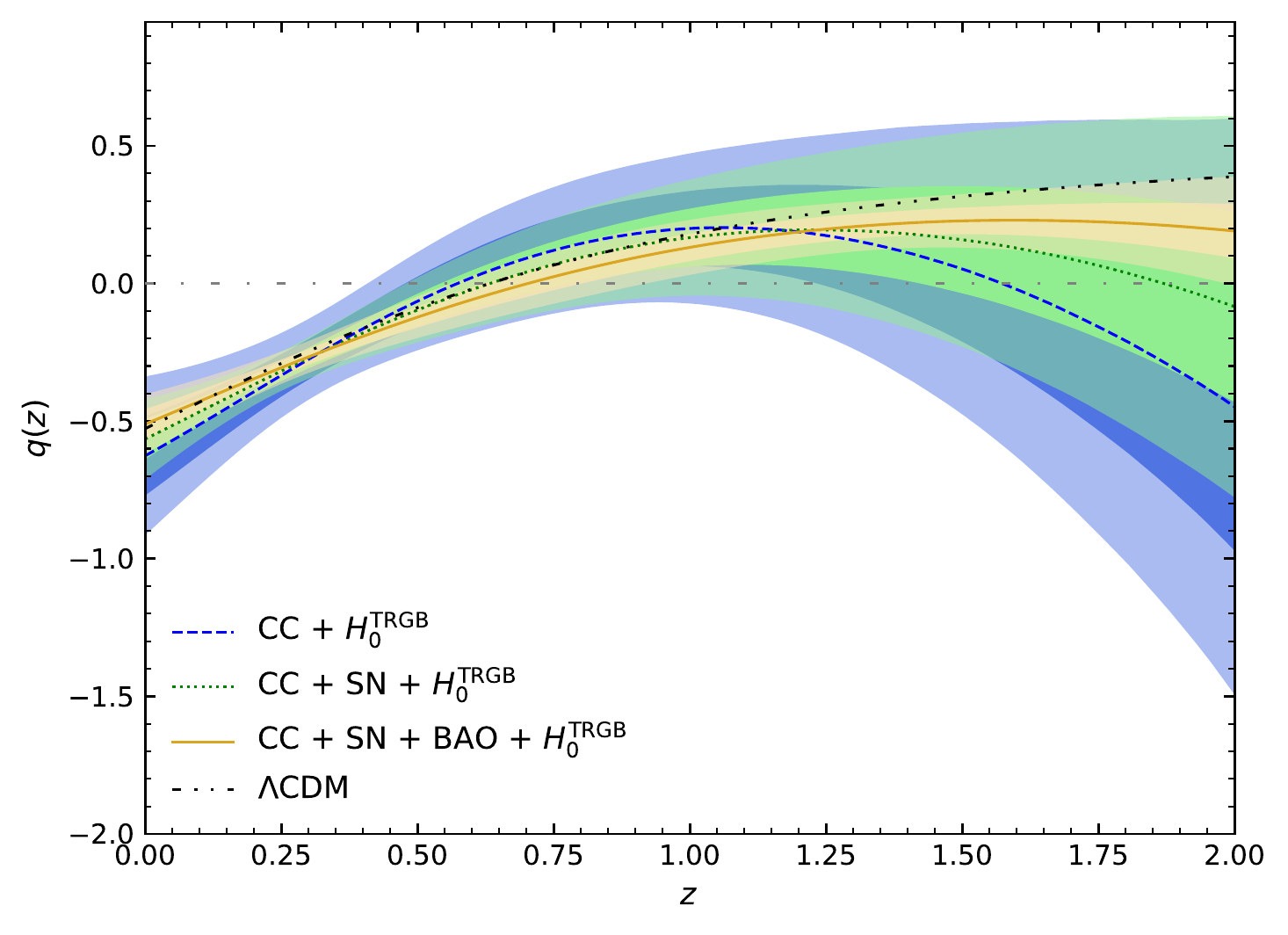}
    \caption{\label{fig:qz_squaredexp_cauchy}
    GP reconstructions of $q(z)$ with the squared exponential kernel functions, along with the $\Lambda$\gls{cdm} prediction. Permission for use of this figure was kindly provided by the authors of Ref.~\cite{Briffa:2020qli}.
    }
\end{center}
\end{figure}

Any deviation from $\mathcal{L}^{(1)}(z) = 0$ represents a deviation from $\Lambda$\gls{cdm}, which makes $\mathcal{L}^{(1)}(z)$ a good diagnostic over which to assess the behavior of the concordance model. The $\mathcal{L}^{(1)}(z)$ diagnostic is presented in Fig.~\ref{fig:L1_squaredexp}.
In Fig.~\ref{fig:H_squaredexp}, the square exponential kernel GP reconstructions are shown for the redshift range of the full data set. In all cases, the \gls{bao} data reduce the $1\sigma$ and $2\sigma$ uncertainties at higher redshifts since the other data sets do not feature points in that regime. In fact, in the cases of \gls{cc} and \gls{cc}+\gls{sn}, the $\Lambda$\gls{cdm} theoretical prediction only deviates into the $2\sigma$ uncertainty region for these high values of redshift, and only outside of both when the \gls{bao} data are included. The \gls{bao} data are dependent on the concordance model of cosmology, and so one would expect it to produce issues of this kind since the other two data sets are independent of cosmological models. The remainder of the reconstruction regions remain in close range of the $\Lambda$\gls{cdm} prediction. This is further exposed by the diagnostic consistency test shown in Fig.~\ref{fig:L1_squaredexp} where a number of regions mark slight deviations from $\Lambda$\gls{cdm}. However, it is the \gls{bao} data set that exposes this deviation at high redshifts. Another interesting feature of these diagnostic tests is that as with the $H_0$ reconstructions in Table~\ref{tab:se_kernel}, the $H_0^{\rm R}$ prior brings about the largest deviation of the reconstructions.
One could also infer the GP reconstructed deceleration parameter, given by
\begin{equation}
    q(z)=(1+z)\frac{H'(z)}{H(z)}-1\,.
\end{equation}
We illustrate the GP reconstruction of $q(z)$ in Fig.~\ref{fig:qz_squaredexp_cauchy}.

We finally discuss and present the application of the above cosmologically model-independent data for the GP reconstruction of a general $f(T)$ function without assuming a specific form of the Lagrangian. We start by expressing the Friedmann equation in terms of redshift dependence alone. We therefore convert the Lagrangian derivative term so that
\begin{equation}
    f_T = \frac{\mathrm{d}f/\mathrm{d}z}{\mathrm{d}T/\mathrm{d}z} = \frac{f'(z)}{T'(z)}\,,
\end{equation}
 where $f'(z) = \mathrm{d}f/\mathrm{d}z$ and $T'(z)= -12HH'$ are redshift derivatives. The immediate follow up becomes the issue of handling $f'(z)$ in the analysis which is tackled in this analysis through the central difference method as
\begin{equation}
    f'(z_i) \simeq\lim_{\Delta z \to 0} \frac{f(z_{i+1}) - f(z_{i-1})}{z_{i+1} - z_{i-1}}\,,\label{eq:centr_diff_meth}
\end{equation}
since this will produce smaller uncertainties $\mathcal{O}(\Delta z^2)$ rather than $\mathcal{O}(\Delta z)$ which occur for the forward and backward difference methods, where $\Delta z = z_{i+1} - z_{i-1}$. By substituting into the Friedmann equation~\eqref{Friedmann_1B} (taking the mapping $f(z) \rightarrow -T + f(z)$, as in Ref.~\cite{LeviSaid:2020mbb,Ren:2022aeo}) and simplifying, this method produces a numerical propagation equation for $f(z)$ given by
\begin{equation}\label{prop_eq_f_T}
    f(z_{i+1}) = f(z_{i-1}) + 2\left(z_{i+1} - z_{i-1}\right) \frac{H'(z_i)}{H(z_i)} \left(3H^2(z_i) + \frac{f(z_i)}{2} - 3 H_0^2 \Omega_{\rm m0} \left(1+z_i\right)^3\right)\,,
\end{equation}
where the propagation equation parameters $H_0$ and $\Omega_{\rm m0}$ are selected from the corresponding GP reconstruction within the P18 and $H_0$ priors, respectively. While superior in terms of having lower associated uncertainties, the central differencing requires two initial conditions to be employed which we form as follows:
\begin{enumerate}
    \item Friedmann equation boundary condition: Evaluating the Friedmann Eqs.~\eqref{Friedmann_1B} -- \eqref{Friedmann_2B} at $z=0$ gives
    \begin{equation}\label{f_T_boundary_condition}
        f(z=0) \simeq -16\pi G \rho_{m}^0 + 6 H_0^2 = -6H_0^2\left(\Omega_{\rm m0} - 1\right)\,,
    \end{equation}
    where we have imposed that $\Lambda$\gls{cdm} dominates at present times, i.e. $f_T(z=0) \simeq 0$. This again relies on the same parameters as the propagation equation itself;
    \item The second boundary condition can be obtained by using the forward differencing method through
    \begin{equation}\label{f_T_2nd_boundary_condition}
        f'(z_i) \simeq \lim_{\Delta z \to 0}\frac{f(z_{i+1}) - f(z_{i})}{z_{i+1} - z_{i}}\,,
    \end{equation}
    that leads to the equation
    \begin{equation}\label{f_T_boundary_condition_2}
        f(z_{i+1}) = f(z_{i}) + 2 \left(z_{i+1} - z_{i}\right) \frac{H'(z_{i})}{H(z_{i})} \left[3H^2(z_{i}) + \frac{f(z_{i})}{2} - 3H_0^2 \Omega_{\rm m0} \left(1+z_{i}\right)^3\right]\,,
    \end{equation}
    which straightforwardly leads to the necessary second boundary condition.
\end{enumerate}
The propagation of the $f(T)$ function is complemented by its associated \gls{mcmc} error propagation which produces the 1$\sigma$ and 2$\sigma$ uncertainties.

\begin{figure}[t!]
\begin{center}
    \includegraphics[width=0.495\columnwidth]{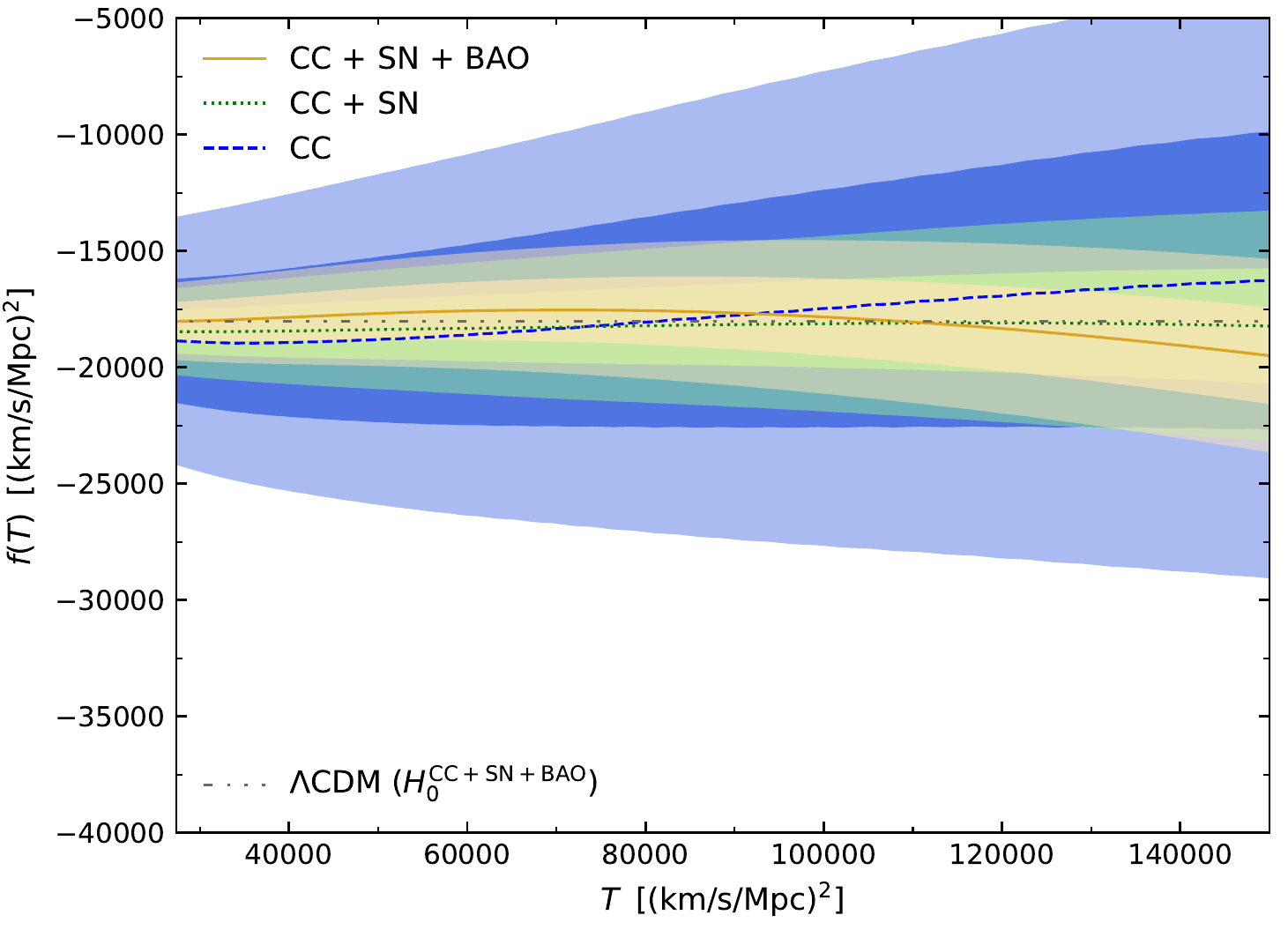}
    \includegraphics[width=0.495\columnwidth]{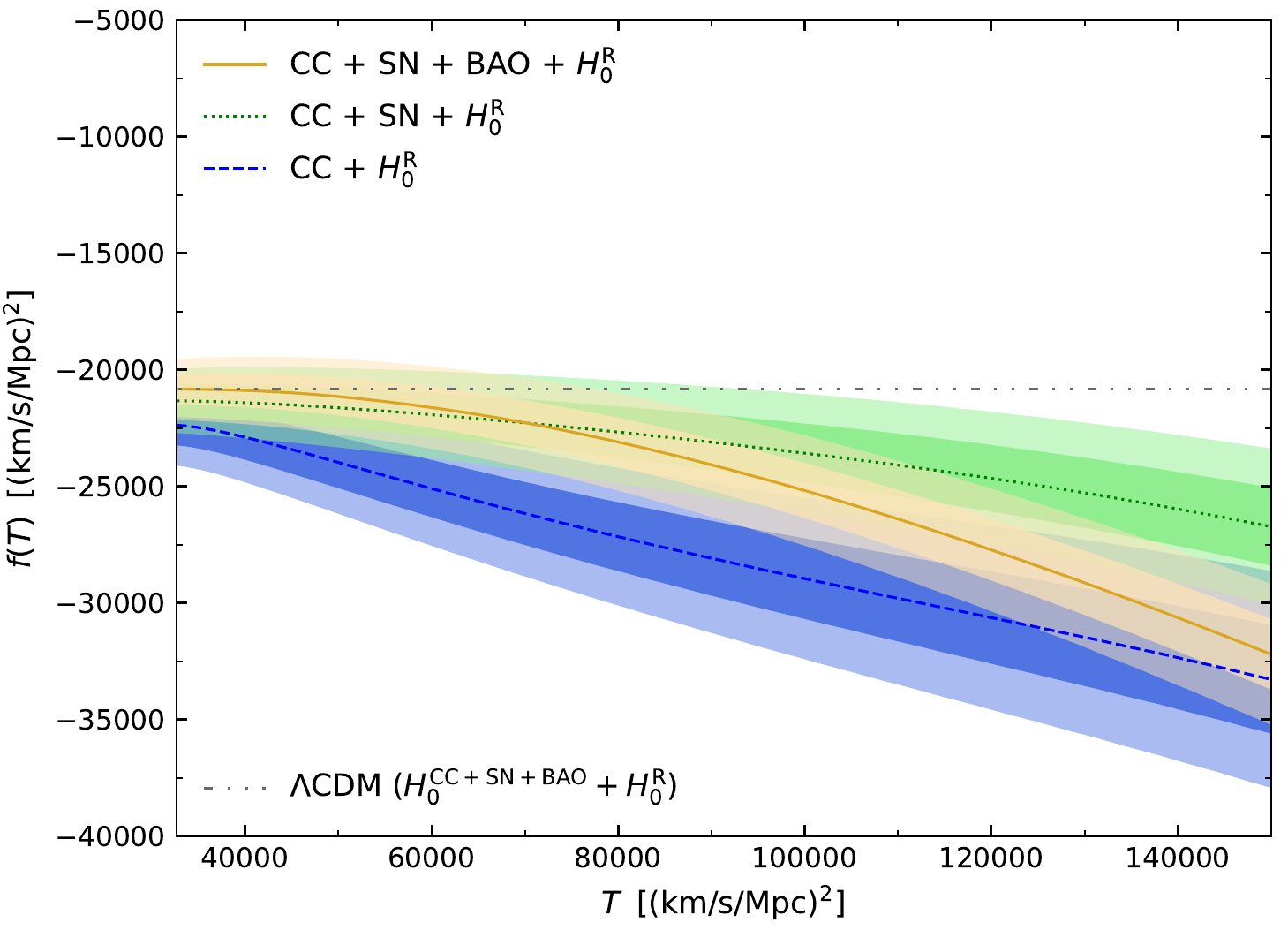}
    \includegraphics[width=0.495\columnwidth]{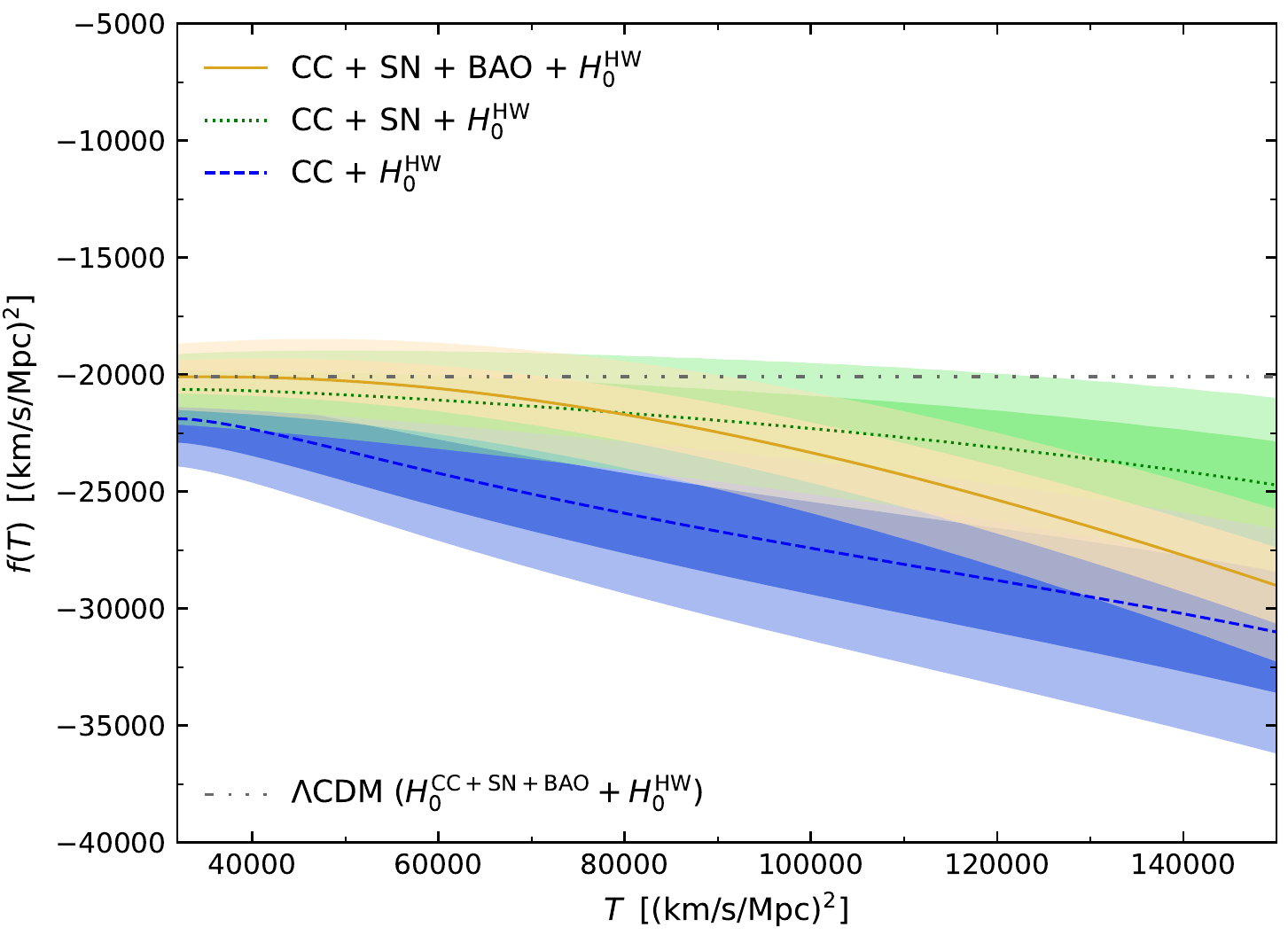}
    \includegraphics[width=0.485\columnwidth]{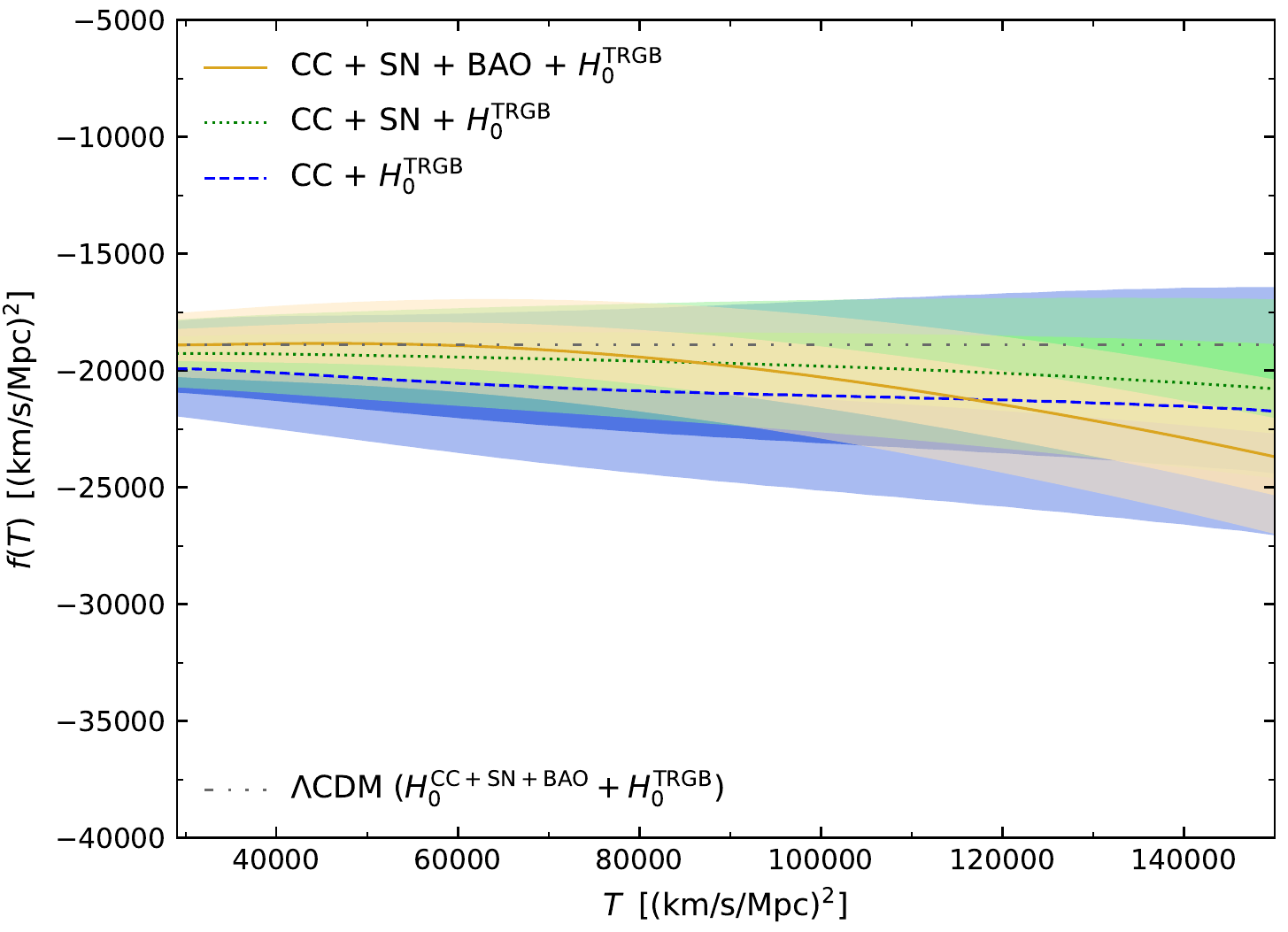}
    \caption
    {GP reconstructions of $f(T)$ with the squared exponential kernel function expressed in the Supplementary annexes (Supplementary 6). The data sets along with the different $H_0^{}$ priors are indicated in each respective panel. Permission for use of this figure was kindly provided by the authors of Ref.~\cite{Briffa:2020qli}.
}
\label{fig:fT_squaredexp}
\end{center}
\end{figure}

Together the propagation equation in Eq.~\eqref{prop_eq_f_T} along with the boundary conditions in (i) and (ii) can express the redshift dependent Lagrangian $f(z)$ in terms of $z$ in a (cosmological) model-independent way. Similarly, the corresponding torsion scalar can be associated with each of the redshift values in question through the Hubble parameter relation in Eq.~\eqref{eq:Tor_sca_flrw}. In this way, the Lagrangian function $f(T)$ can be plotted as a function of the torsion scalar $T$.

The $f(T)$ reconstructions against the torsion scalar are illustrated\footnote{We here reproduce the same figure as reported in Ref.~\cite{LeviSaid:2020mbb}, therefore $f(T)<0$ in this figure.} in Fig.~\ref{fig:fT_squaredexp} where the $1\sigma$ and $2\sigma$ regions are shown in every case. We mention that the $\Lambda$\gls{cdm} paradigm appears as a constant in these plots with a value of $f(T)\rightarrow-6H_0^2(\Omega_{\rm m0}-1)$, which is denoted by horizontal lines in Fig.~\ref{fig:fT_squaredexp}. One could easily observe that the $\Lambda$\gls{cdm} scenario lies inside each and every reconstructed region. Nevertheless, the GP reconstruction procedure shows a slight tendency of $f_T$ to negative values, i.e. to $f(T)$ forms that are slightly decreasing functions of $T$. This is the main result of this section since the aforementioned feature needs to be taken into account in the $f(T)$ model building.

In the case where no prior is used for the GP reconstruction, the $f(T)$ evolution remains within the 2$\sigma$ confidence region of $\Lambda$\gls{cdm} for the breadth of the evolution interval, and mostly within the 1$\sigma$ region for the data set combinations. The furthest propagated line to the $\Lambda$\gls{cdm} is the Hubble parameter reconstructed from the combined data that includes the \gls{bao} data set. On the other hand, once the priors are included the situation changes drastically, with the $H_0^{\rm R}$ prior favouring a slight deviation from $\Lambda$\gls{cdm} over a portion of the cosmic evolution for all data sets. The same situation, but to a lesser extent, occurs for the $H_0^{\rm HW}$ prior, with the $H_0^{\rm TRGB}$ prior being the only one to not affect this propagation in a significant way. These $f(T)$ propagations show that the data sets alone favour an $f(T)$ that only slightly deviates from $\Lambda$\gls{cdm} within the redshift region being probed, while the propagations that do contain literature priors prefer a stronger deviation from $\Lambda$\gls{cdm}, however the latter is inside the allowed regions.

\subsubsection{Machine learning for teleparallel cosmology} \label{subsection:ML_p}

In order to use machine learning in a specific deep learning neural network, we need to classify \gls{tg} cosmological models using the surveys described in Sec.~\ref{sec:cosmology_in_TG}. To perfom these analyses we require the architecture described in the Supplementary annexes (Supplementary 7). Following this, we begin by training several neural networks with different data and different activation functions. Next, we will generate new data with the trained networks and merge this data in a single data frame. Once we have all our new data in a single data frame, we split it in a training set and a test set. By evaluating the model with the test data, we can decide whether the neural network architecture is appropriate for this task or not.

For the first training we use a small data set of \gls{cc}, which has information about the redshift ($z$) and the Hubble parameter ($H(z)$). In this coding we use Tensorflow V2\footnote{\url{https://www.tensorflow.org/?hl=es-419}} and Keras\footnote{\url{https://keras.io}}. The data consists of two columns and 17 rows. Due the small size of this data set, it is standard to not split them in test and train sets. First, we extract the data and normalize the values of $H(z)$. This is done because it is easier for the training to work with small numbers. The architecture of our first network consists in a sequential model with a single layer of just one neuron. This is because we want the output to be a single number for each prediction. We compile this sequential model with the optimizer RMSprop\footnote{\url{https://keras.io/api/optimizers/rmsprop/}} with a standard value of learning rate of $0.003$ for this type of algorithm.

According to the natural language processing on Adam optimiser, and the loss is the mean squared error. The training of this data can be done with a small number of epochs, say 100. To generate non-linear data we must use a different architecture. By including layers with a certain activation function we can train non-linear models. The architecture for the following networks consists of two hidden layers, one with 20 neurons and the other with 12. Each of them have the same activation function, and an output layer with a single neuron. This kind of architecture with several hidden layers is also known as Deep Neural Networks.

\begin{figure}
\centering
\includegraphics[width=0.6\textwidth,origin=c,angle=0]{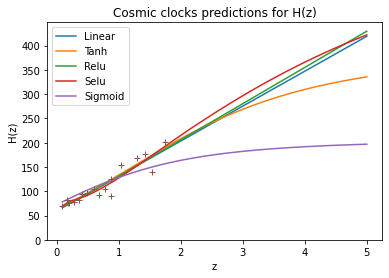}
\caption{Deep learning training for \gls{cc} compilation up to $z=5$. The observations from the sample are denote by `+'. }
\label{fig:DL_training_CC}
\end{figure}

The first non-linear model that we train uses the rectified linear unit (ReLU) as activation function. This function is defined as $ f(x) = \max(0,x)$. Due to the nature of our data we do not expect very different predictions compared to the linear model. For the rest of the neural networks we will use the Adam optimizer\footnote{\url{https://keras.io/api/optimizers/}} with a learning rate of $0.003$ and the mean squared error as loss. We train the network for 300 epochs and again make store predictions for values up to 5 to avoid missing less obvious behaviours such as appear in partial time-series problems. We repeat the previous process for different activation functions. We decided to use the hyperbolic tangent, the Scaled Exponential Linear Unit and the sigmoid function as our activation functions. The predictions made in each case could be different for each training because we are predicting for values of $z$ far from the highest value used for training. The predictions for all the trained neural networks are shown in Fig.~\ref{fig:DL_training_CC}.

\begin{figure}
\centering
\includegraphics[width=0.65\textwidth,origin=c,angle=0]{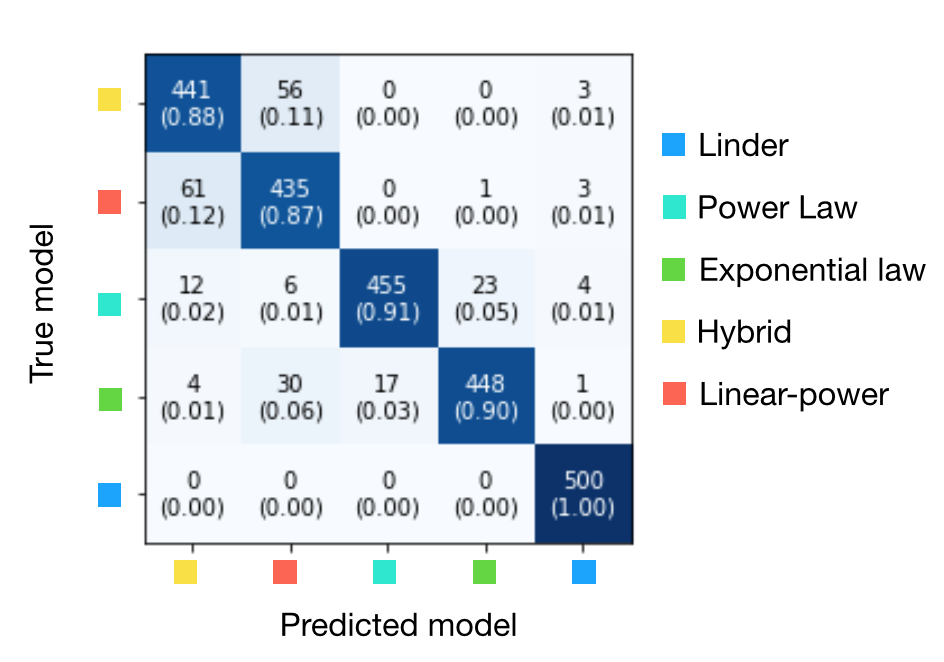}
\caption{Example of classifications of $f(T)$ models described in the compendium Table~\ref{tab:compendium} using \gls{cc} data.}
\label{fig:DL_training_CC2}
\end{figure}

Building on this base, we can introduce our cosmological models and analyze them through the lens of the deep neural network that has since been constructed. To do this, we consider another deep neural network consisting of two hidden layers each with 32 neurons using the ReLU activation function, together with an output layer with the softmax activation function and 5 neurons. This architecture is widely used for nonbinary classification problems such as the current one where we want to measure the degree to which a model is approximated. Between the last hidden layer and the output layer we include a dropout of $0.2$, in order to reduce the over fitting for regularization. This is important since overfitting can become a serious issue otherwise. We use the Adam optimizer with a learning rate of $0.003$ and for the loss the sparse categorical cross entropy. For this example we use the accuracy as metrics. We train our model with a batch size of 80 and a validation split of 0.2 to evaluate if we are over-fitting. The fitting is done for 500 epochs. The result of this process is shown in Fig.~\ref{fig:DL_training_CC2} where the classification output is more clearly shown.

\begin{figure}
\centering
\includegraphics[width=0.45\textwidth,origin=c,angle=0]{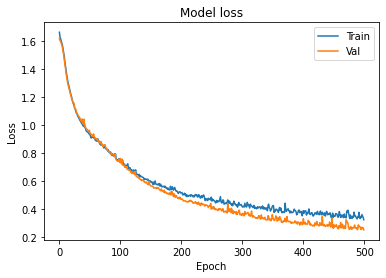}
\includegraphics[width=0.45\textwidth,origin=c,angle=0]{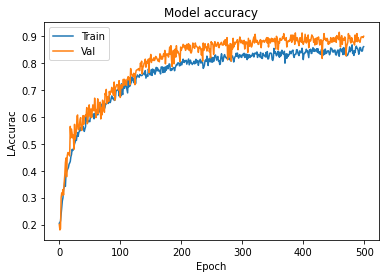}
\caption{Metrics of the trained classifier for \gls{cc} compilation up to $z=5$.}
\label{fig:DL_training_CC2a}
\end{figure}

The lower value of the loss for the validation set and the higher accuracy is a good signal that we are not over fitting the data. To test the classifier we use a confusion matrix. The rows of the matrix are the true labels and the columns are the predicted labels. The closer the confusion matrix is to the identity, the better the classifier is. We make predictions with all the values from the initial data-set and insert the results in the confusion matrix. This matrix that we study contains five different $f(T)$ models. In Fig.~\ref{fig:DL_training_CC2a} we show the percentage between the predicted and true model. We notice that Linder model (see Table~\ref{tab:compendium}) seems to be better identify and preferred by \gls{cc} observations forecasted up to $z=5$, in comparison to the other four models.

\subsection{Perspectives in observational cosmology}
Over the last few years, we have witnessed great advances in precision cosmology in terms of our understanding of the Universe. In addition to known forms of matter, \gls{gr} and an ansatz for the space-time metric as \gls{flrw} spacetime, two contributions remain a mystery and need to be added accounted for in observations, namely: the \gls{cdm} to explain the formation of cosmic structures, and a cosmological constant to account for the late-time cosmic acceleration. The drastically increasing precision and wealth of observations and the surprising findings may call for a revision of the cosmological model hypothesis. On this matter, addressing the dynamics of the cosmic acceleration can be achieved by postulating modifications of \gls{gr} acting only on the largest cosmological scales. Comparing these modified models against surveys does not only provide potential hints to test new physics: it is itself a way to validate the assumptions on which the standard paradigm is formulated. Model dependent supposition are usually made in the analysis of raw data, and certain curtailment which might be compatible with the usual supposition might not be fulfilled in modified scenarios. The study of the different modified theories helps to identify these flaws and make the method of comparison between theory and observations as model independent as much as possible. Studying the predictions of different cosmological models and constraining them with observations provides a way to validate the foundations of the $\Lambda$\gls{cdm} model and quantify deviations from the constituent suppositions. The analysis of self-consistent models allows the examination of a given physical set-up within computational restrictions. This supposition will become important as the cosmological and astrophysical wealth and quality of data increase.

In the near future, there will be next-generation experiments, space-missions and facilities on Earth, that will be of key importance to confirm or to falsify models beyond the standard one proposed at present and in the past years \cite{Abdalla:2022yfr}. Between these models, the \gls{tg} theories are starting to gain ground because they can easily address the $H_0$ and $S_8$ tensions and simultaneously explain the acceleration of the Universe due to the \gls{de}. Unfortunately, at the moment, these theories have only been used for cosmology at the background level and very poorly explored with the current available data and, for example, a complete full analysis, taking into account the effect of the perturbations, is completely missing. The topic of the \gls{tg} models is therefore a ``work in progress" field, that is quickly growing and trying to fill the gap with the other beyond-the-standard-model scenarios extensively studied and compared to the data, we have in the literature. Many works appeared in the last few years going in this direction (see Table~\ref{Table-TG-criteria}), and we expect many more coming out in the near future. For this reason it is important to list here which are the experiments expected in the next few years that could have a big impact in the \gls{de}/DM field, improving significantly the sensitivity we have on the current cosmological parameters, and helping us shaping plans for the kind of work we need to do in the future to make this \gls{tg} field more robust and supported by the data.

For example, a 50\% improvement on the current Hubble constant measurements is expected by the local distance ladder observations, that will reach a precision of 1\%~\cite{Riess:2020xrj}. A better $H_0$ measurement is predicted also for the gravitational time delays experiments, expected to achieve a $\sim 1.5\%$ precision~\cite{Birrer:2020jyr}, and for the standard sirens (\gls{gw}SS)~\cite{Schutz:1986gp,Holz:2005df,Chen:2017rfc,DiValentino:2018jbh,Palmese:2019ehe}, the \gls{gw} analogue of astronomical standard candles, that will reach a $2\%$ precision.

New improved \gls{cmb} data will be possibly released by the ground-based experiments like \gls{cmb}-S4~\cite{Abazajian:2019eic,Abazajian:2019tiv} and Simons Observatory~\cite{Koopman:2020gkh}, and satellite experiments like \gls{litebird}~\cite{Suzuki2018}, \gls{core}~\cite{DiValentino:2016foa}, \gls{pixie}~\cite{Kogut2011}, \gls{prism}~\cite{Delabrouille:2019thj}, and \gls{pico}~\cite{Hanany:2019lle}. Of great importance will be the Euclid space-based survey mission~\cite{Laureijs:2011gra}, that will be able to release an important variety of cosmological probes, starting from the gravitational lensing, to the \gls{bao}, to the galaxy clustering, and the multi-purpose radio-interferometer \gls{ska},~\footnote{~\url{https://www.skatelescope.org}} making measurements of the 21 cm line in neutral hydrogen (HI).

Moreover, starting in 2022 it is planned the galaxy survey Rubin Observatory Legacy Survey of Space and Time (LSST~\cite{Ivezic:2008fe}), and in 2025 the powerful heir of the \gls{hst}, i.e. the Roman Space Telescope (formerly known as WFIRST~\cite{akeson2019wide}). Future \gls{bao} data are expected also from galaxy spectroscopic surveys such as \gls{desi}~\cite{Levi:2013gra}, or radio telescopes such as \gls{bingo}~\cite{Wuensche:2019znm}.

All these experiments combined together are expected to shed light on the current cosmological tensions, giving us the possibility of ruling out large portions of the parameters space, and therefore, of the scenarios beyond-the-standard-model, that now compete for fitting the data.


\clearpage

\section{Conclusion} \label{sec:conclusion}

In this Review of teleparallel gravity, we have discussed the foundations of this approach to constructing gravitational theories, together with the observational and precision cosmology consequences of some of the most popular forms of teleparallel theories. We started with a survey of the historical context in which \gls{tg} was born (into from the early work by Einstein, Weitzenb\"{o}ck and others). In Sec.~\ref{sec:introduction}, we also point out new results that are first being reported here together with corrections to the literature where incorrect results may have appeared. This section also contains our \textit{review convention} which would be advantageous to use in future works since we have made a significant effort to represent all theories and the vast majority of their results in this convention.

We begin the technical details of the formulation of teleparallel geometry through the position of metric-affine geometry in Sec.~\ref{sec2:affine} which was formulated in the later years after Einstein's original work. In this part of the Review, we see how \gls{tg} forms part of a much larger trinity of gravity that involves curvature but also non-metricity. Together, these three geometries, through their respective connections, can fully utilize the components of a general gravitational connection. We then focus our attention solely on the teleparallel geometry manifestation of this trinity in Sec.~\ref{sec3:torsional} where we delve into the details of how \gls{tg} can be sourced through a \textit{gauge theory of translations}. We also go into how the \textit{minimal coupling prescription} is needed to couple matter to this geometry, as well as the symmetries and transformations within the formulation. We close this section with a new discussion on the relationship of teleparallel geometries to other branches of physics such as continuum mechanics where we lay out the way in which this has been used to describe crystal structures.

After laying out the geometric foundations to \gls{tg} and its relationship to the broader class of metric-affine theories, we apply these concepts to the \textit{teleparallel equivalent of general relativity} in Sec.~\ref{sec4:TEGRNGR} also, similar to \gls{gr}, holds a special place in teleparallel theories as the best behaved gravitational model in \gls{tg}. Indeed, we also discuss some aspects related to attempts to quantize \gls{tegr} and their relationship to similar approaches in \gls{gr}. On the other hand, there have been a plethora of proposals to modify \gls{tegr} which we examine in Sec.~\ref{sec5:extended} where we present the best studied and most promising teleparallel theories of gravity. We here tackle a number of issues related to fully describing the ideas behind \textit{good tetrad-spin connection pair} which is crucial for determining the correct field equations of a theory. We also point out progress in some of these modifications and extensions to \gls{tegr} in terms of their spin connection contribution. In particular, we flesh out the impact of the antisymmetric field equations which are related to the local Lorentz transformation degrees of freedom within these theories. Finally, we expand on the scalar-tensor sector of the landscape of teleparallel theories and focus our attention on the new and interesting BDLS theory (teleparallel analogue of Horndeski gravity) which have been gaining popularity as a way to revive standard Horndeski gravity, which has become severely constrained due to its prediction on the speed of propagation of gravitational waves (in many of its models).

Taking these theories, we explore possible cosmologies in their backgrounds through various approaches in Sec.~\ref{sec:cosmology_in_TG}. By and large, we explore flat homogeneous and isotropic backgrounds but are not limited to this. Moreover, we also focus on $f(T)$ cosmology since this is the most studied in the literature, but again we also analyze other scenarios. It is worth noting that the majority of works in the literature in \gls{tg} have been associated with works in this regime, namely background cosmology. In this context, we first explore possible reconstruction methods, which have shown promise in many works in the literature (even outside of \gls{tg}). This is a convenient way of designing models in the various classes of gravity theories which can then be probed against other phenomena. We also explore the wide range of works that make use of Noether symmetries as an approach to again design particular models of gravity that adhere to particular symmetry setups. At a higher level, we present dynamical systems and its approach in determining important generic behaviors of particular classes of cosmology models and whether they are consistent with our Universe. In this section, we apply these approaches to some of the theories of gravity being surveyed in Sec.~\ref{sec5:extended}. The level to which we probe these regimes of theories depends on the amount of work in the literature on these branches of theories with \gls{tg}.

\gls{tg} poses new challenges in the perturbative sector as well since the tetrad and spin connection approach means that we cannot utilize the traditional mechanics of perturbation theory in the metrical formulation of gravity theories. In Sec.~\ref{sec:cosmo-pert} we discuss these crucial points by opening on the traditional 3+1 decomposition of the metric and how this can be translated to teleparallel geometry. Together with a discussion about the \gls{svt} decomposition, we then explore how this can be applied within this framework. We also discuss the idea of gauge transformations which are important for relating results between gauge setups. While we use this pathway to cosmological perturbation theory in a number of theoretical setups, the analysis is much more general and lays out the general approach one should take to this kind of study. Indeed, in Sec.~\ref{sec9:GW}, we use this approach to probe the polarization structure of \gls{gw} in the various scenarios that it has been studied in the literature. Here, we find a rich structure of polarizations which manifest some of the propagating \gls{dof} of each theory, and their subclasses.

In the final part of our Review on \gls{tg}, we explore the observational sector of the theory. Firstly, in Sec.~\ref{sec6:astrophysics} we analyze works related to astrophysics and how the theories from Sec.~\ref{sec5:extended} can impact those physical process and observational measurements. Indeed, to probe this regime of physics we first need to understand very well how the tetrad spin connection pair can be formulated in a good way to best describe the physical systems that we know from metrical approaches to gravity. We open with long discussion on these technical details together with an explanation of the problem of how a generalized Birkhoff theorem may be conceptualized in \gls{tg}. On the observational side of things, we work out both the classic Solar System tests and their extension to the so-called gravito-magnetic effect, as well as some details of the \gls{ppn} analysis as applied to \gls{tg}. We then move to the idea of compact objects in \gls{tg} which is poorly understood. By this, we mean that the literature contains some analyses on the topic but which could be better expanded to explore other regime more related to astrophysical measurements. Moreover, due to the good tetrad-spin connection pair paradigm, some early works may need correcting, as we explain in this section. In addition, we discuss the open problems related to this topic such as the issue about interpreting black holes singularities and the issue of wormholes. The section also contains a discussion of the impact of some gravitational models in galactic rotation curve analyses which are important to understanding well the issue of astrophysical dark matter.

Finally, we explore the all important topic of observational cosmology together with the results and general investigation of precision cosmology in Sec.~\ref{sec10:Observation}. We do this by discussing firstly the state-of-the-art of cosmology and the growing issues being posed by cosmological tensions. We also discuss the state-of-the-art in terms of cosmography. Interestingly, we have one of the few surveys that details the way in which community codes need to be used to correctly explore modified gravity theories (rather than interpreting modified gravity as an effective fluid in a dynamical \gls{de} fashion). In this background, we then collect the most promising models being introduced in the theories in Sec.~\ref{sec5:extended} and then promoted in the background and perturbative sections that ensue in a compendium survey for convenience. We explore the literature for these models both in terms of observational constraints and also their restrictions coming from precision cosmology in the community codes we discussed earlier. This section also entails a significant discussion on possible applications of machine learning in modified gravity in general, and \gls{tg} in particular, such as in the use of Gaussian processes to find confidence regions for cosmological proposals for $f(T)$ gravity models.

In this work, we have expanded on proposed solutions to the problems related to finding good tetrad-spin connection pairs that allow us to probe modified teleparallel gravity. We have also corrected a number of small mistakes in the literature, and discussed places where the literature has been incorrect in more significant ways. We explain a lot of this in Sec.~\ref{sec:corrections}, and through the remainder of the Review. Below, we also point out open problems and issues in the teleparallel community that need further attention for future work.

\begin{itemize}
    \item \textbf{Strong coupling in $f(T)$ Gravity} -- Strong coupling is a pathology of the disparity between \gls{dof} that appear in a full Hamiltonian analysis and the \gls{dof} that appear at linear perturbative level for particular spacetimes. These pathologies may cause perturbations that are strongly coupled to potentially arbitrarily higher order perturbations which is problematic for making finite calculations. We describe the issue further in Sec.~\ref{ssec:Action_and_field_equations} as well as in other places throughout the $f(T)$ gravity analysis. It is not clear without a full Hamiltonian analysis of other theories whether this issue pervades other formulations of \gls{tg} for these specific spacetimes. However, we know that certain more nuanced setups which can, with certainty, escape this issue such as $f(T,B) = f(-T+B) = f(\lc{R})$ gravity. It is important to highlight that strong coupling is not a problem of a theory or the model that populate it but of the solutions that can be probed using traditional perturbation theory. It may be the case, that the issue of strong coupling points to the need for a new approach to perturbation theory. \gls{tg} is a real alternative to the curvature-based geometry of \gls{gr}, and thus it is not totally surprising that certain approaches breakdown such as the way that the linearized equations are derived. Thus, new approaches to taking perturbations in $f(T)$ should be investigated for potential curing of the strong coupling problem, particularly in the case of cosmological perturbations.
    \item \textbf{Good tetrad-spin connection pairs} -- In this Review, we have made an active effort to fully immerse readers into the covariant formulation of \gls{tg}. To this end, we introduce the spin connection and its integral importance in Secs.~\ref{sec2:affine}, \ref{sec3:torsional}, and \ref{subsec:goodtetrad}. Here, we also describe how the antisymmetric field equations are related to the spin connection components. Thus, in \gls{tg} we find additional six equations from these antisymmetric equations of motion which are related to the Local Lorentz degrees of freedom. We give examples of this in Sec.~\ref{subsec:goodtetradsph} but more work needs to be done in order to better understand other examples in the plethora of \gls{tg} theories, as well as to ground a more routine procedure to finding these good tetrad-spin connection components.
    \item \textbf{Quantum gravity} -- The quantization of \gls{gr} and other gravitational theories has been a long-standing programme in gravitational research. As is well known in the community, there are many attempts at constructing a quantum completion of \gls{gr}, which had had various successes. In Sec.~\ref{ssec:quantumhighdim}, we detail some possible insights that \gls{tegr} may have in resolving the problems of quantizing \gls{gr}. However, many details remain to be worked out and a fully fledged quantum completion of \gls{tegr} remains unknown in the literature. Moreover, attempts at using the mechanisms of modified \gls{tg} to construct teleparallel quantum theories of gravity remain poorly studied and would be an interesting problem to have tackled. In this vein, loop quantum gravity also uses very similar analytical tools as \gls{tg} to approach the quantum gravity problem. It would be interesting to use similar routes in \gls{tg} and to explore the differences that may emerge between the two approaches.
    \item \textbf{Black holes} -- Black holes are a fundamental part of astrophysics research from individual black hole systems to understanding the dynamics of galaxies. In Sec.~\ref{sec:blackholes}, we discuss the foundational concepts behind black hole physics in \gls{tg} from their definition to their impact on the photon sphere which has become a detectable phenomenon in recent years. However, a fully fleshed out description continues to remain an open problem. It is crucial that a fully fledged route is opened to black hole physics for the theory to continue to progress beyond toy model solutions and into more complicated and physical scenarios.
    \item \textbf{Compact stars} -- Another important aspect to astrophysics is that of compact stars which are ubiquitous throughout astronomy and have a huge impact on observational astronomy such as in the measurements of pulsars. Better exploring the inner mechanics of these stars in \gls{tg}, both in terms of foundations and in the context of simulations, is crucial to having a competitive description of these all important stars in \gls{tg}.
    \item \textbf{Galactic simulations} -- \gls{gr} necessitates \gls{cdm} to accurately describe the dynamics of galaxies and clusters of galaxies. In Sec.~\ref{sec:galaxy}, we detail how the galactic dynamics may be explained using a particular form of $f(T)$ gravity. It is not clear if this can also be achieved in other $f(T)$ models or other \gls{tg} theories. It would be interesting to pursue this problem further and to also expand the analysis to include simulations for clusters of galaxies.
    \item \textbf{Teleparallel cosmology codes} -- Many cosmological surveys interpret extended theories of gravity as an effective fluid in the cosmological context. This is accurate at background level and can help in many implementations of modified gravity in community codes related to cosmology. However, the perturbations that one determines from the gravitational section, for these terms, and that determined by the code through the fluid interpretation do not equate to each other. While interesting a dynamical dark energy approach, a full implementation of perturbations from \gls{tg} in these codes in terms of modified Boltzmann equations and perturbation modules would be extremely advantageous to moving the framework forward in the observational sector, and particularly for the precision cosmology approach. We explain this prospect further in Sec.~\ref{sec10:Observation}.
\end{itemize}

In this Review, we have attempted to give a snapshot of the current state-of-the-art of \gls{tg} together with possible future directions which are necessary for the topic to progress forward both in terms of its foundations as well as in terms of observational predictions. \gls{tg} is a very novel approach to modified gravity which naturally produces problems that require novel solutions. It will be interesting to see how the community advances over the next decade of research.
\clearpage

\subsection*{Acknowledgements}
SB and MH were supported by the Estonian Research Council grants PRG356 ``Gauge Gravity"  and by the European Regional Development Fund through the Center of Excellence TK133 ``The Dark Side of the Universe". SB is also supported by JSPS Postdoctoral Fellowships for Research in Japan and KAKENHI Grant-in-Aid for Scientific Research No. JP21F21789. KFD acknowledges support by the Hellenic Foundation for Research and Innovation (H.F.R.I.) under the “First Call for H.F.R.I. Research Projects to support Faculty members and Researchers and the procurement of high-cost research equipment grant” (Project Number: 2251). The work was also supported by Nazarbayev University Faculty Development Competitive Research Grant No. 11022021FD2926.
CE-R acknowledges the Royal Astronomical Society as FRAS 10147 and DGAPA-PAPIIT-UNAM Project IA100220.
JLS and JM would also like to acknowledge funding support from Cosmology@MALTA which is supported by the University of Malta. JLS would also like to acknowledge funding from ``The Malta Council for Science and Technology'' in project IPAS-2020-007.
EDV acknowledges the support of the Addison-Wheeler Fellowship awarded by the Institute of Advanced Study at Durham University.
This research was partially carried out using computational facilities procured through the European Regional Development Fund, Project No. ERDF-080 ``A supercomputing laboratory for the University of Malta'' and the ``CosmoNag project'' through Tochtli ICN-UNAM cluster.
The authors would like to acknowledge Laur J\"{a}rv for kindly preparing Fig.~\ref{fig:RTQ}. The authors would like to thank Jos\'{e} Beltran Jimenez for very useful discussions regarding the strong coupling problem in $f(T)$ gravity. All the authors would like to acknowledge the thorough and significant work performed by the anonymous referees during the preparation of this review. This article is based upon work from COST Action CA21136 Addressing observational tensions in cosmology with systematics and fundamental physics (CosmoVerse) supported by COST (European Cooperation in Science and Technology).
\clearpage

\appendixtitleon
\appendixtitletocon
\begin{appendices}
\section{Variations in Teleparallel Gravity}
\label{App:variations}

In this section we will briefly collect the variations of some basic and some more complicated quantities of teleparallel theory, in order for the reader to familiarize themselves with the notation. Apart from that we will calculate the gauge current and finally we will discuss the variations of specific theories, i.e. the $f(T,B,\phi,X)$, $f(T,B,T_G,B_G)$ and the teleparallel analogue of Horndeski gravity (BDLS).

\subsection{Basic teleparallel quantities}
Let us start by showing some necessary and useful identities
\begin{subequations}
\begin{align}
    \frac{\partial e^B{}_{\nu}}{\partial e^A{}_{\mu}} &= \delta^B_A\delta^\mu_{\nu}\,,\label{Eq:1st_der_tetrad}\\[0.5ex]
    \frac{\partial E_B{}^{\nu}}{\partial e^A{}_{\mu}} &= -E_B{}^{\mu}E_A{}^{\nu}\,,\label{Eq:1st_der_inv_tetrad}\\[0.5ex]
    \frac{\partial e }{\partial e^A{}_{\mu}} &= eE_A{}^{\mu}\,,\label{Eq:1st_der_det}\\[0.5ex]
    \frac{\partial g^{\alpha\beta}}{\partial e^A{}_{\mu}} &= -g^{\mu\beta}E_A{}^{\alpha} - g^{\mu\alpha}E_A{}^{\beta}\label{Eq:1st_der_inv_metric}\,.
\end{align}
\end{subequations}
Using the above identities, one can easily obtain the variations of basic quantities with respect to the tetrad, $\delta_e$, such the inverse tetrad, $E_A{}^{\mu}$, the determinant of the tetrad, $e$, a general metric and its inverse, $(g_{\mu\nu},g^{\mu\nu})$, as well as the torsion tensor, $T^\alpha{}_{\mu\nu}$ and the torsion vector, $T^{\mu}$. All of the above vary as follows
\begin{subequations}
\begin{alignat}{2}
    \delta_e E_A{}^{\mu} & =\: & & -E_B{}^{\mu} E_A{}^{\nu} \delta e^B{}_{\nu}\,,\\[0.5ex]
    \delta_e e& =\: & & \delta \, \textrm{det}(e^A{}_\mu)= e\, E_A{}^{\mu} \delta e^A{}_{\mu}\,,\\[0.5ex] \label{var_determinant}
    \delta_e g_{\mu\nu} & =\: & & \eta_{AB}\left( e^A{}_{\mu}\delta e^B{}_{\nu}+ e^A{}_{\nu} \delta e^B{}_{\mu}\right)\,,\\[0.5ex]
    \delta_e g^{\mu\nu} & =\: & & -\left(g^{\mu\alpha}E_A{}^{\nu}+g^{\nu\alpha}E_A{}^{\mu}\right)\delta e^A{}_{\alpha}\,,\\[0.5ex]
    \delta_e T^{\alpha}_{\hphantom{\alpha}\mu\nu} & =\: & & -E_A{}^{\alpha}T^{\beta}{}_{\mu\nu}\delta e^A{}_{\beta} + 2E_A{}^\alpha\delta_e \Gamma ^A{}_{[\nu\mu]} \\[0.5ex]
    & =\: & & -E_A{}^{\alpha}T^{\beta}{}_{\mu\nu}\delta e^A{}_{\beta}+E_A{}^{\alpha} \Big[\partial_\mu\delta e^A{}_{\nu}-\partial_{\nu}\delta e^A{}_{\mu}+\omega^A{}_{B\mu}\delta e^{B}{}_\nu-\omega^A{}_{B\nu}\delta e^{B}{}_\mu\Big]\,, \\[0.5ex]
    \delta _e T^{\mu} & =\: & & -\left( E_A{}^{\mu}T^{\lambda}+g^{\mu\lambda}T_A +T^{\lambda}{}_A{}^{\mu}\right)\delta e^A{}_\lambda \nonumber \\[0.5ex]
    & \: & & +g^{\mu\nu} E_A{}^{\lambda} \left( \partial _{\lambda} \delta e^A{}_{\nu} - \partial _{\nu} \delta e^A{}_{\lambda} +\omega^A{}_{B\lambda}\delta e^{B}{}_\nu-\omega^A{}_{B\nu}\delta e^{B}{}_\lambda\right)\,.
\end{alignat}
\end{subequations}
In what follows, we consider these variations familiar, in order to move to more complicated expressions. It is crucial to see how the torsion scalar varies with respect to the tetrad. Thus, using the above quantities, we have
\begin{equation}\label{var_tor_scalar}
    \delta_e T = \frac{1}{4}\delta (T^{\mu\nu\alpha}T_{\mu\nu\alpha})+\frac{1}{2}\delta (T^{\mu\nu\alpha}T_{\nu\mu\alpha}) - \delta (T^{\mu}T_{\mu})\,,
\end{equation}
where
\begin{subequations}
\begin{align}\label{var_tor_scalar1}
    \delta(T^{\mu\nu\alpha}T_{{\mu\nu\alpha}}) & = 4T_\mu{}^{\nu\alpha}E_A{}^\mu(\partial _\nu \delta e^A{}_{\alpha}+\omega ^A{}_{B\nu}\delta e^B{}_\alpha)-4T^{\mu\nu\alpha}T_{\mu\nu\beta}E_A{}^\beta \delta e^A{}_\alpha \,, \\[0.5ex] \label{var_tor_scalar2}
    \delta(T^{\mu\nu\alpha}T_{\nu\mu\alpha}) &= 2(T^{\beta\nu\mu} -T^{\mu\nu\beta})T_{\nu\mu\alpha}E_A{}^{\alpha} \delta e^A{}_{\beta} +2 (T^\mu{}_\nu{}^{\beta} - T^\beta{}_{\nu}{}^\mu)E_A{}^\nu (\partial _{\mu} \delta e^A{}_\beta + \omega^A{}_{B\mu}\delta e^B{}_{\beta}) \,, \\[0.5ex] \label{var_tor_scalar3}
    \delta(T^{\mu}T_{\mu}) &= -2(-T^\beta T^\alpha{}_{\beta\mu}+T^\alpha T_{\mu})E_A{}^\mu \delta e^A{}_\alpha -2 (T^\mu E_A{}^\beta - T^\beta E_A{}^\mu)(\partial_{\mu}\delta e^A{}_\beta + \omega^A{}_{B\mu}\delta e^B{}_\beta )\,.
\end{align}
\end{subequations}
By replacing Eqs.~\eqref{var_tor_scalar1}, \eqref{var_tor_scalar2}, \eqref{var_tor_scalar3} in Eq.~\eqref{var_tor_scalar}, integrating by parts and neglecting the boundary terms, we get
\begin{equation}\label{var-tor-scalar}
    e \,\delta _e T = 2e\left(\frac{1}{e}\partial_{\mu}(eS_A{}^{\lambda\mu}) - T^{\sigma}{}_{\mu A}S_{\sigma}{}^{\mu\lambda} +\, \omega ^B{}_{A \nu} S_B{}^{\nu\lambda} \right) \delta e^A{}_{\lambda}\,.
\end{equation}

Similarly, the variation of the torsion tensor with respect to the spin connection $\omega^A{}_{B\mu}$ reads from Eq.~\eqref{torsion_tensor}
\begin{equation}
    \delta _{\omega} T^\alpha{}_{\mu\nu} = \delta _{\omega} (E_A{}^\alpha T^A{}_{\mu\nu}) = E_A{}^\alpha \delta_{\omega}T^A{}_{\mu\nu} = E_A{}^\alpha e^B{}_\nu\delta \omega ^A {}_{B\mu} - E_A{}^\alpha e^B{}_\mu\delta \omega ^A {}_{B\nu}\,,
\end{equation}
and thus, the torsion vector becomes
\begin{equation}
    \delta _{\omega} T_{\mu} = \delta _\omega T^\alpha{}_{\alpha\mu} = E_A{}^\alpha e^B{}_\nu\delta \omega ^A {}_{B\alpha} - \delta \omega ^B {}_{B\nu}\,.
\end{equation}
It is worth noticing that the spin connection has to be purely inertial, since if one varies the teleparallel action with respect to arbitrary spin connection coefficients, the associated equations of motion will lead to the vanishing of the torsion tensor \cite{Golovnev:2017dox}. However, if one assumes a spin connection in the inertial class only, $\omega ^A{}_{B\mu} = \Lambda^A{}_C \partial _{\mu} (\Lambda^{-1}) ^C{}_B$, effectively it means that there exists a frame in which $\omega = 0 $ (Weitzenb\"{o}ck gauge), but simultaneously one is allowed to perform a local Lorentz transformation $\Lambda ^A{}_B$ that belongs to the Lorentz group and which produces the above connection.

\subsection{Calculation of the gauge current}
\label{App:Gauge_Curr_def}
By definition, the gauge current can be written as
\begin{equation}
    \udt{J}{A}{\mu} = -\frac{1}{e} \frac{\partial\left(eT\right)}{\partial \udt{e}{A}{\mu}}= -T\dut{E}{A}{\mu} - \frac{\partial T}{\partial \udt{e}{A}{\mu}}\,,
\end{equation}
where Eq.~\eqref{Eq:1st_der_det} was used for the tetrad determinant derivative. Now, using Eq.~\eqref{Torsion_scalar}, this can be expanded to
\begin{alignat}{2}
    \udt{J}{A}{\mu} & =\: & & -T\dut{E}{A}{\mu} - \frac{1}{4} \frac{\partial \udt{T}{B}{\sigma\bar{\mu}}}{\partial \udt{e}{A}{\mu}} \dut{T}{B}{\sigma\bar{\mu}} - \frac{1}{4} \frac{\partial \dut{T}{B}{\sigma\bar{\mu}}}{\partial \udt{e}{A}{\mu}} \udt{T}{B}{\sigma\bar{\mu}} -\frac{1}{2} \frac{\partial \udt{T}{B}{\sigma\rho}}{\partial \udt{e}{A}{\mu}} \udt{T}{\rho\sigma}{B}\nonumber\\[0.5ex]
    & \: & & - \frac{1}{2} \udt{T}{B}{\sigma\rho} \frac{\partial \udt{T}{\rho\sigma}{B}}{\partial \udt{e}{A}{\mu}} + \frac{\partial \udt{T}{\rho\sigma}{\rho}}{\partial \udt{e}{A}{\mu}} \dut{T}{\mu\sigma}{\mu} + \udt{T}{\rho\sigma}{\rho} \frac{\partial \dut{T}{\bar{\mu}\sigma}{\bar{\mu}}}{\partial \udt{e}{A}{\mu}}\,,\label{Eq:App_gauge_current_def}
\end{alignat}
where the torsion tensor expression in Eq.~\eqref{eq:torsion_tensor} means that the derivatives in the second and fourth terms can be determined by
\begin{equation}
    \frac{\partial \udt{T}{B}{\sigma\bar{\mu}}}{\partial \udt{e}{A}{\mu}} = \udt{\omega}{B}{A\sigma} \delta^{\mu}_{\bar{\mu}} - \udt{\omega}{B}{A\bar{\mu}} \delta^{\mu}_{\sigma}\,,
\end{equation}
while the derivatives in the third, fifth, sixth and seventh term turn out to respectively be represented by
\begin{subequations}
\begin{alignat}{2}
    \frac{\partial \dut{T}{B}{\sigma\bar{\mu}}}{\partial \udt{e}{A}{\mu}}& =\: & &\frac{\partial}{\partial \udt{e}{A}{\mu}} \left(\eta_{B\bar{B}} \udt{T}{\bar{B}}{\bar{\sigma}{\lambda}} g^{\sigma\bar{\sigma}}g^{\bar{\mu}\lambda}\right)\nonumber\\[0.5ex]
    & =\: & &-\dut{T}{B}{\mu\bar{\mu}}\dut{E}{A}{\sigma} - \dut{T}{BA}{\bar{\mu}} g^{\mu\sigma} - \dut{T}{B}{\sigma\mu}\dut{E}{A}{\bar{\mu}} - \dudt{T}{B}{\sigma}{A} g^{\mu\bar{\mu}} + \eta_{B\bar{B}} \udt{\omega}{\bar{B}}{A\bar{\sigma}} g^{\sigma\bar{\sigma}} g^{\mu\bar{\mu}}\nonumber\\[0.5ex]
    & \: & &- \eta_{B\bar{B}} \udt{\omega}{\bar{B}}{A\lambda} g^{\sigma\mu} g^{\mu\lambda}\,,\\[0.5ex]
    \frac{\partial \udt{T}{\rho\sigma}{B}}{\partial \udt{e}{A}{\mu}} & =\: & & \frac{\partial}{\partial \udt{e}{A}{\mu}}\left(\dut{E}{B}{\bar{\mu}} g^{\sigma\bar{\sigma}} \dut{E}{\bar{B}}{\rho} \udt{T}{\bar{B}}{\bar{\sigma}\bar{\mu}}\right)\nonumber\\[0.5ex]
    & =\: & & - \udt{T}{\rho\sigma}{A} \dut{E}{B}{\mu} - \udt{T}{\mu\sigma}{B}\dut{E}{A}{\rho} - \udt{T}{\rho\mu}{B}\dut{E}{A}{\sigma} - \udt{T}{\rho}{AB}g^{\mu\sigma}\nonumber\\[0.5ex]
    &\: & & + \dut{E}{\bar{B}}{\rho}\left(\dut{E}{B}{\mu} g^{\sigma\bar{\sigma}}\udt{\omega}{\bar{B}}{A\bar{\sigma}} - \dut{E}{B}{\bar{\mu}} g^{\sigma\mu} \udt{\omega}{\bar{B}}{A\bar{\mu}}\right)\,,\\[0.5ex]
    \frac{\partial \udt{T}{\rho\sigma}{\rho}}{\partial \udt{e}{A}{\mu}} & =\: & & \frac{\partial}{\partial \udt{e}{A}{\mu}} \left(\dut{E}{B}{\rho} g^{\sigma\bar{\sigma}} \udt{T}{B}{\bar{\sigma}\rho}\right)\nonumber\\[0.5ex]
    & =\: & & - \udt{T}{\mu\sigma}{A} - \udt{T}{\rho\mu}{\rho} \dut{E}{A}{\sigma} - \udt{T}{\rho}{A\rho} g^{\mu\sigma} + \dut{E}{B}{\mu} g^{\sigma\bar{\sigma}} \udt{\omega}{B}{A\bar{\sigma}} - \dut{E}{B}{\rho} g^{\sigma\mu} \udt{\omega}{B}{A\rho}\,,\\[0.5ex]
    \frac{\partial \dut{T}{\bar{\mu}\sigma}{\bar{\mu}}}{\partial \udt{e}{A}{\mu}} & =\: & & \frac{\partial}{\partial \udt{e}{A}{\mu}} \left( g_{\sigma\bar{\sigma}} \udt{T}{\bar{\mu}\bar{\sigma}}{\bar{\mu}}\right)\nonumber\\[0.5ex]
    & =\: & & -\udt{T}{\mu}{\sigma A} + \dut{E}{B}{\mu} \udt{\omega}{B}{A\sigma} - \delta^{\mu}_{\sigma} \dut{E}{B}{\rho} \udt{\omega}{B}{A\rho}\,,
\end{alignat}
\end{subequations}
where the derivatives in Eqs.~\eqref{Eq:1st_der_tetrad}, \eqref{Eq:1st_der_inv_tetrad} and \eqref{Eq:1st_der_inv_metric}) were used throughout. By combining these derivative contributions in Eq.~\eqref{Eq:App_gauge_current_def}
\begin{equation}
    \dut{J}{A}{\mu} = -T\dut{E}{A}{\mu} + 2\udt{\omega}{B}{A\sigma} \dut{S}{B}{\mu\sigma} + 2\udt{T}{B}{\sigma A} \dut{S}{B}{\sigma\mu}\,,\label{Eq:App_gauge_current_sim}
\end{equation}
where a long but straightforward series of simplifications were taken together with the expressions for the contortion tensor in Eq.~\eqref{Eq:Contortion_def} and superpotential in Eq.~\eqref{Eq:Superpotential_def}.

\subsection{Example teleparallel theories}
Here we will give the variations of teleparallel quantities that appear in three well-known modifications of teleparallel gravity; the $f(T,B,\phi,X)$, the $f(T,B,T_G,B_G)$ theory presented in Sec.~\ref{Sec:GB_theories} and the BDLS theory in Sec.~\ref{sec:BDLS}.

\subsubsection{$f(T,B,\phi,X)$}

Apart from the \textit{bare} quantities that we discussed so far, one can encounter those in arbitrary functions, such as $f(T), f(T,B)$ and so on. For this reason, we thought it would be useful to consider a general function of the form $f(T,B,\phi,X)$ in order to discuss its variations. Thus, varying the Lagrangian
\begin{equation}
    L = e f(T,B,\phi,X)\,,
\end{equation}
with respect to the tetrad we get
\begin{equation}\label{var_f(T,B,phi,X)}
    \delta _e (ef(T,B,\phi,X)) = f(T,B,\phi,X)\delta _e e + e f_T \delta _e T+ e f_B \delta _e B + e f_X \delta _e X\,,
\end{equation}
where $f_i$ with $i = {T,B,X}$ is the partial derivative of $f$ with respect to $i$. The first term is easily obtained by Eq.~\eqref{var_determinant} and the second one by
\begin{equation}
    e f_T \delta T = -2 \left[e \left(\partial _\mu f_T \right)S_A{}^{\mu\beta} + \partial _{\mu} \left(e S_A{}^{\mu\beta} \right)f_T - e f_T T^\sigma{}_{\mu A}S_\sigma {}^{\beta \mu} \right]\delta e^A{}_{\beta}\,.
\end{equation}
For the third term we have $B = 2\partial _{\mu}(eT^\mu)/e$ and thus
\begin{equation}\label{var_B}
    ef_B\delta _e B = -\left[Bf_B+2 T^\mu \partial_\mu f_B\right]\delta _e e - 2e(\partial_\mu f_B)\delta_e T^\mu\,.
\end{equation}
The first term of the right-hand side is trivial; the second term gives
\begin{alignat}{2}
    -2e(\partial_\mu f_B)\delta_e T^\mu& =\: & & \Big[2e(\partial_\mu f_B)\Big(T_Ag^{\mu\beta}+T^\beta E_A{}^\mu+T^\beta{}_A{}^\mu+\omega^{\beta}{}_{A}{}^{\mu}-\omega^\sigma{}_{A\sigma} g^{\mu\beta}\Big)\nonumber\\[0.5ex]
    &\: & & +2\partial_\mu\Big(eE_A{}^\mu (\partial^\beta f_B)\Big)-2\partial_\nu\Big(eE_A{}^\beta (\partial^\nu f_B)\Big)\Big]\delta e^A{}_\beta\,.
\end{alignat}
Thus, Eq.~\eqref{var_B} will be
\begin{alignat}{2}
    e\,f_B\delta_e B& =\: & &e\Big[(\partial_\mu f_B)\Big(S_A{}^{\mu\beta}+K^{\mu\beta}{}_A+E_C{}^\beta E_A{}^\sigma g^{\lambda\mu}\partial_\sigma e^C{}_\lambda-E_C{}^\sigma E_A{}^\lambda g^{\mu\beta}\partial_\lambda e^C{}_\sigma\Big)\nonumber\\[0.5ex]
    &\: & &+E_A{}^\mu \lc{\nabla}_\mu \lc{\nabla}^\beta f_B-E_A{}^\beta \lc{\square}f_B-Bf_B\, E_A{}^\beta\Big]\delta e^A{}_\beta \,,
\end{alignat}
where the relation $e\lc{\nabla}_\mu Y^\mu = \partial _\mu(eY^\mu)$ and the definitions of the superpotential and the contortion were used. Using several identities, it is possible to show that all the terms appearing in the first line of the above equation is $2(\partial_{\mu} f_B)S_{A}{}^{\lambda\mu}$ (see~\cite{Bahamonde:2015zma} for further details). Thus, the variations of the boundary term contribution are
\begin{equation}
e\,f_B\delta_e B=
	2e \Big(
 E_{A}{}^{\nu}\lc{\nabla}^{\lambda}\lc{\nabla}_{\mu}f_{B}-E_{A}{}^{\lambda}\lc{\Box} f_{B}-
	 \frac{1}{2}Bf_{B}E_{A}{}^{\lambda}+(\partial_{\mu}f_{B})S_{A}{}^{\lambda\mu}
	 \Big) \delta e^{A}{}_\lambda \,.
\end{equation}

Finally, the kinetic term in Eq.~\eqref{var_f(T,B,phi,X)} will give
\begin{equation}
    ef_X\delta _e X = -\frac{1}{2}e f_X (\partial_\mu \phi) (\partial_\nu \phi) \delta _e g^{\mu\nu} =e f_X (\partial_\mu \phi) (\partial_\nu \phi) g^{\mu\beta} E_A{}^\nu \delta e^A{}_{\beta} \,.
\end{equation}

\subsubsection{$f(T,B,T_G,B_G)$ theory}
The action of the theory is given by
\begin{equation}
 \mathcal{S}_{f(T,B,T_{G},B_{G})} = \frac{1}{2 \kappa ^2} \int
\dd^4x \,e \,f(T,B,T_G,B_G)+ \mathcal{S}_{\rm m}\,,
\end{equation}
and its variation will be
\begin{equation}
    \delta \mathcal{S}_{f(T,B,T_G,B_G)} = \frac{1}{2\kappa ^2}\int \dd ^4x \left[ f\delta e+e f_T \delta T + e f_B \delta B + e f_{T_G} \delta T_G + e f_{B_G}\delta B_G\right] + \delta \mathcal{S}_{\rm m}\,,
\end{equation}
where
\begin{subequations}
\begin{align}
    f \delta e &= e f E_A{}^{\beta} \delta e ^A{}_{\beta} \,,\\[0.5ex]
    e f_T \delta T &= -2 \left[e \left(\partial _\mu f_T \right)S_A{}^{\mu\beta} + \partial _{\mu} \left(e S_A{}^{\mu\beta} \right)f_T - e f_T T^\sigma{}_{\mu A}S_\sigma {}^{\beta \mu} \right]\delta e^A{}_{\beta}\,, \\[0.5ex]
    e f_B \delta B &= \left[ 2 e E_A{}^{\nu} \lc{\nabla}^\beta \lc{\nabla}_\mu f_B -2 e E_A{}^\beta \lc{\Box}f_B - B e f_B E_A{}^{\beta} - 2 e \left(\partial_{\mu}f_B\right)S_A{}^{\mu\beta}\right] \delta e^A{}_{\beta} \,,\\[0.5ex]
    e f_{T_G} \delta T_G &= \Big[ \partial _\mu \left(E_H{}^{\mu}E_B{}^{\beta}\eta_{AL}(Y^{B[LH]}-Y^{H[LB]} +Y^{L[BH]})\right) + T_{IAB}E_H{}^{\beta} \left( Y^{B[IH]}-Y^{H[IB]}+Y^{I[B H]}\right)-\nonumber \\[0.5ex]
    &\,\,\,\,- 2 e f_{T_G}\delta ^{MBCD}_{IJKL}E_D{}^{\beta}K^J{}_M{}^{I}K_{EB}{}^K \partial _A(K^L{}_C{}^{E}) \Big] \delta e^A{}_{\beta}\,, \\[0.5ex]
    e f_{B_G} \delta B_G &= -\Big[ \partial_{\mu}\left((P^{B[LH]}-P^{H[LB]} +P^{A[BH]})\eta_{AL}E_H{}^{\mu}E_B{}^{\beta}\right) + T_{IAB}E_H{}^\beta \left(P^{B[IH]}-P^{H[IB]}+P^{I[BH]}\right) - \nonumber \\[0.5ex]
    &\,\,\,\, - \delta ^{MBCD}_{IJKL}e E_D{}^\beta (\partial _M f_{B_G})K^J{}_B{}^{I}(\partial _A K^L{}_C{}^{K}) + e (\partial _\mu f_{B_G})(E_A{}^\beta B^\mu _G - E_A{}^\mu B^\beta _G) +\nonumber \\[0.5ex]
    &\,\,\,\,+ ef_{B_G}B_GE_A{}^{\beta}\Big] \delta e^A{}_{\beta}\,.
\end{align}
\end{subequations}
In the above, we have defined
\begin{align}
    X^A{}_{IJ} = \frac{\partial T_G}{\partial K_A{}^{IJ}}\,,\quad Y^B{}_{IJ} = e f_{T_G}X^B{}_{IJ} - 2 \delta ^{CABD}_{ELKJ}\partial _{\mu}\left( ef_{T_G} E_D{}^{\mu} K^L{}_C{}^{E}K_{IA}{}^{K}\right)\,,
\end{align}
as well as
\begin{align}
    B^A_G &= \epsilon_{IJKL}\epsilon^{BCDA} \left(\frac{1}{2}K^J{}_{B}{}^I R^{KL}{}_{CD} + K^J{}_B{}^I K_{FC}{}^K K^L{}_D{}^F \right) \nonumber \\[0.5ex]
    &= \delta^{ABCD}_{IJKL} K^J{}_B{}^I\left( (K^L{}_C{}^K)_{,D} +K_{CD}{}^M K^L{}_M{}^K\right)\,
\end{align}
and
\begin{align}
    P^B{}_{IJ} = &e E_M{}^{\mu} (\partial _{\mu} f_{B_G}) \Big[ \left( (K^L{}_C{}^K)_{,D}+K_{CD}{}^P K^L{}_P{}^Q \right)\delta ^{MBCD}_{IJKL} + \eta _{PJ} \delta ^{MDPB}_{QCKL} K^C{}_D{}^Q K^L{}_I{}^K +\nonumber \\[0.5ex]
    &+ \delta ^{MPCD}_{KLIJ}K^L{}_P{}^K K_{CD}{}^B \Big] - \delta ^{ACBD}_{KLIJ}\partial _{\sigma} \left( eE_D{}^\sigma E_A{}^\mu (\partial_{\mu}f_{B_G})K^L{}_C{}^K\right)\,.
\end{align}
The equations of motion are given in Sec.~\ref{Sec:GB_theories} Eq.~\eqref{eq:fTTG}. Note that the index G in the above equations is not an index since it refers to the Gauss-Bonnet Teleparallel invariant.

\subsubsection{Teleparallel analogue of Horndeski theory}

By varying the action~\eqref{BDLS_Action} with respect to the tetrads, we find that
\begin{equation}
 \delta_{e}\mathcal{S}_{\rm BDLS}= e\mathcal{L}_{\rm Tele}E_A{}^{\mu}\delta e^A{}_\mu+e\delta_{e} \mathcal{L}_{\rm Tele}+e\sum_{i=2}^{5}\mathcal{L}_iE_A{}^{\mu}\delta e^A{}_\mu+e \delta_{e} \sum_{i=2}^{5}\mathcal{L}_{i}+2\kappa^2e\Theta_A{}^\mu\delta e^{A}{}_\mu=0\,.\label{deltaS}
\end{equation}
The variations of $\delta_{e}\sum_{i=2}^{5} \mathcal{L}_i$ give the standard Horndeski field equations whereas the variations $\delta_{e} \mathcal{L}_{\rm Tele}$ are related to the extra terms coming from \gls{tg}. After doing several computations, one finds that the field equations can be written as
\begin{alignat}{2}
& \: & &2(\partial_{\lambda}G_{\rm Tele,T})S_{A}\,^{\mu\lambda}+2e^{-1}\partial_{\lambda}(e S_{A}\,^{\mu\lambda})G_{\rm Tele,T}-2G_{\textrm{Tele},T}T^{\sigma}\,_{\lambda A}S_{\sigma}\,^{\lambda\mu}+2G_{\rm Tele,T}\omega^{B}{}_{A\nu}S_{B}{}^{\mu\nu}\nonumber\\[0.5ex]
&\: & &-\phi_{;A}\Big[G_{\rm Tele,X}\phi^{;\mu}-G_{\rm Tele,I_2}v^\mu -2G_{\rm Tele,J_1}a^{\mu}a_{J}\phi^{;J} +G_{\rm Tele,J_3}v_I t_{K}{}^{\mu I}\phi^{;K}-2G_{\rm Tele,J_5} t^{I\mu K}t_{IJK}\phi^{;J}\nonumber\\[0.5ex]
&\: & &+2G_{\rm Tele,J_6}t_{ILK}t^\mu{}_M{}^I\phi^{;K}\phi^{;L}\phi^{;M}-2G_{\rm Tele,J_8}t_{IJK}t^{IJ}{}^{\mu}\phi^{;K}-G_{\rm Tele,J_{10}}a^J \phi^{;I}\Big(\epsilon_{\mu JCD}t_I{}^{CD}+\epsilon_{IJCD}t^{\mu CD}\Big) \Big]\nonumber\\[0.5ex]
&\: & &+\frac{1}{3}\Big[M^I(\epsilon_{IB}{}^{CD}
E_C{}^{ \mu} T^{B}{}_{ AD}
-\epsilon_{IB}{}^{ CD}
E_D{}^{ \mu} \omega^B{}_{ AC})+e^{-1}\partial_\nu\Big(eM^I\epsilon_{IA}{}^{CD} E_C{}^{ \nu}E_D{}^{ \mu}\Big)\Big]\nonumber\\[0.5ex]
&\: & &-N^I(E_I{}^{ \mu} \omega^\rho{}_{ A \rho}	- \omega^\mu{}_{AI} - T^\mu{}_{AI}	- v_A	E_I{}^{ \mu})+e^{-1}\partial_\nu\Big(eN^I(E_A{}^{ \nu } E_I{}^{ \mu} - E_A{}^{ \mu } E_I{}^{ \nu})\Big)\nonumber\\[0.5ex]
&\: & &-O^{IJK}H_{IJKA}{}^{\mu}+e^{-1}\partial_\nu\Big(eO^{IJK}L_{IJKA}{}^{\mu\nu}\Big)- \mathcal{L}_{\rm Tele}E_A{}^{\mu}+2E_A{}^{\nu}g^{\mu\alpha}\sum_{i=2}^{5}\mathcal{G}^{(i)}{}_{\alpha\nu}=2\kappa^2\Theta_A{}^{\mu}\,,
\end{alignat}
where we have defined
\begin{subequations}
\begin{alignat}{2}
    M^{I}& =\: & &2G_{\rm Tele,T_{\rm ax}} a^I+2G_{\rm Tele,J_1}\phi^{;I}\phi^{;J}a_{J}+G_{\rm Tele,J_{10}}\epsilon_{A}{}^{I}{}_{CD}\phi^{;A}\phi^{;J}t_{J}{}^{CD}\,,\label{M}\\[0.5ex]
N^I& =\: & &2G_{\rm Tele,T_{\rm vec}} v^I+G_{\rm Tele,I_2}\phi^{;I}+2G_{\rm Tele,J_2}\phi^{;I}\phi^{;J}v_{J}+G_{\rm Tele,J_3}\phi^{;K}\phi^{;J}t^{I}{}_ {KJ}\,,\label{N}\\[0.5ex]
O^{IJK}& =\: & &G_{\rm Tele,J_3}\phi^{;J}\phi^{;K}v^{I}+2G_{\rm Tele,J_5}\phi^{;L}\phi^{;J}t^{I}{}_{L}{}^{K}+2G_{\rm Tele,J_6}\phi^{;J}\phi^{;K}\phi^{;L}\phi^{;M}t^{I}{}_{LM}\nonumber\\[0.5ex]
&\: & &+2G_{\rm Tele,J_8}\phi^{;L}\phi^{;K}t^{IJ}{}_{L}+G_{\rm Tele,J_{10}}\epsilon_{AB}{}^{JK} \phi^{;A}\phi^{;B}\phi^{;I}\,, \\[0.5ex]
H_{IJKA}{}^\mu & =\: & & \frac{\partial t_{IJK}}{\partial e^A{}_{\mu}} \,,\\[0.5ex]
L_{IJKA}{}^{\mu\nu} & =\: & & \frac{\partial t_{IJK}}{\partial e^A{}_{\mu,\nu}}\,.
\end{alignat}
\end{subequations}
In addition, the variation of the scalars appearing in $\mathcal{L}_{\rm Tele}$ is given by
\begin{subequations}
\begin{alignat}{2}
    eG_{\rm Tele,T_{\rm ax}} \delta_e T_{\rm ax}& =\: & &2eG_{\rm Tele,T_{\rm ax}} a^I\delta_e a_I\,,\label{Taxvar}\\[0.5ex]
    eG_{\rm Tele,T_{\rm vec}} \delta_e T_{\rm vec}& =\: & &2eG_{\rm Tele,T_{\rm vec}} v^I\delta_e v_I\,,\\[0.5ex]
    eG_{\rm Tele,I_2} \delta_e I_2& =\: & &eG_{\rm Tele,I_2}\phi^{;I} \delta_e v_{I}-eG_{\rm Tele,I_2}v^\mu \phi_{;A} \delta_e e^{A}{}_\mu\,,\\[0.5ex]
    eG_{\rm Tele,J_1} \delta_e J_1& =\: & &2eG_{\rm Tele,J_1}\phi^{;I}\phi^{;J}a_{J} \delta_e a_{I}-2eG_{\rm Tele,J_1}a_{J}a^{\mu}\phi^{;J}\phi_{;A} \delta_e e^{A}{}_\mu \,,\\[0.5ex]
    eG_{\rm Tele,J_3} \delta_e J_3& =\: & &eG_{\rm Tele,J_3}\phi^{;K}\phi^{;J}t^{I}{}_ {KJ}\delta_e v_{I}+eG_{\rm Tele,J_3}\phi^{;J}\phi^{;K}v^{I}\delta_e t_{IJK} \nonumber\\[0.5ex]
    &\: & &+eG_{\rm Tele,J_3}v_I t_{K}{}^{\mu I}\phi^{;K}\phi_{;A} \delta_e e^{A}{}_\mu\,, \\[0.5ex]
    eG_{\rm Tele,J_5} \delta_e J_5& =\: & &2eG_{\rm Tele,J_5}\phi^{;L}\phi^{;J}t^{I}{}_{L}{}^{K}\delta_e t_{IJK}-2eG_{\rm Tele,J_5} t^{I\mu K}t_{IJK}\phi^{;J}\phi_{;A}\delta_e e^{A}{}_\mu\,,\\[0.5ex]
    eG_{\rm Tele,J_6} \delta_e J_6& =\: & &2eG_{\rm Tele,J_6}\phi^{;J}\phi^{;K}\phi^{;L}\phi^{;M}t^{I}{}_{LM}\delta_e t_{IJK}+2eG_{\rm Tele,J_6}t_{ILK}t^\mu{}_M{}^I\phi^{;K}\phi^{;L}\phi^{;M}\phi_{;A} \delta_e e^{A}{}_\mu\,,\\[0.5ex]
    eG_{\rm Tele,J_8} \delta_e J_8& =\: & &2eG_{\rm Tele,J_8}\phi^{;L}\phi^{;K}t^{IJ}{}_{L}\delta_e t_{IJK}-2eG_{\rm Tele,J_8}t_{IJK}t^{IJ}{}^{\mu}\phi^{;K}\phi_{;A} \delta_e e^{A}{}_\mu\,,\label{J8var}\\[0.5ex]
    G_{\rm Tele,J_{10}} \delta_e J_{10}& =\: & &eG_{\rm Tele,J_{10}}\epsilon_{A}{}^{I}{}_{CD}\phi^{;A}\phi^{;J}t_{J}{}^{CD}\delta_e a_I+eG_{\rm Tele,J_{10}}\epsilon_{AB}{}^{JK}A^B \phi^{;A}\phi^{;I}\delta_e t_{IJK}\nonumber\\[0.5ex]
    & \: & &-eG_{\rm Tele,J_{10}}a^J \phi^{;I}\phi_{;A}\Big(\epsilon_{\mu JCD}t_I{}^{CD}+\epsilon_{IJCD}t^{\mu CD}\Big) \delta_e e^{A}{}_\mu\,.
\end{alignat}
\end{subequations}
By varying the action \eqref{BDLS_Action} with respect to the scalar field we obtain the modified Klein Gordon equation
\begin{equation}
\lc{\nabla}^\mu\Big(J_{\mu\rm -Tele}+\sum_{i=2}^{5}J^{i}_\mu\Big)=P_{\phi-\rm Tele}+\sum_{i=2}^{5}P_{\phi}^i\,,
\end{equation}
where $J_{\mu\rm -Tele}$ and $P_{\phi-\rm Tele}$ are defined as
\begin{subequations}
\begin{alignat}{2}
J_{\mu\rm -Tele}& =\: & & -G_{\rm Tele,X}(\lc{\nabla}_\mu\phi)+G_{\rm Tele,I_2} v_\mu+2G_{\rm Tele,J_1}a_\mu a^\nu\lc{\nabla}_\nu \phi-G_{\rm Tele,J_3}v_\alpha t_{\mu}{}^{\nu\alpha}(\lc{\nabla}_\nu\phi)\nonumber\\[0.5ex]
& \: & & -2G_{\rm Tele,J_5}t^{\beta\nu\alpha}t_{\beta\mu\alpha}(\lc{\nabla}_\nu\phi)+2G_{\rm Tele,J_8}t^{\alpha\nu}{}_{\mu}t_{\alpha\nu}{}^{\beta}(\lc{\nabla}_\beta\phi)-2G_{\rm Tele,J_6}t^{\nu\alpha\beta}t_{\mu}{}^{\sigma}{}_\nu(\lc{\nabla}_\alpha\phi)(\lc{\nabla}_\beta\phi)(\lc{\nabla}_\sigma\phi)\,,\nonumber\\[0.5ex]
&\: & & -G_{\rm Tele,J_{10}} a^\nu (\lc{\nabla}_\alpha \phi) (\epsilon^{\mu}{}_{\nu\rho\sigma}t^{\alpha\rho\sigma}+\epsilon^{\alpha}{}_{\nu\rho\sigma}t^{\mu\rho\sigma})\,,\label{Jtele2}\\[0.5ex]
P_{\phi-\rm Tele}& =\: & & G_{\rm Tele,\phi}\,.\label{PTele2}
\end{alignat}
\end{subequations}
For more details about the derivation of these equations, see \cite{Bahamonde:2020cfv}. Using $\lc{R}=-T+B$, one finds that $P_{\phi}^i$ is given by~\cite{Capozziello:2018gms}
\begin{subequations}\label{Pphi}
	\begin{align}
	P_{\phi}^{2} &= G_{2,\phi}\,,\\[0.5ex]
	P_{\phi}^{3} &= \lc{\nabla}_{\mu}G_{3,\phi}\lc{\nabla}^{\mu} \phi \,,\\[0.5ex]
	P_{\phi}^{4} &= G_{4,\phi}(-T+B) + G_{4,\phi X} \left[ (\lc{\square} \phi)^2 - (\lc{\nabla}_{\mu}\lc{\nabla}_{\nu}\phi)^2\right]\,,\\[0.5ex]
	P_{\phi}^{5} &= -\lc{\nabla}_{\mu}G_{5,\phi} \lc{G}^{\mu\nu}\lc{\nabla}_{\nu}\phi - \frac{1}{6}G_{5,\phi X}\left[(\square \phi)^3 - 3 \square \phi (\lc{\nabla}_{\mu}\lc{\nabla}_{\nu} \phi)^2 + 2 (\lc{\nabla}_{\mu}\lc{\nabla}_{\nu}\phi)^3 \right]\,,
	\end{align}
\end{subequations}
and $J^{i}_\mu$ is defined as~\cite{Capozziello:2018gms}
\begin{subequations}\label{Jis}
\begin{alignat}{2}
	J_{\mu}^{2} & =\: & & -\mathcal{L}_{2,X}\lc{\nabla}_{\mu}\phi \,,\\[0.5ex]
	J_{\mu}^{3} & =\: & & -\mathcal{L}_{3,X}\lc{\nabla}_{\mu}\phi + G_{3,X} \lc{\nabla}_{\mu}X + 2 G_{3,\phi} \lc{\nabla}_{\mu} \phi \,,\\[0.5ex]
	J_{\mu}^{4} & =\: & & - \mathcal{L}_{4,X}\lc{\nabla}_{\mu} \phi +2 G_{4,X}\lc{R}_{\mu\nu}\lc{\nabla}^{\nu} \phi - 2 G_{4,XX}\left(\lc{\square} \phi \lc{\nabla}_{\mu}X - \lc{\nabla}^{\nu} X \lc{\nabla}_{\mu}\lc{\nabla}_{\nu} \phi \right) \nonumber \\[0.5ex]
	& \: & & -2 G_{4,\phi X} (\lc{\square} \phi \lc{\nabla}_{\mu}\phi + \lc{\nabla}_{\mu}X) \,,\\[0.5ex]
	J_{\mu}^{5} & =\: & & -\mathcal{L}_{5,X}\lc{\nabla}_{\mu}\phi - 2 G_{5,\phi}\lc{G}_{\mu\nu} \lc{\nabla}^{\nu} \phi \nonumber \\[0.5ex]
	& \: & & -G_{5,X}\left[ \lc{G}_{\mu\nu}\lc{\nabla}^{\nu} X + \lc{R}_{\mu\nu}\square \phi \lc{\nabla}^{\nu} \phi - \lc{R}_{\nu \lambda} \lc{\nabla}^{\nu} \phi \lc{\nabla}^{\lambda}\lc{\nabla}_{\mu} \phi - \lc{R}_{\alpha \mu \beta \nu}\lc{\nabla}^{\nu}\phi \lc{\nabla}^{\alpha} \lc{\nabla}^{\beta}\phi\right] \nonumber \\[0.5ex]
	& \: & & +G_{5,XX} \Big\{ \frac{1}{2}\lc{\nabla}_{\mu}X \left[(\lc{\square} \phi)^2 - (\lc{\nabla}_{\alpha}\lc{\nabla}_{\beta}\phi)^2 \right]- \lc{\nabla}_{\nu}X\left(\lc{\square} \phi \lc{\nabla}_{\mu}\lc{\nabla}^{\nu}\phi - \lc{\nabla}_{\alpha}\lc{\nabla}_{\mu}\phi\lc{\nabla}^{\alpha}\lc{\nabla}^{\nu}\phi\right)\Big\} \nonumber \\[0.5ex]
	& \: & & +G_{5,\phi X} \Big\{ \frac{1}{2}\lc{\nabla}_{\mu}\phi \left[(\lc{\square} \phi)^2 - (\lc{\nabla}_{\alpha}\lc{\nabla}_{\beta}\phi)^2 \right] + \lc{\square} \phi \lc{\nabla}_{\mu}X -\lc{\nabla}^{\nu}X \lc{\nabla}_{\nu}\lc{\nabla}_{\mu}\phi \Big\}\,.
	\end{alignat}
\end{subequations}
It is easy to see that the equations of motion of well-studied theories can be recovered by choosing the form of the $G_i$ functions; e.g. Brans-Dicke theory can be obtained by setting $G_{\rm Tele} = 0,\, G_2 =\frac{\omega}{\phi}X ,\,G_3 = 0 ,\, G_4 = \phi ,$ and $G_5 = 0.$

\section{\texorpdfstring{$G_{{\rm eff}}$}{} Calculation in \texorpdfstring{$f(T,B)$}{} Gravity} \label{sec:Geff-appendix}

Here we present all the coefficients appearing in Sec.~\ref{f_TB_matter_dens} regarding the effective Newton's constant in $f(T,B)$ gravity and subcases (for more details see Ref.~\cite{Bahamonde:2020lsm})

\begin{subequations}
\begin{alignat}{2}
A_{1} & =\: & & -a^{4}\Upsilon\dot{\Pi}(\dot{\Pi}+12H^{3}f_{TT})\,,\\[0.5ex]
A_{2} & =\: & & -4a^{2}H\Upsilon(2\Pi_{B}\dot{\Pi}-12H^{3}f_{BB}f_{TB}-H\Upsilon f_{T})\,,\\[0.5ex]
A_{3} & =\: & & -16H^{2}\Xi\Upsilon\,,\\[0.5ex]
\nonumber \\[0.5ex]
A_{4} & =\: & & -3a^{6}\dot{\Pi}(\Pi_{B}+\Pi_{T})\Biggl[\Pi'\Big(6\dot{H}^{2}(\Pi_{B}-f_{TB})+6H^{2}\dot{H}(3f_{TB}-\Pi_{B})+18H^{4}(\Pi_{T}-f_{TB})\nonumber \\[0.5ex]
 & \: & & -H^{2}f_{T}-H\dot{f}_{T}\Big)+6H^{2}\Big(\dot{H}(-24H^{3}(f_{TB}-\Pi_{B})(f_{TB}-\Pi_{T})+\Pi_{B}\dot{f}_{T})+H\dot{\Pi}^{2}\nonumber \\[0.5ex]
 & \: & & -12H\dot{H}^{2}(f_{TB}(\Pi_{B}-f_{TB})+\Xi)+H^{2}(2Hf_{T}+\dot{f}_{T})(f_{TB}-\Pi_{T})\Big)\Biggl]\,,\\[0.5ex]
\nonumber \\[0.5ex]
A_{5} & =\: & & 3a^{4}\Biggl(-4H^{4}f_{T}{}^{2}(\Pi_{B}+\Pi_{T})^{2}+48H^{4}f_{TB}{}^{3}\bigl(-12H^{2}(5\Pi_{B}+4\Pi_{T})\dot{H}\nonumber \\[0.5ex]
 & & & +24(\Pi_{B}+\Pi_{T})\dot{H}{}^{2}+H\dot{\Pi}\bigr)+24H^{3}f_{TB}{}^{2}\biggl(24H^{3}\bigl(2\Xi+(4\Pi_{B}+\Pi_{T})(\Pi_{B}\nonumber \\[0.5ex]
 & & & +2\Pi_{T})\bigr)\dot{H}-12H\Bigl(3\Xi+2\bigl(\Pi_{B}{}^{2}+4\Pi_{B}\Pi_{T}+\Pi_{T}{}^{2}\bigr)\Bigr)\dot{H}{}^{2}-H^{2}\bigl(2(\Pi_{B}\nonumber \\[0.5ex]
 & & & +2\Pi_{T})\dot{\Pi}+(\Pi_{B}+\Pi_{T})\dot{f_{T}}\bigr)-\dot{H}\bigl((11\Pi_{B}+10\Pi_{T})\dot{\Pi}+(\Pi_{B}+\Pi_{T})\dot{f_{T}}\bigr)\biggr)\nonumber \\[0.5ex]
 & & & +2Hf_{T}\Biggl(-24H^{3}f_{TB}{}^{2}(\Pi_{B}+\Pi_{T})(3H^{2}-8\dot{H})+(\Pi_{B}+\Pi_{T})\bigl(-H^{2}\Pi_{T}\nonumber \\[0.5ex]
 & & & +\Pi_{B}(7H^{2}-3\dot{H})\bigr)\dot{\Pi}+4H^{2}f_{TB}\biggl(6H^{3}\Bigl(-\Pi_{B}{}^{2}+3\Pi_{B}\Pi_{T}+2\bigl(\Xi+\Pi_{T}{}^{2}\bigr)\Bigr)\nonumber \\[0.5ex]
 &  &  & -12H\bigl(4\Xi+2\Pi_{B}{}^{2}+5\Pi_{B}\Pi_{T}-\Pi_{T}{}^{2}\bigr)\dot{H}+(\Pi_{B}+\Pi_{T})\dot{\Pi}\biggr)\nonumber \\[0.5ex]
 & & & +3H\Bigl(16H^{4}\Pi_{B}\bigl(-\Xi+\Pi_{T}{}^{2}\bigr)+16H^{2}\Pi_{B}\bigl(4\Xi+3\Pi_{B}\Pi_{T}-\Pi_{T}{}^{2}\bigr)\dot{H}\nonumber \\[0.5ex]
 & & & -4\Xi(\Pi_{B}+\Pi_{T})\dot{H}{}^{2}-H(\Pi_{B}+\Pi_{T})^{2}\dot{f_{T}}\Bigr)\Biggr)+3\Pi_{B}\biggl(192H^{6}\Bigl(\Xi\Pi_{B}+\bigl(\Xi\nonumber \\[0.5ex]
 & & & +\Pi_{B}{}^{2}\bigr)\Pi_{T}\Bigr)\dot{H}-96H^{4}\Pi_{T}(3\Xi+2\Pi_{B}\Pi_{T})\dot{H}{}^{2}+(\Pi_{B}+\Pi_{T})\dot{H}\dot{\Pi}{}^{2}\nonumber \\[0.5ex]
 & & & +3H^{2}(\Pi_{B}+\Pi_{T})\dot{\Pi}(-\dot{\Pi}+\dot{f_{T}})-8H^{5}\Bigl(\bigl(-2\Xi+3\Pi_{T}(\Pi_{B}+\Pi_{T})\bigr)\dot{\Pi}+2(\Xi\nonumber \\[0.5ex]
 & & & +\Pi_{B}\Pi_{T})\dot{f_{T}}\Bigr)-4H^{3}\dot{H}\Bigl(\bigl(29\Xi+20\Pi_{B}\Pi_{T}-7\Pi_{T}{}^{2}\bigr)\dot{\Pi}+2\bigl(\Xi-\Pi_{T}{}^{2}\bigr)\dot{f_{T}}\Bigr)\biggr)\nonumber \\[0.5ex]
 & & & +f_{TB}\Biggl(-12H^{3}\Bigl(H^{2}\bigl(4\Xi-8\Pi_{B}{}^{2}-6\Pi_{B}\Pi_{T}-6\Pi_{T}{}^{2}\bigr)-\bigl(29\Xi+11\Pi_{B}{}^{2}\nonumber \\[0.5ex]
 & & & +33\Pi_{B}\Pi_{T}-7\Pi_{T}{}^{2}\bigr)\dot{H}\Bigr)\dot{\Pi}+(\Pi_{B}+\Pi_{T})(4H^{2}-3\dot{H})\dot{\Pi}{}^{2}\nonumber \\[0.5ex]
 & & & +24H^{3}\Biggl(12H(\Pi_{B}+\Pi_{T})(3\Xi+4\Pi_{B}\Pi_{T})\dot{H}{}^{2}+H^{2}\bigl(2\Xi+\Pi_{B}{}^{2}\nonumber \\[0.5ex]
 & & & +3\Pi_{B}\Pi_{T}\bigr)\dot{f_{T}}+\dot{H}\biggl(-24H^{3}\Bigl(\Pi_{B}{}^{3}+\Xi\Pi_{T}+4\Pi_{B}{}^{2}\Pi_{T}+\Pi_{B}\bigl(3\Xi\nonumber \\[0.5ex]
 & & & +2\Pi_{T}{}^{2}\bigr)\Bigr)+\bigl(\Xi-\Pi_{T}{}^{2}\bigr)\dot{f_{T}}\biggr)\Biggr)\Biggr)\Biggr)\,,\\[0.5ex]
\nonumber \\[0.5ex]
A_{6} & =\: & & a^{2}\Biggl(-4H^{2}f_{T}{}^{2}(\Pi_{B}+\Pi_{T})^{2}+864H^{6}(f_{TB})^{3}(3\Pi_{B}+2\Pi_{T})+24H^{3}\Pi_{B}(7\Xi\nonumber \\[0.5ex]
 & & & +6\Pi_{B}\Pi_{T})\dot{\Pi}-\Pi_{B}(\Pi_{B}+\Pi_{T})\dot{\Pi}{}^{2}+4Hf_{T}\biggl(-6H^{3}(f_{TB}-\Pi_{B})\bigl(2\Xi\nonumber \\[0.5ex]
 & & & +(\Pi_{B}-\Pi_{T})\Pi_{T}\bigr)-9H\Bigl((f_{TB})^{2}(\Pi_{B}+\Pi_{T})+\Pi_{B}\bigl(\Xi-\Pi_{T}{}^{2}\bigr)+f_{TB}\bigl(-\Xi\nonumber \\[0.5ex]
 & & & +\Pi_{T}{}^{2}\bigr)\Bigr)\dot{H}+\Pi_{B}(\Pi_{B}+\Pi_{T})\dot{\Pi}\biggr)-12H^{3}f_{TB}{}^{2}\bigl(72H^{3}\Pi_{B}(4\Pi_{B}+5\Pi_{T})\nonumber \\[0.5ex]
 & & & +60H\Xi\dot{H}-(7\Pi_{B}+5\Pi_{T})\dot{\Pi}\bigr)+f_{TB}\Bigl(864H^{6}\bigl(3\Xi\Pi_{B}+\Pi_{B}{}^{3}-\Xi\Pi_{T}\nonumber \\[0.5ex]
 & & & +4\Pi_{B}{}^{2}\Pi_{T}\bigr)+720H^{4}\Xi(\Pi_{B}+\Pi_{T})\dot{H}-12H^{3}\bigl(10\Xi+7\Pi_{B}{}^{2}\nonumber \\[0.5ex]
 & & & +17\Pi_{B}\Pi_{T}\bigr)\dot{\Pi}+(\Pi_{B}+\Pi_{T})\dot{\Pi}{}^{2}\Bigr)+12H^{3}\biggl(-72H^{3}\Pi_{B}\Bigl(\Xi\Pi_{B}+\bigl(\Xi\nonumber \\[0.5ex]
 & & & +\Pi_{B}{}^{2}\bigr)\Pi_{T}\Bigr)-60H\Xi(2\Xi+\Pi_{B}\Pi_{T})\dot{H}+5\Xi(\Pi_{B}+\Pi_{T})\dot{f_{T}}\biggr)\Biggr)\,,\\[0.5ex]
\nonumber \\[0.5ex]
A_{7} & =\: & & 12H^{2}\Xi(12H^{2}(f_{BB}f_{TT}+2\Xi)+\Upsilon f_{T})\,,\\[0.5ex]
\Delta_{1} & =\: & & -a^{4}A_{1}\dot{\Pi}(12H\Pi_{B}\dot{H}-\dot{\Pi})\,,\\[0.5ex]
\Delta_{2} & =\: & & a^{2}\left(-2H\Pi_{B}\dot{\Pi}(6a^{2}A_{2}\dot{H}-5A_{1})+a^{2}A_{2}\dot{\Pi}^{2}+8A_{1}H^{2}(-\Upsilon f_{T}-6\Xi\dot{H})\right),\\[0.5ex]
\Delta_{3} & =\: & & a^{4}A_{3}\dot{\Pi}^{2}-2a^{2}H\Pi_{B}\dot{\Pi}(6a^{2}A_{3}\dot{H}-5A_{2})+8H^{2}\left(a^{2}A_{2}(-\Upsilon f_{T}-6\Xi\dot{H})+3A_{1}\Xi\right)\,,\\[0.5ex]
\Delta_{4} & =\: & & 2H\left(4H\left(a^{2}A_{3}(-\Upsilon f_{T}-6\Xi\dot{H})+3A_{2}\Xi\right)+5a^{2}A_{3}\Pi_{B}\dot{\Pi}\right)\,,\\[0.5ex]
\Delta_{5} & =\: & & 24A_{3}H^{2}\Xi\,,\\[0.5ex]
\Delta_{6} & =\: & & 2a^{4}A_{4}\dot{\Pi}(\dot{\Pi}+12H^{3}f_{TT})\,,\\[0.5ex]
\nonumber \\[0.5ex]
\Delta_{7} & =\: & & 2a^{2}\left(\dot{\Pi}\left(a^{2}A_{5}(\dot{\Pi}+12H^{3}f_{TT})+8A_{4}H\Pi_{B}\right)\right)\nonumber \\[0.5ex]
 & \: & & +96A_{4}H^{4}(f_{BB}f_{TT}+\Xi)+48A_{4}H^{4}f_{TB}(f_{TT}-\Upsilon)-4A_{4}H^{2}\Upsilon f_{T}\,,\\[0.5ex]
\nonumber \\[0.5ex]
\Delta_{8} & =\: & & -2a^{4}A_{6}\dot{\Pi}(-\dot{\Pi}-12H^{3}f_{TT})+8a^{2}A_{5}H\nonumber \\[0.5ex]
 & \: & & +\left(-(H\Upsilon f_{T}-2\Pi_{B}\dot{\Pi})+24H^{3}(f_{BB}f_{TT}+\Xi)+12H^{3}f_{TB}(f_{TT}-\Upsilon)\right)+32A_{4}H^{2}\Xi\,,\\[0.5ex]
\nonumber \\[0.5ex]
\Delta_{9} & =\: & & 2a^{4}A_{7}-\dot{\Pi}(-\dot{\Pi}-12H^{3}f_{TT})+32A_{5}H^{2}\Xi\nonumber \\[0.5ex]
 & \: & & +8a^{2}A_{6}H\left(-(H\Upsilon f_{T}-2\Pi_{B}\dot{\Pi})+24H^{3}(f_{BB}f_{TT}+\Xi)+12H^{3}f_{TB}(f_{TT}-\Upsilon)\right)\,,\\[0.5ex]
\nonumber \\[0.5ex]
\Delta_{10} & =\: & & 8H\left(a^{2}A_{7}\left(-(H\Upsilon f_{T}-2\Pi_{B}\dot{\Pi})+24H^{3}(f_{BB}f_{TT}+\Xi)+12H^{3}f_{TB}(f_{TT}-\Upsilon)\right)+4A_{6}H\Xi\right)\,,\\[0.5ex]
\Delta_{11} & =\: & & 32A_{7}H^{2}\Xi\,.
\end{alignat}
\end{subequations}

\section{PPN parameters for Teleparallel Analogue of Horndeski} \label{app:PPN}

The terms appearing in Table~\ref{table:PPN} are
\begin{subequations}
\begin{alignat}{2}
\HH_{,1}& =\: & & \GGG_{2,X}-2 \GGG_{3,\phi}+\GGG_{\text{Tele},X}\,,\label{H1}\\[0.5ex]
\HH_{,2}& =\: & & \GGG_{\text{Tele},\phi T}\,,\\[0.5ex]
\HH_{,3}& =\: & & \GGG_{\text{Tele},I_2}-2 \GGG_{4,\phi}\,,\\[0.5ex]
\HH_{,4}& =\: & & \GGG_{\text{Tele},T_{\text{vec}}}\,,\\[0.5ex]
\HH_{,5}& =\: & & 2 \GGG_{\text{Tele},T_{\text{vec}}}-\GGG_{\text{Tele},T}+\GG_4\,,\label{H5}\\[0.5ex]
\HH_{,6}& =\: & & 2 \Big(\GGG_{2,X}-2 \GGG_{3,\phi}+\GGG_{\text{Tele},X}\Big) \Big(\left(\GGG_{\text{Tele},\phi T}-\GGG_{\text{Tele},\phi T_{\text{vec}}}\right) \left(\GGG_{2,X}-2 \GGG_{3,\phi}+\GGG_{\text{Tele},X}\right)\nonumber\\[0.5ex]
& \: & & +\left(\GGG_{\text{Tele},I_2}-2 \GGG_{4,\phi}\right) \left(2 \GGG_{4,\phi\phi}-\GGG_{\text{Tele},\phi I_2}\right)\Big)+\left(\GGG_{\text{Tele},I_2}-2 \GGG_{4,\phi}\right){}^2 \left(\GGG_{2,\phi X}-2 \GGG_{3,\phi\phi}+\GGG_{\text{Tele},\phi X}\right)\,,\nonumber\\[0.5ex]\label{H6}
\end{alignat}
\end{subequations}
and
\small{\begin{alignat}{2}
\tilde{\beta}& =\: & & 4 \HH_{,1}^2 \HH_{,3} \Big(\HH_{,2} \left(4 \HH_{,4} \HH_{,5}-2 \HH_{,4}^2-3 \HH_{,5}^2\right)+\HH_{,3} \left(3 \HH_{,4} \HH_{,5}-\HH_{,4}^2+\HH_{,5}^2\right)\Big)-4 \HH_{,1}^3 \HH_{,4} \HH_{,5} \left(\HH_{,4}-2 \HH_{,5}\right)\nonumber\\[0.5ex]
& \: & & +\HH_{,1} \HH_{,3}^3 \Big(8 \HH_{,2} \left(\HH_{,4}-2 \HH_{,5}\right)+\HH_{,3} \left(4 \HH_{,4}+7 \HH_{,5}\right)\Big)+\HH_{,3} \left(-6 \HH_{,2} \HH_{,3}^4+3 \HH_{,3}^5+2 \HH_{,6} \left(\HH_{,5}-2 \HH_{,4}\right){}^2\right)\nonumber\\[0.5ex]
& \: & & +\GG_{4,\phi} \Big(-8 \HH_{,1} \HH_{,3}^3 \left(\HH_{,4}-2 \HH_{,5}\right)+8 \HH_{,1}^2 \HH_{,3} \left(\HH_{,5}^2-\HH_{,4}^2\right)+6 \HH_{,3}^5\Big)\,.\label{betafinal}
\end{alignat}
}
\normalsize \noindent Here, $\GGG_{i}$ and $\GGG_{\text{Tele}}$ denote the constant Taylor coefficients at the background level for $G_i$ and $G_{\rm Tele}$ (see Eqs.~\eqref{HG_2}-\eqref{HG_5} and \eqref{Ltele}).

\section{Equation of State terms for \texorpdfstring{$f(T,B)$}{fTB} gravity} \label{sec:f_T_B_quantities}

We include some definitions that are used in the derivation of $f(T,B)$ cosmology in Sec.~\ref{sec:f(TB)_test} which originated in Ref.~\cite{Escamilla-Rivera:2019ulu}.

\noindent General Taylor expansion model:
\begin{eqnarray}
    w_{x_1}&=& \left[\left(4 A_2+27 A_3+11 A_4\right) \ddot{a}(t)-\left(4 A_2+12 A_3+7 A_4\right) \dot{a}(t)\right]\,, \\
    w_{x_2}&=& \left[4\left(3 A_3+A_4\right) \ddot{a}(t)^2+\dot{a}(t)^2+4 a^{(3)}(t) \left(2 A_3 \dot{a}(t)-A_4 \ddot{a}(t)\right)\right]\,, \\
    w_{x_3}&=& \left[\left(9 A_3+2 A_4\right) a^{(3)}(t) \dot{a}(t)+\ddot{a}(t) \left(\left(4 A_2+3 A_4\right) \ddot{a}(t)+2 \left(2 A_2+9
    A_3+4 A_4\right) \dot{a}(t)\right)\right]\,, \\
    w_{x_4}&=& -\left[\left(4 A_2+9 A_3+6 A_4\right) \ddot{a}(t)+\left(4 A_3+A_4\right) \dot{a}(t)\right]\,, \\
    w_{x_5}&=& 72 \left(4 A_2+12 A_3+7 A_4\right) \dot{a}(t)^5+6 a(t)^3 \dot{a}(t)\left(12 A_3 a^{(3)}(t)+\dot{a}(t)\right)\,, \\
    w_{x_6}&=& 72 \left(3 A_3+A_4\right) a(t)^2 \dot{a}(t)^2 \left[\ddot{a}(t)-a^{(3)}(t)\right]+a(t)^5\left[A_3 B^2+T \left(A_4 B+A_2 T+A_1\right)+A_0\right]\,, \nonumber\\
    &&\\
    w_{x_7}&=& 8 A_2+36 A_3+17 A_4\,, \\
    w_{x_8}&=& a(t)^5 \left[-\left(A_3 B^2+T \left(A_4 B+A_2T+A_1\right)+A_0\right)\right]\,,
\end{eqnarray}

\begin{eqnarray}
    w(z)_1&=& -\frac{A_3 B^2+T \left(A_4 B+A_2 T+A_1\right)+A_0}{(z+1)^5}\,, \\
    w(z)_2&=& \frac{\frac{2}{(z+1)^3}-\frac{96A_3}{(z+1)^5}}{(z+1)^4}\,,\\
    w(z)_3&=& \frac{\left(8 A_2+36 A_3+17 A_4\right)}{(z+1)^{10}}\,, \\
    w(z)_4&=& \frac{\frac{2 \left(4A_2+27 A_3+11 A_4\right)}{(z+1)^3}+\frac{4 A_2+12 A_3+7 A_4}{(z+1)^2}}{(z+1)^7}\,, \\
    w(z)_5&=& \frac{\frac{6 \left(9 A_3+2A_4\right)}{(z+1)^6}+\frac{2 \left(\frac{2 \left(4 A_2+3 A_4\right)}{(z+1)^3}-\frac{2 \left(2 A_2+9 A_3+4
    A_4\right)}{(z+1)^2}\right)}{(z+1)^3}}{(z+1)^4}\,, \\
    w(z)_6&=& \frac{\frac{16 \left(3A_3+A_4\right)}{(z+1)^6}-\frac{24 \left(-\frac{2 A_3}{(z+1)^2}-\frac{2
    A_4}{(z+1)^3}\right)}{(z+1)^4}+\frac{1}{(z+1)^4}}{(z+1)^3}\,, \\
    w(z)_7&=&- \frac{\frac{72 A_3}{(z+1)^4}+\frac{1}{(z+1)^2}}{(z+1)^5}\,, \\
    w(z)_8&=& \frac{\left(3A_3+A_4\right) \left[\frac{2}{(z+1)^3}+\frac{6}{(z+1)^4}\right]}{(z+1)^6}\,, \\
    w(z)_9&=& \frac{4 A_2+12 A_3+7A_4}{(z+1)^{10}}\,,\\
    w(z)_{10}&=& \frac{\frac{4 A_3+A_4}{(z+1)^2}-\frac{2 \left(4 A_2+9 A_3+6A_4\right)}{(z+1)^3}}{(z+1)^7}\,.
\end{eqnarray}

\noindent Power law model:
\begin{eqnarray}
    w_{x_1}&=&\left\{(k-1) a(t)^5 \dddot{a}(t)+48 \dot{a}(t)^6-30 a(t)
    \dot{a}(t)^4 \ddot{a}(t)+a(t)^4 \ddot{a}(t) \left[\multidots{4}{a}(t)+3 (k-1) \dot{a}(t)\right]
    \right. \nonumber \\ && \left. -a(t)^2 \dot{a}(t)^2 \left[21 \ddot{a}(t)^2+2 \dddot{a}(t)
    \dot{a}(t)\right]+a(t)^3 \left[3 \ddot{a}(t)^3+\dot{a}(t) \left(2 \dot{a}(t) \left(\multidots{4}{a}(t)-2 (k-1) \dot{a}(t)\right)
    \right.\right.\right. \nonumber\\&& \left.\left.\left.
    -\dddot{a}(t) \ddot{a}(t)\right)\right]\right\}\,, \\
    w_{x_2}&=& \left[\dot{a}(t)^2-a(t) \ddot{a}(t)\right] \left\{\dot{a}(t)^2-a(t)\left[\ddot{a}(t)+\dot{a}(t)\right]\right\}\,, \\
    w_{x_3} &=& b_0 3^{k+1} (k-1) k \dot{a}(t)^3 \left[a(t)^2
    \dddot{a}(t)-4 \dot{a}(t)^3+3 a(t) \dot{a}(t) \ddot{a}(t)\right] \left[\frac{2 a(t) \ddot{a}(t)+4 \dot{a}(t)^2}{a(t)^2}\right]^k\,, \\
    w_{x_4}&=& \frac{t_0 2^{m+2} 3^m
    (m-1) m \left[\frac{\dot{a}(t)^2}{a(t)^2}\right]^m \left[a(t) \ddot{a}(t)-\dot{a}(t)^2\right]
    +\frac{w_{x_3}}
    {\left[a(t) \ddot{a}(t)+2
    \dot{a}(t)^2\right]^2}}{a(t) \dot{a}(t)}\,, \\
    w_{x_5}&=& \frac{b_0 6^k (k-1) k \dot{a}(t) \left[-a(t)^2 \dddot{a}(t)+4 \dot{a}(t)^3-3 a(t) \dot{a}(t) \ddot{a}(t)\right]
    \left[\frac{a(t) \ddot{a}(t)+2 \dot{a}(t)^2}{a(t)^2}\right]^k}{\left[a(t) \ddot{a}(t)+2 \dot{a}(t)^2\right]^2}\,.
\end{eqnarray}

\begin{eqnarray}
    w(z)_1&=&\left\{-\frac{6 (k-1)}{(z+1)^9}+\frac{2
    \left[\frac{24}{(z+1)^5}-\frac{3 (k-1)}{(z+1)^2}\right]}{(z+1)^7}+\frac{\frac{24}{(z+1)^9}-\frac{\frac{12}{(z+1)^7}-\frac{2
    \left[\frac{2 (k-1)}{(z+1)^2}+\frac{24}{(z+1)^5}\right]}{(z+1)^2}}{(z+1)^2}}{(z+1)^3}-\frac{108}{(z+1)^{12}}\right\}\,, \nonumber\\
    &&\\
    w(z)_2&=& \left[\frac{1}{(z+1)^4}-\frac{\frac{2}{(z+1)^3}-\frac{1}{(z+1)^2}}{z+1}\right]\,, \\
    w(z)_3&=& \left\{\frac{b_0 2^{3 k-1} 3^{k+1} (k-1) k \left[\frac{1}{(z+1)^2}\right]^k}{(z+1)^4}+\frac{t_0 2^{m+2} 3^m (m-1) m \left[\frac{1}{(z+1)^2}\right]^m}{(z+1)^4}\right\}\,.
\end{eqnarray}

\section{Background Cosmology -- Methods and Solutions}

This appendix summarizes some basic methods that are used in Sec.~6 as well as some definitions regarding bouncing solutions.

\subsection{Reconstruction Methods} \label{sec:reconstruction}

The reconstruction method attempts to reconstruct the gravitational Lagrangian based on observations or desired cosmological behaviors stemming from, say, specific choices of $a(t)$ or $H(t)$. In this way, the construction of a Lagrangian becomes somewhat physically motivated. Generally, this approach involves solving partial differential equations which yield the desired Lagrangian \cite{Sotiriou:2008rp,DeFelice:2010aj}. However, they are usually not generally solvable or yield Lagrangian solutions which are extensively complex thereby limiting their applications in other sectors such as smaller scale physics, for instance in galactic dynamics and the solar system scale. In addition, the reconstructed Lagrangian is by no means necessarily viable. While it may recover the observed cosmological late-time background expansion, it may not necessarily conform with early universe observations. Thus, the results from this approach should be treated with caution.

As the general functional form of the Hubble parameter is not known for the whole of the Universe's history, the reconstruction procedure can instead be applied for specific domination epochs. In this way, the approximate dominant behavior of the Lagrangian during those epochs is obtained \cite{Capozziello:2005ku}. Moreover, a more complete description of the gravitational Lagrangian can be achieved by matching the respective sequential approximate behaviors \cite{Nojiri:2006gh,Nojiri:2006be}. Nonetheless, comparisons with observational data especially concerning the evolution of the Hubble parameter can be used as means to reconstruct the Lagrangian in a model independent way. A detailed review on the topic is provided in Sec.~10.

Reconstruction has been applied in an extended number of works, including $f(\lc{R})$ \cite{Nojiri:2008nt,Nojiri:2010wj,Nojiri:2009kx}, $f(\lc{R},\lc{\mathcal{G}})$ \cite{Bamba:2009uf}, $f(\lc{R},\Theta)$ \cite{Momeni:2011am,Houndjo:2011tu,Sharif:2014jpa}, non-local \cite{Koivisto:2008xfa,Deffayet:2009ca,Elizalde:2012ja,Vernov:2012nr,Roobiat:2017wkx}, higher-order curvature contributions \cite{Simon:2004tf}, and unimodular theories \cite{Houndjo:2017jsj,Rajabi:2017alf}. Further information on unimodular and mimetic reconstruction is available in Ref.~\cite{Sebastiani:2016ras}. Therefore, it is only natural to explore this approach in the context of \gls{tg}. For the forthcoming reviewed works, the following main cosmological behaviors have been explored:

(a) \textbf{$\Lambda$\gls{cdm}} -- The popular model which stood for quite a long period of time as the one to match with observations. Fundamentally, the model assumes a contribution from matter (\gls{cdm} and baryonic), radiation and a positive cosmological constant. In most situations which follow, only its late-time behavior is considered which effectively suppresses the contribution arising from radiation leading to the simplified Friedmann equation
\begin{equation}
H^2 = {H_0}^2\left(\Omega_{\rm m0} a^{-3} + \Omega_\Lambda\right)\,.
\end{equation}

(b) \textbf{de Sitter and Einstein static universes} -- When the Hubble parameter is a constant value, depending on its magnitude, the Universe can experience distinct evolutionary behaviors:
\begin{enumerate}[label = (\roman*{})]
\item $H > 0$ (de Sitter): The Universe expands at an accelerating rate and is commonly associated with a positive cosmological constant. This behavior is also used in the context of inflation to generate the required exponential expansion of 60 $e$-folds \cite{Liddle:2003as,Remmen:2014mia,Akrami:2018odb};
\item $H = 0$ (Einstein static): The Universe becomes static. Formulated by Einstein in 1917 \cite{Einstein:1917ce}, a static behavior can be achieved in the presence of a cosmological constant and spatial curvature subject to the conditions
\begin{equation}
    \Lambda = \dfrac{1+3w}{2}\kappa^2 \rho_{0}\,, \quad k = \dfrac{1+w}{2}\kappa^2 \rho_{0}\,,
\end{equation}
where $\rho_{0}$ represents the present matter energy density having associated \gls{eos} $w > -\frac{1}{3}$. Trivially, the static behavior is achieved in vacuum. In the presence of fluids, however, the above conditions imply that the Universe has to be closed. In the context of the reconstruction of general theoretical frameworks, the flatness condition can be maintained in many circumstances.

Naturally, following observations by Hubble, the idea was abandoned. Nonetheless, the model still remains of interest \cite{Goswami:2008fs,Boehmer:2003iv,Clifton:2005at} especially around the very early Universe. Small perturbations can induce an unstable expansion (contraction) leading to an accelerating (Big Crunch) universe \cite{Barrow:2003ni}, an aspect extensively studied in various modified theories of gravity \cite{Boehmer:2007tr,Goheer:2008tn,Shabani:2016dhj,Bohmer:2009fc,Atazadeh:2015zma,Parisi:2007kv}. Thus, the Einstein static universe serves as a possible alternative in realizing inflation without a Big Bang singularity (i.e. the emergent universe) \cite{Ellis:2002we,Ellis:2003qz,Wu:2009ah,Debnath:2008nu,Mulryne:2005ef}.

\end{enumerate}

(c) \textbf{Power-law scale factor} -- A popular simple model is the power-law scale factor $a(t) \propto t^\alpha$ for some constant $\alpha$, commonly associated with a perfect fluid domination era following the identification $\alpha = \frac{2}{3(1+w)}$. The model has been used in various cosmological contexts, ranging from describing late-time acceleration \cite{Sethi:2005au,Cai:2007us,Kaeonikhom:2010vq} to fitting \gls{sn}e~Ia data \cite{Dev:2000du,Dev:2002sz,Nielsen:2015pga}. It has also been applied in inflation, better known as power-law inflation \cite{Abbott:1984fp,Burd:1988ss,Barrow:1990vx,Lucchin:1984yf}. However, the latter is not favourable due to a number of observational and phenomenological issues \cite{Halliwell:1986ja,Liddle:1992wi,Souradeep:1992sm,Stewart:1993bc,Unnikrishnan:2013vga,Akrami:2018odb}.

The reconstruction procedure is not only limited to \textit{a priori} choices of the cosmic expansion. Ultimately, the latter is determined by the different matter constituents present in the universe. In the late-time accelerating universe, this is attributed to the domination of \gls{de}. As outlined in Sec.~6.1.3, \gls{tg} theories can act as a gravitational source which represent \gls{de}. Thus, in the aspect of reconstruction, \textit{a priori} choice of the \gls{de} fluid can be alternative means to reconstruct the gravitational Lagrangian. In turn, if the \gls{de} model is able to match with observations, the respective reconstructed model may yield a viable model at least at a background level.

This approach in the context of \gls{tg} has been mostly considered for Holographic Dark Energy (\gls{hde}) and its extensions, Quantum Chronodynamics (\gls{qcd}) \gls{de} and Pilgrim Dark Energy (\gls{plde}). The respective energy density can be encapsulated in the following form
\begin{equation}
    \rho \propto L^{-s},
\end{equation}
where choices of the IR cut-off length, $L$, and the constant $s$ determine the nature of the \gls{de} fluid:

(a) \textbf{\gls{hde}} $(s = 2)$ -- The theory stems from quantum field theory following the requirement that the quantum zero-point energy density in a region of size $L$ should not exceed the mass of a black hole having the same size. Originally, \gls{hde} was met with various issues including violation of the second law of thermodynamics \cite{Zhang:2005hs,Huang:2004ai}, instabilities \cite{Setare:2007mp,Jawad:2012xy}, and causality \cite{Gao:2007ep,Zhang:2009un} (can be avoided through appropriate choices of $L$ \cite{Gao:2007ep,Zhang:2009un,Feng:2008hk,Trockel:2011sd}), despite agreement with observations \cite{Zhang:2005hs,Chang:2005ph,Zhang:2006qu,Zhang:2007sh,Li:2009bn,Guo:2018ans}. Reconstruction has been applied in numerous works \cite{Jawad:2012xy,Wu:2007tn,Karami:2010aq,Houndjo:2011fb,Zhang:2006av,Setare:2007hq}. Further \gls{hde} extensions include $(m,n)$ corrected \gls{hde} \cite{Ling:2012gu,Ghosh:2014mva} as well as entropy corrections \cite{Karami:2010aq,Ghosh:2014mva,Wei:2009kp}. Meanwhile, extensions such as Tsallis \gls{hde} \cite{Tavayef:2018xwx} or Barrow \gls{hde} \cite{Barrow:2020tzx} have shown the ability to match with SNIa and cosmic chronometers data \cite{Saridakis:2018unr,Anagnostopoulos:2020ctz}.

(b) \textbf{\gls{qcd}} $(s = 1, L = H^{-1})$ -- The model originates from a quantum field theory approach where a Veneziano ghost \cite{Veneziano:1979ec} is constructed as a means to solve the $U(1)_A$ problem in low-energy \gls{qcd}. This ghost has an unphysical nature in Minkowski spacetime but yields an observable effect in dynamical spacetime (like \gls{flrw}) \cite{Urban:2009vy,Urban:2009wb,Urban:2009yg,Ohta:2010in}. In its simplicity, its energy density takes the form $\rho_\text{QCD} \sim \Lambda_\text{QCD}^3 H$ where $\Lambda_\text{QCD} \sim \SI{100}{\mega\electronvolt}$ is some \gls{qcd} mass scale which yields a value comparable to the cosmological constant and hence can resolve the fine tuning problem \cite{Cai:2010uf}. In the standard context, this behaves as a quintessence fluid starting with $w_\text{QCD} = -\frac{1}{2}$ at early-times leading to $w_\text{QCD} \to -1$ at late-times. However, this results in instabilities as the square sound speed is always negative \cite{Cai:2010uf,Ebrahimi:2011js}. \gls{qcd} can also be reconstructed via other sources including $k$-essence \cite{RozasFernandez:2011je} and quintessence scalar fields \cite{Sheykhi:2011nb}.

(c) \textbf{\gls{plde}} $(s \leq 2$ and has phantom-like behaviour) -- \gls{plde} revisits the ideology of \gls{hde} by constructing a \gls{de} model which avoids the formation of black holes \cite{Wei:2012wt}. It can interact with other sources \cite{Sharif:2013ona,Jawad:2014qoa,Sharif:2013lra,Jawad:2015awa} and can be sourced by gravity or scalar fields through reconstruction procedures \cite{Debnath:2014wga,Jawad:2015edc,Sharif:2013sva}. Logarithmic and power-law entropy corrections have also been investigated in literature \cite{Debnath:2014wga,Saha:2015bbx}. 

Viability is crucial as means to obtain reliable models. There is no merit in reconstructing for the Lagrangian which is then unable to match observations or suffer from other pathologies. Thus, due to the large number of diverse reconstructed Lagrangian models, constraints become necessary to identify the physically viable solutions.

A common constraint which appears in literature is the requirement that the reconstructed gravitational Lagrangian hosts vacuum solutions such as Minkowski spacetime. Generally, this condition requires that when the relevant gravitational scalars become null (e.g.~$T \to 0$), the field equations yield Minkowski spacetime. As a consequence, this usually results in having a vanishing gravitational Lagrangian or, equivalently, that it does not emanate any cosmological constant. The vacuum solution, or vanishing Lagrangian condition for the Minkowski spacetime, also appears in curvature-based theories such as $f(\lc{R})$ gravity \cite{delaCruzDombriz:2006fj} as well as in teleparallel contexts \cite{Ferraro:2011ks,Tretyakov:2016uvv,delaCruz-Dombriz:2017lvj,delaCruz-Dombriz:2018nvt,Caruana:2020szx}. Thus, the condition is a physical one rather than based on theoretical considerations.

Teleparallel models which constitute more than one variable such as $f(T,B)$ and $f(T,T_G)$ gravity usually lead to partial differential equations which are not generally solvable hence requiring the use of some functional ansatz. In particular, one popular ansatz is the separable additive type (e.g. $f(T,B) = g(T) + h(B)$) which normally leads to the Friedmann equations to become separable hence reducing the equation to a system of ordinary differential equations. However, whenever this approach leads to such a result, an issue arises since the functions depend on two variables and the reconstruction form might not be unique. This instance occurs whenever an invertible relation between the coordinate time $t$ and the teleparallel gravitational scalars exists, hereby creating an element of freedom of choice~\cite{delaCruzDombriz:2011wn,Bahamonde:2016cul,Zubair:2018wyy,Paliathanasis:2017flf}.

\subsection{Noether symmetry}\label{sec:noether_symm}

The investigation of an extensive number of gravitational models is a daunting task especially without prior knowledge or hint of its possible form. While select criteria to further restrict the possible choices of the Lagrangian can be imposed, there is still an element of freedom in choosing its form. Furthermore, even if the models happen to conform with observations, they are by no means necessarily viable. In addition, their construction may lack physical motivation and therefore raises the question of how can one consistently deem which model is more reliable than another.

In the following, a more geometric approach is explored using Noether symmetries \cite{Noether:1918zz}. Systems exhibiting such symmetries lead to conservation laws usually denoting important physical information of the system such as conservation of energy, linear or angular momentum. In this sense, imposing the existence of Noether symmetries offers a more physically motivating approach in constructing gravitational models \cite{Paliathanasis:2011jq,Dialektopoulos:2018qoe,Tsamparlis:2018nyo}.

However, models originating from Noether symmetries are not necessarily viable as they still need to be confronted against observations and other tests \cite{Will:2001mx}. Furthermore, there is no imposition that the symmetry has to be a Noether symmetry. Galileon \cite{Nicolis:2008in,Deffayet:2009wt,deRham:2010ik,Deffayet:2011gz,Clifton:2011jh} and Hojman symmetry \cite{Hojman:1992anc,Capozziello:2013bma,Wei:2015xax,Paolella:2015gma,Paliathanasis:2016lxr,Wei:2015oua} are two examples of other possible avenues.

In the context of cosmology and modified gravity, Noether's theorem has been studied extensively including $f(\lc{R})$ gravity \cite{Capozziello:2008ch,Capozziello:2008ima,Paliathanasis:2011jq} (in $(n+1)$ dimensions \cite{Vakili:2008ea} or in the presence of a tachyon field \cite{Jamil:2011pv}), $f(\lc{R},\lc{\mathcal{G}})$ gravity \cite{Dialektopoulos:2018qoe} and non-local gravity \cite{Capozziello:1999xs}. As an illustrative example, in $f(\lc{R})$ gravity, the use of Noether symmetry constrains the Lagrangian to be $f(\lc{R}) \propto \lc{R}^{3/2}$ with an associated cosmological behavior \cite{Capozziello:2008ima}
\begin{equation}
a(t) = \sqrt{a_4 t^4 + a_3 t^3 + a_2 t^2 + a_1 t}\,,
\end{equation}
where $a_i$ are constants. This is an interesting exact solution as appropriate parameter fixing may realize viable cosmological behaviors such as transition periods and domination epochs. Other Noether symmetry derived $f(\lc{R})$ models were also checked whether they satisfy the respective viability conditions, but one should also determine whether these models satisfy lower scale tests of gravity such as in the solar system as well as galactic scale physics which are critical to ascertaining an exhaustive model of gravity that can describe different levels of dynamics across the various phenomenological tests available \cite{DeFelice:2010aj}.

In the presence of scalar fields, minimal \cite{Capozziello:2009te,Basilakos:2011rx} and non-minimal \cite{Capozziello:1994du} couplings have been investigated while the relation between Jordan and Einstein frames has been explored in Ref.~\cite{Capozziello:1996xg}. Here, the Noether symmetry approach serves as a means to constrain the form of the potential or coupling of the scalar field.

The role of Noether symmetries can also be associated with quantum cosmology (see Refs.~\cite{Capozziello:1994du,Capozziello:2012hm} and references therein). The Wheeler-de Witt equation \cite{DeWitt:1967yk,wheeler1968wdw} is one attempt to unify the effects of gravity and quantum mechanics. In a cosmological context, it can help to provide the necessary initial conditions to yield a classical, observable Universe provided that the Universe's wave function is obtained. In particular, the Hartle criterion \cite{hartle1986criterion} provides a mechanism as a means to identify when such classical universes are generated. As discussed in Refs.~\cite{Capozziello:1994du,Capozziello:2012hm}, these classical universes correlate to Noether symmetries since the resulting wave function obeys said criterion.

Before examining the works carried out in the context of teleparallel based theories, a brief analytical outlook is outlined that focuses on the Euler-Lagrange equations formulation as opposed to the field theory approach (further details are provided in Refs.~\cite{Capozziello:1996bi,Capozziello:2012hm,Capozziello:2007wc,Dialektopoulos:2018qoe}). For further information and application of the latter, see Refs.~\cite{Wald:1993nt,Bak:1993us,Iyer:1994ys,Aros:1999id,Deruelle:2003ps,Papadimitriou:2005ii}.

We start by considering a Lagrangian $\mathcal{L}(t, q^i, \dot{q}^i)$ which depends on some time parameter $t$ and coordinates $q^i$ with associated velocities $\dot{q}^i$. The Hessian \cite{ARFKEN2013469} determinant of this Lagrangian is then assumed to satisfy the condition $\det \left|\left|\frac{\partial^2\mathcal{L}}{\partial \dot{q}^i \partial \dot{q}^j} \right|\right| \neq 0$. Taking infinitesimal point transformations
\begin{equation}
\bar{t} = t + \epsilon \,\xi(t,q^i) + \mathcal{O}(\epsilon^2)\,, \qquad \bar{q}^i = q^i + \epsilon\, \eta^i(t,q^i) + \mathcal{O}(\epsilon^2)\,,
\end{equation}
where $\epsilon$ represents a small (order) parameter, and $\xi, \eta^i$ represent point transformations of $t$ and $q^i$ respectively, the Lagrangian is said to exhibit a Noether symmetry if the Lagrangian remains invariant under such transformations. The corresponding symmetry is expressed through the Noether symmetry vector
\begin{equation}
\textbf{X} = \xi \partial_t + \eta^i \partial_{q^i}\,.
\end{equation}
Formally, a Lagrangian exhibits a Noether symmetry if there exists a function $g(t,q^i)$ such that the relation
\begin{equation}\label{eq:Noether-condition-relation}
\textbf{X}^{[1]} \mathcal{L} + \mathcal{L} \frac{\dd \xi}{\dd t} = \frac{\dd g}{\dd t}\,,
\end{equation}
holds \cite{Capozziello:1996bi}. Here, $\textbf{X}^{[1]} \equiv \textbf{X} + \eta^{i,[1]} \partial_{\dot{q}^i}$ is the first prolongation of the symmetry generator \textbf{X} and $\eta^{i,[1]} \coloneqq \frac{\dd}{\dd t}(\eta^i - \dot{q}^i \xi) + \ddot{q}^i \xi$. Consequently, each Noether symmetry gives rise to a constant of motion
\begin{equation}
\Sigma_0 = \xi \left(\dot{q}^i \frac{\partial\mathcal{L}}{\partial \dot{q}^i} - \mathcal{L}\right) - \eta^i \frac{\partial \mathcal{L}}{\partial \dot{q}^i} + g\,.
\end{equation}
In simple cases, the function $g$ is chosen to be zero at the expense of restricting the possible number of symmetries the theory can exhibit. Furthermore, the time symmetry yields the energy conservation constraint.

In order to obtain the Noether symmetries of a given gravitational Lagrangian, Eq.~\eqref{eq:Noether-condition-relation} is solved. This relation yields a homogeneous polynomial expression in terms of the velocities which can only be satisfied when the velocity coefficients are zero as none of the symmetry generator components depend on the velocities \cite{Christodoulakis:2013xha}. This leads to an overdetermined system of partial differential equations which, when solved, result into the Noether symmetries. In the special case when the Lagrangian is explicitly time independent, this homogeneous polynomial is of second degree in velocities plus an inhomogeneous term in $q^i$ \cite{Capozziello:1996bi}.

Once a symmetry is obtained, together with the associated constants of motion, the Euler-Lagrange equations can then be solved which, in this case, yields the cosmological evolution. However, this step is not straightforward. As an alternative, the coordinates can then be transformed into a new set where one (or more) coordinates become cyclic \cite{Capozziello:1996bi,Capozziello:2012hm} in the aims of simplifying the problem.

\subsection{Dynamical systems}\label{sec:dyn_sys}

Up to this point, the gravitational formulation can be determined through a choice of the gravitational Lagrangian or arises from particular symmetry considerations. However, it is not necessarily the case that the resulting model is capable of realizing a viable cosmology in terms of observational cosmology \cite{DiValentino:2020vhf}. While the reconstruction approach attempts to tackle this issue via observations or appropriate choices of scale factors, it still has its limitations. Therefore, one asks the following questions
\begin{enumerate}
\item Given a gravitational model, what is the resulting cosmology? Does the behavior change depending on the free parameters?
\item Does the chosen model always realize a particular era or sequence of eras, irrespective of the chosen parameters?
\end{enumerate}
While one may attempt to answer these questions using the aforementioned methods, the reality is that it is a difficult task to achieve. Instead, the dynamical system analysis approach can provide a further insight into these questions.

Dynamical systems allow for the extraction of key main features of the cosmology without solving the evolution equations directly (in an exact form). Thus, it then becomes possible to describe the overall nature of the gravitational theory and henceforth determine whether the model can generate a viable cosmological evolution. This therefore serves as a very useful tool especially in models where it is difficult to extract any cosmological solutions from directly solving the field equations such as in $f(\lc{R})$ gravity. The dynamical systems approach has been applied in \gls{gr} \cite{wainwright1997dynamical,coley2003dynamical} and in a diverse number of curvature-based theories (see \cite{DeFelice:2010aj,Sotiriou:2008rp} and references therein, among others).

To explore the implications of the approach, a short introduction to the method is provided. For further information, see Refs.~\cite{strogatz1994nonlinear,wiggins2003nonlinear,Boehmer:2014vea,Bahamonde:2017ize}. Consider a set of variables $x_i \in \mathbb{R}$, $i = 1, \, \dots{}, \, n$ which evolve \gls{wrt} some time variable $t$ according to the relations
\begin{equation}
    \dot{x}_i = f_i(x_1, \, \dots{}, \, x_n)\,,\label{eq:DSeq111}
\end{equation}
where $f_i(x_1, \, \dots{}, \, x_n)$ are functions describing the evolutionary behavior. Generally, these evolution equations are not easily solvable. However, the solutions may contain special behavioral forms which could be of interest. One way to extract these solutions is by finding the critical (fixed) points of the system, which correspond to solutions when $f_i(x_j) = 0$ or, equivalently, $\dot{x}_i = 0$. As critical points only represent a point-solution of the system, it is then necessary to investigate the respective dynamical behavior to see how the system evolves given a set of initial conditions.

Generally speaking, critical points are said to be stable if any nearby solution stays next to the critical point (but not necessarily evolve towards it) while it is asymptotically stable if the nearby solutions evolve towards the critical point. Otherwise, the point is considered to be unstable. It is precisely these stability conditions which highlight the strength of the dynamical system approach since, without finding the explicit analytical (or numerical) behavior of the system, the overall evolution of the system can still be described by simply following the trajectories of the resulting phase-space. A summary of these characteristics are detailed in Table~\ref{table:critical_points_behavior}.

\begin{table}[H]
    \centering
    \midsepremove
    \begin{tabularx}{\textwidth}{Xc}
    \toprule
    \cellcolor{gris3}\textbf{Eigenvalues of the Jacobian matrix} &
    \cellcolor{gris3}\textbf{Stability} \\ \midrule
    \cellcolor{gris2}All real parts are positive & \cellcolor{gris2}Unstable \\
    \cellcolor{gris1}All real parts are negative & \cellcolor{gris1}Stable \\
    \cellcolor{gris2}At least two eigenvalues have real parts with opposite signs & \cellcolor{gris2}Saddle\\
     \cellcolor{gris1}At least one eigenvalue has a zero real part & \cellcolor{gris1}Linear stability fails/Further analysis\\
    \bottomrule
    \end{tabularx}
    \midsepdefault
    \caption{Generic summary of behavior and stability nature of critical points. }
    \label{table:critical_points_behavior}
\end{table}

Stability can be obtained following a linear stability approach by finding the eigenvalues of the Jacobi matrix of the functions $f_i$ evaluated at the critical point (see also the Hartman-Grobman theorem \cite{Grobman:1959dm,Hartman:1960p,wiggins2003nonlinear}). Depending on the magnitudes of the eigenvalues, then the following stability behaviors are achieved:
\begin{enumerate}
\item If none of the eigenvalues have a zero real part, the points are hyperbolic and are classified as follows:
\begin{enumerate}
\item if they all have a negative real part, then the point is stable (or an attractor),
\item if they all have a positive real part, the point is a repeller,
\item if at least one eigenvalue has a positive real part, then the point is a saddle. These points can either attract or repel trajectories depending on the solutions;
\end{enumerate}
\item Otherwise, the points are non-hyperbolic. In this case, linear stability is not sufficient to determine stability and other approaches need to be considered to determine the behavior (for instance, centre manifold theory and Lyapunov's functions \cite{wiggins2003nonlinear,carr1982centre}).
\end{enumerate}

While the choice of the cosmological phase-space variables used in the approach are not defined following some procedure, they are usually chosen to be dimensionless and encapsulate the time evolution of the Hubble parameter as well as the dynamics of any scalar fields present in the theory under study. The autonomous system is then constructed based on the Friedmann equations. In certain cases, however, the chosen phase-space may be unbounded with the possibility of critical points appearing on the phase-space boundary. These points require other stability analysis techniques to fully understand their behavior and nature. One approach is Poincar\'{e}'s projection method (see Ref.~\cite{Xu:2012jf} and references therein) where a new set of phase-space variables, referred to as Poincar\'{e} variables, are introduced in order to compactify the phase-space. In turn, the critical point behavior of the points appearing at infinity can then be determined.

As is standard, finding an optimal gravitational model can be a tedious task, however, selecting viable cosmological models can be done by considering a sequence of cosmological properties, such as: \cite{Boehmer:2014vea}
\begin{enumerate}[label = (\alph*{})]
    \item An early-time inflationary period with graceful exit (not eternal);
    \item Radiation domination followed by matter domination epochs;
    \item The Universe evolves towards an accelerated expansion era representing the late-time behavior. This is an attractor if it represents the final cosmological state or a saddle if a later cosmological state is reached (for example, the Big Rip scenario).
\end{enumerate}
For the remainder of the section, models satisfying the above three requirements shall be considered as good gravitational models.

\subsection{Bouncing Cosmologies}\label{sec:boun_cosmo}

The Big Bang singularity, a state where the Universe births from a vanishing volume, is a well known problem in cosmology, especially in the context of \gls{gr}. As stipulated by the Penrose-Hawking singularity theorems \cite{Penrose:1964wq,Hawking:1967ju}, the field equations break down at the Big Bang \cite{Brandenberger:1993ef,Novello:2008ra,Brandenberger:2016vhg}. Additionally, the trans-Planckian problem is another important issue. It describes the fact that the origin of modes during inflation come from a time before it when the Universe's physical size is smaller than a Planck length, hence breaking down the classical nature of \gls{gr} \cite{Brandenberger:2012um,Brandenberger:2016vhg}. Thus, motivation for seeking alternative mechanisms to avoid the formation of such a singularity and account for the trans-Planckian problem followed.

One major development is the consideration of so called bouncing cosmologies. Here, the Universe does not necessarily have a beginning or else is described by a behavior which replaces the common Big Bang paradigm. Indeed, the initial singularity is replaced by a cosmological bounce, an era where the Universe experiences a contraction to a non-zero minimal size which then re-expands. Equivalently, if the bounce occurs at some time $t_B$, it can be described via the Hubble parameter through $H(t < t_B) < 0$, $H(t_B) = 0$ and $H(t > t_B) > 0$ (or $\dot{H} > 0$). The converse also represents a bounce, the cosmological turnaround, where the Universe expands to some maximal size after which it starts to contract again (hence corresponding to $\dot{H} < 0$). Alternatively, a singular Hubble parameter can also host a bounce provided that its signature changes pre- and post- the singularity. One such instance is the superbounce scenario described by $a(t) \propto (t_c-t)^{\frac{2}{\tilde{c}^2}}$, where $t_c$ represents the crunch time and $\tilde{c} > \sqrt{6}$ \cite{Koehn:2013upa,Oikonomou:2014yua}.

The nature of a cosmological bounce extends beyond the relatively simple aforementioned notions, with various models constructed to tackle diverse scenarios both at the beginning and the possible end of the Universe. Here, only a brief description of the encountered bouncing scenarios in \gls{tg} is provided. For detailed reviews on bouncing cosmologies, we refer the readers to Refs.~\cite{Novello:2008ra,Battefeld:2014uga,Brandenberger:2016vhg}.
\begin{enumerate}[label = (\alph*{})]
\item \textbf{Symmetric Bounce} -- Described by $a(t) \propto e^{f_0 t^2}, \, f_0 > 0$, it has been used to replace the Big Bang singularity. However, it hosts issues with primordial modes \cite{Bamba:2013fha,Nojiri:2016ygo} hence requiring other processes \cite{Cai:2012va,Cai:2014bea}.
\item \textbf{Cyclic Universe} -- The Universe experiences a continuous series of expansions and contractions which avoids the need of any initial conditions \cite{tolman1934relativity,Steinhardt:2001st}. However, the singularity problem may still persist \cite{Nojiri:2011kd}. This model has been explored in various modified gravity models including $f(\lc{R})$ \cite{Nojiri:2011kd}, massive gravity \cite{Cai:2012ag}, Loop Quantum Cosmology and Brane world models \cite{Steinhardt:2001vw,Steinhardt:2001st,Mukherji:2002ft,Piao:2004hr,Barragan:2009sq}.
\item \textbf{Matter Bounce} -- Originating from Loop Quantum Cosmology, a non-singular cosmology arises due to a contraction era caused by a matter dominated epoch pre-bounce. However, if the pre-bounce contraction era is sourced by a dust fluid, a much larger tensor-scalar ratio arises \cite{Cai:2009fn,Cai:2011tc} together with a red tilted scalar spectrum \cite{Cai:2011tc,Cai:2013kja}. This can be resolved through fluids with \gls{eos} $0 < w \ll 1$ \cite{WilsonEwing:2012pu}. Near the bounce point, the model suffers from BKL instability coming from radiation and anisotropic stresses which can be cured through ekpyrotic processes and scalar fields \cite{Cai:2012va,Cai:2013kja,Cai:2014bea}.
\item \textbf{Past/Future Singularities} -- Commonly referred to as Type I--V singularities, they have been extensively investigated in literature \cite{Nojiri:2004ip,Nojiri:2004pf,Nojiri:2005sx,Cattoen:2005dx,Bamba:2009uf,Capozziello:2009hc,Houndjo:2012ij,Odintsov:2015zza,Odintsov:2015ynk,Dabrowski:2009kg}. One common choice of scale factor ansatz which contains Types I--IV is $a(t) \propto \exp\left[\frac{f_0}{\alpha + 1} (t-t_s)^{\alpha + 1}\right]$ where $t_s$ is the bouncing time, $\alpha \neq 0, \pm 1$ is a constant\footnote{This restriction is such that it excludes the superbounce $(\alpha = -1)$, de Sitter $(\alpha = 0)$ and symmetric bounce $(\alpha = 1)$ scenarios.}, and $f_0 > 0$ is a constant. Briefly, these models are classified as follows with associated range of values for $\alpha$ where applicable:

\begin{enumerate}[label = (\roman*{})]
    \item Type I/Big Rip $(\alpha < -1)$: The Universe expands infinitely in a finite time (i.e. $a(t), \, H(t)$ both diverge) leading to a dissociation of bound structures \cite{Caldwell:2003vq}. Avoidance of the singularity is possible through a generalized Chaplygin gas \cite{GonzalezDiaz:2003bc,BouhmadiLopez:2004me}, quantum effects \cite{Nojiri:2004pf,Nojiri:2005sx,Elizalde:2005ju} or through gravitational modifications \cite{Nojiri:2005vv,Wei:2005fq}. A milder version is the Little Rip ($\alpha > 0$) scenario where the scale factor increases continuously while not approaching the singularity (i.e. $H(t \to \infty) \to \infty$) but still causes a dissolution of bound structures \cite{Frampton:2011sp,Brevik:2011mm,Nojiri:2011kd,Bamba:2012cp}. If the Hubble parameter reaches a finite value at the singularity, a Pseudo-Rip model is produced which describes an asymptotically de Sitter behavior \cite{Frampton:2011aa,Bamba:2012vg,Bamba:2012cp}. While the Little Rip and Pseudo-Rip scenarios are derivatives of the Big Rip cosmology, they are not Type I singularities.
    \item Type II/Sudden ($0 < \alpha < 1$): At the singularity, $a(t)$ is finite but $\ddot{a}$ diverges, leading to a sudden decelerated expansion \cite{Barrow:2004xh,Lake:2004fu} while the Universe still persists beyond the singularity \cite{FernandezJambrina:2004yy,Denkiewicz:2012bz}. A sub-class is the Big Brake singularity where, at the singularity, $a(t)$ is finite, $H \to 0$ and $\ddot{a} \to -\infty$ \cite{Gorini:2003wa}. This describes a universe where the evolution stops.
    \item Type III/Big Freeze ($-1 < \alpha < 0$): At the singularity, $a(t)$ is finite but $\dot{a}$ and $\ddot{a}$ diverge. Originally formulated in Ref.~\cite{Stefancic:2004kb} and later generalized in Ref.~\cite{Nojiri:2005sx}, this bouncing scenario can describe an inflationary epoch as the Hubble radius decreases as it approaches the singularity \cite{Borowiec:2015qrp}.
    \item Type IV/Generalized Sudden ($\alpha > 1$)\footnote{$\alpha$ is commonly defined as $\alpha = \frac{2n+1}{2m+1}$ where $n, \, m \in \mathbb{Z}$, to ensure that all cosmological parameters are well-defined \cite{Nojiri:2015wsa,Odintsov:2015jca,Oikonomou:2015qha,Nojiri:2017ncd}.}: At the singularity, $a(t), \rho$ and $p$ are finite but higher-order derivatives of $H$ diverge. This scenario can describe an inflation graceful exit as the model is unstable and the slow-roll parameters diverge \cite{Oikonomou:2015qha,Nojiri:2015wsa,Odintsov:2015jca,Odintsov:2015ynk}. However, the scalar power spectrum still exhibits problems with primordial modes \cite{Oikonomou:2015qha,Odintsov:2015ynk,Nojiri:2016ygo}.
    \item Type V/$w$-singularity: At the singularity, $a(t)$ is finite, $\rho$ and $p$ vanish but effective \gls{eos} diverges \cite{Dabrowski:2009kg}. While similar to the Big Brake or Type~IV singularity types, its nature is distinct. Additionally, these singularities are weak as they continue to evolve past the singularity \cite{FernandezJambrina:2010ck}. This type emerges if the Taylor expanded scale factor takes on the form $a(t) \sim \sum\limits_{n=3}^\infty (t_s-t)^n$ \cite{FernandezJambrina:2010ck,FernandezJambrina:2010ps}.
\end{enumerate}

\end{enumerate}
Currently, there is significant interest in cosmological models that replace the cosmological singularity (or ``Big Bang'') with a ``Big Bounce'', which denotes a smooth transition from contraction to expansion in order to resolve crucial problems in cosmology. In this direction, theories that analyze the behavior described above are based on more ordinary elements, 3 spatial dimensions, scalar fields and, most importantly, a non-singular bounce that happen at densities below the Planck scale where quantum gravity effects are considered small. Alternatively, this phenomenology can be approached using modified gravity where the corresponding effective stress-energy tensor, which accounts for both the matter fields and the gravitational contribution, must violate the null energy condition. Additionally, avoidance of any instability formation must be guaranteed to ensure a successful bounce. An interesting review of the state-of-the-art of these theories is available in Ref.~\cite{Battefeld:2014uga,Brandenberger:2016vhg}.

In what follows, the aforementioned bouncing solutions shall be explored from the viewpoint of \gls{tg}, namely within the context of $f(T)$, $f(T,B)$, $f(T,T_G)$ and unimodular \gls{tg} theories. Under each model, the existence and formation of such bouncing scenarios shall be explored primarily through the methods of reconstruction and dynamical systems.

\subsubsection{\texorpdfstring{$f(T)$}{} Gravity}\label{sec:bouncing_sols_f_T}

\textbf{(a) $\;$ Using \textit{a priori} $f(T)$ Ansatz} \\

Following Ref.~\cite{Bamba:2012vg}, a selection of $f(T)$ model ansatz were considered to determine whether future bouncing solutions can be realized. During such regimes, $w_\text{eff} \approx w_\text{DE}$ with the latter sourced by torsion. In turn, the conditions for the existence of said bouncing solutions can be more easily determined. For the power-law ansatz $f(T) \propto (-T)^\alpha$, the index classifies the nature of the singularity, being $\alpha < 0$ for Types I, III, $\alpha = \frac{1}{2}$ for Types II, IV (although this does not lead to any cosmological dynamics), and $\alpha > 1$ for Type V. On the other hand, the ansatz $f(T) \propto e^{\lambda T}$ and $f(T) \propto \ln (\gamma T)$ do not host any finite-time singularities. Additionally, the power-law inflation and Little Rip cosmologies are achievable for $\alpha < 0$ or $\alpha = \frac{1}{2}$ whereas the Pseudo-Rip scenario can be realized when $\alpha = \frac{1}{2}$. Furthermore, it was shown that the second law of thermodynamics is satisfied at least for times in the vicinity of the singularity.

The authors further explore the possibility to cure the power-law model from the formation of such future singularities by introducing an additional correction term $T^\beta$. Coupled with the aforementioned $\alpha$ restrictions, the singularities can be cured under the following conditions: $\beta > 1$ for Types I, III, $\beta \neq \frac{1}{2}$ for Types II, IV, and $\beta < 0$ for Type V.

An extension of the former explores whether, given an $f(T)$ ansatz, the singularity can be avoided by an appropriate dynamical \gls{de} behavior \cite{Astashenok:2013kka}. In particular, the avoidance of the Big Rip singularity for the ansatz $f(T) = \alpha (-\frac{T}{6})^\beta$, $\beta > 1$, $\alpha < 0$ was explored. When $|\alpha| \ll (2\beta-1)^{-1}{H_0}^{2(1-\beta)}$, the Friedmann equations lead to $H \sim \rho^{-2\beta}$ at late-times. Setting $p = -\rho - g(\rho)$ with $g(\rho) \sim \rho^\gamma$, the range $\frac{1}{2} < \gamma \leq 1-\frac{1}{2\beta}$ generates the Little Rip cosmology as opposed to Big Rip. \\

\noindent \textbf{(b) $\;$ Bounce Solutions through Reconstruction Approaches} \\

The reconstruction procedure can be useful to determine the necessary conditions for the formation of bouncing scenarios \cite{Astashenok:2013kka}. Taking the presence of a dust fluid as an example, an explicit connection between cosmic time and the Hubble parameter can be established\footnote{To match the signatures as per Ref.~\cite{Astashenok:2013kka}, the transformation $f \to -f$ is necessary.}
\begin{equation}\label{eq:bouncing-fT-recon}
t = -\frac{1}{3}\int \frac{\dd x}{\sqrt{x}} \frac{\dd}{\dd x}\ln \left(2xf_x - f\right) = -\frac{1}{3} \frac{\ln \left(2xf_x - f\right)}{\sqrt{x}}\bigg|_x^{x_0} - \frac{1}{6} \int_{x_0}^x \frac{\ln \left(2xf_x - f\right)}{x^{3/2}} \, \dd x\,,
\end{equation}
where $x = H^2$ and $x_0$ refers to present values. Indeed, it can be noted that:
\begin{enumerate}[label = (\roman*{})]
	\item If the first term converges as $x \to \infty$, a finite-time singularity results; otherwise, a Little Rip arises;
	\item If $x$ approaches a finite value as $t \to \infty$, a Pseudo-Rip occurs.
\end{enumerate}
It is remarked that the quantity $2xf_x - f = 2\kappa^2\rho$ from the $f(T)$ field equations. Thus, the fluid's behavior in conjunction with the above points determines the existence of the singularity. As an example, the model $\rho = 3(x - x_\lambda) e^{-\beta x^{\gamma}}$ generates a finite-time singularity for $\gamma \leq 1/2$ and Little Rip for $\gamma > 1/2$. Correspondingly, an associated reconstructed $f(T)$ function can be derived. In particular, for $\gamma = 1$\footnote{The solution has been corrected from that reported in Ref.~\cite{Astashenok:2013kka}.} we obtain
\begin{equation}
f(x) = 3 \sqrt{\frac{\pi x}{\beta}} (1+2 \beta x_\lambda)\, \text{erf}\left(\sqrt{\beta x}\right)+6 x_\lambda e^{-\beta x}\,,
\end{equation}
which can easily be rewritten in terms of the torsion scalar. It can be observed that when $\beta H^{2\gamma} \ll 1$, the $\Lambda$\gls{cdm} cosmology results. In fact, in this limit, $f(x) \approx 6x + 6x_\lambda$ which is precisely the $\Lambda$\gls{cdm} Lagrangian. This procedure has also been used in Refs.~\cite{Bamba:2012ka,Amoros:2013nxa} with the main crucial difference that $\rho = \rho(H)$ is gravitationally sourced by torsion. Here, reconstructed forms including Little Rip and Big Freeze can be obtained.

Meanwhile, the superbounce cosmology is only realized in the presence of matter fluids leading to $f(T) \propto (-T)^{\frac{3(1+w)}{\tilde{c}^2}}$. For the Loop Quantum Cosmology, ekpyrotic cosmology $a(t) = (a_0 t^2+1)^{\frac{\rho}{2}}$ with $0 < \rho \ll 1$ and $a_0 = \frac{\rho_c}{3\rho^2}$, an analytical solution was not derived. Instead, only the limiting behaviors were obtained: at early-times, $f(T) \sim (-T)^{-3(1+w)}$, whereas at late-times, $f(T) \propto T^{\frac{3(1+w)\rho}{2}}$. It was also found that the models are stable against homogeneous perturbations excluding the presence of any matter perturbations \cite{Odintsov:2015uca,Nojiri:2017ncd}.

Future Type I--IV singularities were explored in Ref.~\cite{Setare:2012vs} following $H(t) = h(t_s-t)^{-\beta} + C$, where $C = 0$ for $\beta > 0$ and non-zero otherwise. In the presence of a perfect fluid, the reconstructed solutions take on the forms as given in Table~\ref{table:fT_bounce_reconst}. The analysis was then extended to explore whether a viscous fluid with bulk velocity $\zeta(\rho)$, i.e. $p = w \rho - 3H\zeta(\rho)$, can avoid the singularity. Mainly, it was observed that: (a)~$\zeta = \text{const.} \neq 0$ cannot avoid it, (b) $\zeta = 3\tau H$ can avoid the Big Rip and possibly the sudden type for $\tau \gg 1$, and (c) $\zeta = \tau(3H)^n$ can avoid Big Rip $(n > 2\beta-1)$ and Big Freeze $(n > -2+\frac{2}{\beta})$ whereas the sudden and Big Brake may be avoided for large $n, \tau$.

\begin{table}[!ht]
	\centering
	\midsepremove
	\begin{tabularx}{\textwidth}{p{4.5cm}X}
	    \toprule
	    \cellcolor{gris1} \textbf{Singularity}& \cellcolor{gris1}\textbf{Reconstruction form}\\ \midrule
		\cellcolor{gris3}Big Rip $(\beta \geq 1)$ & \cellcolor{gris3}$\beta = 1, \, f(T) \propto (-T)^{-\frac{3h(1+w)+2}{2}}$; otherwise, close to singularity, $\sqrt{-T}$ \\
    	\cellcolor{gris1}Big Freeze $(0 < \beta < 1)$ & \cellcolor{gris1}$f(T) \propto (-T)^{\frac{\beta+1}{2\beta}}$ \\
		\cellcolor{gris3}Sudden $(-1 < \beta < 0)$ and Big Brake $(\beta < -1)$ & \cellcolor{gris3}\multirow{2}{*}{Close to singularity, $f(T) \propto \left[\sqrt{-\frac{4T}{9h^2}} - \frac{2C}{h}\right](\sqrt{-6T}-6C)$} \\
		\bottomrule
	\end{tabularx}
	\midsepdefault
	\caption{Reconstruction of $f(T)$ gravity for future Type I--IV singularities investigated in Ref.~\cite{Setare:2012vs} for the evolutionary behaviour $H(t) = h(t_s-t)^{-\beta} + C$, where $C = 0$ for $\beta > 0$ and non-zero otherwise. For the sudden and Big Brake singularities, only the behaviour close to the bounce was found.}
	\label{table:fT_bounce_reconst}
\end{table}

Focusing on Type~IV singularities, using a scalar-tensor representation reconstruction procedure, in vacuum, the reconstructed solution turns out to be $f(T) \propto \sqrt{-T}$ \cite{Abadji:2019cie}\footnote{The field equations in Ref.~\cite{Logbo:2018tni} are incorrect.}. Near the singularity, the Lagrangian can be Taylor expanded to yield a teleparallel Starobinsky-like model $f(T) \sim -T + T^2 + 2\Lambda$ which permits a further examination of the stability of the model. Whereas the slow-roll indices already indicate evolutionary instability leading to a graceful inflation exit, this is further supported through the trace anomaly approach \cite{Bamba:2014zra} as the homogeneous perturbations diverge hence ceasing inflation.

The avoidance of singularities can also be achieved through reconstruction using scale factors tailored to avoid their formation \cite{Astashenok:2013kka}. For instance, the $\Lambda$\gls{cdm} Big Bang singularity can be avoided through the scale factor modification $a(t) = a_0 \sinh^{\frac{2}{3}}\Big(\frac{\sqrt{3\Lambda(t^2+\tau^2)}}{2}\Big)$ which, for \makebox{$t \gg \tau$} guarantees the $\Lambda$\gls{cdm} behavior whereas for $t \ll \tau$, a bounce occurs as opposed to a Big Rip. Assuming the presence of an ideal dust fluid, the reconstructed function is simply $f(T) = \frac{2\rho_0}{{a_0}^3}\left(1-\frac{T}{2\Lambda}\right)$\footnote{By virtue of introducing a new time parameter $\tilde{t} \equiv t^2 + \tau^2$, an analytical solution in terms of $T$ can be derived.}. For the Quasi-Rip scenario, a future bounce is introduced via the scale factor $a(t) = \left(\frac{{t_f}^2+{t_1}^2}{{t_1}^2+(t_f-t)^2}\right)^{\frac{1}{\epsilon}}$ with $t_1 \ll t_f$ and $\epsilon$ is a constant which acts under the influence of a phantom energy density $\rho = \rho_0 a(t)^{\epsilon}$. Finally, the modified cyclic Little Rip scenario $a(t) = e^{\alpha^2 {t_b}^2 \sin^2(t/t_b)}$ was also investigated. However, contrary to what has been reported, the reconstructed $f(T)$ function is not periodic thus requiring a further investigation. \\

\noindent \textbf{(c) $\;$ Formation of Bounce Scenarios viewed from Dynamical Systems} \\

A further extension of the reconstruction approach can be achieved via dynamical systems. As $f(T)$ gravity can be modelled as a one dimensional autonomous system
\begin{equation}
\dot{H} = -\dfrac{1+w}{4}\dfrac{f - 2Tf_T}{f_T + 2T f_{TT}} = 3(1+w) \dfrac{f(H)-Hf_H}{f_{HH}} = -\dfrac{3}{2}\left(1+w_{\rm Tot}\right)H^2 \equiv \mathcal{F}(H)\,,
\end{equation}
a bouncing cosmological behavior can be reconstructed and studied from a dynamical viewpoint.

\begin{figure}[!ht]
	\centering
	\subfigure[Bounce leading to inflation]{\includegraphics[width=0.49\textwidth, height = 6.8 cm]{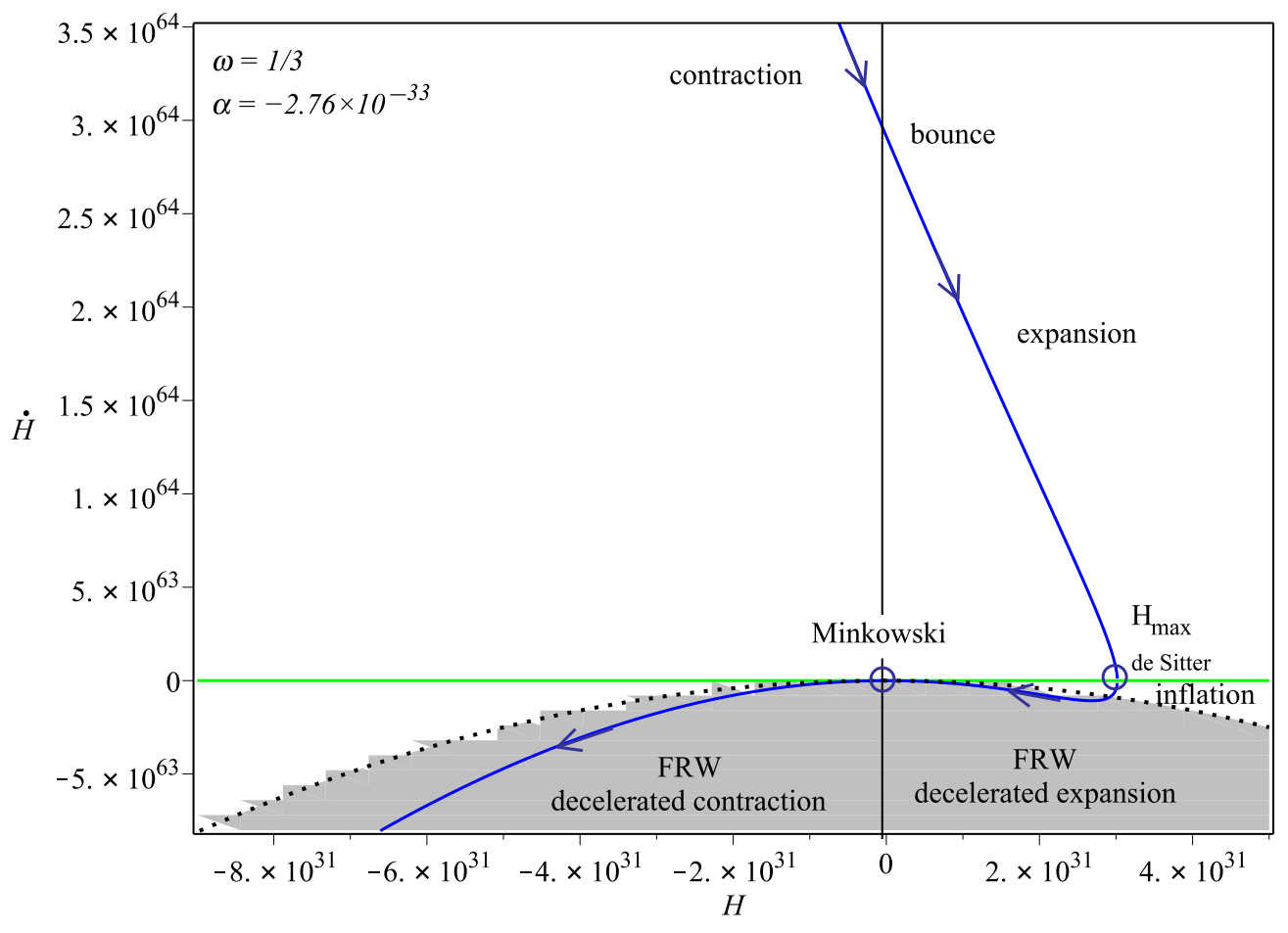}}
\subfigure[Graceful inflation exit to Minkowski]{\includegraphics[width=0.49\textwidth, height = 6.8 cm]{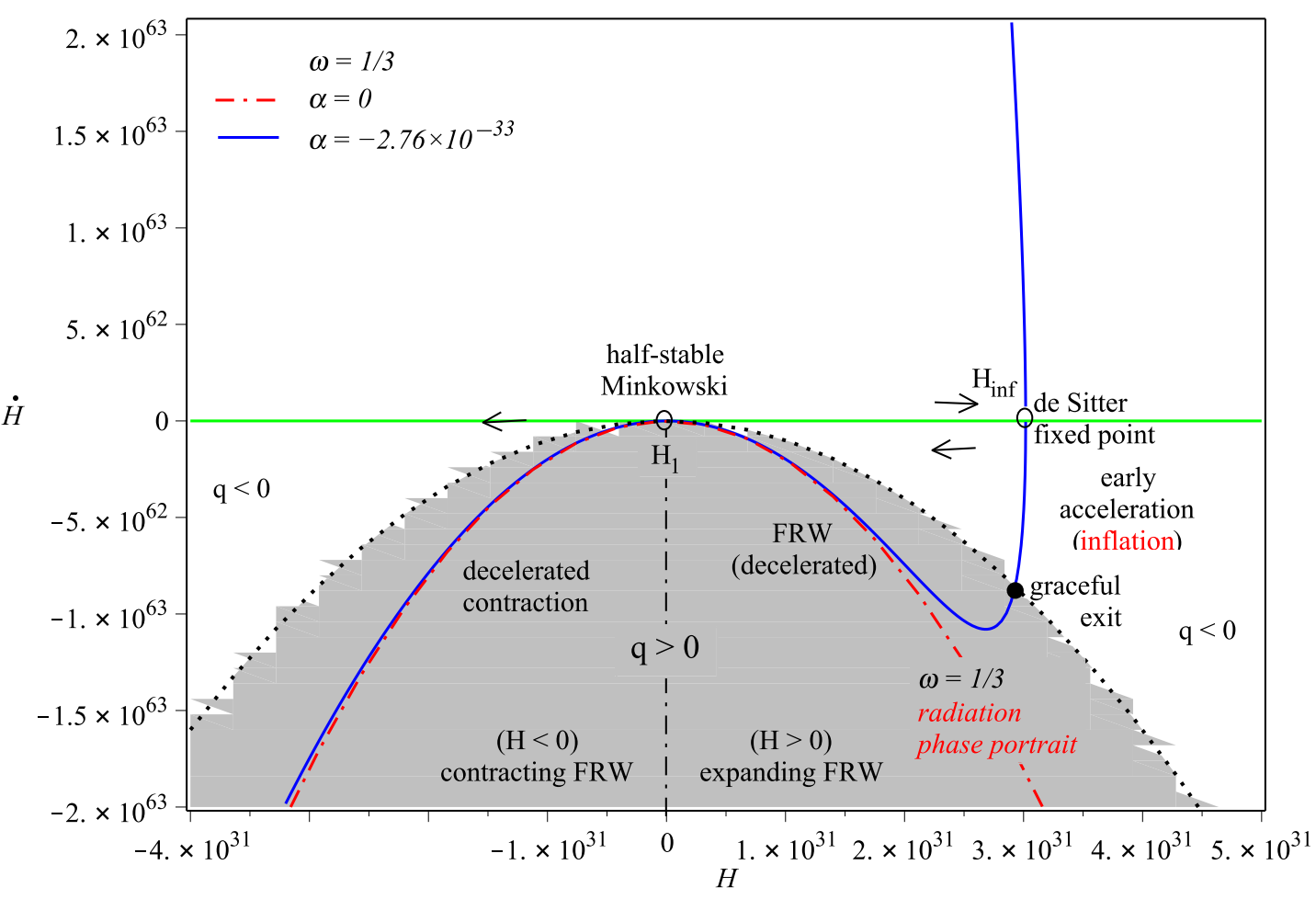}}
	\caption{A phase-space representation of the bouncing cosmology leading to inflation followed by graceful inflation exit for $\alpha = 2.76 \times 10^{-33} < 0$. Permission for use of this figure was kindly provided by the authors of Ref.~\cite{Bamba:2016gbu}.}
	\label{fig:fT_dynamics_boun}
\end{figure}

In Ref.~\cite{Bamba:2016gbu}, a unification of a bounce and late-time radiation/matter domination through the scale factor $a(t) \propto \left[(1+w)(t-t_i)\right]^{\frac{2}{3(1+w)}}e^{\frac{\alpha}{(1+w)(t-t_i)}}$, where $t_i$ represents the bounce time, was investigated. The signature of the constant $\alpha$ determines the nature of the evolution at early-times where $\alpha > 0$ leads to a graceful inflation exit and a bouncing cosmology for $\alpha < 0$. The dynamical behavior of the $\alpha < 0$ model case was studied extensively from which, the cosmological sequence starting with a bounce $\to$ inflation with graceful exit $\to$ decelerating radiation dominated cosmology is realized as presented in Fig.~\ref{fig:fT_dynamics_boun}. Additionally, the resulting maximum temperature during inflation is below that of Grand Unified Theories. Although the model describes a semi-stable Minkowski critical point, this final state is not achieved as the time required is infinite. Thus, the model still needs to be modified to ensure the realization of matter dominated and late-time acceleration phases.

In the case of Type I--IV singularities described by $H(t) = \beta(t-t_s)^\alpha$, $\beta \equiv (1+\alpha)f_0$, the reconstructed $f(T)$ solution was found in Ref.~\cite{ElHanafy:2017xsm}. A summary of the dynamical analysis for each singularity type is given below:
\begin{enumerate}[label = (\roman*{})]
	\item Type I: Existence of a stable Minkowski critical point. $f_0 > 0$ leads to an eternal accelerating contraction while $f_0 < 0$ leads to Big Bang $\to$ inflation (with possible graceful exit) $\to$ decelerated expansion;
	\item Type II: No finite critical points. A sudden singularity $(f_0 > 0)$ or a Big Brake singularity $(f_0 < 0)$ can be realized;
	\item Type III: Stable or semi-stable Minkowski critical point depending on the magnitude of $\alpha$. $f_0 > 0$ is preferred as it generates a transition from deceleration to acceleration which can match $\Lambda$\gls{cdm}. However, the ultimate fate is Minkowski;
	\item Type IV: An initial unstable Minkowski singularity which evolves to a phantom $(f_0 > 0)$ or non-phantom $(f_0 < 0)$ cosmology.
\end{enumerate}

\normalsize Following Ref.~\cite{ElHanafy:2017sih}, use of fixing the boundary term $B = -4\Lambda e^{-1}$ for some constant $\Lambda$ and $e = \det{(\udt{e}{A}{\mu})}$ is discussed. For $\Lambda = 0$, $a(t) \propto (t-t_i)^{\frac{1}{3}}$ i.e. a Type~I singularity, with associated reconstructed function $f(T) \propto (-T)^{\frac{3(1+w)}{2}}$. For $\Lambda \neq 0$, $a(t) = \left({c_2}^2 \Lambda -\frac{9}{4\Lambda} c_1 +2 c_2 \Lambda t+\Lambda t^2\right)^{\frac{1}{3}}$ where $c_1, c_2$ are integration constants\footnote{Eq.~(12) in Ref.~\cite{ElHanafy:2017sih} is missing a factor of $a(t)^3/2$ on the \gls{lhs} hence affecting the form of the scale factor.}. At $t = -c_2$, a bounce $(c_1 < 0)$, turnaround $(c_1 > 0)$ or superbounce $(c_1 = 0)$ occurs. Due to the complicated relationship between $t$ and $T$, the reconstructed $f(T)$ function is not derived.

Recently, a new bouncing scenario called Pseudo-Bang cosmology has been proposed in Ref.~\cite{ElHanafy:2020pek}, which describes a unification of the bounce cosmology $a_B(t) \propto \left[\frac{3}{2} (1+w) \alpha (t-t_B)^2 + 1\right]^{\frac{1}{3(1+w)}}$ and late-time acceleration through $a(t) \propto e^{\beta(t-t_B)} a_B(t)$. The introduction of the exponential term ensures the late-time acceleration for times $t \gg t_B$ while generating a non-symmetrical evolution in the $H$-$\dot{H}$ phase-space plane. The corresponding reconstructed $f(T)$ function is given by
\begin{subequations}
\begin{align}
f(T)_\pm &= \frac{18\rho_0 \left(\mathcal{A} e^{\theta} \text{Ei}(1,\mathcal{B}^+) - \mathcal{A} e^{-\theta} \text{Ei}(1,\mathcal{B}^-) + \frac{2}{9}\alpha \theta e^{-3\gamma\beta \tau_{\pm}}\right)}{\alpha\theta {a_k}^{3\gamma}(2+3\gamma\alpha {\tau_\pm}^2)}\,, \\[0.5ex]
\mathcal{A} &= \gamma\beta \left(\frac{2}{3}\beta+\frac{2}{3}\alpha\tau + \alpha\gamma\beta \tau^2\right)\,,\quad \mathcal{B}^\pm = 3\gamma\beta \tau \pm \theta\,,\\[0.5ex]
\tau_\pm &= \frac{\alpha \pm \sqrt{\alpha^2-6\alpha\gamma\beta^2+2\alpha\beta\gamma\sqrt{-6T}+\alpha\gamma T}}{3\gamma\alpha(\frac{1}{6}\sqrt{-6T}-\beta)}\,,
\end{align}
\end{subequations}
where $a_k \equiv a(t_i)$, $\theta = \beta \sqrt{-6\gamma/\alpha}$ and the function's signature corresponds to the positive and negative $\dot{H}$ branches respectively. Through dynamical systems and observational considerations to set the free parameters, the resulting Pseudo-Bang cosmology is described \cite{ElHanafy:2020pek}.

The Universe starts at the so called Pseudo-Bang singularity described by $a(t) \to 0$, \makebox{$H(t) \to \beta$} as $t \to -\infty$, followed by a first inflationary epoch and a cosmological turnaround which is not of Big Brake type. A transition to an accelerated contraction of Type IV occurs followed by a bounce which avoids both ghost instability and anisotropy problems. Another phantom crossing transitioning to a quintessence regime occurs which avoids the formation of a Big Rip, leading to an inflationary phase with graceful exit without production of monopoles. Ultimately, the standard cosmology of radiation $\to$ matter $\to$ late-time acceleration follows which evolves towards a Pseudo-Rip state.

More recently, Ref.~\cite{Casalino:2020kdr} has proposed an $f(T)$ model which closely resembles the behaviour of loop quantum cosmology, and thus exhibits a bouncing solution. The work contains cosmological perturbations and shows some promising results, but more work is needed to relate it to observations.

\subsubsection{\texorpdfstring{$f(T,B)$}{} Gravity}\label{sec:fTB-gravity-bouncing}

In Ref.~\cite{Caruana:2020szx}, a number of bouncing cosmologies were reconstructed. Since the $f(T,B)$ function cannot be solved in general, various ansatz forms were considered
\begin{align*}
\text{(i)} & \quad g(T) + h(B)\,, & \text{(iv)} &\quad -T + B g(T)\,, \\[0.5ex]
\text{(ii)} &\quad -T+T g(B)\,, & \text{(v)} &\quad -T + \mu \left(\frac{T}{T_0}\right)^\sigma \left(\frac{B}{B_0}\right)^\gamma\,, \\[0.5ex]
\text{(iii)} & \quad B g(T)\,,
\end{align*}
where $\mu, \, \sigma$ and $\gamma$ are constants while $T_0$ and $B_0$ represent the values when the scale factor is unity. Despite the similarity between models~(iii) and (iv), model~(iv) cannot recover \gls{tegr} and hence are fundamentally different. Nonetheless, the reconstructed solutions only differ by a term
\begin{equation}
    \text{Model (iv)} = - T + \frac{B \ln (-T)}{6} + \text{Model (iii)}\,.
\end{equation}
For the considered bouncing scenarios, most ansatz yield an analytical reconstructed solution (see Table~\ref{table:bouncing-recon-fTB}) except for most of model~(v). It is remarked that the reconstructed $Tg(B)$ solution for the matter bounce scenario is complex for the considered domain. Meanwhile, the lack of solutions for the Type I--IV Singularities and Little Rip cosmology scenarios is due to the relationship between $T$ and $B$ as it is not generally invertible.

As the solutions are mostly applicable for periods close to the bounce, the limiting behaviors have been investigated. The reconstructed Lagrangian generally take much simpler forms (see Table~\ref{table:bouncing-recon-fTB-asymptotic}). Starting with the symmetric bounce cosmology, the additive model yields an expanded power-series similar to Ref.~\cite{Escamilla-Rivera:2019ulu} and similar to the quadratic limiting order behavior encountered in $f(\lc{R})$ gravity, \mbox{$f(\lc{R}) \propto 144\alpha -72\lc{R} + \alpha^{-2} \lc{R}^2$} \cite{Bamba:2013fha,Nojiri:2014zqa}.

In the case of the oscillatory bouncing scenario, two asymptotic forms are required as the evolution experiences two distinct bounces: (a) the Big Crunch/Big Bang singularity $H(t) \to \infty$, and (ii) a cosmological turnaround $H(t) \to 0$. In the latter, the higher-order contributions in the additive solution have indices $p \geq 2$ which is a necessary requirement for deceleration after the turnaround \cite{Basilakos:2018arq,Mirza:2017vrk,Zhang:2011qp,Wu:2010xk,Bengochea:2008gz,Nesseris:2013jea,Basilakos:2016xob,Hohmann:2017jao}. For the former, the additive case \mbox{$\mathcal{L}_\text{grav.} \sim T^{3(1+w)} + B^{\frac{5}{2}}$} which ensures the generation of an accelerated contraction \cite{Wei:2011aa,Atazadeh:2011aa,Basilakos:2013rua}.

The matter bounce scenario gives relatively simple forms including $\Lambda$\gls{cdm} and rescaled \gls{tegr} with a logarithmic correction. Finally, for $\alpha < -1$ (Type~I), the models effectively behave as \mbox{$\mathcal{L}_\text{grav.} \sim (-T)^{-\frac{\alpha+1}{2\alpha}} \exp\left(T^{\frac{\alpha+1}{2\alpha}}\right)$} and \mbox{$\mathcal{L}_\text{grav.} \sim B (-T)^{-\frac{3\alpha+1}{2\alpha}} \exp\left(T^{\frac{\alpha+1}{2\alpha}}\right)$} for models~(i) and (iii) respectively. For the remaining cases, a power-law behavior arises. Here, the model~(i) $g(T)$ solution matches with Ref.~\cite{Setare:2012vs}.

The vacuum condition $f(0,0) = 0$, a constraint which is used to host vacuum solutions such as Minkowski spacetime, can be explored to investigate the viability of the solution. This greatly reduces the number of suitable models (see Table~\ref{table:bouncing-recon-fTB-vacuum}). For instance, only model~(i) can satisfy the vacuum constraint for a symmetric bouncing cosmology. On the other hand, practically all superbounce solutions satisfy the constraint. Under the assumption that the universe is not in a vacuum state, models~(i) and (ii) are both suitable choices for the oscillatory cosmology while only model~(ii) works for the matter bounce. Interestingly, only the Type~III singularity naturally satisfies the vacuum constraint in the presence of perfect fluids.

\begin{table}[H]
	\centering
	\footnotesize
	\midsepremove
	\begin{tabularx}{\textwidth}{p{2cm}X}
		\toprule
		\cellcolor{gris3}	\textbf{\large Ansatz} & 	\cellcolor{gris3}\textbf{\large Reconstructed Solution} \\ \midrule
		\multicolumn{2}{l}{	\cellcolor{gris1}\textbf{Symmetric Bounce} $a(t) = A e^{\alpha t^2/{t_*}^2}, \, t_*, \alpha > 0, \, 0 < A < 1$} \\
		\cellcolor{gris3}	$g(T) + h(B)$ & 	\cellcolor{gris3}$g(T) = T_{0} \Omega_{0} A^{-3(1+w)} [e^{-{u_1}^2} + \sqrt{\pi} u_1 \,\mathrm{erf}(u_1)], \, h(B) = c_1 \mathrm{L}\left[-\frac{1}{2}, \frac{3}{2}, -\frac{1}{2}\left(1 + \frac{Bt_{*}^{2}}{12\alpha}\right) \right]$ \\

		\cellcolor{gris1}	$-T+T g(B)$ & 	\cellcolor{gris1}$g_1(B) = 1+\sqrt{\frac{u_2 t_{*}^{2} \pi }{24\alpha}} e^{\frac{u_2 t_{*}^{2}}{24\alpha}} \, \mathrm{erf}\left(\sqrt{\frac{u_2 t_{*}^{2}}{24\alpha}}\right), \, g_2(B) = \sqrt{u_2} e^{\frac{u_2 t_{*}^{2}}{24\alpha}}$ \\

		\cellcolor{gris3}$B g(T)$ & 	\cellcolor{gris3}$g(T) = -\frac{T_{0} \Omega_{0} A^{-3(1+w)}}{6 T} \left[e^{-{u_1}^2} + {u_1}^2 \, \mathrm{Ei}(-{u_1}^2) \right]$ \\ \midrule

		\multicolumn{2}{l}{	\cellcolor{gris1}\textbf{Superbounce} $a(t) \propto \left(t_s - t\right)^{\frac{2}{\tilde{c}^2}}$, $\tilde{c} > \sqrt{6}$} \\
		\cellcolor{gris1}	$g(T) + h(B)$ &	\cellcolor{gris1} $g(T) = \frac{ T_0 \Omega_0}{1-2x} \left(\frac{T}{T_{0}}\right)^{v}$ for $v \neq \frac{1}{2}$, else $g(T) = \frac{\Omega_{0} \sqrt{T_{0}\,T}}{2} \ln(-T)$; $h(B) \propto B^{\frac{1 - 3\gamma}{2}}$ \\

		\cellcolor{gris3}	$-T+T g(B)$ & 	\cellcolor{gris3}$g(B) = 1 + \frac{2(1 - 3\gamma) \Omega_{0}}{4 + 2v (2v + 3\gamma - 5)} \left(\frac{B}{B_0}\right)^{v - 1} + c_2 B^{\frac{1-3\gamma-\sqrt{(3\gamma - 9)(3 \gamma - 1)}}{4}} + c_3 B^{\frac{1-3\gamma + \sqrt{(3\gamma - 9)(3 \gamma - 1)}}{4}}$ \\

		\cellcolor{gris1}	$B g(T)$ & 	\cellcolor{gris1}$g(T) = \frac{\Omega_{0}}{6 (1-v)} \left(\frac{T}{T_{0}} \right)^{v-1}$ for $v \neq 1$, otherwise $g(T) = -\frac{\Omega_{0}}{6} \ln(-T)$ \\

		\cellcolor{gris3}Power-Law & 	\cellcolor{gris3}For $\Omega_0 \neq 0$ and $\sigma \neq 0$, $\sigma + \gamma = v = 1$ and $\mu = \frac{T_{0} (1-\Omega_{0})}{\sigma}$ \\ \midrule

		\multicolumn{2}{l}{	\cellcolor{gris1}\textbf{Oscillatory Bouncing Cosmology} $a(t) = A \sin^2 \left(\frac{Ct}{t_*}\right), A > 1, C > 0$} \\
		\cellcolor{gris3}	$g(T)+h(B)$ &	\cellcolor{gris3} $g(T) = T_0 \Omega_0 A^{-3(1+w)} \bigg[\frac{T^3}{40x^3} \, _2F_1\left(\frac{5}{2},-3 w ;\frac{7}{2}; \frac{T}{2x} \right) - \frac{T^2}{4x^2} \, _2F_1\left(\frac{3}{2},-3 w ;\frac{5}{2}; \frac{T}{2x} \right) + \frac{3T}{2x} \, _2F_1\left(\frac{1}{2},-3 w ;\frac{3}{2}; \frac{T}{2x}\right) + \, _2F_1\left(-\frac{1}{2},-3 w ;\frac{1}{2}; \frac{T}{2x}\right)\bigg]$ \newline 	\cellcolor{gris3}$h(B) = c_4 \left[\sqrt{-B + x} \left(3 B^{2} - 288 B x + 40 x^{2}\right) + \frac{B}{80 \sqrt{5} x^{\frac{7}{2}}}\arctan{\left(\sqrt{\frac{-B + x}{5 x}}\right)}\right] $ \\

		\cellcolor{gris1}	$-T+T g(B)$ & 	\cellcolor{gris1}$g_1(B) = \, _2F_1\left(\frac{5-i \sqrt{15}}{4},\frac{5+i \sqrt{15}}{4};\frac{1}{2};\frac{B-x}{5x}\right)$, $g_2(B) = \sqrt{\frac{-B+x}{5x}} \, _2F_1\left(\frac{7-i \sqrt{15}}{4},\frac{7+i \sqrt{15}}{4};\frac{3}{2};\frac{B-x}{5x}\right)$ \\

		\cellcolor{gris3}	$B g(T)$ & 	\cellcolor{gris3}$w \neq \frac{n}{3}$: $g(T) = -\frac{T_{0} \Omega_{0} A^{-3(1+w)}}{12x} \big\lbrace (1-\frac{T}{2x})^{1+3w} \left[1 + \frac{T}{4x^2}\frac{-T(1+3w)+2x(5+9w)}{2 + 9w (1+w)}\right] + \frac{T}{2x[2 + 9w (1+w)]} - 3 (-1)^{3w} \left(1+w\right) \, \beta\left[\frac{2x}{T}; -3w, 1+3w\right] \big\rbrace$, \newline $w = \frac{n}{3}$: $g(T) = \frac{T_{0} \Omega_{0} A^{-3(1+w)}}{12x} \Big[(n+3) \ln (-T) + \sum\limits_{\substack{k = 0 \\ k \neq 1}}^{n+3} \binom{n+3}{k}\frac{1}{k-1}\left(-\frac{T}{2x}\right)^{k-1} \Big]$ \\ \midrule

		\multicolumn{2}{l}{\cellcolor{gris1}\textbf{Matter Bounce} $a(t) = A \left(\frac{3}{2}\rho_c t^2+1\right)^{\frac{1}{3}}, 0 < \rho_c \ll 1, 0 < A < 1$} \\
		\cellcolor{gris3}	$g(T) + h(B)$ & \cellcolor{gris3}$g(T) = T_{0} \Omega_{0} A^{-3(1+w)} \sqrt{1 - \frac{y}{2}} \left(\frac{y}{2}\,_2F_1\left[\frac{1}{2}, \frac{1}{2} - w,\frac{3}{2}; \frac{y}{2}\right] + \,_2F_1\left[-\frac{1}{2}, -\frac{1}{2}-w; \frac{1}{2}; \frac{y}{2}\right]\right)$, \newline $h(B) \propto \frac{(-B - 6\rho_{c})^{\frac{3}{2}}}{9 \sqrt{-B} \rho_{c}}$ \\

		\cellcolor{gris1}	$-T+T g(B)$ & \cellcolor{gris1}$g_\pm(B) = \left(-\frac{B}{6 \rho_{c}}\right)^{\frac{-1\pm\sqrt{7} i}{4}}\, _2F_1\left[\frac{-1\pm\sqrt{7}i}{4}, \frac{3\pm\sqrt{7} i}{4};1 \pm \frac{\sqrt{7}i}{2};-\frac{B}{6 \rho_{c}}\right]$ \\

		\cellcolor{gris3}	$B g(T)$ & \cellcolor{gris3}$w \neq n$: $g(T) = -\frac{\Omega_0 T_0 A^{-3 (1+w)}}{24 \rho_{c}}\left(1-\frac{y}{2}\right)^{w} \bigg[\frac{2}{y}-\frac{1}{w }-1+ \left(1+\frac{1}{w}\right) \left(\frac{y}{y-2}\right)^{w} \, _2F_1\left[-w ,-w ;1-w ;\frac{2}{y}\right]\bigg]$, \newline $w = 0$: $g(T) = -\frac{\Omega_{0} T_{0}}{24 A^3 \rho_{c}} \left[\frac{2}{y} - \ln \left(\frac{2}{y}-1\right)\right]$, \newline $w \geq 1$: $g(T) = -\frac{\Omega_0 T_0 A^{-3 (1+w)}}{12} \left[\frac{1}{y} +\ln y +\frac{1}{2} (1-w) \left(y-\ln y\right) + \sum\limits_{k = 2}^{w-1} \binom{w-1}{k} \left(-\frac{y}{2}\right)^k \left(\frac{1}{y(1-k)}+\frac{1}{k}\right) \right]$ \\ \midrule

		\multicolumn{2}{l}{\cellcolor{gris1}\textbf{Type I--IV Singularities and Little Rip Cosmology} $a(t) = A \exp\left[\frac{f_0}{\alpha+1}(t-t_s)^{\alpha+1}\right], \, A, f_0 > 0$} \\
		\cellcolor{gris1}	$g(T) + h(B)$ & \cellcolor{gris1}$g(T) = T_{0} \Omega_{0} A^{-3(1+w)} \, _1F_1 \left[-\frac{\alpha}{1 + \alpha}, \frac{1}{1 + \alpha}; -\frac{3f_0 {t_0}^{\alpha + 1} (1+w)}{1+\alpha} \left(\frac{T}{T_0}\right)^{\frac{\alpha + 1}{2 \alpha}} \right]$; no $h(B)$ solution obtained \\
		\cellcolor{gris3}	$Bg(T)$ & \cellcolor{gris3}$g(T) = \frac{T_{0} \Omega_0 A^{-3(1+w)}}{6 T}\, _1F_1\left[-\frac{2 \alpha}{1 + \alpha}, \frac{1-\alpha}{1+\alpha}; -\frac{3f_0 {t_0}^{\alpha + 1} (1+w)}{1+\alpha} \left(\frac{T}{T_0}\right)^{\frac{\alpha + 1}{2 \alpha}} \right]$ \\
		\bottomrule
	\end{tabularx}
	\midsepdefault
	\caption{Summary of the reconstructed $f(T,B)$ functions for the considered bouncing cosmologies investigated in Ref.~\cite{Caruana:2020szx}. Here, $c_1, \dots{}, c_4$ are integration constants, $\Omega_0$ represents the density parameter at $a(t) = 1$, ${u_1}^2 \coloneqq -\frac{3 Tt_{*}^{2} (1+w)}{24\alpha}$, $u_2 \coloneqq -B - \frac{12\alpha}{{t_*}^2}$, $v \coloneqq \frac{3 \gamma (1+w)}{2}$, $x \coloneqq \frac{12 C^2}{t_*^2}$ and $y \coloneqq 1 - \sqrt{1+\frac{T}{\rho_c}}$ while $\beta[x;a,b]$ is the incomplete beta function, $_{2}F_{1}(a,b;c;z)$ and $_{1}F_{1}(a,b;z) = \frac{\Gamma(b)}{\Gamma(a)} \sum\limits_{k=0}^{\infty} \frac{\Gamma(a+k)}{\Gamma(b+k) \, k!} z^k$ represent Gauss's and the confluent hypergeometric functions respectively \cite{buchholz2013confluent}.}
	\label{table:bouncing-recon-fTB}
\end{table}

\begin{table}[H]
	\centering
	\midsepremove
	\small
	\begin{tabularx}{\textwidth}{lXX}
		\toprule
		\cellcolor{gris3}\textbf{\large Ansatz} & \multicolumn{2}{l}{	\cellcolor{gris3}\textbf{\large Asymptotic Form of the Reconstructed Solution}} \\ \midrule
		\multicolumn{3}{l}{	\cellcolor{gris1}\textbf{Symmetric Bounce}} \\
		\cellcolor{gris3}	$g(T) + h(B)$ & \multicolumn{2}{l}{	\cellcolor{gris3}$g(T) = T_{0} \Omega_{0} A^{-3(1+w)} \left(1 + {u_1}^2 - \frac{{u_1}^4}{6} \right)$, $h(B)= c_{1} \left(\frac{17}{14 \pi } - \frac{11 B{t_*}^2}{1260 \pi \alpha} + \frac{B^2 {t_*}^4}{10080 \pi \alpha^{2}}\right) $} \\

		\cellcolor{gris1}	$-T+T g(B)$ & \multicolumn{2}{>{\hsize=\dimexpr2\hsize+2\tabcolsep+\arrayrulewidth\relax}X}{	\cellcolor{gris1}$g_1(B) = -\frac{B{t_*}^2}{12\alpha}$, $g_2(B) = \sqrt{-B-\frac{12 \alpha}{t_{*}^{2}}}$.\newline The matter source takes the form $g(B) = 1 +\frac{T_0 \Omega_0 {t_*}^2 A^{-3(1+w)}}{12\alpha} \ln \left(-B-\frac{12 \alpha}{t_{*}^{2}}\right)$} \\

		\cellcolor{gris3}	$B g(T)$ & \multicolumn{2}{l}{	\cellcolor{gris3}$g(T) = \frac{T_{0} \Omega_{0} A^{-3(1+w)}}{6T}$} \\ \midrule

		\multicolumn{3}{l}{	\cellcolor{gris1}\textbf{Oscillatory Bouncing Cosmology}} \\
		\cellcolor{gris1}	& 	\cellcolor{gris1}\textbf{Close to $H(t) \to 0$} & 	\cellcolor{gris1}\textbf{Close to $H(t) \to \infty$}\\
		\cellcolor{gris3}$g(T)+h(B)$ & \cellcolor{gris3}$g(T) = T_{0} \Omega_{0} A^{-3(1+w)} \left[1 +\frac{3(1+w)T}{2x}\right]$, \newline $h(B) \propto \left(\frac{1}{400 x^3}-245 x^2\right) \sqrt{-B+x}$ &	\cellcolor{gris3} $g(T) = \frac{T_{0} \Omega_{0} A^{-3(1+w)}}{5 + 6w} \left(-\frac{T}{2x}\right)^{3(1+w)}$, \newline $h(B) \propto B^{\frac{5}{2}}$ \\

		\cellcolor{gris1}	$-T+Tg(B)$ & 	\cellcolor{gris1}$g_1(B) = \frac{B}{x}$, $g_2(B) = \sqrt{\frac{-B+x}{5x}}$. \newline The matter source takes on the form \newline $g(B) = 1 - \frac{5T_{0} \Omega_{0} A^{-3(1+w)}}{2x} \ln \left(\frac{2(-B+x)}{5}\right)$ & 	\cellcolor{gris1}$g_1(B) = z^{-\frac{5}{4}}\sin \left[\frac{\sqrt{15}}{4} \ln \left(\frac{2(-B+x)}{5}\right) \right]$, \newline $g_2(B) = z^{-\frac{5}{4}} \cos \left[\frac{\sqrt{15}}{4} \ln \left(\frac{2(B+x)}{5}\right) \right]$. \newline The matter source takes on the form $g(B) = 1+\frac{5\Omega_0 T_0 A^{-3 (1+w)}}{2x(18 w^2 +39w+23)}\left(\frac{2(-B+x)}{5}\right)^{2+3 w}$ \\

		\cellcolor{gris3}$Bg(T)$ & 	\cellcolor{gris3}$g(T) = \frac{T_{0} \Omega_{0} A^{-3(1+w)}}{6T}$ & 	\cellcolor{gris3}$g(T) = \frac{T_{0} \Omega_{0} A^{-3(1+w)}}{12x} \frac{1+3w}{2 + 9w+9w^2} \left(-\frac{T}{2x}\right)^{2+3w}$ \\ \midrule

		\multicolumn{3}{l}{	\cellcolor{gris1}\textbf{Critical Matter Bounce}} \\
		\cellcolor{gris3}	$g(T) + h(B)$ & \multicolumn{2}{l}{	\cellcolor{gris3}$g(T) = T_{0} \Omega_{0} A^{-3(1+w)} \left[1 + \frac{(1+w)}{2} y - \frac{(1+w)(4+w)}{24} y^2\right], \, h(B) \propto \frac{(-B-6 \rho_c )^{3/2}}{9 \sqrt{6} {\rho_c} ^{3/2}}$} \\

		\cellcolor{gris1}	$-T+T g(B)$ & \multicolumn{2}{>{\hsize=\dimexpr2\hsize+2\tabcolsep+\arrayrulewidth\relax}X}{\cellcolor{gris1}In the presence of fluids: $g(B) = 1 - \frac{T_0 \Omega_0 A^{-3(1+w)}}{4 \rho_c} \left[2 - \frac{B}{6 \rho_{c}} \ln{\left(-\frac{B}{6 \rho_{c}}-1\right)}\right]$} \\

		\cellcolor{gris3}	$B g(T)$ & \multicolumn{2}{l}{	\cellcolor{gris3}$g(T) = -\frac{\Omega_{0} T_{0} A^{-3(1+w)}}{12 \rho_{c} y}$} \\ \midrule

		\multicolumn{3}{l}{	\cellcolor{gris1}\textbf{Type I--IV Singularities and Little Rip Cosmology}} \\
		\cellcolor{gris1}$g(T) + h(B)$ & \multicolumn{2}{>{\hsize=\dimexpr2\hsize+2\tabcolsep+\arrayrulewidth\relax}X}{	\cellcolor{gris1}$\alpha > -1$: $g(T) = T_0 \Omega_0 A^{-3(1+w)} \left[1 + \frac{3f_0 \alpha {t_0}^{\alpha + 1} (1+w)}{1 + \alpha} \left(\frac{T}{T_0}\right)^{\frac{\alpha + 1}{2 \alpha}}\right]$, \newline $\alpha<-1$: $g(T) = \frac{T_0 \Omega_0 A^{-3(1+w)}}{3f_0 {t_0}^{\alpha + 1} (1+w)}\left(\frac{T}{T_0}\right)^{-\frac{\alpha + 1}{2 \alpha}} \exp{\left( -\frac{3f_0 {t_0}^{\alpha + 1} (1+w)}{1+\alpha} \left(\frac{T}{T_0}\right)^{\frac{\alpha + 1}{2 \alpha}} \right)}$} \\
		\cellcolor{gris3}	$Bg(T)$ & \multicolumn{2}{>{\hsize=\dimexpr2\hsize+2\tabcolsep+\arrayrulewidth\relax}X}{	\cellcolor{gris3}$\alpha >-1$: $g(T) = \frac{T_0 \Omega_0 A^{-3(1+w)}}{6 T} \left[1 - \frac{6f_0 \alpha {t_0}^{\alpha + 1} (1+w)}{\alpha^2 - 1} \left(\frac{T}{T_0}\right)^{\frac{\alpha + 1}{2 \alpha}} \right]$ \newline $\alpha < -1$: $g(T) = \frac{T_0 \Omega_0 A^{-3 (1+w)} \alpha}{9 T f_0 {t_0}^{\alpha + 1} (1+w)} \left(\frac{T}{T_0}\right)^{-\frac{\alpha + 1}{2 \alpha}} \exp{\left( -\frac{3f_0 {t_0}^{\alpha + 1} (1+w)}{1+\alpha} \left(\frac{T}{T_0}\right)^{\frac{\alpha + 1}{2 \alpha}} \right)}$} \\
		\bottomrule
	\end{tabularx}
	\midsepdefault
	\caption{The leading order asymptotic forms of the reconstructed $f(T,B)$ bouncing solutions near the bounce \cite{Caruana:2020szx}. For the oscillatory case, two limiting forms are obtained as the model undergoes a bounce $(H(t) \to \infty)$ and a cosmological turnaround $(H(t) \to 0)$. Here, ${u_1}^2 \coloneqq -\frac{3 Tt_{*}^{2} (1+w)}{24\alpha}$, $x \coloneqq \frac{12C^2}{{t_*}^2}$ and $y \coloneqq 1 - \sqrt{1+\frac{T}{\rho_c}} $.}
	\label{table:bouncing-recon-fTB-asymptotic}
\end{table}

\begin{table}[H]
	\centering
	\midsepremove
	\begin{tabularx}{\textwidth}{rp{2.55cm}p{2.9cm}p{2.45cm}p{1.65cm}X}
		\toprule
		\cellcolor{gris3}	& 	\cellcolor{gris3}\textbf{Symm.} & 	\cellcolor{gris3}\textbf{Superbounce} & 	\cellcolor{gris3}\textbf{Oscillatory} & 	\cellcolor{gris3}\textbf{Matter} & 	\cellcolor{gris3}\textbf{Type I--IV and Little Rip} \\ \midrule
		\cellcolor{gris1}(i) & 	\cellcolor{gris1}$T_{0} \Omega_{0} A^{-3(1+w)}$ \newline $= \frac{17 c_1}{14 \pi}$ & 	\cellcolor{gris1}Always & 	\cellcolor{gris1}$T_{0} \Omega_{0} A^{-3(1+w)}$ \newline $= 40 c_4 x^{\frac{5}{2}}$ & 	\cellcolor{gris1}\xmark & 	\cellcolor{gris1}$h(0) = 0$ for $-1<\alpha<0$; $T_0 \Omega_0 A^{-3(1+w)} = h(0)$ \newline otherwise \\
		\cellcolor{gris3}(ii) & 	\cellcolor{gris3}\xmark & 	\cellcolor{gris3}Always & 	\cellcolor{gris3}Always & 	\cellcolor{gris3}Always & 	\cellcolor{gris3}N/A \\
		\cellcolor{gris1}(iii) & 	\cellcolor{gris1}\xmark & 	\cellcolor{gris1}$0 < v < 1$ \newline vacuum; always otherwise & 	\cellcolor{gris1}Vacuum & 	\cellcolor{gris1}Vacuum & 	\cellcolor{gris1}$-1<\alpha<0$; vacuum \newline otherwise \\
		\cellcolor{gris3}(iv) & 	\cellcolor{gris3}\xmark & 	\cellcolor{gris3}$v \geq 1$ &	\cellcolor{gris3} \xmark & 	\cellcolor{gris3}\xmark &	\cellcolor{gris3} $-1<\alpha<0$ \\
		\cellcolor{gris1}(v) & 	\cellcolor{gris1}N/A & 	\cellcolor{gris1}Always & 	\cellcolor{gris1}N/A &	\cellcolor{gris1}N/A & 	\cellcolor{gris1}N/A \\
		\bottomrule
	\end{tabularx}
	\midsepdefault
	\caption{The necessary requirements for the reconstructed bouncing $f(T,B)$ models obtained in Ref.~\cite{Caruana:2020szx} to satisfy the vacuum condition $f(0,0) = 0$ for the symmetric, superbounce, oscillatory, critical matter, Type I--IV and Little Rip bouncing cosmologies.}
	\label{table:bouncing-recon-fTB-vacuum}
\end{table}

\subsubsection{\texorpdfstring{$f(T,T_G)$}{} Gravity}

A similar analysis for the bouncing cosmological scenarios considered in $f(T,B)$ gravity has also been performed in the \gls{tegb} extension using the same functional ansatz listed in Sec.~\ref{sec:fTB-gravity-bouncing} \cite{delaCruz-Dombriz:2018nvt}. The reconstructed solutions generally take on a complex form which hinder their practical use (see Table~\ref{table:bouncing-reconstruction-fTTG}). Nonetheless, appropriate limiting behaviors such as those carried out in $f(T,B)$ gravity could greatly reduce their form. Contrary to the $f(T,B)$ gravity power-law ansatz case, except for the Type I--IV bounce scenarios, the TEGB power-law ansatz can host bouncing scenarios although in the absence of fluids. Meanwhile, the reconstructed solutions for the $Tg(T_G)$ model ansatz are only obtained in the absence of fluids, a consequence of the fact that the resulting differential equations take on complicated forms. This feature is strongly present in the Type I--IV bouncing scenarios since there is no general invertible expression between $T$ and $T_G$. As such, a limited number of reconstructed solutions could be obtained. Of note, the considered choices for $\alpha$ stem from the confluent Gauss's hypergeometric function as it is not defined when $\frac{1-3\alpha}{\alpha +1} \neq N \leq 0$. In particular, the special cases $\alpha = \frac{1}{3}$ and $\alpha = 3$ represent Type~II and Type~IV bouncing scenarios respectively whereas the general solution is representative of the Type~I scenario.

The reconstructed models have been also investigated as to whether they are able to host vacuum solutions, i.e. whether the condition $f(0,0) = 0$ is satisfied. Contrary to the $f(T,B)$ case, the constraint is mostly satisfied when the Universe is absent from any matter fluids with the only exception being the Type~III singularity (see Table~\ref{table:bouncing-reconstruction-fTTG-vacuum}).

\begin{table}[H]
	\centering
	\footnotesize
	\midsepremove
	\begin{tabularx}{\textwidth}{lX}
		\toprule
		\cellcolor{gris3}\textbf{\large Ansatz} & 	\cellcolor{gris3}\textbf{\large Reconstructed Lagrangian} \\ \midrule
		\multicolumn{2}{l}{	\cellcolor{gris1}\textbf{Symmetric Bounce} $a(t) = A \exp \left(\frac{\alpha t^2}{{t_*}^2}\right)$, $\alpha > 0$ and $0 < A < 1$} \\
		\cellcolor{gris3}$g(T) + h(T_G)$ & 	\cellcolor{gris3}$h(u_1) \propto e^{u_1} \left[u_1\left(2u_1+1\right)F\left(\sqrt{u_1}\right)-\sqrt{u_1}(u_1+1)\right]$\\

		\cellcolor{gris3}$T g(T_G)$ &	\cellcolor{gris3} $g_1(u_1) = 2\left(8{u_1}^2+14u_1-3\right)+\sqrt{\frac{\pi}{u_1}}\left(16{u_1}^3+36{u_1}^2+3\right)\text{erf}\left(\sqrt{u_1}\right)e^{u_1}$, \newline $g_2(u_1) = \frac{e^{u_1}}{\sqrt{u_1}}\left(16{u_1}^3+36{u_1}^2+3\right)$ \\

		\cellcolor{gris1}$T_G g(T)$ & 	\cellcolor{gris1}$g(u_2) = \dfrac{3 \Omega_0 A^{-3(1+w)} {u_{2,0}}^2}{8T_0} \left[\dfrac{u_2-1}{{u_2}^2}e^{-u_2} + \text{Ei}\left(-u_2\right)\right]$ \\

		\cellcolor{gris3}$-T + \mu (-T)^\beta {T_G}^\gamma$ & 	\cellcolor{gris3}Vacuum with $\{\gamma, \beta\} = \{1,-1\}$ and $\mu = \frac{3}{4}$ \newline For $w = -1$, $\{\gamma, \beta\} = \{-3,2\}$ provided that $\alpha = \frac{2{t_*}^2}{7}$, $\mu = -\frac{9}{20}$ and $\Omega_{\Lambda,0} T_0 = \frac{288}{35}$ \\ \midrule

		\multicolumn{2}{l}{	\cellcolor{gris1}\textbf{Oscillatory Bounce} $a(t) = A \sin^2\left(\frac{Ct}{t_*}\right)$, $A > 1, \; C > 0$} \\
		\cellcolor{gris3}$g(T) + h(T_G)$ &	\cellcolor{gris3} $h(v_1) \propto (8-3 v_1) \sqrt{v_1}-6 v_1(v_1-1) \tan ^{-1}\left(\sqrt{v_1}\right)$ \\

		\cellcolor{gris1}$T g(T_G)$ & 	\cellcolor{gris1}$g(v_1) \propto v_1-\frac{3 {v_1}^3}{8}+\frac{21 {v_1}^5}{640}-\frac{11 {v_1}^6}{1600}+\dots$ \\

		\cellcolor{gris3}$T_G g(T)$ & 	\cellcolor{gris3}$w \neq \frac{n}{3}$: $g(v_2) = \frac{3T_0 \Omega_0}{24w {v_2}^3}\left[\frac{{t_*}^2}{24 C^2}\right]^2 A^{-3(1+w)} (1+v_2)^{1+3 w} \big\lbrace \frac{6 w {v_2}^3}{1+3w} -v_2 \left[1+3w+{v_2}(3w-1)\right] \times $ \newline $ \, _2F_1\left(1,1;1-3 w;-\frac{1}{{v_2}}\right) + (1+{v_2}) \left[1+{v_2}(3 w+5)\right] \, _2F_1\left(1,2;1-3 w;-\frac{1}{{v_2}}\right)\big\rbrace$ \newline $w = \frac{n}{3}$: $g({v_2}) = \frac{3}{4}\left[\frac{{t_*}^2}{24 C^2}\right]^2 T_0 \Omega_0 A^{-3-n} \left[\binom{n+3}{2} \ln {v_2} + \sum\limits_{\substack{k = 0 \\ k \neq 2}}^{n+3} \binom{n+3}{k} \frac{{v_2}^{k-2}}{k-2}\right]$ \\

		\cellcolor{gris1}$-T + \mu (-T)^\beta {T_G}^\gamma$ & 	\cellcolor{gris1}Vacuum with $\{\gamma, \beta\} = \{1,-1\}$ and $\mu = \frac{3}{4}$. \\ \midrule

		\multicolumn{2}{l}{	\cellcolor{gris3}\textbf{Critical Matter Bounce} $a(t) = A \left(\frac{3}{2}\rho_{\text{cr}}t^2+1\right)^{\frac{1}{3}}$, $0< \rho_\text{cr} \ll 1$, $0 < A < 1$} \\
		\cellcolor{gris1}$g(T) + h(T_G)$ &	\cellcolor{gris1} No $h(T_G)$ solution was found \\

		\cellcolor{gris3}$T g(T_G)$ & 	\cellcolor{gris3}$g_1(x) = 1 - \frac{1}{2}(x-1)^2 + \frac{15}{648}(x-1)^4+\frac{1851}{2430}(x-1)^5 + \dots$ \newline $g_2(x) = x - 1 +\frac{1}{6}(x-1)^2-\frac{769 }{648}(x-1)^4 + \frac{1706}{2430}(x-1)^5 + \dots$ \newline where the torsion scalar $T$ is expressed in terms of $T_G$ \\

		\cellcolor{gris1}$T_G g(T)$ & 	\cellcolor{gris1}$w \neq 0$: $g(x) = \frac{3T_0}{64{\rho_\text{cr}}^2} \Omega_0 A^{-3(1+w)} \left(\frac{x}{2}\right)^w \bigg[\frac{2}{(w-1) x}+\frac{1}{w}+4 \left(\frac{x}{x-2}\right)^{-w}\frac{\, _2F_1\left(2-w,-w;3-w;\frac{2}{2-x}\right)}{(w-2) (x-2)^2}$ \newline $-\frac{1}{w}\left(\frac{x}{x-2}\right)^{-w} \,_2F_1\left(-w,-w;1-w;\frac{2}{2-x}\right)\bigg]$ \newline $w = 0$: $g(x) = -\frac{3T_0 \Omega_0}{32A^3 {\rho_\text{cr}}^2} \left[\frac{1}{(x-2)^2}+\frac{1}{x}+\ln \sqrt{\frac{2-x}{x}} \, \right]$ \\

		\cellcolor{gris3}$-T + \mu (-T)^\beta {T_G}^\gamma$ &	\cellcolor{gris3} Vacuum with $\{\gamma, \beta\} = \{1,-1\}$ and $\mu = \frac{3}{4}$. \\ \midrule

		\multicolumn{2}{l}{	\cellcolor{gris1}\textbf{Type I--IV Bounce} $a(t) = A \exp\left[\frac{f_0}{\alpha +1}(t-t_s)^{\alpha+1}\right]$, $A, f_0 > 0$} \\
		\cellcolor{gris3}$g(T) + h(T_G)$ & 	\cellcolor{gris3}No $h(T_G)$ solution was found \\

		\cellcolor{gris1}$T_G g(T)$ & 	\cellcolor{gris1}$\frac{1-3\alpha}{\alpha +1} \neq N$: $g(T) = -\frac{3 \Omega_0 T_0 A^{-3 (1+w)}}{8 T^2} \, _1F_1\left(-\frac{4 \alpha }{\alpha +1};\frac{1-3\alpha}{\alpha +1};-\frac{ 3 f_0 (1+w)}{\alpha +1} \left(-\frac{T}{6{f_0}^2}\right)^{\frac{\alpha +1}{2 \alpha }} \right)$ \newline $\alpha = \frac{1}{3}$: $g(T) = -\frac{3 \Omega_0 T_0 A^{-3 (1+w)}}{8 T^2} \left[\frac{T^2 (1+w)}{16 {f_0}^3} \text{Ei}\left(-\frac{T^2 (1+w)}{16 {f_0}^3}\right)+\exp\left(-\frac{T^2 (1+w)}{16 {f_0}^3}\right)\right]$ \newline $\alpha = 3$: $g(y) = -\frac{3 \Omega_0 T_0 A^{-3 (1+w)}}{8 T^2} \left[\frac{y^3}{2} \text{Ei}\left(-y\right)+ e^{-y} \left(1-\frac{y}{2}+\frac{y^2}{16} \right)\right]$ \newline $\frac{1-3\alpha}{\alpha +1} = N \leq -4$: No general solution is derived. \\

		\cellcolor{gris3}$-T + \mu (-T)^\beta {T_G}^\gamma$ & 	\cellcolor{gris3}Vacuum with $\{\gamma, \beta\} = \{1,-1\}$ and $\mu = \frac{3}{4}$. \\
		\bottomrule
	\end{tabularx}
	\midsepdefault
	\caption{Summary of the reconstructed $f(T,T_G)$ bouncing solutions investigated in Ref.~\cite{delaCruz-Dombriz:2018nvt}. For simplicity, the variables $4u_1 \coloneqq -1 + \sqrt{1-\frac{T_G {t_*}^4}{24\alpha^2}}$, $u_2 \coloneqq -\frac{T (1+w){t_*}^2}{8 \alpha}$, $v_1 \coloneqq 2\Big(1 - \sqrt{1-\frac{T_G {t_*}^4}{48 C^4}}\Big)$, $v_2 \coloneqq \frac{T {t_*}^2}{24C^2}$, $x \coloneqq 1+\sqrt{1+\frac{T}{\rho_\text{cr}}}$ and $y^3 \coloneqq \frac{3T^2 (1+w)^3}{256f_0}$ have been defined whereas $n$ and $N \leq 0$ are integers and the function $F(z)$ represents the Dawson integral. Additionally, as the model~(i) $g(T)$ solutions are identical to those obtained in previous sections, they are not listed in the table.}
	\label{table:bouncing-reconstruction-fTTG}
\end{table}

\begin{table}[!t]
	\centering
	\midsepremove
	\begin{tabularx}{\textwidth}{rp{2.5cm}p{3.5cm}p{3.5cm}X}
		\toprule
		\cellcolor{gris3}	& 	\cellcolor{gris3}\textbf{Symmetric} & 	\cellcolor{gris3}\textbf{Oscillatory} & 	\cellcolor{gris3}\textbf{Matter} & 	\cellcolor{gris3}\textbf{Type I--IV and LR} \\ \midrule
		\cellcolor{gris1}(i) & 	\multicolumn{1}{c}{ \cellcolor{gris1}Vacuum}&	\multicolumn{1}{c}{ \cellcolor{gris1}Vacuum} &	\cellcolor{gris1} Depends on $h(T_G)$ & 	\cellcolor{gris1}$-1 < \alpha < 0$, otherwise only in vacuum \\
		\cellcolor{gris3}(ii) & 	\multicolumn{1}{c}{ \cellcolor{gris3}Vacuum }& 	\cellcolor{gris3}Vacuum, unknown otherwise & 	\cellcolor{gris3}Vacuum, unknown otherwise & 	\cellcolor{gris3}Unknown \\
		\cellcolor{gris1}(iii) &	 \multicolumn{1}{c}{ \cellcolor{gris1}\xmark} & 	\multicolumn{1}{c}{\cellcolor{gris1}\xmark} & \multicolumn{1}{c}{	\cellcolor{gris1}\xmark} & 	\cellcolor{gris1}$-1 < \alpha < 0$ \\
		\cellcolor{gris3}(iv) & 	\multicolumn{1}{c}{\cellcolor{gris3}\xmark }& 	\multicolumn{1}{c}{\cellcolor{gris3}\xmark } & 	\multicolumn{1}{c}{\cellcolor{gris3}\xmark } & 	\cellcolor{gris3}Vacuum or $-1 < \alpha < 0$ \\
		\cellcolor{gris1}(v) & 	\multicolumn{1}{c}{\cellcolor{gris1}\xmark } & 	\multicolumn{1}{c}{\cellcolor{gris1}\xmark } &	\multicolumn{1}{c}{\cellcolor{gris1}\xmark } & 	\cellcolor{gris1}N/A \\
		\bottomrule
	\end{tabularx}
	\midsepdefault
	\caption{The necessary requirements for the reconstructed bouncing $f(T,T_G)$ models obtained in Ref.~\cite{delaCruz-Dombriz:2018nvt} to satisfy the vacuum condition $f(0,0) = 0$ for the symmetric, oscillatory, critical matter, Type I--IV and Little Rip bouncing cosmologies.}
	\label{table:bouncing-reconstruction-fTTG-vacuum}
\end{table}

\subsubsection{Unimodular Teleparallel Gravity}

The unimodular formulation has also been applied in the context of bouncing cosmologies. The Starobinsky inflation model \cite{Starobinsky:1980te,Barrow:1988xh,Odintsov:2015gba,Nojiri:2015sfd} can be described via the Hubble evolution $H(t) = H_i - \frac{M^2(t-t_i)}{6}$, where $t_i$ is the time when inflation starts, $H_i = H(t = t_i)$ and $M \gg 1$, and corresponds to the scale factor $a(t) \propto e^{H_i (t-t_i)-\frac{1}{12} M^2 (t-t_i)^2}$. At early-times, $a(t) \approx e^{H_i (t-t_i)-\frac{1}{6} M^2 t t_i}$. However, the ${t_i}^2$ contribution is retained in Ref.~\cite{Nassur:2016yhc},
\begin{equation}
    a(t) \approx e^{H_i (t-t_i)-\frac{1}{6} M^2 t t_i + \frac{1}{2}M^2 {t_i}^2}\,.
\end{equation}
Transforming in terms of unimodular time coordinates, $t^\prime$, leads to $a(t^\prime) \propto \left(t^\prime - k\right)^\frac{1}{3}$, for some integration constant $k$, resulting in the singular unimodular Hubble parameter $\mathcal{H} = \frac{1}{3 (t^{\prime}-k)}$\footnote{The listed result differs quite considerably from that found in Ref.~\cite{Nassur:2016yhc}. While $a(t^\prime)$ differs from the one listed in Eq.~(55), it agrees with Ref.~\cite{Nojiri:2015sfd} even if the ${t_i}^2$ contribution is included.} . Interestingly, $T$ becomes constant making the Friedmann equation to become algebraic. Thus, a vast class of unimodular $f(T)$ functions can host the singular bounce.

Within the context of a unimodular boundary term fixing $B(t^\prime) = -4\Lambda$ sets $a(t^\prime) = a_B [\beta(t^\prime - {t_B}^\prime)^2+1]^\frac{1}{6}$, where $a_B \equiv a({t_B}^\prime)$, $\beta = \frac{\Lambda}{{a_B}^6}$ which describes a cosmological sequence which passes through Type~IV bounces, an inflationary phase and a standard decelerated cosmology. Despite no pathologies arise near the bounces, the cosmology ends in a Minkowskian fate making it a non-viable scenario~\cite{ElHanafy:2017sih} . Here, the associated reconstructed unimodular $f(T)$ function expressed in terms of time is
\begin{equation}
f(t^\prime) = - \frac{t^\prime}{\sqrt{\beta {t^\prime}^2+1}} \int \frac{\lambda(t^\prime)}{{t^\prime}^2 \sqrt{\beta {t^\prime}^2+1}} \, \dd t^\prime + \frac{2\rho_{m,B}}{\sqrt{\beta {t^\prime}^2+1}} \, _2F_1\left(-\frac{1}{2}, 1+\frac{w}{2}, \frac{1}{2}, -\beta {t^\prime}^2\right)\,,
\end{equation}
where $\lambda(t^\prime)$ is the unimodular Lagrange multiplier and $\rho_{m,B} \equiv \rho_{m}({t_B}^\prime)$.

\subsubsection{Perspectives on bouncing cosmologies in teleparallel cosmology}

The bouncing cosmological description aims to provide an alternative viewpoint to the beginning and late phases of the Universe. In the context of \gls{tg} theories, bouncing cosmological scenarios can be generated either through a specification of the gravitational Lagrangian or via reconstruction methods. Thus, these bouncing cosmologies can be achieved without invoking any exotic matter. For instance, $f(T)$ gravity can host a vast number of bouncing solutions with the possibility of realizing a suitable cosmological sequence, such as the Pseudo-Bang cosmology. Meanwhile, current observations do not lead to the formation of a Type~III singularity in the case of teleparallel \gls{de} whereas avoidance of ghost instabilities have been verified in $f(T)$ gravity models. Meanwhile, $f(T,B)$ gravity can also host a diverse number of bouncing solutions which Lagrangian is able to host vacuum (such as Minkowski) solutions, and can also generate the necessary accelerating contraction prior a Big Crunch. On the other hand, $f(T,T_G)$ gravity can be used to generate bouncing cosmologies, however, only the models hosting a Type~III singularity were found to be compatible with the vacuum constraint. Finally, while the unimodular formulation can host the singular bounce, the theory has not yet been investigated in detail whether it can host other bouncing cosmologies.

It is worth nothing that most of the reported analysis has been conducted within the aspect of reconstruction. An exception lies within $f(T)$ gravity where the use of dynamical systems (some are also reported in Sec.~6.1.3) and its consequences in scalar and tensor power spectra (see Sec.~6.7.2) has been explored. While there has been works which briefly outline the existence of bouncing scenarios in other \gls{tg} formulations (see Secs.~6.3.3, 6.5.3.1 and 6.6.4), there has not yet been an extensive analysis investigating the existence and suitable conditions to realise a viable cosmological bounce. More work has yet to be conducted in this sector.

\subsection{Anisotropic cosmologies in teleparallel theories of gravity} \label{sec:anisotorpic_cosmo}

In the previous sections, the Universe was assumed to be spatially homogeneous and isotropic leading to the extensive use of the \gls{flrw} metric. However, it is known that inhomogeneities and anisotropies do exist in the Universe, making the \gls{flrw} metric insufficient to describe these behaviors unless inhomogeneous and anisotropic effects are included. As discussed in Sec.~7, these can be accounted for through small, linear perturbations in the \gls{flrw} metric. However, the cosmological perturbation procedure also has limitations, for instance when the size of the perturbations are comparable to the background cosmology.

There are also other philosophical arguments to consider. Could the early Universe be initially anisotropic and inhomogeneous which isotropizes and homogenesizes at later times? If inflation did not occur, what other mechanism could have taken place which generated the anisotropies and inhomogeneities? Even though it has been observed that the \gls{cmb} is statistically isotropic and Gaussian \cite{Bunn:1996ut}, anomalies appear at large scales including: (i) a lack of power at largest scales (quadropole contributions), (ii) hemispherical asymmetry and (iii) a preferred alignment at largest scales (quadropole and octopole). These features, which appeared in both \gls{wmap} \cite{Koivisto:2007bp,Pereira:2007yy,Akarsu:2009gz,Koivisto:2008ig,Watanabe:2010fh,Gumrukcuoglu:2007bx} and Planck data \cite{Ade:2015hxq,Akrami:2018odb}, could indicate a departure from this statistical isotropy principle. While investigations whether these are sourced from possible systematics in the experiments (\gls{wmap} \cite{Koivisto:2007bp,Gumrukcuoglu:2007bx}, Planck \cite{Akrami:2018odb}), a deviation from statistical isotropy may still be a reality.

A possible resolution is through the use of Bianchi models \cite{bianchi1897sugli,bianchi1918gruppi}, metrics which are spatially homogeneous but not necessarily isotropic. In this way, these could describe an initial anisotropic cosmology which correctly accounts for the early Universe anisotropies while it isotropizes at late-times leading to the overall observed \gls{flrw} cosmology. While a brief overview of the Bianchi metrics is provided, a detailed account is available in Refs.~\cite{Ellis:1998ct,McCabe:2005gu,Ellis:2006ba,gron2007einstein}.

Given a set of spatial basis vectors \textbf{e}$_i$ obeying the relations
\begin{equation}
    [\textbf{e}_i, \textbf{e}_j] = C^k_{ij} \textbf{e}_k\,,
\end{equation}
where $[\cdot,\cdot]$ is the commutator and $C^k_{ij}$ are structure constants, their associated dual vectors \textbf{w}$^k$ are constructed through
\begin{equation}
    \dd\textbf{w}^k = -\frac{1}{2} C^k_{ij} \textbf{w}^i \wedge \textbf{w}^j\,,
\end{equation}
where $\wedge$ represents the exterior product. This leads to the construction of the Bianchi metric
\begin{equation}
    \dd s^{2}= \dd t^{2}-\gamma_{ij}(t)\textbf{w}^{i}\textbf{w}^{j}\,,
\end{equation}
where $\gamma_{ij}$ is the spatial metric. Nine distinct types of Bianchi models, referred to as Type I--IX (BT-I--BT-IX), can be constructed from the structure constants, which are further classified under two classes:
\begin{enumerate}
    \item Class A: BT-I, II, VI$_0$, VII$_0$, VIII, IX;
    \item Class B: BT-III, IV, V, VI$_h$ and VII$_h$.
\end{enumerate}
The parameter $h$ is an arbitrary non-zero parameter which generates a family of isometries for Types VI and VII.

The Bianchi models have been extensively studied in the context of cosmology. In the case of the background evolution, Collins and Hawking have shown that for ordinary matter universes, most Bianchi models become anisotropic at early-times even if the Universe is very isotropic at late-times \cite{Collins:1972tf}. Furthermore, Wald showed that for a positive cosmological constant, Bianchi universes become isotropic at late-times \cite{Wald:1983ky}. Hu and Parker showed that for a renormalized energy-momentum tensor of a quantised matter field in a BT-I model, particle creation during times of the order of a Planck time could be strong enough to isotropize the Universe at an early-time \cite{Hu:1978zd}. BT-III models have also been studied for late-time isotropy \cite{Akarsu:2009gz}.

In the case of \gls{cmb} anisotropies, Class B models generate a spiral-pattern and hot spot regions, while BT-IX models can affect the quadropole spectrum \cite{Ellis:1998ct,Saadeh:2016sak}. The quadropole effects can also be investigated through other Bianchi spacetimes \cite{Akarsu:2009gz,Koivisto:2007bp,Watanabe:2010fh,Akarsu:2008fj,Pitrou:2008gk,Koivisto:2008ig}.

While cosmological perturbations are mainly considered for the \gls{flrw} cosmology, a similar approach can be considered for Bianchi metrics. As discussed in Refs.~\cite{Pereira:2007yy,Koivisto:2008ig}, the approaches are distinct for non-perfect fluids and the perturbation equations become scale dependent, which introduce a coupling between the scalar and tensor modes leading to a scalar-tensor see-saw mechanism.

\subsubsection{Tetrad and spin connection in anisotropic cosmologies}

For every Bianchi model choice, an associated tetrad needs to be constructed such that it generates the Bianchi metric. However, as discussed in Sec.~5.12, this tetrad does not guarantee that it satisfies the field equations. Adequately, a good tetrad-spin connection pair must be considered to host a consistent set of field equations. In the following, we shall first present and discuss the diagonal tetrad with zero spin connection for the BT-I, BT-III, Kantowski-Sachs and Kasner metrics in the context of the $f(T,B)$ gravity. For the BT-III case, this choice does not constitute a good tetrad-spin connection pair. Thus, we then present a novel set of good tetrad-spin connection pairs for a selection of Bianchi models which have not yet appeared in literature.

\paragraph{Using a diagonal tetrad - not good tetrad-spin connection pair}

Based on the anisotropic cosmological models considered in teleparallel theories of gravity, only a select number of Bianchi type metrics shall be explored in detail, namely BT-I and III. Furthermore, Kantowski-Sachs \cite{Kantowski:1966te} and Kasner \cite{Kasner:1921zz} spacetimes are also discussed in $f(T)$. The BT-III metric in Cartesian coordinates is given to be
\begin{equation}\label{eq:BianchiTypeIII-metric}
\dd s^2 = \dd t^2 - a^2(t) \dd x^2 - e^{-2\alpha x} b^2(t) \dd y^2 - c^2(t) \dd z^2\,,
\end{equation}
where $\alpha$ is a constant parameter and $a(t)$, $b(t)$ and $c(t)$ are the directional scale factors. Trivially, the \gls{flrw} metric is obtained when $a = b = c$ and $\alpha = 0$. On the other hand, BT-I is obtained when $\alpha = 0$ while the locally rotationally symmetric BT-I and Kantowski-Sachs models are obtained when $\alpha = 0$ and $b= c$. Lastly, the Kasner universe is obtained when $\alpha = 0$ and the scale factors reduce to power-law behaviors $a(t) \propto t^{p_1}$, $b(t) \propto t^{p_2}$ and $c(t) \propto t^{p_3}$ with $p_{1,2,3}$ are constants obeying the Kasner relations $\sum\limits_{i = 1}^3 p_i = 1$, and $\sum\limits_{i = 1}^3 {p_i}^2 = 1$.

Next, we choose a tetrad-spin connection pair which yields the metric. One natural choice that would generalize the flat \gls{flrw} tetrad~(6.4) is the diagonal tetrad
\begin{equation}
    \udt{e}{A}{\mu} = \text{diag}\left[1, a(t), e^{-\alpha x} b(t), c(t)\right]\,,\label{diagtetrad}
\end{equation}
and zero spin connection components, for which the torsion scalar and the boundary term become
\begin{subequations}
\begin{align}
    T &= -2\left(H_a H_b + H_a H_c + H_b H_c\right)\,,\\[0.5ex]
    B&=\frac{2 \alpha ^2}{a^2}-2 \left[\dot{H}_a+\dot{H}_b+\dot{H}_c+(H_a+H_b+H_c)^2\right]\,,
\end{align}
\end{subequations}
where $H_a=\dot{a}/a,H_b=\dot{b}/b,H_c=\dot{c}/c$ represent the directional Hubble parameters. Finally, it is convenient to define an average Hubble parameter
\begin{equation}\label{eq:average-Hubble}
    H \equiv \frac{1}{3}\left(H_a + H_b + H_c\right)\,,
\end{equation}
which describes the average expansion of this universe. Consequently, an average scale factor $\tilde{a} \equiv (abc)^{1/3}$ is defined. Naturally, this is related to the average Hubble parameter through the standard relation $H = \dot{\tilde{a}}/\tilde{a}$.
Interestingly, the effect of $\alpha$ is absent in the torsion scalar, which has important consequences to the field equations, which is best illustrated through the context of $f(T,B)$ gravity. Assuming a perfect fluid with anisotropic pressures in each spatial direction, $p_{x,y,z}$, for this choice of tetrad, the field equations of $f(T,B)$ gravity result into

\begingroup
\small
\vspace*{-\baselineskip}
\begin{subequations}
\begin{align}
\kappa^2\rho &= -\frac{f}{2} + f_T \left(T + \frac{\alpha^2}{a^2}\right)+\frac{B f_B}{2}+ \dot{f}_B \left(H_a+H_b+H_c\right)\,,\label{bianchi111} \\[0.5ex]
-\kappa^2 p_x &= -\frac{f}{2} -f_T \left[(H_b+H_c)^2+H_a (H_b+H_c)+\dot{H}_b+\dot{H}_c\right]- \left(H_b + H_c\right) \dot{f}_{T}+\frac{Bf_B}{2} +\ddot{f}_B\,, \\[0.5ex]
-\kappa^2 p_y &= -\frac{f}{2} - f_T \left[(H_a+H_c)^2+H_b (H_a+H_c)+\dot{H}_a+\dot{H}_c\right]- \left(H_a + H_c\right) \dot{f}_T+\frac{Bf_B}{2} +\ddot{f}_B\,, \\[0.5ex]
-\kappa^2 p_z &= -\frac{f}{2} - f_T \left[(H_b+H_a)^2+H_c (H_b+H_a)+\dot{H}_b+\dot{H}_a-\frac{\alpha ^2}{a^2}\right]- \left(H_a + H_b\right) \dot{f}_T+\frac{Bf_B}{2} +\ddot{f}_B\,, \\[0.5ex]
0 &=\alpha \left(H_a - H_b\right) f_T\,, \label{eq:fieldeq-BianchiTypeIII-offdiag2}
\end{align}
\end{subequations}
\endgroup
\normalsize and there is an additional equation from the antisymmetric part, namely
\begin{equation}
    \frac{1}{2}\alpha(\dot{f}_T+\dot{f}_B)=0\,. \label{eq:fieldeq-BianchiTypeIII-offdiag1}
\end{equation}
The respective $f(T)$ field equations reported in Refs.~\cite{Rodrigues:2012qua,Rodrigues:2014xam} are also recovered.

The last (antisymmetric) equation shows the problem of the choice of the tetrad not being compatible with a zero spin connection, since the only way to solve it is either $\alpha=0$ (which recovers BT-I), or $f=f(-T+B)$ which is just $f(\lc{R})$, or $\dot{T} = \dot{B} = 0$. In the latter case, however, this reduces to $\Lambda$\gls{cdm}. On the other hand, BT-I, Kantowski-Sachs, locally rotationally symmetric and Kasner cosmologies do not host this issue as the antisymmetric equation Eq.~\eqref{eq:fieldeq-BianchiTypeIII-offdiag1} is naturally satisfied.

At least in the context of $f(T)$ gravity, there has been an attempt to obtain a consistent set of field equations \cite{Rodrigues:2012wt}. However, it has been unsuccessful since an incompatible tetrad-spin connection pair was considered. A resolution of this issue is explored in the following part.

\paragraph{Using a good tetrad-spin connection pair}\label{sec:anisotropyteleparallel}

As we saw in the previous section, for BT-III, if one uses the diagonal tetrad in Eq.~\eqref{diagtetrad} in the Weitzenb\"{o}ck gauge, the antisymmetric field equations in $f(T)$ gravity constrain the model to be \gls{tegr} (or to become BT-I). This is related to the choice of the tetrad-spin connection pair that is not compatible with the symmetries of the system. In this section, we will derive a good tetrad-spin connection pair for $f(T,B)$ gravity for different types of Bianchi models. Let us first start with the following metric
\begin{equation}
    \dd s^2=\dd t^2-\sigma_1(t,x,y,z)\dd x^2-\sigma_2(t,x,y,z)\dd y^2-\sigma_3(t,x,y,z)\dd z^2\,,
\end{equation}
where $\sigma_i$ are arbitrary functions of the Cartesian coordinates. This metric is in the form of BT-I, BT-III, BT-V and BT-VI. Let us then start by assuming a zero spin connection and start with following diagonal tetrad
\begin{equation}
    \udt{e}{A}{\mu} = \text{diag}(1, \sigma_1, \sigma_2, \sigma_3)\,,\label{diagtetrad2}
\end{equation}
which is not a good tetrad-spin connection pair choice. Then, we can perform a Lorentz transformations $e'{}^A{}_\mu= \Lambda^A{}_B e^B{}_\mu$ with the three Euler $\alpha_i$ angles which explicitly gives us the following rotated tetrad
\begin{subequations}
\begin{alignat}{2}
\mathbf{e}'^0& =\: & &\dd t\,,\\[0.5ex]
\mathbf{e}'^1& =\: & & \sigma_1\cos \alpha_2 \cos \alpha_3 \, \dd x -\sigma_2(\sin \alpha_1 \sin \alpha_2 \cos \alpha_3+\cos \alpha_1 \sin \alpha_3) \, \dd y\nonumber\\[0.5ex]
& \: & &+ \sigma_3(\sin \alpha_1 \sin \alpha_3-\cos \alpha_1 \sin \alpha_2 \cos \alpha_3) \, \dd z\,, \\[0.5ex]
 \mathbf{e}'^2& =\: & & \sigma_1 \cos \alpha_2 \sin \alpha_3 \, \dd x + \sigma_2(\cos \alpha_1 \cos \alpha_3-\sin \alpha_1 \sin \alpha_2 \sin \alpha_3) \, \dd y \nonumber\\[0.5ex]
 & \: & &-\sigma_3(\cos \alpha_1 \sin \alpha_2 \sin \alpha_3+\sin \alpha_1 \cos \alpha_3) \, \dd z\,, \\[0.5ex]
 \mathbf{e}'^3& =\: & & \sigma_1\sin \alpha_2 \, \dd x + \sigma_2\sin \alpha_1 \cos \alpha_2 \, \dd y +\sigma_3 \cos \alpha_1 \cos \alpha_2 \, \dd z\,,
\end{alignat}
\end{subequations}
where $\alpha_i$ can depend on all the coordinates $(t,x,y,z)$. If we replace this tetrad into the $f(T,B)$ gravitational field equations~(5.80), we get that the antisymmetric equations become
\begin{subequations}
\begin{alignat}{2}
0& =\: & &(\dot{f}_B+\dot{f}_T) \Big[\sigma_3 \left(\sigma_1 \left(\alpha_{2,y} \sin \alpha_1+\alpha_{3,y} \cos \alpha_1 \cos \alpha_2\right)-\sigma_{2,x}\right) \nonumber\\[0.5ex]
&\: & &+\sigma_2 \left(\sigma_1 \left(\alpha_{2,z} \cos \alpha_1-\alpha_{3,z} \sin \alpha_1 \cos \alpha_2\right)-\sigma_{3,x}\right)\Big] \,,\label{anti1} \\[0.5ex]
0& =\: & &(\dot{f}_B+\dot{f}_T) \left(\sigma_1 \left(\sigma_2 \left(\alpha_{1,z}+\alpha_{3,z} \sin \alpha_2\right)-\sigma_{3,y}\right)-\sigma_3 \left(\sigma_2 \left(\alpha_{2,x} \sin \alpha_1+\alpha_{3,x} \cos \alpha_1 \cos \alpha_2\right)+\sigma_{1,y}\right)\right)\,, \label{anti22} \\[0.5ex]
0& =\: & &(\dot{f}_B+\dot{f}_T) \left(\sigma_1 \left(\sigma_3 \left(\alpha_{1,y}+\alpha_{3,y} \sin \alpha_2\right)+\sigma_{2,z}\right)+\sigma_2 \left(\sigma_3 \left(\alpha_{2,x} \cos \alpha_1-\alpha_{3,x} \sin \alpha_1 \cos \alpha_2\right)+\sigma_{1,z}\right)\right)\,. \label{anti3}
\end{alignat}
\end{subequations}
Here, commas denote derivatives. Clearly if $\dot{f}_B+\dot{f}_T\neq0$ (which is a theory different to $f(\lc{R})$ gravity), the system is difficult to solve. It is also easy to see that if $\alpha_i=0$, one cannot have a solution for the system for $\dot{f}_B+\dot{f}_T\neq0$. Then, we need the functions $\alpha_i$ in order to solve the antisymmetric equations and then to find a good tetrad-spin connection pair. Since the majority of Bianchi models with a diagonal metric form only depend on the time coordinate and only one extra Cartesian coordinate, we will concentrate on three specific types of metrics:
\begin{enumerate}
    \item All metric functions depending on $t$ and $x$: $\sigma_1=\sigma_1(t,x)$, $\sigma_2=\sigma_2(t,x)$ and $\sigma_3=\sigma_3(t,x)$
    \item All metric functions depending on $t$ and $y$: $\sigma_1=\sigma_1(t,y)$, $\sigma_2=\sigma_2(t,y)$ and $\sigma_3=\sigma_3(t,y)$
    \item All metric functions depending on $t$ and $z$: $\sigma_1=\sigma_1(t,z)$, $\sigma_2=\sigma_2(t,z)$ and $\sigma_3=\sigma_3(t,z)$
\end{enumerate}
BT-I, BT-III and BT-V are subcases of 1. and BT-VI is a subcase of 3. For the three cases mentioned above, it is possible to solve the antisymmetric equations~\eqref{anti1}--\eqref{anti3} by choosing
\begin{subequations}
\begin{eqnarray}
    \textrm{Case 1:}&&\quad \cos\alpha_1=\frac{\sigma_{2,x} \sigma_{3}+\sigma_{2} \sigma_{3,x}}{ \sigma_{1} \sigma_{2}}\,,\quad \alpha_2= z\,,\quad\alpha_3=\frac{\pi}{2}\,.\\[0.5ex]
    \textrm{Case 2:}&&\quad \cos\alpha_2=\frac{\sigma_{1,y} \sigma_{3}+\sigma_{1} \sigma_{3,y}}{ \sigma_{2} \sigma_{3}}\,,\quad \alpha_3=  x\,,\quad \alpha_1=\pi\,.\\[0.5ex]
    \textrm{Case 3:}&&\quad \sin\alpha_2=-\frac{\sigma_{1,z} \sigma_{2}+\sigma_{1} \sigma_{2,z}}{ \sigma_{1} \sigma_{3}}\,,\quad \alpha_3= y\,,\quad \alpha_1=\frac{\pi}{2}\,.
\end{eqnarray}
\end{subequations}
Using these results, we find that a good tetrad-spin connection pair for different Bianchi models with a zero spin connection, with the following tetrads:
\begin{subequations}
\begin{eqnarray}
\textrm{\underline{Bianchi Type I:}}&& \dd s^2=\dd t^2-a(t)^2\,\dd x^2 - b(t)^2\,\dd y^2-c(t)^2\,\dd z^2 \,, \\[0.5ex]
&&e'{}^A{}_\mu=\left(
\begin{array}{cccc}
 1 & 0 & 0 & 0 \\
 0 & 0 & 0 & c\\
 0 & a\cos z & -b \sin z & 0 \\
 0 & a \sin z & b \cos z & 0
\end{array}
\right)\,.\\ \nonumber\\[0.5ex]
\textrm{\underline{Bianchi Type II:}}&&\dd s^2=\dd t^2-a(t)^2 (\dd x- K z \dd y)^2- b(t)^2\dd y^2- c(t)^2\dd z^2\,,\\[0.5ex]
&&e'{}^A{}_\mu=\left(
\begin{array}{cccc}
 1 & 0 & 0 & 0 \\
 0 & 0 & -b & 0 \\
 0 & a & -K z a & 0 \\
 0 & 0 & 0 & c \\
\end{array}
\right)\,.\\ \nonumber \\[0.5ex]
\textrm{\underline{Bianchi Type III:}}&& \dd s^2=\dd t^2-a(t)^2\,\dd x^2 -b(t)^2 e^{-2 \alpha x}\,\dd y^2 -c(t)^2\,\dd z^2 \,, \\[0.5ex]
&&e'{}^A{}_\mu=\left(
\begin{array}{cccc}
 1 & 0 & 0 & 0 \\
 0 & 0 & \frac{\alpha bc}{a}e^{-\alpha x} & c \sqrt{1-\frac{\alpha ^2 c^2}{ a^2}} \\
 0 & a\cos z & -be^{-\alpha x} \sin z \sqrt{1-\frac{\alpha ^2 c^2}{a^2}} & \frac{\alpha c^2 \sin z}{ a} \\
 0 & a\sin z& be^{-\alpha x} \cos z \sqrt{1-\frac{\alpha ^2 c^2}{ a^2}} & -\frac{\alpha c^2 \cos z}{ a} \\
\end{array}
\right)\,.\label{bianchi3}
\\ \nonumber \\[0.5ex]
\textrm{\underline{Bianchi Type V:}}&&\dd s^2=\dd t^2-a(t)^2\,\dd x^2 -e^{-2 m x}\Big(b(t)^2\,\dd y^2 +c(t)^2\,\dd z^2 \Big)\,, \\[0.5ex]
&&e'{}^A{}_\mu=\left(
\begin{array}{cccc}
 1 & 0 & 0 & 0 \\
 0 & 0 & \frac{2 m bc}{ a} e^{-2 m x} & c e^{-m x} \sqrt{1-\frac{4 m^2 c^2 e^{-2 m x}}{ a^2}} \\
 0 & a \cos z & -b e^{-m x} \sin z \sqrt{1-\frac{4 m^2 c^2 e^{-2 m x}}{ a^2}} & \frac{2 m c^2 e^{-2 m x} \sin z}{ a} \\
 0 & a \sin z& b e^{-m x} \cos z \sqrt{1-\frac{4 m^2 c^2 e^{-2 m x}}{a^2}} & -\frac{2 m c^2 e^{-2 m x} \cos z}{ a} \\
\end{array}
\right). \nonumber\\ \\[0.7ex]
\textrm{\underline{Bianchi Type VI:}}&& \dd s^2=\dd t^2-a(t)^2 e^{-2 m z}\,\dd x^2 -b(t)^2 e^{2 n z}\,\dd y^2 -c(t)^2\,\dd z^2 \,, \\[0.5ex]
&&e'{}^A{}_\mu=\left(
\begin{array}{cccc}
 1 & 0 & 0 & 0 \\
 0 & a e^{-m z} \cos y \sqrt{\frac{c^2-b^2 (m-n)^2 e^{2 n z}}{c^2}} & \frac{b^2 (n-m) e^{2 n z} \cos y}{ c} & c \sin y \\
 0 & a e^{-m z} \sin y \sqrt{\frac{c^2-b^2 (m-n)^2 e^{2 n z}}{c^2}} & \frac{b^2 (n-m) e^{2 n z} \sin y}{ c} & -c \cos y \\
 0 & \frac{ab (m-n) e^{z (n-m)}}{c} & b e^{n z} \sqrt{\frac{ c^2-b^2 (m-n)^2 e^{2 n z}}{ c^2}} & 0 \\
\end{array}
\right)\,.\nonumber \\ \\[0.7ex]
\textrm{\underline{Bianchi Type IX:}}&&\dd s^2=\dd t^2-a(t)^2 (\dd x- \cos y\, \dd z)^2-b(t)^2 \left(\dd y^2+ \sin ^2y \, \dd z^2\right)\,, \\[0.5ex]
&&e'{}^A{}_\mu=\left(
\begin{array}{cccc}
 1 & 0 & 0 & 0 \\
 0 & 0 & -b \cos x & b \sin x \sin y \\
 0 & a & 0 & -a\cos y\\
 0 & 0 & b \sin x & b \cos x \sin y \\
\end{array}
\right)\,.
\end{eqnarray}
\end{subequations}
Here, we have also used the same idea for other Bianchi models having cross terms in the metric. All of these tetrads are new results that have not been reported in the literature. It is also possible to make a Lorentz transformation to switch on the spin connection and rewrite the above tetrads in another way. It is important to mention that these good tetrad-spin connection pairs were found in $f(T,B)$ gravity but, it can be shown that they are also good pairs for other more general modified teleparallel theories such as those with scalar fields $f(T,B,\phi,X)$ with $\phi=\phi(t)$.

Let us now re-derive the equations for the BT-III with the correct good tetrad-spin connection given by Eq.~\eqref{bianchi3} with a zero spin connection. For this spacetime the torsion scalar and the boundary term becomes
\begin{subequations}
\begin{align}
T&=-\frac{2 \alpha ^2}{a^2}-2 (H_a H_b+H_a H_c+H_b H_c)\,,\\[0.5ex]
B&=-2 \left[\dot{H}_a+\dot{H}_b+\dot{H}_c+(H_a+H_b+H_c)^2 \right]\,,
\end{align}
\end{subequations}
where we clearly see that $\alpha$ now appears in $T$ and does not appear in $B$. The cosmological equations then yield
\begingroup
\small
\vspace*{-\baselineskip}
\begin{subequations}
\begin{align}
\kappa^2\rho &= -\frac{f}{2} + f_T \left(T + \frac{\alpha^2}{a^2}\right)+\frac{B f_B}{2}+ \dot{f}_B \left(H_a+H_b+H_c\right)\,, \\[0.5ex]
-\kappa^2 p_x &= -\frac{f}{2} -f_T \left[\frac{\alpha^2}{a^2} + (H_b+H_c)^2+H_a (H_b+H_c)+\dot{H}_b+\dot{H}_c\right]- \left(H_b + H_c\right) \dot{f}_{T}+\frac{Bf_B}{2} +\ddot{f}_B\,, \\[0.5ex]
-\kappa^2 p_y &= -\frac{f}{2} - f_T \left[\frac{\alpha^2}{a^2} + (H_a+H_c)^2+H_b (H_a+H_c)+\dot{H}_a+\dot{H}_c\right]- \left(H_a + H_c\right) \dot{f}_T+\frac{Bf_B}{2} +\ddot{f}_B\,, \\[0.5ex]
-\kappa^2 p_z &= -\frac{f}{2} - f_T \left[(H_b+H_a)^2+H_c (H_b+H_a)+\dot{H}_b+\dot{H}_a\right]- \left(H_a + H_b\right) \dot{f}_T+\frac{Bf_B}{2} +\ddot{f}_B\,, \\[0.5ex]
0 &=\alpha \left(H_a - H_b\right) f_T\,. \label{E21bianchi}
\end{align}
\end{subequations}
\endgroup
\normalsize We can notice a big difference between these field equations and the ones derived using the incorrect tetrad~\eqref{bianchi111}--\eqref{eq:fieldeq-BianchiTypeIII-offdiag2}, which is that one does not have the antisymmetric equation~\eqref{eq:fieldeq-BianchiTypeIII-offdiag1}. This equation is the one which constrains the model to be $\alpha=0$ or $f_{TT}=0$ (\gls{gr}) or $\dot{T}=0$ (\gls{gr} again). Exactly as in \gls{gr} (and \gls{tegr}), from Eq.~\eqref{E21bianchi} one notices that one needs a fluid having a linear momentum with components $T_{01}=T_{10}$ in order to construct a BT-III with $H_a\neq H_b$ (or equivalently $a\neq b$). One interesting feature here is that for $f=F(B)$ gravity (without having \gls{gr} in the background), one can study BT-III with $a\neq b$ for an anisotropic fluid without requiring to have a fluid with linear momentum. This feature cannot be achieved in \gls{gr}.

This explicit example related to BT-III shows that it is very important to be careful choosing the correct tetrad-spin connection pair when one is studying certain non-trivial geometries. It is important to mention that the vast majority of the literature related to anisotropic cosmology in \gls{tg} did not consider this important point, and for this, many of them give incorrect conclusions about these models. As a future work, it would be interesting to re-analyze different Bianchi models using the correct tetrads shown here.

\subsubsection{Reconstruction method} \label{sec:anisotropy-reconstruction}

Reconstructing the gravitational Lagrangian in anisotropic universes has been vastly investigated in curvature-based theories \cite{Hossienkhani:2014cja,Zubair:2016way,Mishra:2017sdq,Chakraborty:2018thg}. In contrast to the \gls{flrw} case, the reconstruction approach is not straightforward due to an anisotropic metric and matter fluid, requiring further considerations and assumptions prior to solving for the unknown gravitational Lagrangian. Even if a cosmological evolution is assumed, one then has to consider whether this behavior is observed in particular directions or as an average expansion which may lead to distinguishable directional expansions. In what follows, various known cosmologies shall be examined in detail.

\paragraph{de Sitter Spacetime}

In the first scenario, each individual direction is assumed to evolve following a de Sitter expansion behavior. However, each direction may not necessarily be expanding at the same rate, otherwise it reduces to the \gls{flrw} cosmology. This approach has been considered in Refs.~\cite{Rodrigues:2012qua,Fayaz:2014swa,Fayaz:2015yka} using BT-I where the directional scale factors are taken to be in the form
\begin{equation}
a(t) \propto e^{H_{a,0}t}\,, \quad b(t) \propto e^{H_{b,0}t}\,, \quad c(t) \propto e^{H_{c,0}t}\,,
\end{equation}
where $H_{a,b,c,0}$ are constants. Note that the torsion scalar is now constant, $T_0$. Assuming an isotropic fluid with constant \gls{eos}, the following scenarios are obtained:
\begin{enumerate}
\item $w = -1$: The fluid behaves as a cosmological constant leading to the condition
\begin{equation}
\kappa^2\rho = -\frac{f(T_0)}{2} + T_0 f_T(T_0)\,,
\end{equation}
where $\rho$ is now constant in time. Choosing an $f(T)$ function leads to an algebraic relation between $H_{a,b,c,0}$ which, however, constrains the geometry to reduce to the \gls{flrw} one. Alternatively, treating the equation as an ordinary differential equation gives the class of functions which naturally host the de Sitter behavior. The latter has been considered in Refs.~\cite{Fayaz:2014swa,Fayaz:2015yka} where the matter fluid is assumed to behave as \gls{hde}, for which the $f(T)$ Lagrangian reduces to a \gls{tegr} rescaling.
\item $H_{a,0} + H_{b,0} + H_{c,0} = 0$: This implies that a decelerated expansion occurs along at least one spatial direction.
\end{enumerate}

The second scenario focuses on when the average Hubble parameter is a constant, $H_0$. As investigated in Ref.~\cite{Rodrigues:2013iua} for the locally rotationally symmetric BT-I cosmology $(b = c)$, this assumption imposes the condition $ab^2 \propto e^{3H_0 t}$ which is insufficient to reconstruct the Lagrangian. To this end, the anisotropic fluid is parametrized to behave as a cosmological constant at late-times through
\begin{equation}
    p_x = w_x \rho\,, \quad p_y = p_z = (w_x + \delta)\rho\,,
\end{equation}
where $\delta$ is the skewness parameter and $w_x$ is chosen to be $-1$. In the special case of \gls{tegr}, an analytical form for the directional scale factors are obtained,
\begin{equation}
\label{eq:anisotropy-TEGR-case-scalesolns}
    a(t) \propto e^{H_0 t} \left(\frac{\gamma}{H_0} + 3\lambda e^{-3H_0t}\right)^{-\frac{2}{3}}\,, \quad b(t) = e^{H_0 t} \left(\frac{\gamma}{H_0} + 3\lambda e^{-3H_0t}\right)^{\frac{1}{3}}\,,
\end{equation}
where $\gamma, \lambda$ are integration constants. Consequently, the skewness parameter takes on the following form
\begin{equation}
    \delta = -\frac{27{H_0}^2\lambda^2 e^{-3H_0 t}}{\gamma\left(\gamma e^{3H_0 t}+6H_0 \lambda\right)}\,.
\end{equation}
Evidently, this becomes a realistic model for a de Sitter late-time cosmology since, at late-times, $A \simeq B$ and $\delta \to 0$, meaning this Universe becomes isotropized.

From these considerations, the reconstruction method is then applied to obtain a functional form of the $f(T)$ Lagrangian with the assumption that the directional scale factors take on the previously obtained forms Eq.~\eqref{eq:anisotropy-TEGR-case-scalesolns}. Careful re-examination of the work reveals that the continuity equation does not lead to the ordinary differential equation $1+f_T - f_{TT} = 0$ as reported in Ref.~\cite{Rodrigues:2013iua}, but reduces to an identity relation. In other words, the continuity equation is insufficient to constrain the Lagrangian. Further considerations of the fluid behavior, either through $w_x$ or $\delta$, may be sufficient to reconstruct the $f(T)$ function.

\paragraph{Power-law behavior}

Throughout the works in Refs.~\cite{Rodrigues:2012qua,Paliathanasis:2016vsw,Fayaz:2014swa,Fayaz:2015yka,Sharif:2014fqa}, a power-law behavior has been considered for reconstruction. First, the standard power-law behavior
\begin{equation}
    a(t) \propto t^{p_1}\,, \quad b(t) \propto t^{p_2}\,, \quad c(t) \propto t^{p_3}\,,
\end{equation}
where $p_{1,2,3}$ are constants, is considered. This simple ansatz can host the Kasner metric provided that $p_{1,2,3}$ obey the Kasner relations. However, the introduction of an arbitrary $f(T)$ Lagrangian has important consequences to the existence of the Kasner metric or the form of the Kasner relations. Similar considerations appear in other modified theories of gravity \cite{Barrow:2005dn,Clifton:2006kc,Toporensky:2016kss,Muller:2017nxg}. To illustrate this, starting from the torsion scalar, which is given to be
\begin{equation}\label{eq:Tscalar_anisotropic_power-law}
    T = -2 t^{-2}\left(p_1 p_2 + p_1 p_3 + p_2 p_3\right)\,,
\end{equation}
is zero for a standard Kasner universe. In the absence of matter fluids, the Kasner cosmology is a solution to the field equations if $f(0) = 0$ provided that $|f_T(0)|$, $|f_{TT}(0)| < \infty$\footnote{A more general consideration can be taken, for example $|f_T(0)| \to \infty$ but $|Tf_T|_{T \to 0} < \infty$ but this has not been investigated in detail.}. This allows for an extensive class of functions which can host the Kasner cosmology, including (i) power-law $f(T) \propto (-T)^n$ for $n > 0$, and (ii) exponential $f(T) = -T - Te^{-pT}$, $p$ is some constant \cite{Paliathanasis:2016vsw,Skugoreva:2017vde}.

Alternatively, a Kasner metric is obtained through the modification of Kasner's relations \cite{Paliathanasis:2016vsw}. For instance, in the absence of matter fluids and taking $f(T) \propto (-T)^n$ for $n > 0$, the field equations are satisfied if $T = 0$, which lead to the following modified Kasner relations:
\begin{subequations}
\begin{align}
&n > 0: & &\sum\limits_{i = 1}^3 p_i = 2n - 1\,, & &\sum\limits_{i=1}^3 {p_i}^2 = (2n-1)^2\,, \label{eq:anistropy-modified-KasnerI} \\[0.5ex]
&n > 1: & &\sum\limits_{i = 1}^3 p_i = C_0\,, & &\sum\limits_{i=1}^3 {p_i}^2 = {C_0}^2\,, \label{eq:anistropy-modified-KasnerII}
\end{align}
\end{subequations}
where $C_0$ is some arbitrary constant. It can be observed that the standard Kasner universe is recovered only for $n \geq 1$ by setting $C_0 = 1$, while a non-standard one is obtained for $0 < n < 1$.

For non-Kasner cosmologies, a perfect isotropic fluid was considered in Refs.~\cite{Rodrigues:2012qua,Fayaz:2014swa,Fayaz:2015yka,Sharif:2014fqa} including constant \gls{eos} $w$, Ricci Dark Energy (RDE), New Agegraphic Dark Energy (NADE) and \gls{hde} (including their PLE corrections) fluid types. For the constant \gls{eos} case, the reconstructed Lagrangian is
\begin{equation}
    f(T) = \frac{2\kappa^2\rho_0}{-1+(1+w)(p_1+p_2+p_3)} \left(\frac{-T}{2\chi}\right)^{n}, \quad \chi \equiv p_1 p_2 + p_1 p_3 + p_2 p_3 \neq 0\,.
\end{equation}
Here, an expanding universe is obtained when $\chi < 0$ which imposes the extra necessary condition $(1+w)\beta = 2n$, $n \in \mathbb{N}$, in order to maintain a real Lagrangian. Furthermore, the field equations give rise to two possible scenarios \cite{Rodrigues:2012qua}:
\begin{enumerate}
\item $w(p_1+p_2+p_3) = 1$: Yields an anisotropic expansion within an isotropic fluid;
\item $p_1 = p_2 = p_3$: Reduces to \gls{flrw} cosmology.
\end{enumerate}
For the remaining fluids, the reconstruction procedure was applied within an locally rotationally symmetric BT-I cosmology where the scale factors take on either standard power-law or pole-like forms satisfying\footnote{The index $h$ as seen in Refs.~\cite{Fayaz:2014swa,Fayaz:2015yka} has been redefined in order to simplify the results. The original solutions are recovered via the transformation $h \to -h$. In the case of Ref.~\cite{Sharif:2014fqa}, the transformations $h \to -m h$ and $m \to 1/m$ need to be applied.}
\begin{equation}
    a(t) = b(t)^\frac{1}{m}, \text{ with } b(t) = b_0(t_s-t)^{mh}\,,
\end{equation}
where $b_0, m, h$ are constants and $t_s \geq t$ represents the time of a future singularity. For these cases, the $f(T)$ Lagrangian was reconstructed and are summarized in Table~\ref{table:anisotropy-power-law-reconstruction}. It is remarked that in Ref.~\cite{Fayaz:2014swa}, the correct reconstructed solution is the pole-like one but the solution remains identical for the non-pole-like case. Furthermore, the solutions for PLECHDE and PLECNADE as given in Ref.~\cite{Sharif:2014fqa} are based on assuming the future event horizon and the conformal time computed based on the expansion of $b(t)$ as opposed to the average scale factor. Overall, it is observed that the models reduce to power-law type models.

\begin{table}[!ht]
\centering
\midsepremove
	\begin{tabular}{ll}
		\toprule
		\cellcolor{gris3}	\textbf{Fluid Type} & 	\cellcolor{gris3}\boldmath{$f(T)$} \\ \midrule
		\cellcolor{gris1}	RDE & 	\cellcolor{gris1}$-\frac{\alpha \kappa ^2 (2 m+1) \left[h (4 m+2)-3\right]}{3 h m (m+2)} T$ \\
		\cellcolor{gris3}	PLECHDE & 	\cellcolor{gris3}$-\frac{6 \delta^2 (h m-1)^2}{\gamma}T-\frac{2 \beta (1-h m)^{\alpha}}{\alpha -1}\left(-\frac{T}{\gamma}\right)^{\frac{\alpha}{2}}$ \\
		\cellcolor{gris1}	PLECNADE &	\cellcolor{gris1} $-\frac{6 {b_0}^2 \eta^2 (h m-1)^2}{2 h m-1} \left(-\frac{T}{\gamma}\right)^{1-h m} + \frac{2 {b_0}^\alpha (hm-1)^{\alpha} \beta}{1-\alpha(1- h m)} \left(-\frac{T}{\gamma}\right)^{\frac{\alpha}{2} (1-h m)}$ \\
		\bottomrule
	\end{tabular}
	\midsepdefault
	\caption{Summary of the reconstructed $f(T)$ Lagrangian for an isotropic fluid in the case of power-law (or pole-like) directional scale factors as derived in Refs.~\cite{Fayaz:2014swa,Fayaz:2015yka,Sharif:2014fqa}. Here, $\gamma \coloneqq 2h^2 m (m+2)$ while the constant $\delta$ appears in the RDE fluid, $\eta$ for NADE, $\alpha$ and $\beta$ for the PLE corrections of both \gls{hde} and NADE. The \gls{hde} and NADE reconstructed models derived in Ref.~\cite{Sharif:2014fqa} are recovered by setting $\beta \to 0$.}
	\label{table:anisotropy-power-law-reconstruction}
\end{table}

\paragraph{\texorpdfstring{$\Lambda$}{}CDM behavior}

The $\Lambda$\gls{cdm} cosmology has been studied in Refs.~\cite{Sharif:2011bi,Amir:2015wja} for a perfect isotropic matter fluid with constant \gls{eos} $w$. Here, the cosmology is built on the principle that the average expansion is $\Lambda$\gls{cdm}, meaning it obeys the same evolution equation $H^2 = {H_0}^2\left(\Omega_{\rm m0} \tilde{a}^{-3} + \Omega_\Lambda\right)$. However, since $\tilde{a}$ and $H$ are now the average quantities, each spatial direction may now evolve differently from one another without each individually exhibiting a behavior similar to the $\Lambda$\gls{cdm} cosmology.

To reconstruct the Lagrangian, the matter fluid is expressed in terms of the torsion scalar as
\begin{equation}
    \rho = \rho_0 \left(\frac{J-T- 9{H_0}^2 \Omega_\Lambda}{9{H_0}^2 \Omega_{\rm m0}}\right)^{1+w}\,,
\end{equation}
where $\rho_0$ is some constant and $J \coloneqq {H_a}^2 + {H_b}^2 + {H_c}^2$. Then, using the Friedmann equations lead to the solution
\begin{equation}\label{eq:anisotropy-LCDM-reconstruction}
    f(T) = \sqrt{-T} \int\limits^{-T}_0 \frac{\kappa^2 \rho_0}{x^{3/2}} \left(\frac{J-x- 9{H_0}^2 \Omega_\Lambda}{9{H_0}^2 \Omega_{\rm m0}}\right)^{1+w} \; \dd x\,.
\end{equation}
Here, the solution has been generalized for an arbitrary \gls{eos}. It is remarked that the signature of the $f_T$ coefficient of the Friedmann equation in Refs.~\cite{Sharif:2011bi,Amir:2015wja} is incorrect while the $\sin \theta$ contributions in Ref.~\cite{Amir:2015wja} should not appear. Evidently, contrary to the \gls{flrw} case, an analytical expression cannot be obtained because $J$ cannot be solely expressed in terms of $T$. Nonetheless, prior knowledge of the directional scale factors would resolve this issue.

\paragraph{Unification of matter and dark energy behaviors}

The reconstruction for a unification of matter and \gls{de} domination epochs has been explored in Ref.~\cite{Rodrigues:2012qua}. In particular, the model follows a similar approach to the one encountered in Ref.~\cite{Nassur:2015zba} but is now applied to the anisotropic BT-I cosmology. Here, each directional scale factor evolves as $e^{h_i(t) \ln t}$ where each $h_i(t)$ is a slowly varying function chosen to be
\begin{equation}\label{eq:anisotropy-slowlyvar}
    h_i(t) = \frac{h_{i,1} + h_{i,2} \zeta t^2}{1+ \zeta t^2}\,,
\end{equation}
where $h_{i,1}, h_{i,2}$ and $\zeta$ are positive constants. This choice signifies a directional decelerating era for $0 < h_i < 1$ and an accelerating era for $h_i > 1$. Contrary to the \gls{flrw} case, however, the average expansion behavior of this universe may be different. This can be observed from the deceleration parameter
\begin{equation}
    q \sim -1 + \frac{3}{h_x + h_y + h_z} \equiv -1 + \frac{1}{\tilde{h}}\,,
\end{equation}
with $\tilde{h}$ representing the average of the slowly varying parameters. On average, deceleration is obtained when $0 < \tilde{h} < 1$ while acceleration for $\bar{h} > 1$, which implies that at least one spatial direction could have an opposite expansion behavior from the average one.

Due to the complexity of the resulting field equations, an isotropic fluid as well a Kantowski-Sachs metric is assumed. As $h_i$ are slowly varying $(\dot{h}_i \sim \ddot{h}_i \sim 0)$, the reconstructed Lagrangian takes on the form\footnote{The solution obtained in Ref.~\cite{Rodrigues:2012qua} is not recovered.}
\begin{equation}\label{eq:anisotropy-generalexp-reconstruction}
    f(T) \propto \exp \left\lbrace -\frac{1}{2}(1+w) \left[\ln \left(1+\zeta \Psi_0(T)\right) (h_{x,2}+2 h_{y,2})+(h_{x,1}+2 h_{y,1}) \ln \left(\frac{\Psi_0(T)}{1+\zeta \Psi_0(T)}\right)\right]\right\rbrace \,,
\end{equation}
where $\Psi_0(T)$ is the real root to the polynomial
\begin{subequations}
\begin{align}
0&= \zeta^2 T {\Psi_0}^3 + (2\zeta T + 2\zeta^2 X){\Psi_0}^2 + (T +2 \zeta Z)\Psi_0 +2 Y\,, \\[0.5ex]
X &\equiv h_{x,2} h_{y,2} + h_{x,2} h_{z,2} + h_{y,2} h_{z,2}, \hspace{2mm} Y \equiv h_{x,1} h_{y,1} + h_{x,1} h_{z,1} + h_{y,1} h_{z,1}\,,
 \\[0.5ex]
Z &\equiv h_{x,1} h_{y,2} + h_{x,2} h_{y,1} + h_{x,1} h_{z,2} + h_{x,2} h_{z,1} + h_{y,1} h_{z,2} + h_{y,2} h_{z,1}\,.
\end{align}
\end{subequations}
During early- and late-times, the Lagrangian takes on a more simplified form:
\begin{subequations}
\begin{align}
&\text{At early-times:} & \hspace{-2cm}f(T) \propto \Psi_0(T)^{-\frac{1+w}{2} (h_{x,1}+2 h_{y,1})}\,, \\[0.5ex]
&\text{At late-times:} &\hspace{-2cm} f(T) \propto \Psi_0(T)^{-\frac{1+w}{2} (h_{x,2}+2 h_{y,2})}\,.
\end{align}
\end{subequations}
Observe that at early-times, the reconstructed form depends on the constants $h_{i,1}$ which highlight the early-time dominant behavior as seen from Eq.~\eqref{eq:anisotropy-slowlyvar}. Likewise, the late-time reconstructed form is dependent on $h_{i,2}$ as expected.

\subsubsection{Noether's symmetry approach} \label{sec:anistro_noether}

The application of Noether symmetry for anisotropic cosmologies has been explored in curvature-based theories. Searching for the existence of such symmetries allow for the simplification of the field equations while exploring the nature of the anisotropic cosmology, for instance, whether the latter isotropizes at late-times. Some works include, but are not limited to, Refs.~\cite{Capozziello:1996ay,Jamil:2012zm,Kucukakca:2012zm,Camci:2007zz,Sharif:2015oxa,Shamir:2017jim}. In the context of \gls{tg}, Refs.~\cite{Aslam:2013coa,Motavalli:2016tsz,Tajahmad:2019lin} are reviewed.

Starting with $f(T)$ gravity, Ref.~\cite{Motavalli:2016tsz} considers a Kantowski-Sachs spacetime in polar coordinates. However, the chosen diagonal tetrad ${e^A}_{\mu} = \text{diag}\left(1,a,b,b \sin \theta\right)$ proves to be problematic as the spin connection becomes non-zero which is not considered in the work. This is due the fact that the antisymmetric field equations were not checked for consistency. Thus, the resulting Noether symmetries and cosmologies may not be entirely correct.

In Ref.~\cite{Aslam:2013coa}, a BT-I metric was considered with a configuration space $\mathbb{Q} = \lbrace t, a, b, c, T\rbrace$. The Noether constraint imposes the functional Lagrangian to $f(T) = -T + m T^n$ for some constants $m$ and $n$ which, depending on the magnitude of $n$ ($n = 1$ or $n \neq 1$), different Noether symmetries are obtained. However, it is remarked that Eq.~(22) in Ref.~\cite{Aslam:2013coa} is missing a factor of $T$ for the $f_{TT}$ coefficients which may introduce more symmetries than those reported. While the resulting cosmology was not analyzed for $n \neq 1$, the directional scale factors have been solved for $n = 1$ (i.e. \gls{tegr}) when the matter fluid is absent. However, the scale factors do not host a simple analytic form and hence were not analyzed in detail.

Besides, the anisotropic scenario, Ref.~\cite{Aslam:2013coa} explores the \gls{flrw} limit where only a number of Noether symmetries remain. However, these do not match those investigated in Sec.~6.1.2 due to how the configuration space and the Noether condition are treated. When the \gls{flrw} limit is considered, the configuration space dimension $\mathbb{Q}$ decreases, adequately reducing the number of partial differential equations arising from the Noether condition. Thus, in order to correctly recover the system of equations, the \gls{flrw} limit must be taken before the Noether condition is considered. In fact, the partial differential equation \cite{Aslam:2013coa}
\begin{equation}
    c \,\eta_{1,a} + b \,\eta_{4,a} = 0\,,
\end{equation}
introduces the constraint $\tilde{a}\, \eta_{1,\tilde{a}} = 0$, where $\tilde{a}$ now becomes the standard isotropic \gls{flrw} scale factor since $a = b = c = \tilde{a}$. This extra constraint does not appear in the standard $f(T)$ \gls{flrw} Noether equations.

Next, \gls{tegr} in the presence of a scalar-electromagnetic field coupling is considered \cite{Tajahmad:2019lin}
\begin{equation}
\mathcal{L} = -\frac{T}{2\kappa^2} - \frac{1}{2}\phi_{,\mu} \phi^{,\mu} + V(\phi) + \frac{1}{4}f^2(\phi) F_{\mu\nu}F^{\mu\nu}\,.
\end{equation}
As discussed in Sec.~6.5.2.1, due to the action's equivalence with its curvature analogue, the same Noether symmetries are expected. An locally rotationally symmetric BT-I model $(b = c)$ with $a = b^m$ for some constant $m$ was then considered where the Noether condition constrains the potential and gauge kinetic functions to take the following forms:
\begin{equation}
V(\phi) \propto \left(c_1 e^{\mu \phi} - c_2 e^{-\mu \phi}\right)^2\,, \quad f(\phi) \propto \left(c_1 e^{\mu \phi} - c_2 e^{-\mu \phi}\right)^n\,,
\end{equation}
where $\mu = \frac{m+2}{2\sqrt{4m+2}}$ and $n = \lbrace \frac{2-m}{2+m}, \frac{m}{2+m} \rbrace$ is a parameter which value depends on the choice of the electromagnetic vector potential. While a general cosmological behavior was not explored, four special cases were examined in detail: (i) $c_1 \neq 0, c_2 = 0$, (ii) $c_1 = 0,\,c_2 \neq 0$ and (iii) $c_1 = c_2 = 1/2$ (iv) $c_1 = -c_2 = \frac{1}{2}$. With appropriate fine-tuning of parameters, the resulting cosmology can match with observations and also in the last case some interesting potential forms arise that could account for a unified dark matter model.

\subsubsection{Cosmological stability and dynamical system approach}

In Refs.~\cite{Skugoreva:2017vde,Paliathanasis:2017htk}, the background cosmological stability of a Kasner universe in the context of $f(T)$ gravity was investigated in the case of vacuum. In Ref.~\cite{Paliathanasis:2017htk}, a perturbative approach on the background Kasner cosmology was considered, which focuses on perturbing the directional scale factors $\textbf{A} = \lbrace a, b, c\rbrace$ via
\begin{equation}
\textbf{A} = \textbf{A}_0\left(1 + \varepsilon \textbf{A}_\varepsilon\right) + \mathcal{O}(\epsilon^2)\,,
\end{equation}
where $\textbf{A}_0 \propto \left(t^{p_1}, t^{p_2}, t^{p_3} \right)$ is the background Kasner cosmology, $\varepsilon$ is an infinitesimal parameter and $\textbf{A}_\varepsilon = \left( a_\varepsilon, b_\varepsilon, c_\varepsilon \right)$ represent the directional scale factor perturbations. The stability is determined based on whether $\textbf{A}_\varepsilon$ grows or decays with time. On the other hand, a numerical integration on the field equations is performed in Ref.~\cite{Skugoreva:2017vde} to investigate the stability of the Kasner cosmology. In both works, the power-law ansatz $f(T) = -T - f_0 T^n$, where $f_0$ is some constant and $n \in \mathbb{N}$ is considered while the exponential model $f(T) = -T - f_1\left(1- e^{-pT^m}\right)$, where $f_1, p$ and $m$ are constants is only considered in Ref.~\cite{Paliathanasis:2017htk}.

Starting with the power-law case, as discussed in Sec.~\ref{sec:anisotropy-reconstruction}, a Kasner solution only exists when $T = 0$ or $T^{n-1} = \frac{1}{f_0 (1-2n)}$. The $T = 0$ case is an exact solution to the field equations and therefore stability is not considered. For the remaining case, at least for $n \geq 2$, the Kasner solution is unstable and evolves towards a de Sitter attractor, which only exists when $\text{sign}(f_0) = (-1)^n$. While this unstable point is stated to be a saddle point in Ref.~\cite{Paliathanasis:2017htk}, it actually represents a source in an expanding universe \cite{Skugoreva:2017vde}.

For the exponential model case, the Kasner solution is a hyperbolic point (hence cannot be stable) for $m = 1$ while for $m > 1$, the behavior is identical to the power-law case discussed previously. However, the existence of de Sitter attractors it not explored for this case.

Meanwhile, the dynamical behavior of a BT-I cosmology has been investigated in Refs.~\cite{Paliathanasis:2017htk,Skugoreva:2019bwt,Tretyakov:2021cgb} for the $f(T) = -T - f_0 T^n$ ansatz. Starting with Ref.~\cite{Paliathanasis:2017htk} in the absence of matter fluids, for $n = 2$, the critical points $({H_a}^\star, {H_b}^\star, {H_c}^\star)$
\begin{equation}
    P_{\pm} = \pm \left(\frac{1}{6\sqrt{f_0}}, \frac{1}{6\sqrt{f_0}}, \frac{1}{6\sqrt{f_0}}\right)\,, \quad P_0 = \left(0,0,0\right)\,,
\end{equation}
are obtained. $P_{\pm}$ correspond to a \gls{flrw} de Sitter cosmology while $P_0$ represents a static universe. Furthermore, $P_-$ is unstable, $P_0$ is a saddle point and $P_+$ is stable, meaning that a BT-I cosmology can isotropize towards a de Sitter late-time behavior. While the full critical point analysis has not been explored for arbitrary $n$, a de Sitter attractor solution only exists when $\text{sign}(f_0) = (-1)^n$.

In Ref.~\cite{Skugoreva:2019bwt}, the effect of matter fluids is also considered. The chosen dimensionless phase-space variables are obtained by dividing Eq.~\eqref{bianchi111} with $9H^2f_T$ to yield
\begin{equation}
    \tilde{x} := \frac{T}{9H^2}\,, \quad \tilde{y} := \frac{f_0(1-2n)T^{n-1}-1}{f_T}\,, \quad \tilde{r} := \frac{\kappa^2 \rho}{9H^2f_T}\,.
\end{equation}
As the dynamical variable $\tilde{r}$ appears in the Friedmann equations, it can be used to further constrain the parameter space leaving $(\tilde{x},\tilde{y})$ as the free dynamical variables. This gives rise to a number of critical points, summarized in Table~\ref{table:anisotropy-dynamics-critpts}, which can be associated to distinct cosmological behaviors.

\begin{table}[!ht]
	\small
	\centering
	\midsepremove
	\begin{tabularx}{\textwidth}{llp{5.5cm}X}
		\toprule
		\cellcolor{gris3}	& 	\cellcolor{gris3}\boldmath{$(\tilde{x}^\star,\tilde{y}^\star)$} & 	\cellcolor{gris3}\textbf{Stability} & 	\cellcolor{gris3}\textbf{Cosmological Behavior} \\ \midrule
		\cellcolor{gris1}	$P_1$ & 	\cellcolor{gris1}$\left(0, \frac{1-2n}{n}\right)$ & 	\cellcolor{gris1}Unstable node $-1 < w \leq w_\text{cr.}$ \newline Saddle $w_\text{cr.} < w \leq 1$ \newline Unstable $w = w_\text{cr.}$ & 	\cellcolor{gris1}Anisotropic asymptotic behavior where each directional scale factor evolves as $(t-t_0)^{p_{i,1}} e^{\frac{p_{i,2} (t-t_0)^{2-\alpha}}{2-\alpha}}$ as $t \to t_0$, and $\sum p_{i,1} = \frac{n}{1-w(n-1)}$, $\sum {p_{i,1}}^2 = \left(\sum p_{i,1}\right)^2$ and $p_{i,2} \neq 0$ \\
		\cellcolor{gris3}	$P_2$ & 	\cellcolor{gris3}$\left(-\frac{2}{3}, \frac{1-2n}{n}\right)$ & 	\cellcolor{gris3}Unstable node $w_\text{cr.} < w \leq 1$ \newline Saddle $-1 < w < w_\text{cr.}$ \newline Unstable $w = w_\text{cr.}$ &	\cellcolor{gris3} Isotropic asymptotic power-law behavior \newline $a(t) \propto (t-t_0)^{\frac{2n}{3(1+w)}}$ as $t \to t_0$ \\
		\cellcolor{gris1}	$P_3$ & 	\cellcolor{gris1}$\left(-\frac{2}{3}, -1\right)$ &	\cellcolor{gris1} Stable node $-1 < w < 1$ & 	\cellcolor{gris1}Isotropic asymptotic power-law behavior \newline $a(t) t^{\frac{2}{3(1+w)}}$ as $t \to \infty$ \\ [2ex]
		\cellcolor{gris3}	$P_4$ & 	\cellcolor{gris3}$\left(-\frac{2}{3}, 0\right)$ & 	\cellcolor{gris3}Stable node $-1 < w \leq 1$ & 	\cellcolor{gris3}Isotropic de Sitter solution $H = H_0$ with ${T_0}^{n-1} = \frac{1}{f_0(1-2n)}$ \\
		\cellcolor{gris1}	$P_5$ & 	\cellcolor{gris1}$\left(x^*, \frac{1-2n}{n}\right)$ & 	\cellcolor{gris1}Unstable, only exists for $w = w_\text{cr.}$ & 	\cellcolor{gris1}Anisotropic asymptotic power-law behavior with $a_i \propto (t-t_0)^{p_i}$ as $t \to t_0$ with $\sum p_i = 2n-1$ \\
		\bottomrule
	\end{tabularx}
	\midsepdefault
	\caption{The summary of possible critical points for the $f(T) = -T - f_0 T^n$, $n > 1$, model Lagrangian obtained in Ref.~\cite{Skugoreva:2019bwt}, which describe various cosmological scenarios. The nature of the critical points are also highlighted. Here, $\alpha = \frac{1+w}{1-w(n-1)}$ and $w_\text{cr.} \coloneqq \frac{1}{2n-1}$ have been defined.}
	\label{table:anisotropy-dynamics-critpts}
\end{table}

\normalsize Evidently, the critical points describe both isotropic and anisotropic behaviors, and since certain isotropic critical points can be stable, isotropization of this universe can be achieved at late-times. Furthermore, the de Sitter isotropic solution is again recovered but, contrary to Ref.~\cite{Paliathanasis:2017htk}, its stability is further emphasized by the \gls{eos} of the matter fluid.

For the special cases $n = 2, 3$, the overall cosmological dynamical behavior has been analyzed in detail where, it has been observed that the behavior changes depending on the signature of $f_0$. Starting with the $n = 2$ case:
\begin{enumerate}
\item $f_0 > 0$: The universe starts at an anisotropic non-standard singularity where $T$ finite but $\dot{T}$ is divergent, which later isotropizes to either the power-law behavior $(P_3)$ or de Sitter $(P_4)$.
\item $f_0 < 0$: A non-standard singularity where $T$ approaches a point where $f_T = 0$ is identified. The following scenarios are obtained:
\begin{enumerate}
\item Universe begins at either one of the critical points $P_{1,2,5}$ which later evolves to the $P_3$ behavior;
\item The Universe starts and finishes with the aforementioned non-standard singularity, which evolution causes either a cosmological bounce or turnaround;
\item Starts at the critical point $P_1$ which later evolves to the non-standard singularity.
\end{enumerate}
\end{enumerate}
For the $n = 3$ case, $f_0 > 0$ only leads to the behavior described in 2(a) while the remaining cases are all possible for $f_0 < 0$. Therefore, it is possible to realize a late-time accelerating behavior for appropriate parameter choices.

\subsubsection{Perspective on anisotropic cosmologies in teleparallel cosmology}

Within \gls{tg}, realization of various cosmological behaviors with possible late-time isotropization has been observed. However, further investigations still need to be performed especially since the main attention has been on the background cosmological behavior. Resolving the $H_0$ tension problem is extremely important for any future model of cosmology, however, an anisotropic cosmology has not yet been shown to be a realistic path in producing the necessary dynamics for this since it would lead to multiple expansion profiles in the different orientations. Meanwhile, only one \gls{tg} extension, that of $f(T)$ gravity, has been explored for anisotropic cosmology\footnote{While work on $f(T,T_G)$ gravity has been carried out in Ref.~\cite{Sharif:2016oqy}, for $f(T,T_G) \propto T_G$, the listed field equations still give rise to cosmological dynamics despite the fact that the Lagrangian is expressed in terms of a linear boundary term.}. Additionally, due to the spin connection choice, only the BT-I anisotropic cosmology has been extensively explored. In light of the newly presented good tetrad-spin connection pairs (at least for the $f(T,B)$ class), the remaining BT cosmologies may now be explored. All in all, a substantial amount of research still remains to be explored for anisotropic cosmologies.

\subsection{Reconstructed Solutions} \label{App:reconstruct_sols}

This section contains the reconstructed solutions which are derived for some extended teleparallel gravity scenarios in Sec.~6.

\begin{table}[H]
    \centering
    \midsepremove
    \small
    \begin{tabular}{lccl}
    \toprule
    \cellcolor{gris3}\textbf{Ansatz} &\cellcolor{gris3}\boldmath{$a(t)$} & \cellcolor{gris3}\boldmath{$w$} & \cellcolor{gris3}\textbf{Reconstructed} \boldmath{$f(T)$} \textbf{Lagrangian} \\ \midrule
  \cellcolor{gris2}\multirow{-1}{*}{$j = 1$} & \cellcolor{gris2}\multirow{-1}{*}{$\left(A e^{\lambda t} + Be^{-\lambda t}\right)^{\frac{2}{3}}$} & \cellcolor{gris2}-- & \cellcolor{gris2}$\frac{3 T}{8 \lambda ^2} \, _2F_1\left(\frac{1}{2},-w;\frac{3}{2};-\frac{3 T}{8 \lambda ^2}\right)- \, _2F_1\left(-\frac{1}{2},-w;\frac{1}{2};-\frac{3 T}{8 \lambda ^2}\right)$ \\
    \cellcolor{gris1} & \cellcolor{gris1} & \cellcolor{gris1}$0$ & \cellcolor{gris1}$1-3y^2 + \frac{y^3}{2} + 3y \ln y$ \\
    \cellcolor{gris1} & \cellcolor{gris1} & \cellcolor{gris1}$-\frac{1}{2}$ & \cellcolor{gris1}$(1+2y)\sqrt{1-y} + 3y \ln \left(\frac{1+\sqrt{1-y}}{\sqrt{y}}\right)$ \\
  \cellcolor{gris1}\multirow{-2}{*}{$j = \frac{s^2}{H^2}$} & \cellcolor{gris1}\multirow{-2}{*}{$\frac{1}{s}\left(me^{st} - ne^{-st}\right) + p$} & \cellcolor{gris1}$\frac{1}{2}$ & \cellcolor{gris1}$(35+388y-156y^2+58y^3-10y^4)\sqrt{1-y} - 315y \ln \left(\frac{1+\sqrt{1-y}}{\sqrt{y}}\right)$ \\
   \cellcolor{gris1} & \cellcolor{gris1} & \cellcolor{gris1}$1$ & \cellcolor{gris1}$1-15y^2+10y^3-5y^4+\frac{3}{2}y^5 - \frac{y^6}{5} + 6y \ln y$\\
    \bottomrule
    \end{tabular}
    \midsepdefault
    \caption{Summary of the reconstructed $f(T)$ functions derived from the jerk parameter $j(t)$ following Ref.~\cite{Chakrabarti:2019bed}. Here, $A, B, \lambda, m, n, s$ and $p = \frac{1}{2}$ are constants and $y \coloneqq \frac{1}{s}\sqrt{\frac{-T}{6}}$.}
    \label{table:fT_recon_jerk}
\end{table}

\begin{table}[H]
\centering
\midsepremove
\begin{tabularx}{\textwidth}{p{2.4cm}X}
\toprule
 \cellcolor{gris3}\textbf{Evolution} & \cellcolor{gris3}\textbf{Reconstructed Lagrangian} \\ \midrule
 \cellcolor{gris1}& \cellcolor{gris1}Any solution of the equation $f-T(3f_B+2f_T) = 0$ evaluated at the corresponding \\
\cellcolor{gris1}&\cellcolor{gris1}constant $T$ and $B$ values. Treating the equation as a partial differential equation \\
\multirow{-2}{*}{\cellcolor{gris1}de Sitter} &\cellcolor{gris1} leads to $f(T,B) = \sqrt{-T} g\left(\frac{3}{2} T-B\right)$ with $g(x)$ representing an arbitrary function. \\
\cellcolor{gris1}&\cellcolor{gris1}For the $f(T,B) = -T + f_2(B)$ ansatz, $f_2(B) = -\frac{B}{3} \ln B$.\\ \midrule 
\multicolumn{2}{l}{\cellcolor{gris3}Model I: $f(T,B) = f_1(T) + f_2(B)$} \\ \midrule
\cellcolor{gris1}Power-law & \cellcolor{gris1}$f_1(T) = \frac{2 \kappa ^2 \rho_0}{3 h (1+w)-1} \left(\frac{T}{T_0}\right)^{\frac{3}{2} h (1+w)}\,, \ \ \ \ f_2(B) \propto B^{\frac{1-3h}{2}}$ \\
\cellcolor{gris2}& \cellcolor{gris2}$f_1(T) = T_0 \Omega_{w0} \left(-\frac{\Omega_\Lambda}{\Omega_{\rm m0}}\right)^{1+w}\left[\, _2F_1\left(-\frac{1}{2},-w;\frac{1}{2};\frac{T}{T_0 \Omega_\Lambda}\right)+ \frac{T}{T_0 \Omega_\Lambda} \, _2F_1\left(\frac{1}{2},-w;\frac{3}{2};\frac{T}{T_0 \Omega_\Lambda}\right)\right],$\\
\multirow{-2}{*}{\cellcolor{gris2}$\Lambda$CDM} & \cellcolor{gris2}$f_2(B) \propto \sqrt{\left|1+\frac{B}{3\Lambda}\right|} - \frac{B}{3\Lambda} \arctan\sqrt{\left|1+\frac{B}{3\Lambda}\right|}$ \\
\cellcolor{gris1}Phantom \newline Domination & \multirow{2}{*}{\cellcolor{gris1}$f_1(T) = -2b_0 - \frac{b_1 T}{3{h_0}^2} + \sqrt{\frac{2}{3}} \frac{2b_2 (1+m) (-T)^{\frac{5}{2}}}{15 {h_0}^2}\,, \ \ \ \ f_2(B) \propto B^{\frac{m+3}{2m}}$} \\ \midrule
\multicolumn{2}{l}{\cellcolor{gris3}Model II: $f(T,B) = -T + f_2(B)$} \\ \midrule
\cellcolor{gris1}Power-law & \cellcolor{gris1}$d_1 B^{\frac{1-3h}{2}} + \frac{hB}{3h+1}\ln \big(\frac{B}{B_0}\big) + \frac{4 \kappa^2 \rho_0 (3h-1)}{[3h(1+w)-2][3h(2+w)-1]}\big(\frac{B}{B_0}\big)^{\frac{3h(1+w)}{2}}$ \\
\cellcolor{gris2}$\Lambda$CDM & \cellcolor{gris2}$2\Lambda + d_2 \bigg[\sqrt{\left|1+\frac{B}{3\Lambda}\right|} - \frac{B}{3\Lambda} \arctan\sqrt{\left|1+\frac{B}{3\Lambda}\right|}\bigg]$ for dust matter \\
\cellcolor{gris1}Phantom \newline Domination & \cellcolor{gris1}$2b_0 + \frac{B \ln B}{m-3}\left(1-\frac{b_1}{3{h_0}^2}\right) + \sqrt{\frac{2}{3}}\frac{16 b_2 (m+1) (-B)^{\frac{5}{2}}}{45 {h_0}^5 (m+3)^{\frac{3}{2}}(4m-3)} + d_3 B^{\frac{m+3}{2m}}\,,$ \newline solution not applicable for $m = -3, \frac{3}{4}$ and $3$ \\
\bottomrule
\end{tabularx}
\midsepdefault
\caption{The reconstructed $f(T,B)$ Lagrangian for the evolution behaviors considered in Refs.~\cite{Bahamonde:2016cul,Paliathanasis:2017flf,Zubair:2018wyy}. For the phantom domination case, $H = h_0 e^{m\widetilde{N}}$ and $\rho = b_0 + b_1 e^{2m\widetilde{N}} + \frac{96(m+1)}{5}b_2 e^{5m\widetilde{N}}$, where $b_0, b_1, b_2, h_0$ and $m$ are constants and $\widetilde{N}$ represents the $e$-folding parameter. The constants $d_1, d_2, d_3$ are integration constants.}
\label{table:reconstruction-fTB}
\end{table}

\begin{table}[H]
\centering
\footnotesize
\midsepremove
\begin{tabularx}{\textwidth}{lX}
\toprule
\cellcolor{gris3}\textbf{\large Ansatz} & \cellcolor{gris3}\textbf{\large Reconstructed Lagrangian} \\ \toprule
\cellcolor{gris2}\textbf{de Sitter} &\cellcolor{gris1}In vacuum, any solution to the equation $f - 2T f_T - T_G f_{T_G} = 0$ evaluated at the constant de Sitter $T$ and $T_G$ values \\ \midrule
\cellcolor{gris2}\textbf{\boldmath{$\Lambda$}\gls{cdm}}&\cellcolor{gris2} \\ 
\midrule
\cellcolor{gris1} & \cellcolor{gris1}\footnotesize{$g(T)= T_0 \Omega_{w0} \left(-\frac{\Omega_\Lambda}{\Omega_{\rm m0}}\right)^{1+w}\left[\, _2F_1\left(-\frac{1}{2},-w;\frac{1}{2};\frac{T}{T_0 \Omega_\Lambda}\right)+ \frac{T}{T_0 \Omega_\Lambda} \, _2F_1\left(\frac{1}{2},-w;\frac{3}{2};\frac{T}{T_0 \Omega_\Lambda}\right)\right]$}\\
\cellcolor{gris1}\multirow{-2}{*}{$g(T) + h(T_G)$} &\cellcolor{gris1}$h(y) \propto (1-3 y)^{2/3} \Big[(40y+24)\sqrt{1-y}+ 5 \sqrt{6} \, _2F_1\left(\frac{1}{2},\frac{2}{3};\frac{5}{3};\frac{3y-1}{2}\right) (y^2-1)\Big]$ \\
\cellcolor{gris2}$T g(T_G)$ & \cellcolor{gris2}Only vacuum solutions: $g_1(y) = 1 + \frac{y^2}{3} - \frac{y^4}{3} + \dots{}$ and $g_2(y) = 1 - \frac{y^3}{3} - \frac{13y^4}{16} + \dots{}$ \\ 
\cellcolor{gris1} & \cellcolor{gris1}For $w \neq n \in \mathbb{Z}^+$, \newline $g(T) = -\frac{3 \Omega_{w0} T_0}{8 T^2}\left(\frac{\Omega_{\rm m0} T_0}{T-\Omega_\Lambda T_0}\right)^{-(1+w)} \Big[1+\frac{1+w}{1-w} \, _2F_1\left(1,2;2-w;\frac{\Omega_\Lambda T_0}{T}\right)\Big]$, \\
\multirow{-3}{*}{\cellcolor{gris1}$T_G\, g(T)$} & \cellcolor{gris1}otherwise solution depends on $w$. \\ 
\cellcolor{gris2}$-T + \mu (-T)^\beta {T_G}^\gamma$ & \cellcolor{gris2}For $w = 0$, $\{\gamma, \beta\}=\{0,0\}, \{1,-2\}$, otherwise $\{\gamma, \beta\}=\{0,1\},\{1,-1\}$. \\ 
\end{tabularx}
	\begin{tabularx}{\textwidth}{llX}
		\midrule
		\multicolumn{3}{l}{\cellcolor{gris3}\textbf{Power-law} $a(t) = \left(\frac{t_s-t}{t_0}\right)^\alpha$, $\alpha \neq 0$ and $t_0 > 0$}\\ \midrule
		\cellcolor{gris1}& \cellcolor{gris1} & \cellcolor{gris1}$g(T) = \frac{1}{2} \Omega_{w0} \sqrt{T T_0}\ln\left(\frac{T}{T_0}\right)$ for $1-3 \alpha (1+w) = 0$ and \\
		\cellcolor{gris1}& \multirow{-2}{*}{\cellcolor{gris1}$\alpha \neq 1$} & \cellcolor{gris1}$ g(T) = \frac{\Omega_{w0} T_0}{1-3 \alpha (1+w)} \left(\frac{T}{T_0}\right)^{\frac{3\alpha(1+w)}{2}}$ for $1-3 \alpha (1+w) \neq 0$; $h(T_G) \propto {T_G}^{\frac{1-\alpha}{4}}$ \\
		\multirow{1}{*}{\cellcolor{gris1}$g(T) + h(T_G)$} & \cellcolor{gris1}& \cellcolor{gris1}For finite $h(0), h_{T_G}(0), h_{T_G T_G}(0)$, same solution as $\alpha \neq 1$. If $h_{T_G T_G}$ diverges \\
		\cellcolor{gris1}& \multirow{-1}{*}{\cellcolor{gris1}$\alpha = 1$}& \cellcolor{gris1}but $T_G h_{T_G T_G}|_{T_G \to 0} = \gamma$, $g(T)$ permits an extra particular solution $\frac{8\gamma T^2}{9}$.\\
		\cellcolor{gris1}&\cellcolor{gris1}&\cellcolor{gris1}In both cases, $h(T_G)$ is not uniquely defined. \\
		\cellcolor{gris3}& \cellcolor{gris3} & $g(T_G) = c_1 {T_G}^{m_-}+c_2 {T_G}^{m_+} + A \left(\frac{T_{G}}{T_{G,0}}\right)^{\frac{3(1+w)\alpha-2}{4}},$ \cellcolor{gris3} \\
		\cellcolor{gris3}& \multirow{-2}{*}{\cellcolor{gris3}$\alpha \neq 1$} & \cellcolor{gris3}where $m_\pm \equiv \frac{1}{8} \left(3-\alpha \pm \sqrt{\alpha^2 -22\alpha+25}\right),$ $A \equiv \frac{4 (1-\alpha) \Omega_{w0}}{3 \alpha^2 \left(3w^2+7 w+4\right)-\alpha (21 w+19)+6}$ \\
		\multirow{1}{*}{\cellcolor{gris3}$T g(T_G)$} & \cellcolor{gris3} &\cellcolor{gris3}Lagrangian behaves as rescaled \gls{tegr} with $g(T_G)$ not uniquely defined. \\
		\cellcolor{gris3}& \cellcolor{gris3}&\cellcolor{gris3}For $T_G = 0$, $g(T_G)$ satisfies: (a) for $\alpha(1+w) = 2$, $0 = g + T_G g^\prime$ and $\beta = 48(g^\prime + 2T_G g^{\prime\prime})$; \\
		\cellcolor{gris3}&\multirow{-2}{*}{\cellcolor{gris3}$\alpha = 1$}& \cellcolor{gris3}(b) for $3\alpha(1+w) = 2$, $\beta = g + T_G g^\prime$ and $0 = g^\prime + 2T_G g^{\prime\prime}$. \\
		\cellcolor{gris3}&\cellcolor{gris3}&\cellcolor{gris3} Here, $\beta \equiv \Omega_{w0} {t_0}^{5+3w}$ and primes denote derivatives \gls{wrt} $T_G$. \\
		\cellcolor{gris1} & 	\cellcolor{gris1}$\alpha \neq 1$& 	\cellcolor{gris1}$\beta + 2\gamma = 1$ and $\gamma \neq \frac{\alpha-1}{3\alpha-1}$. In the presence of a perfect fluid, $3(1+w)\alpha = 2$.\\
		\multirow{-2}{*}{\cellcolor{gris1}$-T + \mu (-T)^\beta {T_G}^\gamma$}& \cellcolor{gris1}$\alpha = 1$& \cellcolor{gris1}$w = -\frac{1}{3}$ and $\gamma > 0$ \\
	\end{tabularx}
	\begin{tabularx}{\textwidth}{lX}
		\cellcolor{gris3} &\cellcolor{gris3}$g(T) = \frac{3 \Omega_{w0}}{4 T_0} \ln \left(\frac{T}{T_0}\right)$ for $4-3 \alpha (1+w) = 0$ and \\
		\multirow{-2}{*}{\cellcolor{gris3}$T_G g(T)\ \ \ \ \ \ \ \ $}& \cellcolor{gris3}$ g(T) = -\frac{3 \Omega_{w0} T_0}{2T^2 [4-3 \alpha (1+w)]}\left(\frac{T}{T_0}\right)^{\frac{3\alpha (1+w)}{2}}$ for $4-3 \alpha (1+w) \neq 0$ \\
		\bottomrule
	\end{tabularx}
	\midsepdefault
	\caption{The reconstructed $f(T,T_G)$ Lagrangian for the various cosmological behaviors derived in Refs.~\cite{delaCruz-Dombriz:2017lvj,delaCruz-Dombriz:2018nvt}. Here, $c_1, c_2$ are integration constants and $y \equiv \sqrt{1+\frac{4 T_G}{3{\Omega_\Lambda}^2 {T_0}^2}}$.}
	\label{table:reconstruction-fTTG}
\end{table}

\begin{table}[H]
\centering
\midsepremove
\begin{tabularx}{\textwidth}{lX}
\toprule
\cellcolor{gris3}\textbf{Scenario} & \cellcolor{gris3}\boldmath{$f(T,\Theta)$} \\ \midrule
\cellcolor{gris1}$\Lambda$\gls{cdm} & \cellcolor{gris1}$-T + 2\Lambda + \sqrt{-T} f_1\left(\frac{\Theta}{T}\right)$ \\
 \cellcolor{gris3}& \cellcolor{gris3}Any solution to $Tf_T - \frac{f}{2} + \Theta f_\Theta = \kappa^2 \Theta$ evaluated at constant \\
\multirow{-2}{*}{\cellcolor{gris3}de Sitter}&\cellcolor{gris3}de Sitter $T$ and $\Theta$ values \\
\cellcolor{gris1}Einstein Static Universe & \cellcolor{gris1}Any solution to $Tf_T - \frac{f}{2} + \Theta f_\Theta = 0$ with $T = \Theta = 0$ \\
\cellcolor{gris3}Single Perfect Fluid & \cellcolor{gris3}$\frac{2 \Theta \kappa ^2}{1+5 w}+\sqrt{-T} \, f_2\left(\Theta (-T)^{\frac{1+w}{3w-1}}\right)$ \\
\cellcolor{gris1}Chaplygin Gas & \cellcolor{gris1}$e^{-\frac{B(\Theta)}{2}}\bigg\lbrace f_3\left(T e^{B(\Theta)}\right) - \frac{1}{2}{\displaystyle\int_0^\Theta} \frac{\rho(x) \dd x}{A-\rho(x)^2}\big[T e^{B(\Theta)-\frac{1}{2}B(x)} + 2\rho(x) e^{\frac{B(x)}{2}}\big]\bigg\rbrace$ \\
\cellcolor{gris3}Two Fluid Scenario & \cellcolor{gris3}$-T + \sqrt{-T} f_4\left(\frac{(-T)^{3/2}\left(-64\rho_{p,0} + \rho_{q,0} \Theta^2\right)}{\rho_{q,0}}\right)$ \\
\cellcolor{gris1}Massless Scalar Field & \cellcolor{gris1}$\frac{\kappa^2 \Theta}{3} + \sqrt{-T} \, f_5(T \Theta)$ \\
\bottomrule
\end{tabularx}
\midsepdefault
\caption{Summary of the reconstructed $f(T,\Theta)$ functions for the various cosmological models considered in Ref.~\cite{Momeni:2014jja}. Here, $\Lambda$ represents the standard $\Lambda$\gls{cdm} cosmological constant, $f_i$ are arbitrary functions, $w$ is the \gls{eos} of the perfect fluid, $\rho_{p,q,0}$ represent present energy density values of the phantom and non-phantom fluids respectively, $A$ is a constant and $B(x) \equiv \int_0^x \frac{\rho(y) \dd y}{A-\rho(y)^2}$. The de Sitter condition is derived in the presence of a dust fluid. For the two fluid scenario, the \gls{eos}s were assumed to be $w_p = -\frac{7}{3}$ and $w_q = \frac{1}{3}$ whereas the massless scalar field was assumed to behave as a stiff fluid.}
\label{table:reconstruction-fTT-Momeni}
\end{table}

{
\begin{table}[H]
\centering
\midsepremove
\begin{tabularx}{\textwidth}{lX}
\toprule
\cellcolor{gris3}\textbf{Evolution} & \boldmath{\cellcolor{gris3}$P_1(\Phi)$} \\ \midrule
\cellcolor{gris1}Unification & \cellcolor{gris1}$\widetilde{k} \sqrt{\frac{\Phi}{g_0 \Phi + g_1}} + \frac{\lambda}{8g_0}\bigg[\Phi- \frac{g_1}{g_0}\sqrt{\frac{g_0 \Phi}{g_0 \Phi + g_1}} \ln \left(g_0 \Phi^{\frac{1}{2}} \left[1 + \sqrt{\frac{g_0 \Phi + g_1}{g_0 \Phi}}\right]\right)\bigg]$ \\
\cellcolor{gris3} & \cellcolor{gris3}$\widetilde{k} \sqrt{\Phi} + \frac{3\lambda \Phi^2}{3h_f} + \frac{k_0 \sqrt{\Phi}}{2}\left[2 \arctan(1-k_1 \sqrt{\Phi}) - 2 \arctan(1+k_1 \sqrt{\Phi}) \right.$ \\
\multirow{-2}{*}{\cellcolor{gris3}Transition}&\cellcolor{gris3}$+\ln \left(\sqrt{h_i} + (q h_i h_f)^{\frac{1}{4}} \sqrt{2\Phi} + \sqrt{q h_f} \Phi\right)- \ln \left(\sqrt{h_i} - (q h_i h_f)^{\frac{1}{4}} \sqrt{2\Phi} + \sqrt{q h_f} \Phi\right) \big]$ \\
\bottomrule
\end{tabularx}
\midsepdefault
\caption{The reconstructed $P_1(\Phi)$ function for the $\lambda \neq 0$ case obtained in Ref.~\cite{Nassur:2015zba}. The limiting case $\lambda = 0$ (i.e. $P_2 = -\frac{1}{2}$) reduces to $P_1 \propto (-T)^{-\frac{1}{4}}$ hence leading to the expected result. Here, $\widetilde{k}$ is an integration constant, $k_0 \equiv \frac{\lambda \sqrt{2} (h_i - h_f)}{2q^3 {h_f}^{\frac{7}{4}} {h_i}^{\frac{1}{4}}}$ and $k_1 \equiv \left(\frac{4qh_f}{h_i}\right)^{\frac{1}{4}}$.}
\label{table:reconstruction-fTT-Nassur}
\end{table}
}

\section{Cosmological Perturbations in flat FLRW Teleparallel Gravity}

This section is devoted in presenting the most important quantities needed for the cosmological perturbations.

\subsection{Background}

The non-zero components of the torsion tensor and superpotential, and the torsion and boundary term in the background (flat \gls{flrw}) are
\begin{subequations}
\begin{align}
    T^{i}{}_{0j} & = H\delta^{i}{}_{j}\,,\\[0.5ex]
    S_{i}{}^{0j} & = -H\delta^{i}{}_{j}\,,\\[0.5ex]
    T & = -6H^{2}\,,\\[0.5ex]
    B & = -6(3H^{2}+\dot{H})\,.
\end{align}
\end{subequations}
The matter content is fully conserved giving the standard conservation equation for a perfect fluid
\begin{equation}
\lc{\nabla}_{\nu}\Theta_{\mu}{}^{\nu}:\,\dot{\rho}+3(\rho+P)=0\,.
\end{equation}

\subsection{Tensor perturbations}

The non-zero components of the torsion tensor and the superpotential are
\begin{subequations}
\begin{align}
\delta T^{i}{}_{0j} & = \frac{1}{2}\dot{h}_{ij}\,,\\[0.5ex]
\delta T^{i}{}_{jk} & = \frac{1}{2}(\partial_{j}h_{ik}-\partial_{k}h_{ij})\,,\\[0.5ex]
\delta S_{0}{}^{0i} & = 0\,,\\[0.5ex]
\delta S_{i}{}^{0j} & = \frac{1}{4}\dot{h}_{ij}\,,\\[0.5ex]
\delta S_{i}{}^{jk} & = -\frac{1}{4a^{2}}(\partial_{j}h_{ik}-\partial_{k}h_{ij})\,,
\end{align}
\end{subequations}
while the scalars are
\begin{equation}
\delta T=0\,,\quad\delta B=0\,.
\end{equation}

\subsection{Vector and pseudo-vector perturbations}

The non-zero components of the vectorial and pseudo-vectorial perturbations for the torsion tensor and the superpotential are
\begin{subequations}
\begin{alignat}{2}
\delta T^{0}{}_{0i} & =\: & & a\dot{\beta}_{i}\,,\\[0.5ex]
\delta T^{i}{}_{0j} & =\: & & 2\partial_{i}\dot{h}_{j}-\frac{1}{a}\partial_{j}b_{i}-\epsilon_{kij}\dot{\sigma}_{k}\,,\\[0.5ex]
\delta T^{0}{}_{ij} & =\: & & a(\partial_{i}\beta_{j}-\partial_{j}\beta_{i})\,,\\[0.5ex]
\delta T^{i}{}_{jk} & =\: & & 2(\partial_{i}\partial_{j}h_{k}-\partial_{i}\partial_{k}h_{j})+\epsilon_{ijl}\partial_{k}\sigma_{l}-\epsilon_{ikl}\partial_{j}\sigma_{l}\,,\\[0.5ex]
\delta S_{0}{}^{0i} & =\: & & -\frac{1}{2a^{2}}\Big[2aH(b_{i}-\beta_{i})+\epsilon_{ilk}\partial_{k}\sigma_{l}\Big]\,,\\[0.5ex]
\delta S_{i}{}^{0j} & =\: & & -\frac{1}{2a}\Big[\frac{1}{2}\Big(\partial_{i}(b_{j}+\beta_{j}-a\dot{h}_{j})+\partial_{j}(b_{i}-\beta_{i}-a\dot{h}_{i})\Big)\Big]\,,\\[0.5ex]
\delta S_{0}{}^{ij} & =\: & & -\frac{1}{4a^{3}}\Big[\partial_{i}(b_{j}-\beta_{j}+2a\dot{h}_{j})-\partial_{j}(b_{i}-\beta_{i}+2a\dot{h}_{i})-2a\epsilon_{lij}\dot{\sigma}_{l}\Big]\,,\\[0.5ex]
\delta S_{i}{}^{jk} & =\: & & -\frac{1}{2a^{2}}\Big[\delta_{im}\epsilon_{kjl}\partial_{l}\sigma_{m}+\delta_{ij}\Big(2aH(b_{k}-\beta_{k})-a\dot{\beta}_{k}-2\partial^{2}h_{k}\Big)\nonumber \\[0.5ex]
 & \: & & -\delta_{ik}\Big(2aH(b_{j}-\beta_{j})-a\dot{\beta}_{j}-2\partial^{2}h_{j}\Big)-2\delta_{il}\partial_{k}\partial_{l}h_{j}+2\delta_{kl}\partial_{i}\partial_{j}h_{l}\Big]\,,
\end{alignat}
\end{subequations}
and the perturbations related to the torsion and boundary term scalars are
\begin{equation}
\delta T = 0\,,\quad \delta B = 0\,.
\end{equation}

\subsection{Scalar and pseudo-scalar perturbations}
\label{scalarappendix}

The components of the torsion tensor and the superpotential for scalar and pseudo-scalar perturbations up to first order are
\begin{subequations}
\begin{align}
\delta T^{0}{}_{0i} & = \partial_{i}(a\dot{\beta}-\phi)\,,\\[0.5ex]
\delta T^{i}{}_{0j} & = \partial_{i}\partial_{j}(\dot{h}-a^{-1}b)-\epsilon_{lij}\partial_{l}\dot{\sigma}-\dot{\psi}\delta_{ij}\,,\\[0.5ex]
\delta T^{0}{}_{ij} & = 0\,,\\[0.5ex]
\delta T^{i}{}_{jk} & = \delta_{ij}\partial_{k}\psi-\delta_{ik}\partial_{j}\psi+\delta_{il}(\epsilon_{klm}\partial_{j}\partial_{m}\sigma-\epsilon_{jlm}\partial_{k}\partial_{m}\sigma)\,,\\[0.5ex]
\delta S_{0}{}^{0i} & = -\frac{H}{a}\partial_{i}\Big(b-\beta-(aH)^{-1}\psi\Big)\,,\\[0.5ex]
\delta S_{i}{}^{0j} & = \Big[(2H\phi+\dot{\psi})\delta_{ij}+\frac{1}{2}\partial_{i}\partial_{j}(\dot{h}-a^{-1}b)-\frac{1}{2}\partial^{2}(\dot{h}-a^{-1}b)\delta_{ij}\Big]\,,\\[0.5ex]
\delta S_{0}{}^{ij} & = \frac{1}{2a^{2}}\epsilon_{ijk}\partial_{k}\dot{\sigma}\,,\\[0.5ex]
\delta S_{i}{}^{jk} & = \frac{1}{2a^{2}}\Big[\delta_{ik}\partial_{j}\Big(2aH(b-\beta)+\phi-\psi-a\dot{\beta}\Big)-\delta_{ij}\partial_{k}\Big(2aH(b-\beta)+\phi-\psi-a\dot{\beta}\Big)\Big]\,,
\end{align}
\end{subequations}
and the perturbations up to first order to the scalar torsion and boundary term become
\begin{subequations}
\begin{alignat}{2}
\delta T & =\: & & 4H\Big(3H\phi+3\dot{\psi}+\frac{1}{a}\partial^{2}b-\partial^{2}\dot{h}\Big)\,,\\[0.5ex]
\delta B & =\: & & -\Big[H\left(\frac{1}{a}\partial^{2}(6\beta-10b)-6(6\dot{\psi}+\dot{\phi}-2\partial^{2}\dot{h}+6H\phi)\right)+\frac{2}{a}\partial^{2}(\dot{\beta}-\dot{b})+\frac{2}{a^{2}}\partial^{2}(2\psi-\phi)\nonumber \\[0.5ex]
 & \: & & +2(\partial^{2}\ddot{h}-6\dot{H}\phi-3\ddot{\psi})\Big]\,.
\end{alignat}
\end{subequations}
Then, the perturbation conservation equations become
\begin{subequations}
\begin{align}
\lc{\nabla}_{\mu}\Theta_{0}{}^{\mu} & =\delta\dot{\rho}+3H(\delta p+\delta\rho)+\frac{\partial^{2}v(p+\rho)}{a}-3\dot{\psi}(p+\rho)+\partial^{2}\dot{h}(p+\rho)=0\,,\\[0.5ex]
\lc{\nabla}_{\mu}\Theta_{i}{}^{\mu} & =\partial_{i}\Big[\delta p+(\rho+p)\Big(4aH(b+v-\beta)+\phi+a(\dot{b}-\dot{\beta}+\dot{v})\Big)+a(\dot{\rho}+\dot{p})(v+b-\beta)\Big]=0\,.
\end{align}
\end{subequations}

\subsection{Sub-horizon limit in the Newtonian gauge}
\label{appendix:Newtonian-gauge}

The torsion scalar, the boundary term and the arbitrary functions of them, become in the Newtonian gauge
\begin{subequations}
\begin{align}
\delta T &\simeq -\frac{4H}{a}\left(k^{2}b-3aH(\psi+\phi)\right)\,,\\[0.5ex]
\delta B &\simeq -\frac{2k^{2}}{a^{2}}(2abH-2\psi+\phi)\,,\\[0.5ex]
\delta f_{T} &\simeq -\frac{2k^{2}}{a^{2}}\left(2abH\left(f_{TB}+f_{TT}\right)+f_{TB}(\phi-2\psi)\right)\,,\label{deltafT}\\[0.5ex]
\delta f_{B} &\simeq -\frac{2k^{2}}{a^{2}}\left(2abH(f_{BB}+f_{TB})+f_{BB}(\phi-2\psi)\right)\,.\label{deltafB}
\end{align}
\end{subequations}

\section{Boltzmann equations analysis}

At the present time, the \gls{cmb} dominates the constraints on the standard model of cosmology. This era of precision cosmology can be seen through plots of the current status of the power spectra coming from the \gls{cmb}. Furthermore, this spectrum carries relevant information on both the early- and late-time Universe. We can take advantage of modified versions of the \texttt{CLASS} code\footnote{Another example of numerical code is \texttt{hiCLASS}, which can compute predictions for modified models in Horndeski gravity. However, our new code can consider any \textit{additional} modification as an effective fluid derived from the numerical solution of the field equation in a generic way.} to analyze the perturbations. This implementation allows the analysis of structure formation since each power spectrum is related to the curvature, baryonic matter, dark matter, etc. We start with the perturbations obtained for $\Lambda$\gls{cdm} and afterwards we carry out a modified version that includes any modified/extended theory as an effective-like fluid. The effects will be seen as deviations in the power spectrum. These perturbations are described by equations for density contrast and velocity divergence in the synchronous gauge and valid for a perfect fluid (see Sec.~7 for examples on the procedure through which these perturbations are implemented)~\cite{Ma:1995ey,Hu:1998kj}
\begin{subequations}
\begin{align}
    \dot{\delta}_{i} + 3\mathcal{H}(c^2_{s,\rm eff} - w_{i})\left[\delta_{i}+3\mathcal{H}(1+w_{i})\frac{v_{i}}{k}\right]+(1+\omega_{i})k v_{i} + 3\mathcal{H} \dot{w}_i \frac{v_i}{k} &=-3(1+w_i)\dot{h}\,,\label{eq:idelta_cmb}\\[0.5ex]
    \dot{v}_i + \mathcal{H}(1-3c^2_{s,\rm eff})v_{i} &= \frac{k \delta_i c^2_{s,\rm eff}}{1+w_{i}}\label{eq:iv_cmb}\,.
 \end{align}
\end{subequations}
The derivatives denoted by dots are \gls{wrt} the conformal time, $\mathcal{H}$ is the conformal Hubble parameter, $c^2_{s,\rm eff}$ is the effective sound speed, $v_i$ is the velocity of the $i$th fluid, $\delta_i$ are the perturbations of each matter component, i.e baryons, photons, radiation, cold dark matter, and $w_i$ is the \gls{eos} for each component and $k$ is the wave mode. One way to implement these equations of motion is through the Parameterized Post-Friedmann \cite{Baker:2012zs} approach in the \texttt{CLASS} code, which is written in the programming language C and requires no specific version of the compiler; it has flexibility because it is not hard coded, it has a clear structure and a dynamical allocation of all indices. Furthermore, it controls accuracy as all accuracy parameters are grouped within a single structure; and finally it is faster than other codes.

\begin{figure}
\centering
    \includegraphics[width= 0.6\textwidth]{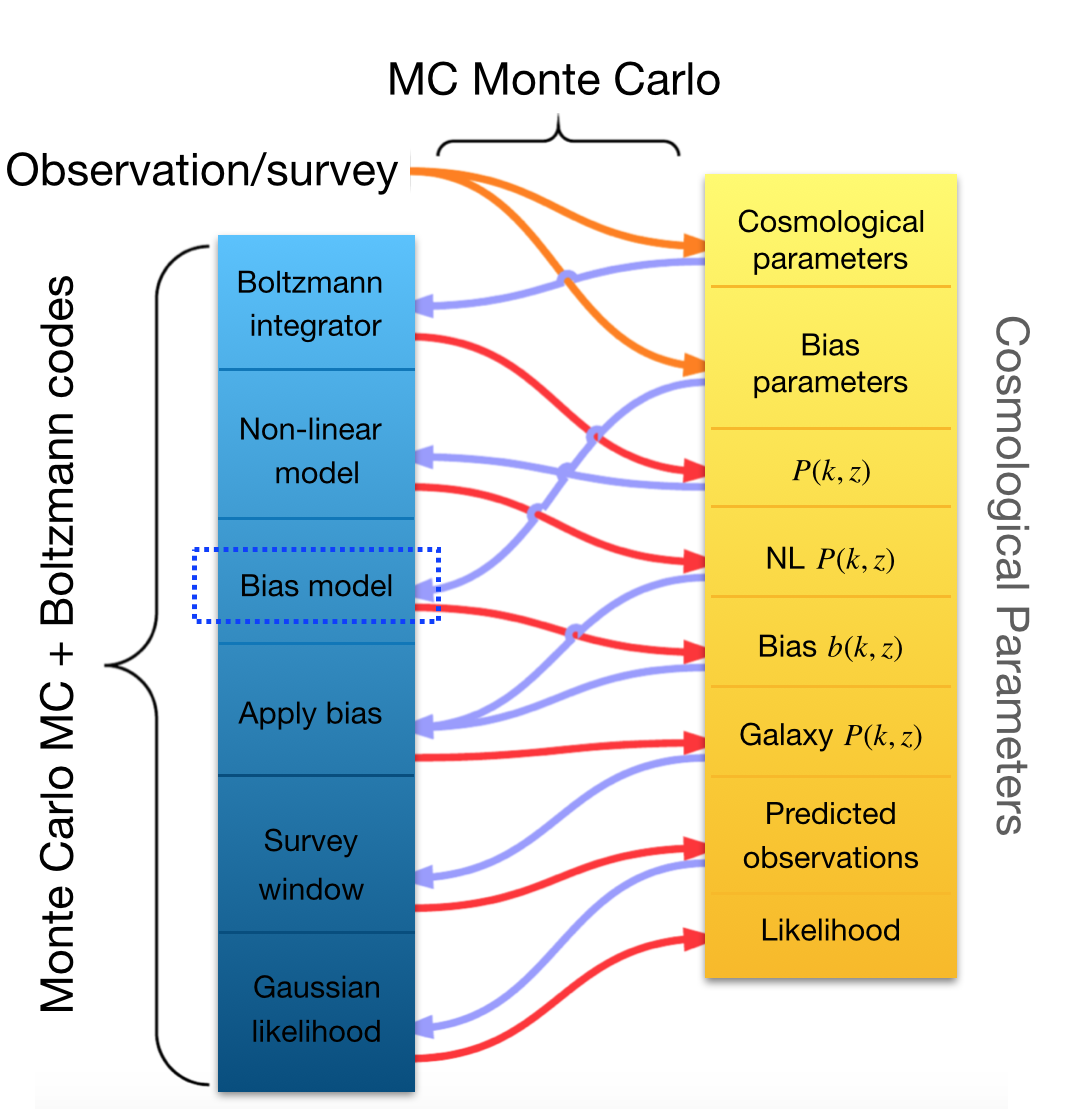}
    \caption{A schematic diagram with \gls{mcmc} + Boltzmann process. This algorithm develops a perturbation analysis for cosmological models, e.g for a galaxy power spectrum likelihood. The dashed box indicates how to insert the model to be analyzed.
    }
    \label{fig:monte_class1}
\end{figure}

\begin{figure}
\centering
    \includegraphics[width= 0.8\textwidth]{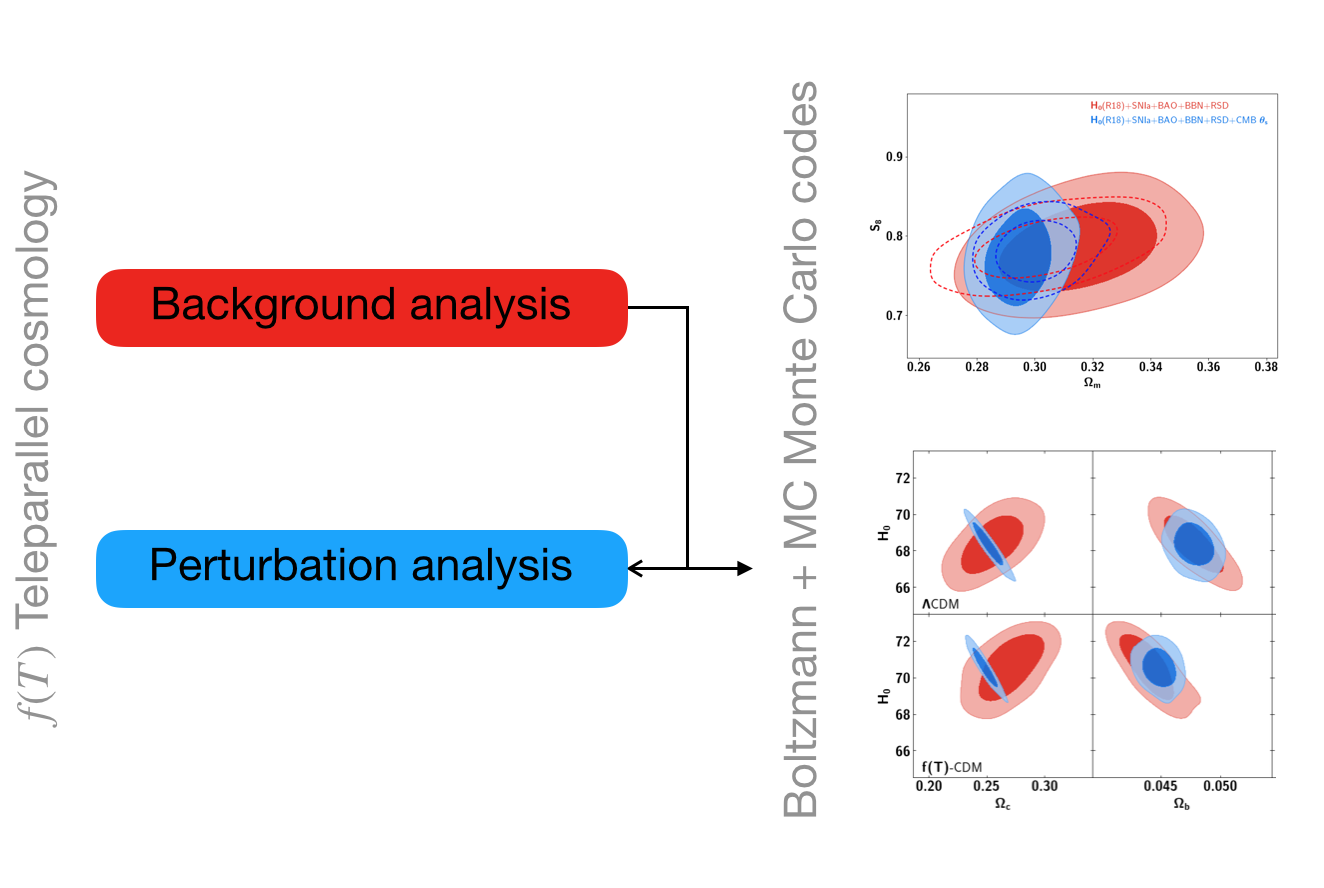}
    \caption{A schematic diagram to test $f(T)$ according to Fig.~\ref{fig:monte_class1}.
This algorithm develops as follows:
\textit{Top:} a background analysis entry, which includes solely the modified Friedmann equations analysis. These equations are given in Sec.~6.1. \textit{Bottom:} A scalar perturbation analysis entry, which includes the analysis at first order on the perturbed modified Friedmann equations. The equations related to his blue box are given in Sec.~7.6 after setting $f(T,B)=f(T)$ presented in Sec.~7.6.
Finally, the likelihood results are presented in probability confidence regions for their specific cosmological parameters, e.g $H_0$, $\Omega_{\rm m0}$ and $S_8$.}
    \label{fig:monte_class_FT}
\end{figure}

\begin{figure}
\centering
    \includegraphics[width= 0.85\textwidth]{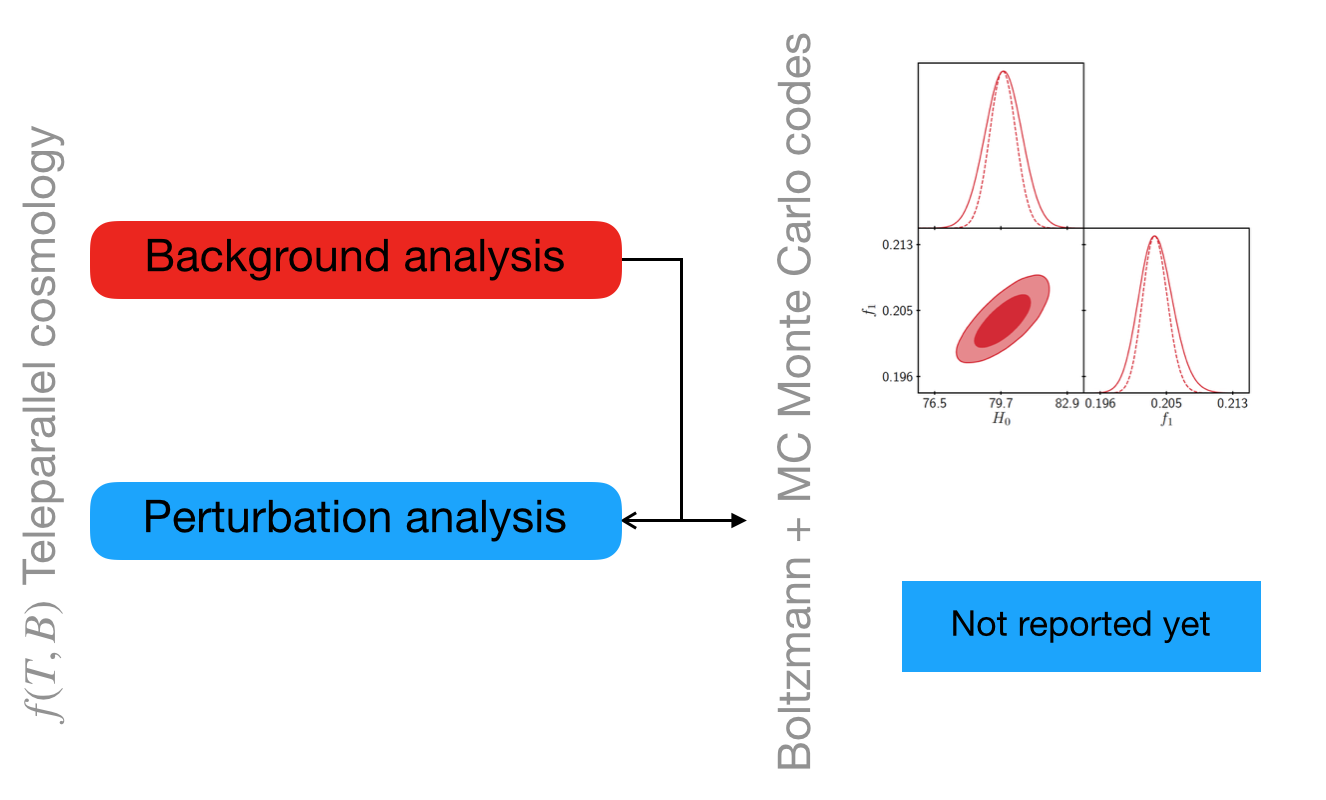}
    \caption{A schematic diagram to test $f(T,B)$ according to Fig.~\ref{fig:monte_class1}.
This algorithm develops: \textit{Top:} a background analysis with the equations presented in Sec.~6.2. \textit{Bottom:} a perturbation analysis is presented in Sec.~7.6. Finally, the likelihood results are presented as CL regions for their specific cosmological parameters, e.g $H_0$, $\Omega_{\rm m0}$. We remark that tests for the perturbation analysis are still an ongoing challenge.}
    \label{fig:monte_class_FTB}
\end{figure}

In order to work with \texttt{CLASS}, one has to express quantities and equations in terms of the specific code units. The units for the wavenumbers $k$ and the Hubble parameter in the conformal time $\mathcal{H}$ are 1/Mpc, while the conformal time $\tau$ has the units of Mpc. Also, we employ the units from Sec.~1.3. Finally, the default \gls{cmb} multipoles ($C_{l}$'s) are dimensionless. A schematic architecture is depicted in Fig.~\ref{fig:monte_class1} and the process to implement \gls{tg} models is illustrated (with their equations) in the top left panels of Figs.~\ref{fig:monte_class_FT}--\ref{fig:monte_class_FTB}.

\begin{figure}
\centering
    \includegraphics[width= 0.7\textwidth]{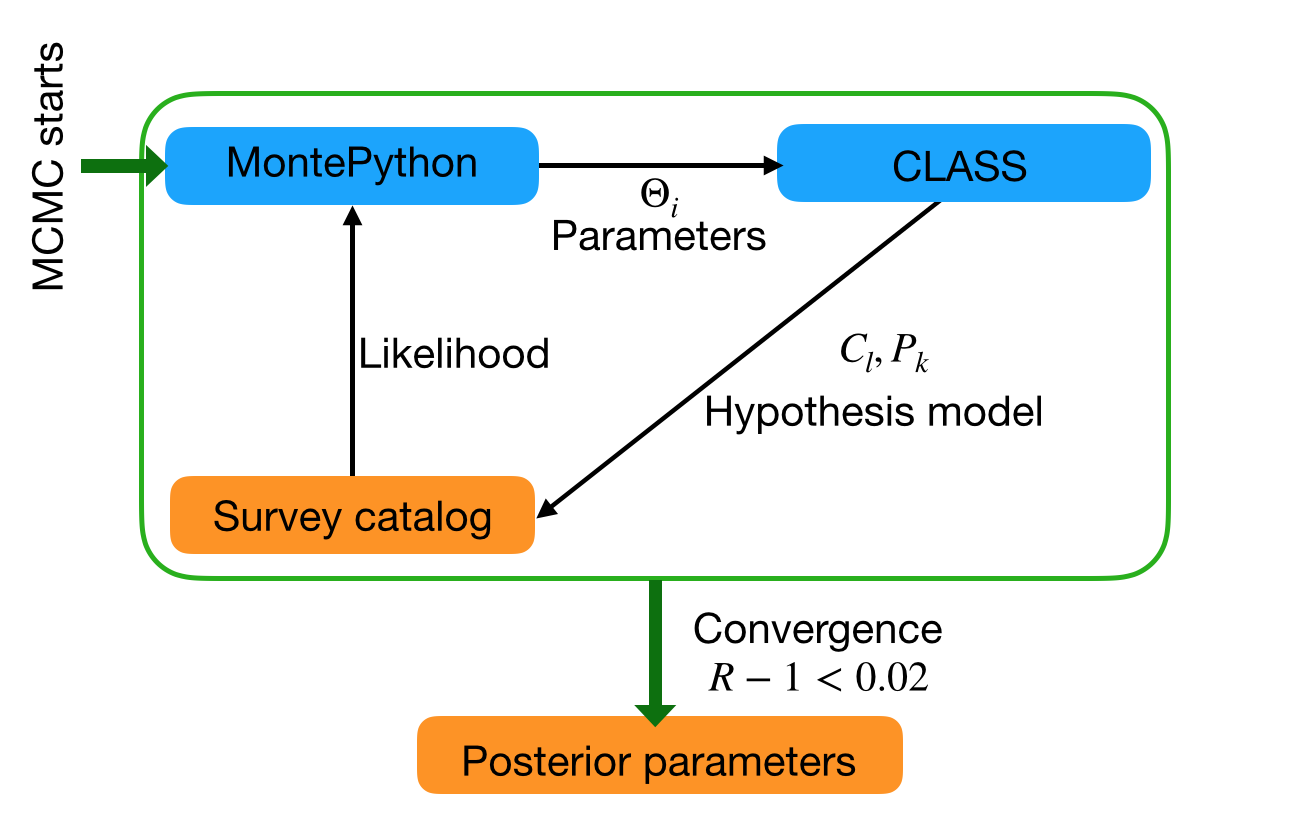}
    \caption{\gls{mcmc} code diagram for cosmology.  We start the test run by asking the code to take a sample of the model parameter which is used to simulate or predict data using that specific version of the model. We extract the $C_l$ and $P_k$ with CLASS for a random combination of parameters chosen by MontePython. Afterwards, we compare this theoretical model with the data of the survey catalog, repeating many times the steps and each time we have a value for the likelihood. Finally, once reached the convergence $R-1<0.02$, compute the posterior parameters.
    The blue boxes indicate the numerical codes, the orange boxes the samples generated and the green box denotes the repeating process.
    }
    \label{fig:monte}
\end{figure}


\subsection{A note on Boltzmann equations at numerical code level}

From Eq.~\eqref{eq:iv_cmb}, we can rewrite the full perturbed system of equations that includes the full Boltzmann equations: photons, neutrinos, baryons, \gls{cdm}, plus the perturbed Einstein equations in Sec.~7.7.2 as
\begin{subequations} \label{system}
\begin{align}
    \frac{ d \Theta_0}{d\eta} + k \Theta_1 & =-\frac{d \Phi}{d\eta}\,, \\[0.5ex]
    \frac{d \Theta_1}{d\eta} - \frac{k}{3} \Theta_0 &= - \frac{k}{3} \Phi \\[0.5ex]
    \frac{d \delta}{d\eta} + i k v & = - 3 \frac{d \Phi}{d\eta}\,, \\[0.5ex]
    \frac{ d v}{d\eta}+ \frac{1}{a} \frac{d a}{d\eta} &= i k \Phi\,,\\[0.5ex]
    k^2 \Phi + 3 \frac{1}{a} \frac{d a}{d\eta} \left(\frac{d \Phi}{d\eta} + \frac{1}{a} \frac{d a}{d\eta} \Phi \right) &= 4 \pi G a^2 \left[\rho_m \delta + 4 \rho_r\Theta_0\right]\,,
\end{align}
\end{subequations}
where the time variable is the conformal time ($d\eta = dt/a $, where $a$ is the scale factor) and $k$ is the comoving wavenumber. The primary components of the photon distribution that affect the matter variables are the monopole and dipole moments denoted by $\Theta_0$ and $\Theta_1$ respectively. The matter fluctuations are characterized by the overdensity $\delta$ and the irrotational peculiar velocity $v$. Or by each component (and for numerical code purposes):
\begin{subequations}
\begin{align}
\dot{\delta}  & = - ku -3\dot{\Phi} \,, \label{eq:continuity} \\[0.5ex]
\dot{u} & = - \frac{\dot{a}}{a}u + k \Psi\,,\\[0.5ex]
\dot{\delta_b} & = -ku_b -3\dot{\Phi}\,, \\[0.5ex]
\dot{u_b} & = -\frac{\dot{a}}{a}u_b + k \Psi + k c_s^2 \delta_b + \frac{\dot{\tau}}{R} [ u_b-3\Theta_1] \,, \\[0.5ex]
\dot{\Theta}_0 & = -k \Theta_1 - \dot{\Phi}\,, \\[0.5ex]
\dot{\Theta}_1 & = \frac{k}{3} \left[ \Theta_0-2\Theta_2+ \Psi \right] + \dot{\tau} \left[ \Theta_1 - \frac{u_b}{3} \right]\,, \\[0.5ex]
\dot{\Theta}_2 & = \frac{k}{5} \left[ 2 \Theta_1 - 3 \Theta_3 \right] + \dot{\tau} \left[ \Theta_2 - \frac{\Pi}{10} \right]\,, \\[0.5ex]
\dot{\Theta}_l & = \frac{k}{2 l +1} \left[ l \Theta_{l-1} - (l+1) \Theta_{l+1} \right] + \dot{\tau} \Theta_l\,,\quad l>2 \,,\label{eq:theta_l} \\[0.5ex]
\dot{\Theta}_{Pl} & = \frac{k}{2 l +1} \left[ l \Theta_{P(l-1)} - (l+1) \Theta_{P(l+1)} \right] + \dot{\tau} \left[ \Theta_{Pl}-\frac{\Pi}{2} (\delta_{l,0}+\frac{\delta_{l,2}}{5}) \right] \,, \\[0.5ex]
\dot{\mathcal N}_0 & = -k \mathcal N_1 - \dot{\Phi} \,,\\[0.5ex]
\dot{\mathcal N}_1 & = \frac{k}{3} \left[ \mathcal N_0-2\mathcal N_2+ \Psi \right]\,, \\[0.5ex]
\dot{\mathcal N}_l & = \frac{k}{2 l +1} \left[ l \mathcal N_{l-1} - (l+1) \mathcal N_{l+1} \right]\,,\quad l>1 \,, \\[0.5ex]
k^{2}\Phi+3 \frac{\dot{a}}{a} \left( \dot{\Phi}- \frac{\dot{a}}{a} \Psi \right) & = 4 \pi G a^2 \left[ \rho_m \delta_m + 4 \rho_r \Theta_{r0} \right]\,,
\end{align}
\end{subequations}
where dot denotes derivatives \gls{wrt} conformal time $\eta$, $\delta$ ($\delta_b$) and $u$ ($u_b$) are the density and velocity perturbations for the dark matter (baryons), $\Theta_l$ and $\Theta_{Pl}$ are the photon temperature and polarization multipole moments, ${\mathcal N}_l$ are the multipole moments of the (massless) neutrino temperature, and $\Pi = \Theta_2+\Theta_{P0}+\Theta_{P2}$. The baryon-to-photon ratio is given by $R = 3\rho_b/(4\rho_{\gamma})$, $\tau$ is the Thomson scattering optical depth and $c_s$ is the baryon sound speed. The subscripts $r$ and $m$ refer to the density-weighted sum of all radiation and matter components, and $\rho_r$ and $\rho_m$ is the mean density in each of these components.

\section{Surveys and cosmological data}

The interest in understanding the physics of the late Universe can be well comprehended by using observational data from: \gls{sn}eIa luminosity distances from the Pantheon compilation \cite{Scolnic:2017caz}, \gls{bao} from redshift surveys~\cite{Gil-Marin:2016wya,Delubac:2014aqe,Gil-Marin:2014sta}, and high-$z$ measurements of $H(z)$ from galaxy ages, the so-called Cosmic Chronometers (\gls{cc})~\cite{Moresco:2016mzx}.

To carry out cosmological tests on the free parameters that can be considered in the \gls{tg} theories described in Sec. 10.3, we are going to consider the numerical solutions over theoretical parameters imposed by each model and determine the specific cosmological parameters for each model in addition to $\Omega_{\rm m0}$ and $H_0$. Publicly available Boltzmann codes, such as \texttt{CLASS} \cite{classcode2}, along with \gls{mcmc}-like methods, as implemented in \texttt{Monte Python} \cite{MontePythoncode2} can constrain the models using joint data sets, for instance, with \gls{cc} + \gls{sn}eIa + \gls{bao}.

The survey compilations considered for these studies described in this Review are provided in the following sections.


\subsection{Galaxy ages compilations}

Modified gravity constraints have been studied through measurements related to the galaxy ages. Currently, this kind of observable includes a sample of 51 observations in the redshift range $0.07< z < 2.0$~\cite{Moresco:2016mzx}. A calibration using stellar evolution from the BC03 model \cite{Bruzual:2003tq} of this sample was presented in Ref.~\cite{Magana:2017nfs} reducing this sample to 31 data points. The normalized parameter $h(z)$ can be computed by considering the values from \texttt{Planck} 2018 and \gls{shoes}+\gls{holicow}. In this sample, 31 data points from passive galaxies are included, and 20 data points are estimated from \gls{bao} data under a $\Lambda$\gls{cdm} prior. However, \gls{bao} data points can be computed by using the $r_s$ at the drag epoch (which is when photon pressure stops supporting gravitational instability in baryons and occurs slightly after the decoupling of photons) from PL18 .

The free parameters of the cosmological model are determined through $\chi^2$-statistics through
\begin{equation}
    \chi_H^2 = \sum_{i=1}^{\# \rm data}\frac{\left[H\left(z_i,\mathbf{x}\right) - H_{\text{obs}}(z_i)\right]^2}{\sigma^2_H(z_i)}\,,
\end{equation}
where $H_{\rm obs}(z_i)$ is the observed value at $z_i$, $\sigma_H(z_i)$ are the observational errors, and $H\left(z_i,\mathbf{x}\right)$ is the value of a theoretical Hubble parameter $H$ for the same $z_i$ with the specific parameter vector $\mathbf{x}$.

\subsection{Pantheon type Ia supernovae compilation} 

The distance modulus data can be added to the analysis as a late-time observable by considering its current sample denoted by the latest Type Ia \gls{sn} compilations \cite{Scolnic:2017caz} and it contains 1048 \gls{sn}eIa in the redshift range $0.01<z<2.26$. In the below analyses, we also make use of the preceding SNLS-\gls{sdss} \gls{jla} \cite{Betoule:2014frx}, consisting of 740 \gls{sn}eIa. The constraining power of this kind of \gls{sn} sample is such that these observations can be used as standard candles. This can be implemented through the use of the distance modulus
\begin{equation}
    \mathcal{F}(z,A)_\text{theo}:=5\log_{10}\left[D_L(z,A)\right]+\mu_0\,.
\end{equation}
The nuisance parameter $\mu_0$ encodes the Hubble parameter and the absolute magnitude $M$ (which is a consatnt \cite{Benisty:2022psx}), and has to be marginalized over and $D_L(z,A)$ is the luminosity distance given by
\begin{equation}
    D_L(z,A):=(1+z)\int_0^z{\frac{c\, \mathrm{d}z'}{H(z',A)}}\,,
\end{equation}
and $A =\{w_0,w_a\}$ is the vector with the free cosmological parameters to be fitted. We notice that the factor $c/H_0$ can be rewritten in terms of $\mu_0$. Furthermore, we can write $\Delta\mathcal{F}(A)=\mathcal{F}_{\text{theo}}-\mathcal{F}_{\text{obs}}$, using for this purpose the distance modulus $\mathcal{F}_{\text{obs}}$ associated with the observed magnitude. At this point it may be thought that a possible $\chi_{\text{SNeIa}}^2$ is given by
\begin{equation}
    \chi_{\text{SNeIa}}^2(A)=\left(\Delta\mathcal{F}(A)\right)^{T}\cdotp C_{\text{SNeIa}}^{-1}\cdotp \Delta\mathcal{F}(A)\,,
\end{equation}
where $C_{\text{SNeIa}}$ is the covariance matrix of the data set (which contains the covariance errors on the cosmological parameter constraints). This equation can be seen to contain the nuisance parameter $\mu_0$, which turns to be a function of the Hubble constant, the speed of light $c$, and the \gls{sn}eIa absolute magnitude. This log-likelihood can also be expressed by marginalizing $\chi_{\text{SNeIa}}^2$ analytically \gls{wrt} $\mu_0$, from which we obtain a new $\chi_{\text{SNeIa}}^2$ estimator
\begin{equation}\label{app_eq_likelihood_SN}
    \chi_{\text{SNeIa}}^2(A) = \left(\Delta\mathcal{F}(A)\right)^{T} C_{\text{SNeIa}}^{-1} \Delta\mathcal{F}(A)+\ln{\frac{S}{2\pi}}-\frac{k^2(A)}{S}\,,
\end{equation}
where $S = \sum_{ij} C_{{\text{SNeIa}}_{ij}}^{-1}$ is the sum of all the components of $C_{SNeIa}^{-1}$. This equation gives a measure of how close the predicted model is to the observation values of the data set for particular sets of model parameter values, which is also weighted by the covariance matrix through
\begin{equation}
    k(A)={\left(\Delta\mathcal{F}(A)\right)^{T}\cdotp C_{\text{SNeIa}}^{-1}}\,.
\end{equation}

For this sample we consider the nuisance parameter $M$ inside the sample, which we can choose from a statistical analysis of the $\Lambda$\gls{cdm} model with a fixed $H_0$ from the late Universe measurements.


\subsection{Baryon acoustic oscillations compilation}

According to fluctuations in the density of the visible baryonic matter of the Universe, the \gls{bao} caused by acoustic density waves in the primordial plasma of the early Universe can be used as a tool to constrain cosmological parameters. Currently, their sample is characterized by a set of 15 transverse measurements which are obtained from the angular diameter distance in a quasi model-independent approach. This can be done by computing the 2-point angular correlation function tracers via $D_A (z; r_{\text{drag}})$ \cite{Nunes:2020hzy}. The sample spans a redshift range $0.11\lesssim z \lesssim 2.225$. These kind of observations contribute to important features by comparing the data of the sound horizon at present, to the sound horizon at the time of recombination (extracted from the \gls{cmb} anisotropy data). The \gls{bao} distances are given by
$d_{z} \equiv \frac{r_{s}(z_{d})}{D_{V}(z)}$, with $r_{s}(z_{d}) = \frac{c}{H_{0}} \int_{z_{d}}^{\infty} \frac{c_{s}(z^\prime)}{E(z^\prime)} \mathrm{d}z^\prime$ and
$r_{s}(z_{d})$ being the comoving sound horizon at the baryon drag epoch, $c$ the light velocity, $z_{d}$ is the
drag epoch redshift and $c^{2}_{s} = c^2/3[1+(3\Omega_{b0}/4\Omega_{\gamma 0})(1+z)^{-1}]$ the sound speed, with $\Omega_{b0}$ and $\Omega_{\gamma 0}$ the present values of baryon and photon energy density parameters, respectively. The dilation scale is given by
\begin{equation}\label{app_eq:comov_vol}
    D_{V}(z,\Omega_{\rm m}; A) = \left[ \frac{c\, z (1+z)^2 D_{A}^2}{H(z,\Omega_{\rm m}; A)} \right]^{1/3}\,,
\end{equation}
where $D_{A}$ is the angular diameter distance
\begin{equation}
    D_{A}(z,\Omega_{\rm m}; A) = \frac{1}{1+z} \int_{0}^{z}
\frac{c \, \mathrm{d}z^\prime}{H(z^\prime, \Omega_{\rm m};A)} \,,
\end{equation}
where $A=\{w_0,w_a\}$. Note that $D_V$ is the comoving volume in which the density of objects remains constant with redshift, $D_A$ is the angular distance in the sky those oscillations have. Through the comoving sound horizon, the distance ratio $d_{z}$ is related to the expansion parameter $h$ (defined such that {$H(z) \doteq 100 h(z)$)} and the physical densities $\Omega_{\rm m}$ and $\Omega_{\rm b}$.

The Alcock-Paczynski distortion parameter measures how flattened or elongated a cluster of galaxies appears along the line of sight for incorrect Hubble parameters, and is given by
\begin{equation}
    F(z,A)=(1+z)\frac{D_{A}(z,A)H(z,A)}{c}\,.
\end{equation}
We need this quantity to ensure that angular and radial clustering match constraints in the various cosmological models. Here we use \gls{bao} data to connect \gls{sn}eIa (Pantheon) to \gls{cmb} data (PL18). This can be achieved by calibrating the $D_A$ from \gls{bao} with the $D_{L}$ from \gls{sn} in a cosmology-independent way and we define the $\chi^2$-statistics as
\begin{equation}
    \chi^2_{\text{BAO}}=(\Delta\mathcal{F}_{\text{BAO}})^{T}\cdotp C_{\text{BAO}}^{-1}\cdotp\Delta\mathcal{F}_{\text{BAO}}\,,
\end{equation}
where $\Delta\mathcal{F}_{\text{BAO}}$ is the difference between the observational data and the resulting value for $\Theta$, and $C_{\text{BAO}}^{-1}$ is the inverse of the covariance matrix mentioned before.


\subsection{Planck legacy 2018}

It is standard to consider \gls{cmb} data via the so-called shift parameters when examining the evolution of the cosmological background. The set of shift parameters is related to the first peak of the anisotropic \gls{cmb} power spectrum. For this purpose, we adopt the low-$\ell$ and high-$\ell$ likelihoods from Ref.~\cite{Aghanim:2018eyx} and set three quantities that give information about the position of the first peak in the temperature angular power spectrum through the ratio between its position in the model one wants to analyze and that of a $\Lambda$\gls{cdm} model. We construct the $\chi^2_{\text{CMB}}$ estimator as
\begin{equation}
    \chi^2_{\text{CMB}}=(\Delta\mathcal{F}^{\text{CMB}})^{T} C_{\text{CMB}}^{-1}\Delta\mathcal{F}^{\text{CMB}}\,,
\end{equation}
where $\Delta\mathcal{F}^{\text{CMB}}$ is a vector constructed by the position of the first peak. Also, with the \texttt{Planck} data Legacy 2018 data we can obtain tight limits on the so-called shift parameters
\begin{equation}
    R(A)=\sqrt{\Omega_{\rm m} H_0^2}\frac{r(z_*,A)}{c}\,, \quad l_a(A)=\pi\frac{r(z_*,A)}{r_s(z_*,A)}\,, \quad \omega_{\rm b}=\Omega_{\rm b} h^2\,,
\end{equation}
where $r(z,\,\Theta)$ is the comoving distance at a redshift $z$, and $r_s(z_,\,\Theta)$ is the comoving sound horizon, which read as follows
\begin{equation}\label{soundhorizon}
    r(z,\,A)=\int^{z}_{0}\frac{c}{H(z',A)}\mathrm{d}z'\,,\quad r_s(z_,A)=\int^{\infty}_{z}\frac{c_s(z')}{H(z',A)}\mathrm{d}z'\,.
\end{equation}
The comoving distance, and the comoving sound horizon are evaluated at the photon-decoupling redshift $z_*$, specified by
\begin{equation}
    z_*=1048\left[1+0.00124(\Omega_{\rm b} h^2)^{-0.738}\right]\left[1+g_1(\Omega_{\rm m} h^2)^{g_2}\right]\,,
\end{equation}
where
\begin{equation}
g_1=0.0783(\Omega_{\rm b} h^2)^{-0.238}\left[1+39.5(\Omega_{\rm b} h^2)^{-0.763}\right]^{-1}\,,\quad
g_2=0.560\left[1+21.1(\Omega_{\rm b} h^2)^{1.81}\right]^{-1}\,.
\end{equation}
Further, here $z_*$ is the photon-decoupling redshift. We also need to consider that the sound speed $c_s(z)$ is given by the expression
\begin{equation}
    c_s(z)=\frac{c}{\sqrt{3(1+\hat{R}_b(1+z)^{-1})}}\,,
\end{equation}
where
\begin{equation}
    \hat{R}_b=31500\,\Omega_{\rm b} h^2\left(\frac{T_{\rm CMB}}{2.7}\right)^{-4}\,.
\end{equation}
The shift parameters depend on the position of the \gls{cmb} acoustic peaks, which are functions of the geometry of the model considered. Therefore, they can be used to discriminate between different models or different values of the free parameters, $A$ which includes $\Omega_{\rm b}$ in this case.

\subsection{\texorpdfstring{$H_0$}{} values}

As described in the introductory part of this section, there has been an decrease in the reported values of $H_0$ coming from \gls{cmb} data \cite{Aghanim:2018eyx} using the $\Lambda$CDM model against model independent measurements of this parameter. We recall that the \texttt{Planck} Collaboration reported a low value of $67.4 \pm 0.5 \,{\rm km\, s}^{-1} {\rm Mpc}^{-1}$ \cite{Aghanim:2018eyx} while the \gls{des} Collaboration gave a similar value of $67.4^{+1.1}_{-1.2} \,{\rm km\, s}^{-1} {\rm Mpc}^{-1}$ \cite{PhysRevD.98.043526}. Consequently, these predicted values of $H_0$ contrast with the cosmology independent estimates from late-time observations, where the highest of these values is given by $H_0^{\rm R} = 74.22 \pm 1.82 \,{\rm km\, s}^{-1} {\rm Mpc}^{-1}$ \cite{Riess:2019cxk}. Another late-time $H_0$ value was reported by the \gls{holicow} Collaboration \cite{Wong:2019kwg} with a value of $H_0^{\rm HW} = 73.3^{+1.7}_{-1.8} \,{\rm km\, s}^{-1} {\rm Mpc}^{-1}$. Another value was reported via the tip of the red giant branch method, which is one of the most accurate and precise means of measuring distances to nearby galaxies, where $H_0^{\rm TRGB} = 69.8 \pm 1.9 \,{\rm km\, s}^{-1} {\rm Mpc}^{-1}$ \cite{Freedman:2019jwv}. Although this method has been used for over thirty years to infer high-precision distances to a large
number of nearby galaxies, it is only recently that the \gls{tgrb} method has been applied directly to the determination of $H_0$ \cite{Mould:2008ua,Freedman:2020dne}.

The $\chi^2$-statistic for any $H_0^i$ prior is given by
\begin{equation}
    \chi_{H_0^i}^2 = \frac{\left[H\left(z=0,\mathbf{x}\right) - H_{0,\text{obs}}^i\right]^2}{\sigma^2_{H_0^i}}\,,
\end{equation}
where $H_{0,\text{obs}}^i$ denotes the observed value of $H_0$ from the $i^{\rm th}$ cosmic probe, $\sigma^2_{H_0^i}$ is the corresponding observational error, and $H\left(z=0,\mathbf{x}\right)$ is the theoretical value of $H_0$ with the specific model parameter vector $\mathbf{x}$.

\section{Bayesian recipe for cosmology}

Usually, it turns out to be very difficult to numerically compute the posterior distribution and the corresponding best-fits of the cosmological model parameters.
For these non-standard setups, the available numerical tools play an important role during the parameter estimation task. There exists several options to carry out this work;
nevertheless here we will focus only on
\gls{mcmc} with the Metropolis Hastings algorithm.

The purpose of a Monte Carlo algorithm is to construct sequences of data points, which are the so-called \textit{chains}, in a parameter space to evaluate a Bayesian posterior. A \gls{mcmc} simulation can be assigned to certain algorithms that use random number generators to approximate a specific quantity. Moreover, a succession $X_1,\,X_2,\,\ldots, X_{n}$ of elements of a set composed by $n$ elements, is a Monte Carlo set while the conditional distribution of $X_{n+1}$ given $X_1,\,\ldots,\,X_n$ depends solely on $X_n$, i.e., a \gls{mcmc} is a process where we can compute subsequent steps based on the information given by the previous data points. A relevant property of a \gls{mcmc} is that it converges to a stationary scenario where successive elements of the chain are samples from the distribution; therefore it converges to the posterior $P(\theta|\mathcal{D},\mathcal{H})$, where $\theta$ are the free cosmological parameters of the model, $\mathcal{D}$ is the data and $\mathcal{H}$ the hypothesised model. In this way, we can estimate the standard quantities of interest. The number of points required to obtain better estimates in \gls{mcmc} tends to scale linearly with the number of parameters of the model, so that, this method becomes much faster as the dimensionality increases. The density is approximated by a set of delta functions given by
\begin{equation}
    p(\theta|\mathcal{D},\mathcal{H})\simeq \frac{1}{N}\sum_{i=1}^N \delta(\theta-\theta_i)\,,
\end{equation}
where $\Theta_i$ denotes the observable data and with $N$ being the number of points in the chain. The posterior mean can be calculated using
\begin{equation}
    \langle\theta\rangle=\int \theta P(\theta,\mathcal{H}|\mathcal{D}) \dd\theta \simeq \frac{1}{N}\sum_{i=1}^N\theta_i \,,\label{eq:proba}
\end{equation}
where the integral intervals can be assumed to be as $[a,b]$, which can be divided into $n$ equally spaced points depending on the observational sample and size of the nested, and where $\simeq$ is used since the samples $\theta_i$ are generated out of the posterior by numerical construction. Then, we can estimate any integrals using Simpson’s rule as
\begin{equation}
    \langle f(\theta)\rangle =\int f(\theta) P(f(\theta),\mathcal{H}|\mathcal{D}) \dd f(\theta) \simeq\frac{1}{N}\sum_{i=1}^N f(\theta_i)\,.
\end{equation}
In a \gls{mcmc} routine it is necessary to generate a new point $\theta_{i+1}$ from the previous point $\theta_i$. However, as expected, we need a criterion for accepting (or refusing) this new point, depending on whether it turns out to fit the data better for our model or not. If this new step is worse than the previous one, we may accept it with a certain probability, since it could be the case that, if we only accept steps with better probability, we could be converging into a local maximum in our parameter space and, therefore, not completely mapping all of the points. The standard adopted algorithm that contains this information is the Metropolis-Hastings algorithm. In this kind of algorithm it is necessary to start from a random pivot point $\theta_i$, with an associated posterior probability $p_i=p(\theta_i|\mathcal{D},\mathcal{H})$. Afterwards, we need to propose a candidate variable $\theta_c$ from a proposed distribution $q(\theta_i,\theta_c)$ used as a generator of new random points in the sequence. Then, the probability of accepting the new point can be defined by
\begin{equation}\label{eq:pacceptance}
    p(\text{acceptance}):=\text{min}\left[1,\frac{p_cq(\theta_c,\theta_i)}{p_iq(\theta_i,\theta_c)}\right]\,.
\end{equation} 
With a symmetric proposed distribution the algorithm reduces to the Metropolis algorithm
\begin{equation}
    p(\text{acceptance})=\text{min}\left[1,\frac{p_c}{p_i}\right]\,. \label{eq:post2}
\end{equation}
The full algorithm can therefore be detailed using the following steps:
\begin{enumerate}
\item Choose a random initial condition $\theta_i$ in the parameter space and compute the posterior distribution using Eq.~\eqref{eq:pacceptance}.
\item Generate a new candidate from a proposal distribution in the parameter space and compute
the corresponding posterior distribution using Eq.~\eqref{eq:post2}.
\item Accept (or not) the new point with the help of the Metropolis-Hastings algorithm using Eq.~\eqref{eq:pacceptance}.
\item If the point is not accepted, repeat the previous point in the chain.
\item Repeat steps 2--4 until you have a large enough chain (according to the Gelman–Rubin criteria).
\end{enumerate}
A schematic of this process is shown in Supplementary 3. Once the cosmological model is established, we need a statistical code to estimate the free parameters of our model. Fortunately there are several publicly available \gls{mcmc} codes that can make this task easy to handle. In this Review and in the literature the \texttt{Monte Python} code is employed, which is a \gls{mcmc} code for cosmological parameter extraction \cite{MontePythoncode}. This code contain likelihoods of the most recent cosmological experiments, and interfaces with the Boltzmann code Cosmic Linear Anisotropy Solving System (\texttt{CLASS}) \cite{classcode} observables. The code has several sampling methods available: Metropolis-Hastings, Nested Sampling (through MultiNest), EMCEE (through CosmoHammer) and Importance Sampling. To adapt our theories to the code, we must consider Gaussian likelihoods for each data compilation with the following form
\begin{equation}
    \mathbb{L}\propto \exp\left[-\sum_i \frac{(F(z_i)-F_i)^2}{2\sigma_i}\right]\,,
\end{equation}
where $F(z_i)$ is the theoretical value related to
the observation $F_i$, and $\sigma_i$ are the corresponding errors associated to each measurement. In our estimation $F(z_i)$ is given by the definitions of $\chi^2$ (see Supplementary 4) for each cosmological observable. This function contains the background evolution and equation of state equations described for the \gls{tg} theories.

\section{Gaussian processes for cosmology}

Gaussian processes (\gls{gp}) generalize the concept of a Gaussian distribution for a finite set of data points to a continuous distribution over a specified range for a function (see Refs.~\cite{10.5555/971143,10.5555/1162254} for a comprehensive review on the topic). Thus, given a set of Gaussian distributed data points, \gls{gp} provide an iterative process by which to produce the most likely underlying continuous function that describes the data together with its associated confidence bounds without assuming a prescribed form of the function \cite{Seikel:2013fda}. In the current era of precision cosmology with multiple state-of-the-art surveys, \gls{gp} has been exhaustively applied in cosmology \cite{Shafieloo:2012ht,Seikel:2013fda,Cai:2015zoa,Cai:2015pia,Wang:2017jdm,Zhou:2019gda,Cai:2019bdh,Mukherjee:2020vkx,Gomez-Valent:2018hwc,Zhang:2018gjb,Aljaf:2020eqh,Li:2019nux,Liao:2019qoc,Yu:2017iju,Escamilla-Rivera:2021rbe,LeviSaid:2021yat,Colgain:2021ngq,Dhawan:2021mel}, particularly to infer the late-time dynamics of the Universe.

As previously mentioned, \gls{gp} extends the idea of a Gaussian distribution by defining a mean function $\mu(z)$ together with a two-point covariance function $\mathcal{C}(z,z')$ to form a GP as a continuous curve
\begin{equation}
	\xi(z)\sim\mathcal{GP}\left(\mu(z),\mathcal{C}(z,z')\right)\,,
\end{equation}
together with its associated error regions $\Delta\xi(z)$, thus resulting in $\xi(z) \pm \Delta \xi(z)$. Here, $z'$ represents other points in the training data set. Without losing generality, the mean can be set to zero for all points in the reconstruction, since each point is not very sensitive to this. For the redshifts $z^*$ of the reconstruction at which we do not have data points, we can define a kernel function for the covariance such that $\mathcal{C}\left(z^*,z^{*'}\right) = \mathcal{K}\left(z^*,z^{*'}\right)$, which will produce the reconstruction at those points. The kernel will embody all the information about the strength of the correlations between these reconstructed values as well as the amplitude of the deviations from the mean \cite{Gomez-Valent:2018hwc}. The only generic property about this is that it must be a symmetric function. On the other hand, for observational data points $\tilde{z}$ where observations exist, we have access to the associated errors and covariance matrix $\mathcal{D}\left(\tilde{z},\tilde{z}'\right)$ between the points, so that the covariance can be written as $\mathcal{C}\left(\tilde{z},\tilde{z}'\right) = \mathcal{K}\left(\tilde{z},\tilde{z}'\right) + \mathcal{D}\left(\tilde{z},\tilde{z}'\right)$ which will give information about the kernel. Lastly, observational points and reconstructed points will be correlated by the kernel alone through $\mathcal{C}\left(z^*,\tilde{z}'\right) = \mathcal{K}\left(z^*,\tilde{z}'\right)$ \cite{Busti:2014aoa}.

There exists a number of kernel function choices \cite{10.5555/1162254}, where some behave slightly better in certain situations, which remains an open discussion in the literature \cite{Seikel:2013fda}. In this Review, we will only be considering the square exponential kernel function, specified by
\begin{equation}\label{eq:square_exp}
	\mathcal{K}\left(z,\tilde{z}\right) = \sigma_f^2 \exp\left[-\frac{\left(z-\tilde{z}\right)^2}{2l_f^2}\right]\,,
\end{equation}
where $\sigma_f$ and $l_f$ are the kernel hyperparameters which relate the strength and scope of the correlations between the reconstructed data points, respectively \cite{2012JCAP...06..036S}. Here, the \textit{hyperparameters} are statistical parameters rather than model parameters. In this setting, hyperparameters model the functional behaviour of the underlying data points without any knowledge of the physical processes being undertaken. The hyperparameter $l_f$ can be considered as the typical correlation length scale of the independent variable, while the signal variance $\sigma_f$, can be thought of as the typical variation scale of the dependent variable.

The observational data being input into a \gls{gp} then appears as a subset of Gaussian points that can be extended through the \gls{gp} approach. This is achieved by maximizing the likelihood of the \gls{gp} producing the observational data by estimating the underlying functional behavior for different values of the hyperparameters for particular kernel instances.

Furthermore, the \gls{gp} approach is model-independent in the context of a physical model and instead assumes a particular statistical kernel which dictates the correlation between the reconstructed points. The \gls{gp} approach is also non-parametric in terms of physical models.

For a Gaussian prior for both the data and the random functions, one can marginalize over the space of functions and obtain the logarithm of the marginal likelihood via a Bayesian approach
\begin{alignat}{2}
    \ln \mathbb{L}& =\: & &-\frac{1}{2}\sum_{i,j=1}^N\big\{\left[y_i-\mu(x_i)\right]\left[\mathcal{K}(x_i,x_j)+\mathcal{C}(x_i,x_j)\right]^{-1}[y_j-\mu(x_j)]\nonumber\\
    & \: & &-\frac{1}{2}\ln\vert \mathcal{K}(x_i,x_j)+\mathcal{C}(x_i,x_j)\vert\big\}-\frac{N}{2}\ln2\pi\,.
\end{alignat}
The optimal hyperparameter values of $\sigma_f$ and $l_f$ are then inferred via the maximisation of the logarithm of the marginal likelihood.

This reconstruction method can also be implemented for the reconstruction of the derivatives of a function which has been \gls{gp} reconstructed, since the derivatives will again be a \gls{gp}. The discussed \gls{gp} algorithm is nicely implemented in a publicly available \texttt{Python} code \texttt{GaPP} (\gls{gp} in Python) \cite{gappcode,2012JCAP...06..036S}, which reconstructs a continuous function for any given data set along with its derivative functions.

\section{Machine learning for cosmology}

Bayesian methods remain the preferred choice compared to the information criterion and \gls{gp} reported in the literature \cite{Escamilla-Rivera:2020fxq}. A complete Bayesian method for model selection is highly involved computationally and usually suffers from multi modal posteriors and free parameter degeneracies. The calculation of the required evidence, in order to obtain a final result, is an immensely time consuming process which makes it very expensive computationally.

The study of \gls{lss} necessitates the use of computer simulations in order to address the issue of constraining cosmological parameters using Bayesian methods. Due to the computational complexity of these simulations, some studies remain computationally unfeasible for the foreseeable future. It is at this point that computational techniques such as machine learning can have a number of important uses, even as for example for trying to understand the evolution of our Universe.

On the issue of \gls{lss}, on a macroscopic level, galactic structure form clusters which are some of the largest known objects to populate the Universe. On even larger scales, filamentary structures are observed which are composed of galactic superclusters that are billions of light years in term of their length scale. This is the \gls{lss} of the Universe, and by measuring its properties and structure, we can test different cosmological models of the evolution of the Universe. These structures are often described by a matter density field, or by its statistical properties through the matter power spectrum. These can be measured in a variety of ways: \gls{cmb} photons travel through the \gls{lss} and thus interacts with it. This leaves an imprint on the photon properties which is then used to infer the structure and growth of the \gls{lss}. Large-scale galaxy surveys and intensity mapping experiments directly sample the matter density field, allowing us to build up an image of the \gls{lss} with unparalleled precision. Surveys of galaxy clusters—whether from optical surveys, X-ray, or through their Sunyaev-Zeldovich interaction with the CMB—sample the extreme peaks of the density field. Sampling the \gls{lss} through its greatest non-linearities gives a powerful lever arm on the physics that drive its evolution.

The scheme behind the machine learning is based on considering a neural network with a complex combination of several neurons organised in nested layers. Each of these neuron implements a function that is parametrized by a set of weights which can be denoted by $W$. Each layer of a neural network thus transforms one input vector - or tensor depending the dimension of the system - to another through a differentiable function. Theoretically, given a neuron $n$ that has an input vector and the choice of an activation function $A_n$, the output of the neuron can be calculated as
\begin{subequations}
\begin{align}
    h^{<t>}&=A_n(h^{<t-1>}\cdot W_{h}+x^{<t>}\cdot W_{x}+b_{a})\,,\\
    y^{t}&=A_n(h^{t}\cdot W_{y}+b_{y})\,,
\end{align}
\end{subequations}
where $b$ is the bias, $ h^{<t>}$ is called the hidden state, $A_n$ is the activation function, $x^{<t>}$ is the input data and $y^{t}$ is the output.

We introduce a data set in order to train this array and therefore the architecture can learn to finally give an output set of data. For example: the network can learn the distribution of the distance moduli in the \gls{de} models, then feed the astrophysical samplers (surveys) to the network to reconstruct the \gls{de} model and then discriminate the most probable model\footnote{In this text review we are employing a recurrence neural network. There are several approaches in this machine learning field e.g. in Ref.~\cite{Ntampaka:2019udw} and references therein. }. Interesting works in that these directions have been developed in Ref.~\cite{Escamilla-Rivera:2019hqt,Munoz:2020gok}.

While neural networks can learn complex nested representations of the data sample, allowing them to achieve high performance levels, it also bounds our understanding of the model learned by the network itself. The choice of an architecture \cite{Ntampaka:2019udw} can have an important relevant influence on the performance of the neural network. Some methods have to be designed by evaluating the number and type of layers, and the number and the size of the filters used per layer. A good way to select these choices is through experimentation, which we can do since we have different data sets from observational cosmology. We can select the size of the network, which depends on the number of training tests, because networks with a large number of cosmological parameters are likely to be over-fitted if not enough training tests are available.

A growing interest in these kind of algorithms has put forward new opportunities for data-driven cosmological analyses, but also the adoption of machine learning approaches also opens new challenges for the community. In our case, we consider deep learning, a method that can interpret the results when the data are highly complex for standard model developments. A number of proposals in this area have been suggested to explore the deep learning methods for measurements of cosmological parameters from density fields \cite{Schmelzle:2017vwd} and for future large-scale photometric surveys \cite{Charnock:2016ifh}.

\begin{figure}
\centering
\includegraphics[width=0.85\textwidth,origin=c,angle=0]{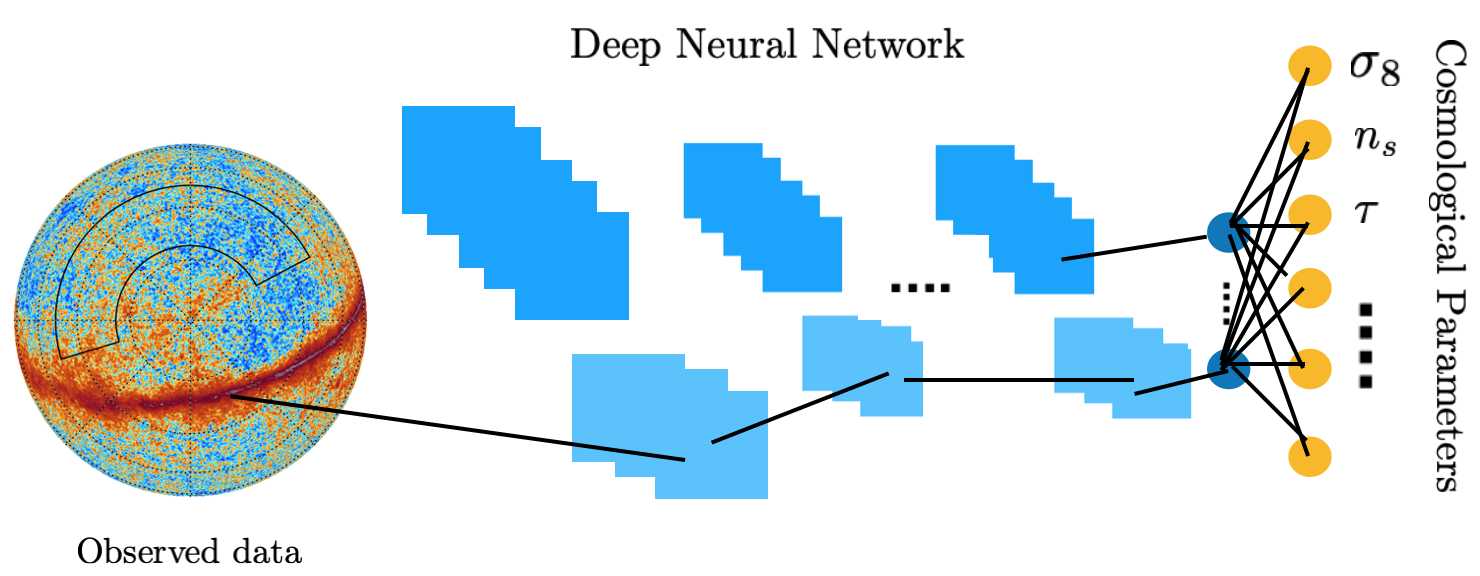}
\caption{A deep learning architecture for a cosmological model. On the left we have an example of an observed data as an input, on the middle the architecture of an optimal deep neural network and finally, as output results, on the right the cosmological parameters that describe the cosmological model. Permission for use of this figure was kindly provided by the authors of Ref.~\cite{Escamilla-Rivera:2020fxq}.}
\label{fig:DL_cosmo}
\end{figure}

To start to train an astrophysical or a cosmological sample, we need to design an architecture with a function of neural network, that can have many unstable points and a local minima. This architecture makes the optimisation process very difficult, but in real scenarios, high levels of noise degrade the training data and results in optimisation scenarios with more local minima points and increases the difficulty in training the neural network. It is desirable to start by optimising the neural network using noise-free data which typically yield smoother scenarios. As an example, in Fig.~\ref{fig:DL_cosmo} we present a standard network using an image of a cosmological simulation (the data) and then we divided an array of several layers to finally, extract the output cosmological parameters value. Each neuron uses a Bayesian process to compute the error propagation as it is done in the standard inference analyses.

We can describe a brief but effective recipe to develop a recurrence neural network with a Bayesian computation training in the following steps:
\begin{itemize}
        \item \textbf{Step 1 - Design of the the neural network} -- For a recurrence neural network method we can select values that have one layer and a specific number of neurons.
        \item \textbf{Step 2 - Organising the data sample} -- We need to sort the samples from lower to higher redshift $z$ in the observations. Afterwards, we can re-arrange our data using a number of steps that compares the same points of the data samples, e.g. if a \gls{sn} sample contains approximately 1000 data points, then our array should be of the same dimension of the data sample.
        \item \textbf{Step 3 - Bayesian training} -- It is important to choose a model of regularization. With a Bayesian standard method to calculate the evidence, the algorithm can calculate errors via regularization methods. Finally, over the cost function we can use Adam optimiser.
        \item \textbf{Step 4 - Training the full architecture} -- It is required to consider a high number of epochs (e.g for a sampler as Pantheon \gls{sn}, one can use around 1000 epoch per layer). After the training, we need to read the model, which now will be the new initial model and apply the architecture using the same hyperparameters and re-run the algorithm until we got the final convergence. The result of this step is the construction of the confidence regions.
        \item \textbf{Step 5 - Calculation of the modulus distance $\mu(z)$ for each cosmological model} -- Using $H(z)$, we can compute $\mu(z)$ by using a specific dark energy \gls{eos} in terms of $z$ and then integrate them.
        \item \textbf{Step 6 - Calculations of best-fits} -- The output values can be obtained by using the training data as a simulated sample. For this case we use the publicly codes \texttt{CLASS} and \texttt{Monte Python} to constrain the models as it is standard as it is done in the standard Bayesian cosmology case.
\end{itemize}

\end{appendices}

\bibliographystyle{utphys}
\bibliography{references}
\clearpage

\end{document}